\documentclass[hyper]{JHEP3}

\input{epsf}
\usepackage{epsfig}
\usepackage{amssymb}
\usepackage{amsmath}
\usepackage{amsfonts}
\usepackage{textgreek,upgreek}
\usepackage{amsbsy}
\usepackage[all]{xy}
\usepackage{amsmath}

\usepackage{amssymb,amscd}
\usepackage{mathrsfs}
\usepackage{amsmath,amsthm}
\usepackage{bbm}
\usepackage{mathtools}
\def\t{{ \sf t}}
\def\d{{\mathrm d}}

\def\tt{{\mathfrak t}}

\def\EE{{\sf E}}
\def\veps{\varepsilon}
\def\o{{\mathbf o}}

\def\H{{\mathcal H}}
\def\Q{{\mathcal Q}}

\def\FF{{\mathcal  F}}

\def\RR{{\mathcal R}}

\def\uu{u}

\def\p{{\eurm p}}

\def\VV{{\eusm V}}

\def\w{w}
\def\Chi{\chi}

\def\be{\begin{equation}}
\def\ee{\end{equation}}

\def\Re{{\mathrm{Re}}}
\def\Im{{\mathrm{Im}}}
\def\hat{\widehat}

\def\tilde{\widetilde}
\def\fN{{\mathfrak N}}
\def\frak{\mathfrak}
\def\h{\widehat}

\def\D{{\mathcal D}}
\def\SSS{{\mathcal S}}
\def\SIgma{\Sigma}

\def\V{{\mathcal V}}
\def\O{{\mathcal O}}

\def\Bbb{\mathbb}

\def\red{{\mathrm{red}}}

\def\d{{\mathrm d}}

\def\R{{\mathbb R}}
\def\C{{\mathbb C}}
\def\U{{\mathcal U}}
\def\D{{\mathcal D}}
\def\[{\bigl [}

\def\]{\bigr ]}

\def\CP{{\mathbb{CP}}}
\def\N{{\mathcal N}}
\def\T{{\mathcal T}}
\def\E{{\mathcal E}}

\def\Z{{\mathbb Z}}
\def\ZZ{{\mathcal Z}}

\def\CC{{\mathcal C}}

\def\L{{\mathcal  L}}
\def\t{\widetilde }
\def\h{\widehat}

\def\V{{\mathcal V}}
\def\I{{\mathcal I}}

\def\B{{\mathfrak{B}}}

\def\M{{\mathcal M}}

\def\H{{\mathcal H}}

\def\trho{\text{\textmu}}

\def\VV{{\Bbb V}}
\def\vol{{\mathrm{vol}}}

\def\RR{{\mathcal R}}

\def\tilde{\widetilde}

\def\bar{\overline}

\font\teneurm=eurm10 \font\seveneurm=eurm7  \font\fiveeurm=eurm5
\newfam\eurmfam
\textfont\eurmfam=\teneurm \scriptfont\eurmfam=\seveneurm
\scriptscriptfont\eurmfam=\fiveeurm

\font\teneusm=eusm10 \font\seveneusm=eusm7 \font\fiveeusm=eusm5
\newfam\eusmfam
\textfont\eusmfam=\teneusm \scriptfont\eusmfam=\seveneusm
\scriptscriptfont\eusmfam=\fiveeusm

\font\tencmmib=cmmib10 \skewchar\tencmmib='177
\font\sevencmmib=cmmib7 \skewchar\sevencmmib='177
\font\fivecmmib=cmmib5 \skewchar\fivecmmib='177
\newfam\cmmibfam
\textfont\cmmibfam=\tencmmib \scriptfont\cmmibfam=\sevencmmib
\scriptscriptfont\cmmibfam=\fivecmmib

\def\bar{\overline}
\def\tilde{\widetilde}
\def\M{{\mathcal M}}
\def\Id{\boldsymbol{\mathrm{Id}}}
\def\Q{{\mathcal Q}}
\def\fId{\mathfrak{I}\mathfrak{d}}
\def\fS{\mathfrak{S}}

\def\im{\mbox{Im }}
\def\mod{{\rm mod}}

\def\half{\frac{1}{2}}

\renewcommand{\Im}{{\rm Im }}
\renewcommand{\Re}{{\rm Re }}

\def\one{{\hbox{ 1\kern-.8mm l}}}

\def\vol{{\rm vol\,}}
\def\p{\partial}

\def\be{\bar{e}}

\def\half{\frac{1}{2}}

\def\p{\partial}

\def\Tr{{\rm Tr}}

\def\be{ \begin{equation} }
\def\ee{ \end{equation}}

\def\fg{\mathfrak{g}}
\def\fm{\mathfrak{m}}
\def\fr{\mathfrak{r}}
\def\fu{\mathfrak{u}}
\def\fv{\mathfrak{v}}

\def\End{{\rm End}}
\def\Hom{{\rm Hom}}
\def\Hop{{\rm Hop} }
\def\I{{\rm i}}
\def\i{{\rm i}}

\def\Sh{{\rm Sh}}

\def\fa{\mathfrak{a}}

\def\fc{\mathfrak{c}}
\def\fd{\mathfrak{d}}
\def\fe{\mathfrak{e}}
\def\fE{\mathfrak{E}}

\def\fF{\mathfrak{F}}
\def\lieg{\mathfrak{g}}
\def\fg{\mathfrak{g}}

\def\fh{\mathfrak{h}}

\def\fm{\mathfrak{m}}

\def\fr{\mathfrak{r}}
\def\fs{\mathfrak{s}}
\def\ft{\mathfrak{t}}

\def\fr{\mathfrak{r}}
\def\fs{\mathfrak{s}}
\def\ft{\mathfrak{t}}
\def\fu{\mathfrak{u}}
\def\fv{\mathfrak{v}}
\def\fw{\mathfrak{w}}
\def\fx{\mathfrak{x}}

\def\fz{\mathfrak{z}}
\def\fA{\mathfrak{A}}
\def\fB{\mathfrak{B}}
\def\fC{\mathfrak{C}}
\def\fD{\mathfrak{D}}
\def\fE{\mathfrak{E}}
\def\fI{\mathfrak{I}}
\def\fL{\mathfrak{L}}
\def\fM{\mathfrak{M}}
\def\fR{\mathfrak{R}}
\def\fS{\mathfrak{S}}
\def\fT{\mathfrak{T}}
\def\fU{\mathfrak{U}}

\def\fV{\mathfrak{V}}
\def\fW{\mathfrak{W}}

\def\im{\mbox{Im }}
\def\mod{{\rm mod}}
\def\Tr{{\rm Tr}}

\def\IC{\mathbb{C}}

\def\IM{\mathbb{M}}

\def\IP{\mathbb{P}}
\def\IQ{\mathbb{Q}}
\def\IR{{\mathbb{R}}}
\def\IS{{\mathbb{S}}}

\def\IV{{\mathbb{V}}}
\def\IZ{{\mathbb{Z}}}

\def\CA{{\cal A}}
\def\CB{{\cal B}}
\def\CC {{\cal C}}
\def\CD {{\cal D}}
\def\CE {{\cal E}}
\def\CF {{\cal F}}
\def\CG {{\cal G}}
\def\CH {{\cal H}}
\def\CI {{\cal I}}
\def\CJ {{\cal J}}
\def\CK {{\cal K}}
\def\CL {{\cal L}}
\def\CM {{\cal M}}
\def\CN {{\cal N}}
\def\CO {{\cal O}}
\def\CP {{\cal P}}
\def\CR {{\cal R}}
\def\CV {{\cal V}}
 \def\v{{\mathbf v}}
\def\CW {{\cal W}}
\def\CX {{\cal X}}
\def\CO {{\cal O}}

\def\CE {{\cal E}}
\def\CG {{\cal G}}
\def\CH {{\cal H}}
\def\CI {{{\cal I}}}
\def\CB {{\cal B}}
\def\CQ {{\cal Q}}
\def\CS {{\cal S}}
\def\CT {{\cal T}}
\def\CU {{\cal U}}
\def\CX {{\cal X}}
\def\CY {{\cal Y}}

\def\afty{{$A_{\infty}$}}
\def\DET{{\rm Det}}

\newcommand\fro{{\overline{\underline{\Omega}}}}
\def\fVac{ { \fV \fa\fc} }
\def\Rvtx{ R^{\rm int}}
\def\IntfcTimes{ \boxtimes }

\title{Algebra of the Infrared: String Field Theoretic Structures
 in Massive $\CN=(2,2)$  Field Theory In Two Dimensions}

\author{Davide Gaiotto,$^1$  Gregory W. Moore,$^2$  and Edward Witten$^3$\\
$^1$Perimeter Institute for Theoretical Physics\\
31 Caroline Street North, ON N2L 2Y5, Canada\\
 $^2$ NHETC and Department of Physics and Astronomy,
Rutgers University,\\
Piscataway, NJ 08855--0849, USA\\
$^3$ School of Natural Sciences, Institute for Advanced Study, \\
Princeton, NJ 08540, USA\\
\\
{\rm dgaiotto@gmail.com, gmoore@physics.rutgers.edu, witten@ias.edu} }

\abstract{We introduce a ``web-based formalism'' for
describing the category of half-supersymmetric boundary conditions
in $1+1$ dimensional massive field theories with $\CN=(2,2)$ supersymmetry and unbroken
$U(1)_R$ symmetry. We show that the category can be completely constructed
from  data available in the far infrared, namely, the vacua, the
central charges of soliton sectors, and the spaces of soliton states
on $\IR$, together with certain ``interaction and boundary emission amplitudes.''
These amplitudes are shown to satisfy a system of algebraic constraints
related to the theory of $A_\infty$ and $L_\infty$ algebras.
The web-based formalism  also gives a method of finding the BPS states
for the theory on a half-line and on an interval. We  investigate half-supersymmetric
interfaces between theories and show that they have, in a certain sense, an associative ``operator product.''
We derive a categorification of wall-crossing formulae.
The example of Landau-Ginzburg theories is described in depth drawing
on ideas from Morse theory, and its interpretation in terms of supersymmetric quantum mechanics.
In this context we show that the web-based category is equivalent to a version of the Fukaya-Seidel
\afty-category associated to a holomorphic Lefschetz fibration, and we describe  unusual local
operators that appear in massive Landau-Ginzburg theories.
We indicate potential applications to the theory of surface defects in theories of class S
and to the gauge-theoretic approach to knot homology. }

\begin{document}

\section{Introduction}\label{sec:Introduction}

\subsection{Preliminaries}

This paper is devoted to the study of massive two-dimensional theories with $(2,2)$ supersymmetry.
The supersymmetry operators of positive spacetime chirality
are denoted $Q_+, \bar Q_+$ and those of negative chirality by $Q_-,\bar Q_-$.
(The adjoint of an operator $\O$ is denoted $\bar\O$.) It will be important that
there is an unbroken   $U(1)$ $R$-symmetry, whose
generator we call $\FF$ or ``fermion number.'' Supersymmetry generators
of $\FF=+1$ are $Q_-$ and $\bar Q_+$, and those of $\FF=-1$ are $Q_+$ and $\bar Q_-$.
In Minkowski space with metric
$\d \ell^2=-\d t^2+\d x^2$, the  supersymmetry algebra is
\begin{align}\{Q_+,\bar Q_+\}& = H+P \cr \{Q_-,\bar Q_-\}&=H-P \cr \{Q_+,  Q_-\}&=\bar Z \cr\{\bar Q_+,\bar Q_-\}&=  Z,\label{trrt}\end{align}
with other anticommutators
vanishing.  Here $H\sim -i\partial_t$ and $P\sim -i\partial_x$ are the energy and momentum and $Z$ is a central charge, which commutes
with the whole algebra and with all local operators.

Typically, we consider a theory with a finite
set $\VV$ of vacua (in some applications, one allows infinitely many vacua) in each of which there is a mass gap.    Because $Z$ is central,
in the $ij$ sector, which is defined as  the
space of states that interpolate from a vacuum $i$ at $x\to -\infty$ to a vacuum $j$ at $x\to +\infty$,  $Z$ is equal to a fixed complex number $z_{ij}$.  Cluster decomposition implies that for
$i,j,k\in \VV$, $z_{ij}+z_{jk}=z_{ik}$, and therefore there are complex numbers $W_i$ (unique up to a common additive constant) such that
$W_i-W_j=z_{ij}$, $i,j\in\VV$.  $W_i$ is called the value of the superpotential in vacuum $i\in\VV$.

A large supply of massive $\N=2$  theories with $U(1)$ $R$-symmetry
 can be constructed as Landau-Ginzburg (LG) models with chiral superfields $\phi_1,\dots,\phi_n$ valued in $\IC^n$,  and a suitable superpotential function $W(\phi_1, \dots,\phi_n)$.  (Any superpotential at all leads to a theory with a  $U(1)$ $R$-symmetry. Generically such a theory is massive and we usually call these theories massive $\N=2$ theories, leaving the $R$-symmetry understood.) Many of our considerations apply to more general LG models with
general K\"ahler target space $X$ and holomorphic Morse function $W$, but our considerations are already quite nontrivial
for the case $X=\IC$, and we restrict attention to $X =\IC^n$ in this introduction.
Since a model defined in some other way may have a description as an effective LG theory at low energies,
it may be that for some purposes this type of example is universal.  In this paper, we describe a framework that we believe applies
generally, but on some key points we rely on knowledge of LG models to infer what structure to expect.

Our goal is really to understand what additional information beyond the vacua and their central charges  is needed in order to describe the supersymmetric
states of a massive $\N=2$ theory.  The most elementary extra needed information concerns the BPS soliton states
in the $ij$ sector. For useful background and further references
see \cite{Cecotti:1992qh,Cecotti:1992rm,Hori:2003ic}.  Let $\Q_{ij}=Q_- -\zeta_{ij}^{-1} \bar Q_+$, with
$|\zeta|=1$.
Then for a state of $P=0$ in the $ij$ sector, \begin{equation}\label{tolf}\{\Q_{ij},\bar\Q_{ij}\}=2(H-\Re\,(\zeta_{ij}^{-1}z_{ij})).\end{equation}
A standard argument shows that   BPS states --
states annihilated by $\Q_{ij}$ and $\bar \Q_{ij}$
-- can exist only if $H=|z_{ij}|$ and $\zeta_{ij}=z_{ij}/|z_{ij}|$.
Such states come in supermultiplets consisting of a pair of states with $\FF=f,f+1$
for some $f$.  Using cluster decomposition, it can be shown that the values of $f$ mod $\Z$ depend only on the vacua  $i$ and $j$.   The
number of BPS multiplets for a given value of $f$ is the most basic observable that goes beyond a knowledge of the set $\VV$ of vacua and
the corresponding superpotential values $W_i$.  In this paper, we write $\RR_{ij}$ for the space of BPS solitons in the $ij$ sector.

As reviewed in Section \S \ref{subsec:MorseComplexRealLine}, in an LG model, a classical approximation to $\RR_{ij}$ is the space $R_{ij}$  of solutions of a certain supersymmetric soliton equation.
In general, $R_{ij}$ does not necessarily give a basis for the space $\RR_{ij}$ of quantum BPS soliton states.
To determine $\RR_{ij}$, one must in general compute certain instanton corrections to the classical soliton spectrum.  The instantons are solutions of
a certain nonlinear partial differential equation that we will call the $\zeta$-instanton equation.\footnote{For some prior
work on this equation, see \cite{Witten1993,Fan:2007ba,Guffin}.}  
 For LG models, the construction of this paper can be
developed taking as the starting point either the classical space $R_{ij}$ of BPS solitons or the corresponding quantum-corrected space $\RR_{ij}$.
However, the construction is probably easier to understand if one starts with the classical space $R_{ij}$, so we will use that language in this
introduction.  For an abstract massive $\N=2$ model that is not presented as an LG model (and more generally is not presented with
anything one would call a classical limit), there is no space $R_{ij}$ of classical solitons and one has to make the construction in terms of the
space $\RR_{ij}$ of quantum solitons, that is, BPS states in the $ij$ sector.

\subsection{Branes}

We can get a much richer story by considering also half-BPS branes.  We consider our theory
on a half-plane $\H$ defined by $x\geq 0$ with a boundary
condition at $x=0$ determined by a brane $\B$.  We suppose that, for some complex number $\zeta_\B$ of modulus 1,
 the brane $\B$ is invariant under the supersymmetry
$\Q_\B=Q_- -\zeta_\B^{-1}\bar Q_+$ and its adjoint $\bar\Q_\B$.  We also generally assume that $\zeta_\B$ does not coincide
with any of the $\zeta_{ij}$.  (This assumption keeps us away from  walls at which jumping phenomena occur.)

A basic question about such a brane is as follows.
If we formulate a massive $\N=2$ theory on the half-plane $\H$ with the brane $\B$ at $x=0$ and a vacuum $i\in\VV$ at $x=\infty$,
then what supersymmetric states are there, and with what values of the fermion number $\FF$?
We write $\E_i(\B)$ or just $\E_i$
for the space of such states.  The spaces $\E_i(\B)$ depend on the brane $\B$ and on additional microscopic details of the theory.
(In a Landau-Ginzburg theory, as explained in Section \S \ref{subsec:MorseComplexHalfLine}, $\E_i(\B)$ can be determined by solving the classical soliton equation with boundary
conditions determined by $\B$ and computing instanton corrections. The relevant instantons are solutions of the $\zeta_\B$-instanton equation obeying
certain boundary conditions.)

\begin{figure}
 \begin{center}
   \includegraphics[width=3in]{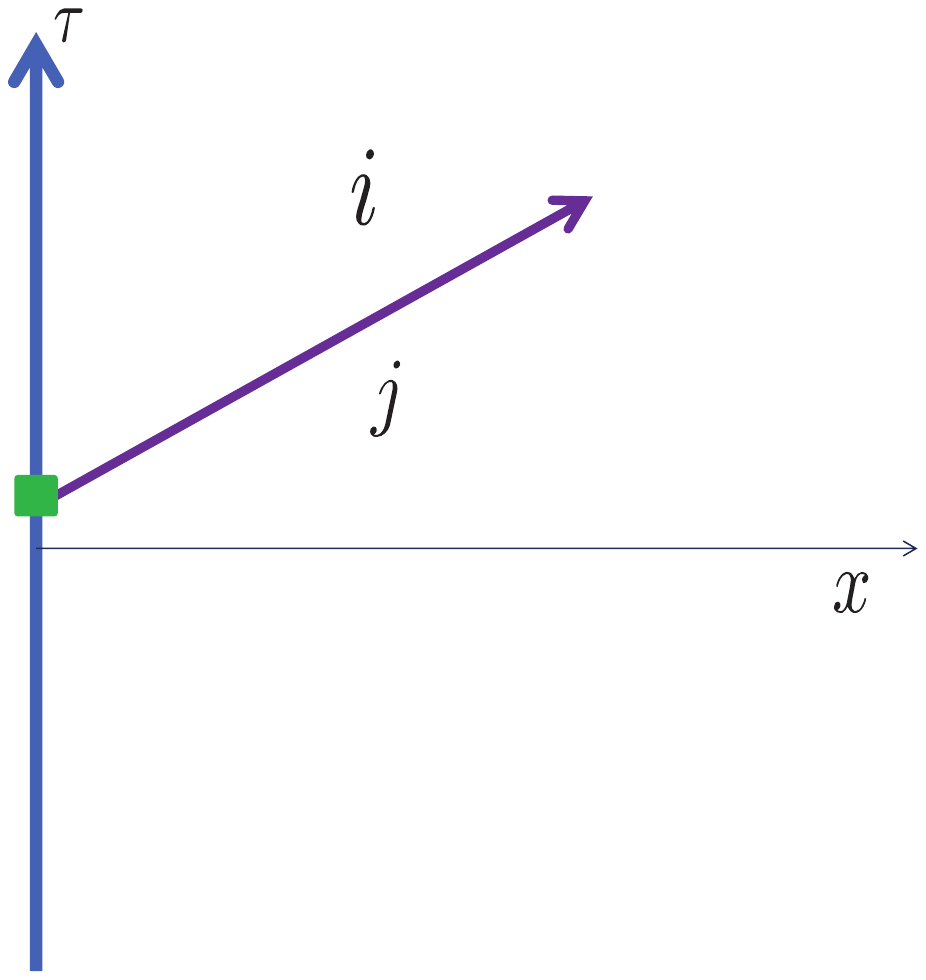}
 \end{center}
\caption{\small An $ij$ soliton emitted from the boundary.  This process is supersymmetric if the soliton is emitted at the proper angle.}
 \label{firstone}
\end{figure}

The assumption that $\zeta_\B$ does not coincide with any of the $\zeta_{ij}$ ensures that one cannot make a supersymmetric state consisting of a BPS soliton at rest in the presence of the brane $\B$.
However, if we transform to Euclidean signature (by letting $t=-i\tau$ so that the metric on $\R^2$ becomes $\d s^2=\d \tau^2+\d x^2$), there can potentially exist a BPS configuration consisting of $\B$ together with a boosted or more
precisely (in Euclidean signature) a rotated soliton.  To understand why, recall that $\zeta_{ij}$ was defined so that an $ij$ BPS
 soliton at rest at fixed $x$ -- so that its world line is
a straight line in the $\tau$ direction -- is invariant under $\Q_{ij}=Q_- -\zeta_{ij}^{-1}\bar Q_+$.  If we rotate the $x-\tau$ plane by an angle $\varphi$,
then $\Q_{ij}$ transforms to $\hat\Q_{ij}= e^{-\I \varphi/2}Q_- -\zeta_{ij}^{-1} e^{\I \varphi/2}\bar Q_+$.  If we pick $e^{\I \varphi}=\zeta_{ij}/\zeta_{\B}$, then $\hat\Q_{ij}$
is a multiple of $\Q_\B$.  Hence a soliton whose worldline is a straight line at an angle $\varphi_{ij}=\mathrm{Arg}(\zeta_{ij}/\zeta_{\B})$ to the $\tau$-axis
preserves the same supersymmetry as the brane $\B$.  Accordingly, one can ask (Figure \ref{firstone}) whether there is a supersymmetric
coupling by which $\B$ emits a BPS solition of type $ij$ at an angle $\varphi_{ij}$ to the vertical.  The answer to this question is not determined
in any elementary way by any data we have mentioned so far.  All we can say from the point of view of low energy effective field theory is that
in general, the answer depends on the choice of a soliton state in $R_{ij}$, and on the assumed initial and final states in $\E_j(\B)$ and $\E_i(\B)$.
Thus the answer can be summarized by a linear transformation\footnote{In the abstract formulation of
\S \ref{sec:RepWeb}, $T_{ij\B}$  is generalized to to a multisoliton emission amplitude  $\CB$
(an element of the vector space \eqref{eq:Rbd-def}) that satisfies a Maurer-Cartan equation (eqn. \eqref{eq:boundary-amp}). }
\begin{equation}\label{zomp}T_{ij\B}: \E_j(\B)\to \E_i(\B)\otimes R_{ij}. \end{equation}
(In a Landau-Ginzburg theory,  this linear transformation can be determined, in principle by solving the $\zeta$-instanton equation with suitable boundary conditions near the brane and at infinity.  This is explained
in Section \S \ref{zetawebs}.)

\subsection{Supersymmetric $(\B_\ell,\B_r)$ Strings}

At first sight, it may not be obvious that the amplitude for a supersymmetric brane to emit a supersymmetric soliton at an angle is related to anything
that is usually studied in a supersymmetric theory.
To see that it is, replace the half-plane $x\geq 0$ with a strip $0\leq x\leq L$, where we can take $L$ to be much
greater than the Compton wavelength of any particle in any of the vacua $i\in \VV$.  Let $\B_\ell$ and $\B_r$ be a pair of mutually BPS
branes, meaning that $\zeta_{\B_\ell}=\zeta_{\B_r}$, so that $\Q_{\B_{\ell}}=\Q_{{\B_r}}$; we denote them as $\Q_\B$.
Use $\B_\ell$ and $\B_r$ to define boundary conditions at $x=0$ and $x=L$, respectively.  By a supersymmetric $(\B_\ell,\B_r)$ state, we mean
a state of this system of zero energy or equivalently a state annihilated by $\Q_\B$ and its adjoint.

What is the space of these supersymmetric
states?  There is an obvious approximation for large $L$.  Far from $x=0$ and from $x=L$, the system must be exponentially close to one of the
vacua $i\in \VV$.  For given $i$, near $x=0$, the system is in some state in $\E_i(\B_\ell)$ and near $x=L$, it is in some state  in $\E_i(\B_r)$.
The mass gap means that to a good approximation, these two states can be specified independently and thus a large $L$ approximation to the
space of supersymmetric $(\B_\ell,\B_r)$ states is given by
\begin{equation}\label{info}\oplus_{i\in \VV}\E_i(\B_\ell)\otimes \E_i(\B_r).  \end{equation}

\begin{figure}
 \begin{center}
   \includegraphics[width=3in]{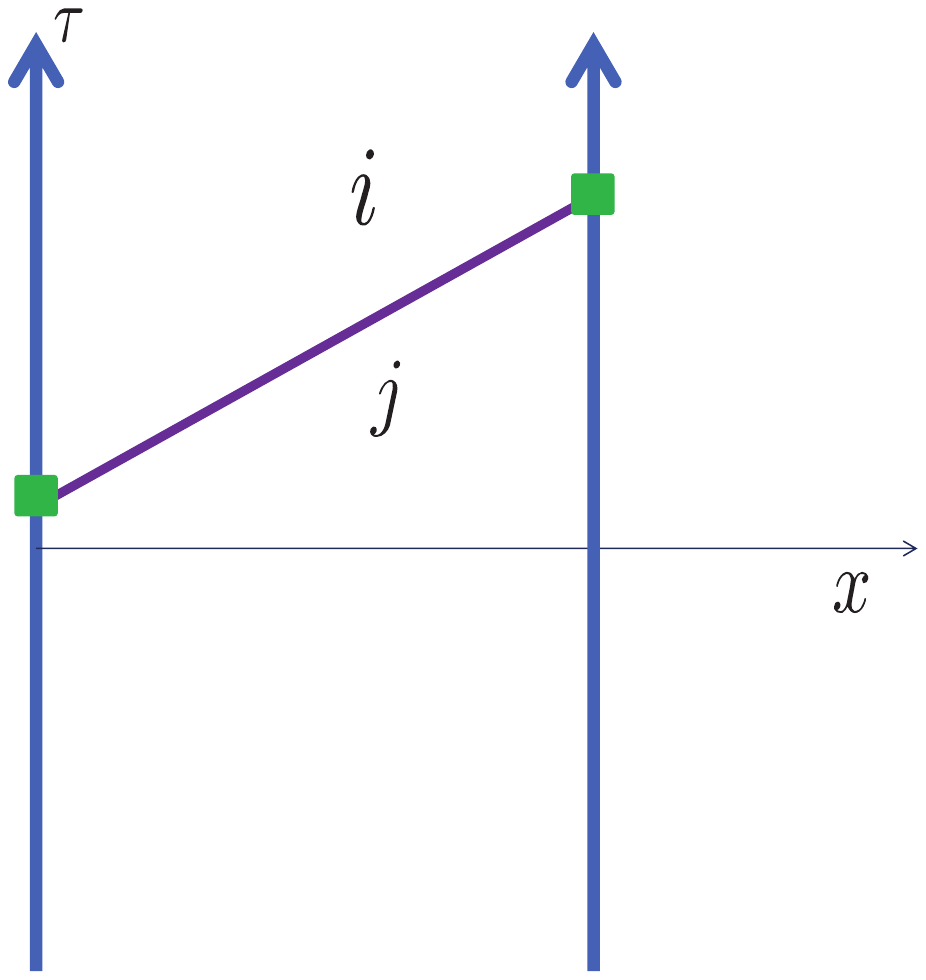}
 \end{center}
\caption{\small  A supersymmetric soliton exchanged between the two branes on the left  and the right.}
 \label{secondone}
\end{figure}

In general, however, eqn. (\ref{info}) only gives an approximation to the space of supersymmetric $(\B_\ell,\B_r)$ states. The states just described
have very nearly zero energy, but they may not have precisely zero energy. The reason for
this is that, rather as in supersymmetric quantum mechanics \cite{Witten:1982im}, instanton corrections can lift some approximate zero energy states
away from zero energy, though only by an exponentially small amount.
A simple instanton in this context is a process in which brane $\B_\ell$ emits a BPS soliton at an angle and that
soliton is absorbed by $\B_r$ (Figure \ref{secondone}).  The angle at which the soliton propagates in this figure is determined by supersymmetry,
and therefore it is reasonable to expect that the instanton sketched in the figure has precisely 1 real modulus, which one can think of as the ``time'' at which the soliton is exchanged.

In supersymmetric quantum mechanics, an instanton that depends on only 1 real modulus -- the time of the instanton event -- gives a correction
to the matrix element of a supercharge of fermion number $\FF=1$.  (An anti-instanton that depends on 1 real modulus similarly
corrects the matrix element of an $\FF=-1$ supercharge.)  More generally, in the field of an instanton that depends on $k$ real moduli, there are
$k$ fermion zero modes, so to get a non-zero amplitude, one must insert operators that shift $\FF$ by $k$ units.  In massive
LG models,  the same relationship holds
between the dimension of instanton moduli space and the violation of fermion number, as explained in Section \S \ref{zetawebs}.

Given the facts stated in the last paragraph, the instanton of Figure \ref{secondone} describes a process in which $\FF$ changes by 1 and thus this instanton can contribute
to the matrix element of $\Q_\B$ between initial and final states that have zero energy in the approximation of eqn. (\ref{info}).
In other words, such instantons can shift some approximately supersymmetric states away from zero energy. They must be taken into
account in order to determine the supersymmetric $(\B_\ell,\B_r)$ states.

\begin{figure}
 \begin{center}
   \includegraphics[width=4in]{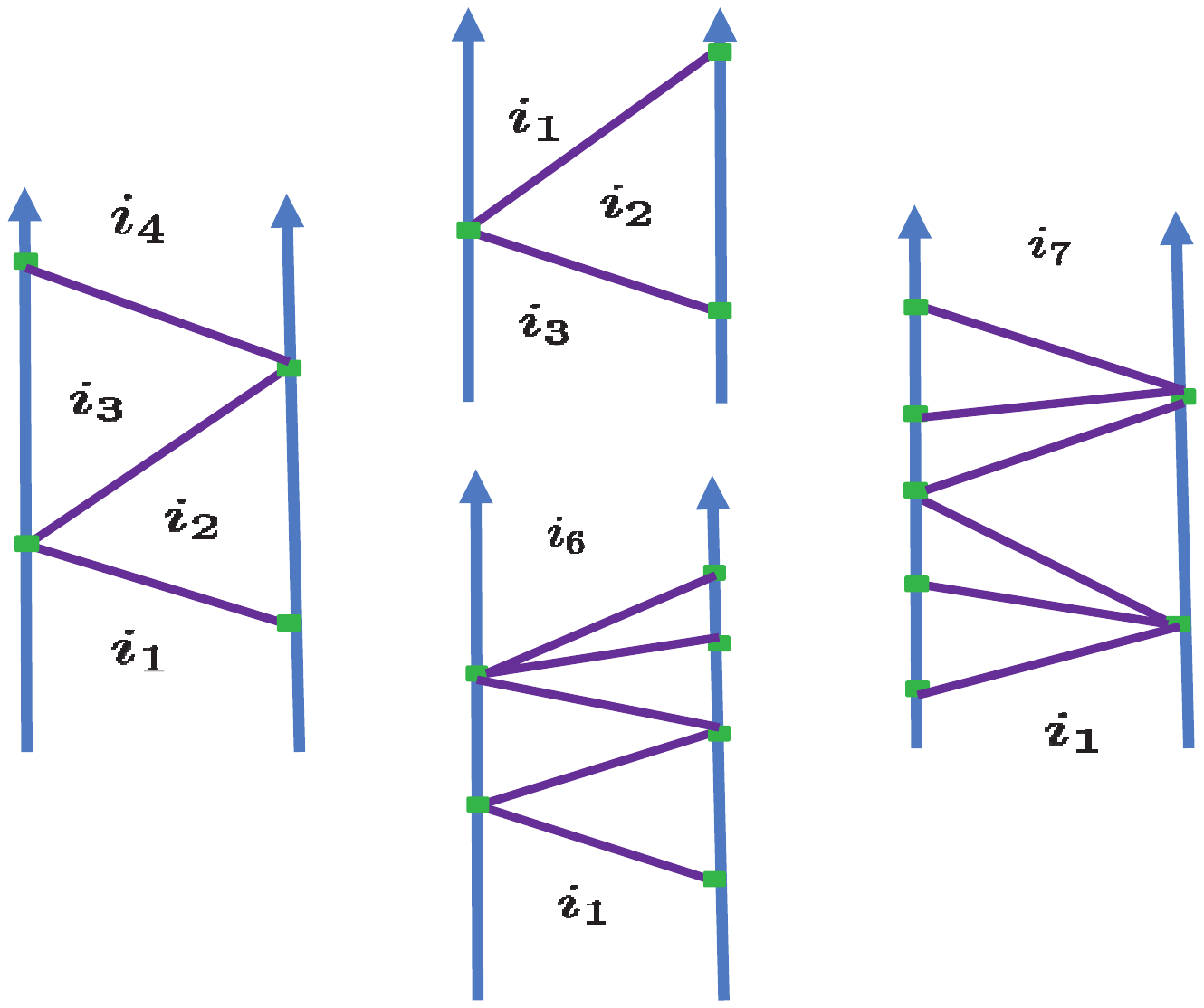}
 \end{center}
\caption{\small This figure shows a variety of rigid strip instantons constructed using  boundary
vertices only. }
 \label{thirdone}
\end{figure}

Once one gets this far, it is not hard to see that additional types of supersymmetric instantons might also be relevant.  First of all, there
might  be  amplitudes for the branes $\B_\ell$ and/or $\B_r$ to emit simultaneously  two or more supersymmetric solitons at suitable angles.
(In a massive Landau-Ginzburg model, such a multiple emission event is again computed by a solution of the
$\zeta$-instanton equation with suitable boundary
conditions.  We consider  again  the solutions  that have only a single real modulus corresponding to  overall time translations.)
If so, when the theory is formulated on a strip,
many additional types of supersymmetric instantons are possible.  In particular, the instantons indicated in Figure
\ref{thirdone} all depend on only a single real modulus -- the overall time of the tunneling event -- since the angles are fixed by supersymmetry.
Therefore, these instantons must
 all must be taken into account to determine which states of the $(\B_\ell,\B_r)$ system are precisely supersymmetric.  We will
say that an instanton in the strip is ``rigid'' if it has no moduli except the one associated to time translations.  So the instantons depicted in
Figure \ref{thirdone} are all rigid.  (Later in this paper, we will make a distinction between ``rigid'' and ``taut'' webs of BPS states on $\R^2$ or on
a half-plane, but in the present
context of instantons on a strip, this distinction does not arise.)

\subsection{Bulk Vertices}

\begin{figure}
 \begin{center}
   \includegraphics[width=4.5in]{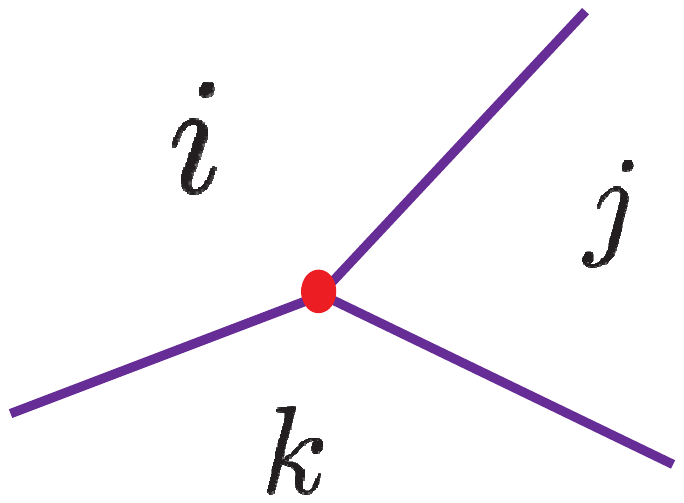}
 \end{center}
\caption{\small A ``bulk'' vertex that involves a coupling of three BPS solitons.  The vacua involved are $i,j,k\in\VV$,
and the solitons that emanate from the vertex are respectively of types $ij,$ $jk$, and $ki$.}
 \label{fourthone}
\end{figure}

This is still far from the whole story.  Certain ``closed string'' processes  must also be considered.   Once one realizes that there can be supersymmetric ``boundary''
vertices in which BPS solitons are emitted from a brane, it is natural to wonder if similarly there can be ``bulk'' vertices involving the coupling
of BPS solitons.\footnote{Examples of such bulk vertices have been constructed in \cite{Carroll:1999wr,Gibbons:1999np}.
No analogous explicit examples of boundary vertices are known.}  The most basic example is a trilinear coupling of three BPS solitons (Figure \ref{fourthone}), with  each soliton  emitted at an angle
$\varphi_{ij}=\mathrm{Arg}\,\zeta_{ij}/\zeta_{\B}$, as above.  Low energy effective field theory allows this possibility.  (In a Landau-Ginzburg theory, such
a coupling arises from a solution of the $\zeta$-instanton equation with suitable asymptotic conditions.)

A cubic bulk vertex has at least 2 real moduli, corresponding to the position of the vertex in $\R^2$.
For our present discussion, the relevant
case is that these are the only moduli (otherwise, we will not be able to make a rigid instanton in the strip).   We say that a bulk vertex is
rigid if it has only the 2 real moduli associated to spacetime translations.
Is a rigid cubic bulk vertex, if it exists, relevant to the problem of understanding the supersymmetric
$(\B_\ell,\B_r)$ states?  One may think the answer is ``no'' because once a bulk vertex is included, even a rigid one,
the number of real moduli will be at least 2.

\begin{figure}
 \begin{center}
   \includegraphics[width=3.5in]{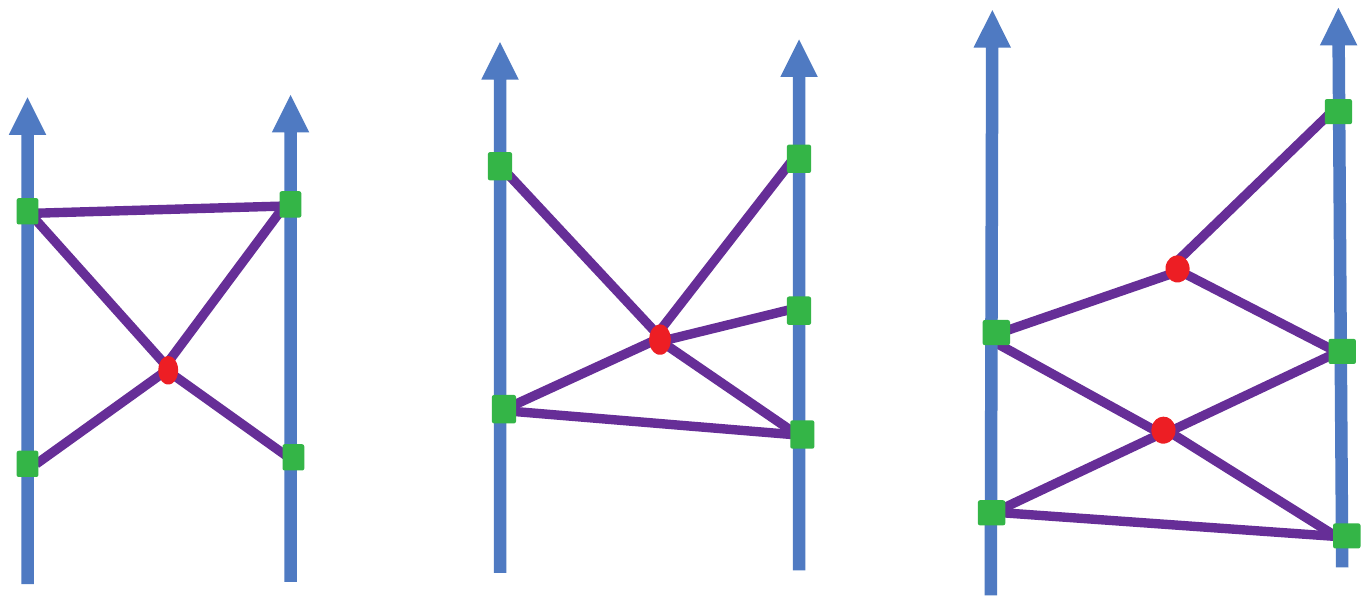}
 \end{center}
\caption{\small Rigid strip instantons whose construction makes use of bulk vertices.   We assume
the bulk vertices have no moduli except the ones associated to spacetime translations.}
 \label{fifthone}
\end{figure}
However, it is not hard to construct strip instantons that depend on only 1 real modulus even though they involve
bulk vertices that individually would depend on 2 real moduli each.  Some examples are shown in Figure \ref{fifthone}.  These web instantons must all
be included to determine  which $(\B_\ell,\B_r)$ states have precisely zero energy.

In short, to answer the seemingly simple question of finding the supersymmetric states on an interval in a massive theory,
we need a full understanding of all bulk and boundary vertices involving couplings of BPS solitons, and how they can be put  together to
make what we will call a ``web'' of BPS solitons.

\subsection{An Algebraic Structure}

At this point, matters may seem bewilderingly complicated.  However, there is a hidden simplification: the data that we have described can be combined into a rich algebraic structure that makes things tractable.  This structure is the real topic of the present paper.

\begin{figure}
 \begin{center}
   \includegraphics[width=5.5in]{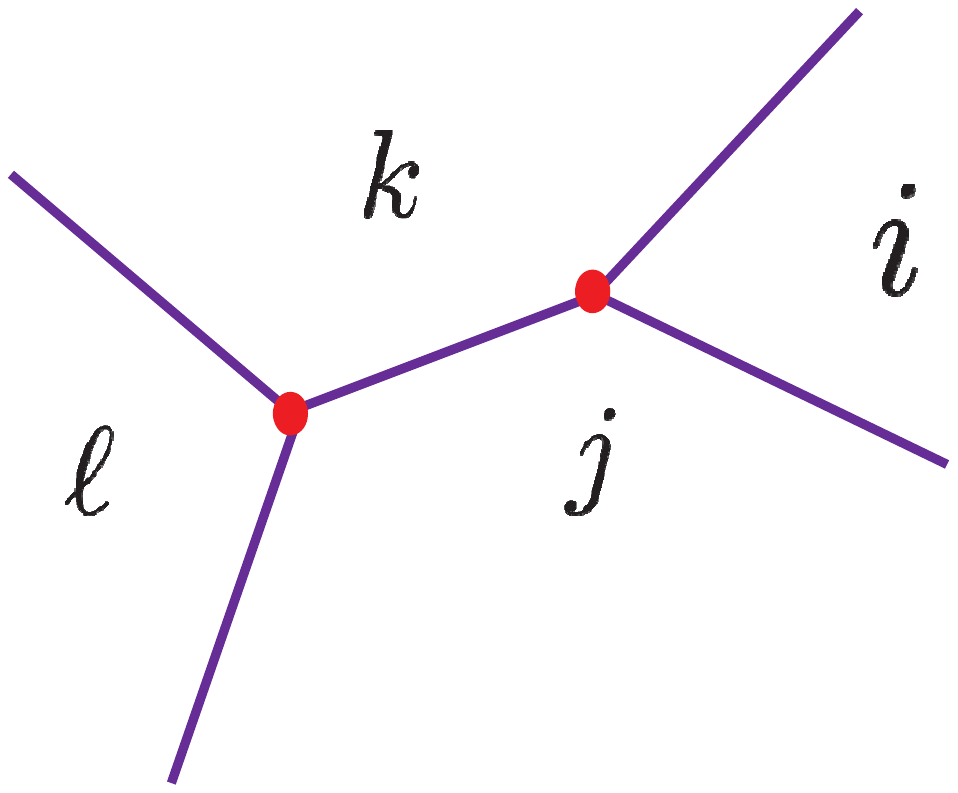}
 \end{center}
\caption{\small An ``end'' of the moduli space of solutions of the $\zeta$-instanton equation corresponding to two widely
separated vertices of type $ijk$ and $kj\ell$.  Assuming the individual vertices have only the obvious moduli associated to spacetime translations,
this component of the moduli space is three-dimensional.}
 \label{seventhone}
\end{figure}

To illustrate the basic idea, we start  with a cubic vertex involving vacua $i,j,k$ and another cubic vertex involving vacua $k,j,l$.
We assume that each vertex has only the 2 moduli associated to spacetime translations.  If the vertices are far apart, we can make an approximate
solution involving all four vacua $i,j,k,l$ by gluing together the $jk$ soliton that emerges from one vertex with the $kj$ soliton that emerges from
the other vertex (Figure \ref{seventhone}).  After slightly adjusting the output of this gluing operation, one gets a family of solutions of the $\zeta$-instanton equation
with a three-dimensional moduli space that we will call $\M$.  (One expects that index theory and ellipticity of the LG instanton equation ensure that this adjustment
can be made.)

To the extent that we can identify the moduli from the figure, two of them are associated to spacetime translations and the third is the distance $d$
between the two vertices.  So if we rely entirely on this figure, it looks like  $\M$ is a copy of $\R^2\times \R_+$, where $\R^2$ parametrizes
the position of, say, the $ijk$ vertex, and $\R_+$  is the half-line $d\geq 0$.

At least in the context of LG models, $\R^2\times \R_+$  cannot be the correct answer, since $\R_+$ has a boundary at $d=0$.
Because of the superrenormalizable nature of the LG theory, there will be no such boundary in the moduli space of
solutions of the $\zeta$-instanton equation.  (A technical statement is that the $\zeta$-instanton equation is a linear equation plus lower order nonlinear
terms. The linear equation with target space $X=\IC^n$ does not admit ``bubbling'' and the superpotential
is a lower order term which does not change that property.)
For a generic superpotential, any family of solutions can be continued, with no natural boundaries or singularities, and with ends that arise
only when something goes to infinity.  In a massive LG theory, the scalar fields cannot go to infinity
(since the potential energy grows when they
do), so all that can go to infinity
are the vertices.
  This means that the ``ends'' of the moduli space
have a semiclassical picture in terms of a soliton web, as in Figure \ref{seventhone}. Moreover, this is also true for the reduced moduli
space $\M'=\M/\R^2$ that is obtained by dividing out by spacetime translations.
%
%
%\footnote{In this paper we leave the notion of an
% ``end'' of a moduli space somewhat heuristic. If the moduli space is $\IR$ it has
% two ends, and if it is $\IR^d$ with $d>1$ then it has one end. For a formal definition
% one could start with the Wikipedia entry End(topology). For more discussion about possible
% ``ends'' of moduli spaces in QFT see Sections \S\S \ref{whyindeed},\ref{impanom}, and \ref{zetafan} below. }
%
%

In Figure \ref{seventhone},  the reduced moduli space, to the extent that we can understand it from the figure, is a copy of $\R_+$ with
one visible end for $d\to\infty$.  But the moduli space cannot just end at $d=0$.  It has to continue somehow. Because a one-manifold
without boundary
that has at least one noncompact end is a  copy of $\R$, which has two ends, it must continue to infinity with a second end.
A correct view of the figure is that it gives a good approximate picture of a family of solutions of the $\zeta$-instanton equation when  $d$ is large.
When $d$ is not large, the semiclassical picture of the solution given in Figure \ref{seventhone} is not valid.  But the reduced moduli $\M'$ must
have a second ``end'' that again has a semiclassical interpretation.

\begin{figure}
 \begin{center}
   \includegraphics[width=5.5in]{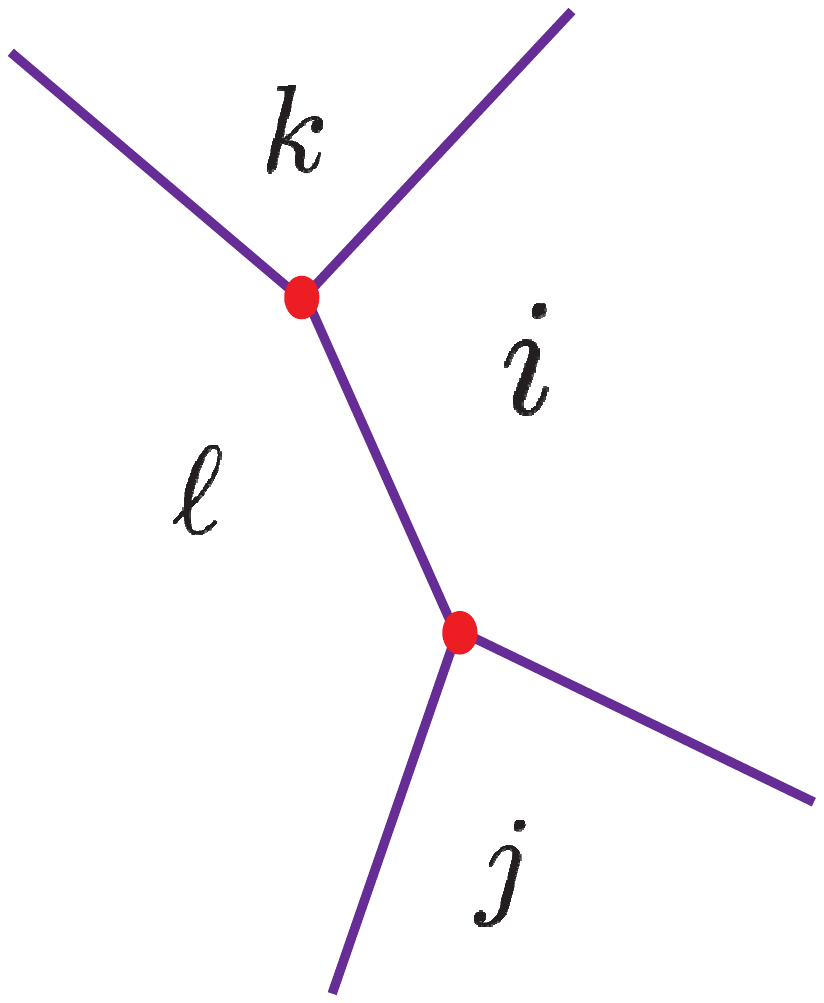}
 \end{center}
\caption{\small The component of moduli space that has one end depicted in Figure \protect\ref{seventhone}
%
%\ref{seventhone}
%
must have a second end.
The second end might be as depicted here.  Note that the ``fans'' of vacua at infinity are the same in this figure and in Figure
\protect\ref{seventhone}
%
%\ref{seventhone},
%
%
so they can appear as parts of the same moduli space.}
 \label{eighthone}
\end{figure}
Low energy effective field theory is not powerful enough to predict what this second end will be.  In general, there are different possibilities.
In the case at hand, a natural possibility (Figure \ref{eighthone}) is that in
addition to the $ijk$ and $kjl$ vertices that we started with, there are also solutions of the $\zeta$-instanton equation corresponding to
$ijl$ and $lki$ vertices.    The reason that  the soliton webs shown in figs. \ref{seventhone} and  \ref{eighthone} can both appear as part of the same moduli space
is that they connect to the same ``web'' of BPS solitons at infinity.

In this situation, if we represent an $ijk$ vertex by a symbol $\beta_{ijk}$, we see that there is some sort of relation between the products
$\beta_{ijk}\beta_{kjl}$ and $\beta_{ijl}\beta_{klj}$.  What is this relationship precisely?

The answer turns out to be that the bulk vertices are part of  an algebraic structure known as an $L_\infty$ algebra. (More precisely,
they define a solution of the Maurer-Cartan equation in an $L_\infty$ algebra.  Such a solution can be used to deform
an $L_\infty$ algebra to a new $L_\infty$ algebra.)
This structure often appears in closed-string theory (for instance, see \cite{Zwiebach:1992ie,Zwiebach:1997fe}), and in related areas of mathematics.
 The vertices associated to emissions from a brane can similarly be used to define what in
open-string theory and related areas is sometimes called an $A_\infty$ algebra.    We expect that these algebraic structures are universal for massive $\N=2$ theories, though in motivating them,
we have made essential use of LG models.  The reason that we expect so is that these algebraic structures are well-adapted to answering
our basic question of how to determine the spectrum of supersymmetric states in the presence of branes.

\subsection{Categorical Wall-Crossing Formula}
The spaces $\CR_{ij}$ of quantum BPS soliton states and the spaces of quantum ground states $\CE_i(\fB)$ on the half line
are objects of independent interest. As the parameters of the underlying theory are varied,
these spaces of ground states are expected to jump across certain walls of marginal stability.
The standard theory of wall-crossing constrains the variation of the Witten indices of such spaces of states:
the BPS and framed BPS degeneracies
\be
\mu_{ij} = \Tr_{\CR_{ij}} (-1)^F \qquad \qquad \fro(\fB,i) = \Tr_{\CE_i(\fB)} (-1)^F
\ee
The framed BPS degeneracies jump across walls where $\zeta^\fB$ aligns to some $\zeta_{ij}$.
The form of the jump is universal \cite{Gaiotto:2011tf}:
\be
S_{ij}:~\fro(\fB,j) \to \fro(\fB,j) + \fro(\fB,i) \mu_{ij}
\ee
 with all other $\fro(\fB,k)$ remaining unchanged.

The BPS degeneracies jump across walls where $\zeta_{ij}$ aligns to some $\zeta_{jk}$. The form of the jump is universal \cite{Cecotti:1992rm}
\be \label{eq:cswall}
\mu_{ik} \to \mu_{ik} + \mu_{ij} \mu_{jk}
\ee
 with all other $\mu_{kt}$ remaining unchanged. This formula can be derived directly by requiring compatibility with
 the framed BPS wall-crossing formula: the relation
\be
S_{ij}[\mu] S_{ik}[\mu] S_{jk}[\mu] = S_{kj}[\mu'] S_{ik}[\mu'] S_{ij}[\mu']
\ee
 implies that the $\mu$ and $\mu'$ degeneracies are related as in \eqref{eq:cswall}.

In Sections \ref{sec:CatTransSmpl} and \ref{sec:GeneralParameter} we address the problem of providing a categorification of such wall-crossing formulae, i.e. we describe the categorical data which
should be added to the vector spaces $\CR_{ij}$ and $\CE_i(\fB)$ in order to allow for a universal description of how such vector spaces
(and the categorical data itself) jump across walls of marginal stability.

The categorical data which has to be added to the $\CE_i(\fB)$ to describe their wall-crossing properties
essentially coincides with the amplitudes for the emission of BPS solitons from the
boundary condition $\fB$, organized into an object of an appropriate category.
The categorical wall-crossing of $\CE_i(\fB)$ is encoded in a ``mutation'' of that category.

The categorical wall-crossing of the $\CR_{ij}$ is determined again by requiring compatibility with the
categorical framed BPS wall-crossing formula. The existence of a categorical BPS wall-crossing formula
is related to the observation that mutations form, in an appropriate sense,
a representation of the braid group.

\subsection{A More Detailed Summary}

In this introduction, we have omitted several subjects that are treated in considerable detail in the main text.
These include interfaces  between theories, as well as bulk and boundary local operators in massive $\N=2$ theories.

The curious reader who wants to learn more detail, but is daunted by the
length of the present paper, is referred to  
\cite{FloridaLectures}. These are lecture notes that summarize 
the entire paper from a broad perspective,
and can serve as a detailed introduction.

\section{Webs}\label{sec:Webs}

In the previous section we have motivated from
qualitative physical considerations the concept of
webs associated to a massive two-dimensional supersymmetric
QFT. In this section we abstract that idea and discuss
in some detail a purely mathematical construction.

\subsection{Plane Webs}\label{planewebs}

We begin with webs in the plane $\IR^2$, which
we sometimes identify with $\IC$.

The definition of a web depends on
some data. We fix a finite
set $\IV$ called the \emph{set of vacua}.
Typical elements are denoted $i,j,\dots \in \IV$.
We also fix a set of \emph{weights} associated
to these vacua which are  complex numbers $\{z_i\}$,
that is, we fix a map $z:\IV \to \IC$. We assume that
$z_{ij}\not=0$ for $i\not=j$.  The pair
$(\IV,z)$ will be called \emph{vacuum data}.
The following definition is absolutely fundamental to
our formalism:

\bigskip
\noindent
\textbf{Definition:} A \emph{plane web}
is a graph in $\IR^2$,   together with a
labeling of the \emph{faces} (i.e. connected components
of the complement of the graph) by vacua
such that the labels across  each edge are different
and moreover, when oriented with
$i$ on the left and $j$ on the right the edge is
straight and parallel to the complex number
$z_{ij}:= z_i - z_j$. We take plane webs
to have all vertices of valence at least three.
In Section \S  \ref{subsec:ExtendedWebs} we define
a larger class of \emph{extended plane webs}
which have two-valent vertices. In Section \S
\ref{sec:LocalOpsWebs} we will introduce a further
generalization to \emph{doubly extended plane webs}
by allowing certain zero-valent vertices.

\begin{figure}[htp]
\centering
\includegraphics[scale=0.3,angle=0,trim=0 0 0 0]{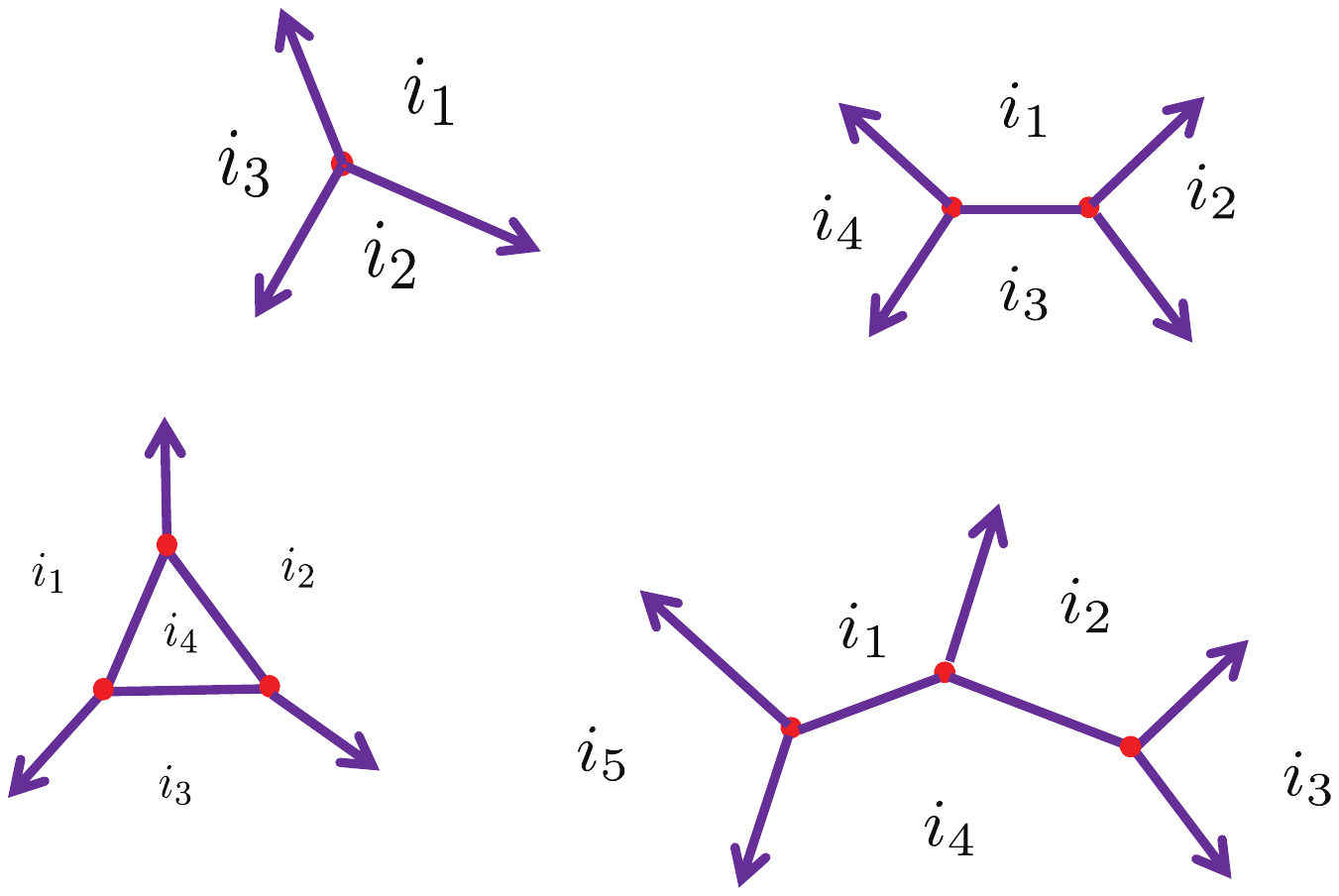}
\caption{Some examples of plane webs.    }
\label{fig:FIRST-WEB-EXAMPLES}
\end{figure}
\begin{figure}
 \begin{center}
   \includegraphics[width=3in]{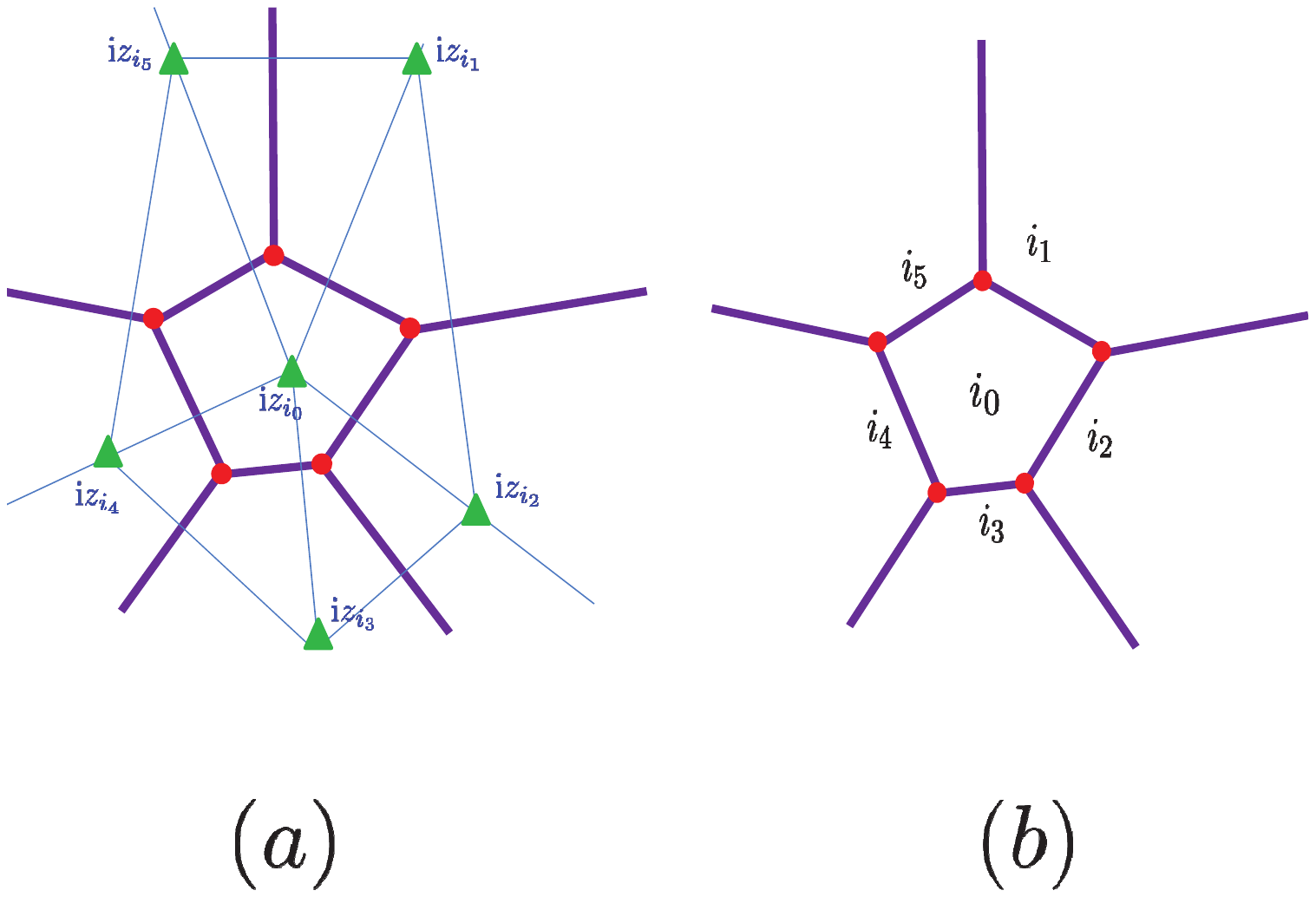}
 \end{center}
\caption{(a) A configuration of weights $z_i$ which is not
convex. The green triangles indicate the points  $\I z_i$.
The dual graph gives an example of a web shown in $(b)$. Note that the
  corresponding web has a loop.  }
 \label{fig:CONVEXWEIGHTS}
\end{figure}

Some examples of webs are shown in Figure \ref{fig:FIRST-WEB-EXAMPLES}.
We make a number of remarks on some basic properties that
immediately follow from this simple definition:

\begin{enumerate}

\item  Note that the edges do not have
an intrinsic orientation. If we reverse
the orientation of an edge then $j$ is on the left
and $i$ is on the right and then the oriented
edge is parallel to $z_{ji}= - z_{ij}$.
Edges which go to infinity are called \emph{external
edges} and the remaining edges are \emph{internal edges}.
In section
\S \ref{sec:RepWeb} we give external edges
a canonical outward orientation.

\item At each vertex of a plane web
 the labels in the angular regions in the
 \emph{clockwise direction} define a \emph{cyclic fan of vacua},
which is, by definition, an ordered set $\{ i_1, \dots, i_s\}$
so that the phases of   $z_{i_k, i_{k+1}}$,
with $k$ understood modulo $s$,
form a clockwise ordered collection of points on the
unit circle. Put differently
\be
\Im z_{i_{k-1},i_{k} } \overline{z_{i_{k },i_{k+1} } } > 0
\ee
for all $k$. We generally denote a cyclic fan of vacua
by $I=\{ i_1, \dots, i_s\}$ and we  say that $I$ has length $s$.

\item A useful intuition is obtained by thinking of the
edges as strings under a tension given by $z_{ij}$.
Then at each vertex we have a no-force condition:
\be\label{eq:zero-force}
z_{i_1,i_2} + z_{i_2,i_3} + \cdots + z_{i_n,i_1} = 0
\ee
It follows that the edges emanating from any vertex
cannot lie in any half plane.

\item For plane webs the graph must be connected.
Moreover, for  a fixed set of weights $\{ z_i \}$
there are only a finite number of plane webs.
\footnote{This finiteness property is one advantage
of the requirement that all vertices have valence
bigger than two.}
We can prove these statements with a useful argument which
we will call the \emph{line principle}:
If we consider any oriented line in the
plane which does not go through a vertex of the web
then we encounter an ordered set of vacua given
by the labels of the regions intersecting the line. The line principle
says that no vacuum can appear twice.  To prove this note that we can orient
all the edges which intersect the line to point into
one half-plane cut out by the line. Then a vacuum cannot
appear twice since if it did in the sequence $\{ i, j_1,\dots, j_k, i\}$
then, on the one hand, the sum of the tensions $z_{i,j_1} + \cdots z_{j_k,i} =0$,
but on the other hand all the terms in the sum point into the same half-plane,
which is impossible. Therefore there are only a finite number of
possible sequences of vacua. This implies that there are only
a finite number of possible vertices. Other corollaries
of the line principle are that  no vertex
can appear twice within any given web and there are at least three
external edges.

 \item  A sequence of weights $z_{i_k}$ is associated to a cyclic fan of
 vacua $I=\{ i_1, \dots, i_s\}$ if and only if they are the clockwise ordered
vertices of a convex polygon in the complex plane.  The topology of a web $\fw$ is
captured by the decomposition of the polygon $P_\infty$ associated to $I_\infty(\fw)$
into the polygons $P_v$ associated to the $I_{v}(\fw)$. This can be seen by noting
that an internal edge of the web connecting vertices $v_1$ and $v_2$ corresponds
to a   shared edge of the polygons $P_{v_1}$ and $P_{v_2}$. On the other hand,
each external edge of the web is associated to a single vertex $v$ and corresponds
to an external boundary of $P_\infty$.   See, for example,
 Figure \ref{fig:CONVEXWEIGHTS}.
 Indeed, the   decomposition can be identified with a dual graph to the web.
This provides an alternative, intuitive explanation of many properties of the webs.
The paper  \cite{Kapranov:2014uwa} of Kapranov, Kontsevich, and Soibelman
emphasizes this dual viewpoint and suggests that it is the proper formulation
for generalizing the structures we find to higher dimensional field theories.

\item  A corollary of the above remark is that for a given
set of weights $\{ z_i\} $  there
 is a web with a closed loop if and only if
there is a sequence of weights
 $\{ z_{i_1},\dots, z_{i_s}  \}$ which are vertices of a
 convex polygon such that there is a weight $z_{i_0}$ in
 the interior of the polygon. The existence of a web
 with a closed loop implies that there are positive
 numbers $\lambda_{\alpha}$ with
\be
\lambda_1 z_{i_1,i_0} +\cdots + \lambda_s z_{i_s,i_0} =0
\ee
and hence
\be\label{eq:cvx-1}
z_{i_0} = t_1 z_{i_1,i_0} +\cdots + t_s z_{i_s,i_0}
\ee
with $t_\alpha = \lambda_\alpha/\sum_\beta \lambda_\beta $.
Conversely, if \eqref{eq:cvx-1} holds with $\Im(z_{i_{k-1},i_0}\overline{z_{i_{k},i_0} })>0$
for all $k$ then it is easy to show that $\{i_{k-1},i_k,i_0 \}$ is a cyclic
fan of vacua so the web shown in Figure   \ref{fig:CONVEXWEIGHTS}(b) exists.

\item \emph{Some notation}: We will denote a web
(or rather its ``deformation type'' - defined below) by a gothic ``w,''
which looks like  $\fw$.
The set of vertices is denoted by $\CV(\fw)$, and it has
order $V(\fw)$. Similarly, the set of internal edges
is $\CE(\fw)$ and has order $E(\fw)$. At each vertex
$v\in \CV(\fw)$ there is a cyclic fan of vacua $I_{v}(\fw)$.
The cyclic fan of vacua at infinity is denoted $I_\infty(\fw)$.

\item We have the relation
 $V(\fw) - E(\fw) + F(\fw) = 1$, where $F(\fw)$ is the
 number of bounded faces (and hence the number of internal loops).
 This follows since if we add a vertex at infinity then
 the web triangulates $S^2$.

\item In the applications to Landau-Ginzburg theories in Sections
\S\S \ref{lgassuper}-\ref{subsec:LG-Susy-Interface} below the vacua $\IV$ will be the critical points
of the superpotential and the vacuum weights $z_i$ are essentially the critical
values of of the superpotential. The precise relation, as determined by
equation \eqref{hopeful} and Figure \ref{fig:BOOSTEDSOLITON} below is
$z_i = \zeta\overline{W_i}$, where $W_i$ is the critical value
of the superpotential and $\zeta$ is a phase, introduced in Section \S \ref{lgassuper}.

\item Finally, we note that in the application to knot-homology described in
\S \ref{subsec:KnotHomology} we will need to relax the constraint that $\IV$
is a finite set. In general, if $\IV$ is infinite one can choose weights $z_i$
leading to pathologies. (For example, if the weights have an accumulation
point in the complex plane, there will be infinite numbers of webs with the
same fan at infinity.) In Section \S \ref{subsec:TwistedMasses} we describe
a class of models where $\IV$ is infinite, but for which our theory still applies.
The knot homology examples belong to this class.

\end{enumerate}

\begin{figure}[htp]
\centering
\includegraphics[scale=0.3,angle=0,trim=0 0 0 0]{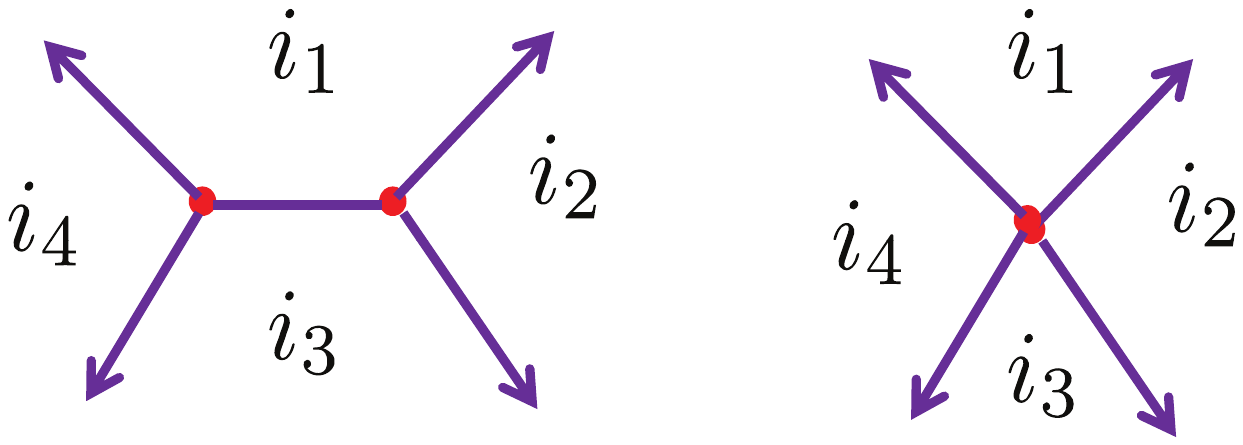}
\caption{The two webs shown here are considered to be   different
deformation types, even though the web on the left can clearly degenerate
to the web on the right.   }
\label{fig:DIFFERENT-DEFORMATION-TYPE}
\end{figure}

A plane web has a \emph{deformation type}: This is an equivalence
class under translation and/or scaling of the lengths of some
subset of the internal edges. This scaling must of course be
compatible with the constraints that define a web: In terms of
the string model of the web mentioned above we are allowed to
stretch and translate the strings, but we must not rotate them,
 and we must  maintain the no-force condition.  In a deformation
type no edge is allowed to be scaled to zero size. See Figure
\ref{fig:DIFFERENT-DEFORMATION-TYPE}. The set of webs with fixed
deformation
type $\fw$   is naturally embedded as
a cell $\CD(\fw) \subset (\IR^2)^{V(\fw)}$ by considering the $(x,y)$
coordinates of all the vertices of the web. The (internal) edge conditions
impose $E(\fw)$ linear relations on these coordinates, together with
inequalities requiring that each edge have positive length.
When the vacuum weights $z_i$ are in general position the edge conditions
will be independent equations and then
 $\CD(\fw)$ will be a convex cone of dimension
\be\label{eq:Def-Dim}
d(\fw):= 2 V(\fw) - E(\fw).
\ee
We will sometimes refer to $d(\fw)$ as the \emph{degree} of the web.
Note that there is a free action of translations
on the set of webs of a given deformation type, so $d(\fw) \geq 2$.
We will refer to the quotient $\CD_r(\fw)$ of the moduli space $\CD(\fw)$ by the translation group
as the {\it reduced} moduli space. Thus, provided the weights are in general position, the dimension of
the reduced moduli space,  called the
reduced dimension, is  $d_r(\fw):=2 V(\fw) - E(\fw)-2$.

For generic configurations of weights $\{ z_i \} $ the
 boundary of the closure $\bar \CD(\fw)$ of $\CD(\fw)$ in $\IR^{2 V(\fw)}$
consists of $d(\fw)-1$ dimensional cells where some edge inequality is saturated.
Thus at each boundary cell two or more vertices of $\fw$ collapse to a single point
$p$ and $\fw$ reduces to a simpler web $\fw_1$ with a marked vertex $v$ at $p$.
In a small neighbourhood of such boundary cell $\fw$ can be recovered from $\fw_1$ by replacing $v$ with an
infinitesimally small copy of a second web
$\fw_2$ formed by the collapsing vertices and edges. The cyclic fan $I_\infty(\fw_2)$ coincides with the cyclic fan $I_p(\fw_1)$.

In order to formalize the relation between $\fw$, $\fw_1$ and $\fw_2$ we introduce the key construction of   \emph{convolution of webs}:

\bigskip
\noindent
\textbf{Definition:} Suppose $\fw$ and $\fw'$ are two plane
 webs and there is a vertex $v\in \CV(\fw)$ such that
 \be
 I_v(\fw) = I_\infty(\fw').
 \ee
 We then define $\fw*_v \fw'$ to be the deformation type
 of a web obtained by cutting out a small disk around $v$
 and gluing in a suitably scaled and translated copy of
 the deformation type of $\fw'$. It is important
 that we only use a deformation type here. In general
 the external edges of $\fw'$
 do not necessarily meet at a single point when continued
 inward. However we can deform $\fw'$ so that the edges
 literally fit with those of $\fw$, provided we take the
 disk sufficiently small. This and similar statements can be proven trivially from the
 linear nature of the constraints imposed on the positions of the vertices by the topology of a web.

\begin{figure}[htp]
\centering
\includegraphics[scale=0.3,angle=0,trim=0 0 0 0]{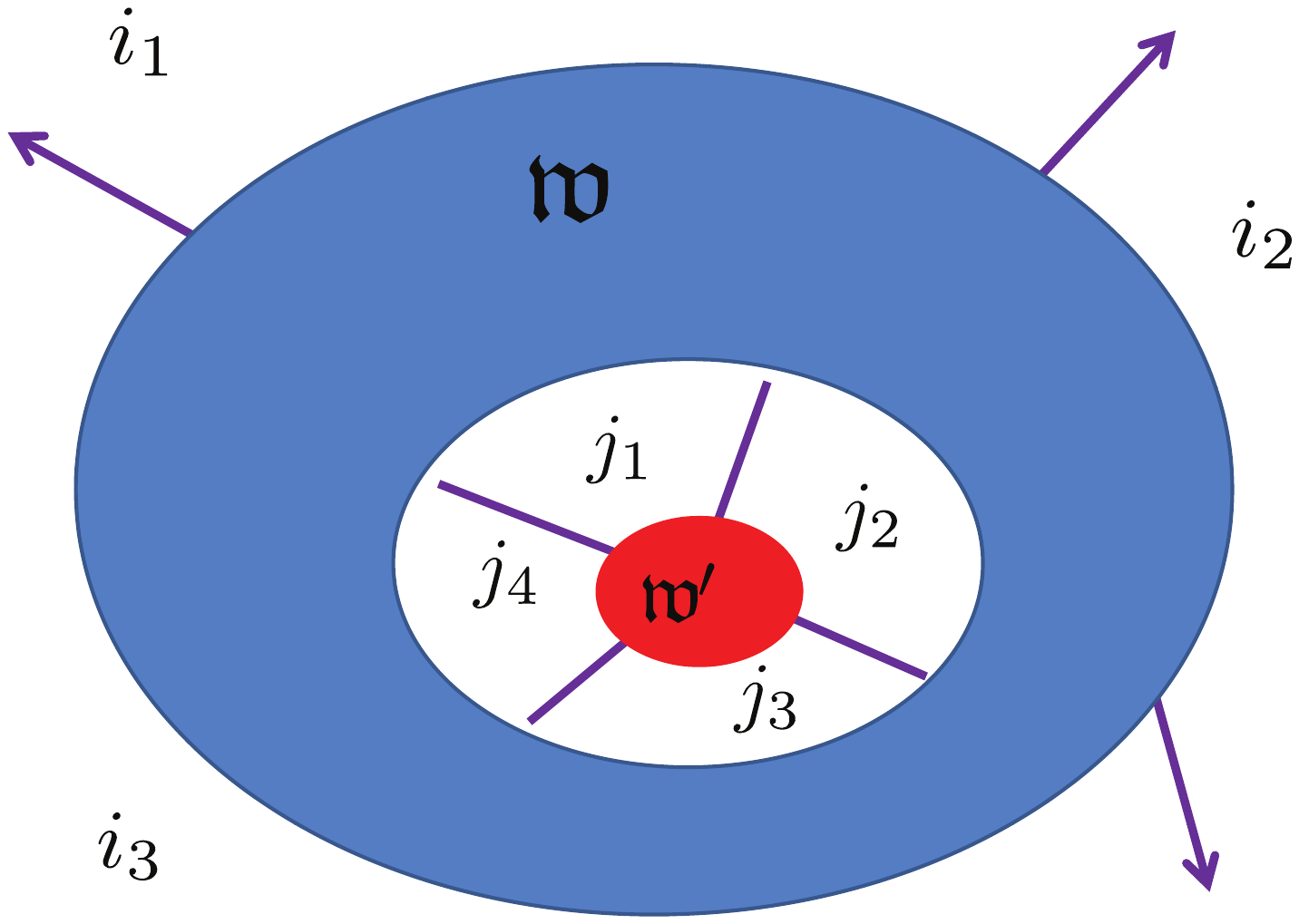}
\caption{Illustrating the convolution of a web $\fw$ with
internal vertex $v$ having a fan $I_v(\fw)=\{j_1,j_2,j_3,j_4\}$ with
a web $\fw'$ having an external fan $I_\infty(\fw') =\{j_1,j_2,j_3,j_4\}$.     }
\label{fig:CONVOLUTION}
\end{figure}

 The procedure is illustrated in Figure \ref{fig:CONVOLUTION}.
 When writing convolutions below we always
 put the ``container web'' on the left.

One easily verifies the relations
\be
E(\fw *_v \fw') = E(\fw) + E(\fw')
\ee
\be
V(\fw *_v \fw') = V(\fw) + V(\fw')-1
\ee
and hence we have the important relation
\be\label{eq:expct-dim-conv}
d(\fw *_v \fw') = d(\fw) + d(\fw')-2
\ee
showing that we can take any sufficiently small representative $\fw'$ in $\CD_r(\fw')$ and
insert it into any given representative $\fw$ in $\CD_r(\fw')$.

With these results at hand, it is now clear that, for generic weights $\{ z_i \}$,  the top dimension
boundary cells of $\bar \CD(\fw)$ are in one-to-one correspondence with pairs $(\fw_1, \fw_2)$
such that $\fw = \fw_1 *_v \fw_2$ and $d(\fw_1)=d(\fw)-1$. In the neighbourhood of each such
boundary cell we have a local isomorphism
between $\CD(\fw)$ and $\CD(\fw_1) \times \CD_r(\fw_2)$.

We now introduce some special classes of webs which will
be of the most use to us:

\bigskip
\noindent
\textbf{Definition:} A \emph{rigid web} is a web with $d(\fw)=2$.
A \emph{taut web} is a web with $d(\fw)=3$ and a
\emph{sliding web} is a web with $d(\fw)=4$.

A rigid web must have $E(\fw)=0$ hence $V(\fw)=1$
 and hence is just a single vertex. Using
  $V(\fw) - E(\fw) + F(\fw) = 1$ and eliminating $E(\fw)$
 we have
 \be
 V(\fw) = d(\fw)-1 + F(\fw)
 \ee
 and hence a taut web has at least two vertices, a sliding web
 at least three vertices, and so forth.

Let $\CW$ be the free abelian group generated by
\emph{oriented} deformation types of webs.  By ``oriented''
we mean that we have chosen an orientation $o(\fw)$ of
the cell $\CD(\fw)$. Henceforth the notation
$\fw$ will usually refer to such an oriented deformation type,
rather than a specific web.
In $\CW$ the object $-\fw$ is the
oriented web with the opposite orientation to $\fw$.
Henceforth, when working with $\CW$ we will assume the vacuum
weights are in generic position. We return to this assumption
in Section \S  \ref{subsec:SpecialVacWt} below.

We now define a convolution operation
\be
*: \CW \times \CW \to \CW
\ee
by defining $\fw_1 *_v \fw_2 =0$ if $I_\infty(\fw_2) \not= I_v(\fw_1)$
and then setting
\be\label{eq:conv-def}
\fw_1 * \fw_2 := \sum_{v\in \CV(\fw_1)} \fw_1 *_v \fw_2.
\ee
Note that, because of the line principle, at most one term on
the right hand side of \eqref{eq:conv-def}
can be nonzero. Moreover, in order for this to be well-defined on
$\CW$ we must orient $\CD(\fw_1*\fw_2)$. If $o(\fw)$ is an orientation
of $\CD(\fw)$, thought of as a top-degree form, we can use the
freely-acting translation symmetry to define a
``reduced orientation'' by
\be
o_r(\fw):=\iota(\frac{\p}{\p y}) \iota(\frac{\p}{\p x})  o(\fw)
\ee
and then we define
\be
o(\fw_1*\fw_2):= o(\fw_1) \wedge o_r( \fw_2) .
\ee
(This uses the product structure near the boundary of the cell where
$\fw_2$ shrinks to a single vertex.)

Since taut webs have
a one-dimensional reduced cell we can and will choose a standard orientation
for all taut webs to be
the orientation with tangent vector in the direction of
increasing size. That is, the moduli of the taut web can be taken to be
an overall position $x,y$ together with a scaling modulus $\ell$. We
take the orientation to be $dx dy d\ell$.
Now we can define the \emph{taut element} $\ft\in \CW$ to be
the sum of all oriented taut webs with standard
orientation:
\be\label{eq:taut-planar}
\ft := \sum_{d(\fw)=3} \fw.
\ee

Including the orientation data, we arrive at our final characterization of the
generic codimension one boundaries of $\bar \CD(\fw)$: a typical web
$\fw$ looks like a convolution $\fw_1 *_v \fw_2$ where $\fw_2$ is a
taut web and the orientation of $\fw$ is written as $o(\fw) = o(\fw_1) \wedge d \ell_2$,
with $\ell_2$ oriented towards the interior of the cell $\bar \CD(\fw)$.
Applying this picture to the case where $\fw$ is a sliding web we
 note that   $\fw_1$ is a taut web as well, and the natural orientation
$dx dy d\ell_1 d\ell_2$ might or might not agree with the orientation of $\fw$.
We should thus write $\fw = \pm \fw_1 *_v \fw_2$. Looking carefully at the global structure of the moduli
spaces of sliding webs, we deduce our first result:

\bigskip
\noindent
\textbf{Theorem}: We have
\be\label{eq:Taut-sq-zer}
\ft* \ft=0.
\ee

\begin{figure}[htp]
\centering
\includegraphics[scale=0.3,angle=0,trim=0 0 0 0]{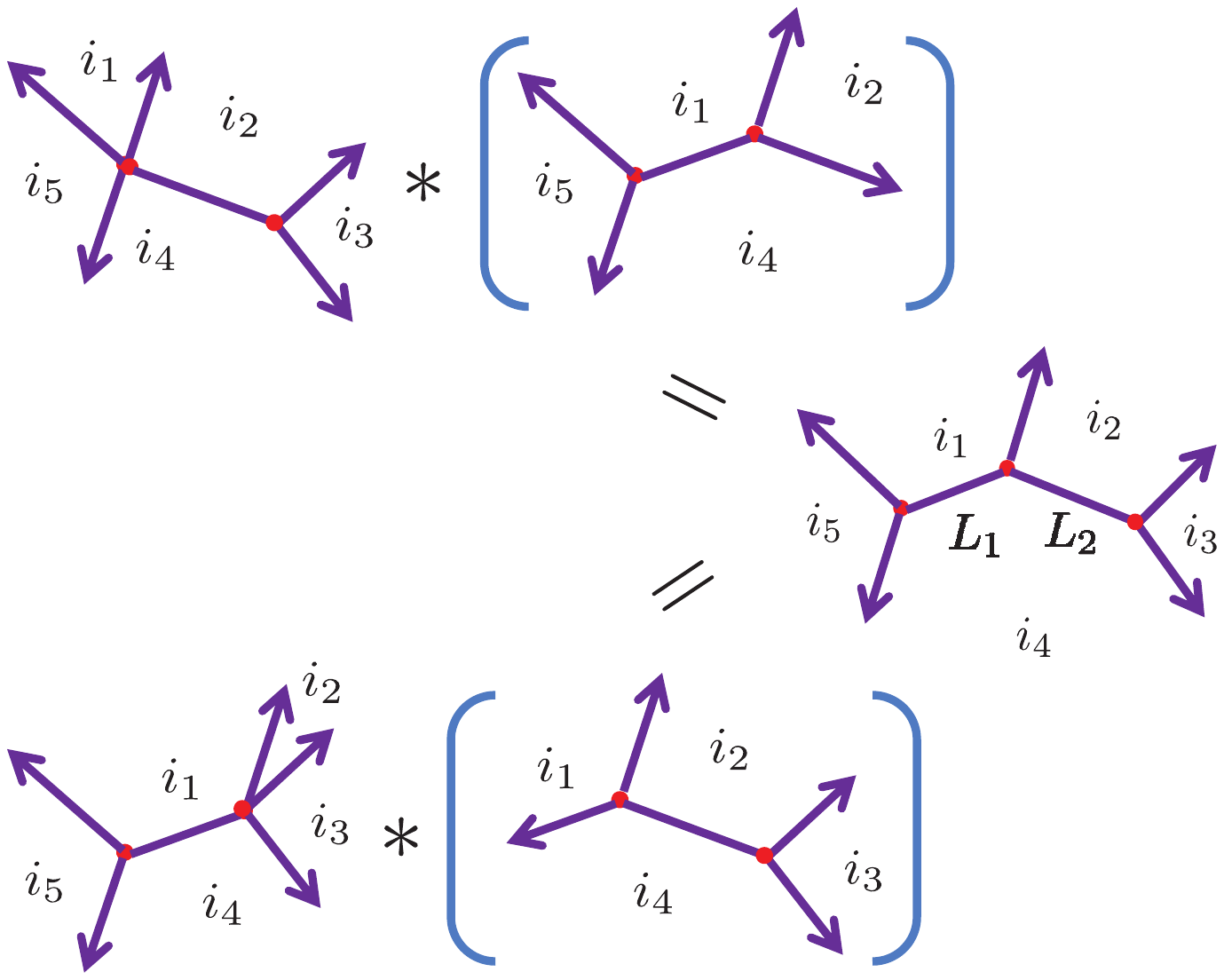}
\caption{The two boundaries of the deformation type of the sliding
web shown on the right correspond to different convolutions
shown above and below. If we use the lengths $L_1,L_2$
of the edges as coordinates then the orientation from the
top convolution is $dL_2 \wedge d L_1$. On the other
hand the orientation from the bottom convolution is
 $dL_1 \wedge d L_2$ and hence the sum of these two
 convolutions is zero. This is the key idea in the demonstration
 that $\ft*\ft=0$.   }
\label{fig:TAUT-SQUARE}
\end{figure}

\begin{figure}[htp]
\centering
\includegraphics[scale=0.5,angle=0,trim=0 0 0 0]{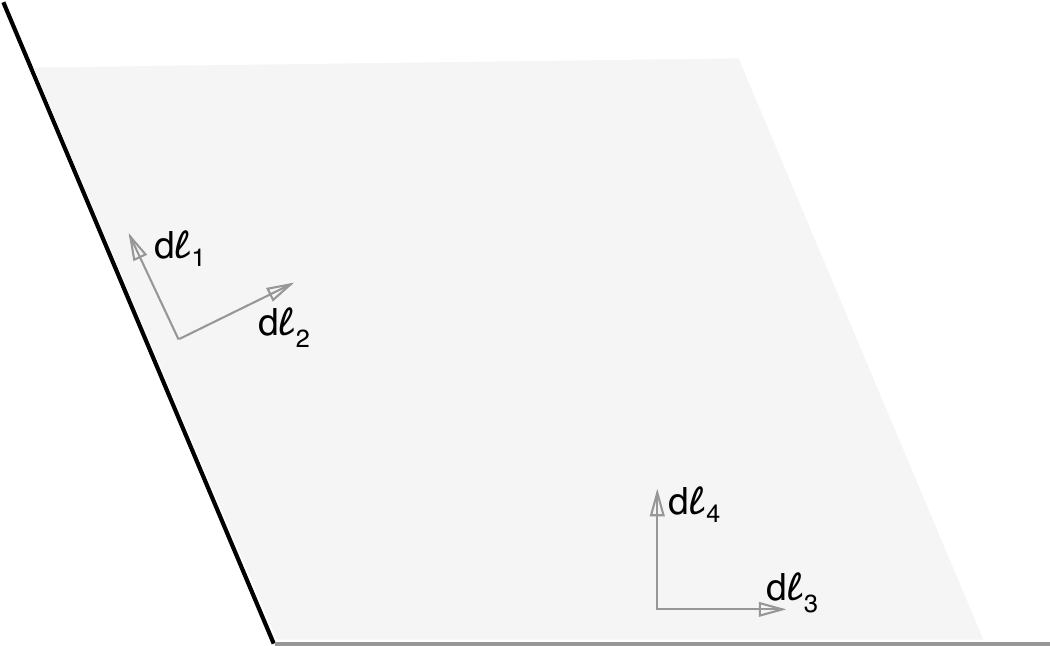}
\caption{A graphical proof that the two boundaries of the reduced moduli space of a sliding web are associated to opposite orientations. It is drawn as a cone since the
 moduli space of plane webs inherits a metric from the embedding into $\IR^{2V}$, and
with this metric the two boundaries are not orthogonal in general. }
\label{fig:TAUT-SQUAREBIS}
\end{figure}

\emph{Proof}:  Every element $\fw*\fw'$ in the convolution
is a sliding web, since reduced dimension is additive.
The reduced moduli space of a sliding web is a two-dimensional cone.
Up to a linear transformation it has boundary:
\be
\p \IR_+^2 =\left(  \IR_{>0} \times \{0\}\right)\amalg \left( \{0\} \times \IR_{>0}\right) \amalg \{ (0,0) \}.
\ee
Therefore, the terms can be grouped into pairs, each pair contributing
to the same deformation type. If the two boundaries are represented
 by convolutions of taut webs $\fw_1 * \fw_2$ and $\fw_3*\fw_4$ respectively,
 the corresponding orientations $d \ell_1 d\ell_2$ and $d \ell_3 d \ell_4$ of the reduced cell
 are opposite to each other. Thus
\be
\fw_1 * \fw_2 +  \fw_3*\fw_4 = 0.
\ee
This concludes the proof.  A concrete example illustrating the above
argument is shown in Figures \ref{fig:TAUT-SQUARE} and \ref{fig:TAUT-SQUAREBIS}.

%

%\textbf{Remark}: Although the convolution of plane webs is not associative, it is still true that
%for any web $\fw$
%\begin{equation}
%(\fw * \ft)*\ft=0
%\end{equation}
%Indeed, the difference $(\fw* \ft) * \ft - \fw*(\ft * \ft)$ consists of terms where two taut webs $\fw_1$ and $\fw_2$ are inserted separately
%at two vertices $v^1$ and $v^2$ of $\fw$. Each such terms appears twice in the sum, either from $(\fw*_{v^1} \fw_1) *_{v^2} \fw_2$
%or from $(\fw*_{v^2} \fw_2) *_{v^1} \fw_1$. The two contributions have opposite orientations and cancel out against each other.
%The nilpotent differential
%\begin{equation}
%Q_\ft : \fw \to \fw * \ft
%\end{equation}
%has a simple geometric interpretation: if we assemble all moduli spaces of webs with a given $I_\infty$ into a single cell complex,
%$Q_\ft$ becomes the adjoint of the standard boundary operator on the cell complex.

\subsection{Half-Plane Webs}

\textbf{Definition}:

a.) Let $\CH \subset \IR^2$ be a half-plane, whose boundary is not parallel to any of the $z_{ij}$.
A  half-plane web   in $\CH$ is
a graph in the half-plane,
which allows some vertices, but no edges, to be subsets of the boundary.
The boundary vertices have valence of at least one.
We apply the same rule as for plane webs: Label connected components
of the complement of the graph by vacua so that if the
 edges are oriented with $i$ on the left and
$j$ on the right then they  are parallel to $z_{ij}$.

b.) A   half-plane fan (often, we will just say, ``fan'')
 is an ordered sequence of vacua $\{i_1, \dots, i_n\}$ so
that the rays from the origin through   $z_{i_k, i_{k+1}}$ are ordered
clockwise for increasing $k$ and   $z_{i_k, i_{k+1}}\in \CH$.
\footnote{When we write $z_{ij}\in \CH$ for a
general half-plane $\CH$  we mean that if we rigidly translate $\CH$ to $\CH'$ so that
the origin is on its boundary then $z_{ij}\in \CH'$. We will use this slightly
sloppy notation again later in the paper.}

\textbf{Remarks}:

\begin{enumerate}

\item Unlike plane webs, half-plane webs need not be connected.

\item Let $\fu$ denote a typical half-plane web. There are now two different kinds of
vertices, the boundary vertices $\CV_\p(\fu)$ and the interior vertices
$\CV_i(\fu)$ with cardinalities  $V_\p(\fu)$ and $V_i(\fu)$, respectively.

\item
We will consider $\CV_\p(\fu)$ to be an \emph{ordered set} and we
will use a uniform ordering convention for all half-planes $\CH$ which
is invariant under rotation. To this end we choose a direction $\p_{\parallel}$ along $\p \CH$
so that if $\p_{\perp}$ is the outward normal to $\CH$ then $\p_\perp \wedge \p_{\parallel}  $ is
the standard orientation of the $(x,y)$ plane $\IR^2$, namely $\frac{\p}{\p x}\wedge \frac{\p}{\p y}$.
\footnote{Thus the mnemonic is ``Outward-Normal-First'' = ``One Never Forgets.''}
Now, our ordering of the
boundary vertices
\be\label{eq:bdry-order}
\CV_\p(\fu)= \{ v^\p_1,\dots, v^\p_n\}
\ee
is that reading from left to right proceeds in the direction of $\p_{\parallel}$.
In particular, if $\CH_L$ is the \emph{positive} half-plane $x \geq x_0$,  (with boundary
on the left) then  $v^\p_1,\dots, v^\p_n$ is a sequence of vertices with
decreasing ``time'' $y$, while for the \emph{negative} half-plane $\CH_R$, $x\leq x_0$, (with boundary on
the right) the sequence of vertices is in order of increasing time.

\item  We denote general half-plane fans by $J$, reserving $I$ for cyclic fans.
  We will denote
the half-plane fan at infinity by $J_{\infty}(\fu)$. Similarly, if $v \in \CV_\p(\fu)$
there is a half-plane fan $J_v(\fu)$.

\end{enumerate}

\begin{figure}[htp]
\centering
\includegraphics[scale=0.3,angle=0,trim=0 0 0 0]{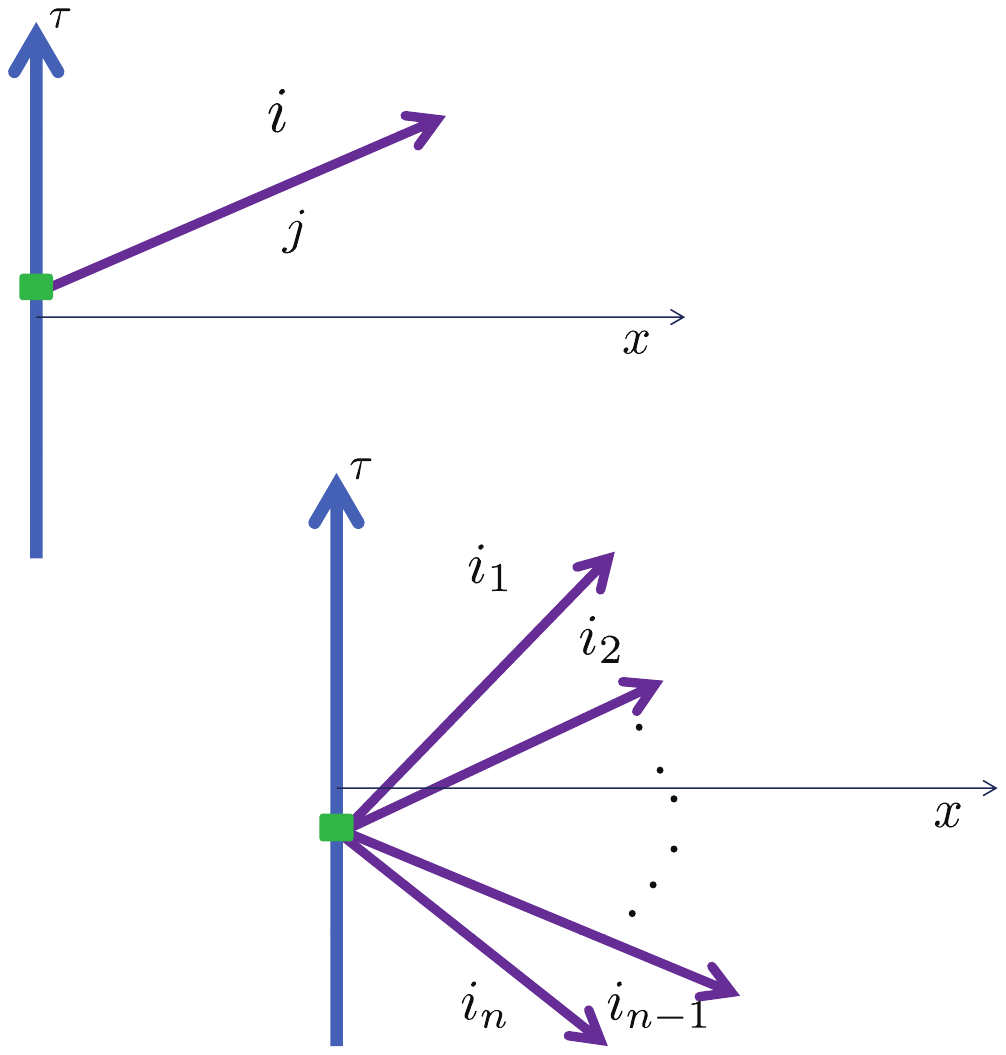}
\caption{Two examples of rigid positive-half-plane webs.   }
\label{fig:BNDRY-WEB-1}
\end{figure}

\begin{figure}[htp]
\centering
\includegraphics[scale=0.3,angle=0,trim=0 0 0 0]{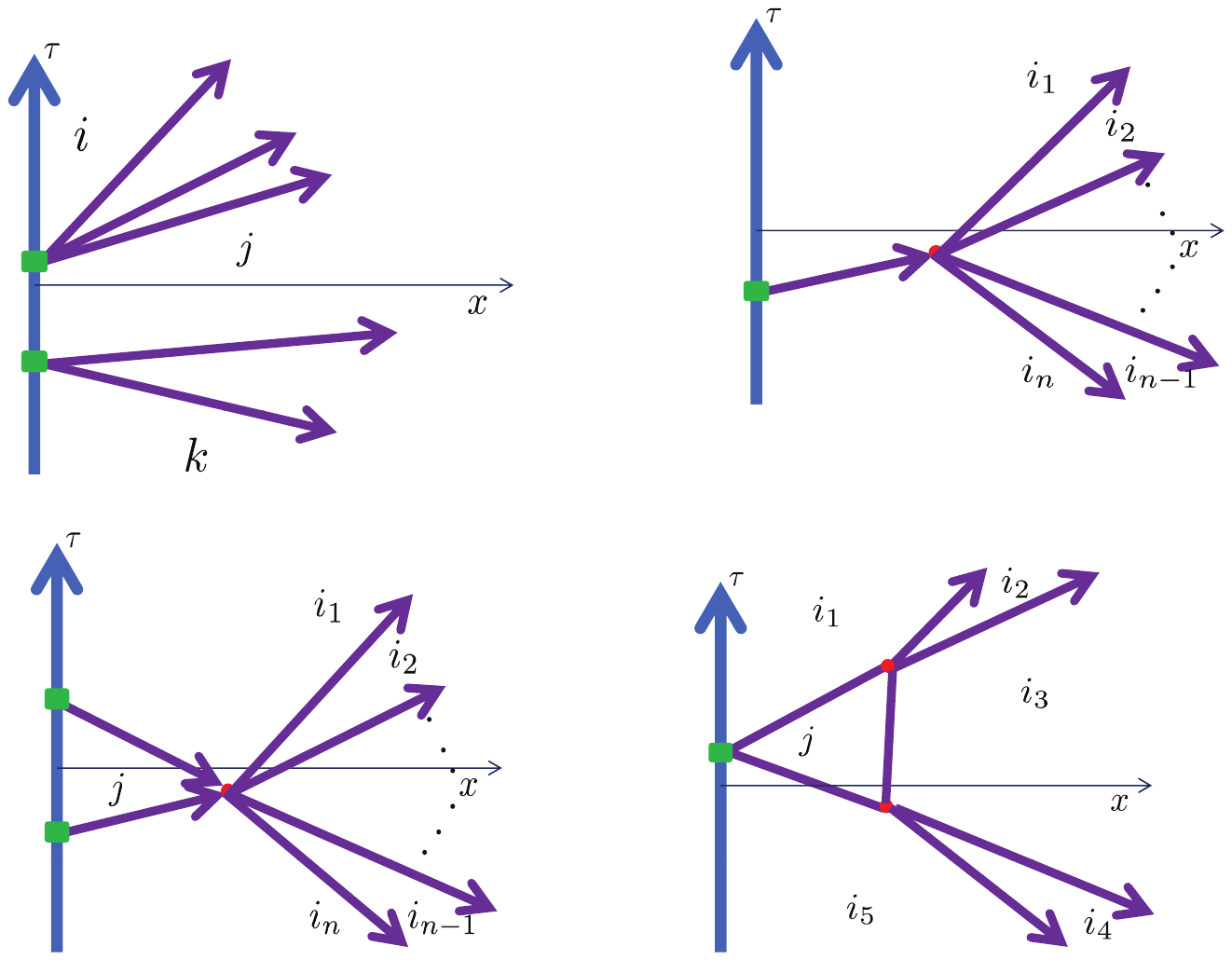}
\caption{Four examples of taut positive-half-plane webs   }
\label{fig:HALFPLANE-TAUTWEB}
\end{figure}

\begin{figure}[htp]
\centering
\includegraphics[scale=0.3,angle=0,trim=0 0 0 0]{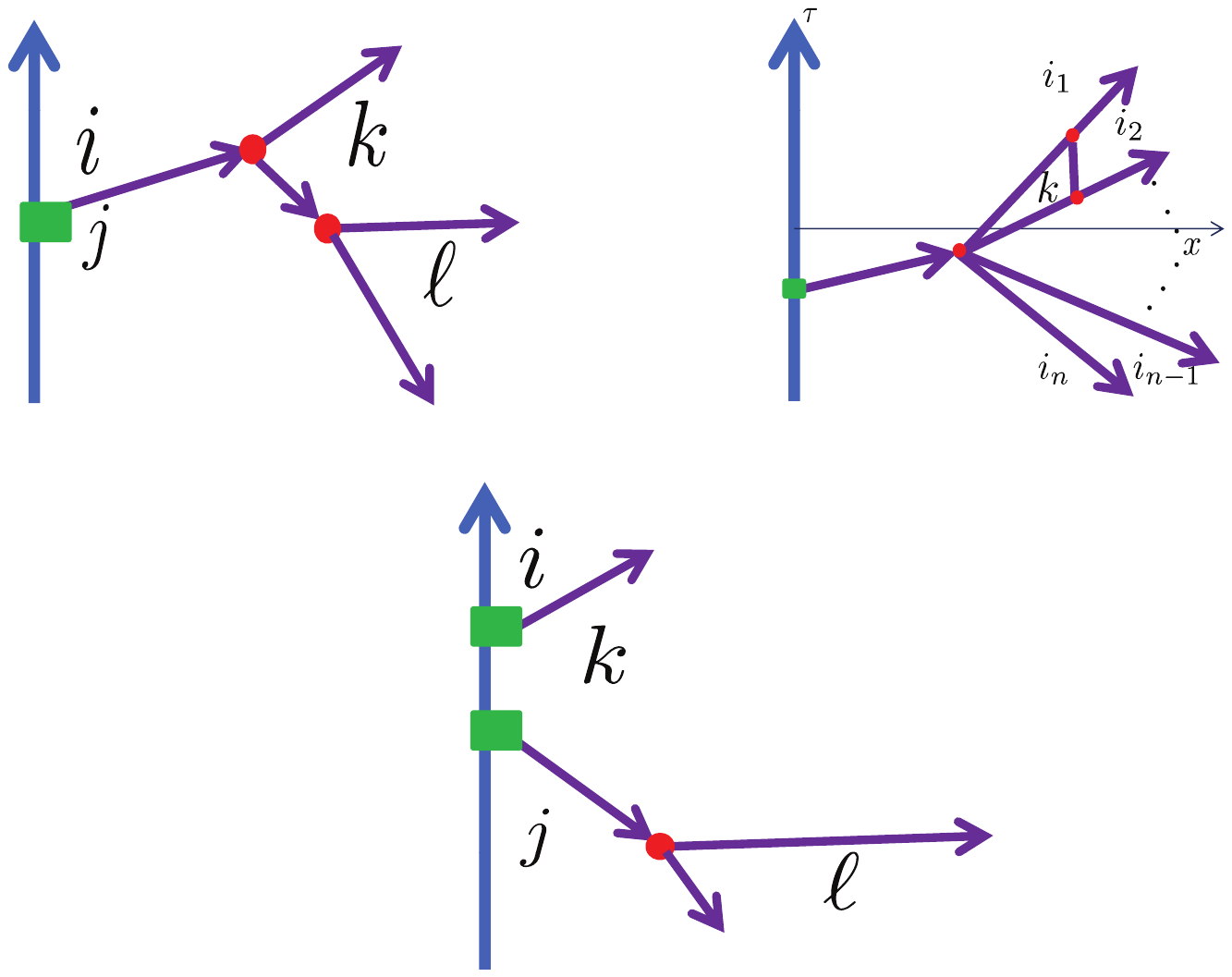}
\caption{Three examples of sliding half-plane webs   }
\label{fig:HALFPLANE-SLIDINGWEB}
\end{figure}

We can again speak of a deformation type of a half-plane web $\fu$.
The set of webs of a given deformation type is denoted  $\CD(\fu)$. It has
dimension:
\be
d(\fu) := 2 V_i(\fu) + V_\p(\fu) - E(\fu).
\ee
Now translations parallel to the boundary of $\CH$ act freely on
$\CD(\fu)$ and hence $d(\fu)\geq 1$.
Once again we define half-plane webs to be \emph{rigid, taut,} and \emph{sliding}
if $d(\fu) = 1,2,3$, respectively. Similarly, we can define oriented deformation type
in an obvious way
and consider the free abelian group $\CW_{\CH}$ of oriented deformation types
of half-plane webs in the
  half-plane $\CH$. Some examples where $\CH= \CH_L$ is the positive half-plane  are shown in Figures
\ref{fig:BNDRY-WEB-1}, \ref{fig:HALFPLANE-TAUTWEB}, and \ref{fig:HALFPLANE-SLIDINGWEB}.


We can again ask how a half plane web $\fu$ can degenerate near the boundary of the closure
$\bar \CD(\fu)$ in $\IR^{2V_i(\fu) + V_\partial(\fu)}$. We have now two types of boundary cells:
either the collapsing vertices of $\fu$ come together to a point in the interior of $\CH$ or they come together
to a point in the boundary $\CH$. Correspondingly, we can define two kinds of convolution.

 If $\fu$ and $\fu'$ are two half-plane
fans,  $v^\p \in \CV_\p(\fu)$, and $J_{v^\p}(\fu) = J_{\infty}(\fu')$  then
\be
\fu *_{v^\p} \fu'
\ee
is obtained by cutting out a small half-disk around $v^\p$ and gluing in
a small copy of $\fu'$. For this operation $V_i$ is additive as is $E$
but $V_\p(\fu *_{v^\p} \fu') = V_\p(\fu) + V_\p(\fu') -1$, so
the dimension behaves like:
\be
d(\fu *_{v^\p} \fu') = d(\fu) + d(\fu')-1.
\ee
We can extend $*_{v^\p}$ to an operation
\be\label{eq:bbstar}
*: \CW_{\CH} \times \CW_{\CH} \to \CW_{\CH}
\ee
by defining $\fu *_{v^\p} \fu'=0$ if $J_{v^\p}(\fu) \not= J_\infty(\fu')$
and then taking
\be
\fu * \fu' := \sum_{v^\p \in \CV_\p(\fu)} \fu *_{v^\p} \fu'.
\ee
Once again, at most one term in this sum can be nonzero.
To define the orientation of $\CD(\fu*\fu')$ we again
introduce a reduced orientation $o_r(\fu) := \iota(\p_{\parallel}) o(\fu)$
by contracting with the vector field $\p_{\parallel}$ described above
\eqref{eq:bdry-order}   and defining
\be
o(\fu*\fu') := o(\fu) \wedge o_r(\fu').
\ee

Similarly, if $v \in \CV_i(\fu)$ is an interior vertex  then we can convolve
with a plane-web $\fw$ to produce a deformation type
$\fu *_v \fw$ with orientation $o(\fu) \wedge o_r(\fw)$.   Now $V_\p$ and $E$ are additive but
$V_i(\fu *_v \fw) = V_i(\fu) + V(\fw) -1$ and hence we now have
\be
d(\fu *_{v} \fw) = d(\fu) + d(\fw)-2.
\ee
Again we can define
\be\label{eq:bistar}
*: \CW_{\CH} \times  \CW \to \CW_{\CH}
\ee
by defining $\fu *_{v} \fw =0$ if $I_{v }(\fu) \not= I_\infty(\fw)$
and then
\be
\fu * \fw = \sum_{v  \in \CV_i(\fu)} \fu *_{v } \fw.
\ee
Once again, by the line principle, at most one term in this
sum can be nonzero.

Thus all the boundary cells of $\bar \CD(\fu)$ are associated to some container half plane web
$\fu_1$ with either a marked interior vertex $v$ resolved to a small taut plane web $\fw_2$,
or a marked boundary vertex $v^\p$ resolved to a small taut half-plane web $\fu_2$.
We can write either $\fu = \fu_1 *_{v} \fw_2$ or $\fu = \fu_1 *_{v^\p} \fu_2$
at each boundary cell.

We now define the half-plane taut element
\be
\ft_{\CH} := \sum_{d(\fu)=2} \fu
\ee
There is one scale modulus $\ell$ so that as $\ell$ increases the
web gets bigger. The canonical orientation of taut elements is
then $dy_{\parallel} \wedge d\ell$ where $\p_{\parallel} = \frac{\p}{\p y_{\parallel}}$.
Since there are now two kinds of taut elements we henceforth denote the
planar taut element \eqref{eq:taut-planar} by $\ft_p$.
We now have:

\bigskip
\noindent
\textbf{Theorem:} Let $\ft_p$ be the taut element for planar
webs and $\ft_{\CH}$ the taut element for the half-plane $\CH$.
Then, combining the two convolutions \eqref{eq:bbstar} and
\eqref{eq:bistar}
\be\label{eq:hp-wb-id}
\ft_{\CH} * \ft_{\CH} + \ft_{\CH} * \ft_p =0.
\ee

\begin{figure}[htp]
\centering
\includegraphics[scale=0.3,angle=0,trim=0 0 0 0]{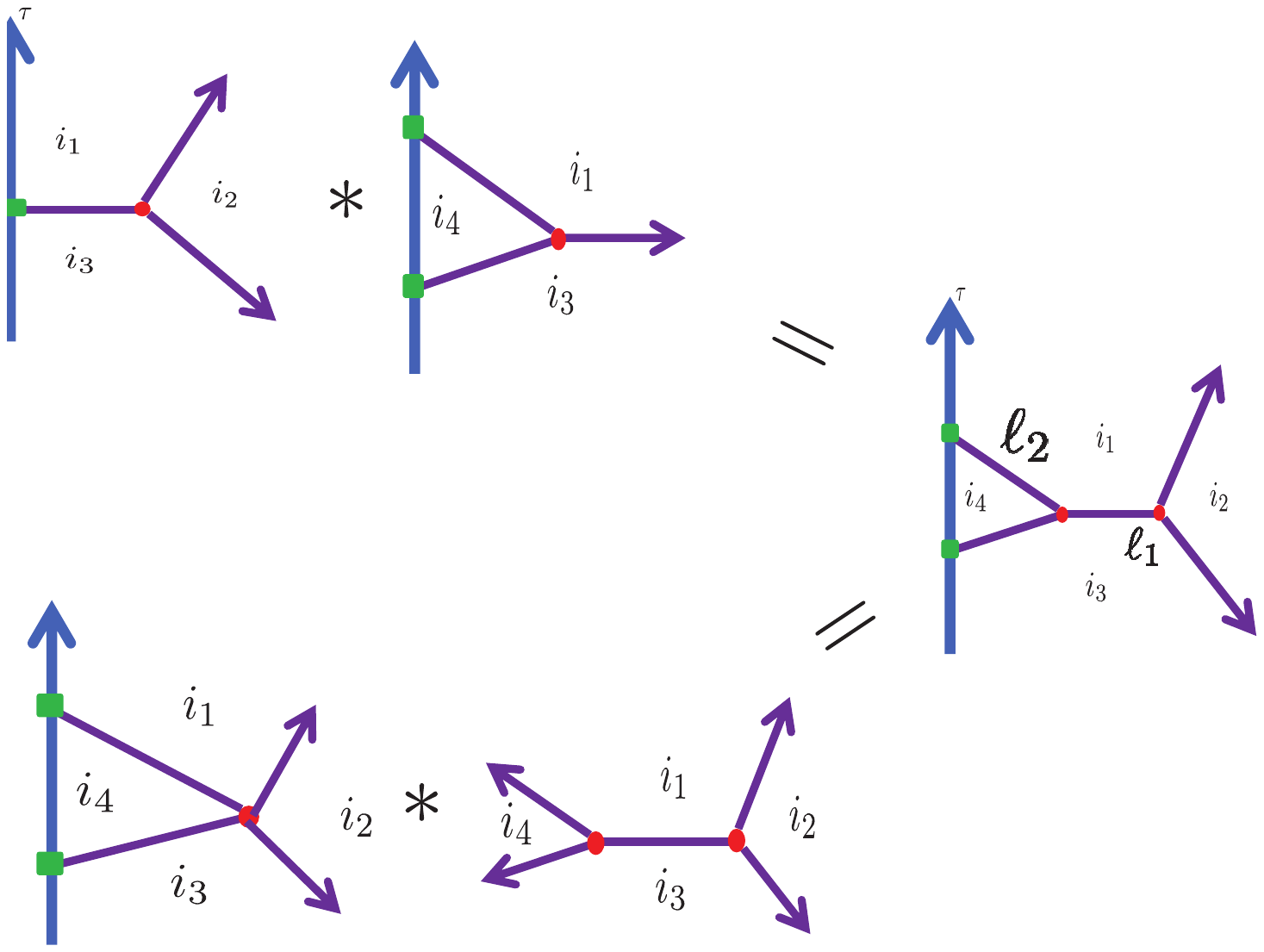}
\caption{An example of the identity on plane and half-plane
taut elements. On the right is a sliding half-plane web. Above is
a convolution of two taut half-plane webs with orientation
$dy \wedge d\ell_1 \wedge d\ell_2$. Below is a convolution
of a taut half-plane web with a taut plane web. The orientation is
$dy \wedge d\ell_2 \wedge d\ell_1$.  The two convolutions
determine the same deformation type but have opposite orientation, and
hence cancel. }
\label{fig:BLK-BDRY-WEBIDENT}
\end{figure}

\emph{Proof}: The idea of the proof is essentially the same
as in the proof of \eqref{eq:Taut-sq-zer}.  The reduced moduli spaces of sliding half-plane webs are still two-dimensional cones, and have paired boundary cells which induce opposite orientations. Thus all terms in \ref{eq:hp-wb-id} cancel out in pairs. An example of the argument is shown in Figure \ref{fig:BLK-BDRY-WEBIDENT}.

\bigskip
\noindent
\textbf{Remark}: Since half-plane webs are not connected one
might wonder whether we should introduce a new operation of
time-convolution in the identity \eqref{eq:hp-wb-id}. Certainly terms
appear which can be interpreted as time-convolutions but
these are properly accounted for by the first term
of \eqref{eq:hp-wb-id}. On the other hand,  will see in \S
\ref{subsec:Strip-Webs} below that when we replace the
half-plane by a strip we do need to introduce a separate
time-convolution operation.

%\textbf{Remark}: A few simple manipulations  We can define the nilpotent differential
%\begin{equation}
%Q_{\ft_p} : \fu \to \fu * \ft_p
%\end{equation}
%on half plane webs as we did for plane webs. The convolution of half webs
%is compatible with this differential: \footnote{This formula does {\it not} hold if any of the half-plane webs are replaced by plane webs, due to an extra term on the right hand side}
%\begin{equation}
%(\fu_1 * \fu_2) * \ft_p = \fu_1*(\fu_2 * \ft_p) + (-1)^{d(\fu_2)-1} (\fu_1 * \ft_p)*\fu_2
%\end{equation}
%A few more manipulations
%\begin{equation}
%(\fu * \ft_\CH)* \ft_\CH = \fu * (\ft_\CH* \ft_\CH) = - \fu * (\ft_\CH* \ft_p)= - (\fu * \ft_\CH)* \ft_p - (\fu * \ft_p)* \ft_\CH
%\end{equation}
%show us that
%\begin{equation}
%Q_{\ft_p} + Q_{\ft_\CH} : \fu \to \fu * \ft_p+ \fu * \ft_\CH
%\end{equation}
%is a nilpotent differential. Again, this is the adjoint to the boundary operation on a cell complex of all half-plane webs with a given $J_\infty$.

\subsection{Strip-Webs}\label{subsec:Strip-Webs}

We now consider webs in the strip $[x_{\ell}, x_r]\times \IR$.
Again, we assume that the boundary of the strip is not parallel to any of the $z_{ij}$.
Strip-webs are defined similarly to half-plane webs: We allow some vertices
but no edges to lie on the boundary of the strip.
Now there are two connected components of the boundary of the
strip so the boundary vertices are decomposed as a disjoint
union of two sets $\CV_{\p} = \CV_{\p,L} \amalg\CV_{\p,R}$.
Every strip web is associated to a certain choice of vacua in the far future
($y\to +\infty$) and in the far past ($y\to -\infty$). We can refer to them as the
future and past vacua of the strip-web, respectively.

Once again we can speak of deformation type.
We denote a generic strip-web, or rather an oriented deformation type,  by $\fs$.
The dimension of the space of strip-webs of fixed deformation type is
\be
d(\fs) := 2 V_i(\fs) + V_\p(\fs) - E(\fs).
\ee
Again time translation acts freely on the set $\CD(\fs)$ and hence
$d(\fs) \geq 1$.
As overall rescaling is not a symmetry of the problem, the moduli spaces $\CD(\fs)$ are not cones anymore.

\bigskip
\noindent
\textbf{Definition:} We  define
\emph{taut strip-webs} to be those with $d(\fs)=1$ and \emph{ sliding
strip-webs} to be those with $d(\fs)=2$. In other words
there is no distinction between rigid and taut strip-webs.

\bigskip
\noindent
\textbf{Remark}: The above definition might be surprising
since we did not introduce rigid strip webs. The source of
the distinction is the presence or lack of scaling symmetry.
When the geometry in which the webs live has a scaling symmetry,
such as the plane or half-plane, we distinguish between
rigid and taut webs. Otherwise rigid and taut webs are indistinguishable,
and have no reduced modulus.
 \bigskip

We can define as usual the closure $\bar \CD(\fs)$ of $\CD(\fs)$ in $\IR^{2 V_i(\fs) + V_\p(\fs)}$.
This introduces three kinds of boundary cells, corresponding to a collection of vertices collapsing to a point in the interior
or on either boundary
of the strip. Correspondingly, we have now three kinds of convolution. Recall that the ``container web'' is written
on the left. First, we can convolve
strip webs with planar webs so that
\be
d(\fs *_{v_i} \fw) = d(\fs) + d(\fw)-2
\ee
and $o(\fs*_{v_i}\fw)= o(\fs) \wedge o_r(\fw)$.
Next, we can convolve a strip web $\fs$ with
a positive
half-plane web with vertices on the left boundary as
\be
 \fs *_{v^\p} \fu_L
\ee
where $v^\p \in \CV_{\p,L}(\fs)$. Similarly, if $v^\p \in \CV_{\p,R}(\fs)$ and $\fu_R$ is a
web in the negative-half-plane (so it has boundary vertices on the right) then
we write
\be
 \fs *_{v^\p} \fu_R.
\ee
We have
\be
d(\fs *_{v^\p} \fu) = d(\fs) + d(\fu) - 1
\ee
with orientation  $o(\fs) \wedge o_r(\fu)$
in both cases. The
 reduced orientation is defined with the vector field $\p_{\parallel}$
defined in \eqref{eq:bdry-order}.

An important aspect in what follows is that we can introduce another
operation on strip-webs namely time concatenation: If $\fs_1$ and $\fs_2$
are two strip webs such that the future vacuum of $\fs_2$ coincides with the past vacuum of $\fs_1$ we define
\be
\fs_1 \circ \fs_2
\ee
to be the deformation type of a strip web where $\fs_1$ and $\fs_2$ are
disconnected and separated by a line at fixed time, with $\fs_1$ in the
future of $\fs_2$.
If the future vacuum of $\fs_2$ and the past vacuum of $\fs_1$ differ,
we define the concatenation $\fs_1 \circ \fs_2$ to be
zero. Note that when $\fs_1 \circ \fs_2$ is nonzero then
\be
d(\fs_1 \circ \fs_2) =  d(\fs_1) + d(\fs_2).
\ee

Because of the assumption that none of the $z_{ij}$ points in the direction along the strip,
no connected strip-web may have an arbitrarily large extension along the strip. The only webs which
can grow to arbitrarily large size have at least two disconnected components, and can thus be written as the
concatenation of simpler webs.

Proceeding as before we define the   free abelian group $\CW_{S}$ generated
by oriented deformation types of strip-webs. We extend $*$ in the usual
way to define operations $*: \CW_{S} \times \CW  \to \CW_{S}$
and $*: \CW_{S} \times \CW_{L,R}  \to \CW_{S}$
where $\CW_{L}$ is the group of positive-half-plane webs (with boundary on the left)
and $\CW_{R}$ is the group of negative-half-plane webs (with boundary on the right).
We introduce the taut element for the strip:
\be
\ft_s := \sum_{d(\fs)=1} \fs
\ee
with $\fs$ oriented towards the future. That is, choosing any boundary vertex $v^\p$ with
$y$-coordinate $y^\p $ the orientation is $o(\fs) = d y^\p$.

%\cg{The conventions on the orientation of $\p_{\parallel}$ for left or right boundaries do no appear to affect
%the convolution identity: the translation modulus of the taut half-plane webs is stripped off in the convolution
%and only the scaling modulus matters}
%

It is natural now to study the moduli spaces of sliding webs. There are two possible topologies:
the closure of the reduced moduli space
$\bar \CD_r(\fs)$ for a sliding web $\fs$ can be either a segment or a half-line.
\footnote{In the case of extended webs introduced below there are some exceptional cases where
the moduli space of sliding webs can be $\IR$. Nevertheless there are two ends with opposite
orientation and the convolution identity holds.}
At each of the boundaries
at finite distance
$\fs$ can be written as the convolution of an appropriate taut strip-web and a taut plane or half-plane web.
In all cases, the convolution gives an orientation pointing away from the boundary.
On the other hand, the semi-infinite end of a half-line moduli space is associated to a {\it concatenation} of two taut strip-webs.
If we denote the coordinates on the moduli space of two taut strip-webs $\fs_1$ and $\fs_2$ as $y_1$ and $y_2$,
the orientation of $\fs_1 \circ \fs_2$ is
\begin{equation}
o(\fs_1 \circ \fs_2) = dy_1 \wedge dy_2 = - dy_1 \wedge d(y_1 - y_2)
\end{equation}

In the conventions where $\fs_1$ in the future of $\fs_2$, $y_1 - y_2$ is the natural coordinate on the reduced moduli space, increasing towards
infinity. Thus the concatenation gives an orientation on the half line towards the origin.

%
%\begin{figure}[htp]
%\centering
%\includegraphics[scale=0.3,angle=0,trim=0 0 0 0]{STRIPWEB2-eps-converted-to.pdf}
%\caption{In this example with only three vacua there are no interesting
%interior webs, but we see the interplay between the various terms in
%the convolution identity, and in particular the need for the
%time-concatenation.  }
%\label{fig:STRIPWEB2}
%\end{figure}
%

Thus we have the

\bigskip
\noindent
\textbf{Theorem}: Let $\ft_p$ be the planar taut element and $\ft_L$ and $\ft_R$
the taut elements in the positive and negative half-planes, respectively,
 and $\ft_s$ the strip taut
element. Then
\be\label{eq:Strip-Web-Ident}
 \ft_s*\ft_L + \ft_s * \ft_R + \ft_s * \ft_p +\ft_s \circ \ft_s =0.
\ee
\begin{figure}[htp]
\centering
\includegraphics[scale=0.3,angle=0,trim=0 0 0 0]{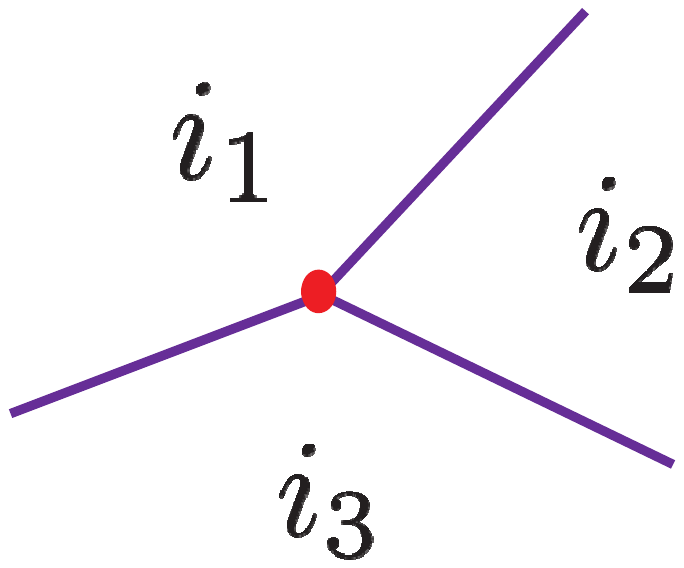}
\caption{The only vertex in a theory with three vacua. }
\label{fig:THREEVACUA-VERTEX}
\end{figure}
\begin{figure}[htp]
\centering
\includegraphics[scale=0.3,angle=0,trim=0 0 0 0]{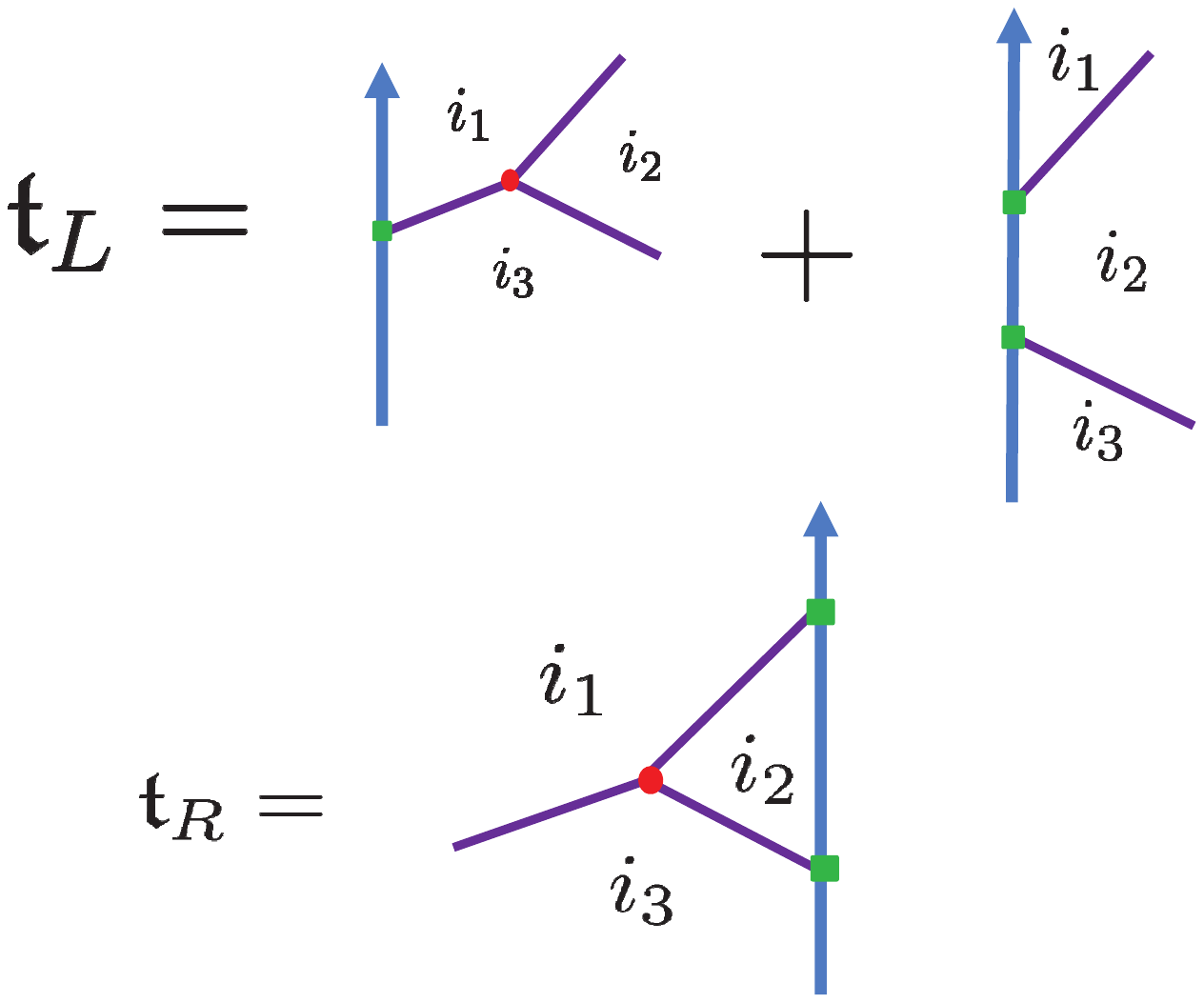}
\caption{The positive and negative half-plane taut elements are illustrated here. Letting $y$ denote
the $y$-coordinate of any boundary vertex and $\ell$ the internal edge length
$\ft_L$ has orientation $-dy d \ell$ and $\ft_R$ has orientation $dy d \ell$.  }
\label{fig:THREEVACUA-TAUTLR}
\end{figure}
\begin{figure}[htp]
\centering
\includegraphics[scale=0.3,angle=0,trim=0 0 0 0]{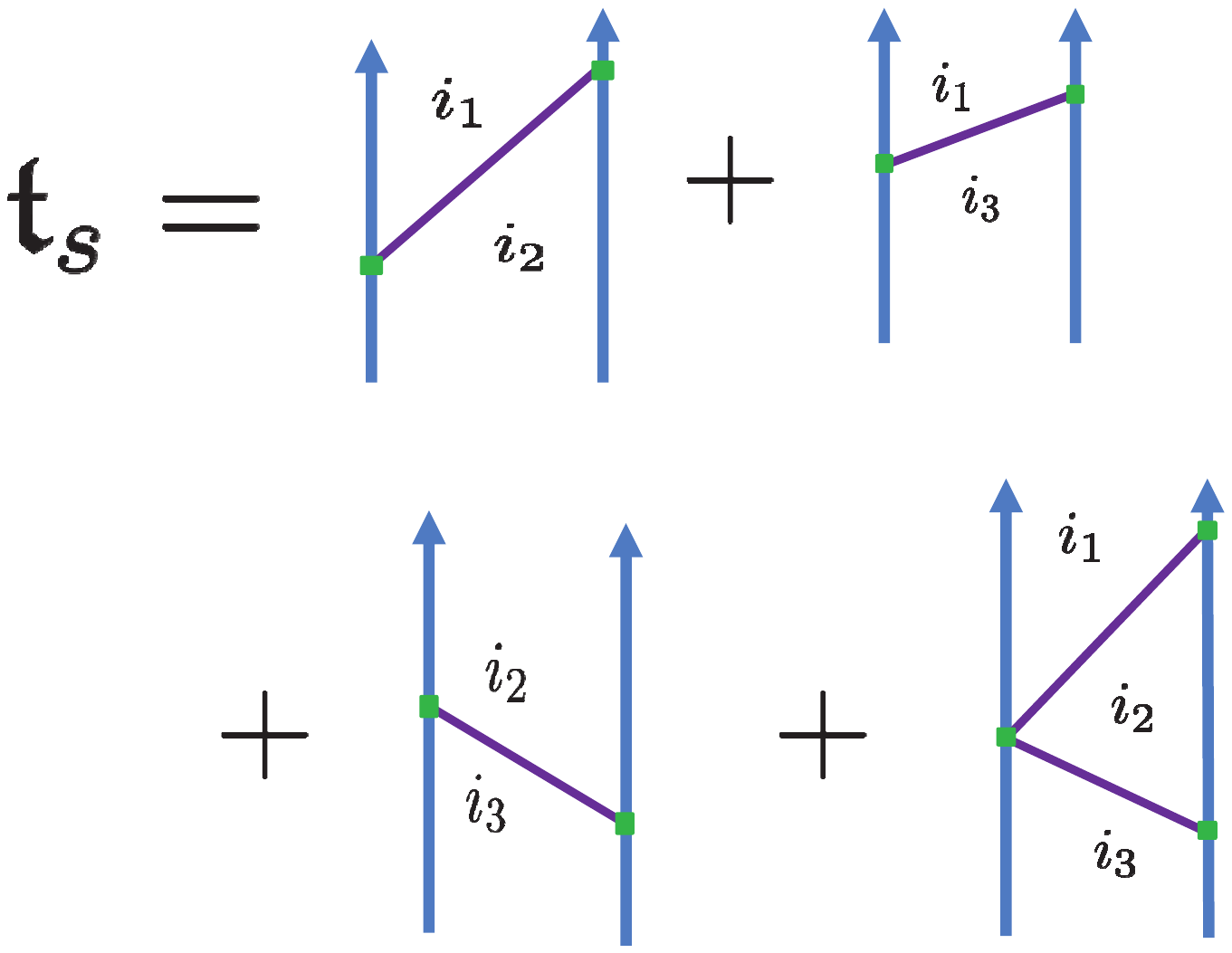}
\caption{The taut element on the strip with three vacua. Letting $y$ denote the
$y$ coordinate of any edge vertex the orientation is $dy$  }
\label{fig:THREEVACUA-TAUTSTRIP}
\end{figure}
\begin{figure}[htp]
\centering
\includegraphics[scale=0.3,angle=0,trim=0 0 0 0]{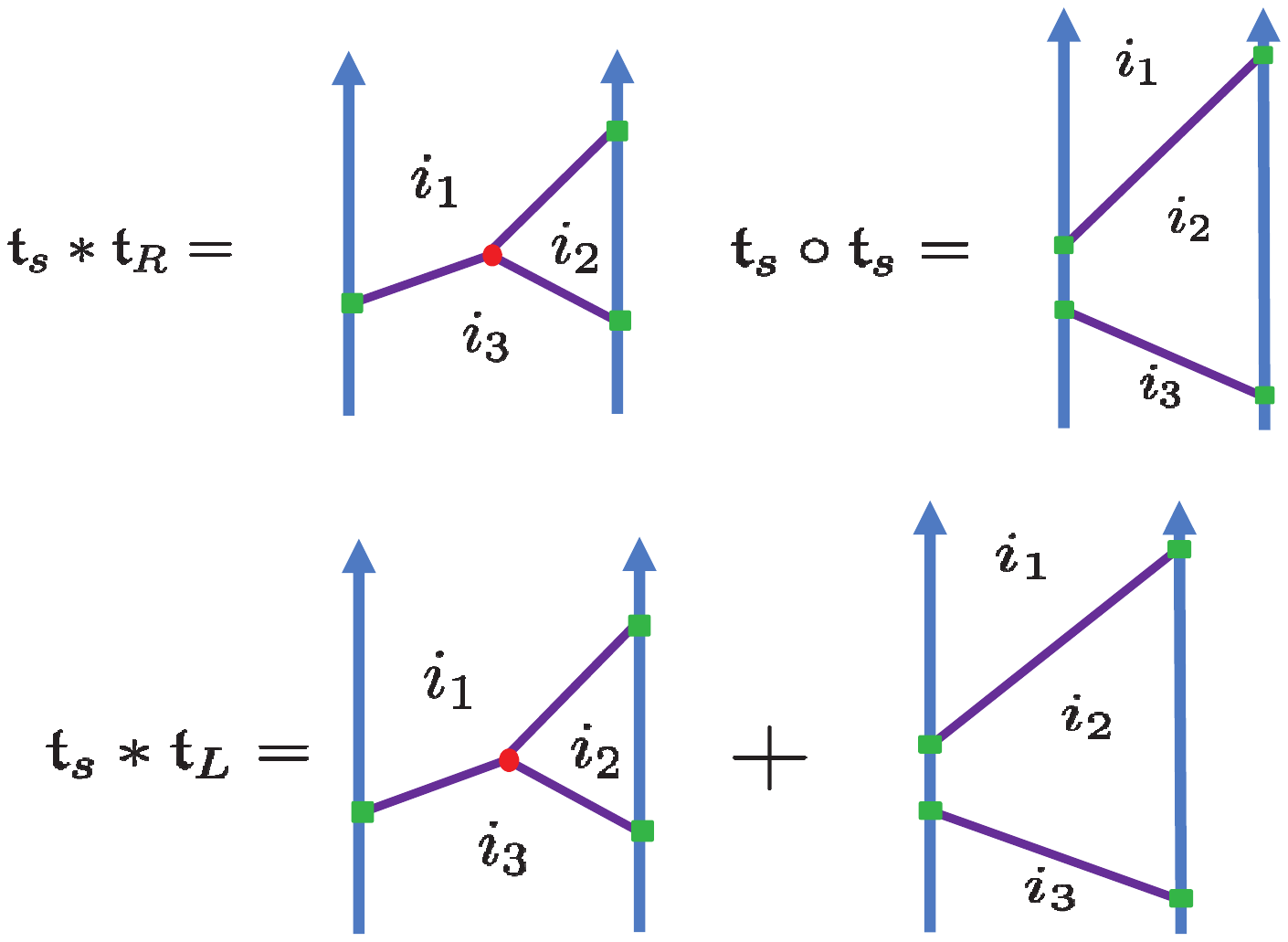}
\caption{The various terms in the convolution identity on the strip. In
this simple example with three vacua $\ft_s * \ft_p=0$. The orientations of the
three terms are $-dy d\ell$  on the
first line and $+dy d \ell$ on the second line.   }
\label{fig:THREEVACUA-CONVOLUTIONS}
\end{figure}

\bigskip
\noindent
\textbf{Example} Suppose there are three vacua in $\IV$. Then two
of the $\Re z_{ij}$ have the same sign, and without loss of generality
we will assume that $\Re z_{12} > 0 $ and $\Re z_{23} > 0 $. Thus
the only planar vertex is of the form shown in
Figure \ref{fig:THREEVACUA-VERTEX}.  The taut element
$\ft_L$ for the positive-half plane then has two summands
 while the taut element for the negative half-plane has a
single summand as shown in Figure \ref{fig:THREEVACUA-TAUTLR}.
The taut element on the strip is shown in Figure
\ref{fig:THREEVACUA-TAUTSTRIP}. The various convolutions
are illustrated in
\ref{fig:THREEVACUA-CONVOLUTIONS} and cancel.
%

%\textbf{Remark}: We could have defined the time-concatenation map $\circ$ already in
%the half-plane case but it was unnecessary since all half-plane webs which are sliding
%can already be considered as convolutions of taut half-plane webs. However, in the
%case of the strip, sliding strip-webs which arise from time-concatenation  cannot
%be reinterpreted as convolutions of taut or rigid strip webs. Hence the
%fourth term in \eqref{eq:Strip-Web-Ident} is necessary.
%\cg{ I dislike this remark. Two typical half plane webs cannot be concatenated, even if the vacua agree. Concatenation is simply not a natural operation for half-plane webs}

%\textbf{Remark}: We can define the nilpotent differential
%\begin{equation}
%Q_{\ft_p} : \fs \to \fs * \ft_p
%\end{equation}
%on strip-webs as well. Furthermore, the convolution of strip-webs and half webs
%is compatible with this differential:
%\begin{equation}
%(\fs * \fu) * \ft_p = \fs*(\fu * \ft_p) + (-1)^{d(\fu)-1} (\fs * \ft_p)*\fu
%\end{equation}
%and so is the concatenation of strip-webs.
%We can also define anti-commuting differentials
%\begin{equation}
%Q_{\ft_{L,R}} : \fs \to \fs * \ft_{L,R}
%\end{equation}
%such that $Q_{\ft_p} + Q_{\ft_{L}} + Q_{\ft_{R}}$ is nilpotent, adjoint to the
%boundary operator on the cell complex of strip-webs. Operators $t_s \circ$ and $\circ t_s$
%can be added to include the contributions from boundaries at infinity with a taut element in the future or in the past.

\subsection{Extended Webs}\label{subsec:ExtendedWebs}

A small generalization of the webs defined above will turn out to play a role below.
In some circumstances it will be useful to relax slightly the restriction that interior vertices should be at least tri-valent and boundary vertices at least univalent.
In particular, we will consider the following two generalizations: First,
we can allow two-valent interior vertices and second, we can allow zero-valent
boundary vertices. These generalizations weaken somewhat the finiteness properties of webs.
Nevertheless, the number of webs of a given degree $d$  is still finite since the addition of such
vertices increases $d$ by one. This is  sufficient to keep our formulae sensible.

We will refer to this larger class of webs as \emph{extended webs} when the distinction is important.
Extended webs satisfy most of the same properties as standard webs.
The definitions and properties of deformation types, orientation and convolution will all hold true for extended webs.
The taut elements for the extended webs satisfy the same convolution identities as the taut elements for standard webs.
The whole algebraic structure defined in the next section \ref{sec:AlgebraicStructures} persists as well if we consider extended webs.

\subsection{Special Configurations Of Vacuum Weights}\label{subsec:SpecialVacWt}

At several times in our discussion above we required the set of vacuum weights
$\{ z_i \}$ to be in general position. Nevertheless, there are special configurations
of weights which will be of some importance in the discussion of certain homotopies
in Section \S \ref{subsec:CompThrIntfc} and in the discussion of
wall-crossing in Section \S \ref{sec:GeneralParameter} below.

We can consider $\{ z_i \}$ to define a point in $\IC^{  \IV  } - \Delta$,
where $\IC^{  \IV  }$ is the space of maps $\IV \to \IC$ and
$\Delta$ is the large diagonal where $z_i = z_j$ for some pair $i\not=j$.
Within this space are two subspaces of special weights. They are generically
of real codimension one but have complicated self-intersections of higher codimension.

The first special codimension one subspace is defined by weights such that  some triple of
weights $z_i,z_j, z_k$ for three distinct vacua $i,j,k$ become colinear:
\be\label{eq:MS-WALL}
\Im z_{ij} \overline{z_{jk}} =0.
\ee
We call these   \textit{walls of marginal stability}.
Generic one-parameter families of weights will cross such walls. When this happens
the set of cyclic vacua and the set of webs changes discontinuously. For example,
with four vacua we can pass from a set of webs which are all tree graphs to a set
of webs with loops. We will discuss some consequences of such wall-crossings in
Section \S \ref{sec:GeneralParameter}, and especially in Section \S \ref{subsec:MS-WC} below.

\begin{figure}[htp]
\centering
\includegraphics[scale=0.3,angle=0,trim=0 0 0 0]{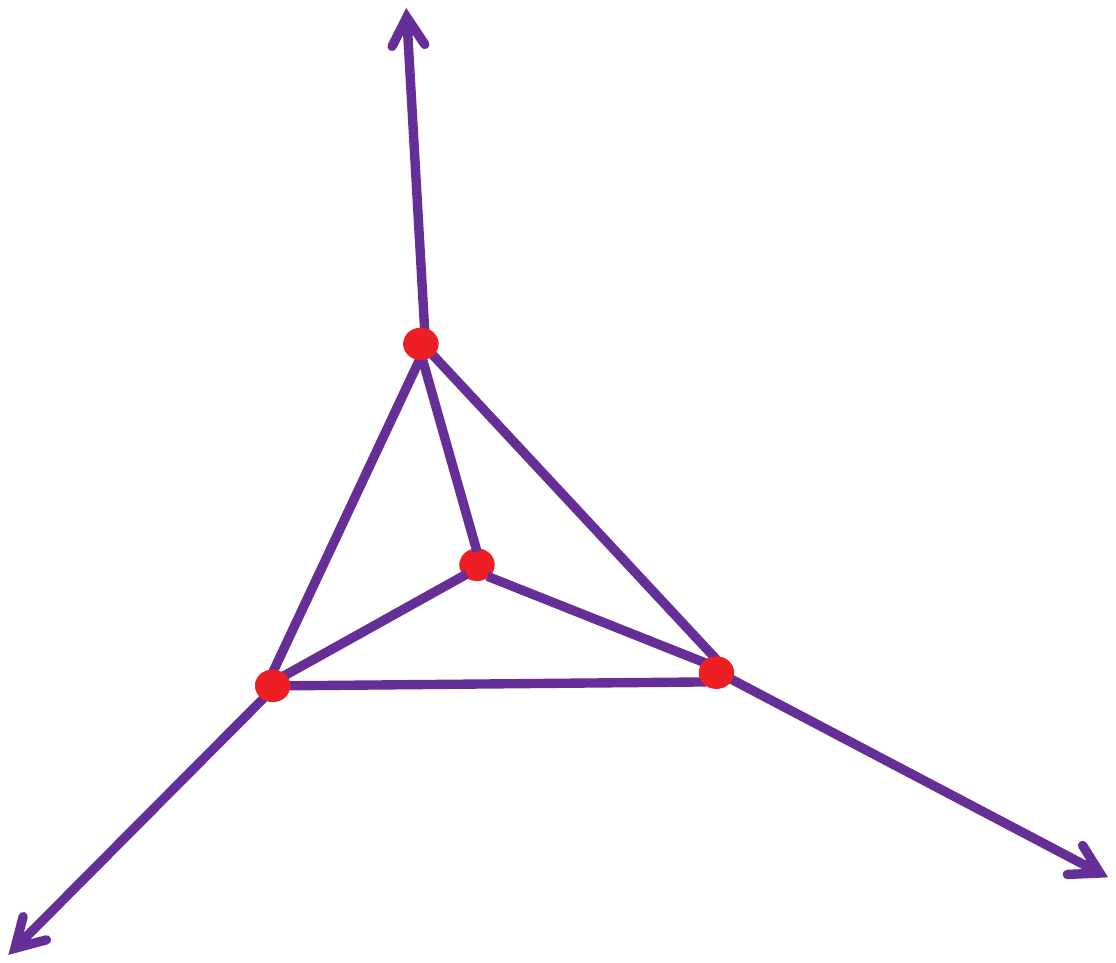}
\caption{An example of an exceptional web. There is only one reduced
modulus corresponding to overall scaling. Therefore $\dim \CD(\fw)=3$. Nevertheless
$V=4$ and $E=6$ so $d(\fw)= 2V-E = 2$.   }
\label{fig:EXCEPTIONALWEB}
\end{figure}

A  more subtle special configuration of weights is one for which there
exist \emph{exceptional webs}. These are, by definition, webs such that
\be\label{eq:DefExcWeb}
D(\fw):=\dim \CD(\fw) > d(\fw).
\ee
Such webs can arise because, for some configurations of vacuum weights
there can be webs where the edge constraints are not all independent.
We say that some edge constraints are ineffective. Let us decompose $z_{ij}$
into real and imaginary parts $z_{ij} = u_{ij} + \I v_{ij}$. If an edge $e$
is of type $ij$ and  has vertices $(x^{(1)}_{e}, y^{(1)}_{e} )$ and  $(x^{(2)}_{e}, y^{(2)}_{e} )$
then the edge constraints are a set of linear equations
\be
u_{ij(e)} (y^{(2)}_{e} - y^{(1)}_{e}) - v_{ij(e)} (x^{(2)}_{e} - x^{(1)}_{e})  = 0
\ee
which we can write as $M(\fw) L(\fw)  =0$ for a matrix $M(\fw)$ of edge constraints and a vector $L(\fw)$ of vertex coordinates. The real codimension
one \emph{exceptional walls} in $\IC^{\IV}-\Delta$ are defined by the loci where the rank
of $M(\fw)$  drops from $E(\fw)$ to $E(\fw)-1$. Of course, the closure of the exceptional walls will have many
components, intersecting in places where the rank drops further.

An example of an exceptional web is shown in Figure \ref{fig:EXCEPTIONALWEB}. This is plainly a
taut web, so the dimension of is deformation space is three, but $d(\fw)=2$!
When such phenomena arise we will distinguish the true dimension
$D(\fw)$ from $d(\fw)$ by calling the latter the \emph{expected dimension}.
We use the terminology of index theory because, as we will see in
Section \S \ref{zetawebs}, this literally does correspond to an issue in
index theory.

The example  shown in Figure \ref{fig:EXCEPTIONALWEB} requires at least six
vacua. If we hold all but one fixed and vary the last then it is clear that
any small perturbation will destroy the web. However, a generic one-parameter
family of weights nearby this configuration will have a point admitting
such an exceptional web. Further triangulating one of the triangles in
Figure \ref{fig:EXCEPTIONALWEB} reduces the expected dimension by $1$ and in
this way  we can produce examples
of exceptional webs with arbitrarily small and negative expected dimension.
In general, if a deformation type $\fw$ is exceptional so
that $\nu:= D(\fw) - d(\fw) >0$ then generic $\nu$-parameter families
of weights $\{ z_i \}$ will intersect the loci of such webs. We will discuss the
consequences of exceptional walls in the wall-crossing story in Section \S
\ref{subsec:X-WEB-WC}.

We can now be more precise about the meaning of ``generic weights'' or
``general position'' used  both above and below.
This term implies that the weights are not on walls
of marginal stability and do not admit exceptional webs.

%
%\cg{Perhaps say special things that happen on half-planes and strips if needed.
%We do need $\fe_{\p}$ in Section \ref{sec:GeneralParameter}. }
%

%
\begin{figure}[htp]
\centering
\includegraphics[scale=0.3,angle=0,trim=0 0 0 0]{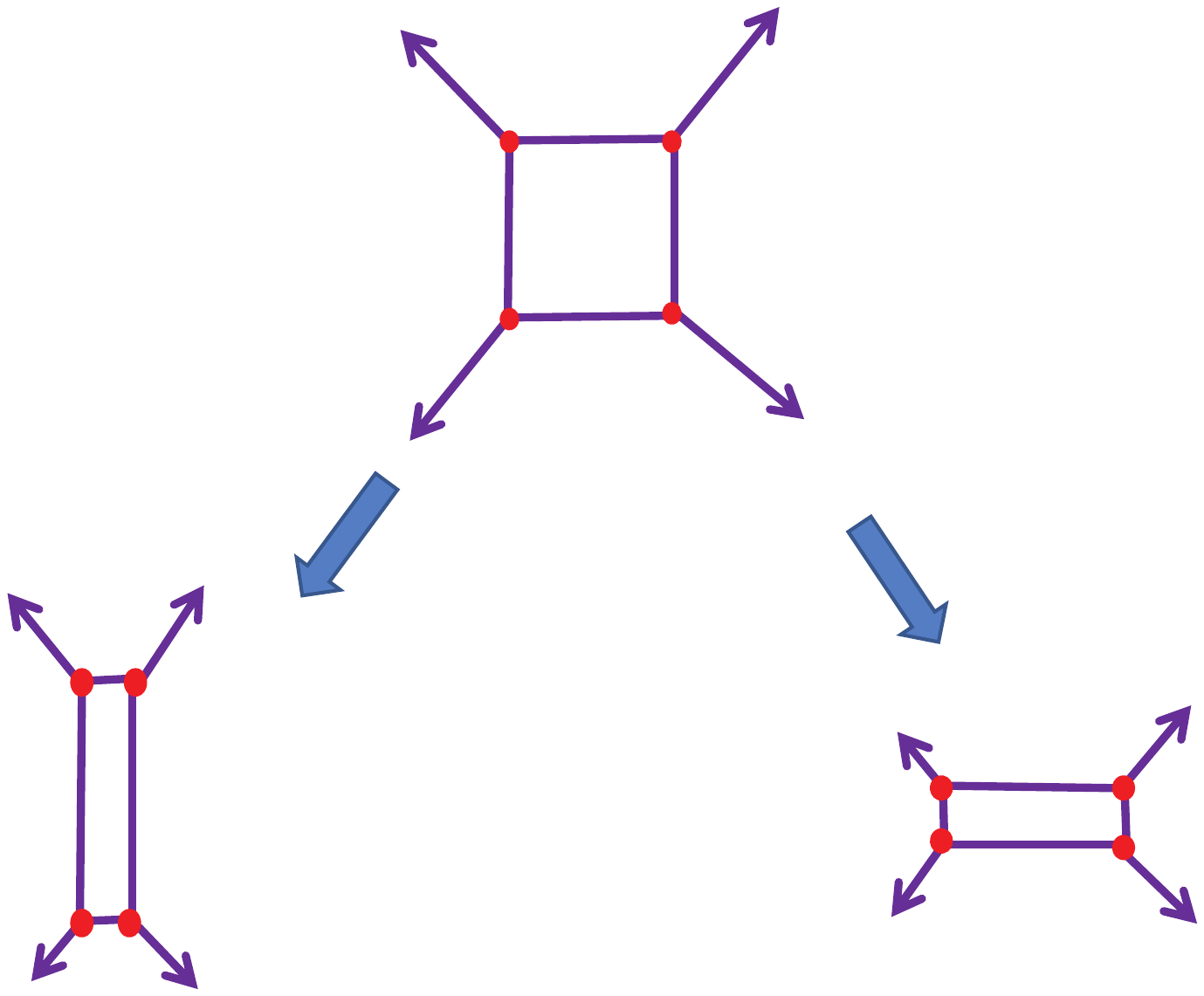}
\caption{The sliding web shown here has two degenerations, neither of which is
of the form $\fw_1*\fw_2$.   }
\label{fig:ULTRAEXCEPTIONAL}
\end{figure}

\textbf{Remarks}

\begin{enumerate}

\item  Finally, we remark that there are certain high codimension
configurations of weights where webs can degenerate in ways which are not
described in terms of convolution at a single vertex. An example is shown in
Figure \ref{fig:ULTRAEXCEPTIONAL}.

\item There are also exceptional vacuum configurations and webs
for the half-plane and strip geometries,  and these will
be used in   Section \S \ref{sec:GeneralParameter}.

\end{enumerate}

\section{Tensor Algebras Of Webs And Homotopical Algebra}\label{sec:AlgebraicStructures}

In this section we consider algebraic operations defined by webs on various tensor algebras
such as
\footnote{We take the tensor algebra without a unit, i.e.
we do not include the ground ring $\IZ$.}
\be\label{eq:TensorCW}
T\CW  := \CW \oplus \CW^{ \otimes 2}   \oplus \CW^{\otimes 3}  \oplus \cdots
\ee
and its analogs for $\CW_{\CH}$ and $\CW_{S}$.
Here we   take the graded tensor product using the Koszul rule. The
grading of a web such as $\fw,\fu,\fs$ will be given by the dimension
$d(\fw), d(\fu), d(\fs)$, respectively. We will find various algebraic
structures familiar from applications of homotopical algebra to string field
theory. While these algebraic structures emerge naturally from thinking
about webs the reader should be aware that the $L_\infty$ and $A_\infty$ algebras
which are used to make contact with the physics only make their
appearance when we come to Section \S \ref{sec:RepWeb}.

\subsection{$L_\infty$ And Plane Webs}

The convolution of webs  is not associative:
$(\fw_1*\fw_2)*\fw_3 - \fw_1*(\fw_2 * \fw_3)$ consists of terms where $\fw_2$ and $\fw_3$ are are glued in at distinct
vertices of $\fw_1$. One could readily write down a tower of associativity relations for
some generalized convolution operations, which insert multiple webs at distinct vertices of a single container web.
It turns out that for our applications we only need   an operation
\begin{equation}\label{eq:TensorT-def}
T(\fw): \CW^{\otimes V(\fw)} \to \CW
\end{equation}
which replaces {\it all} the vertices of a plane web with other webs. More precisely, we
define $T(\fw)$ as follows:

Given an \emph{ordered} collection of $\ell=V(\fw)$ plane webs $\{ \fw_1, \dots, \fw_\ell \}$, we seek some permutation
$\sigma$ of  an   ordered set of vertices $\{ v_1,\dots, v_\ell\}$ of $\fw$ (with any ordering),
such that $I_{v_{\sigma(a)} }(\fw) = I_\infty(\fw_a)$ for $a=1,\dots, \ell$.
If the permutation does not exist, i.e. if
the arguments cannot be inserted into $\fw$ (saturating all the vertices exactly once) then we set $T(\fw)[\fw_1, \dots, \fw_\ell]=0$.
If the permutation exists, it is unique, since a given cyclic fan of vacua can appear at most once in $\fw$. We then define
$T(\fw)[\fw_1, \dots, \fw_\ell]$ to be the oriented deformation type obtained by gluing in $\fw_a$ in small disks cut out around the
vertices $v_{\sigma(a)}$ of $\fw$. The orientation is given by $o(\fw) \wedge o_r(\fw_1) \wedge \cdots \wedge o_r(\fw_\ell)$.
This is the only place the ordering of  $\{ \fw_1, \dots, \fw_\ell \}$ is used. In particular, $T(\fw)$ is graded symmetric,
exactly as we would expect from manipulating graded elements $\fw_a$ of degree $d(\fw_a)$ with the Koszul rule.
(Since $d$ and $d_r$ differ by two the sign is the same.) Now, we regard  $\{\fw_1, \dots, \fw_\ell \}$ as a
monomial in $\CW^{\otimes \ell}$ and extend by linearity to define \eqref{eq:TensorT-def}.
Finally, we can  extend $T(\fw)$ to a map $T(\fw): T\CW \to \CW$,
by setting $T(\fw): \CW^{\otimes n} \to \CW$ to be zero unless $n = V(\fw)$.

It is useful to recall at this stage
the definition of \emph{$n$-shuffles}. If $S$ is an
ordered set then an $n$-shuffle of $S$ is an ordered disjoint
  decomposition into $n$ ordered subsets
\be
S = S_1 \amalg S_2 \amalg \cdots \amalg S_n
\ee
where the ordering of each summand $S_\alpha$ is inherited
from the ordering of $S$ and the $S_\alpha$ are allowed
to be empty. Note that the ordering of the
sets $S_\alpha$ also matters so that $S_1\amalg S_2$
and $S_2 \amalg S_1$ are distinct 2-shuffles of $S$.
For an ordered set $S$ we let ${\rm Sh}_n(S)$ denote
the set of distinct $n$-shuffles of $S$.
We can count $n$-shuffles by successively asking
each element of $S$  which set $S_\alpha$ it belongs to.
Hence there are  $n^{\vert S\vert}$ such shuffles.

We are now ready to formulate a useful compatibility relation between the $*$ and $T$ operations:
\be\label{eq:T-and-star}
T(\fw * \fw')[\fw_1, \dots, \fw_{n} ] =
\sum_{{\rm Sh}_2(S)} \epsilon ~ T(\fw) [ T(\fw')[S_1], S_2]
\ee
where we sum over 2-shuffles $S= S_1 \amalg S_2$ of the ordered
set $S=\{ \fw_1, \dots, \fw_{n}  \}$ and we understand that
$T(\fw)[\emptyset] = 0$. The sign $\epsilon$ in the sum keeps track
of the web orientations, and it is determined as follows.  We let
 $o_r(S_\alpha)$ be the ordered product of reduced orientations of the
$ \fw_i$ in $S_{\alpha}$ and define
\be
\epsilon = \frac{o_r(S_1) \wedge o_r(S_2) }{o_r(\fw_1) \wedge \cdots \wedge o_r(\fw_n)}:= \epsilon_{S_1,S_2}.
\ee
exactly as we would expect from manipulating graded elements $\fw$ of
degree $d(\fw)$ with the Koszul rule.

% That is
%we sum over ordered disjoint decompositions
%
%\be
%\{ \fw_1, \dots, \fw_{n}  \} = S_1 \amalg S_2
%\ee
%
%where each of $S_1$ and $S_2$ retains the ordering
%on the LHS.

Next we extend the map $\fw \to T(\fw)$ to be a linear map by  setting
 $T(\fw_1 + \fw_2) := T(\fw_1) + T(\fw_2)$. It now makes sense to speak of
   $T[\ft]$, which will play a particularly important role for us.
The relation \eqref{eq:T-and-star} is bilinear in $\fw$ and
$\fw'$. Summing $\fw$ and $\fw'$ separately
over taut elements and applying   $\ft*\ft=0$ it follows
that, for any ordered set $S$ (i.e. for any  monomial in $T\CW$):
\be\label{eq:L-infty-interior}
\sum_{{\rm Sh}_2(S) } \epsilon_{S_1,S_2} ~ T(\ft) [ T(\ft)[S_1], S_2] = 0.
\ee

We can interpret these relations as defining a version of
an ``$L_\infty$ algebra.'' To make this clear and to
lighten the notation let us denote by $b_n$ (``closed brackets'') the
restriction of $T(\ft)$ to   $\CW^{\otimes n}$,
so
\be
b_n: \CW^{\otimes n} \rightarrow \CW
\ee
has degree $\deg(b_n) = 3 -2n$ and satisfies the identities
\be\label{eq:Linf1}
\sum_{{\rm Sh}_2(S)} \epsilon_{S_1,S_2} b_{n_2}(b_{n_1}(S_1), S_2) = 0
\ee
where $n_i= \vert S_i\vert$, $i=1,2$.

\bigskip
\noindent
\textbf{Remarks}:

\begin{enumerate}

\item The degrees and associativity relations \eqref{eq:Linf1} coincide with the notion of $L_\infty$ algebra which appears in other areas
of physics, such as closed-string field theory \cite{Zwiebach:1992ie,Gaberdiel:1997ia}.
 The degrees and signs used in the mathematical literature are slightly different. Our definitions
are known as the $L_\infty[-1]$ relations. (For a relation of these relations to the more standard
$L_\infty$ relations see \cite{MehtaZambon}.)

\item It is worth noting that $(\fw_1*\fw_2)*\fw_3 - \fw_1*(\fw_2 * \fw_3)$
can be interpreted as webs with $\fw_2$ and $\fw_3$ inserted into
distinct vertices of $\fw_1$ and is therefore graded symmetric in $\fw_2$ and
$\fw_3$. This is precisely the definition of a (graded) ``pre-Lie-algebra,'' thus
making contact with the papers \cite{ChapotonLivernet,LadaMarkl,OudomGuin,Bandiera}.

\item Note that since every taut web has at least two
vertices  the differential $b_1$ is
always identically zero. In technical terms
these algebras are ``minimal'' \cite{KellerIntroduction}.

\item
Moreover, $b_n(\fw_1, \dots, \fw_n) $ is a web with
\be
V = n + \sum_{i=1}^n ( V(\fw_i) -1)
\ee
vertices. Since we can grade  $\CW$ by the number of vertices
it follows that $b_2$ is a nilpotent multiplication.

%\item  It is worth noting that there is a dual formulation
%of these identities. We consider the supercommutative algebra generated
%by elements $x_\fw$ of degree $d(\fw)$. Then to each $\fw$ we associate
%a vector field
%
%\be
%\hat V(\fw) = \left( \sum_{I_v(\fw) = I_\infty(\fw_v)} \epsilon \prod x_{\fw_v} \right)\frac{\p}{\p x_{\fw}}
%\ee
%
%then the vector field $\hat V = \hat V(\ft)$ simply squares to zero.
%
%\cg{I'm not sure this remark is correct. Check it.}
%\cg{It is wrong as it is. It can be made right, but at the cost of expanding it quite a bit. Probably not worth the effort.}
%

\item The $T$ operation also satisfies a natural associativity relation.
\be\label{eq:Interior-Web-Associativity}
T( T(\fw)[\fw_1 \otimes \cdots \otimes \fw_n ])[\tilde \fw_1 \otimes \cdots \otimes \tilde \fw_N]
= \sum_{{\rm Sh}_n(S)} \epsilon ~
T(\fw)\left[ T(\fw_1)[S_1], T(\fw_2)[S_2], \dots, T(\fw_n)[S_n] \right]
\ee
where the sum over $n$-shuffles refers to the ordered set
\be
S = \{ \tilde \fw_1 ,  \cdots , \tilde \fw_N \}
\ee
The sign $\epsilon$ in the sum keeps track as usual of the webs orientations.
This relation will not play an important role in the following.
%, and itis determined as follows.  We let
% $o_r(S_\alpha)$ be the ordered product of reduced orientations of the
%$\tilde \fw_i$ in that set. Then
%
%\be
%\epsilon = \frac{ o(\fw) \wedge \left( o_r(\fw_1) \wedge o_r(S_1) \right) \cdots \wedge \left( o_r(\fw_n) \wedge o_r(S_n)\right) }{
%o(\fw) \wedge o_r(\fw_1) \wedge \cdots \wedge o_r(\fw_n) \wedge o_r(\tilde \fw_1)
%\wedge \cdots \wedge o_r(\tilde \fw_N) }
%\ee
%
% Again, because of the line
%principle, at most one term in \eqref{eq:Interior-Web-Associativity}
%can be nonzero.

%This relation is only nontrivial when $n=V(\fw)$ and
%
%\be
%N=\sum_{i=1}^n V(\fw_i).
%\ee
%
%Note that $N = V(T(\fw)[\fw_1 \otimes \cdots \otimes \fw_n ])$.

%The sum is over \emph{$n$-unshuffles} of $\{ \tilde \fw_1 ,  \cdots , \tilde \fw_N \} $.
%By this we mean an ordered
%  disjoint decompositions into $n$ ordered  sets:
%
%\be
%\{ \tilde \fw_1 ,  \cdots , \tilde \fw_N \} = S_1 \amalg S_2 \amalg \cdots \amalg S_n
%\ee
%
%where the sets $S_\alpha$ retain the ordering of $\{ \tilde \fw_1 ,  \cdots , \tilde \fw_N \} $
%and the ordering of the sets $S_1, \dots, S_n$ matters, so that $S_1 \amalg S_2$ is
%considered different from $S_2 \amalg S_1$, etc.

\end{enumerate}

\subsubsection{Examples Of Web Algebras }

We can describe the $L_\infty[-1]$-algebra fairly explicitly if there are $n$
vacua with weights
$z_i$ which are in the set of
 extremal points of a convex set. We can enumerate the
 vertices by ordering the vacua so that $\{1,\dots, n\}$ is
 a cyclic fan of vacua. Then there is an  $n$-valent vertex
$\fw_{12\cdots n}$ and we can make all other  $(n-j)$-valent vertices by deleting
$j$ vacua from the cyclic fan $\{1,\dots, n\}$ to form cyclic fans with
smaller numbers of vacua.
 We must have at least $3$ vacua so there are in all:
\be
\sum_{j=0}^{n-3} {n \choose j} = 2^n - \half (n^2 + n + 2)
\ee
different vertices. We denote these by $\fw_{I}$ where $I$ is a cyclic fan
of vacua, which, in these examples is just a cyclically
ordered subset of $\{ 1, \dots, n\}$ with at least three elements.

We can make all taut elements by ``resolving'' the vertices. These will be
enumerated by pairs of cyclic fans  of vacua $I_1, I_2$ which are
compatible in the sense that they are of the form:
\be
I_1 = \{ i_1, i_2, \dots, i_k \} \qquad I_2 = \{ i_k, i_{k+1}, \dots, i_1 \}
\ee
and, if we denote
\be
I_1*I_2 = \{ i_1, i_2, \dots, i_k  , i_{k+1}, \dots, i_1 \}
\ee
then $I_1*I_2$ is also a cyclic fan.
Then we have
\be
\ft = \sum \fw_{I_1; I_2}
\ee
where we sum over such compatible pairs of fans.
All the taut webs have exactly two vertices and therefore
(for such convex configurations of vacuum weights) the higher
products $b_n=0$ for
$n>2$.

The only nonzero products of vertex webs is
\be
b_2(\fw_{I_1}, \fw_{I_2}) = \fw_{I_1;I_2}
\ee
However, if a taut web $\fw_{I_1;I_2}$ has vertices with $I_1$ or $I_2$ of length
greater than $3$ then it can also define products of non-vertex webs which have
$I_\infty(\fw) = I_1$ or $=I_2$. Note also that no taut web has any
vertex with $I_v = \{ 1,2,\dots, n\}$ and so any web with
$I_\infty(\fw) =\{ 1,2, \dots, n \}$ such as the
vertex $\fw_{1,2,\dots, n}$, and all its resolutions, will be in the
annihilator of $b_2$. We  have an $(n-2)$-step
nilpotent algebra. That is, $(n-2)$ applications of $b_2$ will always
vanish.

When the weights $z_i$ of the vacua are not extremal points of a convex
set then the algebras can be more complicated and the higher products
$b_n$ can be nonzero.

\subsection{Algebraic Structures From Half-Plane Webs}\label{subsec:AlgRel-HalfPlane}

There are three obvious generalizations of $T$ to half-plane webs:
we can either replace all interior vertices of some half-plane webs $\fu$ with
plane webs, replace all boundary vertices with half-plane webs, or both.
The latter operation is the composition of the former two: we can first replace the interior vertices,
then the boundary vertices (if we try the opposite, we create new interior vertices at the first step).
It is instructive to discuss all three possibilities. The first case, discussed in
\S \ref{subsubsec:HalfPlane-L}, shows  how half-plane webs
provide an $L_\infty$-module for the $L_\infty$-algebra of plane webs. Then, in
\S \ref{subsubsec:HalfPlane-A} we show that inserting half-plane webs into half-plane
webs defines an $A_\infty$-algebra structure. When we combine the two operations
we end up with a set of identites we call the $LA_\infty$-identities in
\S \ref{subsubsec:HalfPlane-LA}. Finally in \S \ref{subsubsec:HalfPlane-CLA}
we give a conceptual interpretation of the $LA_\infty$-identities in terms
of an $L_\infty$-morphism between the $L_\infty$ algebra of plane webs and
the $L_\infty$ algebra of Hochschild cochains on the $A_\infty$ algebra of
half-plane webs.

\subsubsection{$L_\infty$-Modules}\label{subsubsec:HalfPlane-L}

The first possibility - replacing just interior vertices -  is not our essential goal,
 but it is instructive. We define a multilinear map $T_i[\fu]:  T\CW \to \CW_\CH$
which is, as before, zero unless there is some permutation $\sigma$ which matches the arguments $\fw_{a}$
of a monomial $\{ \fw_1, \dots, \fw_\ell\}$  to the
 interior vertices of $\fu$ in the sense that
 $I_{v^i_{\sigma(a)} }(\fu) = I_\infty(\fw_{a}) $, in which case
it is the simultaneous convolution with orientation
$o(\fu) \wedge o_r(\fw_1) \wedge \cdots \wedge o_r(\fw_\ell)$.
This map has a simple relation to convolutions:
\begin{align}\label{eq:Ti-and-star}
T_i(\fu * \fw)[\fw_1, \dots, \fw_{n} ] &=
\sum_{{\rm Sh}_2(S)} \epsilon ~ T_i(\fu) [ T(\fw)[S_1], S_2] \cr
T_i(\fu * \fu')[\fw_1, \dots, \fw_{n} ] &=
\sum_{{\rm Sh}_2(S)} \epsilon' ~ T_i(\fu) [S_1]* T_i(\fu')[S_2]
\end{align}
with the usual definition $S = \{\fw_1, \dots, \fw_{n} \}$.

The $\epsilon$ signs keep track of the relative web orientations on the two sides of the equations.
It is important to observe that
the signs arise from the reorganization of a product of \emph{reduced} web orientations.
It would thus be incorrect to  assume glibly
that $\epsilon$   coincides with the Koszul rule: we defined the degree of a web as the dimension of the unreduced moduli space. The correct sign rule could be denoted as the ``reduced Koszul rule'': treat the symbols as if they had
degree given by the reduced dimension of moduli spaces. This subtlety was invisible for bulk webs, for which the reduction of moduli space removes two dimensions, but it is important for half-plane webs. The prime on
the $\epsilon$ in the second equation of \eqref{eq:Ti-and-star} takes into account that we must bring $\fu'$
across the monomial $S_1$ in the tensor algebra using the reduced Koszul rule.

If we plug our second theorem $\ft_{\CH} * \ft_{\CH} + \ft_{\CH} * \ft_p =0$ into \eqref{eq:Ti-and-star}, we get a neat relation
\be\label{eq:L-infty-modone}
\sum_{{\rm Sh}_2(S) } \epsilon_{S_1,S_2} ~ T_i(\ft_\CH) [ T(\ft_p)[S_1], S_2]  + \epsilon'_{S_1,S_2} ~ T_i(\ft_\CH) [S_1]* T_i(\ft_\CH)[S_2]= 0
\ee
This identity can be used to define $\CW_{\CH}$ as a ``right-module for the $L_\infty$-algebra $\CW$.''

In general,  if $\CL$ is an
$L_{\infty}$-algebra with products $b_n^{\CL}$ of degree $3-2n$,
then a left $L_{\infty}$-module $\CM$ is a graded $\IZ$-module
with operations $b_n^{\CM}: \CL^{\otimes n} \otimes \CM \to \CM$ defined for $n\geq 0$
and of degree $1-2n$
so that the analog of the $L_{\infty}$-identities holds, i.e. for all $S=\{ \ell_1, \dots, \ell_s\}$
and $m\in \CM$:
\be\label{eq:Lifty-module}
\sum_{{\rm Sh}_2(S)} \epsilon b_{s_2 +1}^{\CM}(b^{\CL}_{s_1}(S_1),S_2; m)
+
\sum_{{\rm Sh}_2(S)} \epsilon b_{s_1 }^{\CM}(S_1 ; b^{\CM}_{s_2}(S_2; m))
= 0
\ee
In the present case, the   module operations for a set $S=\{ \fw_1, \dots, \fw_n\}$ are
 \be
 \fu \mapsto \fu*T_i(\ft_{\CH})[S]
 \ee
  The module relations are
\be\label{eq:L-infty-modtwo}
\sum_{{\rm Sh}_2(S) } \epsilon_{S_1,S_2} ~ \fu * T_i(\ft_\CH) [ T(\ft_p)[S_1], S_2]  + \epsilon'_{S_1,S_2} ~ (\fu *T_i(\ft_\CH) [S_1])* T_i(\ft_\CH)[S_2]= 0
\ee
and can be proven from \eqref{eq:L-infty-modone} by convolving with $\fu$.
The only nontrivial step is the observation that
\be
(\fu_1*\fu_2)*\fu_3 - \fu_1*(\fu_2 * \fu_3)
\ee
 is (reduced Koszul)
graded symmetric in $\fu_2$ and $\fu_3$. That is, once again we use the property
that convolution defines a pre-Lie algebra structure.

\subsubsection{$A_\infty$-Algebras}\label{subsubsec:HalfPlane-A}

We can define the second natural multilinear map $T_\CH[\fu]:  T\CW_\CH \to \CW_\CH$, by inserting
half-plane webs at all boundary vertices of $\fu$.
In contrast to   the interior vertices, the boundary vertices will always be time ordered, so we define the
map to be zero on $\fu_1 \otimes \cdots \otimes \fu_\ell$ unless the arguments match the boundary vertices,
$J_{v^\p_a}(\fu) = J_{\infty}(\fu_a)$ for $a=1,\dots, \ell$,
\emph{in that order}. (Recall from equation \eqref{eq:bdry-order}
that we have chosen an ordering of the boundary vertices.)
When this is satisfied we glue in to get
 a new deformation type in the usual way with the orientation
\be
o(\fu)\wedge o_r(\fu_1) \wedge \cdots \wedge o_r(\fu_\ell).
\ee

Before stating the compatibility of $T_{\CH}$ with $*$
 it is useful at this point to define a notion of   \emph{ordered $n$-partitions}.
If $P$ is an ordered set we define an ordered $n$-partition of $P$ to be an ordered disjoint
  decomposition into $n$ ordered subsets
\be
P = P_1 \amalg P_2 \amalg \cdots \amalg P_n
\ee
where the ordering of each summand $P_\alpha$ is inherited
from the ordering of $P$ and all the elements of $P_\alpha$ precede all elements of $P_{\alpha+1}$  inside $P$.
We allow the $P_\alpha$ to be the empty set.
For an ordered set $P$ we let ${\rm Pa}_n(P)$ denote
the set of distinct $n$-partitions of $P$. If $p=\vert P\vert$ there are
${n+p-1 \choose p}$ such partitions.

Now we can state the compatibility:
\begin{align}\label{eq:Th-and-star}
T_\CH(\fu * \fw)[\fu_1, \dots, \fu_{n} ] &=
\epsilon ~\left( T_\CH(\fu) [\fu_1, \dots, \fu_{n} ] \right) * \fw - \sum_{m=1}^n \epsilon ~ T_\CH(\fu) [\fu_1, \dots, \fu_m * \fw,\dots  \fu_{n} ] \cr
T_\CH(\fu * \fu')[\fu_1, \dots, \fu_{n} ] &=
\sum_{{\rm Pa}_3(P)} \epsilon ~ T_\CH(\fu) [P_1,T_\CH(\fu')[P_2],P_3]
\end{align}
In the second identity we have introduced a sum over ordered $3$-partitions of an ordered set $P$ of
half-plane webs. As before, we take $T_\CH[\fu][P]=0$ if $P=\emptyset$.

Combining this with $\ft_{\CH} * \ft_{\CH} + \ft_{\CH} * \ft_p =0$ we arrive at the relation
\be\label{eq:A-infty}
\begin{split}
\epsilon_1
\left( T_\CH(\ft_\CH) [\fu_1, \dots, \fu_{n} ] \right)* \ft_p &
- \sum_{m=1}^n \epsilon_2 ~ T_\CH(\ft_\CH) [\fu_1, \dots, \fu_m * \ft_p,\dots  \fu_{n} ]\\
& + \sum_{{\rm Pa}_3(P)} \epsilon_3 ~ T_\CH(\ft_\CH) [P_1,T_\CH(\ft_\CH)[P_2],P_3]= 0.\\
\end{split}
\ee
where
\be
\epsilon_1 = (-1)^{\sum_s d_r(\fu_s)} \qquad \epsilon_2 = (-1)^{\sum_{s=1}^m d_r(\fu_s)} \qquad
\epsilon_3 = (-1)^{P_1} := (-1)^{ \sum_{\fu \in P_1} d_r(\fu)} .
\ee

We can interpret \eqref{eq:A-infty}
as the  the standard axioms for an $A_\infty$ algebra structure on $\CW_{\CH}$.  To make this clear and to
lighten the notation let us denote by $a_n$ (``open brackets'') the
restriction of $T_\CH(\ft_\CH)$ to $\CW_\CH^{\otimes n}$ for $n>1$ and
the operation
\begin{equation}\label{eq:HalfPlane-a1}
a_1(\fu) =T_\CH(\ft_{\CH})[\fu] - (-1)^{d_r(\fu)} \fu * \ft_p
\end{equation} for $n=1$.
The first of the \afty-relations demands that  $a_1(a_1(\fu))=0$.
This works out to be
\be
T_\CH(\ft_\CH)(T_\CH(\ft_\CH)[\fu]) - (-1)^{d_r(\fu)} T_\CH(\ft_\CH)[\fu*\ft_p] +
(-1)^{d_r(\fu)} (T_\CH(\ft_\CH)[\fu])*\ft_p - (\fu * \ft_p)*\ft_p = 0
\ee
Thus to match to \eqref{eq:A-infty} we also need to check that $(\fu * \ft_p)*\ft_p=0$.
 Although convolution is not associative,
%
% it is still true that for any web $\fu$
%\begin{equation}
%(\fu * \ft_p)*\ft_p=0
%\end{equation}
%Indeed,
the difference $(\fu* \ft_p) * \ft_p - \fu*(\ft_p * \ft_p)$ consists of terms where two taut webs $\fw_1$ and $\fw_2$ are inserted separately
at two vertices $v^1$ and $v^2$ of $\fu$. Each such terms appears twice in the sum, either from $(\fu*_{v^1} \fw_1) *_{v^2} \fw_2$
or from $(\fu*_{v^2} \fw_2) *_{v^1} \fw_1$. The two contributions have opposite orientations and cancel out against each other.

Moving on to the higher identities, the
\be
a_n: \CW_\CH^{\otimes n} \rightarrow \CW_\CH \qquad n \geq 1
\ee
have degree $\deg(a_n) = 2-n$ and satisfy the identities
\be\label{eq:Ainf1}
\sum_{{\rm Pa}_3(P)} (-1)^{P_1}  a_{p_1+p_3+1}(P_1,a_{p_2}(P_2), P_3) = 0
\ee
where $p_i= \vert P_i\vert$, $i=1,2,3$.

\subsubsection{The $LA_\infty$-Identities}\label{subsubsec:HalfPlane-LA}

Finally, we can consider the combined operation
\be\label{eq:CombOp}
T(\fu): T\CW_{\CH} \otimes T \CW \to  \CW_{\CH}
\ee
as
\be\label{eq:tensor-u-def}
\begin{split}
T(\fu)[\fu_1, \dots, \fu_n; \fw_1, \dots, \fw_m]
% & :=
%(\fu *_{\CV_\p(\fu)} \{ \fu_1, \dots, \fu_n \} )*_{\CV_i(\fu)} \{ \fw_1, \dots, \fw_m \}\\
&
:= \epsilon~ T_\CH\left(T_i(\fu)[\fw_1, \dots, \fw_m] \right)[\fu_1, \dots, \fu_n]\\
\end{split}
\ee
We included a sign, to convert the orientation
\be
o(\fu)\wedge \left( o_r(\fw_1) \wedge \cdots \wedge o_r(\fw_m) \right) \wedge
\left( o_r(\fu_1) \wedge \cdots \wedge o_r(\fu_n) \right)
\ee
in the right hand side to the natural orientation for the order of the arguments on left hand side
\be
o(\fu)\wedge\left( o_r(\fu_1) \wedge \cdots \wedge o_r(\fu_n) \right) \wedge \left( o_r(\fw_1) \wedge \cdots \wedge o_r(\fw_m) \right).
\ee

%The associativity relation for the tensor operation of half-plane webs is
%
%\be
%\begin{split}
%&
%(T(\fu)[\fu_1, \dots, \fu_n; \fw_1, \dots, \fw_m])\left[\tilde \fu_1, \dots, \tilde \fu_N; \tilde\fw_1,
%\dots, \tilde\fw_M \right] \\
%& = \sum_{\ell_1 + \cdots + \ell_n = N} \sum_{{\rm Sh}_{n+m}(S)} \epsilon ~
%T(\fu)[ T(\fu_1)[\tilde u_1, \dots, \tilde u_{\ell_1};\tilde S_1],
%T(\fu_2)[\tilde u_{\ell_1+1}, \dots, \tilde u_{\ell_1+\ell_2};\tilde S_2], \dots \\
%& \dots, T(\fu_n)[\tilde u_{N-\ell_n+1}, \dots, \tilde u_{N};\tilde S_n];
%T(\fw_1)[\tilde S_{n+1}], \dots, T(\fw_m)[\tilde S_{n+m}] ]\\
%\end{split}
%\ee
%
%where we sum over $(n+m)$-shuffles of $S =  \{ \tilde\fw_1, \dots, \tilde\fw_M\}$.
%The sign is determined as usual.

Again, we note that
convolution and $T$ interact well together:
\be\label{eq:compat-1}
T(\fu * \fu')[P;S] =
\sum_{{\rm Sh}_2(S), {\rm Pa}_3(P)} \epsilon ~    T(\fu)[ P_1,
T(\fu')[P_2 ; S_1 ], P_3; S_2]\\
\ee
where $S$ is the set of plane web arguments and $P$ is the set of half-web arguments.
The sign is given by the usual reduced Koszul rule.

%The sign can be written as:
%
%\be\label{eq:UseSign}
%\epsilon = (-1)^{d_r(\fu')(d_r(\fu_1) + \cdots + d_r(\fu_r)) } (-1)^{d_r(S_1)(d_r(\fu_{r+n'+1}) +
%\cdots + d_r(\fu_{n''} ) )} \epsilon_{S_1,S_2} .
%\ee
%
%where $d_r$ is the reduced dimension.
%This will be used in deriving \afty-relations below.

Similarly
\be\label{eq:compat-2}
T(\fu * \fw)[P; S] =
\sum_{{\rm Sh}_2(S)}   \epsilon_{S_1,S_2} ~   T(\fu)[ P;
T(\fw)[  S_1 ],  S_2]
\ee

Combining \eqref{eq:compat-1} and \eqref{eq:compat-2} with
\eqref{eq:hp-wb-id} we get some nontrivial algebraic
identities
\be\label{eq:big-rel}
 \sum_{{\rm Sh}_2(S), {\rm Pa}_3(P)} \epsilon ~  T(\ft_{\CH})[P_1,
T(\ft_{\CH})[P_2 ; S_1 ], P_3; S_2]   +
\sum_{{\rm Sh}_2(S)}   \epsilon ~  T(\ft_{\CH})[ P;
T(\ft_p)[  S_1 ],  S_2] = 0  .
\ee

We will refer to this hybrid equation as an
``$LA_\infty$'' identity. (This is not standard terminology.)
It  has a somewhat refined algebraic meaning, which we will decode presently.
Before that, we would like to remark that both $T_i$ and $T_\CH$ can be recovered from
$T$ by filling either kinds of slots with the sum over all rigid plane webs $\fr$ or all rigid half-plane webs $\fr_\CH$:
\begin{align}
T_i[S] &= \sum_n T[\fr_\CH^{\otimes n};S] \cr
T_\CH[P] &= \sum_n \frac{1}{n!} T[P,\fr^{\otimes n}]
\end{align}
and we can similarly fill in the slots of \eqref{eq:big-rel} to get the corresponding equations \eqref{eq:L-infty-modone} and \eqref{eq:A-infty}.

\subsubsection{Conceptual Meaning Of The $LA_\infty$-Identities}\label{subsubsec:HalfPlane-CLA}

We now give a conceptual interpretation of the $LA_\infty$ identities \eqref{eq:big-rel}.
It is useful to organize the equations \eqref{eq:big-rel} by first considering the special cases where $S$
has cardinality  $0,1,2$. Then the general structure will become clear. Correspondingly,
we can decompose the taut element $\ft_\CH$ according to the number of
interior vertices in the taut webs:
\be
\ft_{\CH} = \ft_{\CH}^{(0)} + \ft_{\CH}^{(1)} + \ft_{\CH}^{(2)} + \cdots
\ee
Note that the only taut half-plane webs with no interior vertices
have precisely two boundary
vertices and therefore $\mu:= T(\ft_{\CH}^{(0)})$ is simply
a multiplication map
\be
\mu: \CW_{\CH} \times \CW_{\CH} \rightarrow \CW_{\CH}.
\ee
When \eqref{eq:big-rel} is restricted to $S=\emptyset$ we
learn that  $\mu$ is an associative
multiplication, up to sign:
\be
\mu(\mu(\fu_1, \fu_2), \fu_3) + (-1)^{d(\fu_1)-1} \mu(\fu_1,\mu( \fu_2, \fu_3))=0
\ee
and therefore $\tilde \mu(\fu_1,\fu_2) = (-1)^{d(\fu_1)-1} \mu(\fu_1, \fu_2)$ is
strictly associative.

Next, for a fixed planar web $\fw$
let
\be
\mu^{(1)} := T(\ft^{(1)})[\cdots; \fw]:  T\CW_{\CH} \to \CW_{\CH}
\ee
Recall that a Hochschild cochain  on an algebra $\CA$ is simply a
collection of linear maps $F_n: \CA^{\otimes n} \to \CA$, $n\geq 1$,  or
equivalently an element of $CC^\bullet(\CA):=\Hom(T\CA, \CA)$. Therefore for
 a fixed $\fw$ we may view $\mu^{(1)}$ as a Hochschild cochain on the
 associative algebra  $\CW_{\CH}$.
Then, taking $S=\{ \fw \}$, equation \eqref{eq:big-rel}  becomes
\be\label{eq:HochschildDiff}
\begin{split}
0 & = \tilde{\mu}( \fu_1, \tilde{\mu}^{(1)}(\fu_2, \dots, \fu_n) ) + \\
& + \sum_{r=0}^{n-1} (-1)^r \tilde{\mu}^{(1)}(\fu_1 , \dots, \fu_r , \tilde{\mu}(\fu_{r+1}, \fu_{r+2}), \fu_{r+3}, \dots, \fu_n) \\
& + (-1)^{n} \tilde{\mu}(\tilde{\mu}^{(1)}(\fu_1, \dots, \fu_{n-1}), \fu_n) \\
& =: B^{(1)}(\tilde{\mu}^{(1)})(\fu_1, \dots, \fu_n)  \\
\end{split}
\ee
where $\tilde \mu^{(1)}$ is related to $\mu^{(1)}$ by signs in a way analogous to the relation
of $\mu$ and $\tilde\mu$.
In the last line of \eqref{eq:HochschildDiff} we have recognized that the previous
lines define the Hochschild differential $B^{(1)}$ on the Hochschild
complex. Thus, our identity \eqref{eq:big-rel} when $\vert S \vert = 1$ simply says that $\mu^{(1)}$ is
a Hochschild cocycle. In order to discuss the  cases $\vert S \vert \geq 2 $
we introduce the notation
\be
\tilde\mu^{(n)}(S) :=\epsilon_n T(\ft^{(n)})[\cdots; S]:  T\CW_{\CH} \to \CW_{\CH}
\ee
where  $S= \{ \fw_1, \dots, \fw_n \} $,  $n \geq 1$, and $\epsilon_n$ is again
an appropriate sign redefinition.
Then the identity for $\vert S \vert = 2$ reads
\be
\tilde\mu^{(1)}(b_2(\fw_1, \fw_2)) = B^{(1)} \tilde\mu^{(2)}(\fw_1, \fw_2) + B^{(2)}(\tilde\mu^{(1)}(\fw_1), \tilde\mu^{(1)}(\fw_2))
\ee
where   $b_2$ is the multiplication on $\CW$ defined by $T(\ft_p)$, and
  we have introduced the Hochschild bracket $B^{(2)}$ on the Hochschild complex $CC^\bullet(\CW_{\CH})$.

Quite generally the Hochschild bracket on $CC^\bullet(\CA)$
may be defined  on two cochains $F,G$ of degree $n,m$ by
\be
\begin{split}
B^{(2)}(F,G) & :=  F\circ G - (-1)^{(\vert F \vert - 1) (\vert G \vert -1)} G\circ F \\
F\circ G & :=  \sum_r \epsilon F(\fu_1, \dots, \fu_r , G(\fu_{r+1},\dots, \fu_{r+m}), \dots \fu_{n+m-1})
\end{split}
\ee
where the sign $\epsilon$  can be found in many papers. See, for examples,
\cite{KellerKontsevichQuantization,Abbaspour}. In general, the Hochschild complex is a
differential graded Lie algebra with operations $B^{(1)}$ and $B^{(2)}$.
A differential graded Lie algebra can be considered to be   a special
case of an $L_\infty$-algebra, so we can speak of the $L_\infty$ algebra
of a Hochschild complex.

To summarize these facts concisely we need the general notion of an $L_\infty$
morphism between two $L_\infty$ algebras. See equation \eqref{eq:Lin-Morph} below.
% $\CA_1$ and $\CA_2$. Such a morphism is a
%map $f: T\CA_1 \to \CA_2$ so that
%
%\be
%\sum_s \sum_{\Sh_s(S)} \epsilon m_s^{\CA_2}\left( f(S_1),\dots, f(S_s) \right)
%= \sum_{\Sh_2(S)} \epsilon f_{s_2+1}\left(m_{s_1}^{\CA_1}(S_1),S_2) \right)
%\ee
%
%where on each side $\epsilon$ represents suitable signs.

Now, returning to our example,  we consider the map
\be
\tilde\mu: T\CW \to CC^\bullet(\CW_{\CH})
\ee
defined by
\be
\fw_1 \otimes \cdots \otimes \fw_n \mapsto T(\ft_{\CH}[\cdots; \fw_1 \otimes \cdots \otimes \fw_n]).
\ee
Then, separating \eqref{eq:big-rel} into the cases $S_1 = \emptyset$, or $S_2=\emptyset$,
or both $S_1, S_2$ are nonempty, and   grouping together the terms $S_1 \amalg S_2$ with $S_2 \amalg S_1$ in
the latter case
we see that the equation \eqref{eq:big-rel} may be concisely summarized
as the statement   that
\emph{ $\tilde\mu$ is an $L_\infty$-morphism from the
$L_\infty$-algebra of planar webs to the $L_\infty$-algebra of the Hochschild complex of $\CW_{\CH}$.}
This is the conceptual meaning of the $LA_\infty$ identities.

\subsection{Bimodules And Strip Webs}\label{subsec:Bimodule-StripWeb}

Starting from a strip web $\fs$, we can define three elementary operations $T_{i, L, R}(\fs)$ which replace either all interior,
 left boundary or right boundary vertices
with plane, positive half-plane or negative half-plane webs, respectively. (Recall that the positive half-plane
$\CH_L$ has boundary on the left.)
The definitions of these operations are completely parallel to the definitions given in the previous two sub-sections.
Due to our choice of ordering of boundary vertices, the arguments of $T_{L}(\fs)$ should be ordered left to right in decreasing time,
and right to left in decreasing time for  $T_{R}(\fs)$.

We can then define appropriate composite operations
\begin{align}
T_{L,i}(\fs)[P;S] &:= \epsilon T_L\left(T_i(\fs)[S]\right)[P] \cr
T_{i,R}(\fs)[S;P'] &:= T_R\left(T_i(\fs)[S]\right)[P'] \cr
T_{L,R}(\fs)[P;P'] &:= T_R\left(T_L(\fs)[P]\right)[P'] \cr
T(\fs)[P;S;P'] &:=\epsilon'  T_R\left(T_L\left(T_i(\fs)[S]\right)[P]\right)[P']
\end{align}
where $\epsilon,\epsilon'$ are (reduced, as always) Koszul signs for reordering the
arguments.

We will now just sketch some of the various algebraic structures which follow from
the strip web identity
\eqref{eq:Strip-Web-Ident} which we quote here:
\be
 \ft_s*\ft_L + \ft_s * \ft_R + \ft_s * \ft_p +\ft_s \circ \ft_s =0.
\ee
These structures will involve the notion of an \afty-module.

 In general, if $\CA$ is an
\afty-algebra with products $m_n^{\CA}$ then a left \afty-module $\CM$ is a graded $\IZ$-module
with operations $m_n^{\CM}: \CA^{\otimes n} \otimes \CM \to \CM$ defined for $n\geq 0$
and of degree $1-n$
so that the analog of the \afty-identities holds, i.e. for all $P=\{ a_1, \dots, a_p\}$
and $m\in \CM$:
\be\label{eq:afty-module}
\sum_{{\rm Pa}_3(P)} (-1)^{p_1} m_{p_1+ p_3 +1}^{\CM}(P_1, m_{p_2}^{\CA} (P_2), P_3; m)
+
\sum_{{\rm Pa}_2(P)} (-1)^{p_1} m_{p_1}^{\CM}(P_1; m_{p_2}^{\CM}(P_2;m))
= 0
\ee
Similarly, one can define right \afty-modules as well as bimodules. When we include
interior operations $T_i$ then there will be $L_\infty$ maps to the $L_\infty$ algebra
of  Hochschild cochains with values these modules.

As the simplest example let us consider $\CH_L$, the positive half-plane and denote $\ft_L := \ft_{\CH_L}$
and $\CW_L := \CW_{\CH_L}$.  Proceeding as in the previous section the
identity \eqref{eq:Strip-Web-Ident} implies:
\be\label{eq:Strip-L-Mod}
\begin{split}
 & \sum_{{\rm Pa}_3(P)}\epsilon T_{L}(\ft_s)[P_1, T_{\CH_L}(\ft_L)[P_2],P_3] + \cr
 &+ \epsilon T_{L}(\ft_s)[P ] * \ft_R + \epsilon T_{L}(\ft_s)[P ] * \ft_p- \sum_{m=1}^n \epsilon T_{L}(\ft_s)[\fu_1, \dots, \fu_m * \ft_p,\dots  \fu_{n} ]+ \cr &+\sum_{{\rm Pa}_2(P)}\epsilon ~ T_{L}(\ft_s)[P_1] \circ T_{L}(\ft_s)[P_2] =0
\end{split}
\ee
where, as usual $P= \{ \fu_1, \dots  \fu_{n}\}$.

One can interpret the equations \eqref{eq:Strip-L-Mod} as defining a left \afty-module structure
on $\CM = \CW_{S}$ for the \afty-algebra $\CA = \CW_L$ defined using $T_{\CH_L}(\ft_L)$.
  To see this define
\be
m_0^{\CM}(\fs) := (-1)^{d_r(\fs)} \fs*(\ft_p + \ft_R)
\ee
and, for $n\geq 1$,  $m_n^{\CM}: \CW_L^{\otimes n} \otimes \CW_{S} \rightarrow  \CW_{S}$ by
\be
m_p^{\CM}(P;s) = T_{L}(\ft_s)[P]\circ \fs
\ee
In verifying the module relations we find that $(m_0^{\CM})^2=0$ for reasons
analogous to those mentioned above. Then we compose the LHS of
\eqref{eq:Strip-L-Mod} with $\circ \fs$ and use
\be
(-1)^{d_r(\fs)} \left( T_{L}(\ft_s)[P]*(\ft_p + \ft_R) \right)\circ \fs =
\left( T_{L}(\ft_s)[P]\circ \fs\right) *(\ft_p + \ft_R) -
T_{L}(\ft_s)[P]\circ \left( \fs * (\ft_p + \ft_R) \right)
\ee
to recast these equations into the left-module conditions.

%Up to some simple manipulations, this relation defines a left $A_\infty$ module for the $A_\infty$ algebra we associated to $T_\CH(\ft_L)$,
%with higher operations
%\begin{equation}
%T_{L}(\ft_s)[\fu_1, \dots  \fu_{n}]\circ : T\CW_\CH \otimes \CW_\CS \to \CW_\CS
%\end{equation}
%and first operation $\fs \to (-1)^{d(\fs)-1} \fs * (\ft_p + \ft_R)$.

In a similar fashion, $T_{R}(\ft_s)$ gives a right $A_\infty$ module for the $A_\infty$ algebra we associated to $T_{\CH_R}(\ft_R)$
and $T_{L,R}(\ft_s)$ gives an $A_\infty$ bi-module, with left and right actions given by the two $A_\infty$ algebras
we associated to $T_{\CH_L}(\ft_{L})$ and $T_{\CH_R}(\ft_{R})$.

As for the interior operation, $T_i(\ft_s)$ will satisfy relations such that the operations $* T_i(\ft_L)+ * T_i(\ft_R)+ *T_i(\ft_s)$ define a
right $L_\infty$ module. The three $T_{L,i}(\ft_s)$, $T_{i,R}(\ft_s)$, $T(\ft_s)$ satisfy lengthy axioms, which
essentially define some left, right or bi-module for the $\mu$ operations (as in Section
\S \ref{subsubsec:HalfPlane-CLA})  associated to either boundary, together with
$L_\infty$ maps from the $L_\infty$-algebra of planar webs to the $L_\infty$-algebra of the Hochschild complexes of the modules,
compatible with the maps defined before.

\section{Representations Of Webs}\label{sec:RepWeb}

\textbf{Definition}: Fix a set of vacua $\IV$ and weights
$\{ z_i \}_{i \in \IV} $. A \emph{representation of webs} is
 a pair $\CR = ( \{ R_{ij} \}, \{ K_{ij} \} )$ where

a.) $R_{ij}$ are $\IZ$-graded $\IZ$-modules defined for
all ordered pairs $ij$ of distinct vacua.

b.) $K_{ij}$ is a degree $-1$ symmetric perfect  pairing

\be
K_{ij}: R_{ij} \otimes R_{ji} \to \IZ.\label{whichone}
\ee

\bigskip

By degree $-1$ we mean that $K_{ij}(r_{ij}, r_{ji}')$ is
only nonzero when $\deg(r_{ij}) + \deg(r_{ji}')= + 1$
so that the integer $K_{ij}(r_{ij}, r_{ji}')$ is degree zero.
The pairing is symmetric in the sense that
\begin{equation}\label{eq:K-Symm}
K_{ij}(r_{ij}, r_{ji}') =K_{ji}(r_{ji}',r_{ij})
\end{equation}
As the degrees of the arguments differ by one unit, the symmetry of $K_{ij}$
makes sense. Property $(b)$  of $K_{ij}$ is motivated by
the realization in Landau-Ginzburg models explained near
equation \eqref{elz} above. Note that since $K_{ij}$ is nondegenerate,
$R_{ij}$ and $R_{ji}$ have the same rank. The property that it is
a perfect pairing will be
used in several points of the development, for example, in the derivation
of equations \eqref{eq:SijFactor-def-fs},\eqref{eq:SijFactor-def-ps} below.
Often we will drop the subscripts and just write to $K$ for the pairing when no confusion can arise.

\subsection{Web Representations And Plane Webs}\label{subsec:WebRepPlane}

Given a representation of webs, for every cyclic fan of vacua
$I = \{ i_1, i_2, \dots, i_n\}$ we form
\be
R_I := R_{i_1,i_2} \otimes R_{i_2,i_3} \otimes \cdots \otimes R_{i_n,i_1}
\ee
Note that to write this formula we needed to choose a place to start the cyclic sequence.
Different choices in the definition of $R_I$ are related by a canonical isomorphism
because  the Koszul rule gives  a canonical isomorphism
\be
 R_{i_1,i_2} \otimes R_{i_2,i_3} \otimes \cdots \otimes R_{i_n,i_1}
 \cong
  R_{i_2,i_3} \otimes R_{i_3,i_4} \otimes \cdots \otimes R_{i_n,i_1} \otimes R_{i_1,i_2}.
\ee
We will sometimes refer to $R_I$ as a \emph{representation of a fan}.

Now we collect the representations of all the vertices by forming
\be\label{eq:Rint-def}
\Rvtx := \oplus_{I} R_I
\ee
where the sum is over all cyclic fans of vacua. We want to define a map
\be
\rho(\fw): T \Rvtx  \to \Rvtx
\ee
with properties akin to $T(\fw)$.

\begin{figure}[htp]
\centering
\includegraphics[scale=0.3,angle=0,trim=0 0 0 0]{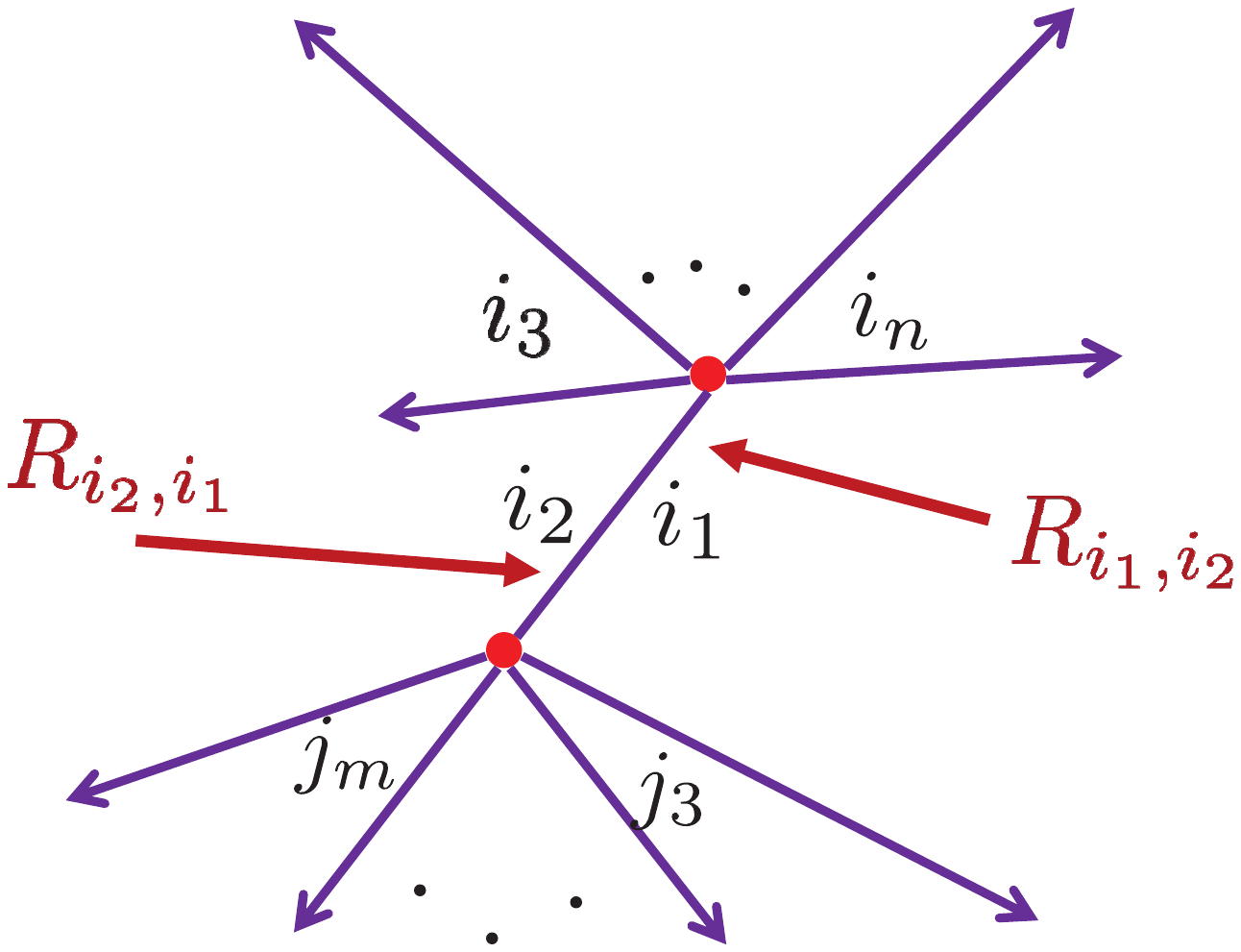}
\caption{The internal lines of a web naturally pair
spaces $R_{i_1,i_2}$ with $R_{i_2,i_1}$ in a web representation,
as shown here.    }
\label{fig:WEBEDGE}
\end{figure}

As for $T(\fw)$, we take $\rho(\fw)[r_1, \dots, r_n]$ to be zero unless
$n = V(\fw)$ and there exists an order $\{v_1 \dots v_n \}$ for the vertices of $\fw$
such that $r_a \in R_{I_{v_a}}$. If such an order exists, we will define our map
\be\label{eq:web-rep-1}
\rho(\fw): \otimes_{v \in \CV(\fw)}  R_{I_v(\fw)} \to R_{I_\infty(\fw)}
\ee
as the application of the contraction map $K$ to all internal edges of the web.
Indeed, if an edge joins two vertices $v_1, v_2 \in \CV(\fw)$ then if
$R_{I_{v_1}(\fw)}$ contains a tensor factor $R_{ij}$ it follows that
$R_{I_{v_2}(\fw)}$ contains a tensor factor $R_{ji}$ and these two
factors can be paired by $K$ as shown in Figure \ref{fig:WEBEDGE}.

In order to define $\rho(\fw)$ unambiguously, we need to be very precise about the
details of the contraction: Since $K$ has odd degree, the order of the contractions will
affect the sign of the result! It is useful to denote by $K_e$ the pairing which we will apply to the
tensor factors associated to the edge $e$. Then we are attempting to make sense of the overall sign of an expression such as
\begin{equation}\label{eq:tempK}
\otimes_{e \in \CE(\fw)} K_e \circ \otimes_{a=1}^n r_a
\end{equation}
where each of the $r_a \in R_{I_{v_a}(\fw)}$ is a linear combination of tensor products $r_{i_1,i_2} \otimes \dots \otimes r_{i_m, i_1}$ if
$I_{v_a}(\fw) = \{ i_1, i_2, \dots, i_m\}$.

Given a specific order of the edges in $\CE(\fw)$ and vertices in $\CV(\fw)$, the meaning of \ref{eq:tempK} is clear:
we shuffle the symbols around using the Koszul rule until each $K_e$ is followed by the two tensor factors it is supposed to contract, and then we
execute all the contractions. We are left with a sequence of residual tensor factors, which can be reordered again with the Koszul rule until they agree with the order in $R_{I_\infty(\fw)}$.
The final result depends on the initial order we picked for the vertices and edges of $\fw$ in an obvious way: by the Koszul rule for permuting the $r_a$ or the degree $-1$ $K_e$ symbols among themselves.
Our aim is to define a graded-symmetric operation. As the $r_a$ appear in the product in the same order as the arguments of $\rho(\fw)$, this is automatically true.

The only remaining subtlety is to relate the order for the edges of $\fw$ and the orientation $o(\fw)$
in such a way that
\begin{equation}
\rho(-\fw) = - \rho(\fw)
\end{equation}
We can do so if we remember that the moduli space of deformations of a web is given by a locus in $\IR^{2 V(\fw)}$ cut locally by a linear constraint for each edge of the web.
We can easily describe a vector field transverse to some edge constraint. For example, we can define $\p_e$ by acting with a clockwise rotation on the coordinates of the two endpoints of the edge
(the choice of origin for the rotation is immaterial). If we have some order for the edges, we can define an orientation for $\fw$ from the canonical orientations $dx_v dy_v$ in each $\IR^2$ factor as
\begin{equation}
\prod_{e \in \CE(\fw)} \p_e \circ \prod_{v \in \CV(\fw)} dx_v dy_v
\end{equation}
where $\circ$ means we contract the poly-vector field on the left with
the differential form on the right.

We are finally ready to give a complete definition: if the arguments are compatible with the vertices of the web
\begin{equation}
\rho(\fw)[r_1, \dots, r_n] = \frac{o(\fw)}{\prod_{e \in \CE(\fw)} \p_e \circ \prod_{v \in \CV(\fw)} dx_v dy_v} \otimes_{e \in \CE(\fw)} K_e \circ \otimes_{a=1}^n r_a
\end{equation}
where we use the same ordering of edges for the product over $\p_e$ in the denominator and for
the product over $K_e$. Otherwise, $\rho(\fw)[r_1, \dots, r_n]=0$.
This map has degree $- E(\fw)$.

The analogy between $T$ and $\rho$ extends to the interplay with the convolution operation.
\be\label{eq:rho-and-star}
\rho(\fw * \fw')[r_1, \dots, r_{n} ] =
\sum_{{\rm Sh}_2(S)} \epsilon_{S_1,S_2} ~ \rho(\fw) [ \rho(\fw')[S_1], S_2]
\ee
where we sum over 2-shuffles of the ordered
set $S=\{ r_1, \dots, r_{n}  \}$.
Once again we define $\rho(\fw)[\emptyset]=0$.
The sign $\epsilon_{S_1,S_2}$ keeps track as usual of the Koszul signs encountered in the
shuffling of the arguments $r_i$.
The only subtlety in checking this relation is the overall sign of each term in the left and right hand sides.
It is useful to observe that we can order the edges of $\fw * \fw'$ by listing all edges of $\fw$ first, then all edges of $\fw'$.
Then the order of the $K$ factors on the two sides of the equation is the same, the order of the $\p_e$ vector fields
in the denominators is the same, the orientations in the numerators coincide and the reshuffling of the
arguments is accounted for by the $\epsilon$ sign.

On the left hand side, the overall position of the
second web is removed (convolution uses the reduced orientation of $\fw'$) from the numerator,
the position of the insertion vertex in $\fw$ is removed from the denominator. As the overall position of the second web
is identified naturally with the position of the insertion vertex, this does not introduce any extra sign.

Now we extend $\rho$ linearly by defining $\rho(\fw + \fw') := \rho(\fw) + \rho(\fw')$.
 In close analogy to the previous section we
 can plug $\ft_p * \ft_p=0$ into the relation \eqref{eq:rho-and-star}
 and arrive at the axioms of an $L_\infty$ algebra $\rho(\ft): T\Rvtx \to \Rvtx$:
\be\label{eq:L-infty-rho}
\sum_{{\rm Sh}_2(S) } \epsilon_{S_1,S_2} ~ \rho(\ft_p) [ \rho(\ft_p)[S_1], S_2] = 0
\ee
The main difference between this algebra and the web algebra $\CW$ we encountered before is that $\Rvtx$ may have a rich subspace of degree $2$, which
allows us to discuss solutions to the Maurer-Cartan equation for the $L_\infty$ algebra:

\bigskip
\noindent
\textbf{Definition:} An \emph{interior amplitude} is an element $\beta \in \Rvtx$
of degree $+2$
so that if we define $e^\beta \in T \Rvtx \otimes \IQ$ by
\be
e^\beta := \beta + \frac{1}{2!} \beta \otimes \beta + \frac{1}{3!} \beta \otimes \beta\otimes\beta + \cdots
\ee
then
\be\label{eq:bulk-amp}
\rho(\ft_p)(e^\beta) = 0.
\ee

Note that any taut summand $\fw$ in $\ft$ has $2V(\fw)  - E(\fw) = 3$
and then $\rho(\fw)$ has degree $-E(\fw)$, so evaluated on $\beta^{\otimes V(\fw)}$ we get an
element of degree $3$. Thus, \eqref{eq:bulk-amp} is a nontrivial identity consisting of a sum
of elements of degree $3$.

\bigskip
\noindent
\textbf{Definition:} A \emph{Theory} $\CT$ consists of a set of vacuum data  $(\IV,z)$,
a representation of webs $\CR=(\{ R_{ij}\}, \{ K_{ij}\} )$ and an interior amplitude $\beta$.
If we want to talk about Theories with different data we can
write $\CT(\IV,z,\CR,\beta)$.
In the remainder of this section we will assume we are working within a specific Theory.

An interesting property of the Maurer-Cartan equation \eqref{eq:bulk-amp} for an $L_\infty$ algebra
is that a solution can be used to ``shift the origin'' of the algebra.
If we define
\begin{equation}
\rho_\beta(\fw)[r_1, \dots, r_\ell] := \rho(\fw)[r_1, \dots, r_\ell ,e^\beta]
\end{equation}
then we claim that $\rho_{\beta}(\ft_p):T\Rvtx \to \Rvtx$ satisfies:
\be\label{eq:L-infty-rhoz}
\sum_{{\rm Sh}_2(S) } \epsilon_{S_1,S_2} ~ \rho_\beta(\ft_p) [ \rho_\beta(\ft_p)[S_1], S_2] = 0 .
%= \sum_{{\rm Sh}_2(S) } \epsilon_{S_1,S_2} ~ \rho(\ft) [ \rho(\ft)[S_1,e^\beta], S_2,e^\beta] =0.
\ee
To prove this note that
 the 2-shuffles of the ordered set $\tilde S = S \cup \{ \beta, \dots, \beta\}$ with $n$ copies of $\beta$
appended at the right end of $S$
include ${n \choose k}$ copies of decompositions of the
form
\be
\tilde S =  \left(S_1\cup \{ \beta, \dots, \beta\} \right) \amalg \left( S_2 \cup \{ \beta, \dots, \beta\} \right)
\ee
with $k$ $\beta$'s in the first summand and $n-k$ $\beta$'s  in the second.
\footnote{We are being slightly sloppy here about the difference between union and
disjoint union. Consider the initially appended $\beta$'s as distinct and only identify
them after we apply $\rho(\ft_p)[\tilde S_1]$, etc. }
Now we multiply the $L_\infty$ axiom for $\rho(\ft_p)$ applied to $\tilde S$ by $\frac{1}{n!}$ and sum over $n$.
Thanks to the above remark the sum can be rearranged to give the
left-hand-side of the the $L_\infty$ axioms for $\rho_\beta(\ft_p)$ applied to $S$.
Thus far the argument applies to \emph{any}  element $\beta \in \Rvtx$.
To see what is special about interior amplitudes
note that while we defined $\rho(\fw)[S]=0$ for $S=\emptyset$, we have $\rho_\beta(\fw)[\emptyset]\not=0$
in general! Hence, for general $\beta$, the term with $S_1=\emptyset$ will contribute an extra
``source term'' in the identities. However, if $\beta$ is an interior amplitude then we can drop this term
and just sum over shuffles with $S_1 \not= \emptyset$ to recover the standard $L_\infty$ relations.

\bigskip
\noindent
\textbf{Remarks}:

\begin{enumerate}

\item
The $\rho$ and $T$ operations are compatible:
\be\label{eq:rhoT-1}
\rho( T(\fw)[\fw_1 , \cdots , \fw_n ])(r_1, \cdots , r_N)
= \sum_{{\rm Sh}_n(S) } \epsilon ~
\rho(\fw)\left[ \rho(\fw_1)[S_1], \rho(\fw_2)[S_2], \dots, \rho(\fw_n)[S_n] \right]
\ee
where $S=\{ r_1, \cdots , r_N\}$.
This equation is clearly analogous to the $TT$ associativity relation. In a sense, $\rho$ behaves as a representation for the
algebraic structure defined by $T$, hence our terminology.

\item The origin of the term ``shift the origin'' is from the analogy to string field theory.
Our space $\Rvtx$ is analogous to the space of closed-string fields, and solutions of the
Maurer-Cartan equation are analogous to on-shell backgrounds (at tree level). Now, there is an
identity
\be
\rho_{\beta}(S,e^{\beta'}) = \rho(S, e^{\beta + \beta'} )
\ee
which shifts the origin of the space of string fields.

\end{enumerate}

\subsubsection{Isomorphisms Of Theories}\label{subsubsec:IsomTheory}

It is worth giving a careful definition of an isomorphism
between two Theories  $\CT^{(1)}$ and $\CT^{(2)}$. First
of all, we require that there be a bijection
\be
\varphi: \IV^{(1)} \to \IV^{(2)}
\ee
so that the weights are mapped into each other. That is,
viewing the vacuum weight as a map $z: \IV \to \IC$ we have
\be
\varphi^*( z^{(2)}) = z^{(1)}
\ee

It will be convenient to ``trivialize'' $\varphi$ so that
we identify $ \IV^{(1)} = \IV^{(2)} = \IV$. Then $\varphi$
is a bijection of $\IV$ with itself. Because we will discuss
successive composition of interfaces from the right it will be convenient to
write the action from the \emph{right} so
\be
i \mapsto i\varphi
\ee
and the condition on the weights is
\be\label{eq:WtCond}
z^{(2)}_{i\varphi} = z^{(1)}_i\qquad\qquad \forall i\in \IV
\ee

Next, for   every distinct pair of vacua $(i,j)$ we have an isomorphism
of graded $\IZ$-modules:
\be
\varphi_{ij}: R^{(1)}_{ij} \to R^{(2)}_{i\varphi, j \varphi}
\ee
such that
\be\label{eq:Kpullback}
(\varphi_{ij}\otimes \varphi_{ji})^*(K^{(2)}_{i\varphi, j \varphi})=
K^{(1)}_{i , j  }
\ee

Finally, for   any cyclic fan of vacua $I$ we let $I\varphi$ be the image cyclic
fan of vacua (it is cyclic thanks to \eqref{eq:WtCond}). Then the $\varphi_{ij}$ induce an isomorphism
$\varphi_I: R^{(1)}_I \to R^{(2)}_{I\varphi} $ and we require that
\be\label{eq:BetaPres}
\varphi_I(\beta^{(1)}_I) = \beta^{(2)}_{I\varphi}
\ee

These three conditions define an \emph{isomorphism of Theories}.

\textbf{Remarks}:

\begin{enumerate}

\item If $\varphi^{(12)}: \CT^{(1)}\to \CT^{(2)}$ is an isomorphism
and $\varphi^{(23)}: \CT^{(2)}\to \CT^{(3)}$ is another isomorphism
then $\varphi^{(12)}\varphi^{(23)}$ is an isomorphism
$\CT^{(1)}\to \CT^{(3)}$.

\item Automorphisms are isomorphisms of a Theory with itself, and
these always form a group. Note that a non-identity automorphism
must still induce the identity permutation on $\IV$. For, suppose
that $i\varphi\not= i$ for some $i$. Then \eqref{eq:WtCond} implies
$z_{i} = z_{i\varphi}$. But,   setting $j:=i\varphi$
we see that this clashes with the condition on vacuum data that
  $z_{ij}\not=0$ for all $i\not=j$. The maps $\varphi_{ij}$ can still be
nontrivial so a Theory can still have a nontrivial automorphism group.
In the text we make use of some nontrivial isomorphisms which are not
automorphisms.

\end{enumerate}

\subsection{Web Representations And Half-Plane Webs}\label{subsubsec:WebRep-Halfplane}

In   Section \S \ref{subsec:BraneCat}  below we will introduce
an abstract notion of the Lefschetz thimbles which, in the context of  Landau-Ginzburg
theory define special branes in the theory associated to each of the
vacua. (See Section \S \ref{lgassuper} below.)  This motivates the following

\bigskip
\noindent
\textbf{Definition}: Fix a set of vacua $\IV$.
We define \emph{Chan-Paton data} to be  an assignment $i \to \CE_i$ of
a graded $\IZ$-module  to each vacuum $i\in \IV$. The modules $\CE_i$
are often referred to as \emph{Chan-Paton factors}.
\bigskip

Now fix a half-plane $\CH$.
If $J= \{ j_1, \dots, j_n\} $ is a half-plane fan in $\CH$ then we define
\be \label{eq:RJ}
R_J(\CE):= \CE_{j_1} \otimes R_{j_1,j_2} \otimes \cdots \otimes R_{j_{n-1},j_n} \otimes \CE_{j_n}^*.
\ee
and the counterpart to \eqref{eq:Rint-def} is
\be\label{eq:Rbd-def}
R^\p(\CE) := \oplus_J R_J(\CE)
\ee
where we sum over all half-plane fans in $\CH$.
%
%When the dependence on the
%choice of Chan-Paton spaces is important, as it is in Section
%\S \ref{sec:CategoriesBranes},  we will write $R_J(\CE)$ and $R^\p(\CE)$
%for \eqref{eq:RJ} and \eqref{eq:Rbd-def}, respectively.
%

We are ready to define the web-representation analogue of $T(\fu)$ defined in
\eqref{eq:CombOp}, namely  a map
\be
\rho(\fu): TR^\p(\CE) \otimes T\Rvtx \to TR^\p(\CE)
\ee
graded symmetric on the second tensor factor.
As usual, we define the element
\be\label{eq:Rho-rp-r}
\rho(\fu)[r^\p_1, \dots, r^\p_m;r_1, \dots, r_n]
\ee
by contraction.
In the equations below we will abbreviate this to $\rho(\fu)[P;S] $ where
\be
P = \{r^\p_1, \dots, r^\p_m\} \qquad  S= \{ r_1, \dots, r_n \}.
\ee
We define $\rho(\fu)[P;S] $  to be zero  unless the following conditions hold:
\begin{itemize}
\item The numbers of interior and boundary vertices
of $\fu$ match the number of arguments of either type: $V_\p(\fu) = m$ and $V_i(\fu)=n$.
\item The boundary arguments match in order and type those of the boundary vertices: $r^\p_a \in R_{J_{v^\p_a}(\fu)}$ (Recall these are
ordered from left to right in the order described in \eqref{eq:bdry-order}.).
\item We can find an order of the interior vertices $\CV_i(\fu)=\{v_1, \dots, v_n \}$ of $\fu$ such that they match the order and type of the interior arguments: $r_a \in R_{I_{v_a}(\fu)}$.
\end{itemize}

If the above
 conditions hold, we will simply contract all internal lines with $K$ and contract the Chan Paton elements of consecutive pairs of
$r^\p_a$ by the natural pairing $\CE_{j} \otimes \CE_j^* \to \IZ$.
We keep track of signs as before, building an orientation for $\fu$ from the orientation on $\IR^{2V_i(\fu) + V_\p(\fu)}$, denoting the coordinates of interior vertices as
$x_v, y_v$ and boundary vertices as $y^a_\parallel$. The edge vector fields $\p_e$ can be built as before, adjusting them so that the boundary vertices remain on the boundary.
\begin{align}\label{eq:bdy-rho-signs}
\rho(\fu)[r^\p_1, \dots, r^\p_m;r_1, \dots, r_n] &= \frac{o(\fu)}{\left[ \prod_{e \in \CE(\fu)} \p_e \right]\circ\left[\prod_{a=1}^m dy^a_\parallel\right]\left[  \prod_{v \in \CV_i(\fu)} dx_v dy_v\right] } \cr
&\left[\otimes_{e \in \CE(\fu)} K_e \right] \left[\prod_{a=1}^m \partial_{\theta_a} \right]\circ \left[\otimes_{a=1}^m \theta_a r^\p_a \right] \left[ \otimes_{a=1}^n r_a\right]
\end{align}
The ordering of the products over $dy^a_\parallel$ and $\p_{\theta_a}$ follows that specified in
\eqref{eq:bdry-order}.

In order for the signs to follow as closely as possible the conventions in $T[\fu]$, we introduced $m$
auxiliary degree $-1$ variables $\theta_a$, to be contracted with dual $\partial_{\theta_a}$ to get the final result.
The $\theta_a$ produce useful signs as they are brought across the $r^\p_a$ by the Koszul rule.
The use of $\theta_a r^\p_a$ mimics the use of reduced orientations in the definition of $T$.
Omitting the $\theta_a$ auxiliary variables in $\rho$ would have the same effect as
replacing
\begin{equation}
\prod_a o_r(\fu_a) \to \prod_a \partial_{y_\parallel^a} \prod_a o(\fu_a)
\end{equation}
in $T$, giving rise to somewhat less pleasing sign rules in the various associativity identities.

\begin{figure}[htp]
\centering
\includegraphics[scale=0.3,angle=0,trim=0 0 0 0]{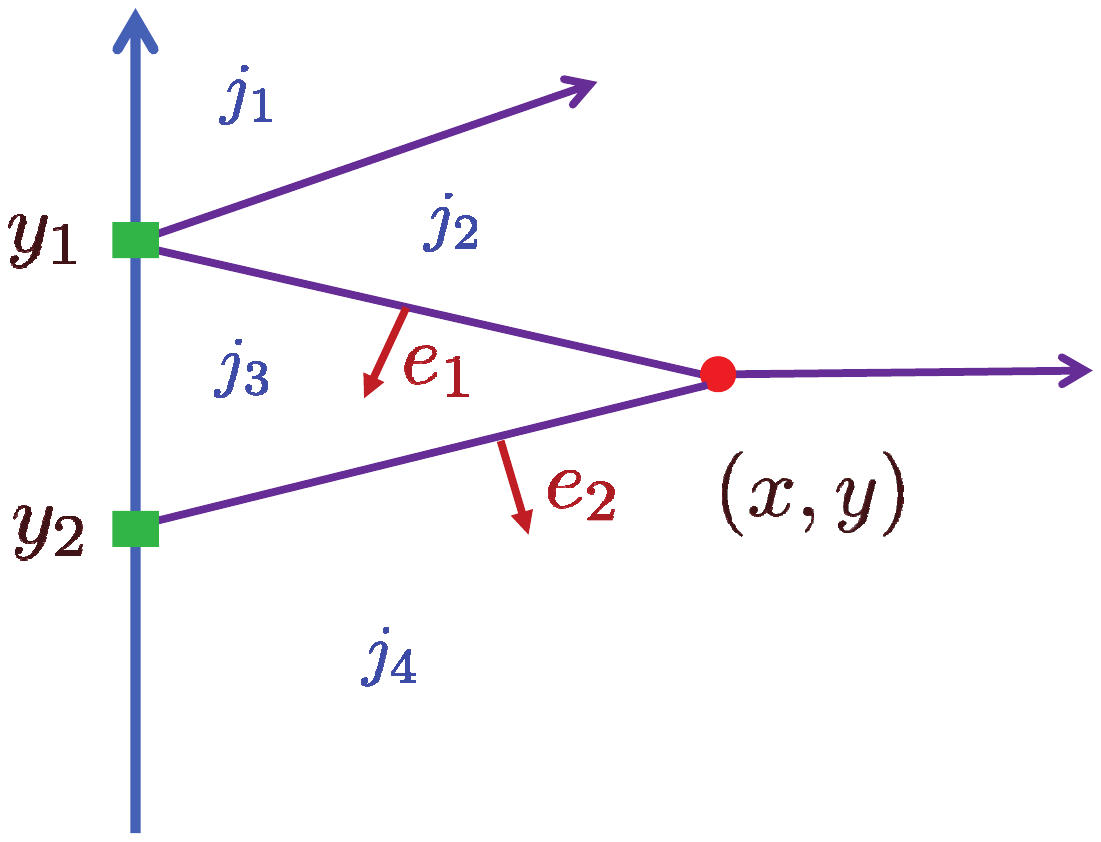}
\caption{A typical half-plane web. The signs for the contraction are fixed as
explained in the example.    }
\label{fig:EXAMPLE-CONTRACTION}
\end{figure}

\bigskip
\bigskip
\noindent
\textbf{Example}: As an example of how the formalism works consider the half-plane
web shown in Figure  \ref{fig:EXAMPLE-CONTRACTION}. This half-plane web is taut
and hence has a canonical orientation $o_r(\fu)$     oriented towards larger
webs. Therefore $o_r(\fu) = [dx] = [dy_1] = [-dy_2]$. (Note that the web gets
larger if we increase $x$ or $y_1$ but smaller if we increase $y_2$.) Now we can take
$\p_{\parallel} = - \frac{\p}{\p y_1}$ so we get $o(\fu) = [dy_1 dy_2]$.
Similarly, $\p_{e_1} \wedge \p_{e_2} = - \frac{\p}{\p y} \wedge \frac{\p}{\p x} $.
Thus, the prefactor on the first line of \eqref{eq:bdy-rho-signs} works out to
\be
\frac{o(\fu)}{ \p_{e_1} \wedge \p_{e_2} (-dy_1) (-dy_2) (dx dy)} = \frac{[dy_1 dy_2]}{-[dy_1 dy_2]} = -1
\ee
Now, $\rho(\fu)$ can only be nonzero on sums of vectors of the form
$r_1^\p \otimes r_2^\p \otimes r$ where
\be
\begin{split}
r_1^\p & \in \CE_{j_1}\otimes R_{J_1} \otimes\CE_{j_3}^* \qquad J_1 = \{ j_1, j_2,j_3 \} \\
r_2^\p & \in \CE_{j_3} \otimes R_{J_2} \otimes \CE_{j_4}^* \qquad J_2 = \{  j_3,j_4 \} \\
r & \in R_I \qquad\qquad\qquad\qquad I = \{ j_2, j_4,j_3 \} \\
\end{split}
\ee
Moreover it suffices to consider monomials of definite degree:
\be
\begin{split}
r_1^\p &  = \varepsilon_{j_1} r_{j_1j_2} r_{j_2j_3} \varepsilon_{j_3}^* \\
r_2^\p & = \varepsilon_{j_3}' r_{j_3j_4} \varepsilon_{j_4}^* \\
r & = r_{j_2j_4} r_{j_4j_3} r_{j_3j_2}  \\
\end{split}
\ee
Therefore
\be
\begin{split}
\rho(\fu)[r_1^\p, r_2^\p; r] & = - K_{e_1} K_{e_2} \p_{\theta_1} \p_{\theta_2} (\theta_1 r_1^\p) (\theta_2 r_2^\p) r \\
& = (-1)^{1 + \vert r_1^\p \vert} K_{j_2j_3} K_{j_3j_4}   r_1^\p  r_2^\p r \\
\end{split}
\ee
From here on, we simply apply the Koszul rule. The net result is
\be
\rho(\fu)[r_1^\p, r_2^\p; r] = \kappa \varepsilon_{j_1} \otimes r_{j_1j_2} \otimes r_{j_2j_4} \otimes \varepsilon_{j_4}^*
\in \CE_{j_1} \otimes R_{J_{\infty}} \otimes \CE_{j_4}^*
\ee
where $J_\infty = \{ j_1,j_2,j_4 \}$ and $\kappa$ is a scalar given by
\be
\begin{split}
\kappa & = (-1)^s \cdot \left( \varepsilon_{j_3}^*(\varepsilon_{j_3}')\right)\cdot \left( K_{j_3j_4}(r_{j_3j_4},r_{j_4j_3})\right)\cdot
\left( K_{j_2j_3}(r_{j_2j_3}, r_{j_3j_2}) \right)\\
s & = 1 + \vert r_1^\p\vert + \vert r_{j_2j_4} \vert (\vert r_{j_4 j_3} \vert + \vert r_{j_3j_2} \vert ) + \vert r_{j_2j_3} \vert. \\
\end{split}
\ee
\bigskip
\bigskip

With this definition in hand, we can check that $\rho$ behaves just like $T$ as far as convolutions are involved.
Since the combinatoric structure is the same as for $T$, we can focus on the signs.
First, we can look at:
\be\label{eq:rho-compat-2}
\rho(\fu * \fw)[P;S] =
\sum_{{\rm Sh}_2(S)}   \epsilon_{S_1,S_2} ~   \rho(\fu)[P;
\rho(\fw)[  S_1 ],  S_2].
\ee
The orientations in the numerators appear in the same way on both sides of the equation.
The $K$ factors in $\rho(\fw)$ on the right hand side are inserted to the right of the $\theta_a r^\p_a$ factors in $\rho(\fu)$ and need to be brought to the
left in order to match the left hand side. This reproduces the reduced Koszul rule.
We also need to bring the $\partial_e$ factors in the denominator to the left of the $dy^a_\p$,
but this cancels against the sign to bring the $K$ factors to the left of the $\partial_{\theta_a}$.

Next, we can look at
\be\label{eq:rho-compat-1}
\rho(\fu * \fu')[P;S] = \\
\sum_{{\rm Sh}_2(S), {\rm Pa}_3(P)} \epsilon ~    \rho(\fu)[ P_1,
\rho(\fu')[P_2 ; S_1 ], P_3; S_2].
\ee
To compare the right hand side to the left hand side, we need to transport the $K \partial_\theta$ block in $\rho(\fu')$, together with the $\theta$
in front of it, to the left of the $P_1$ arguments $\theta_a r^\p_a$ in $\rho(\fu)$. We also need to transport the arguments
 $r_a$ of $S_1$  to the right of the
$P_3$ arguments $\theta_a r^\p_a$. This reproduces the reduced Koszul rule.
All denominator manipulations needed to reorganize the $\partial_e$ and $dy^a$ give signs which cancel out against the
identical manipulations on the $K$ and $\partial_{\theta_a}$.

Plugging in the usual convolution identities for taut elements, we derive the $LA_\infty$ relation for $\rho[\ft_\CH]$
analogous to \eqref{eq:big-rel}:
\be\label{eq:big-rel-rho}
  \sum_{{\rm Sh}_2(S), {\rm Pa}_3(P)} \epsilon ~  \rho(\ft_{\CH})[P_1,
\rho(\ft_{\CH})[P_2 ; S_1 ], P_3; S_2] \\
 +
\sum_{{\rm Sh}_2(S)}   \epsilon ~  \rho(\ft_{\CH})[ P;
\rho(\ft_p)[  S_1 ],  S_2] = 0  .
\ee

The most important consequence of these identities is that if we are given an interior amplitude $\beta$,
we   immediately receive an $A_\infty$ algebra with operations
\be\label{eq:R-AFTYALG}
\rho_\beta(\ft_\CH): TR^\p(\CE) \to R^\p(\CE)
\ee
defined by
\begin{equation}
\rho_\beta(\ft_\CH)[r^\p_1, \dots, r^\p_n] := \rho(\ft_\CH)[r^\p_1, \dots, r^\p_n;e^\beta]
\end{equation}
This is the main object of interest for us.
A useful point of view on this derivation is that because $\rho(\ft_p)[e^\beta]=0$,
any convolution of the form $\fu*\ft_p$ will give zero when inserted into $\rho_\beta$:
applying the convolution identities to $e^\beta$ we get
\begin{equation}
\rho(\fu * \ft_p)[P,e^\beta] =  \rho(\fu)[P;
\rho(\ft_p)[e^\beta],e^\beta]
\end{equation}

We are ready for the the half-plane analog of the interior amplitude:

\bigskip
\noindent
\textbf{Definition}

\noindent
a.) A \emph{boundary amplitude} in a Theory $\CT$ is an element $\CB \in R^\p(\CE)$
of degree $+1$ which solves the Maurer-Cartan equations
\be\label{eq:boundary-amp}
\sum_{n=1}^\infty \rho_\beta(\ft_\CH)[\CB^{\otimes n}] = 0.
\ee

\noindent
b.) A \emph{Brane} in a  Theory $\CT$ is a pair $\fB=(\CE,\CB)$
of Chan-Paton data $\CE$, together with a compatible  boundary amplitude $\CB$.
\footnote{We will often simply refer to a Brane $\fB$ by its boundary
amplitude $\CB$ when the Chan-Paton data are understood.}

\bigskip

We remark that equation \eqref{eq:boundary-amp} is a sum of
elements of degree $2$. Note that
 we can also define formally \footnote{The reader is cautioned about a possible notational confusion. Late on, we will introduce an identity element $\Id$.
 Given a multilinear function $f$, the first term in the expansion
 of some $f(X,\frac{1}{1-\CB},Y)$ is $f(X,Y)$, {\it not} $f(X,\Id,Y)$ }
\begin{equation}
\frac{1}{1-\CB} = \sum_{n=0}^{\infty} \CB^{\otimes n}
\end{equation}
and write the equation as
\be\label{eq:boundary-amp2}
\rho_\beta(\ft_\CH)[\frac{1}{1-\CB}] = 0.
\ee
%

%\bigskip
%\noindent
%\textbf{Definition} A \emph{Brane} in some given theory consists of a choice of Chan-Paton factor together with a
%boundary amplitude for the corresponding $A_\infty$ algebra.
%
%\cg{The definition of boundary amplitude already has the CP
%factors in it. I think these are the same and we should
%just have one definition here.}
%

\bigskip
\noindent
\textbf{Remarks}:
\begin{enumerate}

\item In conformal field theory the term ``brane'' is often used for
conformally invariant boundary conditions consistent with a given
conformal field theory $\CC$. These branes form a category. In our
context we think of a boundary amplitude as a boundary condition
and indeed in the context of Landau-Ginzburg theories, as described in
Sections \S\S \ref{lgassuper}-\ref{halfspace} below, we will see that boundary conditions
indeed provide a boundary amplitude.
We will see that, for a fixed Theory $\CT$, the boundary amplitudes,
or equivalently the different Branes, also form a category.

\item The higher operations
\begin{equation}
\rho_\beta(\ft_\CH)[P;S] := \rho(\ft_\CH)[P;S,e^\beta]
\end{equation}
still satisfy $LA_\infty$ relations. They will not play a further role for us.

\item As for $\beta$, we can use $\CB$ to ``shift the origin'' in the $A_\infty$ algebra.
The operations $\rho^\CB_\beta(\ft_\CH): TR^\p(\CE) \rightarrow R^\p(\CE)$ defined by
\begin{equation}
\rho^\CB_\beta(\ft_\CH)[r^\p_1, \dots, r^\p_n]  = \rho_\beta(\ft_\CH)[\frac{1}{1-\CB}, r^\p_1,\frac{1}{1-\CB} \dots,\frac{1}{1-\CB}, r^\p_n,\frac{1}{1-\CB}]
\end{equation}
again satisfy the $A_\infty$-relations if $\CB$ is a boundary amplitude. The proof is similar
to that of \eqref{eq:L-infty-rhoz}.
We will identify this $A_\infty$ algebra in \S \ref{subsec:BraneCat}
 with an $A_\infty$ algebra of endomorphisms $\mathrm{Hom}(\CB,\CB)$ (see \eqref{eq:BraneMultiplications}).

%\item It is easy to promote the Chan-Paton factors from graded vector spaces to complexes $(\CE_i, d_i)$:
%define $d^\p$ to coincide with $d_i$ when acting on $\CE_i$ and add
%\begin{equation}
%r^\p \to d^\p r^\p + (-1)^{{\rm deg}(r^\p)} r^\p d^\p
%\end{equation}
%to the first operation in the $A_\infty$ algebra.
%Alternatively, one can simply include boundary vertices with no outgoing edges in the taut webs, i.e. fans with a single vacuum. Then one can add the $d_i$ as the component of $\CB$
%associated to the empty fans.
%
%\cg{I don't understand the last sentence here. Why is $\CB$ part of the differential? How
%do we define "taut" if there are bivalent vertices on the boundary? }
%

 \end{enumerate}

\subsection{Web Representations And Strip-Webs}\label{subsec:WebRepStrip}

Now we will explore what implications a representation of webs
has when combined with strip webs.
Suppose we are given
a web representation $\CR = (\{ R_{ij} \}, K)$
 and Chan-Paton factors $\CE_{L,i}$ and $\CE_{R,i}$ for the left and right boundaries of the strip.
We will denote the fans for the left boundary as $J$ and the fans for the right boundary
as $\tilde J$, with corresponding spaces $R_J(\CE_L)$ and $R_{\tilde J}(\CE_{R})$.
The direct sum over positive- and negative- half-plane fans with these Chan-Paton spaces
will be denoted as $R^\partial_L(\CE_L)$ and $R^\partial_R(\CE_R)$, respectively.

\bigskip
\noindent
\textbf{Definition:} We define the  space of
\emph{approximate ground states} to be
\be\label{eq:ELRdef}
\CE_{LR} := \oplus_{i \in\IV} \CE_{L,i} \otimes \CE_{R,i}^*
\ee
and a typical element is denoted by $g$.

Given a strip web $\fs$ we plan to define an operation
\begin{equation}\label{eq:Strip-Contract}
\rho[\fs]: TR^\p_L(\CE_L) \otimes T\Rvtx \otimes \CE_{LR} \otimes TR^\p_R(\CE_R) \to \CE_{LR}
\end{equation}
As usual, we take this to be zero unless all the arguments are compatible with the appropriate vertices of the strip web
and defined by a familiar formula otherwise:
\begin{align}
&\rho(\fs)[r^\p_1, \dots, r^\p_m;r_1, \dots, r_n;g;\tilde r^\p_1, \dots, \tilde r^\p_s] = \cr
&\frac{o(\fu)}{\left[ \prod_{e \in \CE(\fs)} \p_e \right]\circ\left[\prod_{a=1}^m dy^a_\parallel\right]\left[  \prod_{v \in \CV_i(\fs)} dx_v dy_v\right] \left[\prod_{a=1}^s d\tilde y^a_\parallel\right]} \cr
&\left[\otimes_{e \in \CE(\fs)} K_e \right] \left[\prod_{a=1}^m \partial_{\theta_a} \right]\left[\prod_{a=1}^s \partial_{\tilde \theta_a} \right]
\circ \left[\otimes_{a=1}^m \theta_a r^\p_a \right] \left[ \otimes_{a=1}^n r_a\right] \otimes g \left[\otimes_{a=1}^s \tilde \theta_a \tilde r^\p_a \right]
\end{align}
Recall that, reading from left to right the $r_j^\p$ are inserted on the boundary in order of decreasing $y$ while the
$\tilde r_j$ are inserted in order of increasing time.

\begin{figure}[htp]
\centering
\includegraphics[scale=0.3,angle=0,trim=0 0 0 0]{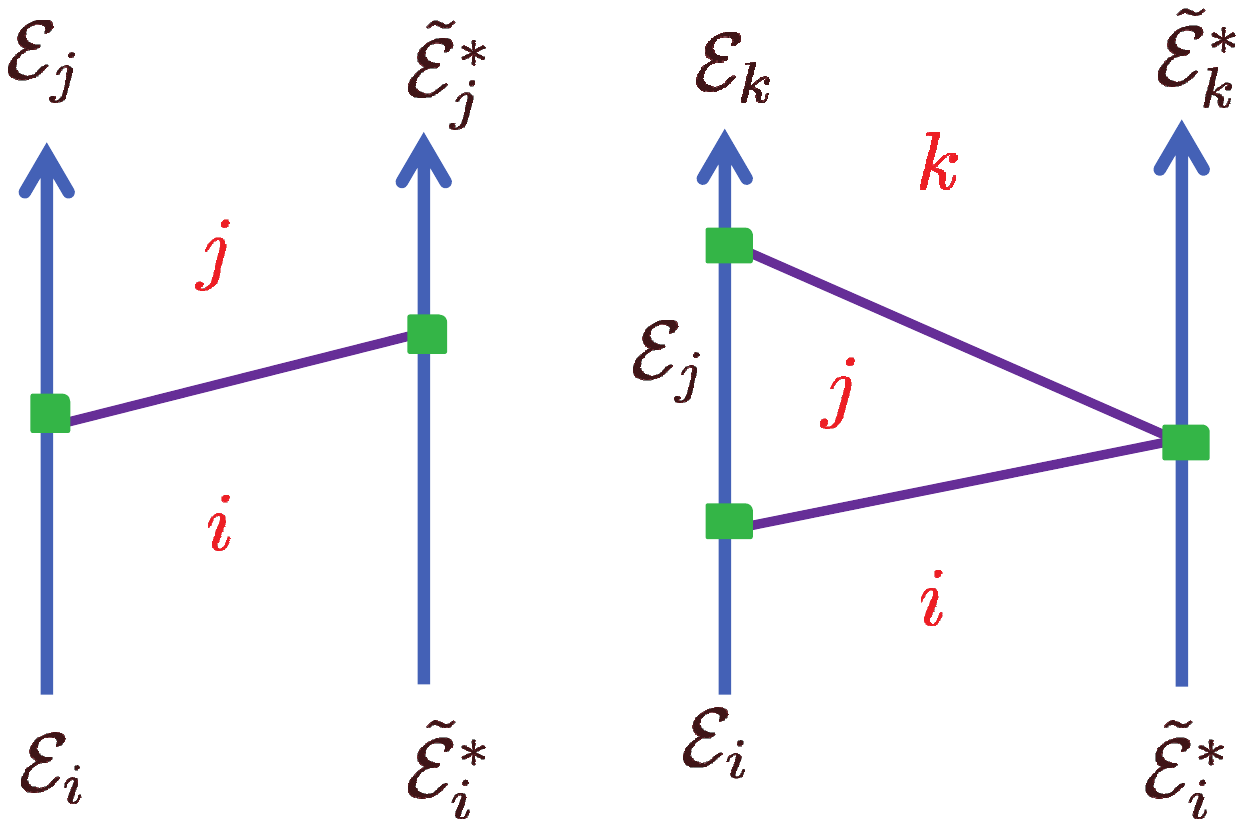}
\caption{Strip webs whose contractions are described in the text.  }
\label{fig:STRIPWEB1}
\end{figure}

\bigskip
\noindent
\textbf{Example}: The contraction associated with the strip-web
on the left in Figure \ref{fig:STRIPWEB1} maps
\be
R_{ji}(\CE) \otimes (\CE_i \otimes \tilde \CE_i^*) \otimes R_{ij}(\tilde \CE) \to \CE_j \otimes \tilde \CE_j^*
\ee
It operates on a typical primitive tensor via
\be
(v_j \otimes r_{ji} \otimes v_i^*) \otimes (v_i' \otimes \tilde v_i^{*'})
\otimes (\tilde v_i \otimes \tilde r_{ij} \otimes \tilde v_j^*)
\mapsto \pm (v_i^*\cdot v_i') ( \tilde v_i^{*'}\cdot \tilde v_i) K(r_{ji},\tilde r_{ji}) v_j \otimes \tilde v_j^*
\ee
where the superscript $*$ indicates a vector is in $\CE^*$ and the   sign is determined by the Koszul rule.
Similarly, the   strip-web
on the right in Figure \ref{fig:STRIPWEB1} maps
\be
\biggl( R_{kj}(\CE) \otimes   R_{ji}(\CE)\biggr) \otimes
(\CE_i \otimes \tilde \CE_i^*) \otimes R_{ijk}(\tilde \CE) \to \CE_k \otimes \tilde \CE_k^*.
\ee
It operates on a typical primitive tensor via
\be
\begin{split}
(v_k \otimes r_{kj} \otimes v_j^*) \otimes (v_j' \otimes r_{ji} \otimes v_i^{*'})\otimes (v_i \otimes \tilde v_i^{*})
& \otimes (\tilde v_i' \otimes \tilde r_{ij}\otimes \tilde r_{jk} \otimes \tilde v_k^*)\\
\mapsto \pm (v_j^*\cdot v_j') (v_i^{*'}\cdot v_i)
&
( \tilde v_i^{*}\cdot \tilde v_i') K(r_{kj},\tilde r_{jk})
K(r_{ji},\tilde r_{ij})  v_k \otimes \tilde v_k^*\\
\end{split}
\ee
where the sign is determined by the Koszul rule.

\bigskip

The full map $\rho(\fs)$ satisfies the same compatibility relations with convolutions as $T(\fs)\circ$, which combined with the convolution identities for the taut elements
tell us that $\rho(\ft_s)$ satisfies the same lengthy algebraic relations as $T(\ft_s)\circ$ ,
described in Section \S \ref{subsec:Bimodule-StripWeb}.
%
%\cg{refer to appropriate formulae. Appendix?}
%\cg{Yeah, we should probably write them out.}
%

Now let us  select a specific choice of interior amplitude
$\beta$, together with left and right  boundary
amplitudes $\CB_L$ and  $\CB_R$, respectively,  and
define an  operator
$d_{LR}: \CE_{LR} \to \CE_{LR}$   by the equation
\begin{equation}\label{eq:strip-diff-def}
d_{LR}: g \mapsto \rho(\ft_s)[\frac{1}{1-\CB_L};e^\beta; g; \frac{1}{1-\CB_R}].
\end{equation}
The $\ft_s \circ \ft_s + \cdots =0$ identity \eqref{eq:Strip-Web-Ident} reduces to the crucial nilpotency
\begin{equation}
d_{LR}^2=0
\end{equation}
essentially because all other terms in the convolution identity give zero when evaluated
 on $\frac{1}{1-\CB_L}$, $e^\beta$, and $ \frac{1}{1-\CB_R}$. That is, $d_{LR}$ is a differential on the complex $\CE_{LR}$.

We believe these considerations amply justify the
following

\bigskip
\noindent
\textbf{Definition:} The \emph{complex of ground states} associated to a left and right brane in a given theory
is $(\CE_{LR}, d_{LR})$. The cohomology of this complex gives us the \emph{exact ground states} for this system.

\bigskip
\noindent
\textbf{Remarks}:
\begin{enumerate}

\item When our formalism is applied to physical theories and physical branes
the above definition coincides with the
physical notion of groundstates, thus realizing one of the primary objectives of the
introductory section \S \ref{sec:Introduction}.

\item We can define operations
\begin{equation}
\rho_{\beta,R}(\ft_s)[P;g] := \rho(\ft_s)[P;e^\beta; g; \frac{1}{1-\CB_R}]
\end{equation}
These endow $\CE_{LR}$ with the structure of an $A_\infty$ left module for $\rho_{\beta}(\ft_L)$.
Similarly,
\begin{equation}\label{eq:rho-beta-strip}
\rho_{\beta}(\ft_s)[P;g;P'] := \rho(\ft_s)[P;e^\beta; g; P']
\end{equation}
defines an $A_\infty$ bimodule structure on $\CE_{LR}$.

%\item If the Chan-Paton factors are promoted to complexes $(\CE_{L,i}, d_{L,i})$, $(\CE_{R,i}, d_{R,i})$,
%we should add $\oplus_{i \in \IV} d_{L,i} \otimes 1 + 1 \otimes d^*_{R,i}$ to the differential $d_{L,R}$.
 \end{enumerate}

\subsection{On Degrees, Fermion Numbers And R-Symmetry}\label{subsec:OnDegrees}

Throughout this section, and in later sections, we define the $R_{ij}$ as graded vector spaces,
with a $\IZ$ valued degree which determines the Grassmann parity of objects and allows us to use
the Koszul rule in our manipulations.

The integral grading of objects such as the complex of ground states $\CE_{LR}$, the
$R^{\mathrm{int}}$ and $R^\partial[\CE]$ spaces used in defining the interior and boundary amplitudes
should be canonically well-defined, as these objects are expected to be in correspondence to objects in a physical theory
which have a well-defined, integral grading given by the conserved R-charge.

On the other hand, the individual $R_{ij}$ and $\CE_i$ spaces are expected to be in correspondence
with objects in a physical theory for which the definition of R-charge is possibly not integral and ambiguous,
due to contributions from boundary terms at infinity. Concretely, the R-charge operators $\hat q_{ij}$ and $\hat q_i$ on $R_{ij}$ and $\CE_i$
are defined up to a constant shift
\begin{equation}\label{eq:shift}
\hat q_{ij} \to \hat q_{ij} + f_i - f_j  \qquad  \hat q_i \to \hat q_i - f_i
\end{equation}
which leaves the R-charges of $\CE_{LR}$, $R^{\mathrm{int}}$ and $R^\partial[\CE]$ invariant.

When we attempt to associate a web representation to a certain physical theory, we can always select some choice of $f_i$
such that the R-charges are integral, and can be used to define integral degrees. Such a choice, though, it is not unique, and
may break some symmetry of the theory. Different choices are related by shifts with integral $f_i$.

As changes in degrees affect the Koszul rules, a shift in degree in the $R_{ij}$ will lead to sign changes in the definition of $\rho$. Furthermore, it may affect the
signs in the MC equations for interior and boundary amplitudes. In order for our algebraic structures to behave well under degree shifts, we would like to
be able to reabsorb such signs into sign redefinitions in the $\CE_{LR}$, $R^{\mathrm{int}}$ and $R^\partial[\CE]$ spaces and in the $K$ pairing.

More precisely, we would like to define a new web representation in terms of some ${}^\vee R_{ij}$ isomorphic, perhaps not canonically, to the degree-shifted
$R_{ij}^{[f_i - f_j]}$, and CP factors ${}^\vee \CE_i$ isomorphic, perhaps not canonically, to the degree-shifted
$\CE_i^{[-f_i]}$ such that we have canonical isomorphisms
\begin{equation}
\CE_{LR} \cong {}^\vee \CE_{LR} \qquad  R^{\mathrm{int}} \cong {}^\vee R^{\mathrm{int}} \qquad R^\partial[\CE] \cong {}^\vee R^\partial[\CE]
\end{equation}
which intertwine between $\rho$ defined by the original representation, and ${}^\vee \rho$ defined by the new representation and map interior and boundary amplitudes for the original representation
to interior and boundary amplitudes for the new representation.

There is a natural, physical way to find such maps, but the story has an unexpected twist: in order to relate
naturally $\rho$ and ${}^\vee \rho$, we need to also act with an automorphism of the web algebra,
i.e. a linear map $f_{\CW}: \fw \to \fw$ which commutes with all web convolution operations.
Such a map will {\it not}, in general, preserve the taut element and thus will {\it not}
map interior amplitudes to interior amplitudes, except in some special cases we will describe below,
unless we generalized the notion of taut element and interior amplitude slightly.

Lets first describe our degree-shift maps. The maps will act as $\pm 1$ on each summand $\CE_i\otimes \CE_i^*$, $R_I$, $R_J[\CE]$.
Consider some one-dimensional graded vector spaces $V_i$ of degree $f_i$ and their duals $V^*_i$, with a canonical isomorphism $V^*_i \otimes V_i \cong \IZ$.
Define
\begin{equation}
 {}^\vee R_{ij} = V_i \otimes R_{ij} \otimes V^*_j \qquad  {}^\vee \CE_i = \CE_i \otimes V^*_i
\end{equation}

We can focus on $R^{\mathrm{int}}$ and the plane web representation. The same analysis holds for half-plane and strip web representations.
Consider the $ {}^\vee R_I$. We can apply the canonical isomorphism $V^*_i \otimes V_i \cong \IZ$
to relate canonically
\begin{equation}
 {}^\vee R_{i_1 \cdots i_n} \cong V_{i_1} \otimes R_{i_1 \cdots i_n} \otimes V^*_{i_1}
\end{equation}
We can then define a canonical isomorphism  ${}^\vee R_{i_1 \cdots i_n}\cong R_{i_1 \cdots i_n}$ by Koszul-commuting
$V_{i_1}$ all the way to the right and applying the canonical isomorphism. It is easy to see that such canonical isomorphism
makes a neat commutative diagram with the isomorphisms $R_{i_1 i_2\cdots i_n} \cong R_{ i_2\cdots i_n i_1}$ and
${}^\vee R_{i_1 i_2\cdots i_n} \cong {}^\vee R_{ i_2\cdots i_n i_1}$ defined in \ref{subsec:WebRepPlane}.

We should define ${}^\vee K$ as well. Of course, ${}^\vee R_{ij} \otimes {}^\vee R_{ji}$ is canonically
isomorphic to $V_i \otimes R_{ij} \otimes R_{ij} \otimes V^*_i$. As the degrees of the middle factors add up to $1$, there is no sign to pay to bring $V_i$ all the way to the right
and apply the canonical isomorphism again to $R_{ij} \otimes R_{ij}$. Thus we can take ${}^\vee K$ to coincide with $K$ up to this canonical isomorphism.

Lets compose ${}^\vee \rho$ with the canonical isomorphisms: we take the arguments $r_a$ in $R_{I_a}$,
map them canonically to elements in ${}^\vee R_{I_a}$ and do our contractions with ${}^\vee K$,
which means we contract the $R_{ij}$ elements with $K$ and the $V_i$, $V^*_i$ pairwise
according to the same pattern.

Effectively, the only difference between ${}^\vee \rho$ and $\rho$ is the composition of a bunch of
canonical ``pair creation'' maps $\IZ \to V_i^* \otimes V_i$ and ``annihilation'' maps $V^*_i \otimes V_i \cong \IZ$,
along a pattern dictated by the topology of the web. It is easy to see that the chain of contractions produces a loop for every internal face of the web.
Thus ${}^\vee \rho$ and $\rho$ differ by a factor of $\prod_{i \in \mathrm{faces}[\fw]} (-1)^{f_i}$.

We can absorb the difference into a linear map
\begin{equation}
\fw \to f[\fw] \fw \equiv \left[ \prod_{i \in \mathrm{faces}[\fw]} (-1)^{f_i }  \right] \fw
\end{equation}
Thus the web representation transforms canonically under the degree shift combined with the action of this
linear map on the space of webs.

The sign $f[\fw] $ has a striking property:
\begin{equation}
f[\fw_1 *_v \fw_2] = f[\fw_1] f[\fw_2]
\end{equation}
as convolution does not create new internal faces. Thus the map commutes with all web algebraic operations.
We can call a collections of numbers with such property a {\it cocycle} for the web algebra.

It should be clear that the taut element $\ft$ is not invariant under twisting by
a general cocycle $\sigma[\fw]$:
\footnote{A notable exception is a case of vacuum weights which
form a convex polygon: the special cocycles are trivial because
there are no internal faces. Half-plane and strip taut webs may have internal faces bounded by
one of the boundaries of the space, but the extra sign can be reabsorbed in a re-definition of
$R^\partial[\CE]$. This will be important in later examples}
\begin{equation}
\ft \to \ft_\sigma = \sum_{\mathrm{taut} \fw} \sigma[\fw] \fw.
\end{equation}

On the other hand, the twisted taut element $\ft_\sigma$ is still nilpotent,
and we could extend our definition of theory by replacing $\ft$ with $\ft_\sigma$
in our definition of interior amplitudes, etc. Although we will suppress this possibility in the remainder of the paper,
it is likely relevant to concrete applications.

Our final statement is that degree shifts $f_i$ in the $R_{ij}$ relate canonically
a theory associated to a cocycle $\sigma$ and a theory associated to a cocycle
$\sigma f$.

\begin{figure}[htp]
\centering
\includegraphics[scale=0.3,angle=0,trim=0 0 0 0]{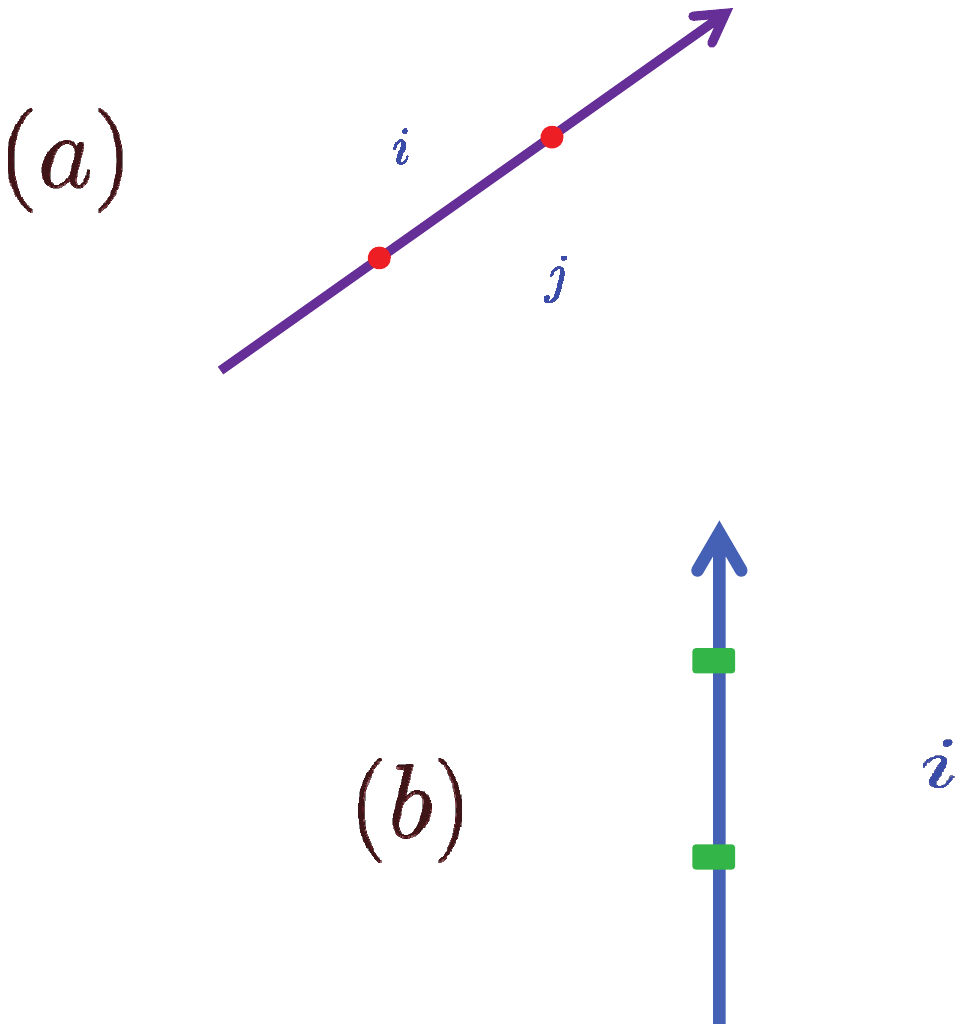}
\caption{In figure (a) we show an extended taut planar web. The contribution of this
web to the equation for an interior amplitude shows that such an (extended) interior
amplitude can be used to define a differential on $R_{ij}$. Similarly, in figure (b)
we show a taut extended half-plane web. Its contribution to the Maurer-Cartan equation
for the corresponding $A_\infty$ algebra shows that a component of the (extended)
boundary amplitude defines a differential on $\CE_i$.   }
\label{fig:EXTENDED-DIFFERENTIAL}
\end{figure}
\begin{figure}[htp]
\centering
\includegraphics[scale=0.3,angle=0,trim=0 0 0 0]{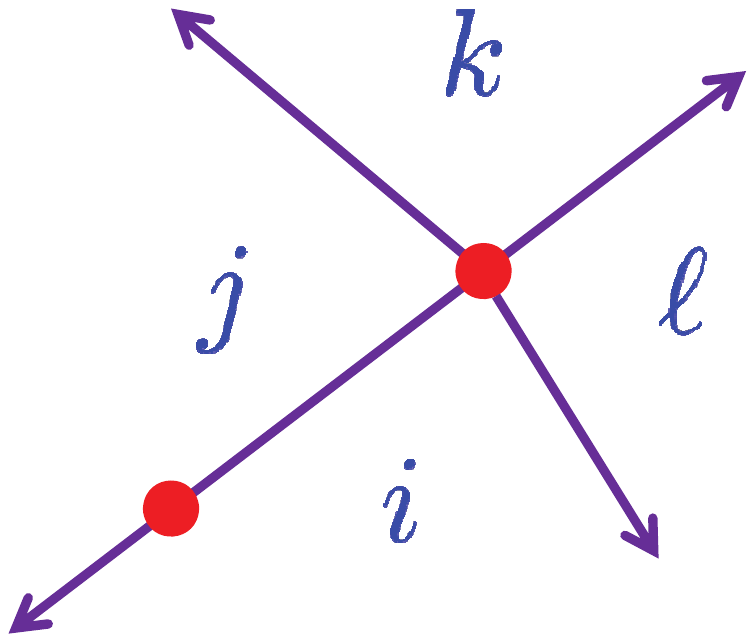}
\caption{A bivalent vertex can be added to any leg of any vertex to produce a
taut extended web, as shown here. }
\label{fig:EXTENDED-DIFFERENTIAL-2}
\end{figure}

\subsection{Representations Of Extended Webs}\label{subsec:RepExtendedWebs}

We can extend the definition of web representations to extended webs.
For plane webs, we have new fans available, with two vacua only,
and associated vector spaces
\begin{equation}
R_{(ij)} = R_{ij} \otimes R_{ji}
\end{equation}
The interior amplitude includes now a component $\beta_{ij}$ in $R_{(ij)}$. We can use $K$ to ``raise an index'' of $\beta_{ij}$
to define a degree $1$ map
\begin{equation} Q_{ij}: R_{ij} \to R_{ij}
\end{equation}
by
\be
Q_{ij}(r_{ij}) := (1\otimes K_{23}) ( \beta_{ij}\otimes r_{ij}).
\ee
where the subscript $23$ means that $K$ is contracting the second and third factors in $R_{ij}\otimes R_{ji}\otimes R_{ij}$.
The equation satisfied by $\beta$ implies that $Q_{ij}$ is a nilpotent differential, making $R_{ij}$ into a complex.
This is illustrated in Figure \ref{fig:EXTENDED-DIFFERENTIAL}(a).
\footnote{To give a little more detail: The interior amplitude identity says that $K_{23}(\beta_{ij}\otimes \beta_{ij})=0$,
where again the subscript $23$ indicates which factors the $K$ acts upon.
Then, using similar notation,  to check $Q_{ij}^2=0$ we need to verify $K_{23}( \beta_{ij} \otimes K_{45}(\beta_{ij}\otimes r_{ij}))=0$.
Since $K_{23}$ and $K_{45}$ act on different spaces and hence (anti)commute
we can first contract $K_{23}(\beta_{ij}\otimes \beta_{ij})$ to get zero. }
As the two-valent interior vertex only appears in taut plane webs with two vertices, the equations for $\beta$
differ from the standard case only by terms where some $Q_{ij}$ acts on an external $ij$ leg of $\beta$.
The relevant kinds of taut webs are illustrated in Figure \ref{fig:EXTENDED-DIFFERENTIAL-2}.

For half-plane extended webs, we have a single-vacuum fan available, and
associated vector spaces
\begin{equation}
R_{i}(\CE) := \CE_i \otimes \CE_{i}^*
\end{equation}
associated to a half-plane ``web'' consisting of one vertex on the boundary, with no ingoing lines.
When working with extended webs   the definition of $R^\p(\CE)$ in \eqref{eq:Rbd-def} now
reads
\be\label{eq:extwebrep}
R^\p(\CE) = \oplus_i \CE_i \otimes \CE_i^* \oplus \oplus_{z_{ij}\in \CH}
\left( \CE_i \otimes R_{ij}\otimes \CE_j^* \right) \oplus \cdots
\ee
When speaking of elements of $R^\p(\CE)$ we refer to the new summand $\oplus_i \CE_i \otimes \CE_i^*$
in the definition of $R^\p(\CE)$  as the \emph{scalar part}.
Thus the boundary amplitude includes now a scalar part $Q_i$ in $R_{i}(\CE)$. The Maurer-Cartan equation
 now includes a taut web with two zero-valent vertices and this contribution requires
 the scalar part $Q_i$ to be a differential on $\CE_i$, making the Chan-Paton factors into a complex. See
Figure \ref{fig:EXTENDED-DIFFERENTIAL}(b).  Moreover,
 the zero-valent boundary vertices only appear in taut half-plane webs with two vertices,
so the equations for $Q_i + \CB$ differ from the equations for $\CB$ with
unextended webs only by an anti-commutator between some $Q_i$ and $\CB$.

It is also interesting to observe that we can define an element $\Id_i$ as the
canonical identity element in   $R_{i}$. Then we set
\be\label{eq:GdId}
\Id := \oplus_i \Id_i ~.
\ee
 The $\Id$ element
behaves as a graded identity for the $A_\infty$ algebra $\rho_\beta(\ft_\CH)$, i.e.
$\rho_\beta(\ft_{\CH})[\Id] =0$, (since there is no taut web with a
single boundary vertex)  while, using the conventions of
\eqref{eq:bdy-rho-signs},
\be\label{eq:GdId-p}
\rho_\beta(\ft_{\CH})[\Id,r]=r \qquad\qquad \rho_\beta(\ft_{\CH})[r,\Id] = (-1)^{\vert r\vert} r ,
\ee
while $\rho_\beta(\ft_{\CH})[P_1,\Id, P_2] =0$ if both $P_1,P_2$ are nonempty
(simply because there are no taut webs of the appropriate kind).   This feature alone can make extended webs useful.

%
%\cg{With graded identity I mean that I think $\rho_\beta(\ft_\CH)[\Id,r]$ and $\rho_\beta(\ft_\CH)[r,\Id]$ have opposite signs! We should parse the %definitions
%to see which one is $r$, and which $-r$.}
%

For extended strip webs, the new vertices only appear in simple taut (=rigid) webs consisting of a single zero-valent vertex
on either boundary. Thus we should simply add \begin{equation}\oplus_{i \in \IV} \left[ Q_{L,i} \otimes 1 + 1 \otimes Q_{R,i} \right]\end{equation} to the differential $d_{L,R}$.

\subsection{A Useful Set Of Examples With Cyclic Vacuum Weights}\label{subsec:CyclicVacWt}

We will now describe an infinite family of non-trivial examples of Theories and Branes
which will be used again in Sections \S \ref{subsec:VacCat-SUN} and \S \ref{subsec:RotIntfc-TSUN}
to illustrate our formal constructions. As explained in Section \S \ref{subsubsec:RelationTSUN-Physical},
they are also of physical interest.

Fix a positive integer $N$ and consider the vacuum data
\begin{equation}\label{eq:CyclicWt}
\IV^N_\vartheta: z_k = e^{- \I \vartheta - \frac{2 \pi \I}{N} k} \qquad k = 0, \cdots N-1
\end{equation}
For convenience, in this section,
 we choose  a small positive $\vartheta$, so that $z_0$ has the most positive real part among all vacua, and
a small negative imaginary part. The vacua are a regular sequence of points on the unit circle ordered
in the clockwise direction. In particular they form a regular convex polygon and hence there are no webs with loops.
Let us  enumerate the rigid vertices. Note that   if    $i,j,k$ are three successive vacua in a cyclic fan then,
by our conventions, they label regions in the clockwise order and hence $z_{jk}$ rotates counterclockwise through
an   angle less than $\pi$ to point in the direction of $z_{ij}$. It follows that the
corresponding   vertices $z_i, z_j, z_k$ on the unit circle must be clockwise ordered.
From this we can conclude that the rigid vertices are in one-one correspondence with
increasing (reading left to right) sequences of numbers between $0$ and $N-1$. In particular,
the trivalent vertices are labeled by triples of vacua with $0\leq i<j<k\leq N-1$.

In what follows we
 will consider two examples of web representations $\CR$. We will analyze the resulting
  $L_\infty$ MC equation and, for a specific choice of interior amplitude $\beta$ we will
  analyze the  $A_\infty$ MC equations and Branes in these Theories.
While the development is a purely formal illustration of the mathematical constructions
developed above, the two classes of models are meant to correspond to two physical
models, as explained in Section \S \ref{subsubsec:RelationTSUN-Physical}  below.

\subsubsection{The Theories $\CT^N_{\vartheta}$   }

Our first class of Theories, denoted   $\CT^N_{\vartheta}$ have web representations
 such that  $R_{ij}$ is  one-dimensional space, with degree (or ``fermion number'')  $0$ or $1$:
\begin{align}\label{eq:ExpleTN-webrep}
R_{ij} &= \IZ^{[1]} \qquad i<j \cr
R_{ij} &= \IZ \qquad i>j
\end{align}
Moreover, we take  $K$ to be the natural degree $-1$ map $K: \IZ^{[1]} \otimes \IZ \to \IZ$ given by multiplication.
\footnote{One could introduce an extra multiplicative factor in the definition of $K$.
 Invertibility over the integers constrains it to be  $\pm 1$. One could extend it to
be a nonzero rational number and tensor the representations over $\IQ$. When counting the
$\zeta$-webs of Section \S \ref{zetawebs} it is natural to take the multiplicative factor
to be $1$.}

This   restriction on the degree has a neat consequence: the degree of $R_I$ equals $|I|-1$, where $|I|$ is the number of vacua in the fan $I$.
Thus the interior amplitude, which must have degree $2$,
 is concentrated on trivalent vertices only. We can therefore label the independent
 components of the interior amplitude by
 $\beta_{ijk} \in R_{ij} \otimes R_{jk} \otimes R_{ki}$ with $i<j<k$.
To complete the definition of the theory we must choose a specific interior amplitude. Therefore,
let us examine the $L_\infty$ MC equation.

\begin{figure}[htp]
\centering
\includegraphics[scale=0.3,angle=0,trim=0 0 0 0]{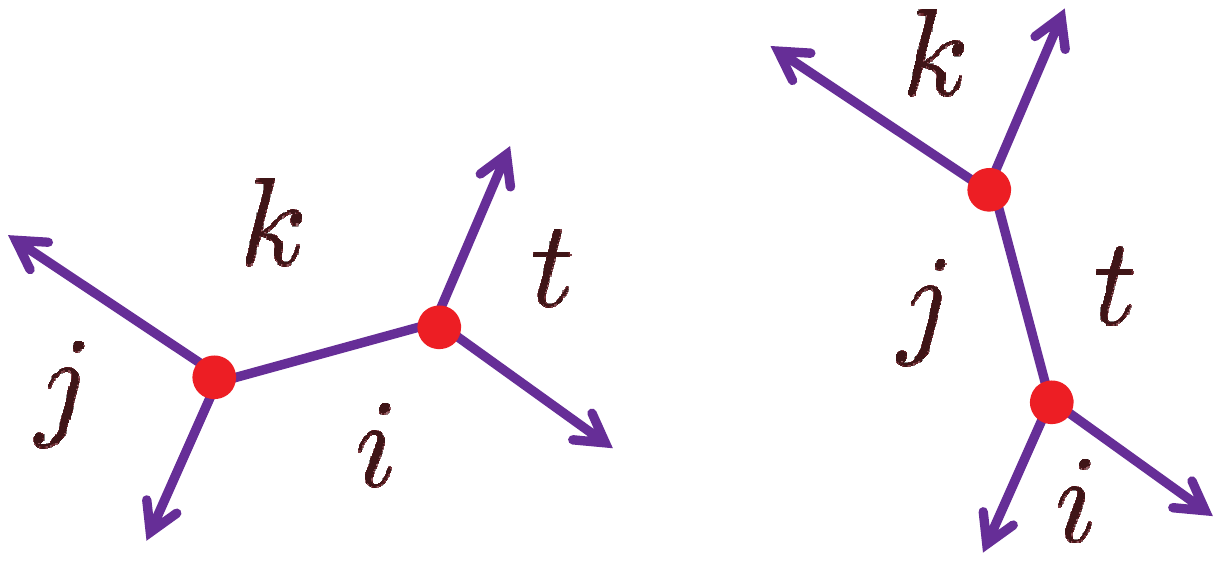}
\caption{The two terms in the component of the $L_\infty$ equations
for $i<j<k<t$.     }
\label{fig:TNEXAMPLE-1}
\end{figure}

The only taut webs with only trivalent vertices have four vacua at infinity. For each
increasing sequence of four vacua, $i<j<k<t$, there are two taut webs corresponding
to the two ways to resolve the 4-valent vertex as shown in Figure \ref{fig:TNEXAMPLE-1}. Therefore
the $L_\infty$ MC equation is a collection of separate equations, one for each such
increasing sequence,  of the form
\begin{equation}
\rho(\ft_p)[\beta_{ijk}, \beta_{ikt}] + \rho(\ft_p)[\beta_{ijt}, \beta_{jkt}] = K_{ik} \circ \left( \beta_{ijk} \otimes \beta_{ikt} \right) + K_{jt} \circ \left( \beta_{ijt} \otimes  \beta_{jkt} \right)=0
\end{equation}
where $K_{ik}$ is the contraction of the $R_{ik}\otimes R_{ki}$ factors and so forth.
We used the fact that for a canonically oriented   taut web $\fw$ with two vertices,
\begin{equation}
\partial_e \circ \left( dx_1 dy_1 dx_2 dy_2 \right) = dx_1 dy_2 d(y_1-y_2) = o(\fw)\end{equation}
to deal with the orientation ratio in $\rho$. (See Section \S \ref{subsec:WebRepPlane} for the
definition of $\partial_e$.)

In order to compute the relative signs in the two terms of the MC equation,
we remember that
\begin{equation}
\beta_{ijk} \otimes \beta_{ikt} \in R_{ij} \otimes R_{jk} \otimes R_{ki} \otimes R_{ik} \otimes R_{kt} \otimes R_{ti}
\end{equation}
and thus $K_{ik}$ has to go through two degree $1$ spaces $R_{ij} \otimes R_{jk}$ to contract $R_{ki} \otimes R_{ik}$, while
\begin{equation}
\beta_{ijt} \otimes  \beta_{jkt}   \in R_{ij} \otimes R_{jt} \otimes R_{ti} \otimes R_{jk} \otimes R_{kt} \otimes R_{tj}
\end{equation}
As $R_{tj}$ has degree $0$, we can carry it through the other factors to the right of $R_{jt}$, and then $K_{jt}$ has to go through a single degree $1$
factor $R_{ij}$ in order to contract $R_{jt} \otimes R_{tj}$. Therefore, if we identify now $\beta_{ijk}$ with an integer $b_{ijk}$
then the $L_\infty$ MC equations becomes the system of quadratic equations:
\be\label{eq:ExpleMC-1}
b_{ijk} b_{ikt} - b_{ijt} b_{jkt} = 0  \qquad  i<j<k<t
\ee
For example, for $N=4$ we have a single equation familiar from the conifold. On the other hand,
for large $N$ there are more equations than variables so it is nontrivial to have solutions
at all, let alone integral solutions.
\footnote{From the relation to Landau-Ginzburg theory discussed in
Section \S \ref{zetawebs} we expect integral solutions.}
Fortunately there is a   simple canonical solution given by
 $b_{ijk} =1$ for all $i<j<k$.  This choice of interior amplitude $\beta$ defines the Theory we will call
  $\CT^N_\vartheta$.

Next, we can look at half-plane webs and Branes. In general  the half-plane taut element
will depend on the relative choice of $\vartheta$ and of the slope for the half-plane $\CH$. We have
already chosen $\vartheta$ to be small and positive and we will now choose $\CH$ to be the positive
half-plane.

\begin{figure}[htp]
\centering
\includegraphics[scale=0.3,angle=0,trim=0 0 0 0]{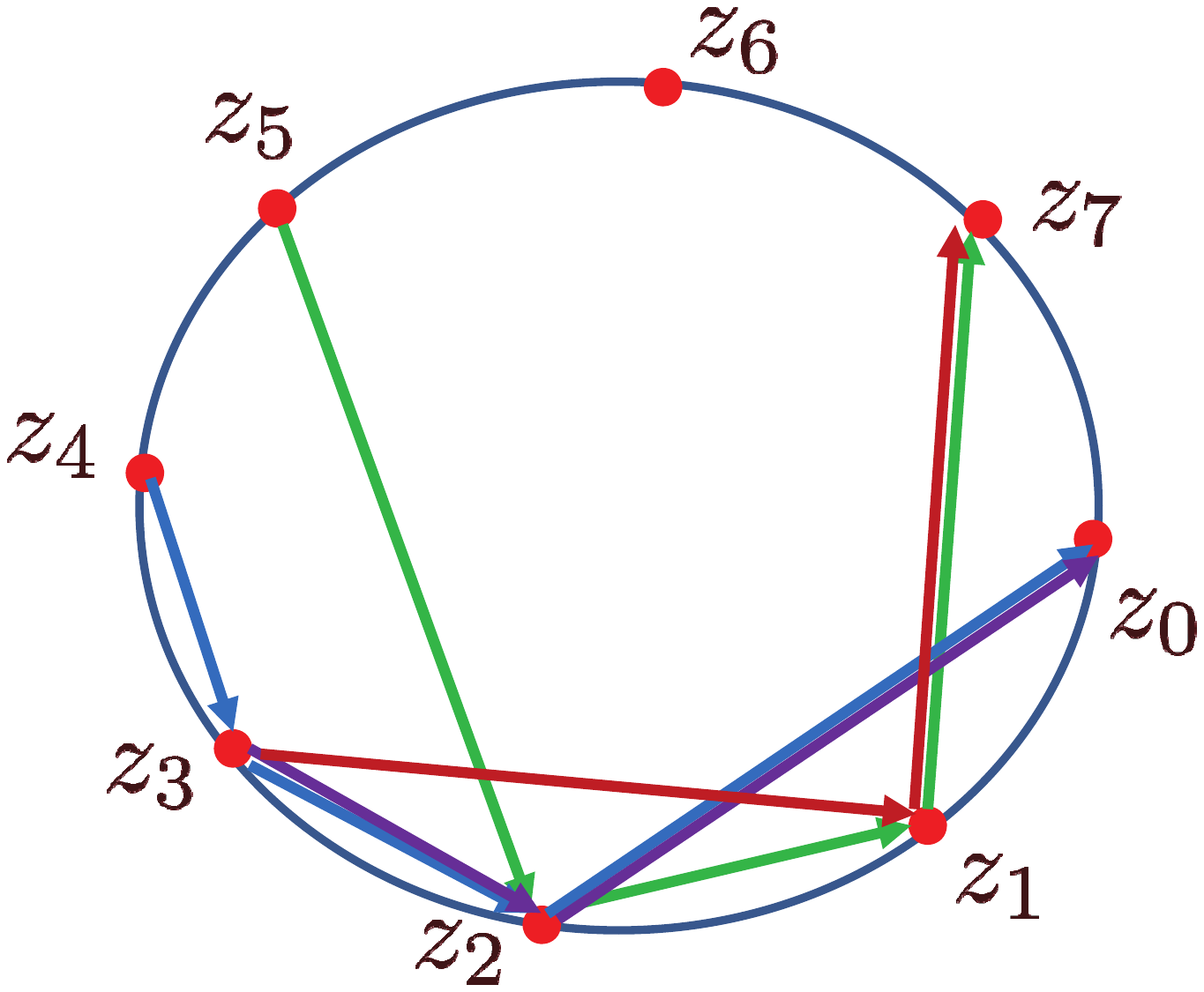}
\caption{Paths defining positive-half-plane fans for the cyclic weights \protect\eqref{eq:CyclicWt}.
%
%\eqref{eq:CyclicWt}.
%
These can
be divided into four types according to whether the vacua in $J = \{ i, \dots, j\}$ are down-type vacua
$i,j \in [0,\frac{N}{2})$ or up-type vacua $i,j\in [\frac{N}{2}, N-1]$. (We use the fact that
$\vartheta$ is small and positive here.)  Reading
the paths in the direction of the arrows gives the sequence of vacua encountered reading the fan in
the counterclockwise direction, and this corresponds to reading the vacua in $J = \{ i, \dots, j\}$
from right to left. Shown here is the case $N=8$.
The vacua $z_0,\dots, z_3$ are lower vacua.
The vacua $z_4,\dots, z_7$ are upper vacua. The green path is of type $u\dots u$. The maroon path is of
type $u\dots d$. The blue path is of type $d \dots u$. Finally the purple path is of type $d\dots d$.     }
\label{fig:TNEXAMPLE-2}
\end{figure}

Let us now enumerate the possible positive half-plane fans $J=\{ j_1, \dots, j_s \}$, where
we recall that reading from left to right we encounter the vacua in the clockwise direction.
The presence of the boundary breaks the cyclic symmetry so that when speaking of half-plane
webs and Branes it is very useful to distinguish   ``upper vacua'' or ``up vacua'' from   ``lower vacua''
or ``down vacua.'' The up vacua have positive imaginary part and the down vacua have negative imaginary
part. Thus, the down vacua correspond to $i$ with $0\leq i < \frac{N}{2}$ and the up
vacua correspond to $i$ with $\frac{N}{2} \leq i < N$.

In order to enumerate the positive half-plane fans we begin by noting
 two points: First, $z_{j_{p}, j_{p+1}}$ has to have positive real part,
and hence $\Re(z_{j_p}) > \Re( z_{j_{p+1}})$. Second, $z_{j_{p}, j_{p+1}}$ must
rotate counterclockwise to point in the direction of $z_{j_{p-1}, j_{p}}$.
Now read the list of vacua counterclockwise, i.e. from bottom to top, and from
\emph{right to left} in $J$. The vacua then define a path of points on the unit circle,
and  the half-plane fans are enumerated
by paths which move to the right so that the successive segments rotate counterclockwise.
One way to classify these paths is the following:
Denoting generic up- and down-type vacua as $u$ or $d$ respectively
note that the segments $dud$ or $uuu$ rotate clockwise and are excluded.
Therefore the half-plane fans $J$ must be sequences of the type
 $\{u_1 \cdots u_2\}$, $\{u_1 \cdots d_2\}$, $\{d_1\cdots u_2\}$, $\{d_1 \cdots d_2 \}$ where in each case
 the ellipsis $\cdots$, if nonempty, is an ordered sequence of down-type vacua. An example is shown in
 Figure \ref{fig:TNEXAMPLE-2}.

Since our interior amplitude is supported on trivalent vertices, to write the $A_\infty$ MC equations we need only
list all the taut half-plane webs with trivalent vertices. This can be done, with some effort.
Indeed, if we associate to each half plane fan the sequence of edges in the weight plane between the corresponding vacua, then
each half-plane web with non-zero representation can be associated to a triangulation of the polygon defined by the sequences of edges for
half-fans at boundary vertices together with the half-fan at infinity.

\begin{figure}[htp]
\centering
\includegraphics[scale=0.3,angle=0,trim=0 0 0 0]{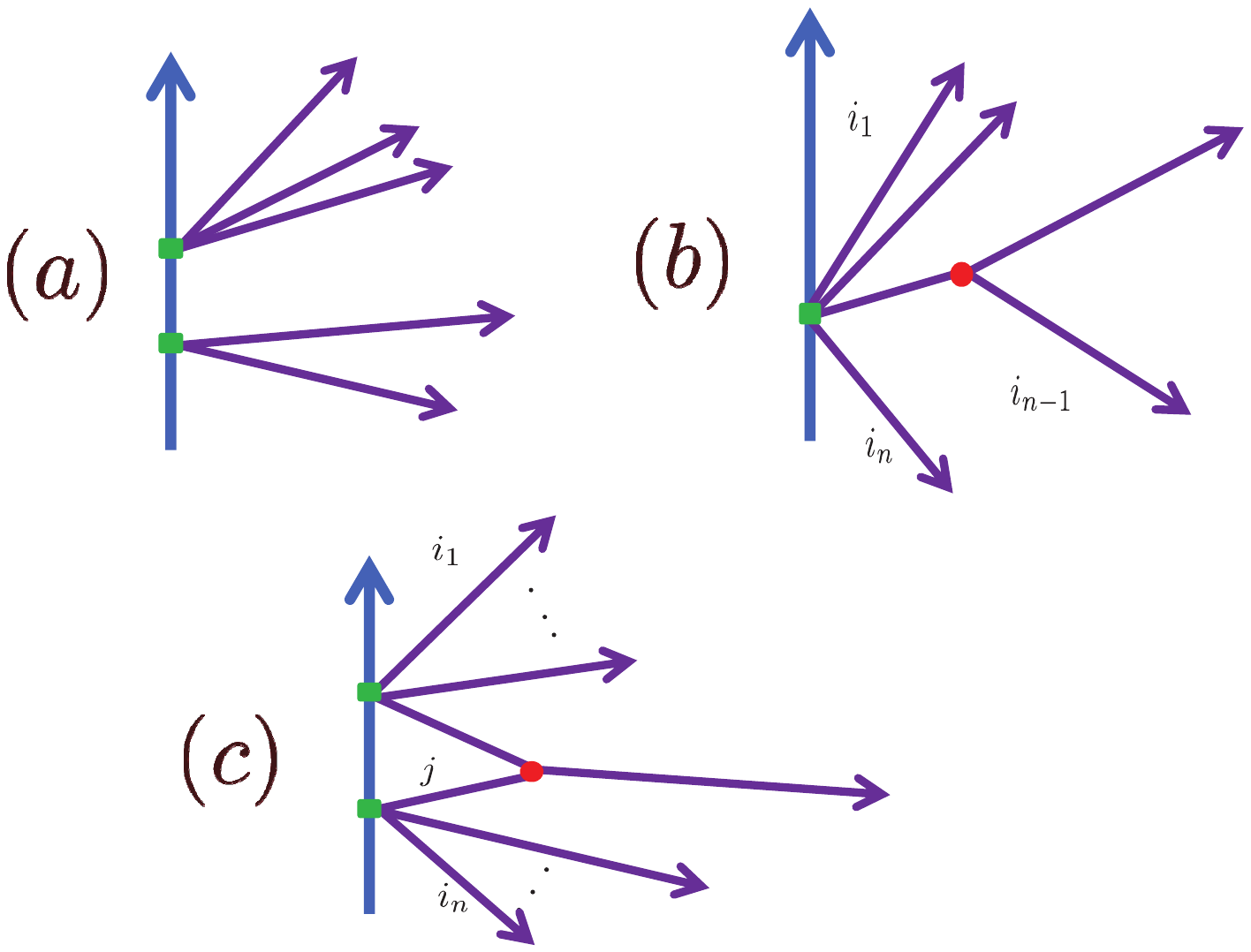}
\caption{The three kinds of taut half-plane webs which can contribute
to the $A_\infty$-Maurer-Cartan equation in the examples with cyclic weights.   }
\label{fig:TNEXAMPLE-3}
\end{figure}

If the half-plane web does not include intermediate upper vacua in the sequence of vacua along the boundary, all we can have is a disconnected taut web  (Figure \ref{fig:TNEXAMPLE-3}(a)) or a
web with a single boundary vertex, an edge of which splits at an interior vertex to give a half-fan at infinity with one extra edge than the half fan at the boundary vertex (Figure \ref{fig:TNEXAMPLE-3}(b)) .
If the half plane web includes a single intermediate upper vacua, the only relevant taut half-plane web has two half-fans and a single interior vertex
at which the last edge of one fan and the first of the next fan join to a new semi-infinite line
(Figure \ref{fig:TNEXAMPLE-3}(c)). Webs with more intermediate upper vacua cannot be taut.
%
%\cg{Need to explain more clearly why these are the only cases.}
%

Solving the MC equation is still a rather formidable task, and therefore (with some later applications in mind), we will constrain the problem further, and
impose a simple but powerful constraint on the degrees assigned to the (nonzero) Chan-Paton factors $\CE_i$: we will choose
the degrees to be decreasing as we move clockwise around the lower vacua and decreasing as we move counterclockwise
around the upper vacua. Moreover, we require a reflection symmetry on the degrees of nonzero Chan-Paton factors. So we take:
\begin{align}\label{eq:CPDEG}
\mathrm{deg} \, \CE_k \quad &= n-k \qquad \qquad 0 \leq k < N/2 \cr
\mathrm{deg} \, \CE_{N-k-1} &= n-k \qquad \qquad 0 \leq k <  N/2 \cr
\end{align}
for some $n$. (The integer $n$ is not really needed here but cannot be shifted away in the
related example \eqref{eq:SLDEF} below.)
%For even N, the Chan-Paton spaces on the two lines of \eqref{eq:CPDEG} correspond
%to distinct vacua.  The same is true for odd $N$ except for the space $\CE_{(N-1)/2}$,
%for which the   two lines agree.

%
\begin{figure}[htp]
\centering
\includegraphics[scale=0.3,angle=0,trim=0 0 0 0]{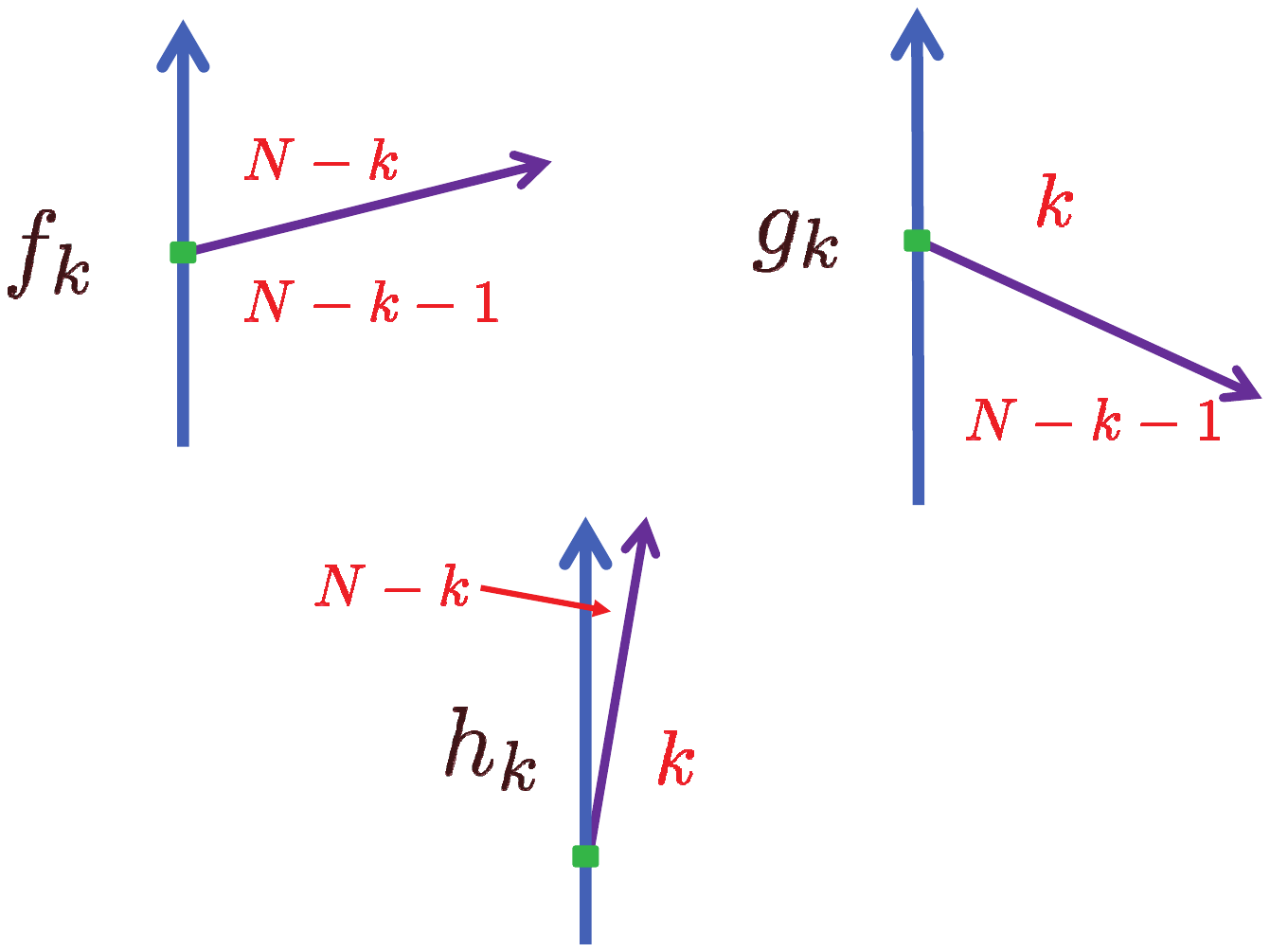}
\caption{Three nontrivial components of a boundary amplitude in
the cyclic theories.  Note that
for small $\vartheta$ the slope $z_{N-k,k}$ is nearly vertical, $z_{k, N-k-1}$ always points
downwards, while $z_{N-k,N-k-1}$ can point upwards or downwards, depending on $k$.   }
\label{fig:TNEXAMPLE-4}
\end{figure}

The restriction \eqref{eq:CPDEG} strongly constrains which half-plane fans can support
a nonzero boundary amplitude $\CB$ (since $\CB$ must have degree $1$).
It will allow a simple analysis of the MC equation for a boundary amplitude with such Chan-Paton factors.
Suppose we consider a component $\CB_{ij} \in R_{\{ i\cdots j\} }(\CE)$.
It is useful to look at first at fans which include two vacua only
so $\CB_{ij} \in \CE_i \otimes R_{ij} \otimes \CE_j^*$.  If $i>j$
then by \eqref{eq:ExpleTN-webrep}  $R_{ij}$ has degree $0$ and hence the degree of $\CE_i \otimes \CE_j^*$
must be $1$. The only two possibilities are $i = N-k$ and $j=k$, with $1 \leq k < N/2$
or $i=N-k$ and $j = N-k-1$, with $1 \leq k < N/2$. In both cases  $i$ is $u$-type.
On the other hand, if
 $i<j$ then by \eqref{eq:ExpleTN-webrep}  $R_{ij}$
 has degree $1$ and hence the degree of $\CE_i \otimes \CE_j^*$ must be zero. Then
 the only possibility is $i = k$, $j = N-k-1$ with $0 \leq k < N/2$.  In this case $i$ is $d$-type.
It turns out that  no half-fans with more than two vacua may contribute to a degree $1$ amplitude.
For example, if we consider $\CE_{N-k} \otimes R_{N-k,\ell}\otimes R_{\ell,k} \otimes \CE_k^* $
with $k<\ell< N-k$ then $\{ N-k, \ell, k \}$ is never a valid half-plane fan. The reason is
that the real part of $z_{N-k,k}$ is $\sin(\vartheta) \sin(2\pi k/N)$ and is arbitrarily small,
and will be smaller than $\Re(z_{\ell,k})$ whenever  $\Re(z_{\ell,k})>0$.
We conclude that the only potentially nonzero components of a boundary amplitude are of the form:
\be\label{eq:Bd-Lc}
\begin{split}
\CB_{N-k,N-k-1} \quad \leftrightarrow\quad &   f_k \in \Hom(\CE_{N-k-1}, \CE_{N-k}) \qquad\qquad 1\leq k < N/2 \\
\CB_{k,N-k-1} \quad \leftrightarrow\quad &   g_k \in \Hom(\CE_{N-k-1},\CE_k) \qquad\qquad 0 \leq k < N/2 \\
\CB_{N-k,k} \quad \leftrightarrow\quad &  h_k \in \Hom(\CE_k, \CE_{N-k} ) \qquad\qquad 1\leq k < N/2 \\
\end{split}
\ee
where in the second column we have interpreted the indicated component of $\CB$  in terms of linear transformations $f_k, g_k, h_k$.
See Figure \ref{fig:TNEXAMPLE-4}.

\begin{figure}[htp]
\centering
\includegraphics[scale=0.5,angle=0,trim=0 0 0 0]{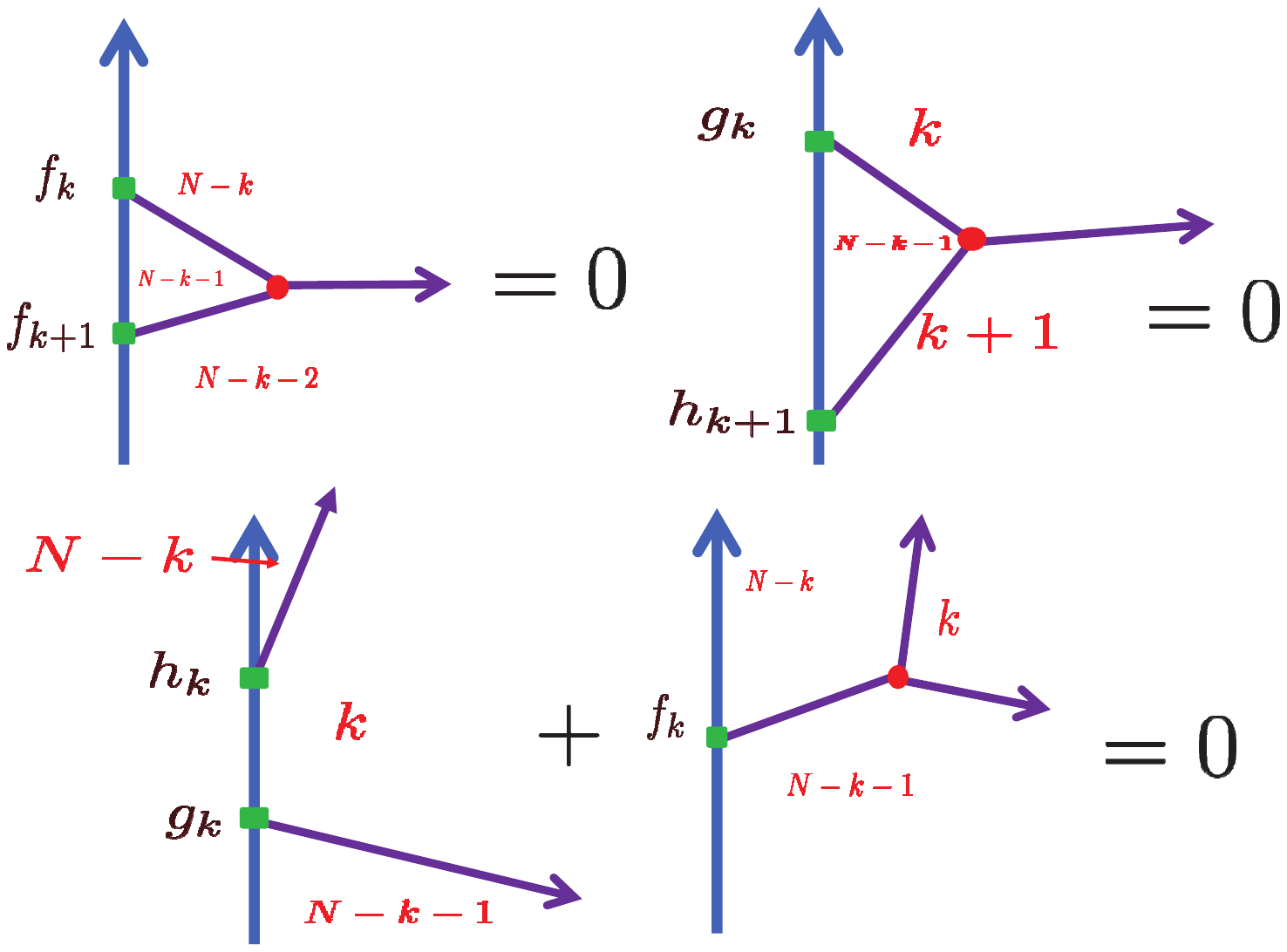}
\caption{Three nontrivial equations in the $A_\infty$ MC equation. }
\label{fig:TNEXAMPLE-6}
\end{figure}

We can now write out the nontrivial components of the $A_\infty$ Maurer-Cartan equation for $\CB$.
We organize them by the type of the half-plane fan $J_{\infty}$.
 The $uu$ component arises from a taut web with a single interior vertex
and two boundary vertices as in Figure \ref{fig:TNEXAMPLE-3}(c). It takes the form
\begin{equation}
\rho(\ft_{\CH} )[\CB_{N-k,N-k-1},\CB_{N-k-1,N-k-2};\beta_{N-k-2,N-k-1,N-k}] =0
\end{equation}
and it tells us that  $f_k f_{k+1} =0$ in the notation of \eqref{eq:Bd-Lc}.

In a similar fashion, the $dd$ component takes the form
\begin{equation}
\rho(\ft_{\CH})[\CB_{k,N-k-1},\CB_{N-k-1,k+1};\beta_{k,k+1,N-k-1}] =0
\end{equation}
and it tells us that if  $g_k h_{k+1} =0$ in the notation of \eqref{eq:Bd-Lc}
(where $k< N/2 -1 $).

Finally, the $udu$ component involves two kinds of webs: a web with two boundary vertices only
as in Figure \ref{fig:TNEXAMPLE-3}(a), and a web with a
single boundary vertex and a single interior vertex as in Figure \ref{fig:TNEXAMPLE-3}(b).
Both webs contribute to a term in the MC equation with a half fan of three vacua, which takes the form $N-k, k, N-k-1$. Thus we need to solve
\begin{equation}
\rho(\ft_{\CH})[\CB_{N-k,k},\CB_{k,N-k-1}] + \rho(\ft_{\CH})[\CB_{N-k,N-k-1};\beta_{k,N-k-1,N-k}]=0
\end{equation}
which tells us that   $f_k = h_k g_k$.
%
%\cg{ Should recheck sign. Show details?}
%

Note that the $f_k f_{k+1} =0$ constraint follows from the other constraints.
Thus the general solution of the $A_\infty$ MC equation subject to the constraint \eqref{eq:CPDEG} is
given by a set of linear transformations $\{ f_k, g_k, h_k \}$ as in \eqref{eq:Bd-Lc} subject
to the two conditions $g_k h_{k+1} =0$
and  $f_k = h_k g_k$ (for values of $k$ for which this makes sense). These equations are illustrated in Figure \ref{fig:TNEXAMPLE-6}.

%If we are willing to ignore issues of torsion and tensor our Chan-Paton spaces with the field $\IQ$ then  we
% can decompose the Chan-Paton vector spaces into subspaces such that the restrictions of the $f_k$, $g_k$, $h_k$ functions
%are either $0$ or the identity, by splitting off kernels and co-kernels of the maps. (Put differently,
%the only invariants of an $n\times m$ integral matrix under the action of $GL(n,\IQ)\times GL(m,\IQ)$ is the rank,
%since $\IQ$ is a field.)
%Thus a Brane obeying our degree restrictions can be essentially split into a direct sum of simpler Branes,
%where we define the direct sum of Branes
%\begin{equation}
%\oplus V_a \otimes \CB_a
%\end{equation}
%by taking the direct sum of the Chan-Paton factors tensored with the coefficient vector spaces $V_a$, and
%assemble together the individual boundary amplitudes acting as the identity on the $V_a$. Together with
%the above restrictions on which boundary amplitudes can be nonzero we see that (over $\IQ$) the category of Branes
%$\fB\fr(\CT,\CH)$ (defined in Section \S \ref{subsec:BraneCat} below) has two kinds of simple objects -
%corresponding to $\CB=0$ and to the nonzero $\CB$ above.
%
%\cg{I'm not sure this claim is correct. What if $g_k$ and $h_k$ are not equal?
%Put differently: In simplifying $f_k, g_k, h_k$ you must use the same change of basis on $\CE_k$ for both
%$g_k$ and $h_k$. Then it is not obvious they can be simultaneously put into the canonical form stated here. }
%

Let us consider two simple types  of solutions.
 The the first   are the  ``thimbles'' $\fT_i$ defined by the Chan-Paton factors
\begin{equation}
\CE(\fT_i)_j = \delta_{ij} \IZ
\end{equation}
with  $f_j=g_j = h_j =0$. The second kind   are the Branes  $\fC_k$, with  $1 \leq k < N/2$.
These are defined by taking Chan-Paton spaces
\begin{equation}\label{eq:Ck-CP-Spaces}
\CE(\fC_k)_{N-k} = \IZ^{[1]} \qquad \CE(\fC_k)_{N-k-1} = \IZ \qquad \CE(\fC_k)_{k} = \IZ
\end{equation}
with all other Chan-Paton spaces $\CE(\fC_k)_j$ equal to zero. For the boundary amplitude we take $f_k, g_k, h_k$ to be
be multiplication by $1$ (and of course $f_j, g_j, h_j$ vanish for $j\not= k$)
and hence $f_j = h_j g_j$ and $g_j h_{j+1}=0$  for all $j$
is satisfied.
The motivation for writing down these branes is that they are generated from
rotational interfaces as described at length in Section \S \ref{subsec:RotIntfc-TSUN} below.

In order to look at the strip we need to define some boundary amplitudes and Branes for the negative half-plane.
One convenient way to do this is to use the $\IZ_N$ symmetry of the model and rotate the half-planes by $\pi$.
We can construct a family of cyclic Theories by simultaneous rotation of all the vacuum weights $z_i \to e^{-\I \phi} z_i$.
As the edges parallel to $z_{ij}$ rotate we continue to associate the same spaces $R_{ij}$ to them, together
with the same interior amplitudes, and thus we obtain a family of Theories.    A rotation
 by a multiple of $2\pi/N$ leaves the set of vacuum weights invariant.  Hence there is an
isomorphism of the rotated planar theory with the original theory.
However, because of the degree assignments in equation  \eqref{eq:ExpleTN-webrep} this isomorphism  involves an
interesting degree shift on the web representation. If $z_i \to \omega^{-d} z_i$ with $\omega = \exp[2\pi \I/N]$ then the
isomorphism acts by $i \to \hat i$ where
$\hat i = (i+d) \mod N$ with $0 \leq \hat i < N$ and the rotated web representation $\Phi^*_d(R_{ij})$ is
related to the old one by
\be
\Phi^*_d(R_{ij}) = R^{[s_{ij}]}_{\hat i \hat j }
\ee
with
\be
[s_{ij}] = \begin{cases}
0 & i<j  \qquad \& \qquad  \hat i   < \hat j \\
0 & i>j \qquad \& \qquad  \hat i > \hat j \\
+1 & i<j \qquad \& \qquad  \hat i > \hat j \\
-1 & i>j \qquad \& \qquad  \hat i < \hat j \\
\end{cases}
\ee
\begin{figure}[htp]
\centering
\includegraphics[scale=0.3,angle=0,trim=0 0 0 0]{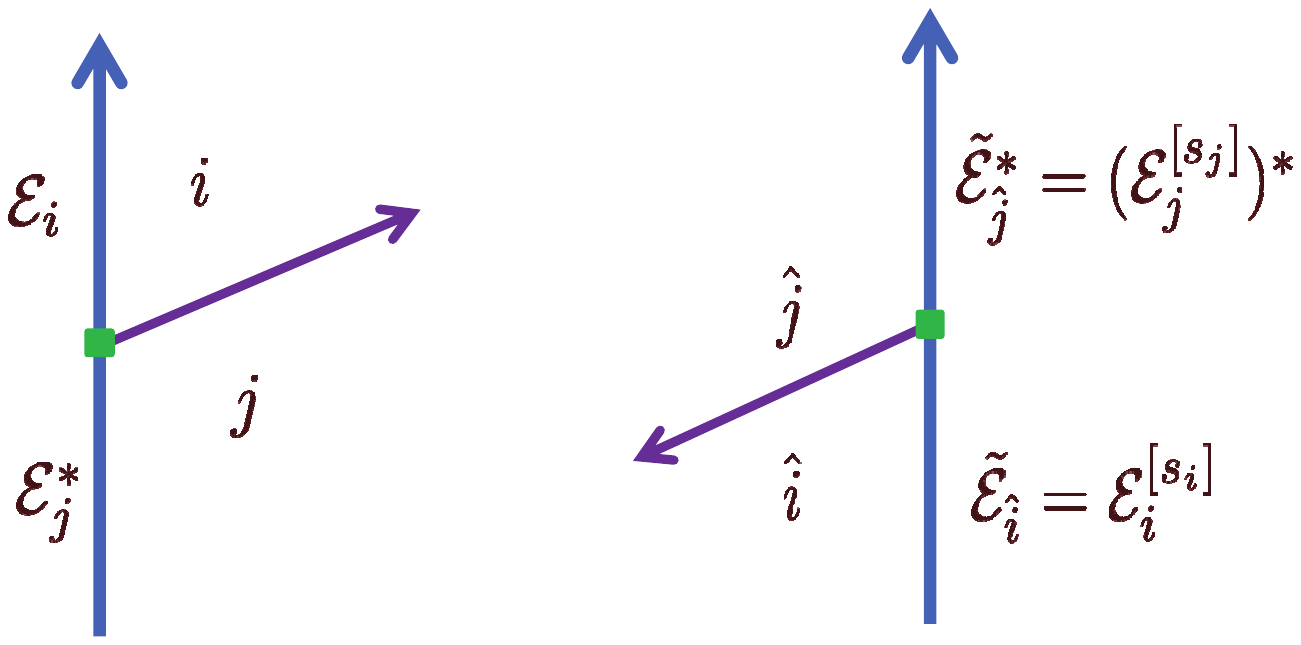}
\caption{Rotating the theory for $N$ even by $\pi$ maps left Branes to right Branes
using the above rule.   }
\label{fig:PIROTATION}
\end{figure}

In general the rotation takes one half-plane theory to another half-plane theory. If $N$ is even
we can use a rotation by $\omega^{-N/2} = -1$ to take the positive half-plane theory to the negative
half-plane theory as in Figure \ref{fig:PIROTATION}.  Our rule for mapping Chan-Paton spaces will be that
\be
\tilde\CE_{\hat i} = \CE_{i}^{[s_i]}
\ee
and the degree shifts are chosen so that there is a degree zero isomorphism
\be
\CE_i \otimes \Phi^*_{N/2}(R_{ij}) \otimes \CE_j^* \cong \tilde \CE_{\hat i}\otimes  R_{\hat i  \hat j} \otimes  \tilde \CE_{\hat j}^*
\ee
and hence the degree-shifts on the Chan-Paton spaces are determined, up to an overall shift,  by
\be\label{eq:DegShft-2}
[s_{ij}] = [s_i] -  [s_j].
\ee
The general solution to \eqref{eq:DegShft-2} is
\be\label{eq:DegShft-3}
s_i = \begin{cases}  s +1  & 0 \leq i < \frac{N}{2} \\
s & \frac{N}{2} \leq i \leq N-1 \\
\end{cases}
\ee
for some $s$.
These maps define an isomorphism of theories as in \ref{subsubsec:IsomTheory}. See also \ref{subsubsec:CyclicIsoms}.

In particular, applying this procedure to the Branes $\fC_k$ produces a collection of Branes
for the right boundary, $\tilde\fC_k$, $1 \leq k < N/2$    with
\be
\CE(\tilde\fC_k)_{N-k} = \IZ^{[s]} \qquad \CE(\tilde\fC_k)_{k-1} = \IZ^{[s-1]} \qquad \CE(\tilde\fC_k)_{k} = \IZ^{[s]}
\ee
and the simplest choice is to take $s=1$.
In the above we have renamed the Brane
$\fC_{k}$ rotated by $\pi$ to be $\tilde\fC_{\frac{N}{2}-k}$.
The components of the boundary amplitude of $\tilde \fC_k$ are obtained
from those of $\fC_{N/2-k}$ and are  $\pm1$.

%%half plane. It is convenient to restrict ourselves to even $N$, so that we can relate such Branes to the ones we defined above by a simultaneous %rotation of $\pi$
%in the plane and in the weight space, up to a simple relabelling of the vacua which exchanges upper and lower vacua and some appropriate degree %shifts.
%
%In particular, we can readily define the right version of the $\fC_k$ Brane: $\tilde\fC_k$ is defined by
% the Chan-Paton factors
%\begin{equation}
%\CE(\tilde\fC_k)_{N-k} = \IZ^{[1]} \qquad \CE(\tilde\fC_k)_{k-1} = \IZ \qquad \CE(\tilde\fC_k)_{k} = \IZ^{[1]}
%\end{equation}

The strip complex for the Branes $\fC_k$ and $\tilde\fC_t$ is non-empty only if $k=t$ or $k=t-1$. In either case, it is a two-dimensional complex,
with differential ``$1$''. For example for $k=t$ we have
the complex of approximate ground states
\be\label{eq:CkCktilde-cplx}
\begin{split}
\CE_{LR} & = \oplus_i \CE(\fC_k)_i \otimes \CE(\tilde\fC_k)_i^* \\
& \cong  \IZ \otimes \IZ^{[-s]} \oplus \IZ^{[1]} \otimes \IZ^{[-s]}\\
\end{split}
\ee
There is only one taut web which contributes to $d_{LR}$ given by
the amplitude $h_k$ and hence  $d_{LR}(m,n) = (0,m)$.
 The cohomology is therefore zero: there are no exact ground states on the strip between $\fC_k$ and $\tilde\fC_t$.
In particular, in a physical manifestation of this example, although there are good approximations to supersymmetric
groundstates on the interval in fact instanton effects break supersymmetry.

\subsubsection{ The Theories $\CT^{SU(N)}_{\vartheta}$ }

There is an interesting, and much richer, variant of the Theories we have described above: we keep the same weights, but define
\begin{align}\label{eq:SUN-Rij}
R_{ij} &= A_{j-i}^{[1]} \qquad i<j \cr
R_{ij} &= A_{N+j-i} \qquad i>j
\end{align}
where $A_\ell$ is the $\ell$-th antisymmetric power of a fundamental representation of $SU(N)$.
\footnote{If we want to work over $\IZ$ or $\IQ$ we should replace $SU(N)$ with $SL(N,\IZ)$
or $SL(N,\IQ)$. We will informally write $SU(N)$, since this is what appears in the main physical
applications.}
We will show now how to define an $SU(N)$-invariant interior amplitude, and thus a family of
Theories $\CT^{SU(N)}_\vartheta$ whose algebraic structures will be $SU(N)$ covariant.
We choose an orientation on the fundamental representation $A_1$, or equivalently,
a nonzero vector in $A_N$, denoted by $\vol$.
It will also be convenient in some formulae to choose an oriented orthonormal basis $\{e_1, \dots, e_N \}$
and, for multi-indices $S = \{ a_1< a_2 < \cdots < a_\ell\}$ the corresponding
  vector $e_S = e_{a_1} \wedge e_{a_2} \wedge \cdots \wedge e_{a_\ell}$. These vectors
  form an orthonormal basis for $A_\ell$.

The  fan spaces $R_I$ are the product of $SU(N)$ representations whose Young tableaux
have a   total of $N$ boxes, and thus  contain a
\emph{unique} $SU(N)$ invariant line: Taking the outer product of the antisymmetric
tensors from each of the factors we antisymmetrize on all the indices.
%
%\cg{Actually, not quite true. Consider $R_{12} \otimes R_{23} \otimes \cdots \otimes R_{N1}$.
%There are two invariant lines for symm. and antisymmetric. But I think this is the only
%exception.}
%If we denote elements in $R_{ij}$ as an antisymmetric tensor with the appropriate number of $SU(N)$ indices, then the $SU(N)$ invariant line in %$R_I$
%is generated by multiples of the totally antisymmetric $\epsilon_{a_i \cdots a_N}$ tensor.

The  pairing $K_{ij}: R_{ij}\otimes R_{ji} \to \IZ$ is uniquely determined by $SU(N)$ invariance to be
%\begin{equation}
%K_{ij} = {N \choose j-i} \epsilon^{a_1 \cdots a_{j-i} b_1 \cdots b_{N+i-j}} \qquad i < j
%\end{equation}
%
\be\label{eq:Kij-TSUN}
K_{ij} ( v_1 \otimes v_2) = \kappa_{ij} \frac{ v_1 \wedge v_2}{\vol}
\ee
where $\kappa_{ij}$ is a nonvanishing normalization factor. For simplicity we
will take the $\kappa_{ij}$ to be given by a single factor $\kappa$ for $i<j$.
Then, by the natural isomorphism $R_{ij}\otimes R_{ji} \cong R_{ji}\otimes R_{ij}$
  $\kappa_{ji}= \pm \kappa $ with a sign determined by $i,j$.

Since the degree assignments of \eqref{eq:SUN-Rij} are the same as those
of \eqref{eq:ExpleTN-webrep} the interior amplitude only has components
on trivalent vertices. We will assume our interior amplitudes are valued
in the invariant line in $R_I$ and therefore, for $i<j<k$ the amplitude
must be of the form
\be\label{eq:beta-TSUN}
\beta_{ijk} = b_{ijk} \sum_{\Sh_3^{ijk} } \frac{ e_{S_1} e_{S_2} e_{S_3}}{\vol} e_{S_1} \otimes e_{S_2} \otimes e_{S_3}
\ee
where $b_{ijk}$ is a scalar, exterior multiplication is understood in the first factor in the
sum on the RHS, and $\Sh_3^{ijk} $ is the set of   3-shuffles of $S=\{1, \dots, N\}$ such that
$\vert S_1 \vert = j-i$ and $\vert S_2 \vert = k-j$.

%and the interior amplitude
%\begin{equation}
%\beta_{ijk} =  \epsilon_{a_1 \cdots a_{j-i} b_1 \cdots b_{k-j} c_1 \cdots c_{N+i-k}}
%\end{equation}
%
%\cg{Say this more invariantly?}
%

Now let us write the $L_\infty$ MC equation. When we compute $K_{ik}(\beta_{ijk} \otimes \beta_{ikt})$
we apply the contraction to a sum over pairs of 3-shuffles $S_1 \amalg S_2 \amalg S_3 \in \Sh_3^{ijk}$ and
$S_1' \amalg S_2' \amalg S_3 ' \in \Sh_3^{ikt}$. In fact, the contraction turns out to be
valued in the $SU(N)$ invariant line in $R_{\{ i,j,k,t \}}$ because the contraction of three epsilon
tensors is proportional to an epsilon tensor. In our notation we have the identity:
\be
\frac{ e_{S_1} e_{S_2} e_{S_3}}{\vol}\frac{ e_{S_1'} e_{S_2'} e_{S_3'}}{\vol}\frac{ e_{S_1'} e_{S_3}  }{\vol}
= \begin{cases}    \frac{ e_{S_1} e_{S_2} e_{S_2'} e_{S_3'} }{\vol} & S_1 \amalg S_2 \amalg S_2' \amalg S_3' \in \Sh_4^{ijkt} \\
0 & {\rm else} \\
\end{cases}
\ee
%
%where $\Sh_4^{ijkt}$ is the set of $4$-shuffles with $\vert S_1\vert = j-i$, $\vert S_2 \vert = k-j$, and $\vert S_2'\vert = t-k$.
where $\Sh_4^{ijkt}$ is a sum over 4-shuffles with lengths $j-i, k-j, t-k, N+i-t$.
Note that one must be careful to contract $R_{ik}\otimes R_{ki} \to \IZ$ in that order.
It therefore follows that
\be
K_{ik}(\beta_{ijk} \otimes \beta_{ikt}) = \kappa b_{ijk} b_{ikt}
\sum_{\Sh_4^{ijkt}}  \frac{ e_{S_1} e_{S_2} e_{S_2'} e_{S_3'} }{\vol} e_{S_1} \otimes e_{S_2}\otimes  e_{S_2'}\otimes e_{S_3'}
\ee

A similar result holds for $K_{jt}(\beta_{ijt} \otimes \beta_{jkt})$, which turns out to be
(when considered as an element of $R_{ij}\otimes R_{jk}\otimes R_{kt} \otimes R_{ti}$)
\be
K_{jt}(\beta_{ijt} \otimes \beta_{jkt})= - \kappa b_{ijt} b_{jkt}
\sum_{\Sh_4^{ijkt}}  \frac{ e_{S_1} e_{S_1'} e_{S_2'} e_{S_3} }{\vol} e_{S_1} \otimes e_{S_1'}\otimes  e_{S_2'}\otimes e_{S_3} .
\ee
The overall minus sign has the same origin as in the second term of \eqref{eq:ExpleMC-1}.
Therefore   the $L_\infty$ MC equations are
\be
  b_{ijk} b_{ikt}  -  b_{ijt} b_{jkt} = 0 \qquad \qquad i<j<k<t.
\ee
These are the same equations as before. For general $N$ they are overdetermined,
and once again we take the canonical solution $b_{ijk}=1$ to define the
Theories $\CT^{SU(N)}_{\vartheta}$.

%If we go back to the MC equation,
%\begin{equation}
%K_{ik} \circ \left( \beta_{ijk} \otimes \beta_{ikt} \right) = {N \choose k-i}\epsilon^{d_1 \cdots d_{k-i} c_1 \cdots c_{N+i-k}}\epsilon_{a_1 \cdots %a_{j-i} b_1 \cdots b_{k-j} c_1 \cdots c_{N+i-k}}\epsilon_{d_1 \cdots d_{k-i} e_1 \cdots e_{t-k} f_1 \cdots f_{N+i-t}}
%\end{equation}
%coincides with $N! \epsilon_{a_1 \cdots a_{j-i} b_1 \cdots b_{k-j} e_1 \cdots e_{t-k} f_1 \cdots f_{N+k-t}}$.
%On the other hand,
%\begin{equation}
%K_{jt} \circ \left( \beta_{ijt} \otimes  \beta_{jkt} \right) =-  {N \choose t-j}\epsilon^{d_1 \cdots d_{t-j} c_1 \cdots c_{N+j-t}}\epsilon_{a_1 %\cdots a_{j-i} d_1 \cdots d_{t-j} f_1 \cdots f_{N+i-t}}\epsilon_{b_1 \cdots b_{k-j} e_1 \cdots e_{t-k} c_1 \cdots c_{N+j-t}}
%\end{equation}
%coincides with $-N! \epsilon_{a_1 \cdots a_{j-i} b_1 \cdots b_{k-j} e_1 \cdots e_{t-k} f_1 \cdots f_{N+k-t}}$.
%The MC equation is thus still satisfied.

We can impose the same restrictions on the degree of Chan-Paton factors as in \eqref{eq:CPDEG}.
The same reasoning as before implies that we can interpret the possible nonzero components
of the boundary amplitude as linear transformations. We give them the same names as before,
but now equation \eqref{eq:Bd-Lc} is generalized to give a set of three   maps
\be\label{eq:Bd-Lc-SUN}
\begin{split}
\CB_{N-k,N-k-1} \quad \leftrightarrow\quad &   f_k \in \Hom(\CE_{N-k-1}, \CE_{N-k}\otimes A_{N-1}) \qquad\qquad 1\leq k < N/2 \\
\CB_{k,N-k-1} \quad \leftrightarrow\quad &   g_k \in \Hom(\CE_{N-k-1},\CE_k\otimes A_{N-2k-1}^{[1]}) \qquad\qquad 0 \leq k < N/2 \\
\CB_{N-k,k} \quad \leftrightarrow\quad &  h_k \in \Hom(\CE_k, \CE_{N-k}\otimes A_{2k} ) \qquad\qquad 1\leq k < N/2 \\
\end{split}
\ee

There are two independent MC equations. The first (see the northeast corner of Figure
\ref{fig:TNEXAMPLE-6})   says that
\be
\begin{split}
K_{26}K_{35} (1\otimes \beta_{k,k+1,N-k-1}) g_k h_{k+1}:\CE_{k+1} & \rightarrow \CE_{N-k-1}\otimes A_{2k+2} \\
& \rightarrow \CE_k \otimes A_{N-2k-1}^{[1]}\otimes A_{2k+2} \\
& \rightarrow \CE_k \otimes A_{N-2k-1}^{[1]} \otimes A_{2k+2} \otimes A_1^{[1]} \otimes A_{N-2k-2}^{[1]} \otimes A_{2k+1} \\
& \rightarrow \CE_k \otimes A_{1} \\
\end{split}
\ee
must vanish. Written out in components this means the following. Choose bases $\{ v_{\alpha_k} \}$
for $\CE_k$ and define matrix elements:
\be
\begin{split}
h_k(v_{\alpha_k}) & := \sum_{ \beta_{N-k}, \vert I \vert = 2k} h_{\beta_{N-k}, I \vert \alpha_k} v_{\beta_{N-k} }\otimes e_I \\
g_k(v_{\beta_{N-k-1} }) & :=\sum_{ \gamma_{k}, \vert I \vert = N-2k-1}   g_{\gamma_{k}, I \vert \beta_{N-k-1}} v_{\gamma_k }\otimes e_I \\
\end{split}
\ee
and the MC equation becomes (assuming $b_{k,k+1,N-k-1}\not=0$ as is true for the canonical interior amplitude):
\be\label{eq:SUN-MC-1}
0 = \sum_{ \beta_{N-k-1}, I_1, I_2 }  \varepsilon_{i,I_1,I_2}
g_{\gamma_{k}, I_2 \vert \beta_{N-k-1}}  h_{\beta_{N-k-1}, I_1 \vert \alpha_{k+1} }
\ee
The sum is over all multi-indices $I_1,I_2$ with $\vert I_1 \vert = 2k+2, \vert I_2\vert = N-2k-1$.
The equation is meant to hold for all $\gamma_k,  \alpha_{k+1}$, and $1\leq i \leq N$.
The factor  $\varepsilon_{i,I_1,I_2} \in \{ 0, \pm 1\} $  comes from contracting 3 epsilon tensors.
The explicit formula  is
\be
\varepsilon_{i,I_1,I_2} := \left( \frac{e_i e_{I_2'} e_{I_1'} }{\vol}\right)  \varepsilon_{I_2} \varepsilon_{I_1}
\ee
where   $I'$ denotes the complementary multi-index to $I$ in  $\{1,\dots, N \}$ and
 $\varepsilon_I := \frac{e_I \wedge e_{I'}}{\vol}$.
%
%\cg{Check and simplify! It is probably useful to dualize the indices $I \to I'$
%on the matrix elements of $f,g$. This simplifies things a little. More importantly, we need to provide intuition about the meaning of the formulae.}
%

The second MC equation says that the sum of the two diagrams on the bottom of Figure
\ref{fig:TNEXAMPLE-6})  must vanish. Thus,
\be
\begin{split}
  h_{k} g_k :\CE_{N-k-1} & \rightarrow \CE_{k}\otimes A_{N-2k-1}^{[1]} \\
& \rightarrow \CE_{N-k} \otimes A_{2k}\otimes A_{N-2k-1}^{[1]} \\
\end{split}
\ee
plus
\be
\begin{split}
  K_{24} (1\otimes \beta) f_k  :\CE_{N-k-1} & \rightarrow \CE_{N-k}\otimes A_{N-1} \\
& \rightarrow \CE_{N-k} \otimes A_{N-1}\otimes A_{N-2k-1}^{[1]} \otimes A_1^{[1]} \otimes A_{2k} \\
& \rightarrow \CE_{N-k} \otimes   A_{N-2k-1}^{[1]} \otimes   A_{2k} \\
\end{split}
\ee
must vanish. When written out in terms of the matrix elements this means that,
for all $v_{\alpha_{N-k-1}}$ and all $v_{\beta_{N-k}}$ and all multi-indices
$\vert I_1 \vert = N-2k-1$ and $\vert I_2 \vert = 2k$ we must have
\be\label{eq:SUN-MC-2}
\xi_{I_1,I_2} f_{\beta_{N-2k}, \hat i\vert \alpha_{N-k-1} }= \sum_{\gamma_k} h_{\beta_{N-k},I_2\vert \gamma_k} g_{\gamma_k, I_1 \vert \alpha_{N-k-1} }
\ee
where $\xi_{I_1,I_2}$ is a constant given by
\be
\xi_{I_1,I_2} = (-1)^{n-k+1}\left( \frac{e_{I_1} \wedge e_i \wedge e_{I_2}}{\vol} \right)
\left(\frac{e_i \wedge e_{\hat i} }{\vol} \right) \kappa b_{k,N-k-1,N-k}
\ee
This constant is zero unless $I_1 \amalg I_2 = \hat i$ for some $i\in \{1,\dots, N\}$, and $\hat i$ is the multi-index of length $N-1$ complementary
to $i$.

The last MC equation follows from the first two, as in the $\CT^N_\vartheta$ theories.
  $f_k$ is a  map $f_k : \CE_{N-k-1} \to \CE_{N-k} \otimes A_{N-1}$.
We identify $A_{N-1}\cong A_1^*$ and then the component of the MC given by the upper left
diagram in Figure \ref{fig:TNEXAMPLE-6} becomes a map
\begin{equation}
[f_k f_{k+1}] : \CE_{N-k-2} \to \CE_{N-k} \otimes A_{2}^*
\end{equation}
obtained by antisymmetrizing the two $A_{1}^*$  indices to an $A_{2}^*$ index.
The antisymmetrization is due to the contraction with $\beta_{N-k-2, N-k-1, N-k}$.
The form of $f_k$ determined above in \ref{eq:SUN-MC-2} shows that $[f_k f_{k+1}]=0$.

Writing down solutions of these $A_\infty$ MC equations is considerably less
trivial than in the $\CT^N_\vartheta$ theories!
It is now natural to impose   a requirement of $SU(N)$ invariance
on the boundary amplitude component so that $\CB_{i,j}$ is  in the invariant subspace of $\CE_i \otimes R_{ij} \otimes \CE_j^*$.
Equivalently, we require that $f_k, g_k, h_k$ be \emph{intertwiners}.

\begin{figure}[htp]
\centering
\includegraphics[scale=0.3,angle=0,trim=0 0 0 0]{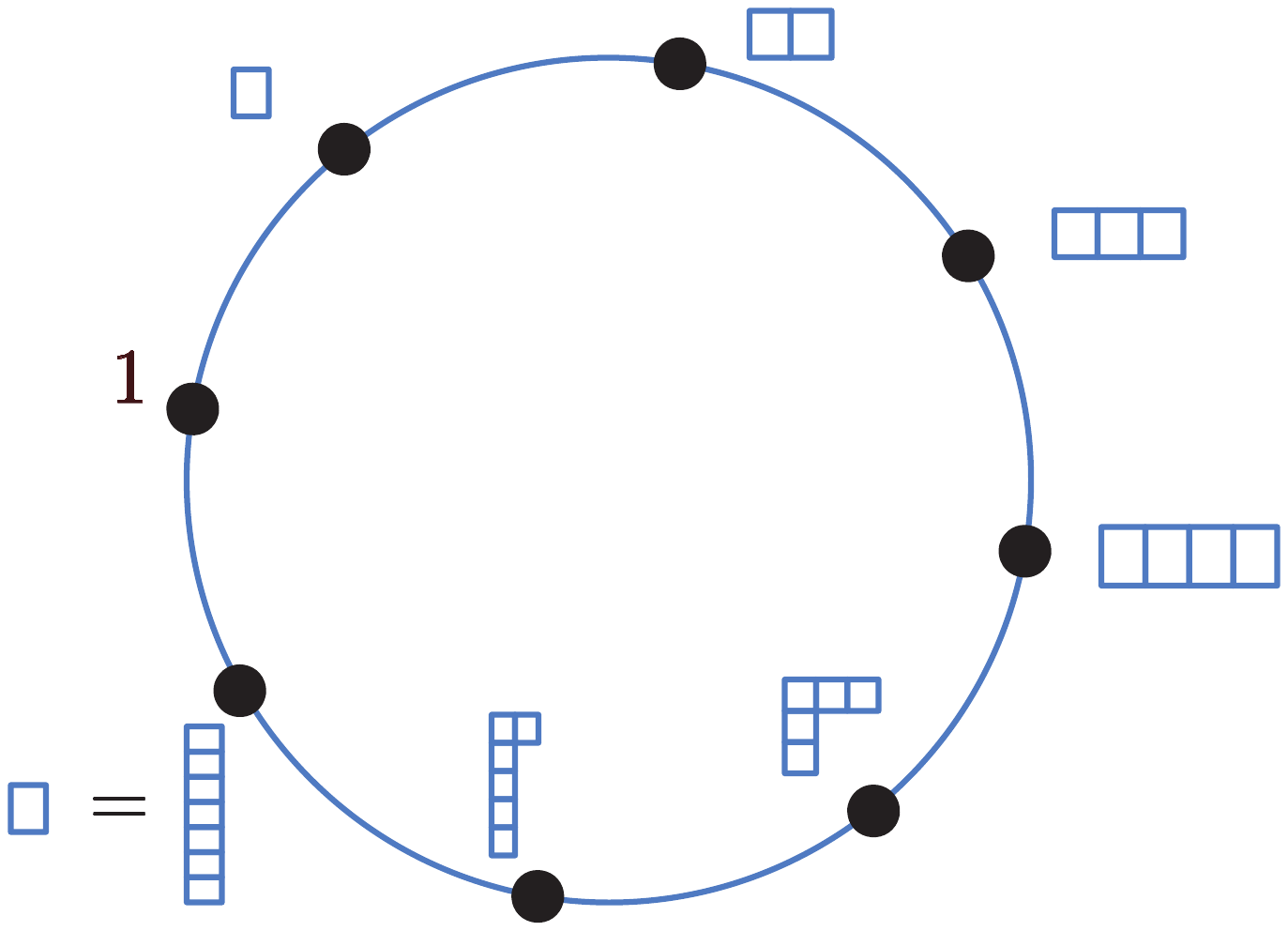}
\caption{The CP factors for the Brane $\fN_n$ in the $\CT^{SU(N)}_\vartheta$
theory, for the case $N=8$ and $n=3$.  }
\label{fig:TSUN-CP-FACTORS}
\end{figure}

Based on this observation and a certain degree of guesswork using the
rotational interfaces discussed in Section \ref{sec:CatTransSmpl} below,
we have found a neat class of Branes $\fN_n$ for this model:
They are
generated from thimbles by using the rotational interfaces. See equation
\eqref{eq:B2hat-CP-TSUN} et. seq.   below.
The Chan-Paton factors of $\fN_n$ are
%
%\begin{align}\label{eq:SLDEF}
%\CE(\fN_n)_k &= L_{2k+1,n-k+1}^{[n-k]} \qquad 0 \leq k < N/2 \cr
%\CE(\fN_n)_{N-k-1} &= L_{1,n-k}^{[n-k]}= S_{n-k}^{[n-k]} \qquad  0 \leq k < N/2 \cr
%
\begin{equation}\label{eq:SLDEF}
\CE(\fN_n)_k = \begin{cases}
L_{2k+1,n-k+1}^{[n-k]} &  0 \leq k < N/2 \\
L_{1,n+k+1-N}^{[n+1-N+k]} &  \frac{N}{2}-1 < k \leq N-1 \\
\end{cases}
\end{equation}
The superscript indicates the degree in which the complex is concentrated,
as usual. Here $L_{\ell,m}$ are the representations of $SU(N)$ labelled by an (upside down) $L$-shaped Young diagram with
 a column of height $\ell$, a row of length $m$ and a total of $\ell+m-1$ boxes. They have dimension
\be
\dim L_{\ell,m} = \frac{(N+m-1)!}{(N-\ell)!(\ell-1)! (m-1)! }\frac{1}{\ell+ m-1}
\ee
Note that $L_{1,m} = S_m$, the $m^{th}$ symmetric power of $A_1$. Note that we
also have $L_{N,m+1}  \cong  S_m$ and $L_{\ell, 1} = A_\ell$.
For the lower-vacua, moving clockwise the $L$ shrinks in width and gets taller.
For the upper-vacua, the representation is always a symmetric power, and the Young diagram
gets longer as the vacua move clockwise. Note that the two sets of cases in
equation \eqref{eq:SLDEF} overlap for $k=(N-1)/2$. The upper representation
is $L^{[n- (N-1)/2]}_{N, 1+n-(N-1)/2}$ and the lower one is $L^{[n- (N-1)/2]}_{1, n-(N-1)/2}$
and these are isomorphic.
In order for the representations to make sense the definition requires us to take
$n$ sufficiently large. In particular,
 $n \geq \left[ \frac{N}{2} \right] + 2 $ will suffice. We will extend it to all integer $n$ in a later section \ref{subsec:RotIntfc-TSUN}.

To define the Brane we must specify the maps
\be\label{eq:SUN-fgh-k}
\begin{split}
 f_k :&  S_{n-k} \to S_{n-k+1} \otimes A_{N-1} \\
 g_k :&  S_{n-k} \to L_{2k+1,n-k+1} \otimes A_{N-2k-1} \\
 h_k :&  L_{2k+1,n-k+1} \to S_{n-k+1} \otimes A_{2k}, \qquad \qquad 0 \leq k < N/2 \\
\end{split}
\ee
such that the MC equations \eqref{eq:SUN-MC-1} and \eqref{eq:SUN-MC-2} are
satisfied. It will be convenient to use the volume form to
perform a partial dualization on $f_k$ and $g_k$ and instead use
the equivalent maps
\be\label{eq:SUN-fghat-k}
\begin{split}
\hat f_k: S_{n-k}\otimes A_1 & \to S_{n-k+1} \\
\hat g_k: S_{n-k}\otimes A_{2k+1} & \to L_{2k+1,n-k+1}\\
\end{split}
\ee

Our choice of Chan Paton factors is such that we have a unique,
non-zero $SU(N)$ invariant line where to choose the maps $f_k$, $g_k$, $h_k$ maps.
Indeed   we note the isomorphism
\begin{equation} \label{eq:ASL}
A_\ell \otimes S_m \cong  L_{\ell+1,m} \oplus L_{\ell,m+1}
\end{equation}
and we choose nonzero intertwiners (projection operators)
\be
\begin{split}
\Pi^1_{\ell,m}: A_\ell \otimes S_m  \to   L_{\ell+1,m}  \\
\Pi^2_{\ell,m}: A_\ell \otimes S_m  \to   L_{\ell,m+1}  \\
\end{split}
\ee
These projection operators are very easily understood. Given a
tensor product of an antisymmetric and symmetric tensor, $t^a \otimes t^s$,
we could antisymmetrize one index of $t^s$ with the indices of $t^a$ to
obtain $\Pi^1_{\ell,m}$ or we could symmetrize one index of $t^a$ with the
indices of $t^s$ to obtain $\Pi^2_{\ell,m}$.

Now we can make $SU(N)$-equivariant maps by declaring:
\be
\begin{split}
\hat f_k &:= \nu_k \Pi^2_{1,n-k} \\
\hat g_k &:= \gamma_k \Pi^2_{2k+1,n-k} \\
& \Pi^1 \circ h_k  := \eta_k 1 \qquad     \Pi^2 \circ h_k =0   \\
\end{split}
\ee
where $\nu_k, \gamma_k$ and $\eta_k$ are scalars.
The map $\hat f_k$ is particularly easy to write when
viewed as a map $S_{n-k} \to S_{n-k+1}\otimes A_1^*$
since it is just given by symmetrization:
\be\label{eq:fk-symm}
f_k: {\rm Sym}(e_{a_1} \otimes \cdots e_{a_{n-k}} ) \rightarrow \nu_k
\sum_b {\rm Sym}(e_{a_1} \otimes \cdots e_{a_{n-k}}\otimes e_b) \otimes e^b
\ee
where $\{ e^b \}$ is a dual basis to $\{ e_a \}$. From this viewpoint
the $[f_k f_{k+1}]=0$ follows easily
as we are antisymmetrizing two indices which had been symmetrized.
For the $[g_k h_{k+1}]=0$ equation we note that the composition
(given by contracting with the interior amplitude) is a map
$L_{2k+3, n-k} \to L_{2k+1,n-k+1}\otimes A_1 $ but working out
the tensor product there is no nonzero intertwiner, by Schur's lemma.
Finally, $f_k$ can be defined in terms of $h_k g_k$, and since the
relevant space of intertwiners is one-dimensional it is always
possible to choose $\nu_k$ appropriately given $\gamma_k$ and $\eta_k$.
To be precise
\be
\Pi^2_{2k+1,n-k}(f_k \otimes 1) = \eta_k \hat g_k
\ee
When the scalars are related in this way the $A_\infty$ MC equations
are solved.

%
%\cg{I don't think this paragraph was quite right.  }
%
%In the same spirit $g_k$ is derived from the natural   projection
%$S_{n-k} \otimes A_{2k+1} \to L_{2k+1,n-k+1}$ and $h_k$ is the natural embedding $L_{2k+1,n-k+1} \to S_{n-k+1} \otimes A_{2k}$.
%It is straightforward to test that this defines a boundary amplitude. We can sketch why the various pieces of the MC equation vanish.
%It should be clear that $[f_k f_{k+1}]=0$, as we are antisymmetrizing two indices which had been symmetrized.
% Similarly,
%$[g_k h_{k+1}]$ is essentially the composition of an injection into the first summand of \ref{eq:ASL} and a projection on the second summand
%and is thus zero. Finally, we have $[h_k g_k] = f_k$.
%

It is not hard to build a sequence of Branes $\bar \fN_n$ based on the conjugate representations $\bar S_m$, $\bar L_{n,m}$.
They have Chan-Paton factors
\be\label{eq:NbarCP-1}
\CE(\bar \fN_n)_k = \begin{cases} \bar L^{[-k-n]}_{N-2k,n+k+1} & 0 \leq k < \frac{N}{2}\\
\bar S^{[k+1-n-N]}_{n+N-k} & \frac{N}{2} \leq k \leq N-1 \\
\end{cases}
\ee
%
%We can pick Chan Paton factors
%\begin{align}
%\CE_{N-k} &= \bar S^{[1-k-n]}_{n+k} \qquad 1 \leq k < N/2+1
%\cr \CE_{k} &= \bar L^{[-k-n]}_{N-2k,n+k+1}\qquad 0 \leq k < N/2
%\end{align}
%\cg{Check ranges of $k$.}

Finally, we can look at strip complex. We take $N$ even so that we can produce Branes for
the negative half-plane using rotation by $180$ degrees, as in the $\CT^N_\vartheta$ theories.
The rotation exchanges upper and lower vacua and, taking into account the degree shift
\eqref{eq:DegShft-3} we produce Branes  $\tilde\fN_n$:
\begin{align}
\CE(\tilde\fN_n)_{k} &= S_{\tilde n +k}^{[\tilde n + k + s +1 ]} \qquad  0  \leq k \leq \frac{N}{2}-1
\cr \CE(\tilde\fN_n)_{N-k-1} &= L_{N-2k-1,\tilde n + k }^{[\tilde n + k + s ]} \qquad 0 \leq k \leq \frac{N}{2}-1
\end{align}
where $\tilde n = n - \frac{N}{2} +1$ and $s$ is an arbitrary degree shift.

The complex for the segment with Branes $\fN_{n_L}$ and $\tilde\fN_{n_R}$ is a somewhat forbidding direct sum of
tensor products of $SU(N)$ representations:
\be\label{eq:FEARSOME}
\begin{split}
\left(  \oplus_{k=0}^{\frac{N}{2}-1} L^{[n_L-k]}_{2k+1, n+1-k} \otimes S^{[\tilde n_R+k+s+1]}_{\tilde n_R + k} \right)
&  \oplus  \left( \oplus_{k=0}^{\frac{N}{2}-1} S^{[n_L-k]}_{n_L-k} \otimes L^{[\tilde n_R+k+s]}_{N-2k-1, \tilde n_R + k} \right) \\
\end{split}
\ee
where the left and right sums come from the    lower and upper vacua, respectively. Using this block form
the differential is schematically of the form
\be
\oplus_{k=0}^{\frac{N}{2}-1}  \begin{pmatrix} 0 & g_k\otimes \tilde h_k  \\ h_k\otimes \tilde g_k  & 0 \\ \end{pmatrix}
\ee
%
%
%\cg{Check if subscripts are $k$ or $k\pm 1$. Review how this squares to $0$ because of MC equations }
%
%Intuitively, the differential injects $S \otimes L$ into something of the form $S \otimes A \otimes S$ and then projects to $L \otimes S$.
The techniques explained in Section \S \ref{subsec:RotIntfc-TSUN} below
(see especially equations \eqref{eq:ProveHopCoho-1} et. seq.) can probably be used to evaluate
the cohomology of \eqref{eq:FEARSOME}.

%Up to numeric pre-factors, we can write
%\begin{align}
%f_k(s_{a_1\cdots a_{n-k}}) &= s_{(a_1\cdots a_{n-k}} \delta_{a_{n-k+1})}^b \cr
%g_k(s_{a_1\cdots a_{n-k}}) &= \Pi_{L_{2k+1,n-k+1}}^{a_1\cdots a_{n-k};b_1 \cdots b_{2k+1}} s_{a_1\cdots a_{n-k}} \delta_{b_{1}}^{[c_{1})} \cdots \delta_{b_{2k+1}}^{c_{2k+1}]} \cr
%h_k(\ell) &= \ell_{(a_1\cdots a_{n-k+1});[b_1 \cdots b_k]}
%\end{align}

\subsubsection{Cyclic Isomorphisms Of The Theories}\label{subsubsec:CyclicIsoms}

In Section \S \ref{subsubsec:IsomTheory} we gave a formal definition
of an isomorphism of Theories. If we relax the constraint that $\vartheta$ be
small and positive we can illustrate that notion with some nontrivial isomorphisms
which will be extremely useful to us in Section \S \ref{subsec:RotIntfc-TSUN} below.

  If we consider the Theories $\CT^N_{\vartheta}$ and
$\CT^{N}_{\vartheta\pm \frac{2\pi}{N}}$ then there are isomorphisms
between them. For example
\be\label{eq:CyclIsom-TN}
\varphi^\pm: \CT^N_{\vartheta}\rightarrow \CT^{N}_{\vartheta\pm \frac{2\pi}{N}}
\ee
satisfies
\be
j \varphi^\pm = j\mp 1 \,\,\mod \,\, N
\ee
The map $\varphi_{ij}^+: R_{i,j} \to R_{i-1,j-1}$ is a map $\IZ \to \IZ$ or
$\IZ^{[1]} \to \IZ^{[1]}$ and consequently has degree zero,
 \emph{except} when $i=0$ or $j=0$. If $i=0$ then
\be
\varphi_{0,j}: R_{0,j} \to R_{N-1,j-1}
\ee
is a map $\IZ^{[1]} \to \IZ$ for all $j=1,\dots, N-1$. This necessarily has degree $-1$.
Similarly,
\be
\varphi_{i,0}: R_{i,0} \to R_{i-1,N-1}
\ee
is a map $ \IZ \to \IZ^{[1]}$ for all $i=1,\dots, N-1$. This necessarily has degree $+1$.
Note that $\varphi_I$ has degree zero for a cyclic fan of vacua.
 It is natural to take $\varphi_{ij}$ to be multiplication by $1$, together with an appropriate degree shift.
Since this map preserves cyclic  ordering, and $b_{ijk}$ all have the same value,  the condition \eqref{eq:BetaPres} is satisfied.

 If we consider the Theories $\CT^{SU(N)}_{\vartheta}$ and
$\CT^{SU(N)}_{\vartheta\pm \frac{2\pi}{N}}$ then there are isomorphisms
between them. Once again
\be\label{eq:CyclIsom-TSUN}
\varphi^\pm: \CT^{SU(N)}_{\vartheta}\rightarrow \CT^{SU(N)}_{\vartheta\pm \frac{2\pi}{N}}
\ee
satisfies
\be
j \varphi^\pm = j\mp 1 \,\,\mod\,\, N
\ee
For $\varphi_{ij}$, so long as $i\not=0,j\not=0$ we have
\be
\begin{split}
\varphi_{ij}: A_{j-i}^{[1]} & \to A_{j-i}^{[1]} \qquad i< j \\
\varphi_{ij}: A_{N+j-i}^{[1]} & \to A_{N+j-i}^{[1]} \qquad i> j \\
\end{split}
\ee
and for $i=0$ or $j=0$ the representations are cunningly chosen so that
\be
\begin{split}
\varphi_{0,j}: A_{j}^{[1]} & \to A_{j} \qquad  j=1,\dots, N-1 \\
\varphi_{i,0}: A_{N-i} & \to A_{N-i}^{[1]} \qquad  j=1,\dots, N-1 \\
\end{split}
\ee
Again, it makes sense to take all of these to be the identity map, up to the appropriate degree shift.

\subsubsection{Relation To Physical Models}\label{subsubsec:RelationTSUN-Physical}

We now briefly explain how the above models arise in a physical context.
The first class of examples, $\CT^N_{\vartheta}$,  is meant to correspond to the physics of the simple LG model with
  target space  $X= \IC$ with Euclidean metric and superpotential
\be\label{eq:TN-Superpot}
W =   \zeta \frac{N+1}{N}\left( \phi - e^{-\I N \vartheta} \frac{\phi^{N+1}}{N+1}\right) .
\ee
The prefactor is introduced so that, with the relation $z_k = \zeta \bar W_k$ the vacuum weights
coincide with those in \eqref{eq:CyclicWt}. Indeed the vacua
are the critical points of the superpotential $\phi_k =  e^{\I \vartheta + \frac{2 \pi \I k}{N}}$.
(Here $\zeta$ is a phase introduced in Section \S \ref{lgassuper}. In those sections we fix a superpotential
and vary $\zeta$.)    The $S$-matrix and soliton spectrum have been worked out in
\cite{Fendley:1992dm}. All the data agree with the theory $\CT^N_{\vartheta}$.

We refer to Section \S \ref{boundary} for a general discussion of boundary conditions in LG models.
For now, we can describe the specialization to this simple model. Following Section
\S \ref{goodclass}, we want $\Im(\zeta^{-1} W) \to + \infty$ at large $\vert\phi\vert$ for branes on a boundary of the positive
half-plane and $\Im(\zeta^{-1} W) \to - \infty$ at large $\vert\phi\vert$ for branes on a boundary of the negative half-plane.
The sign of $\Im(\zeta^{-1} W)$ at large $\vert \phi\vert$ is governed by that of $-\sin((N+1)\arg \phi - N \vartheta)$.
Accordingly, the $\phi$-plane at infinity  is subdivided into a
sequence of $2N+2$ angular sectors of width $\frac{\pi}{N+1}$ with boundaries $\arg\phi = \frac{N}{N+1}\vartheta + \frac{2s}{2N+2}\pi$, $s\in \IZ$.
Typical  boundary conditions are represented by
open curves in the $\phi$ plane whose endpoints go to infinity in sectors labeled by $s\in \IZ$ such that
for a boundary of a positive half-plane we have:
\be\label{eq:LeftBoundaryAsymp}
{\rm Left~~ boundary}: \qquad  \frac{4s-2}{2N+2}\pi + \frac{N}{N+1}\vartheta < \arg \phi < \frac{4s}{2N+2}\pi + \frac{N}{N+1}\vartheta
\ee
and for the boundary of a negative half-plane we have:
\be
{\rm Right~~ boundary}: \qquad  \frac{4s}{2N+2}\pi + \frac{N}{N+1}\vartheta < \arg \phi < \frac{4s+2}{2N+2}\pi + \frac{N}{N+1}\vartheta
\ee

It is instructive to try to draw the $\fC_k$ branes in the $\phi$-plane.
In Section \S  \ref{subsec:RotIntfc-TSUN} we construct a family of Branes for
the positive half-plane $\hat{\fB}_k$ such that $\hat{\fB}_0$ is the left
Lefshetz thimble
\footnote{For a definition see Section \S \ref{thimbles}.}
$\fT_0$  and successive brane are obtained by rotation by $\frac{2\pi}{N}$ in the
space-time plane. It is shown in  Section \S  \ref{subsec:RotIntfc-TSUN}
that for $2 \leq k \leq [ \frac{N}{2} ]$ we have $\hat{\fB}_k = \fC_{k-1}^{[1]}$.
Now equation \eqref{eq:Ck-CP-Spaces} only defines $\fC_k$ for $k<N/2$ so
the rotation procedure extends the definition to larger values of $k$.
It turns out we extend
$\fC_k$ by a sequence of down-vacua thimbles, and $\tilde \fC_k$
by a sequence of up-type thimbles.
When this is done for the $\CT^N_{\vartheta}$ Theory the rotation
operation is periodic and $\hat{\fB}_{N+1} = \hat{\fB}_0$. In order
to draw figures of the Branes we should bear in mind that the
Chan-Paton spaces for LG branes are obtained from intersections with
the \emph{right} Lefshetz thimbles. See equation \eqref{eq:CP-realized} below.
Using this one can deduce that the $\fC_k$ branes are Lagrangians stretching
between sectors labeled $s$ and $s+1$ in \eqref{eq:LeftBoundaryAsymp},
where $s$ depends on $k$.

\begin{figure}[htp]
\centering
\includegraphics[scale=0.25,angle=0,trim=0 0 0 0]{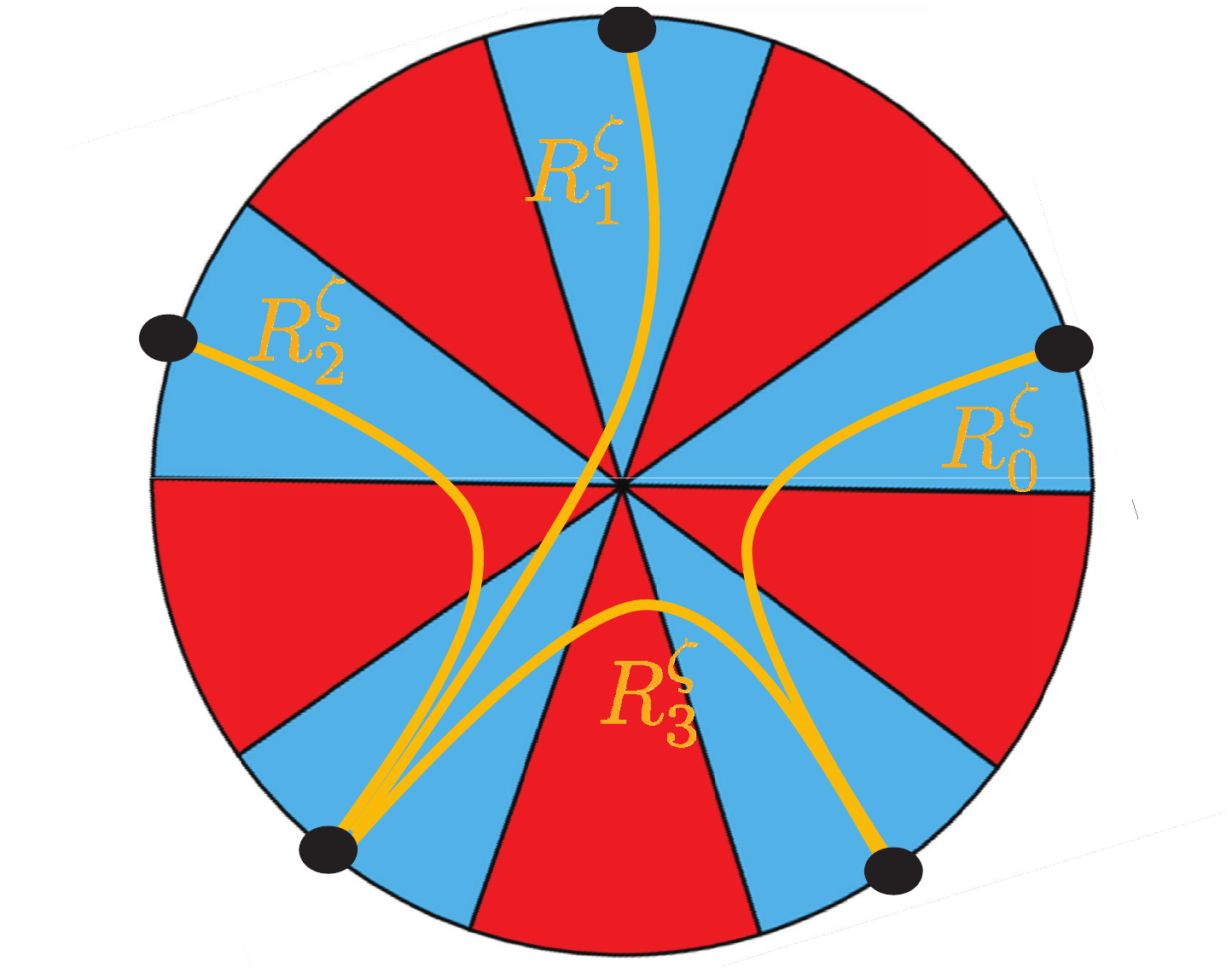}
\caption{This figure illustrates the  stability sectors in the $\phi$-plane
for the $\CT^N_{\vartheta}$ Theory for $N=4$ and small positive $\vartheta$.
Branes for left boundaries must asymptote to infinity in the red regions.
Branes for right boundaries must asymptote to infinity in the blue regions. These
regions are divided equally by the rays along which Lefshetz thimbles asymptote.
The right thimbles are shown in gold.  }
\label{fig:TN-BRANES-C}
\end{figure}
\begin{figure}[htp]
\centering
\includegraphics[scale=0.25,angle=0,trim=0 0 0 0]{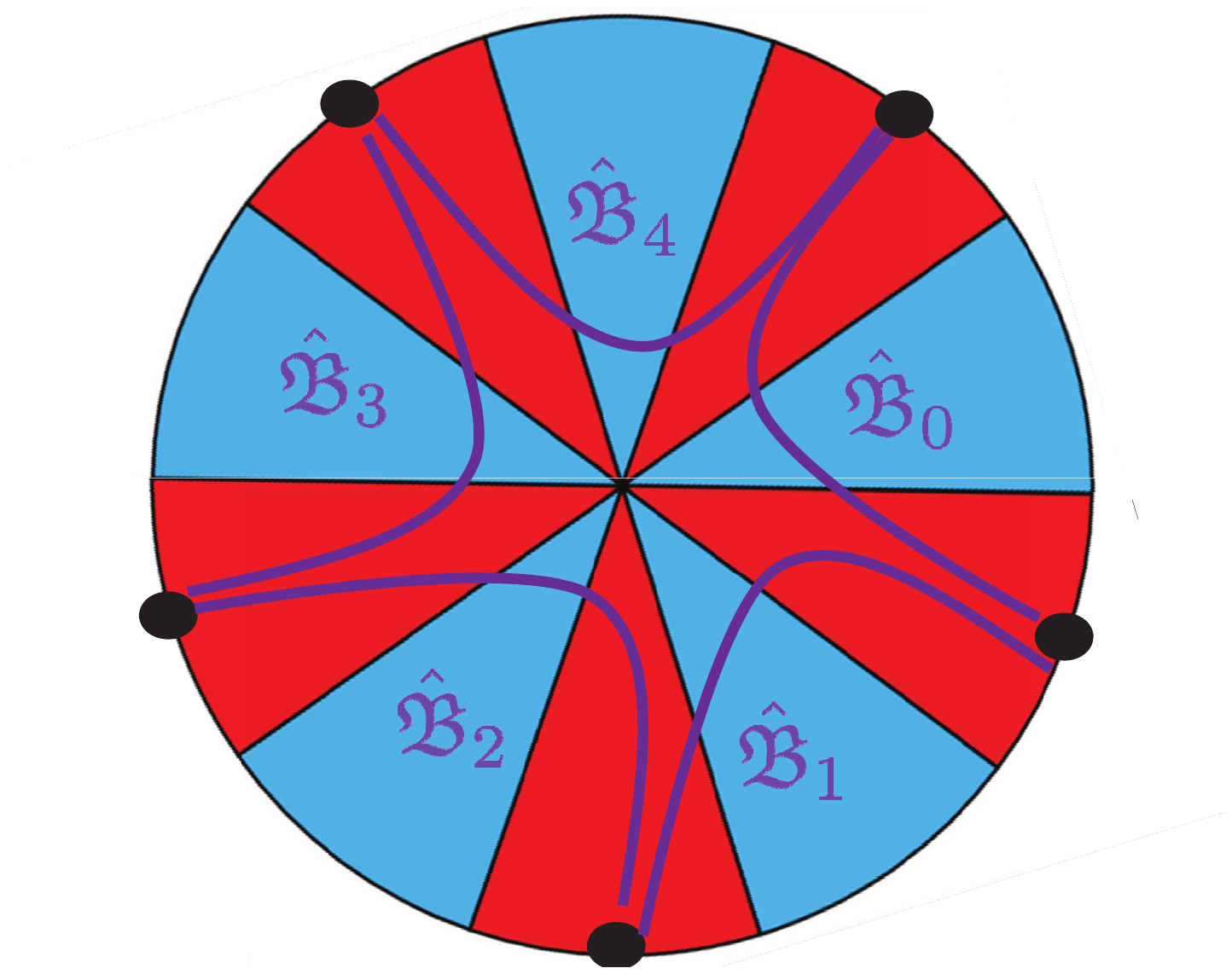}
\caption{A system of elementary Branes for the $\CT^N_\vartheta$ Theory
for $N=4$ and  small positive $\vartheta$. The Brane $\fC_{1}$ is the Brane $\hat\fB[2]$. }
\label{fig:TN-BRANES-D}
\end{figure}

Let us illustrate the procedure for $N=4$. The stability sectors and a
basis (for the relevant relative homology group) of right Lefshetz thimbles
is shown in Figure \ref{fig:TN-BRANES-C}. For $N=4$ the only $\fC_k$ Brane
defined by \eqref{eq:Ck-CP-Spaces} has $k=1$, and corresponds to $\hat{\fB}_2^{[-1]}$.
From the intersection numbers we see that it has the shape of $\hat{\fB}_2$ in
Figure \ref{fig:TN-BRANES-D}. The Brane $\hat{\fB}_1$ is neither a left thimble,
nor a $\fC_k$ Brane. The Branes $\hat{\fB}_3$ and $\hat{\fB}_4$ turn out to be
thimbles in this case. Finally, the Branes are related by $2\pi/5$ rotations
\emph{in the $\phi$-plane}. There is a similar picture for Branes $\tilde\fC_k$ on the
right boundary of a negative half-plane obtained by rotating by $\pi/10$. In
particular, $\tilde \fC_1$ stretches from $\arg\phi=\pi/10$ at infinity
to $\arg\phi= 5\pi/10$ at infinity. We conjecture that the obvious
generalization of \ref{fig:TN-BRANES-D} holds for all values of $N$.
It seems quite likely that the Branes described here are closely related to those
studied in (in the conformal theory) in \cite{Maldacena:2001ky} and in section 6.4 of \cite{Moore:2004yt}.

One can now go on to consider the local operators between these Branes.
This is worked out in detail for several pairs of Branes in the $\fC^N_\vartheta$
Theories in Section \S \ref{subsec:VacCat-SUN}.

The complex of ground states on a segment is built from
a vector space generated by intersections between the left and right branes.
The intersection number, i.e. the Witten index of the
complex of ground states on the segment, is robust under continuous deformation
thanks to the boundary conditions at infinity. Note that $\fC_1$, as depicted in
Figure \ref{fig:TN-BRANES-D} is equivalent to a sum of Lefshetz thimbles
$L^\zeta_1 + L^\zeta_2  +L^\zeta_3 $ while $\tilde \fC_1$ is analogously equivalent to
$R^\zeta_0 + R^\zeta_1 + R^\zeta_3$.
%
%\cg{Well, I should really check that the orientations are correct
%for this to be true...}
%
 The representation in terms of sums of thimbles gives the
complex \eqref{eq:CkCktilde-cplx}, while the brane $\fC_1$ shown in Figure
\ref{fig:TN-BRANES-D} and its analogue for $\tilde \fC_1$ have zero intersection.
These are consistent because  \eqref{eq:CkCktilde-cplx} has no cohomology.

Using Remark 5 near equation \eqref{eq:Strip-HalfPlane} we can go on to compute
the actual spaces of BPS states more generally
once we know both the space of local operators between Branes, and how
the Branes behave under rotation (in the $(x,\tau)$ plane) by $\pi$.

The second class of examples is meant to correspond to the A-model twist of the
affine Toda theory, i.e. the Landau-Ginzburg model whose target space is
the subvariety of $X\subset (\IC^*)^N$ defined by
\be
Y_1 \cdots Y_N = q
\ee
and
\be\label{eq:TSUN-Superpot}
W = Y_1 + \cdots  + Y_N
\ee
In our case $q= e^{-\I \frac{N}{N-1} \vartheta}$. The expected $SU(N)$ global symmetry is not fully manifest
in this LG model. Clearly there is a symmetry by the Weyl group of $SU(N)$, and the
$N$ critical points are permuted by the center of $SU(N)$. The fundamental group of $X$
can be identified with the root lattice of $SU(N)$ and the solitons have a ``winding number'' of
 $\log Y_i$  which is independent of $i$ and  equal to $k/N$ modulo integers for some $k$.
 Thus, the winding numbers can be identified with the weight lattice of $SU(N)$.

According to \cite{Hori:2000ck,Hori:2003ic} the B-model mirror
of this A-model is the supersymmetric   $\IC P^{N-1}$, with B-twist.
This model has manifest $SU(N)$ global symmetry.
\footnote{Indeed, the mirror transformation is a T-duality along the $(\IC^*)^N$
isometries.}
The study of this model goes back some time \cite{D'Adda:1978kp}\cite{Witten:1978bc}.
and the solitons of type $(i,i+1)$ are in the fundamental
representation of $SU(N)$, while those of type $(i,i+k)$ are boundstates of
$k$ fundamental solitons. This justifies nicely our choice of the $R_{ij}$
used in \eqref{eq:SUN-Rij} above.
The exact S-matrix for the $\IC P^N$ model was written in \cite{Koberle}.
The Witten indices $\mu_{ij}$ were  computed in \cite{Cecotti:1992rm,Hori:2000ck}.

A nice geometrical interpretation of the Branes should be available in the
$\IC P^{N-1}$ model in terms of homogeneous vector bundles. We propose that
the thimbles may be identified as
\be
\fT_j = \CO(-j)  \qquad 0\leq j < \frac{N}{2}
\ee
for the down-vacua, and
\be
\fT_{N-j} = \Lambda^{N-2j} TX \otimes \CO(j-N)  \qquad \frac{N}{2} \leq j  \leq N-1
\ee
for the up-vacua. The main justification for this is that the space $Q$-closed
local operators between these thimbles coincides with the results
\eqref{eq:HopCoho-1}-\eqref{eq:HopCoho-3} found below using formal techniques.
The branes $\fN_n$ are obtained from the down-vacua thimbles by rotations, which can be interpreted as B field shifts,
and hence the integer $n$ with the first Chern class on the bundles.
%
%\cg{up to a constant we should determine. Add link to sec 7.9}
%

We should note a few differences between the purely formal presentation of this
section and the physical models. First, in the physical models the fermion number
of the $ij$ soliton sector is
\footnote{Incidentally, this gives an example where the adiabatic formula
for the fermion number, \eqref{explained} below, which is often used in the
literature, is in fact not correct.}
\be
f_{ij} = \{ \frac{i-j}{N} \}
\ee
where, for a real number $x$,  $\{ x \} \in [0,1)$ is the fractional part of $x$.
That is, $x = [x] + \{ x \}$. In general, it is possible to add an exact one-form
to a conserved fermion number current $J^F$ and thereby define a different consistent
fermion number. In the LG model this means that $J^F \to J^F + df$, and if the
function $f$ has values $f(\phi_i) = f_i$ at the critical points of $W$ then
the fermion number $\int_{\IR} *J^F $ is shifted according to  $f_{ij} \to f_{ij} -f_i + f_j$.
Since we wish to apply the Koszul rule it is useful to make such a redefinition
to get integral fermion numbers. One choice which achieves this is  $f_i = i/N$.
In that case  we obtain the fermion number assignments used in
\eqref{eq:ExpleTN-webrep} and \eqref{eq:SUN-Rij} above. Of course, this still
leaves the ambiguity further integral shifts of $f_i$, but those modifications
are dealt with in Section \S \ref{subsec:OnDegrees} above.

Another difference from the
physical models is that we work over $\IZ$. We make a canonical choice of interior
amplitude $\beta$ and it would be very gratifying to know if it corresponds to that
which applies to the physical models.

\section{Categories Of Branes}\label{sec:CategoriesBranes}

\subsection{The Vacuum \afty-Category}\label{subsec:VacCategory}

In this section we would like to discuss the properties of Branes associated
to a   Theory $\CT$ and a half-plane $\CH$.
Since we will be comparing branes with different Chan-Paton factors we should recall the
definition of $R_J$ from Section \S \ref{subsubsec:WebRep-Halfplane}. We
 denote by $R_J(\CE)$ and $R^\p(\CE)$ the $\IZ$-modules built as in \eqref{eq:RJ} from some Chan Paton factors $\CE_i$
and in this section  $R_J$ will denote the ``bare space'' with trivial CP factors
\be \label{eq:RJb}
R_J:=  R_{j_1,j_2} \otimes \cdots \otimes R_{j_{n-1},j_n}.
\ee
In defining the morphisms of the vacuum category below we will make use of the space
$\widehat{R}_{ij}$ defined as the direct sum of $R_J$ over all half-plane fans of the form $J=\{i, \dots, j \}$,
that is:
\be
\widehat{R}_{ij}:= R_{ij} \oplus ~ \left( \oplus_k' R_{ik} \otimes R_{kj} \right)
\oplus  \left( \oplus_{k_1,k_2}' R_{ik_1} \otimes R_{k_1 k_2}\otimes  R_{k_2 j} \right)  \oplus \cdots
\ee
where the prime in the direct sum indicates that we only sum over half-plane fans in $\CH$.
\footnote{Note that, as opposed to $R_{ij}$,
the expression $\widehat{R}_{ij}$   depends on a choice of half-plane $\CH$ and is only defined
for half of the pairs $(i,j)$. It will also be convenient to define $\widehat{R}_{ii}=\IZ$, concentrated
in degree zero. Care should be taken when using this notation to distinguish webs from extended webs.  }
This allows us to write the \afty-algebra defined by \eqref{eq:R-AFTYALG} as
\begin{equation}
R^\p(\CE) = \oplus_{z_{ij} \in \CH } \CE_i \otimes \widehat{R}_{ij} \otimes \CE^*_j.
\end{equation}

Until now, we always extended all our maps to give zero whenever some of the arguments cannot be fit together.
This is self-consistent, but it hides some structure which is sometimes useful to make manifest.
The natural way to discuss operations which only make sense (or are only nontrivial)
 for certain pairs of arguments is to use a categorical language.
In the present case, we need an $A_\infty$ category:

\bigskip
\noindent
\textbf{Definition:} Suppose we are given the data of a
Theory $\CT$ and a half-plane $\CH$.
The associated \emph{Vacuum \afty-category} $\fVac$ has as objects
the vacua $i,j,\dots \in \IV$ while the space of morphisms is given by
%
%\be\label{eq:Hom-DEFS}
%\Hom(j,i) = \begin{cases} \widehat{R}_{ij} &  z_{ij} \in \CH \\
%\IZ & i=j \\
%0 & z_{ij} \notin \CH \end{cases}
%\ee
%
\be\label{eq:Hom-DEFS}
\Hom(j,i) = \begin{cases} \widehat{R}_{ij} &  z_{ij} \in \CH \\
\widehat{R}_{ii}=\IZ & i=j \\
0 & i\not=j \qquad {\rm and} \qquad z_{ij} \notin \CH \end{cases}
\ee
The \afty-compositions $m^{\fVac}$ are given by $  \rho_\beta(\ft_\CH)$ (defined as before,
but without the contractions of Chan Paton factors). See \eqref{eq:VacCatComp} below for
a more precise statement.

% The objects in the category are the vacua in $\IV$. The spaces of morphisms $\Hom(i,j)$
%are the $\widehat{R}_{ij}$. The morphisms are composed through the usual map
%This is our definition of
%
%Notice that $\Hom(i,j)$ is non-trivial only if $z_{ij}  \in \CH$.

We should make a number of remarks about this definition

\begin{enumerate}

\item   Recall that $z_{ij} \in \CH$ means that if $\CH$ is translated
so that the origin is on its boundary then $z_{ij}$ is in the translated
copy of $\CH$.

%    If we use standard webs,
%we will add the graded identity element and its scalar multiples by hand in $\Hom(i,i) = \IZ$ for future convenience.
%
%\cg{I don't understand this sentence which was here:``
%Given our definition of $m^{\fVac}$ we have a non-unital \afty-category.''. Does it mean ``it would have been non-unital if we had not added the %units by hand?''}
%

%\item If we included empty half-fans to accommodate the possibility of
%Chan-Paton complexes, we also have $\Hom(i,i)=\IZ$, concentrated in degree zero, which behave as identities.
%  Thus the vacuum category is akin to an \emph{exceptional collection}.
%The resemblance is not accidental.
%
%\cg{Combine remarks 2 and 3. Remark 3 is a bit   cryptic. Is it the set $\IV$ which
%is akin to an exceptional collection?  Add forward reference where it is
%further explained? }
%

\item
The category depends on   the data $\CT$ and $\CH$. We will generally suppress this dependence in the notation
but if we wish to stress the dependence or distinguish different choices of
data then we will indicate them in the argument and write some or all of the
data by writing   $\fVac(\CT,\CH)$
and so forth.

\item Note that we defined $\fVac$ with the ``bare'' $R_J$.
 We can ``add'' Chan-Paton factors to define a new \afty-category which we will denote as
  $\fVac(\CE)$ (when $\CT,\CH$ are  understood, and as $\fVac(\CT,\CH, \CE) $
  when they are not).  The morphism spaces are determined by
  replacing $\widehat{R}_{ij} \to  \CE_i \otimes \widehat{R}_{ij}\otimes \CE_j^*$
  in \eqref{eq:Hom-DEFS}. To define the \afty-multiplications
  we tensor $m^{\fVac}$ with the obvious contraction operations $\left(\CE_1 \otimes \CE_2^*\right) \otimes \left(\CE_2 \otimes \CE_3^*\right)
\to \left(\CE_1 \otimes \CE_3^*\right)$.

\item  Note that  $\Hom(i,i) = \IZ$ means the complex
 concentrated in degree zero, consisting of scalar multiples of the graded
 identity element $\Id_i$. The definition fits in well with
 \eqref{eq:VacCatComp} if we use extended webs. Recall that the \afty-multiplications
involving $\Id_i$ are defined near \eqref{eq:GdId-p}.

\item There are three interlocking conventions for composition of morphisms,
which we will now spell out somewhat pedantically. First, one can compose morphisms
successively on the left or on the right. Second, one can read an equation from left
to right, or vice versa. Third, the time ordering implicit in successive operations might or might not
agree with the geometrical time ordering of increasing $y$ in the $(x,y)$ plane.
In equation \eqref{eq:Hom-DEFS}, $\Hom(j,i)$ refers to the set of arrows which go
\emph{from} object $j$ \emph{to} object $i$. It is useful to define
\be
\Hop(i,j):=\Hom(j,i)
\ee
where ${\rm Hop}$ is an abbreviation for ${\rm Hom}^{\rm opp}$.
Equations with $\Hop$ should generally be read from right to left.
Including the Chan-Paton factors we define Hom-spaces by using
\be\label{eq:Hop-ij-CP}
\Hop^{\CE}(i,j) := \CE_i \otimes \Hop(i,j) \otimes \CE_j^*
\ee
%
%(when $z_{ij}\in \CH$)
%
 so that the \afty-multiplications
\be
\Hop^{\CE} (i_0,i_1)\otimes \Hop^{\CE}(i_1,i_2) \otimes \cdots \otimes \Hop^{\CE}(i_{n-1},i_n)
\rightarrow \Hop^{\CE}(i_0,i_n)
\ee
are computed from
\be\label{eq:VacCatComp}
\rho_{\beta}(\ft_{\CH})(r_1,\dots, r_n) \in \Hop^{\CE}(i_0,i_n)
\ee
with $r_s \in \Hop^{\CE}(i_{s-1},i_s)$. Successive morphisms are
composed on the left and the successive composition of
arrows should be read from right to left. Now, if $\CH$ is the
positive half-plane $x\geq x_0$ with boundary on the left then
this successive composition of morphisms can be visualized as taking place
forward in ``time'' $y$. However, by the same token, if $\CH$
is the negative half-plane $x\leq x_0$ with boundary on the right
then successive composition on the left and reading from
right to left corresponds to going
\emph{backwards} in time $y$. This leads to some awkwardness
when we discuss web representations and categories associated
to strips and interfaces, since these involves both positive
and negative half-plane webs. (Of course, for the negative
half-plane,  reading the composition of morphisms from left to right
corresponds to going forward in time, but reversing the
time ordering corresponds to composition of morphisms in the \emph{opposite} category. )
We will adopt the convention that successive operations on
$\CE_{LR}$ of \eqref{eq:ELRdef} act on the left and
correspond to transitions forward in time. This makes
$\CE_{LR}$ an $R_L^\p \otimes (R_R^\p)^{\rm opp}$
\afty-module.
(That is, an $R_L^\p - R_R^\p $ bimodule.) Similarly,
$i \to \CE_{L,i} \otimes \CE_{R,i}^*$ defines a
$\fVac(\CE_L) \times \fVac(\CE_R)^{\rm opp}$ module
of \afty-categories, in the sense of Definition 8.14
of \cite{ClayBook2}.

\end{enumerate}

\textbf{Examples}

\begin{enumerate}

\item Consider the top left taut positive  half-plane web in
Figure \ref{fig:HALFPLANE-TAUTWEB}. Choose  arbitrary
elements  $r_1 \in \widehat{R}_{ij}$ and
$r_2 \in  \widehat{R}_{jk}$. In this case the composition $m^{\fVac}(r_1,r_2)$
is   simply $r_1 \otimes r_2 \in \widehat{R}_{ik}$.

\item Now consider the bottom left taut web in
Figure \ref{fig:HALFPLANE-TAUTWEB}. This taut web leads to
a  contribution to $m_2^{\fVac}$ which we can illustrate as follows.
If $r_1 \in R_{i_1, j}$
and $r_2 \in R_{j,i_n}$ then the contribution
of this web to
\be
m_2^{\fVac}(r_1, r_2)   \in {\rm Hop}(i_1, i_n) = \widehat{R}_{i_1, i_n}
\ee
is
computed as follows. The interior vertex has a cyclic fan of vacua
\be
I = \{ i_n,j,i_1, i_2, \dots, i_{n-1} \}.
\ee
Let $\beta_I$ be the component
of $\beta$ in $R_I$ for this cyclic fan. Then we consider
\be
K_{14} K_{23} (r_1\otimes r_2 \otimes \beta_I)
\ee
where   $K_{23}$ contracts  $R_{j, i_n} \otimes R_{i_n,j}\to \IZ$
and then $K_{14}$ contracts   $R_{i_1 j } \otimes R_{j,i_1} \to \IZ$ leaving behind an
element of
\be
R_{i_1, i_2} \otimes \cdots \otimes R_{i_{n-1},i_n} \subset \widehat{R}_{i_1, i_n}
\ee

\item Similarly, the two taut webs on the right in
Figure \ref{fig:HALFPLANE-TAUTWEB} contribute to $m_1$.
The first at order $\beta$ with one contraction of $K$
and the second at order $\beta^2$ with three contractions of $K$.

\end{enumerate}

\subsection{Branes And The Brane Category $\fB\fr$}\label{subsec:BraneCat}

We are now ready to define an $A_\infty$ category of branes $\fB\fr$,
associated to a given choice of Theory $\CT$ and half-plane $\CH$.
The objects of $\fB\fr$ are Branes, i.e. pairs $\fB=(\CE,\CB)$ of
  \emph{some} choice of Chan Paton data $\CE$ together with a compatible
boundary amplitude $\CB$.
Recall that a boundary amplitude $\CB$ is a degree one element
\be\label{eq:BEL}
\CB \in \oplus_{i,j\in \IV} \CE_i \otimes \Hop(i,j) \otimes \CE_j^*
\ee
which satisfies the Maurer-Cartan equation in $\fVac(\CE)$:
\be
\sum_{n=1}^\infty \rho_\beta(\ft_\CH)(\underbrace{\CB, \dots, \CB}_{n\ {\rm times}} ~ ) =0.
\ee

The space of morphisms from $\fB_2=(\CE^2,\CB_2)$, a Brane in $\fVac(\CE^2)$ to
$\fB_1=(\CE^1,\CB_1)$, a Brane in $\fVac(\CE^1)$ is defined to be
\be\label{eq:BHOM}
\Hop(\fB_1, \fB_2) :=
\oplus_{i,j\in \IV} \CE_i^1 \otimes \Hop(i,j) \otimes (\CE_j^2)^*.
\ee
In order to  define the composition of morphisms
\be
\delta_1 \in \Hop(\fB_0, \fB_1), \quad \delta_2 \in \Hop(\fB_1, \fB_2) ,  \dots, \delta_n \in
\Hop(\fB_{n-1}, \fB_n)
\ee
It is useful to observe that the boundary amplitude $\CB$
of a Brane $\fB$ is,  thanks to the definition \eqref{eq:BHOM},
a morphism in  $\Hop(\fB,\fB)$. With this in mind it makes sense to
define the multiplication operations in $\fB\fr$
using the formula
\be\label{eq:BraneMultiplications}
M_n(\delta_1,\dots, \delta_n) :=
   m\left(\frac{1}{1-\CB_0},\delta_1, \frac{1}{1-\CB_1}, \delta_2, \dots, \delta_n, \frac{1}{1-\CB_n} \right).
\ee
where $m$ is the tensor product of $m^{\fVac}$ with the natural contraction on CP spaces.
Note that $M_n(\delta_1,\dots, \delta_n)\in \Hop(\fB_0,\fB_n)$.

After some work (making repeated use of the fact that the $\CB_a$ solve the Maurer-Cartan equation)
one can show that the $M_n$ satisfy the \afty-relations and hence $\fB\fr$ is an \afty-category.
Although this is well-known we provide a simple proof in Appendix \ref{app:cat}, below
equation \eqref{eq:BrCatDef}.

\bigskip
\noindent
\textbf{Remarks}:
\begin{enumerate}

\item As in our notation
for $\fVac$, the \afty-category  $\fB\fr$
implicitly depends on the Theory $\CT(\IV,z,\CR, \beta)$, as well as
the half-plane $\CH$. When we wish
to stress this dependence we will write these as arguments. Note that
$\fB\fr(\CT,\CH)$ does not depend on any specific choice of Chan-Paton spaces $\CE$.

\item We should note that the passage from $\fVac$ to $\fB\fr$
 is quite standard in the theory of homotopical algebra
where a Brane is known as a ``twisted complex.''
See, for examples,
Section 3l, p.43 of \cite{SeidelBook}  or Definition 8.16, p.614 of \cite{ClayBook2}.
Our category $\fB\fr$ would be denoted ${\rm Tw}(\fVac)$ in the mathematics
literature.

\item It will be useful to extend the Brane category to include the extended webs of Sections
\ref{subsec:ExtendedWebs}\ref{subsec:RepExtendedWebs}. If we use extended webs then
the extended web representations \eqref{eq:extwebrep} mean that the morphisms now
have a ``scalar part'' $Q_i$ with $Q_i^2 =0$. Now we define the morphism spaces to be
\be\label{eq:ExtendedWebHom}
\Hop(\fB',\fB) = \left( \oplus_i \CE_i' \otimes \CE_i^*\right)  \oplus
\left( \oplus_{z_{ij}\in \CH} \CE_i' \otimes \widehat{R}_{ij}\otimes \CE_j^*\right)
\ee
and we again refer to the component in $ \oplus_i \CE_i' \otimes \CE_i^* $ as the ``scalar part.''
The scalar part of $\delta\in \Hop(\fB',\fB)$  is just
a collection of linear maps $f_i: \CE(\fB)_i \to \CE(\fB')_i$. Those maps can be composed or applied to \eqref{eq:BHOM}
and hence we can define the multiplications \eqref{eq:BraneMultiplications} by using
the extended webs in the taut element $\ft_{\CH}$ in the definition \eqref{eq:VacCatComp} of $m^{\fVac}$.
We will generally work with this extended Brane category.

\item The category of vacua is naturally a full subcategory of  the category of Branes: each vacuum $i$ maps to
``thimble Branes'' $\fT_i$ (also called simply ``thimbles''). The reason for the name is explained in \S \ref{thimbles}. The
Chan-Paton spaces of $\fT_i$ are defined by $\CE(\fT_i)_j = \delta_{ji} \IZ$, in degree zero. It follows
that $\Hop(\fT_i, \fT_j) = \Hop(i,j)$. Moreover, the
boundary amplitude $\CB_i$ of $\fT_i$ is taken to be zero (even if we used extended webs).  Thus the insertion
of $\frac{1}{1-\fT_i}$ in \eqref{eq:BraneMultiplications} has no effect on the contractions and hence $M_n = m_n^{\fVac}$ on chains of
morphisms between thimbles.

\item The thimbles form an ``exceptional collection,''  from which all other branes, by definition, arise as twisted complexes.
Other exceptional collections will exist in $\fB\fr$, especially the ones obtained from mutations of the thimble collection.
We will discuss these mutations briefly in Section \S  \ref{subsec:Mutations} below.

\item Finally, we note that $M_1^2=0$ and hence each space of morphisms $\Hop(\fB_1, \fB_2)$ is a chain complex.
In the physical applications the cohomology of the differential $M_1$ on this complex is interpreted as BRST-invariant
local operators which can be inserted at the boundary of the half-plane $\CH$ and change the boundary conditions.
See the end of Section \S \ref{subsec:BindPoints}, Remark 5 for further discussion.

\end{enumerate}

\subsection{Homotopy Equivalence Of Branes}\label{subsec:HomEquivBr}

Working in the framework of the extended Brane category we can define a
notion of homotopy equivalence which will be extremely useful in Sections
\S \ref{sec:Interfaces}-\ref{sec:GeneralParameter}.

Because  $\delta \to M_1(\delta)$ for  $\delta \in \Hop(\fB',\fB)$ is a differential
we can consider $M_1$-exact and $M_1$-closed morphisms.
The composition $M_2$ is compatible with $M_1$. We can use $M_1$ and $M_2$ to define useful notions such as homotopy and homotopy equivalence,
treating the branes as an analogue of a chain complex.

We define   two $M_1$-closed morphisms $\delta_{1,2}$ to be \emph{ homotopic} if they differ by an $M_1$-exact morphism, i.e.
\begin{equation}\label{eq:hmtpy-Morph}
\delta_1 \sim \delta_2\qquad  \longleftrightarrow \qquad \delta_1 - \delta_2 = M_1(\delta_3)
\end{equation}
for some $\delta_3$. Similarly, we define  two branes $\fB$ and $\fB'$ to be \emph{ homotopy equivalent}, denoted,
\begin{equation} \fB \sim \fB',  \end{equation}
if there are two $M_1$-closed morphisms $\delta: \fB \to \fB'$ and $\delta': \fB' \to \fB$ which
are inverses up to homotopy. That is:
\begin{equation}\label{eq:hmtpy-Br}
M_2(\delta, \delta') \sim \Id \qquad \qquad M_2(\delta', \delta) \sim \Id.
\end{equation}
Recall that $\Id$ is the graded identity defined in \eqref{eq:GdId}. \footnote{Physically, $\delta$ and $\delta'$ correspond to boundary-changing local operators whose OPE is the identity operator up to
exact boundary operators. Homotopy equivalent branes essentially represent the same D-brane in a topologically twisted physical model.}

If we have two morphisms $\delta_1$ and $\delta_2$ in $\Hop(\fB',\fB)$ and $\Hop(\fB'',\fB')$,
respectively, with scalar parts
given by collections of maps $f_{1,i}$ and $f_{2,i}$, the scalar part of the
composition $M_2(\delta_1, \delta_2)$ is simply given by the composition of the scalar parts $ f_{2,i}f_{1,i}$.
Similarly, if the scalar parts of the boundary amplitudes of $\fB'$ and $\fB$
  are $Q_i$ and $Q'_i$, respectively, and the scalar part of a morphism $\delta: \fB \to \fB'$
is $f_i$, then the scalar part of $M_1(\delta)$ is
\begin{equation}
Q'_i f_i \pm f_i Q_i
\end{equation}
Thus the scalar part of an $M_1$-closed morphism is a collection of chain maps between the Chan Paton factors,
homotopic morphisms have homotopic scalar parts and homotopy equivalent branes have homotopy equivalent Chan Paton factors.

 The following will prove to
be a very useful criterion for homotopy equivalence between branes $\fB_1$ and $\fB_2$ in our discussions in Sections
\S\S \ref{sec:Interfaces},\ref{sec:CatTransSmpl},\ref{sec:GeneralParameter}.
(See, for examples equation \eqref{eq:BfBpHom}   and the discussion at the end of Section \S
\ref{subsec:BindPoints}.)
We consider the special case where the two Branes have the \emph{same} Chan-Paton data:
$\CE(\fB_1)=\CE(\fB_2)$,  and where the  homotopy equivalence can be written
as  a  morphism of the form
$\Id + \epsilon$, where $\epsilon$ is a degree zero element in $  \Hop(\fB_2,\fB_1)$  with no scalar part.
Note that $\Id\in  \Hop(\fB_2,\fB_1)$ makes sense because the Chan-Paton factors
are assumed to be the same.
If we require such a morphism between two branes $\fB_1$ and $\fB_2$ to be $M_1$-closed, we find  the relation:
\begin{equation}\label{eq:HomEqBr}
\begin{split}
M_1(\Id + \epsilon) & := \rho_\beta(\ft_\CH)[\frac{1}{1-  \CB_2},\Id +\epsilon, \frac{1}{1-\CB_1}]\\
& = \rho_\beta(\ft_\CH)[\CB_2,\Id] + \rho_\beta(\ft_\CH)[ \Id,\CB_1] + \rho_\beta(\ft_\CH)[\frac{1}{1-  \CB_2}, \epsilon, \frac{1}{1-\CB_1}]\\
& =   \CB_1 - \CB_2 +  \rho_\beta(\ft_\CH)[\frac{1}{1-  \CB_2}, \epsilon, \frac{1}{1-\CB_1}]=0, \\
\end{split}
\end{equation}
where in the last line we used the condition that $\CB_2$ has degree one
 and the  third term  refers to unextended webs. Recall that
 we can regard $\CB_1$ and $ \CB_2$ to be themselves elements of
$\Hop(\fB_2,\fB_1)$, so the equation \eqref{eq:HomEqBr} makes sense.

We must also guarantee the invertibility up to homotopy \eqref{eq:hmtpy-Br}
of $\Id + \epsilon$. This  is actually automatic in our setup.
Indeed, we can solve $M_2(\Id + \epsilon , \Id - \epsilon') = \Id$ recursively by
\begin{equation}
\epsilon' = \epsilon - M_2(\epsilon, \epsilon')
\end{equation}
The recursion will stabilize after a finite number of iterations thanks to the finiteness properties of the webs involved in $M_2$.
As $M_2$ is associative up to homotopy, $\Id - \epsilon'$ is also a left inverse up to homotopy.
More generally, if we consider a  morphism $f+\epsilon$ with scalar part $f$ and $g$ is
an inverse of $f$ up to homotopy then we can find $\epsilon'$ so that $g-\epsilon'$ is
an inverse of $f+\epsilon$, up to homotopy.

For examples of homotopy equivalent pair of branes, see Sections \S \ref{subsec:SWallIntfc} and \S \ref{subsec:RotIntfc-TSUN}.

In all the constructions of homotopy equivalences in the rest of the paper, we will actually produce both morphisms
$\Id + \epsilon , \Id - \epsilon'$ explicitly and will not rely on these
finiteness properties to argue for the existence of the inverse. Aside from being philosophically more
satisfying, this will be useful because we can then extend the results to cases of interest with
an infinite number of vacua (such as those relevant to knot homology and the 2d/4d wall-crossing-formula).

\subsection{Brane Categories And The Strip}

Given a Theory $\CT$ we can associate two \afty-categories to the strip geometry $[x_{\ell}, x_r] \times \IR$.
We have  the category of Branes
$\fB\fr_L$ associated to the left boundary, controlled by the operation $\rho_\beta(\ft_L)$
and the category of Branes $\fB\fr_R$ associated to the right boundary, controlled by the operation $\rho_\beta(\ft_R)$.

Our strip bimodule operation $\rho_\beta(\ft_s)$
defined in \eqref{eq:rho-beta-strip} above can be given several different useful interpretations.
We will use the concept of a ``module for an \afty-category'' as defined in 8.14 of
\cite{ClayBook2}. In general if $\CA$ is an \afty-category with objects $x_\alpha$ then
a (left) module over $\CA$ consists of a choice of graded $\IZ$-module $\CM(x_\alpha)$
for each object together with a collection of maps
\be\label{eq:ModuleMaps}
m_n^{\CM}: \Hop(x_0,x_1) \otimes \cdots \otimes \Hop(x_{n-1},x_n)\otimes \CM(x_n) \to \CM(x_0)
\ee
which are defined for $n\geq 1$, are of degree $2-n$, and  satisfy the categorical
analog of the identities \eqref{eq:afty-module}. As usual, a bimodule for a pair of \afty-categories
$(\CA , \CB)$ is a module for the \afty-category $\CA \times \CB^{\rm opp}$.

Our first interpretation is that $\rho_\beta(\ft_s)$ can be used to define a bimodule
for the pair of vacuum categories $(\fVac_L , \fVac_R)$. The objects of   $\fVac_L \times \fVac_R^{\rm opp}$
are pairs of vacua $(i,j)$ and we take $\CM(i,j) = \delta_{i,j} \IZ$, with $\IZ$ in degree zero.

As a second interpretation, we can define a bimodule for the pair of \emph{Brane} categories
$(\fB\fr_L,\fB\fr_R)$. The objects of $\fB\fr_L\times \fB\fr_R^{\rm opp}$ are pairs of Branes
and now we take
\be
\CM(\CB_L, \CB_R) := \CE_{LR} = \oplus_{i\in \IV} \CE_{L,i}\otimes \CE_{R,i}^*
\ee
To define the module maps \eqref{eq:ModuleMaps} in this case
we take
%
%\cg{What if the CP factors of the Branes have nothing to do
%with those of $\CE_{LR}$? }
%
\be
\begin{split}
m^{\CM}(\delta_1,\dots, \delta_n;g;\delta'_1,\dots, \delta'_{n'}) & := \\
   \rho_\beta(\ft_s)\biggl(\frac{1}{1-\CB_{L,0}},\delta_1,
    %\frac{1}{1-\CB_{L,1}}, \delta_2,
     \dots, \delta_n, & \frac{1}{1-\CB_{L,n}} ; g; \frac{1}{1-\CB_{R,0}},\delta'_1,
     %\frac{1}{1-\CB_{R,1}}, \delta'_2,
     \dots, \delta'_{n'}, \frac{1}{1-\CB_{R,n'}} \biggr).\\
\end{split}
\ee
where
\be
\delta_1 \in \Hop(\CB_{L,0}, \CB_{L,1}), \dots, \delta_n \in \Hop(\CB_{L, n-1}, \CB_{L,n})
\ee
\be
\delta_1' \in \Hop(\CB_{R,0}, \CB_{R,1}), \dots, \delta_{n'} \in \Hop(\CB_{R, n'-1}, \CB_{R,n'}).
\ee

The maps $\delta_1 \to m^{\CM}(\delta_1;\ \cdot \ ;\ \cdot \   )$ and $\delta'_1 \to m^{\CM}(\ \cdot \ ; \ \cdot \  ;\delta'_1)$ are particularly interesting for us.
Indeed, we have
\begin{equation}
m^{\CM}\left( M_1(\delta_1);\ \cdot \ ;\ \cdot \  \right) =\pm d_{LR} m^{\CM}(\delta_1;\ \cdot \ ;\ \cdot \ ) \pm m^{\CM}(\delta_1;\ \cdot \ ;\ \cdot \  ) d_{LR}
\end{equation}
and a similar equation for $d(\ \cdot \  ; \ \cdot \  ;M_1(\delta'_1))$.

Thus $M_1$-closed morphisms map to chain maps
for the chain complexes of approximate ground states $(\CE_{LR},d_{LR})$, homotopic morphisms map to
homotopic chain maps and homotopy equivalent Branes give rise to homotopy equivalent chain complexes.
In particular, homotopy equivalent Branes have isomorphic spaces of exact ground states on the segment.
This is a special case of a general principle: homotopy equivalent Branes are the ``same'' Brane for most purposes.

We can also use these constructions to interpret the right-Branes as elements of a standard category
along the following lines. Given a right Brane $\CB_R$ we can define a module for the \afty-category
$\fVac_L$ by the assignment
\be\label{eq:funny}
\CM(i) =   \CE_i^*
\ee
Then the module maps \eqref{eq:ModuleMaps} are just
\begin{equation}\label{eq:SeqMap-1}
r_1\otimes \cdots \otimes r_n \otimes g \to \rho_\beta(\ft_s)[r_1, \dots, r_n;g;\frac{1}{1-\CB_R}]
\end{equation}
which thus defines a family (parametrized by $\CB_R$) of   $A_\infty$ modules
 for $\fVac_L$ with $i \to   \CE_i^*$.

Notice that in this formula we have Chan-Paton factors on the right boundary arguments, but not the left boundary arguments.
The data \eqref{eq:SeqMap-1} defines a set of maps
\begin{equation}
\widehat R_{i_0, i_1} \otimes \widehat R_{i_1, i_2} \cdots \widehat R_{i_{n-1}, i_n} \to  \CE_{R,i_0}^* \otimes \CE_{R,i_n}.
\end{equation}
This observation suggests the following definition of a \emph{mapping category}:
\footnote{The term ``mapping category'' is nonstandard. It appears to be
closely related to the notion of a Koszul dual to an \afty-category.
See, for example, \cite{KoszulDual}.}
The objects will be
sets of Chan-Paton data $\CE= \{ \CE_i \}_{i\in \IV}$ and the morphisms $\Hop(\CE,\CE')$ will
be the set of collections of linear maps
\begin{equation}\label{eq:JustMaps}
\widehat R_{i_0, i_1} \otimes \widehat R_{i_1, i_2} \cdots \widehat R_{i_{n-1}, i_n}  \to (\CE'_{i_0})^*\otimes \CE_{i_n}.
\end{equation}
The difference between the collection of maps \eqref{eq:ModuleMaps} and the collection of maps of the
form \eqref{eq:JustMaps} defining a generic morphism in the mapping category is that the latter need not
satisfy the \afty-relations.
Let us denote a collection of such linear maps by $\mathfrak{m}$ and the value on a monomial
of the form   $P =  \widehat r_{i_0, i_1} \otimes \widehat r_{i_1, i_2} \cdots \widehat r_{i_{n-1}, i_n} $
by $\fm[P]$.
Two morphisms $\mathfrak{m}_1\in \Hop(\CE,\CE')$ and $\mathfrak{m}_2\in \Hop(\CE',\CE'')$ are composed
as
\begin{equation}
\left(\mathfrak{m}_1 \circ \mathfrak{m}_2 \right)[P] = \sum_{\mathrm{Pa}_2(P)} \mathfrak{m}_1[P_1]  \mathfrak{m}_2[P_2]
\end{equation}
where on the right-hand-side we contract the spaces $(\CE'_{i})\otimes (\CE'_j)^* $ in the natural way.
This composition is associative, thus making the mapping category an \emph{ordinary} category.

There is also a differential on the morphism spaces of the mapping category given by
\begin{equation}
\mathfrak{d} \mathfrak{m}[P] = \sum_{\mathrm{Pa}_3(P)} \epsilon \, \mathfrak{m}[P_1,m^{\fVac_L}[P_2],P_3].
\end{equation}
It is compatible with the composition. Collections of maps annihilated by the differential make the
assignment $i \to \CE_i$ into an \afty-module for $\fVac_L$.

Now, as noted above, for every Brane $\CB_R \in \fB\fr_R$ we can define an object in the mapping category,
namely the Chan-Paton data of the Brane for the negative half-plane.
Moreover, if we regard the mapping category as a very degenerate
version of an \afty-category then we can define an \afty-functor from $\fB\fr_R^{\rm opp}$ to the mapping category.
The image of morphisms $\delta_1, \cdots \delta_m$ is the collection of maps
\begin{equation}
\fm[r_1, \dots, r_n] =  \rho_\beta(\ft_s)[r_1, \dots, r_n;\cdot ;\frac{1}{1-\CB_{R,0}},\delta_1,
     %\frac{1}{1-\CB_{R,1}}, \delta'_2,
     \dots, \delta_{m}, \frac{1}{1-\CB_{R,m}}]
\end{equation}
The usual convolution identity for strip webs gives the functor property.

This allows us to identify the right Branes as elements of a standard category, if needed.

\subsection{Categorification Of 2d BPS Degeneracies}\label{subsec:Cat-Muij}

There is a very nice way to define the  complexes $\widehat{R}_{ij}$ using matrices of chain complexes.
Suppose there are $N$ vacua so we can identify
$\IV = \{1,\dots, N\}$. Introduce the elementary $N\times N$ matrices $e_{ij}$
with a 1 in the $i^{th}$ row, $j^{th}$ column and zero elsewhere. Then we
can define $\hat R_{ij}$ from  the formal product
\be\label{eq:Cat-KS-prod}
\IZ\cdot \textbf{1}  + \oplus_{z_{ij}\in \CH} \widehat{R}_{ij} e_{ij} = \bigotimes_{z_{ij}\in \CH} (\IZ\cdot \textbf{1} + R_{ij} e_{ij} )
\ee
where $\textbf{1}$ is the $N\times N$ unit matrix and
 in the tensor product we order the factors left to right by the
 clockwise order of the argument of $z_{ij}$. Phase-ordered products of this kind
involving operators, rather than complexes,
have appeared in the work of Cecotti and Vafa
\cite{Cecotti:1992rm} and in the work of Kontsevich and Soibelman
\cite{Kontsevich:2008fj,Kontsevich:2009xt} on wall-crossing.
\footnote{Of course, such phase ordered products have also appeared
in many previous works on Stokes data.}
Our work here can be considered as a ``categorification'' of the
wall-crossing formulae. This will be discussed further in
 the sections on wall-crossing  \S \ref{subsec:Cat-FramedWC} and
 \S \ref{sec:GeneralParameter}.

In the physical context one finds that $R_{ij}$ are complexes of approximate groundstates
and the Witten indices
\be
\mu_{ij} := \Tr_{R_{ij}} (-1)^F
\ee
are known as the (two-dimensional) BPS degeneracies. With $F$ denote the integer fermion number, coinciding with the integer degree
we defined on $R_{ij}$.

They were extensively studied in \cite{Fendley:1992dm,Cecotti:1992rm}.
Since fermion number behaves well under tensor product we can take a trace (on the web representations) of
\eqref{eq:Cat-KS-prod} to obtain
\be\label{eq:2d-CVKS-prod}
  \textbf{1}  + \oplus_{z_{ij}\in \CH} \widehat{\mu}_{ij} e_{ij} = \bigotimes_{z_{ij}\in \CH} (  \textbf{1} + \mu_{ij} e_{ij} ).
\ee
where
\be
\hat \mu_{ij} := {\Tr}_{\widehat{R}_{ij}} (-1)^F.
\ee
The Cecotti-Vafa-Kontsevich-Soibelman wall-crossing formula states that certain continuous deformations of
Theories lead to jumps in the BPS degeneracies $\mu_{ij}$ while the $\widehat{\mu}_{ij}$ remain constant.
In Section \S \ref{sec:GeneralParameter} we will discuss the categorified version of that statement.

\textbf{Remark}: In LG theories, $\hat \mu_{ij}$ can be computed by intersecting infinitesimally rotated Lefschetz thimbles \cite{Cecotti:1992rm}\cite{Hori:2000ck}. See Section \S \ref{subsec:CatSpecGen} for a categorized version of that statement.

\subsection{Continuous Deformations}
%\cg{Polish up}
In this section we elaborate a bit on the meaning of equation.  \eqref{eq:HomEqBr} (repeated here)
\begin{equation}
\CB' - \CB+ \rho_\beta(\ft_\CI)[\frac{1}{1-\CB'};\epsilon(s); \frac{1}{1-\CB}] =0 \end{equation}
This material will not be used later
and the reader should feel free to skip it.

If we are given a brane $\CB$ and some degree zero morphism $\epsilon$
with no scalar part we can solve the constraint
recursively to find some new $\CB'$.
It is interesting to verify that the result of such recursion is indeed a boundary amplitude: we can compute
\begin{align}
\rho_\beta(\ft_\CI)[\frac{1}{1-\CB'}] = &\rho_\beta(\ft_\CI)[\frac{1}{1-\CB}]+ \rho_\beta(\ft_\CI)[\frac{1}{1-\CB'}; \CB' - \CB;\frac{1}{1-\CB}]=  \cr
&= - \rho_\beta(\ft_\CI)[\frac{1}{1-\CB'}; \rho_\beta(\ft_\CI)[\frac{1}{1-\CB'};\epsilon; \frac{1}{1-\CB}] ;\frac{1}{1-\CB}] = \cr
&= \rho_\beta(\ft_\CI)[\frac{1}{1-\CB'};\rho_\beta(\ft_\CI)[\frac{1}{1-\CB'}];\frac{1}{1-\CB'};\epsilon; \frac{1}{1-\CB}] +\cr&+  \rho_\beta(\ft_\CI)[\frac{1}{1-\CB'};\epsilon;\frac{1}{1-\CB} \rho_\beta(\ft_\CI)[\frac{1}{1-\CB}],\frac{1}{1-\CB}]\end{align}
This relation proves recursively that $\rho_\beta(\ft_\CI)[\frac{1}{1-\CB'}]=0$.

If we have a continuous family $\epsilon(s)$, the corresponding family of branes $\CB(s)$ is the exponentiation of an exact deformation.
Start from
\begin{equation}
\dot \CB(s) + \rho_\beta(\ft_\CI)[\frac{1}{1-\CB(s)};\dot \CB(s); \frac{1}{1-\CB(s)} ; \epsilon(s);  \frac{1}{1-\CB}]+  \rho_\beta(\ft_\CI)[\frac{1}{1-\CB(s)};\dot \epsilon(s); \frac{1}{1-\CB}]  =0 \end{equation}
This equation is solved by $\dot \CB(s) = -\rho_\beta(\ft_\CI)[\frac{1}{1-\CB(s)};\dot \epsilon(s); \frac{1}{1-\CB(s)}]$.
We can exclude other solutions recursively, by writing $\dot \CB(s) = r -\rho_\beta(\ft_\CI)[\frac{1}{1-\CB(s)};\dot \epsilon(s); \frac{1}{1-\CB(s)}]$
and plugging into the second term in the equation.
Conversely, the exponentiation of an exact deformation is a family of isomorphisms.
Of course, the notion of isomorphism is sensible even when boundary amplitudes were defined, say, on $\IZ$.

\subsection{Vacuum And Brane Categories For The Theories $\CT^N_{\vartheta}$ And $\CT^{SU(N)}_{\vartheta}$ }\label{subsec:VacCat-SUN}

As an illustration of the above definitions we comment on  $\fVac$ and $\fB\fr$ for the Theories
 $\CT^N_{\vartheta}$ and $\CT^{SU(N)}_{\vartheta}$ described in Section \ref{subsec:CyclicVacWt}.
Again, we will work with very small, positive $\vartheta$.

Let us consider first the morphism spaces $\Hop(i,j)$ of $\fVac$, for either theory. In principle
these could be worked out from  equation \eqref{eq:Cat-KS-prod}, but it is easier to enumerate
the half-plane fans and work out the complexes in special cases. These
divide into 4 cases because (for small positive $\vartheta$)
 there are four distinct kinds of half-plane fans as enumerated below Figure \ref{fig:TNEXAMPLE-2}.
 Recall these depend on whether $i,j$ are upper or lower vacua, determined by
  the sign of the imaginary parts of $z_i$ and $z_j$. Of course we always
  have $\Hop(i,i) = \widehat{R}_{ii}  = \IZ$ in degree zero.
On the other hand, when   $i\not=j$ we have:

\begin{enumerate}

\item  $\{d \dots d \}$.
If both $i$ and $j$ are lower vacua, then $\Hop(i,j)$   is non-empty
only if $i \leq j$.
If $i<j$ the   possible half-plane fans $J$ between $i$ and $j$ can be enumerated
by all strictly increasing sequences of integers beginning with $i$ and ending with $j$.
The phase $\vartheta_{ij}$ of $z_{ij}$ is given very nearly (for $\vartheta\to 0^+$)
by $\tan(\vartheta_{ij}) = \cot(\pi \frac{i+j}{N} )$
so the product \eqref{eq:Cat-KS-prod} simplifies considerably and we may write
\be\label{eq:CatProdSC}
\IZ\cdot \textbf{1}  + \oplus_{0 \leq i < j < \frac{N}{2} } \widehat{R}_{ij} e_{ij} = \bigotimes_{0 \leq i < j < \frac{N}{2} }
 (\IZ\cdot \textbf{1} + R_{ij} e_{ij} )
\ee
where the product on the RHS is ordered left to right by increasing values of $i+j$ (since we also have
$i<j$ the product is then well-defined). In terms of $\widehat{R}_{ij}$   we have
\be\label{eq:Hope-Expl1}
\Hop(i,j) = \begin{cases}
\widehat R_{ij} & 0 \leq i \leq  j <  \frac{N}{2} \\
 0 &  0 \leq j < i < \frac{N}{2}  \\
\end{cases}
\ee

\item $\{u \dots d \}$. Similarly, if  $i=N-k$ is an upper vacuum, so $1\leq k \leq \frac{N}{2}$
 and $0 \leq j< \frac{N}{2}$ is a lower vacuum then
\be\label{eq:Hope-Expl2}
\Hop(N-k,j)  = \begin{cases}  0 & 0 \leq j < k \\
\widehat{R}_{N-k,j}
   &  k \leq j < \frac{N}{2} \\
\end{cases}
\ee
where for $k\leq j$ we have
\be
\widehat{R}_{N-k,j}= \oplus_{\ell=k}^{j} R_{N-k,\ell}\otimes \widehat R_{\ell, j}
\ee

\item $\{ d \dots u \}$. Now if $i$ is a lower vacuum and $j = N-k$ is upper then
\be\label{eq:Hope-Expl3}
\Hop(i,N-k)    = \begin{cases}  0 & k \leq i < \frac{N}{2} \\
\widehat{R}_{i,N-k}  &  0 \leq i < k \\
\end{cases}
\ee
where for $i<k$
\be
\widehat{R}_{i,N-k} :=  \oplus_{\ell=i}^{k-1}   \widehat R_{i,\ell}\otimes R_{\ell, N-k}
\ee

\item $\{ u \dots u \}$. We label a pair of up vacua by $(N-k,N-t)$ with
$1 \leq k,t \leq \frac{N}{2}$. Then
\be\label{eq:Hope-Expl4}
\Hop(N-k, N-t)    = \begin{cases} 0   & t < k   \\
\IZ & t= k \\
 \widehat{R}_{N-k,N-t}  & k\leq \ell \leq s \leq t-1 \\
\end{cases}
\ee
where for $k \leq t-1$:
\be
\widehat{R}_{N-k,N-t} := R_{N-k,N-t} \oplus
 \oplus_{k\leq \ell \leq s \leq t-1 } R_{N-k,\ell}\otimes \widehat{R}_{\ell,s}\otimes   R_{s,N-t}.
\ee

\end{enumerate}

Now let us describe the multiplication operations on $\fVac$, again for both theories.
Since the only taut webs which contribute to $\rho_\beta$ have one or two boundary
vertices the \afty-category $\fVac$   is in fact just a differential graded algebra.
The differential $m_1$ arises from Figure \ref{fig:TNEXAMPLE-3}(b). The multiplication $m_2$ arises from
Figures \ref{fig:TNEXAMPLE-3}(a) and \ref{fig:TNEXAMPLE-3}(c). In \ref{fig:TNEXAMPLE-3}(a)
$J_1,J_2$ share a common d-type vacuum. In Figure \ref{fig:TNEXAMPLE-3}(c) the intermediate
vacuum is an up-type vacuum, $j_s \in [ \frac{N}{2}, N-1]$.

Now let us specialize the discussion to $\fVac(\CT^N_\vartheta)$.
In this case the spaces $R_{ij}$ are very simple and given by \eqref{eq:ExpleTN-webrep}.
The key spaces $\widehat{R}_{ij}$ for a pair of down vacua with $i<j$ is - as we have said above -
a sum over all increasing sequences of integers beginning with $i$ and ending
with $j$. Each sequence   contributes a summand $R_J \cong \IZ^{[|J|-1]}$ to $\widehat R_{ij}$.
Denoting generators
of $\Hop(i,j) $ by $e_J$, where $J$ is a positive half-plane fan, $m_1(e_J)$ is a signed
sum of generators obtained by inserting a down-type vacuum into $J$ in all possible ways.
For example, if $J = \{d_1, \dots, d_s \}$ with $0 \leq d_1 < d_s < \frac{N}{2}$ then
%
%We can denote the corresponding generator as $e_{J}$.
%The differential $m_1$ arises from taut webs with a single boundary vertex
%and must be of the form shown in Figure \ref{fig:TNEXAMPLE-3}(b) above,  which splits a line in two and thus acts on
%$e_J$ with $J=\{ d_1, \dots, d_s\}$ according to
%
\begin{equation}
m_1(e_{d_1 \cdots d_s}) = \sum_{n=1}^{s-1} \sum_{d=d_n+1}^{d_{n+1}-1} (-1)^{n-1} e_{d_1\cdots d_n d d_{n+1}\cdots d_s}
\end{equation}
 The sign is determined by the usual patient commutation of $K_{d_n d_{n+1}}$ through the first
 $(n-1)$ $R_{ij}$ factors in $R_J \otimes R_{d_n d} \otimes R_{d d_{n+1}} \otimes R_{d_{n+1} d_n}$.
 Note that $R_{d_{n+1} d_n}$ has degree zero and can be brought from the far right into the relevant
 place in the product to effect the contraction by $K$ in the definition of $\rho_\beta$.
It is easy to check that $m_1$ is nilpotent. Similar formulae hold for the other three types of
half-plane fans $J$: The operation $m_1$ is a signed sum of all fans where we insert one extra down vacuum.

Now consider the multiplication of $e_{J_1}$ and $e_{J_2}$ with
\be
J_1 = \{ j_1, \dots,j_{s-1},  j_s \} \qquad J_2 = \{j_s, j_{s+1}, \dots ,j_n \}
\ee
If the intermediate vacuum is down-type  $0 \leq j_s < \frac{N}{2}$ then
\be
m_2(e_{J_1}, e_{J_2}) = e_{J_1*J_2}
\ee
with
\be
J_1*J_2 := \{ j_1. \dots, j_{s-1}, j_s , j_{s+1}, \dots, j_n \}
\ee
If the intermediate vacuum   $j_s$ is an up-type vacuum, $j_s \in [ \frac{N}{2}, N-1]$ then
\be
\rho(\ft_{\CH})(e_{J_1}, e_{J_2}) = K_{j_s,j_{s-1}} K_{j_s,j_{s+1}} (e_{J_1} \otimes e_{J_2}\otimes \beta_{j_{s-1},j_s,j_{s+1}})
= e_{J_1\wedge J_2}
\ee
%
%\cg{Is there a sign here? I think so in order for $M_1^2=0$ below.}
%
with
\be
J_1\wedge J_2 := \{ j_1. \dots, j_{s-1},  j_{s+1}, \dots, j_n \}.
\ee
%

%
%The only other non-trivial operation is $m_2$. If $k$ is a lower vacuum,
%\begin{align}
%m_2(e_{i \cdots k}, e_{k \cdots j}) &= e_{i \cdots k \cdots j} \cr
%m_2(e_{i \cdots N-k}, e_{N-k \cdots j}) &= \pm \sum_d e_{i\cdots d \cdots j}
%\end{align}
%

Next let us turn to the Brane category $\fB \fr(\CT^N_\vartheta)$. We will
limit ourselves to the description of some of the Hom-spaces and their cohomology
using the differential $M_1$. As noted above, the $M_1$ cohomology will have the physical interpretation
as the space of $\CQ$-invariant local operators changing boundary conditions of one
Brane into another.

The easiest class of Branes to consider are of course the thimbles $\fT_i$ since
\be
\Hop(\fT_i, \fT_j) = \Hop(i,j)
\ee
and in  this case $M_n = m_n^{\fVac}$. In particular we can study the cohomology
of $m_1$. It is easy to see that the cohomology is nonzero in general.
This happens when the spaces \eqref{eq:Hope-Expl1}-\eqref{eq:Hope-Expl4} consist
of a single summand $R_{ij}$. Then $R_{ij}$ has a definite degree so $m_1$ must
be identically zero, and hence the cohomology is the space $R_{ij}$ itself.
Thus, for example, $\Hop(i,i), \Hop(i,i+1)$, $\Hop(N-k,k)$, $\Hop(k-1,N-k)$ and $\Hop(N-k,N-k-1)$
have $m_1=0$ and are equal to their own cohomologies.
%
%\cg{Also true for $j=i$ right?}
%
On the other hand, in the
other cases we can construct a contracting homotopy so that the cohomology vanishes.
Consider, for example, the case where $i<j$ are two down-type vacua and $i+1<j$.
Then define
\be
\kappa (e_{i d_2 \cdots d_{s-1} j} ) := \begin{cases}  e_{i d_3 \cdots d_{s-1} j } & d_2  = i+1 \\
 0 & d_2 > i+1 \\
 \end{cases}
\ee
One can check that $\kappa m_1 + m_1 \kappa = {\rm Id} $ and hence the cohomology vanishes.
Similarly, for $\Hop(N-k,j)$ with $j>k$ we can define a contracting homotopy
 operator
\be
\kappa (e_{N-k, d_2 \cdots d_{s-1} j} ) := \begin{cases}  e_{N-k , d_3 \cdots d_{s-1} j } & d_2  = k \\
 0 & d_2 > k  \\
 \end{cases}
\ee
and so forth.

In conclusion: \emph{The cohomology of $\Hop(\fT_i,\fT_j)$ is nonzero
 if and only if the only half-plane fan of the form $J=\{i, \dots, j \}$
is in fact $J = \{ i,j \}$. } We leave the calculation of the appropriate homotopy contractions to the enthusiastic reader.

As a test, we present the matrix of Poincar\'e polynomials for the $\Hop(\fT_i,\fT_j)$ for $N=10$:
\begin{equation}\label{eq:Text-PoinPol}
\left(
\begin{array}{cccccccccc}
 1 & y & y (y+1) & y (y+1)^2 & y (y+1)^3 & y (y+1)^4 & y (y+1)^3 & y (y+1)^2 & y (y+1) & y \\
 0 & 1 & y & y (y+1) & y (y+1)^2 & y (y+1)^3 & y (y+1)^2 & y (y+1) & y & 0 \\
 0 & 0 & 1 & y & y (y+1) & y (y+1)^2 & y (y+1) & y & 0 & 0 \\
 0 & 0 & 0 & 1 & y & y (y+1) & y & 0 & 0 & 0 \\
 0 & 0 & 0 & 0 & 1 & y & 0 & 0 & 0 & 0 \\
 0 & 0 & 0 & 0 & 0 & 1 & 0 & 0 & 0 & 0 \\
 0 & 0 & 0 & 0 & 1 & y+1 & 1 & 0 & 0 & 0 \\
 0 & 0 & 0 & 1 & y+1 & (y+1)^2 & y+1 & 1 & 0 & 0 \\
 0 & 0 & 1 & y+1 & (y+1)^2 & (y+1)^3 & (y+1)^2 & y+1 & 1 & 0 \\
 0 & 1 & y+1 & (y+1)^2 & (y+1)^3 & (y+1)^4 & (y+1)^3 & (y+1)^2 & y+1 & 1 \\
\end{array}
\right)
\end{equation}
Setting $y = -1$ we recover the Witten indices of $\Hop(\fT_i,\fT_j)$, i.e. the $\hat \mu_{ij}$:
\begin{equation}\label{eq:Text-PoinPol}
\left(
\begin{array}{cccccccccc}
 1 & -1 & 0 & 0 & 0 & 0 & 0 & 0 & 0 & -1 \\
 0 & 1 & -1 & 0 & 0 & 0 & 0 & 0 & -1 & 0 \\
 0 & 0 & 1 & -1 & 0 & 0 & 0 & -1 & 0 & 0 \\
 0 & 0 & 0 & 1 & -1 & 0 & -1 & 0 & 0 & 0 \\
 0 & 0 & 0 & 0 & 1 & -1 & 0 & 0 & 0 & 0 \\
 0 & 0 & 0 & 0 & 0 & 1 & 0 & 0 & 0 & 0 \\
 0 & 0 & 0 & 0 & 1 & 0 & 1 & 0 & 0 & 0 \\
 0 & 0 & 0 & 1 & 0 & 0 & 0 & 1 & 0 & 0 \\
 0 & 0 & 1 & 0 & 0 & 0 & 0 & 0 & 1 & 0 \\
 0 & 1 & 0 & 0 & 0 & 0 & 0 & 0 & 0 & 1 \\
\end{array}
\right)
\end{equation}
This supports our statement. Equation \eqref{eq:Text-PoinPol}, and
similar examples below follow a pattern that leads to
a conjectural formula for the general case. We leave it as a challenge to the
reader to give a proof of these formulae.

Next we can look at the morphisms between the Branes  $\fC_k$ defined in
Section \S \ref{subsec:CyclicVacWt} and the thimbles.
 Using the formula \eqref{eq:Ck-CP-Spaces} for the Chan-Paton spaces of $\fC_k$
we have
\be\label{eq:CkTj-hop}
\Hop(\fC_k, \fT_j )  = \Hop(N-k,j)^{[1]} \oplus \Hop(N-k-1,j) \oplus \Hop(k,j)
\ee
Consider the multiplications $M_n$ for this pair of Branes.
Since the only webs which contribute to $\ft_{\CH}$ are those shown in Figure
\ref{fig:TNEXAMPLE-3} the only nonzero multiplications $m_n^{\fVac}$ are
$m_1$ and $m_2$ and hence in the category of Branes likewise only $M_1$ and $M_2$
can be nonzero. Moreover $M_2$ coincides with $m_2$. However,
  because $\fC_k$ has nontrivial boundary
amplitudes $\CB(\fC_k)$ with components $\CB_{N-k,N-k-1}$, $\CB_{k,N-k-1}$, and
$\CB_{N-k,k}$ there will be an important difference between $m_1$ and $M_1$.
In particular,
\be
\begin{split}
M_1(\delta) & = m^{\fVac}(\frac{1}{1-\CB(\fC_k)}, \delta, \frac{1}{1-\CB(\fT_j)} ) \\
& = m^{\fVac}(\frac{1}{1-\CB(\fC_k)}, \delta  ) \\
& = m_1 (  \delta  )+ m_2(\CB(\fC_k), \delta)  \\
\end{split}
\ee

Now, in order to analyze the $M_1$-cohomology of \eqref{eq:CkTj-hop}  recall
that $\fC_k$ are only defined when $k$ is a down-type vacuum. Then,
when $j$ is a down-type vacuum we can use   \eqref{eq:Hope-Expl1} et. seq. above
to  find the   morphism spaces:
\begin{equation}
\begin{split}
\Hop(\fC_k, \fT_j ) & =
\begin{cases} \widehat R_{N-k,j}^{[1]} \oplus \widehat R_{N-k-1,j} \oplus \widehat R_{k,j}  & k+1 \leq j < \frac{N}{2} \\
R_{N-k,k}^{[1]} \oplus \IZ  & 0 \leq k  = j < \frac{N}{2} \\
0  & 0 \leq j < k < \frac{N}{2} \\
\end{cases}\\
\end{split}
\end{equation}

If $k=j$ then $\Hop(\fC_k, \fT_k) \cong \IZ^{[1]} \oplus \IZ$ and under
this isomorphism $M_1(x \oplus y ) = y \oplus 0$. The cohomology is
therefore zero. If $k+1 \leq j$ and we write the three components of $\delta $ as
\be
\delta = \delta_{N-k,j} \oplus  \delta_{k,j} \oplus \delta_{N-k-1,j} = (  \delta_{N-k,j} , \delta_{k,j},\delta_{N-k-1,j}  )
\ee
and  then we have
\begin{align}
M_1(\delta_{N-k,j},0,0) &= \left( m_1(\delta_{N-k,j}),0,0 \right) \cr
M_1(0, \delta_{k,j},0)&=\left( m_2(\CB_{N-k,k},\delta_{k,j}),   m_1(\delta_{k,j}),0 \right) \cr
&=\left( h_k m_2(e_{N-k,k},\delta_{k,j}),   m_1(\delta_{k,j}),0 \right) \cr
M_1(0,0,\delta_{N-k-1,j}) &= \left( m_2(\CB_{N-k,N-k-1},\delta_{N-k-1,j}),m_2(\CB_{k,N-k-1},\delta_{N-k-1,j}),  m_1(\delta_{N-k-1,j}) \right) \cr
&= \left(f_k m_2(e_{N-k,N-k-1},\delta_{N-k-1,j}),     g_k  m_2(e_{k,N-k-1},\delta_{N-k-1,j}), m_1(\delta_{N-k-1,j}) \right) \cr
\end{align}
The ordering is chosen so that the upper-triangular structure is clear.
It is a good exercise to show that $M_1^2=0$ (one must use $f_k = h_k g_k$).

Now we claim that there is a homotopy contraction of   $M_1$ of the form
\be
\begin{pmatrix} \kappa & \kappa_{12} & \kappa_{13} \\
0 &\kappa & \kappa_{23} \\
0 & 0 & \kappa
\end{pmatrix}
\ee
For this to be a homotopy contraction we need (we use the property that $h_k$ is
multiplication by $1$ here):
\be
\kappa_{12}(m_1(\delta_{k,j})) + m_1(\kappa_{12}(\delta_{k,j})) +
\kappa(m_2(e_{N-k,k}, \delta_{k,j})) + m_2(e_{N-k,k},\kappa(\delta_{k,j})) =0
\ee
with similar equations for $\kappa_{13}$ and $\kappa_{23}$. One can
solve this using
\be
 \kappa_{12}(e_{k,d_2,d_3,...,j}) = \begin{cases} - e_{N-k,d_3,....,j} & d_2=k+1 \\
 0 & {\rm else} \\
 \end{cases}
\ee
Thus, there are never boundary-condition-changing operators from $\fC_k$ to down-type thimbles.

Turning now to the case of thimbles for up-type vacua we write, for $1\leq t \leq \frac{N}{2}$,
\be\label{eq:CkTj-hop-2}
\begin{split}
\Hop(\fC_k, \fT_{N-t}  ) & = \Hop(N-k,N-t)^{[1]} \oplus \Hop(N-k-1,N-t) \oplus \Hop(k,N-t) \\
\end{split}
\ee
Using \eqref{eq:Hope-Expl1} et. seq. above  we find that this   vanishes when $k>t$.
When $t=k$ only the first summand, namely,
\be
\Hop(N-k,N-k)^{[1]}\cong \IZ^{[1]},
\ee
is nonzero so the complex is concentrated
in a single degree, therefore $M_1=0$ and the cohomology is nontrivial. For $k\leq t-1 $ all three summands
are nonzero and we expect that a nontrivial analysis like that above for down-type thimbles shows there are
no other cases with nonzero cohomology.

In conclusion we have given considerable evidence for the claim that $\Hop(\fC_k, \fT_j )$
only has nontrivial cohomology when $j=N-k$, in which case the
cohomology is one-dimensional.

As a test, we provide the matrix of Poincar\'e polynomials for $\Hop(\fC_k, \fT_j )$ for $N=10$:
\begin{equation}
\left(
\begin{array}{cccccccccc}
 0 & y+1 & (y+1)^2 & (y+1)^3 & (y+1)^4 & (y+1)^5 & (y+1)^4 & (y+1)^3 & (y+1)^2 & y \\
 0 & 0 & y+1 & (y+1)^2 & (y+1)^3 & (y+1)^4 & (y+1)^3 & (y+1)^2 & y & 0 \\
 0 & 0 & 0 & y+1 & (y+1)^2 & (y+1)^3 & (y+1)^2 & y & 0 & 0 \\
 0 & 0 & 0 & 0 & y+1 & (y+1)^2 & y & 0 & 0 & 0 \\
\end{array}
\right)
\end{equation}

We expect that a very similar story holds for
\begin{equation}
\Hop(\fT_i, \fC_k) = \Hop(i,N-k)^{[1]} \oplus \Hop(i,N-k-1) \oplus \Hop(i,k).
\end{equation}
It is straightforward to check that for $i=k $ there are only two nonzero
summands and the complex is isomorphic to $\IZ^{[1]} \oplus \IZ$
with   $M_1(x\oplus y) = y\oplus 0$, and hence the cohomology vanishes.
The other easy case is $i=N-k-1$. Then only the middle summand is nonzero
so we get nonzero cohomology. We expect that for the other values, $i\not= k,N-k-1$,
 a detailed analysis like that
we did above for the other order would show that for the
the cohomology vanishes, but we have not confirmed this.
In any case,  we expect that \emph{the only nonzero cohomology
appears for $\Hop(\fT_{N-k-1}, \fC_k)$, in which case it is one-dimensional. }
Again, the cohomology is limited to half-plane fans of length $2$.

As a test, we provide the matrix of Poincar\'e polynomials for $\Hop(\fT_i, \fC_k)$ for $N=10$:
\begin{equation}
\left(
\begin{array}{cccccccccc}
 (y+1)^2 & y+1 & 0 & 0 & 0 & 0 & 0 & 0 & 1 & \frac{(y+1)^2}{y} \\
 (y+1)^3 & (y+1)^2 & y+1 & 0 & 0 & 0 & 0 & 1 & \frac{(y+1)^2}{y} & \frac{(y+1)^3}{y} \\
 (y+1)^4 & (y+1)^3 & (y+1)^2 & y+1 & 0 & 0 & 1 & \frac{(y+1)^2}{y} & \frac{(y+1)^3}{y} & \frac{(y+1)^4}{y} \\
 (y+1)^5 & (y+1)^4 & (y+1)^3 & (y+1)^2 & y+1 & 1 & \frac{(y+1)^2}{y} & \frac{(y+1)^3}{y} & \frac{(y+1)^4}{y} & \frac{(y+1)^5}{y} \\
\end{array}
\right)
\end{equation}

Finally, we could look at the morphisms $\Hop(\fC_k, \fC_t)$. Each Brane has  Chan-Paton
spaces with nonzero support at three vacua and hence we now get nine summands:
\be
\begin{split}
\Hop(\fC_k, \fC_t) & =   \Hop(N-k,N-t) \oplus \Hop(N-k-1,N-t)^{[-1]} \oplus \Hop(k,N-t)^{[-1]}\\
 & \oplus \Hop(N-k,N-t-1)^{[1]} \oplus \Hop(N-k-1,N-t-1) \oplus \Hop(k,N-t-1) \\
  &\oplus \Hop(N-k,t)^{[1]} \oplus \Hop(N-k-1,t) \oplus \Hop(k,t) \\
 \end{split}
\ee
and the differential is
\be
M_1(\delta) = m_1 (  \delta  )+ m_2(\CB(\fC_k), \delta)  + m_2(\delta, \CB(\fC_t)).
\ee
A systematic analysis of the cohomology would be tedious. So we will limit ourselves to
the cases $t \leq k$.  If $t < k-1$ then $\Hop(\fC_k, \fC_t) =0$. If $t=k-1$
then the morphism space is concentrated in one degree and is $\Hop(N-k,N-k)^{[1]}$.
Therefore the cohomology is nonzero in this case. For $t=k$ we must carry out a
nontrivial computation. There are six nonvanishing morphism spaces
\be
\begin{split}
& \Hop(N-k,N-k) \oplus \Hop(N-k-1,N-k-1) \oplus \Hop(k,k) \\
& \oplus \Hop(N-k,N-k-1)^{[1]} \oplus \Hop(k,N-k-1)
\oplus \Hop(N-k,k)^{[1]} \\
\end{split}
\ee
This space has rank $7$ because
\be
\Hop(N-k,N-k-1)^{[1]} = R_{N-k,N-k-1}^{[1]} \oplus R_{N-k,k,N-k-1}^{[1]}
\ee
has rank $2$. A little computation shows that
\be
  M_1: \begin{pmatrix}
   \delta_{N-k,N-k}\\
   \delta_{N-k-1,N-k-1} \\
   \delta_{k,k}\\
    \delta_{N-k,N-k-1}\\
     \delta_{N-k,k,N-k-1}\\
      \delta_{k,N-k-1}\\
      \delta_{N-k,k}\\
\end{pmatrix}
\mapsto
\begin{pmatrix}
 0 \\ 0 \\ 0\\  \delta_{N-k,N-k} - \delta_{N-k-1,N-k-1}\\
\delta_{k,N-k-1}+\delta_{N-k,k} - \delta_{N-k,N-k-1}\\
  \delta_{k,k}-\delta_{N-k-1,N-k-1}\\  \delta_{N-k,N-k}-\delta_{k,k}  \\
\end{pmatrix}
\ee
A short computation then shows that the cohomology is rank one and
generated by  $(1,1,1,0,0,0,0)$.

In a similar way we conjecture the absence of cohomology when $t>k$.
As a test, we provide the matrix of Poincar\'e polynomials for $\Hop(\fC_k, \fC_t)$ for $N=10$:
\begin{equation}
\left(
\begin{array}{cccc}
 y^2+3 y+3 & \frac{(y+1)^4}{y} & \frac{(y+1)^5}{y} & \frac{(y+1)^6}{y} \\
 y & y^2+3 y+3 & \frac{(y+1)^4}{y} & \frac{(y+1)^5}{y} \\
 0 & y & y^2+3 y+3 & \frac{(y+1)^4}{y} \\
 0 & 0 & y & y^2+3 y+3 \\
\end{array}
\right)
\end{equation}

%In order to express this in a more useful form we choose a natural basis for $\Hop(\fC_k, \fT_j )$.
%Using \eqref{eq:ExpleTN-webrep} and \eqref{eq:Hope-Expl1} et. seq. we can choose a
%basis for  $\widehat R_{N-k,j}^{[1]} $ which can be identified with  $e_{J_1}$ where $J_1$ is a fan of down-type
%vacua with $J_1 = \{ \ell, \dots, j\}$ and $k \leq \ell\leq j $. Similarly for $\widehat R_{N-k-1,j} $
%we choose $e_{J_2} $ with  $J_2 = \{ \ell, \dots, j\}$ and $k+1 \leq \ell\leq j $, and again
%for $\widehat R_{k,j}  $ we have   $e_{J_3}$ where   $J_3 = \{ \ell, \dots, j\}$ and $k \leq \ell\leq j $.
%Then,
%
%\be
%M_1( e_{J_1}\oplus e_{J_2} \oplus e_{J_3} ) = \left( m_1(e_{J_1}) + e_{J_3} \right) \oplus m_1(e_{J_2}) \oplus
%\left( m_1(e_{J_3}) + e_{k,J_2} \right)
%\ee
%
%\cg{Need to check this! Signs? Check it squares to zero.}
%

Now let us consider briefly the $\CT^{SU(N)}_{\vartheta}$ theories of Section \S \ref{subsec:CyclicVacWt}.
The formulae \eqref{eq:Hope-Expl1}-\eqref{eq:Hope-Expl4} apply to this case as well. In this way we find
the following morphism spaces:

For two lower vacua,  $0\leq i<j < \frac{N}{2}$  we have:
\begin{equation}\label{eq:Morph-SUN-1}
\Hop(i,j) = \sum_{n\geq 1}  \sum_{\sum_{s=1}^n d_s =j-i} \bigotimes_{s=1}^n A^{[1]}_{d_s}
\end{equation}
If $i=N-k$ is an upper vacuum and $j$ a lower vacuum, we need $j\geq k$ for a nonzero morphism
 space. In this case we have:
\begin{equation}\label{eq:Morph-SUN-2}
\Hop(N-k,j) =A_{j+k} \oplus  \sum_{n>1}  \sum_{\sum_{s=1}^n d_s =j-k+1} A_{2k-1+d_1}\otimes \bigotimes_{s=2}^n A^{[1]}_{d_s}.
\end{equation}
If $i$ is a lower vacuum, and $j = N-k$ an upper vacuum, similar considerations apply with $i<k$:
\begin{equation}\label{eq:Morph-SUN-3}
\Hop(i,N-k) =A^{[1]}_{N-k-i} \oplus  \sum_{n> 1} \sum_{\sum_{s=1}^n d_s =k-i} \bigotimes_{s=1}^{n-1} A^{[1]}_{d_s} \otimes A_{N-2k+d_n}^{[1]}
\end{equation}
Finally if we have two upper vacua an $t>k$ then
\begin{equation}\label{eq:Morph-SUN-4}
\Hop(N-k,N-t) = A_{N+k-t} \oplus \sum_{n>1} \sum_{\sum_{s=1}^n d_s =t-k+1} A_{2k-1+d_1}\otimes \bigotimes_{s=1}^{n-1} A^{[1]}_{d_s} \otimes A_{N-2t+d_n}^{[1]}
\end{equation}
In all four cases \eqref{eq:Morph-SUN-1}-\eqref{eq:Morph-SUN-4} the sums are over partitions with $d_s>0$.

Let us turn now to the differential $m_1$ in the vacuum category $\fVac(\CT^{SU(N)}_{\vartheta})$.
We use the contraction \eqref{eq:Kij-TSUN} and the interior amplitude \eqref{eq:beta-TSUN} with
$b_{ijk}=b$ for all $i<j<k$. For $\Hop(i,j)$ in \eqref{eq:Morph-SUN-1} $m_1$ is a signed sum of
operations on each of the tensor factors in the product. On a factor of the form $R_{\ell, \ell + d}\cong A_d^{[1]}$
with $d>0$ it acts as an intertwiner:
\be\label{eq:TSUN-diffl}
m_1: A_d \rightarrow \oplus_{d_1+d_2=d} A_{d_1} \otimes A_{d_2}
\ee
Note that for each decomposition $d=d_1+d_2$ there is a canonical intertwiner $\Pi_{d_1,d_2}: A_{d_1} \otimes A_{d_2}\to A_d$
given by the wedge product.
The components $m_1^{(d_1,d_2)}$ of $m_1$ in \eqref{eq:TSUN-diffl} are such that
\be
\Pi_{d_1,d_2}\circ m_1^{(d_1,d_2)} e_S = \kappa b {d \choose d_1} \varepsilon_S e_S
\ee
(Recall that $\varepsilon_S = \frac{ e_S \wedge e_{S'}}{\vol}$ where $S'$ is the complementary
multi-index to $S$.) In formulae
\be
m_1(e_S) = \kappa b \varepsilon_S \sum_{d_1=1}^{d-1} \sum_{\Sh_2(S): \vert S_1\vert = d_1}
\frac{e_{S_1} e_{S_2} e_{S'} }{\vol } e_{S_1} \otimes e_{S_2}
\ee

To describe the multiplication $m_2$ in the vacuum category we need to distinguish between
the two cases where the intermediate vacuum is down-type, as in Figure \ref{fig:TNEXAMPLE-3}(a)
or up-type, as in Figure \ref{fig:TNEXAMPLE-3}(c). In the first case we simply take a tensor
product. In the second case we must use the contraction. Here we are taking a product
\be
m_2: \Hop(i,N-k) \otimes \Hop(N-k,j) \to \Hop(i,j)
\ee
so we must combine the final factors in equation \eqref{eq:Morph-SUN-3} with the
initial factors in equation \eqref{eq:Morph-SUN-2}. The main step is captured by
the map $R_{i,N-k} \otimes R_{N-k,j} \to R_{i,j}$ with $i<j<N-k$,  namely
\be
A^{[1]}_{N-k-i} \otimes A_{k+j} \to A_{j-i}^{[1]}.
\ee
This is simply taking the dual of the wedge product of the duals (up to a factor of
$b\kappa^2$).

Turning now to the Brane category $\fB\fr(\CT^{SU(N)}_{\vartheta})$
we   can use the above complexes to compute the space of boundary-changing
operators between thimbles,  $H^*( \Hop(\fT_i,\fT_j), M_1)$. In the $\CT^{SU(N)}_\vartheta$
theories these will be representations of $SU(N)$. We conjecture the
following

\bigskip
\noindent
\textbf{Conjecture}

a.) If $i,j$ are lower vacua with $i<j$ then
\be\label{eq:HopCoho-1}
H^*( \Hop(\fT_i,\fT_j), M_1) \cong S^{[j-i]}_{j-i}
\ee

b.) If  $i=N-k$ is an upper vacuum and $j$ a lower vacuum with  $j\geq k$
\be\label{eq:HopCoho-2}
H^*( \Hop(\fT_{N-k},\fT_j), M_1) \cong L_{2k,j-k+1}^{[j-k]}
\ee

c.) If $i$ is a lower vacuum, and $j = N-k$ an upper vacuum  with $i<k$:
\be\label{eq:HopCoho-3}
H^*( \Hop(\fT_{i},\fT_{N-k}), M_1) \cong L_{N-2k+1,k-i}^{[k-i]}
\ee
%

%\cg{Say something about remaining case $(N-k,N-t)$? Could it be an irrep with Tableaux of shape $||\_$?}

The above conjecture is easily checked for the simple cases in which
$\Hop$ is concentrated in a single degree, but in general appears to be
an extremely challenging computation. We will deduce equation \eqref{eq:HopCoho-1}
using the rotational interfaces of Section \S \ref{sec:CatTransSmpl}.
See equation \eqref{eq:ProveHopCoho-1} below.

One could contemplate computing the morphisms spaces involving the
Branes $\fN_n$ defined by the Chan-Paton factors \eqref{eq:SLDEF}
and amplitudes \eqref{eq:SUN-fgh-k} et. seq.  We leave this exercise to
the truly energetic reader (with lots of time to spare). Since these Branes
are generated from thimbles by rotational Interfaces (see equation
\eqref{eq:B2hat-CP-TSUN} et. seq.) it is conceivable that arguments
along the lines of \eqref{eq:ProveHopCoho-1} lead to a derivation of these
cohomology spaces. It would also be interesting to see if these results
can be checked using the   $\sigma$-model or Fukaya-Seidel viewpoint
described in Sections \S\S \ref{lgassuper}-\ref{notif}.

\section{Interfaces}\label{sec:Interfaces}

\subsection{Interface Webs}\label{subsec:DomainWallWebs}

\subsubsection{Definition And Basic Properties}

We now consider webs in the presence of an ``Interface,'' a notion we will define
precisely just below equation \eqref{eq:InFcMC}. Roughly speaking, an Interface is a domain wall separating
two Theories.
For simplicity we take the  wall to be localized on the line $D$ described by $x=x_0$ in the $(x,y)$
plane. \footnote{More generally one could rotate our construction
in the plane.} We now consider two sets of vacuum data,
$(\IV^-, z^-) $, associated with the negative half-plane
$x \leq x_0$ and
$(\IV^+, z^+)$ associated with the positive half-plane
$x\geq x_0$. The data $(\IV^\pm, z^\pm, x_0)$ will be
collectively denoted by $\CI$. As in the half-plane case, we assume that
  none of the $z_{ij}^\pm$ are parallel to $D$.

%%%%%%%%%%%%
\bigskip
\noindent
\textbf{Definition}:

a.) An \emph{interface web} is a union of half-plane webs $(\fu^-, \fu^+)$,
with $\fu^-$ a negative half-plane web and $\fu^+$ a positive half-plane
web, where the half-planes share a common
boundary $D$ at $x=x_0$. The webs are determined by  the vacuum data $(\IV^\pm, z^\pm)$, respectively.
\footnote{We stress that at this point $\fu^-$ and $\fu^+$ are
webs, not deformation types of webs.}

b.) An \emph{interface fan} is the union of a fan for the negative half-plane data and a fan for the positive half-plane data.
It will be denoted $J= (J^+, J^-)$ where one of  $J^\pm$ (but not both) is allowed to be the empty set.

\begin{figure}[htp]
\centering
\includegraphics[scale=0.3,angle=0,trim=0 0 0 0]{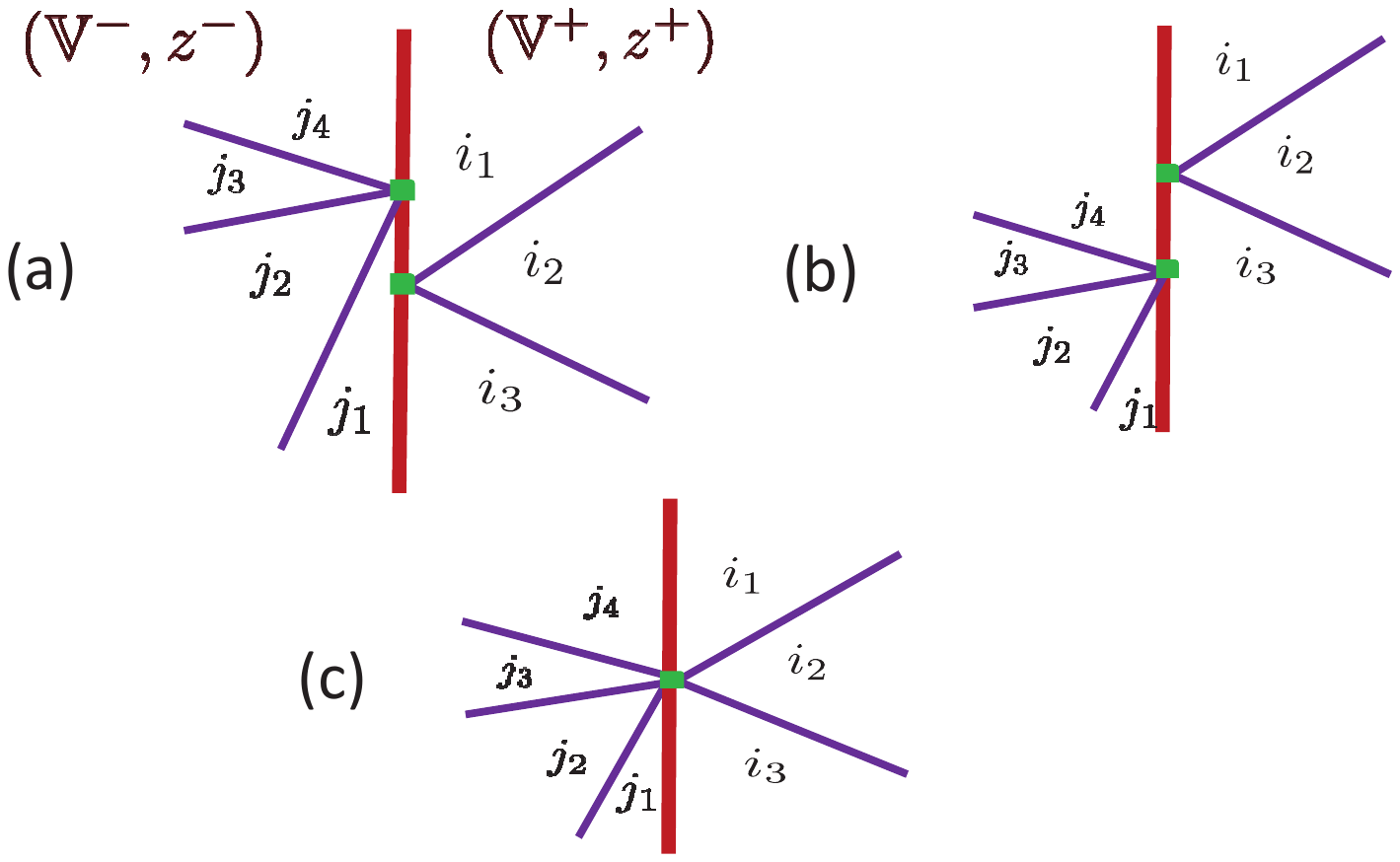}
\caption{The three interface webs shown here have different
deformation types. The webs (a) and (b) are   taut, while (c) is rigid. In all three
webs $J_\infty(\fd) = \{ i_1, i_2, i_3 ;j_1, j_2 , j_3, j_4 \}$. In Figure $(b)$ the
top vertex has $J_v(\fd) = \{  i_1, i_2, i_3 ; j_4 \}$.   }
\label{fig:DOMAINWALLWEB3}
\end{figure}

\textbf{Remarks}:
\begin{enumerate}
\item Let $\fd$ denote a typical interface web.
\footnote{The gothic ``d,'' which looks like $\fd$, is for ``domain wall,''
although in the course of our work that term has been deprecated in favor of ``interface.''}
We divide up the set of vertices of $\fd$ into
the  set of  \emph{wall vertices}  $\CV_\p(\fd)$ located on $D$ and \emph{ interior vertices}
$\CV^{\pm}_i(\fd)$ in either half-plane with cardinalities $V_\p(\fd)$ and $V^{\pm}_i(\fd)$, respectively.
The interior edges are subsets of the negative or positive half-planes,
cannot lie in $D$, and do not go to infinity.
The sets of interior edges are denoted   $\CE^\pm(\fd)$ and have cardinality $E^\pm(\fd)$.

\item We will consider $\CV_\p(\fd)$ to be an \emph{ordered set}.
Our convention is that reading left to right
 we order the vertices from future to past, as we would for a left boundary of a positive half-plane.

\item An interface web has an interface fan at infinity $J_{\infty}(\fd)$.
If $\fd=(\fu^-, \fu^+)$ then  $J_{\infty}(\fd) = \{ J_\infty(\fu^+); J_\infty(\fu^-) \} $.
Similarly, if $v\in\CV_\p(\fd)$ we can define local interface fans   $J_v(\fd)$.

\item Using a standard reflection trick we could  make a precise correspondence between
interface webs and half-plane webs for the ``disjoint union'' of the vacuum data
(a term we will not try to make precise). Note that any
 half-plane web could be seen as an interface web with
trivial vacuum data on one side of $D$. As we will see, interface webs
behave very much like half-plane webs.

\item We can speak of a deformation type of an interface web $\fd$.
In order to avoid confusion, let us stress that the deformation type $\fd$
of an interface  web is \emph{not} just a pair
of deformation types of negative and positive half-plane webs.
The reason is that when vertices of the negative and positive  half-plane webs
coincide deformations must maintain this identification. When they do {\it not} coincide,
deformations must preserve the relative order.
See, for example, Figure \ref{fig:DOMAINWALLWEB3}. In particular note that
one cannot unambiguously combine deformation types of negative and positive
half-plane webs into a deformation type of an interface web. Thus if we
identify $\fd$ with $(\fu^-, \fu^+)$ we must bear in mind that $\fu^\pm$
represent webs, not deformation types, so there is further data specifying how
the webs are combined,
in particular, how their boundary vertices are ordered and/or identified
to form the set of wall vertices of $\fd$. When the vacuum data
are in general position,
the moduli space $\CD(\fd)$ of interface webs of a fixed deformation
type has dimension
\be
d(\fd) := \left( 2 V^-_i(\fd)- E^-(\fd) \right) + \left(2 V^+_i(\fd)- E^+(\fd) \right)  + V_\p(\fd) .
\ee
For nongeneric vacuum data there can be exceptional webs with $\dim \CD(\fd) > d(\fd)$.

\item We define interface webs to be \emph{rigid, taut,} and \emph{sliding}
if $d(\fd) = 1,2,3$, respectively, just as for half-plane webs.
 Similarly, we define oriented deformation type
and consider the free abelian group $\CW_{\CI}$ generated by oriented
deformation types of  interface webs. The taut webs have a canonical
orientation (towards larger webs) and we denote the sum of taut canonically
oriented
interface webs by $\ft^{-,+}_\CI$ or, usually, just $\ft^{-,+}$ when the data $\CI$ is understood.
We can also denote by $\CW_{p}^\pm$ the group of plane webs associated to the data in the positive and negative
half-planes respectively.

\end{enumerate}

There are natural convolution operations inserting elements of $\CW_{p}^\pm$, $\CW_\CI$
at interior vertices in the appropriate half-planes or at wall vertices respectively
and we have the natural

\bigskip
\noindent
\textbf{Theorem}: Let $\ft_\CI^{-,+}$ be the interface taut element and $\ft_{p}^\pm $
the plane taut elements associated to the data in the two half-planes.
It is useful to define the formal sum $\ft_p = \ft_{p}^+ + \ft_{p}^-$.
We have a familiar-looking
convolution identity:
\be\label{eq:interface-ident}
 \ft_\CI^{-,+}*\ft_\CI^{-,+} + \ft_\CI^{-,+} * \ft_{p} =0.
\ee

The proof is closely modeled on that of the
half-plane case \eqref{eq:hp-wb-id}.

\subsubsection{Tensor Algebra Structures}

Turning now to the tensor algebras of webs, we
 also have natural operations associated to the insertion of appropriate webs at all interior vertices on either half plane and/or
at the wall vertices. As usual, the operations involving interior vertices define complicated $L_\infty$-type structures.
The basic operation $T_\p(\fd)$ on $T\CW_{\CI}$, denoted $T_\p(\fd)[\fd_1, \dots, \fd_n]$
 is defined as usual by replacing all wall vertices of $\fd$ on $D$
with appropriate interface webs with $J_{v_a}(\fd) = J_\infty(\fd_a)$. To repeat,
the ordering of vertices $v_a$ is toward decreasing $y$.  This behaves as it did for half-plane webs
and in particular applying the reasoning of \eqref{eq:A-infty} et. seq. the operator
$T_\p(\ft_\CI^{-,+}): T\CW_{\CI} \to \CW_{\CI} $ defines the structure of an $A_\infty$ algebra on $\CW_{\CI}$.

\begin{figure}[htp]
\centering
\includegraphics[scale=0.3,angle=0,trim=0 0 0 0]{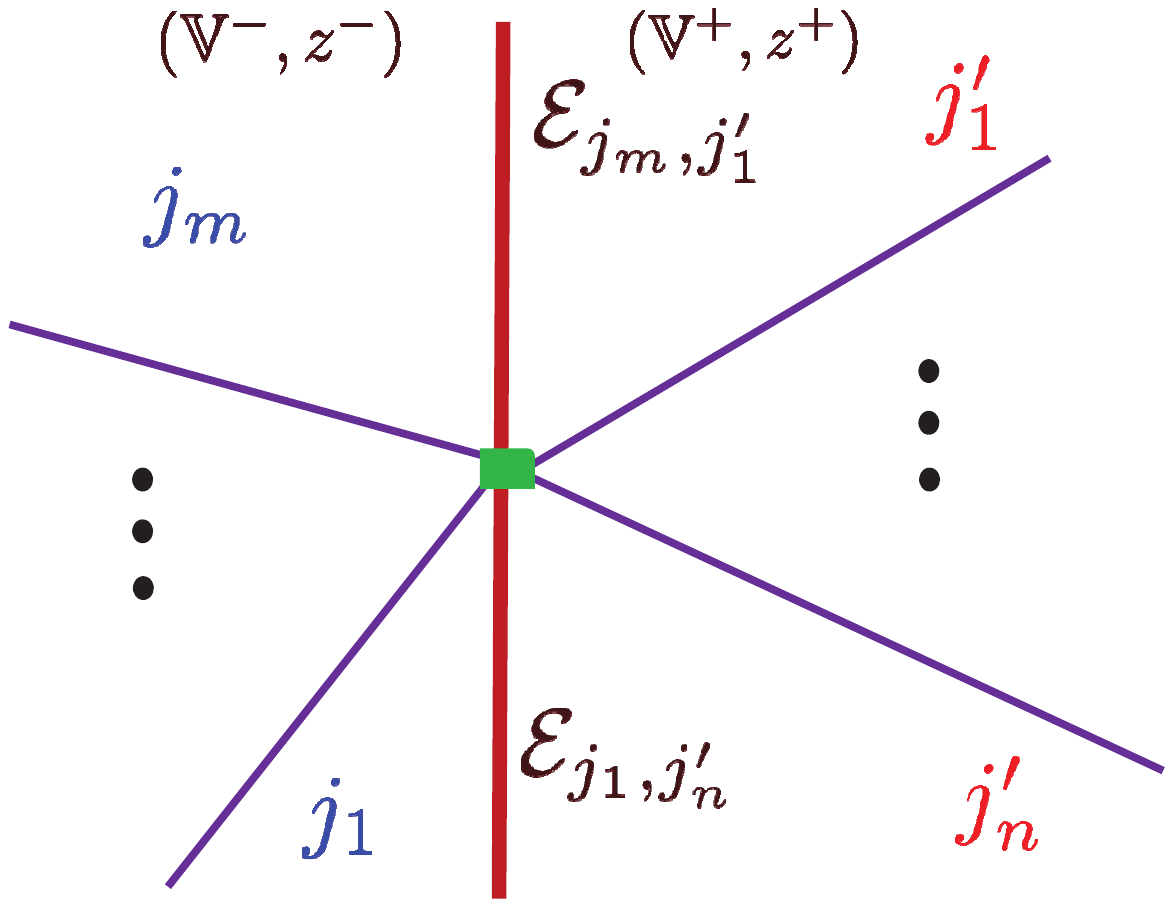}
\caption{Conventions for Chan-Paton factors localized on  interfaces. If representation spaces are attached
to the rays then this figure would represent a typical summand in
$\Hom(j_m j_1', j_1 j_n')$. We order such vertices from left to right using the conventions
of positive half-plane webs.  }
\label{fig:DOMAINWALL-CHANPATON}
\end{figure}

\subsubsection{Web Representations, Interfaces, And Interface Categories }

We now consider the algebraic structures that arise when we
are given a pair of representations of the vacuum data $(\IV^\pm,z^\pm)$.
The discussion closely parallels that for the half-plane theory.

We define a \emph{ representation of interface webs } to be a pair of representations
\be
\CR_{\CI} = \left( (\{ R_{ij}^-\}, \{ K_{ij}^-\}), ( \{ R_{i'j'}^+\} , \{ K_{i'j'}^+\}) \right)
\ee
for the vacuum data $\IV^\pm$. Similarly, we define \emph{Chan-Paton data} for
an interface to be an assignment $(i,i') \to \CE_{i,i'}$ where the
  Chan-Paton factors $\CE_{i,i'}$ are graded $\IZ$-modules. We picture this
  with a vacuum $i$ on the negative half plane and $i'$ on the positive half plane
  with the Chan-Paton factor located on the boundary $D$.
Our convention will be that the wall vertices on $D$ of interface fans
$J = \{ j'_1, \dots j'_n ;j_1, \dots, j_m \}$ will be represented by
the graded $\IZ$-module:
\be \label{eq:RJ-intfc}
R_J(\CE):= \CE_{j_m,j'_1} \otimes R_{j'_1,j'_2}^+ \otimes \cdots \otimes R_{j'_{n-1},j'_n}^+
 \otimes \CE_{j_1,j'_n}^* \otimes R_{j_1,j_2}^- \otimes \cdots \otimes R_{j_{m-1},j_m}^-.
\ee
This is illustrated in Figure \ref{fig:DOMAINWALL-CHANPATON}.
As usual we define the direct sum over all domain wall fans to be:
\be\label{eq:domainR}
R^\p(\CE) := \oplus_J R_J(\CE).
\ee
It is straightforward to define a map $\rho(\fd): TR^{{\rm int},-} \otimes TR^\p(\CE) \otimes TR^{{\rm int},+} \to R^\p(\CE)$
with arguments associated respectively to the interior vertices in the negative half plane,
wall vertices, and positive half plane interior vertices.
Given a Theory $\CT^-$ on the negative half plane and a Theory $\CT^+$ on the positive half-plane
we can define $\rho_\beta(\fd): TR^\p(\CE) \to R^\p(\CE)$ by
\begin{equation}
\rho_\beta(\fd)[r^\p_1, \dots, r^\p_n] = \rho_\beta(\fd)[e^{\beta^-};r^\p_1, \dots, r^\p_n;e^{\beta^+}]
\end{equation}
where $\beta^\pm$ are the interior amplitudes of $\CT^\pm$
and $\beta = (\beta^-;\beta^+)$. Familiar reasoning shows that this defines an
\afty-algebra structure on $R^\p(\CE)$.
%
%As before, this is one of the possible tensor products between the $A_\infty$ algebras associated to the left and right half-plane %web representations.
%

In analogy to the half-plane case we can now define  an
\emph{interface amplitude} to be an element $\CB_{\CI} \in R^\p(\CE)$,
for some $\CE$,
which solves the Maurer-Cartan equations
\be\label{eq:InFcMC}
\sum_{n=1}^\infty \rho_{\beta}(\ft_\CI)(\CB_{\CI}^{\otimes n}) =
\rho_\beta(\ft_\CI)[\frac{1}{1-\CB_{\CI}}]=0
\ee

\bigskip
\noindent
\textbf{Definition:}
An \emph{Interface} is a choice of $D$, a pair of Theories $\CT^\pm$,
 a choice of Chan-Paton data  for the interface, and an interface amplitude.

\bigskip
We generally denote an Interface by a capital Gothic letter, such as $\fI$.
The Chan-Paton data is $\CE(\fI)$ and the interface amplitude is $\CB(\fI)$.
Occasionally we will simply denote an Interface by its interface amplitude $\CB$.

As in the half-plane case, given data $(\IV^\pm, z^\pm,x_0)$ and Chan-Paton spaces $\CE_{ii'}$ we can introduce
a vacuum category $\fVac(\CT^-,\CT^+,\CE)$ with morphisms
\be\label{eq:Inf-Vac-Homs}
\Hom^{\CE}(jj', ii') =\begin{cases}
\CE_{ii'} \otimes \widehat{R}^+_{i'j'} \otimes \CE_{jj'}^*\otimes \widehat{R}^-_{ji}    & \Re(z_{i'j'}) > 0 \qquad
\emph{and} \qquad \Re(z_{ij}) > 0\\
\IZ & i=j \qquad and \qquad i'=j' \\
0 & \qquad {\rm else} \\
\end{cases}
\ee
See Figure \ref{fig:DOMAINWALL-CHANPATON} for a typical summand. The superscripts $\pm$ remind us
that the $\widehat{R}$'s are defined with respect to positive and negative half-planes, respectively.
As before, if we just
take $\CE_{ii'} = \IZ$ for all $i,i'$ then we get the ``bare'' Interface vacuum category
$\fVac(\CT^-,\CT^+)$.

\bigskip

Now, taking all Chan-Paton spaces into account we can define an
\afty-category of Interfaces, denoted $\fB\fr(\CT^-, \CT^+)$,
following closely the definitions of $\fB\fr(\CT)$ in
Section \S \ref{subsec:BraneCat}.  The objects are Interfaces and the
space of morphisms from the Interface $\fI_2$ to the Interface
$\fI_1$ is the natural generalization of
\eqref{eq:BHOM}:
\be
\Hop(\fI_1, \fI_2) := \oplus_{ii',jj'} \CE(\fI_1)_{ii'} \otimes \Hop(ii',jj') \otimes (\CE(\fI_2)_{jj'})^*.
\ee
where $\Hop(ii',jj')$ refers to the morphisms \eqref{eq:Inf-Vac-Homs} of the ``bare'' category with
$\CE_{ii'} = \IZ$ for all $i,i'$.
The \afty-multiplications are given by the natural generalization of  equation \eqref{eq:BraneMultiplications}.
There is no difficulty defining the formalism for
extended webs so, as  in the discussion of \eqref{eq:hmtpy-Morph} et. seq.,  we can
use compositions   $M_1$ and $M_2$ to
define notions of homotopic morphisms and of homotopic Interfaces.

\begin{figure}[htp]
\centering
\includegraphics[scale=0.5,angle=0,trim=0 0 0 0]{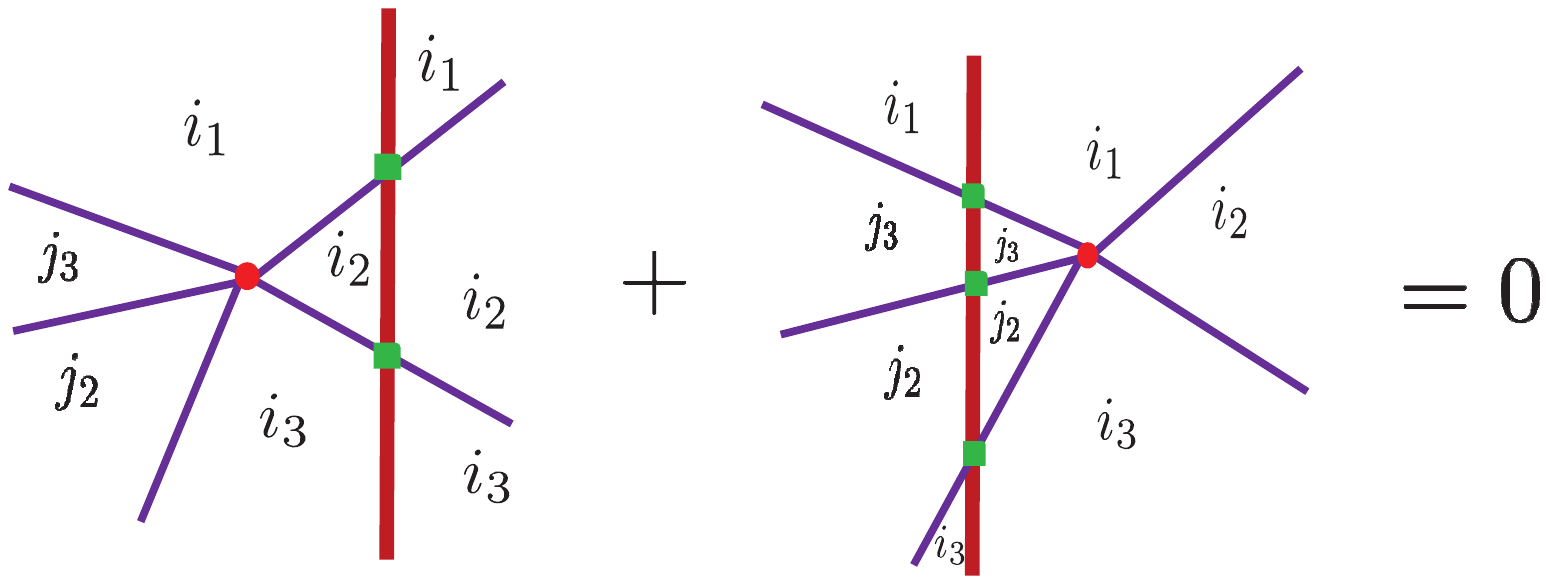}
\caption{Examples of taut interface webs which contribute to the
Maurer-Cartan equation for the identity interface $\fId$ between
a Theory and itself.    }
\label{fig:ID-INTERFACE}
\end{figure}

\subsubsection{Identity And Isomorphism Interfaces}\label{subsubsec:IsomIntfc}

 There is a very simple, universal, and
 instructive example of an interface between a Theory and itself: the \emph{identity interface } $\fId$.
We can pick as Chan-Paton factors $\CE_{ij} = \delta_{ij} \IZ$. With such a choice of CP factors amplitudes are valued in
\be
R^\p(\CE) = \oplus_{z_{ij} \in \CH^+ } \widehat{R}^+_{ij} \otimes \widehat{R}^-_{ji}.
\ee
Recall that the notation $\widehat{R}_{ij}$ implies a choice of half-plane.
We use the positive half-plane for the left factor
and the negative half-plane for the right factor.
To define the interface we take $\CB(\fId)$ to have nonzero component only in summands of the form
$R_{ij}\otimes R_{ji}$ corresponding to the fan $\{i,j;j,i\}$.
The vertex looks like a straight line
of a fixed slope running through the domain wall. The specific component of
$\CB(\fId)$  in $  R_{ij}\otimes R_{ji}$ will be $-K_{ij}^{-1}$, where
$K_{ij}^{-1}$  is defined as follows:

The element  $K_{ij}^{-1}\in R_{ij}\otimes R_{ji}$
uniquely characterized by the property that the map
\be\label{eq:Kinv-def1}
R_{ij} \rightarrow R_{ij}\otimes R_{ji} \otimes R_{ij} \rightarrow R_{ij}
\ee
defined by
\be\label{eq:Kinv-def2}
r \rightarrow K^{-1}_{ij}\otimes r  \rightarrow (1\otimes K_{ji})( K^{-1}_{ij}\otimes r )
\ee
is simply the identity transformation $r\to r$.
\footnote{Warning: If we map $r' \in R_{ji}$ by $r' \to r' \otimes K^{-1}_{ij}$ and then contract on the first
two factors the result is $r' \mapsto (-1)^{F+1} r' $. }
 It is worth expanding $K^{-1}_{ij}$ in terms
of a basis. We   introduce bases $\{ v_\alpha \}$ and
$\{ v_{\alpha'} \}$ for $R_{ij}$ and $R_{ji}$, respectively, where $v_\alpha, v_{\alpha'}$ are assumed  to have definite degree.
Then $K_{\alpha'\alpha} = K_{ji}(v_{\alpha'}, v_\alpha )$
and $K_{\alpha\alpha'} = K_{ij}(v_\alpha, v_{\alpha'})$ are related by $K_{\alpha\alpha'} = K_{\alpha' \alpha}$
since $K$ is symmetric. The element $K^{-1}_{ij}$ defined above is
\be\label{eq:Kinv-def3}
 K^{-1}_{ij} = (-1)^{\deg(v_{\alpha})} K^{\alpha\alpha'} v_{\alpha} \otimes v_{\alpha'},
\ee
where  $K^{\alpha\alpha'} $
is the matrix inverse of  $K_{\alpha'\alpha}$. That is $K^{\alpha\alpha'} K_{\alpha'\beta} = \delta^{\alpha}_{~\beta}$.
Note well that under the natural isomorphism $R_{ij}\otimes R_{ji} \rightarrow R_{ji}\otimes R_{ij}$
we have $K_{ij}^{-1} \rightarrow - K_{ji}^{-1}$. Hence, in this sense, $K_{ij}^{-1}$ is
\emph{antisymmetric}, a fact that will be useful in Sections \S\S \ref{sec:CatTransSmpl} and
\ref{sec:LocalOpsWebs}. Thus, the
component of $\CB(\fId)$ in $R_{ji}\otimes R_{ij}$, where $ji$ is the fan in the negative
half-plane,  is just $K_{ji}^{-1}$.

Now that we have defined $\fId$ let us verify that the interface amplitude
indeed satisfies the Maurer-Cartan equation.
The only non-zero contributions to $\rho_\beta(\ft_\CI)(\frac{1}{1-\CB})$ arise from
 taut webs with a single bulk vertex in either the positive or negative
 half-plane, but not both, as shown in Figure \ref{fig:ID-INTERFACE}. The interior
 vertex is saturated by the interior amplitude $\beta$. These
 vertices form pairs obtained by ``transporting''
 the vertex across the wall $D$. These pairs of taut webs   cancel out together
 in verifying the Maurer-Cartan equation for $\fId$. In slightly more
 detail, suppose that $I= \{i_1, \dots, i_n \}$ is a cyclic fan
 so that the vacuum amplitude $\beta$ has component $\beta_I \in R_I$.
Since (by assumption) none of the $z_{ij}$ for $i,j\in \IV$ is pure imaginary
we can choose to start the fan so that $z_{i_1,i_2}, \dots, z_{i_{m-1},i_m}$ point
into the positive half-plane and $z_{i_m,i_{m+1} }, \dots, z_{i_{n},i_1}$ point
into the negative half-plane. The taut web with the vertex in the negative
half-plane contributes (up to a sign, determined by equation \eqref{eq:bdy-rho-signs})
\be
K_{i_1,i_2} \cdots K_{i_{m-1},i_m} \left( K^{-1}_{i_1,i_2} \otimes \cdots \otimes K^{-1}_{i_{m-1},i_m} \otimes \beta_I \right)
\ee
to the Maurer-Cartan equation, where the $K^{-1}$'s come from $\CB_{\CI}$ while the taut web with the
same vertex in the positive half-plane contributes (up to a sign, determined by equation \eqref{eq:bdy-rho-signs})
\be
K_{i_{m+1},i_m} \cdots K_{i_{n},i_1} \left( K^{-1}_{i_{m+1},i_m} \otimes \cdots \otimes K^{-1}_{i_{1},i_n} \otimes \beta_I \right).
\ee
Both of the webs in Figure \ref{fig:ID-INTERFACE} are taut and hence canonically oriented. If $x$ is the coordinate of the
interaction vertex then the one on the left has $o_r(\fu)= -[dx]$ and the one on the right has $o_r(\fu) = +[dx]$.
We claim all the other sign factors cancel out and hence the two expressions in fact sum to zero. Essentially, this follows from the fact that the
edge vector fields associated to $K_{i_{n},i_{n+1}}$ get contracted with the one-forms for the wall vertex associated to $K^{-1}_{i_{n},i_{n+1}}$,
and the relative order of the vector fields and one forms mimic the relative order of the $K$ and $K^{-1}$ symbols.

From this description it is clear that the existence of this vertex strongly
uses  the fact that the left and right interior amplitudes are assumed to be equal.
   This property of the MC equation
    anticipates a theme which will recur later in the paper:
the existence of interfaces with given properties can encode relations between two theories.
See Section \S \ref{sec:GeneralParameter} for an implementation of this idea.

There is a very useful generalization of the identity Interface.
Suppose there is an isomorphism $\varphi: \CT^{(1)}\to \CT^{(2)}$,
as defined in Section \S \ref{subsubsec:IsomTheory}.
Then we can construct an almost-canonical invertible Interface
\be
\fId^\varphi \in \fB\fr(\CT^{(1)},\CT^{(2)})
\ee
which we will call an \emph{isomorphism Interface}.  The Chan-Paton factors are defined by
\be
\CE(\fId^\varphi)_{i,j} = \delta_{j,i\varphi} \IZ^{[e_i]}
\ee
where $e_i$ is a degree-shift which will be fixed, up to a common
shift $e_i \to e_i + s$,  below. (This ambiguity is the reason we
say the interface is only ``almost'' canonical.)

In order to define the amplitudes we define a set of canonical elements
\be
K^{-1,\varphi}_{ij} \in R^{(1)}_{ij} \otimes R^{(2)}_{j\varphi, i\varphi}
\ee
labeled by pairs of distinct vacua. This element can be defined by
requiring commutativity of the diagram
\be
\xymatrix{ R^{(1)}_{ij} \ar[r]^-{1 \otimes K^{-1,\varphi}_{ji}} \ar[rd]^{\varphi_{ij}} & R^{(1)}_{ij}\otimes R^{(1)}_{ji} \otimes R^{(2)}_{i\varphi, j\varphi}\ar[d]^{K^{(1)}_{ij}\otimes 1} \\
& R^{(2)}_{i\varphi, j\varphi}
}
\ee
This is equivalent to the condition
\be
\xymatrix{ R^{(2)}_{j\varphi,i\varphi} \ar[r]^-{  K^{-1,\varphi}_{ji}\otimes 1}   & R^{(1)}_{ji}\otimes R^{(2)}_{i\varphi,j\varphi} \otimes R^{(2)}_{j\varphi, i\varphi}\ar[d]^{1\otimes K^{(2)}_{i\varphi,j\varphi}} \\
& R^{(1)}_{ji}\ar[lu]_{\varphi_{ji}}
}
\ee
thanks to \eqref{eq:Kpullback}.

In order to give an explicit formula for $K^{-1,\varphi}_{ij}$ we choose bases $v^{(ij)}_\alpha$ for the $R^{(1)}_{ij}$
and similarly for $R^{(2)}_{ij}$. We write linear transformations $v \mapsto v \varphi_{ij}$ so that
the matrix elements relative to a basis are defined by $v_\alpha \varphi = \varphi_{\alpha\beta} w_\beta$.
Then the composition of linear transformations $\varphi_1 \varphi_2$ is represented by the standard matrix
product $(\varphi_1)_{\alpha\gamma} (\varphi_2)_{\gamma\beta} $. With this understood we have the formula
\be
K^{-1,\varphi}_{ij} = (-1)^{\deg(v^{(ij)}_\alpha )} (K^{-1,\varphi}_{ij})^{\alpha,\beta} v^{(ij)}_{\alpha } \otimes v^{(j\varphi,i\varphi)}_{\beta}
\ee
with
\be\label{eq:Kinv-explct}
(K^{-1,\varphi}_{ij})^{\alpha,\beta} = (K^{(1)}_{ij})^{-1,\alpha,\gamma} (\varphi_{ij})_{\gamma,\beta}
\ee
Now the boundary amplitudes for $\fId^{\varphi}$ are valued in
\be
\CE_{i,i\varphi} \otimes \widehat R^{(2)}_{i\varphi,j\varphi} \otimes \CE_{j,j\varphi}^* \otimes \widehat R^{(1)}_{ji}
\ee
and these are taken to be $\CB = K^{-1,\varphi}_{ij}$ up to degree shifts. The degree shifts are
used to ensure that $\CB$ has degree one.
Pictorially we have a bivalent vertex on the domain wall with a straight line going through it
from one half-plane to the other, just as in Figure \ref{fig:ID-INTERFACE}.

The demonstration that these boundary amplitudes satisfy the MC equation is
closely analogous to that of $\fId$.

\subsubsection{Trivial Theories}\label{subsubsec:TrivialTheories}

Once we speak of interfaces it is useful to introduce a formal concept
of a \emph{trivial theory} $\CT_{\rm triv}$. This is a Theory whose vacuum data is a set $\IV$
with a single element $\upupsilon$. The corresponding vacuum weight $z$ is irrelevant.  There are
no planar webs. There is a unique, trivial web representation, as there are no $R_{ij}$. Of course, $\widehat R_{\upupsilon,\upupsilon}=\IZ$.
 However,
there are extended half-plane webs: They are simply a collection of vertices
on the boundary of the half-plane. The   Chan-Paton data consists simply of
a graded $\IZ$-module $\CE$. The boundary amplitude consists entirely of its scalar part
$\CB \in \CE \otimes \CE^*$,  which can be viewed as an operator $Q\in \Hom(\CE)$ of degree one.
The taut web is the case of two boundary vertices so the MC equation simply
says that $Q^2=0$. Thus, giving a Brane for the trivial Theory is equivalent to giving
a chain complex over $\IZ$.

Now, an Interface   between the trivial Theory $\CT^-=\CT_{\rm triv}$ and $\CT$ is a
Brane for the theory $\CT$ on the positive half-plane. An Interface between the trivial Theory and itself
is therefore, once again, a chain complex over $\IZ$.

\subsubsection{Tensor Products Of \afty-Algebras}

 We remark in passing that our representations of interface webs lead
to a nice mathematical construction of tensor products of $A_\infty$-algebras.
In general the problem of defining a tensor product structure on \afty-algebras is
nontrivial and has been discussed, for examples, in
\cite{Amorim,MarklShnider,Loday}. In general there is a moduli space of possible
tensor products. From our present viewpoint, at least for a pair of
 \afty-algebras of the form   $\fVac(\CT^-, \CE^-),\fVac(\CT^+, \CE^+)$
we can choose   our interface Chan-Paton
spaces to be tensor products $\CE_{i,i'} = \CE_i^- \otimes \CE_{i'}^+$
and then our construction defines a canonical tensor product structure on
$R^\p(\CE) \cong (R^\p(\CE^-))^{\rm opp} \otimes R^\p(\CE^+)$. This is a distinguished
 $A_\infty$-algebra structure on the  tensor product, given the vacuum weights.

\begin{figure}[htp]
\centering
\includegraphics[scale=0.5,angle=0,trim=0 0 0 0]{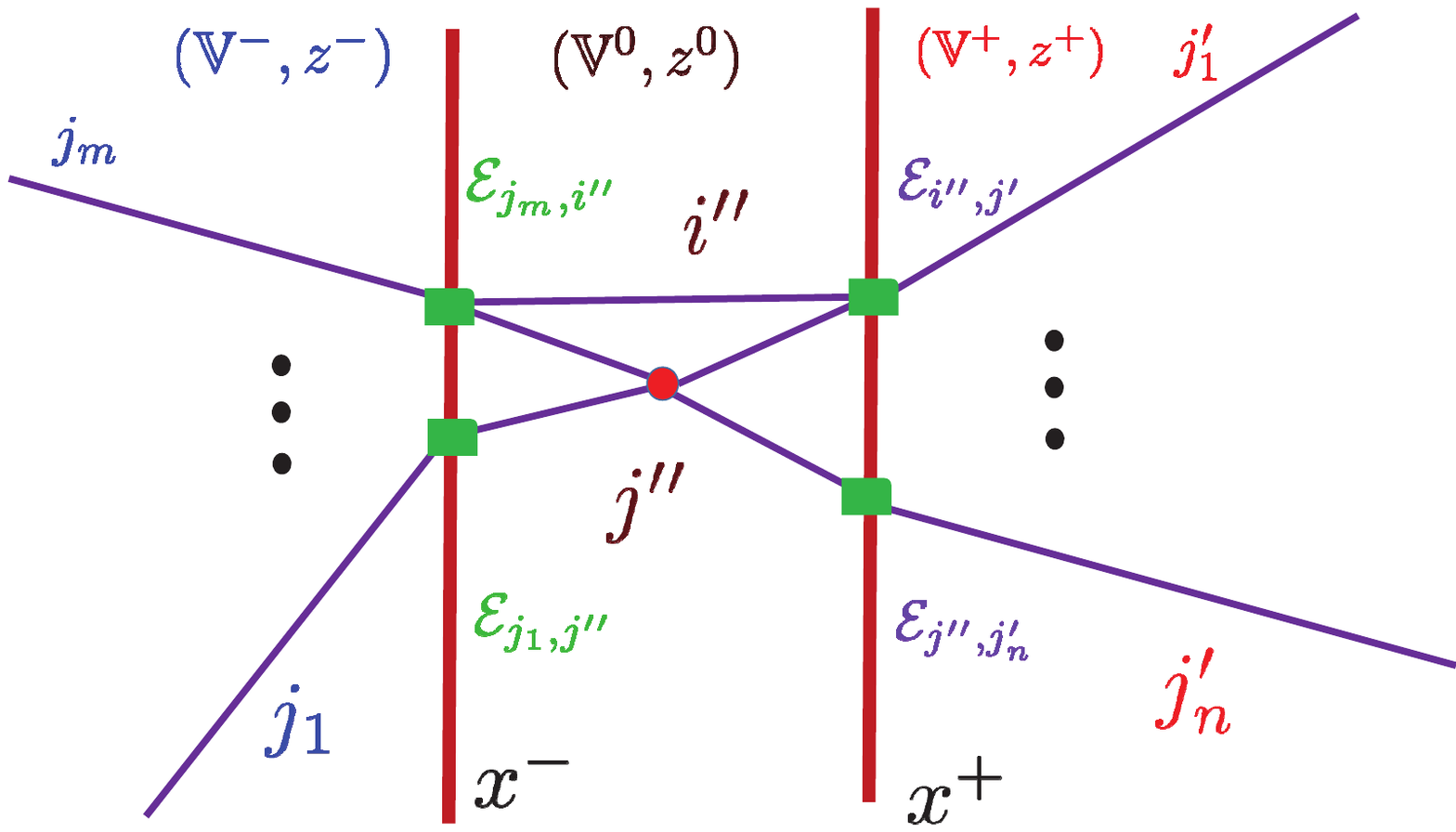}
\caption{An example of a composite web, together with conventions for
Chan-Paton factors. In this web the fan of vacua at
infinity has
$J_{\infty}(\fc) =  \{ j'_1, \dots j'_n; j_1, \dots, j_m \}$
and $\check{J}_{\infty}(\fc)=   \{i''; j'_1, \dots j'_n; j''; j_1, \dots, j_m \}$.
Reading from left to right the indices are in clockwise order.   }
\label{fig:COMPOSITEWEB1}
\end{figure}

\subsection{Composite Webs And Composition Of Interfaces}\label{subsec:ComposeInterface}

The crucial property of Interfaces, which goes beyond the properties of Branes, is that they can be composed.
We will discuss here the composition of two Interfaces. The composition of an Interface and a Brane
is a special case of that. Physically, we are defining a notion of operator product expansion
of supersymmetric interfaces.

The composition of Interfaces is based on a generalization of the strip geometry.
Choose $x^-< x^+$ and define a   tripartite geometry $G_2$ to be the union
of the  negative half-plane
$x\leq x^-$, the strip $x^- \leq x \leq x^+$, and the positive half-plane
$x^+ \leq x$. To these three regions we associate the three vacuum data
 $(\IV^-,z^-)$, $(\IV^0,z^0)$, and $(\IV^+,z^+)$, respectively. See Figure
 \ref{fig:COMPOSITEWEB1}.

By definition a \emph{composite} web for this tripartite geometry is a triplet $\fc = (\fu^-, \fs, \fu^+)$
of half-plane and strip webs on $G_2$ which are based on the appropriate vacuum data in
each connected component. (The definition can clearly be generalized to regions with
multiple strips with data $(\IV^-,z^-),(\IV^0,z^0), \dots, (\IV^n, z^n),(\IV^+,z^+)$.)
Once again, the moduli space of deformation types is not just the product
 $\CD(\fc) \not= \CD(\fu^-) \times \CD(\fs) \times \CD(\fu^+)$ because the deformations follow the
 rules for interface webs between   $(\IV^-,z^-)$ and $(\IV^0,z^0)$ and
 between $(\IV^0,z^0)$, and $(\IV^+,z^+)$.
 \footnote{Once again, $\fu^-, \fs,\fu^+$ are webs, not deformation types. Given
three such deformation types there are several ways to combine them into a deformation type
of a composite web.}
In general we denote the free abelian group generated by oriented deformation types
of composite webs as $\CW_C[\IV^-, \IV^0,  \dots,  \IV^n, \IV^+]$ (with the vacuum
weights understood).
 As for strip webs, the geometry has no scale invariance, and thus reduced moduli spaces are
quotiented by time translations only. Thus the definitions of taut and sliding composite webs are
analogous to those for strip webs:

\bigskip
\noindent
\textbf{Definition}: Assuming that at least two of the Theories
$(\CT^-, \CT^0, \CT^+)$ are nontrivial,   composite webs
with $d(\fc) = 1$ are called \emph{taut} (or \emph{rigid}) and
composite webs with $d(\fc) = 2$ are called \emph{sliding}.
\bigskip

For composite webs there are two senses in which we can speak of the fans of vacua
at infinity. If $\fc = (\fu^-, \fs, \fu^+)$ then we could define
\be
J_\infty(\fc) := \{ J_\infty(\fu^+); J_{\infty}(\fu^-) \}.
\ee
Notice that this has the same structure as boundary vertex fans for an interface
web between vacua $\IV^-$ and $\IV^+$, a fact which will be useful presently.
Sometimes it can be useful to include the past and future vacua
$j^{-}(\fs)$ and $j^+(\fs)$ of $\fs$, respectively. Then we define
\be
\check{J}_\infty(\fc) := \{  j^+(\fs); J_{\infty}(\fu^+);j^-(\fs); J_\infty(\fu^-) \}
\ee
See for example Figure \ref{fig:COMPOSITEWEB1}.

We will now describe the convolution identity for $\CW_C$, the free abelian group generated by the
  oriented deformation types of  composite webs.
Here a novel feature arises and the identity itself involves a tensor operation.
As usual we consider the possible boundaries of deformation types of sliding
composite webs. We encounter again the same phenomenon as for strip webs: some components of the moduli space
of taut composite webs are not segments, but half lines. While the boundaries at finite distance are accounted for
by convolutions, we need a different operation to account for boundaries at infinity.

\begin{figure}[htp]
\centering
\includegraphics[scale=0.5,angle=0,trim=0 0 0 0]{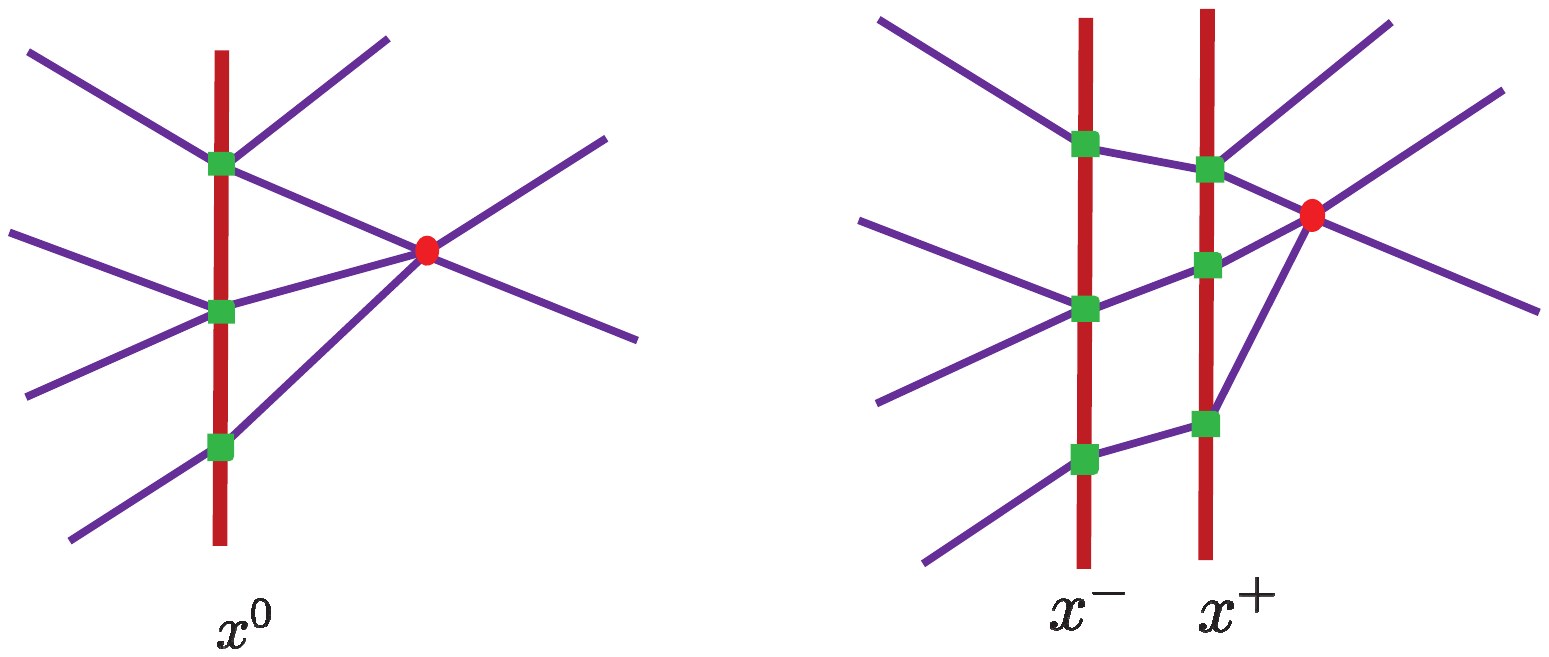}
\caption{The taut interface web on the left can be convolved with
three taut (=rigid) composite webs at the three green wall vertices
to form the sliding composite web on the right. This represents one
type of degeneration in the convolution identity for the taut composite
web. The region at infinity is represented by the limit in which the red
vertex moves off to infinity in the positive half-plane.   }
\label{fig:CMPST-CONV}
\end{figure}

For strip webs,
the new operation was time convolution: a large strip sliding web takes the form of two taut strip webs separated by a long stretch of time.
For composite webs, we can do something similar, but there is an important difference: as a web becomes large in size, the restriction to the central strip
will consist of two or more components of finite extent and well separated in time, but the components in the left and right half-planes may simply grow to arbitrarily large size. See Figure \ref{fig:CMPST-CONV} for an example.

We can make this statement precise by separating the ``bound'' vertices whose distance from the boundaries stabilizes as the web keeps growing
from the ``scaling'' vertices whose distance from the boundaries scales linearly with the distance from the boundaries. The bound vertices form clumps consisting of vertices
whose distance in time remains bounded as the web grows. We can take a sliding web of some large size $L$,
and re-scale all coordinates by a factor of $L$. The intermediate strip is now of very small width and the composite web is well approximated by an interface
web $\fd$ between $\CT^{-}$ and $\CT^{+}$
 whose interior vertices correspond to the scaling vertices of the original composite web and whose boundary vertices correspond to
each of the clumps of bound vertices of the original composite web.

We can put these heuristic pictures on a firm footing by considering a tensor operation
where a composite web is obtained from an interface web $\fd$ by convolving composite webs into the
wall-vertices of the  interface web $\fd$.  To be more precise, suppose $\fd$ is an
%
%
%Conversely, we can recover the re-scaled composite web from $\fd$ by resolving each boundary vertex of $\fd$ into a small composite web. This is %clearly in the same spirit as the tensor operations we defined
%in previous sections. If $\fd$ is any
%
interface web between  $(\IV^-, z^-) $ and $(\IV^+, z^+)$.
We can define an operation
\be
T_{\p}(\fd): T \CW_C \to \CW_C
\ee
whose nonzero values on  monomials $\fc_1 \otimes \cdots \otimes \fc_n$ are obtained
by inserting
  the $\fc_a$  (in the correct time order \footnote{By convention we use the order set by
  the positive half-plane webs. Therefore reading from left to right corresponds to
  vertices with decreasing $y$.} ) into the
  wall vertices $v^\p_a$ of $\fd$ provided $J_{v^p_a}(\fd) = J_{\infty}(\fc_a)$
and provided that the past strip vacuum of $\fc_{a}$ agrees with the future strip vacuum
of $\fc_{a+1}$. We orient the resulting web in the standard way, wedging the reduced orientations
of the arguments in the same order as the arguments themselves. It is easy to check that the
dimensions of the deformation spaces of the webs are related by
\be\label{eq:ModTensOp}
d \left( T_\p(\fd)[\fc_1,\dots, \fc_n] \right) = d(\fd) + \sum_{a=1}^n \left( d(\fc_a) -1 \right).
\ee
If we take into account the positions of the walls it is most natural to take the position $x^0$
of the wall for $\fd$ to be somewhere in the open interval   $x^- <  x^0 < x^+$, and all the composite webs $\fc_a$
have the same positions $(x^-,x^+)$. Since we are defining an operation on deformation types
the precise choice of $x^0$ does not matter.

According to \eqref{eq:ModTensOp} if all the composite webs are taut, so $d(\fc_a)=1$, and
if the interface web $\fd$ is taut, so $d(\fd)=2$, then $T_\p(\fd)[\fc_1,\dots, \fc_n]$ has $d=2$ and is
hence a \emph{sliding} composite web. In this way the  generic \emph{sliding} composite web is associated to a
\emph{taut} interface web, with insertions of \emph{an arbitrary number} of taut composite webs.
We thus claim that the regions at   infinity of the reduced moduli space of  sliding composite webs
are well described by
\be
T_{\p}(\ft^{-,+})\left[ \frac{1}{1-\ft_c} \right]
\ee
where $\ft_c$ is the taut element for composite webs and $\ft^{-,+}$ is the taut
element for interface webs between  $(\IV^-, z^-) $ and $(\IV^+, z^+)$. It is worth
noting that the time convolution of two composite webs is a special case of this operation.
It arises from the taut interface webs with precisely two wall vertices. For another example see
Figure \ref{fig:CMPST-CONV}. Note that this is qualitatively different from all the previous
convolution operations we have encountered because the ``more primitive'' structure plays
the role of the ``container web.''

To write down the full convolution identity for the taut composite webs we should also take into account
other degenerations at finite distance in the reduced moduli space.
To do this let
\be
\ft_{pl} := \ft^-_p + \ft^0_p + \ft^+_p
\ee
be the formal sum of plane web taut elements for the three vacuum data.
Similarly, let
\be
\ft_{\CI} := \ft^{-,0} + \ft^{0,+}
\ee
where   $\ft^{-,0}$ is the taut element associated to the interface webs between
vacuum  data $(\IV^-,z^-)$ and $(\IV^0,z^0)$, while $\ft^{0,+}$ is the taut element associated to the interface webs between
vacuum  data $(\IV^0,z^0)$ and $(\IV^+,z^+)$. The convolution identity for composite webs is
\be\label{eq:Composite-Strip-Web-Ident}
 \ft_c * \ft_{pl} + \ft_c*\ft_{\CI} + T_\p(\ft^{-,+})[\frac{1}{1-\ft_c}]=0.
\ee
Following the discussion of Section \ref{subsubsec:HalfPlane-A} we
can identify this equation as the Maurer-Cartan equation for an $A_\infty$ algebra structure on $\CW_{C}$ with operations $T_\p(\ft^{-,+})$, plus an extra differential
$* \ft_{pl} + * \ft_{\CI}$.

We now consider
a triplet of Theories   $(\CT^-, \CT^0 , \CT^+)$ associated  to the
three regions of $G_2$. Given Intefaces $\fI^{-,0} \in \fB\fr(\CT^-,\CT^0)$
and $\fI^{0,+}\in \fB\fr(\CT^0,\CT^+)$ we want to define a
product Interface, $\fI^{-,0} \IntfcTimes \fI^{0,+}\in \fB\fr(\CT^-,\CT^+)$.

We first determine the Chan-Paton factors of $\fI^{-,0} \IntfcTimes \fI^{0,+}$.
The choice of Theories $(\CT^-, \CT^0 , \CT^+)$  implies a choice of
three representations $(\CR^-, \CR^0 , \CR^+)$
of vacuum data.  The Interfaces $\fI^{-,0}$ and $\fI^{0,+}$ have
Chan-Paton spaces $\CE^{-,0}_{i,i''}$ and $\CE^{0,+}_{i'',i'}$, respectively.
Define the Chan-Paton data for the product Interface between vacua $\CT^-$ and $\CT^+$ as
\be\label{eq:Comb-CP}
\CE(\fI^{-,0} \IntfcTimes \fI^{0,+})_{ii'} := \CE^{-,+}_{ii'}  :=\oplus_{i'' \in \IV^0} \CE_{i, i''}^{-,0}\otimes \CE_{i'', i'}^{0,+}
\ee
Note that $\CE^{-,+}:= \oplus_{i\in \IV^-, i'\in \IV^+} \CE_{ii'}^{-,+}$
 is a generalization of the approximate ground states on the strip of equation \eqref{eq:ELRdef}.

Now, in order to define the interface amplitudes of $\fI^{-,0} \IntfcTimes \fI^{0,+}$ we need
some more preliminaries.  Viewing $\CE_{ii'}^{-,+}$ as Chan-Paton factors for an interface between $\CT^-$
and $\CT^+$ we can   formulate the spaces  $ R^\p(\CE^{-,+})$ using equation \eqref{eq:RJ-intfc}.
For a composite web $\fc$ we follow the usual procedure and define
\be\label{eq:Cmpste-Contract}
\rho_\beta(\fc): TR^\p(\CE^{-,0})   \otimes TR^\p(\CE^{0,+}) \to R^\p(\CE^{-,+}),
\ee
where $\CE^{-,+}$ is given by \eqref{eq:Comb-CP},
by inserting $\beta = (\beta^-, \beta^0, \beta^+)$ into the interior vertices
of $\fc$,  so that
\be\label{eq:rhofc-def}
\begin{split}
\rho_\beta(\fc)[r_1^{-,0}, \dots, r_n^{-,0};r_1^{0,+}, \dots, r_m^{0,+}]:= &
\rho(\fc)[ e^{\beta^-}; r_1^{-,0}, \dots, r_n^{-,0};e^{\beta^0};r_1^{0,+}, \dots, r_m^{0,+};
e^{\beta^+}] \\
& \qquad  \in R^\p(\CE^{-,+}).\\
\end{split}
\ee
\begin{figure}[htp]
\centering
\includegraphics[scale=0.5,angle=0,trim=0 0 0 0]{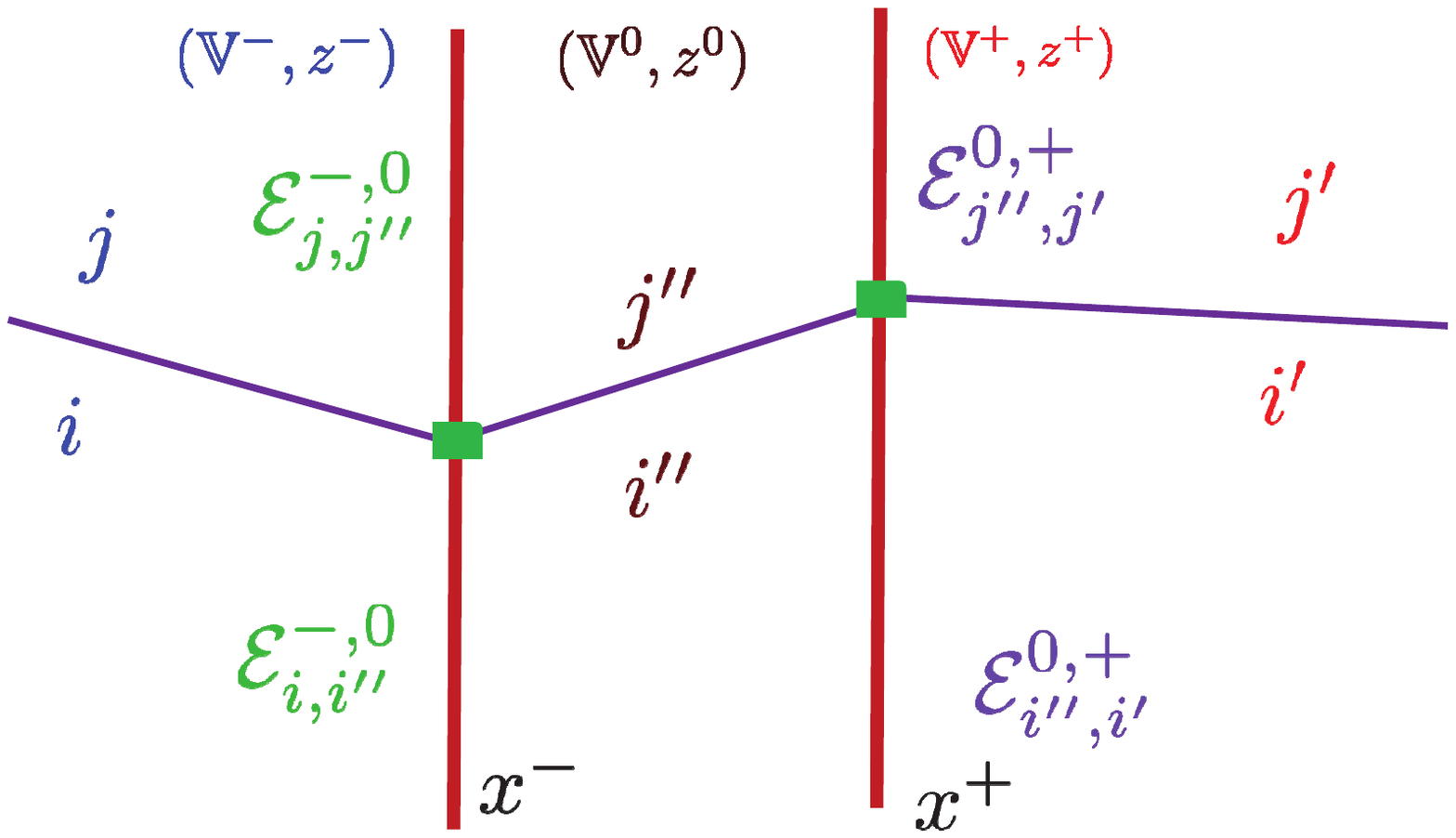}
\caption{A simple taut web is illustrated here. It leads to the contractions
described in the example below.}
\label{fig:COMPOSITE-EXAMPLE}
\end{figure}

\bigskip
\noindent
\textbf{Example}: As an example consider the taut composite web $\fc$ shown
in Figure \ref{fig:COMPOSITE-EXAMPLE}. The action of $\rho_\beta(\fc)$ is
zero on every component of $TR^\p(\CE^{-,0})   \otimes TR^\p(\CE^{0,+}) $
\emph{except} on
\be\label{eq:cpst-expl-1}
\left( \CE^{-,0}_{j,j''} \otimes R^0_{j'',i''} \otimes (\CE^{-,0}_{i,i''})^* \otimes R^-_{i,j} \right) \otimes
\left( \CE^{0,+}_{ j'',j'} \otimes R^+_{j',i'} \otimes (\CE^{0,+}_{i'',i'})^* \otimes R^0_{i'',j''}\right)
\ee
The superscripts on the $R$'s indicates which Theory we are speaking of,
and there is no sum on any of the indices.
The action of $\rho_{\beta}(\fc)$ on this summand uses the contraction
\be
K^0_{j'',i''}:  R^0_{j'',i''}\otimes R^0_{i'',j''}\to \IZ
\ee
together  with the Koszul rule  to map an element of \eqref{eq:cpst-expl-1} to
\be\label{eq:cpst-expl-2}
\left( \CE^{-,0}_{j,j''} \otimes \CE^{0,+}_{ j'',j'} \right)
\otimes R^+_{j',i'}  \otimes \left( \CE^{-,0}_{i,i''}\otimes \CE^{0,+}_{i'',i'}\right)^*
\otimes  R^-_{i,j}
\ee
Now note that \eqref{eq:cpst-expl-2} is a summand of $ R^\p(\CE^{-,+})$.

\begin{figure}[htp]
\centering
\includegraphics[scale=0.5,angle=0,trim=0 0 0 0]{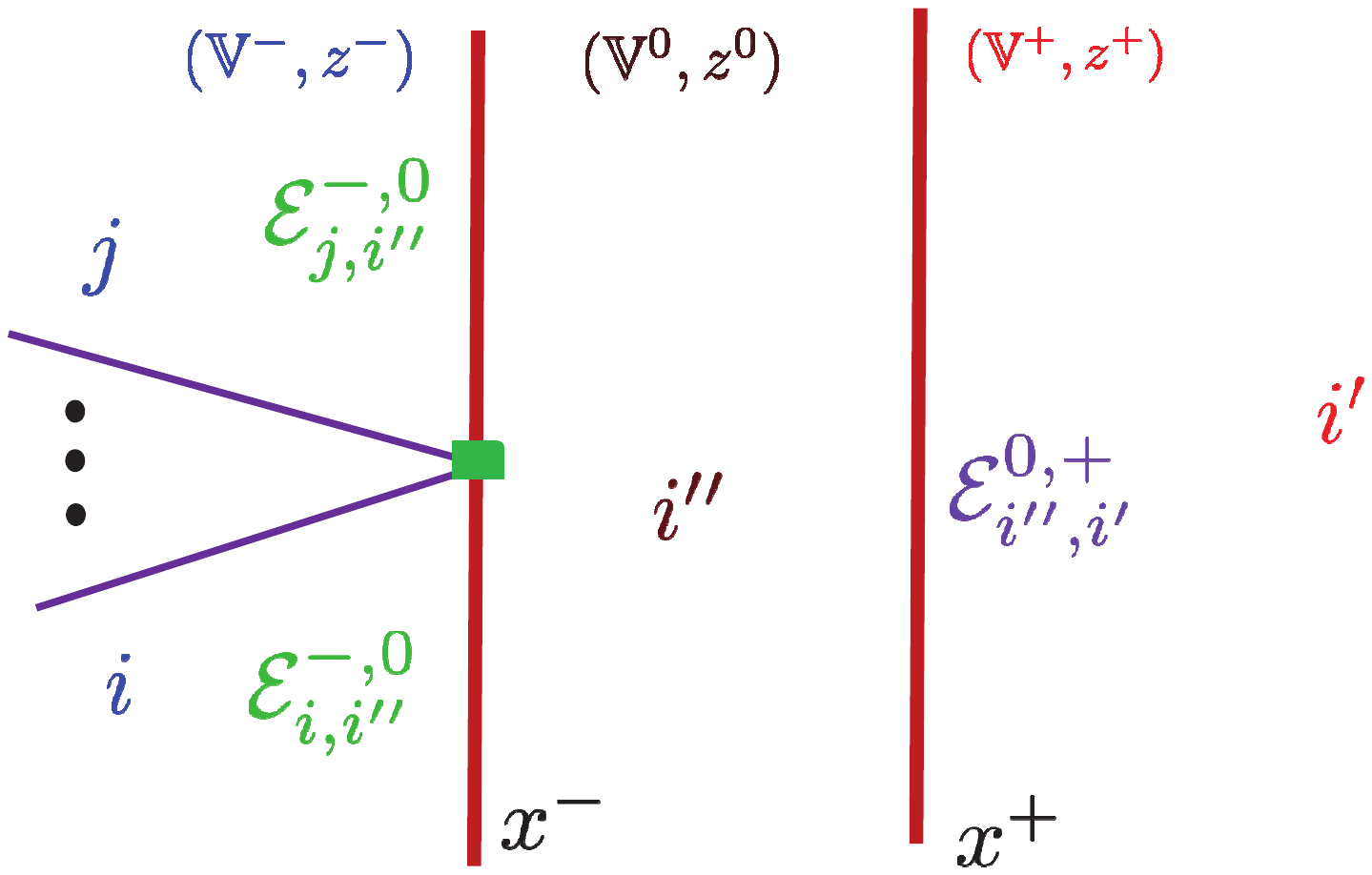}
\caption{A  taut web with no vertices on the $\CT^0,\CT^+$ boundary. }
\label{fig:COMPOSITE-SPECIAL}
\end{figure}

There is a special case of \eqref{eq:rhofc-def}
we must deal with separately, namely when $n=0$ or $m=0$. The reason is
that we can have composite webs with \emph{no} vertices one one of the two interfaces.
See for example the taut web in Figure \ref{fig:COMPOSITE-SPECIAL}. For such webs
$\rho_\beta(\fc)$ will map
\be\label{eq:Cmpste-Special}
\rho_\beta(\fc): TR^\p(\CE^{-,0})     \to R^\p(\CE^{-,+}),
\ee
by taking $\rho_\beta(\fc)[r_1,\dots, r_n; \emptyset]$ to have
a value only in the component:
\be\label{eq:Cmpste-Special-1}
\left( \CE^{-,0}_{j,i''}\otimes \CE^{0,+}_{i'',i'} \right) \otimes \widehat R^+_{i',i'} \otimes \left(\CE^{-,0}_{i,i''}\otimes \CE^{0,+}_{i'',i'}\right)^* \otimes \
 \widehat R^-_{i,j}
 \cong
 \left(  \CE^{-,0}_{j,i''}\otimes  \widehat R^0_{i'',i''} \otimes  (\CE^{-,0}_{i,i''})^* \otimes\widehat R^-_{i,j} \right)
  \otimes \left( \CE^{0,+}_{i'',i'} \otimes (\CE^{0,+}_{i'',i'})^*\right)
\ee
with a value given by
\be\label{eq:Cmpste-Special-2}
\rho_\beta(\fc)[r_1,\dots, r_n] \otimes \Id_{i'',i'}
\ee
where $\rho_\beta(\fc)[r_1,\dots, r_n]$ is just the contraction for an interface web between $\CT^-$ and $\CT^0$. We make
a similar definition with webs that have no vertices on the $(\CT^-,\CT^0)$ boundary.

Now we can define the interface amplitude of $\fI^{-,0}\IntfcTimes \fI^{0,+}$.
Suppose that $\fI^{-,0}$ and $\fI^{0,+}$ have interface amplitudes  $\CB^{-,0}$ and $\CB^{0,+}$,
respectively. We claim that
\begin{equation}\label{eq:InterfaceComp}
\CB(\fI^{-,0}\IntfcTimes \fI^{0,+}):=
%\CB^{-,0} \IntfcTimes \CB^{0,+} :=
\rho_\beta(\ft_c)\left[ \frac{1}{1-\CB^{-,0}}; \frac{1}{1-\CB^{0,+}}\right]
\end{equation}
satisfies the Maurer Cartan equation for an interface amplitude between the theories $\CT^-$ and $\CT^+$
with Chan-Paton spaces \eqref{eq:Comb-CP}.
To prove  this  claim  we first note that
\be\label{eq:RepTdel}
\rho_\beta(\ft^{-,+})\left[ \frac{1}{1- \CB(\fI^{-,0} \IntfcTimes \fI^{0,+}) } \right] =
\rho_\beta\left( T_\p(\ft^{-,+})\left[ \frac{1}{1-\ft_c} \right]\right)\left[ \frac{1}{1- \CB^{-,0}  },
\frac{1}{1-  \CB^{+,0}}\right]
\ee
This forbidding identity has a simple meaning. On the right hand side we are computing the amplitude of
composite webs produced by inserting $\ft_c$ in all possible ways in $\ft^{-,+}$.
On the left hand side we compute the amplitude for the individual $\ft_c$ sub webs first, and then insert that in $\ft^{-,+}$.
Finally, we apply the convolution identity \eqref{eq:Composite-Strip-Web-Ident}
and use the fact that $\beta$ is an interior amplitude and
$\CB^{-,0}$ and $\CB^{0,+}$ are interface amplitudes, thus establishing that \eqref{eq:InterfaceComp}
is an interface amplitude.

It follows from the the above discussion that Interfaces can be composed.
In fact, the   product $\IntfcTimes$ can be extended to define an \afty\ bi-functor from the
Cartesian product of
 Interface categories $\fB\fr(\CT^-,\CT^0) \times \fB\fr(\CT^0,\CT^+)$ to the Interface category
$\fB\fr(\CT^-,\CT^+)$. That means that if we have

\begin{enumerate}

\item   A   sequence of
interface amplitudes   $\fI_0^{-,0},\dots, \fI_{n}^{-,0}$
in $\fB\fr(\CT^-,\CT^0)$ together with morphisms  $\delta_1,\dots, \delta_n$   between
 them, and

\item  similarly, we have interface amplitudes    $\fI_0^{0,+},\dots, \fI_{n'}^{0,+}$
 in  $\fB\fr(\CT^0,\CT^+)$, together with morphisms
   $\delta_1',\dots, \delta_{n'}'$ between them,

\end{enumerate}

\noindent
then we can produce an element:
\be
\nu(\delta_1,\dots, \delta_n; \delta_1', \dots, \delta'_{n'})\in \Hop\left(\fI_0^{-,0}\IntfcTimes \fI_0^{0,+}, \fI_n^{-,0}\IntfcTimes \fI_{n'}^{0,+}\right)
\ee
such that the \afty-relations are satisfied separately in the two sets of arguments.
The element  $\nu(\delta_1,\dots, \delta_n; \delta_1', \dots, \delta'_{n'})$  is defined by   $\rho_\beta(\ft_c)$:
\begin{equation}\label{eq:bifunctormap}
\rho_\beta(\ft_c)[\frac{1}{1-\CB_0^{-,0}}, \delta_1, \cdots, \delta_n \frac{1}{1-\CB_{n}^{-,0}}; \frac{1}{1-\CB_0^{0,+}}, \delta'_1, \cdots,\delta_{n'}' \frac{1}{1-\CB_{n'}^{0,+}}]
\end{equation}
This extends the composition of Interfaces to a full $A_\infty$ bi-functor.
(We have not written out the full details of a proof that this is in fact a
bi-functor.)

%
%\cg{Need to comment on special cases $n=0$ or $n'=0$}
%
%
\begin{figure}[htp]
\centering
\includegraphics[scale=0.3,angle=0,trim=0 0 0 0]{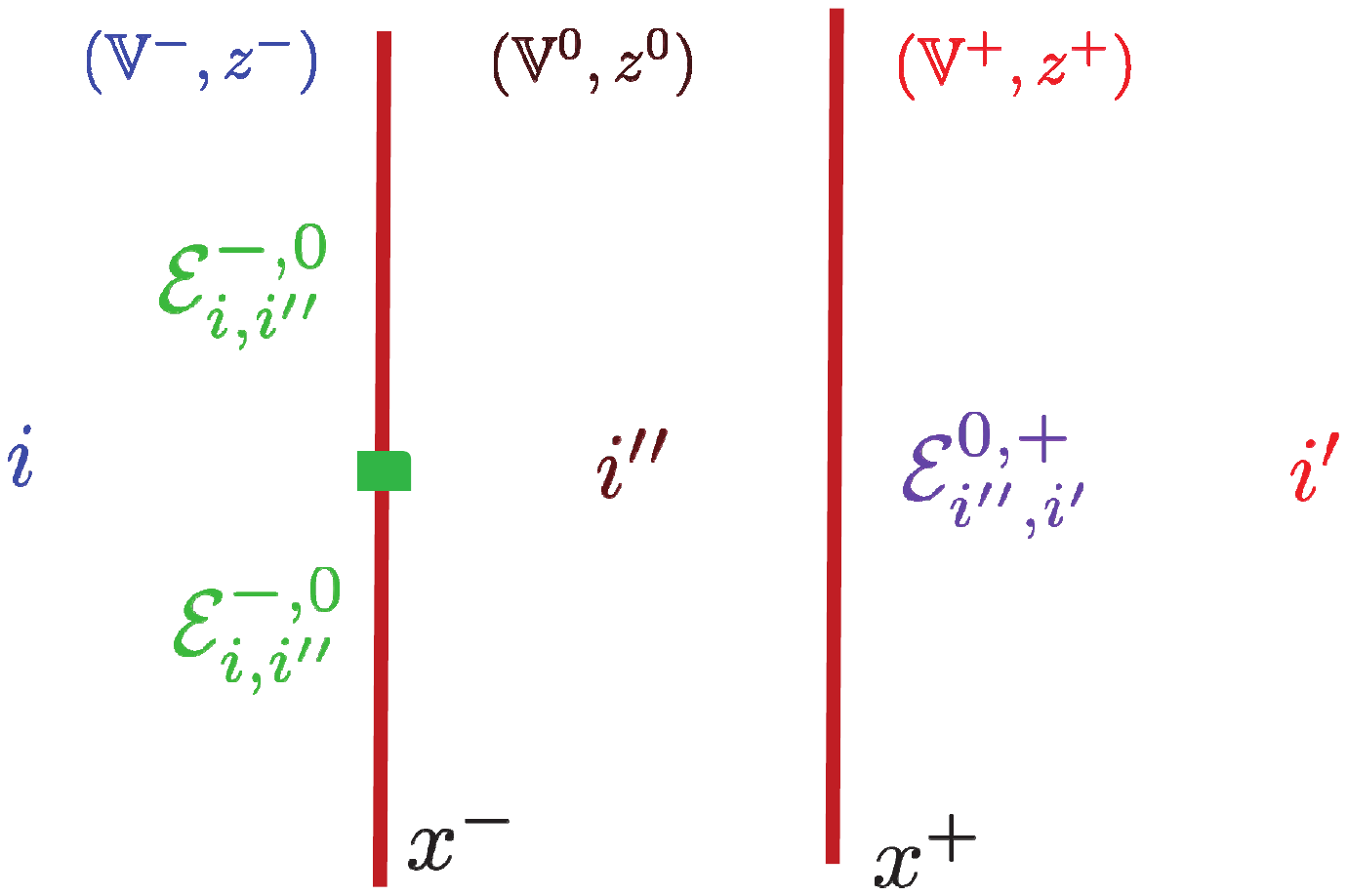}
\caption{The (extended) taut web shown here contributes to  $\nu(\Id;\emptyset)$.}
\label{fig:CPST-MAP-IDENT}
\end{figure}

If we specialize the above discussion to $n=1$ and $n'=0$ or $n=0$ and $n'=1$ then we
obtain an  interesting interplay with notions of homotopy equivalent
branes and interfaces. In particular, even though $\fI^{-,0}\IntfcTimes\fI^{0,+}$ is not a
 bilinear operation, we claim that if $\fI^{-,0}_0$ is homotopy equivalent to $\fI^{-,0}_1$
 then $\fI^{-,0}_0\IntfcTimes \fI^{0,+}$ is homotopy equivalent to $\fI^{-,0}_1\IntfcTimes \fI^{0,+}$.
There is a completely parallel result for homotopy equivalences of $\fI^{0,+}$ holding  $\fI^{-,0}$ fixed.
To prove this we note that  we have identities like the commutativity (up to sign) of the diagram:
\be\label{eq:bifunctor-2}
\xymatrix{\Hop(\fI^{-,0}_0, \fI^{-,0}_1)  \ar[r]^-\nu \ar[d]^{M_1^{-,0}}  &
\quad \Hop(\fI^{-,0}_0\IntfcTimes\fI^{0,+} , \fI^{-,0}_1\IntfcTimes\fI^{0,+}  )  \ar[d]^{M_1^{-,+} } \\
\Hop(\fI^{-,0}_0, \fI^{-,0}_1) \ar[r]^-\nu   & \quad \Hop(\fI^{-,0}_0\IntfcTimes\fI^{0,+} , \fI^{-,0}_1\IntfcTimes\fI^{0,+}  ) \\
}
\ee
where $\nu$ is the map obtained by specializing \eqref{eq:bifunctormap} to $n=1$ and $n'=0$.
Of course there is a similar identity for   $n=0$ and $n'=1$. This follows by applying
the representation of the identity \eqref{eq:Composite-Strip-Web-Ident} to the sequence of arguments
\be
e^{\beta^-} ; \frac{1}{1-\CB_0^{-,0}}, \delta, \frac{1}{1-\CB_1^{-,0}};e^{\beta^0};  \frac{1}{1-\CB^{0,+}}; e^{\beta^+}
\ee
where $\delta \in \Hop(\fI^{-,0}_0, \fI^{-,0}_1)$.
It follows from \eqref{eq:bifunctor-2}
that $\nu$ maps an $M_1$-closed or exact morphism  between    $\fI^{-,0}_0$  and $\fI^{-,0}_1$ to
an $M_1$-closed or exact morphism between  $\fI^{-,0}_0\IntfcTimes\fI^{0,+}$ and $\fI^{-,0}_1\IntfcTimes\fI^{0,+}$.
Now note that if $\Id$ is the graded identity element of equation \eqref{eq:GdId} then $\nu(\Id;\emptyset) = \Id$.
To prove this note that the only taut webs which can contribute to
\be
\rho_\beta(\ft_c)[\frac{1}{1-\CB^{-,0}}, \Id ,   \frac{1}{1-\CB^{-,0}}; \frac{1}{1-\CB^{0,+}} ]
\ee
are those with a single vertex on the boundary between $\CT^-$ and $\CT^0$, and a single vacuum $i''$
in the region $x^- \leq x \leq x^+$, as shown in Figure \ref{fig:CPST-MAP-IDENT}. Now, using
the definitions \eqref{eq:Cmpste-Special}-\eqref{eq:Cmpste-Special-2} we can check that the sum
over such taut webs gives $\nu(\Id;\emptyset) = \Id$. Similarly, $\nu(\emptyset;\Id) = \Id$ comes
from taut webs with a single vertex on the $\CT^0,\CT^+$ boundary.

Now, since $\nu$ is an \afty-functor, if $\delta, \delta'$ define a homotopy equivalence between $\fI_0^{-,0}$ and $\fI_1^{-,0}$ then
\footnote{In this equation we got lazy about the signs.}
\be
\begin{split}
  M_2(\nu(\delta), \nu(\delta'))  & = \nu(M_2(\delta,\delta')) \pm \nu(M_1(\delta),\delta')\pm  \nu(\delta, M_1(\delta'))
  \pm M_1(\nu(\delta,\delta')) \\
& =  \nu(M_2(\delta,\delta')) \pm M_1(\nu(\delta,\delta')) \\
& =  \nu(\Id + M_1(\delta'') ) \pm M_1(\nu(\delta,\delta')) \\
 & = \Id + M_1(\nu(\delta'') \pm \nu(\delta,\delta')) \\
\end{split}
\ee
In the first line we used the definition of an \afty-functor. In the second line we used the hypothesis that $\delta, \delta'$ are closed,
in the third line we used the hypothesis that they define a homotopy equivalence.  This finally completes the proof  that homotopy equivalence
is nicely compatible with $\IntfcTimes$.

In the special case where $\CT^-$ is the  trivial Theory, we have a useful result: each Interface in $\fB\fr(\CT^0,\CT^+)$ gives us an $A_\infty$ functor
from $\fB\fr(\CT^0)$ to $\fB\fr(\CT^+)$. This will be important for us later in Section \S \ref{sec:CatTransSmpl}
so let us spell it out a bit more. If $\fI^{0,+}$ is a fixed Interface we define an \afty-functor by declaring that on objects
\be\label{eq:Intfc-Functor}
\CF_{\fI^{0,+}}(\fB) := \fB\IntfcTimes \fI^{0,+}
\ee
and if  $\delta_1,\dots, \delta_n$ is a composable set of morphisms between Branes
 $\fB_0 ,\dots, \fB_{n} $ in $\fB\fr(\CT^0)$ then
\be\label{eq:Ifc-F2}
\CF_{\fI^{0,+}}(\delta_0, \dots, \delta_n) := \delta \in \Hop(\CF_{\fI^{0,+}}(\fB_0), \CF_{\fI^{0,+}}(\fB_n))
\ee
is given by
\be\label{eq:Ifc-F3}
\delta =
\rho_\beta(\ft_c)[\frac{1}{1-\CB_0 }, \delta_1, \cdots, \delta_n \frac{1}{1-\CB_{n} }; \frac{1}{1-\CB^{0,+}} ]
\ee
and we claim, moreover, that the \afty-relations defining an \afty-functor are satisfied:
\be\label{eq:afty-functor}
\begin{split}
& \sum_k \sum_{{\rm Pa}_k(P)} \rho_{\beta^+}(\ft^+_{\CH})\left( \CF_{\fI^{0,+}}(P_1),\dots,  \CF_{\fI^{0,+}}(P_k)\right)\\
& = \sum_{{\rm Pa}_3(P)} \epsilon_{P_1,P_2,P_3}  \CF_{\fI^{0,+}}\left( P_1, \rho_{\beta^0}(\ft^0_{\CH})(P_2), P_3 \right)  \\
\end{split}
\ee
where $P = \{ \delta_1, \dots, \delta_n \}$,  $\CF_{\fI^{0,+}}(\emptyset)=0$, $\epsilon_{P_1,P_2,P_3} $ is an
appropriate sign, and we note that $\ft^{-,+}=\ft^+_{\CH}$ is the taut element of the Theory $\CT^+$ in the positive half-plane
while $\ft^{-,0} = \ft^0_{\CH}$ is the taut element of the Theory $\CT^0$ in the positive half-plane.

Moreover, suppose that $\psi \in \Hop(\fI_1^{0,+}, \fI_2^{0,+})$ is a morphism between Interfaces.
Then we claim that there is an \afty-natural transformation $\tau(\psi)$ between the corresponding
functors $\CF_{\fI_1^{0,+}}$ and $\CF_{\fI_2^{0,+}}$. That is, for every $\fB \in \fB\fr(\CT^0)$
we can define
\be
\tau(\psi)_{\fB} \in \Hop(\CF_{\fI_1^{0,+}}(\fB), \CF_{\fI_2^{0,+}}(\fB))
\ee
so that, if $\delta_1, \dots, \delta_n$ is a composable sequence of morphisms between
Branes $\fB_0, \dots, \fB_n $ in $\CT^0$, then
\be
\xymatrix{\CF_{\fI_2^{0,+}}(\fB_n) \ar[r]^{\tau(\psi)_{\fB_n}} \ar[d]^{\CF_{\fI_2^{0,+}}(\delta_0,\dots, \delta_n) } &
\CF_{\fI_1^{0,+}}(\fB_n)\ar[d]^{\CF_{\fI_1^{0,+}}(\delta_0,\dots, \delta_n) } \\
\CF_{\fI_2^{0,+}}(\fB_0) \ar[r]^{\tau(\psi)_{\fB_0}}  & \CF_{\fI_1^{0,+}}(\fB_0)\\
}
\ee
is a commutative diagram. The formula for $\tau(\psi)_{\fB}$ is just
\be
\tau(\psi)_{\fB} := \rho_\beta(\ft_c) \left[ \frac{1}{1-\CB}; \frac{1}{1-\CB_1^{0,+}}, \psi , \frac{1}{1-\CB_2^{0,+}}\right].
\ee
where $\CB$ is the boundary amplitude of $\fB$ and $\CB_1^{0,+}$, $\CB_2^{0,+}$ are
the interface amplitudes of $\fI_1^{0,+}$, $\fI_2^{0,+}$, respectively.
Moreover, a natural transformation $\tau$ between an Interface and itself is homotopic to the identity if, for every
$\fB$, $\tau_\fB = \Id + M_1(\delta)$. Two \afty-functors can be regarded as homotopy equivalent
if they are related by two natural transformations whose composition is homotopic to the identity.
Now
\be
M_2(\tau(\psi_1)_{\fB}, \tau(\psi_2)_{\fB}) =\tau(M_2(\psi_1,\psi_2))_{\fB}+ \cdots ,
\ee
where the extra terms in $\cdots$ involve $M_1$. This simply follows from the
observation that $\tau(\psi)_{\fB} = \nu(\emptyset; \psi)$ (compare equation
\eqref{eq:bifunctormap}) and is part of the statement that $\nu$ is an \afty-functor
in its second set of arguments. Hence a homotopy equivalence between Interfaces $\fI_1^{0,+}$ and $\fI_2^{0,+}$ leads to
a homotopy equivalence of the corresponding  \afty-functors.

If both $\CT^-$ and $\CT^+$ are trivial, the above results reduce to the previous results for the strip.
For example,   equation \eqref{eq:Strip-Web-Ident} is a special case of  equation \eqref{eq:Composite-Strip-Web-Ident}:
The only nontrivial taut element in $\ft^{-,+}$ consists of two boundary vertices, and the
composition operation then gives concatenation of strip-webs. As noted previously, the Chan-Paton space
\eqref{eq:Comb-CP} becomes the space of approximate ground states \eqref{eq:ELRdef}. Moreover
equation \eqref{eq:Cmpste-Contract} is equivalent to equation \eqref{eq:Strip-Contract},
when we bring $\CE_{LR}$ from the LHS to $\CE_{LR}^*$ on the RHS of the latter equation.
The differential $d_{LR}$ on the complex of approximate ground states
can be thought as a solution of a trivial MC equation, where only the composition of the Chan-Paton factors remain interesting.

\begin{figure}[htp]
\centering
\includegraphics[scale=0.3,angle=0,trim=0 0 0 0]{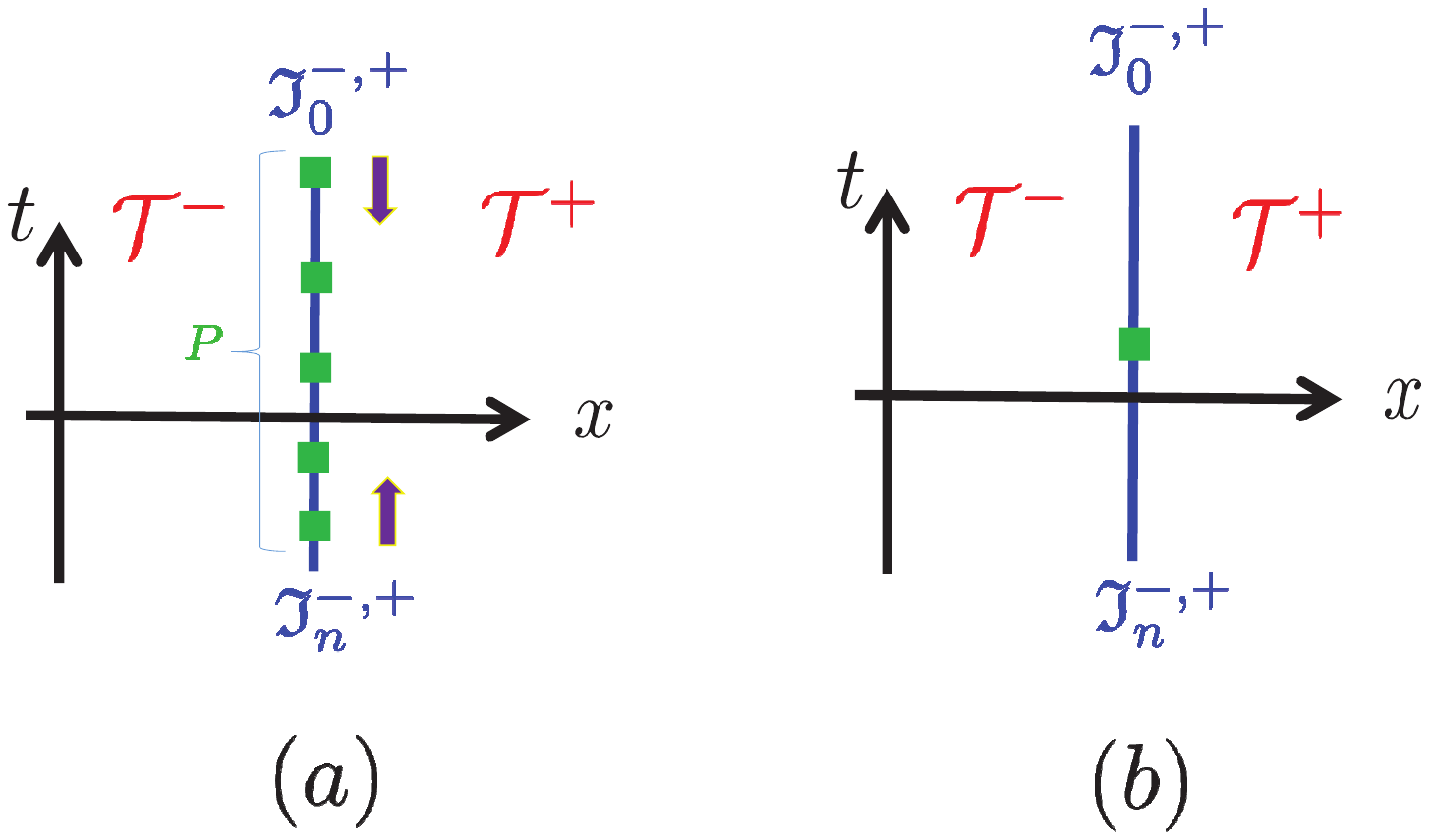}
\caption{Illustrating the \afty-multiplications on the local operators on
an Interface. Figure (a) shows a number of local operators between different
Interfaces. The vertical purple arrows indicate that they are to be multiplied.
The result is a single local operator between the initial and final Interfaces,
as illustrated in Figure (b). }
\label{fig:InterfaceBiFunctor-8}
\end{figure}
\begin{figure}[htp]
\centering
\includegraphics[scale=0.3,angle=0,trim=0 0 0 0]{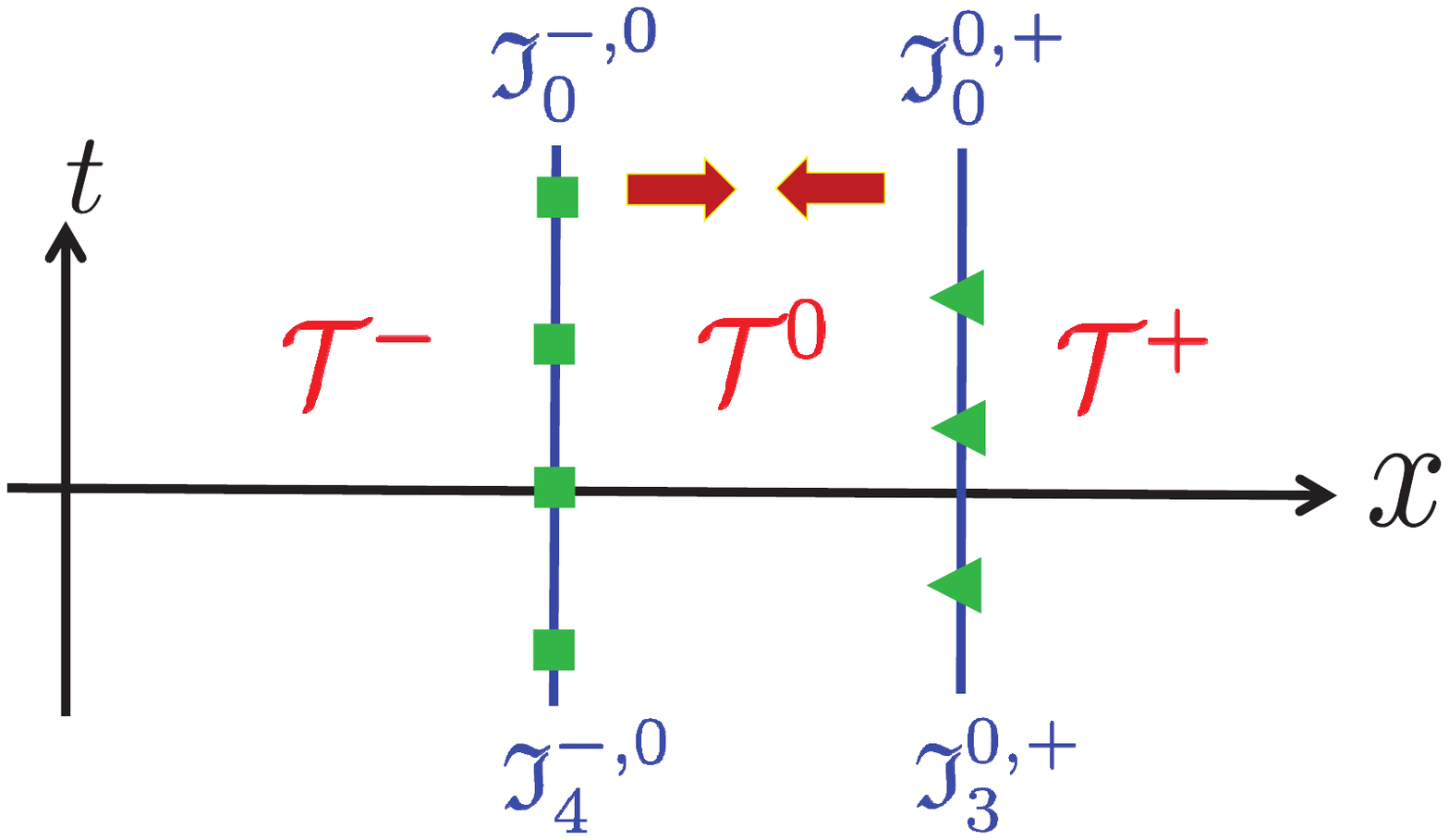}
\caption{Illustrating the bifunctor $\nu$. Here the green squares illustrate
$4$ local operator insertions $\delta_1,\delta_2,\delta_3, \delta_4$ between Interfaces from Theory $\CT^-$ to Theory $\CT^0$.
Similarly, the green triangles represent $3$ local operator insertions $\delta_1',\delta_2', \delta_3'$ between
Interfaces from Theory $\CT^0$ to Theory $\CT^+$. The maroon arrows indicate that the two interfaces
with their local operator insertions are being moved together (adiabatically).  }
\label{fig:InterfaceBiFunctor-4}
\end{figure}
\begin{figure}[htp]
\centering
\includegraphics[scale=0.3,angle=0,trim=0 0 0 0]{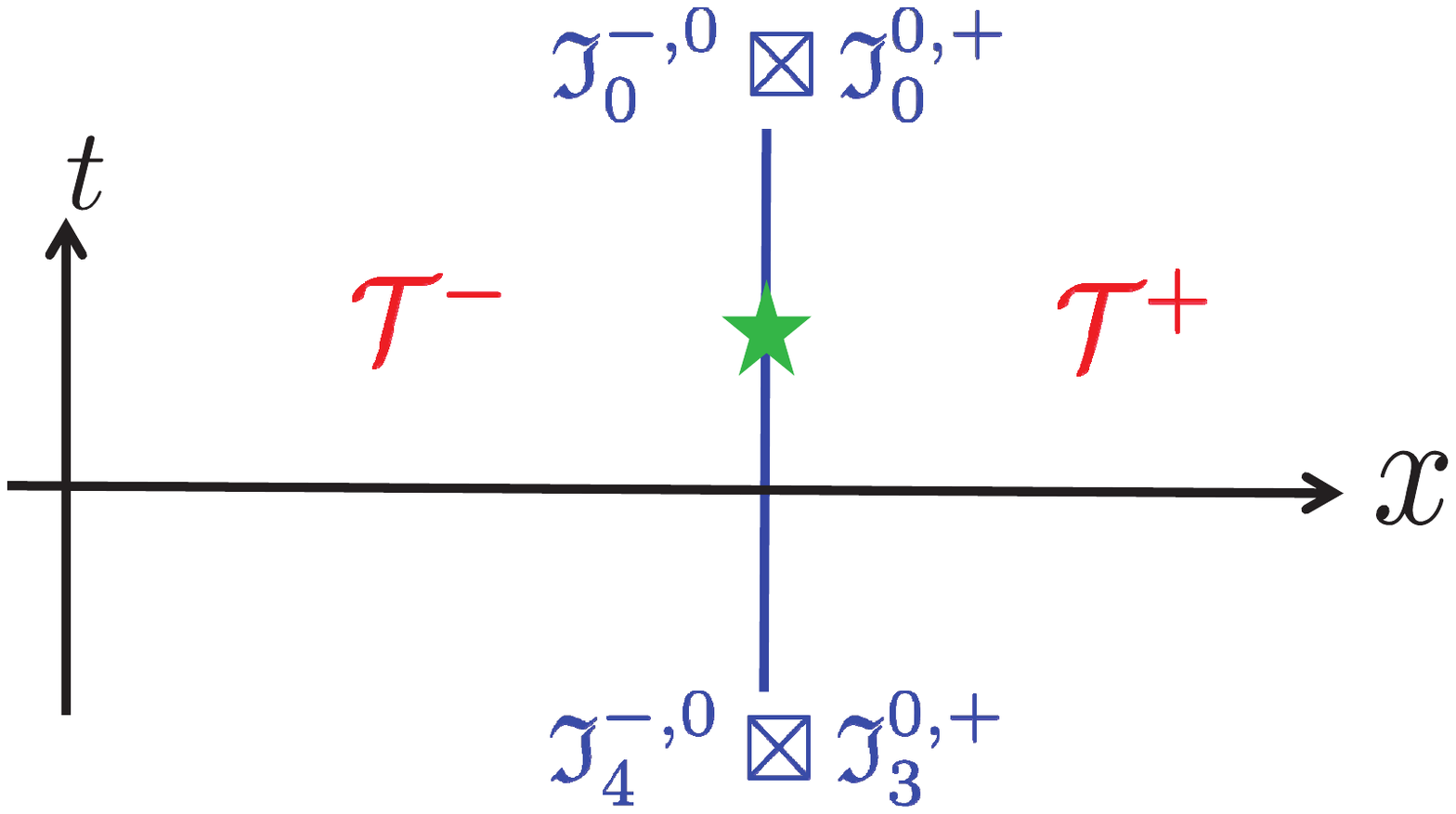}
\caption{The result of the process described in Figure \protect\ref{fig:InterfaceBiFunctor-4} is
a single local operator, described by the green star, between the products of the initial and final
Interfaces. This is the operator $\nu(\delta_1,\delta_2,\delta_3, \delta_4;\delta_1',\delta_2', \delta_3')$.  }
\label{fig:InterfaceBiFunctor-5}
\end{figure}
\begin{figure}[htp]
\centering
\includegraphics[scale=0.3,angle=0,trim=0 0 0 0]{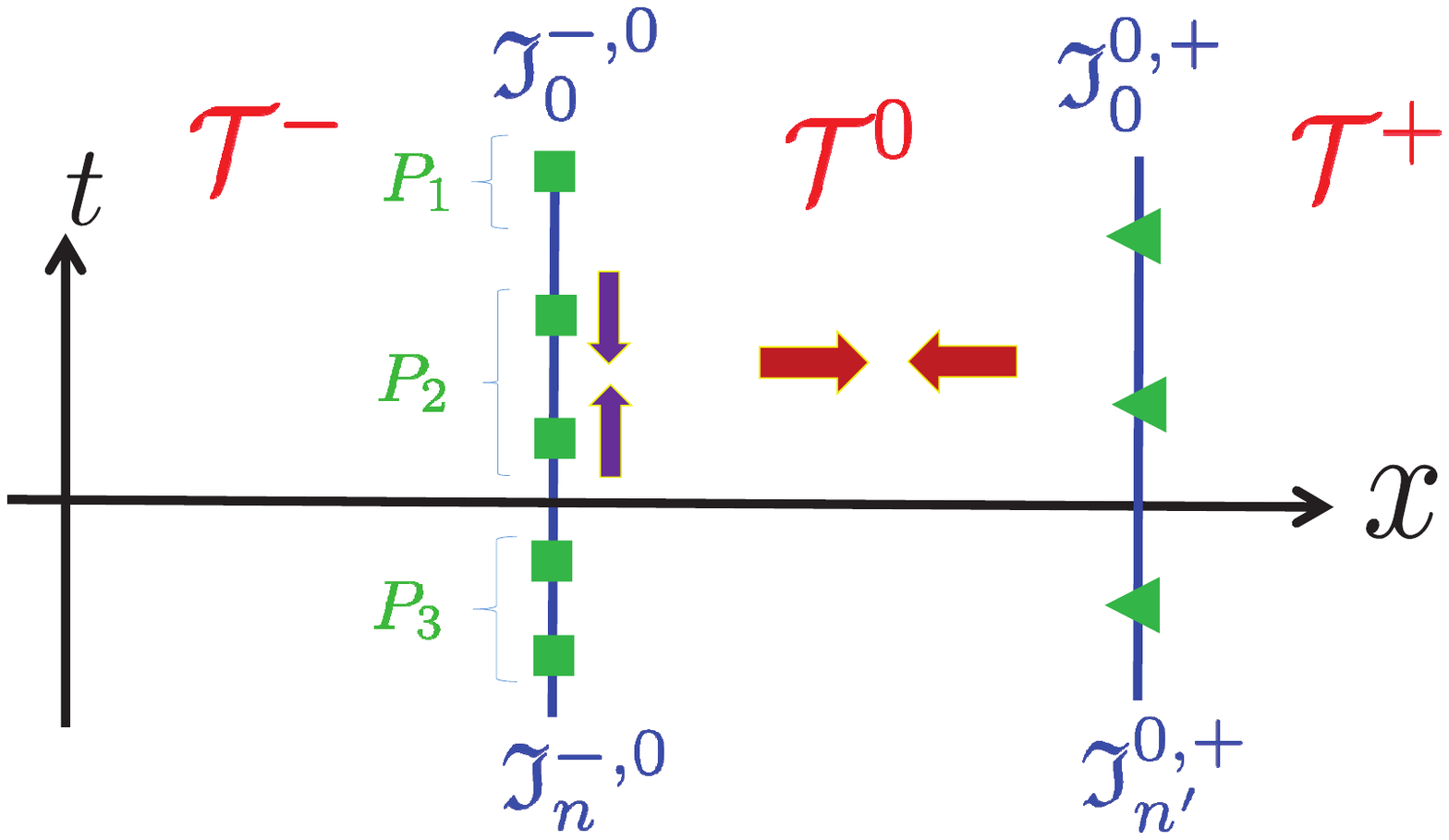}
\caption{This figure represents one side of the equation stating that $\nu$ is a bifunctor.
We first take the ``operator product'' of the ordered set $P_2$ of local operators on
Interfaces between $\CT^-$ and $\CT^0$, as indicated by the vertical purple arrows, and then
apply the Interface product, as indicated by the horizontal maroon arrows. We sum over all
decompositions of the local operators on the left Interface into $P_1 \amalg P_2 \amalg P_3$. }
\label{fig:InterfaceBiFunctor-6}
\end{figure}
\begin{figure}[htp]
\centering
\includegraphics[scale=0.3,angle=0,trim=0 0 0 0]{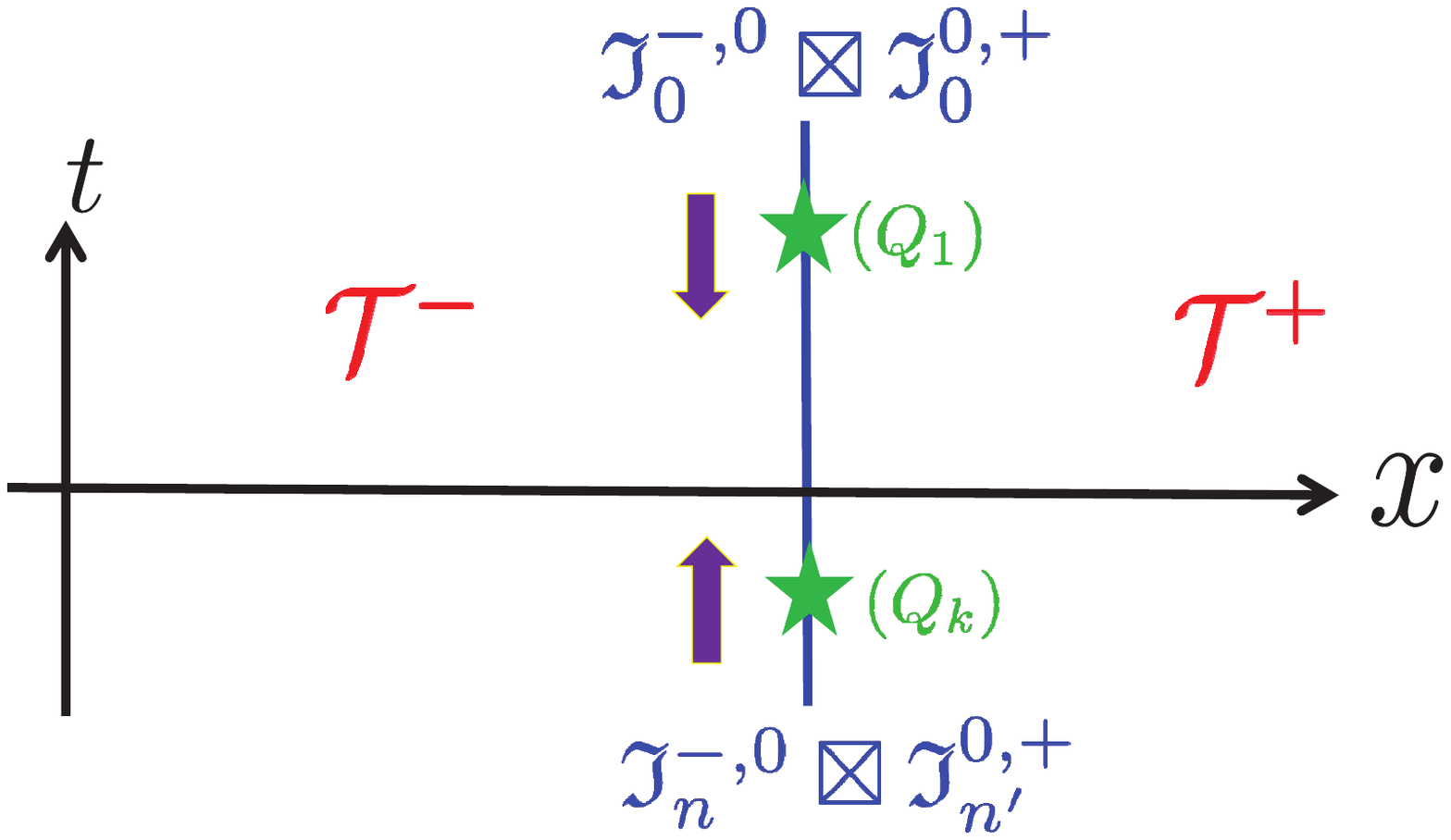}
\caption{This figure represents the other side of the equation stating that $\nu$ is a bifunctor.
We consider all ordered decompositions $Q_1\amalg \dots \amalg Q_k$ of local operators
on the left Interface (keeping $\delta_1',\dots, \delta_{n'}'$ fixed). We apply the Interface
product separately to these collections to produce the Interfaces and local operators
indicated by $\bigstar(Q_1),\dots, \bigstar(Q_k)$. Then we take the product of these local
operators, as indicated by the vertical purple arrows.   }
\label{fig:InterfaceBiFunctor-7}
\end{figure}

\bigskip
\noindent
\textbf{Remarks}:

\begin{enumerate}

\item  In principle we should keep track of the  positions $x^{-,0}$ and $x^{0,+}$ of
the original Interfaces as well as the position $x^{-,+}$ of the final product
Interface. Given the translation invariance and homotopy equivalence we can be a bit
sloppy about this, but it is relevant to the sense in which the product is associative.
We take that issue up in the next Section \S \ref{subsec:CompThrIntfc}.

\item It might help to restate some of the above formal expressions in more physical
terms. First of all, the basic \afty-product on local operators on Interfaces
is illustrated in Figure \ref{fig:InterfaceBiFunctor-8}.
Physically the product $\fI^{-,0}\IntfcTimes \fI^{0,+}$ is a kind of operator product of supersymmetric
interfaces. In the physical models there is no ``Casimir force'' between the Interfaces,
even with the insertions of (a suitable class of) local operators. Therefore they can
 be adiabatically brought together   as illustrated in Figure
\ref{fig:InterfaceBiFunctor-4} to produce a new Interface with a single local operator,
as illustrated in Figure \ref{fig:InterfaceBiFunctor-5}. An illustration of the statement
that $\nu$ is an \afty-functor (for fixed $\delta_1',\dots, \delta_{n'}'$) is shown in
Figure \ref{fig:InterfaceBiFunctor-6} and Figure \ref{fig:InterfaceBiFunctor-7}. Of course,
an analogous statement can also be made holding $\delta_1,\dots, \delta_{n}$ fixed.

\end{enumerate}

\subsection{Composition Of Three Interfaces}\label{subsec:CompThrIntfc}

We can now consider a geometry $G_3^\eta$ with three interfaces, set at
$x=-L$, $x=\eta L$, $x=L$, $-1 < \eta < 1$, with vacuum data $(\IV^\alpha, z^\alpha)$,
$\alpha\in \{ -,0,1,+ \}$,
in the negative half plane, the two strips and the positive half plane respectively.
The composite webs in this geometry have essentially the same properties
as in the case with two interfaces, with a convolution identity of the same general form.

Following through the same derivation, we arrive at  the statement that
given four Theories $\CT^-$, $\CT^0$, $\CT^1$, $\CT^+$ we obtain a
triple $A_\infty$ functor from $\fB\fr(\CT^-,\CT^0) \times \fB\fr(\CT^0,\CT^1)\times \fB\fr(\CT^1,\CT^+)$
to $\fB\fr(\CT^-,\CT^+)$, composing three consecutive Interfaces $\fI^{-,0}$, $\fI^{0,1}$, $\fI^{1,+}$
to a single Interface we can denote as
\be
\fI^{-,+}_{\eta} := (\fI^{-,0} \fI^{0,1} \fI^{1,+})_\eta .
\ee
It has boundary
amplitude
\be\label{eq:etaprod}
\CB(\fI^{-,+}_{\eta}):=  \rho_{\beta}(\ft_c)\left( \frac{1}{1- \CB^{-,0}},
\frac{1}{1-\CB^{0,1}}, \frac{1}{1-\CB^{1,+}} \right)
\ee
where $\ft_c$ is the taut composite element in $G_3^\eta$.

Thus we get a family of triple compositions, parameterized by $\eta$, and we may obviously wonder how would they compare to the
repeated compositions $(\fI^{-,0}\IntfcTimes \fI^{0,1}) \IntfcTimes\fI^{1,+}$ and $\fI^{-,0}\IntfcTimes (\fI^{0,1} \IntfcTimes \fI^{1,+})$.
We want to argue that there is an homotopy equivalence between any pair of interfaces $(\fI^{-,0} \fI^{0,1} \fI^{1,+})_\eta$ and
$(\fI^{-,0} \fI^{0,1} \fI^{1,+})_{\tilde \eta}$ and that moreover there are  well-defined limits such that:
\begin{align}
(\fI^{-,0} \fI^{0,1} \fI^{1,+})_{\eta \to -1} &= (\fI^{-,0}\IntfcTimes \fI^{0,1}) \IntfcTimes\fI^{1,+} \cr
(\fI^{-,0} \fI^{0,1} \fI^{1,+})_{\eta \to 1} &= \fI^{-,0}\IntfcTimes(\fI^{0,1}\IntfcTimes \fI^{1,+})
\end{align}
It then follows that  $(\fI^{-,0}\IntfcTimes \fI^{0,1}) \IntfcTimes\fI^{1,+}$
and $\fI^{-,0}\IntfcTimes(\fI^{0,1}\IntfcTimes \fI^{1,+})$ themselves are homotopy equivalent.

\subsubsection{Limits}

We can start from the analysis of the $\eta \to -1$ limit of the triple composition.
We need to study the fate of a composite taut web in $G_3^\eta$ when $\eta$ is sent to $-1$.
The taut element itself may jump in the process. As there are only finitely many possible taut web topologies,
as we make $\eta$ sufficiently close to $-1$, the taut element will ultimately stabilize.

After the taut element has stabilized, we can analyze the problem in the same way as we did for composite webs.
As the left strip shrinks, some vertices will remain at finite distances from the boundaries, while the distance from the left boundary of
some other ``bound'' vertices will scale as $1+\eta$. The bound vertices will form clumps at finite locations along the
left boundary. The whole web is well-approximated  by a composite taut web in the $G_2$ geometry with vacua $\IV^-$, $\IV^1$, $\IV^+$,
with boundary vertices on the left boundary replaced by tiny taut webs in a local $G_2$ geometry with vacua $\IV^-$, $\IV^0$, $\IV^1$.

If $\fc$ is any composite web in the $G_2$ geometry with vacua $\IV^-$, $\IV^1$, $\IV^+$,
so $\fc \in \CW_C[\IV^-,\IV^1,\IV^+]$ then
we can define an operation
\be
T_{\p}(\fc): T \CW_C[\IV^-, \IV^0, \IV^1] \to \CW_C[\IV^-, \IV^0, \IV^1, \IV^+]
\ee
whose nonzero values on  monomials $\fc_1 \otimes \cdots \otimes \fc_n$ are obtained
by inserting the $\fc_a$  into the left boundary vertices $v^\p_a$ of $\fc$ provided $J_{v^p_a}(\fc) = J_{\infty}(\fc_a)$
and provided that the past strip vacuum of $\fc_{a}$ agrees with the future strip vacuum
of $\fc_{a+1}$. We orient the resulting web in the standard way, wedging the reduced orientations
of the arguments in the same order as the arguments themselves.

\begin{figure}[htp]
\centering
\includegraphics[scale=0.5,angle=0,trim=0 0 0 0]{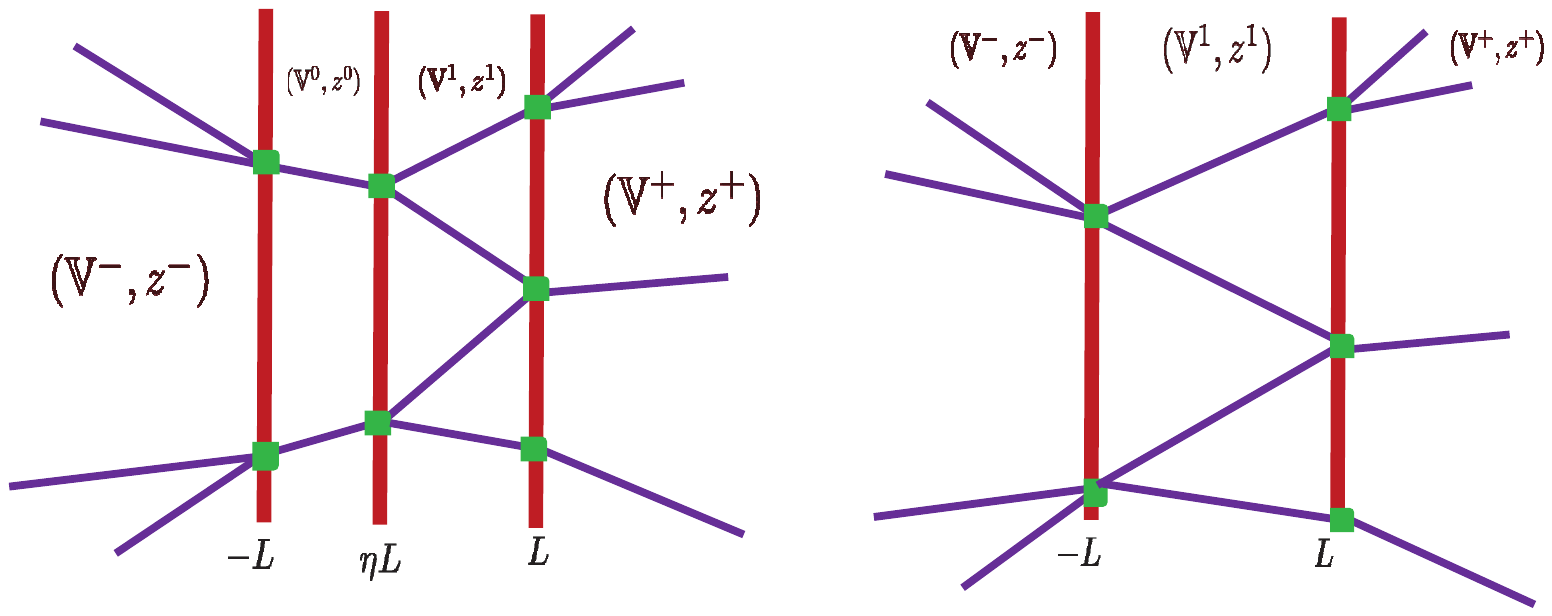}
\caption{In the limit that $\eta\to -1$ the taut composite web on the left
degenerates to that on the right. We can view this as a contribution of
$T_{\p}(\ft^{-,1,+})[\ft^{-,0,1}, \ft^{-,0,1}]$ to $\ft_\eta^{-,0,1,+}$.   }
\label{fig:HMTPY-CMPST}
\end{figure}

Thus with these definitions, if $\ft_\eta^{-,0,1,+}$ is the taut element in $G_3^\eta$ and $\ft^{-,1,+}$ and $\ft^{-,0,1}$ the taut elements for
the two $G_2$ geometries respectively,
\begin{equation}
\lim_{\eta\to -1} \ft_\eta^{-,0,1,+} = T_\p[\ft^{-,1,+}](\frac{1}{1-\ft^{-,0,1} })
\end{equation}
See, for example, Figure \ref{fig:HMTPY-CMPST}.

Now we can compute
\begin{equation} (\CB^{-,0} \CB^{0,1} \CB^{1,+})_{\eta \to -1}  = \rho_\beta(\lim_{\eta\to -1} \ft_\eta^{-,0,1,+}  ) [\frac{1}{1-\CB^{-,0}};\frac{1}{1-\CB^{0,1}};\frac{1}{1-\CB^{1,+}}]\end{equation}
from the representations of webs and interface amplitudes.
We can write
\begin{align} \rho_\beta(T_\p[\ft^{-,1,+}]&(\frac{1}{1-\ft^{-,0,1}  })) [\frac{1}{1-\CB^{-,0}};\frac{1}{1-\CB^{0,1}};\frac{1}{1-\CB^{1,+}}] =\cr
&\rho_\beta(\ft^{-,1,+})[\frac{1}{1-\rho_\beta(\ft^{-,0,1})[\frac{1}{1-\CB^{-,0}};\frac{1}{1-\CB^{0,1}}]};\frac{1}{1-\CB^{1,+}}]
\end{align}

Thus
\begin{equation}
(\fI^{-,0} \fI^{0,1} \fI^{1,+})_{\eta \to -1} = (\fI^{-,0} \IntfcTimes\fI^{0,1})\IntfcTimes \fI^{1,+}
\end{equation}
A similar analysis holds for $\eta \to 1$.

At this point, we are left with the task of proving the homotopy equivalence of the triple compositions for different values $\eta_{p,f}$ of $\eta$.
The Chan-Paton factors are independent of $\eta$. If we define $\fI_{p,f} = (\fI^{-,0} \fI^{0,1} \fI^{1,+})_{\eta_{p,f}}$
then, according to \eqref{eq:HomEqBr}, we need to find $\delta_h\in R^\p(\CE^{-,+})$ such
that
\begin{equation}\label{eq:BfBpHom}
\CB_f - \CB_p+ \rho_\beta(\ft^{-,+} )[\frac{1}{1-\CB_f};\delta_h; \frac{1}{1-\CB_p}] =0 \end{equation}
where
\be
\CE^{-,+}_{i_-,i_+} := \oplus_{i_0,i_1} \CE^{-,0}_{i_-,i_0} \otimes \CE^{0,1}_{i_0,i_1}\otimes \CE^{1,+}_{i_1,i_+}
\ee

As the interface amplitudes $\CB_{p,f}\in \Hop(\fI_{p},\fI_{f})$ are computed directly from the
corresponding composite taut elements $\ft_{p,f}^{-,0,1,+}$
we will first try to find  a similar identity for the difference between these two taut elements.

\subsubsection{Homotopies}

How can we compare the taut elements for different values of $\eta$? One way would be to vary $\eta$ continuously, and study
the special values at which the taut element jumps. This is a somewhat subtle but interesting analysis, and we will come back to it
at the very end of the section. Here we will use a different, more effective strategy, which produces precisely the desired result.

\begin{figure}[htp]
\centering
\includegraphics[scale=0.5,angle=0,trim=0 0 0 0]{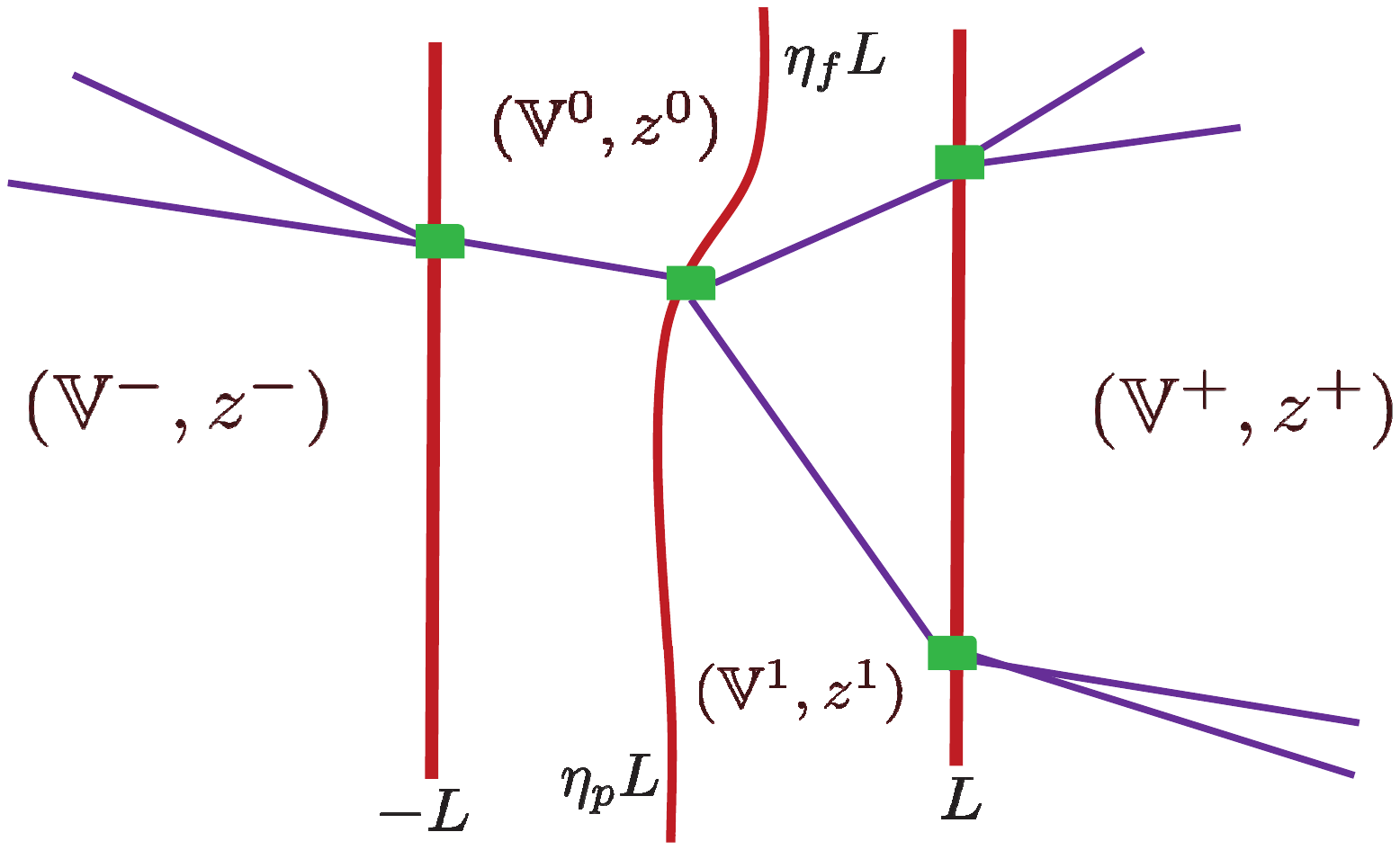}
\caption{A smooth adiabatic variation of $\eta$ as a function of $y$ can relate the
products of interfaces at different values of $\eta$.    }
\label{fig:ETAVARY}
\end{figure}

We can encode the problem in a simple geometric setup: a time-dependent setup where $\eta$ depends very slowly on the time direction.
We want $\eta(y)$ to vary slowly enough that the deviation from the vertical of the slope of the boundary between the $\IV^0$ and $\IV^1$ at any $y$
does not affect the local interface taut element between $\IV^0$ and $\IV^1$. In particular the slope should remain close enough to vertical so that it never crosses the slopes of the weights $z^0_{ij}$ and $z^1_{i'j'}$ for any $i,j,i',j'$.
We also want the variation of $\eta$ to be restricted to some compact region $y\in [y_p,y_f]$
with fixed values $\eta_p$ and $\eta_f$ in the past and the future, respectively.  We will call such
functions $\eta(y)$ \textit{tame}.

As soon as these conditions are met, we can consider composite webs in the time-dependent geometry.
\footnote{Actually, we could consider the positions of all three interfaces to be time-dependent.
The resulting discussion would be similar to what we give here.}
The definitions for composite webs from Section \S \ref{subsec:ComposeInterface} have
straightforward generalizations, with one important exception noted in the next
paragraph.  The only slightly new point is that the wall vertices between
$\CT^0$ and $\CT^1$ do not sit at a definite value of $x$ since the boundary between
these theories is $y$-dependent. Nevertheless, one can define deformation types of
time-dependent composite webs. The only new point is that wall vertices between
$\CT^0$ and $\CT^1$ must slide along the wall $x = \eta(y) L$.  If a composite web $\fc$ has $n$
wall-vertices on the boundary between $\CT^0$ and $\CT^1$ and $\fd_1, \dots, \fd_n$
are $n$ interface webs between $\CT^0$ and $\CT^1$ then the convolution
$T[\fc](\fd_1, \dots, \fd_n)$ makes sense as a composite web in the time-dependent
geometry. Moreover, we claim that if $\eta(y)$ satisfies the above conditions
then the deformation type of $T[\fc](\fd_1, \dots, \fd_n)$ only depends on
the deformation types of $\fc$ and $\fd_1, \dots, \fd_n$. To prove this we note that
there are only a finite number of possible webs $\fc, \fd_1,\dots, \fd_n$. But then note that
the  homotopy which straightens out the boundary between $\CT^0,\CT^1$ to a vertical line in the neighbourhood of each
interface web will change
the edge lengths by an amount which can be made arbitrarily small by making the slope of
the boundary arbitrarily close to vertical.

The one new point in the definitions for composite webs in time-dependent geometries is that
 we must change the definitions of rigid, taut and sliding webs
from what we used in Section \ref{subsec:ComposeInterface}.
Webs in the time-dependent geometry  with \emph{no moduli} are \emph{rigid} or \emph{taut}, and
webs with a \emph{single modulus} are called \emph{sliding}. An example of a sliding web in a time-dependent
geometry is a rigid composite web for $\eta = \eta_f$ whose support lies in $y\geq y_f$ or, similarly,
a rigid composite web for $\eta = \eta_p$ whose support lies in $y\leq y_p$.

We can now repeat our usual exercise: We define $\ft[\eta(y)]$  to be the taut (= rigid) element in
 $\CW_C$ in the time-dependent geometry determined by $\eta(y)$ and find its convolution identity by
 examining the   end-points of the moduli spaces of sliding webs in this time-dependent
geometry. We encounter standard boundaries at finite distance: some edge inequalities get saturated, some subset of vertices collapse to a point
in the interior or at any of the interfaces. These endpoints are enumerated by convolutions of appropriate taut elements.

Boundaries at infinite distance are also rather standard. Much as for the time-independent geometry we can consider directions along which a web grows to large size.
These boundaries are accounted for by a standard tensor operation, inserting taut composite webs at the boundary vertices of a taut interface web.
In order for the result to have only one modulus in the time-dependent geometry, exactly one of the composite webs must be localized in the compact region where $\eta$ varies.
%
%\cg{why is that? Better write the analog of equation
%\eqref{eq:ModTensOp}. }
%
The others will be sliding along the regions of constant $\eta$ in the past or future. Thus these endpoints are enumerated by the usual
$T_\p(\ft^{-,+})$ operation with $\ft^{-,+}$ being the taut element for interface webs between $\IV^-$ and $\IV^+$,
acting on a collection of taut webs in the far past or future, together with a single rigid/taut web stuck somewhere in the region
$y_p \leq y \leq y_f$.

The only new terms are very simple: a single taut web for the $G_3^{\eta_p}$ geometry inserted far in the past, or a
 taut web for the $G_3^{\eta_f}$ geometry inserted far in the future. We are now ready to write the convolution
 identity for the taut element $\ft[\eta(y)]$ in the time-dependent geometry.
Let $\ft_{pl} = \ft_p^- + \ft_p^0 + \ft_p^1 + \ft_p^+$ denote the sum of the planar
taut elements for the four theories. Similarly,
let   $\ft_\p$ denote the sum of the   taut interface
elements for the three boundaries between pairs of theories:
 $\ft_\p = \ft^{-,0} + \ft^{0,1} + \ft^{1,+}$,
and let $\ft_c^{p,f}$ be the taut composite elements for the initial and
final  $G_3^{\eta_{p,f}}$ geometries. If $\ft_c^f$ is understood to be
 made of webs with all edges and vertices in the region $y\geq y_f$ and
likewise  $\ft_c^p$ has all lines and vertices in the region $y \leq y_p$,
then we can consider these as elements of the web group of the time-dependent
geometry. With this understanding we have  the   convolution identity
\begin{equation} \label{eq:fth}
\ft[\eta(y)] * \ft_{pl} + \ft[\eta(y)] * \ft_\p + \ft_c^f - \ft_c^p + T_\p(\ft^{-,+})[\frac{1}{1-\ft_c^f};\ft[\eta(y)]; \frac{1}{1-\ft_c^p}] =0.
\end{equation}
The $\ft_c^f - \ft_c^p$ could
be absorbed into the $T_\p(\ft^{-,+})$ term if we include the empty
web (no vertices nor edges) into $\ft[\eta(y)]$.

Now we combine equation \eqref{eq:fth} together with a representation of the
composite webs and apply the result to
\be
e^{\beta^-};  \frac{1}{1-\CB^{-,0}}; e^{\beta^0}; \frac{1}{1-\CB^{0,1}} ; e^{\beta^1 } ; \frac{1}{1-\CB^{1,+}} ; e^{\beta^+}
\ee
The first two terms of \eqref{eq:fth} give zero. The next two terms give $\CB_f - \CB_p$ using the
definition \eqref{eq:etaprod}, and, using identities analogous to \eqref{eq:RepTdel}, the last term in \eqref{eq:fth} gives
\be
\rho_{\beta}(\ft^{-,+})\left[ \frac{1}{1-\CB_f} ; \delta[\eta(y)] ; \frac{1}{1-\CB_p} \right]
\ee
with
\begin{equation}
\delta[\eta(y)] := \rho_\beta(\ft[\eta(y)])[\frac{1}{1-\CB^{-,0}};\frac{1}{1-\CB^{0,1}};\frac{1}{1-\CB^{1,+}}]
\end{equation}
finally leading to the desired relation \eqref{eq:BfBpHom}. Recall from
 \eqref{eq:HomEqBr} that \eqref{eq:BfBpHom} implies that the morphism  $\Id + \delta[\eta(y)]$ (or just $\delta[\eta(y)]$ if we include the empty
web into $\ft[\eta(y)]$) from $\CB_p$ to $\CB_f$  is $M_1$-closed.

Next, we need to show that $\Id + \delta[\eta(y)]$ has a closed inverse up to homotopy. In principle, we could cheat a bit.
The morphism $\delta[\eta(y)]$ can have a non-trivial scalar part, from composite webs with no external edges.
As the vacua are naturally ordered, in the sense that for any pair of vacua only one of the two can be to the future of the other,
this scalar contribution is upper triangular and thus the scalar part of $\Id + \delta[\eta(y)]$ is invertible. We could build the required inverse recursively, as described in Section \ref{subsec:BraneCat}. However, there is a much better way to proceed.

In order to prepare some tools which will be useful later, we will instead prove directly that the closed morphism
$\Id + \delta[\eta(y)]$
 associated to the $\eta(y)$ deformation and the closed morphism $\Id + \delta[\eta(-y)]$
associated to the time-reversed deformation are inverse of each other up to homotopy.

The first observation is that if we have two continuous deformations $\eta_1(y)$ and $\eta_2(y)$,
with the same value $\eta_{1,p}=\eta_{2,f}$ in the past of $\eta_1$ and in the future of $\eta_2$, we can build a ``shifted time composition''
$\eta_1 \circ_T \eta_2$ where $\eta_1$ is placed at some large time $T$ after $\eta_2$. If $T$ is sufficiently large,
it is easy to see that
\begin{equation}\label{eq:mult-homtpy}
\Id + \delta[\eta_1 \circ_T \eta_2] = M_2(\Id + \delta[\eta_1],\Id + \delta[\eta_2])
\end{equation}
where $M_2$ is computed in the Interface category $\fB\fr(\CT^-,\CT^+)$, and $\delta[\eta_1 \circ_T \eta_2]$, $\delta[\eta_1], $
and $\delta[\eta_2]$   are computed
in different time-dependent geometries.

In order to establish \eqref{eq:mult-homtpy} we observe that
 the most general rigid web in the composite geometry can be approximated by a large interface web with a boundary vertex
resolved into a rigid web in the region of $\eta_1$ variation, a boundary vertex resolved into a rigid web in the
region of $\eta_2$ variation
and any number of boundary vertices resolved to rigid webs in the regions of constant $\eta$.  Thus we have:
\begin{equation}\label{eq:mult-homtpy-2}
\ft[\eta_1 \circ_T \eta_2] = \ft[\eta_1] + \ft[\eta_2] + T_\p(\ft^{-,+})[\frac{1}{1-\ft_c^{1,f}};\ft[\eta_1];\frac{1}{1-\ft_c^{1,p}};\ft[\eta_2]; \frac{1}{1-\ft_c^{2,p}}]
\end{equation}
where $\ft_c^{1,f}$ is the taut composite element for the value $\eta_{1,f}$ and we recall that
$\ft_c^{1,p} =   \ft_c^{2,f}$. This equation is to be thought of as valued in the web group of
the time-dependent geometry described by $\eta_1 \circ_T \eta_2$.
Now, again using identities similar to \eqref{eq:RepTdel} we learn that \eqref{eq:mult-homtpy-2} implies \eqref{eq:mult-homtpy}.
Thus, the desired result
\be
\fI^{-,+}_{\eta_p} \sim \fI^{-,+}_{\eta_f}
\ee
will follow by proving that
\be\label{eq:HomToShow}
\Id + \delta[\eta(y) \circ_T \eta(-y)] \sim \Id.
\ee
We will prove \eqref{eq:HomToShow} in the next Section.

\subsubsection{Homotopies Of Homotopies}\label{subsubsec:Homotopy-homotopy}

The required identity \eqref{eq:HomToShow} will  follow
from a more general result: Suppose that $\eta_1(y)$ and $\eta_2(y)$ define two tame
time-dependent geometries such that $\eta_1(y) = \eta_2(y) = \eta_f$ for $y\geq y_f$
and $\eta_1(y) = \eta_2(y) = \eta_p$ for $y\leq y_p$. Suppose moreover that
$\eta(y;s)$ is a homotopy between these functions. Then we claim that
\be\label{eq:hom-to-hom}
\delta[\eta_1(y)] \sim \delta[\eta_2(y)]
\ee
are homotopic morphisms in $\fB\fr(\CT^-,\CT^+)$. When writing $\eta(y;s)$
we take $s\in \IR$ and assume that $\frac{\p}{\p s} \eta(y;s)$ has
compact support in some interval $(s_1,s_2)$ with $\eta(y;s)= \eta_1(y)$
for $s\leq s_1$ and $\eta(y;s)= \eta_2(y)$ for $s\geq  s_2$. We will also
assume that the homotopy $\eta(y;s)$ is tame for all fixed $s$ and is furthermore generic.

We will establish \eqref{eq:hom-to-hom} by studying how the taut element
$\ft[\eta(y;s)]$ jumps as we vary $s$. Some of the ideas we introduce here
will be very useful in Section \S \ref{sec:GeneralParameter} on wall-crossing
as well as in Sections \S   \ref{whynot} and \S \ref{notif} on Landau-Ginzburg models.

The basic idea is to allow deformations in the geometry  $G_3[s]:=G_3^{\eta(\cdot;s)}$
determined by $\eta(y;s)$ in our notion of ``deformation type''. We will call this enlarged
notion ``deformation type up to homotopy''
or just $h$-type, for brevity. Thus, we allow the usual translations and
dilation of internal edges at fixed $s$, but also we allow the parameter $s$ to be adjusted.
For readers who demand more precision we will spell this out more
formally.  Less fastidious readers can skip to the examples below equation \eqref{eq:h-dim} below.

Define a \emph{continuous family of webs $\fw[s]$}, labeled by $s_- \leq s \leq s_+$ to be
a family of webs such that

\begin{enumerate}

\item For each fixed $s\in [s_-,s_+]$, $\fw[s]$ is a web in the geometry $G_3[s]$.

\item As $s$ varies the vertices of $\fw[s]$ in the plane vary continuously,  no
edge shrinks to zero length, and no boundary vertices collide.

\end{enumerate}

  For a fixed $s\in \IR$ let ${\rm Web}[G_3[s]]$
denote the set of all webs in the geometry $G_3[s]$. According to the definition in
 Section \S \ref{planewebs}, a deformation type of a web in $G_3[s]$ is a subset   $\CD(\fw)\subset {\rm Web}[G_3[s]]$,
 defined by an equivalence relation under translation and dilation of internal edges.
 We now consider the union
 \be
 {\rm WEB}:= \cup_{s\in \IR } {\rm Web}[G_3[s]]
 \ee
 and define a
\emph{deformation type up to homotopy} (or, just an \emph{$h$-type}, for brevity)
 to be an equivalence class of webs $\CD^h \subset {\rm WEB}$ with the following two equivalence
 relations:

\begin{enumerate}

\item If $\fw_a, \fw_b \in {\rm Web}[G_3[s]] $ for the same value of $s$ then,
if they define the same deformation type within
${\rm Web}[G_3[s]]$, they are equivalent.

\item   If
there exists a continuous family of webs $\fw[s]$, $s_-\leq s\leq s_+$ interpolating
between $\fw_-$ and $\fw_+$, then the web $\fw_- \in {\rm Web}[G_3[s_-]] $
  is equivalent to the web $\fw_+ \in {\rm Web}[G_3[s_+]] $.

\end{enumerate}

As usual, we will let $\CD^h(\fw)$ denote the set of deformation types up to homotopy
of any given web $\fw \in {\rm Web}[G_3[s]] $ for some $s$.

Of course, there is a projection $\pi: {\rm WEB} \to \IR$ given by the $s$-coordinate so
an $h$-type $\CD^h$ will be fibered over a connected subset of some interval $[s_-,s_+]\in \IR$.
See, Figure \ref{fig:SLIDING-MODULISPACE} below. The
fiber of the projection $\pi: \CD^h \to  [s_-,s_+]$ above some
$s\in [s_-,s_+]$  will be a subset of $\IR^{2V_i(\fw) + V_\p(\fw) }$
given by the data of the vertices of the web. We can therefore regard
 $\CD^h(\fw) \subset \IR^{2V_i(\fw) + V_\p(\fw) +1}$. This motivates us to define the
\emph{expected $h$-dimension} (or just $h$-dimension, for brevity) by
\be\label{eq:h-dim}
d^h(\fw) := 2V_i(\fw) + V_\p(\fw) +1 - E(\fw).
\ee
The set $\CD^h(\fw)$ is orientable.
Oriented $h$-types are denoted by $\fw^h$. If $ \fw \in {\rm Web}[G_3[s]] $ for some $s$
then we let $\fw^h$ denote the induced $h$-type with orientation $o(\fw^h)=o(\fw)\wedge ds$.
It can happen that the projection of $\CD^h$ under $\pi$ is a single point $s_*\in \IR$.
In this case we take the orientation to be $o(\fw^h)=ds$.
Oriented $h$-types again generate a free abelian group $\CW^h$.

We now consider the convolution identity in $\CW^h$. As usual we examine
the moduli space of sliding $h$-types, i.e. those with $d^h(\fw)=1$.
What are the possible boundary regions? Provided that $\eta(y;s)$ is tame and generic
these can be listed as follows:

\begin{figure}[htp]
\centering
\includegraphics[scale=0.3,angle=0,trim=0 0 0 0]{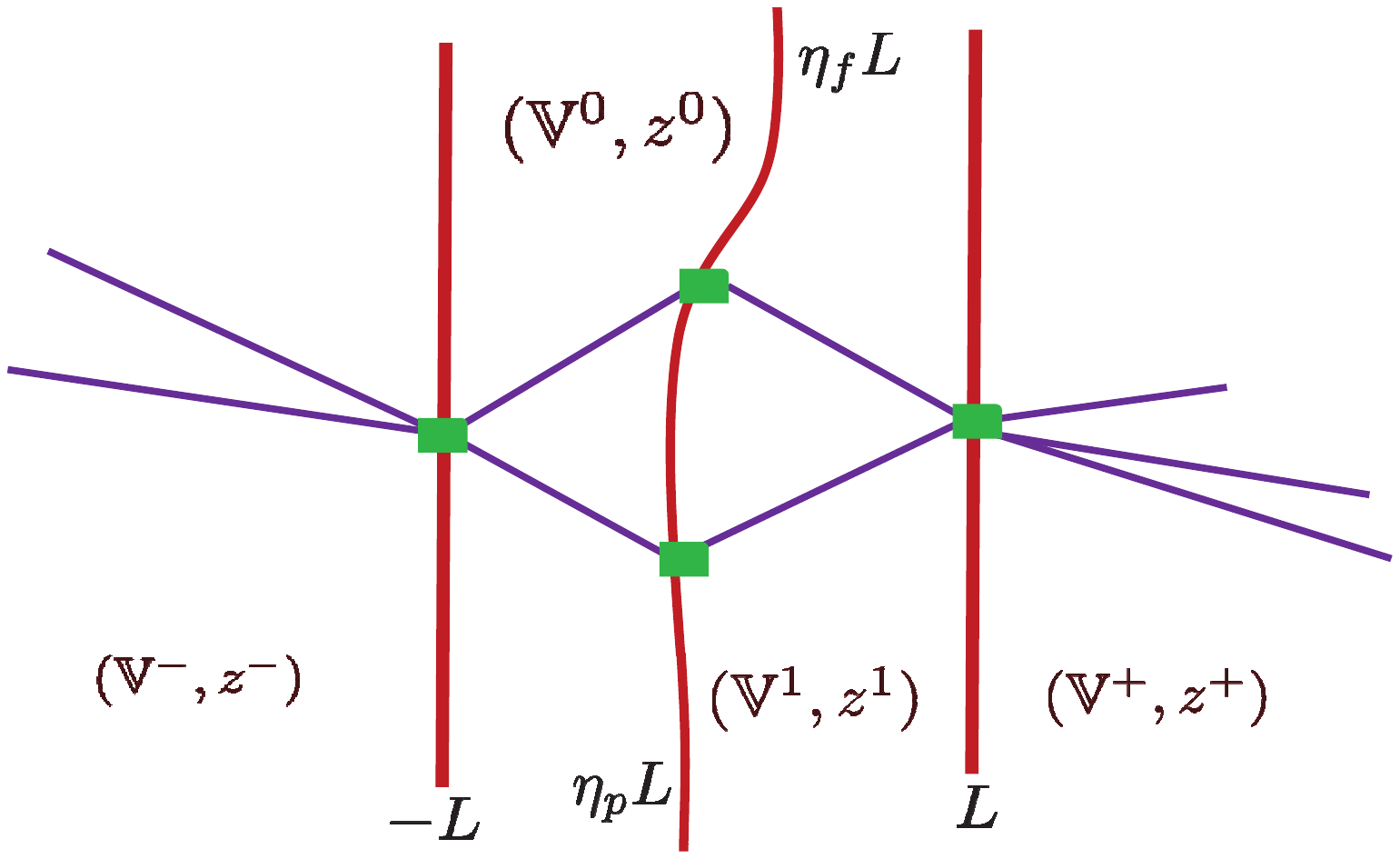}
\caption{At fixed $s$ the web shown here
 has expected and true dimension zero: There are four boundary vertices and four internal edges.
 If the edge constraints can be satisfied then the equations are all independent. At fixed
values of $s$ the web cannot be deformed at all. If it exists it will only
exist for a finite set of possible $y$-coordinates of the left vertex.
However, if it exists,  for tame and generic homotopies $\eta(y,s)$ if we deform $s$ we
can deform the web so that it will  generically define a deformation
type up to homotopy of expected and true dimension $d^h = 1$. This is a typical contribution
to both $\ft^h[\eta_1(y)]$ and $\ft^h[\eta_2(y)]$.}
\label{fig:ETAVARY-TRYTWO-1}
\end{figure}
\begin{figure}[htp]
\centering
\includegraphics[scale=0.3,angle=0,trim=0 0 0 0]{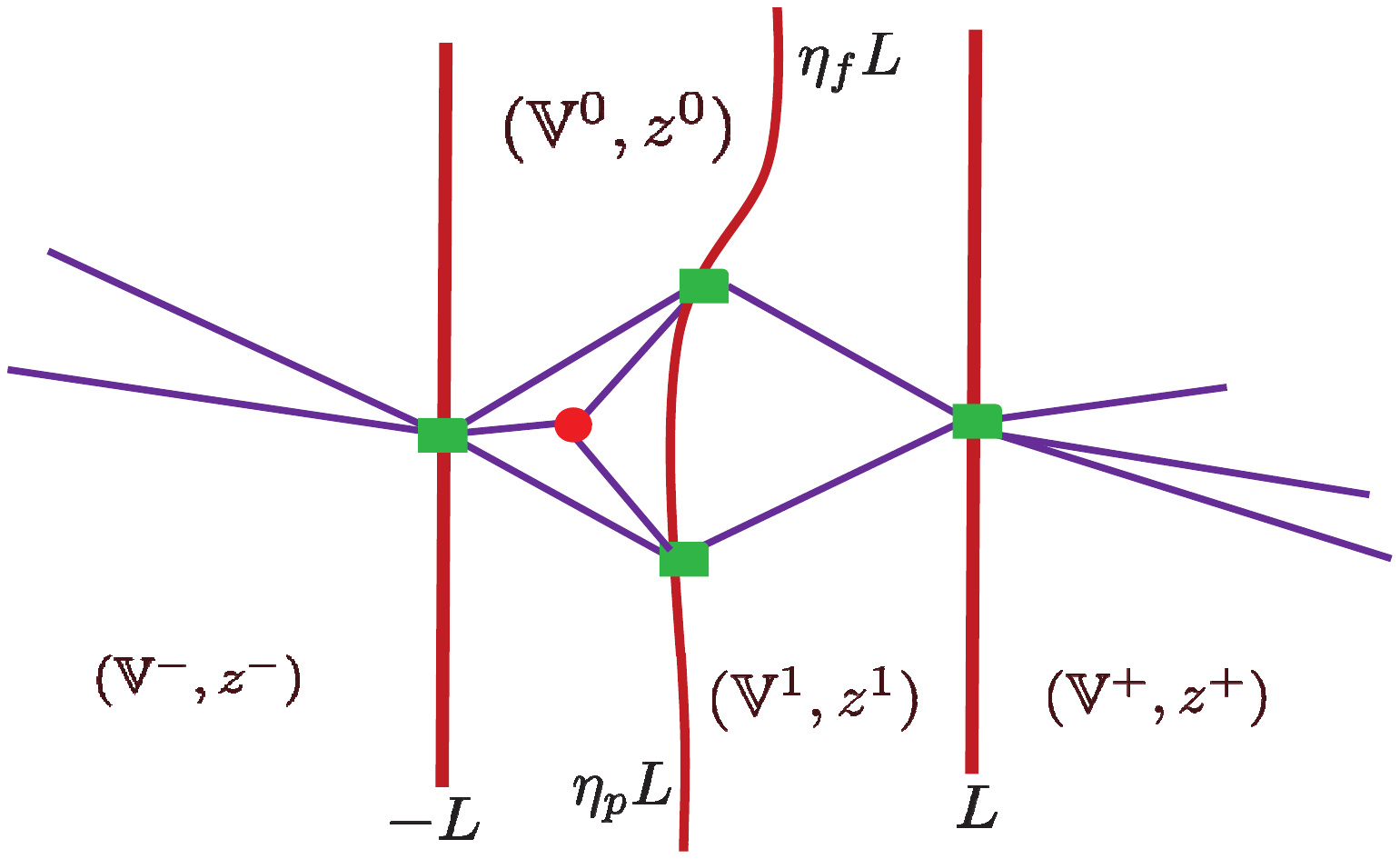}
\caption{This web will only exist at special values of $s$. It has dimension $2V_i + V_\p -E=2+4 -7=-1$
at fixed $s$ and $h$-dimension $d^h = 0$.      }
\label{fig:ETAVARY-TRYTWO-2}
\end{figure}
\begin{figure}[htp]
\centering
\includegraphics[scale=0.3,angle=0,trim=0 0 0 0]{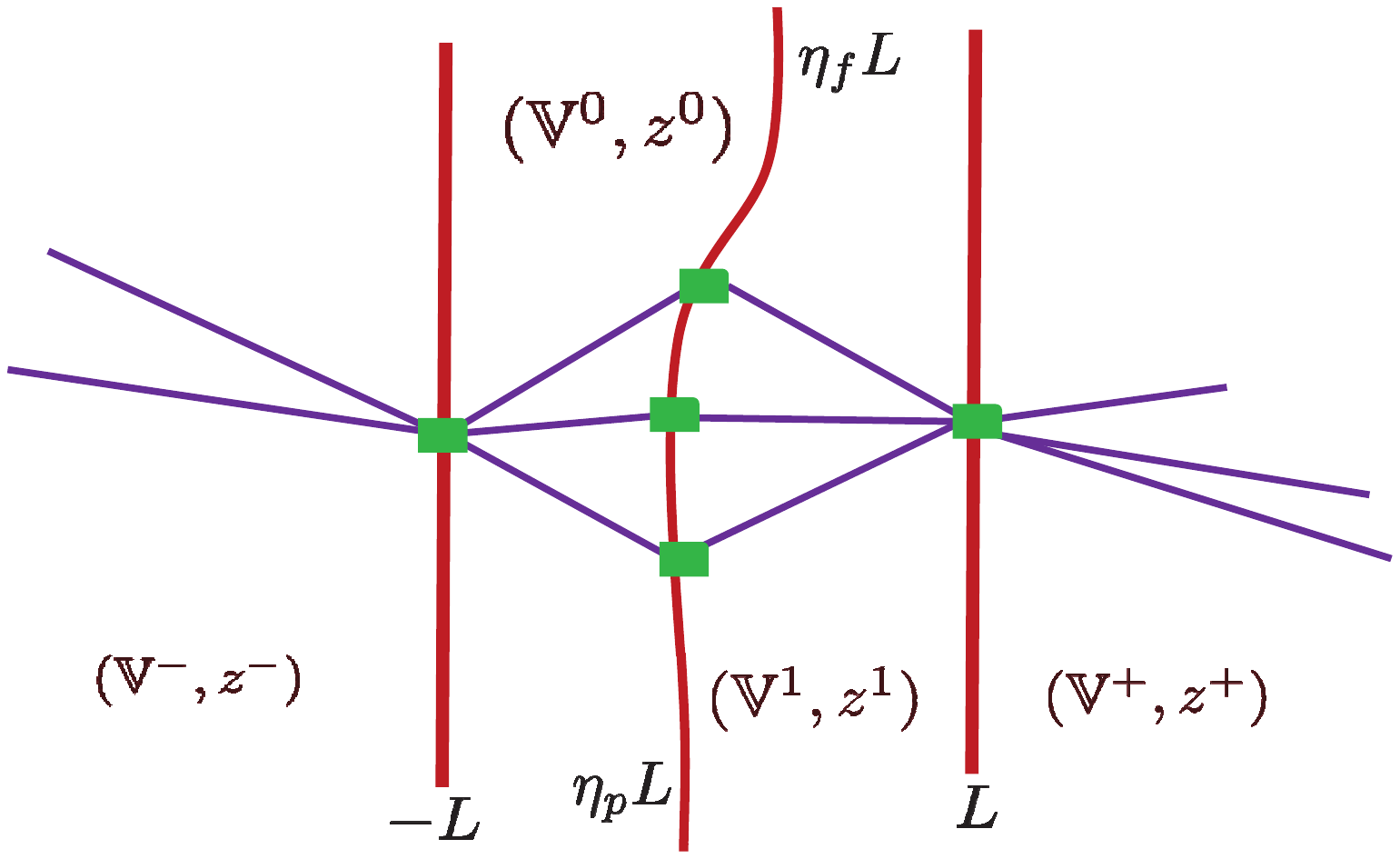}
\caption{This web will only exist at special values of $s$. It has dimension $ V_\p -E=5 -6=-1$
at fixed $s$ and $h$-dimension $d^h = 0$.     }
\label{fig:ETAVARY-TRYTWO-3}
\end{figure}
%

%
%\begin{figure}[htp]
%\centering
%\includegraphics[scale=0.3,angle=0,trim=0 0 0 0]{ETAVARY-EXCPL-3-eps-converted-to.pdf}
%\caption{This web will only exist at special values of $s$. It has dimension $2V_i + V_\p -E=2+3-6=-1$
%at fixed $s$ and $h$-dimension $d^h = 0$.    }
%%\label{fig:ETAVARY-EXCPL-3}
%\end{figure}
%
%
\begin{figure}[htp]
\centering
\includegraphics[scale=0.3,angle=0,trim=0 0 0 0]{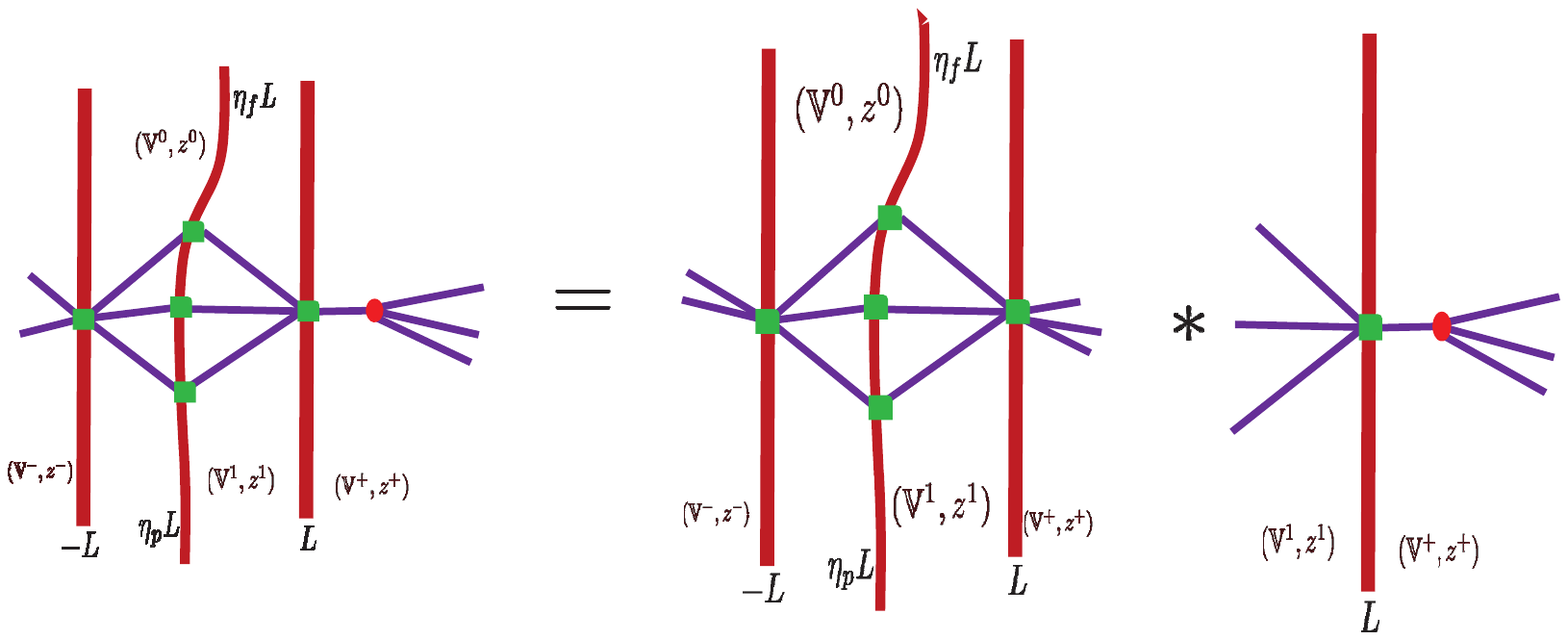}
\caption{Here we show a typical example of a contribution $\fe^h* \ft_\p$ to the convolution identity.
The sliding $h$-type on the left projects to a single value $s_* \in \IR$.   }
\label{fig:ETAVARY-TRYTWO-5}
\end{figure}
\begin{figure}[htp]
\centering
\includegraphics[scale=0.3,angle=0,trim=0 0 0 0]{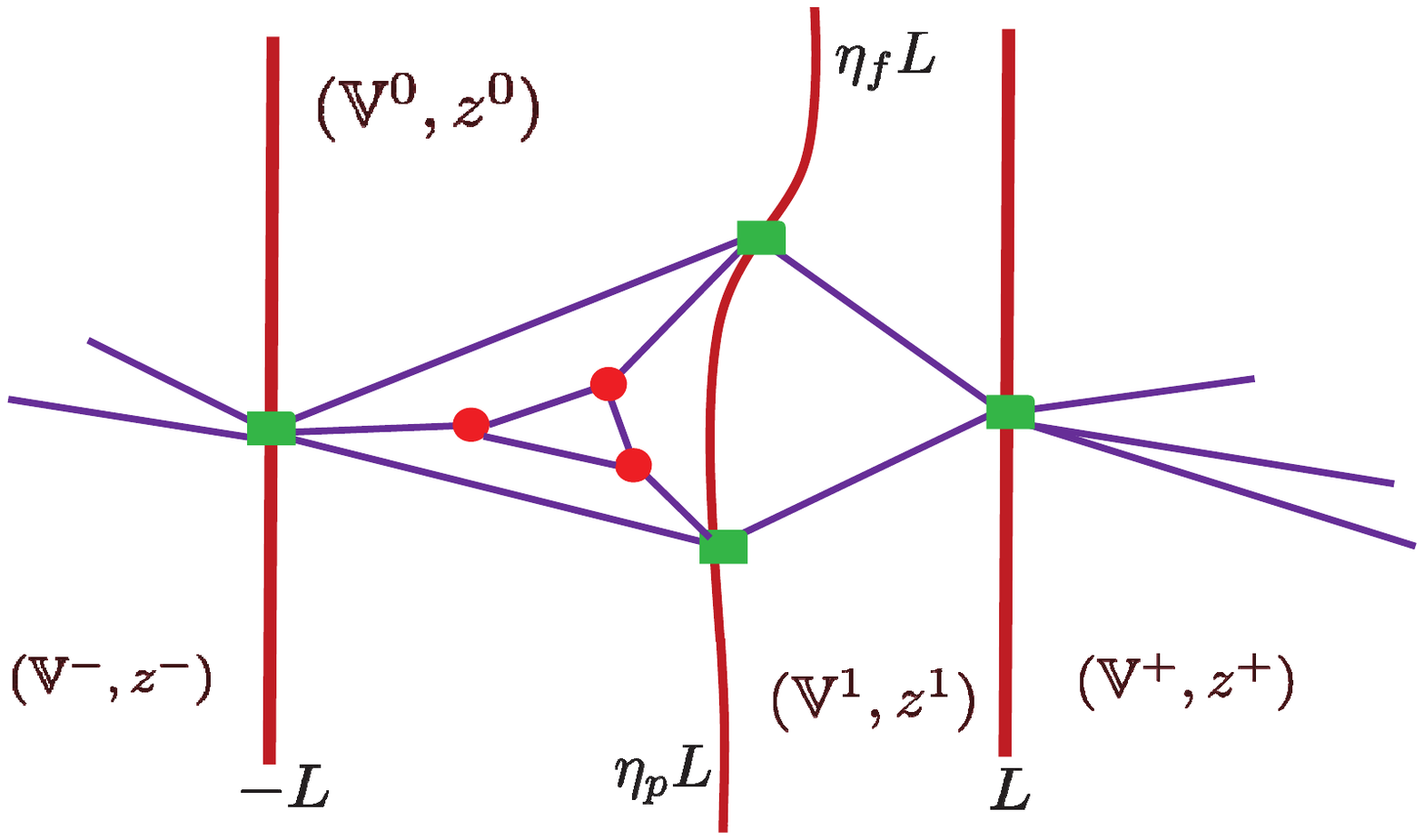}
\caption{The web shown here has expected dimension $d=0$ at fixed $s$ and $h$-dimension $d^h=1$.
If it exists it will project to an interval of the $s$-line. At the boundary of the interval
the inner triangle shrinks and the figure degenerates to an exceptional web which only exists
at a fixed value $s_*$. The nearby sliding $h$-types represent contributions to $\fe^h* \ft_{pl}$
in the convolution identity.     }
\label{fig:ETAVARY-TRYTWO-6}
\end{figure}
\begin{figure}[htp]
\centering
\includegraphics[scale=0.3,angle=0,trim=0 0 0 0]{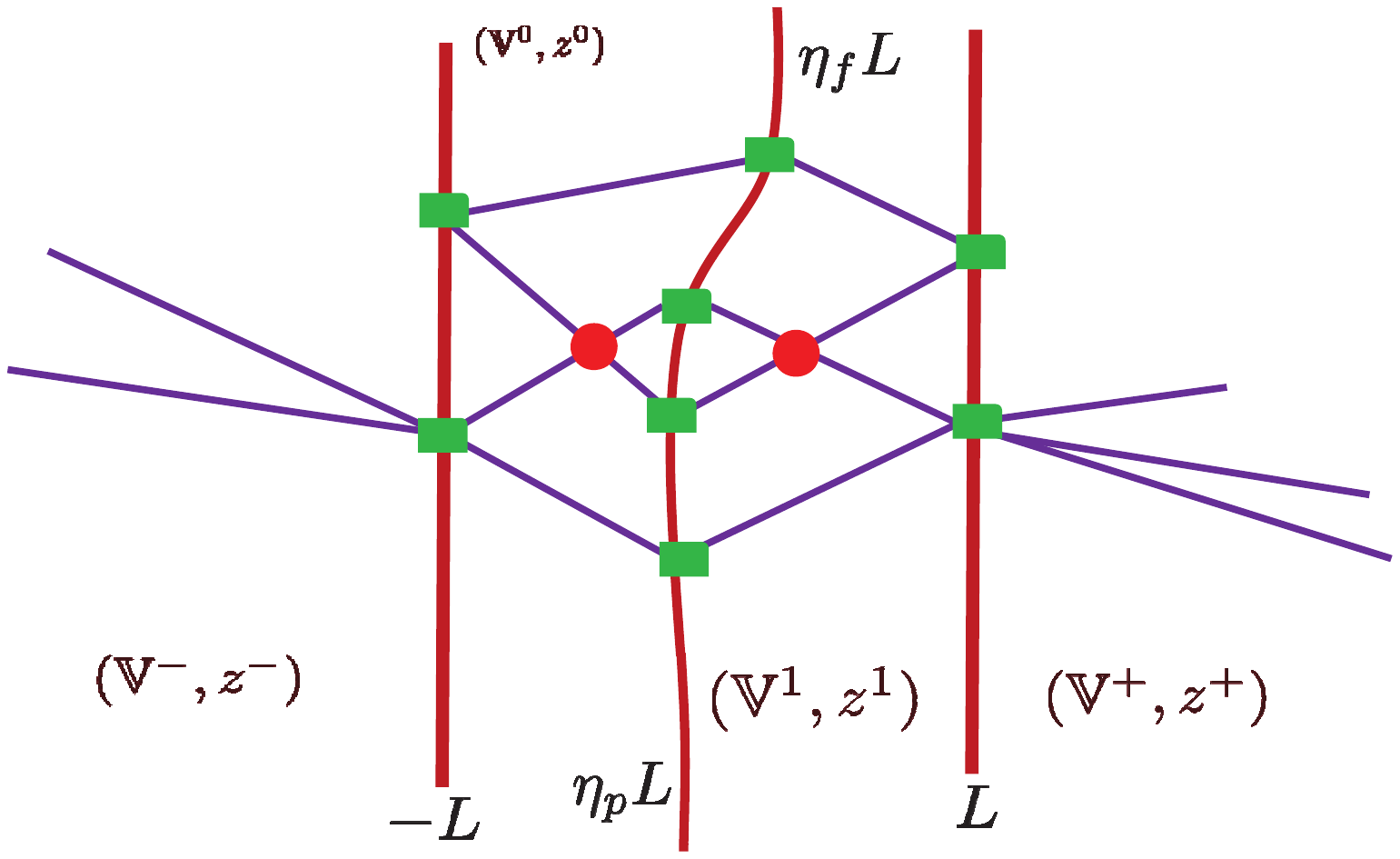}
\caption{The web shown here has expected dimension $d=0$ at fixed $s$ and $h$-dimension $d^h=1$.
As $s$ varies the two ``causal diamonds'' can merge and exchange places, leading to a component
of the moduli space of sliding webs which is a circle.     }
\label{fig:ETAVARY-TRYTWO-7}
\end{figure}
\begin{figure}[htp]
\centering
\includegraphics[scale=0.3,angle=0,trim=0 0 0 0]{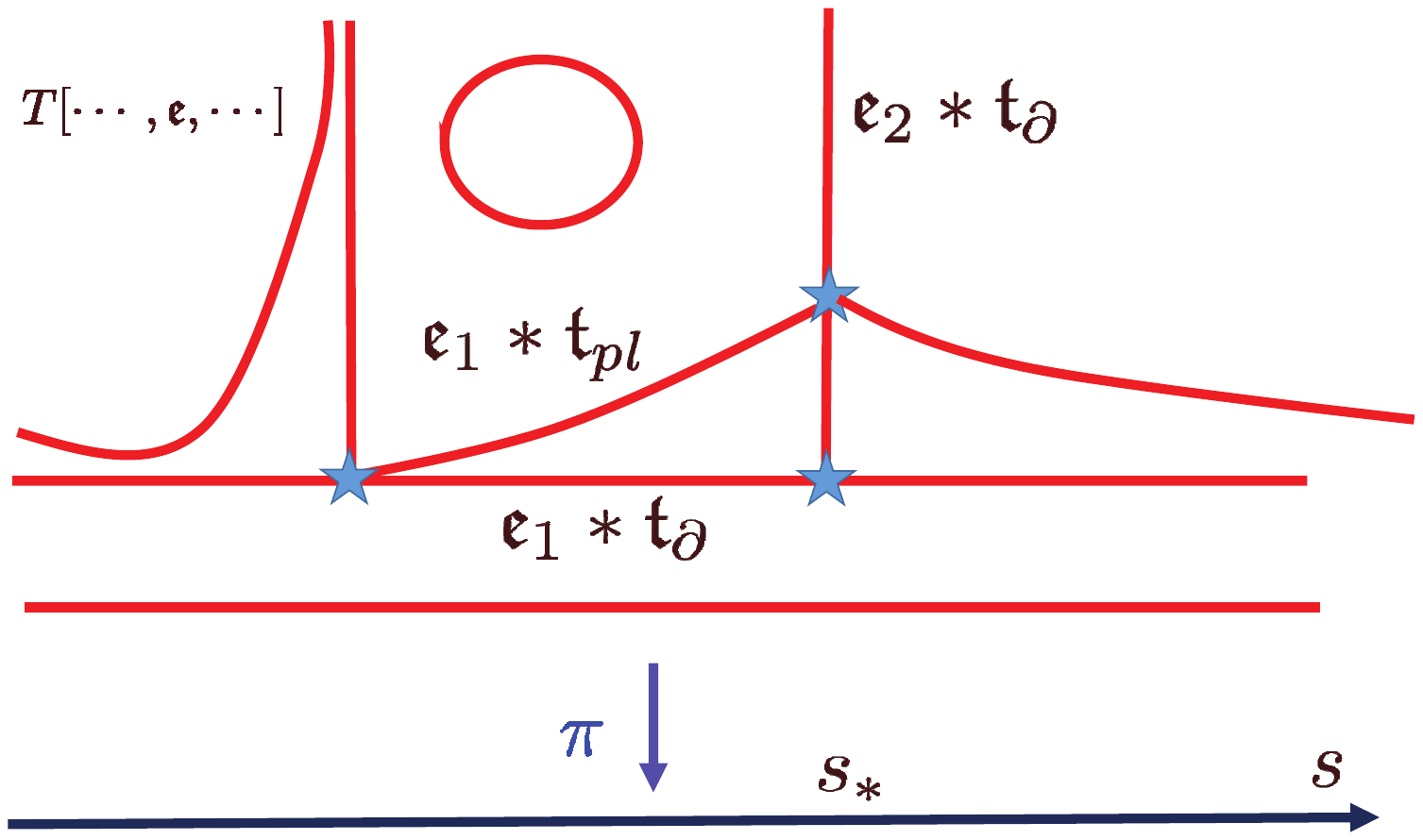}
\caption{A schematic picture of the set of sliding $h$-types. The horizontal axis at the bottom is
the $s$-line but no meaning is assigned to the vertical direction.   }
\label{fig:SLIDING-MODULISPACE}
\end{figure}

\begin{enumerate}

\item First, for $s\leq s_1$ or $s\geq s_2$ any of the summands in $\ft^h[\eta(y;s)]$ are
sliding. Recall that we assume $\eta(y;s)$ has nontrivial $y$-dependence, so the taut elements
are the same as the rigid elements. That is, they have $d=0$. For fixed $s$ the
moduli space of such deformation types is either empty or a finite set. (It could
be identified, for example, with the finite set of $y$ values of the vertex on the left boundary.)
On  the other hand, the image under
$\pi$ of such an $h$-type is a semi-infinite interval containing $(-\infty, s_1]$ or
$[s_2, +\infty)$. The boundaries at infinity of these webs contributes $\ft^h[\eta_1(y)] - \ft^h[\eta_2(y)] $ to the convolution
identity, where $\ft^h[\eta_1(y)]$, $\ft^h[\eta_2(y)]$ denote sums of  oriented $h$-types of webs.
 For a typical example, see Figure \ref{fig:ETAVARY-TRYTWO-1}.  We are interested in the
\emph{difference} $\ft^h[\eta_1(y)] - \ft^h[\eta_2(y)] $. It can be nonzero because some webs
can appear and disappear by shrinking to zero and then violating edge constraints, or by
blowing up to infinity.

\item It can happen that, at some special values of $s$, exceptional webs exist.
See, for example Figures \ref{fig:ETAVARY-TRYTWO-2}
and \ref{fig:ETAVARY-TRYTWO-3}.  Now at those special values
of $s$, $d(\fw)  =-1$ and $d^h(\fw)=0$. Such webs will exist only at
isolated values $s_*$. Such exceptional configurations can be convolved with taut elements
$\ft_p$ or $\ft_\p$ to produce $h$-types of $h$-dimension one. The case of a convolution
with $\ft_\p$ is   illustrated in Figure \ref{fig:ETAVARY-TRYTWO-5}. This $h$-type projects
under $\pi$ to a single point in the $s$-line. If we denote by $\fe^h$ the $h$-types
of all the exceptional webs of $h$-dimension $0$ then this represents a contribution
of $\fe^h*\ft_\p$ to the convolution identity.

\item  By contrast in Figure  \ref{fig:ETAVARY-TRYTWO-6}
we show an $h$-type of $h$-dimension $1$ which will project to an interval in the $s$-line.
The boundary of this $h$-type contributes to one of the terms
$\fe^h*\ft_{pl}$ in the convolution identity.

\item   Finally, there are boundary regions at infinity: a very large exceptional sliding web, which can be described
as usual by the tensor operation $T_\p(\ft^{-,+})[\frac{1}{1-\ft_c^f};\fe^h; \frac{1}{1-\ft_c^p}]$.

\item There can also be contributions to the moduli space of sliding $h$-types which are circles with
no boundary. This is illustrated in Figure     \ref{fig:ETAVARY-TRYTWO-7}. Altogether, the moduli space
of sliding $h$-types can be schematically pictured as shown in Figure  \ref{fig:SLIDING-MODULISPACE}.

\end{enumerate}

Putting these various boundaries together we can write the resulting convolution identity as
\footnote{We will not try to give a formal definition of a convolution $\CW^h \times \CW \to \CW^h$,
and so forth for all possible webs. }
\begin{equation}\label{eq:HomHomConv}
\ft^h[\eta_1(y)] - \ft^h[\eta_2(y)] +\fe^h* \ft_{pl} + \fe^h * \ft_\p +
 T_\p(\ft^{-,+})[\frac{1}{1-\ft_c^f};\fe^h; \frac{1}{1-\ft_c^p}] =0
\end{equation}
where, again,  $\fe^h$ includes the sum of oriented $h$-types of all possible exceptional webs encountered along the deformation.
%
%\cg{Where we choose to put the $ds$ in our definitions is related to how we orient the summands in $\fe^h$.}
%

If we apply a web representation to equation \eqref{eq:HomHomConv}, and again use identities of the form \eqref{eq:RepTdel}
to evaluate $\rho_\beta(T(\ft^{-,+}))$  then
\begin{equation}
\delta[\eta_1(y)] - \delta[\eta_2(y)] +M_1\left( \rho_\beta(\fe^h)[\frac{1}{1-\CB^{-,0}};\frac{1}{1-\CB^{0,1}};\frac{1}{1-\CB^{1,+}}]\right)=0
%\delta[\eta_1(y)] - \delta[\eta_2(y)] +M_1( \rho_\beta(\fe^h) )=0.
\end{equation}
This finally concludes our proof of equation \eqref{eq:hom-to-hom}.

\begin{figure}[htp]
\centering
\includegraphics[scale=0.3,angle=0,trim=0 0 0 0]{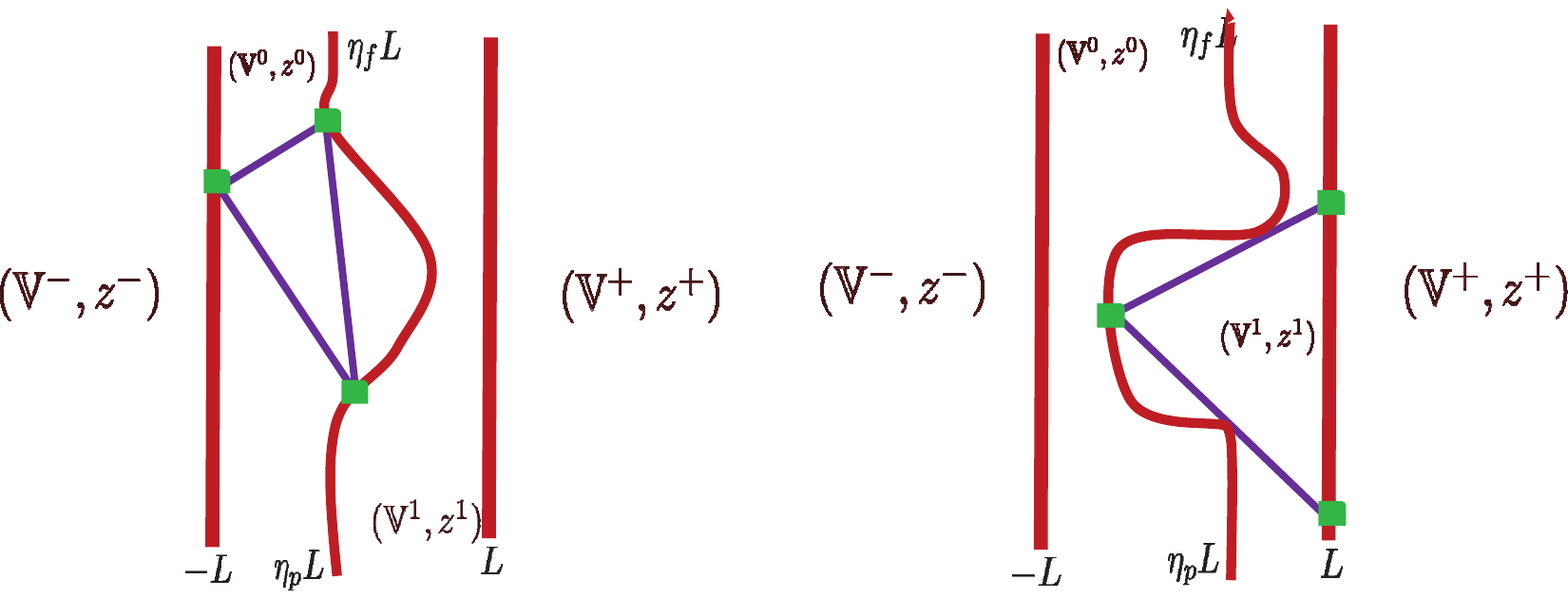}
\caption{If the homotopy $\eta(y;s)$ is not tame then configurations can occur which do not
fit into our convolution identity. On the left we have sliding webs. On the right we have an exceptional
web with $d^h=0$. It could be decorated to make a sliding $h$-type.     }
\label{fig:ETAVARY-EXCPL-2}
\end{figure}

\bigskip
\textbf{Remarks}

\begin{enumerate}

\item We stress that the assumption that we have a tame homotopy is crucial to our argument above.
Otherwise configurations such as those shown in Figure \ref{fig:ETAVARY-EXCPL-2} can occur which
do not fit into our convolution identity.

\item All these results are easily generalizable to the composition of any number of interfaces.

\item
Notice that we could have attempted the approach based on deformation type up to homotopy  in order to compare the triple
compositions for different values of $\eta$, by considering instead a variation
$\eta(s)$ which is not encoded as a time-dependent configuration, but as a family of
$G_3$ geometries. The problem with such an approach is that the corresponding convolution identity would look like
\begin{equation}
\fe' * \ft_p + \fe' * \ft_\p + \ft_c^f - \ft_c^p + T_\p(\ft^{-,+})[\frac{1}{1-\ft_c(s)};\fe'; \frac{1}{1-\ft_c(s)}] =0
\end{equation}
where the tensor operation involving some exceptional web in $\fe'$ picks the taut element $\ft_c(s)$ for the value of $s$ at which
the exceptional web exists. This convolution identity is clearly less useful than \ref{eq:fth}.
%In general, encoding the deformations of a parameter into a space-time dependence is a more effective strategy unless
%the deformed problem has no translation symmetry.
%
%\cg{I don't understand the last sentence of this paragraph. gm. Feb. 10, 2014}
%

\item  We have found that Theories and Interfaces produce a very interesting mathematical structure
which should, perhaps, be called an \afty-2-category. The objects (``zero-morphisms'') are Theories.
The space of (one-)morphisms between two Theories $\CT^-$ and $\CT^+$, is just the category
of Interfaces: $\Hom(\CT^-,\CT^+) = \fB\fr(\CT^-,\CT^+)$. The morphisms between two objects
of this category $\fI^{-,+}_1$ and $\fI^{-,+}_2$ are the two-morphisms. Rather than associativity
we have \afty-type axioms for the composition of morphisms.
%
%\cg{I didn't write out or check all the axioms. It could be quite a chore.}
%
%\cg{There is an extra layer of structure from the notion of homotopy equivalence.
%If we mod out by homotopy equivalence do we get a strict 2-category?}
%

\end{enumerate}

%We have accumulated a collection of definitions: we have Theories, we have an $A_\infty$ category of interfaces between
%each pair of theories, a composition functor which is associative up to an isomorphism. This is the data of a bicategory \cg{to be distinguished from %a 2-category where the
%composition is just associative, I think.}
%
%\cg{Give a reference where this term is defined/used?
%Otherwise, we should explain the difference from a 2-category.
%There are strict and non strict 2-categories. I wonder if you
%mean a nonstrict 2-category?.}
%

\subsection{Invertible Interfaces And Equivalences Of Theories}

The existence of the identity $\fId$ interface allows us to make an obvious definition: an interface $\fI\in \fB\fr(\CT^-,\CT^+)$ has
a right inverse $\tilde \fI\in \fB\fr(\CT^+,\CT^-)$
if $\fI\IntfcTimes \tilde \fI = \fId$. This is an extremely strong condition. For example it implies that the
Chan-Paton spaces satisfy
\be
\sum_{i' \in \IV^+} \CE_{ii'}^{-+} \otimes \tilde \CE_{i'j}^{+-} = \delta_{ij}\IZ
\ee

To give a nontrivial example of inverses, suppose
that we have isomorphisms $\varphi^{(12)}: \CT^{(1)}\to \CT^{(2)}$
and $\varphi^{(23)}: \CT^{(2)}\to \CT^{(3)}$. Then define
$\varphi^{(13)}:= \varphi^{(12)}\varphi^{(23)}$. We claim that
\be\label{eq:isom-intfc-grplw}
\fI^{\varphi^{(12)}} \IntfcTimes \fI^{\varphi^{(23)}} = \fI^{\varphi^{(13)}}
\ee
The key identity needed to establish this claim is that
\be
K^{(2)}_{i \varphi^{(12)}, j\varphi^{(12)}} \left( K^{-1,\varphi^{(12)} }_{ji}
\otimes K^{-1,\varphi^{(23)} }_{j\varphi^{(12)},i\varphi^{(12)}}\right) =
K^{-1,\varphi^{(13)} }_{ji}
\ee
With some patience this can be proven using \eqref{eq:Kinv-explct} and
\eqref{eq:Kpullback}.

We will need a more flexible notion of invertibility of Interfaces. It turns out
to be much more useful to define a  right-inverse up to homotopy if $\fI\IntfcTimes \tilde \fI$ is homotopy equivalent to the identity interface $\fId$.
 Similar definitions hold for left inverses. Because of associativity up to homotopy, a left inverse and a right inverse up to homotopy are equal up to homotopy.

\textbf{Definition}: We will refer to an interface which has right and left inverse up to homotopy as an {\it invertible interface}.

A good example of invertible Interfaces which are only invertible up to homotopy are the
rotation Interfaces $\fR[\vartheta_{\ell}, \vartheta_r]$ discussed at length in Section
\ref{sec:CatTransSmpl}.

The existence of an invertible Interface between two Theories $\CT^-$ and $\CT^+$ implies a strong relation between the Theories.
It defines a functor between the categories of Branes or Interfaces which is invertible up to natural transformations.
Concretely, it allows one to identify the spaces of exact ground states between branes of one theory and the other:
sandwiching the interface and its inverse between two branes and using associativity we get a quasi-isomorphism between the
complex of ground states for the two branes and the complex of ground states for their image in the other theory.
In Section \S \ref{sec:GeneralParameter} we will pursue this idea further to define a notion of equivalence of
Theories.

\section{Categorical Transport: Simple Examples}\label{sec:CatTransSmpl}

In the previous Section \S \ref{sec:Interfaces} we have constructed an \afty-2-category
of Interfaces. The arguments used in the construction employed (geometric) homotopies of the interface
geometries and their relation to (algebraic) homotopy equivalences  of Interfaces. A crucial result
was that the composition of consecutive Interfaces along the $x$-axis is associative up to homotopy equivalence.
We can therefore consider compositions of several Interfaces, closely spaced along the
$x$-axis. The resulting composition is well-defined up to homotopy equivalence.
This suggests consideration of a third kind of homotopy, namely homotopies
of the data used to define Theories and Interfaces. Thus, without trying to make the
notion too precise at the moment, we can imagine continuous
 families of vacuum weights $z_i$,   contractions $K_{ij}$,   interior amplitudes $\beta$
and so forth. Moreover, we could generalize the set of vacua $\IV$ to be
a discrete (possibly branched) cover over some space of parameters.
In the same spirit we could generalize web representations $R_{ij}$ to bundles of $\IZ$-modules, etc.
Let us denote the relevant parameter spaces for these data generically as $\CC$.
The results of Section \S \ref{sec:Interfaces} thus suggest that given a continuous path
$\wp$ from   some interval $[s_\ell, s_{r}] \subset \IR$ to $\CC$ there might be a way to ``map''
the Theory $\CT^\ell$ at $s_\ell$ to the Theory $\CT^r$ at $s_r$ so that the
Branes for the positive half-plane   with data $\wp(s_{\ell})$ are ``coherently mapped'' to Branes
for the positive half-plane   with data $\wp(s_r)$.

To be slightly more precise, for a continuous map $\wp$ as above, suppose we have a
definite law for constructing $\CT^r$ given $\CT^\ell$. Then we are aiming to define an \afty-functor
\be
\CF(\wp): \fB\fr(\CT^\ell, \CH ) \rightarrow \fB\fr(\CT^r,\CH)
\ee
where $\CH$ is, say, the positive half-plane.
The functor $\CF(\wp)$ is meant to be a categorical version of parallel transport.
Thus, it should be an \afty-equivalence of categories, and moreover, if $\wp_1$
and $\wp_2$ are two paths which can be composed then there
should be an invertible natural transformation
between  $\CF(\wp_1 \circ \wp_2)$ and the composition of the \afty-functors
$\CF(\wp_1)$ and $\CF(\wp_2)$. Of course, if $\wp$ is the constant path then
$\CF(\wp)$ should be the identity functor. We will refer to such a family of
functors as a \emph{categorical parallel transport law}, or just \emph{categorical transport}
for brevity.   We aim to show, furthermore,
that the ``connection'' defining this transport law is in fact a flat connection.
That is, if $\wp_1$ is homotopic to $\wp_2$ in an appropriate space of
parameters $\CC$ then there is an invertible natural transformation between
$\CF(\wp_1)$ and $\CF(\wp_2)$.

 Given the above motivation our general strategy
for constructing the functors $\CF(\wp)$ will be to regard the variation $\wp(s)$ as
a \emph{spatially dependent variation of parameters} allowing us to construct families of
Interfaces $\fI[\wp]$ which satisfy homotopy properties analogous to $\CF(\wp)$.
In particular, we require two key properties:

\begin{enumerate}

\item \emph{Parallel transport}: If $\wp_1$ and $\wp_2$ are composable paths then
there must be a homotopy equivalence of Interfaces:
\be \label{eq:PT-Intf1}
\fI[\wp_1]\IntfcTimes \fI[\wp_2] \sim \fI[\wp_1 \circ \wp_2].
\ee

\item \emph{Flatness}: If $\wp_1 \sim \wp_2$ are homotopy equivalent, in
a suitable sense, then correspondingly there must be a homotopy equivalence
of Interfaces:
\be \label{eq:PT-Intf2}
\fI[\wp_1] \sim  \fI[ \wp_2].
\ee

\end{enumerate}

Given a family of Interfaces satisfying \eqref{eq:PT-Intf1} and \eqref{eq:PT-Intf2}
we can then   invoke the construction of equations \eqref{eq:Intfc-Functor}, \eqref{eq:Ifc-F2},
and \eqref{eq:Ifc-F3} to produce the corresponding flat parallel transport \afty-functors.

In this Section and in Section \S \ref{sec:GeneralParameter} we will make some of these general
ideas much more precise for certain variations of the data used to define Theories and Interfaces.
For further discussion see the introduction to Section \S \ref{sec:GeneralParameter}.

\subsection{Curved Webs And Vacuum Homotopy}\label{subsec:CurvedWebs}

One of the most important physical examples of  the variations of parameters  described
above are variations of vacuum weights.  Thus we consider
paths of weights $\wp: \IR \to \IC^{  \IV  } - \Delta$, where $\Delta$ is the large
diagonal. If $\wp$ is continuous and nonconstant only within some finite interval
we will call it a \emph{vacuum homotopy}.
Such a collection of maps $z_i: \IR \to \IC$ for $i\in \IV$ can be used to define a set of
spatially-dependent vacuum weights $z_i(x)$. If the maps are continuously differentiable
and suitably generic then in the context of Landau-Ginzburg
theories such collections of maps can be used to define interesting supersymmetric interfaces.
This is described in more detail in Section \S \ref{subsec:LG-Susy-Interface} below.

The notion of webs for spatially-dependent
weights $z_i(x)$  still makes sense: They are again graphs in $\IR^2$ but now the edges are allowed
to be smooth non-self-intersecting curves. The connected components of the complement of the graph
are labeled by vacua and edges separate regions labeled by pairs of distinct vacua. If $z_i(x)$ are
all continuously differentiable in the neighborhood of a point $x_0$ then we can orient the
$ij$ edges in a strip centered on $x_0$ and the tangent
vector to an edge at a point $(x,y)$ in this strip, with $i$ on the left and $j$ on the right,
is parallel to $z_{ij}(x):= z_i(x)-z_j(x)$. Such webs are called \emph{curved webs}.
\footnote{We will assume that $z_i(x)$ are sufficiently generic that edges have
transverse intersections, except at special points called ``binding points,''
defined below.}
In fact, when discussing homotopies of paths it is useful to generalize still further
and consider spacetime dependent weights $z_i(x,y)$ with the natural generalized
definition of curved webs. We will see many examples of such curved webs below.
See, for examples, the figures in Section \S \ref{subsec:BindPoints}.

In the remainder of Section \ref{sec:CatTransSmpl}
we will discuss a very simple class of curved webs such that
\be\label{eq:SpinningWeights}
z_i(x) = e^{-\I \vartheta(x)} z_i
\ee
where $\vartheta: \IR \to \IR$ is a smooth function with compact support for $\vartheta'$.
We will call these \emph{spinning weights} because they are all related by a uniform
(albeit $x$-dependent) rotation. If we wish to distinguish their webs from general
curved webs we call them \emph{spinning webs}. We will already find quite a rich set of
phenomena in this case.

One advantage of the spinning webs is that we can consider the interior amplitude
$\beta_I$ to be $x$-independent.  In a general composite web with several
interfaces the interior amplitude can vary discontinuously across the interfaces.
However, the particular family of ``spinning'' vacuum weights \eqref{eq:SpinningWeights}
have the property that the set of cyclic fans $I$ does not change with $x$. Therefore
there is  a natural choice of interior amplitude where we take $\beta_I$ to be
$x$-independent. In the remainder of Section \S \ref{sec:CatTransSmpl} we make that choice.
In Section \S \ref{sec:GeneralParameter} we consider more general situations.

There is another very nice way to motivate the study of spinning vacuum weights
and describe that in the next two subsections.

\subsection{Rotation Interfaces }\label{subsec:RotInts}
%
%\cg{The word "mutations" does not appear to be used anywhere below!}
%

Given a choice of
 a Theory $\CT$ and a half-plane $\CH$, we have defined a category of Branes $\fB\fr[\CT, \CH]$. It is natural to wonder how
 this category depends on the choice of $\CH$ for fixed $\CT$. Certainly it is literally unchanged
  if we apply a translation to $\CH$. On the other hand, there is an interesting change
  if we apply a rotation to $\CH$. It is useful to denote the phase of the normal
 to the boundary pointing into $\CH$ as $\vartheta$ and the corresponding half-plane (defined up to translation)
as $\CH_\vartheta$. The corresponding  category of Branes will be denoted $\fB\fr_\vartheta$.
The notion of half-plane webs is ill-defined for those special values of $\vartheta$   such that the boundary of the half-plane aligns with
 $z_{ij}$ for some $i,j\in \IV$. We define an \emph{ $S_{ij}$-ray} to be the ray in the complex plane
\footnote{The terminology is motivated by the relation to the theory of spectral networks
\cite{Gaiotto:2011tf,Gaiotto:2012rg,Gaiotto:2012db}. The relation to
spectral networks is discussed in more detail in Section \S \ref{subsec:CatSpecNet} below. }
 through the angle $e^{\I\vartheta_{ij}}$ such that the canonically
  oriented boundary has direction $-z_{ij}$. In formulae, the $S_{ij}$-ray is the ray through $e^{\I\vartheta_{ij}}$ such that
\be\label{eq:Sij-Ray-def}
\Re\left( e^{-\I \vartheta_{ij}} z_{ij} \right) = 0  \qquad \& \qquad
\Im\left( e^{-\I \vartheta_{ij}} z_{ij} \right) > 0 .
\ee
Thus we have a well-defined notion of $\fB\fr_\vartheta$ when $e^{\I \vartheta}$ is in the complement of the union,
over all pairs $(i,j)$ of distinct vacua, of the $S_{ij}$-rays.
%
%\cg{Important definition! Check signs carefully.}
%

We would like to define an \afty-equivalence of the categories of Branes $\fB\fr_\vartheta$
for different values of $\vartheta$.
%
%\cg{Actually, the linear interpolation gives a \emph{canonical} functor between the
% categories. The more general continuous interpolations from $\vartheta_\ell$ to $\vartheta_r$
% give an identification up to \afty-equivalence.}
% %
The Branes associated to the $\CH_\vartheta$ half-plane for some Theory $\CT$ with vacuum data $(\IV,z)$
can be re-interpreted as Branes associated to the positive half-plane for a Theory $\CT^\vartheta$.
To define  $\CT^\vartheta$ we choose the same set of vacua $\IV$, but now the weights are rotated:
\begin{equation}z_j^\vartheta := e^{- \I \vartheta} z_j. \end{equation}
The interior amplitude $\beta$ and the web representation $\CR$ can be taken to be the same.
In other words, $\fB\fr_\vartheta[\CT] = \fB\fr_0[\CT^\vartheta]$ and we can focus on relations between Branes for rotated Theories
in the positive half plane. The \afty-equivalences we seek will be given by the functors associated (via equations
\eqref{eq:Intfc-Functor} et. seq.) to a family of invertible Interfaces $\fR[\vartheta_\ell, \vartheta_r]$
between any pair of rotated Theories $\CT^{\vartheta_\ell}$ and $\CT^{\vartheta_r}$. Here we must regard
$\vartheta_{\ell}$ and $\vartheta_r$ as belonging to the real numbers. Although the
Theories $\CT^{\vartheta}$ only depend on $\vartheta ~\mod 2\pi$ there are interesting
monodromy phenomena associated with interpolations that have $\vert\vartheta_{\ell} - \vartheta_r \vert \geq 2\pi$.
(See Section \S \ref{subsubsec:Monodromy} below.)

The definition of $\fR[\vartheta_\ell, \vartheta_r]$
appears in equation \eqref{eq:RigRotInt-Def} at the end of Section \S  \ref{subsubsec:Monodromy}
and uses the key definition \eqref{eq:defIth} below.
We now give some motivation for the somewhat elaborate definitions which follow.

%As individual branes will not necessarily
%come back to themselves as $\vartheta \to \vartheta+\pi$, we take $\vartheta$ to be valued in $\IR$ from now on. The definition
%of the interface can be found in equation  It is only defined up to homotopy equivalence.
%
%\cg{Again, should distinguish between linear and general interpolation.}
%

As discussed above, we want our Interfaces to behave as a categorical version of parallel transport:
the composition $\fR[\vartheta_1, \vartheta_2] \IntfcTimes \fR[\vartheta_2, \vartheta_3]$ should be homotopy equivalent
to $\fR[\vartheta_1, \vartheta_3]$, and   $\fR[\vartheta, \vartheta]$ should be the identity interface $\fId$.
%
%\cg{Refer here to categorification of parallel transport
%defined by spectral networks.}
%
Once the problem is approached from the point of view of the composition of interfaces, the variation of $\vartheta$ becomes naturally tied to the space direction.
In order to define an interface $\fR[\vartheta_\ell, \vartheta_r]$ we could then hope that there is a simple
definition when $\vartheta_{\ell}$ and $\vartheta_r$ are infinitesimally close and then
imagine subdividing the interval $[\vartheta_\ell, \vartheta_r]$ into
a very large number of small sub-intervals $[\vartheta_k, \vartheta_{k+1}]$, and use the definition
 of the infinitesimal interfaces. This line of reasoning naturally leads to the idea that
 given any   continuous interpolation $\vartheta(x)$ with $\vartheta(x) = \vartheta_\ell$ for $x<-L$
and $\vartheta(x) = \vartheta_r$ for $x>L$ there is a corresponding Interface $\fI[\vartheta(x)]$.
In Section \S \ref{subsubsec:DefiningAmplitudes} we will indeed define such an interface. (See  equation \eqref{eq:defIth} below.)
We will need to define Chan-Paton factors and amplitudes.
The main tool is to use representations of spinning webs.  We will also use curved webs with space-time dependent weights
to show that homotopy equivalent interpolations will give us homotopy equivalent interfaces.

Before giving the complete set of rules to deal with such curved webs, we gain some
intuition by looking at a special case which can be reduced to standard webs with straight edges.
Choosing real lifts so that
$\vartheta_\ell>\vartheta_r$ and $\vert \vartheta_\ell - \vartheta_r\vert \leq \pi $  we can consider
the smooth interpolation
 $\vartheta(x) = -x$ (where the sign is chosen for future convenience) defined on the interval
 $-\vartheta_\ell \leq  x \leq -\vartheta_r$. The advantage of this is that if
  we apply the exponential map $u+ \I v := e^{-\I x + y}$ taking the strip
  $-\vartheta_\ell \leq  x \leq -\vartheta_r$ in the $x + \I y$ plane
  to a wedge, denoted $\CH[\vartheta_{\ell}, \vartheta_r]$, in the $u + \I v$ plane,  then curves $x(s)+\I y(s)$ with tangent
  $ \frac{d}{ds}(x(s)+\I y(s))= e^{\I x(s) } z_{ij}$
  in the $x + \I y$ plane are mapped to curves $u(s)+\I v(s)$ satisfying
  $ \frac{d}{ds}(u(s)+\I v(s))=-\I   z_{ij}$. These will be straight rays (or line segments) parallel
  to $-\I z_{ij}$ inside   the wedge $\CH[\vartheta_{\ell}, \vartheta_r]$.   Therefore, a curved  web
 on the strip will map to a web  with straight edges in the wedge $\CH[\vartheta_{\ell}, \vartheta_r]$.

Now, given a left Brane $\fB$ for $\CT^{\vartheta_\ell}$ and a right Brane $\tilde \fB$ for $\CT^{\vartheta_r}$,
we can define a complex of approximate ground states by sandwiching an $\fR[\vartheta_\ell, \vartheta_r]$ Interface
between the two Branes, i.e. by looking at the composition
$(\fB \fR[\vartheta_\ell, \vartheta_r] \tilde \fB)_{\eta}$ computed with a $G_3$ geometry as in Section \S \ref{subsec:CompThrIntfc}.
Recall that an Interface between a trivial Theory and itself is just a chain complex. We could equally well
consider the chain complex of approximate groundstates  (recall the definition from Section \S \ref{subsec:WebRepStrip})
for the strip with $\fB\IntfcTimes\fR[\vartheta_\ell, \vartheta_r]$ on the left and $\tilde\fB$
on the right or that with $\fB$ on the left and $ \fR[\vartheta_\ell, \vartheta_r]\IntfcTimes \tilde\fB$ on the right.
All of these chain complexes are homotopy equivalent. Our main heuristic is that these chain complexes should
also be computed by a natural generalization of the complex of approximate groundstates associated to webs in
the wedge geometry. In the next Section \S \ref{subsec:WedgeWebs} we will describe that generalization, and
then the requirement that there be a  homotopy equivalence to the chain complexes $(\fB \fR[\vartheta_\ell, \vartheta_r] \tilde \fB)_{\eta}$
will help us figure out how to compute the Chan-Paton factors for $\fR[\vartheta_\ell, \vartheta_r]$.

\subsection{Wedge Webs}\label{subsec:WedgeWebs}

A wedge geometry consists of the conical region $\CH[\vartheta_\ell,\vartheta_r]$ of $\IR^2$ included between the two rays $\vartheta_\ell$ and $\vartheta_r$, $\vartheta_\ell>\vartheta_r$,
clockwise from $\vartheta_\ell$. A wedge web is a web with vertices which may lie in the interior, on the two boundary rays or at the origin of the wedge.
The interior vertices of a wedge web are associated to standard interior fans and the left and right boundary vertices to half-plane fans for $\CH_{\vartheta_\ell}$
and $\CH_{\pi + \vartheta_r}$ respectively. The possible vertex at the origin is associated to a wedge fan, a sequence of vacua compatible with edges lying in the wedge.
The same type of wedge fan labels the external edges of the web, i.e. the fan at infinity. It is convenient to include the trivial wedge fan, with a single vacuum and no edges.

We can define as usual deformation types of wedge webs $\fv$, moduli spaces of deformations, orientations, etc.
We can define the convolution and tensor operations taking a wedge web as a container and inserting appropriate plane or half-plane webs
or another wedge web at the vertex at the origin, as long as the fans match.

The wedge geometry has a scaling symmetry and thus a taut wedge web has a single modulus associated to the scale transformations,
oriented towards larger webs. The wedge taut element $\ft_w$ satisfies the usual type of convolution identity
\begin{equation}\label{eq:WedgeConv}
\ft_w * \ft_w + \ft_w * \ft_{pl} + \ft_w * \ft_{\p,\ell} + \ft_w * \ft_{\p,r} =0
\end{equation}
In the first term we are convolving summands from $\ft_w$ into summands from $\ft_w$ at the
origin of the wedge.

Given a choice of Theory $\CT$ and of Chan Paton factors $\CE_i$, $\tilde \CE_i$ at the two boundary rays, we can define in a standard way a representation of wedge webs. We can associate to a wedge fan $\{i_1, \dots, i_n\}$ a vector space $\CE_{i_1} \otimes R_{i_1, i_2}\otimes \cdots R_{i_{n-1},i_n}\otimes \tilde \CE_{i_n}^*$ and collect all these vector spaces into a single
\begin{equation}
R_{w}^{\vartheta_\ell,\vartheta_r} [\CE,\tilde \CE] =  \oplus_{z_{ij}\in \CH[\vartheta_\ell,\vartheta_r]} \CE_{i}\otimes \widehat{R}_{ij}^{\vartheta_\ell,\vartheta_r} \otimes \tilde \CE_j^*
\end{equation}
with
\be \label{eq:KSprod}
\IZ\cdot \textbf{1}  + \oplus_{z_{ij}\in \CH[\vartheta_\ell,\vartheta_r]} \widehat{R}_{ij}^{\vartheta_\ell,\vartheta_r} e_{ij} := \bigotimes_{z_{ij}\in \CH[\vartheta_\ell,\vartheta_r]} (\IZ\cdot \textbf{1} + R_{ij} e_{ij} )
\ee

As usual, given a wedge web $\fv$ we have an associated multilinear map:
\begin{equation}
\rho_\beta(\fv): TR^{\p, \ell}[\CE] \otimes R_{w}^{\vartheta_\ell,\vartheta_r} [\CE,\tilde \CE] \otimes TR^{\p, r}[\tilde\CE]  \to R_{w}^{\vartheta_\ell,\vartheta_r} [\CE,\tilde \CE]
\end{equation}

The convolution identity tells us that $\rho_\beta(\ft_w)$ defines an $A_\infty$ bimodule, satisfying the same type of $A_\infty$ axioms as
the strip-web operation $\rho_\beta(\ft_s)$ did, but with a larger vector space $R_{w}^{\vartheta_\ell,\vartheta_r} [\CE,\tilde \CE]$ instead of the
$\CE_{LR}$ we found for the strip. In particular, for any pair of Branes $\fB$ and $\tilde \fB$ in the $\fB\fr_{\vartheta_\ell}$ and $\fB\fr_{\vartheta_r+\pi}$ categories
we have a chain complex $R_{w}^{\vartheta_\ell,\vartheta_r} [\CE,\tilde \CE]$ with differential
\begin{equation}
g \to \rho_\beta(\ft_w)[\frac{1}{1-\CB},g, \frac{1}{1-\tilde \CB}].
\end{equation}
This complex should be thought of as describing local operators placed at the tip of the wedge.

If we want the chain complex $R_{w}^{\vartheta_\ell,\vartheta_r} [\CE,\tilde \CE]$
to be homotopy equivalent to $\left( \CB \fR[\vartheta_\ell, \vartheta_r] \tilde \CB\right)_\eta$,
as suggested by the exponential map,
we are led to the idea that the Chan Paton factors of the interface $\fR[\vartheta_\ell, \vartheta_r]$ should coincide with the
$\widehat{R}_{ij}^{\vartheta_\ell,\vartheta_r}$.
The exponential map relates the straight edges of a wedge web which go to the origin or to infinity in the wedge geometry
to external edges of the curvilinear web which go to infinity in the Euclidean time direction $y$, either sitting in the
 far past or far future at values of $x$ such that $e^{\I \vartheta_{ij}(x)}$ lies on an $S_{ij}$ ray for the
 Theory $\CT$.

This suggests that the Chan Paton factors for the $\fR[\vartheta_\ell, \vartheta_r]$ interface should be built
from individual factors of $R_{ij}$ associated to such ``vertical'' external edges. With this hint
we   are now ready to propose the full set of rules for spinning webs.

\subsection{Construction Of Interfaces For Spinning Vacuum Weights}

Let us now return to the general case of spinning vacuum weights
of the form \eqref{eq:SpinningWeights}, determined by a generic, smooth
function $\vartheta: \IR \to \IR$ so that $\vartheta'$ has compact support.

In what follows we choose some $L$ so that the support of $\vartheta'$ is
in $(-L, L)$. We set $\vartheta(x)=\vartheta_{\ell}$ for $x\leq -L$
and $\vartheta(x) = \vartheta_r$ for $x \geq L$.
Moreover, we assume that none of the complex numbers $e^{-\I\vartheta_{\ell}}z_{ij}$,
$e^{-\I\vartheta_{r}}z_{ij}$ is pure imaginary for $i\not=j$.
Our goal is to define an Interface
\be
\fI[\vartheta(x)] \in \fB\fr(\CT^{\vartheta_{\ell}}, \CT^{\vartheta_{r}})
\ee
with the flat parallel transport properties spelled out in equations
\eqref{eq:PT-Intf1} and \eqref{eq:PT-Intf2}.
To implement \eqref{eq:PT-Intf2} we define $\vartheta^1(x) \sim \vartheta^2(x)$ to be homotopic
if the functions $e^{-\I \vartheta^a(x)}$, $a=1,2$, define homotopic maps from the real line into the circle.

\begin{figure}[htp]
\centering
\includegraphics[scale=0.25,angle=0,trim=0 0 0 0]{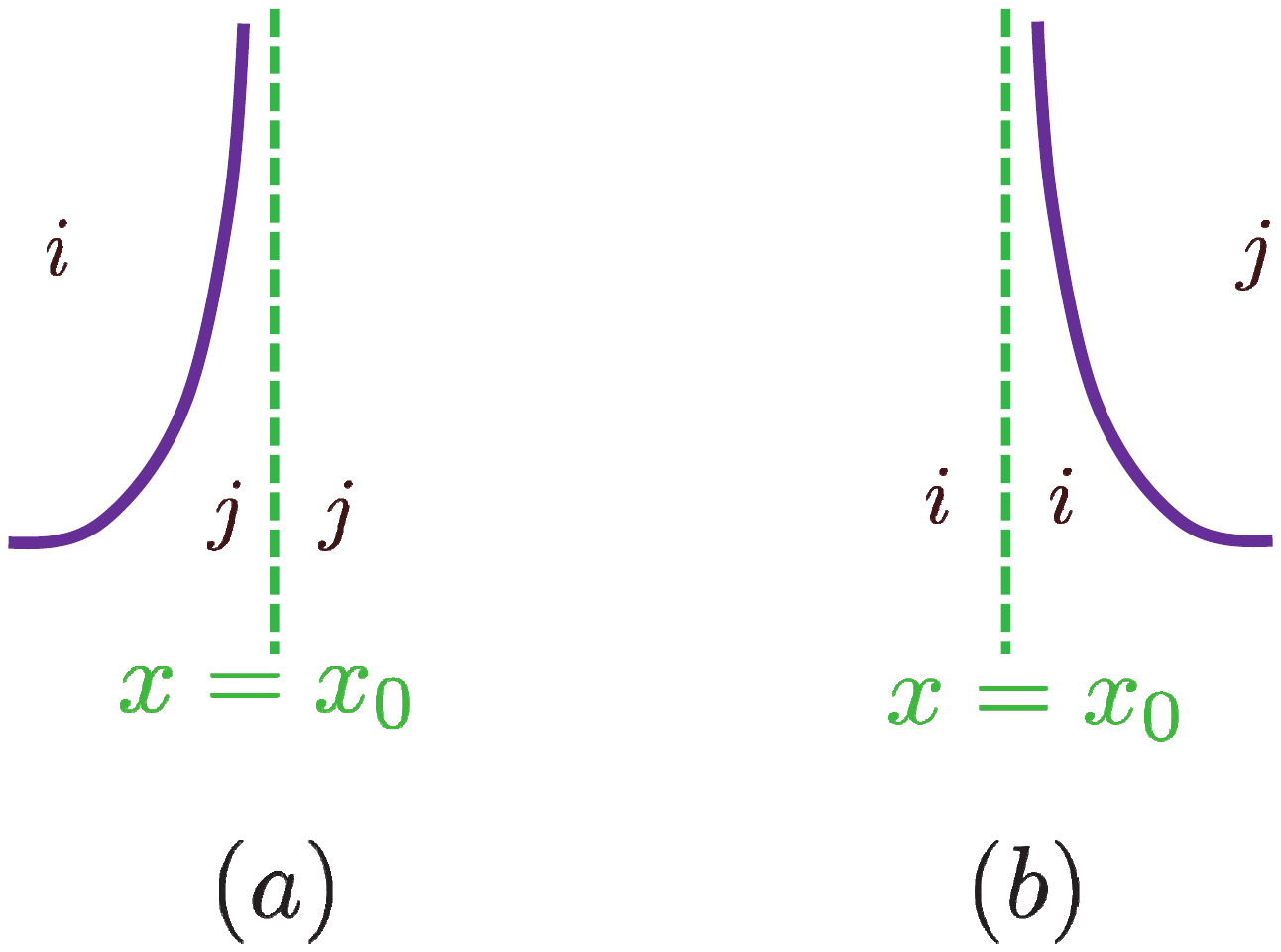}
\caption{Near a future stable binding point $x_0$ of type $ij$ the edges separating
 vacuum $i$ from $j$ asymptote to the dashed green line $x=x_0$ in the future. Figures (a) and (b)  show
 two possible behaviors of such lines. The phase $e^{-\I \vartheta(x)} z_{ij}$ rotates through the
 positive imaginary axis in the counterclockwise direction.}
\label{fig:FUTURESTABLE-1}
\end{figure}
\begin{figure}[htp]
\centering
\includegraphics[scale=0.25,angle=0,trim=0 0 0 0]{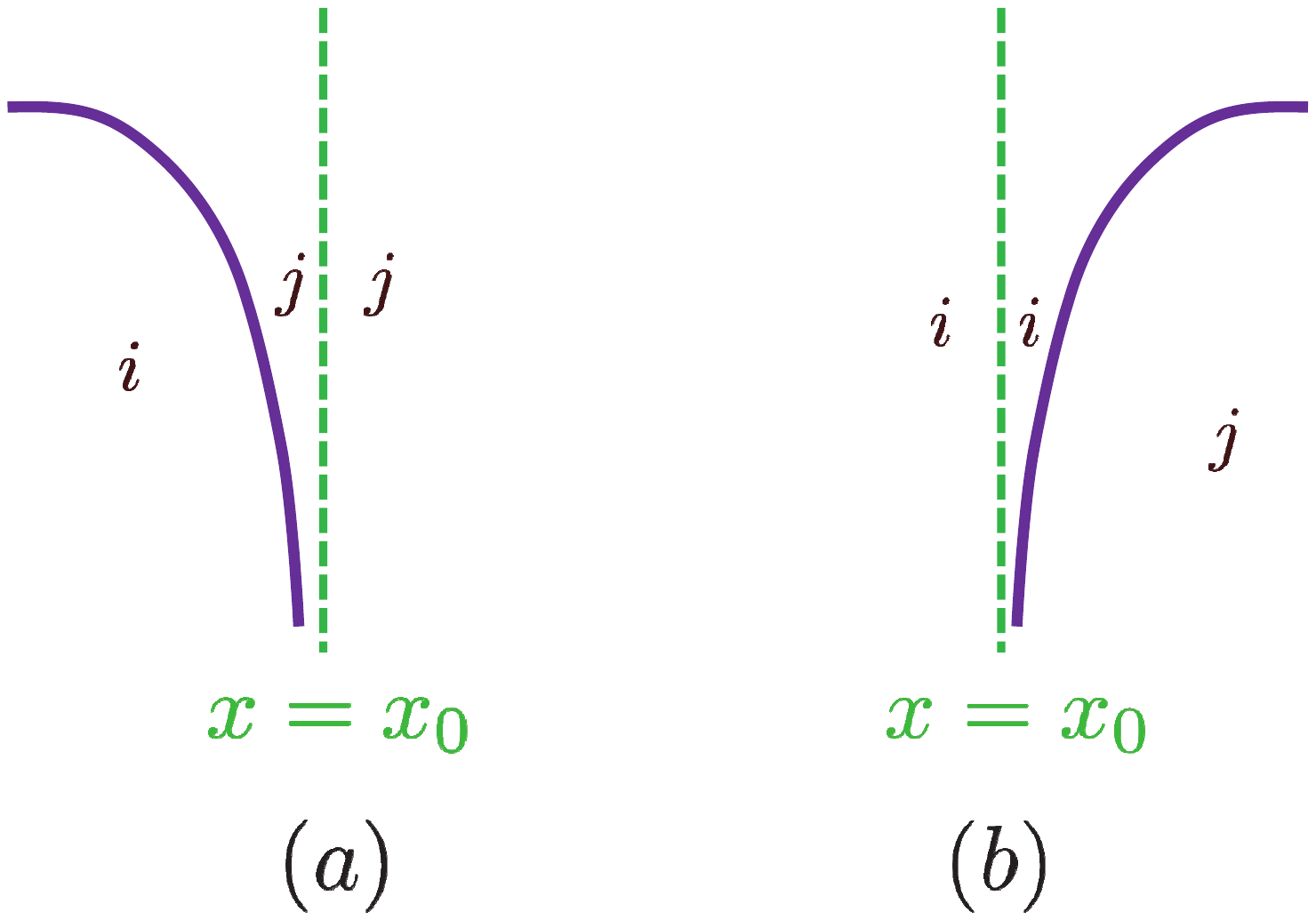}
\caption{Near a past stable binding point $x_0$ of type $ij$ the edges separating
 vacuum $i$ from $j$ asymptote to the dashed green line $x=x_0$ in the past. Figures (a) and (b)  show
 two possible behaviors of such lines. The phase $e^{-\I \vartheta(x)} z_{ij}$ rotates through the
 positive imaginary axis in the clockwise direction.}
\label{fig:PASTSTABLE-1}
\end{figure}

\subsubsection{Past And Future Stable Binding Points  }\label{subsec:BindPoints}

Let us first describe some special properties of the spinning webs.
They share some characteristics of plane webs,  of interface
webs for data $(\IV^-,z^-) = (\IV, z^{\vartheta_\ell})$ and $(\IV^+,z^+) = (\IV, z^{\vartheta_r})$,
and also of composite and strip webs. In general, curved webs will be denoted by $\fz$.
All of the vertices $v$ in $\fz$  will be equipped with a cyclic fan of vacua $I_v(\fz)$
drawn from $\IV$, just as for plane webs. Nevertheless,  there are
two very different kinds of external edges.
To see this let us consider what the external edges
of the web might look like. The external edges with support
in the regions $x\geq L$ or $x\leq  -L$ extend to $x \to + \infty$ or $x\to -\infty$ and define
positive and negative half-plane fans $J^+$ and $J^-$, respectively. On the other hand,
a novel aspect of curved webs is that there can also be vertical external
edges supported in the region $-L < x < L$. We now
describe this important phenomenon in some detail.

%In a web, edges separating vacua $i$ and $j$ are unoriented.
%
%\footnote{As noted before, external edges carry a natural orientation
%pointing to infinity.}
%
%However in the following discussion
%it will be convenient to orient them such that, if the tangent direction
%has a positive $x$-component, then the vacuum $i$ is on the left and $j$ is on
%the right. We will say such an edge is \emph{of type $ij$}.
%
%
%

At any fixed
$x$ we can apply the line principle to a vertical line through $x$, and hence
the edges will be graphs of functions over certain intervals. We define an
``edge of type $ij$'' to be an edge that separates vacua $i$ and $j$.
(There is no distinction between an edge of type $ij$ and type $ji$.)
Locally, there is a parametrization of the edge with tangent vector
oriented so that vacuum $i$ is on the left and $j$ is on the right.
With this parametrization
$\frac{d}{ds}(x_{ij}(s) + \I y_{ij}(s)) = e^{-\I \vartheta(x)} z_{ij}$
and hence locally the edge is the graph of a function $y_{ij}(x)$ satisfying
\be
\frac{dy_{ij}}{dx} = \tan(\alpha_{ij}-\vartheta(x))
\ee
where the phase $\alpha_{ij}$ is defined by
 $z_{ij} := \vert z_{ij} \vert e^{\I \alpha_{ij}}$. Note that
 $\alpha_{ji} = \alpha_{ij} + \pi$. This differential equation is
singular at those values of $x$ for which $\alpha_{ij}-\vartheta(x) \in \frac{\pi}{2}+\pi\IZ$.
Suppose that when $x$ is near $x_0$ we have
\be\label{eq:BindingPoint-def}
\alpha_{ij}-\vartheta(x) = \frac{\pi}{2} +  \frac{(x-x_0)}{\kappa} + \CO((x-x_0)^2) + 2 \pi n
\ee
for some   $n\in \half \IZ $ and some nonzero real number $\kappa$.
\footnote{We are using here the assumption that $\vartheta(x)$ is suitably generic.}
Then near $x_0$ we must have
\be
y_{ij}(x) \cong  -\kappa\log \vert x-x_0 \vert + const + \cdots
\ee

The local behavior of edges separating vacuum $i$ from $j$ in the neighborhood
of $x=x_0$ (but not at $x=x_0$) is governed by four cases according to whether $n$
in \eqref{eq:BindingPoint-def} is integer or half-integer and the
sign of $\kappa$. It is useful to make the following definition:
\footnote{Once again the terminology here is motivated by the theory of spectral networks
\cite{Gaiotto:2011tf,Gaiotto:2012rg,Gaiotto:2012db}. As discussed
in Section \S \ref{subsec:LG-Susy-Interface} and \S \ref{subsec:CatSpecNet} below,
the ``binding points'' represent places where supersymmetric domain walls in $1+1$
dimensional theories (or, more generally, on surface defects) can form boundstates
with two-dimensional BPS particles.
}

\bigskip
\noindent
\textbf{Definition:} Given vacuum weights of the form \eqref{eq:SpinningWeights}
 we  define a point $x_0 \in \IR$ to be a
\emph{binding point of type $ij$}  if $\vartheta$ satisfies \eqref{eq:BindingPoint-def}
with some \underline{integer} $n$.  It is called a \emph{future stable binding point} if
$\kappa>0$ and a \emph{past stable binding point} if $\kappa<0$. We denote the set of
future stable binding points of type $ij$ by $\curlywedge_{ij}$ and the set of past stable
binding points of type $ij$ by $\curlyvee_{ij}$. If $x_0$ is a (future- or past-  stable)
binding point of type $ij$ we call the line $x=x_0$ in the $(x,y)$-plane a
\emph{(future- or past-  stable) binding wall of type $ij$}.

\bigskip
\textbf{Remarks}:

\begin{enumerate}

\item
Note well that the set   $\curlyvee_{ij}~ \cup ~\curlywedge_{ij}$ of
all binding points of type $ij$ can be characterized as precisely
those positions $x_0$ on the real line where $e^{\I \vartheta(x_0)}$ defines
an $S_{ij}$-ray. (Recall the definition of $S_{ij}$-rays in  \eqref{eq:Sij-Ray-def}.)

\item
The differential equation for $y_{ij}$ is also singular when $n$ is a half-integer
$n\in \half + \IZ$ in \eqref{eq:BindingPoint-def}.
In this case $x_0$ is a binding point of type $ji$.

\item The sign of the derivative $\vartheta'(x_0)$ determines whether the point is
past or future stable. As $x$ \underline{increases} past $x_0$, the complex number
 $z_{ij}(x)$ goes through the positive imaginary axis in the \underline{counter-clockwise}
 direction for a future stable binding point, and in the \underline{clockwise} direction for a
 past stable binding point.

\end{enumerate}

\begin{figure}[htp]
\centering
\includegraphics[scale=0.25,angle=0,trim=0 0 0 0]{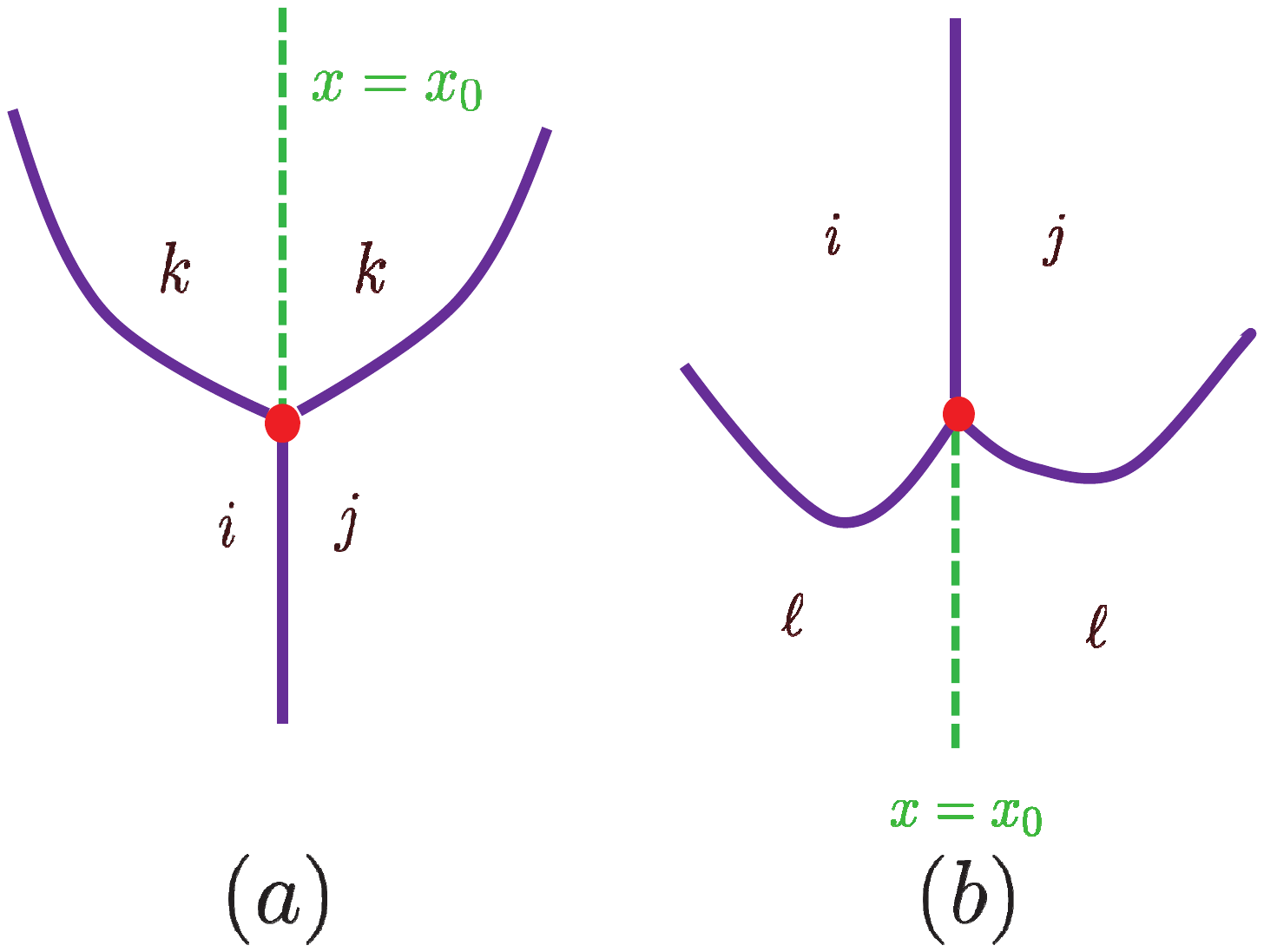}
\caption{When a vertex for a line separating vacua $(i,j)$ has an $x$-coordinate
which is a binding point $x_0$ of type $ij$ then it can be ``frozen''. In Figure (a) the vertex
cannot move off the binding wall $x=x_0$ if $x_0$ is a future-stable binding point, since the figure
cannot smoothly deform to Figure \protect \ref{fig:FUTURESTABLE-1}.
 In Figure (b) the vertex
cannot move off the binding wall $x=x_0$ if $x_0$ is a past-stable binding point, since the figure
cannot smoothly deform to Figure \protect\ref{fig:PASTSTABLE-1}.
On the other hand, in Figure (a)
the vertex is not frozen if $x_0$ is past stable and in Figure (b) the vertex is not
frozen if the $x_0$ is future stable. }
\label{fig:FROZENVERTEX-4}
\end{figure}
\begin{figure}[htp]
\centering
\includegraphics[scale=0.25,angle=0,trim=0 0 0 0]{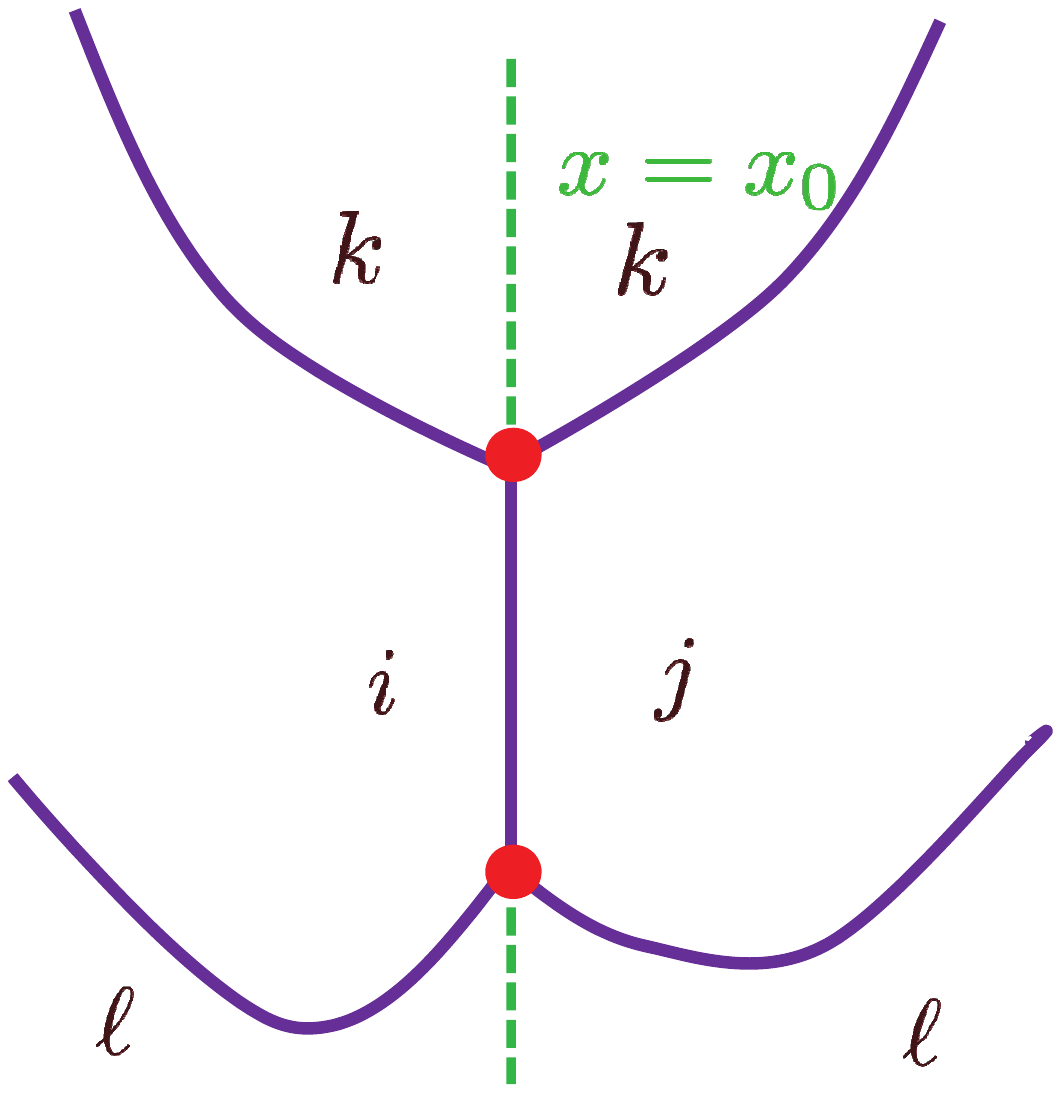}
\caption{A vertical internal edge at an $ij$ binding point. The internal edge
can be deformed away from the binding wall $x=x_0$.  }
\label{fig:FROZENVERTEX-3}
\end{figure}

Putting these remarks together we see that edges of type $ij$
have a behavior in the neighborhood of a binding point $x_0$ of type $ij$
of the form shown in   Figure \ref{fig:FUTURESTABLE-1}  for future stable
binding points and shown in  Figure  \ref{fig:PASTSTABLE-1} for past stable
binding points.

It is also possible for an $ij$-edge to sit within the
vertical line $x=x_0$. This is illustrated in Figures  \ref{fig:FROZENVERTEX-4}
and  \ref{fig:FROZENVERTEX-3}. As explained in the caption of
Figure \ref{fig:FROZENVERTEX-4} a vertex can be \emph{frozen} in the
sense that it cannot be translated in the $x$ direction.
%
%\cg{Davide says we only need frozen edges and not frozen vertices. This is
%correct, but I find the frozen vertices at the ends of frozen edges to be
%useful.}
%
Such vertices are connected to frozen external edges.  We therefore
split the vertices of $\fz$ into \emph{free} and \emph{frozen}
subsets $\CV(\fz) = \CV_{\rm free}(\fz) \cup \CV_{\rm frozen}(\fz)$.
In an analogous way the external edges can be divided up as
$\CE^{\rm ext}(\fz) = \CE^{\rm ext}_{\rm free}(\fz) \cup \CE^{\rm ext}_{\rm frozen}(\fz)$.

Now we can define the fan at infinity $I_\infty(\fz)$ for a curved web $\fz$. As we have said,
edges extending outside $-L < x < L$ define negative- and positive- half-plane fans $J^-, J^+$,
respectively.  Suppose these two fans are $J^- = \{ j_1, \dots, j_m \}$
and $J^+ = \{ j_1', \dots, j_n'\}$ as in Figure \ref{fig:DOMAINWALL-CHANPATON}. Then
the vacua in the regions encountered moving from $x=-L$ to $x=+L$ will stabilize, for sufficiently
large positive $y$, to a set
\be
\CJ^+ = \{ j_m, f_1, \dots, f_s , j_1' \}
\ee
while the vacua encountered moving from $x=+L$ to $x=-L$ will stabilize, for sufficiently
large negative $y$, to a set
\be
\CJ^- = \{ j_n', i_1, \dots, i_t , j_1 \}.
\ee
Therefore, reading left to right, the vacua encountered in a clockwise traversal at infinity are
\be\label{eq:curv-I-infty}
I_\infty(\fz) = \{ \CJ^+, J^+, \CJ^-, J^- \}.
\ee

We are now ready to define the oriented deformation type of a curved web.
Given a curved web $\fz$ we can deform it by varying the positions of the
vertices subject to the edge constraints and subject to the condition that
the web does not change topology, i.e., that no edge collapses to zero length.
Free vertices contribute two degrees of freedom but frozen vertices only contribute
one (it can be taken as the $y$ coordinate of the vertex). Thus the  expected dimension of the deformation space
of a generic curved web $\fz$ is
\be\label{eq:CurvedExpDim-1}
d(\fz) = 2 V_{\rm free}(\fz) + V_{\rm frozen}(\fz) - E^{\rm int}(\fz).
\ee
This can also be written as:
\be\label{eq:CurvedExpDim-2}
d(\fz) = 2 V(\fz)  - E^{\rm int}(\fz) - E^{\rm ext}_{\rm frozen}(\fz).
\ee
since each frozen vertex is uniquely associated with a frozen edge.
As usual, we are assuming $\vartheta(x)$ is sufficiently generic when we make
this definition. There will sometimes be exceptional webs $\fe$ where some of the
edge constraints are ineffective and the dimension of the deformation space is
larger than $d(\fe)$. In addition to this, the above discussion assumed that
the set of vertices $\CV(\fz)$ is nonempty. Actually, we can have curved webs
with no vertices at all, and in fact these will play an important role below.
(See, for examples, Figure  \ref{fig:CURVEDTAUT-1} and Figures \ref{fig:SPCLAMP-3}-
\ref{fig:SPCLAMP-7} below.)  In
that case the expected dimension is the true dimension of the deformation space
and is simply the number of components of the web.

Unlike the deformation spaces we have considered until now the moduli spaces
$\CD(\fz)$ do not have  piecewise linear boundaries. Nevertheless, they
are cells, and for generic $\fz$ they will have a dimension $d(\fz)$. We can
give them an orientation to define an oriented deformation type $\fz$ and
consider the free abelian group $\CW_{\rm curv}$ generated by the oriented
deformation types of curved webs. When we form the tensor algebra $T\CW_{\rm curv}$
we give $\fz$ the degree $d(\fz)$.

Curved webs have a translation symmetry in the $y$-direction but no scaling
symmetry. In this sense they are very much like strip webs and composite webs, and indeed can
be thought of as a continuous version of composite webs where many closely
spaced interfaces have been joined together. We therefore adopt the definitions
of Section \S \ref{subsec:ComposeInterface} and define $\fz$ to be
\emph{rigid} or \emph{taut} if $d(\fz)=1$ and to be \emph{sliding} if
$d(\fz) =2 $. The canonical orientation for the taut webs is $dy$ where $y$ is a
measure of the $y$ position of the web. We denote the taut element in $\CW_{\rm curv}$
by $\ft$

\begin{figure}[htp]
\centering
\includegraphics[scale=0.25,angle=0,trim=0 0 0 0]{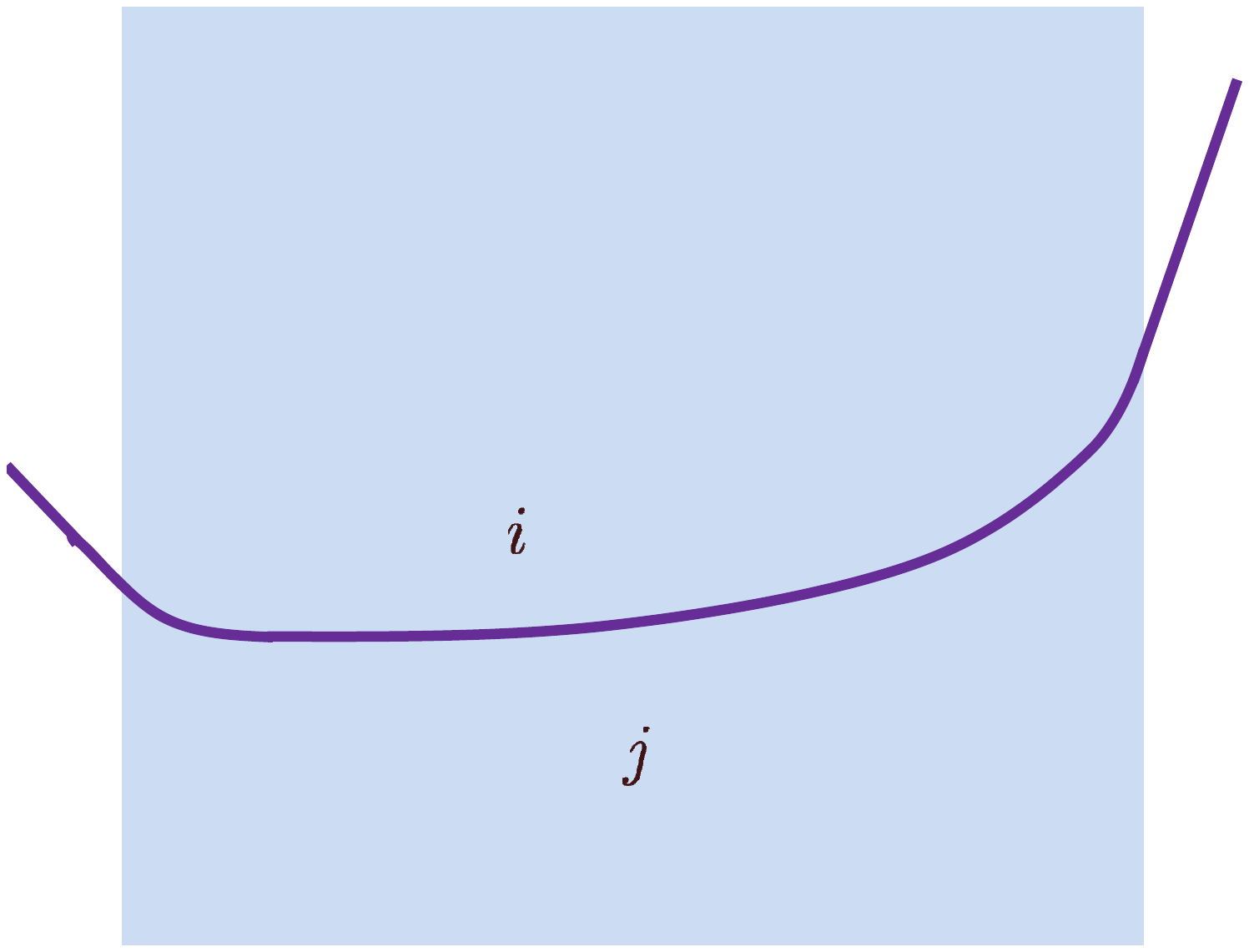}
\caption{An example of a taut curved web consisting of a single free
  edge of type $ij$. This contributes two external free edges. In this and similar
  figures the  light blue shaded region indicates the support of $\vartheta'(x)$.   }
\label{fig:CURVEDTAUT-1}
\end{figure}
%
%
%\begin{figure}[htp]
%\centering
%\includegraphics[scale=0.25,angle=0,trim=0 0 0 0]{CURVEDTAUT-2-eps-converted-to.pdf}
%\caption{Two   examples of taut curved webs in the presence of a
% binding point of type $ij$. In Figure (a)   $x=x_0$ is a future stable binding wall. An
%  edge of type $ij$ can come in from the negative half-plane. It is not
%possible to have a single edge of type $ij$ in the positive half-plane. In Figure (b)
%$x=x_0$ is a past stable binding wall. Now it is possible to have a taut web consisting
%of a single edge of type $ij$ in the positive half-plane but we cannot have a single edge
%of type $ij$ in the negative half-plane.  }
%\label{fig:CURVEDTAUT-2}
%\end{figure}
%
%
\begin{figure}[htp]
\centering
\includegraphics[scale=0.25,angle=0,trim=0 0 0 0]{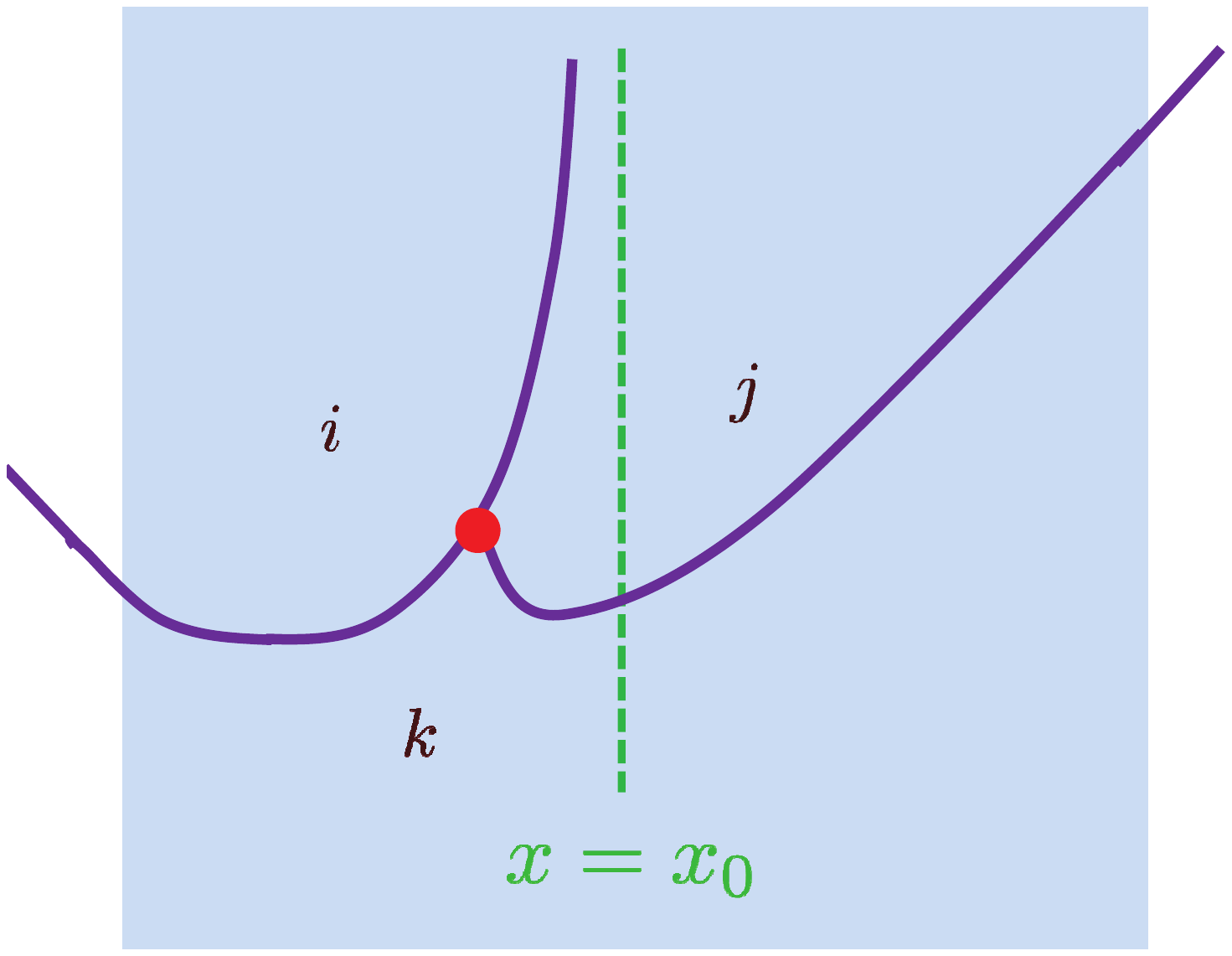}
\caption{An example of a curved sliding web. If $x=x_0$ is a future stable binding wall of
type $ij$ then the vertex can move across the line.   }
\label{fig:CURVEDTAUT-3}
\end{figure}
\begin{figure}[htp]
\centering
\includegraphics[scale=0.25,angle=0,trim=0 0 0 0]{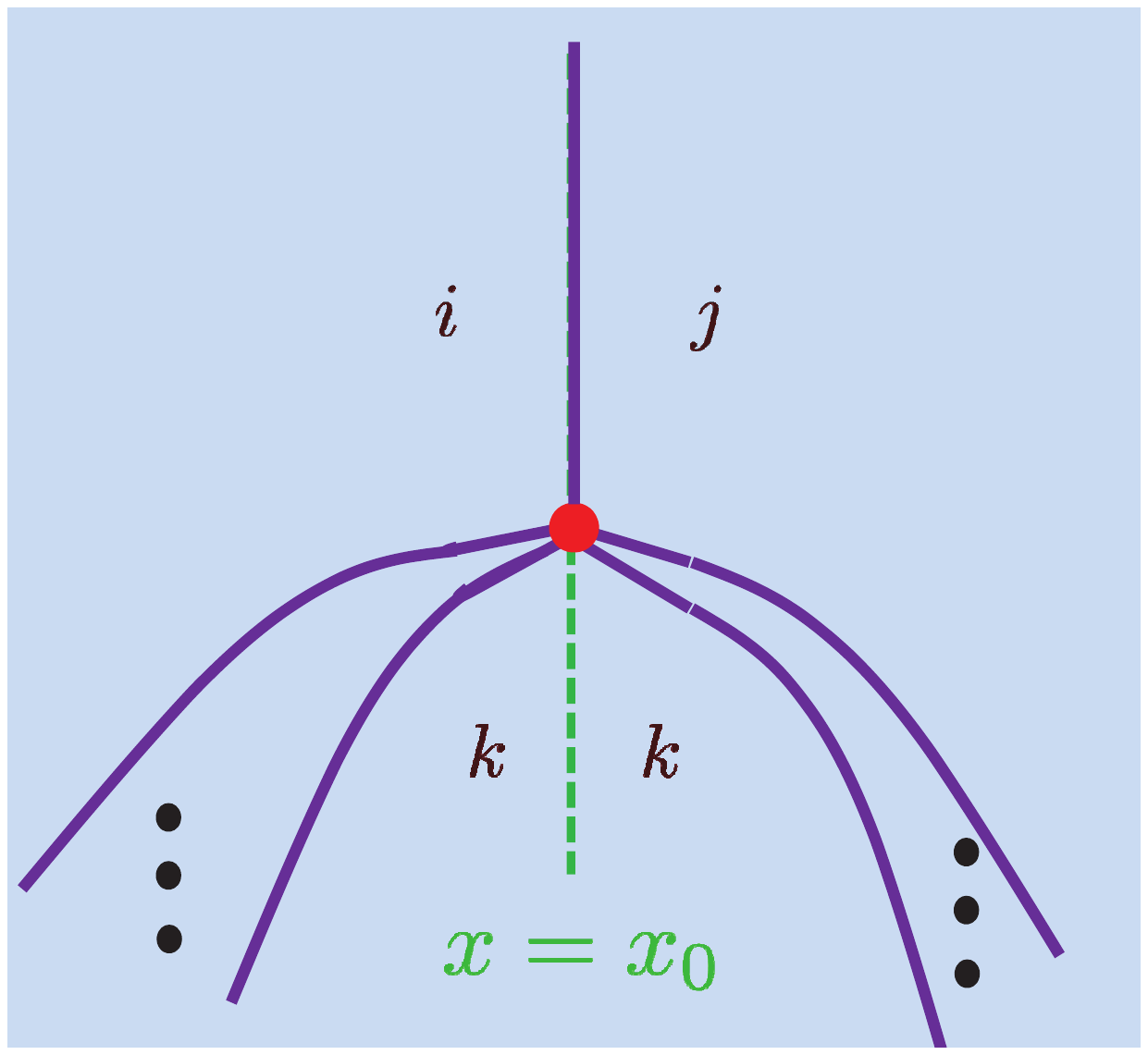}
\caption{An example of a curved taut web in the case that $x=x_0$ is a \underline{past} stable binding wall of
type $ij$. The edge of type $ij$ is a frozen external edge; the only degree of freedom
corresponds to moving the vertex vertically along the $ij$ binding wall, and hence the web is taut.
Similarly, if an $ij$ edge extends vertically to
$-\infty$ at a \underline{future} stable binding point then the external edge is frozen  and the web is a taut web. }
\label{fig:CURVEDTAUT-4}
\end{figure}

\bigskip
\noindent
\textbf{Examples}

\begin{enumerate}

\item If there is no binding point of type $ij$ then an edge of type  $ij$   can stretch from $x=-\infty$ to $x=+\infty$
with no vertices. This is a taut curved web $\fz$. See Figure \ref{fig:CURVEDTAUT-1}.

\item If there is a future stable binding point of type $ij$ then an edge of type $ij$ can come in from the negative half-plane
or from the positive half plane, as shown in Figure \ref{fig:FUTURESTABLE-1}. A curved with with a single component,
as shown in either Figure \ref{fig:FUTURESTABLE-1}(a) or Figure \ref{fig:FUTURESTABLE-1}(b) is a taut curved web.

\item Similarly if there is a past stable binding point of type $ij$ then an edge of type $ij$ can come in from
the negative or positive half-plane, as shown in Figure \ref{fig:PASTSTABLE-1}. A curved with with a single component,
as shown in either Figure  \ref{fig:PASTSTABLE-1}(a) or Figure \ref{fig:PASTSTABLE-1}(b) is a taut curved web.

\item Any vertex of the Theories $\CT^{\vartheta_\ell}$ or $\CT^{\vartheta_r}$ defines a \underline{sliding} web. See Figure \ref{fig:CURVEDTAUT-3}.

\item However, if a vertex of the Theories  $\CT^{\vartheta_\ell}$ or $\CT^{\vartheta_r}$ has an edge of type $ij$
then it might also define a taut curved web located on binding walls of type $ij$.
The case of a past stable binding wall is shown  in Figure \ref{fig:CURVEDTAUT-4}.

\item Finally, it is possible to have a completely rigid curved web with no moduli at all. These correspond
to entire lines of type $ij$, parallel to the $y$-axis and located at $ij$ binding points. They will play
some role in Section \ref{sec:LocalOpsWebs} below.

\end{enumerate}

Let us now discuss what kinds of convolution are possible with curved webs. Every web $\fw$ for the
vacuum data $(\IV, z^{\vartheta_\ell})$ can be continued to a web for any $(\IV, z^{\vartheta(x)})$ simply by a rotation,
until it becomes a web for $(\IV, z^{\vartheta_r})$.
If $\fz$ is a curved web it makes sense to consider the convolution $\fz*_v \fw$,
declaring the convolution to be nonzero when the fan coincide and
rotating $\fw$ by the appropriate angle before inserting it at $v$.

Now suppose that $\fd$ is an interface web between $\CT^{\vartheta_{\ell}}$ and $\CT^{\vartheta_{r}}$
and $\fz_1,\dots, \fz_n$ are a collection of curved webs. Then we can define a curved
web  $T_\p(\fd)[\fz_1, \cdots, \fz_n] \in \CW_{\rm curv}$.  We think of the interface as placed somewhere
in the region $-L< x< L$ and we replaced each of the
wall vertices in $\fd$ with the sequence of curved webs $\fz_1, \cdots, \fz_n$,
defining the convolution to be zero when the curved webs do not fit in properly
with the set of wall vertices of $\fd$.  In particular the operation is zero
unless the left and right half-fans of the $\fz_a$ are compatible with the
fan for the corresponding vertex of $\fd$. Moreover it is zero unless
the (transpose of the)
past fan $\CJ^-(\fz_a)$ coincides with the future fan  $\CJ^+(\fz_{a+1})$, including the specific location for
each vertical external edge. The vertical external edges of consecutive arguments can be connected into
a single internal edge. It is important to observe that one always connects frozen vertical edges of some curved web
with un-frozen vertical edges of another curved web.

%
%As each binding point of type $ij$
% is either past- or future-stable, connecting these vertical  external edges
%into an internal edge does not change the degree of the result.
%
One can show that  the moduli space of deformations of $T_\p(\fd)[\fz_1, \cdots, \fz_n]$ is locally the product of the deformation
space of $\fd$ times the product of reduced moduli spaces of the $\fz_a$. In particular,
\be
d(T_\p(\fd)[\fz_1, \cdots, \fz_n])  = d(\fd) + \sum_{a=1}^n (d(\fz_a) -1 )
\ee
just like equation \eqref{eq:ModTensOp}.
Thus, the operation is defined on the oriented deformation types of these webs.

Now we can write the convolution identity for the taut element $\ft$ in $\CW_{\rm curv}$.
Let $\ft_{pl} = \ft^{\vartheta_\ell} + \ft^{\vartheta_r}$  be the formal sum of
taut elements for the Theories $\CT^{\vartheta_{\ell}}$ and $\CT^{\vartheta_{r}}$
respectively. Similarly, let $\ft_{\CI}$ be the taut interface element between
$\CT^{\vartheta_{\ell}}$ and $\CT^{\vartheta_{r}}$. Then, examining the ends of the
moduli spaces of sliding webs, as usual, produces
\be\label{eq:Rotating-Web-Ident}
\ft * \ft_{pl} + T_\p(\ft_\CI)[\frac{1}{1-\ft}]=0
\ee
This is to be compared with equation \eqref{eq:Composite-Strip-Web-Ident}. The difference
is that all vertices of $\ft$ are now interior.

\subsubsection{Defining Chan-Paton Spaces And Amplitudes}\label{subsubsec:DefiningAmplitudes}

Let us now consider a representation of webs $\CR$. The usual definition of representations of webs
only makes use of the set of vacua $\IV$ and not the weights, so it essentially carries over immediately to the case of
curved webs, up to an important subtlety concerning frozen external edges. Frozen external edges impose an extra edge constraint
on the deformation space of a curved web $\fz$ and thus a good definition of $\rho(\fz)$ should include some extra degree $-1$ map $\check K_{\check e}$,
to be defined momentarily, for each frozen external edge $\check e$:
\begin{align}\label{eq:web-rep-2}
\rho(\fz)[r_1, \dots, r_n] = & \frac{o(\fw)}{\left[\prod_{\check e \in \CE^{\rm ext}_{\rm frozen}(\fz)} \p_{\check e} \prod_{e \in \CE(\fw)} \p_e \right] \circ \prod_{v \in \CV(\fw)} dx_v dy_v} \cdot \cr & \cdot
\left[\otimes_{\check e \in  \CE^{\rm ext}_{\rm frozen}(\fz)} \check K_{\check e}\otimes_{e \in \CE(\fw)} K_e \right]\circ \otimes_{a=1}^n r_a
\end{align}

We would like to define some interface Chan-Paton data $\CE_{j,j'}$ so that
$\rho(\fz)$, for $\fz$ a web with $I_{\infty}(\fz)$ of the form \eqref{eq:curv-I-infty}
can be understood as a map
\be\label{eq:Int-CP-curved-1}
\rho(\fz)~: \otimes_{v \in \CV(\fz)} R_{I_v(\fz)} \to  \CE_{j_m,j_1'} \otimes \widehat R^+_{j_1',j_n'} \otimes
(\CE_{j_1,j_n'})^* \otimes \widehat R^-_{j_1, j_m}
\ee
whose output is an interface vertex for some Interface in
$\fB\fr(\CT^{\vartheta_{\ell}}, \CT^{\vartheta_r})$.

The definitions of $\CE_{j,j'}$ and $\rho(\fz)$ should be compatible with the convolution operation $T_\p(\fd)[\fz_1, \cdots, \fz_n]$:
\be\label{eq:rhoT-curved}
\rho(T_\p(\fd)[\fz_1, \cdots, \fz_n])(S)
= \sum_{{\rm Sh}_{n+2}(S) } \epsilon ~
\rho(\fd)\left[S_1;\rho(\fz_1)[S_2],\dots, \rho(\fw_n)[S_{n+1}];S_{n+2} \right]
\ee
That means that the contraction of vertical external edges on the left hand side of this equation should
match the contraction of $\CE_{j,j'}^*$ and $\CE_{j,j'}$ on the right hand side. We need an independent summand in
$\CE_{j_m,j_1'}$for any possible $\CJ^+(\fz)$ in order to encode the representation data attached to vertical edges and, dually,
a summand in $(\CE_{j_1,j_n'})^*$ for any possible $\CJ^-(\fz)$ in order to encode the representation data attached to past vertical edges.

We define the relevant Chan-Paton data using a construction very similar to the product rule construction of $\widehat R_{ij}$ in
equation \eqref{eq:Cat-KS-prod} above:

For each  binding point $x_0$ of type $ij$   introduce a matrix
with chain-complex entries. It depends on whether $x_0$ is future-stable or past stable:
\be\label{eq:SijFactor-def-fs}
S^f_{ij}(x_0) := \IZ\cdot \textbf{1} + R_{ij} e_{ij}\qquad \qquad {\rm Future\ stable\ }\  ij\quad  {\rm binding\ point}
\ee
\be\label{eq:SijFactor-def-ps}
S^p_{ij}(x_0) := \IZ\cdot \textbf{1} + R_{ji}^* e_{ij}\qquad \qquad {\rm Past\ stable\ }\  ij\quad  {\rm binding\ point.}
\ee
We will refer to $S_{ij}(x_0)$  as a \emph{categorified $S_{ij}$-factor},
or just as an \emph{  $S_{ij}$-factor}, for short. The future and past stable $S$-factors
are related by $S_{ij}^p = (S_{ji}^f)^{tr, *}$.

We define
\be\label{eq:TautCurvedCP}
 \oplus_{j,j'\in \IV} \CE_{j,j'} e_{j,j'} := \bigotimes_{i\not=j}
\bigotimes_{x_0 \in \curlyvee_{ij} ~ \cup \curlywedge_{ij} } S_{ij}(x_0)
\ee
where the tensor product on the RHS of \eqref{eq:TautCurvedCP} is ordered from left to right by
increasing values of $x_0$ and $S_{ij}(x_0)$ is the future- or past-stable factor as appropriate
to the binding point. (Note that the rule \eqref{eq:TautCurvedCP} reduces
to the product for wedges \eqref{eq:KSprod} if $\vartheta(x) = -x$.)

Given this definition, we can associate each $\CJ^+(\fz)$ to a summand in $\CE_{j,j'}$ given as an ordered tensor product
with a factor of $R_{ij}$ for every vertical external unfrozen $ij$ edge
and a factor of $R^*_{ji}$ for every vertical external frozen $ij$ edge. These two choices differ by one unit of degree,
which matches the extra edge constraint for frozen edges and is crucial in defining a degree one boundary amplitude from the taut curved element $\ft$ below.

Dually, we can associate each $\CJ^-(\fz)$ to a summand in $(\CE_{j_1, j_n'})^*$ given again as an ordered tensor product with a factor of $R_{ij}$
for every vertical external unfrozen $ij$ edge and a factor of $R^*_{ji}$ for every vertical external frozen $ij$ edge. To see this
take the dual and transpose of the RHS of  \eqref{eq:TautCurvedCP}. This takes the transpose of the
matrix units $e_{ij}$ and takes $R_{ij} \to R_{ij}^*$. The factors are now ordered
with decreasing values of $x_0$. Now use the relation $S_{ij}^p = (S_{ji}^f)^{tr, *}$.

We have encoded the full fan at infinity $I_\infty(\fz)$ for any curved web $\fz$ into
a summand of the interface factor $R_J(\CE)$ of equation \eqref{eq:RJ-intfc}.

Recall  that $K_{ij}: R_{ij}\otimes R_{ji} \to \IZ$ is a \emph{perfect pairing}
and since $K_{ij}$ has degree $-1$ we can define a degree $-1$ isomorphism of $\IZ$-graded modules:
\be\label{eq:Kcheck-def}
\check K_{ij}: R_{ij} \rightarrow R_{ji}^*
\ee
by $\check K_{ij}(r_{ij})(\cdot ):= K_{ij}(r_{ij}, \cdot )$. In terms of
the notation in  equation \eqref{eq:Kinv-def3} we have
\be
\check K_{ij}(v_{\alpha}) = \sum_{\alpha'} K_{\alpha\alpha'} v_{\alpha'}^*
\ee
We identify the map $\check K_{\check e}$ in \ref{eq:web-rep-2} associated to a frozen $ij$ external edge $\check e$
with $\check K_{ij}$.

In order to complete our definition of $\rho(\fz)$, we only need to define carefully the vector field $\p_{\check e}$ associated to the corresponding
external edge constraint, in such a way that the compatibility condition \ref{eq:rhoT-curved} holds true. We define $\p_{\check e}$
to be a translation of the frozen vertex in the positive $x$ direction. When we contract two vertical external edges at a future-stable binding point,
this coincides with the $\p_e$ vector field for the resulting internal edge. This agrees with the absence of relative sign in
$\check K_{ij}(r_{ij}) \cdot r_{ji}:= K_{ij}(r_{ij},r_{ji} )$. On the other hand, when we contract two vertical external edges at a past-stable binding point,
this has the opposite orientation to the $\p_e$ vector field for the resulting internal edge. This agrees with the relative sign in
$r_{ji} \cdot \check K_{ij}(r_{ij}):= - K_{ij}(r_{ij},r_{ji} )$.

Now that we have defined the Chan-Paton spaces \eqref{eq:TautCurvedCP} for the
Interface we turn to the definition of the interface amplitude. This will be
defined by using taut curved webs.
Recall from Section \S \ref{subsec:CurvedWebs} that we are taking the interior amplitude
$\beta$ to be constant. In particular we take the same interior amplitudes $\beta_I$
for $\CT^{\vartheta_{\ell}}$ and for $\CT^{\vartheta_r}$.
\footnote{Note that the interior amplitude is \emph{not} defined
as a solution of an $L_\infty$ Maurer-Cartan equation $\rho(\ft)(e^\beta) = 0$,
where $\ft$ is the curved taut element! }
With this understood, for any curved web $\fz$
 we can define the operator $\rho_\beta(\fz)$ following the usual insertion of $e^\beta$.
We will be particularly interested in the special element $\rho_{\beta}^0(\fz)$ given by
inserting $\beta$ into every interior vertex of $\fz$.
\be\label{eq:CB-cpt}
\rho^0_{\beta}(\fz) = \rho(\fz)[e^\beta]
\ee
which is valued in \eqref{eq:Int-CP-curved-1}, and has degree $2V(\fz) - E^{\rm int}(\fz)  - E^{\rm ext}_{\rm frozen}(\fz)$.
It follows from \eqref{eq:CurvedExpDim-2} that  if $\fz$ is a taut web then
\eqref{eq:CB-cpt} indeed has degree $1$, as befits an interface amplitude.

\bigskip
\noindent
\textbf{Example}:  An example, which will be useful to us later, is given in Figure \ref{fig:CURVEDTAUT-4}.
Here a vertex of the theory $\CT^{\vartheta_\ell}$ has been
rotated so that the $ij$ edge goes to the future at a past stable binding point of type $ij$. Suppose the
cyclic fan of the vertex is $I$. It can be written as
\be
I = \{j,\dots, k, \dots, i \}= J^+ * J^-
\ee
where $J^+ = \{ j, \dots, k\}$
is a positive half-plane fan for $\CT^{\vartheta_r}$ and  $J^- = \{ k, \dots, i \}$ is a negative half-plane
fan for $\CT^{\vartheta_{\ell}}$ and the amalgamation $ J^+ * J^-$ is regarded as a cyclic fan.
Now,  the interior amplitude $\beta_I$ is a degree two
element of $R_I$ and (after a cyclic transformation) it will be convenient to regard this as
\be
\beta_I \in   R_{ij} \otimes \left( R_{J^+}\otimes R_{J^-}\right) .
\ee
We wish to interpret $\rho^0_\beta(\fz)$ as an interface amplitude valued in
$R_{J}(\CE)$. The Chan-Paton data for $\rho^0_\beta(\fz)$ are defined by
\eqref{eq:TautCurvedCP} and the relevant factors for this web are   $\CE_{ij} = R_{ji}^*$
(coming from \eqref{eq:SijFactor-def-ps}) and $\CE_{kk}^* = \IZ$. Therefore the interface amplitude
must be valued in
\be
\rho^0_\beta(\fz)\in  \CE_{ij} \otimes R_{J^+} \otimes \CE_{kk}^* \otimes R_{J^-}\cong R_{ji}^* \otimes
\left( R_{J^+}   \otimes R_{J^-}\right).
\ee
%
%Let $\check K_{ij}: R_{ij} \to R_{ji}^*$ be the degree $-1$ map induced by $K_{ij}$.
%
%\footnote{Referring the notation below equation \eqref{eq:Kinv-def2} we
%have $\check K_{ij}(v_\alpha) = \sum_{\alpha'} K_{\alpha'\alpha}v_{\alpha'}^*$.}
%
We identify the interface amplitude as
\be
\rho_{\beta}^0(\fz) := (\check K_{ij}\otimes 1)(\beta_I)
\ee
for this particular taut web $\fz$. Note that it indeed has degree $1$.
We do not need any extra sign on the right hand side, as $i_{\p_{\check e}} (dx \wedge dy) = dy$.

\bigskip
\bigskip

\begin{figure}[htp]
\centering
\includegraphics[scale=0.15,angle=0,trim=0 0 0 0]{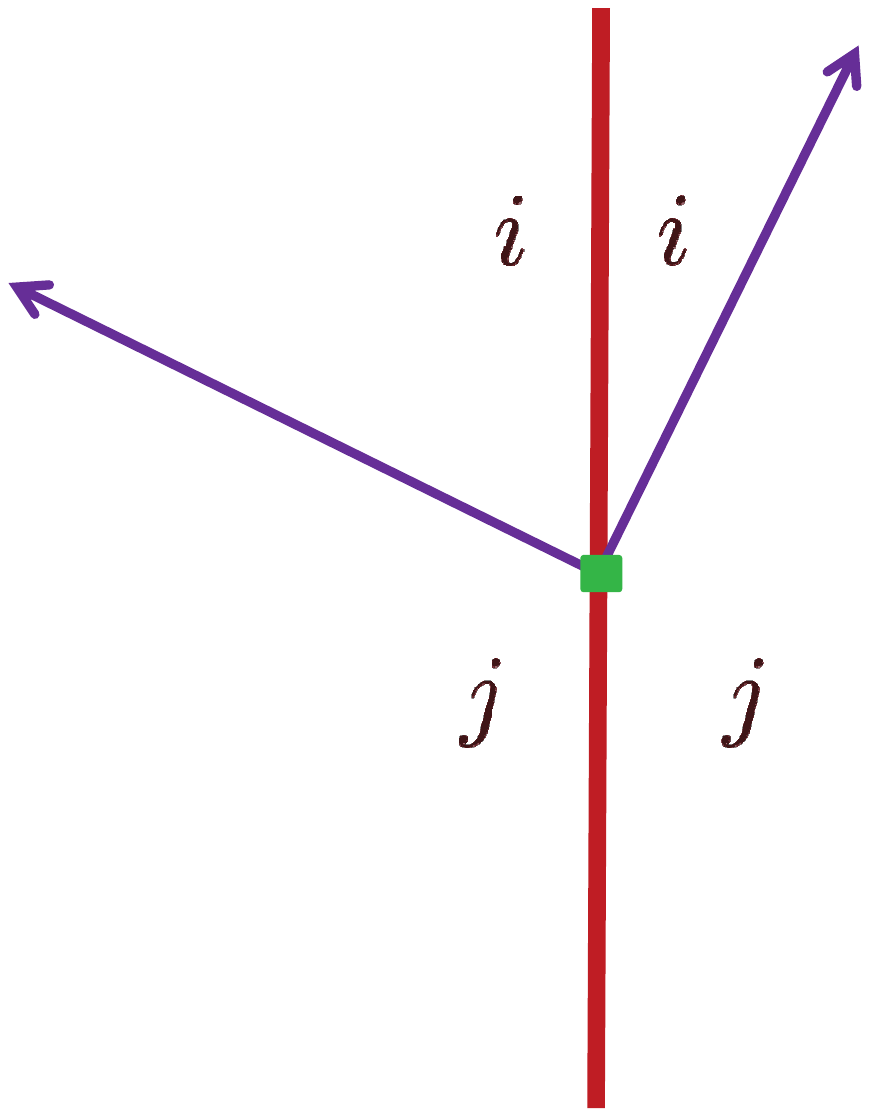}
\caption{The taut curved web of Figure \protect\ref{fig:CURVEDTAUT-1}
 leads to a nonzero interface
amplitude for the rigid interface web shown here.    }
\label{fig:SPECIALAMP-0}
\end{figure}
\begin{figure}[htp]
\centering
\includegraphics[scale=0.15,angle=0,trim=0 0 0 0]{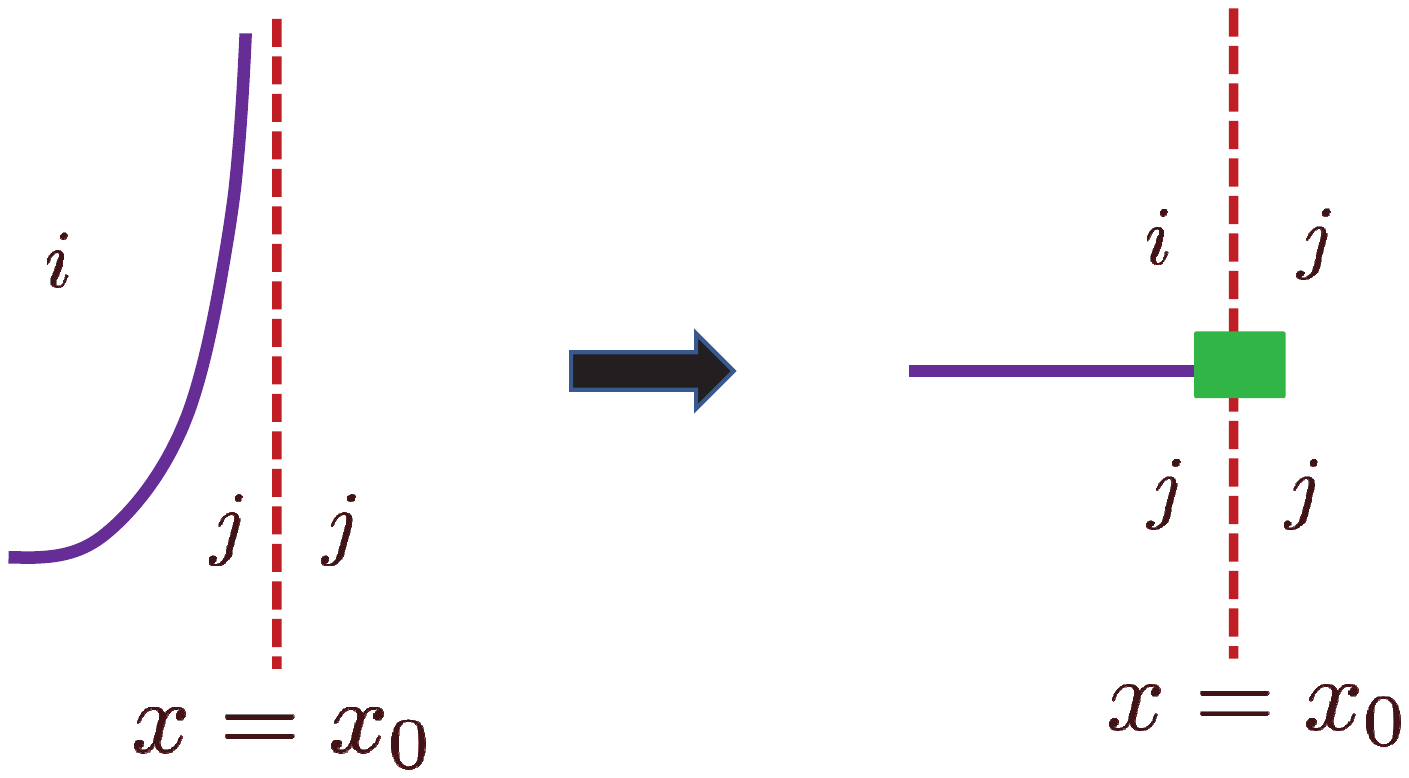}
\caption{The taut curved web of Figure \protect\ref{fig:FUTURESTABLE-1} (a)
leads to a nonzero interface
amplitude for the rigid interface web shown here.    }
\label{fig:SPCLAMP-3}
\end{figure}
\begin{figure}[htp]
\centering
\includegraphics[scale=0.15,angle=0,trim=0 0 0 0]{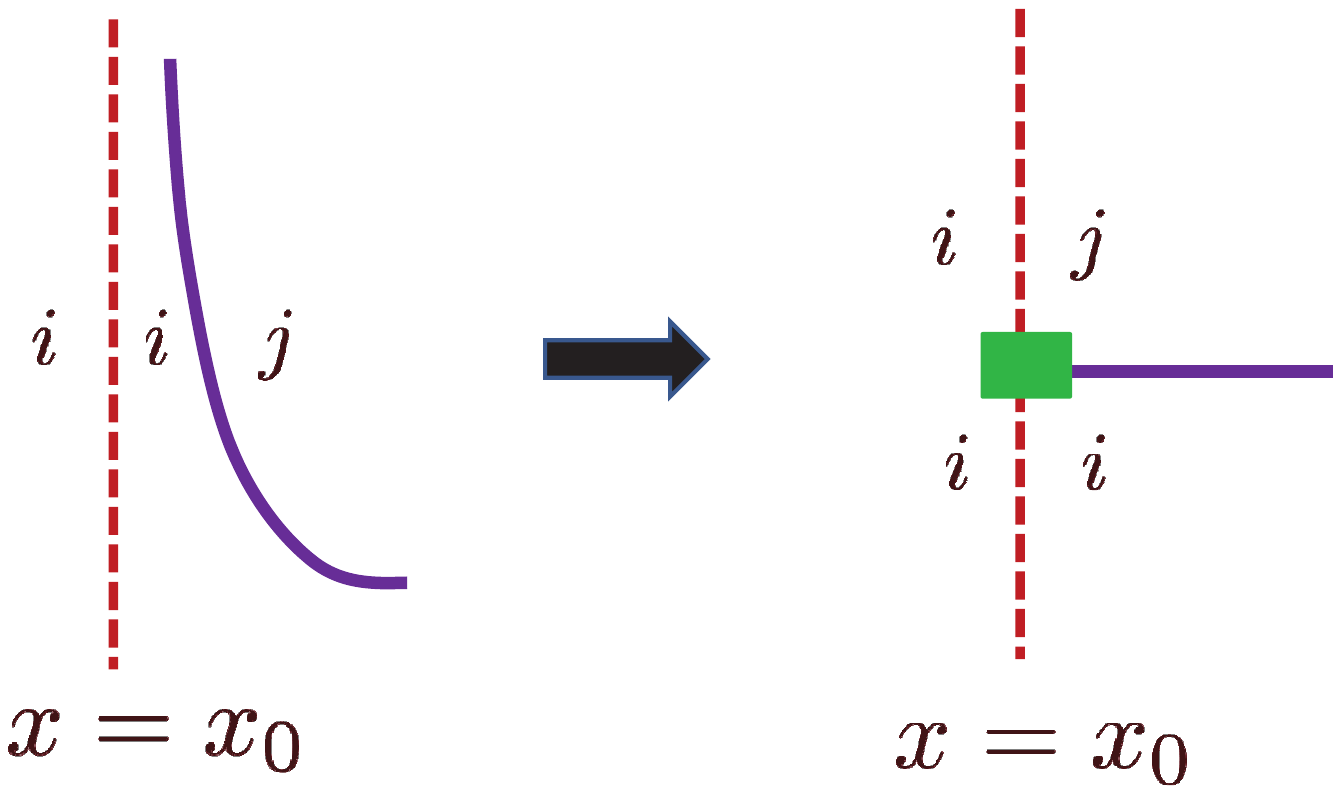}
\caption{The taut curved web of Figure \protect\ref{fig:FUTURESTABLE-1}(b)
leads to a nonzero interface
amplitude for the rigid interface web shown here.    }
\label{fig:SPCLAMP-4}
\end{figure}
\begin{figure}[htp]
\centering
\includegraphics[scale=0.15,angle=0,trim=0 0 0 0]{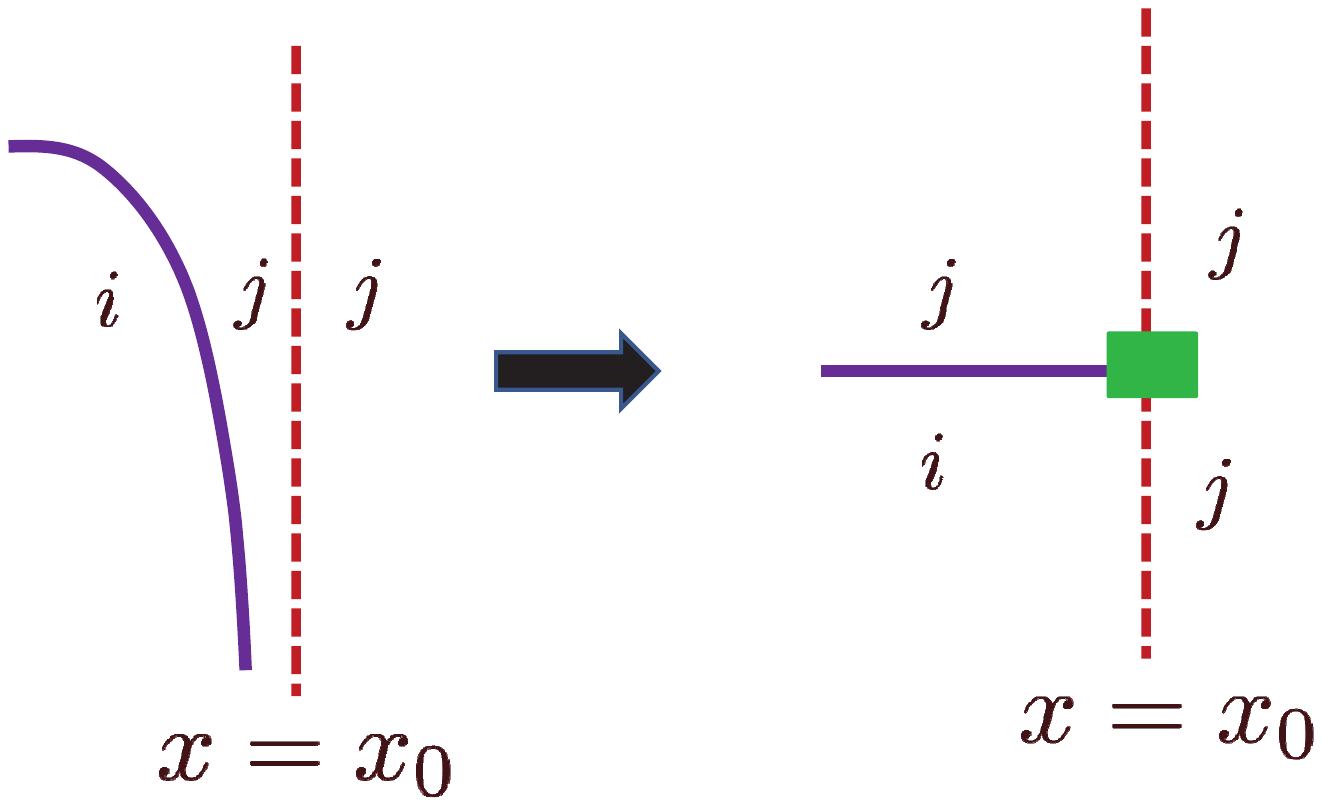}
\caption{The taut curved web of Figure \protect\ref{fig:PASTSTABLE-1}(a)
leads to a nonzero interface
amplitude for the rigid interface web shown here.    }
\label{fig:SPCLAMP-5}
\end{figure}
\begin{figure}[htp]
\centering
\includegraphics[scale=0.15,angle=0,trim=0 0 0 0]{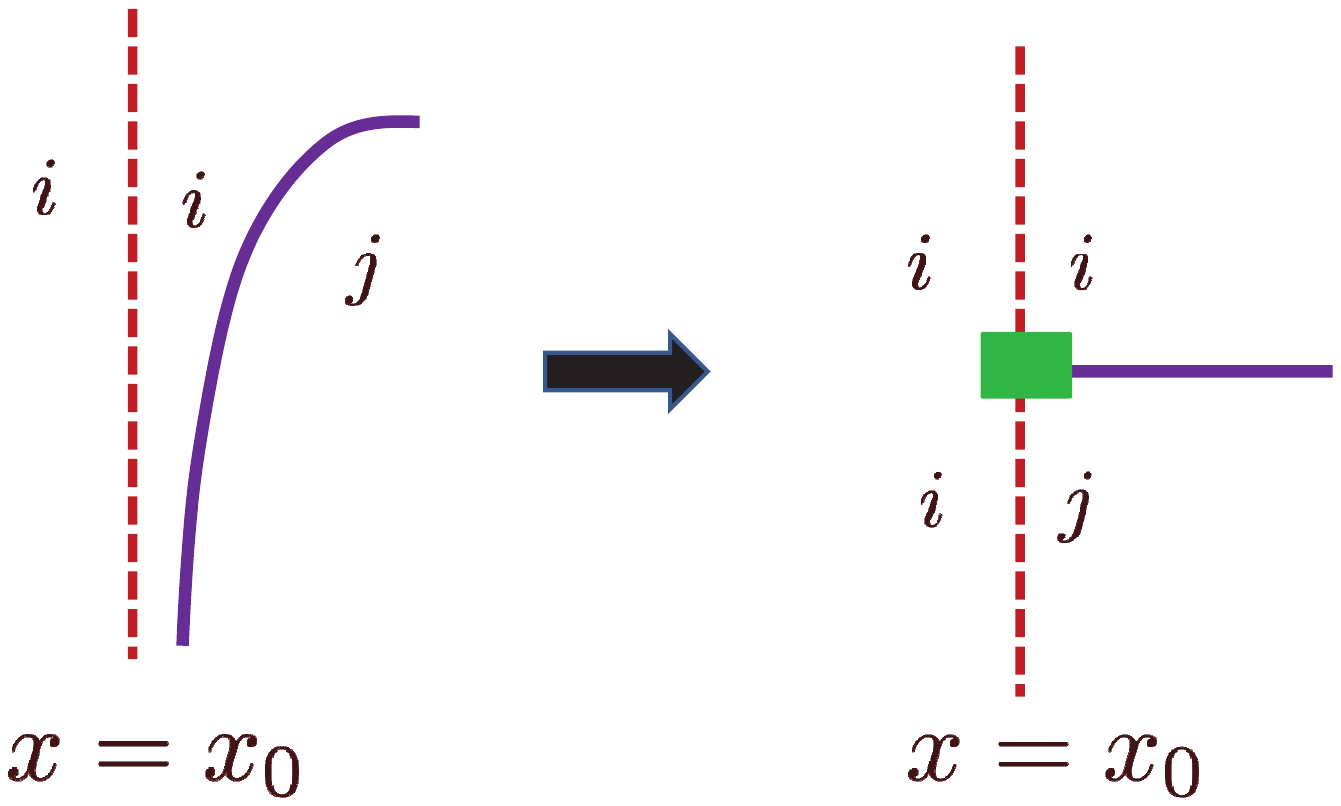}
\caption{The taut curved web of Figure \protect\ref{fig:PASTSTABLE-1}(b)
leads to a nonzero interface
amplitude for the rigid interface web shown here.    }
\label{fig:SPCLAMP-6}
\end{figure}
\begin{figure}[htp]
\centering
\includegraphics[scale=0.25,angle=0,trim=0 0 0 0]{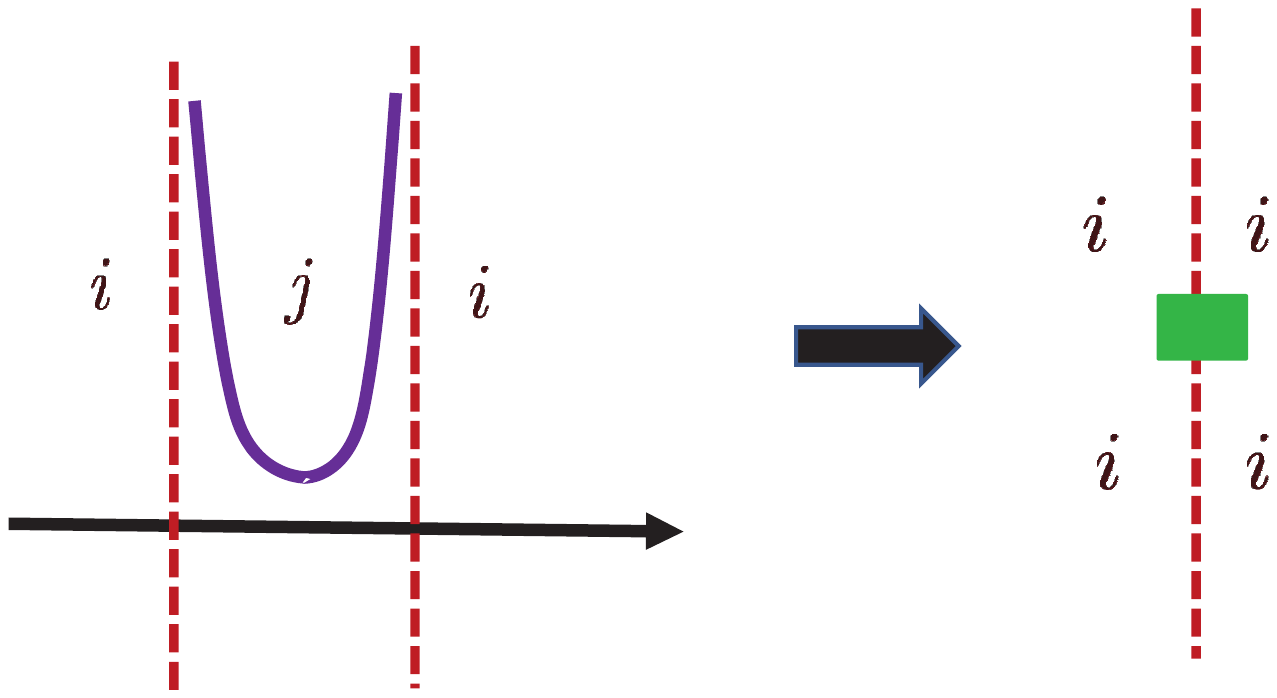}
\caption{A taut curved web with no vertices between future stable $ij$
and $ji$ binding points.   }
\label{fig:SPCLAMP-7}
\end{figure}
\begin{figure}[htp]
\centering
\includegraphics[scale=0.25,angle=0,trim=0 0 0 0]{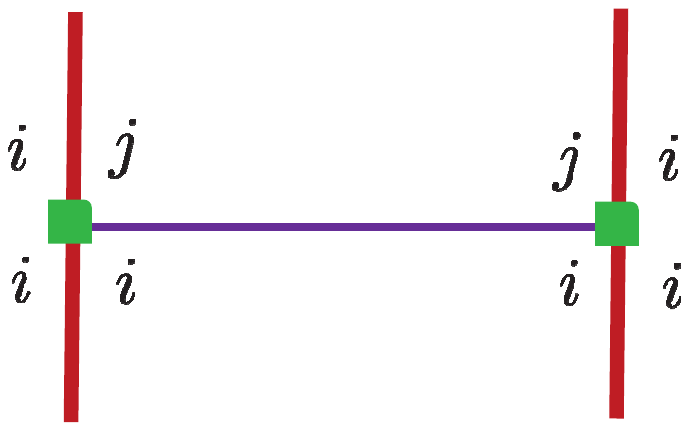}
\caption{The amplitude associated with the taut curved web of Figure
\protect\ref{fig:SPCLAMP-7} is defined by the interface product shown here.  }
\label{fig:TWOVACUA-INTFC-COMP2}
\end{figure}

The above discussion has assumed that the taut curved web $\fz$ has vertices. However,
as we have already noted, it is possible to have taut curved webs with no vertices.
These require special consideration when defining $\rho^0_{\beta}(\fz)$ (again, considered
as an interface amplitude for $\fI[\vartheta(x)]$). We must consider a few cases here.

\begin{enumerate}

\item The most basic case is for taut webs of the form shown in Figure \ref{fig:CURVEDTAUT-1}.
In the case of Figure \ref{fig:CURVEDTAUT-1} there are no binding walls so
the Chan-Paton data defined by  \eqref{eq:TautCurvedCP} give simply
$\CE_{k,\ell} = \delta_{k,\ell}\IZ$ for all $k,\ell\in \IV$. The taut curved web
of Figure \ref{fig:CURVEDTAUT-1} contributes to $\rho^0_{\beta}(\fz)$ as an  amplitude
associated with the rigid interface web shown in Figure \ref{fig:SPECIALAMP-0}.
 The interface amplitude associated with such a web must be
valued in $R_{ji}\otimes R_{ij}$ (where we put the negative half-plane factor first) and we take it to be
\be\label{eq:LocallyTrivial-1}
\rho^0_{\beta}(\fz):= K_{ji}^{-1}.
\ee
This choice can be shown to be required by demanding the
 composition property of Interfaces \eqref{eq:PT-Intf1} or by the
 Maurer-Cartan equation for the $S$-wall Interfaces discussed in
 Section \S \ref{subsec:SWallIntfc} below. Note that the interface amplitudes
 for $\fId$ are a special case of this equation.

\item In addition the future- and past-stable taut curved webs shown in
Figures \ref{fig:FUTURESTABLE-1} and \ref{fig:PASTSTABLE-1} lead to basic
amplitudes shown in Figures \ref{fig:SPCLAMP-3}-\ref{fig:SPCLAMP-6}.
By insisting that the amplitudes define solutions of the Maurer-Cartan
equation for the future- and past-stable S-wall interfaces $\fS_{ij}^{f,p}$
described in Section \S \ref{subsec:SWallIntfc} below we derive the following:

\begin{itemize}

\item
The amplitude of Figure \ref{fig:SPCLAMP-3}, $\CB^{ij}_{jj}\in \CE_{ij}\otimes R_{ji} \cong R_{ij}\otimes R_{ji}$
is
\be\label{eq:SpAmp-1}
\CB^{ij}_{jj}= - K_{ij}^{-1}
\ee

\item
The amplitude of Figure \ref{fig:SPCLAMP-4}, $\CB^{ij}_{ii}\in \CE_{ij}\otimes R_{ji} \cong R_{ij}\otimes R_{ji}$
is
\be\label{eq:SpAmp-2}
\CB^{ij}_{ii}=   K_{ij}^{-1}
\ee

\item
The amplitude of Figure \ref{fig:SPCLAMP-5}, $\CB^{jj}_{ij}\in R_{ij} \otimes \CE_{ij}^* \cong R_{ij}\otimes R_{ji}$
is
\be\label{eq:SpAmp-3}
\CB^{jj}_{ij}=  (-1)^F K_{ij}^{-1}= K^{\alpha\alpha'} v_{\alpha}\otimes v_{\alpha'}
\ee

\item
The amplitude of Figure \ref{fig:SPCLAMP-6}, $\CB^{ii}_{ij}\in R_{ij} \otimes \CE_{ij}^* \cong R_{ij}\otimes R_{ji}$
is
\be\label{eq:SpAmp-4}
\CB^{ii}_{ij} =  -(-1)^F K_{ij}^{-1}= - K^{\alpha\alpha'} v_{\alpha}\otimes v_{\alpha'}
\ee

\end{itemize}

\item There can also be taut curved webs trapped between two binding points. These
should be defined so that the composition property \eqref{eq:PT-Intf1} holds with
an equality. One example, which will be of significance later in Section
\S \ref{sec:LocalOpsWebs}, is shown in Figure \ref{fig:SPCLAMP-7}.
The result is an amplitude   $\CB^{ii}_{ii}\in \End(\CE_{ii})$ where the Chan-Paton space $\CE_{ii}$
is derived from multiplying
\be\label{eq:SpAmp-5}
\begin{pmatrix} \IZ & R_{ij} \\  0 & \IZ \\ \end{pmatrix} \begin{pmatrix} \IZ & 0 \\  R_{ji} & \IZ \\ \end{pmatrix}
\ee
and hence $\CE_{ii} \cong \IZ \oplus R_{ij}\otimes R_{ji}$. We can define $\CB^{ii}_{ii}$
from the interface product   $\IntfcTimes$ of an Interface $\fI_1$ of Figure \ref{fig:SPCLAMP-4}
with an Interface $\fI_2$ of Figure \ref{fig:SPCLAMP-3} (with $i$ and $j$ switched).
 The  relevant diagram is shown in Figure \ref{fig:TWOVACUA-INTFC-COMP2}.
The contraction in this diagram involves:
\be\label{eq:SpAmp-6}
K_{ji} : (\CE(\fI_1)_{ij} \otimes R_{ji})\otimes (R_{ij}\otimes \CE(\fI_2)_{ji}) \rightarrow \CE_{ij}(\fI_1)\otimes \CE(\fI_2)_{ji} \cong R_{ij}\otimes R_{ji}
\ee
and the value is
\be\label{eq:SpAmp-7}
K_{ji}( (K_{ij}^{-1})\otimes (-K_{ji}^{-1}) ) = K_{ij}^{-1}
\ee
We view the amplitude $\CB^{ii}_{ii}$  as a map
\be\label{eq:SpAmp-8}
\IZ \cong \CE(\fI_1)_{ii} \otimes \CE(\fI_2)_{ii} \to \CE(\fI_1)_{ij}\otimes \CE(\fI_2)_{ji} \cong R_{ij}\otimes R_{ji}
\ee
taking $1$ to $K_{ij}^{-1}$ and annihilating $R_{ij}\otimes R_{ji}$.
 It is indeed a degree one differential on $\CE(\fI_1\IntfcTimes \fI_2)_{ii}$,
as required by the Maurer-Cartan equation.

\item
%Other taut curved webs with no vertices can be similarly constructed using the interface
%product and the elementary ones we have described above.
As we noted above, there can be
completely rigid curved webs with no moduli (these are vertical lines at binding points).
These can be ``added'' to generic curved webs at otherwise empty binding points to get new webs with the same number of moduli.
They contribute to $\rho_\beta(\cdot)$ an extra tensor factor acting as the identity on the corresponding
 $R_{ij}$ or $R^*_{ji}$ factors in $\CE_{j,j'}$. It is important to include these contributions in the total curved taut element $\ft$.

\end{enumerate}

\bigskip
\bigskip

We have now completely defined $\rho^0_\beta(\ft)$.
The operator $\rho_\beta(\fz)$ is compatible by construction with convolutions and the
tensor operation $T_\p$, and thus the convolution identity gives us
\begin{equation}\label{eq:CrvdIntAmp}
\rho_\beta(\ft_\CI)[\frac{1}{1-\rho^0_\beta(\ft)}]=0
\end{equation}
In other words, we have the key observation that
\emph{ $\rho^0_\beta(\ft)$ is an interface amplitude and, together with
\eqref{eq:TautCurvedCP}, it defines an Interface
\be\label{eq:defIth}
 \fI[\vartheta(x)] \in \fB\fr(\CT^{\vartheta_{\ell}}, \CT^{\vartheta_r}).
\ee
associated to a function $\vartheta(x)$ defining spinning vacuum weights
\eqref{eq:SpinningWeights}.}

As a special case, note that if $\vartheta(x)=\vartheta$ is constant then $\fI[\vartheta(x)]= \fId$ is the identity Interface
in $\fB\fr(\CT^{\vartheta}, \CT^{\vartheta})$.  Since, by assumption $e^{-\I \vartheta} z_{ij}$
is never imaginary the Chan-Paton factors are indeed given by $\CE_{ij}= \delta_{ij} \IZ$. The amplitude $\rho_{\beta}^0(\ft)$
is defined by \eqref{eq:LocallyTrivial-1}, which is also the definition of the amplitudes for $\fId$.

The above construction can be generalized to discuss curved webs in the presence of interfaces by making $\vartheta(x)$
piecewise differentiable and placing interfaces at positions $x$ where $\vartheta'(x)$
has a discontinuity. For example, suppose we have Interfaces $\fI^{-,\vartheta_{\ell} }\in \fB\fr(\CT^{-}, \CT^{\vartheta_{\ell} })$ at $x_{\ell}$
and $\fI^{\vartheta_r,+}\in \fB\fr(\CT^{\vartheta_r} , \CT^{+})$ at $x_r$. Then we can consider
curved webs in the strip $x_{\ell} \leq x \leq x_r$ given by $\vartheta(x)$ interpolating between $\vartheta_{\ell}$ and $\vartheta_r$.
We could repeat our above discussion and define an interface web $\fI^{\rm strip}[\vartheta(x)] \in \fB\fr(\CT^{-}, \CT^{+})$.
On the other hand, we could instead extend $\vartheta(x)$ to a function on the real line to define
$\fI[\vartheta(x)]$ as in equation \eqref{eq:defIth} above. Then,
using a $G_3$-geometry as in Section \ref{subsec:CompThrIntfc}
we can consider $(\fI^{-,\vartheta_{\ell} }  \fI[\vartheta(x)]   \fI^{\vartheta_r,+})_{\eta}$
as in \eqref{eq:etaprod}. We claim these Interfaces are all homotopy equivalent.
An appropriate space-time dependent geometry can realize an
homotopy equivalence with any setup where the interpolation happens throughout the whole region in between the locations of
$\fI^{-,\vartheta_{\ell} }$ and $\fI^{\vartheta_r,+}$. This leads, in particular, to the connection to the wedge geometries sketched above.

\subsubsection{Verification Of Flat Parallel Transport}\label{subsubsec:Verification}

Now that we have defined $\fI[\vartheta(x)]$ let us verify the two key properties
\eqref{eq:PT-Intf1} and \eqref{eq:PT-Intf2} for defining flat parallel transport.

First, the   composition property \eqref{eq:PT-Intf1} is straightforward.
Suppose $\vartheta^1(x)$  smoothly interpolates from $\vartheta^{-}$ to $\vartheta^0$ and
$\vartheta^2(x)$ smoothly interpolates from $\vartheta^0$ to $\vartheta^+$. Then, on the one
hand, we have defined Interfaces  $\fI[\vartheta^1(x)]\in \fB\fr(\CT^{\vartheta^{-}}, \CT^{\vartheta^{0}}) $ and
$\fI[\vartheta^2(x)]\in \fB\fr(\CT^{\vartheta^{0}}, \CT^{\vartheta^{+}}) $ which can be
composed as in equation \eqref{eq:InterfaceComp}. On the other hand, the functions can be concatenated
to define a smooth interpolation $\vartheta^1\circ \vartheta^2(x)$ from $\vartheta^{-}$ to $\vartheta^+$
and we wish to show:
\be\label{eq:trspt-1}
\fI[\vartheta^1(x)]\IntfcTimes \fI[\vartheta^2(x)] \sim \fI[\vartheta^1\circ \vartheta^2(x)]
\ee
where $\sim$ means homotopy equivalence. The definition \eqref{eq:TautCurvedCP} as
an ordered product along the real line shows that the Interfaces on the left- and
right-hand sides of \eqref{eq:trspt-1} have identical Chan-Paton spaces.
As for the amplitude, if we cut a taut curved web (contributing the the interface
amplitude of $\fI[\vartheta^1\circ \vartheta^2(x)]$) into two pieces along
a vertical line $x=x_0$, then there is a corresponding taut composite web used in the
definition of the interface amplitude of
$\fI[\vartheta^1(x)]\IntfcTimes \fI[\vartheta^2(x)] $ and the two amplitudes match.
Next, as we have seen, in the case of taut curved webs with no
vertices the amplitudes are defined in terms of elementary ones so that the
composition property holds.
In fact, often, one can replace the homotopy equivalence in \eqref{eq:trspt-1} by
an equality sign. In general, we should write a homotopy equivalence because
$\IntfcTimes$ is only associative up to homotopy equivalence.

Thanks to the composition property the general Interface $\fI[\vartheta(x)]$
can be decomposed as a product of elementary Interfaces. We discuss these elementary
Interfaces in detail in Sections \S\S \ref{subsec:LocTriv} and \ref{subsec:SWallIntfc} below.

Next, we need to study the behaviour of $\fI[\vartheta(x)]$ under homotopy of $\vartheta(x)$.
Accordingly, let  $\vartheta(x,y)$ be an homotopy
interpolating between two functions $\vartheta_f(x)$ at very large positive $y$ and $\vartheta_p(x)$ at very large negative $y$
with the same endpoints $\vartheta_{\ell,r}$. Again, our choice of argument is intentional: we interpret the homotopy as an actual space-time dependent configuration.

We can define curved webs with  space-time dependent weights $e^{-\I \vartheta(x,y)}z_i$  in an obvious way. Their properties are somewhat similar to the ones we looked at to prove associativity of composition of interfaces in Section \S \ref{subsec:CompThrIntfc}.
 Looking at sliding webs, we get the convolution identity closely analogous to equation \eqref{eq:fth}
\begin{equation}\label{eq:ST-DeptWT-Hom}
\ft_{st} * \ft_{pl} +\ft^f - \ft^p +  T_\p(\ft_\CI)[\frac{1}{1-\ft^f};\ft_{st}; \frac{1}{1-\ft^p}] =0
\end{equation}
where $\ft^{p,f}$ are the   taut elements for curved webs with the past and future spinning weights
$e^{-\I \vartheta_p(x)}z_i$, $e^{-\I \vartheta_f(x)}z_i$, respectively,  and $\ft_{st}$ is now the space-time dependent taut element
(not including the empty web).  The notations $\ft_{pl}$ and $\ft_{\CI}$ are as in \eqref{eq:Rotating-Web-Ident}.

Applying a web representation, we find an identity of the form of equation \eqref{eq:HomEqBr}
and therefore using the discussion of that result we conclude that   $\delta[\vartheta(x,y)] := \rho^0_\beta(\ft_{st})$
defines a closed morphism $\Id + \delta[\vartheta(x,y)]$
between $\fI[\vartheta_f(x)]$ and $\fI[\vartheta_p(x)]$.

Next, we need to show that the closed morphisms $\Id + \delta[\vartheta(x,y)]$ and $\Id + \delta[\vartheta(x,-y)]$ are inverse up to homotopy. We can use the same strategy as for the proof of associativity of composition of interfaces up to homotopy: show that given a continuous interpolation $\vartheta(x,y,s)$ between two homotopies $\vartheta^1(x,y)$ and $\vartheta^2(x,y)$ the closed morphism $\Id + \delta[\vartheta(x,y,s)]$ varies by an exact amount.
This follows from a convolution identity for curved web homotopies similar to the ones we have already discussed above.

To summarize, we have proven that   given two homotopic maps $\vartheta^1(x)$ and $\vartheta^2(x)$ with the same endpoints
the corresponding Interfaces $\fI[\vartheta^1(x)]$ and $\fI[\vartheta^2(x)]$ are homotopy equivalent.

\subsubsection{Rigid Rotations And Monodromy}\label{subsubsec:Monodromy}

We can now deliver on a promise made in Section \S \ref{subsec:RotInts} and define a
canonical Interface $\fR[\vartheta^\ell, \vartheta^r]$ between Theories $\CT^{\vartheta_\ell}$ and
$\CT^{\vartheta_r}$. Namely, choose lifts of $\vartheta_{\ell}, \vartheta_r$ to $\IR$ so that
$\vartheta_{\ell} >  \vartheta_r$ and  $\vert \vartheta_{\ell} -  \vartheta_r\vert \leq \pi$.
Then, as in Section \S \ref{subsec:RotInts} we can use
\be
\vartheta(x)= \begin{cases} \vartheta_{\ell} & x \leq - \vartheta_\ell \\
-x &  -\vartheta_{\ell} <  x < - \vartheta_r \\
 \vartheta_{r} & x >  - \vartheta_\ell \\
\end{cases}
\ee
to define
\be\label{eq:RigRotInt-Def}
\fR[\vartheta_\ell, \vartheta_r] := \fI[\vartheta(x)].
\ee
Now, it follows from \eqref{eq:trspt-1} that
\be\label{eq:Cps-LinInt}
\fR[\vartheta_{1},\vartheta_{2}]\IntfcTimes \fR[\vartheta_{2},\vartheta_{3}] =
\fR[\vartheta_{1},\vartheta_{2}]
\ee
as desired.

If we try to extend the above discussion to intervals larger than $2\pi$
then we encounter the interesting phenomenon of monodromy.
 Consider the function
\be\label{eq:Cat-Monod}
\vartheta(x)= \begin{cases} \vartheta_*  & x \leq - \vartheta_* \\
-x &  -\vartheta_{*} <  x < - \vartheta_* + 2\pi \\
 \vartheta_{*}-2\pi  & x >  - \vartheta_\ell \\
\end{cases}
\ee
The invertible interface $\fI[\vartheta(x)]$ in this case defines an \afty-functor
from $\fB\fr(\CT^{\vartheta_*})$ to itself.  Note that there is precisely one binding
wall of type $ij$ for each pair distinct pair of vacua $(i,j)$. This can be viewed as a monodromy
transformation on the category of Branes. Indeed, we can see that the Interface for
rigid rotation through an angle $2\pi$ is \underline{not} equivalent to the identity
Interface using the discussion of equations \eqref{eq:SpAmp-5}-\eqref{eq:SpAmp-8} above.
The cohomology of $\CE_{ii}$ is the quotient of $R_{ij}\otimes R_{ji}$ by the one dimensional
line spanned by $K_{ij}^{-1}$, and is in general nontrivial.  See Section \S \ref{subsec:CatSpecGen} below
for further discussion.

\subsubsection{The Relation Of Ground States To Local Operators}

The rigid rotation Interfaces can be used to describe the precise relation between
the complex of ground states on an interval and the complex of local operators on
a half-plane.

 Let us return to the motivation in Sections \S \ref{subsec:RotInts} and \ref{subsec:WedgeWebs}.
There is a   very useful special case of the exponential map,  namely when the wedge
has opening angle $\pi$. In this case we find that if $\fB_1, \fB_2 \in \fB\fr(\CT^\vartheta)$
then $\Hop(\fB_1,\fB_2)$ with differential $M_1$ is literally the same as the complex
 of approximate groundstates on the interval with left-Brane $\fB_1$ and right-Brane
$\fB_2[\pi]:=\fB_2 \IntfcTimes \fR[\vartheta, \vartheta+\pi]$, with differential \eqref{eq:strip-diff-def}.
Indeed, note that on an interval of $\pi$ for every unordered pair of vacua there will be
precisely one binding wall. (It is important to define the rotation Interface so that $\vartheta$ increases,
and hence all the binding walls are past-stable.)  The Chan-Paton factors of $\fR[\vartheta, \vartheta+\pi]$
provide all the relevant half-plane fans and we conclude that
\be\label{eq:Strip-HalfPlane}
H^*(\Hop(\fB_1,\fB_2),M_1) \cong H^*(\CE_{LR}(\fB_1,\fB_2[\pi]), d_{LR}).
\ee
This result will be very useful in Section \S \ref{subsec:RotIntfc-TSUN} below.
Physically, it states that the space of BPS states between branes on an interval is
isomorphic to the local boundary-changing operators on a half-plane.
For more discussion of the relation to local observables see Section
\S \ref{sec:RemarksLocalObs} below.
\begin{figure}[htp]
\centering
\includegraphics[scale=0.25,angle=0,trim=0 0 0 0]{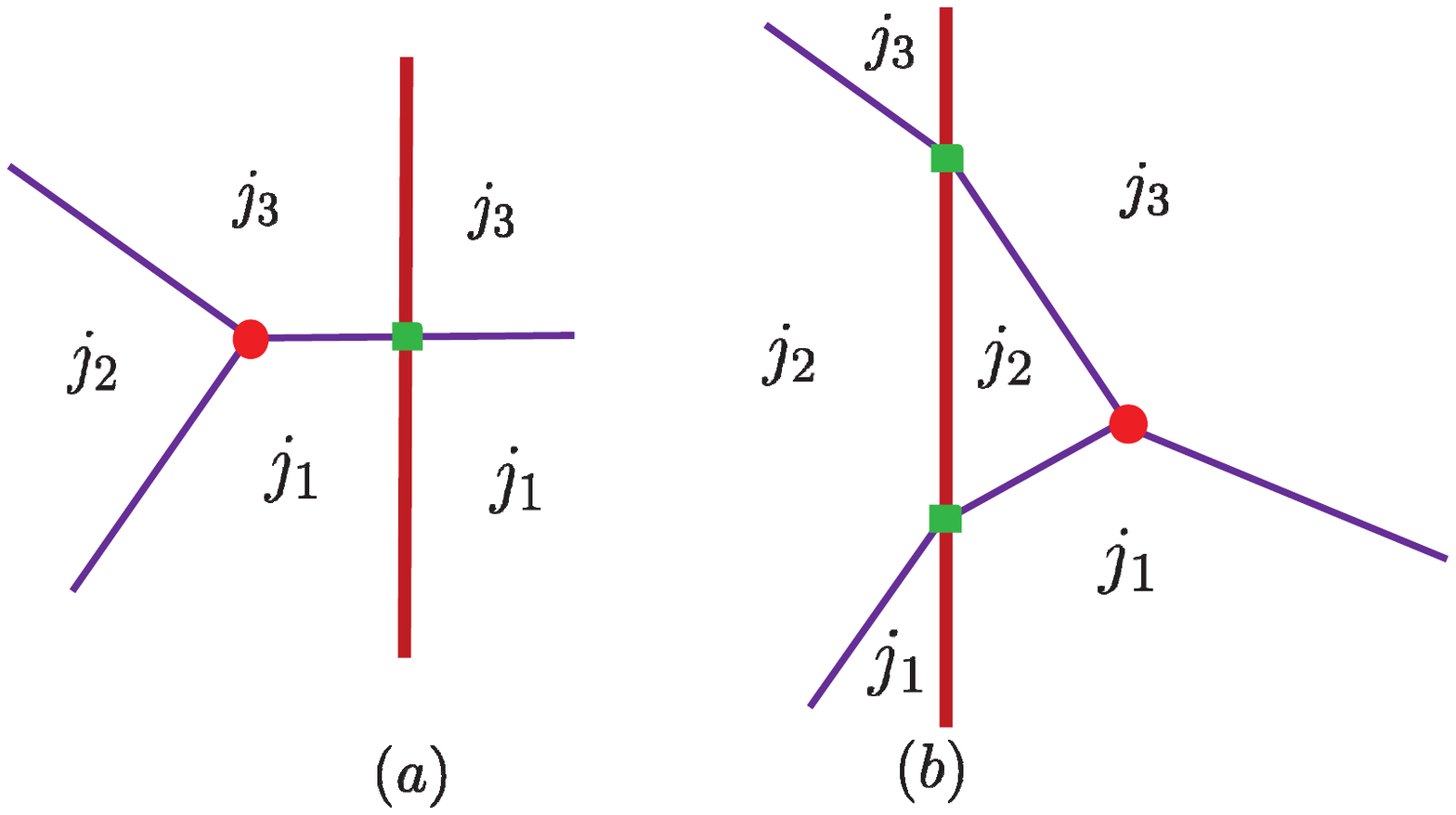}
\caption{Two contributions to the Maurer-Cartan equation for an interface defined by
locally trivial transport, such as that induced by taut webs of the form
of Figure \protect\ref{fig:CURVEDIDENT-MC}.
   }
\label{fig:CURVEDIDENT-MC}
\end{figure}

\subsection{Locally Trivial Categorical Transport}\label{subsec:LocTriv}

Let us consider a function $\vartheta(x)$ interpolating from
$\vartheta_\ell$ to $\vartheta_r$ with $\vartheta_{\ell}=\vartheta_r = \vartheta_*$.
We assume, moreover that there is a homotopy $\vartheta(x,y)$ to the constant
function $\vartheta(x)=\vartheta_*$ such that  $\vartheta(x,y)$
has no binding points (as a function of $x$ at fixed $y$, for all $y$).
Applying the discussion of \eqref{eq:ST-DeptWT-Hom} et. seq. we conclude
that  $\fI[\vartheta(x)]$ is homotopy equivalent to the identify
interface $\fId$ on $\CT^{\vartheta_*}$.

More generally when the function $\vartheta(x)$
has no binding points we say that $\fI[\vartheta(x)]$ defines a
\emph{locally trivial categorical transport}.
The Interface and its associated \afty-functor (via equation \eqref{eq:Intfc-Functor} )
are particularly simple in this case.
Since there are no binding walls equation \eqref{eq:TautCurvedCP} says that
the Chan-Paton factors are identical to those of the identity interface
$\fId$, namely $\CE_{ij}= \delta_{i,j}\IZ$.

We will say that $\vartheta(x)$ and its associated Interface $\fI[\vartheta(x)]$
are \emph{simple} if the  only curved taut webs
look like those in Figure \ref{fig:CURVEDTAUT-1}
(one for each pair of distinct vacua $(i,j)$). The interface amplitudes for
such a simple Interface
are then those given by equation \eqref{eq:LocallyTrivial-1}.
The demonstration that the Maurer-Cartan equation is satisfied,
that is, the demonstration that  \eqref{eq:CrvdIntAmp} holds in this case, is
very similar to that used to show that the identity Interface $\fId$
between a theory and itself satisfies the Maurer-Cartan
equation. Indeed, we should compare  Figure \ref{fig:ID-INTERFACE}
with Figure \ref{fig:CURVEDIDENT-MC}.
The main difference is that when we move an interior vertex across the
interface it gets rotated in order to be compatible with Figure
\ref{fig:SPECIALAMP-0}. Algebraically, the demonstration that the Maurer-Cartan
equation is satisfied is identical to the case of $\fId$. Nevertheless,
we should not identify it with $\fId$ because it is in general
an Interface between different theories $\CT^{\vartheta_{\ell}}$ and
$\CT^{\vartheta_r}$.

In general if $\vartheta(x)$ has no binding points but is not simple then the
Chan-Paton factors are still given by $\CE_{ij}= \delta_{i,j}\IZ$
but in principal there could be exceptional taut webs leading to
different interface amplitudes.
%
%\cg{Given an example by deforming Figure \ref{fig:EXCEPTIONALWEB} so it
%only has translation modulus in the y direction. But why is this homotopy
%equivalent to simple locally trivial transport? }
%
In this case we can divide up the region of support into a union of
small regions $[x_i, x_{i+1}]$ so that $\vartheta(x)$ is simple
in each region. Then, invoking equation \eqref{eq:trspt-1} we learn that
locally trivial transport is always homotopy
equivalent to simple locally trivial transport.

Using the discussion of \eqref{eq:Intfc-Functor} et. seq. we see that
a locally trivial Interface $\fI[\vartheta(x)]$ defines an \afty-functor
$\CF_{\vartheta(x)}: \fB\fr(\CT^{\vartheta_{\ell}}) \to \fB\fr(\CT^{\vartheta_{r}})$.
However, thanks to the very simple Chan-Paton data, it preserves the
vacuum subcategory whose objects are just the thimbles $\fT_i$. That is,
we can think of locally trivial transport as induced from a simpler functor
$\CF_{\vartheta(x)}: \fVac(\CT^{\vartheta_{\ell}}) \to \fVac(\CT^{\vartheta_{r}})$.

Note well that in the above discussion we have strongly used the fact
that there are no binding walls. If, on the other hand,
the rotation of some edge from $e^{-\I\vartheta_{\ell} }z_{ij}$ to $e^{-\I\vartheta_{r} }z_{ij}$
passes through the positive imaginary axis then one of the vertices of $\CT^{\vartheta_\ell}$
cannot be transported through the region of support of $\vartheta'(x)$ to produce the
corresponding vertex of  $\CT^{\vartheta_r}$.    The Maurer-Cartan equation \eqref{eq:CrvdIntAmp} for the simple
interface amplitudes \eqref{eq:LocallyTrivial-1}  will fail in this case. The next subsection
\S \ref{subsec:SWallIntfc} explains in detail how the correct Interface $\fI[\vartheta(x)]$
corrects the simple amplitudes to produce a solution of the
Maurer-Cartan equation.

\subsection{S-Wall Interfaces}\label{subsec:SWallIntfc}

As we mentioned above, thanks to the composition property
\eqref{eq:trspt-1} the general Interface $\fI[\vartheta(x)]$
can be decomposed into elementary factors. These consist of
locally trivial parallel transport together with the   ``S-wall Interfaces''
that are described in detail in this Section.  The \emph{$S$-wall Interfaces}
are defined (up to homotopy equivalence) by functions $\vartheta(x)$
which interpolate from  $\vartheta_{ij}\pm \epsilon$ to
$\vartheta_{ij}\mp \epsilon$, where $e^{\I \vartheta_{ij}}$ defines
an $S_{ij}$-ray and $\epsilon$ is a sufficiently small positive number.
\footnote{Note that while the variation in $\vartheta_{ij}$ is small
we have said nothing about how large the region of support of $\vartheta'(x)$
is. This can be changed, up to homotopy equivalence, and could be taken to
be very large or very small.}

To be concrete, we define an Interface $\fS^{p}_{ij}$ (up to homotopy equivalence) by
choosing $\vartheta(x)$ to interpolate from $\vartheta_{ij}-\epsilon$ to
$\vartheta_{ij}+\epsilon$ on some interval $(x_0-\delta, x_0 + \delta)$.
Here $\delta$ is a positive number (and not a morphism!) and, for
  definiteness, we choose a linear interpolation with $\epsilon \ll \delta$.
The region of support of $\vartheta'(x)$ contains a \emph{past stable} binding
point $x_0$ of type $ij$ and no other binding points. The vacuum weights
$e^{-\I \vartheta(x)} z_k $ all rotate clockwise through an angle $2\epsilon$
and  $e^{-\I \vartheta(x)} z_{ij}$ rotates clockwise through the positive imaginary
axis.

Similarly, we will define an Interface $\fS^{f}_{ij}$ (up to homotopy equivalence) by
choosing $\vartheta(x)$ to interpolate from $\vartheta_{ij}+\epsilon$ to
$\vartheta_{ij}-\epsilon$ on some interval $(x_0-\delta, x_0 + \delta)$.
The region of support of $\vartheta'(x)$ contains a \emph{future stable} binding
point $x_0$ of type $ij$ and no other binding points. The vacuum weights
$e^{-\I \vartheta(x)} z_k $ all rotate counter-clockwise through an angle $2\epsilon$
and  $e^{-\I \vartheta(x)} z_{ij}$ rotates counter-clockwise through the positive imaginary
axis.

The name  \emph{S-wall Interfaces} is apt because the path in the complex plane $e^{-\I \vartheta(x)} z_{ij}$  crosses
an $S_{ij}$-ray (see equation \eqref{eq:Sij-Ray-def}).

According to the definition
\eqref{eq:TautCurvedCP} the Chan-Paton data for these Interfaces are given by
\be\label{eq:SWP-CP}
\CE(\fS^{p}_{ij})_{kl} = \begin{cases} R_{ji}^* & (k,l) = (i,j) \\
\delta_{k,l} \IZ & {\rm else} \\
\end{cases}
\ee
\be\label{eq:SWM-CP}
\CE(\fS^{f}_{ij})_{kl} = \begin{cases} R_{ij} & (k,l) = (i,j) \\
\delta_{k,l} \IZ & {\rm else} \\
\end{cases}
\ee
\begin{figure}[htp]
\centering
\includegraphics[scale=0.35,angle=0,trim=0 0 0 0]{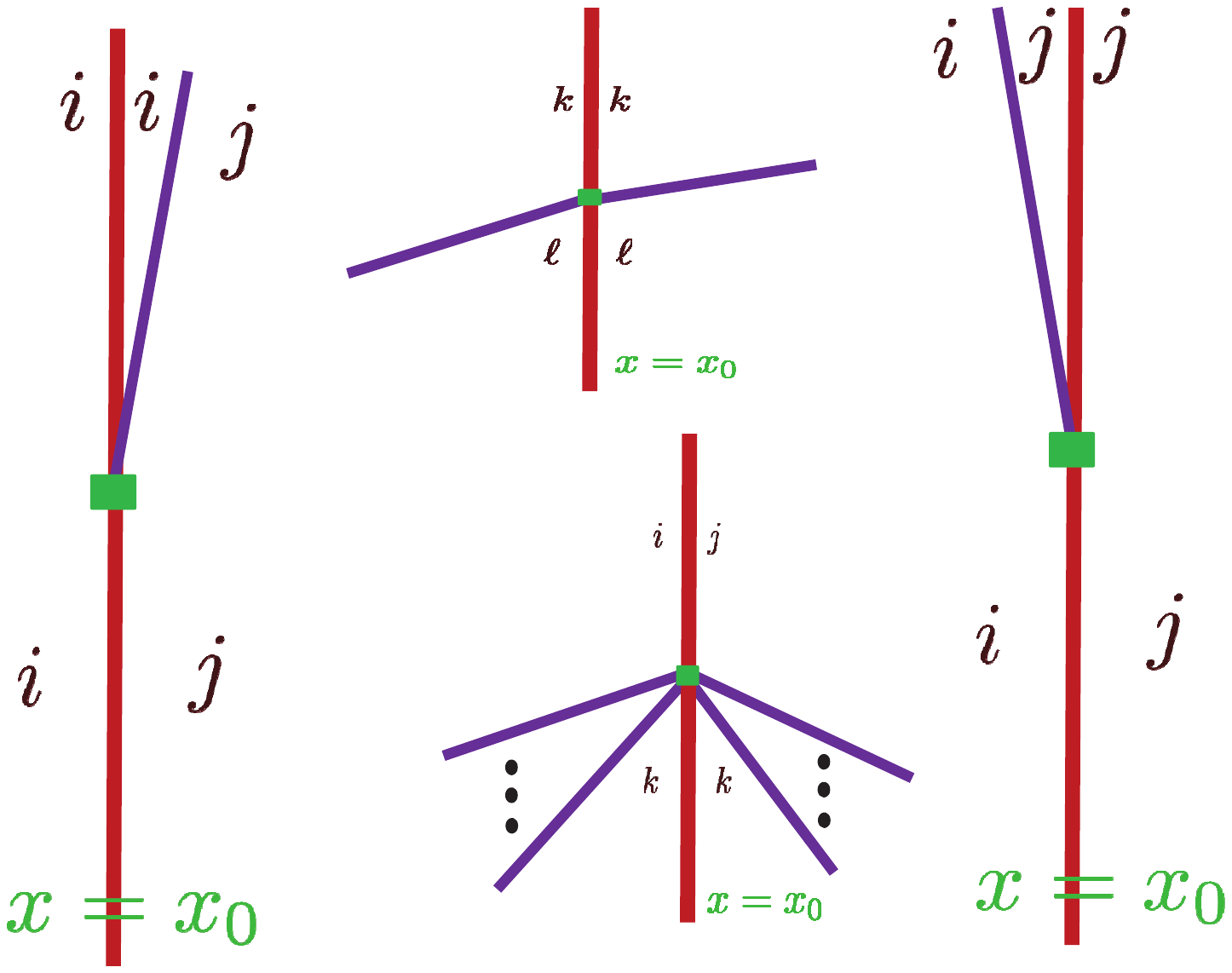}
\caption{This figure shows the nonzero components of the interface
amplitude for   $\fS^{p}_{ij}$. For every pair of vacua $(k,\ell)$ there
is an amplitude of the form shown in the top middle. The acute angle between
the lines in the negative and positive half-planes is $2\epsilon$. The left and right
figures show new amplitudes relative to the locally trivial case. The acute angle here is $\epsilon$.
The amplitudes for all the above three cases are given up to sign by $K^{-1}$, suitably interpreted.
See equations \protect\eqref{eq:LocallyTrivial-1}, \protect\eqref{eq:SpAmp-3}, and \protect\eqref{eq:SpAmp-4} for the
precise formulae.
The lower middle figure is a new amplitude associated to any interior vertex with an $ij$ edge
pointing to the future.
If $\beta_I$ is the interior amplitude associated with that vertex then the corresponding
interface amplitude is $(\check K \otimes 1)(\beta_I)$.  }
\label{fig:SWALL-AMP-PLUS}
\end{figure}
\begin{figure}[htp]
\centering
\includegraphics[scale=0.35,angle=0,trim=0 0 0 0]{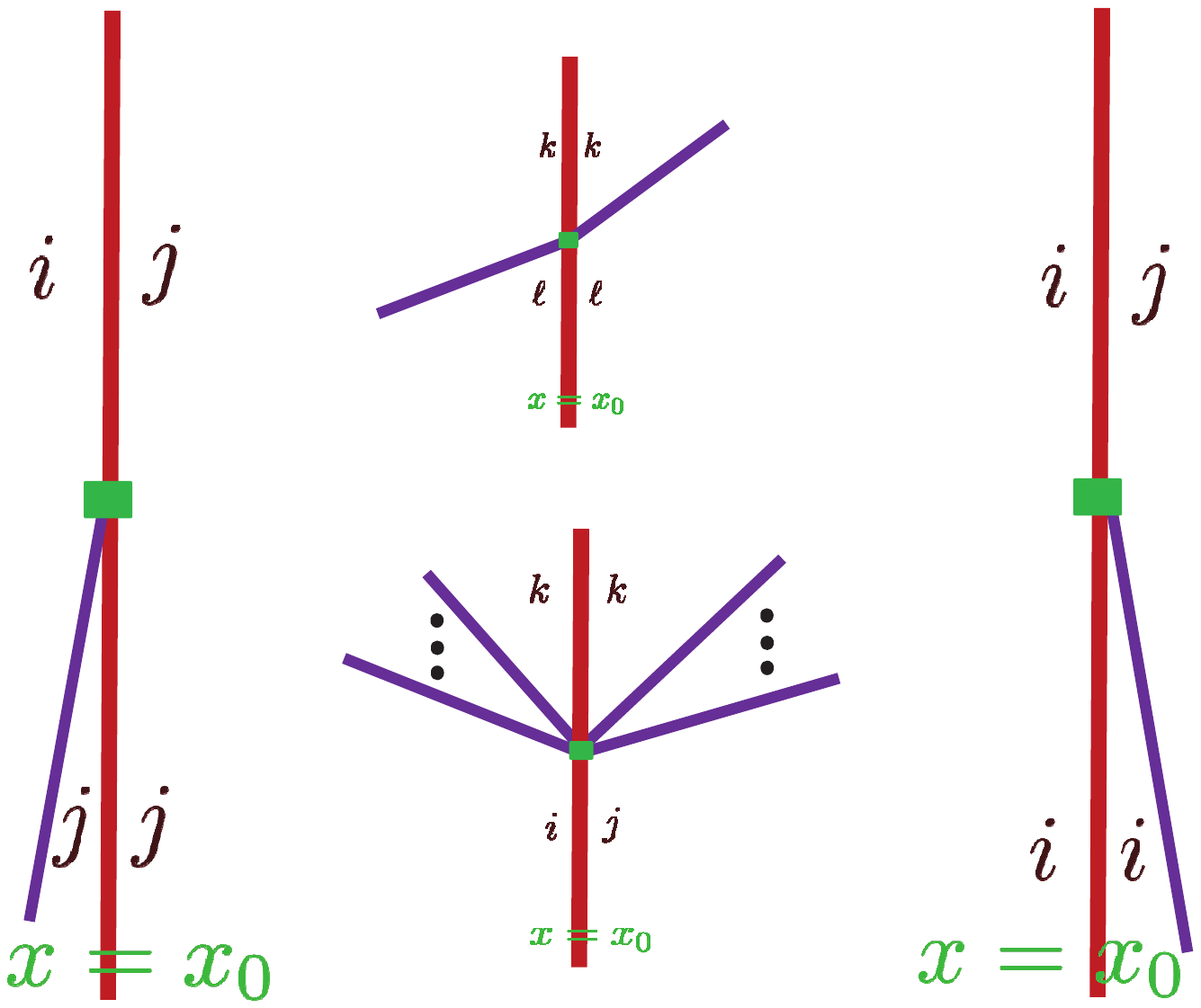}
\caption{This figure shows the nonzero components of the interface
amplitude for   $\fS^{f}_{ij}$ analogous to those for  $\fS^{p}_{ij}$.   }
\label{fig:SWALL-AMP-MINUS}
\end{figure}

The interface amplitudes have already
been described in Section \S \ref{subsubsec:DefiningAmplitudes} above. We summarize
the   nonzero amplitudes in Figure \ref{fig:SWALL-AMP-PLUS} for $\fS^{p}_{ij}$  and
in Figure \ref{fig:SWALL-AMP-MINUS} for $\fS^{f}_{ij}$.

\begin{figure}[htp]
\centering
\includegraphics[scale=0.25,angle=0,trim=0 0 0 0]{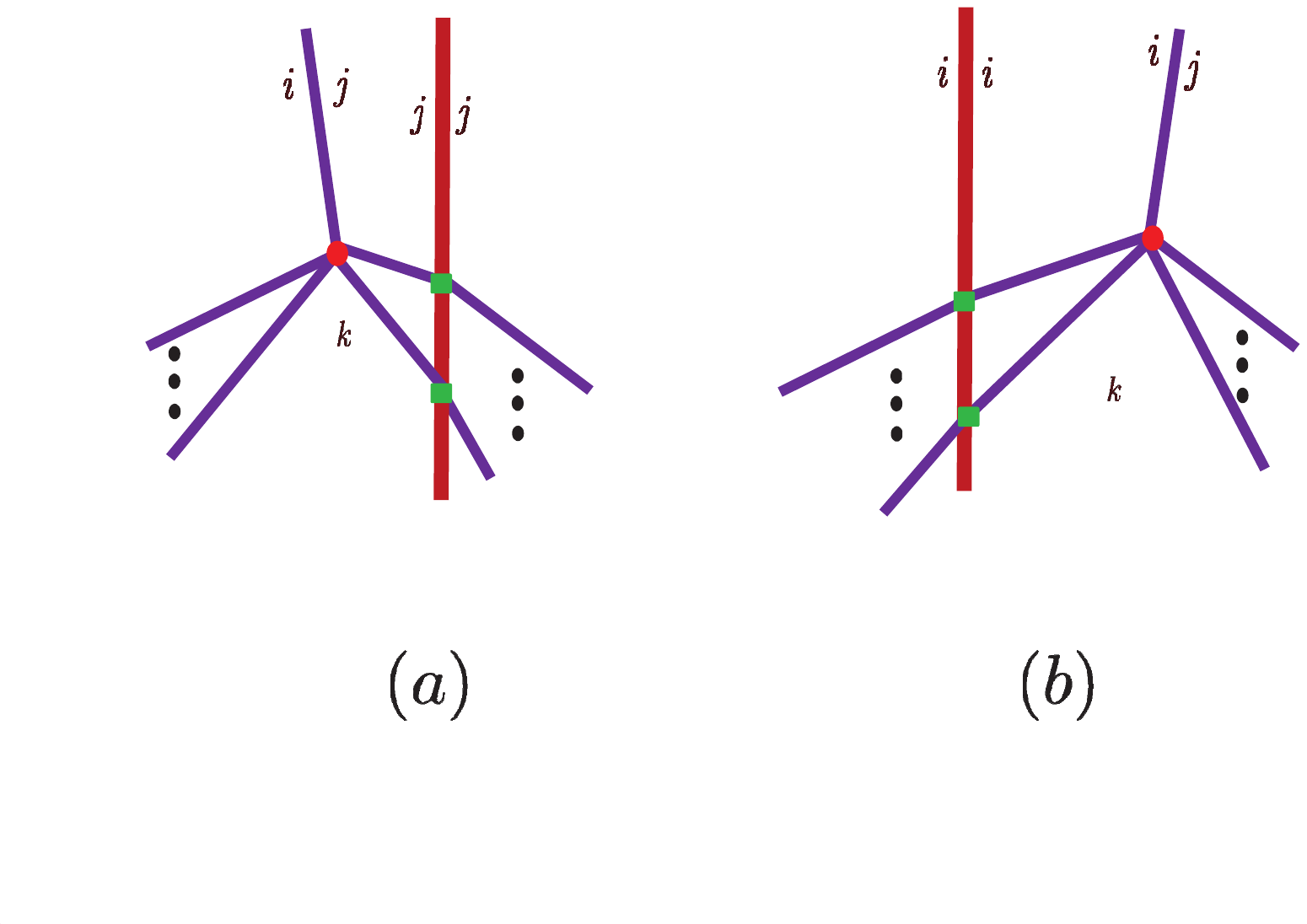}
\caption{These two contributions to the MC equation for $\fS^{p}_{ij}$ are analogous
to canceling contributions for $\fId$ but cannot cancel in this case because now the amplitudes
are valued in different spaces, as explained in the text.    }
\label{fig:SWALL-MC-1}
\end{figure}
\begin{figure}[htp]
\centering
\includegraphics[scale=0.25,angle=0,trim=0 0 0 0]{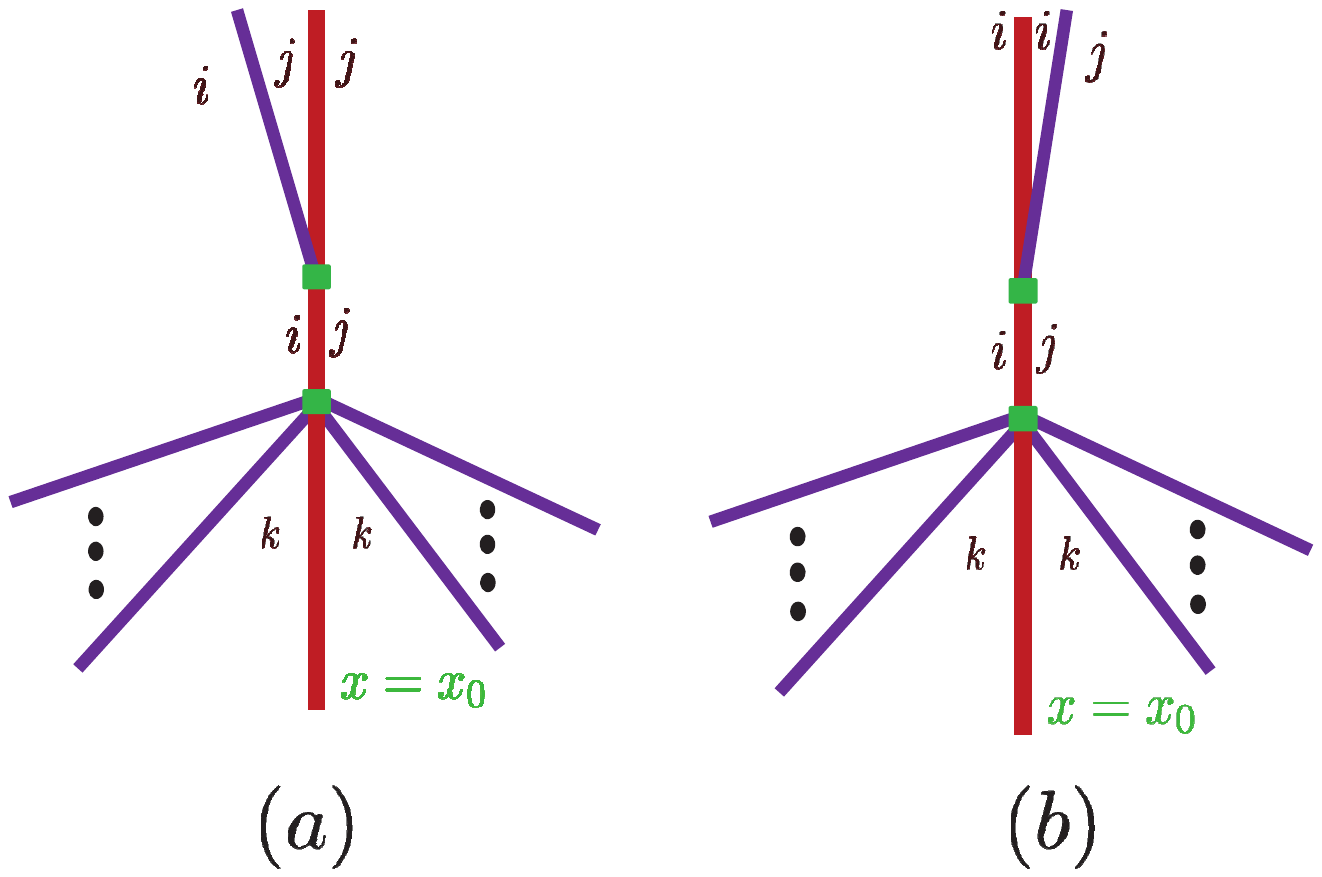}
\caption{Contributions (a) and (b) to the MC equation for $\fS^{p}_{ij}$
cancel those of Figure \protect\ref{fig:SWALL-MC-1}(a)
and Figure \protect\ref{fig:SWALL-MC-1}(b),
 respectively.   }
\label{fig:SWALL-MC-2}
\end{figure}
\begin{figure}[htp]
\centering
\includegraphics[scale=0.25,angle=0,trim=0 0 0 0]{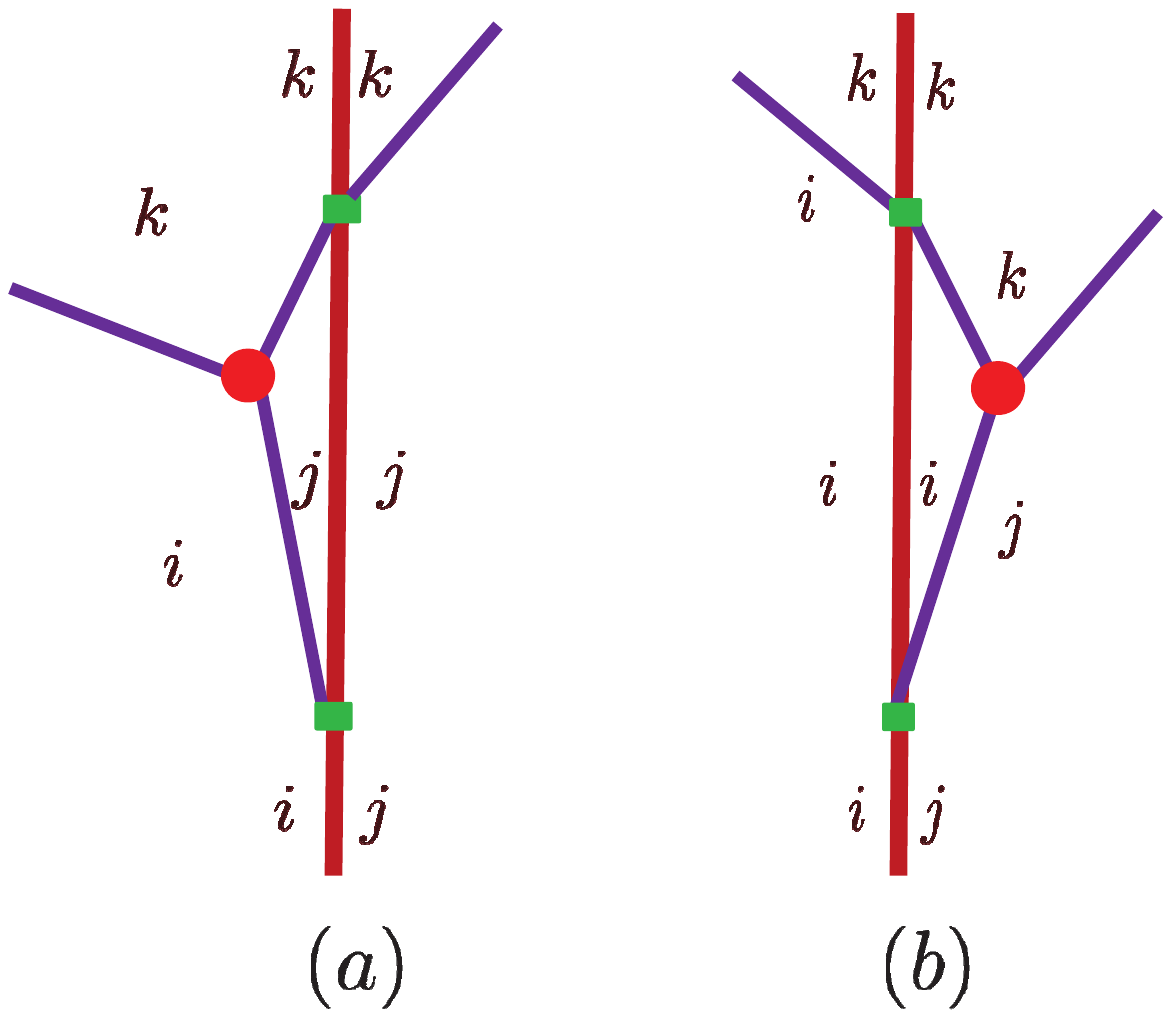}
\caption{Contributions (a) and (b) to the MC equation for $\fS^{p}_{ij}$
cancel.     }
\label{fig:SWALL-MC-3}
\end{figure}
\begin{figure}[htp]
\centering
\includegraphics[scale=0.25,angle=0,trim=0 0 0 0]{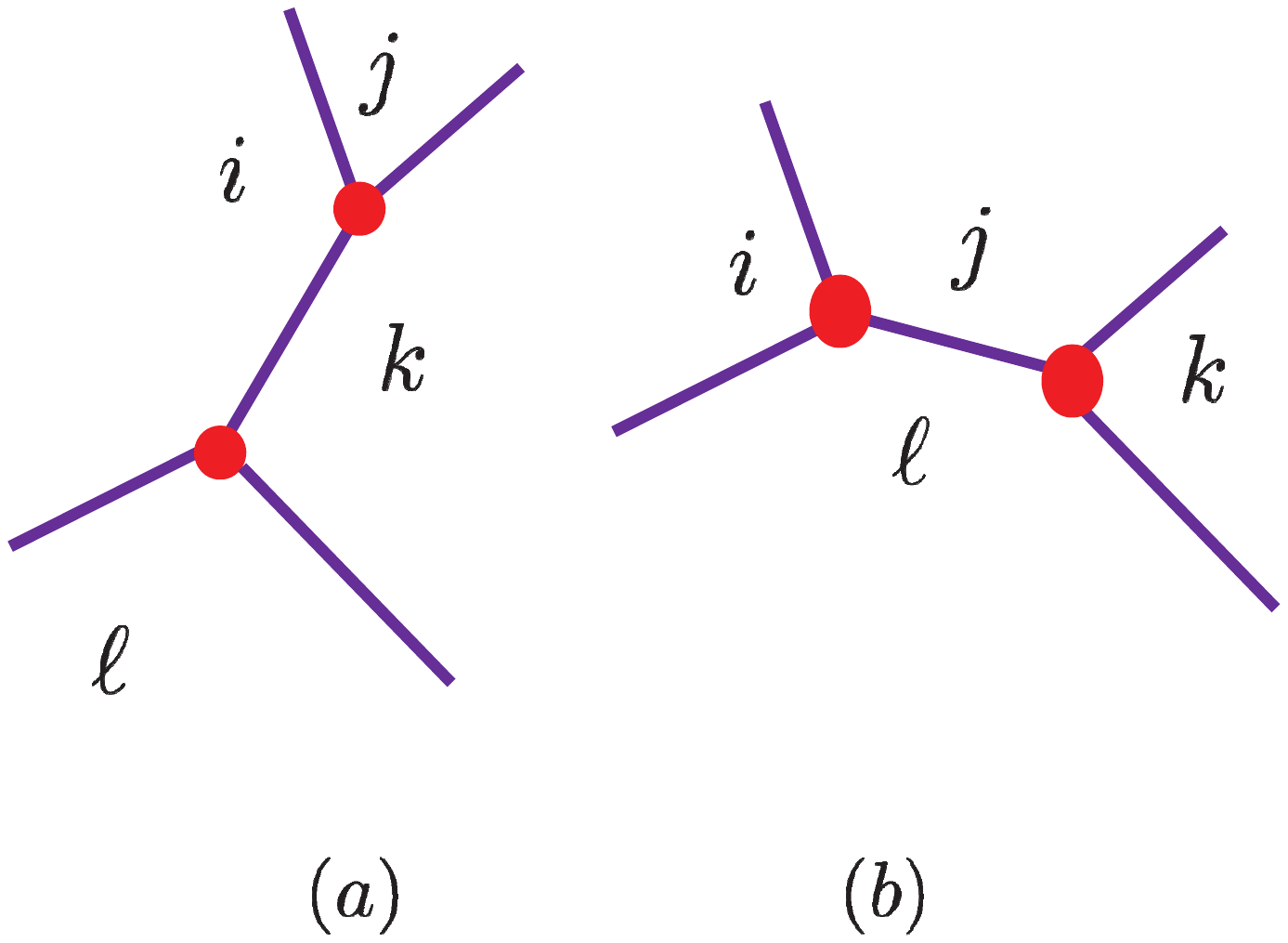}
\caption{ A cancelling pair in the Maurer-Cartan equation for the
Theory $\CT^{\vartheta_{ij}-\epsilon}$.    }
\label{fig:SWALL-MC-5}
\end{figure}
\begin{figure}[htp]
\centering
\includegraphics[scale=0.25,angle=0,trim=0 0 0 0]{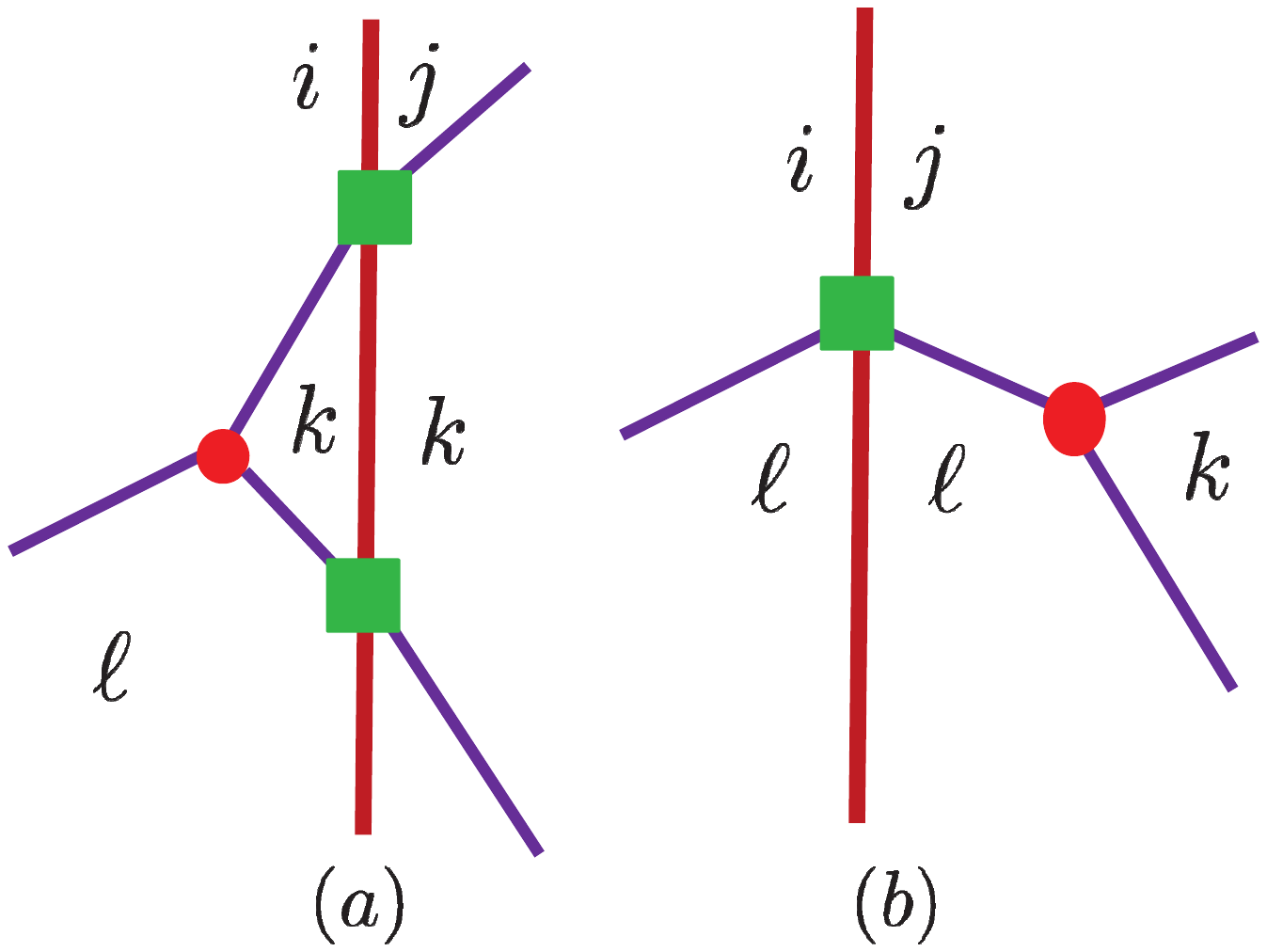}
\caption{A component of the $L_\infty$ MC equation of Figure \protect\ref{fig:SWALL-MC-5}
has a corresponding component for the $A_\infty$ MC equation for $\fS^p_{ij}$ shown here.     }
\label{fig:SWALL-MC-4}
\end{figure}

It is instructive to check explicitly the claim
that the interface amplitude satisfies the Maurer-Cartan equation \eqref{eq:CrvdIntAmp}.
We will do so for $\fS^p_{ij}$, and the check for $\fS^f_{ij}$ is very similar.
For webs not involving $ij$ lines the check is identical to the verification of the
Maurer-Cartan equation for the identity Interface. However, there are some new taut
webs that arise because $\CE_{ij}$ is nonzero and because there are new vertices
involving $ij$ lines.

First consider interior vertices with an external $ij$ line that goes to the
future. When this vertex is ``moved'' through the Interface the lines rotate
and we obtain taut interface webs such as those shown in
Figure \ref{fig:SWALL-MC-1}. The amplitude for Figure \ref{fig:SWALL-MC-1}(a)
is valued in $\CE_{ii}\otimes \widehat R^+_{ik}\otimes \CE_{kk}^* \otimes \widehat R^-_{ki}$
while that for Figure \ref{fig:SWALL-MC-1}(b) is valued in
$\CE_{jj}\otimes \widehat R^+_{jk}\otimes \CE_{kk}^* \otimes \widehat R^-_{kj}$.
The extra rotation, compared to the identity Interface, has led to amplitudes valued in
different spaces which therefore cannot cancel.  However, thanks to the ``new'' interface amplitudes
for $\fS^{p}_{ij}$ there are also two new taut webs shown in Figure \ref{fig:SWALL-MC-2}(a)
and Figure  \ref{fig:SWALL-MC-2}(b).  The amplitude for Figure \ref{fig:SWALL-MC-1}(a)
cancels that for Figure \ref{fig:SWALL-MC-2}(a) and similarly the amplitude
for Figure \ref{fig:SWALL-MC-1}(b) cancels that for Figure \ref{fig:SWALL-MC-2}(b).
In fact, demanding this cancellation gives the derivation of the basic amplitudes
\eqref{eq:SpAmp-3} and \eqref{eq:SpAmp-4} above.

Next, consider interior vertices of $\CT^{\vartheta_{\ell}}$ with an external $ij$ line that goes to the
past. Some new taut webs constructed with such a vertex in either half-plane
are shown in Figure \ref{fig:SWALL-MC-3} (for the case of a trivalent vertex).
Working patiently and carefully with all the sign conventions we have explained one can check that these
two amplitudes do indeed cancel.

Finally, recall that in the identity for the taut planar
element $\ft_{pl} * \ft_{pl}=0$ the terms canceled in pairs. For each such pair
involving a vertex with an $ij$ edge we can construct two taut webs.
For example, consider the component of the Maurer-Cartan equation
shown in Figure \ref{fig:SWALL-MC-5}. This is an identity for the interior
amplitudes $\beta_I$. There is a corresponding pair of taut interface webs
shown in  Figure \ref{fig:SWALL-MC-4}. Since the vertices with $ij$ lines
in Figure \ref{fig:SWALL-MC-4} are defined by $\check K(\beta)$ the
$A_\infty$ Maurer-Cartan equation for the Interface will be satisfied
if $\beta$ satisfies the corresponding $L_\infty$ Maurer-Cartan equation.
This completes the verification of the Maurer-Cartan  equation for the Interface
$\fS^p_{ij}$.

\begin{figure}[htp]
\centering
\includegraphics[scale=0.25,angle=0,trim=0 0 0 0]{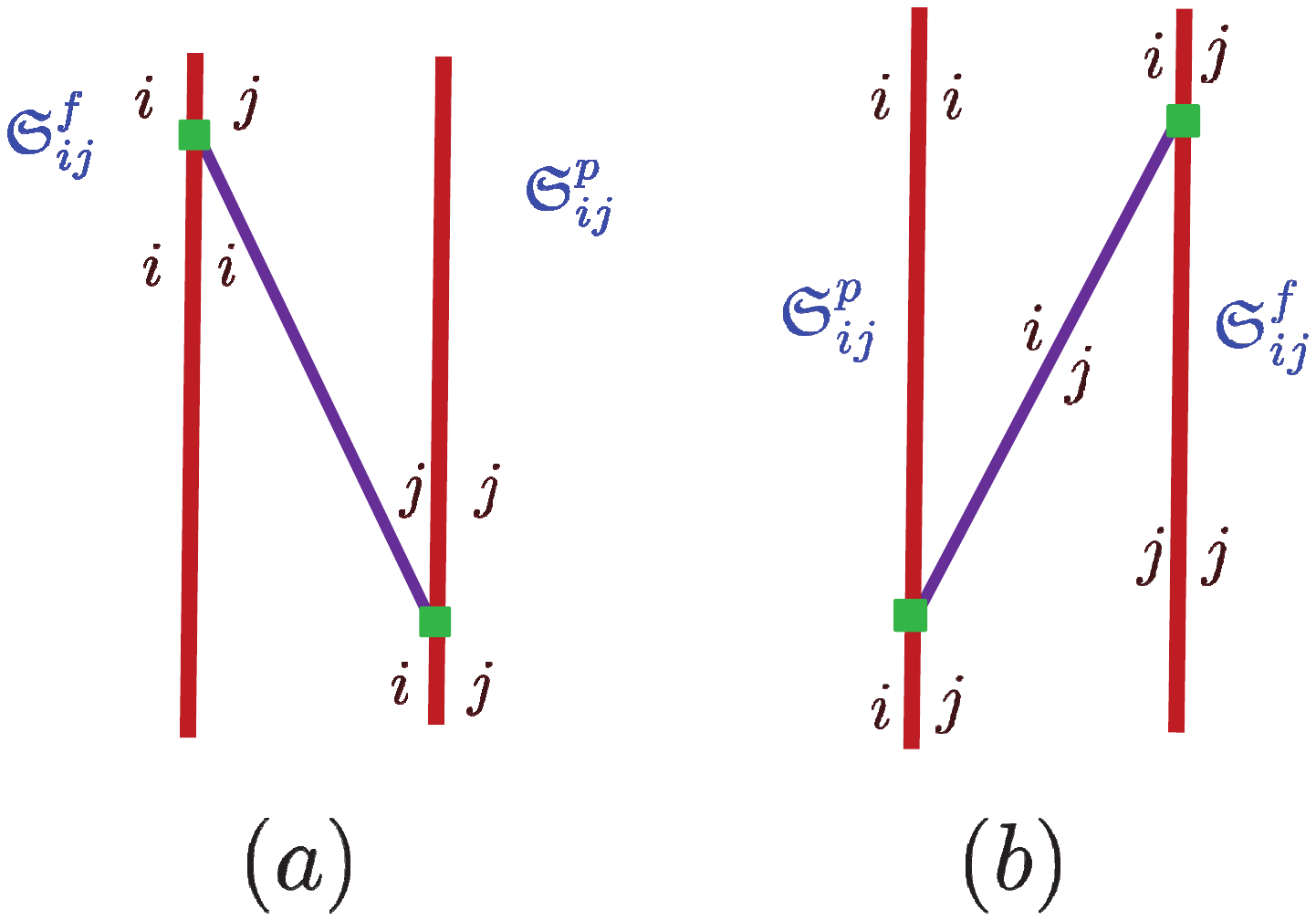}
\caption{The two composite webs shown here lead to differentials
on the $\CE_{ij}$ Chan-Paton space of $\fS^{f}_{ij}\boxtimes \fS^p_{ij}$
and $\fS^{p}_{ij}\boxtimes \fS^f_{ij}$, respectively. In each case there is
a chain homotopy of the identity morphism to zero so that the cohomology
of $\CE_{ij}$ vanishes.    }
\label{fig:SPL-SMN}
\end{figure}

The general arguments we gave in Section \S \ref{subsubsec:Verification} imply
that if we compose past and future S-wall Interfaces the result is homotopy
equivalent to the identity Interface. Nevertheless, it is instructive to
examine this homotopy equivalence in some detail and we turn to this next. Let
\be
\CT^+ = \CT^{\vartheta_{ij}+\epsilon} \qquad \qquad \CT^- = \CT^{\vartheta_{ij}-\epsilon}
\ee
\be
\fI^{fp} = \fS^f_{ij}\boxtimes \fS^p_{ij}\in \fB\fr(\CT^+,\CT^+)\qquad \qquad \fI^{pf} = \fS^p_{ij}\boxtimes \fS^f_{ij}\in \fB\fr(\CT^-,\CT^-)
\ee
Then we claim that
\be\label{eq:TwoComp-1}
\fI^{fp} \sim \fId_{\CT^+} \qquad\& \qquad  \fI^{pf} \sim \fId_{\CT^-}.
\ee

The Chan-Paton data of the Interfaces $\fI^{fp}$ and $\fI^{pf}$ are the
same and are  easily computed
from equations \eqref{eq:Comb-CP}, \eqref{eq:SWP-CP}, and \eqref{eq:SWM-CP},
with the result
\be\label{eq:SWPM-CP}
\CE( \fI^{fp} )_{kl} = \CE( \fI^{pf} )_{kl} = \begin{cases} \CE^f_{ij} \oplus \CE^p_{ij} & (k,l) = (i,j) \\
\delta_{k,l} \IZ & {\rm else}. \\
\end{cases}
\ee
Here we have denoted $\CE^f_{ij}:= \CE(\fS^f_{ij})_{ij} \cong R_{ij}$ and
$\CE^p_{ij} := \CE(\fS^p_{ij})_{ij} \cong R_{ji}^*$.
The first check of a homotopy equivalence to the identity Interface is that the
cohomology of the Chan-Paton space of type $ij$ should vanish. In fact, the
interface product $\boxtimes$ leads to a nontrivial differential associated with the
taut composite webs shown in Figure \ref{fig:SPL-SMN}. We explain this
in detail for $\fI^{fp}$. The taut composite web of Figure \ref{fig:SPL-SMN}(a)
gives an amplitude valued in
\be
\CE(\fI^{fp})_{ij}\otimes \CE(\fI^{fp})_{ij}^* \cong  \End(\CE(\fI^{fp})_{ij}).
\ee
Now the endomorphisms of $\CE(\fI^{fp})_{ij}$ can be organized into block matrices
using the summands $R_{ij}$ and $R_{ji}^*$ so that
elements are valued in the block matrix:
\be
\begin{pmatrix} \End(R_{ij}) & \Hom(R_{ij},R_{ji}^*) \\
\Hom(R_{ji}^*, R_{ij}) & \End(R_{ji}^*) \\  \end{pmatrix}
\ee
With this understood, the amplitude defined by Figure \ref{fig:SPL-SMN}(a) is
\be\label{eq:diffl-1}
\CB(\fI^{fp})^{ij}_{ij} = \begin{pmatrix} 0 & 0 \\
(-1)^F K_{ij}^{-1}  & 0 \\  \end{pmatrix}
\ee
Note that $(-1)^F K_{ij}^{-1}= K^{\alpha\alpha'} v_{\alpha}\otimes v_{\alpha'} \in \Hom(R_{ji}^*, R_{ij})$ really has degree $+1$
since the matrix elements are only nonzero when $\deg(v_\alpha) + \deg(v_{\alpha'}) = +1$.
\footnote{
It is a nice and rather subtle exercise to check that the product Interface does indeed
satisfy the MC equation.}

Now, to exhibit a homotopy equivalence $\fI^{fp} \sim \fId$ we need to produce four morphisms:
\be
\begin{split}
\delta_1 & \in \Hop(\fI^{fp},\fId) \\
\delta_2 & \in \Hop(\fId,\fI^{fp}) \\
\delta_3 & \in \Hop(\fI^{fp},\fI^{fp} ) \\
\delta_4 & \in \Hop(\fId,\fId) \\
\end{split}
\ee
such that
\be\label{eq:Morph-eq-1}
M_2(\delta_1,\delta_2) = \textbf{Id}_{\fI^{fp}} + M_1(\delta_3)
\ee
\be\label{eq:Morph-eq-1}
M_2(\delta_2,\delta_1) = \textbf{Id}_{\fId} + M_1(\delta_4)
\ee

First of all, we take  $(\delta_1)^{kk}_{kk} = (\delta_2)^{kk}_{kk} = 1$ for all $k\in \IV$.
Now, because $\fI^{fp}$ has a CP space
\be
\CE_{ij} = \CE^f_{ij} \oplus \CE^p_{ij} \cong R_{ij} \oplus R_{ji}^*
\ee
the identity morphism $\textbf{Id}_{\fI^{fp}}$ has a component of type $(ij,ij)$,
namely, the identity transformation   $\textbf{Id}_{\CE_{ij}}\in \Hom(\CE_{ij})$ that
is impossible to produce   from $M_2(\delta_1, \delta_2)$. It is impossible
to produce $\textbf{Id}_{\CE_{ij}}$ simply because if
$\delta_1 \in \Hop(\fI^{fp},\fId_{\CT^+})$ has an $ij$ line in the future then
it must have a $jj$ line in the past but if
$\delta_2 \in \Hop(\fId_{\CT^+}, \fI^{fp})$ has an $ij$ line in the past it must
have an $ii$ line in its future. Therefore, no composition of $\delta_1$ and $\delta_2$
can product an element of $\Hom(\CE_{ij})$.
Therefore the $(ij,ij)$ component of $\textbf{Id}_{\fI^{fp}}$ must come from
$M_1(\delta_3)$. We choose $\delta_3$ to have only one nonzero component, valued in $\End(\CE_{ij})$,
and  given by
\be\label{eq:diffl-2}
(\delta_3)^{ij}_{ij} = \begin{pmatrix} 0 & (-1)^F K_{ji}  \\
0  & 0 \\  \end{pmatrix}
\ee
where we interpret $(-1)^F K_{ji}= (-1)^{v_{\alpha'}} K_{\alpha'\alpha} v_{\alpha'}^*\otimes v_{\alpha}^*
\in  \CE^p_{ij} \otimes (\CE^f_{ij})^* $.
Then, when computing $M_1(\delta_3)$ we meet
\be\label{eq:M1-cpt}
 \rho(\ft_\CI)(\CB, \delta_3) + \rho(\ft_\CI)(\delta_3, \CB).
\ee
Using $\CB^{ij}_{ij}$ above, and taking proper care of signs
\footnote{ Define the sign of the dual so that $v_{\alpha}^* \cdot v_{\beta} = \delta_{\alpha,\beta}$.}
one finds that the first summand in \eqref{eq:M1-cpt} gives $\textbf{Id}_{\CE^f_{ij}}$ and the
second gives $\textbf{Id}_{\CE^p_{ij}}$ so the sum gives the desired identity morphism on $\CE(\fI^{fp})_{ij}$.
 Thus, $\delta_3$ provides an explicit chain homotopy equivalence of the
identity morphism on $\CE_{ij}$ to the zero morphism.

Since $\delta_3$ is nonzero there are induced vertices of type $(ii,ij)$ and $(ij,jj)$ in
$M_1(\delta_3)$ but these can be cancelled against $M_2(\delta_1,\delta_2)$ by choosing
\be\label{eq:dlt-cpt-1}
(\delta_2)^{ii}_{ij} = + \textbf{Id}_{\CE^f_{ij}}
\ee
\be\label{eq:dlt-cpt-2}
(\delta_1)^{ij}_{jj} = - \textbf{Id}_{\CE^p_{ij}}
\ee
Now we must check that the other vertices of $\CB(\fI^{fp})$ do not
lead to new components of $M_1(\delta_3)$ or new components of
$M_2(\delta_1, \delta_2)$ or $M_2(\delta_2, \delta_1)$ which
would complicate the homotopy equivalence. There are several potential contractions which
would considerably complicate the discussion, but happily
 they all give zero because, after careful examination, they all involve
 contractions of   $(\CE^f_{ij})^*$ with $\CE^p_{ij}$ or $(\CE^p_{ij})^*$ with $(\CE^f_{ij})$,
and these contractions vanish. Now is easy to check that
\be
M_2(\delta_2, \delta_1) = \textbf{Id}_{\fId_{\CT^+}}
\ee
so we can take $\delta_4 =0$.

\begin{figure}[htp]
\centering
\includegraphics[scale=0.25,angle=0,trim=0 0 0 0]{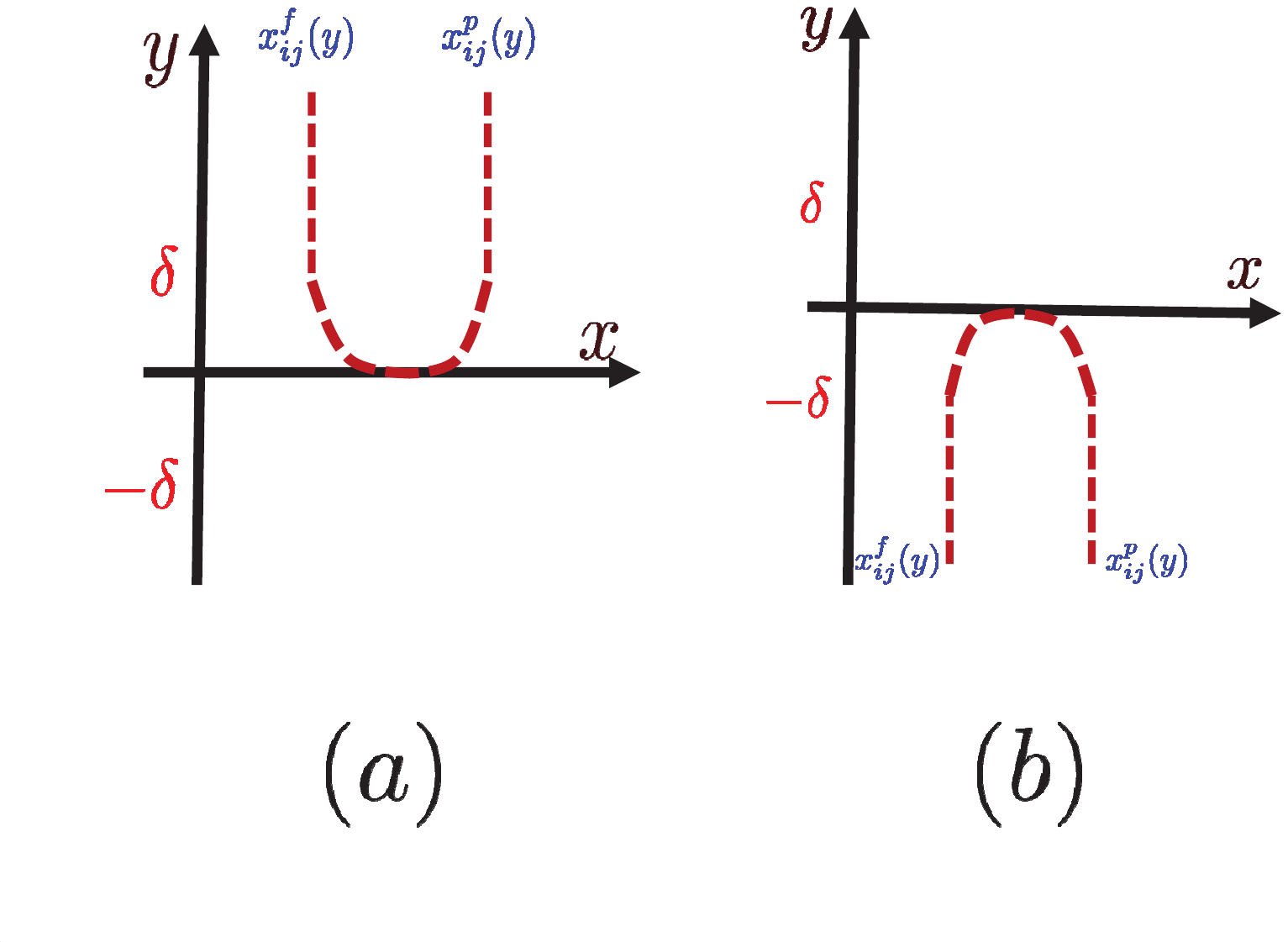}
\caption{The dashed curves show the
behavior of binding points under a homotopy between $\vartheta^{fp}(x)$
and the constant. In $(a)$  the homotopy $\vartheta(x,y)$ has the property that for $y\leq -\delta$
it is just the constant $\vartheta(x,y) = -\epsilon$ and for $y\geq \delta$ it is $\vartheta(x,y)=\vartheta^{fp}(x)$.
As $y$ decreases from $\delta$ to $0$, the binding points approach each other and annihilate.
In $(b)$ we show the location of the binding points for the time-reversed homotopy $\vartheta(x,-y)$.  }
\label{fig:ST-HOMEQV-2}
\end{figure}
\begin{figure}[htp]
\centering
\includegraphics[scale=0.25,angle=0,trim=0 0 0 0]{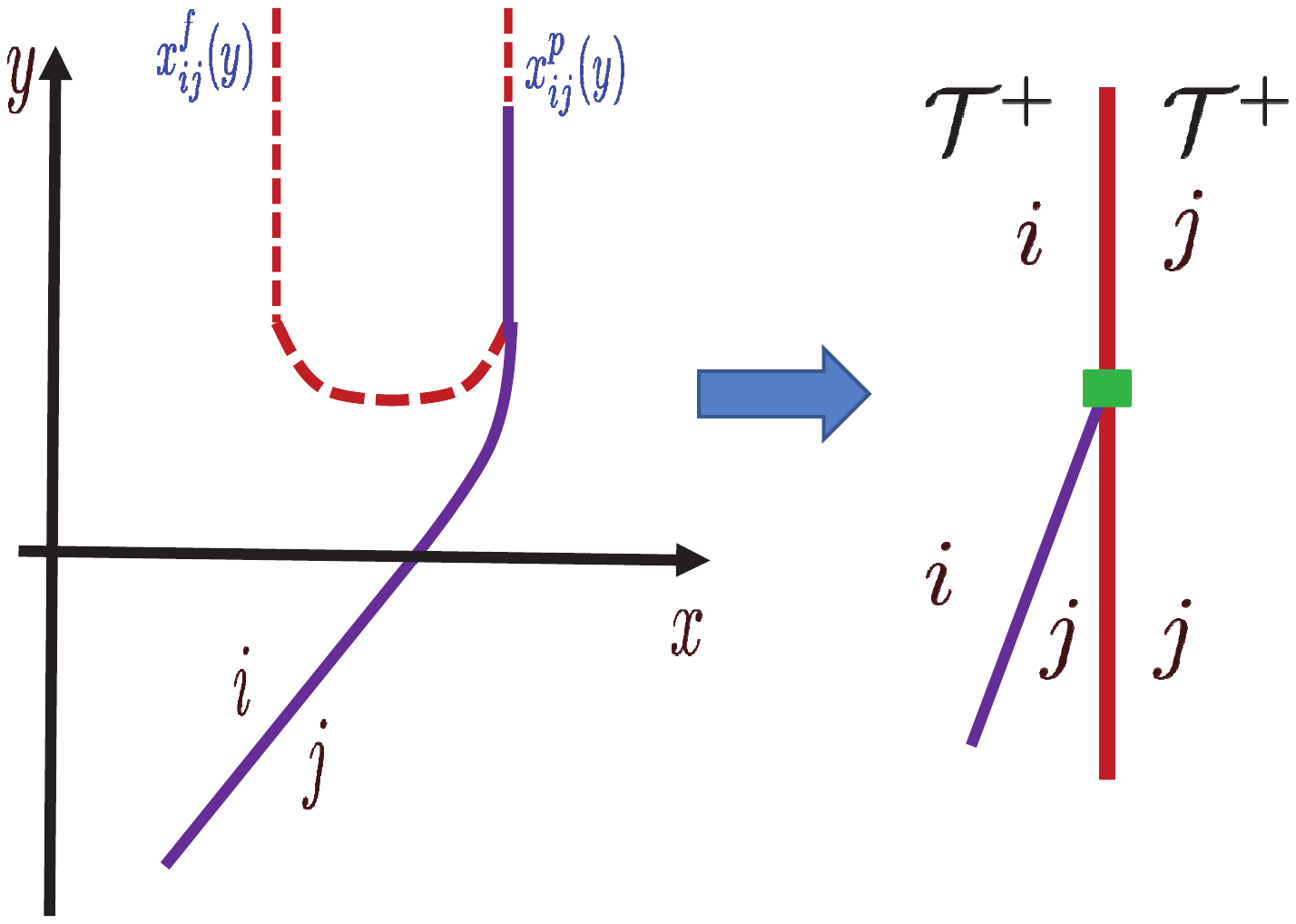}
\caption{The taut (= rigid) web on the left produces a space-time curved web
which contributes to the component $(\delta_1)^{ij}_{jj}$ in the morphism
describing the homotopy equivalence of $\fI^{fp}$ and $\fId_{\CT^+}$.   }
\label{fig:ST-HOMEQV-3}
\end{figure}
\begin{figure}[htp]
\centering
\includegraphics[scale=0.25,angle=0,trim=0 0 0 0]{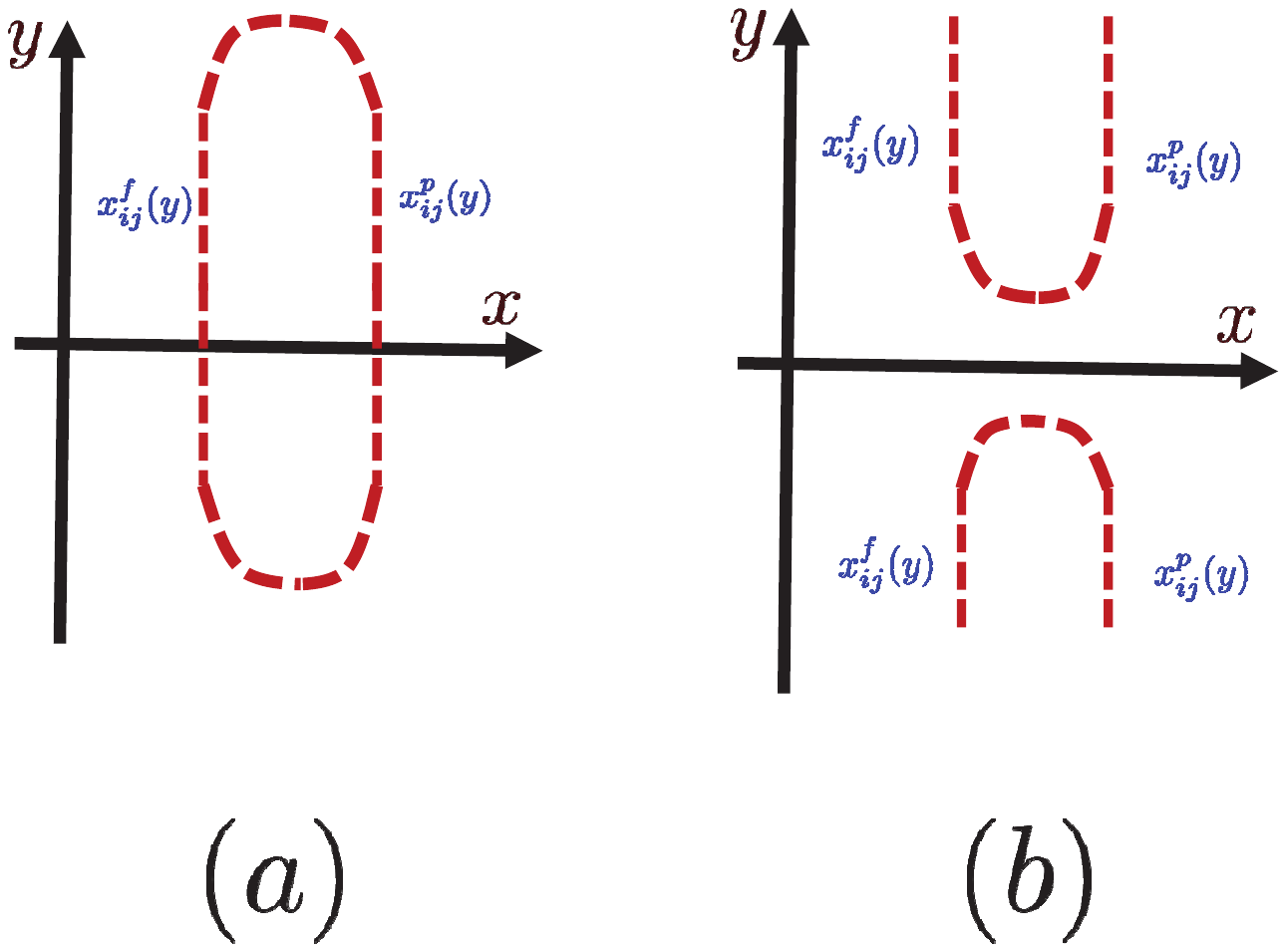}
\caption{A vacuum homotopy $\tilde \vartheta(x,y)$ used to compute
$M_2(\delta_2,\delta_1)$ leads to past and future binding points
 shown in Figure $(a)$. We do not expect
any exceptional webs when we consider a homotopy $\tilde \vartheta(x,y;s)$ to the
constant function.
On the other hand, a vacuum homotopy $\check \vartheta(x,y)$ used to compute
$M_2(\delta_1,\delta_2)$ leads to past and future binding points
 shown in Figure $(b)$. In the corresponding homotopy
$\check \vartheta(x,y;s)$ to the constant the two dashed curves must
start far apart, then merge and turn into two parallel dashed lines.
At some point $s=s_*$ there will be an exceptional web, illustrated in
Figure \protect\ref{fig:ST-HOMEQV-5} leading to the
morphism $\delta_3$.  }
\label{fig:ST-HOMEQV-4}
\end{figure}
\begin{figure}[htp]
\centering
\includegraphics[scale=0.25,angle=0,trim=0 0 0 0]{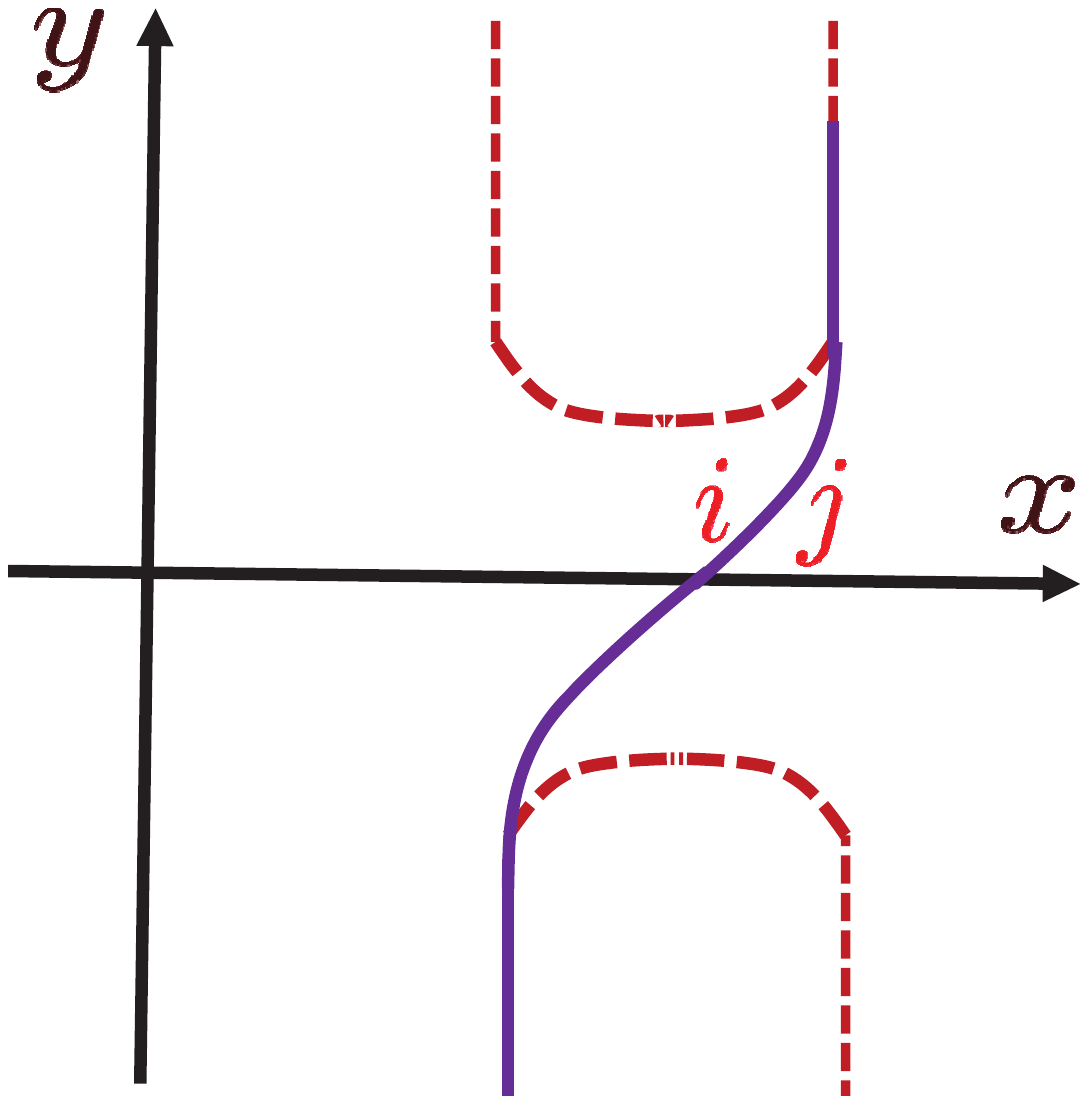}
\caption{When the concatenation time $\circ_T$ is positive it is impossible
to draw an $ij$ line such as that shown here. In the homotopy $\check\vartheta(x,y;s)$
there will be a critical value of $s$ where an $ij$ line of the type shown here
will exist. This is an exceptional web $\fe$ that leads to the morphism $\delta_3$.  }
\label{fig:ST-HOMEQV-5}
\end{figure}

Finally, it is instructive to see how the general argument of
Section \S \ref{subsubsec:Verification} produces the above explicit
homotopy equivalence of $\fI^{fp}$ with the identity Interface.
Let $\vartheta^{fp}(x)$ define a vacuum homotopy corresponding
to $\fI^{fp}$. Thus, on some interval, $z_{ij}(x)$ rotates from
$\I e^{-\I \epsilon}$ to $\I e^{+\I \epsilon}$ and then back to
$\I e^{-\I \epsilon}$, where $\epsilon>0$. Let $\vartheta(x,y)$
be a homotopy of $\vartheta^{fp}(x)$ to the constant path, so
that, for $y\geq \delta$ we have $\vartheta(x,y)=\vartheta^{fp}(x)$
and for $y\leq - \delta$ we have $\vartheta(x,y) = \epsilon$.
For example, as $y$ decreases from $\delta$ to $-\delta$, the
path $z_{ij}(x)$ could rotate more and more slowly so that at $-\delta$
it becomes constant. See Figure \ref{fig:ST-HOMEQV-2} for an illustration
of how the future and past $ij$ binding points evolve. Note that at some
intermediate time, say $y=0$ the vacuum weight $z_{ij}(x;y)$, as a function
of $x$ fails to rotate past the positive imaginary axis. Then the past and future
binding points annihilate.

As described in Section \S \ref{subsubsec:Verification}, the
morphism $\delta_1$ can be constructed from  $\rho^0_{\beta}(\ft_{st})$
where $\ft_{st}$ is the taut element in the space-time curved webs
described by $\vartheta(x,y)$.
In particular, the nontrivial component of equation \eqref{eq:dlt-cpt-2}
arises from the taut (= rigid) web shown in Figure \ref{fig:ST-HOMEQV-3}.
One can similarly derive the nontrivial component \eqref{eq:dlt-cpt-1}
for $\delta_2$. Finally, when we compute $M_2(\delta_1, \delta_2)$
and $M_2(\delta_2, \delta_1)$ we should use equation \eqref{eq:mult-homtpy},
valid for concatenations $\circ_T$ with a large time interval $T$. In the case
of $M_2(\delta_2,\delta_1)$ we have a spacetime configuration with
binding points evolving as in Figure \ref{fig:ST-HOMEQV-4}(a). This
can be homotoped to the constant function without producing any exceptional
webs. On the other hand, when
computing $M_2(\delta_1,\delta_2)$ we use Figure \ref{fig:ST-HOMEQV-4}(b).
In the homotopy $\check \vartheta(x,y;\epsilon)$ to the function
$\vartheta^{fp}(x)$ the dahsed line of binding points merges and then turns
into two parallel dashed lines. When the concatenation time $T$ is positive
it is impossible to draw an $ij$ line, and when the two components are too close
it is again impossible to draw an $ij$ line. At a critical value $s_*$
there will be an exceptional web $\fe$ such as that shown in Figure \ref{fig:ST-HOMEQV-5}
  leading to $\delta_3 = \rho^0_{\beta}(\fe)$,
and producing an amplitude of the type \eqref{eq:diffl-2}.

%In fact, we can go further and produce the explicit homotopy equivalence
% between $\fS^{p}_{ij} \IntfcTimes \fS^{f}_{ij}$ and the identity $\fId$ by taking some clues from
%a space-time dependent geometry. It has scalar components which map $\CE_{ii}$ of the two into each other and $\CE_{ij}$ to zero.
%The other components consist of a one-valent vertex with an $ij$ or $ji$ external edge, with the map
%$\CE_{ii} \otimes R_{ij} \otimes \CE_{ij}^*$ being the identity in the second factor of $\CE_{ij}$ and
%the map $\CE_{ij} \otimes R_{ji} \otimes \CE_{ii}^*$ being the identity in the first factor of $\CE_{ij}$.
%
%\cg{this paragraph is hard to understand. It should be possible to write out the
%explicit homotopy equivalence. }
%\cg{More figures}

\subsection{Categorification Of Framed Wall-Crossing}\label{subsec:Cat-FramedWC}

As we have mentioned, the Interfaces $\fS^{p,f}_{ij}$ may be regarded as a categorification of the ``$S$-factors''
which play an important role in the theory of spectral networks
\cite{Gaiotto:2011tf,Gaiotto:2012rg,Gaiotto:2012db,MOORE_FELIX}. The relation to
spectral networks is discussed in more detail in Section \S \ref{subsec:CatSpecNet} below.
In order to recover wall-crossing formulae for \emph{framed} BPS states
we  replace $R_{ij}$ by its Witten index $\mu_{ij}$ to obtain two-dimensional
soliton BPS counts. For an Interface $\fI^{-,+}$ we define the \emph{framed BPS degeneracies}
to be the Witten indices of the Chan-Paton factors
\be\label{eq:FramedBPS-def}
\fro(\fI^{-,+}, ij'):= \Tr_{\CE( \fI^{-,+})_{ij'}} (-1)^{F}
\ee
To compare with \cite{Gaiotto:2011tf,Gaiotto:2012rg,Gaiotto:2012db,MOORE_FELIX}, note that the role
of the line defect is played by $\fI^{-,+}$ and the IR charge, usually denoted $\gamma_{ij'}$  is
here simply the pair $ij'$.

To illustrate the relation to wall-crossing let us consider a vacuum homotopy
that crosses and $S_{ij}$-wall.  For $x_1<x_2$
define $\fI[x_1,x_2]$ to be $\fI[\vartheta(x;x_1,x_2)]$ where
\be
\vartheta(x;x_1,x_2) = \begin{cases} \vartheta(x_1) &   x \leq x_1 \\
\vartheta(x) &  x_1 \leq x \leq x_2 \\
\vartheta(x_2) & x \geq x_2 \\
\end{cases}
\ee
The family of interfaces satisfies $\fI[x_1,x_3] \sim \fI[x_1,x_2]\IntfcTimes \fI[x_2,x_3]$ for
$x_1<x_2<x_3$. In particular, if $x_{ij}$ is a binding point of type $ij$ then
\be\label{eq:CatSWC-1}
\fI[x_1, x_{ij} + \delta] \sim \fI[x_1,x_{ij}- \delta]\IntfcTimes \fI[x_{ij}-\delta,x_{ij} + \delta] =
\fI[x_1,x_{ij}- \delta]\IntfcTimes \fS^{p,f}_{ij}
\ee
where we choose $\fS^{p,f}_{ij}$ depending on whether $x_{ij}$ is past or future stable, respectively.
This is the categorified $S$-wall-crossing formula.

If we consider the Chan-Paton data to be a matrix of complexes then we have the homotopy equivalence
of matrices of chain complexes:
\be\label{eq:CatSWC-2}
\CE( \fI[x_1, x_{ij} + \delta]) \sim
\CE(\fI[x_1,x_{ij}- \delta]) \CE(  \fS^{p,f}_{ij})
\ee
To relate this to the standard wall-crossing formula note that if we
take the Witten index of the matrix of Chan-Paton spaces we produce
the generating function for framed BPS degeneracies of the Interface:
\be
F[\fI[x_1,x_2] ] := {\rm Tr}_{\CE(\fI[x_1,x_2])}(-1)^F =    \sum_{k,\ell} \fro(\fI[x_1,x_2], k\ell) e_{k,\ell}.
\ee
This matrix-valued function will be continuous in $x_1,x_2$ for locally trivial transport, but when $x$ crosses a binding point
of type $ij$ we have framed wall-crossing formula:
\be\label{eq:CatSWC-3}
F \mapsto \begin{cases} F \cdot (\textbf{1} + \mu_{ij} e_{ij} ) &   x_{ij} \in \curlywedge_{ij} \\
F \cdot (\textbf{1} - \mu_{ij} e_{ij} ) &   x_{ij} \in \curlyvee_{ij} \\
\end{cases}
\ee
In the second line we have used the degree $-1$ isomorphism of $R_{ji}^*$ with $R_{ij}$ to
identify the Witten index of $R_{ji}^*$ with $-\mu_{ij}$. Equation \eqref{eq:CatSWC-3} is precisely the
framed wall-crossing formula of \cite{Gaiotto:2011tf}.

%
%\cg{Important paragraph. Explain better.}
%
%The interfaces $\fS_{ij}^\pm$ are a further categorifications of the $S$-wall factors of \cite{Gaiotto:2012rg,Gaiotto:2012db}.
%The $S$-wall factors controlled how the framed BPS degeneracies, i.e. the Witten index of Chan paton factors for branes and interfaces,
%jumped due to wall-crossing as $\vartheta$ was varied. The interfaces $\fS_{ij}^\pm$ and more generally $\fR[\vartheta, \vartheta']$
%give a full categorical description of the wall-crossing of framed BPS states as $\vartheta$ is varied.
%

\subsection{Mutations}\label{subsec:Mutations}

Categorical transport by  $\fS^{p,f}_{ij}$  makes contact with the
theory of mutations and exceptional collections in category theory.
The relation between mutations of exceptional collections and
D-branes in Landau-Ginzburg models has been discussed at length
in \cite{Zaslow:1994nk,Hori:2000ck,SeidelBook}. We briefly make
contact with these works.

In general there is no natural order on the set of vacua $\IV$ because
the vacuum weights $z_i$ are points in the complex plane. If we choose
a direction in the plane, say parallel to a complex number $\zeta$, then
that direction defines a height function on the plane and, so long as $\zeta$
is not parallel to any of the $z_{ij}$ for $i,j\in \IV$ we can order the
vacua by increasing (or decreasing) height. Note this is the same as the
condition that $\zeta$ is not orthogonal to any $S_{ij}$ ray. Considering
$\zeta$ to the normal direction to a half-plane, the corresponding thimbles $\fT_i$
can now be ordered:
\be
\fT_i < \fT_j \qquad \Leftrightarrow \qquad \Re(\zeta^{-1} z_{ij} ) > 0 .
\ee
Note that with such an ordering we can write
\be
\Hop(\fT_i, \fT_j ) =  \begin{cases} \widehat{R}_{ij} & \fT_i < \fT_j  \\
\widehat{R}_{ii}\cong \IZ & i = j \\
 0  &  \fT_i > \fT_j  \\  \end{cases}.
\ee
where $\widehat{R}_{ij}$ is defined with respect to a half-plane with
inward-pointing normal vector $\zeta$.
Since the thimbles generate
the category of Branes, they form what is known in category theory as an
\emph{exceptional collection}, and the vacuum category $\fVac(\CT,\CH)$ is
an \emph{exceptional category} as defined in Appendix \ref{app:cat} below.
%
%\cg{Well, the actual definition uses Ext groups, or hom in the
%derived category. Is this really exactly the same? }
%

Now consider rotating $\zeta$ or, equivalently, fix $\zeta = +1$, take  the positive half-plane $\CH^+$, and
consider a family of spinning weights $e^{-\I \vartheta(x)} z_i$. When $e^{\I \vartheta(x)}$
passes through an $S_{ij}$ wall   an ordered pair of thimbles $(\fT_i, \fT_j)$ exchanges
its ordering and the old exceptional collection is no longer an exceptional collection in the new Theory.
A natural question is whether one can construct a new exceptional collection of
Branes in the new Theory out of the old exceptional collection of the old Theory. The answer to this question
is ``yes,'' and the procedure is called a \emph{mutation}. We will explain how mutations
work in our formalism.

If our path of Theories  crosses a future stable $S_{ij}$ wall then
  interface product with the $S$-wall Interface $\fS^{f}_{ij}$ defines, as usual, an \afty-functor
$\fB\fr(\CT^{\vartheta_{ij}+\epsilon}) \rightarrow \fB\fr(\CT^{\vartheta_{ij}-\epsilon})$.
The original Theory, $\CT^{\vartheta_{ij}+\epsilon}$ has $\Re(z_{ij})>0$ so $\fT_i < \fT_j$.
So, the original exceptional collection is
\be\label{eq:Old-EC-1}
\dots, \fT_i, \fT_j, \dots \qquad\qquad {\rm Future \  stable}
\ee
and the final Theory, $\CT^{\vartheta_{ij}-\epsilon}$ has $\Re(z_{ij})< 0$.
Similarly, if the path of Theories crosses a past stable $S_{ij}$ wall then
we use $\fS^{p}_{ij}$ to define an \afty-functor
$\fB\fr(\CT^{\vartheta_{ij}- \epsilon}) \rightarrow \fB\fr(\CT^{\vartheta_{ij}+\epsilon})$.
The original Theory, $\CT^{\vartheta_{ij}-\epsilon}$ has $\Re(z_{ij})<0$ so $\fT_i > \fT_j$.
So, the original exceptional collection is
\be\label{eq:Old-EC-2}
\dots, \fT_j, \fT_i, \dots \qquad\qquad {\rm Past \  stable}
\ee
and the final Theory, $\CT^{\vartheta_{ij}+\epsilon}$ has $\Re(z_{ij})>0$.

\begin{figure}[htp]
\centering
\includegraphics[scale=0.25,angle=0,trim=0 0 0 0]{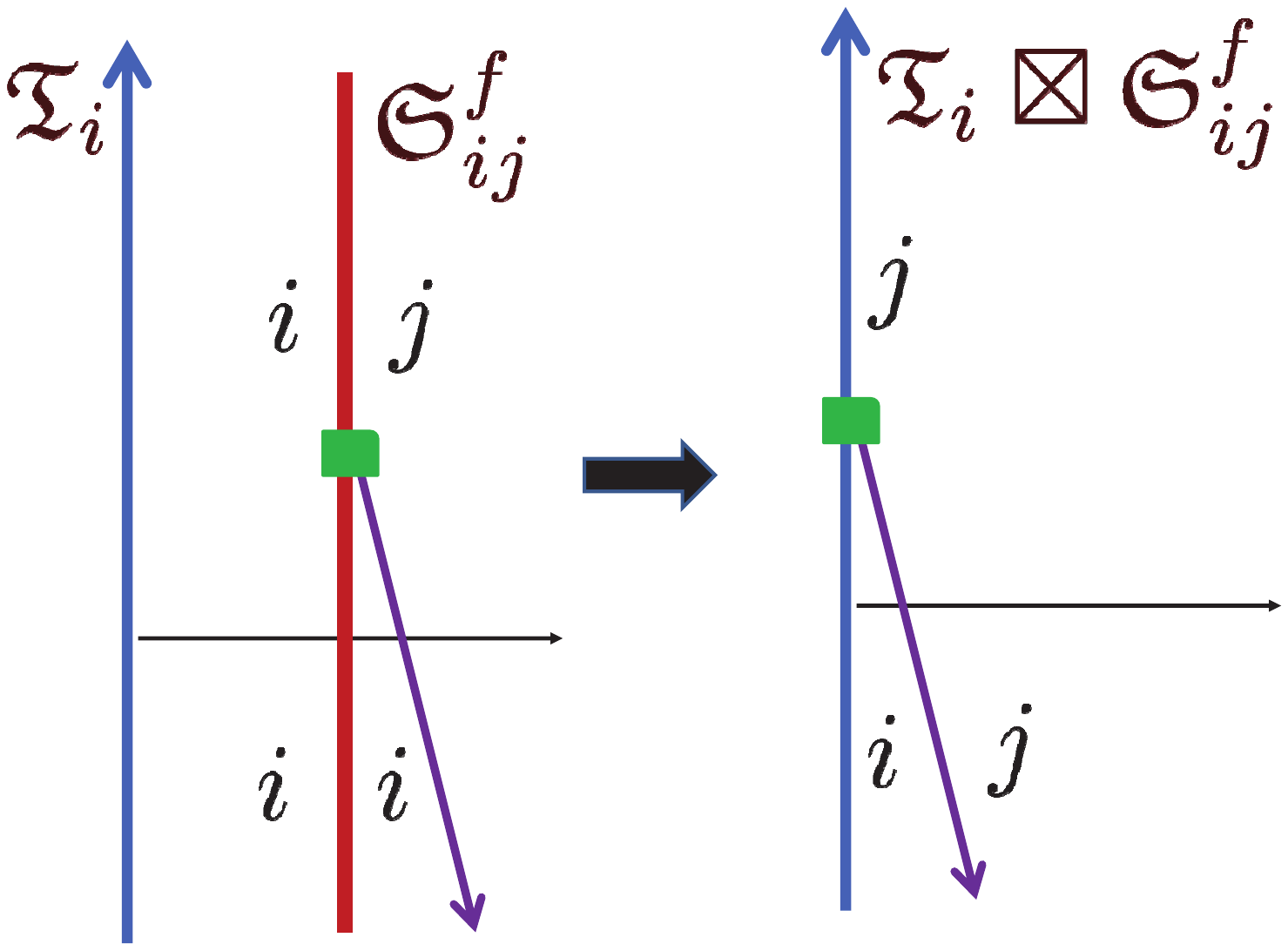}
\caption{The Brane $\fT_i\boxtimes \fS^f_{ij}$ has one nonvanishing
boundary amplitude. Under the isomorphism $\CE_i\otimes \widehat{R}_{ji}\otimes \CE_j^* \cong R_{ij}\otimes R_{ji}$ it is
$K_{ij}^{-1}$. }
\label{fig:MUTATION1}
\end{figure}
\begin{figure}[htp]
\centering
\includegraphics[scale=0.25,angle=0,trim=0 0 0 0]{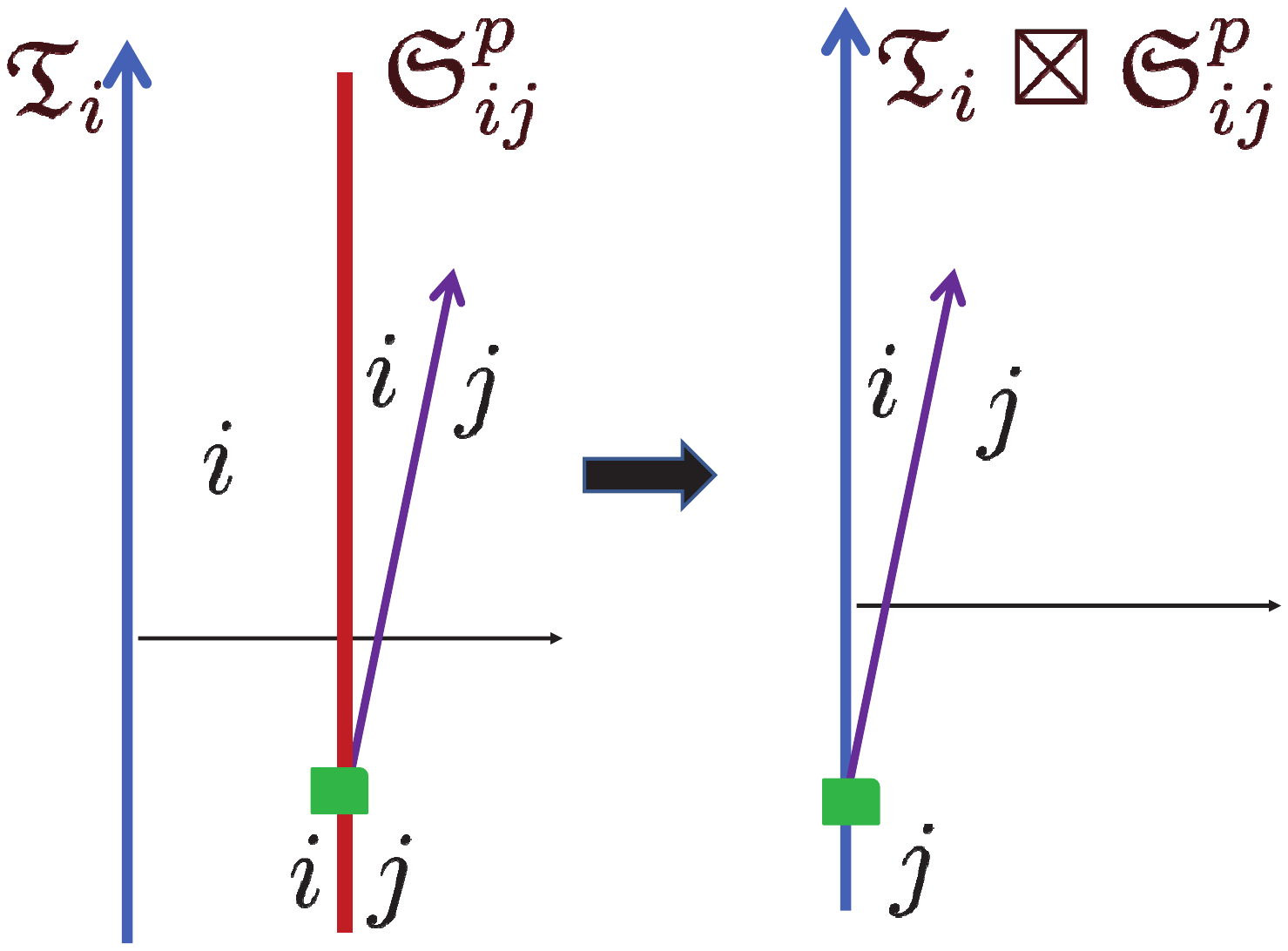}
\caption{The Brane $\fT_i\boxtimes \fS^p_{ij}$ has one nonvanishing
boundary amplitude. Under the isomorphism $\CE_i\otimes \widehat{R}_{ij}\otimes \CE_j^* \cong R_{ij}\otimes R_{ji}$ it is
$-(-1)^FK_{ij}^{-1}$. }
\label{fig:MUTATION2}
\end{figure}

Let us examine the action of the $S$-wall functors on the exceptional collections of the
original Theory in these two cases. It is easy to see that
\be
\fT_k \IntfcTimes \fS^{p,f}_{ij}  = \fT_k \qquad k \not= i
\ee
since none of the amplitudes in Figures \ref{fig:SWALL-AMP-MINUS} or \ref{fig:SWALL-AMP-PLUS}
can contract with $\fT_k$. On the other hand, $\fT_i \IntfcTimes \fS^{p,f}_{ij}$ is nontrivial.
Indeed we can compute the Chan-Paton factors:
\be
\begin{split}
\CE\left(\fT_k \IntfcTimes \fS^{p}_{ij}\right)_\ell &=  \CE(\fT_k)_{\ell}  \oplus \delta_{k,i} R_{ji}^* \otimes \CE(\fT_j)_{\ell} \\
\CE\left(\fT_k \IntfcTimes \fS^{f}_{ij}\right)_\ell &=  \CE(\fT_k)_{\ell}  \oplus \delta_{k,i} R_{ij} \otimes \CE(\fT_j)_{\ell} \\
\end{split}
\ee
Moreover, each of the Branes $\fT_i \IntfcTimes \fS^{p,f}_{ij}$ has a single nonvanishing boundary
amplitude, illustrated in Figures \ref{fig:MUTATION1} and \ref{fig:MUTATION2}.

Let us consider the case of crossing an $S_{ij}$ wall in the future-stable direction. We now
would like to introduce a new generating set of Branes, replacing the old exceptional collection
\eqref{eq:Old-EC-1} by the new collection of Branes
\be\label{eq:NewColl-1}
\dots, \fT_j, \fT_i\boxtimes\fS^f_{ij},\dots
\ee
In the language of Appendix \ref{app:cat} this corresponds to a left-mutation at $j$.
Similarly, when crossing an $S_{ij}$-wall in the past-stable direction we
would like to introduce a new generating set of Branes, replacing the old exceptional collection
\eqref{eq:Old-EC-2} by the new collection of Branes
\be\label{eq:NewColl-2}
\dots,  \fT_i\boxtimes\fS^p_{ij},\fT_j, \dots
\ee
In the language of Appendix \ref{app:cat} this corresponds to a right-mutation at $j$.
The idea is that the ``missing'' Brane $\fT_i$ can be expressed as a boundstate of
the Branes $\fT_j$ and $\fT_i\boxtimes\fS^{f,p}_{ij}$ by condensing local boundary
operators. One way of expressing this, often found in the literature, is to relate
the three Branes by an exact triangle. We explain this momentarily.

On general grounds, mutations of exceptional collections are expected to provide a representation of the braid group, up to homotopy.
Such a braid group representation is intimately connected to the theory of categorical wall-crossing we develop in Section \ref{sec:GeneralParameter}.

Before writing our exact triangles we first note that
the category of Branes forms a module over the category of $\IZ$-modules.
Given any $\IZ$-module $V$ and any Brane $\fB$ we define $V\otimes \fB$ to be the Brane which has
Chan-Paton spaces
\be
\CE(V\otimes \fB)_i := V\otimes \CE(\fB)_i
\ee
Then the amplitude of $V\otimes \fB$ is supposed to live in
\be
\oplus_{z_{ij}\in \CH}  \CE(V\otimes \fB)_i\otimes \widehat{R}_{ij} \otimes
\CE(V\otimes \fB)_j^*
= (V\otimes V^*)\otimes  \oplus_{z_{ij}\in \CH}  \CE(  \fB)_i\otimes \widehat{R}_{ij} \otimes
\CE(  \fB)_j^*
\ee
and   we   take the amplitude (up to an appropriate sign) to be ${\rm Id}_V \otimes \CB$ where $\CB$ is
the amplitude of $\fB$. The hom-spaces of the category of Branes satisfy
\be
\Hop(V_1\otimes \fB_1, V_2 \otimes \fB_2) = V_1\otimes \Hop(\fB_1,\fB_2)\otimes V_2^*.
\ee

Now, consider a path crossing an $S_{ij}$-wall in the future-stable direction
 and consider the triple of Branes $R_{ij}\otimes \fT_j, \fT_i, \fT_i \IntfcTimes \fS^{f}_{ij}$   We compute the hom-spaces
 for the positive half-plane in the new Theory $\CT^{\vartheta_{ij}-\epsilon}$:
\be\label{eq:Mut-HomSpce-1}
\Hop(R_{ij}\otimes \fT_j,\fT_i) = R_{ij}\otimes \widehat{R}_{ji}
\ee
\be\label{eq:Mut-HomSpce-2}
\Hop( \fT_i \IntfcTimes \fS^{f}_{ij} , \fT_i) = \Hop(i,i) \oplus R_{ij}\otimes \widehat{R}_{ji}
\ee
\be\label{eq:Mut-HomSpce-3}
\Hop(R_{ij} \otimes \fT_j, \fT_i \IntfcTimes \fS^{f}_{ij}) =R_{ij}\otimes \widehat{R}_{ji} \oplus  R_{ij}\otimes R_{ij}^*
\ee
Each of the summands above contains a canonical element. In \eqref{eq:Mut-HomSpce-1}
we have the element $K_{ij}^{-1}$,  in \eqref{eq:Mut-HomSpce-2} the first summand has the identity
and the second has $K^{-1}$, and
in    \eqref{eq:Mut-HomSpce-3} the first summand has $K^{-1}$ and the second summand
has the identity $\textbf{Id}_{R_{ij}}$.   Using the definition \eqref{eq:BraneMultiplications}
(contraction with the taut element) we can check the exact triangle of morphisms
\be\label{eq:S-minus-triangle}
\xymatrix{ R_{ij} \otimes \fT_j     & & \fT_i \ar[ll]^{\rm K^{-1} } \ar[ld]^{K^{-1} } \\
& \fT_i \IntfcTimes \fS^{f}_{ij} \ar[lu]^{\textbf{Id}_{R_{ij}} }  \\
}
\ee
is a commutative diagram.

Similarly, for $\fT_i \IntfcTimes \fS^{p}_{ij}$ we compute (again with $z_{ij}$ in the
positive half-plane)
\be
\begin{split}
\Hop(\fT_i, R_{ji}^* \otimes \fT_j) & = \widehat{R}_{ij} \otimes R_{ji}\\
\Hop(\fT_i, \fT_i \IntfcTimes \fS^{p}_{ij}) & = \Hop(i,i) \oplus   \widehat{R}_{ij}R_{ji} \\
\Hop(\fT_i \IntfcTimes \fS^{p}_{ij}, R_{ji}^* \otimes \fT_j) & =  \widehat{R}_{ij}\otimes R_{ji} \oplus R_{ji}^*\otimes R_{ji} \\
\end{split}
\ee
Once again, using suitable canonical elements from the summands we can construct the
exact triangle:
\be\label{eq:S-plus-triangle}
\xymatrix{ \fT_i & & R_{ji}^* \otimes \fT_j   \ar[ll]^{ K^{-1} } \ar[ld]^{\textbf{Id}_{R_{ji}^*}  } \\
& \fT_i \IntfcTimes \fS^{p}_{ij} \ar[lu]^{K^{-1} }  \\
}
\ee
These are the kinds of exact triangles that appear in discussions of mutations of exceptional collections
found in the literature.

When comparing with the general discussion of Appendix \ref{app:cat} we should note that
some of the factors $R_{ij}$ above should really be viewed as  Hop spaces $\widehat{R}_{ij}$.
However, near an $S_{ij}$-wall these two spaces can be identified.

\subsection{Categorical Spectrum Generator And Monodromy}\label{subsec:CatSpecGen}

Let us now return to \eqref{eq:Cat-Monod}.
We define $\fR[\vartheta,\vartheta-\pi]$ to be the \emph{categorical spectrum generator}.
The name is apt because in this case $\vartheta(x)$ has a unique future stable binding point for each
pair of vacua with $z_{ij}$ in a suitable half-plane. Taking $\vartheta$ to be small
 we can rewrite \eqref{eq:TautCurvedCP} as
\be\label{eq:TautCurvedCP-sg}
 \oplus_{j,j'\in \IV} \CE_{j,j'} e_{j,j'} := \bigotimes_{\Re(z_{ij})>0 }
  S_{ij}(x_{ij} )
\ee
where the ordering in the product from left to write is the clockwise ordering of the phases of $z_{ij}$.
If we consider the Witten index of this product we produce precisely the spectrum
generator as defined in \cite{Gaiotto:2009hg,Gaiotto:2011tf}. In the case of
2d Landau-Ginzburg models this is precisely the matrix $S$ defined long ago
by Cecotti and Vafa. (See equation (2.11) in  \cite{Cecotti:1992rm}.)

In our case we have the general result that
\be
\fR[\vartheta, \vartheta-2\pi] \sim  \fR[\vartheta, \vartheta-\pi]\IntfcTimes \fR[\vartheta-\pi, \vartheta-2\pi]
\ee
Examining the Chan-Paton factors for the Interface $\fR[\vartheta, \vartheta-2\pi]$ motivates the interpretation
of $\fR[\vartheta, \vartheta-2\pi]$ as a categorified  version of Cecotti and Vafa's ``monodromy''  $S S^{tr,-1}$
(which in turn is motivated by the monodromy of the cohomology of a Milnor fiber in singularity theory).
\footnote{We can also regard $\fR[\vartheta, \vartheta-2\pi]$ as a categorified  version of a Stokes matrix.
We will not pursue that very interesting direction in the present paper.}
Indeed, if $\vartheta(x) = - x$ for $x\in [0,2\pi]$ then all the $S$-walls are future stable and, for every
pair $ij$ with $i\not=j$ there will be precisely two future stable binding points $x_{ij}$ and $x_{ji}$
in the interval of length $2\pi$, and moreover
$\vert x_{ij} - x_{ji} \vert = \pi$. Again,   with a suitable choice of half-plane $\CH$ (or choosing
$\vartheta$ to be small) we can write the Chan-Paton
factors of $\fR[\vartheta, \vartheta-2\pi]$ as
\be
\cdots \otimes (\IZ \textbf{1} \oplus R_{ij} e_{ij} ) \otimes \cdots \otimes (\IZ \textbf{1} \oplus R_{ji} e_{ji} ) \otimes \cdots
= \IS \otimes \IS^{\rm opp}
\ee
where $\IS$ is the clockwise phase-ordered product for $z_{ij}$ in one half-plane and $\IS^{\rm opp}$ is the clockwise phase-ordered product
in the opposite half-plane. Now, if we take the Witten index to decategorify $\IS \to S$ then we map the factors in $\IS$ via
$ (\IZ \textbf{1} \oplus R_{ij} e_{ij} ) \to  ( \textbf{1} + \mu_{ij} e_{ij} )$.   Now we need the relation
\be
\mu_{ji} = - \mu_{ij}^*
\ee
which follows from the existence of the degree $-1$ pairing $K_{ij}$. (See also the discussion in Landau-Ginzburg theory in Section \S \ref{quantumbps} below.)
If we define the fermion number to be integral (using the gauge freedom discussed in Sections \S
\ref{subsec:OnDegrees} and \S \ref{subsubsec:RelationTSUN-Physical}) then $\mu_{ij}$ is real and hence
if $\IS \to S$ then $\IS^{\rm opp} \to S^{tr,-1}$.

There is a known relation between properties of the UV theory
and the eigenvalues of the matrix $S S^{tr,-1}$ \cite{Cecotti:1992rm}.
If the massive theory flows from a UV SCFT, as is the case for $\CT^N$,
the eigenvalues take the form $\exp 2 \pi \I q_a$, where $q_a$ are the charges of UV
B-model operators under the R-charge broken by the massive deformation of the SCFT.
If the UV theory is asymptotically free, as is the case for $\CT^{SU(N)}$,
$S S^{tr,-1}$ typically has Jordan blocks. In Section \S \ref{subsec:RotIntfc-TSUN} below
we will construct some rotation Interfaces for $\CT^N$ and $\CT^{SU(N)}$ and we will see that
indeed a sufficiently high power of the rotation Interface is trivial in the former case, but
not in the latter case. Moreover, since   $\fR[\vartheta, \vartheta-2\pi]$ is a categorical
lift of $S S^{tr,-1}$ it is natural to wonder if it somehow reconstructs some other properties of the UV theory.
Indeed we will see that it is an essential ingredient in the construction of local operators
(on the plane) in Section \S \ref{subsec:TraceInterface}.

It would be interesting understand whether there is a categorical generalization of the
relation of the eigenvalues of $S S^{tr,-1}$ to R-charges. Perhaps this can be done
by introducing a notion of an ``eigen-interface'' for $\fR[\vartheta, \vartheta-2\pi]$
under interface product $\boxtimes$, but we will leave this idea for future work.

\subsection{Rotation Interfaces For The Theories $\CT^N_{\vartheta}$ And $\CT^{SU(N)}_{\vartheta}$ }\label{subsec:RotIntfc-TSUN}

In this section we construct some interfaces in the Theories $\CT^{N,SU(N)}_{\vartheta}$
which give a very useful construction of nontrivial Branes from the simple thimbles. (Much of the discussion can be developed in
parallel for the two families of Theories $\CT^{N}_{\vartheta}$ and $\CT^{SU(N)}_{\vartheta}$. So we will
delay separating the cases as long as possible.)   We will reveal how one could discover the Branes
$\fC_k$ and $\fN_n$ of the Theories $\CT^N_{\vartheta}$ and $\CT^{SU(N)}_{\vartheta}$, respectively.
(See  Section \S \ref{subsec:CyclicVacWt} above.)   This construction also leads to a very neat
computation of the space of boundary-condition-changing operators $H^*(\Hop(\fB_1,\fB_2),M_1)$
for certain pairs of Branes, thus justifying several claims made in Section \ref{subsec:VacCat-SUN}.

Recall that the Theories $\CT^{N,SU(N)}_{\vartheta}$   are based on the vacuum weights
(making a slight change of notation from \S \ref{subsec:CyclicVacWt}):
\be\label{eq:thet-vc-wt}
z_j^\vartheta := e^{-\I \vartheta - \frac{2\pi \I}{N} j}
\ee
Here $j$ is an integer modulo $N$ and we will always choose it to be in the
fundamental domain $0 \leq j \leq N-1$. As described in section \S \ref{subsubsec:CyclicIsoms},
there are nontrivial isomorphisms
\be
\varphi^\pm: \CT^{N,SU(N)}_{\vartheta}\rightarrow \CT^{N,SU(N)}_{\vartheta \pm \frac{2\pi}{N} }
\ee
and the corresponding isomorphism Interfaces will be denoted $\fId^{\pm}:= \fId^{\varphi^\pm}$.
In particular, we have $j \varphi^\pm = (j\mp 1) \mod N$ and
\be
\CE(\fId^+)_{j,k} = \begin{cases} \delta_{k,N-1} \IZ^{[1]} & j=0 \\
\delta_{k,j-1} \IZ & 1\leq j \leq N-1 \\
\end{cases}
\ee
The choice of degree shift here is the simplest one such that the boundary
amplitudes
\be
K^{-1,\varphi^+}_{ij} \in \CE_{i,i-1} \otimes R_{i-1,j-1}\otimes \CE_{j,j-1}^* \otimes R_{ji}
\ee
all have degree one. In $\CT^N_{\vartheta}$ this is just $\pm 1\in \IZ^{[1]}$ and in
$\CT^{SU(N)}_{\vartheta}$ it is of the form
\be
\pm \sum_{I} \varepsilon_I \left( e_I \otimes e_{I'}\right)^{[1]}
\ee
where the sum is over multi-indices of length $\vert i-j\vert$ and
$\varepsilon_I$ is a sign defined under \eqref{eq:SUN-MC-1}.

We can work out $\fId^-$ similarly. It is useful to think of the Chan-Paton data as
a matrix with chain complexes as entries and in these terms we have
\be
\CE(\fId^+) = \IZ^{[1]} e_{0,N-1} \oplus \bigoplus_{j=1}^{N-1} \IZ e_{j,j-1}
\ee
\be
\CE(\fId^-) = \IZ^{[-1]} e_{N-1,0} \oplus \bigoplus_{j=0}^{N-2} \IZ e_{j,j+1}
\ee
%
%\be
%\CE(\fId^-)_{j,k} = \begin{cases} \delta_{j,N-1} \IZ^{[-1]} & k=0 \\
%\delta_{j,k-1} \IZ & 1\leq k \leq N-1 \\
%\end{cases}
%\ee
%

We now consider rotation Interfaces $\fR[\vartheta_{\ell}, \vartheta_r]$ relating the
Theories $\CT^{N,SU(N)}_{\vartheta}$ for different values of $\vartheta$. Thus, $\vartheta(x)$
varies linearly and the binding walls are determined by the equations
\be\label{eq:bw-1}
\Re\left(z^\vartheta_{jk}\right) = 2 \sin\left( \vartheta + \frac{\pi}{N}(k+j)\right)
\sin\left(  \frac{\pi}{N}(k-j)\right) = 0
\ee
\be\label{eq:bw-2}
\Im\left(z^\vartheta_{jk}\right) = 2 \cos\left( \vartheta + \frac{\pi}{N}(k+j)\right)
\sin\left(  \frac{\pi}{N}(k-j)\right) > 0
\ee

To be more specific we consider the Theories
\begin{equation}
\CT^+ := \CT^{N,SU(N)}_{\epsilon} \qquad \CT^- := \CT^{N,SU(N)}_{\frac{2 \pi}{N} - \epsilon}
\end{equation}
where $\epsilon$ is a small positive phase with $\epsilon \ll \frac{\pi}{N}$.
(Actually,   $0<\epsilon<\frac{\pi}{ N}$ will already suffice.)

We now consider functors between the categories of Branes  $\fB\fr(\CT^+)$ and
$\fB\fr( \CT^-)$. They will be induced by Interfaces as discussed in equation
\eqref{eq:Intfc-Functor} et. seq.  There are four natural ways to relate these Theories.
Indeed, for clockwise rotations, we can rotate $\CT^N_{\epsilon}$ into $\CT^N_{\frac{2\pi}{N}-\epsilon}$, or rotate $\CT^N_{\frac{2\pi}{N}-\epsilon}$ into $\CT^N_{\frac{2 \pi}{N} + \epsilon}$
and then act with a symmetry interface $\fId^-$ to bring it back to $\CT^N_{\epsilon}$. For counterclockwise rotations, we can
rotate $\CT^N_{\frac{2\pi}{N}-\epsilon}$ into $\CT^N_{\epsilon}$, or rotate $\CT^N_{\epsilon}$ into $\CT^N_{- \epsilon}$ and then act with a symmetry interface
$\fId^+$ to bring it back to $\CT^N_{\frac{2\pi}{N}-\epsilon}$.

We can thus define
\begin{align}\label{eq:Iplmn-intfc}
\tilde \fI^{+-} & := \fR[\epsilon,\frac{2 \pi}{N} - \epsilon] \in \fB\fr(\CT^+,\CT^-) \cr
\tilde \fI^{-+} & := \fR[\frac{2 \pi}{N} - \epsilon,\frac{2 \pi}{N} + \epsilon] \IntfcTimes\fId^-  \in \fB\fr(\CT^-,\CT^+)\cr
\fI^{+-} & := \fR[\epsilon, - \epsilon] \IntfcTimes\fId^+   \in \fB\fr(\CT^+,\CT^-) \cr
\fI^{-+} & := \fR[\frac{2 \pi}{N} - \epsilon,\epsilon] \in \fB\fr(\CT^-,\CT^+). \cr
\end{align}

Next, we work out the effect of convolution by these Interfaces on the Chan-Paton factors of Branes.
We begin by computing the binding walls.

Consider the case of $\fI^{-+}$.
Equation \eqref{eq:bw-1}  is equivalent to
 $\vartheta+ \frac{\pi}{N}(k+j) = n \pi$, with $n\in \IZ$,. But,
 given the range of $\vartheta$, we must
have $\vartheta= \frac{\pi}{N}$ and hence $k+j = N-1$ and $n=1$. Then
the positivity constraint \eqref{eq:bw-2} implies $k<j$, and hence  $0\leq k < \frac{N-1}{2}$.
Similar results hold for the other three cases with minor variations. We have binding walls of type $jk$
where, roughly speaking, $j$ is an upper vacuum and $k$ is the lower vacuum
vertically below it.

There is a small subtlety in this computation because all the binding walls of type $jk$
with $j+k=N-1$ are at the same value of $x$,
thanks to the very symmetric choice of vacuum weights we made for  these examples.
However, the matrices $e_{j,k}$ for $j+k=N-1$ and $0\leq k < \frac{N-1}{2}$ all
commute with one another  and hence the product \eqref{eq:TautCurvedCP}
is unambiguous. Indeed, any small deformation of the vacuum weights will split the walls,
leading to a multiple convolution of the walls $\fS^{f}_{j,k}$ for these values of $j,k$. The
different orderings of the convolutions will be homotopy equivalent. Similar remarks apply to the
other three Interfaces.

Since the product of two matrices of the type
$e_{j,k}$ for $j+k=N-1$ and $0\leq k < \frac{N-1}{2}$ is zero
 \eqref{eq:TautCurvedCP} simplifies to the lower triangular matrix
\be
\CE(\fI^{-+}) = \IZ \textbf{1} \oplus \bigoplus_{\frac{N-1}{2}< j \leq N-1} R_{j,N-j-1} e_{j,N-j-1}
\ee
Explicitly, for $N=2,3,4,5$ we have
\be\label{eq:fmp-cp-N2}
\CE(\fI^{-+}) =
\begin{pmatrix}
\IZ & 0 \\
R_{1,0} & \IZ \\
\end{pmatrix}
\ee
\be
\CE(\fI^{-+}) =
\begin{pmatrix}
\IZ & 0 & 0  \\
0 &  \IZ & 0   \\
R_{2,0} & 0 & \IZ \\
\end{pmatrix}
\ee
\be
\CE(\fI^{-+}) =
\begin{pmatrix}
\IZ & 0 & 0 & 0  \\
0 &  \IZ & 0 & 0   \\
0 & R_{2,1} & \IZ & 0 \\
R_{3,0} & 0  & 0  & \IZ  \\
\end{pmatrix}
\ee
\be
\CE(\fI^{-+}) =
\begin{pmatrix}
\IZ & 0 & 0 & 0 & 0 \\
0 &  \IZ & 0 & 0  & 0  \\
0 &  0 & \IZ & 0 & 0    \\
0 & R_{3,1} & 0 & \IZ & 0 \\
R_{4,0} & 0  & 0  & 0 & \IZ   \\
\end{pmatrix}
\ee

The case of $\fI^{+-}$ is slightly more elaborate. The binding walls of $\fR[\epsilon,-\epsilon]$ are
obtained from \eqref{eq:bw-1} with $\vartheta=0$ and hence  $j+k=N$. Again $n=1$ and  then
\eqref{eq:bw-2} implies $1\leq k < \frac{N}{2}$.  We must then convolve with the isomorphism
Interface $\fId^+$. The net result is
\be
\CE(\fI^{+-}) = \IZ^{[1]}e_{0,N-1} \oplus_{j=1}^{N-1} \IZ e_{j,j-1} \oplus_{1\leq k < \frac{N}{2}} R_{N-k,k} e_{N-k,k-1}
\ee
Again, to get a feel for these, we list the cases $N=2,3,4,5$:
\be\label{eq:fpm-cp-N2}
\CE(\fI^{+-}) =
\begin{pmatrix}
0 & \IZ^{[1]}  \\
  \IZ & 0  \\
\end{pmatrix}
\ee
\be
\CE(\fI^{+-}) =
\begin{pmatrix}
0 & 0 &  \IZ^{[1]}  \\
 \IZ & 0  & 0  \\
R_{2,1} &   \IZ & 0  \\
\end{pmatrix}
\ee
\be
\CE(\fI^{+-}) =
\begin{pmatrix}
0 & 0 & 0 &  \IZ^{[1]}  \\
 \IZ & 0 & 0 & 0   \\
0 &   \IZ & 0 & 0   \\
R_{3,1} & 0    & \IZ & 0 \\
\end{pmatrix}
\ee
\be
\CE(\fI^{+-})  =
\begin{pmatrix}
0 & 0 & 0 & 0 &  \IZ^{[1]}  \\
  \IZ & 0 & 0  & 0 & 0  \\
  0 & \IZ & 0 & 0 & 0    \\
0 & R_{3,2} &  \IZ & 0 & 0 \\
R_{4,1} & 0  & 0  & \IZ & 0   \\
\end{pmatrix}
\ee

Using these formulae it easily follows that if $\fB\in \fB\fr(\CT^-)$ then
\be
\CE\left(\fB \IntfcTimes \fI^{-+}\right)_j =
\begin{cases} \CE(\fB)_j \oplus \CE(\fB)_{N-j-1} \otimes R_{N-j-1,j} & 0 \leq j < \frac{N-1}{2} \\
\CE(\fB)_j & \frac{N-1}{2}\leq j \leq N-1 \\
\end{cases}
\ee
and similarly if $\fB\in \fB\fr(\CT^+)$
then the Chan-Paton factors change by
\be
\CE\left(\fB \IntfcTimes \fI^{+-}\right)_j =
\begin{cases} \CE(\fB)_{j+1} \oplus \CE(\fB)_{N-j-1} \otimes R_{N-j-1,j+1} & 0 \leq j < \frac{N}{2}-1 \\
\CE(\fB)_{j+1} & \frac{N}{2}-1 \leq j \leq N-2 \\
\CE(\fB)_0^{[1]} & j=N-1 \\
\end{cases}
\ee

Entirely analogous remarks apply to the Interfaces $\tilde\fI^{\pm\mp}$. The main
difference is that since $\vartheta(x)$ increases in the rotation Interfaces
the binding walls are past stable. In this way we find that if $\fB\in \fB\fr(\CT^+)$
then
\be
\CE\left(\fB \IntfcTimes \tilde\fI^{+-}\right)_j =
\begin{cases} \CE(\fB)_j \oplus \CE(\fB)_{N-j-1} \otimes R_{j,N-j-1}^* & 0 \leq j < \frac{N-1}{2} \\
\CE(\fB)_j & \frac{N-1}{2}\leq j \leq N-1 \\
\end{cases}
\ee
and if $\fB\in \fB\fr(\CT^-)$ then
\be
\CE\left(\fB \IntfcTimes \tilde\fI^{-+}\right)_j
= \begin{cases} \CE(\fB)_{N-1}^{[-1]} & j=0 \\
\CE(\fB)_{j-1} \oplus \CE(\fB)_{N-j-1} \otimes R_{j-1,N-j-1}^* & 1 \leq j < \frac{N}{2} \\
\CE(\fB)_{j-1} & \frac{N}{2} \leq j \leq N-1 \\
\end{cases}
\ee
\begin{figure}[htp]
\centering
\includegraphics[scale=0.35,angle=0,trim=0 0 0 0]{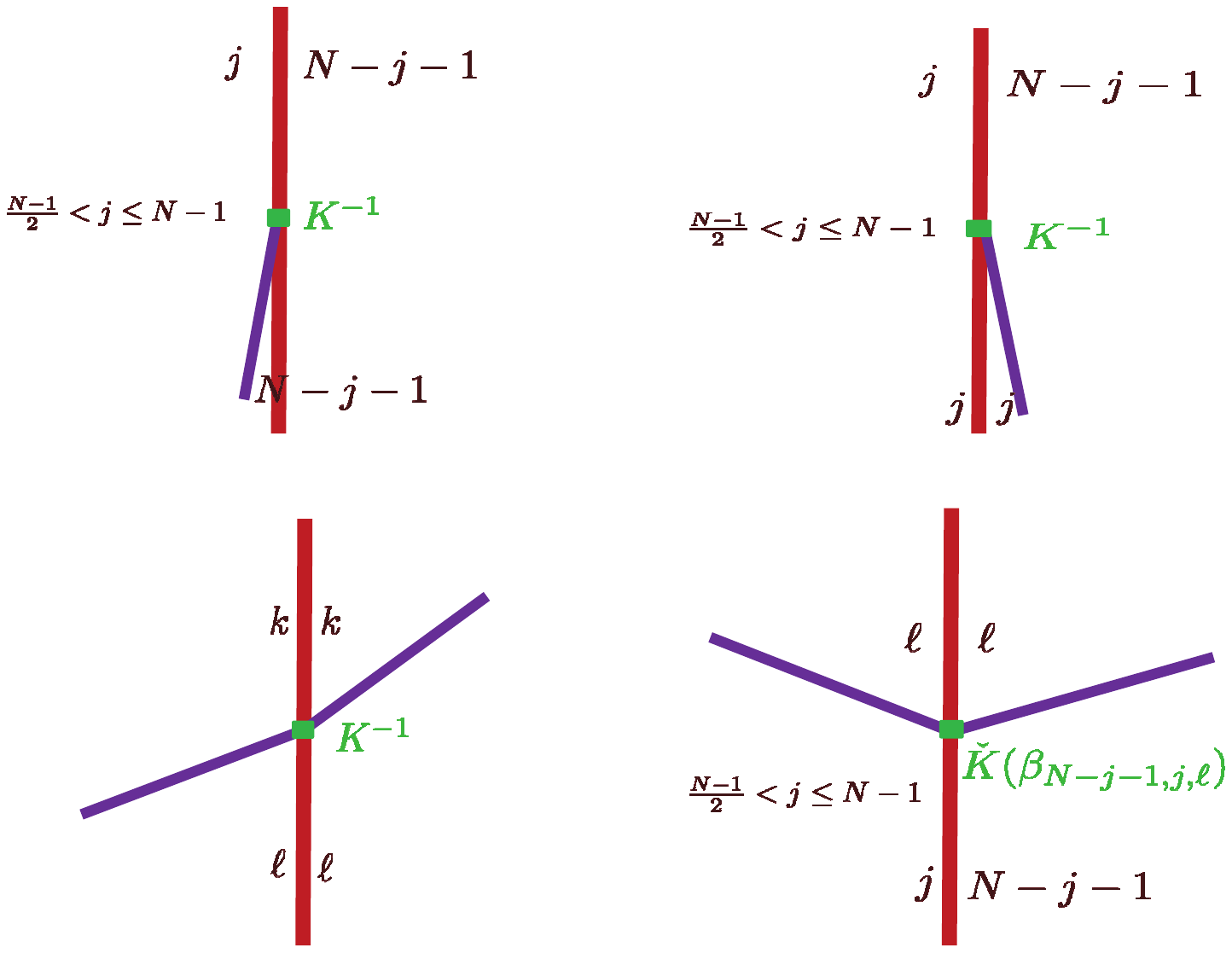}
\caption{Nonzero boundary amplitudes for $\fI^{-+}$.  }
\label{fig:IMINUSPLUS-AMP}
\end{figure}
\begin{figure}[htp]
\centering
\includegraphics[scale=0.35,angle=0,trim=0 0 0 0]{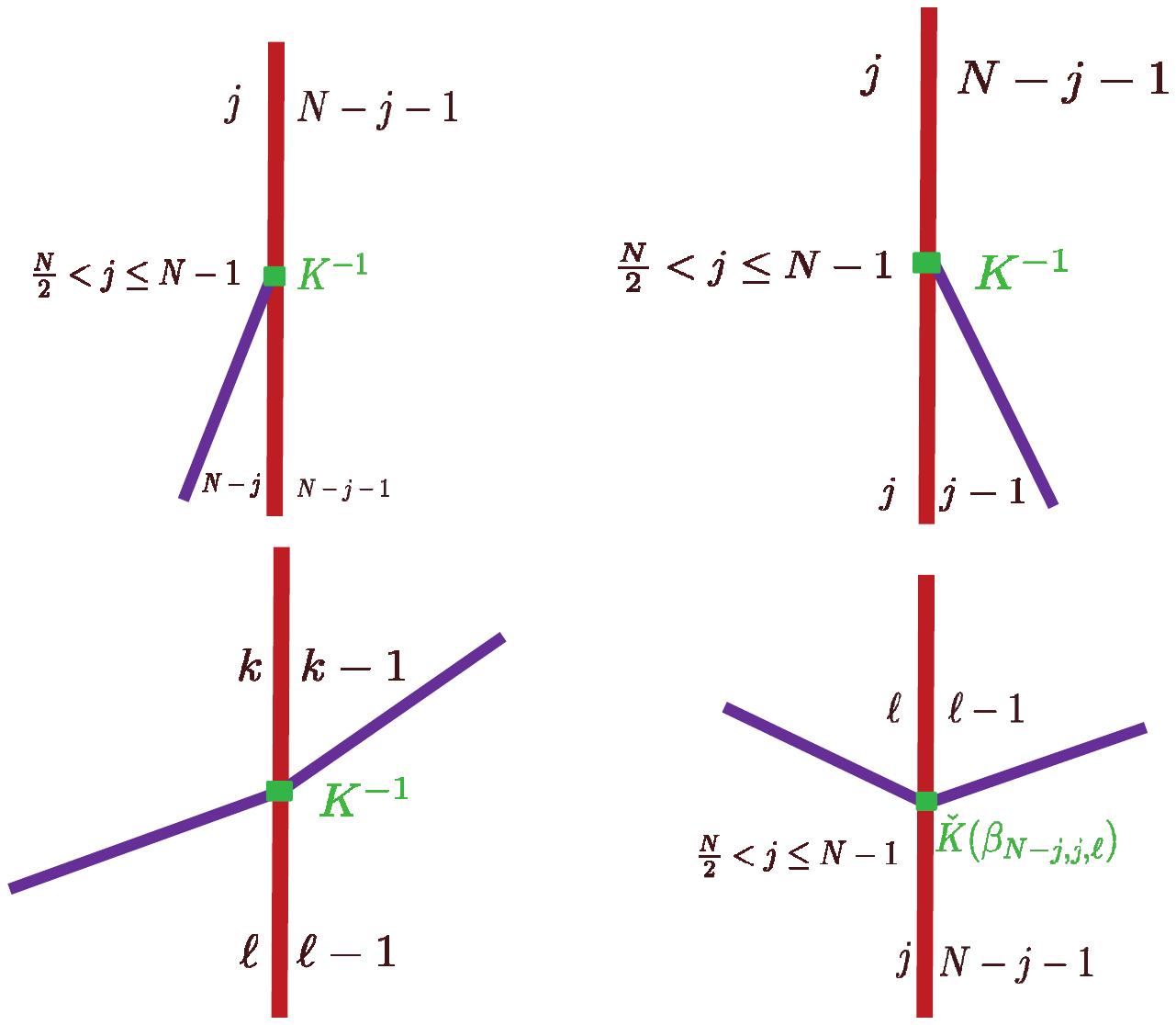}
\caption{Nonzero boundary amplitudes for $\fI^{+-}$.   }
\label{fig:IPLUSMINUS-AMP}
\end{figure}

The boundary amplitudes for $\fI^{-+}$ and $\fI^{+-}$ are shown in
Figures \ref{fig:IMINUSPLUS-AMP} and \ref{fig:IPLUSMINUS-AMP}, respectively.
Similar results hold for $\tilde\fI^{-+}$ and $\tilde\fI^{+-}$.

Now, successive application of these Interfaces generates a sequence of
Branes in $\CT^-, \CT^+$ starting with one Brane in either Theory. To be specific,
suppose $\fB \in \fB\fr(\CT^+)$. We then generate a sequence of Branes
$\fB[n] \in \fB\fr(\CT^+)$ and $\fB[n+\half] \in \fB\fr(\CT^-)$ with $n\in \IZ$
by setting $\fB[0]:=\fB$ and then defining recursively, for $n\geq 0$
\be\label{eq:rec-up}
\fB[n+\half]:= \begin{cases} \fB[n] \IntfcTimes\fI^{+-} & n \in \IZ    \\
\fB[n] \IntfcTimes\fI^{-+} & n \in \IZ + \half \\
\end{cases}
\ee
while for $n\leq 0$ we take
\be\label{eq:rec-down}
\fB[n-\half]:= \begin{cases} \fB[n] \IntfcTimes\tilde\fI^{+-} & n \in \IZ    \\
\fB[n] \IntfcTimes\tilde \fI^{-+} & n \in \IZ + \half \\
\end{cases}
\ee

Note that for $n$ positive we are thus taking successive convolutions (well-defined up
to homotopy equivalence)
\be
\fI^{+-}\IntfcTimes \fI^{-+}\IntfcTimes \fI^{+-}\cdots
\ee
and similarly for $n$ negative with $\tilde\fI^{\pm\mp}$. We do not get anything
interesting by alternating $ \fI^{\pm\mp}$ and $\tilde\fI^{\pm\mp}$ because
of the homotopy equivalences:
\be\label{eq:fmp-fpm-he}
\tilde \fI^{+-}\IntfcTimes \fI^{-+} \sim \fId_{\CT^+} \qquad \qquad
 \fI^{-+}\IntfcTimes\tilde \fI^{+-}  \sim \fId_{\CT^-}
\ee
and
\be\label{eq:fpm-fmp-he}
 \fI^{+-}\IntfcTimes \tilde \fI^{-+} \sim \fId_{\CT^+} \qquad \qquad
 \tilde\fI^{-+}\IntfcTimes \fI^{+-}  \sim \fId_{\CT^-}
\ee
\begin{figure}[htp]
\centering
\includegraphics[scale=0.35,angle=0,trim=0 0 0 0]{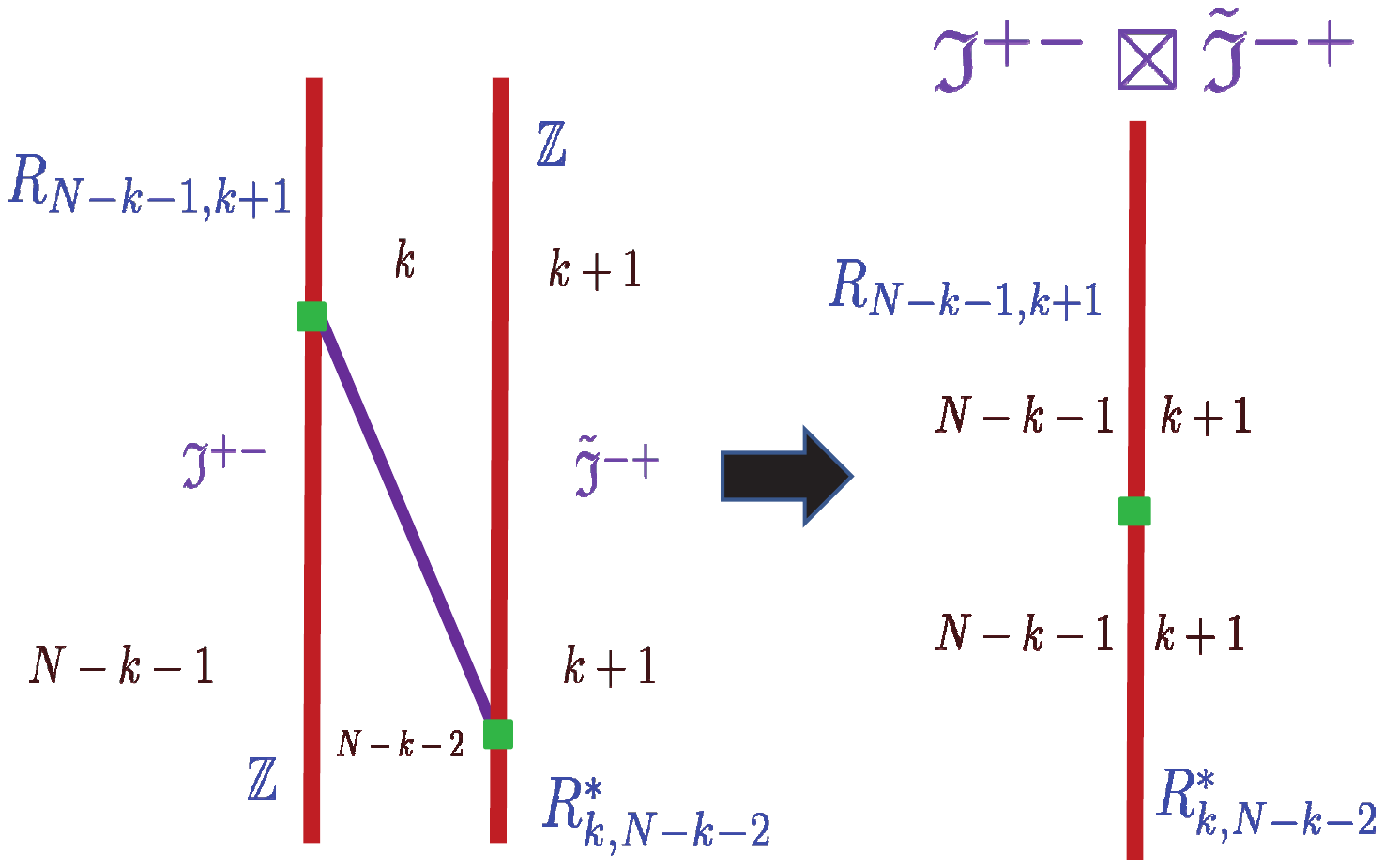}
\caption{A boundary amplitude for $ \fI^{+-}\IntfcTimes \tilde \fI^{-+}$  contributing to the extended web
shown on the right arises from the convolution of boundary amplitudes with the taut web shown
on the left. This defines a differential which eliminates the Chan-Paton factors not present in the
identity Interface, upon taking cohomology.   }
\label{fig:IPM-MP-HE}
\end{figure}
The homotopy equivalences \eqref{eq:fmp-fpm-he} are straightforward given our
previous discussion on categorical parallel transport. The equivalences
\eqref{eq:fpm-fmp-he} require more discussion. A short computation show that
the Chan-Paton data for $ \fI^{+-}\IntfcTimes \tilde \fI^{-+}$ is
\be
\CE( \fI^{+-}\IntfcTimes \tilde \fI^{-+}) = \IZ \textbf{1} \oplus \bigoplus_{0 \leq k < \frac{N}{2}-1}
\left( R_{k,N-2-k}^* \oplus R_{N-k-1,k+1} \right) e_{N-k-1,k+1}
\ee
Plainly,  the Chan-Paton data differ from those of the identity Interface.

A glance at Figure \ref{fig:IPM-MP-HE} shows that the problematic chain complexes
$\CE( \fI^{+-}\IntfcTimes \tilde \fI^{-+})_{N-k-1,k+1}$ have a nonzero differential.
Indeed, using the symmetry isomorphism and $\check K$ we have a degree zero isomorphism:
\be
R_{k,N-2-k}^* \cong R_{k+1, N-k-1}^{[-1]}
\ee
and with this understood the differential given by Figure \ref{fig:IPM-MP-HE}
acts on $R_{k+1, N-k-1}^{[-1]} \oplus R_{N-k-1,k+1}$ as $(r^{[-1]},s) \mapsto (0, r)$.
Thus, the cohomology of the problematic terms in the Chan-Paton data vanishes.

Of course, the above quasi-isomorphism would be a simple consequence of the homotopy equivalence
\eqref{eq:fpm-fmp-he}, but, as we have seen with the homotopy equivalence of $\fS^f_{ij}\boxtimes \fS^p_{ij}$
with the identity Interface, explained at length in Section \S \ref{subsec:SWallIntfc}, more
discussion is needed to establish a homotopy equivalence. The argument in this case is
very similar to that for $\fS^f_{ij}\boxtimes \fS^p_{ij}$.

Let us now study the sequence of Branes $\fB[n]$ for some simple choices of $\fB=\fB[0]$.
It is already quite interesting for thimbles.
To begin, suppose that $\fT_\ell\in \fB\fr(\CT^+)$ is a thimble with $1\leq \ell \leq \frac{N}{2}$,
i.e. a down-vacuum thimble. Then one can easily show that
\be
\CE(\fT_\ell\IntfcTimes \fI^{+-})_k = \delta_{k,\ell-1}\IZ
\ee
and moreover the boundary amplitudes are all zero. To see this note that since
the boundary amplitudes of $\fT_\ell$ are all zero the only possible boundary
amplitude in the convolution would use the amplitude in the upper right of
Figure \ref{fig:IPLUSMINUS-AMP}, but for $\ell$ in the range $1\leq \ell \leq \frac{N}{2}$
there is no such nonzero amplitude.  It follows that we have
\be\label{eq:Thim-Thim-1}
\fT_\ell\IntfcTimes \fI^{+-} = \fT_{\ell-1} \qquad 1\leq \ell \leq \frac{N}{2}
\ee
\be\label{eq:Thim-Thim-2}
\fT_\ell\IntfcTimes \fI^{-+} = \fT_{\ell} \qquad\fT_\ell\in \fB\fr(\CT^-),\quad  0\leq \ell \leq \frac{N-1}{2}.
\ee
\be\label{eq:Thim-Thim-3}
\fT_\ell\IntfcTimes \tilde\fI^{+-} = \fT_{\ell} \qquad\fT_\ell\in \fB\fr(\CT^+),\quad  0\leq \ell \leq \frac{N-1}{2}.
\ee
\be\label{eq:Thim-Thim-4}
\fT_\ell\IntfcTimes \tilde\fI^{-+} = \fT_{\ell+1} \qquad\fT_\ell\in \fB\fr(\CT^-),\quad  0\leq \ell \leq \frac{N}{2}-1.
\ee
where the results \eqref{eq:Thim-Thim-2}-\eqref{eq:Thim-Thim-4} are obtained in an entirely analogous fashion.
\begin{figure}[htp]
\centering
\includegraphics[scale=0.35,angle=0,trim=0 0 0 0]{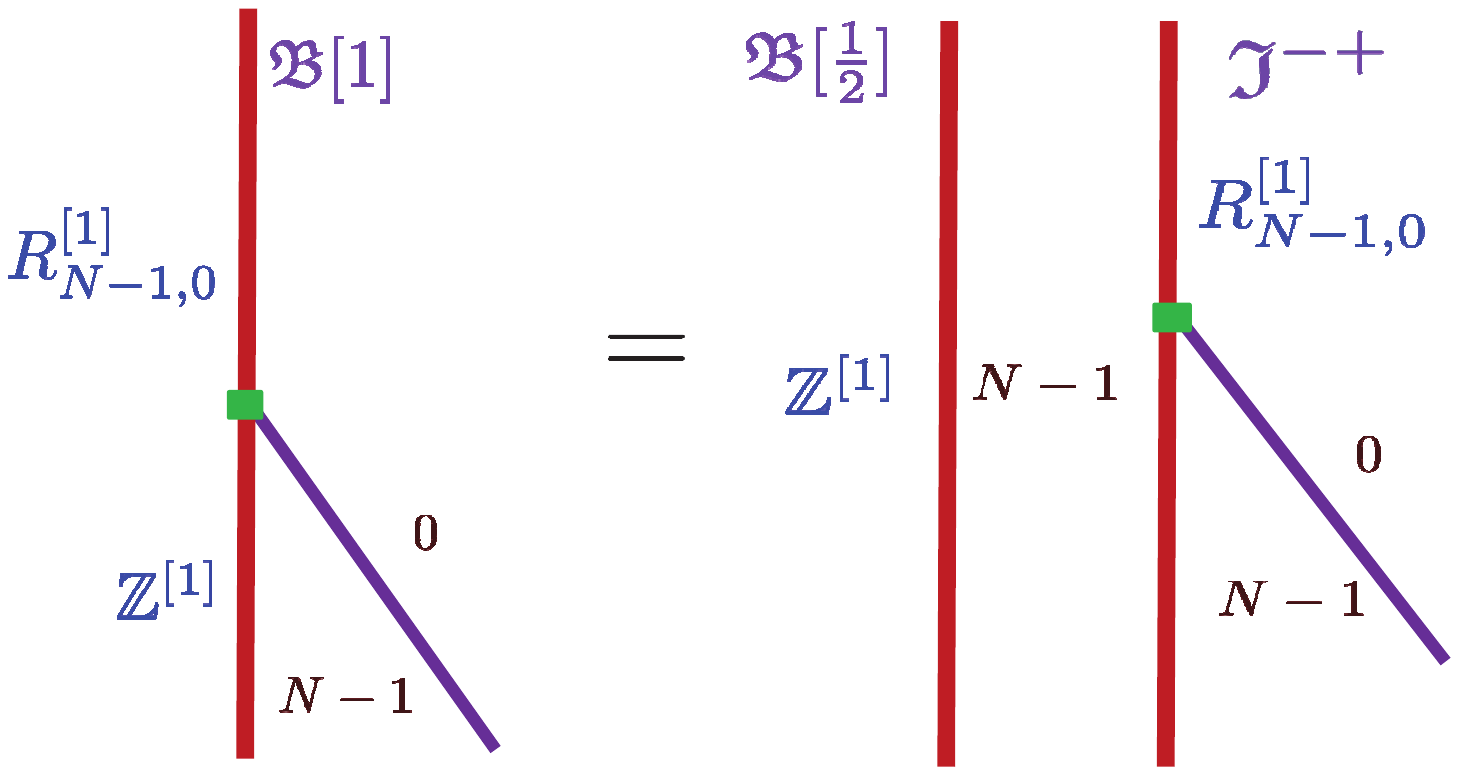}
\caption{This figure shows how a nonzero boundary amplitude can be generated from
the thimble.   }
\label{fig:B1-AMP}
\end{figure}

Equations \eqref{eq:Thim-Thim-1} and \eqref{eq:Thim-Thim-2} cannot be applied to thimbles for upper vacua nor to the
case of $\ell=0$. Let us focus on the latter case and consider the
 more nontrivial sequence of Branes  produced when we take $\fB[0]=\fT_0$.
Then $\fB[\half] = \fT_{N-1}^{[1]}$ is a shifted thimble. However, at the next step we find
\be
\CE(\fB[1])_j = \begin{cases} R_{N-1,0}^{[1]} & j = 0 \\
\IZ^{[1]} & j= N-1 \\
0 & {\rm else} \\
\end{cases}
\ee
Moreover, $\fB[1]$ now acquires a nonzero amplitude for the fan $J = \{ 0, N-1\}$
given by $K^{-1}_{N-1,0} \in R^{[1]}_{N-1,0} \otimes R_{0,N-1} \otimes (\IZ^{[1]})^*$.
See Figure \ref{fig:B1-AMP}.

Proceeding to compute $\fB[\frac{3}{2}] = \fB[1]\IntfcTimes \fI^{+-}$ we get
\be
\CE(\fB[\frac{3}{2}])_j = \begin{cases} R_{N-1,1}^{[1]} & j = 0 \\
\IZ^{[1]} & j= N-2 \\
R_{N-1,0}^{[2]} & j= N-1\\
0 & {\rm else} \\
\end{cases}
\ee
Moreover, computing the boundary amplitudes
we find three fans have nontrivial amplitudes They can be
interpreted as $K^{-1}$ for fans $\{0, N-2\}$ and $\{N-1,N-2\}$
and $\check K(\beta_{0,1,N-1})$ for $\{N-1,0\}$.
(These come from using Figure \ref{fig:IPLUSMINUS-AMP},
upper right with $j=N-1$, lower left with $k=0$, $\ell=N-1$, and
lower right, with $\ell=0$, $j=N-1$, respectively.)

\begin{figure}[htp]
\centering
\includegraphics[scale=0.35,angle=0,trim=0 0 0 0]{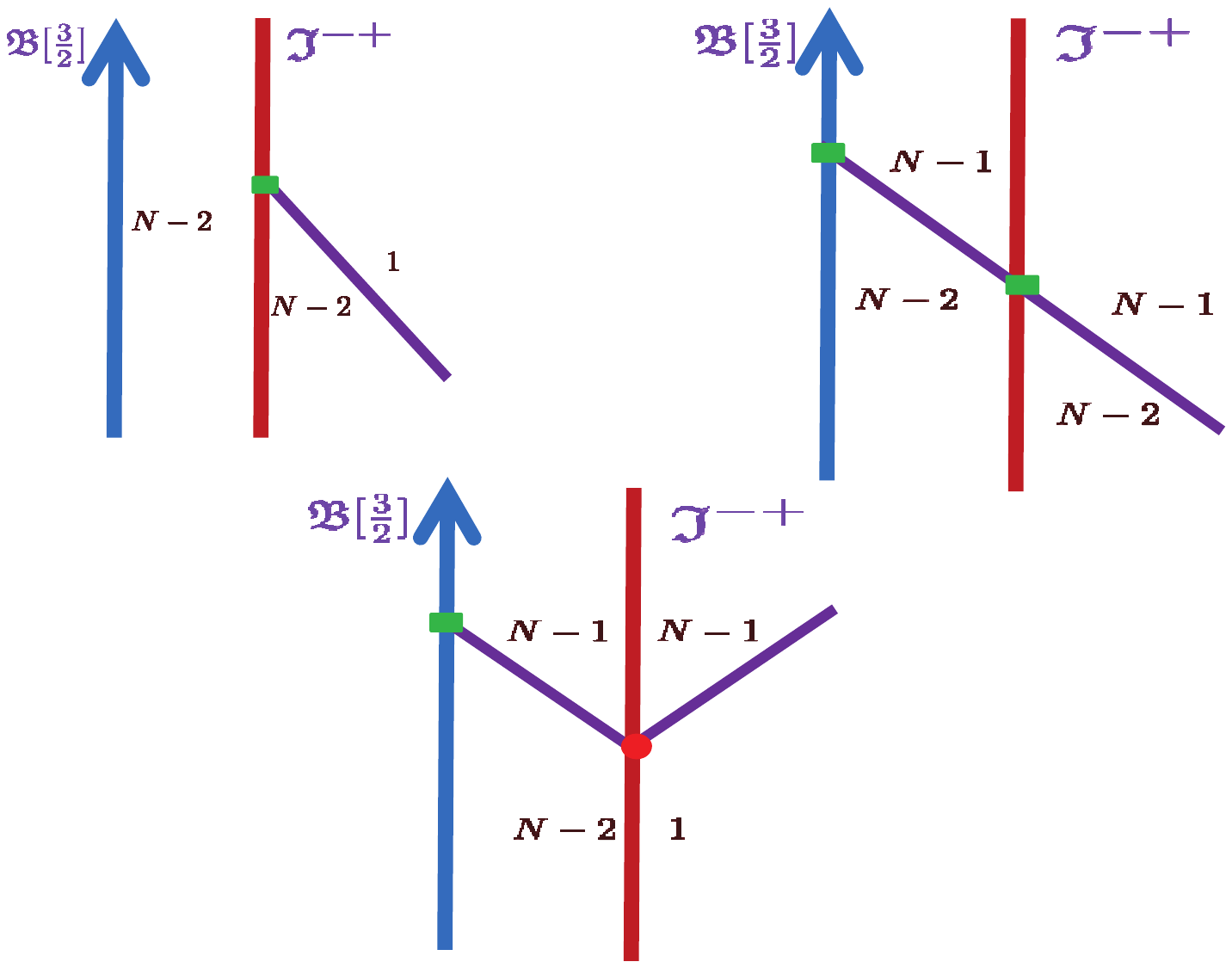}
\caption{ These amplitudes are common to $\fB[2]$ and $\hat{\fB}[2]$.    }
\label{fig:B2-AMP-1}
\end{figure}
\begin{figure}[htp]
\centering
\includegraphics[scale=0.35,angle=0,trim=0 0 0 0]{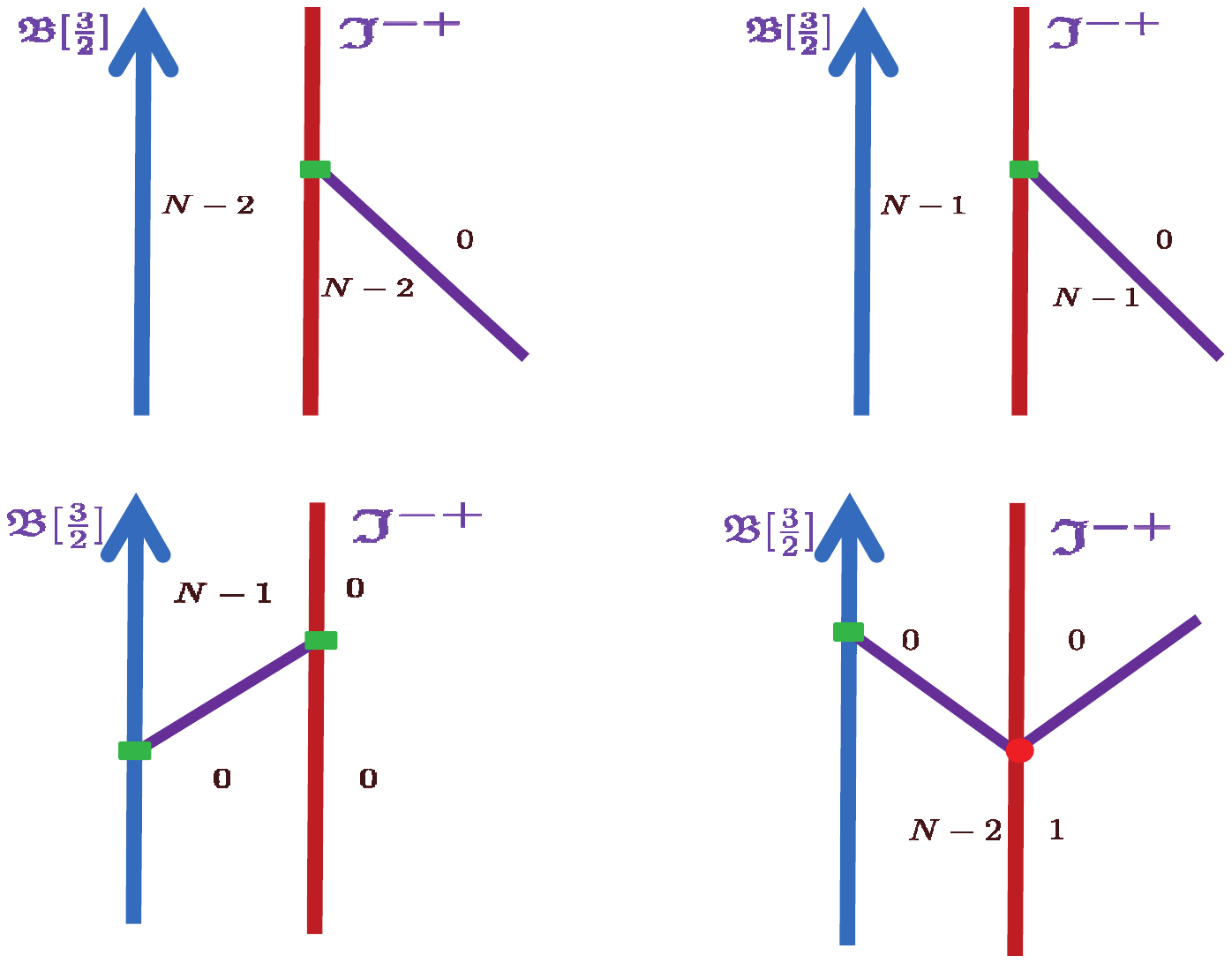}
\caption{ These amplitudes are present for $\fB[2]$ but not for $\hat{\fB}[2]$.    }
\label{fig:B2-AMP-2}
\end{figure}

If we move on to $\fB[2]= \fB[\frac{3}{2}] \IntfcTimes \fI^{-+}$ then
\be\label{eq:B2-CP}
\CE(\fB[2])_j = \begin{cases} R_{N-1,1}^{[1]} \oplus R_{N-1,0}^{[2]} \otimes R_{N-1,0} & j = 0 \\
R_{N-2,1}^{[1]} & j=1 \\
\IZ^{[1]} & j= N-2 \\
R_{N-1,0}^{[2]} & j= N-1\\
0 & {\rm else} \\
\end{cases}
\ee
The nontrivial amplitudes are illustrated in Figures \ref{fig:B2-AMP-1} and \ref{fig:B2-AMP-2}.
Note that $\{N-1,0\}$ is a positive-half-plane fan for
the Theory $\CT^-$ but $\{0,N-1\}$ is a positive-half-plane fan for
the Theory $\CT^+$. Thus the ``emission line'' in the lower left of Figure \ref{fig:B2-AMP-2}
does not continue into the positive-half-plane.

We would now like to replace $\fB[2]$ with a simpler, but homotopy equivalent,
Brane, denoted by $\hat{\fB}[2]$.
At this stage we must distinguish between the Theories $\CT^{N}_{\epsilon}$ and $\CT^{SU(N)}_{\epsilon}$
since we need to use special properties of the $R_{ij}$. We first discuss the case of $\CT^{N}_{\epsilon}$.
We then return and pick up the thread for $\CT^{SU(N)}_{\epsilon}$ at this point.

For $\CT^{N}_{\epsilon}$  equation \eqref{eq:B2-CP} simplifies to
\be\label{eq:B2-CP-TN}
\CE(\fB[2])_j = \begin{cases} \IZ^{[1]} \oplus \IZ^{[2]} & j = 0 \\
\IZ^{[1]} & j=1 \\
\IZ^{[1]} & j= N-2 \\
\IZ^{[2]} & j= N-1\\
0 & {\rm else} \\
\end{cases}
\ee
We want to eliminate the $j=0$ Chan-Paton space so we define $\hat{\fB}[2]$ to have Chan-Paton
data:
\be\label{eq:B2hat-CP-TN}
\CE(\hat\fB[2])_j = \begin{cases}
\IZ^{[1]} & j=1 \\
\IZ^{[1]} & j= N-2 \\
\IZ^{[2]} & j= N-1\\
0 & {\rm else} \\
\end{cases}
\ee
with the same boundary amplitudes as in Figure \ref{fig:B2-AMP-1}.

Now we describe the homotopy equivalence $\hat\fB[2]\sim \fB[2]$.
Note that with nonempty positive-half-plane fans $0$ is always in the future.
Moreover, $\{0,i\}$, $i=1,N-2,N-1$ are all positive-half-plane fans for $\CT^+$.
To construct the homotopy equivalence we need closed morphisms
\be\label{eq:d1-home}
\delta_1 \in \Hop(\hat\fB[2],\fB[2]) = \Hop(\hat\fB[2],\hat\fB[2])
\ee
\be\label{eq:d2-home}
\delta_2 \in \Hop(\fB[2],\hat\fB[2]) = \Hop(\hat\fB[2],\hat\fB[2])\oplus \CD
\ee
\be
\CD:=\bigoplus_{i=1,N-2,N-1} \CE(\fB[2])_0 \otimes \widehat R_{0,i} \otimes (\CE(\hat\fB[2])_i)^*
\ee
such that the products $M_2(\delta_1,\delta_2)$ and $M_2(\delta_2,\delta_1)$ are homotopy
equivalent to $\Id$:
\be
\begin{split}
M_2(\delta_1,\delta_2) & = \Id_{\hat\fB[2]} + M_1(\delta_3) \\
M_2(\delta_2,\delta_1) & = \Id_{ \fB[2]} + M_1(\delta_4). \\
\end{split}
\ee
It will be useful below to compare the boundary amplitudes of $\fB[2]$ and $\hat\fB[2]$
and write
\be
\CB(\fB[2]) = \CB(\hat\fB[2]) + \Delta\CB
\ee

The multiplications $M_k$ are computed using the taut half-plane webs, and the only
ones with \underline{at least} two boundary vertices in fact have \underline{at most}
two boundary vertices. (See, for example, Figure \ref{fig:TNEXAMPLE-3} for the
case of unextended webs.) This property considerably simplifies the computation of $M_2$.
To begin we take $\delta_1 = \Id_{\hat\fB[2]}$ and $\delta_2 = \Id_{ \hat\fB[2]}\oplus 0$,
where the direct sum refers to the decomposition in \eqref{eq:d2-home}.
The equation $M_2(\delta_1,\delta_2)  = \Id_{\hat\fB[2]}$ works nicely with $\delta_3=0$.
On the other hand,
since $\fB[2]$ has a nonzero Chan-Paton space for $i=0$ and $\hat\fB[2]$ does not,
$M_2(\delta_2,\delta_1)$ cannot possibly reproduce $\Id_{\CE(\fB[2])_0}$. Therefore,
$\delta_4\in \Hop(\fB[2],\fB[2])$ must be nonzero. We take it to have only
nonzero scalar component in $\End(\CE(\fB[2])_0)$. Then the scalar component
of the differential is simply given by matrix multiplication
\be
M_1(\delta_4) = \CB_{00} \delta_4 + \delta_4 \CB_{00}
\ee
where $\CB_{00}\in \End(\CE(\fB[2])_0)$ is the boundary amplitude induced by the
lower left diagram of Figure \ref{fig:B2-AMP-2}. Writing vectors in $\CE(\fB[2])_0$
in the form $r_1 \oplus r_2$ where $r_1\in \IZ^{[1]}$ and $r_2 \in \IZ^{[2]}$ we
easily compute that the boundary amplitude is the linear transformation:
\be
\CB_{00}: r_1 \oplus r_2 \mapsto 0 \oplus r_1^{[1]} .
\ee
Therefore, we can take $\delta_4$ to be the chain-homotopy inverse
\be
\delta_4:  r_1 \oplus r_2 \mapsto r_2^{[-1]} \oplus 0
\ee
so that the scalar component of $M_1(\delta_4)$ is the missing
component $\Id_{\CE(\fB[2])_0}$. There will be further
contributions from $M_1(\delta_4)$, producing amplitudes with
positive half-plane fans of type $\{ 0, 1\}$, $\{0,N-2\}$, and $\{0,N-1\}$.
They are given (essentially) by $\Delta \CB$, and can be cancelled by adding
(essentially) $-\Delta\CB$ to $\delta_2$.  Now we have established
the required homotopy equivalence $\hat\fB[2]\sim \fB[2]$.

In what follows we will need to employ repeatedly a maneuver very similar to what
we just explained: We will find a pair of Branes $\fB$ and $\hat \fB$
all of whose Chan-Paton spaces are identical except for one vacuum $j_*$
(the vacuum $j_*=0$ in the example above) and whose boundary amplitudes
are identical for all amplitudes not involving this distinguished vacuum.
Moreover, we have
\be
\CE(\fB)_{j_*} = \CE(\hat \fB)_{j_*} \oplus V \oplus V^{[1]}
\ee
and there is a boundary amplitude in $\Hop(\fB,\fB)$ in
$\End(\CE(\fB)_{j_*} )$ taking
\be
v_1 \oplus v_2 \oplus v_3 \rightarrow v_1 \oplus 0 \oplus v_2^{[1]}
\ee
In this case we can find a homotopy equivalence $\hat \fB \sim \fB$,
exactly as in the above example. This will allow us to replace $\fB$
by the simpler Brane $\hat\fB$. We call this the \emph{cancellation lemma}
below.

% and
%
% \be
% \delta_2 = \Id_{ \fB[2]} - \Delta\CB
%\ee
%
%Then, thanks to the Maurer-Cartan equation $M_2(\delta_1, \delta_2) = \Id_{\hat\fB[2]}$
%and $M_2(\delta_2, \delta_1)\sim  \Id_{ \fB[2]}$.
%
%
%
%Now note that
%
%\be
% \Hop(\fB[2], \fB[2]) = \Hop(\hat\fB[2],\hat\fB[2])\oplus
% \left( \CE(\fB[2])_0 \otimes   (\CE( \fB[2])_0)^*\right)
% \oplus \CD
%\ee
%
%so we can compare boundary amplitudes for $\fB[2]$ and $\hat\fB[2]$ and define
%
%\be
%\CB(\fB[2]) := \CB(\hat \fB[2]) \oplus \Delta \CB .
%\ee
%
%Now we take
%\cg{Check this. I suspect the basic argument is pretty general and actually applies
%in all the cases we use here.}
%
%
%A special case of the above result shows that the ``scalar component'' of $\Delta \CB$
%in ${\rm End}(\CE(\fB[2])_0)$ is a differential of the form $(r_1,r_2) \mapsto (0,r_1)$,
%so  the cohomology of $\CE(\fB[2])_0$ is zero. This is the surrogate argument we will
%use below in lieu of exhibiting actual homotopy equivalences in analogous maneuvers in
%this section.
%
%\cg{again, this is embarrassing...}
%

Returning to our sequence of Branes $\fB[n]$ in the Theory $\CT^N_{\vartheta}$ we
next observe that $\hat\fB[2]$ is the nontrivial Brane $\fC_{k=1}^{[1]}$ of
equation \eqref{eq:Ck-CP-Spaces} above.
We now proceed inductively. Suppose that
$k<\frac{N}{2}$ and that $\fB[k]$ is homotopy equivalent to $\hat\fB[k]=\fC_{k-1}^{[1]}$.
Therefore
\be\label{eq:Bkhat-CP-TN}
\CE(\hat\fB[k])_j = \begin{cases}
\IZ^{[1]} & j=k-1 \\
\IZ^{[1]} & j=N-k \\
\IZ^{[2]} & j=N-k+1 \\
0 & {\rm else} \\
\end{cases}
\ee
so we compute
\be\label{eq:Bkhat-CP-TN-2}
\CE(\hat\fB[k]\IntfcTimes\fI^{+-})_j = \begin{cases}
\IZ^{[1]}\oplus\IZ^{[2]} & j=k-2\\
\IZ^{[1]} & j=k-1 \\
\IZ^{[1]} & j=N-k-1 \\
\IZ^{[2]} & j=N-k \\
0 & {\rm else} \\
\end{cases}
\ee
Once again there is a component of the boundary amplitude which acts as a
differential on the Chan-Paton space with $j=k-2$ and eliminates it upon
passing to cohomology. Our cancellation lemma allows us to replace this
Brane with a homotopy equivalent Brane $\hat\fB[k+\half]$ with
\be\label{eq:Bkhat-CP-TN-3}
\CE(\hat\fB[k+\half] )_j = \begin{cases}
%\IZ^{[1]}\oplus\IZ^{[2]} & j=k-2\\
\IZ^{[1]} & j=k-1 \\
\IZ^{[1]} & j=N-k-1 \\
\IZ^{[2]} & j=N-k \\
0 & {\rm else} \\
\end{cases}
\ee
Now we compute again
\be\label{eq:Bkhat-CP-TN-4}
\CE(\hat\fB[k+\half]\IntfcTimes\fI^{-+} )_j = \begin{cases}
\IZ^{[1]}\oplus\IZ^{[2]} & j=k-1\\
\IZ^{[1]} & j=k  \\
\IZ^{[1]} & j=N-k-1 \\
\IZ^{[2]} & j=N-k \\
0 & {\rm else} \\
\end{cases}
\ee
and again the cancellation lemma gives us
\be
\hat\fB[k+\half]\IntfcTimes\fI^{-+} \sim \hat\fB[k+1] = \fC_{k}^{[1]}.
\ee
completing the inductive step.
%
%\cg{IS THIS REALLY KOSHER? DO THE BOUNDARY AMPLITUDES BEHAVE AS ASSUMED IN THE LEMMA?}
%

The inductive step works until we produce $\hat\fB[k] \in \fB\fr(\CT^+)$ for
$k=[\frac{N}{2}]$. For simplicity assume first that $N$ is even.
Then we compute
\be\label{eq:Bkhat-CP-TN-5}
\CE(\hat\fB[\frac{N}{2}]\IntfcTimes\fI^{+-})_j = \begin{cases}
\IZ^{[1]}\oplus\IZ^{[2]} & j=\frac{N}{2}-2\\
\IZ^{[1]} & j=\frac{N}{2}-1 \\
\IZ^{[2]} & j=\frac{N}{2} \\
0 & {\rm else} \\
\end{cases}
\ee
The cancellation lemma produces a homotopy equivalent Brane with Chan-Paton
factors
\be\label{eq:Bkhat-CP-TN-6}
\CE(\hat\fB[\frac{N}{2}+\half] )_j = \begin{cases}
%\IZ^{[1]}\oplus\IZ^{[2]} & j=\frac{N}{2}-2\\
\IZ^{[1]} & j=\frac{N}{2}-1 \\
\IZ^{[2]} & j=\frac{N}{2} \\
0 & {\rm else} \\
\end{cases}
\ee
Then we compute again
\be\label{eq:Bkhat-CP-TN-5}
\CE(\hat\fB[\frac{N}{2}+\half]\IntfcTimes\fI^{-+})_j = \begin{cases}
\IZ^{[1]}\oplus\IZ^{[2]} & j=\frac{N}{2}-1\\
%\IZ^{[1]} & j=\frac{N}{2}-1 \\
\IZ^{[2]} & j=\frac{N}{2} \\
0 & {\rm else} \\
\end{cases}
\ee
which, by the cancellation lemma, is homotopy equivalent to
a shifted thimble! That is, we have $\hat\fB[\frac{N}{2}+1] = \fT^{[2]}_{N/2}$.

Now using equations \eqref{eq:Thim-Thim-1} and \eqref{eq:Thim-Thim-2} we can continue the procedure to
produce a sequence of thimbles of down-type vacua, until we get to $\fT^{[2]}_{0}$.

It follows from the above discussion that
\be\label{eq:PeriodicPower}
\fT_\ell \IntfcTimes \left( \fI^{+-} \IntfcTimes \fI^{-+} \right)^{N+1} \sim \fT_\ell \qquad \qquad 0 \leq \ell \leq \frac{N}{2}
\ee
a similar story holds if $N$ is odd. In this case the induction works until we get to $\hat \fB[\frac{N-1}{2}]$.
Then $\hat \fB[\frac{N+1}{2}]$ has two Chan-Paton fators $\IZ^{[1]}$ and $\IZ^{[2]}$ at $j=(N-1)/2$ and $j=(N+1)/2$,
respectively, and $\hat \fB[\frac{N+3}{2}] = \fT^{[2]}_{(N-1)/2}$.

We have not worked out the sequence of Branes generated by thimbles for
upper vacua.

We next turn our attention to the Theory $\CT^{SU(N)}_{\epsilon}$.
We begin with the sequence of Branes $\fB[k]$, $k\geq 0$ generated by the
thimble $\fT_0$ in the Theory $\CT^{SU(N)}_{\epsilon}$. We have already described
the general story up to the Brane   $\fB[2]$, as in equation \eqref{eq:B2-CP}. Now, however we have
\be
\begin{split}
\CE(\fB[2])_0 & = R^{[1]}_{N-1,1} \oplus R^{[2]}_{N-1,0} \otimes R_{N-1,0} \\
& = A_2^{[1]} \oplus A_1^{[2]} \otimes A_1 \\
\end{split}
\ee
Next  $A_1^{[2]} \otimes A_1 \cong A_2^{[2]} \oplus S_2^{[2]}$. It is natural to expect
that the differential
computed by the lower left diagram of Figure \ref{fig:B2-AMP-2} maps $A_2^{[1]} \to A_2^{[2]}$
as a degree shift, since we know the amplitudes are all $SU(N)$-covariant.
We will assume this to be the case and proceed, although we have not checked the
boundary amplitudes in detail.

Passing to cohomology we eliminate the two summands of $A_2$ and using the
cancellation lemma we claim a homotopy equivalence to $\hat\fB[2]$ with
\be\label{eq:B2hat-CP-TSUN}
\CE(\hat\fB[2])_j = \begin{cases}
S_2^{[2]} & j= 0 \\
A_3^{[1]} & j=1 \\
\IZ^{[1]} & j= N-2 \\
A_1^{[2]} & j= N-1\\
0 & {\rm else} \\
\end{cases}
\ee
Referring to equation \eqref{eq:SLDEF} we identify these as the Chan-Paton factors of
$\fN_1^{[1]}$.  We expect that the boundary
amplitudes coincide with those of $\fN_1^{[1]}$, although we have not checked in detail.
In this and similar equations below we must interpret
\be\label{eq:Lneg-m}
\begin{split}
L_{n,m}=0 & \qquad n>1, m \leq 0 \\
L_{1,0}= & S_0 \cong \IZ \\
S_{m} = 0 & \qquad m< 0 \\
\end{split}
\ee

Again, we can proceed inductively. Suppose that $\fB[n] \sim \hat\fB[n]$ where
$\hat\fB[n+1] = \fN_n^{[1]}$. It is useful to employ the isomorphism
$L_{N,m+1} \cong S_m$ and rewrite \eqref{eq:SLDEF} as
\be\label{eq:NnCP-1}
\CE(\fN_n)_j = \begin{cases}
L^{[n-j]}_{2j+1, n+1-j}  & 0 \leq j < \frac{N-1}{2} \\
S^{[n+j+1 - N]}_{n+j+1 - N} &  \frac{N-1}{2}\leq j \leq N-1 \\
\end{cases}
\ee
The boundary amplitudes were described in section \S \ref{subsec:CyclicVacWt}.
Now it is straightforward to compute
\be\label{eq:NnCP-3}
\CE(\fN_n\IntfcTimes \fI^{+-})_j = \begin{cases}
L^{[n-j-1]}_{2j+3, n-j} \oplus S^{[n-j]}_{n-j} \otimes R_{N-j-1,j+1} & 0 \leq j < \frac{N}{2}-1 \\
S^{[n+j+2 - N]}_{n+j+2 - N} &   \frac{N}{2}-1 \leq j \leq N-1 \\
\end{cases}
\ee
Thus, using \eqref{eq:SUN-Rij} and \eqref{eq:ASL} we learn that for $0\leq j < \frac{N}{2}-1$
\be\label{eq:cancel-arg}
\CE(\fN_n\IntfcTimes \fI^{+-})_j \cong L^{[n-j-1]}_{2j+3, n-j}\oplus L^{[n-j]}_{2j+2, n-j+1}\oplus
L^{[n-j]}_{2j+3, n-j}
\ee
Using the cancellation lemma we claim a homotopy equivalence to $\hat\fB[n+\frac{3}{2}]$ with
\be\label{eq:NnCP-2}
\CE(\hat\fB[n+\frac{3}{2}])_j = \begin{cases}
L^{[n-j]}_{2j+2, n-j+1}   & 0 \leq j < \frac{N}{2}-1 \\
S^{[n+j+2 - N]}_{n+j+2 - N} &   \frac{N}{2}-1 \leq j \leq N-1 \\
\end{cases}
\ee
Then a similar argument shows that $\hat\fB[n+\frac{3}{2}]\IntfcTimes \fI^{-+} \sim \fN_{n+1}^{[1]}$
thus completing the inductive step. Unlike the case of the Theory $\CT^N_\epsilon$, the sequence
does not simplify and there is no periodicity as we increase $n$.
Physically, $n$ is related to a first Chern class of an equivariant bundle on $\IC\IP^{N-1}$,
and no such periodicity is expected. See Section \S \ref{subsubsec:Powers} below for further
discussion of this non-periodicity.

If, instead, we use the interfaces $\tilde\fI^{+-}, \tilde \fI^{-+}$ we find that $n$ is reduced by
successive composition. If $\fB[0] = \fN_n$ with $n> \frac{N}{2}$ then the recursion
relations \eqref{eq:rec-down}
give a sequence of Branes $\hat \fB[\ell]$ for $\ell<0$ and $\ell \in \IZ + \half$. With
this sequence we find for $\ell$ negative integral (and not too negative)  $\hat \fB[\ell] \sim \fN_{n+\ell}$:
\be
\fN_n \IntfcTimes \left(\tilde\fI^{+-}\IntfcTimes \tilde \fI^{-+}\right)^{\vert \ell\vert}  \sim \fN_{n+\ell}
\ee
At $n-\vert \ell\vert= \frac{N}{2}-1$ the Chan-Paton factor
becomes $\IZ$ for $j=\frac{N}{2}$. Proceeding to lower $N$ the usual cancellation
argument in expressions like \eqref{eq:cancel-arg} produces zero.
Making use of  equation \eqref{eq:Lneg-m} note that
equation \eqref{eq:NnCP-1} makes sense for $n \geq -1 $ and we can proceed to reduce $n$ until
we get to $\fN_{-1} = \fT_0^{[-1]}$. We can then continue the recursion \eqref{eq:rec-down}
using \eqref{eq:Thim-Thim-3} and \eqref{eq:Thim-Thim-4} to define $\fN_{n}$ for lower values of $n$ in
terms of down-type thimbles. The process then stops when we   arrive at $\fN_{-1-N/2} = \fT^{[-1]}_{N/2}$
(taking  $N$ to be even, for simplicity).  At this point we recall the Branes $\bar \fN_n$ of section \S \ref{subsec:CyclicVacWt}
with
\be\label{eq:NbarCP}
\CE(\bar \fN_n)_j = \begin{cases} \bar L^{[-j-n]}_{N-2j,n+j+1} & 0 \leq j < \frac{N}{2}\\
\bar S^{[j+1-n-N]}_{n+N-j} & \frac{N}{2} \leq j \leq N-1 \\
\end{cases}
\ee
which makes sense for $n\geq -N/2$. Note that $\bar \fN_{-N/2}= \fT_{N/2}^{[1]}$.
The usual arguments then show that
\be
\bar \fN_n\IntfcTimes \left(\tilde\fI^{+-}\IntfcTimes \tilde \fI^{-+}\right)^\ell\sim \bar \fN_{n+\ell}
\ee
and hence we can continue to define $\fN_{n}$ for values below $-N/2-1$ by taking
\be\label{eq:NtoNbar}
\fN_{-\frac{N}{2}-n} \cong \bar \fN_{-1-\frac{N}{2}+n}^{[-2]}\qquad n\geq 1
\ee
(We have not checked the above equations at the level of boundary amplitudes.)

The relation of the Branes $\fN_n$ to thimbles for a certain range of $n$
allows us to fill in a gap from Section \ref{subsec:VacCat-SUN}, namely the
proof of equation \eqref{eq:HopCoho-1}. The key idea is that, thanks to the
\afty-bifunctor of Section \S \ref{subsec:ComposeInterface} the complex
between two Branes $\Hop(\fB_1, \fB_2)$ is quasi-isomorphic to the complex
obtained by composition with an invertible interface. If we apply that to the
present case then we can write, for the case of two lower vacua $i,j< N/2$
\be\label{eq:ProveHopCoho-0}
\begin{split}
\Hop(\fT_i, \fT_j) & = \Hop(\fN_{-1-i}^{[1]}, \fN_{-1-j}^{[1]}) \\
& =_{q.i.} \Hop(\fN_{-1 }^{[1]}, \fN_{-1-j+i }^{[1]}) \\
& = \Hop( \fT_0,\fN_{-1+i-j}^{[1]} ) \\
\end{split}
\ee
where $=_{q.i}$ means we have a quasi-isomorphism.

Now assume for simplicity that $N$ is even.
We next use the observation of equation \eqref{eq:Strip-HalfPlane} to say that,
\be\label{eq:ProveHopCoho-1}
\begin{split}
H^*(\Hop(\fT_i, \fT_j),M_1) & \cong H^*(\CE_{LR}(\fT_0,\fN_{-1+i-j}^{[1]}[\pi] ), d_{LR}) \\
\end{split}
\ee
Recall from the discussion of equation \eqref{eq:Strip-HalfPlane} that we should rotate
in the direction of increasing $\vartheta$, so we should compose with
$(\tilde\fI^{+-}\boxtimes \tilde \fI^{-+})^{N/2}$ to rotate the Brane by $\pi$
and hence
\be
\fN_{-1+i-j}^{[1]}[\pi] =  \fN_{-1+i-j-N/2}^{[1]}.
\ee
The complex of groundstates is very simple when one of the Branes is a thimble.
In this case there is no differential and the cohomology is the Chan-Paton factor itself.
In our case
\be\label{eq:ProveHopCoho-2}
H^*(\CE_{LR}(\fT_0, \fN_{-1+i-j-N/2}^{[1]}, ), d_{LR}) \cong  \left(\CE(\fN_{-1+i-j-N/2}^{[1]})_{N/2}\right)^*
\ee
Note that although we have the thimble for the vacuum $i=0$ on the left side of the strip, thanks to
the rotation by $\pi$ we should take the Chan-Paton space with vacuum $N/2$ on the right-Brane.
Next, assuming that $j>i$ we can use equation \eqref{eq:NtoNbar} to say
\be\label{eq:ProveHopCoho-3}
\CE(\fN_{-1+i-j-N/2}^{[1]})_{N/2} = \CE(\bar\fN_{j-i-N/2}^{[-1]})_{N/2}
\ee
Finally, using \eqref{eq:NbarCP} we have
\be\label{eq:ProveHopCoho-4}
\CE(\bar\fN_{j-i-N/2}^{[-1]})_{N/2} = \bar S^{[i-j]}_{j-i}
\ee
Putting together equations \eqref{eq:ProveHopCoho-0}-\eqref{eq:ProveHopCoho-4} we
finally arrive at a proof of equation \eqref{eq:HopCoho-1}.

We expect that similar manipulations allow a computation of the cohomologies
of the groundstate complexes such as \eqref{eq:FEARSOME} and the spaces of local operators
in \eqref{eq:HopCoho-2} and \eqref{eq:HopCoho-3}.

\subsubsection{Powers Of The Rotation Interface}\label{subsubsec:Powers}

It is rather interesting to examine the powers of the Interface $\fI^{+-+} := \fI^{+-}\boxtimes\fI^{-+}$
that corresponds to a rotation by $2\pi/N$ in the worldvolume of the Theory. We will discuss this
at the level of Chan-Paton factors, without investigating the boundary amplitudes.

The Chan-Paton data of $\fI^{+-+}$ is the matrix of complexes:
\be\label{eq:CPpmp}
\begin{split}
\CE(\fI^{+-+}) & = R^{[1]}_{N-1,0} e_{0,0} \oplus \bigoplus_{j=0}^{N-2} \IZ e_{j+1,j} \oplus \IZ^{[1]}e_{0,N-1} \\
&  \oplus \bigoplus_{1\leq j < (N-1)/2} R_{N-1-j,j} e_{N-j,j} \oplus \bigoplus_{0\leq j < (N-2)/2} R_{N-1-j ,j+1} e_{N-1-j,j} \\
\end{split}
\ee
Explicitly, for $N=2$ this is
\be\label{eq:epmp-2}
\CE(\fI^{+-+}) = \begin{pmatrix} R_{1,0}^{[1]} & \IZ^{[1]} \\ \IZ & 0 \\ \end{pmatrix}
\ee
and for $N=3$,
\be\label{eq:epmp-3}
\CE(\fI^{+-+}) = \begin{pmatrix} R_{2,0}^{[1]} & 0 & \IZ^{[1]} \\ \IZ & 0 & 0 \\
R_{2,1} & \IZ & 0  \\ \end{pmatrix}
\ee

The equation \eqref{eq:PeriodicPower} suggests that for $\CT^N_{\vartheta}$ the $(N+1)^{th}$ power
is homotopy equivalent to a (shifted) Identity interface. Indeed, one easily checks that
the third power of \eqref{eq:epmp-2} is just
\be
\begin{pmatrix}
\IZ^{[2]} \oplus \IZ^{[2]}\oplus \IZ^{[3]} &  \IZ^{[2]}\oplus \IZ^{[3]}  \\
 \IZ^{[2]}\oplus \IZ^{[3]} & \IZ^{[2]} \\
\end{pmatrix}
\ee
and is quasi-isomorphic to $\IZ^{[2]} \textbf{1}_2$. Similarly, a check by hand shows that
the fourth power of \eqref{eq:epmp-3} is quasi-isomorphic to $\IZ^{[2]} \textbf{1}_3$,
and we conjecture that for all $N$, $(\fI^{+-+})^{\boxtimes (N+1)}$ is homotopy equivalent
to the isomorphism Interface given by a degree shift of $2$.
 There is a   simple  intuitive explanation in the LG theory   \ref{eq:TN-Superpot}
for this result. The  effect of convolution with $\fI^{+-}$ or $\fI^{-+}$
on geometric branes simply rotates the sectors at infinity by one unit and deforms the
geometric brane accordingly by a rigid rotation by $2\pi/(N+1)$ in the $\phi$ plane.
\footnote{Do not confuse this with the origin of the Interfaces from rotations by
$2\pi/N$ in the $(x,\tau)$ plane.}
As there are $2N+2$ sectors, the $(N+1)$ power of the
interface $\fI^{+-}\fI^{-+}$ acts geometrically on the branes by rotating it by $2 \pi$ back to itself in the $\phi$ plane.

In fact, the characteristic polynomial
of $\CE(\fI^{+-+})$ is given by the remarkable formula:
\be\label{eq:CharPoly}
\det( x \textbf{1}_N - \CE(\fI^{+-+}))=x^N + \sum_{j=1}^{N-1} R_{j} x^j + 1
\ee
where we interpret a shift by $[1]$ as a minus sign and we use the property that $R_{a,b} = R_{a-b}$ only
depends on the difference $a-b$. For a proof see Appendix \S \ref{app:ProveCharPol}

For the Theory $\CT^N_{\vartheta}$ we have $R_j = \IZ$, and hence the ``eigenBranes'' of
$\fI^{+-+}$ have eigenvalues given by the $N$ nontrivial $(N+1)^{th}$ roots of unity.
This proves that $(\fI^{+-+})^{N+1}$ is homotopy equivalent to
the identity (up to an even degree shift).

We can also apply equation \eqref{eq:CharPoly} to the Theory $\CT^{SU(N)}_{\vartheta}$. Now
 we have $R_j = A_{N-j}$ so, at the level of the Witten index we can factorize equation
\eqref{eq:CharPoly} to get the character
\be\label{eq:FactorCharPoly}
 \prod_{i=1}^{N} (x+ t_i)
\ee
where $t= {\rm Diag}\{ t_1,\dots, t_N \}$  is a generic element in the
diagonal Cartan subgroup of $SU(N)$. It is natural to
suspect that the  ``eigen-Branes'' can be interpreted in the $\IC\IP^{N-1}$ $B$-model as
Dirichlet branes located at the $N$ fixed points of the natural $SU(N)$ action on the homogeneous
coordinates, with eigenvalue $t_i$. The result \eqref{eq:FactorCharPoly} will be very
useful when we discuss local operators in Section \S \ref{subsec:LocOp-TSUN} below.

\section{Categorical Transport And Wall-Crossing}\label{sec:GeneralParameter}

\subsection{Preliminary Remarks}\label{subsec:CatTransPrelim}

We now return to the general situation discussed at the beginning of Section \S \ref{sec:CatTransSmpl}. In
Section \S \ref{sec:CatTransSmpl} we considered in detail categorical transport of Brane categories
associated to paths of weights $\wp$ given by spinning weights \eqref{eq:SpinningWeights}. In this section we consider more
general vacuum homotopies. In particular we will consider three kinds of vacuum homotopies:

\begin{enumerate}

\item  Vacuum homotopies $\{ z_i(s) \}$ which are more general than \eqref{eq:SpinningWeights}
but do not cross the real codimension one walls of special webs described in
Section \S \ref{subsec:SpecialVacWt}. In this
case the webs behave in a very similar way to those of \eqref{eq:SpinningWeights}.
We will call these \emph{tame vacuum homotopies}. They are discussed in Section \S \ref{subsec:TameVacHmtpy}.

\item Vacuum homotopies $\{ z_i(s) \}$ which cross the exceptional walls described in
 \S \ref{subsec:SpecialVacWt} above.
 This is discussed in
Section \S \ref{subsec:X-WEB-WC}.

\item Vacuum homotopies $\{ z_i(s) \}$ which cross walls of marginal stability
described in \S \ref{subsec:SpecialVacWt}.  This is discussed in Section
\S \ref{subsec:MS-WC}.

\end{enumerate}

In Section \S \ref{sec:CatTransSmpl} we constructed Interfaces $\fI[\vartheta(x)]\in \fB\fr(\CT^\ell, \CT^r)$.
Given $\CT^\ell$ there was a canonical choice for $\CT^r$ given by taking ``constant'' web representation $\CR$
and interior amplitude $\beta$. In this section we will see that the more general paths of weights listed
above make a canonical determination of $\CT^r$ given $\CT^\ell$ somewhat more problematical. The reason for
this is that there can be \emph{wall-crossing} phenomena associated to the data $(\CR,\beta)$ used to define
a Theory. In particular, we will see that if $\wp(s)$ crosses an exceptional wall then the $L_\infty$ algebra
of closed webs will in general change because the taut element will in general change. Therefore, in general
the interior amplitude will change. If $\wp(s)$ crosses
a wall of marginal stability then the  set of cyclic fans will change and hence $\Rvtx$ must change.
Indeed, in general when crossing a wall of marginal stability
both the interior amplitude and the web representation $\CR$ will change.
The rules for the discontinuity of $\CR$ lead to a categorification of the
Cecotti-Vafa-Kontsevich-Soibelman wall-crossing formula.

We will now make the notion of a ``change of Theory'' somewhat more precise.
Given a vacuum homotopy $\wp: \IR \to \IC^{\IV}-\Delta$ we say that a family
of Theories $\CT(s)$ is continuously defined over $\wp$ if for all $i\not=j$
the $R_{ij}$ form a continuous vector bundle with connection over $\wp$ such that
$K_{ij}$ and $\beta$ are parallel-transported. If we can and do trivialize the bundle with
connection then $R_{ij}, K_{ij}, \beta$ are all constant. As mentioned above,
when the path $\wp(s)$ crosses walls with special configurations of weights
there will be obstructions to defining a continuous family of theories over that path.
Instead, we can only define \emph{piecewise-continuous} paths of Theories over $\wp$.
Naively, the only discontinuities are located at the walls of special weights described in
Section \S \ref{subsec:SpecialVacWt}. We will assume this for the moment, but
that assumption will need to be revised for reasons described in Remark 2 below.
 A formula for the discontinuity
of $\CT$ is a \emph{wall-crossing formula}. Let $\CT^-$ denote the Theory just before
the wall and let $\CT^+$ denote the Theory just after the wall. Thus, a wall-crossing
formula is, in its simplest incarnation, just
 a prescription for determining $(\CR^+,\beta^+)$ from $(\CR^-,\beta^-)$.

Given such a wall-crossing rule, if we have path of vacuum weights $\wp$ then, given $\CT^\ell$
we can construct a corresponding piecewise-continuous path $\CT(s)$ of Theories. The wall-crossing
rule should then be constrained by requiring that the path $\CT(s)$ behave suitably with
respect to homotopy and concatenation of paths of vacuum weights $\wp$. Heuristically speaking, we want to define
a ``flat connection on Theories.'' However a little thought quickly shows such a parallel transport rule must be defined on a
suitable equivalence class of Theories. In order to motivate the relevant notion of equivalence let
us say a little more about how we propose to approach the wall-crossing formula.

As in Section \S \ref{sec:CatTransSmpl} our theme will be to interpret the variation of parameters $\CT(s)$
as spatial-variation of parameters, so we will have spatially dependent vacuum weights  $\wp(x)$ and
spatially-dependent data of Theories $\CT(x)$.   Then it is quite natural to interpret a discontinuity of theories
across some point $x_*$ in terms of a suitable ``wall-crossing Interface''
  $\fI^{\rm wc}\in \fB\fr(\CT^-,\CT^+)$.

Let us make this slightly more precise. We assume that
there exist  $x_{\ell}$ and $x_r$ so that  $\wp(x)$ is constant for $x \leq x_{\ell}$
and $x \geq x_r$. Choose such points and let $z^\ell: \IV \to \IC$ and $z^r: \IV \to \IC$ be the corresponding
weight functions in these regions.  Then there should be a corresponding piecewise-continuous family of
Theories $\CT(x)$ interpolating between $\CT^\ell$ and $\CT^r$ together with an Interface
\be\label{eq:defIwpth}
\fI[\CT(x)] \in \fB\fr(\CT^\ell, \CT^r)
\ee
generalizing equation \eqref{eq:defIth}. As before, such a family of Interfaces allows us to define
a functor of Brane categories $\CF: \fB\fr(\CT^\ell) \to \fB\fr(\CT^r)$ and hence define a categorical
transport law on Brane categories.

When trying to construct the relevant Interfaces we will keep in mind the following
three useful guiding principles:

\begin{enumerate}

\item The Interfaces for paths such that  $z_{ij}(x)$ is never pure imaginary should
already define functors between the vacuum categories $\fVac(\IV, z^\ell)$ and $\fVac(\IV,z^r)$.
In particular the Chan-Paton factors of $\fI[\CT(x)]$ will be   $\CE_{ii'} = \delta_{ii'} \IZ$.

\item We must have properties \eqref{eq:PT-Intf1} and
\eqref{eq:PT-Intf2}: First, if there are  two piecewise-continuous
families of Theories  $\CT^1(x)$ and $\CT^2(x)$ that can be concatenated
at a point of continuity then
\be\label{eq:Tame-CatParTrspt}
\fI[\CT^1(x)] \IntfcTimes \fI[\CT^2(x)] \sim \fI[\CT^1 \circ \CT^2(x) ] .
\ee
Second, let   $\fE$ be the exceptional set of weights in $\IC^{\IV}-\Delta$ described in
Section \S \ref{subsec:SpecialVacWt}. That is, the subset $\{ z_i \}$ where some subset
of three or more weights is colinear, or where there are exceptional webs.
Then, if $\wp^p(x)$ and $\wp^f(x)$ are paths in $\IC^{\IV}-\Delta- \fE $ homotopic
in $\IC^{\IV}-\Delta- \fE $ through a homotopy keeping fixed
$(\IV, z^\ell)$ and $(\IV, z^r)$ for $x\leq x^\ell$ and $x\geq x^r$
and $\CT^p(x)$ and $\CT^f(x)$ are corresponding paths of Theories then
\be\label{eq:HE-INTFC}
\fI[\CT^p(x)] \sim \fI[\CT^f(x)]
\ee%
are homotopy equivalent Interfaces. As before, given such Interfaces we
 have a notion of flat parallel transport along $\wp$ from the category of Branes $\fB\fr(\CT^\ell)$
to the category of Branes $\fB\fr(\CT^r)$, generalizing what was constructed in Section
\S \ref{sec:CatTransSmpl}.

\item Because the underlying physical theory is  rotationally invariant the
Interfaces  should come in a $\vartheta$-dependent family, intertwined by the rotation interfaces.
To be more precise, for any path $\vartheta(x)$ of real numbers from $0$ to $\vartheta$
we can concatenate the path $\wp(x)$ from
$\{ z_i^\ell \}$ to $\{ z_i^r \}$ with a path $e^{-\I \vartheta(x)} z_i^r $ from $z_i^r$  to
$e^{-\I \vartheta} z_i^r $. Alternatively we can concatenate the path $e^{-\I \vartheta(x)} z_i^\ell $
with the path $e^{-\I \vartheta} \wp(x)$. The homotopy $z_i(x) e^{-\I \vartheta(y)}$ shows
that these two vacuum homotopies are homotopic and hence it follows from \eqref{eq:Tame-CatParTrspt}
that
\footnote{The notation $\fR[0, \vartheta]$ is slightly ambiguous since the interface actually depends on
the initial Theory, just like in equation \eqref{eq:defIwpth}.
 The two appearances of this Interface in \eqref{eq:fI-rot-intertwine} are hence slightly different.
 Also the notation $e^{-\I \vartheta}\CT(x)$ means that we take the same continuous family of $(\CR,\beta)$
 but the vacuum weights are $e^{-\I \vartheta}z_i(x)$. }
\be\label{eq:fI-rot-intertwine}
\fR[0, \vartheta] \IntfcTimes \fI[e^{-\I \vartheta} \CT(x) ] \sim \fI[\CT(x)] \IntfcTimes \fR[0, \vartheta].
\ee
More geometrically, we can imagine rotating the plane by angle $\vartheta$ and defining interfaces
for the rotated Theories  for vacuum homotopies defined along the rotated $x$-axis. Of course,
 this should not essentially change the parallel
transport, and that is what equation \eqref{eq:fI-rot-intertwine} is meant to express. Note in particular that if we
set $\vartheta = 2\pi$ then $ \fI[e^{-2\pi \I } \CT(x) ] = \fI[\CT(x)]$ are literally equal, but
$\fR[0,2\pi]$ might well be nontrivial.

%$\fR[\vartheta, \vartheta']$,
%periodic as $\vartheta \to \vartheta+2\pi$.
%
%\cg{Up to homotopy equivalence?}
%
%\item The composition of such interfaces and the standard tame variation away from the special values of the
%parameters will give some invertible interfaces $\CI[\CT[x]]$,
%which are compatible up to homotopy equivalence with compositions and homotopies of families of theories.
%
%\cg{This is too vague. Write out more details.}
%

\end{enumerate}

We can now say what our notion of equivalence of Theories will be. We say that Theories $\CT^1$ and $\CT^2$
are \emph{equivalent Theories} if there exists a \emph{periodic} family of invertible Interfaces
$\fI^{\vartheta}$ between $\CT^{1,\vartheta}$ and $\CT^{2,\vartheta}$ which intertwine with the rotational
Interfaces in the sense of equation \eqref{eq:fI-rot-intertwine}.
The auto-equivalences of a Theory with itself form a kind of ``gauge symmetry'' of the ``flat connection
on Theories.'' We should only hope to define parallel transport up to such ``gauge symmetry.'' As a simple
example, the dependence of equation \eqref{eq:defIwpth} on $x_\ell$ and $x_r$ is only up to equivalence of
Theories in this sense.

Conditions 1,2,3 above on the Interfaces $\fI[\CT(x)]$ are certainly rather restrictive. We do not
know if they are defining properties.

In the remainder of
Section \S \ref{sec:GeneralParameter} we will construct Interfaces and rules for constructing $\CT(x)$
for paths $\wp$ which cross exceptional walls and walls of marginal stability. We will see that
the existence of such Interfaces imposes  strong constraints on how the interior amplitude and the web representation
can vary along the family.  The problem of finding the Interfaces
is over-determined and thus invertibility of the Interfaces, compatibility with rotations, homotopy invariance, etc.
 give constraints on the family $\CT(x)$.
This phenomenon is a categorical version of the derivation of the wall-crossing formula for the $\mu_{ij}$ from the
properties of framed BPS degeneracies
under deformations of parameters
\cite{Gaiotto:2010be,Gaiotto:2011tf,MOORE_FELIX}.
%

%Ideally, a ``categorical wall-crossing formula'' would determine the evolution $\CT(x)$ of the representation data
%and interior amplitude
%in terms of the evolution of the vacuum data $\wp(x)$. There are simple reasons for which this aspiration may be overly ambitious:
%we will demonstrate momentarily that it is easy to produce interfaces
%between pairs theories with the same vacuum data such that the three properties we are after are all satisfied.
%Clearly, we can only hope to determine $\CT(x)$ up to the action of such interfaces.
%
%\cg{Try to restate the above three paragraphs a bit more sharply. Where do we deliver on the three promises made
%above? }
%

\bigskip
\textbf{Remarks}

\begin{enumerate}

\item A more ambitious formulation of a wall-crossing formula is to give an
 $L_\infty$ morphism $\gamma^{\rm web}$ between the planar web algebras determined by $(\IV^\pm, z^\pm)$
compatible with an $L_\infty$ morphism $\gamma$ between the $L_\infty$-algebras associated
with the Theories $\CT^\pm$. We will, in fact, do this for crossing exceptional walls.
One should probably go further and  construct  an ``$LA_\infty$ morphism'' of ``$LA_\infty$  algebras.''
which is compatible with a   functor $\CF: \fB\fr(\CT^-) \to \fB\fr(\CT^+)$. We have not done that.
It is not clear to us if such data is uniquely determined by giving families of Interfaces
\eqref{eq:defIwpth}.

\item There is a natural notion of ``inner auto-equivalence'' between Theories $\CT^\pm$ which have the same
vacuum data and representation of webs, but interior amplitudes which differ by an exact amount:
\be \label{eq:inner1}
\beta^+ = \beta^- + \rho_{\beta_-}(\ft)[\epsilon]
\ee
where $\epsilon$ is a degree $1$ element in $\Rvtx$ supported on a single fan $I_\epsilon$.
Because of the line principle, $\rho(e^{\beta^+}) = \rho_{\beta_-}(e^{\rho_{\beta_-}(\epsilon)})=0$
if $\rho(e^{\beta^-})=0$. It is straightforward to map the Brane categories of the two Theories into each other,
simply by shifting boundary amplitudes in a similar fashion:
\be\label{eq:inner2}
\CB^+ = \CB^- + \rho_{\beta_-}(\ft_\p)[\frac{1}{1-\CB^-};\epsilon]
\ee
This transformation can also be implemented by a family of interfaces $\fI_\epsilon^\vartheta$ which differs from the identity interfaces
only by a shift by $\epsilon$ of the boundary amplitude. It is possible to show that these interfaces are truly invertible (i.e. not just up to homotopy)
and commute with rotation interfaces. Indeed, the two sides of the commutation relation
\ref{eq:fI-rot-intertwine} for these interfaces only differ by an exact term added to the boundary amplitude.
It is also possible to recast these relations in the form of an $LA_\infty$ algebra isomorphism. \footnote{In order to prove these statements,
it is useful to observe that the amplitude for a web defined in the presence of an $\fI_\epsilon^\vartheta$ interface which includes an insertion of
$\epsilon$ at an interface vertex is identical to the amplitude for a web with the same geometry defined in the absence of the $\fI_\epsilon^\vartheta$ interface.
By this identification, the MC equation for $\fI_\epsilon^\vartheta$ becomes \ref{eq:inner1}, the definition of $\fB\IntfcTimes \fI_\epsilon^\vartheta$
maps to \ref{eq:inner2} and \ref{eq:fI-rot-intertwine} maps to the convolution identity for curved webs.
}

\item Inner auto-equivalences play a role in the relation between Theories and concrete physical theories:
although we expect to have a direct map from physical theories to Theories, the image of the map
may change by inner auto-equivalences as we vary the parameters of the underlying physical theory,
even if the corresponding vacuum homotopy does not cross the exceptional set $\fE$. These jumps will occur at
co-dimension one walls whose position depends on the detail of the underlying theory, possibly including D-term
deformations. We will therefore call these \emph{phantom walls}.
Because of the possibility of phantom walls, we should really map physical theories to equivalence classes of Theories
up to inner auto-equivalences. Correspondingly, in physical applications any ``categorical wall-crossing formula'' should be understood
up to inner auto-equivalences. It is also possible to envision another class of phantom walls, across which
the $R_{ij}$ themselves may change to a homotopy equivalent complex, which would require one to quotient
the space of Theories further in order to define a robust map from physical theories. We leave open the problem
to identify which type of equivalences between theories should be associated to the most general possible phantom walls.

\item We can elaborate further on the possibility of phantom walls in the context of physical theories such as the Landau-Ginzburg
theories we discuss in Sections \S\S \ref{lgassuper}-\ref{subsec:LG-Susy-Interface}.
Suppose we are given a one-parameter family of superpotentials $W(\phi;s)$, say with $s\in \IR$.
Following, Remark 9 of Section \S \ref{planewebs} we obtain a vacuum homotopy $\{ z_i(s) \}$.
In general, there will be isolated points $s_*$ where $W_{s_*}$ admits exceptional $\zeta$-instantons
with fan boundary conditions. (See Section \S \ref{zetawebs} below for a discussion
of $\zeta$-instantons.) Such exceptional
instantons will have a moduli space whose formal dimension (given by the
index $\iota(L)$,  discussed in Section \S \ref{tonzo} below) is $1$. This means
that the amplitude associated with the path integral with fan boundary conditions
(as discussed in Section \S \ref{webscot}) will define an element $\gamma \in \Rvtx$
with fermion number $+1$. It can be inserted into some taut webs in $\ft_{s_*}$ to produce an element
$\rho_{\beta}[\ft_{s_*}](\gamma)$ of fermion number $+2$. This will contribute to a jump in the
interior amplitude $\beta$ as $s$ passes through $s_*$. In terms of $\zeta$-instantons,
as $s\to s_*$ the size of the relevant $\zeta$-web
that can accomodate the exceptional instanton at a vertex will grow to infinity.
This is an infrared phenomenon associated with working on a noncompact spacetime;
it has no counterpart
to invariants associated with topological field theory integrals defined on compact manifolds.

\item We should note that these are by no means the most general continuous families we could consider.
An important variation on the above ideas involves replacing the vacua $\IV$ with the
fibers of a branched covering $\pi: \Sigma \to C$, where $C$ is a space of Theories.
This is the setup appearing in the 2d4d wall-crossing formula of \cite{Gaiotto:2011tf}.
It should be possible to extend the ideas of the present paper to the more general setting
of a branched cover, but, beyond
some remarks in Section \S \ref{subsec:CatSpecNet}, that lies beyond the scope of this paper.

\end{enumerate}

\subsection{Tame Vacuum Homotopies}\label{subsec:TameVacHmtpy}

We define a  \emph{tame vacuum homotopy} to be a vacuum homotopy   $\{ z_i(x)\} $  such that:

\begin{enumerate}

\item  For all $x$ the set of weights $\{ z_i(x) \}$ is in general position, in the sense of Section
\S \ref{subsec:SpecialVacWt}.

\item  Each taut web in ${\rm WEB}[x]$ (the set of plane webs determined by
the vacuum weights $\{ z_i(x) \}$ )
fits into a continuous family of webs,  in the sense defined in Section
\S \ref{subsubsec:Homotopy-homotopy}. Moreover, no taut web is created as $x$ varies.

\end{enumerate}

Given these criteria we can speak of a single web group $\CW$
and there is a continuously varying planar taut element $\ft_{pl}(x)$.
We can also define curved webs precisely as in Section \S \ref{subsec:CurvedWebs} and
hence we can define the curved taut element $\ft$ to be the sum of oriented deformation types of curved taut webs, i.e.
those with expected dimension $d=1$. We can then write a convolution identity for $\ft$.
We make the assumptions contained in the paragraph containing equation \eqref{eq:defIwpth}.
In particular, the vacuum data $(\IV,z^\ell)$ and $(\IV, z^r)$ define web groups $\CW^\ell$ and $\CW^r$.
 We can define  $\ft_{pl} = \ft^{\ell}_{pl} + \ft^{r}_{pl}$
to be the formal sum of the planar taut elements in the web groups.
Similarly, we let $\ft^{\ell,r}$ denote the taut interface
element for an interface separating vacuum data   $(\IV, z^\ell)$ and $(\IV, z^r)$.  Then we have the analogue
of equation \eqref{eq:Rotating-Web-Ident}
\be\label{eq:Curved-Web-Ident}
\ft * \ft_{pl} + T_\p(\ft^{\ell,r})[\frac{1}{1-\ft}]=0.
\ee

We choose the   representations $\CR^\ell$ and $\CR^r$ of the webs
determined by $(\IV, z^\ell)$ and $(\IV, z^r)$ to be the same, indeed we can think
of a constant, i.e. $x$-independent representation of the vacuum data $(\IV, z(x))$. For a
tame vacuum homotopy the set of cyclic fans is constant so $\Rvtx$ is constant.  Moreover, since the
taut element $\ft_{pl}(x)$ varies continuously it makes sense to speak of  an $x$-independent
interior amplitude $\beta$. We can therefore define the contraction operation $\rho_{\beta}$ on curved webs and
then \eqref{eq:Curved-Web-Ident} implies that
\begin{equation}\label{eq:TameIntfcAmp}
\rho_\beta(\ft^{\ell,r})[\frac{1}{1-\rho^0_\beta(\ft)}]=0
\end{equation}
where the superscript $0$ on  $\rho^0_\beta(\ft)$ indicates that the interior amplitude $\beta$ is inserted at all
vertices of the  \emph{curved} taut element $\ft$. It follows that $\rho_\beta^0(\ft)$ can be regarded as
an interface amplitude, where the Chan-Paton factors for the  interface are given once again by the formula
\eqref{eq:TautCurvedCP}, repeated here:
\be\label{eq:TautCurvedCP-2}
 \oplus_{j,j'\in \IV} \CE_{j,j'} e_{j,j'} := \bigotimes_{i\not=j}
\bigotimes_{x_0 \in \curlyvee_{ij} ~ \cup \curlywedge_{ij} } S_{ij}(x_0),
\ee
where we just take $x_0$ in the interval $(x^\ell, x^r)$.
Strictly speaking we should define binding points and binding walls in the more general context
of tame vacuum homotopies, but the definitions of Section \ref{subsec:BindPoints} are essentially
the same and will not be repeated.

Therefore, to a tame vacuum homotopy and a choice of points $x_{\ell}, x_r$, with corresponding
Theories $\CT^\ell$ and $\CT^{r}$ determined by a constant web representation $\CR$ and vacuum
amplitude $\beta$ we can construct an Interface between the theories.

We can now imitate closely the ideas used in Section \S \ref{subsec:CompThrIntfc}. If we are
given a tame homotopy $\wp(x,y)$ between two tame vacuum homotopies $\wp^p(x)$ and $\wp^f(x)$
then there is a set of weights $z_i(x,y)$ which we can regard as space-time dependent with $z_i^p(x)$
in the far past and $z_i^f(x)$ in the far future.  Curved webs again make sense with these
space-time dependent weights and we can use them, together with a constant web representation and
interior amplitude to construct a closed invertible morphism $\Id + \delta[\wp(x,y)]$
between $\fI[\wp^p(x)] $ and $\fI[\wp^f(x)]$. Again we can define the time-concatenation $\circ_T$ of two
such homotopies and we claim that
\be\label{eq:shainenti}
\Id + \delta[ \wp^1 \circ_T \wp^2 ] = M_2(\Id + \delta[\wp^1], \Id + \delta[\wp^2])
\ee
Finally, as in Section \ref{subsubsec:Homotopy-homotopy} a homotopy of homotopies $\wp(x,y;s)$ with
fixed vacuum weights $z^\ell$ for $x\leq x_\ell$ and $z^r$ for $x \geq x_r$, and fixed $\wp^p(x)$ in
the far past and $\wp^f(x)$ in the far future defines a homotopy equivalence between the
morphisms $\Id + \delta[ \wp^p(x)]$ and  $\Id + \delta[ \wp^f(x)]$. It then follows from
\eqref{eq:shainenti} that homotopies between vacuum homotopies lead to homotopy-equivalent
Interfaces thus checking \eqref{eq:HE-INTFC} for tame homotopies between tame
vacuum homotopies.  In a similar way we can also check equation \eqref{eq:Tame-CatParTrspt}
for concatenation of tame vacuum homotopies.

\begin{figure}[htp]
\centering
\includegraphics[scale=0.3,angle=0,trim=0 0 0 0]{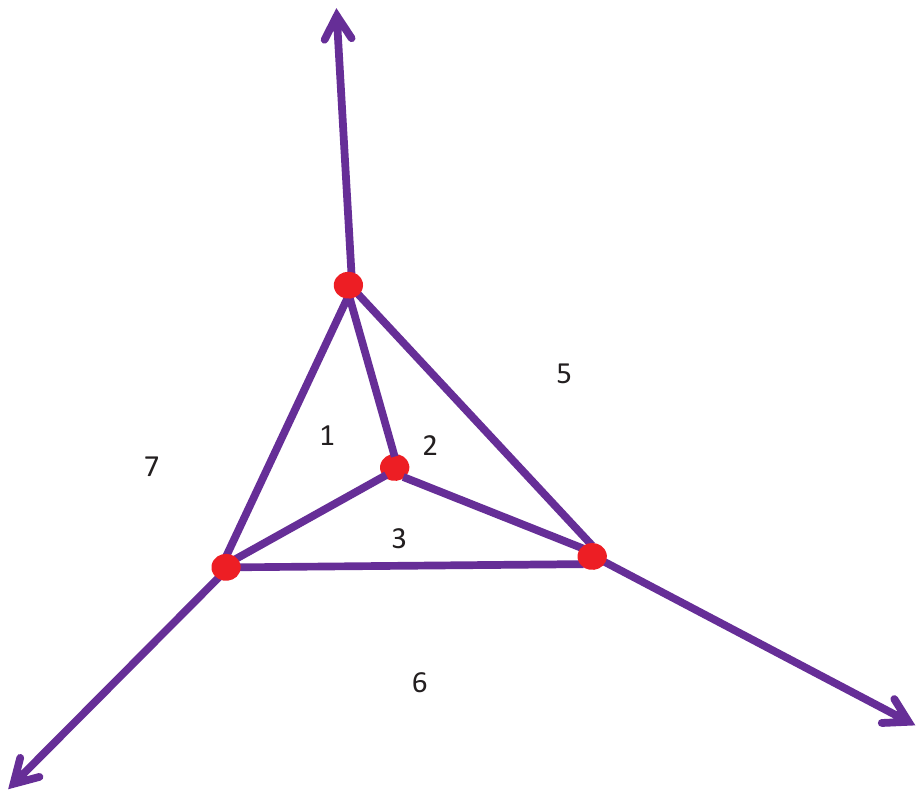}
\caption{An exceptional web which appears only at $s=s_*$ in a family of webs
defined by $\{ z_i(s)\}$, $s\in \IR$.   }
\label{fig:XWC-1}
\end{figure}
\begin{figure}[htp]
\centering
\includegraphics[scale=0.3,angle=0,trim=0 0 0 0]{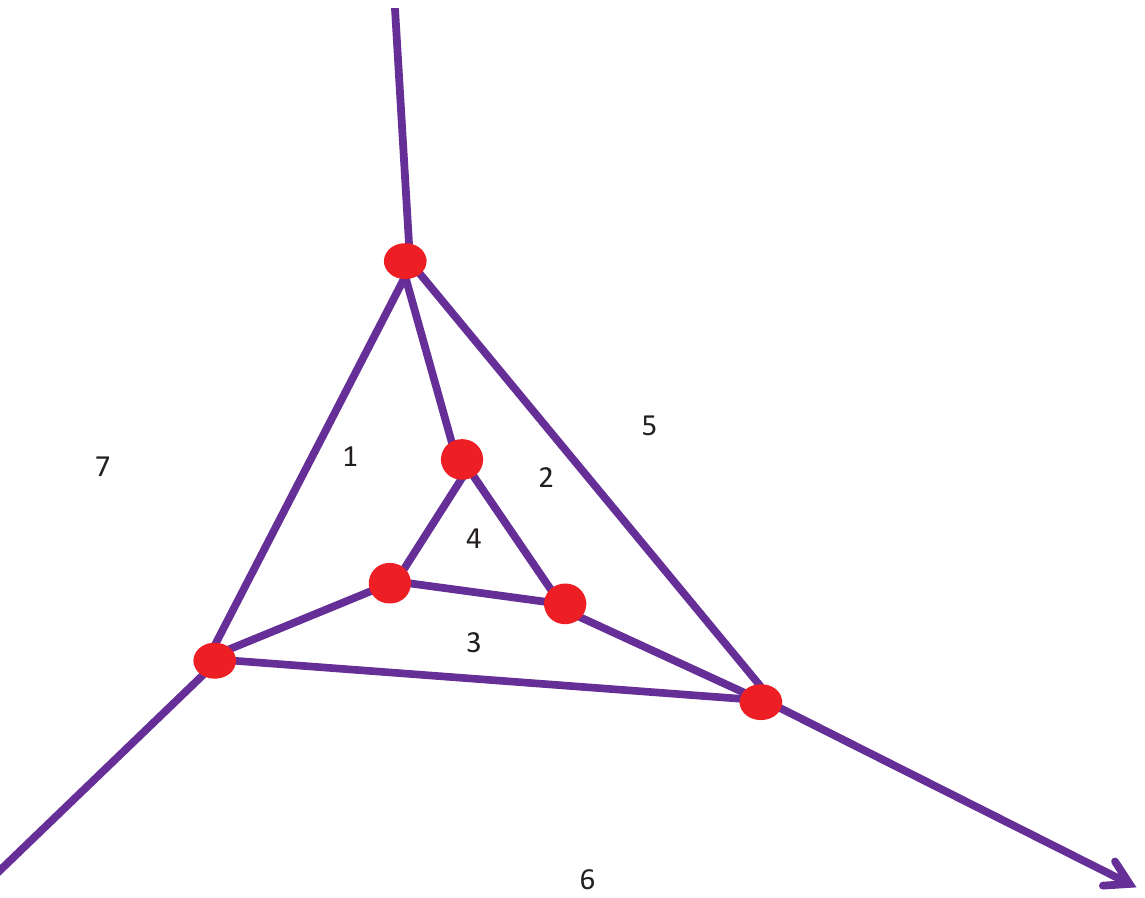}
\caption{A non-exceptional web which can degenerate as $s\to s_*$ to the exceptional
web shown in Figure \protect\ref{fig:XWC-1}.
%
% \ref{fig:XWC-1}.
%
   }
\label{fig:XWC-2}
\end{figure}
\begin{figure}[htp]
\centering
\includegraphics[scale=0.3,angle=0,trim=0 0 0 0]{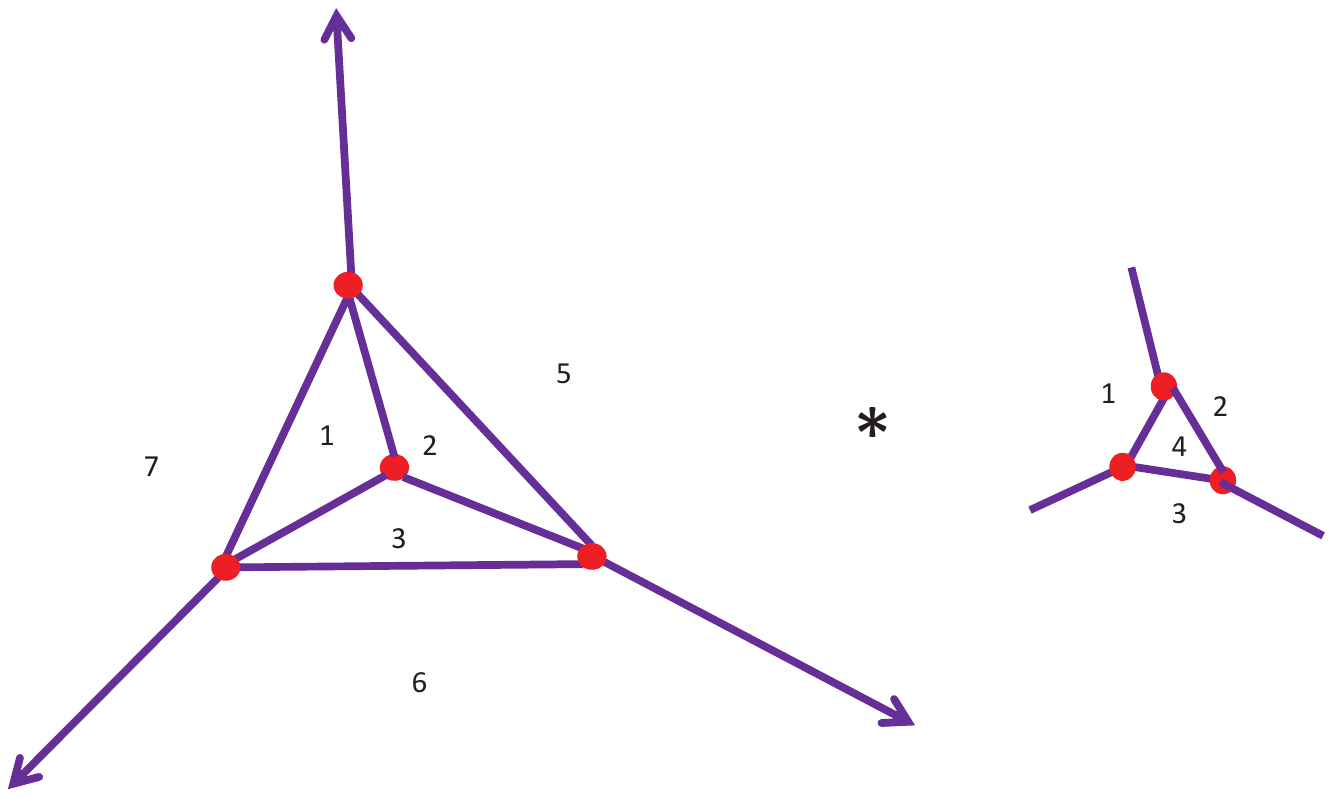}
\caption{The taut web of Figure \protect\ref{fig:XWC-2} disappears
for $s>s_*$, and near $s=s_*$ its $h$-type can be written as a convolution of an
exceptinol web with a taut web.    }
\label{fig:XWC-3}
\end{figure}
\begin{figure}[htp]
\centering
\includegraphics[scale=0.3,angle=0,trim=0 0 0 0]{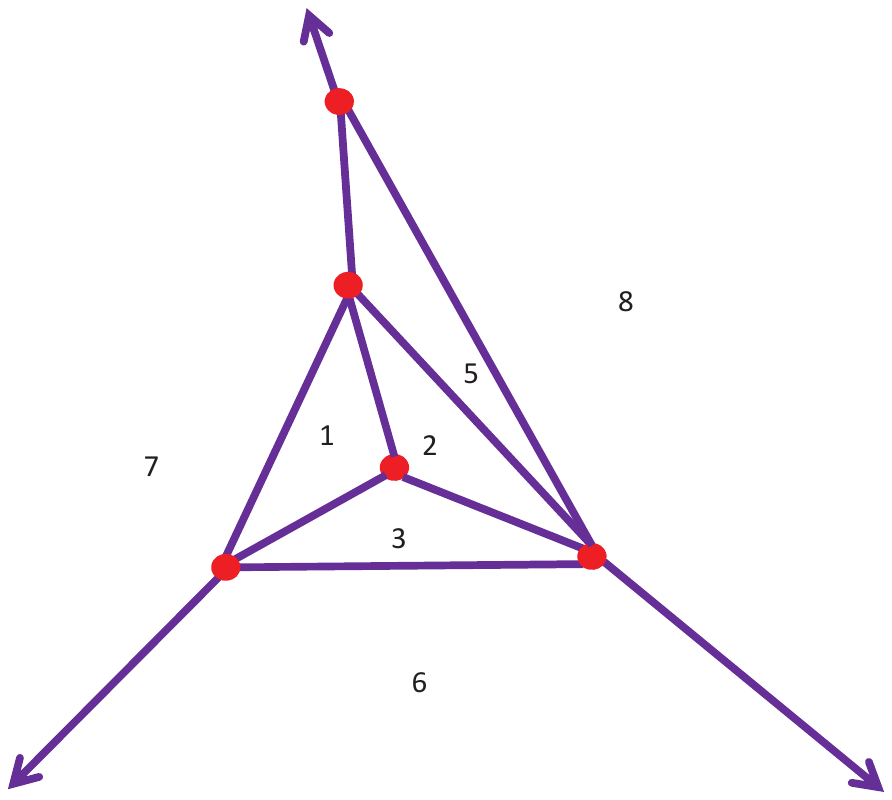}
\caption{Another exceptional web in the exceptional class of the web shown in
Figure \protect\ref{fig:XWC-1}.  }
\label{fig:XWC-4}
\end{figure}
\begin{figure}[htp]
\centering
\includegraphics[scale=0.3,angle=0,trim=0 0 0 0]{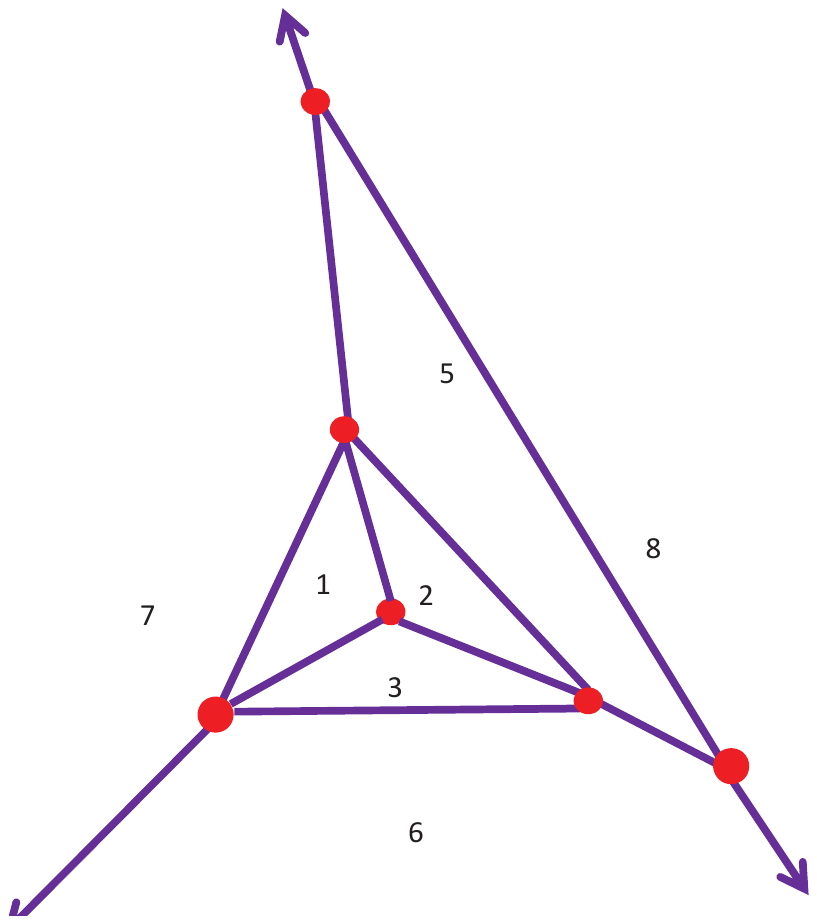}
\caption{An exceptional sliding web that will appear in the convolution
$\fe*\ft$. To see this convolve the vertex with fan $\{ 2,5,8,6,3\}$ of
Figure \protect\ref{fig:XWC-4} with a suitable taut web.    }
\label{fig:XWC-5}
\end{figure}
\begin{figure}[htp]
\centering
\includegraphics[scale=0.3,angle=0,trim=0 0 0 0]{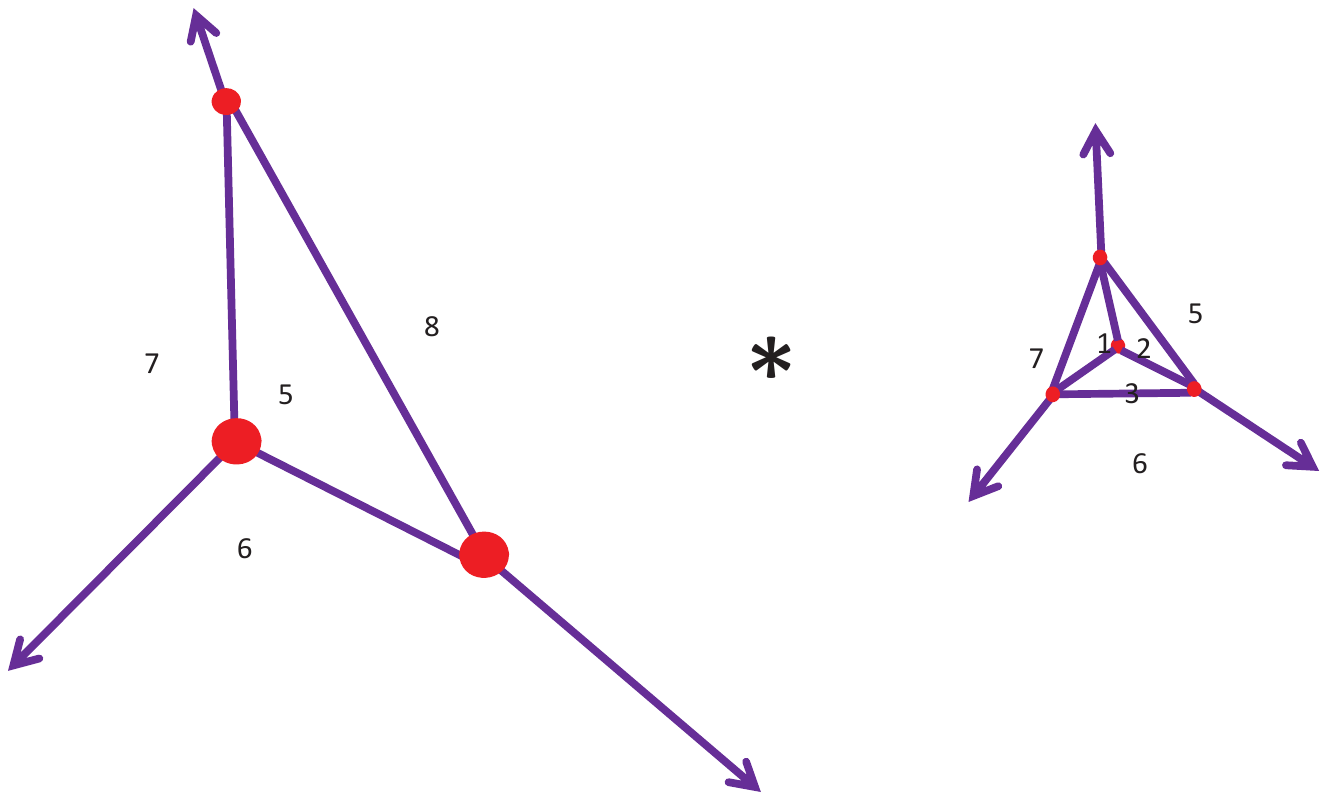}
\caption{Terms in $\ft*\fe$ such as this cancel the exceptional sliding webs such
as those shown in Figure \protect\ref{fig:XWC-5}.    }
\label{fig:XWC-6}
\end{figure}
\begin{figure}[htp]
\centering
\includegraphics[scale=0.3,angle=0,trim=0 0 0 0]{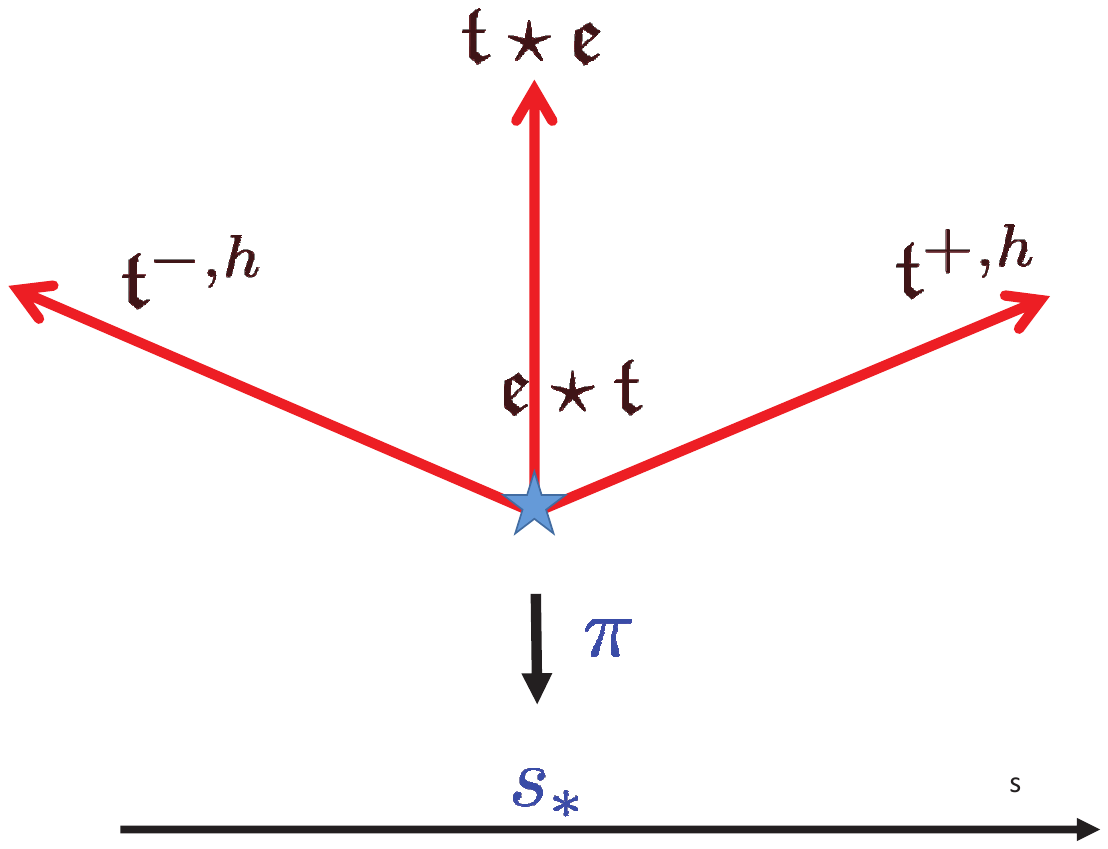}
\caption{A typical component of the moduli space of sliding $h$-types for the family
$\{ z_i(s) \}$. (Three dimensions for translation and dilation have been factored out.)
It can happen that the only nonzero component is just $\ft^+$ or $\ft^-$, as could happen for
the case of Figure \protect\ref{fig:XWC-1}.
%
%\ref{fig:XWC-1}.
%
Or it can happen that there are several branches
meeting at $s=s_*$, as for the case of Figure \protect\ref{fig:XWC-4}.
%
% \ref{fig:XWC-4}.
%
 Comparing the boundaries of this
dimension one complex leads to the convolution identity for exceptional webs.   }
\label{fig:XWC-7}
\end{figure}

\subsection{Wall-Crossing From Exceptional Webs}\label{subsec:X-WEB-WC}

We now come to a different kind of path of vacuum weights where $\wp(x)$ crosses
a wall of exceptional webs. For simplicity assume first that exceptional webs exist
only at a single point $x_*$.  It will be useful to consider first an abstract family of weights $\{z_i(s)\}$,
with $s\in \IR$ and consider the $h$-types of webs for this family rather than curved webs.
(Recall the definition of $h$-type in Section \S \ref{subsubsec:Homotopy-homotopy}.)
 The exceptional webs appear only at $s=s_*$.
We will describe how such families lead to $L_\infty$-morphisms of the $L_\infty$ algebras
$(\CW,T(\ft))$ of Section \S \ref{sec:AlgebraicStructures}
and $(\Rvtx,\rho_\beta(\ft))$ of Section \S \ref{subsec:WebRepPlane},  as well as $A_\infty$-morphisms
of the $A_\infty$ algebras defined by $\ft_\p$. Then, when we consider the family as an
$x$-dependent family of weights we describe the wall-crossing in terms of a suitable Interface.

\subsubsection{$L_\infty$-Morphisms And Jumps In The Planar Taut Element}

Let us begin with an example.

Suppose that the family of weights $\{z_i(s)\}$ admits
an exceptional web such as that shown in Figure \ref{fig:XWC-1} at $s=s_*$. If we
study the behavior of webs in the neighborhood of $s_*$ several things can happen.
The vertex with fan $I=\{1,2,3\}$ will continue to exist for all $s$, but it will typically
happen that   it will not ``fit'' into any larger triangle with fan $I_\infty = \{5,6,7\}$.
When this happens there are several different subcases:

\begin{enumerate}

\item It   can happen that the  exceptional web of Figure    \ref{fig:XWC-1} is not a degeneration of any web
that exists for $s\not= s_*$. Such a web then has no effect on the $L_\infty$ algebras
$(\CW,T(\ft))$ and $(\Rvtx, \rho_\beta(\ft))$.

\item  It can also happen that there
is a vacuum with weight $z_4(s)$ so that the web of Figure \ref{fig:XWC-1}
can be viewed as a degeneration of a nearby non-exceptional web, such as that shown
in  Figure \ref{fig:XWC-2}. Here, some set of edge constraints are effective for $s\not=s_*$ but
become ineffective, or linearly dependent, at $s=s_*$. Geometrically, some set of edges and vertices
shrinks to a single vertex. Generically, we will have a triangle shrink to a single vertex,
reducing the contribution of this set of edges and vertices
 to the expected dimension from $3$ to $2$. The result is an
exceptional web. In this case there are two further subcases we must consider:

\item  It can happen
that such nonexceptional degenerating webs exist both for $s<s_*$ and $s>s_*$. Again, when
this happens there might or might not be a difference in the
$L_\infty$ algebras $(\CW,T(\ft))$ and $(\Rvtx, \rho_\beta(\ft))$ defined by webs
for $s<s_*$ and $s>s_*$. This can be understood once we understand the next and
last case.

\item On the other hand, it can also happen that
the degenerating web of Figure \ref{fig:XWC-2} exists for $s>s_*$, but not for $s<s_*$, or
vice versa. Note that the web of Figure \ref{fig:XWC-2} is a \emph{taut} web. If it exists
for $s>s_*$ and not for $s<s_*$ (say), then there must be a change in the taut element
and hence a change in the $L_\infty$ algebras $(\CW,T(\ft))$ and $(\Rvtx, \rho_\beta(\ft))$.
We are most interested in this fourth case.

\end{enumerate}

In order to understand how the algebras $(\CW,T(\ft))$ and $(\Rvtx, \rho_\beta(\ft))$
change in the fourth case above
let us first note that the web of Figure \ref{fig:XWC-2} can be written as a convolution
as in Figure \ref{fig:XWC-3}. Thus, a convolution of an exceptional web with a non-exceptional
web can be non-exceptional. Moreover, given the rule \eqref{eq:expct-dim-conv}, in such a case
the convolution of a taut exceptional web with a taut non-exceptional web will be a \emph{taut}
non-exceptional web. This example suggests that we can write the change in the taut element
$\ft^+ - \ft^-$,  where $\ft^\pm$ are the taut elements for $s>s_*$ and $s<s_*$ respectively,
in terms of $\fe*\ft$ where $\fe$ is the sum of oriented exceptional taut webs and $\ft$ is the sum of
taut webs that do not change from $s>s_*$ to $s<s_*$.

There is a problem with expressing $\ft^+ - \ft^-$ in terms of $\fe*\ft$. The problem arises
from the fact that there can be further exceptional taut webs such as that shown in Figure \ref{fig:XWC-4}.
Indeed, we will refer to the set of exceptional webs obtained by shrinking the same small triangle
as an \emph{exceptional class} of webs. Generically $\fe$ will be the sum over one exceptional class.
Quite similarly to the case of Figure \ref{fig:XWC-1}, the web of Figure \ref{fig:XWC-4}, and indeed
  every web in the exceptional class, can be viewed as a degeneration
of a nonexceptional taut web obtained by taking the convolution of the $\{1,2,3\}$ vertex
as in Figure \ref{fig:XWC-3}. However, the problem is that the convolution $\fe*\ft$ will also
typically contain exceptional
sliding webs. An example is  shown in Figure \ref{fig:XWC-5}. Such terms must clearly be cancelled
off from $\fe*\ft$ since $\ft^+ - \ft^-$ contains no exceptional webs. We can do this by noting that
Figure \ref{fig:XWC-5} can also be degenerated by
shrinking the exceptional triangular web, producing a boundary in the form of
a convolution as shown in Figure \ref{fig:XWC-6}. This suggests that we should subtract $\ft*\fe$
from $\fe*\ft$.
Note that $\ft*\fe$ is always exceptional, and hence will always produce exceptional sliding webs.

We now generalize the above example by considering the $h$-types of the webs defined by $\{ z_i(s)\}$.
 The moduli space of sliding $h$-types
(that is, $h$-types of $h$-dimension $4$) will have typical components that
(for the doubly-reduced moduli space)  look like Figure \ref{fig:XWC-7},
which the reader should compare with Figure \ref{fig:SLIDING-MODULISPACE}. Comparing the boundaries
of this space leads to the convolution identity for jumps in the taut element due to exceptional webs:
\be\label{eq:excptl-conv}
\ft^+ - \ft^- = \fe * \ft - \ft* \fe
\ee
where, again, $\fe$ is the sum of exceptional taut webs at $s=s_*$ and $\ft$ is the sum of taut
webs which do not change across $s_*$. In the next paragraph we explain that the taut element
$\ft$ on the right-hand side can be either $\ft^+$ or $\ft^-$.

At this point we need some properties of exceptional webs. Let us call the difference between
the dimension of the moduli space of a web and its expected dimension, that is,
$D(\fw) - d(\fw)$, the \emph{excess} dimension. In a generic one-parameter family of webs
the excess dimension will only jump by $\pm 1$. Thus, for example, in a family where
the web of Figure \ref{fig:XWC-2} degenerates to Figure \ref{fig:XWC-1} the excess dimension
jumps by $+1$. Let us call the fans at infinity $I_\infty$ for the
exceptional webs \emph{exceptional fans}. In any web, the set of  local fans $I_v(\fw)$, for
$v\in \CV(\fw)$ will contain at most one exceptional fan. Otherwise the excess dimension would
jump by more than $\pm 1$. Moreover, no local vertex $I_v$ of an exceptional web
can be an exceptional fan, since resolving such a web would change the excess dimension by more than $\pm 1$.
It follows that the taut webs which do jump across $s_*$ cannot have exceptional fans as local vertex fans.
Therefore, we can replace $\ft$ on the right-hand-side of \eqref{eq:excptl-conv} by either $\ft^+$ or $\ft^-$.

Given the change in the taut element \eqref{eq:excptl-conv} how can we express the change
in the $L_\infty$ algebras $(\CW,T(\ft))$ and $(\Rvtx, \rho_\beta(\ft))$?  We will
focus on $\Rvtx$, which is somewhat simpler and just remark on the former case at the
end of this section. It is natural to try to relate the two $L_\infty$ algebras for $s>s_*$
and $s<s_*$ using an $L_\infty$-morphism.

Recall that, in general, given two $L_\infty$ algebras $(\CL^\pm, b^\pm)$ an $L_\infty$ morphism
$\gamma: (\CL^-, b^-) \to (\CL^+,b^+)$ is a map $\gamma: T\CL^- \to \CL^+$ such that, for all
monomials $S \in T\CL^-$ we have
\be\label{eq:Linfty-Morph}
\sum_k \sum_{\Sh_k(S)} \epsilon b^+(\gamma(S_1), \dots, \gamma(S_k)) = \sum_{\Sh_2(S)} \epsilon \gamma(b^-(S_1),S_2)
\ee
where $\epsilon$ are signs following from the Koszul rule. See Appendx \ref{App:HomotopicalAlgebra} below
for more precise definitions.
It is easy to show that, given an $L_\infty$ morphism $\gamma$ and a
 solution $\beta^-$ of the $L_\infty$ MC equation for $(\CL^-,b^-)$ we automatically get a solution
\be\label{eq:xwc-1}
\beta^+ = \gamma(e^{\beta^-})
\ee
of the MC equation for $(\CL^+,b^+)$.

Now we claim that
\be\label{eq:xwc-2}
\gamma = \textbf{1} + \rho[\fe]
\ee
is an $L_\infty$-morphism, where $\textbf{1}$ is the identity on $\Rvtx$ and
vanishes on the higher tensors $\left(\Rvtx\right)^{\otimes n}$ with $n>1$.
To prove this first note that if it is an $L_\infty$ morphism then we must have
\be\label{eq:xwc-3}
\begin{split}
\beta^+ & = \gamma(e^{\beta^-}) \\
& = \beta^- + \rho[\fe](e^{\beta^-}) \\
& = \beta^- + \rho_{\beta^-}^0[\fe].  \\
\end{split}
\ee
Again, $\beta^+$ and $\beta^-$ will only differ on summands $R_I$ where $I$ is an exceptional fan. As
we have just explained, these are
never the fans at vertices of an exceptional web so we may write $\rho_{\beta^-}^0[\fe] = \rho_{\beta^+}^0[\fe]$
and hence it is also true that $\beta^- = \beta^+ - \rho_{\beta^+}^0[\fe]$.
Note that \eqref{eq:xwc-3} is compatible with equation \eqref{eq:excptl-conv} because
\be\label{eq:xwc-4}
\begin{split}
\rho[\ft^+](e^{\beta^+}) & = \rho[\ft^-](e^{\beta^+}) + \rho[\fe*\ft](e^{\beta^+})- \rho[\ft*\fe](e^{\beta^+})\\
& = \rho[\ft^-](\rho_{\beta^-}^0[\fe])  + \rho_{\beta^+}[\fe](\rho^0_{\beta^+}[\ft])     - \rho_{\beta^+}[\ft](\rho^0_{\beta^+}[\fe]) \\
& = 0 \\
\end{split}
\ee
To get to the last line the first and third terms of the second line cancel and the middle term vanishes,
after using the definition of an interior amplitude.

Now to prove that \eqref{eq:xwc-2} is in fact an $L_\infty$ morphism
 recall that taut webs can have at most one exceptional fan
as a local fan $I_v$  at its vertices. Therefore, in \eqref{eq:Linfty-Morph} on the
left-hand-side $\rho[\fe]$ can appear at most linearly. (It might appear linearly
through the expansion of $e^{\beta^+}$ using \eqref{eq:xwc-3} or it might act on the arguments of $S$.)
This, together with the properties of $\textbf{1}$ simplifies the sum over $k$-shuffles considerably
and the required identity follows from applying a representation of webs to the convolution
identity \eqref{eq:excptl-conv}.

\bigskip
\textbf{Remarks}

\begin{enumerate}

\item The final arguments using equations \eqref{eq:xwc-3} and \eqref{eq:xwc-4} made use of the
finiteness properties of $\IV$ and the line principle. In general we would like to have more general
arguments since some of the main applications, namely knot homology and categorified spectral networks
will not enjoy those finiteness principles. We expect that in general there will be an $L_\infty$ morphism
to express the change of the interior amplitude.

\item Let us return briefly to discuss the change in the $L_\infty$ algebra of planar webs $(\CW, T[\ft])$.
It would be preferable to describe the jump in this algebra and then apply web representations to obtain
the jump in $(\Rvtx, \rho_{\beta}[\ft])$. Roughly speaking, we expect that there will be an equation
of the form
\begin{equation}
T[\ft_+](e^\fg) = \fg * \ft_-
\end{equation}
for some object $\fg$ generalizing $\fr + \fe$ (where $\fr$ is the rigid element).
Some further thought suggests that in order to give a direct geometric meaning to such a formula, and in particular to $\fg$ itself,
we need to cook up a setup which is translation invariant, but not scale invariant: $\fg$ has degree number $2$!
For example, we could let the slope of an edge depend on its length, so that long edges
are controlled by the  vacuum data for $s>s_*$ and short edges by the  vacuum data for $s<s_*$.
Then the set of rigid webs in such a setup would give us a degree $2$ object $\fg$.
Sliding webs in such a setup would give the desired convolution identity:
large web endpoints of moduli spaces will look like a large taut web in $\ft_+$ with
all vertices solved to $\fg$ rigid webs, while small web endpoints will look like a rigid web in $\fg$
with a single vertex resolved into a taut web in $\ft_-$.
The advantage of this complicated construction is that it would probably work in situations with weaker
finiteness properties. The disadvantage is that the prescription seems somewhat \emph{ad hoc}
and unphysical.

\item Returning to the path $\wp(x)$ defining $x$-dependent weights and curved webs, when $x$ passes
through $x_*$ we will introduce in Section \S \ref{subsubsec:InterfaceExceptWall} below
an Interface whose ($A_\infty$)  Maurer-Cartan equation is equivalent
to the condition \eqref{eq:xwc-3} above. Then, if there are several values of $x$ where $\wp(x)$ passes
through an exceptional wall we simply take the convolution of the Interfaces.

\end{enumerate}

\begin{figure}[htp]
\centering
\includegraphics[scale=0.3,angle=0,trim=0 0 0 0]{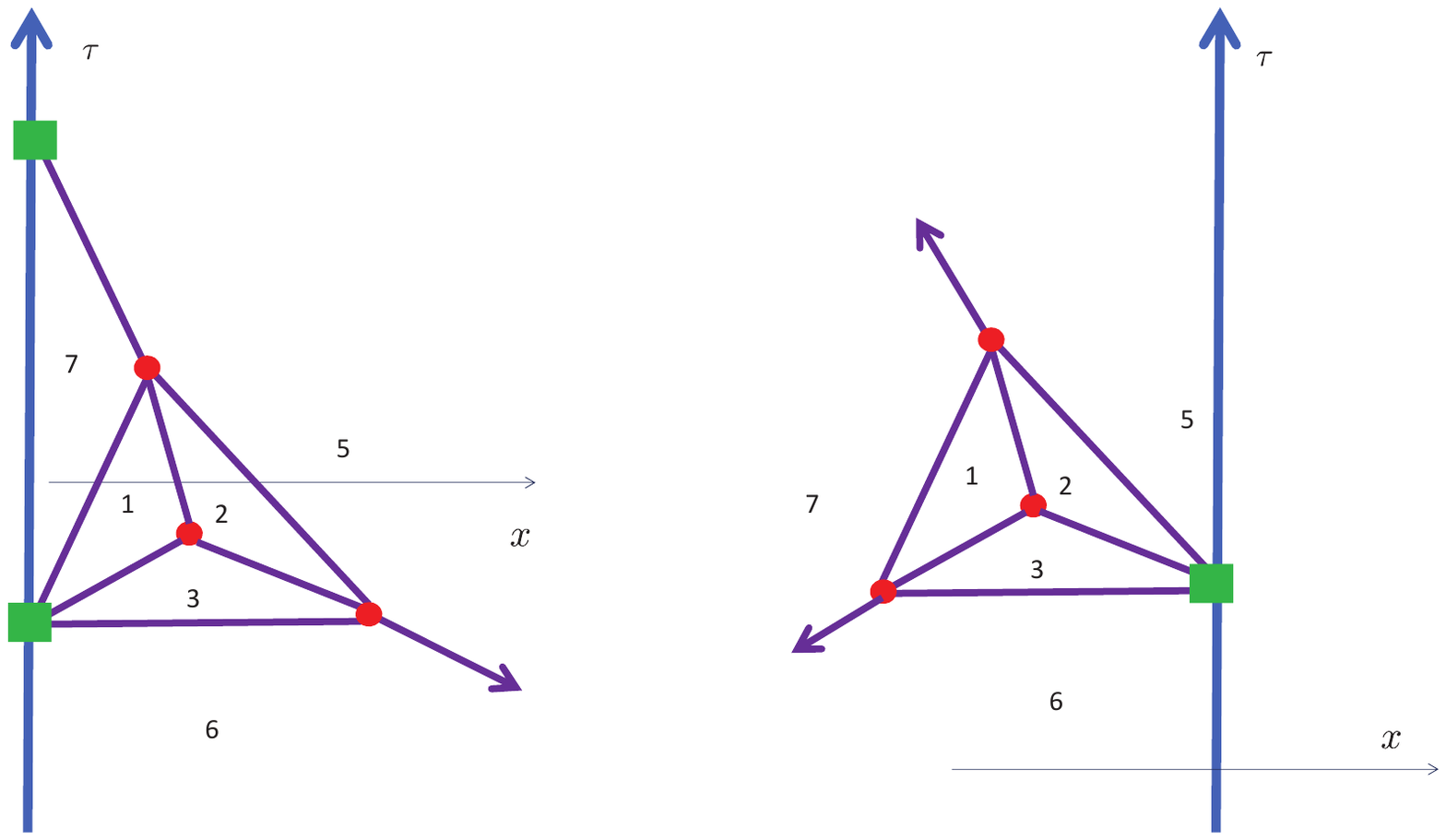}
\caption{Exceptional half-plane webs for the positive and negative
half-planes, whose existence follows from Figure \protect\ref{fig:XWC-1}.
%
%\ref{fig:XWC-1}.
%
   }
\label{fig:XWC-8}
\end{figure}

\subsubsection{$A_\infty$-Morphisms And Jumps In The Half-Plane Taut Element}

Let us fix a half-plane $\CH$, for example the positive or negative half-plane, and continue
to consider the family of weights $\{ z_i(s) \}$ with exceptional half-plane webs appearing at $s=s_*$,
and for no other value of $s$. It is possible for both plane and half plane exceptional webs to appear
at the same value $s=s_*$. Indeed, for any class of planar exceptional webs we can make half-plane exceptional webs by taking an
extremal vertex of the planar exceptional and interpreting it as a boundary vertex. See,
for example, Figure \ref{fig:XWC-8}. In this way we can construct taut exceptional half-plane webs
from taut planar webs. We will denote the sum of oriented taut half-plane exceptional webs at $s=s_*$ by $\fe_\p$
and the sum of oriented taut plane exceptional webs, if present at $s=s_*$, by $\fe$.

If we consider the moduli space of sliding $h$-types of half-plane webs we derive the convolution identity:
\begin{equation}\label{eq:xwc-5}
\ft_\p^+ - \ft_\p^- = \fe_\p * \ft_{pl} + \fe_\p * \ft_\p - \ft_\p * \fe - \ft_\p * \fe_\p.
\end{equation}
Adopting the usual arguments based on finiteness and the line principle, it does not matter whether
we take $\ft_{pl}^\pm$ or $\ft_\p^\pm$ on the right hand side.
%
%\cg{I didn't check that very carefully. gm. Feb. 18, 2014}
%

Let us now suppose that $\CB^-$ is a boundary amplitude for the positive half-plane for a
Theory $\CT^-$ with vacuum weights $z_i(s_-)$ with $s_-<s_*$. We must also choose $\CR$ and Chan-Paton
data $\CE$. Now, holding $\CR$ and $\CE$ fixed, consider a Theory $\CT^+$ for $z_i(s_+)$ with $s_+>s_*$.
We can construct a new solution $\CB^+$ to the MC equation of $\CT^+$ if we set
\begin{equation}\label{eq:xwc-6}
\CB^+ := \CB^- + \rho_\beta(\fe_\p)[\frac{1}{1-\CB}].
\end{equation}
Again, we are using heavily the finiteness principle to insure that this expression is well-defined, and
independent of the choice of $\CB^\pm$ or $\beta^\pm$ on the right hand side. To verify it one must
take
\be\label{eq:xwc-7}
\rho(\ft^+_\p)\left[ \frac{1}{1-\CB^+}; e^{\beta^+} \right]
\ee
and expand everything in terms of amplitudes and webs for $s<s_*$ using \eqref{eq:xwc-3},\eqref{eq:xwc-5}
and \eqref{eq:xwc-6}.  After a few lines of computation,
using the finiteness properties and the fact that $\CB^-$ and $\beta^-$ are boundary and interior amplitudes
one finds that \eqref{eq:xwc-7} is indeed zero.

An obvious way to extend this analysis to more general situations would be to think in terms of an
$A_\infty$ morphism $\gamma_\p$ from the $A_\infty$-algebra $(R^\p, \rho_{\beta^-}(\ft_\p^-))$ to
the $A_\infty$-algebra $(R^\p, \rho_{\beta^+}(\ft_\p^+))$, mapping $\CB^-$ to $\CB^+$.
In the case with a single class of exceptional webs, the morphism is $\gamma_\p = \textbf{1} + \rho_\beta(\fe_\p)$,
in close analogy to the case of planar webs.
%
%\cg{Here we are using $\rho_\beta(\fe_\p)$ but before we used $\rho(\fe)$. So there seems to
%be some inconsistency in the claims.}
%

\bigskip
\textbf{Remarks}

\begin{enumerate}

\item We  could probably extend the above discussion to define an  $LA_\infty$ morphism from
$\rho(\ft_\p^-)$ to $\rho(\ft_\p^+)$, which coincides in the simple case with
$\textbf{1} + \rho_\beta(\fe_\p)$. (See Appendix \S \ref{subsec:LA-ALG} for the
definition of an $LA_\infty$-morphism.)

\item As in the planar case we could give a geometric meaning to these general structures by using the same trick to break scale invariance,
considering half-plane webs with edges whose slope depends on the length. Rigid webs in such
a setup would define an element $\fg_\p$ mapped by a web representation to $\gamma_p$
and satisfying automatically the required axioms.

\end{enumerate}

\begin{figure}[htp]
\centering
\includegraphics[scale=0.3,angle=0,trim=0 0 0 0]{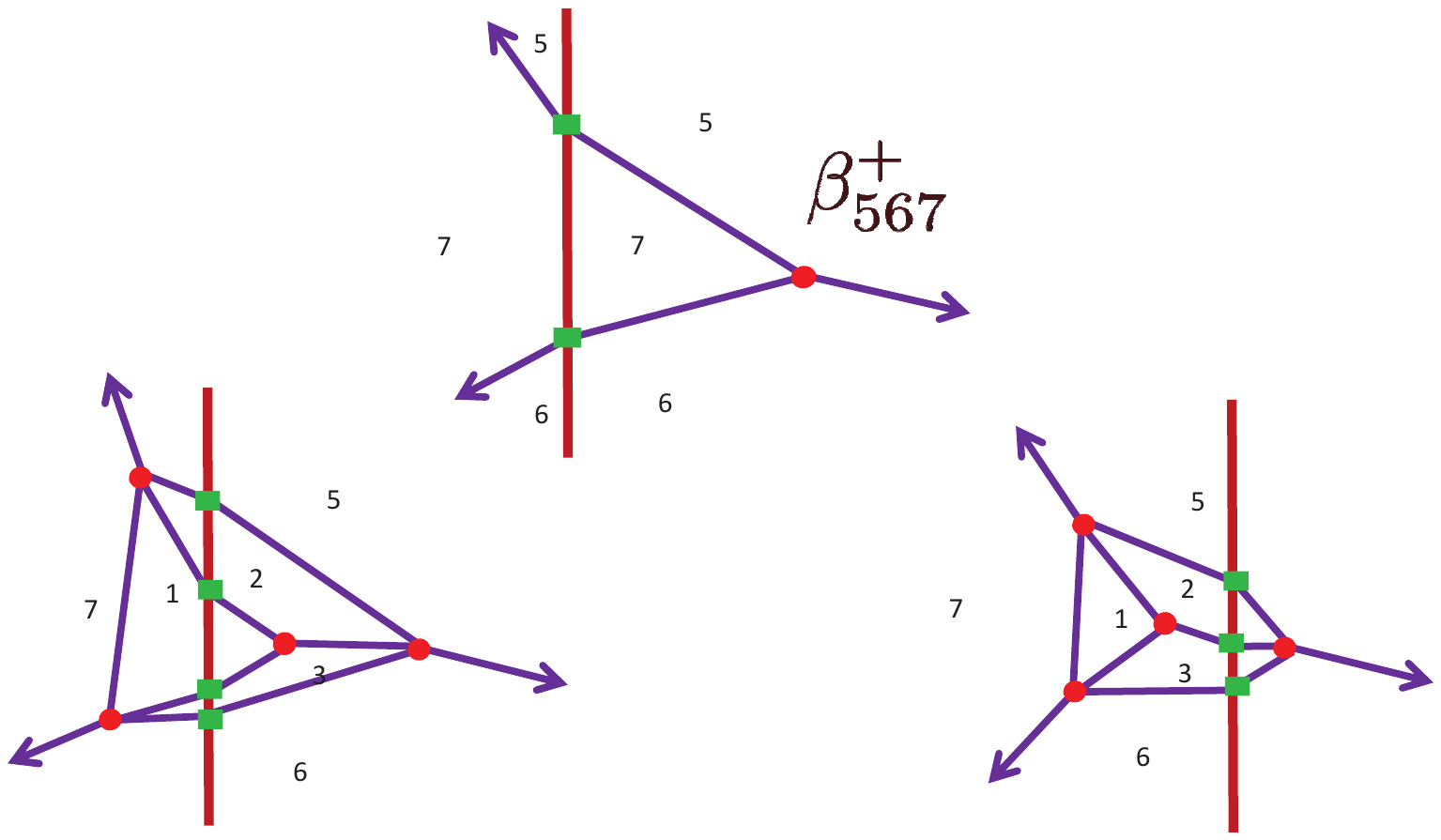}
\caption{We consider an interface with weights $\{ z_i(s_*+\epsilon) \}$ in the positive half-plane
and weights $\{ z_i(s_*-\epsilon) \}$ in the negative half-plane.  In the upper center we show a typical
interior amplitude which is discontinuous at $s_*$ because of the exceptional web of Figure \protect\ref{fig:XWC-1}.
%
% \ref{fig:XWC-1}.
%
There is a unique corresponding taut interface web obtained by placing a vertical slice through
an adjusted version of the exceptional web. Two (out of three) possible places for the placement of this
vertical slice are shown in the lower left and lower right. Only one of the three possible vertical lines will actually
admit a solution to the edge constraints.   }
\label{fig:XWC-9}
\end{figure}

\subsubsection{An Interface For Exceptional Walls}\label{subsubsec:InterfaceExceptWall}

We will now construct an Interface $\fI^{\rm exc}$ whose Maurer-Cartan equation is equivalent to the
discontinuity \eqref{eq:xwc-3} of the interior amplitude and which induces a functor
on brane categories reproducing the discontinuity \eqref{eq:xwc-6} of boundary amplitudes.

We continue to consider the continuous family  $\{ z_i(s) \}$ of vacuum weights crossing
an exceptional wall at $s_*$. For simplicity we  restrict ourselves to the generic situation
where a single class of exceptional webs appear at the jump locus, and use all the necessary
finiteness constraints.

Our Interface $\fI^{\rm exc}$ will separate two Theories with vacuum weights $\{ z_i(s_* - \epsilon) \}$
in the negative half-plane and $\{ z_i(s_* + \epsilon) \}$ in the positive half-plane.
The Interface will be formally the same as the identity Interface $\fId$. That is the Chan-Paton
factors are simply $\delta_{i,j} \IZ$ and the nonzero interface amplitudes are all $2$-valent and take
the value $K_{ij}^{-1} \in R_{ij}\otimes R_{ji}$.
The main difference from the  case of the identity Interface $\fId$, is that the interface taut element is now
 more interesting and the Maurer-Cartan equation satisfied by $\fI^{\rm exc}$ is more subtle.

Quite generally, given any continuous family of weights $\{ z_i(s)\}$,
by taking a vertical slice through a planar web $\fw$  with weights $z_i(s_0)$,  where the slice does not go
through any of the vertices of $\fw$, one can  make a corresponding interface web $\fw^{\rm ifc}$ (adding 2-valent vertices where the lines
intersect the vertical slice). One could then deform the interface web so the slopes have weights
$z_i(s_0 + \epsilon)$  in the positive half-plane and $z_i(s_0 - \epsilon)$  in the negative half-plane.
This procedure will never change the expected dimension:  We always add a boundary vertex and
an internal edge so $d(\fw) = d(\fw^{\rm ifc})$.
 In general, the procedure also does not change the true dimension: $D(\fw) = D(\fw^{\rm ifc})$.
Thus, in general, if we apply the procedure to planar taut webs we get interface sliding
webs. However, in the special case when we apply this procedure to an \emph{exceptional} web
with $s_0=s_*$   the resulting interface web is non-exceptional: $D(\fw^{\rm ifc}) = d(\fw^{\rm ifc})$.
In particular, if we apply the procedure to an exceptional taut planar web we then produce a taut interface web.
We claim, moreover, that for each deformation class of exceptional taut web at $s=s_*$ the procedure will yield a
\emph{unique} deformation class of taut interface web separating weights $z_i(s_0 \pm \epsilon)$.
See Figure \ref{fig:XWC-9}.

It then follows that the Interface $\fI^{\rm exc}$ has a more subtle Maurer-Cartan equation than that of $\fId$.
Since some interior amplitudes $\beta_I$ are discontinuous we cannot apply the simple argument of
Figure \ref{fig:ID-INTERFACE}. But we know from equation \eqref{eq:xwc-3} that $\beta_I$ will
only be discontinuous when $I$ is an exceptional fan.
These extra terms are precisely compensated by the taut interface webs such as those shown in
Figure \ref{fig:XWC-9}! The Maurer-Cartan equation thus becomes
\begin{equation}
\beta^+ - \beta^- = \rho_\beta^0[\fe]
\end{equation}
and thus the MC equation for $\fI^{\rm exc}$ is equivalent to the discontinuity condition \eqref{eq:xwc-3},
as was to be shown.

In a similar way, if we try to compose our Interface with a Brane, the only non-trivial taut (=rigid) composite webs will be in
one-one correspondence with exceptional taut half-plane webs through a similar procedure of introducing a vertical slice.
Thus the \afty-functor $\CF_{\fI^{\rm exc}}$ defined in \eqref{eq:Intfc-Functor} implements the discontinuity equation
\begin{equation}
\CB^+ - \CB^- = \rho_\beta(\fe_\p)[\frac{1}{1-\CB}].
\end{equation}

The Interface $\fI^{\rm exc}$ we have just constructed represents the discontinuity of Theories for crossing an exceptional wall.
It is thus similar to the Interfaces $\fS^{p,f}_{ij}$ for crossing $S$-walls. In a way analogous to the general Interface
for spinning webs of Section \ref{sec:CatTransSmpl}, we can use the concatenation property
\eqref{eq:Tame-CatParTrspt} to define Interfaces for more general paths
$\wp(x)$ which can cross several exceptional walls by taking suitable compositions of the Interfaces such as $\fI^{\rm exc}$ with
interfaces for tame vacuum homotopies. At this point one should engage
in an extensive discussion of homotopy equivalence, well-definedness of concatenation of homotopies up to
homotopy equivalence etc., but we will not spell out the details here.
It is useful to point out, however, that when working with homotopies of homotopies
special codimension two loci in the space of weights $\IC^{\IV}-\Delta$ can become important. In particular, one
one should treat with care points where two distinct co-dimension one walls of exceptional weights intersect.

Another loose end which we leave to the reader's imagination is to show that the Interface $\fI^{\rm exc}$
satisfies the MC equations also in the more general setup we defined with $L_\infty$
morphisms and non-scale invariant configurations. This can be done with a setup where the edge slopes depend on
their length, but only on the positive half-plane. Then the MC equation for the Interface with $\beta^+$ realized by rigid webs
represents a large sliding web in the setup. The other endpoints can be represented by convolutions with standard planar taut elements, and these terms will vanish.
Similar interpolations show that the $A_\infty$ morphism matches the composition with the trivial interface.

\begin{figure}[htp]
\centering
\includegraphics[scale=0.3,angle=0,trim=0 0 0 0]{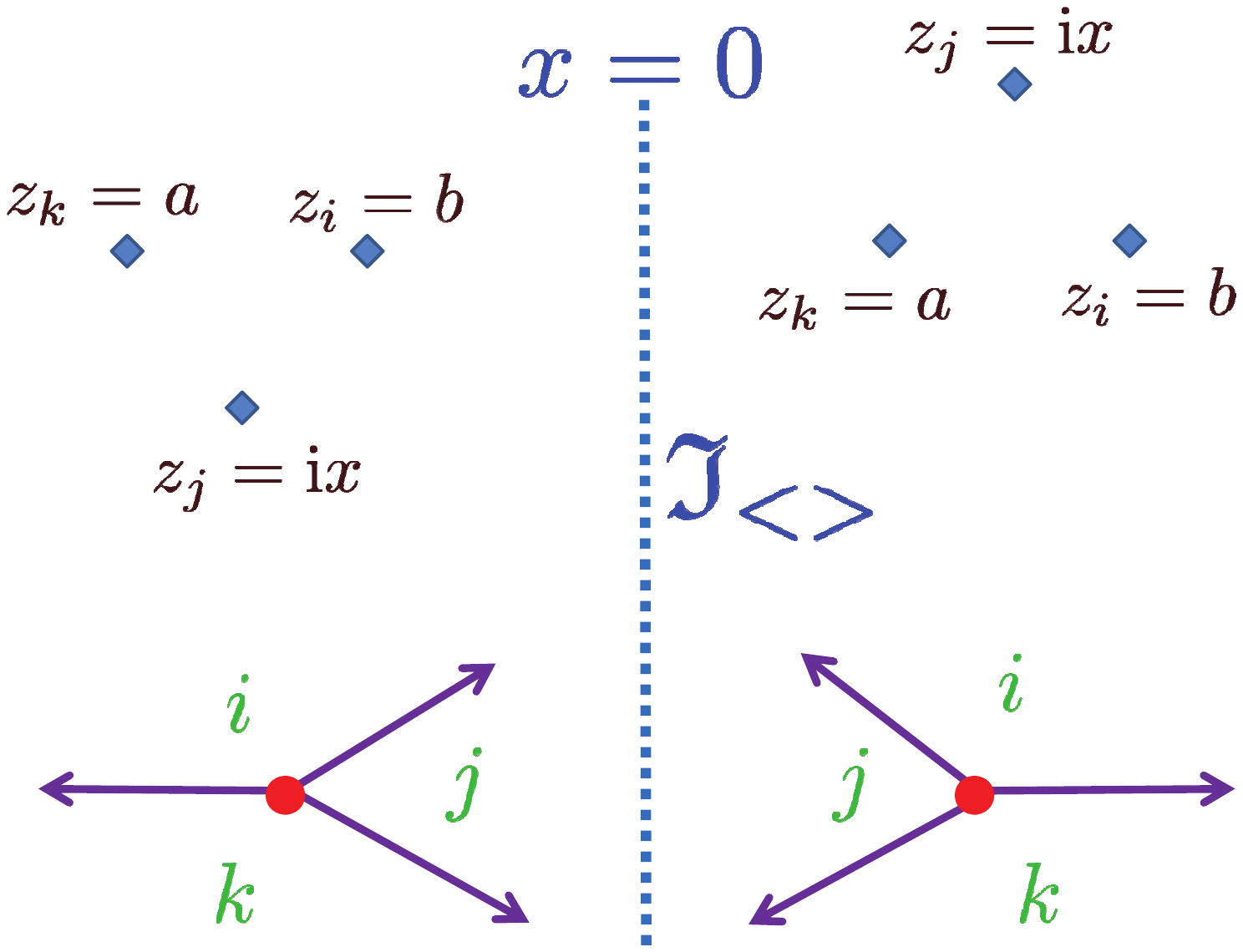}
\caption{An example of a continuous path of vacuum weights crossing a wall of
marginal stability. Here $z_k=a$ and $z_i=b$ with $a,b$ real and $a<0<b$.
They do not depend on $x$, while $z_j(x) = \I x$. We show typical vacuum weights
for negative and positive $x$ and the associated trivalent vertex. All other vacuum
weights are assumed to be independent of $x$. As $x$ passes
through zero the vertex  degenerates with $z_{jk}(x)$  and $z_{ij}(x)$ becoming real.
Note that with this path of weights the $\{i,j,k\}$ form a \emph{positive} half-plane
fan in the negative half-plane, while $\{k,j,i\}$ form a \emph{negative} half-plane
fan in the positive half-plane. If we choose $x_\ell < 0 < x_r$ there is an
associated interface $\fI_{<>}$. (We suppress the dependence on $x_\ell, x_r$ in the
 notation.) The only vertices are divalent vertices. These are all the standard   amplitude $K^{-1}$
familiar from the identity Interface $\fId$, except for $\alpha^{--}_{<>}\in R^{(2)}_{ik}\otimes R^{(1)}_{ki}$. }
\label{fig:CAT-CVWC-1}
\end{figure}
\begin{figure}[htp]
\centering
\includegraphics[scale=0.3,angle=0,trim=0 0 0 0]{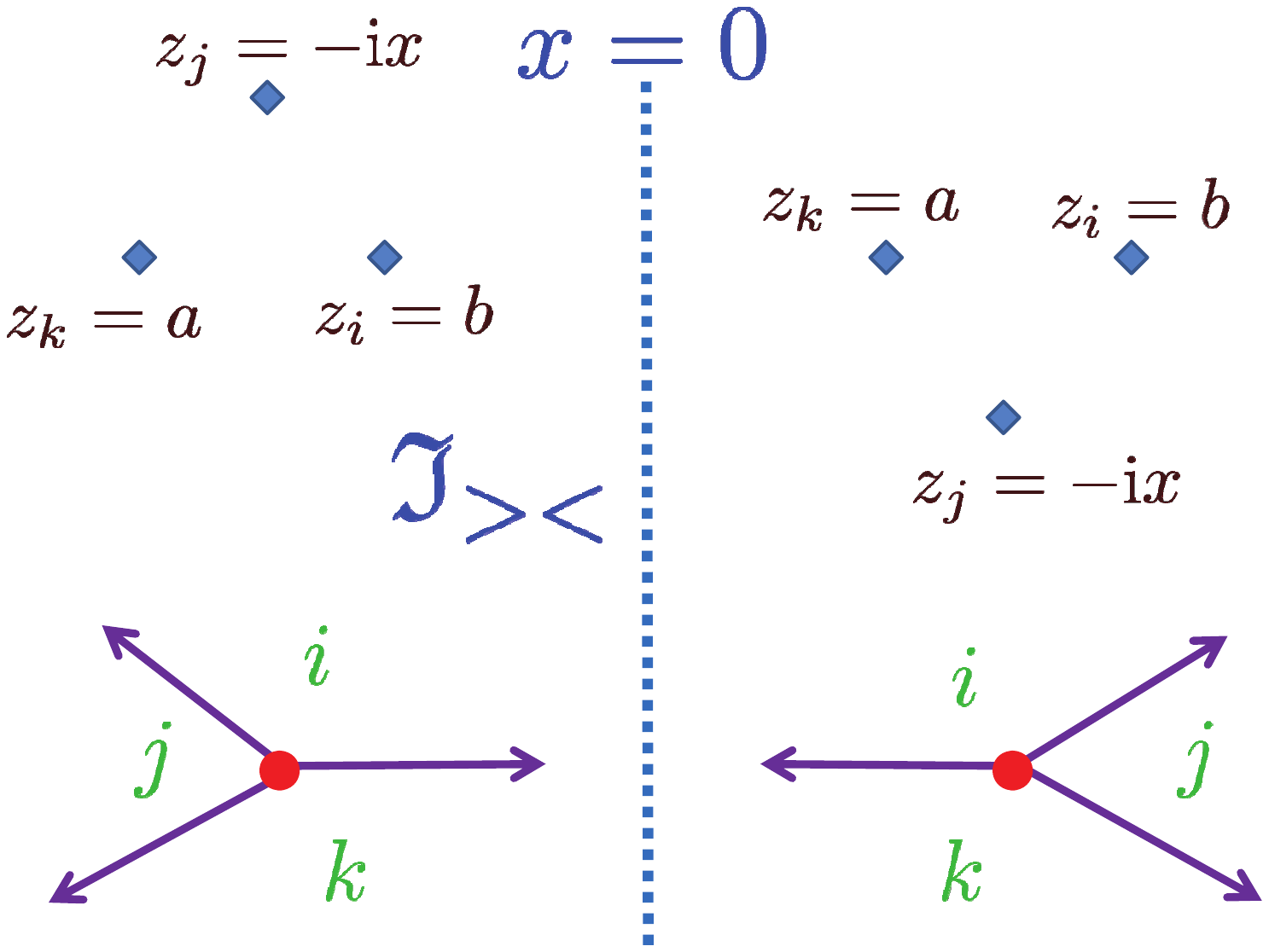}
\caption{In this figure the path of weights shown in Figure \protect\ref{fig:CAT-CVWC-1}
%
%\ref{fig:CAT-CVWC-1}
%
 is reversed.
Again,  $z_k=a$ and $z_i=b$ with $a,b$ real and $a<0<b$, but now   $z_j(x) = - \I x$.
We show typical vacuum weights
for negative and positive $x$ and the associated trivalent vertex. All other vacuum
weights are assumed to be independent of $x$.
Note that with this path of weights the $\{i,j,k\}$ form a \emph{positive} half-plane
fan in the positive half-plane, while $\{k,j,i\}$ form a \emph{negative} half-plane
fan in the negative half-plane. In order to define an interface we choose initial
and final points for the path $-x_r < 0 < - x_\ell$ so that, after translation,
it can be composed with the path defining $\fI_{<>}$. The interface $\fI_{><}$
has several nontrivial vertices. See Figure \protect\ref{fig:CAT-CVWC-8}.
%
%\ref{fig:CAT-CVWC-8}.
%
   }
\label{fig:CAT-CVWC-2}
\end{figure}
\begin{figure}[htp]
\centering
\includegraphics[scale=0.3,angle=0,trim=0 0 0 0]{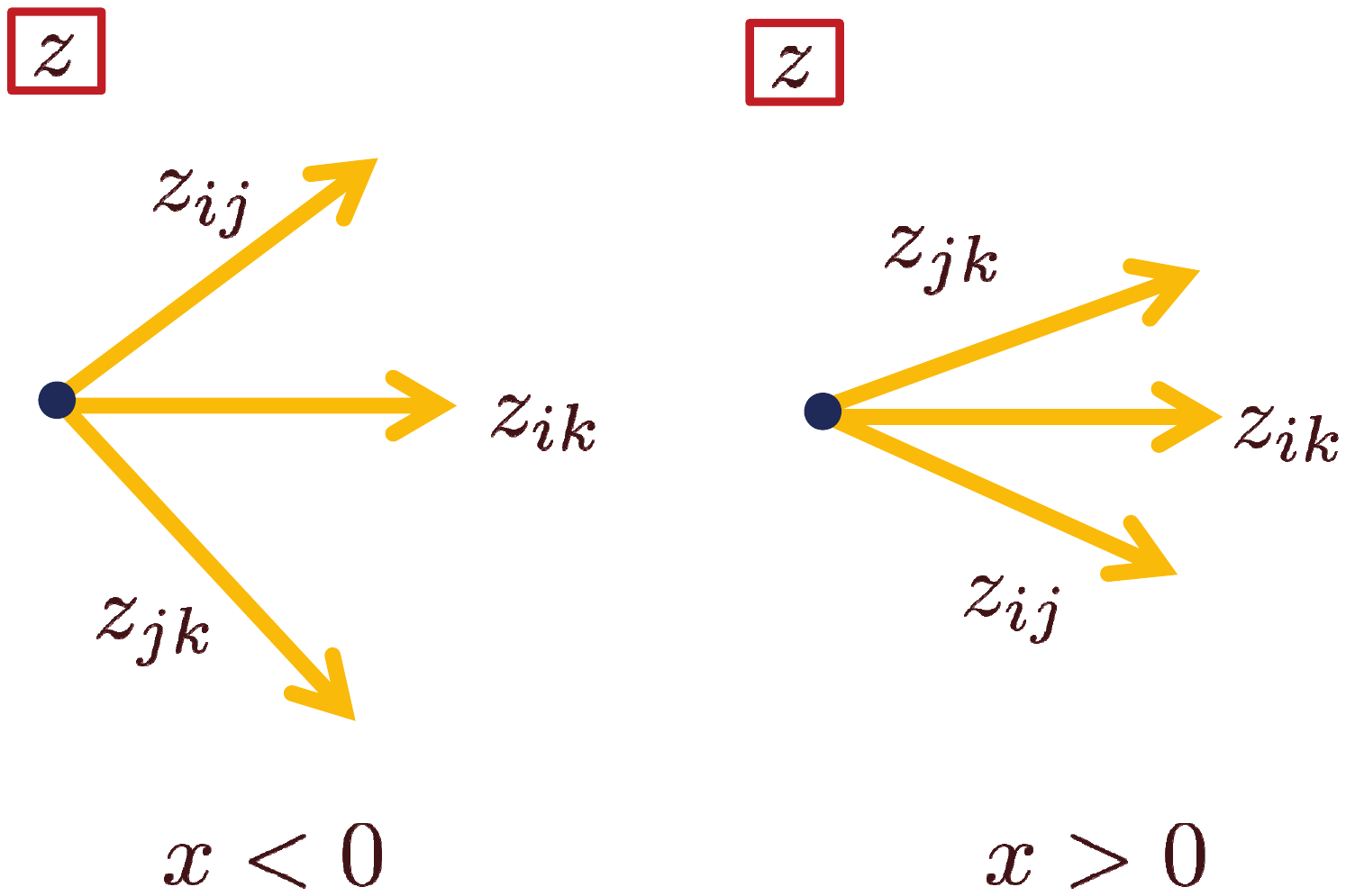}
\caption{For the path of vacuum weights in Figure \protect\ref{fig:CAT-CVWC-1}
%
%\ref{fig:CAT-CVWC-1}
%
we have BPS rays crossing as in the standard marginal stability
analysis of the two-dimensional wall-crossing formula.   }
\label{fig:CAT-CVWC-BPSRAYS}
\end{figure}
\begin{figure}[htp]
\centering
\includegraphics[scale=0.3,angle=0,trim=0 0 0 0]{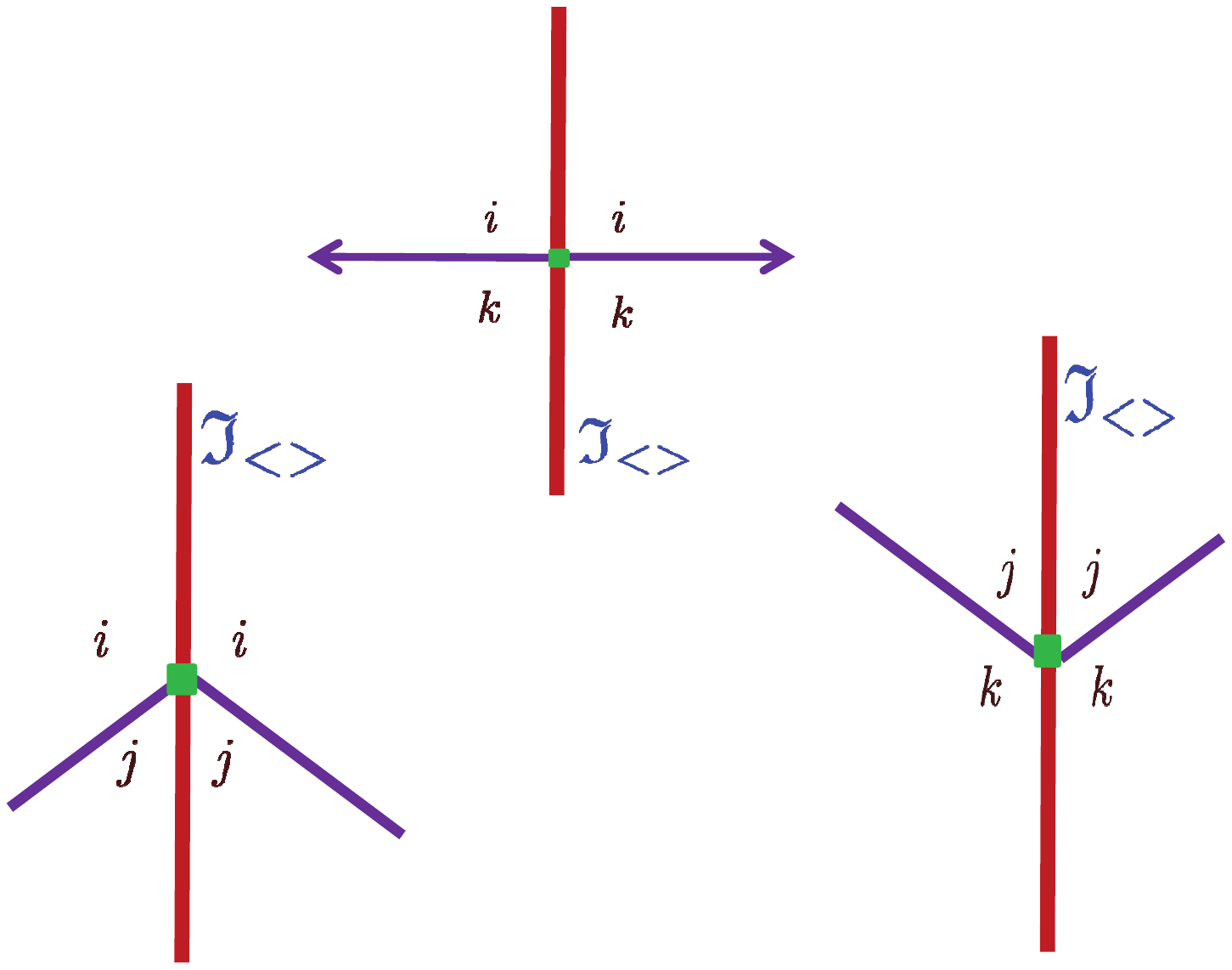}
\caption{Vertices for the interface amplitude $\fI_{<>}$ described in
  Figure \protect\ref{fig:CAT-CVWC-1}.
  %\ref{fig:CAT-CVWC-1}.
  %
The bottom left and right amplitudes
are the standard $K^{-1}$-type of the Interface $\fId$,  but the middle amplitude is a
nontrivial amplitude $\alpha^{--}_{<>} \in R^{(2)}_{ik}\otimes R^{(1)}_{ki}$ in the
wall-crossing identity.    }
\label{fig:CAT-CVWC-6}
\end{figure}
\begin{figure}[htp]
\centering
\includegraphics[scale=0.3,angle=0,trim=0 0 0 0]{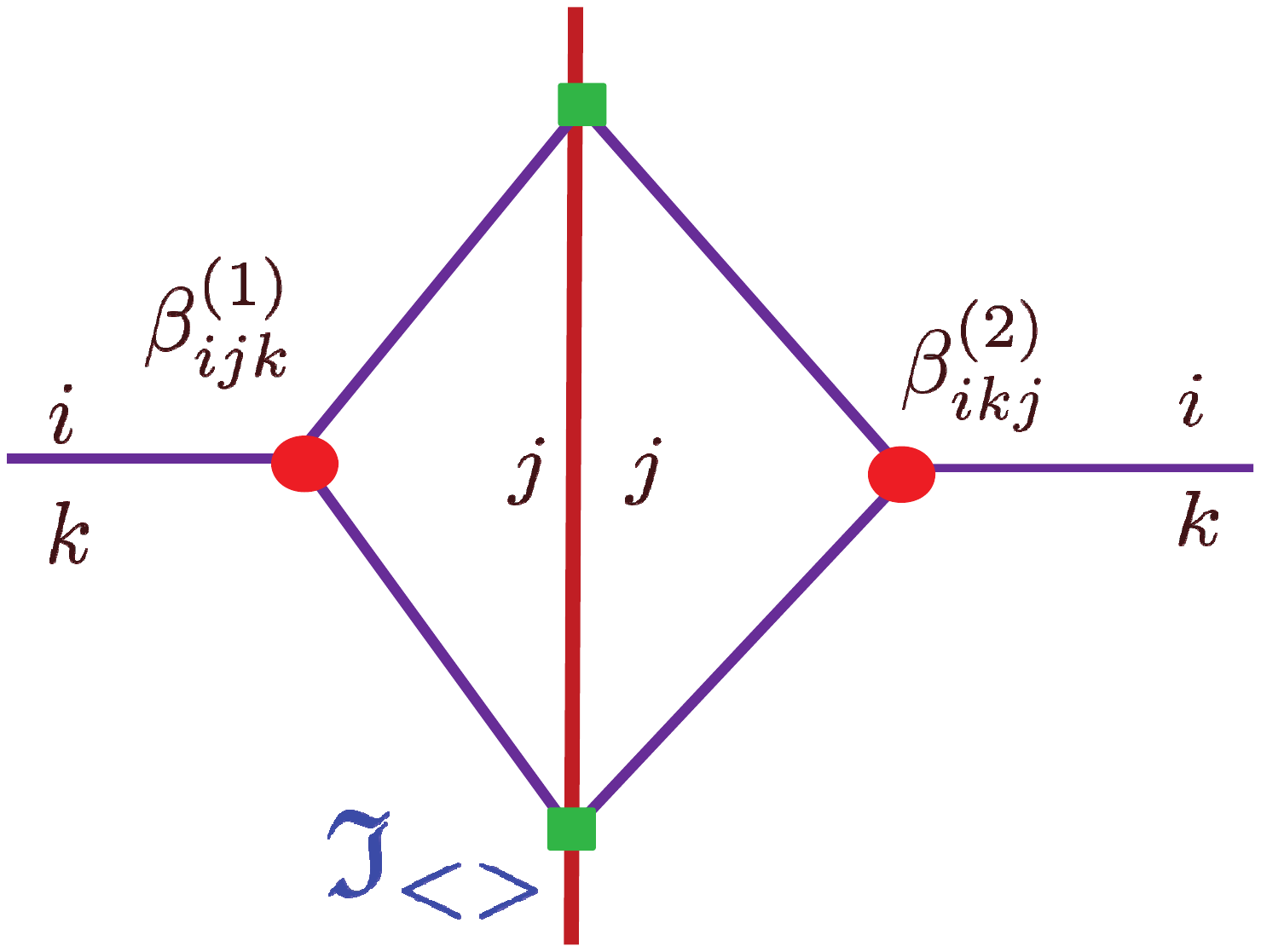}
\caption{There is one nontrivial taut interface web in the MC equation
for $\fI_{<>}$ which only involves vacua $i,j,k$.   }
\label{fig:CAT-CVWC-16}
\end{figure}
\begin{figure}[htp]
\centering
\includegraphics[scale=0.3,angle=0,trim=0 0 0 0]{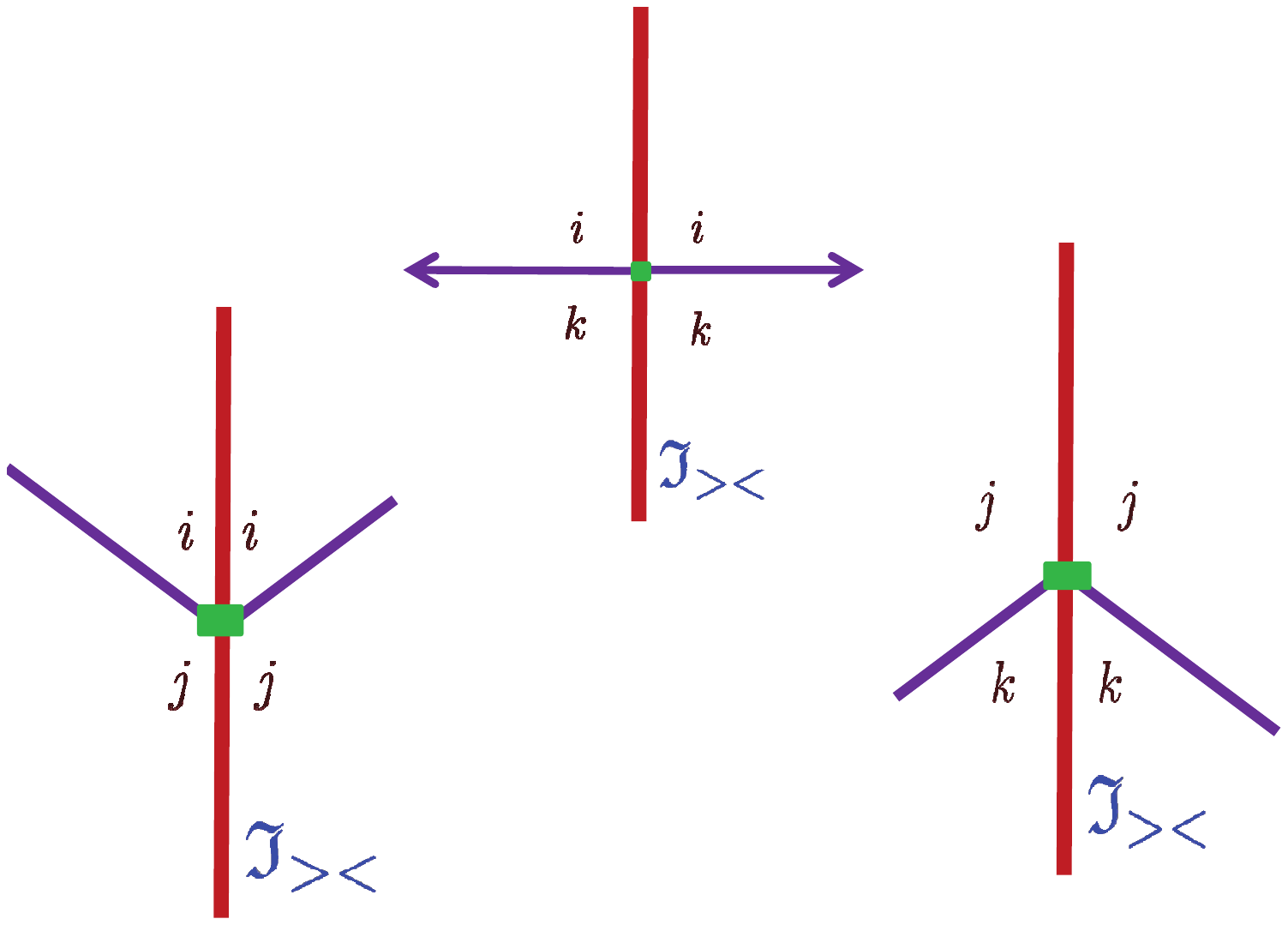}
\caption{Vertices for the interface amplitude $\fI_{><}$ described in
  Figure \protect\ref{fig:CAT-CVWC-2}.
  %
  %\ref{fig:CAT-CVWC-2}.
  %
   The bottom left and right amplitudes
are the standard $K^{-1}$-type of the Interface $\fId$, but the middle amplitude is a
nontrivial amplitude $\alpha^{--}_{><} \in R^{(1)}_{ik}\otimes R^{(2)}_{ki}$ in the
wall-crossing identity.   }
\label{fig:CAT-CVWC-7}
\end{figure}
\begin{figure}[htp]
\centering
\includegraphics[scale=0.3,angle=0,trim=0 0 0 0]{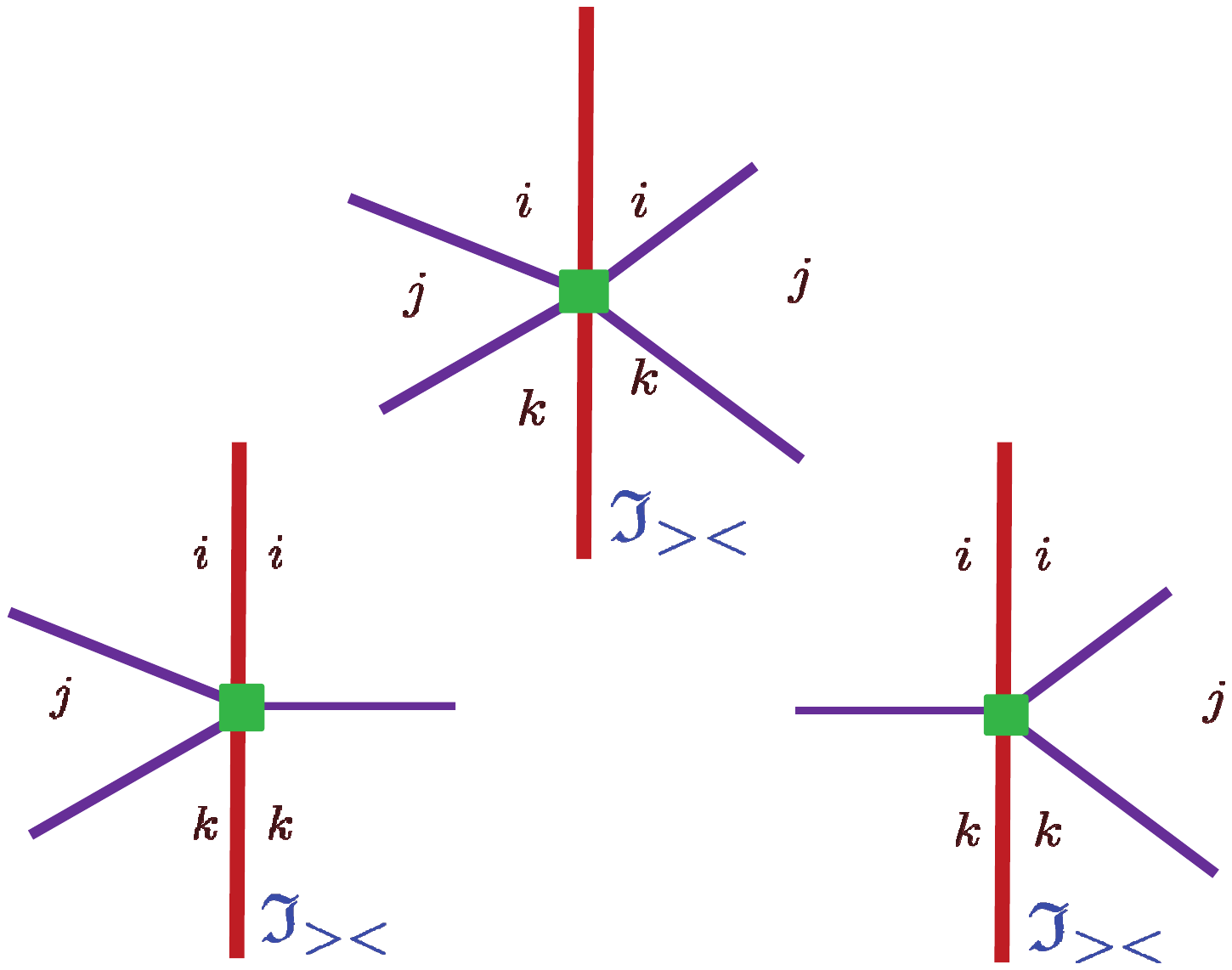}
\caption{There are three other vertices for the interface $\fI_{><}$ described in
  Figure \protect\ref{fig:CAT-CVWC-2},
  %
  %\ref{fig:CAT-CVWC-2},
  %
  shown here. The lower left is an amplitude
  $\alpha^{>-}_{><}\in R^{(1)}_{ik}\otimes R_{kj}\otimes R_{ji}  $. The lower right
  is an amplitude $\alpha^{-<}_{><}\in R_{ij}\otimes R_{jk}\otimes R^{(2)}_{ki}$. The middle
  amplitude is $\alpha^{><}_{><}\in R_{ij}\otimes R_{jk}\otimes R_{kj}\otimes R_{ji}$.      }
\label{fig:CAT-CVWC-8}
\end{figure}
\begin{figure}[htp]
\centering
\includegraphics[scale=0.3,angle=0,trim=0 0 0 0]{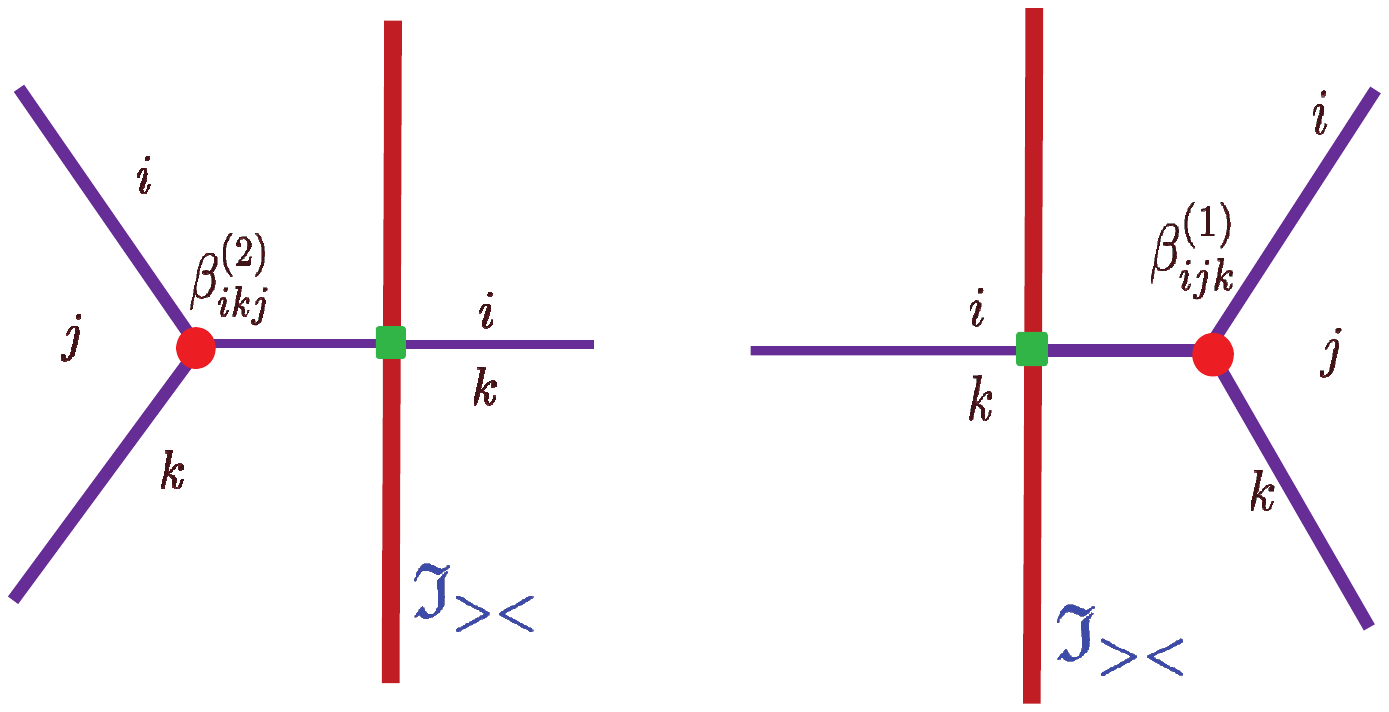}
\caption{There are two nontrivial taut interface webs in the MC equation
for $\fI_{><}$ which only involves vacua $i,j,k$.   }
\label{fig:CAT-CVWC-18}
\end{figure}
\begin{figure}[htp]
\centering
\includegraphics[scale=0.3,angle=0,trim=0 0 0 0]{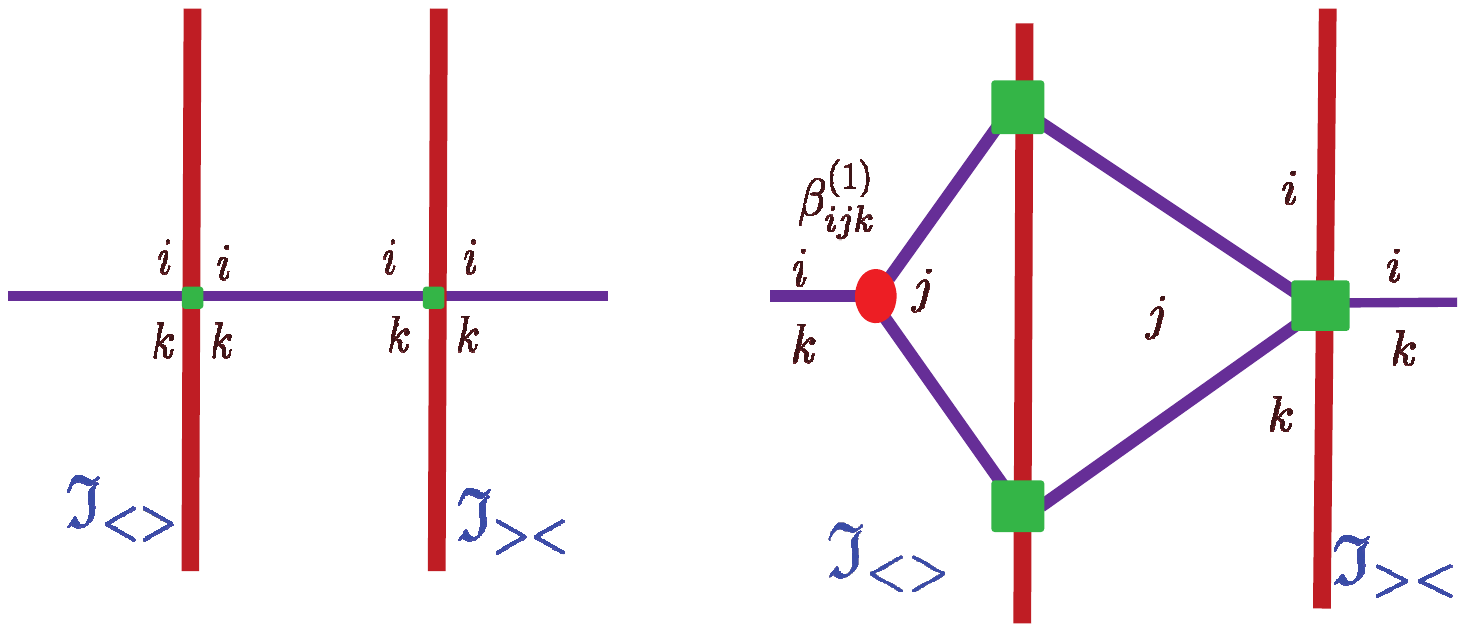}
\caption{The first nontrivial equation in the identity $\fI_{<>}\IntfcTimes \fI_{><} \sim \fId$. The simplest
possibility is to take the amplitude to be $K^{(1),-1}_{ik}$.    }
\label{fig:CAT-CVWC-10}
\end{figure}
\begin{figure}[htp]
\centering
\includegraphics[scale=0.3,angle=0,trim=0 0 0 0]{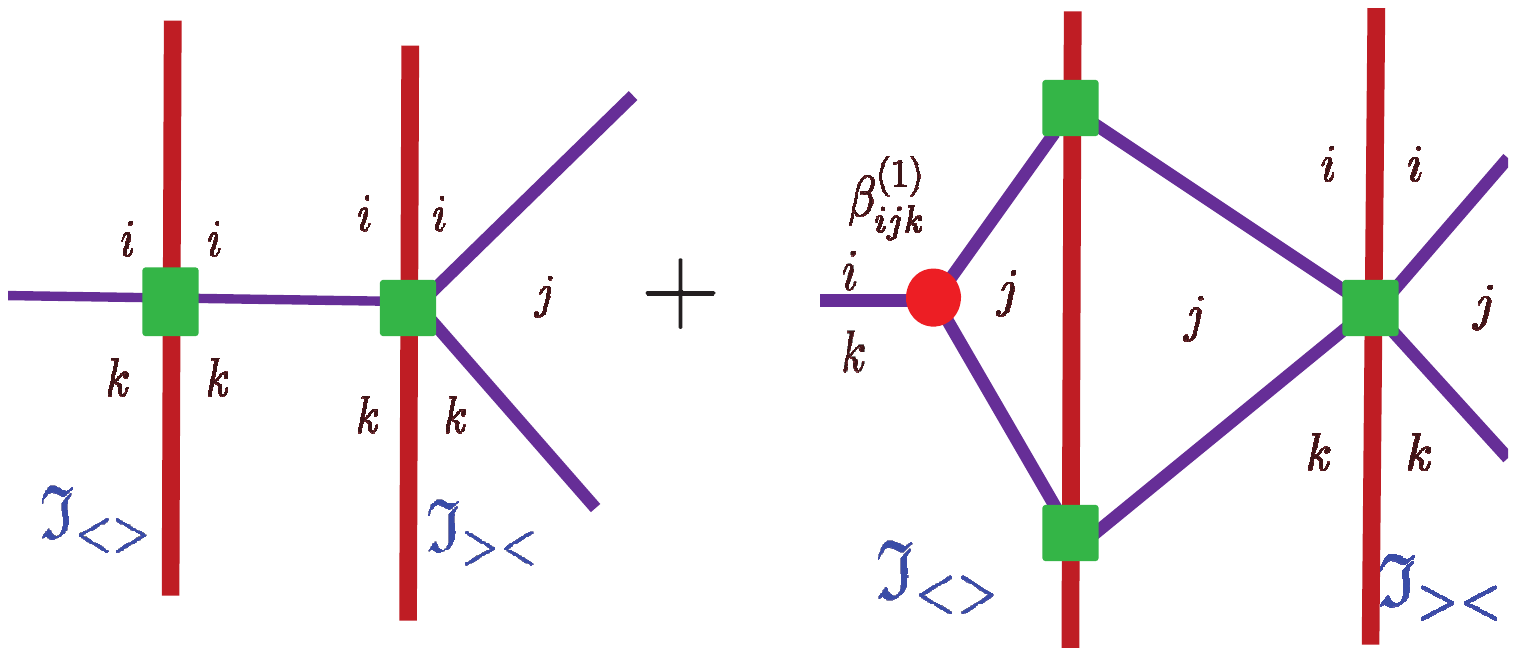}
\caption{The second nontrivial equation in the identity $\fI_{<>}\IntfcTimes \fI_{><} \sim \fId$. The simplest
possibility is to take the amplitude to vanish.    }
\label{fig:CAT-CVWC-12}
\end{figure}
\begin{figure}[htp]
\centering
\includegraphics[scale=0.3,angle=0,trim=0 0 0 0]{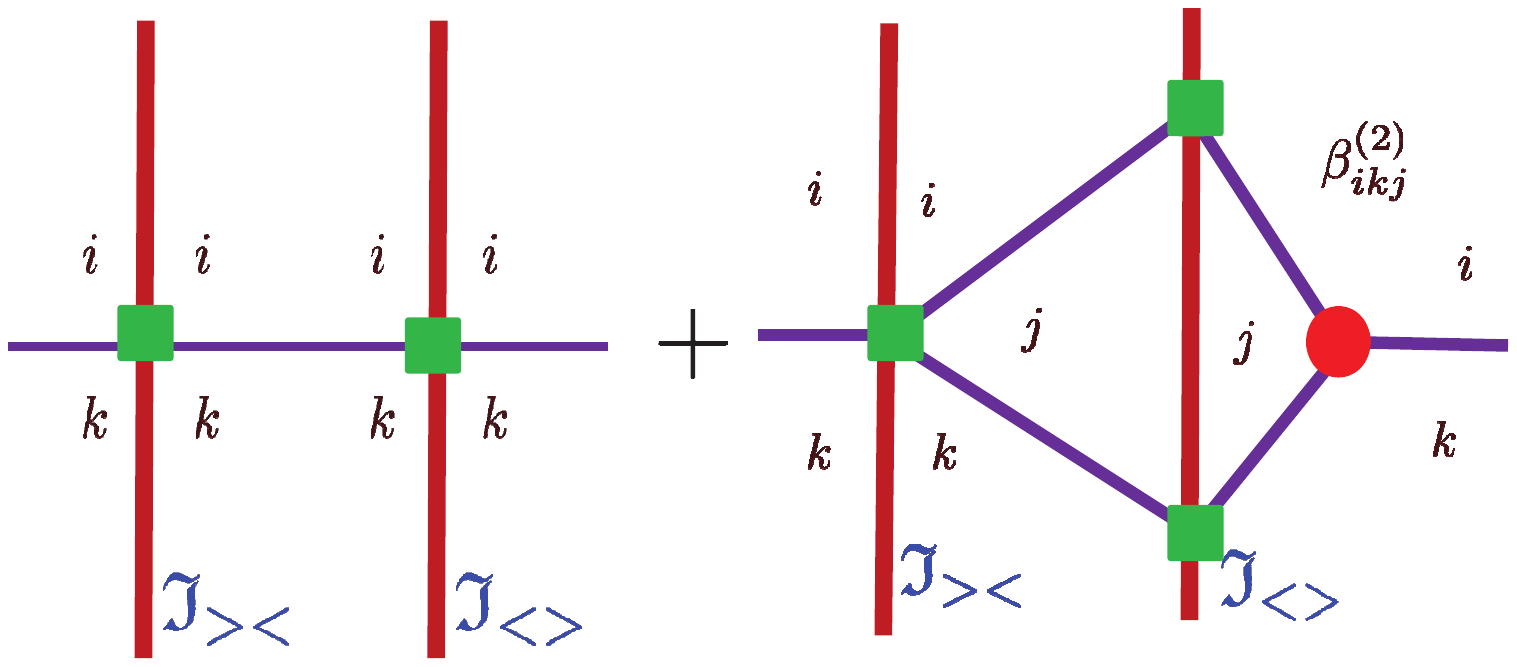}
\caption{The first nontrivial equation in the identity $\fI_{><} \IntfcTimes \fI_{<>}\sim \fId$. The simplest
possibility is to take the amplitude to be $K^{(2),-1}_{ik}$.    }
\label{fig:CAT-CVWC-14}
\end{figure}
\begin{figure}[htp]
\centering
\includegraphics[scale=0.3,angle=0,trim=0 0 0 0]{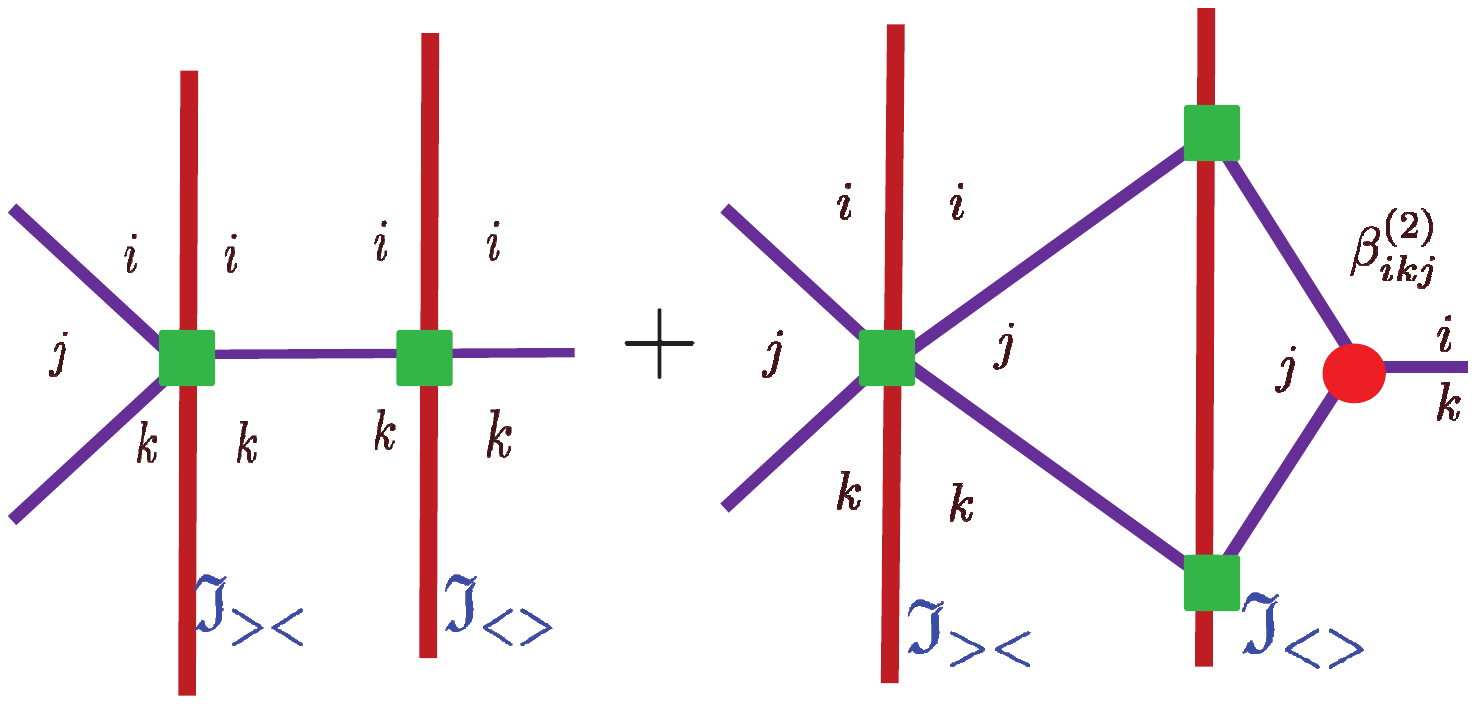}
\caption{The second nontrivial equation in the identity $\fI_{><} \IntfcTimes \fI_{<>}\sim \fId$. The simplest
possibility is to take the amplitude to vanish.    }
\label{fig:CAT-CVWC-13}
\end{figure}

\subsection{Wall-Crossing From Marginal Stability Walls }\label{subsec:MS-WC}

One of the most interesting wall-crossing phenomena occurs when the path of
vacuum weights goes through a wall of marginal stability, such as
equation \eqref{eq:MS-WALL}. In this section we examine some important examples
of such wall-crossing, but we do not give a completely general wall-crossing
prescription.
%
%\cg{But fall short of giving a general result???}
%

One way to cross such a wall is illustrated in Figures \ref{fig:CAT-CVWC-1}
and \ref{fig:CAT-CVWC-2}. While we have chosen a very concrete set of weights
our analysis applies to general configurations where the fans behave as described in the
captions, and so long as none of the $z_{ij}, z_{jk}, z_{ki}$ become
pure imaginary. If our Theory has more than three vacua we
 assume that all other vacuum weights are constant and just the $i,j,k$
``subsector'' of the Theory is changing.

Let us begin by recalling the well-known standard
 Cecotti-Vafa-Kontsevich-Soibelman result for
this situation. Referring to the path of Figure \ref{fig:CAT-CVWC-1} we have
the standard transformation of BPS rays shown in Figure \ref{fig:CAT-CVWC-BPSRAYS}.
Thus the equality of phase-ordered products \eqref{eq:2d-CVKS-prod} where $\CH$ is
the positive half-plane gives:
\be
(1+\mu_{ij}^- e_{ij}) (1+\mu_{ik}^- e_{ik}) (1+\mu_{jk}^- e_{jk}) =
(1+\mu_{jk}^+ e_{jk}) (1+\mu_{ik}^+ e_{ik}) (1+\mu_{ij}^+ e_{ij})
\ee
and so we obtain Cecotti and Vafa's wall-crossing result:
\be\label{eq:W-indx-wc}
\begin{split}
\mu_{ij}^- & = \mu_{ij}^+ \\
\mu_{jk}^-  & = \mu_{jk}^+ \\
\mu_{ik}^- + \mu_{ij}^- \mu_{jk}^-   & = \mu_{ik}^+. \\
\end{split}
\ee
Of course, the inverse transformation is obtained by considering
the path shown in Figure \ref{fig:CAT-CVWC-2}.

At a minimum, a ``categorification'' of the wall-crossing formula \eqref{eq:W-indx-wc}
should describe the discontinuous change in the web representation $\CR$ of
the Theories defined by the weights of Figure \ref{fig:CAT-CVWC-1} at $x_\ell<0$ and $x_r>0$.
As we have seen with the paths involving
exceptional webs we should also allow for a change in the interior amplitude and
indeed this is quite necessary in the present case since the set of cyclic fans
must change by replacing  $\{ i,j,k\}$ with  $\{ i, k, j\}$.

The simplest hypothesis for how $\CR$ changes,  which is
compatible with the change of Witten indices \eqref{eq:W-indx-wc},
is that $R_{ik}, R_{ki}, K_{ik}, K_{ki}$  change while all other representation
spaces and contractions remain unchanged. Similarly, the component
$\beta_{ijk}$ of the interior amplitude can only exist on one side of the
wall while $\beta_{ikj}$ can only exist on the other. We assume all other interior
amplitudes are unchanged.

We will again seek to characterize an Interface
which implements (via the discussion of Section \ref{subsec:ComposeInterface}) the desired
$A_\infty$ functor between Brane categories. The Interface $\fI_{<>}$ relates the Theory with
vacuum weights at $x_\ell$, web representation,
$R_{ik}^{(1)}, R_{ki}^{(1)}, K_{ik}^{(1)}, K_{ki}^{(1)}$, and interior amplitude $\beta^{(1)}_{ijk}$
on the left and the Theory with
vacuum weights at $x_r$, web representation
$R_{ik}^{(2)}, R_{ki}^{(2)}, K_{ik}^{(2)}, K_{ki}^{(2)}$ and interior amplitude $\beta^{(2)}_{ikj}$ on the right.
The interface   $\fI_{><}$ is then defined by the choice of path in Figure \ref{fig:CAT-CVWC-2} beginning at
$-x_r$ and ending at $-x_\ell$.
%
%In order to fix ideas we consider the family of
%vacuum weights in Figure \ref{fig:CAT-CVWC-1} and choose $x_{\ell} < 0  < x_r$.
%The vacuum weights at $x_{\ell}$ together with the representation data $R_{ik}^{(1)}, R_{ki}^{(1)}, K_{ik}^{(1)}, K_{ki}^{(1)}$
%and interior amplitude $\beta^{(1)}_{ijk}$
% %determine a Theory $\CT^\ell$, while the vacuum weights
%at $x_{r}$ together with the representation data $R_{ik}^{(2)}, R_{ki}^{(2)}, K_{ik}^{(2)}, K_{ki}^{(2)}$
%and interior amplitude $\beta^{(2)}_{ikj}$
%determine a Theory $\CT^r$. We will seek to construct Interfaces
%
Thus, we seek to define Interfaces:
\be
\fI_{<>} \in \fB\fr(\CT^\ell, \CT^r) \qquad\qquad \&  \qquad\qquad \fI_{><} \in \fB\fr(\CT^r, \CT^\ell)
\ee
(where the  notation is meant to remind us how the half-plane fans are configured in
the negative and positive half-planes).
Now, the essential statement constraining these Interfaces is that, after a suitable translation
of an Interface to the left or right so that they can be composed, the
composition of the Interfaces should be homotopy equivalent to the identity Interface:
\be\label{eq:Cat-WC-Form1}
\fI_{<>}\IntfcTimes \fI_{><} \sim \fId_{\CT^\ell} \qquad \qquad \&  \qquad \qquad  \fI_{><}\IntfcTimes \fI_{<>} \sim \fId_{\CT^r}.
\ee
The Interfaces only depend on  $x_\ell< 0$ and $x_r >0$ through composition with invertible Interfaces.

We now construct such Interfaces $\fI_{><}$ and $\fI_{<>}$.
The simplest hypothesis is that the Chan-Paton data of the Interfaces $\fI_{<>}$ and $\fI_{><}$ is identical to that of $\fId$
and we will adopt these.
As explained in Figure \ref{fig:CAT-CVWC-6} the Interface $\fI_{<>}$ has amplitudes coinciding
with those of the Identity interface $\fId$, except for
\be\label{eq:wc-amp-1}
\alpha^{--}_{<>} \in R^{(1)}_{ki}\otimes R^{(2)}_{ik}
\ee
where the notation is again meant to be suggestive of the picture. The Mauer-Cartan equation
for this interface is illustrated in Figure \ref{fig:CAT-CVWC-16}. It constrains the interior
amplitudes through the condition:
\be
K_{ij}\otimes K_{jk}\left( \beta^{(1)}_{ijk}\otimes \beta^{(2)}_{ikj} \right) = 0
\ee
where we have used repeatedly the defining properties of $K^{-1}$. There is no constraint
on $\alpha^{--}_{<>}$ from taut webs involving only vacua $i,j,k$. Of course, if there are
other vacua then this amplitude might well be involved in other components of the MC equation.

Similarly, as explained in Figure
\ref{fig:CAT-CVWC-7} and \ref{fig:CAT-CVWC-8} the Interface $\fId_{><}$ has a more intricate
set of amplitudes. In Figure  \ref{fig:CAT-CVWC-7} we have
\be\label{eq:wc-amp-2}
\alpha^{--}_{><} \in R^{(2)}_{ki}\otimes R^{(1)}_{ik}
\ee
and in Figure \ref{fig:CAT-CVWC-8} we have
\be\label{eq:wc-amp-3}
\begin{split}
\alpha^{-<}_{><} & \in R_{ij}\otimes R_{jk}\otimes R^{(2)}_{ki}\\
\alpha^{>-}_{><} & \in R^{(1)}_{ik}\otimes R_{kj}\otimes R_{ji} \\
\alpha^{><}_{><} & \in R_{ij}\otimes R_{jk}\otimes R_{kj}\otimes R_{ji} \\
\end{split}
\ee
Once again, the notation is meant to be a mnemonic for the picture.
There are now two nontrivial components to the MC equation, illustrated in
Figure \ref{fig:CAT-CVWC-18}. These lead to equations
\be
K^{(2)}_{ik}\left(  \beta^{(2)}_{ikj}\otimes \alpha^{--}_{><} \right) = 0
\ee
\be
K^{(1)}_{ik}\left( \alpha^{--}_{><}\otimes  \beta^{(1)}_{ijk}   \right) = 0
\ee

In order to investigate \eqref{eq:Cat-WC-Form1} we work out the nontrivial amplitudes of the composition of the
two Interfaces. For $\fI_{<>}\IntfcTimes \fI_{><}$ there are two nontrivial amplitudes. The first, illustrated
by Figure \ref{fig:CAT-CVWC-10} is given by
\footnote{In these, and similar formulae below we have not attempted to get the
relative signs in the equations right.}
\be\label{eq:I12-amp1}
K^{(2)}_{ik}\left( \alpha^{--}_{<>} \otimes \alpha^{--}_{><}\right)
+ K_{ij}\otimes K_{jk} \left( \beta^{(1)}_{ijk} \otimes \alpha^{>-}_{><}\right) \in R^{(1)}_{ik}\otimes R^{(1)}_{ki}.
\ee
In this formula, and in the similar ones to follow we have used the basic defining
property \eqref{eq:Kinv-def1}, \eqref{eq:Kinv-def2} of $K^{-1}$ several times.
The second nontrivial amplitude, illustrated by Figure \ref{fig:CAT-CVWC-12}, is given by
\be\label{eq:I12-amp2}
K^{(2)}_{ik}\left( \alpha^{--}_{<>} \otimes \alpha^{-<}_{><}\right)
+ K_{ij}\otimes K_{jk} \left( \beta^{(1)}_{ijk} \otimes \alpha^{><}_{><}\right) \in R_{ij}\otimes R_{jk}
\otimes  R^{(1)}_{ki}.
\ee

Similarly, for  $\fI_{><}\IntfcTimes \fI_{<>}$ there are likewise two nontrivial amplitudes. The first, illustrated
by Figure \ref{fig:CAT-CVWC-14} is given by
\be\label{eq:I21-amp1}
K^{(1)}_{ik}\left( \alpha^{--}_{><} \otimes \alpha^{--}_{<>}\right)
+ K_{ij}\otimes K_{jk} \left(\alpha^{-<}_{><}\otimes  \beta^{(2)}_{ikj} \right) \in R^{(2)}_{ik}\otimes R^{(2)}_{ki}.
\ee
The second nontrivial amplitude, illustrated by Figure \ref{fig:CAT-CVWC-13}, is given by
\be\label{eq:I21-amp2}
K^{(1)}_{ik}\left( \alpha^{>-}_{><} \otimes \alpha^{--}_{<>}\right)
+ K_{ij}\otimes K_{jk} \left(\alpha^{><}_{><}\otimes  \beta^{(2)}_{ikj} \right) \in R_{kj}\otimes R_{ji}\otimes
R^{(2)}_{ik}.
\ee

In order to illustrate the categorified wall-crossing we will content ourselves with
constructing a consistent pair of Interfaces $\fI_{<>}$ and $\fI_{><}$ satisfying all
the above criteria. We will not try to construct the most general Interface consistent
with all the criteria. In this spirit we will therefore try to construct these Interfaces
so that equation \eqref{eq:Cat-WC-Form1} is satisfied with equality, rather than homotopy
equivalence. This leads to the four equations:
\be\label{eq:I12-amp11}
K^{(2)}_{ik}\left( \alpha^{--}_{<>} \otimes \alpha^{--}_{><}\right)
+ K_{ij}\otimes K_{jk} \left( \beta^{(1)}_{ijk} \otimes \alpha^{>-}_{><}\right)  = K^{(1),-1}_{ik}
\ee
\be\label{eq:I12-amp21}
K^{(2)}_{ik}\left( \alpha^{--}_{<>} \otimes \alpha^{-<}_{><}\right)
+ K_{ij}\otimes K_{jk} \left( \beta^{(1)}_{ijk} \otimes \alpha^{><}_{><}\right) =0
\ee
\be\label{eq:I21-amp11}
K^{(1)}_{ik}\left( \alpha^{--}_{><} \otimes \alpha^{--}_{<>}\right)
+ K_{ij}\otimes K_{jk} \left(\alpha^{-<}_{><}\otimes  \beta^{(2)}_{ikj} \right) = K^{(2),-1}_{ik}
\ee
\be\label{eq:I21-amp21}
K^{(1)}_{ik}\left( \alpha^{>-}_{><} \otimes \alpha^{--}_{<>}\right)
+ K_{ij}\otimes K_{jk} \left(\alpha^{><}_{><}\otimes  \beta^{(2)}_{ikj} \right) =0
\ee
\begin{figure}[htp]
\centering
\includegraphics[scale=0.3,angle=0,trim=0 0 0 0]{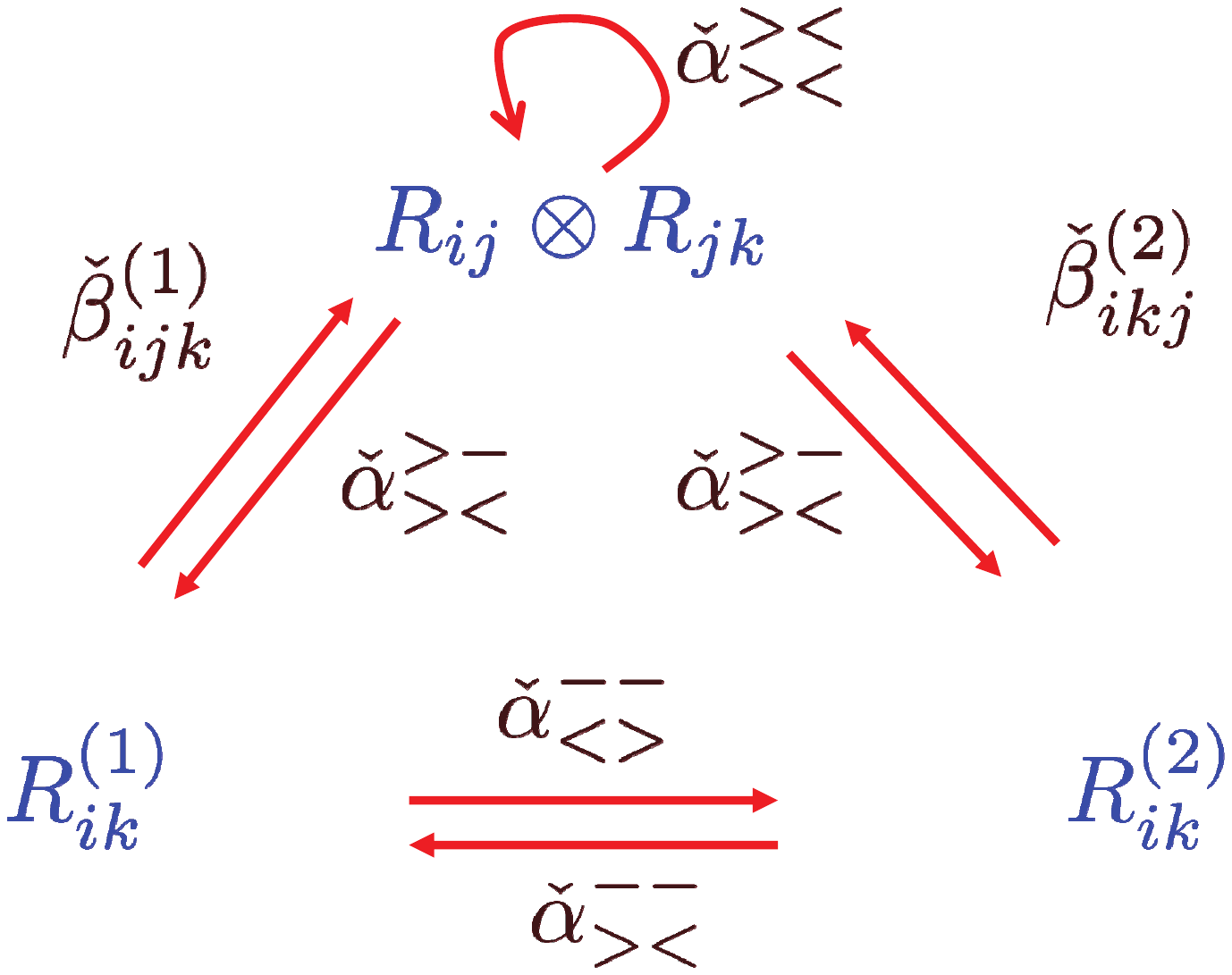}
\caption{A quiver-like figure illustrating the various linear transformations
appearing in the categorified wall-crossing formula.   }
\label{fig:CAT-CVWC-19}
\end{figure}

The conditions \eqref{eq:I12-amp11}-\eqref{eq:I21-amp21} are rather opaque.
They can be considerably simplified by using the property that $K$ is a
\emph{nondegenerate pairing} to define a degree minus one isomorphism
$R_{ji} \to R_{ij}^*$. In this way we can reinterpret the amplitudes
\eqref{eq:wc-amp-1}-\eqref{eq:wc-amp-3} as 7  linear transformations between
three different vector spaces:
\be
\begin{split}
\check \alpha^{--}_{<>} & \in \Hom(R^{(1)}_{ik} , R^{(2)}_{ik} ) \\
\check \alpha^{--}_{><} & \in \Hom(R^{(2)}_{ik} , R^{(1)}_{ik} ) \\
\check \alpha^{-<}_{><} & \in \Hom(R^{(2)}_{ik} , R_{ij}\otimes R_{jk} ) \\
\check \alpha^{>-}_{><} & \in \Hom(R_{ij}\otimes R_{jk}, R^{(1)}_{ik}) \\
\check \alpha^{><}_{><} & \in \Hom(R_{ij}\otimes R_{jk}, R_{ij}\otimes R_{jk}) \\
\check \beta^{(1)}_{ijk} & \in \Hom(R^{(1)}_{ik},  R_{ij}\otimes R_{jk}) \\
\check \beta^{(2)}_{ikj} & \in \Hom(R_{ij}\otimes R_{jk}, R^{(2)}_{ik}) \\
\end{split}
\ee
In these terms the Maurer-Cartan equations become 3 simple conditions on the linear transformations:
\footnote{Our convention here is that subsequent composition of linear transformations
are written on the \emph{right}.}
\be\label{eq:MC-cvwc}
\begin{split}
\check \beta^{(1)}_{ijk}\check \beta^{(2)}_{ikj} & = 0 \\
\check \alpha^{--}_{><}\check \beta^{(1)}_{ijk}& = 0 \\
\check \beta^{(2)}_{ikj}\check \alpha^{--}_{><} & = 0,\\
\end{split}
\ee
while the equivalence of the composition with the identity Interface, equations
 \eqref{eq:I12-amp11}-\eqref{eq:I21-amp21} become the four somewhat more tractable  equations:
\be\label{eq:he-cvwc}
\begin{split}
\check \alpha^{--}_{<>}\check \alpha^{--}_{><} +  \check \beta^{(1)}_{ijk} \check \alpha^{>-}_{><} & = \Id_{R^{(1)}_{ik}} \\
\check \alpha^{--}_{><}\check \alpha^{--}_{<>} +  \check \alpha^{-<}_{><} \check \beta^{(2)}_{ikj}  & = \Id_{R^{(2)}_{ik}} \\
\check \alpha^{--}_{<>}\check \alpha^{-<}_{><} +  \check \beta^{(1)}_{ijk} \check \alpha^{><}_{><} & = 0 \\
\check \alpha^{>-}_{><}\check \alpha^{--}_{<>} +  \check \alpha^{><}_{><} \check \beta^{(2)}_{ikj}  & = 0. \\
\end{split}
\ee
It helps to draw a quiver-like diagram to represent the linear transformations and their constraints
as shown in Figure \ref{fig:CAT-CVWC-19}.

There will be a moduli space of solutions to these constraints.
Some general facts are readily deduced. For example it is
 an easy exercise to show from these equations that $\check \alpha^{--}_{<>}\check \alpha^{--}_{><} $
and $\check \alpha^{--}_{><}\check \alpha^{--}_{<>}$ are projection operators onto subspaces $V_1 \subset R^{(1)}_{ik}$ and
$V_2\subset R^{(2)}_{ik}$ and that $\check \alpha^{--}_{<>}$ is an isomorphism from $V_1$ to $V_2$. We may therefore
take
\be
R^{(1)}_{ik} = V\oplus W_1 \qquad R^{(2)}_{ik} = V \oplus W_2
\ee
Therefore $\check \beta^{(1)}_{ijk} \check \alpha^{>-}_{><}$ and $\check \alpha^{-<}_{><} \check \beta^{(2)}_{ikj} $
are orthogonal projectors onto $W_1$ and $W_2$, respectively. We will not try to give the
most general solution to the constraints. The simplest  solution
of all our constraints is obtained when
\be
R_{ij}\otimes R_{jk} = W_1 \oplus W_2 \oplus U
\ee
and then to take $\check \alpha^{><}_{><}=0$ and
\footnote{There is a slight abuse of notation here. $P_V$ here denotes the
projection to $V$ composed with the identity map to the subspace $V$ in the
codomain. We have suppressed this in an attempt to keep the equations readable.}
\be
\begin{split}
\check \alpha^{--}_{><} = \check \alpha^{--}_{<>} & = P_V \\
\check \alpha^{-<}_{><} =  \check \beta^{(2)}_{ikj} & = P_{W_2} \\
\check \beta^{(1)}_{ijk} = P_{W_1}^{[1]} \qquad & \qquad    \check \alpha^{>-}_{><}   = P_{W_1}^{[-1]} \\
\end{split}
\ee
The superscripts in the last line indicate a degree-shift. Indeed, when passing
from the amplitude $\alpha$ to the linear transformation $\check\alpha$ we must
use the degree $-1$ isomorphism of $R_{ij} \to R_{ji}^*$, and so on. Therefore,
$\check \alpha^{--}_{><}$, $ \check \alpha^{--}_{<>}$,
$\check \alpha^{-<}_{><}$, and $  \check \beta^{(2)}_{ikj}$  all have degree $0$, etc.

In terms of physics, $V$ represents $ik$ solitons which are unchanged by
the wall-crossing, while $W_1$ and $W_2$ are sets of solitons which are
gained or lost during the wall-crossing. Those subspaces are isomorphic to
subspaces of $R_{ij}\otimes R_{jk}$ (and indeed correspond to boundstates).
Thanks to the degree assignments of $\check \alpha$ and $\check \beta$ we see that $W_1$ and $W_2$
contribute with opposite signs in computing the index on $R_{ij}\otimes R_{jk}$
(while $U$ is a subspace which contributes zero)
and in this sense we can say, informally, that the categorified Cecotti-Vafa-Kontsevich-Soibelman
wall-crossing formula is
\be
\begin{split}
R^{(2)}_{ik} - R^{(1)}_{ik} & =  \left(R_{ij}\otimes R_{jk}\right)^+ - \left(R_{ij}\otimes R_{jk}\right)^-\\
& = \left( R_{ij}^+ - R_{ij}^-\right)\otimes \left( R_{jk}^+ - R_{jk}^- \right) \\
\end{split}
\ee
where the superscript $\pm$ on the right hand side refers to the sign of $(-1)^F$.

It is time to stop and assess our results. We have given an explicit description of a pair of ``minimal'' wall-crossing interfaces
$\fI_{<>}$ and $\fI_{><}$, which exist as long as the web representations before and after wall-crossing are related in a natural way,
as described by the above decomposition. We have not  checked that $\fI_{<>}$ and $\fI_{><}$ intertwine with rotation interfaces, nor
that one can encode the relation between the two theories enforced by $\fI_{<>}$ and $\fI_{><}$ into an $L_\infty$ (or better, an $LA_\infty$) morphism.
We leave these problems to future work.

In the context of LG theories, as described in Section \S \ref{quantumbps}
the $R_{ik}$ spaces are generated by certain solitons interpolating between the critical points of the superpotential associated to the vacua $i$ and $k$.
At a wall-crossing, such solitons only appear and disappear generically when the critical point $j$ hits the soliton, splitting it into
 $ij$ and $jk$ solitons.
The subspaces $V$, $W_1$ and $W_2$ should be generated by solitons which respectively are not hit by $j$, or are hit when approaching the wall
in parameter space from either side. It should also be possible to test our solution for the jump in interior amplitudes for LG theories.
We leave that problem, as well, to future work.

We expect our proposal for the wall-crossing of the $R_{ij}$ spaces to hold universally for massive $(2,2)$ theories, in the sense that
the ``true'' wall-crossing interfaces should always factor through our $\fI_{<>}$ or $\fI_{><}$, up to inner auto-equivalences
or other equivalences associated to phantom walls (See Remarks 2,3 and 4 at the end of Section \S \ref{subsec:CatTransPrelim}.)

\section{Local Operators And Webs}\label{sec:LocalOpsWebs}

This section develops some formalism for discussing local operators on the plane,
in the context of the web formalism.

We have already identified the local boundary operators
on the half-plane  between two Branes $\fB_1$ and $\fB_2$
with $\Hop(\fB_1, \fB_2)$. More precisely, using the first $A_\infty$ multiplication $M_1$,
$\Hop(\fB_1, \fB_2)$ is a complex whose cohomology is meant to be the space of $\Q$-cohomology
classes of local operators preserving
suitable supersymmetries. As explained in Section \S \ref{generalities} below, the physical
context for these operators is the ``A-model with superpotential.'' As explained in
Section \S \ref{openlocal}, this space of local operators includes both order and disorder
operators and is slightly unusual in discussions of Landau-Ginzburg models.

Now, in Section \S \ref{closedstrings} we show that further new ideas are needed to
discuss local operators in the bulk. This proves to be the case in the approach from the
web-based formalism as well. We should stress one point: In the Landau-Ginzburg model
it is quite natural to look at the Jacobian ideal $\IC[\phi^I]/(dW)$ (or its generalizations
with curved target space). This is the chiral ring for the B-twisted model, and is
\emph{not} the local operators relevant to the ``A-model with superpotential.'' The latter
has a subspace of local operators given by the DeRham cohomology of the target, although these
are only the order operators, and in principle there will be other, disorder operators,
in the space of local operators.

\subsection{Doubly-Extended Webs And The Complex Of Local Operators On The Plane}

The guiding principle for generalizing the complex $\Hop(\fB_1,\fB_2)$ to the
case of operators on the plane will be the relation to the complex of groundstates
provided by the exponential map
\be\label{eq:uvxy}
u+ \I v = e^{-\I x + y} .
\ee
Recall that, for boundary operators, the spinning webs, \eqref{eq:SpinningWeights}, with
uniform rotation $\vartheta(x)=-x$ on an interval of length $\pi$ map to half-plane
webs with a marked point on the boundary at $u=v=0$. This point corresponds to the
far past $y\to -\infty$ on the strip. The relation of the complex of local operators
to the complex of groundstates is summarized in equation \eqref{eq:Strip-HalfPlane}.

We now imitate the above discussion for closed strings. Accordingly, we will map
the infinite cylinder with coordinates $(x,y)$ and $x \sim x + 2\pi$ to the plane
using the exponential map \eqref{eq:uvxy}. Equivalently, we can consider periodic
webs on the strip in the $(x,y)$ plane with $x_{\ell} \leq x \leq x_{\ell} + 2\pi$.
We again consider curved spinning webs with uniform rotation $\vartheta(x) = -x$.
The complex of groundstates on the strip will be given by the trace of the matrix
of  Chan-Paton factors of the rotation Interface $\fR[\vartheta_{\ell}, \vartheta_{\ell}-2\pi]$.
(See the discussion in Section \S \ref{subsec:WedgeWebs}, and further development in
Section \S \ref{subsec:TraceInterface} below.)  All binding points are future stable
and, from equation \eqref{eq:TautCurvedCP} see see that each \emph{cyclic} fan of vacua
will fit on the cylinder. We thus might expect the complex of groundstates
to be simply the complex $R^{\rm int} = \oplus_I R_I$ we have met before. This is not
quite right since the constant vacuum $i$, corresponding to the ``fan'' $\{ i \}$ is
also an approximate groundstate. Therefore, for each vacuum $i$ we introduce
a module $R_i \cong \IZ$, in degree zero, and we define
\begin{equation}\label{eq:LocalOpComplex}
\begin{split}
R_c & := \left[ \oplus_{i \in \IV} R_i \right] \oplus R^{\rm int} \\
 &  =\left[ \oplus_{i \in \IV} R_i \right] \oplus \left[\oplus_{I} R_I \right]\\
\end{split}
\end{equation}

The complex $R_c$ defined in equation \eqref{eq:LocalOpComplex} is nicely in accord
with the MSW complex of semiclassical twisted ground states discussed in Section
\S \ref{twclosed} below. The summands $R_i$ correspond to states in the constant
vacuum $\phi_i$ that sits at a critical point of the superpotential. The summands $R_I$
where $I$ has length greater than one correspond, for large radius of the cylinder, to the fans of
solitons.

\begin{figure}[htp]
\centering
\includegraphics[scale=0.3,angle=0,trim=0 0 0 0]{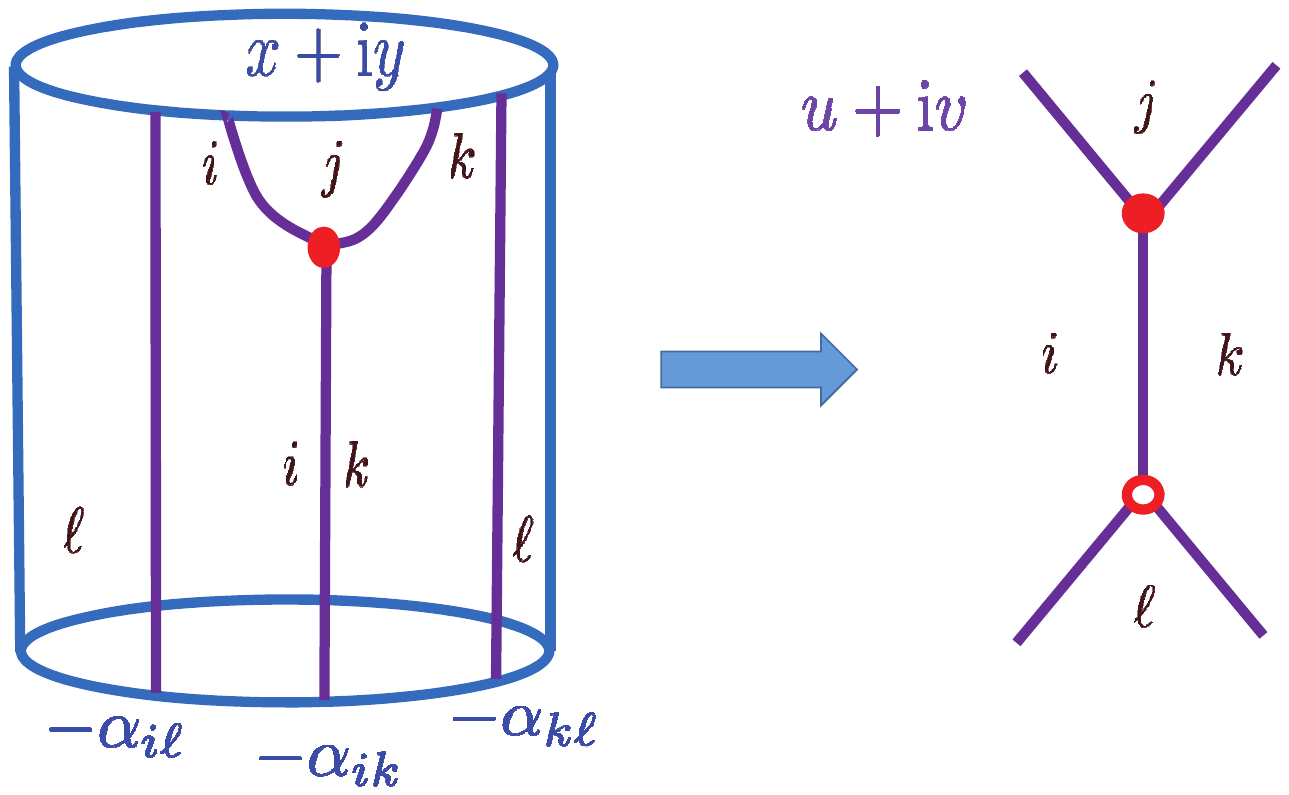}
\caption{This figure represents the relation between a taut curved web on the $x+\I y$ cylinder and a taut
extended web on the $u+ \I v$ plane.  The open red circle
is the origin of the plane and corresponds to the $y\to -\infty$ limit of the cylinder
under the map $u + \I  v = e^{-\I x + y}$. The vertical lines extending from $y=-\infty$ to $+\infty$
at $x= - \alpha_{i \ell}$ and $x= - \alpha_{i\ell}$ have no moduli, and the vertex can only move
vertically at $x=-\alpha_{ik}$. Taut webs are used to define a differential on the complex $R_c$.
Contraction with the taut web shown here takes an element
$r\in R_{\{ i,k,\ell\} }$  to $\rho_{\beta}[\ft](r)$ which in this case is just $\rho[\ft](r \otimes \beta_{ijk})$.
The cylindrical picture is meant to motivate this operation as a transition amplitude from an approximate ground state in the
far past of the cylinder to a state in the future.    }
\label{fig:LOCALOPERATORS-5}
\end{figure}
\begin{figure}[htp]
\centering
\includegraphics[scale=0.3,angle=0,trim=0 0 0 0]{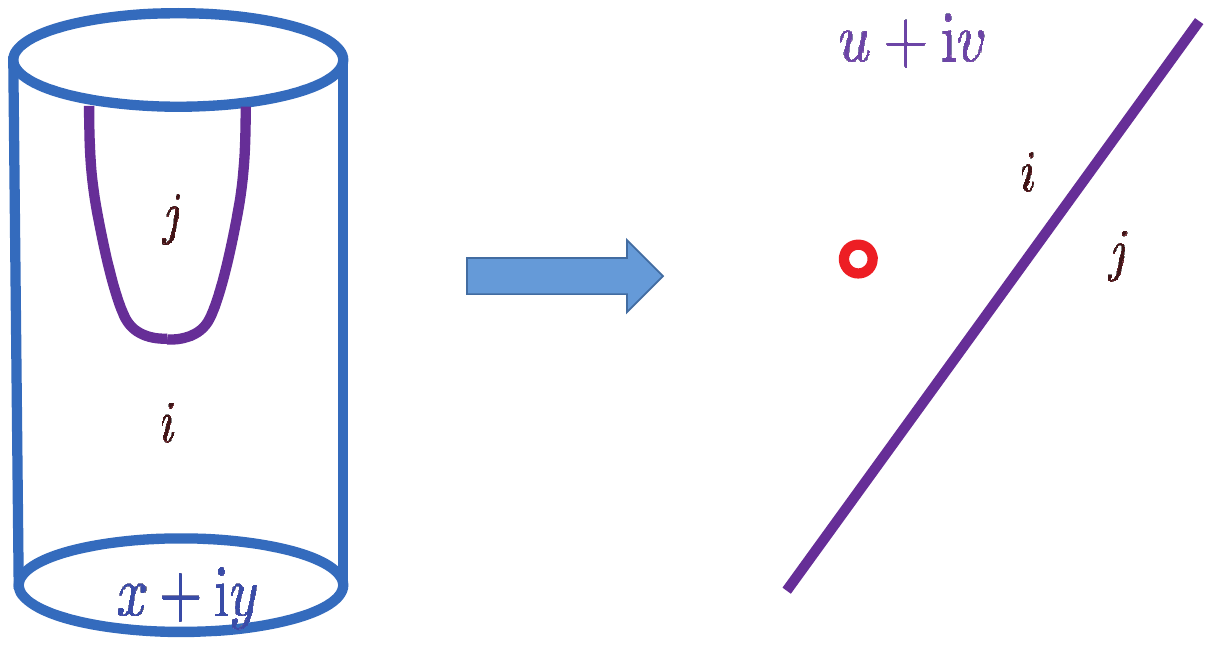}
\caption{The web on the cylinder has one modulus and hence is taut as a curved web. It can be considered as a map
from a state in $R_i$ to a state in $R_{ij}\otimes R_{ji}$. It is natural to give this map the amplitude $K_{ij}^{-1}$ and
indeed that completes the operation of Figure \protect\ref{fig:LOCALOPERATORS-5}
%
%\ref{fig:LOCALOPERATORS-2}
%
 to a differential. The image under the exponential map
shows that we should broaden our notion of extended webs to doubly-extended webs by including a new kind of vertex in the
faces of the webs.   }
\label{fig:LOCALOPERATORS-3}
\end{figure}

Now, we would like to define a differential $d_c$ on $R_c$ to make it into a complex.
We follow the lead of the complex of approximate ground states defined in Section
\S \ref{subsec:WebRepStrip} above. We should contract incoming states at $y=-\infty$
on the strip with all taut webs, saturating all boundary and interior vertices with
boundary and interior amplitudes, as in equation \eqref{eq:strip-diff-def}.

The image under the exponential map \eqref{eq:uvxy} of a taut spinning periodic web will be one of two
types, illustrated in Figures \ref{fig:LOCALOPERATORS-5} and \ref{fig:LOCALOPERATORS-3}.
In the first type, there is a fan of vacua $I$ at $y\to -\infty$ of length larger than one.
In the second type, the fan at $y\to -\infty$ consists of a single vacuum $\{ i \}$.
In the first case the image of the taut curved web in the $(u+\I v)$-plane is a taut web with
one vertex at the origin, as shown in Figure \ref{fig:LOCALOPERATORS-5}. If $I$ is the fan
of vacua at $y\to -\infty $ and if $r_I \in R_I$, then we define
\be\label{eq:dc-def1}
d_c(r_I):= \rho_\beta[\ft_{pl}](r_I),
\ee
where $\ft_{pl}$ is the taut planar element on the  $(u+\I v)$-plane. Thus, all vertices
except for the one at the origin are saturated with the interior amplitude $\beta$. The second
type of taut spinning periodic web will lead to a map
\begin{equation}\label{eq:ExtDiff}
R_i  \rightarrow  \oplus_{j\not=i} R_{ij} \otimes R_{ji}.
\end{equation}
To define this map return to Figure \ref{fig:SPCLAMP-7} and use equations
\eqref{eq:SpAmp-6}-\eqref{eq:SpAmp-8} to write
\be\label{eq:dc-def2}
d_c(\phi_i ):=  \oplus_{j\not=i} K_{ij}^{-1}
\ee
where $\phi_i$ is a generator of $R_i\cong \IZ$ defined below. The crucial
property $d_c^2=0$ will follow from our discussion of ``doubly-extended
webs'' below.

The cohomology $H^*(R_c, d_c)$ is to be identified with the space of local
operators. Recall that under the isomorphism $R_{ij}\otimes R_{ji} \rightarrow
R_{ji} \otimes R_{ij}$, the element $K_{ij}^{-1}$ is \underline{antisymmetric}.
It therefore follows that
\be\label{eq:UnitOp}
\textbf{1}:= \oplus_{i\in \IV} \phi_i
\ee
is always closed and defines a canonical element of the cohomology.
This element simply corresponds to the unit operator.

\begin{figure}[htp]
\centering
\includegraphics[scale=0.3,angle=0,trim=0 0 0 0]{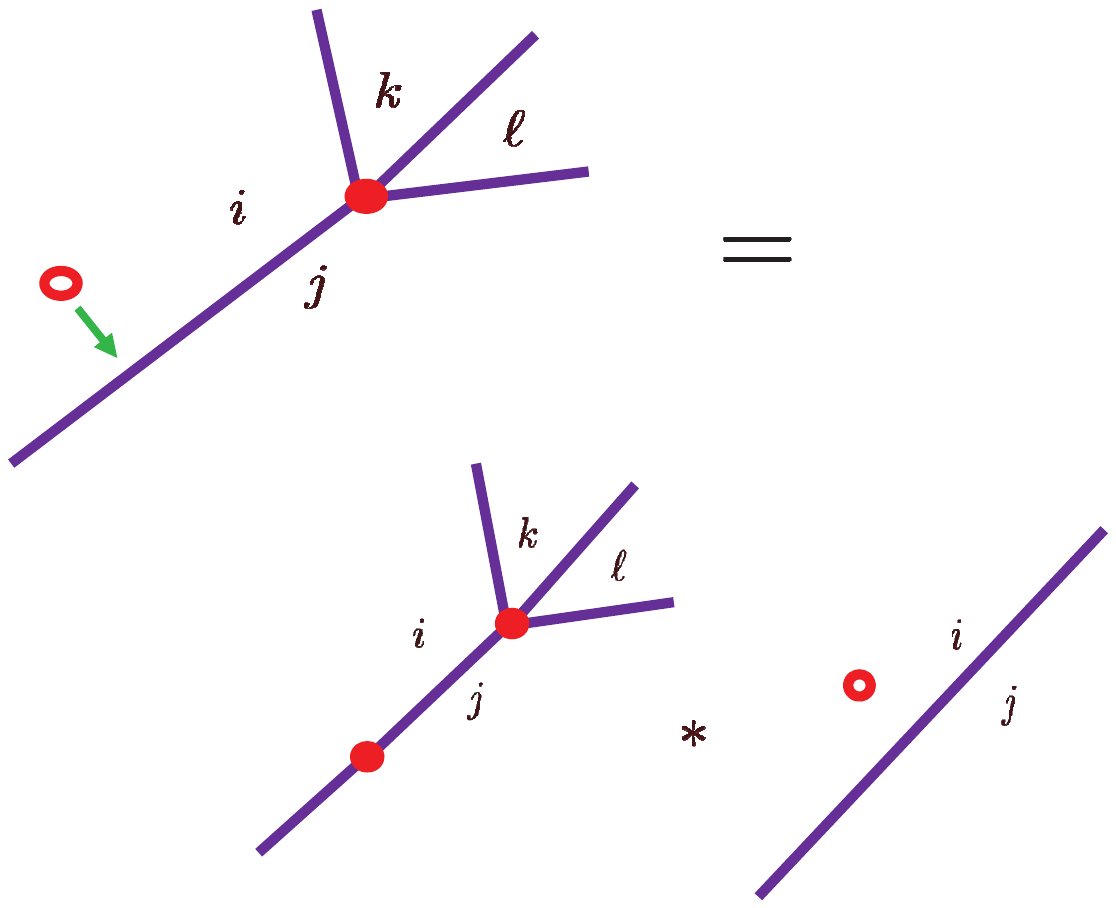}
\caption{The pictorial demonstration of a convolution for doubly-extended webs.
The web with the closed vertex in vacuum $i$ has four moduli - two for each vertex
and is therefore a sliding web. Near one of the boundaries of its moduli space
it can be expressed as a convolution of two taut webs.     }
\label{fig:LOCALOPERATORS-4}
\end{figure}
\begin{figure}[htp]
\centering
\includegraphics[scale=0.3,angle=0,trim=0 0 0 0]{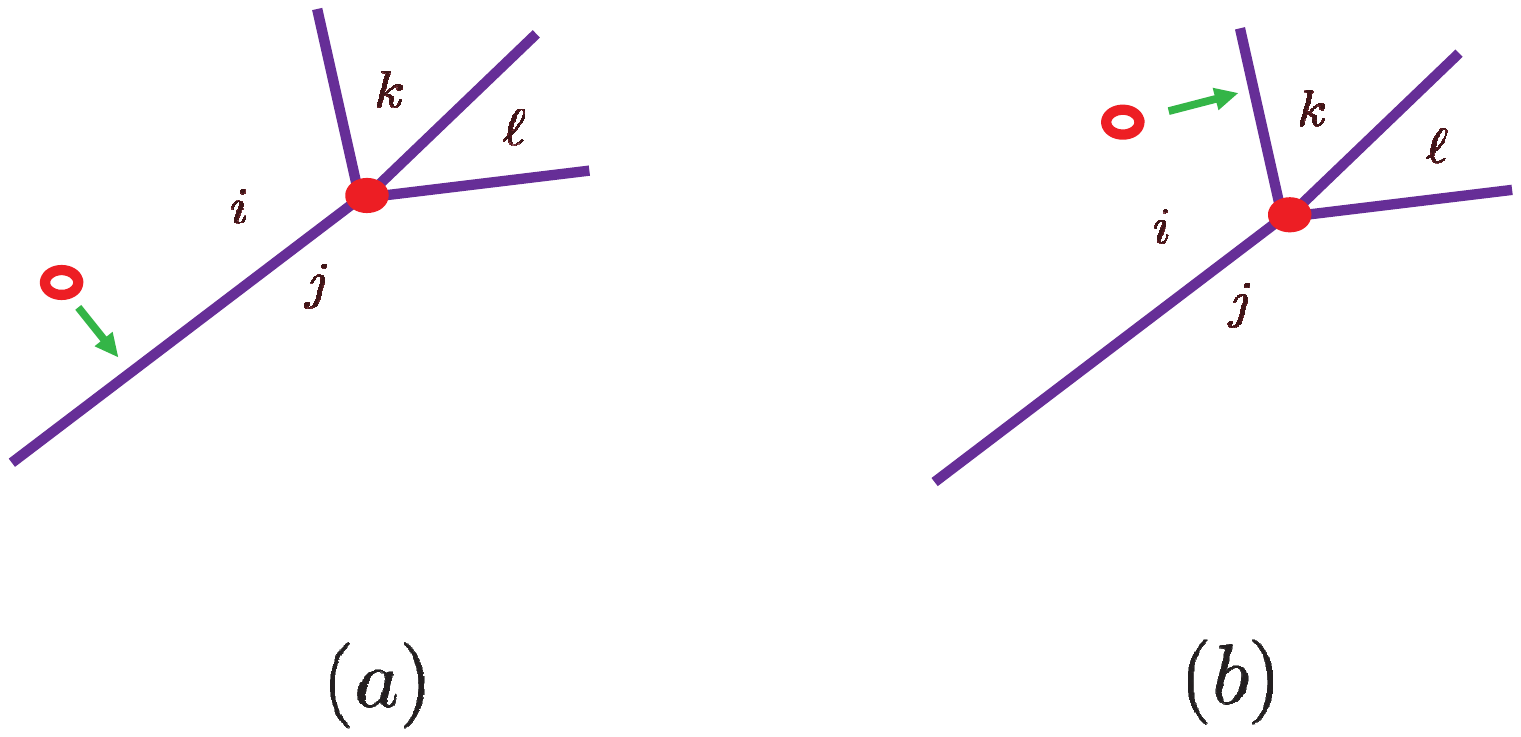}
\caption{The two boundaries illustrated in (a) and (b) correspond to two boundaries
leading to cancelling contributions in the contribution of $d(d(\phi_i))$ to the
summand of $R_c$ with fan $I=\{i,k,\ell,j\}$.   }
\label{fig:LOCALOPERATORS-6}
\end{figure}

We now show that $R_c$ is indeed a complex. In fact, it is an $L_\infty$ algebra,
extending the $L_\infty$ algebra structure on the set of interior vectors $R^{\rm int}$.
To this end we introduce a notion of \emph{doubly-extended webs}. These are plane webs,
defined as before, but now we introduce a new kind of vertex, called a \emph{closed vertex}
that can be inserted into the \underline{interior} of the faces of an ordinary
extended plane web. These closed vertices are denoted by open circles in
Figures \ref{fig:LOCALOPERATORS-4} and \ref{fig:LOCALOPERATORS-6}. We can now define
oriented deformation type and convolution in straightforward ways. Each closed vertex
adds two moduli. The boundaries of the moduli space include ends where a closed vertex
approaches a bounding edge of a face. Given a web $\fw$, let $\tilde \fw$ be the
new doubly-extended web where a closed vertex has been inserted into some face at a
point $(u_o,v_o)$. Then the orientations of the two webs are related by
\be
o(\tilde \fw) = o(\fw) \wedge (du_o dv_o)
\ee
All the convolution identities work as for (extended) plane webs.
An illustration of an important convolution involving a closed vertex
is shown in Figure \ref{fig:LOCALOPERATORS-4}. An example of cancelling
ends in a convolution identity is shown in Figure \ref{fig:LOCALOPERATORS-6}.

With a representation of webs we can \underline{define} generators $\phi_j$
of $R_j$, $j\in \IV$,  by
\be\label{eq:Extd-Linfty}
\rho(\tilde\fw)[S_1, \phi_j, S_2] := \delta_{i,j} \rho(\fw)[S_1,S_2]
\ee
where $\fw$ and $\tilde \fw$ are related as before and the closed vertex in $\tilde\fw$ is
inserted in a face marked with the vacuum $i$. Here $S_1, S_2 \in TR_c$.
The generator $\phi_i$ of $R_i$ is the same as that used in equation \eqref{eq:dc-def2},
as one can check by carefully comparing orientations in Figure \ref{fig:LOCALOPERATORS-4}.
The demonstration of the $L_\infty$ algebra structure on $R_c$ completely parallels
that used before for $R^{\rm int}$. In particular, the demonstration that $d(d(\phi_i))=0$
follows from the consideration of ends of moduli space such as those shown in
Figure \ref{fig:LOCALOPERATORS-6}.

\subsection{Traces Of Interfaces}\label{subsec:TraceInterface}

There is a useful, alternative perspective on the cylinder geometry with $\vartheta(x) = -x$. Up to an homotopy,
we can deform the $\vartheta(x)$ profile so that the variation happens on a small scale compared to the size of the circle.
Thus the geometry reduces to a cylinder with essentially constant vacuum weights, and an interface $\fR[2\pi,0]$
inserted at $x=0$. The complex of approximate ground states is essentially the same as $R_c$, though the differential
will only be chain-homotopic to $d_c$.

This is a special case of a construction which is available every time we have an interface $\CI\in \fB\fr(\CT,\CT)$
between a Theory $\CT$ and itself.
Let us define the \emph{trace of the Interface $\CI$}, denoted $\Tr(\CI)$, to be the complex of approximate groundstates on the cylinder
with $\CI$ running along the axis of the cylinder. Thus, the underlying $\IZ$-module of the complex is
\be
\Tr(\CI) = \oplus_{i\in \IV} \CE(\CI)_{ii}
\ee
To define the differential we consider   ``periodic webs.''  These are  webs in $\IR \times S^1$
where the edges  have constant slope. In other words, if we unroll the cylinder by cutting along a vertical line
to get a strip, then the edges of the webs are straight lines in the strip.
Periodic webs are a close analogue to strip webs or composite webs on a strip geometry.
Taut webs have two moduli (one of which is translation along the axis of the cylinder).
\footnote{In the absence of the Interface $\CI$,
the line principle shows the only webs would be unions of closed loops wrapping around the cylinder.}
The taut element $\ft_{c}$ satisfies a convolution identity
\begin{equation}
\ft_c \circ \ft_c + \ft_c * \ft_{pl} +  \ft_c * \ft_{\CI}  =0
\end{equation}
and hence $d_c = \rho_\beta[\ft_c](  \frac{1}{1-\CB})$, where $\CB$ is the boundary
amplitude of the Interface, is a differential.

In general we claim that $\Tr(\CI \IntfcTimes \fR[0, \pm 2\pi] )$ is homotopy equivalent to the
complex of groundstates on the strip $x \sim x+2 \pi$ with $\CI$ inserted and with
spinning vacuum weights, spinning
by  $\vartheta(x) = -x$. In particular, taking $\CI$ to be the identity Interface,
and the definition \eqref{eq:TautCurvedCP} we see that the Chan-Paton data of
$\fR[\vartheta, \vartheta-2\pi]$ are given by
\be
\oplus_{i\in \IV}  (R_i \oplus \widehat{R}'_{ii} ) e_{ii}   \oplus
\oplus_{i<j} \widehat{R}^+_{ij} e_{ij} \oplus \oplus_{i>j} \widehat{R}^-_{ij} e_{ij}
\ee
where $\widehat{R}'_{ii}$ is the set of cyclic fans with vacuum $i$ at $x=0$.
(See equation \eqref{eq:SpAmp-5} above for the case of two vacua.)  The trace of this matrix of complexes
recovers the complex $R_c$ of equation \eqref{eq:LocalOpComplex}. Thus, given the
results of the previous section,
the space of local operators in a Theory can be described in terms of the trace of
$\fR[\vartheta, \vartheta-2\pi]$.

The above construction can be generalized.
 If we have a sequence of Interfaces    $\CI_i \in \fB\fr(\CT^i, \CT^{i+1})$   at locations
$x_i$, $i=1,\dots, n$ along the periodic direction $x \to x+2 \pi$, interpolating between
a periodic sequence of Theories $\CT^i$ then we can consider the trace of the product
\be\label{eq:TraceProductInterface}
\Tr( \CI_1 \IntfcTimes \cdots \IntfcTimes \CI_n)  = \oplus_{j_1, \cdots j_n} \otimes_{i=1}^n\CE(\CI_i)_{j_i,j_{i+1}}
\ee
There will be many different homotopy-equivalent differentials. If we consider again the
taut element for periodic composite webs $\ft_c$ (generalizing that used above for a
single Interface) then the convolution identity generalizes to
\begin{equation}
\ft_c \circ \ft_c + \ft_c * \ft_{pl} + \sum_{i=1}^n \ft_c * \ft_{\CI_i} =0
\end{equation}
and therefore $d_c = \rho_\beta[\ft_c](\otimes_{i=1}^n \frac{1}{1-\CB_i})$
defines a differential on \eqref{eq:TraceProductInterface}.

\subsection{Local Operators For The Theories $\CT^N$ and $\CT^{SU(N)}$ }\label{subsec:LocOp-TSUN}

We comment briefly on the computation of the cohomology of $R_c$ for
the two examples of Theories  $\CT^N$ and $\CT^{SU(N)}$ discussed
throughout this text.

Let us consider first $\CT^N$. According to Section \S \ref{subsubsec:RelationTSUN-Physical}
this is meant to coincide with the $A$-model  with target space $X=\IC$ and
superpotential \eqref{eq:TN-Superpot}. For the strict $A$-model the target space
cohomology $H^*_{DR}(X)$ has a single operator in degree zero corresponding to the
unit operator. Nevertheless, as discussed in Section \S \ref{closedstrings} the full
space of local operators can in principle include disorder operators, and
$H^*_{DR}(X)$ is only a subspace of the space of local operators. In fact,
for the $\CT^N$ Theories, the cohomology of $R_c$ is one-dimensional,
and spanned by the unit operator $\textbf{1}$  as the following computation shows.

The complex $R_c$ takes the form
\be
\oplus_i R_i \rightarrow \oplus_{i<j} R_{ij}\otimes R_{ji} \rightarrow \oplus_{i<j<k} R_{\{i,j,k\}} \rightarrow \cdots
\rightarrow \oplus_{i  }  R_{\widehat{i}} \rightarrow R_{\{0,1,\dots, N-1\} } \rightarrow 0
\ee
where in degree $N-2$, $\widehat{i}$ denotes the fan that omits $i$ from $ \{0,\dots , N-1\}$.
Using \eqref{eq:ExpleTN-webrep} this becomes
\be
\oplus^N \IZ\rightarrow \oplus^{{ N \choose 2} } \IZ^{[1]} \rightarrow \cdots
\rightarrow \oplus^{{ N \choose j} } \IZ^{[j-1]}\rightarrow \cdots \rightarrow \IZ^{[N-1]} \rightarrow 0
\ee
The Witten index is thus automatically $1$. We can in fact do better and compute the cohomology
as follows. We can identify the
complex with a subspace of a Grassmann algebra $\CG = \IZ[\theta_0, \dots, \theta_{N-1}]/\CI$
where the ideal $\CI$ is generated by $\theta_i \theta_j + \theta_j \theta_i =0$, for all $i,j$
and the $\theta_i$ have degree $+1$. We identify $R_c$ with the subspace of $\CG$ of elements
of degree at least one and then shift the degree by $-1$.  To see this, identify a generator of
$R_{i_1,i_2} \otimes \cdots R_{i_{k-1},i_k} \otimes R_{i_k, i_1}$, where
$i_1< i_2< \cdots < i_k $ with $\theta_{i_1} \theta_{i_2} \cdots \theta_{i_k}[-1]$.
Then, for any fan, the differential acts by diagrams like those of Figure \ref{fig:LOCALOPERATORS-5}.
Since the interior amplitude is only nonzero for 3-valent vertices with $b_{ijk} = 1$ for
$i<j<k$ it is easy to see that the differential $d_c$ is the same as the action of
multiplication by $\Theta = \theta_0 + \cdots + \theta_{N-1}$. There is thus a clear chain-homotopy
inverse between the zero map and the projection operator onto on elements of $R_c$ of positive degree.
It is given by:
\be
\kappa(r) := \begin{cases}  \left(
\frac{\p}{\p \theta_0} + \cdots + \frac{\p}{\p \theta_{N-1}} \right) r  & \qquad \deg(r)>0 \\
0 & \qquad \deg(r) = 0 \\
\end{cases}
\ee
Thus,
\be
\left(\kappa d_c + d_c \kappa\right)(r) := \begin{cases}
N r  & \qquad \deg(r)>0 \\
N r - {\rm tr}(r) \textbf{1} & \qquad \deg(r) = 0 \\
\end{cases}
\ee
Here $\textbf{1} = \sum_{i\in \IV} \phi_i$ is the unit operator discussed above.
In this example $\textbf{1} = \sum_{i=0}^{N-1} \theta_i[-1]$. Moreover, if $r=\sum x_i \theta_i[-1]$
is of degree zero we define ${\rm tr}(r)= \sum_i x_i$. Therefore the cohomology is generated by
the unit operator.
\footnote{This argument only suffices to determine the cohomology up to $N$-torsion. From
examples we find that the cohomology is in fact isomorphic to $\IZ$ in degree zero. In physics
the space of local operators is a vector space over $\IC$, but in the web formalism one
can work over $\IZ$. This raises the interesting question of whether there can be torsion in
the cohomology of $R_c$, and what its physical meaning would be, if any. We leave that for
another time.}

Turning now to the   $\CT^{SU(N)}$ Theory the physical expectation from
Section \S \ref{subsubsec:RelationTSUN-Physical} is that it should
correspond to an A-model with superpotential $W=\sum_{i=1}^N Y_i$
 on the space $\Xi$ defined by
 $\Xi=\{(Y_1,\dots, Y_N) \vert Y_1 \cdots Y_N = q \}\subset (\IC^*)^N$
 where $q\not=0$. By \cite{Hori:2000kt} this should be mirror to the
 B-model on $\IC \IP^{N-1}$. We use this mirror dual pair to check
 our proposal for the local operators using the   complex $R_c$ constructed from the representation
of webs described in equation \eqref{eq:SUN-Rij} above.

In general, the
B-model with target space $X$ has a space of local operators
\be
\oplus_{p,q} H^p(X, \Lambda^q T^{1,0}X).
\ee
In the present case
we compute the cohomology, as a representation of ${\rm su}(N)$  to be:
\be\label{eq:CPN-coho}
H^p(X, \Lambda^q T^{1,0}\IC \IP^{N-1} )= \begin{cases}
L_{N-q,q+1} & p = 0 \\
0 & p>0 \\
\end{cases}
\ee
where the $L_{n,m}$ were defined in Section \S \ref{subsec:CyclicVacWt}.
The result is in fact very intuitive. Cohomology classes with $p=0$
can be represented by global sections
\be
C_{i_1\dots i_q}^{j_1\dots j_q} X^{i_1} \cdots X^{i_q} \frac{\p}{\p X^{j_1}}\wedge \cdots \wedge \frac{\p}{\p X^{j_q}},
\ee
where $[X^1: \cdots : X^N]$ are homogeneous coordinates. The $SU(N)$ tensor $C_{i_1\dots i_q}^{j_1\dots j_q}$
is totally symmetric in $i_k$ and totally antisymmetric in $j_k$ and therefore in $A_{N-q} \otimes S_q$. Now
we must identify by the image of holomorphic vector fields and this requires $C_{i_1\dots i_q}^{j_1\dots j_q}$
to be traceless. Referring to the decomposition \eqref{eq:ASL}, the traceless part is the second summand.
The cohomology is one-dimensional in degree $q=0$ and isomorphic to the adjoint in degree $q=1$.
Indeed, the one-dimensional cohomology should be a Lie algebra of global symmetries of the theory
on \emph{a priori} grounds and that is the ${\rm su}(N)$ symmetry of the present example.

We leave it as an interesting and nontrivial challenge to reproduce \eqref{eq:CPN-coho} from the cohomology of $R_c$, as
defined in equation \eqref{eq:LocalOpComplex}.

We can perform one nontrivial check on this identification by examining
the character-valued index. Let  $t = {\rm Diag}(t_1, \dots, t_N ) $ be   a generic diagonal element of $SU(N)$.
It acts naturally on the homogeneous coordinates of $\IC\IP^{N-1}$ thereby inducing an action on $H^p(X, \Lambda^q T^{1,0}X)$.
By the Atiyah-Bott fixed point formula we have
\footnote{
Incidentally, there is an elegant argument to recover the representations of \eqref{eq:CPN-coho} from
this formula. Consider
\be
U_N(z,x) := \frac{1}{z} \frac{\prod_{j=1}^N (1 + x z/t_j) }{\prod_{j=1}^N (1-t_j/z)} dz
\ee
By equating the residue at $z=\infty$ with the sum of the finite residues,  using the
generating functions for characters of symmetric and antisymmetric representations,
and using the decomposition \eqref{eq:ASL} one can reproduce \eqref{eq:CPN-coho}.}
\be\label{eq:ABsum}
\begin{split}
F_N(x) & :=\sum_{q=0}^{N-1} x^q \sum_{p=0}^{N-1} (-1)^p {\rm Tr}_{H^p(X, \Lambda^q T^{1,0}\IC \IP^{N-1} )}  t \\
& =  \sum_{i=1}^N \frac{\prod_{j\not=i} (1 + x t_i/t_j) }{\prod_{j\not=i} (1-t_j/t_i)}\\
\end{split}
\ee
In particular, the character-valued index is
\be\label{eq:FN-SIMP}
F_N(-1) = \sum_{i=1}^N \frac{\prod_{j\not=i} (1 - t_i/t_j) }{\prod_{j\not=i} (1-t_j/t_i)}
= (-1)^{N-1}(t_1 \cdots t_N)^{-1} \sum_{i=1}^N t_i^N = (-1)^{N-1} \sum_{i=1}^N t_i^N
\ee
where in the second equality we used the property that $t\in SU(N)$.

On the other hand, by direct computation of the character-valued index of $R_c$, using the
characters of the anti-symmetric representations $R_{ij}$ we find
\be\label{eq:Amazing}
{\Tr}_{R_c}(-1)^F = N - N \prod_{i=1}^N t_i + (-1)^{N-1} \sum_{i=1}^N t_i^N = (-1)^{N-1} \sum_{i=1}^N t_i^N
\ee
where in the second equality we used the property that $t\in SU(N)$. We checked equation \eqref{eq:Amazing} for
 $N=2,3,4,5,6$, but giving a direct proof of this equation looks difficult. Fortunately the methods of
 Section \S \ref{subsubsec:Powers} above can be used to give a proof: The Interface $\fI^{+-+}$ of that
 section implements a rotation by $2\pi/N$, and hence its $N^{th}$ power gives the full Interface for
 rotation by $2\pi$. The eigenvalues of the character-valued index of the Chan-Paton data of this Interface
 follow from \eqref{eq:FactorCharPoly}, and are simply $(-t_i)$, $i=1,\dots, N$.
 Using the relation of the complex $R_c$ to the
 trace of the Interface explained in Section \S \ref{subsec:TraceInterface} we arrive at equation
 \eqref{eq:Amazing}.
 \footnote{Actually, there is a disagreement by an $N$-independent minus sign. We have not sorted out the explanation of this
 sign.}

Note that the cohomology is much larger than the naive DeRham cohomology of $\Xi \cong (\IC^*)^{N-1}$ that one
might associate to the $A$-model on $\Xi$. Indeed,
the dimension of the $H^*(R_c, d_c)$ for $\CT^{SU(N)}$ is given by ${}_2 F_1(1-N,N+1,1;-1)$,
and this grows with $N$ far more rapidly than $2^{N-1}$. Thus, there are many disorder operators.
Indeed, we can already see the need for disorder operators for the case $N=2$ discussed in detail
in Section \S \ref{moremirror} below.

\section{A Review Of Supersymmetric Quantum Mechanics And Its Relation To Morse Theory}\label{review}
\def\uu{u}

In Sections \S\S \ref{lgassuper}-\ref{subsec:LG-Susy-Interface} below we will sketch our main physical application of the
formalism we have developed. That application is based in turn on standard ideas about the
interpretation of Morse theory in terms of supersymmetric quantum mechanics \cite{Witten:1982im}.
While this material is well-known, and is nicely reviewed, for example, in
\cite{Hori:2003ic}, we would like to emphasize several key points which are of
particular importance in our application.

\subsection{The Semiclassical Approximation}

We start by reviewing supersymmetric quantum mechanics and its relation to Morse theory
(see \cite{Witten:1982im} and section 10 of \cite{Hori:2003ic}),
since much of our subject can be developed in close parallel to this.  Much of this
material may be familiar to many readers, but in section \ref{whyindeed} we explain a point that may be less familiar
and that is crucial background for the present paper.

We begin with a Riemannian manifold $M$ of dimension $n$, with local coordinates $\uu ^a$, $a=1,\dots,n$,
a metric tensor $g_{ab}$, and a smooth real-valued function $h$, called the superpotential.
A critical point of $h$ is a point at which its gradient vanishes,
$\partial h/\partial \uu ^a=0$, $a=1,\dots,n$, and a critical point is called nondegenerate if at this point the matrix of
second derivatives\footnote{In general, to define the second derivative of a function $h$ on a Riemannian manifold $M$,
we need to use the affine connection of $M$; the only natural second derivative is  $D^2h/D \uu ^a D\uu ^b$, where $D/D \uu ^a$ is a covariant
derivative. But at a critical point, $D^2 h/D \uu ^a D\uu  ^b$ reduces to the more naive $\partial^2 h/\partial \uu ^a\partial \uu ^b$, a fact
that we incorporate in some formulas below.}
 $\partial^2 h/\partial \uu ^a\partial \uu ^b$ is nondegenerate.  (This matrix is sometimes called the Hessian.)
%
%A critical point $p$ has an ``index'' denoted by $n_p$ and defined to be
%the number of negative eigenvalues.
%
We will assume that $h$ is a Morse function.
A Morse function is simply a smooth function such that
all critical points $\phi_i$ are nondegenerate.
%
%with only finitely many critical points $\phi_1,\dots,\phi_s$, all of which
%are nondegenerate.
%
For the moment, we assume $M$ to be compact,
 in which case the number
of critical points of $h$ is finite,
but sometimes one wishes to relax the compactness
assumption and also allow infinitely many critical points.

From this data, we construct a supersymmetric quantum mechanics model, describing maps from $\Bbb R^{1|2}$ --
a supersymmetric worldline with a real coordinate $t$ and odd coordinates $\theta$ and $\bar\theta$ -- to $M$.
The supersymmetry algebra\footnote{We write $\Q$ and $\bar \Q$ (rather than $Q$ and $\bar Q$) for the supercharges
of the quantum mechanical model, since this will be more convenient in discussing the generalization to two-dimensional
LG theories.} is generated by the odd vector fields on $M$
\begin{equation}\label{tolzo} \Q=\frac{\partial}{\partial\bar\theta}+i\theta\frac{\partial}{\partial t}, ~~~\bar \Q=
\frac{\partial}{\partial\theta}+i\bar\theta\frac{\partial}{\partial t};\end{equation}
the only nonzero anticommutator of these operators is
\begin{equation}\label{olzo}\{\Q,\bar \Q\}=2 H ,~~~ H =i\partial_t. \end{equation}
This supersymmetry algebra admits a group $U(1)$ of outer automorphisms called the $R$-symmetry group, generated by a
charge $\FF$ that assigns the values $1$ and $-1$ to $\theta$ and $\bar\theta$, respectively.  This will be a symmetry of the models
we consider.  The supersymmetry algebra commutes in the $\Z_2$-graded sense with the operators
\begin{equation}\label{tolzob} D=\frac{\partial}{\partial\bar\theta}-i\theta\frac{\partial}{\partial t}, ~~~\bar D=
\frac{\partial}{\partial\theta}-i\bar\theta\frac{\partial}{\partial t},\end{equation}
which are used in writing Lagrangians.

To construct a supersymmetric model that describes maps from $\Bbb{R}^{1|2}$ to $X$,
 we promote the local coordinates $\uu ^a$ on $M$ to  superfields $X^a(t,\theta,\bar\theta)=\uu ^a(t)
+ \I \bar\theta \psi^a(t)+\I \theta \bar\psi^a(t)+\bar\theta\theta F^a(t)$, where $\psi^a$ and $\bar\psi^a$   are Fermi fields with $\FF=1$ and $\FF=-1$, respectively, and the $F^a$ are auxiliary fields.\footnote{When working with complex superalgebras we let $*$ denote the complex anti-linear
involution which acts on odd variables according to the rule $(\theta_1\theta_2)^* = \theta_2^* \theta_1^*$.
Our notation is such that $\bar \theta = (\theta)^*$, and so on. Hence $X^i$ is a real superfield when $F^i$ is real.
We are using the same letter $\Q$ for an operator on fields and for an odd vector field on a supermanifold,
so the supersymmetry transformation rule is $[\I \Q, X^i] = \CQ X^i$. Unfortunately, our conventions differ
from those in \cite{Hori:2003ic} by an exchange of $\psi \leftrightarrow \bar\psi$. }
It follows from this that
\begin{align}\label{zolt}\{\Q,\bar\psi^a\}& = i\dot \uu ^a+F^a\cr
                                    \{\bar \Q,\psi^a\}& = -i\dot \uu ^a+F^a.\end{align}

One takes the action to be
\begin{equation}\label{murky}I=\frac{1}{\lambda}\int \d t\,\d^2\theta\, \left(\frac{1}{2}
g_{ab}(X^k)D X^a\bar D X^b - h(X^c).\right)\end{equation}
This action describes a supersymmetric $\sigma$-model in which the target space is $M$.
Perturbation theory is a good approximation if $\lambda$ is small, but $\lambda$ can be eliminated from the formulas by
rescaling $g_{ab}$ and $h$ (and to avoid clutter we do so in what follows).
After integrating over $\theta$ and $\bar\theta$, the action
becomes\footnote{Here $\int d^2 \theta \bar\theta\theta = \int d\theta d\bar\theta ~\bar\theta \theta = +1$.}
\begin{equation}\label{olt}I=\int\d t\left(\frac{1}{2}g_{ab}\dot \uu ^a\dot \uu ^b+ig_{ab}\bar\psi^a\frac{D}{D t}\psi^b+\frac{1}{2}g_{ab}F^aF^b
-F^a\partial_a h +\bar\psi^a\psi^b\frac{D^2h}{D \uu ^aD \uu ^b} + \dots\right),\end{equation}
where the covariant derivative $D/Dt$ is defined using the pullback of the Levi-Civita connection of $M$, and we omit four-fermi
terms.
One can eliminate the auxiliary field $F^a$ via its equation of motion
\begin{equation}\label{olz}F^a=g^{ab}\partial_b h,\end{equation}
and  for the ordinary potential energy, one finds
\begin{equation} V(\uu ^k)=\frac{1}{2}|\nabla h|^2=\frac{1}{2}g^{ab}\partial_a h \partial_b h. \end{equation}
A classical ground state is therefore a critical point of $h$.  Since we have assumed that $h$ is a Morse function,
there is a finite set $\IV$ of such critical points.

The fermion mass term that arises in expanding around a critical point can be read off from the action:
\begin{equation}\label{heffalo} H_{\bar \psi \psi}=\half [\psi^a, \bar\psi^b] \frac{\partial^2 h}{\partial \uu ^a\partial \uu ^b}.\end{equation}
(The bracket appearixng here is an ordinary commutator, not a graded commutator, and accounts for an important
normal ordering convention to preserve supersymmetry.)
Since we assume that $h$ is a Morse function, the fermion mass matrix
$m_{ab}= \partial^2 h/\partial \uu ^a\partial \uu ^b$ is nondegenerate at each critical point.
In Morse theory the number of negative eigenvalues of the Hessian at a critical point $p$
is called the \emph{Morse index}. We denote it by $n_p$.
%
%\be
%n_p := \dim\{ v\in T_pM \vert  {\rm sign}(m) \cdot  v = - v \}.
%\ee
%
Thus the number of negative
 eigenvalues of the mass matrix is $n_p$.    Similarly the bosonic mass squared matrix that arises in expanding around $p$
is positive-definite.  So in expanding around a given critical point,
all bosonic and fermionic modes are massive.  Hence, from the standpoint of perturbation theory, there is precisely one minimum energy
state $\Phi_p$  for every critical point $p\in\IV$.  In perturbation theory, this state has zero energy and is annihilated by the
supercharges $Q$ and $\bar Q$.  Indeed, the supersymmetry algebra
\begin{equation}\label{pozzo} \{ \Q,\bar \Q\}=2H,~~ \Q^2=\bar \Q^2=0,\end{equation}
implies that eigenstates
of $H$ with nonzero eigenvalue come in pairs, and therefore the fact that in expanding around the critical point $p$
one finds only a unique ground state $\Phi_p$ implies that in perturbation theory, $\Phi_p$ is annihilated by $\Q$, $\bar \Q$, and $H$,
so it is  a supersymmetric state of zero energy.   (Beyond perturbation theory, as we discuss in section \ref{zyindeed},
nonperturbative effects can modify this statement.)

The states in the $\sigma$-model are not functions on $M$, but differential forms on $M$.  To see this, we observe that
a state of smallest fermion number must be annihilated by the $\bar\psi^a$ operators, but may have an arbitrary dependence on the
bosonic variables $\uu ^a$.  So letting $|\Omega\rangle$ denote a state annihilated by all $\bar\psi^a$ and independent of the $\uu ^a$,
a state of minimum possible
fermion number is $f(\uu )|\Omega\rangle$, where $f(\uu )$ is an arbitrary function on $M$.  A state whose fermion number is greater
by $n$ is then $\sum_{a_1\dots a_n}f_{a_1a_2\dots a_n}(\uu )\psi^{a_1}\psi^{a_2}\dots \psi^{a_n}|\Omega\rangle$, where
in the language of differential geometry, $\sum_{a_1\dots a_n}f_{a_1a_2\dots a_n}(\uu )\d \uu ^{a_1}\d \uu ^{a_2}\dots \d \uu ^{a_n}$ is called an $n$-form
on $M$.  Thus quantum states in the $\sigma$-model correspond to differential forms on $M$ and, up to an additive constant (which
is the fermion number we assign to the state $|\Omega\rangle$), the
fermion number of a state is the degree of the corresponding differential form.  In differential geometry, it is customary
to define the degree of a differential form on a $d$-manifold to vary from $0$ to $d$.  However, the theory (\ref{murky})
has a symmetry $\psi\leftrightarrow \bar\psi$ (``charge conjugation,'' which in differential geometry is called the Hodge star operator on differential forms).
$\FF$ is odd under this symmetry at the classical level, and to maintain this property quantum mechanically,
 we subtract an overall constant $-d/2$ and say that an $n$-form corresponds
to a state of fermion number $\FF = -d/2+n$.

To determine the fermion number of the low energy state $\Phi_p$  associated to a given
critical point $p$, we need to determine which modes
of $\psi$ and $\bar\psi$ annihilate $\Phi_p$.  All we need to know is that for a real number $m$
and  a single pair of fermion modes $\psi$, $\bar\psi$, the
operator $H_0=\frac{m}{2}[\psi,\bar\psi]$ has {\it (i)} an eigenstate $|\negthickspace\downarrow\rangle$ annihilated by $\bar\psi$ with $H_0=-\frac{m}{2}$
and  {\it (ii)} an eigenstate $|\negthickspace\uparrow\rangle$
annihilated by $\psi$ with $H_0=+\frac{m}{2}$.  For $m>0$, the ground state of $H_0$ is
annihilated by $\bar\psi$ but for $m<0$, it is annihilated by $\psi$.  So the number of modes of $\psi$ that
annihilate the ground state of the fermion mass operator $H_{\bar\psi\psi}$ of eqn. (\ref{heffalo}) is equal to the number
of negative eigenvalues of the  matrix $\partial^2 h/\partial \uu ^a\partial \uu ^b$, or in other words, the Morse
index of the critical point $p$.  So if $p$ has Morse index $n_p$, then $\Phi_p$ is an $n_p$-form.

Thus if $n_+$ and $n_-$ are the number of positive and negative eigenvalues of the fermion mass matrix
in expanding around a given critical point $p$ (so $d=n_++n_-$ and the Morse index of $p$ is $n_p=n_-$), then the fermion number of $\Phi_p$ is
\begin{equation}\label{hoobo} f_p=-\frac{1}{2}\left(n_+-n_-\right). \end{equation}

A standard argument using the supersymmetry algebra (\ref{pozzo}) shows that the space of supersymmetric states --
states annihilated by $\Q,\bar \Q, $ and $H$ -- can be naturally identified with the cohomology of $\Q$ (the kernel of $\Q$ divided
by its image).  Here $\Q$ is an operator mapping $n$-forms to $n+1$-forms and obeying $\Q^2=0$.  In differential geometry,
there is a standard operator with this property, the exterior derivative $\d$.
In differential geometry, it is usual written $\d=\d \uu ^a\partial_{\uu ^a}$, which in our language would be $\psi^a\partial_{\uu ^a}$.  Hence
$\{\d,\bar\psi^a\}=g^{ab}\partial_{\uu ^b}$.  Recalling that in canonical quantization, $\dot \uu ^a$ maps to $-\i g^{ab}\partial_{\uu ^b}$, eqn.
(\ref{zolt}) tells us that $\{\Q,\bar\psi^a\}=g^{ab}(\partial_{\uu ^b}+\partial_bh)$.  So $\Q$ does not coincide with $\psi^a\partial_{\uu ^a}=\d$; rather,
\begin{equation}\label{hozz} \Q=\psi^a(\partial_{\uu ^a}+\partial_a h) =e^{-h}\d e^h. \end{equation}
Thus $\Q$ does not coincide with the exterior derivative $\d$, but rather is conjugate to it.

This means that the cohomology of $\Q$ is naturally isomorphic to the cohomology of $\d$, which is usually called the de Rham
cohomology of $M$: the cohomology of $\Q$ is obtained from that of $\d$ by multiplying by the operator $e^{-h}$.
In particular, the number of states of precisely zero energy with fermion number $-d/2+n$ is the corresponding Betti number
$b_n$ (defined as the rank of the de Rham cohomology for $n$-forms) and does not depend on the
choice of $h$.  By contrast, the number of zero energy states found in perturbation theory for given $n$
is the number of critical points of $h$ of Morse index $n$
and definitely does depend on $h$.  (For example, if $M$ is the circle $0\leq \varphi\leq 2\pi$, the Morse function $h=\cos k\varphi$
has $2k$ critical points, half with index 0 and half with index 1.)  So there will have to be nonperturbative effects that in general
eliminate some of the vacuum degeneracy.

Since $\bar \Q$ is the adjoint of $\Q$, it follows from (\ref{hozz}) that
\begin{equation}\label{ozz} \bar \Q=\bar\psi^a(-\partial_{\uu ^a}+\partial_a h) = e^h\d^\dagger e^{-h},\end{equation}
where $\d^\dagger$ is the adjoint of $\d$ in the standard $L^2$ metric on
differential forms.  The Hamiltonian $H=\{\Q,\bar \Q\}/2$ definitely does depend upon $h$,
though the number of its zero energy states for each value of $\FF$ does not.

\subsection{The Fermion Number Anomaly}\label{fermanom}

The next topic we must understand is the fermion number anomaly.  For this computation, we transform to Euclidean
signature via $t=-i\tau$.  The (linearized) Dirac equation for $\psi$ and $\bar\psi$ becomes
\begin{equation}\label{zoob} L\psi=0=L^\dagger\bar\psi,\end{equation}
with
\begin{align}\label{moob}(L\psi)^a&=\frac{D\psi^a}{D\tau}-g^{ab}\frac{D^2h}{D \uu ^b D \uu ^c}\psi^c \cr
                                         (L^\dagger\bar\psi)^a&=-\frac{D\bar\psi^a}{D\tau}-g^{ab}\frac{D^2h}{D \uu ^b D\uu ^c}\bar\psi^c.\end{align}
We consider expanding around a path $\ell\subset M$  that starts at one critical point $q$ in the far past and ends at another
critical point $p$ in the far future.

Let $n_q$ and $n_p$ be the Morse indices of these critical points.
We want to compute a vacuum-to-vacuum amplitude, that is, a transition
between the  initial state $\Phi_q$, of $\FF= -d/2+n_q$, and the final state $\Phi_p$, of fermion number
$\FF=-d/2+n_p$.  The fermion numbers of the initial and final states differ by $n_p-n_q,$ so the amplitude
must vanish unless one inserts operators that carry a net fermion number $n_p-n_q$.

As usual, the mechanism for this is that the index\footnote{There is a slight clash in the standard
terminology here;
the ``index'' of an operator should not be confused with
the ``Morse index'' of a critical point of a function.} of the operator $L$, which is defined
as  the number of zero-modes of $L$ minus the number of
zero-modes of its adjoint $L^\dagger $, is equal to $n_p-n_q$.  Let us verify directly that this is true.  Since the index is invariant
under smooth deformations (which preserve the mass gap at infinity),
it suffices to consider the case that the Levi-Civita connection of $M$ is trivial along $\ell$
and that the fermion mass matrix is diagonal along $\ell$.  Thus it suffices to consider the case of a single pair of fermions $\psi$,
$\bar\psi$ with
   \begin{align}\label{mooby}L\psi(\tau)&=\frac{\d\psi(\tau)}{\d\tau}-w(\tau)\psi(\tau) \cr
                                         L^\dagger\bar\psi(\tau)&=-\frac{\d\bar\psi(\tau)}{\d\tau}-w(\tau)\bar\psi(\tau)\end{align}
   where in  the one-component case, we abbreviate $D^2h/D \uu ^2$ as $w$.
   We assume that the function $w$ is nonzero for $\tau\to \pm \infty$.
 The equations $L\psi=0$ and $L^\dagger\bar\psi=0$ imply respectively
 \begin{align}\label{ooby} \psi(\tau)& = C\exp\left(\int_0^\tau\d\tau' w(\tau')  \right) \cr
                                         \bar\psi(\tau)& = C'       \exp\left(-\int_0^\tau\d\tau' w(\tau')  \right) , \end{align}
                                         with constants $C,C'$. If $w$ has the same sign for $\tau>>0$ as for $\tau<<0$,
                                         then neither solution is square integrable.  If $w$ is positive for $\tau<<0$ and negative for $\tau>>0$,
 then $\psi$ has a normalizable zero-mode but not $\bar\psi$; if $w$ is negative for $\tau<<0$ and positive for $\tau>>0$,
 then $\bar\psi$ has a normalizable zero-mode but not $\psi$.  In all cases, the index of the $1\times 1$ operator $L$ for a single
 pair $\psi,\bar\psi$
 is the contribution of this pair to $n_p-n_q$.  Summing over all pairs, the index of $L$ equals $n_p-n_q$,
 as expected.

 The index $\iota(L)$ of the  operator $L$ always determines the difference between the number of $\psi$ and $\bar\psi$ zero-modes,
 but in fact generically one of these numbers vanishes and the other equals $|\iota(L)|$.  For example, if $\iota(L)\geq0$, generically
 there are no $\bar\psi$ zero-modes, and the space of $\psi$ zero-modes has dimension $\iota(L)$; if $\iota(L)\leq 0$, generically
 there are no $\psi$ zero-modes and the space of $\bar\psi$ zero-modes has dimension $-\iota(L)$.  The explicit calculation in the last
 paragraph shows that these statements are always true in the $1\times 1$ case; in fact, they hold generically.  Informally,
   $\iota(L)$ is a regularized difference between the number of variables and the number of conditions in
 the equation $L\psi=0$.  So for example if $\iota(L)>0$, the equation $L\psi=0$ is analogous to a finite-dimensional linear problem with
 $\iota(L)$ more variables than equations and generically has a space of solutions precisely of dimension $\iota(L)$.

\subsection{Instantons And The Flow Equation}\label{instflow}

In general, a Morse function on $M$ has too many critical points to match the de Rham cohomology of $M$, so there
must be nonperturbative effects that shift some of the perturbatively supersymmetric states $\Phi_q$ away from zero energy.
For this to happen, the supercharges $\Q$ and $\bar \Q$, instead of annihilating the $\Phi_q$, must have nonzero matrix
elements $\langle \Phi_p|\Q|\Phi_q\rangle$ or $\langle \Phi_p|\bar \Q|\Phi_q\rangle$, for distinct critical points $p$ and $q$.

So we have to analyze tunneling events that involve transitions between two critical points $q$ and $p$.  As a preliminary,
we evaluate the action for
a trajectory $\uu (\tau)$ that starts at $q$ for $\tau\to -\infty$ and ends at $p$ for $\tau\to+\infty$.
For such a trajectory, in Euclidean signature,
 and with the auxiliary fields eliminated via (\ref{olz}), the bosonic part of the action (\ref{olt}) is
 \begin{align}\label{durf}I=&\frac{1}{2}\int_{-\infty}^\infty\d \tau \,\left(g_{ab}\frac{\d \uu ^a}{\d\tau}\frac{\d \uu ^b}{\d\tau}
  +g^{ab}\partial_a h\,\partial_b h\right)
\cr =&\frac{1}{2}\int_{-\infty}^\infty\d\tau\,g_{ab}\left(\frac{\d \uu ^a}{\d\tau}\pm g^{ac}\partial_ch\right)\left(\frac{\d \uu ^b}{\d\tau}\pm g^{be}\partial_e h\right)\mp
(h(p)-h(q)), \end{align}
after integrating by parts.  The action is therefore minimized by a trajectory that obeys
\begin{equation}\label{zobb}\frac{\d \uu ^a}{\d\tau}-g^{ab}\partial_b h=0, ~~~~   h(p)>h(q),\end{equation}
or
\begin{equation}\label{zoobo} \frac{\d \uu ^a}{\d\tau}+g^{ab}\partial_b h=0, ~~~~~h(p)<h(q). \end{equation}
(For a given sign of $h(p)-h(q)$, only one of these equations may have a solution, since the left
hand side of eqn. (\ref{durf}) is non-negative. If $h(q)=h(p)$, 	with $q\not=p$, neither equation has a solution.)
These equations are called gradient flow equations; we can write them
\begin{equation}\label{oob}\frac{\d\vec \uu }{\d\tau}=\pm \vec\nabla h, \end{equation}
where the ``flow'' vector $\d\vec \uu /\d\tau$ has components $\partial_\tau \uu ^a$, and the ``gradient'' vector $\vec\nabla h$ has
components $g^{ab}\partial_b h$.

The gradient flow equations have another interpretation.  The Lorentz signature supersymmetry transformations (\ref{zolt}) can be
transformed to Euclidean signature via $t=-i\tau$, so that $i\dot \uu ^a=i\d \uu ^a/\d t$ is replaced by $-\d \uu ^a/\d\tau$.
After also eliminating the auxiliary fields, the supersymmetry transformations in Euclidean signature read
\begin{align}\label{zolto}\{\Q,\bar\psi^a\}& = -\frac{\d \uu ^a}{\d\tau}+g^{ab}\partial_bh\cr
                                    \{\bar \Q,\psi^a\}& = \frac{\d \uu ^a}{\d\tau}+g^{ab}\partial_b h.\end{align}
The condition for a trajectory to be $\Q$-invariant is that $\{\Q,\bar\psi^a\}$ vanishes for that trajectory.
So $\Q$-invariant trajectories obey
\begin{equation}\label{olto} \frac{\d \uu ^a}{\d \tau}=g^{ab}\partial_b h, \end{equation}
and we call these ascending gradient lines since the flow (for increasing $\tau$) is in the direction of steepest ascent for $h$.
And $\bar \Q$-invariant trajectories are similarly descending gradient flow lines, obeying
\begin{equation}\label{bolto} \frac{\d \uu ^a}{\d\tau}=-g^{ab}\partial_b h. \end{equation}

Now let us discuss the moduli space $\M_{qp}$ of ascending gradient flow lines from $q$ in the past to $p$ in the future.
The tangent space of $\M_{qp}$ at the point corresponding to a given solution of the ascending flow equation (\ref{olto})
is the space of solutions of the linear equation found by linearizing the ascending
flow equation around the given solution.  The linearization of the ascending flow equation is simply the equation $L\psi=0$,
where $L$ is the fermion kinetic operator defined in eqn. (\ref{mooby}).  As explained in section \ref{fermanom}, the index of $L$
is $\iota(L)=n_p-n_q$, and generically, when this number is nonnegative, it equals the dimension of the kernel of $L$ or in other
words of the tangent space of $\M_{qp}$.  Generically (that is, for a generic metric $g_{ab}$ on $M$ and a generic Morse function $h$),
$\M_{qp}$ is a smooth (not necessarily connected) manifold of dimension $n_p-n_q$, assuming that this number is positive.
For this reason, one calls $n_p-n_q$ the expected dimension of $\M_{qp}$.  If the expected dimension is negative, generically
$\M_{qp}$ is empty.

To understand what happens when $n_q=n_p$, we first need the following general comment.  As long as $p\not=q$,
an ascending flow from $q$ to $p$ cannot be invariant under time translations, so there is always a free action of the group $\Bbb R$
of time translations on $\M_{qp}$.  So for $p\not=q$, we can define a reduced moduli space $\M_{qp,\mathrm{red}}$ as the quotient $\M_{qp}/\Bbb R$.  $\M_{qp}$ is a fiber bundle over $\M_{qp,\mathrm{red}}$ with fibers $\Bbb R$.
The expected dimension of $\M_{qp,\mathrm{red}}$ is $n_p-n_q-1$.  So if $n_p=n_q$ and $p\not=q$, the expected dimension of
$\M_{qp,\mathrm{red}}$ is $-1$.  This means that generically $\M_{qp,\mathrm{red}}$ is empty, in which case $\M_{qp}$ is likewise
empty.  If instead $p=q$, the only ascending or descending flow from $q$ to $p$ is the trivial flow in which $\uu ^a(\tau)$ does not
depend on $\tau$. (This is an easy consequence of the fact that the right hand side of (\ref{durf}) vanishes for such a flow.)
So $\M_{pp}$ is always a point.

One last comment along these lines is that even if $\M_{qp}$ has a positive expected dimension, it is empty if $h(p)\leq h(q)$.
This follows easily from (\ref{durf}), whose right hand side would be non-positive for an ascending flow from $q$ to $p$.

We conclude with a more elementary way to determine the dimension of $\M_{qp}$.  Near a nondegenerate critical point $m\in M$, we can find Riemann normal
coordinates $\uu ^a$ such that the metric tensor is just $\sum_a (\d \uu ^a)^2+\O(\uu ^2)$, and
\begin{equation}\label{mubo} h = h_0+\frac{1}{2}\sum_a f_a \uu _a^2, \end{equation}
where the $f_a$ are all nonzero and the number of negative $f_a$ equals the Morse index at $m$.  The flows \eqref{olto} near $m$ look like
\begin{equation}\label{ubo} \uu ^a(\tau)=\sum_a c_a e^{f_a \tau},\end{equation}
so the flows that depart from $m$ for $\tau\to -\infty$ (or approach $m$ for $\tau\to +\infty$) are those in which $c_a=0$ for $f_a<0$ (or $f_a>0$).
The space of all gradient flows has dimension $d$ (since a flow is determined by its value at a specified time).  After imposing $n_q$ conditions
to ensure that a flow starts at $q$ and $d-n_p$ conditions to ensure that a flow ends at $p$, we find that the expected dimension of the space of flows
from $q$ to $p$ is $n_p-n_q$.

\subsection{Lifting The Vacuum Degeneracy}\label{zyindeed}

Now we want to see how instantons can lift the vacuum degeneracy that is present at tree level.
We consider the matrix element of the supercharge $\Q$ between perturbative vacuum states $\Phi_q$ and $\Phi_p$:
\begin{equation}\label{mox}\langle \Phi_p|\Q|\Phi_q\rangle. \end{equation}
For this matrix element to be nonzero, the fermion number of $|\Phi_p\rangle$ must exceed that of $|\Phi_q\rangle$ by 1.
In other words, the Morse index of $p$ exceeds that of $q$ by 1.
To compute the matrix element, we perform a path integral over trajectories that begin at $q$ at $\tau=-\infty$ and end at $p$ at
$\tau=+\infty$.  Since $\Q$ is a conserved quantity (or since $\Phi_p$ and $\Phi_q$ both have zero energy in the approximation that
is the input to this computation), the time at which $\Q$ is inserted does not matter.

The action for trajectories from $q$ to $p$
is $\Q$-exact modulo an additive constant that depends only on the values of the superpotential at the critical points $p$ and $q$:
\begin{equation}\label{drof} I =\{\Q,V\}+\frac{1}{\lambda}(h(p)-h(q)) \end{equation}
This statement is a supersymmetric extension of eqn. (\ref{durf}).
In eqn. (\ref{drof}),  $V$ is proportional to $1/\lambda$.
By rescaling $V$, one goes to arbitrarily weak coupling while only changing $I$ by $\Q$-exact terms.
So the path integral with insertion of arbitrary $\Q$-exact operators, such as $\Q$ itself, can be computed in the weak coupling limit.

Since $\Q$ is a symmetry of the action and obeys $\Q^2=0$, the path integral with insertion of arbitrary $\Q$-invariant operators -- such
as $\Q$ itself -- localizes on $\Q$-invariant fields.  As we deduced from eqn. (\ref{zolto}), a $\Q$-invariant field is a solution of the
ascending gradient flow equation.  These are the appropriate instantons in our problem,
and the desired matrix element can be computed as a sum of instanton contributions.   Since the Morse indices of $p$ and $q$ differ by 1,
the moduli space $\M$ of solutions of the gradient
flow equation is 1-dimensional and the reduced moduli space $\M_{\mathrm{red}}$ is a finite set of points.  The desired matrix element
can be computed by summing over those points.

There is, however, a subtlety in the computation.  To see why, it helps to restore the loop counting parameter $\lambda$ in the original
action (\ref{murky}).  When we do this, $\Q$, $\bar \Q$, and $H$ are all proportional to $1/\lambda$ if expressed in terms of classical
variables $\uu , \dot \uu ,\psi,$ and $\bar\psi$.  For example
\begin{equation}\label{zurky}\Q=\frac{1}{\lambda}\left(g_{ab}\frac{\d \uu ^b}{\d\tau}-\partial_ah\right)\psi^a. \end{equation}
The supersymmetry algebra $\{\Q,\bar \Q\}=2H$ is obeyed (with no factor of $\lambda$) since the canonical commutators are
proportional to $\lambda$.

In computing a transition from $\Phi_q$ to $\Phi_p$ with an insertion of $\Q$ (or any other $\Q$-invariant operator),
there will be a factor of $\exp(-(h(p)-h(q)/\lambda)$ that comes from the value of the classical action for the instanton trajectory.
We will use the phrase ``reduced matrix element'' to refer to a matrix element for a transition from $\Phi_q$ to $\Phi_p$ with
this elementary factor of $\exp(-(h(p)-h(q))/\lambda$ removed.

Though $\Q$ is of order $1/\lambda$, its  reduced matrix element is of order 1 and comes from a 1-loop computation around the classical
instanton trajectory.  There are two ways to see this.  First, if we try to do a leading order calculation in the instanton field,
we immediately get 0, since the instanton is a solution of the gradient flow equation $g_{ab}\frac{\d \uu ^b}{\d\tau}-\partial_ah=0$,
and $\Q$ is proportional to the left hand side of this equation.  So the reduced matrix
element must come from a 1-loop calculation.  Second, an elegant calculation explained in section 10.5.1 of \cite{Hori:2003ic})
shows directly that the reduced matrix is of order $\lambda^0$.

This is shown not by doing the 1-loop computation
\footnote{It turns out that a direct calculation is actually quite tricky.}
 but by  an interesting shortcut that avoids the need for such a calculation.
Instead of computing the reduced matrix element of $\Q$, we pick any function $f$ that has different values at the critical points $p$ and $q$
and compute the reduced matrix element of the commutator $[\Q,f]$.  (For example, we could take $f=h$, and this is the choice actually
made in \cite{Hori:2003ic}.)
 This contains
the same information, for the following reason.  When we compute the matrix element of an equal time commutator
\begin{equation}\label{toldo} \langle \Phi_p| (\Q f - f \Q) |\Phi_q\rangle,\end{equation}
we can as usual assume that the operator that is inserted on the left is inserted at a slightly greater time than the one that is inserted
on the right.  But as $\Q$ is a conserved quantity, the time at which it is inserted does not matter and we can take this to be much
greater or much less than the time at which $f$ is inserted.  So
\begin{equation}\label{noldox}\langle \Phi_p|[\Q,f]|\Phi_q\rangle= \langle\Phi_p|(\Q(\tau)f(\tau')-f(\tau)\Q(\tau')|\Phi_q\rangle, \end{equation}
where we can take $\tau-\tau'$ to be very large. (Only the difference $\tau-\tau'$ matters, since the initial and final states have zero
energy.)   When we evaluate the right hand side of (\ref{noldox}) by integrating over instanton
moduli space, the instanton must occur at a time close to the time at which $\Q$ is inserted.  This means that $f$ is inserted in the initial
or final state $\Phi_p$ or $\Phi_q$.  To lowest order in $\lambda$, we set $f$ to $f(q)$ or $f(p)$ in the initial or final state.
(In a moment, it will be clear that higher order terms are not relevant.)  So
\begin{equation}\label{oldox} \langle\Phi_p|\Q|\Phi_q\rangle =\frac{1}{f(q)-f(p)}\langle\Phi_p|[\Q,f]|\Phi_q\rangle. \end{equation}
What we have gained from this is that $[\Q,f]=\partial_af\psi^a$ is independent of $\lambda$, so the right hand side of (\ref{oldox})
can be evaluated classically.

The actual calculation is explained in \cite{Hori:2003ic}.  We insert $[\Q,f]$ at, say, $\tau=0$.
The instanton trajectory is $u^a(\tau)=u^a_{\mathrm{cl}}(\tau-v)$, where $v$ is a collective
coordinate over which we must integrate.  The corresponding classical fermion zero-mode is
$\psi_{\mathrm{cl}}^a(\tau)= \partial_v u^a_{\mathrm{cl}}(\tau-v).$ At the classical level in an instanton field, we simply set $[\Q,f](0)=\partial_a f\psi^a|_{\tau=0}$ to
its value with $u^a(\tau)$ set equal to the classical trajectory $u^a_{\mathrm{cl}}(\tau-v)$ and $\psi^a$ set equal to $\psi^a_{\mathrm{cl}}$,
both evaluated at $\tau=0$.  Thus $[\Q,f](0)=\partial_v u^a_{\mathrm{cl}}(-v)\partial_a f(u(-v)) =\partial_v f(u(-v))$.  This must be integrated over $v$
and multiplied by the ratio of fermion and boson determinants.   By time translation symmetry, these determinants do not depend on $v$, so we can
integrate over $v$ first, giving $\int_{-\infty}^\infty \d v\,\partial_v f(u(-v))=  f(q)-f(p) $.  Inserting this in (\ref{oldox}) the factor of $f(q)-f(p)$ cancels,
and we find that the contribution of a given instanton to the reduced matrix element is simply the ratio of fermion and boson determinants.
The boson and fermion determinant are equal up to sign because of a pairing of nonzero eigenvalues by supersymmetry, so their ratio gives a factor of $\pm 1$.
In more detail, the ratio of determinants is
\begin{equation}\label{molo} \frac{\det' (L)}{(\det'( L^\dagger L))^{1/2}} ,\end{equation}
where $\det'$ is a determinant in the space orthogonal to the zero-modes.  The numerator in (\ref{molo}) is real, since $L$ is a real operator,
 but is not necessarily positive.  The denominator is positive since $L^\dagger L$ is non-negative and is positive-definite once the zero-mode is removed.
 The cancellation of numerator and denominator up to sign occurs because for a real operator $L$, $\det'(L)=\det'(L^\dagger)$ so $\det'(L^\dagger L)=\det'(L^\dagger)\det'(L)=(\det'(L))^2$.

Actually, as usual, the path integral on $\R$ is not a number but a transition amplitude between initial and final states.
This is clear in eqn. (\ref{moto}), where reversing the sign of the initial or final state $\Phi_q$ or $\Phi_p$ would certainly reverse
the sign of the matrix element $m_{qp}$.  In the mathematical theory, one says that the regularized fermion path integral
 is not a number but
a section of a real determinant line bundle that can be trivialized by choosing the signs of the initial and final states.  We give an
introduction to this point of view in Appendix \ref{sec:SQMSigns}, but here we give a less technical explanation.

Restoring the factor of $\exp(-(h(p)-h(q))/\lambda)$, $\Q$ acts on the states of approximately zero energy by
\begin{equation}\label{moto}\Q\Phi_q =\sum_{p|n_p=n_q+1}\exp\left(-(h(p)-h(q))/\lambda\right)m_{qp}\Phi_p,\end{equation}
where the sum runs over all critical points $p$ of Morse index $n_q+1$. The matrix element
 $m_{qp}$ vanishes if there are no flows from $q$ to $p$ and
receives a contribution of $+1$ or $-1$
for each ascending flow line from $q$ to
$p$.   We will describe this loosely by saying that $m_{qp}$ is computed
by ``counting'' the instanton trajectories from $q$ to $p$.  We always understand that ``counting'' means ``counting with signs.''

Alternatively, denoting as $M$ a matrix that multiplies $\Phi_q$ by
$\exp(-h(q)/\lambda)$, we find that  $\hat \Q=M^{-1}\Q M$ acts by
\begin{equation}\label{bloto}\hat \Q\Phi_q= \sum_{p|n_p=n_q+1}m_{qp}\Phi_p,\end{equation}
Thus matrix elements of $\hat\Q$ are reduced matrix elements of $\Q$.
Of course, the cohomology of $\Q$ is naturally isomorphic to that of $\hat\Q$, but eqn. (\ref{bloto}) has the advantage of making it manifest
that the cohomology is defined over $\Bbb Z$.  We call the complex with basis $\Phi_q$ and differential $\hat\Q$ the MSW (Morse-Smale-Witten)
complex.

Now let us see what we can say about the sign of the fermion path integral, based only on very general ideas.
It is simplest to assume first that the target space $M$ is simply-connected.  The fermion kinetic operator
$L$ makes sense for expanding around an arbitrary path from $q$ to $p$, not necessarily a classical solution, and likewise it makes sense to discuss
the sign of the measure in the fermion path integral for an arbitrary path.  For a particular path, there is no natural way to pick the sign of the fermion
measure, but it makes sense to ask that the sign of the fermion measure should vary continuously as we vary the path.  For simply-connected $X$,
any two paths from $q$ to $p$ are homotopic, and therefore the sign of the fermion measure for any such path is uniquely determined up to an overall
choice of sign that depends only on the choice of $q$ and $p$ and not on a particular path between them.

From this, it seems that the reduced matrix element $m_{qp}$ in (\ref{bloto}) or (\ref{moto}) is well-defined up to  an overall sign
that is the same for all trajectories from $q$ to $p$, say up to multiplication by
$(-1)^{f(p,q)}$ where $f(p,q)$ equals 0 or 1 for each pair $p,q$.  However, there is one more ingredient to consider and this is cluster decomposition.
Let $p,q$, and $r$ be three critical points.  Consider a path from
 $q$ to $r$ that consists of a path that first travels from $q$ to $p$ and then, after a long
time, travels from $p$ to $r$.  (Such a trajectory is called a broken path and plays a further role that we will explain in section \ref{whyindeed}.)
Cluster decomposition -- that is, the condition that the fermion measure should factorize naturally in this situation -- gives the constraint
$(-1)^{f(p,q)}(-1)^{f(r,p)}=(-1)^{f(r,q)}$.  This implies that $(-1)^{f(p,q)}=(-1)^{a(p)+a(q)}$ for some function $a$ on the set of $\IV$ of critical points.
But a factor of $(-1)^{a(p)+a(q)}$ in the reduced matrix element can be eliminated by multiplying the state $\Phi_s$, for any critical point $s$,
by $(-1)^{a(s)}$.  In short, for simply-connected $M$, the fermion measure in all sectors is uniquely determined up to signs that reflect the
choices of sign in the initial and final state wavefunctions.

In case $M$ is not simply-connected, the principles we have used do not necessarily give a unique answer because in general the answer is not
unique.  The signs of the fermion measure in the different sectors might
 not be uniquely determined (even after allowing for the freedom to change the
external states) because one is free to twist the theory by considering differential forms on $M$ valued in a flat real line bundle, rather than ordinary differential
forms.  This will change the signs of the various transition amplitudes.
(If one wishes to weight the different sectors of the path integral by arbitrary complex phases, as opposed to arbitrary minus signs as assumed
above, one would find that the construction is unique up to the possibility of considering differential forms on $M$ valued in a flat complex line bundle.)

The attentive reader might notice one subtlety that was ignored in the above discussion.  If $M$ is simply-connected, we can interpolate between
any two paths from $q$ to $p$.  However, if $\pi_2(M)\not=0$, there are different homotopy types of such interpolations.  If different ways to interpolate
between one path and another would give different signs for the fermion measure, we would say that the theory has a global anomaly and is inconsistent.
This does not happen in the class of supersymmetric quantum mechanical models considered here (basically because the theory of differential forms
can be defined on any manifold $M$), but it does happen in other classes of supersymmetric quantum mechanical models.

In the simple remarks just made, we have explained, just from general principles, that the path integral for trajectories from $q$ to $p$
is well-defined as a transition amplitude from  $\Phi_q$ to $\Phi_p$.  But we have not given a recipe to compute the overall sign of this
transition amplitude.  The reader interested in this should consult Appendix \ref{sec:SQMSigns}.

\subsection{Some Practice}\label{warmup}

\begin{figure}
 \begin{center}
   \includegraphics[width=3.5in]{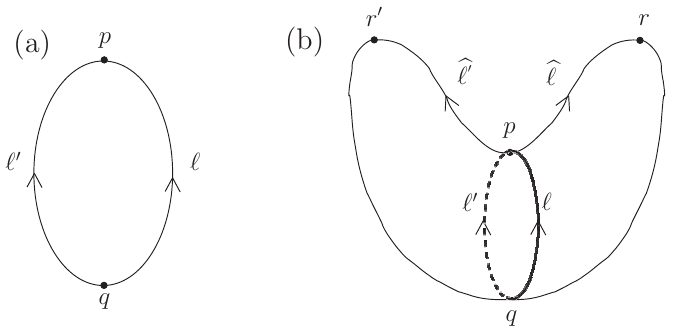}
 \end{center}
\caption{\small Some example of flow lines in Morse theory.  (a) A circle $N=S^1$ is embedded
in $\Bbb R^2$ in such a way that the ``height'' function is a perfect Morse function $h$, with only the minimum possible
set of critical points -- a maximum $p$ and a minimum $q$. (b) A two-sphere $M=S^2$ is embedded in $\Bbb R^3$
in such a way that the height function is {\it not} a perfect Morse function.  It has two local maxima $r$ and $r'$, a saddle
point $p$, and a minimum $q$. The arrows on the flow lines show the directions along which $h$ increases. }
 \label{Morse}
\end{figure}

To help orient the reader and as background for section \ref{whyindeed},
we will here give some simple examples involving flow lines and Morse theory.
  In our first example,
we take the target space of the $\sigma$-model to be
 $N=S^1$ with a Morse function $h$ that has only two critical points -- a maximum $p$ and a minimum $q$.
Accordingly, the supersymmetric quantum mechanics with this target space and Morse function has precisely two
states whose energy vanishes in perturbation theory -- a state $\Phi_q$ that corresponds to a 0-form and a state $\Phi_p$ that corresponds to a 1-form.  Since the cohomology of $S^1$
has rank 2, the states $\Phi_p$ and $\Phi_q$ must survive in the exact quantum theory as supersymmetric states of precisely
zero energy. Since $h$ has the minimum number of critical points needed to reproduce the cohomology of $M$,
it is called a perfect Morse function.
Because $\hat\Q$ increases the fermion number by 1,  its only possibly non-zero matrix element is
\begin{equation} \hat\Q\Phi_q=m_{qp}\Phi_p, \end{equation}
where $m_{qp}$ is a sum over ascending flow lines from $q$ to $p$, weighted by the sign of the fermion determinant.
However, since $\Phi_q$ and $\Phi_p$ must remain at precisely zero energy, we expect $m_{qp}=0$.  Concretely,
as sketched in Figure \ref{Morse}(a), there are two steepest descent or ascent
trajectories from $p$ to $q$, labeled $\ell$ and $\ell'$ in the figure.

Each of them, if properly parametrized by the Euclidean time $\tau$, gives a solution of the equation for ascending gradient
flow from $q$ to $p$.   Each of these trajectories contributes $\pm 1$ to $m_{qp}$.  The expected result $\Q\Phi_q=0$
arises because the two trajectories contribute with opposite signs: whatever orientation we pick at $p$,
the orientation of time translations along one of the two trajectories will agree with it, along the other trajectory
will disagree with it. (Again, see Appendix \ref{sec:SQMSigns} for a technical description of the signs.)

For a slightly more elaborate example, we consider $M=S^2$, but now  with a decidedly non-perfect Morse function
that has two local maxima $r$ and $r'$, a saddle point $p$, and a minimum $q$ (Figure \ref{Morse}(b)).
The MSW complex now has rank 4, generated by the 2-forms $\Phi_r$ and $\Phi_{r'}$, the 1-form $\Phi_p$, and the 0-form $\Phi_q$.
{\it A priori}, the possible matrix elements of $\Q$ are
\begin{align}\label{apr} \hat\Q\Phi_q& = m_{qp}\Phi_p \cr
                                      \hat\Q\Phi_p& =m_{pr}\Phi_r+m_{pr'}\Phi_r' .\end{align}
As shown in Figure \ref{Morse}(b), there are two flow lines $\ell$ and $\ell'$ from $q$ to $p$.  However, they cancel
just as in the previous example.  There is, however, just a single flow line $\hat\ell$ or $\hat\ell'$ from $p$ to $r$ or $r'$,
so no cancellation is possible.  We can pick the signs of the states so that $m_{pr}=m_{pr'}=1$ and then we have
  \begin{align}\label{apro} \hat\Q\Phi_q& = 0 \cr
                                      \hat\Q\Phi_p& =\Phi_r+\Phi_r' .\end{align}
The cohomology of $S^2$ is therefore of rank 2, generated by a 0-form $\Phi_q$ and a 2-form that we can take to
be $\Phi_r$ or $\Phi_{r'}$.

The result (\ref{apro}) makes it clear that in this example $\Q^2=0$.  Considerations of fermion number force $\Q^2$ to annihilate
all the basis vectors except $\Phi_q$, but $\Q^2\Phi_q=0$ since $\Q\Phi_q=0$.

\subsection{Why $\Q^2=0$}\label{whyindeed}

We now want to explain in terms of Morse theory and gradient flow lines why $\Q^2=0$ in general
(See \cite{Hutchings} for a fairly accessible rigorous explanation.)  The explanation gives
important motivation for many of the constructions in the present paper.

Consider in general a target space $M$ with Morse function $h$.  Let $q$ be a critical point of Morse index $n$.
The general form of $\hat\Q\Phi_q$ from eqn. (\ref{bloto}) is
\begin{equation}\label{nox}\hat\Q\Phi_q=\sum_{p_i|n_{p_i} =n+1}m_{qp_i}\Phi_{p_i}, \end{equation}
where the sum runs over critical points $p_i$ of index $n+1$ and $m_{qp_i}$ is the usual sum of contributions $\pm 1$
from ascending flows from $q$ to $p_i$.  And similarly
the general form of $\hat\Q^2\Phi_q$ is
\begin{equation}\label{monox}\hat\Q^2\Phi_q=\sum_{r_\alpha|n_{r_\alpha}=n+2~}\sum_{p_i|n_{p_i} =n+1}
m_{qp_i}m_{p_ir_\alpha}\Phi_{r_\alpha},
\end{equation}
where $r_\alpha$ ranges over the critical points of index $n+2$.  So the statement that $\Q^2\Phi_q=0$ amounts
to the statement that for each $\alpha$,
\begin{equation}\label{wonox} \sum_i m_{qp_i}m_{p_i r_\alpha}=0. \end{equation}

The only possibly dangerous case is the case that there is, for some $i$, an ascending gradient trajectory from $q$ to $p_i$
and also an ascending gradient trajectory from $p_i$ to $r_\alpha$.  (Otherwise, either $m_{qp_i}$ or $m_{p_i r_\alpha}$ vanishes
for all $i$.)  So let us assume this to be the case.  Rather as in Figure \ref{Morse}(b), let $\ell$ be an ascending trajectory from $q$
to $p_i$ and let $\hat\ell$ be an ascending trajectory from $p_i$ to $r_\alpha$.   (We assume that as usual these flows depend on no
moduli except those associated to time translations.)
  We can make an approximate ascending gradient flow
trajectory from $q$ to $r_\alpha$ as follows.
Start at $q$ at $\tau=-\infty$.  After lingering near $q$ until time $\tau_1$, flow from $q$ to $p_i$
along the trajectory $\ell$.  Linger near $p_i$ until some much later time $\tau_2$, and then flow to $r_\alpha$ along the trajectory
$\hat\ell$.  For $\tau_2-\tau_1>>0$, this gives a very good approximate solution of the flow equation, depending on the two parameters
$\tau_2$ and $\tau_1$.

Index theory and the theory of differential equations
can be used to show that these approximate solutions can be corrected to give a family of exact solutions of the flow
equations, also depending on 2 parameters.\footnote{Schematically, if we linearize the problem of correcting the approximate solution $\uu_0$ to an exact solution we find
an inhomogeneous linear problem
\begin{equation}
L \, \delta \uu = - L \uu_0
\end{equation}
The obstruction to inverting $L$ on the space of normalizable fluctuations is controlled by the
kernel of $L^\dagger$, which is generically empty.}

Since $n_{r_\alpha}-n_q=2$ by hypothesis, 2 is the expected dimension of the moduli space $\M_{q r_\alpha}$ of ascending
flows from $q$ to $r_\alpha$.   What we have found is a component $\M^*_{q r_\alpha}$ of $\M_{q r_\alpha}$ that
has  this expected dimension.  ($\M_{q r_\alpha}$
is not necessarily connected; in general, $\M^*_{q r_\alpha}$ is one of its connected components.)

As usual, the  group $\Bbb R$ of time translations acts on $\M^*_{q r_\alpha}$; the quotient is a 1-manifold
$\M^*_{q r_\alpha,\mathrm{red}}$.
For generic metric $g_{ij}$ on $M$ and Morse function $h$, the 2-manifold $\M^*_{q r_\alpha}$ and the 1-manifold
$\M^*_{q r_\alpha,\mathrm{red}}$ are smooth manifolds without boundary.   This follows again from general considerations about
index theory and differential equations.

A smooth 1-manifold without boundary is either a circle or a copy of $\Bbb R$.  However, $\M^*_{q r_\alpha,\mathrm{red}}$ is not compact,
since it has an ``end'' corresponding to $\tau_2-\tau_1\to\infty$.  Therefore, $\M^*_{q r_\alpha,\mathrm{red}}$ must be a copy of $\Bbb R$.
And since $\Bbb{R}$ has two ends, $\M^*_{q r_\alpha,\mathrm{red}}$ must have a second end, in addition to the one that we know about.

The end of $\M^*_{q r_\alpha,\mathrm{red}}$ that we know about is sometimes called a broken trajectory or a broken flow line.
It corresponds to the limit of a gradient flow line from $q$ to $r_\alpha$ that breaks up into a widely separated pair
consisting of a flow  from $q$ to another critical critical point $p_i$ followed by a flow from $p_i$ to $r_\alpha$, where the Morse
index increases by 1 at each step.

It again follows from very general considerations that any end of any component of $\M_{qr_\alpha,\mathrm{red}}$ corresponds to a broken
trajectory from $q$ to $r_\alpha$.    So in addition to the broken trajectory from $q$ to $r_\alpha$ that appears at the end
of  $\M^*_{q r_\alpha,\mathrm{red}}$ that we know about, $\M^*_{q r_\alpha,\mathrm{red}}$ must have a second end that corresponds to a second
broken trajectory from $q$ to $r_\alpha$.

The sum on the left hand side of eqn. (\ref{wonox}) can be regarded as
 a sum over contributions from broken trajectories. The contribution of each
broken trajectory is $\pm 1$.  The mechanism by which
the sum always vanishes is that broken trajectories appear in pairs, corresponding to the two ends of a 1-dimensional reduced
moduli space such as $\M^*_{qr_\alpha,\mathrm{red}}$. Careful attention to signs show that
they do indeed provide canceling contributions. (See Appendix \ref{sec:SQMSigns}.)

To conclude, we should elaborate on the claim that an end of a 1-dimensional reduced moduli space must correspond to a broken
trajectory.  Roughly speaking, an end of a moduli space of solutions of a differential equation corresponds in general
to an ultraviolet effect (something blows up at short distances or times), a large field effect (some fields go to infinity), or an infrared effect
(something happens at large distances or times).   An example of an ultraviolet effect is the shrinking of an instanton to zero
size in four-dimensional gauge theory.  This has no analog in our problem, because at short times the gradient
flow equation reduces to $\d \uu ^a/\d \tau=0$ (the term $g^{ab}\partial_b h$ in the equation is subleading at small times), and the
solutions of this equation do not show any ultraviolet singularity. If $M$ is compact, we  do not have to worry about $\uu ^a$ becoming large;
 if $M$ is not compact (as will actually be the case in our main application),  it is necessary to analyze this possibility.  Finally, as we are
considering a massive theory, the only interesting  effect that is possible at long times is a broken trajectory.
In a massive theory
in Euclidean signature, long times do not come into play unless the trajectory becomes broken.

A final comment is that actually, this subject is one area in which rigorous mathematical theorems
are illuminating for physics.  From a physicist's point of view, perturbation theory gives an approximation
to the space of supersymmetric ground states of supersymmetric quantum mechanics, and the inclusion of instantons
gives a better approximation.  Does inclusion of instantons give the exact answer, or could there be
nonperturbative corrections that near
the classical limit are even smaller than instantons?  One answer to this question is that the rigorous theorems,
described in \cite{Hutchings}, show that inclusion of instantons gives the exact answer for the space of supersymmetric states.

\subsection{Why The Cohomology Does Not Depend On The Superpotential}\label{whynot}

In equation (\ref{hozz}), we showed that the supercharge $\Q$ is conjugate to the de Rham exterior derivative $\d$.
This implies that the cohomology of $\Q$ is canonically isomorphic to the de Rham cohomology, which is defined
without any choice of metric $g$ or superpotential $h$.  Hence, the cohomology of $\Q$ does not depend on the choice of $g$ or $h$.

However, this sort of proof is not available when we get to quantum field theory with spacetime dimension $>1$.
We will give another explanation here that does generalize.  This explanation relies on counting of gradient flow
trajectories. (For a much more precise account,  see Section 4
of \cite{Hutchings}.)

Before beginning the technical explanation, let us explain the physical framework that should make one expect such an explanation to exist.
According to eqn. (\ref{drof}), the Euclidean action of supersymmetric quantum mechanics is $\{\Q,V\}$ plus a surface term, where
\begin{equation}\label{murr}
V=\frac{1}{2\lambda}\int_{-\infty}^\infty \d \tau g_{ab} \bar\psi^a \left(\frac{\d u^b}{\d \tau} - g^{bc}\frac{\partial h}{\partial u^c}\right).  \end{equation}
The definition of $V$ makes sense if $g$ and $h$ have an explicit $\tau$-dependence, rather than being functions of $u^a$ only, and in this
more general situation, we can generalize (\ref{drof}) to the supersymmetric action
\begin{equation}\label{merrier} I=\{Q,V\}+\frac{1}{\lambda}\int_{-\infty}^\infty \d \tau\frac{\partial}{\partial \tau} h(u;\tau). \end{equation}
The fact that it is possible to give $h$ and $g$ an explicit time-dependence while maintaining $\Q$-invariance can be regarded as an explanation of the
technical construction that we will describe in the rest of this section.  This technical construction has numerous analogs that are important
in the rest of this paper, notably in section \ref{notif} where we explain the relation between the Fukaya-Seidel category
and the web-based construction of the abstract part of this paper.

\def\Cr{{\mathcal C}}
\def\Vr{{\mathcal V}}
\def\Ur{{\mathcal U}}
\def\Hr{{\mathcal H}}
Here is the technical explanation.
To compare the cohomology of $\Q$ computed using one metric and superpotential $g$ and $h$ to that computed
using another pair $g',h'$, we proceed as follows.  Let $\Cr$ be the set of critical points of $h$ and $\Cr'$ the set of critical points of $h'$.
In  the classical limit, the system based on $g,h$ has a supersymmetric state $\Phi_p$ for each $p\in \Cr$, furnishing a basis
 of the space $\V$ of approximately supersymmetric states (the MSW complex).
 Similarly, the system based on $g',h'$ has an
approximately supersymmetric state $\Phi_{p'}$ for each $p'\in\Cr'$, furnishing a basis of the analogous space $\Vr'$.  As in eqn. (\ref{bloto}), by counting gradient flow trajectories for the Morse function $h$ with metric $g$, we define a normalized differential
$\h \Q$ acting on $\Vr$, and similarly by counting  trajectories for $g',h'$, we define a normalized differential $\h\Q'$ acting on
$\Vr'$. We write $\Hr$ and $\Hr'$ for the cohomology of $\h\Q$ and $\h\Q'$, respectively.
 We want to define a degree-preserving
linear map $\Ur:\Vr\to \Vr'$ that will establish an isomorphism between $\Hr$ and $\Hr'$.

A degree $d$ linear transformation $\Ur:\Vr\to\Vr'$ will induce a map  $\h\Ur:\Hr\to\Hr'$ if it is a ``chain map,'' meaning that
\begin{equation}\label{suffo} \h\Q'\Ur=(-1)^d \Ur\h\Q.\end{equation}
This ensures that if $\psi\in \Vr$ represents a cohomology class of $\h\Q$, meaning that $\h\Q\psi=0$, then $\Ur\psi$
represents a cohomology class of $\h\Q'$, since $\h\Q'\Ur\psi=\Ur\h\Q\psi=0$.  Moreover the class of $\Ur\psi$ only
depends on the class of $\psi$, since if we replace $\psi$ by $\psi+\h\Q\chi$, then $\Ur\psi$ is replaced by $\Ur\psi
+\Ur\h\Q\chi=\Ur\psi+(-1)^d\h\Q'(\Ur\chi)$.  We define $\h\Ur:\Hr\to\Hr'$ by saying that if a class in $\Hr$ is represented
by a state $\psi$, then $\h\Ur$ maps this state to the class of $\Ur(\psi)$.
To show that such a $\hat\Ur:\Hr\to\Hr'$ is an isomorphism, we will want a map $\Ur':\Vr'\to\Vr$ in the opposite direction, obeying
the analog of (\ref{suffo}) and inducing an inverse on cohomology.
%moreover obeying a pair of relations
%
%\be\label{eq:UUp}
%\begin{split}
%\Ur' \Ur & = 1 + \hat \Q T + T \hat Q \\
%\Ur \Ur' & = 1 + \hat \Q' T' + T' \hat Q' \\
%\end{split}
%\ee
%
%Given \eqref{eq:UUp} we have $\hat\Ur'\hat\Ur=1$.

To construct $\Ur$,
 we first pick an interpolation from the pair $g,h$ to the pair $g'$, $h'$.  We do this
by letting $g$ and $h$ depend on a real-valued ``time'' coordinate $\tau$, such that $(g(\tau),h(\tau))$ approach $(g,h)$ for $\tau\to-\infty$
and approach $(g',h')$ for $\tau\to+\infty$.  We can assume that $g(\tau)$ and $h(\tau)$ are nearly (or even exactly) independent of $\tau$
except near some time $\tau_0$.  Now we consider the gradient flow equation with a time-dependent metric
and superpotential:
\begin{equation}\label{luffo} \frac{\d u^a}{\d\tau}= g^{ab}(u;\tau)\frac{\partial h(u;\tau)} {\partial u^b}.  \end{equation}
The boundary conditions are that $u(\tau)\to p\in \Cr$ for $\tau\to -\infty$ and $u(\tau)\to p'\in \Cr'$ for
$\tau\to +\infty$.
The index determining the expected dimension of the space of solutions of this equation is simply the difference
between the Morse indices $n_p$ and $n_{p'}$  of $p$ and $p'$.  (We know that this is the answer
if $g=g'$ and $h=h'$, and the index does not change
under continuous evolution of $g', h', $ and $p'$.)  So the index is 0 if $p$ and $p'$ have the same Morse index.
For a generic interpolation, the moduli spaces have dimensions equal to
 their expected dimensions; we consider an interpolation that is generic
in this sense.
So the index zero condition means that the moduli space consists of finitely many points.  (Since $g$ and $h$ are time-dependent,
there is no time translation symmetry to force the existence of a modulus.)
These points correspond to solutions in which $u^a(\tau)$ is almost constant except
near some time $\tau=\tau_0$.
We let $u_{pp'}$ be the ``number'' of such solutions (as usual weighting each solution with the sign of the fermion determinant),
and define $\Ur:\Vr\to\Vr'$ by
\begin{equation}\label{fuffo}\Ur\Phi_p=\sum_{p'|n_{p'}=n_p}u_{pp'}\Phi_{p'}.\end{equation}
The condition that $n_{p'}=n_p$ means that $\Ur$ preserves the grading of $\Vr$ and $\Vr'$.
We will need to know that $\Ur$ defined by \eqref{fuffo} is nonzero, and in fact satisfies
\eqref{suffo}. This will be justified at the end of this section.

The left and right hand sides of the equation  $\h\Q'\Ur=(-1)^d\Ur\h\Q$ both increase the degree (or grading) by 1.
So to prove this identity,
we have to look at moduli spaces of time-dependent gradient flows with $n_{p'}=n_p+1$.   The reasoning we need is very similar to the proof
that $\h\Q^2=0$ in section \ref{whyindeed}.  Let $\M$ be a component of the moduli space of solutions of the time-dependent
gradient flow equation, interpolating from $p$ in the past to $p'$ in the future.  $\M$ is a one-manifold without boundary,
and so is either a copy of $S^1$, with no ends at all, or a copy of $\IR$, with two ends.  In the following, only the case that $\M$
is a copy of $\IR$ is relevant.

Just as in section \ref{whyindeed}, an end of $\M$ is a broken path, in which a solution breaks up into two pieces localized
at widely different times.  The building blocks of a broken path in the present context are of the following types:

{\it (A)} One ingredient
is familiar from section \ref{whyindeed}.  In the far past or the far future, where the equation (\ref{luffo}) has no explicit
time-dependence, we may have a solution of the time-independent gradient flow equation that interpolates between two
critical points of $h$ or two critical points of $h'$ with Morse index differing by 1.  Let us say that such a solution is of type
{\it (A$_-$)} if the transition occurs in the past and of type {\it (A$_+$)} if it occurs in the future.

{\it (B)}  The new ingredient in the present context  is a solution that interpolates from
a critical point of $h$ to a critical point of $h'$, near the time $\tau_0$.

Ends of $\M$ correspond to broken paths of two possible types:

{\it (i)}  One type consists of a trajectory of type {\it (A$_-$)} in the far past, interpolating between two critical points of $h$,
followed by a trajectory of type {\it (B)} at $\tau\cong\tau_0$, interpolating from a critical point of $h$ to one of $h'$.

{\it (ii)}  The other type consists of a trajectory of type {\it (B)} at $\tau\cong\tau_0$, interpolating  from a critical point of $h$
to one of $h'$, followed by a trajectory of type {\it (A$_+$)} interpolating to another critical point of $h'$.

Conversely, every broken path of either of these types arises at one end of one component of $\M$, and this component
has a second end that is either of the same type or of opposite type.  This is  true for reasons similar to what we explained
in showing that $\hat\Q^2=0$.

A broken path of type {\it (i)} contributes $\pm 1$ (depending on the sign of the fermion determinant) to a matrix element of
$\Ur\h\Q$,  and a broken path of type {\it (ii)} contributes $\pm 1$ to the  matrix element of $\h\Q'\U$ between the same initial
and final states.
If $\M$ has two ends that are both of type {\it (i)} or both of type {\it (ii)}, then the   corresponding contributions to $\Ur\h\Q$ or
to $\h\Q' \Ur$ cancel.  On the other hand, if $\M$ has one end of each type, then these ends make equal contributions to
$\Ur\h\Q$ and to $\h\Q'\Ur$. Both statements follow from the observation that the sign of a contribution
is equivalent to a choice of orientation of the corresponding component of $\M$, and all contributions are oriented canonically towards the future.
After summing these statements over all components of $\M$, we arrive at the desired identity (\ref{suffo}).

\begin{figure}
 \begin{center}
   \includegraphics[width=4.4in]{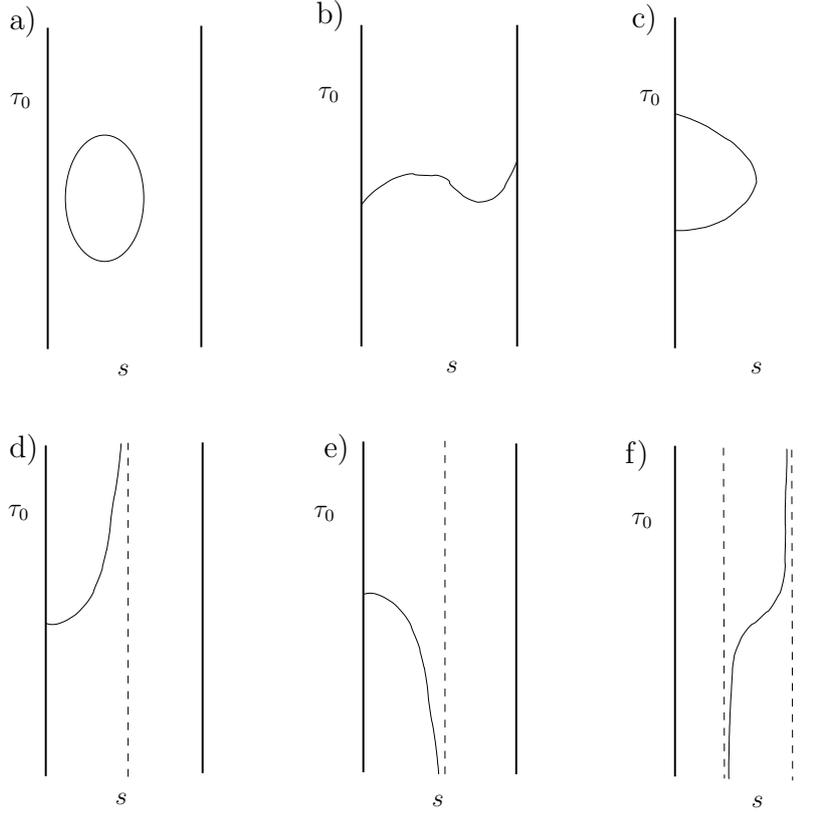}
 \end{center}
\caption{\small   Some possibilities for a component of the moduli space $\M$ of solutions
to (\protect\ref{luffo-p}) that exist for \emph{some} $s$ in the case where $n_p=n_{p'}$.
(Some additional possibilities are omitted).
The picture is only schematic: while the value of $s$
at which a solution exists is precisely defined, the corresponding value of $\tau_0$ which characterizes
the solution is not really well-defined. What is well-defined
is only whether a sequence of solutions goes to $\tau_0=\pm\infty$. A component of
 $\M$ might be compact and without boundary, as in (a).
Otherwise, it has two boundaries and/or ends.  Each boundary or end contributes to one of the four terms in
equation (\protect\ref{mizzo}).
%
%eqn. (\ref{mizzo})
%
 and
the two boundaries and/or ends of $\M$ make compensating contributions to this identity.   In (d), (e), and (f), the vertical dotted
lines represent values of $s$ at which there is an exceptional gradient flow solution contributing to the matrix $\EE$.  These are flows that reduce the Morse index by 1.  }
 \label{sompos}
\end{figure}

One question about this is whether the map $\Ur:\Vr\to\Vr'$ (and hence possibly the induced map $\hat\U$ on cohomology) depends
on the specific choice of a generic time-dependent interpolation from $g,h$ to $g',h'$ (we call an interpolation generic if the
moduli spaces have their expected dimensions). In general,  the counting (with signs) of the solutions
of an elliptic differential equation is invariant under continuous variations of the parameters in the equation, as long as solutions
cannot go to infinity. If we could assume in the present context that solutions cannot go to infinity, then the numbers $u_{pp'}$ would
be independent of the choice of a generic interpolation and $\Ur$ would
likewise not depend on the interpolation.

This is actually not so in general.  If $\Ur_0$ and $\Ur_1$ are the maps determined by two different generic interpolations, then in general
the relationship between them is not $\Ur_0=\Ur_1$ but rather
\be\label{mizzo}\Ur_1-\Ur_0 =\hat \Q'\EE-\EE\hat \Q,\ee
where $\EE$ is a linear transformation $\EE:\Vr\to\Vr'$ that reduces the degree by 1.
This is enough to ensure that the induced maps
on cohomology are equal, $\h\Ur_1=\h\Ur_0$.  The correction terms on the right hand side of eqn. (\ref{mizzo}) should be expected, for
the following reason.  The action of our system is $\Q$-exact up to a surface term (eqn. (\ref{merrier})), and when we change the interpolation
from $g,h$ to $g',h'$ (without changing $g$ or $h$ at $\tau=\pm\infty$) we change the action by a $\Q$-exact term.  After integration by parts,
this results in contributions in which $\Q$ acts on initial and final states, and the transition amplitude changes by $\h\Q'\EE-\EE\h\Q$ for some $\EE$
(as usual the shift from $\Q$ to $\h\Q$ and $\h\Q'$ results from absorbing the surface terms in the action in the normalization of the initial and final states).

Technically, $\EE$ can be found as follows. Given two generic  interpolations from $g,h$ to $g',h'$, we first select an interpolation between the two interpolations.
This means that we choose a metric $g(u;\tau,s)$ and superpotential $h(u;\tau,s)$ that depends not only on the time but on another
parameter $s$, with $0\leq s\leq 1$, such that the restriction to $s=0$ or to $s=1$ gives the two interpolations that we want to compare.
Now we look for solutions of the gradient flow equation in $\tau$ at a fixed value of $s$
\begin{equation}\label{luffo-p} \frac{\d u^a}{\d\tau}= g^{ab}(u;\tau,s)\frac{\partial h(u;\tau,s)} {\partial u^b} \end{equation}
flowing from a critical point $p$ in the past to a critical point $p'$ in the future in such a way that the Morse index is reduced by 1: $n_{p'}=n_p-1$.
For fixed $s$, the expected dimension of the moduli space is $-1$, meaning that for generic $s$ there are no solutions (and there are in fact no
solutions at $s=0$ or $s=1$ since we have assumed $g(u;\tau,s)$ and $h(u;\tau,s)$ to be generic at $s=0,1$).  However, by allowing
$s$ to vary  -- or in other words including $s$ as an additional variable -- we
increase the expected dimension by 1.  The expected dimension of the moduli space of  solutions flowing from $p$ to $p'$
{\it at some unspecified value of $s$}  is 0, and we define an integer $e_{pp'}$ as the ``number'' of such solutions,
weighted by the sign of an appropriate fermion determinant. (We address the sign issues here briefly in
Appendix \ref{sec:SQMSigns}.)

The matrix $\EE$ is defined by
\begin{equation}\label{zelf} \EE\Phi_p=\sum_{p'|n_{p'}=n_{p}-1 }e_{pp'}\Phi_{p'}. \end{equation}
Let also $\Ur_0:\Vr\to \Vr'$ and $\Ur_1:\Vr\to \Vr'$ be the maps defined via eqn. (\ref{luffo-p}) at $s=0$ and $s=1$. We claim that $\Ur_0$, $\Ur_1$,
and $\EE$ satisfy (\ref{mizzo}).

As usual, to justify the claim we analyze the moduli spaces of solutions of the gradient flow equation.
We consider a matrix element of eqn. (\ref{mizzo}) from $\Phi_p$ to $\Phi_{p'}$, where $n_p=n_{p'}$.  The moduli space of gradient flows from $p$ to $p'$ at some unspecified value of $s$
is 1-dimensional.  Some illustrative possibilities for what a component $\M$ of this moduli space might look like are indicated in Figure \ref{sompos}.
As usual (Figure \ref{sompos}(a)), $\M$ might be compact and without boundary, but such a component does not contribute to the discussion.
If $\M$ is not of this type, then it has two boundaries or ends that may be either at $s=0$, $s=1$, $\tau=-\infty$, or $\tau=+\infty$.
Boundaries or ends of $\M$ of the four possible types contribute to the four terms in eqn. (\ref{mizzo}), and the two ends of $\M$ make
canceling contributions in this identity.

 We already know that an endpoint of $\M$
at $s=0$ or $s=1$ contributes to $\Ur_0$ or $\Ur_1$.  What remains is to explain why $\M$ can have an end at $\tau_0=\pm \infty$ and why
such ends correspond to matrix elements of $\h\Q'\EE$ or $\EE\h\Q$.  For example, suppose that $\h\Q'\EE$ has a matrix element from $\Phi_p$
to $\Phi_{p'}$, where $n_{p'}=n_p$.  This means that $\EE$ has a matrix element from $\Phi_{p}$ to $\Phi_{q'}$, where $q'$ is a critical point of $h'$ with $n_{q'}=n_p-1$,
and $\h\Q'$ has a matrix element from $\Phi_{q'}$ to $\Phi_{p'}$.  The matrix element of $\EE$ comes from a gradient flow from $p$ to $q'$ (for the
time-dependent superpotential  $h(u;\tau,s)$) that exists
at some value $s=s_0$ (this solution is localized near some time $\tau=\tau_0$).
The matrix element of $\h\Q'$ comes from a flow from $q'$ to $p'$
(for the time-independent superpotential $h'$)
that exists at generic $s$ (and any $\tau$).  We can try to convert
the ``broken path'' $p\to q'\to p'$ into an exact gradient flow that interpolates from $p$ to $q'$ near time $\tau_0$ and then from $q'$ to $p'$
at some much later time $\tau_1$.  For very large $\tau_1-\tau_0$, we can certainly make a very good approximate solution like this.  However,
in contrast to examples that were considered before, general considerations of index theory do {\it not} predict that this approximate solution
can be corrected to an exact solution at the same value of $s$.  The reason is that the initial flow from $p$ to $q'$
has virtual dimension $-1$, meaning that the linearization of the gradient flow equation near this trajectory is not surjective; the gradient flow equation that this soluton satisfies has one more equation than unknown
and a generic perturbation of the equation causes the solution not to exist.  A generic perturbation can be made by either changing $s$ or including
the second flow near time $\tau_1>>\tau_0$.   However, the fact that the index is $-1$ means that the space of potential obstructions to deforming
a solution is 1-dimensional, so we can compensate for existence of the second flow at very large $\tau_1$ by perturbing $s$ slightly away from $s_0$.
For $\tau_1\to\infty$ (so that the perturbation by the second flow goes to zero), we must take $s\to s_0$ (so that the perturbation by $s$ goes
to $0$). This is why a matrix element of
$\h\Q'\EE$ or $\EE\h\Q$ corresponds to an infinite end of the moduli space, as indicated in Figure \ref{sompos}(d,e,f).

Now that we know that the map induced on cohomology by a generic interpolation from $g,h$ to $g',h'$ does not depend on
the interpolation, it is straightforward to show that this map is invertible.
Just as before, we pick a time-dependent interpolation from $g',h'$ back to $g,h$ and use the counting of trajectories
to define a map $\Ur':\Vr'\to\Vr$ in the opposite direction.  We can compute the product map $\Ur'\Ur:\Vr\to\Vr$ by
considering an interpolation from $g,h$  to itself in which we first interpolate from $g,h$ to $g',h'$ near some time $\tau_0$
and then interpolate back to $g,h$ near some much later time $\tau_1$.    Thus the product $\Ur'\Ur$ is computed
by counting the trajectories for some interpolation from $g,h$ to itself.  As we have just explained, the map $\Ur'\Ur$ will
in general depend on the interpolation, but the induced map $\h\Ur'\h\Ur$ on cohomology will not.
So we can compute it for the trivial, time-independent interpolation from $g,h$ to itself.  The map on cohomology associated
to the trivial interpolation is certainly the identity, so it follows that in general $\h\Ur'\h\U=1$.

\section{Landau-Ginzburg Theory As Supersymmetric Quantum Mechanics}\label{lgassuper}

Now we turn to our real topic -- massive theories in two dimensions.
Our purpose is to give a concrete realization of the abstract algebraic structures described in
Sections \S \ref{sec:Webs} - \S \ref{sec:GeneralParameter} in the context of massive
Landau-Ginzburg theories.

Our analysis will proceed roughly in the opposite order as in the abstract context.
As discussed in the introduction \ref{sec:Introduction}, our starting point is the complex of ground states
for a two-dimensional theory compactified on a strip with supersymmetric boundary conditions.
In a limit where the segment is made very long, we expect to be able to reconstruct this complex in terms of
a web representation and interior and boundary amplitudes which encode properties of the same theory on the plane
and on the left and right half-planes. The advantage of working with LG theories is that we can formulate our questions in the
language of Supersymmetric Quantum Mechanics and Morse theory. As a result, the complex of ground states on the strip,
the web representation data, the interior and boundary amplitudes will all be defined in terms of counting problems
for solutions of certain differential equations on the strip, plane and half planes.

In this section we review the basic data required to define a massive LG theory, some supersymmetric boundary conditions,
the relation to Supersymmetric Quantum Mechanics and Morse theory and the corresponding differential equations.
We also explain the relation between the supersymmetric boundary conditions discussed in this paper and the Fukaya-Seidel
category or more precisely the Fukaya category of the superpotential.

\subsection{Landau-Ginzburg Theory}\label{subsec:LG-Theory}

Let us recall the basic data needed to formulate a $1+1$ dimensional
LG theory with $\CN =(2,2)$ supersymmetry. We
require a K\"ahler manifold $X$, together with a
holomorphic superpotential  $W: X \to \IC$.  $X$ has a K\"ahler form $\omega$, making
it a symplectic manifold,
and a corresponding K\"ahler metric.
A vacuum state corresponds to a critical point of $W$, and at such a critical
point, the fermion mass matrix is the matrix of second derivatives of $W$, also
called the Hessian matrix.
So an LG theory is massive precisely if the Hessian matrix is nondegenerate at every critical
point of $W$.  In this case, we say that $W$ is a Morse function in the holomorphic sense.
(Using the Cauchy-Riemann equations, this is equivalent to the condition that any nontrivial real
linear combination of $\mathrm{Re}\,W$ and $\mathrm{Im}\,W$ is a Morse function in the ordinary real sense.)
Since we assume that the theory is massive in every vacuum, and in addition the soliton states that interpolate
between different vacua will also be massive, there is a characteristic length scale
$\ell_W$  beyond which
the theory should always be, in some sense, close to a vacuum
configuration.    As usual, we write $\Bbb V$ for the set of vacua or equivalently the set of critical points of $W$.

There are many familiar ways to formulate the standard $1+1$-dimensional Landau-Ginzburg
 model associated to the above data.
A slightly less familiar approach will
be convenient for us. We will formulate the LG model
as a special case of the supersymmetric quantum mechanics construction of section \ref{review},
but now with an infinite-dimensional target space.\footnote{The ability to do this uses $(2,2)$ supersymmetry in
an essential way.  A two-dimensional $\sigma$-model with $(1,1)$ supersymmetry actually cannot be viewed as an infinite-dimensional
version of the construction reviewed in section \ref{review}.  For example, such a $\sigma$-model in general does not have an additively conserved
fermion number.}
This construction will make manifest not all four supersymmetries of the $(2,2)$ model, but only a subalgebra
consisting of two supercharges whose anticommutators generate time translations but not spatial translations.
Such a subalgebra is not uniquely determined.  It will be important to have an unbroken $U(1)_R$ symmetry that
acts on this superalgebra.     This will
be an axial $U(1)$ charge normalized so that the negative-chirality supercharge
$Q_-$ and the positive chirality supercharge   $\bar Q_+$ both have
axial $U(1)_R$-charge $+1$ while the positive-chirality supercharge
  $Q_+$ and the negative-chirality supercharge  $\bar Q_-$  have axial $U(1)_R$-charge $-1$.
We will refer to this axial $U(1)_R$ charge as ``fermion number.''

As already explained in section \ref{sec:Introduction}, a subalgebra satisying these conditions depends on the choice
of a complex number $\zeta$ of modulus 1.   The subalgebra that we make
manifest via the quantum mechanical construction is generated by
\be \label{manifest} \CQ_{\zeta}:=Q_- -\zeta^{-1}\bar Q_+,~~\bar{\CQ}_{\zeta} :=\bar Q_- -\zeta   Q_+. \ee
The nonzero anticommutators are
\be \label{anifest}\{\CQ_{\zeta},\bar\CQ_{\zeta}\}=2\CH -2\mathrm{Re}(\zeta^{-1}Z), \end{equation}
where $\CH$ is the Hamiltonian of the quantum field theory and $Z$ is the central charge.
We will call this a small subalgebra of the supersymmetry algebra.

To present the two-dimensional LG model as an abstract quantum mechanical model,
we formulate it on a two-manifold of the form $\Bbb{R}\times D$,
where $\Bbb R$ is parametrized by the ``time,'' and $D$ is a 1-manifold that represents ``space.''
It could be $D=\IR$, or
the half-lines $D=[x_\ell, \infty)$ or $D=(-\infty, x_r]$
or the interval $D=[x_\ell, x_r]$.

The target space of the supersymmetric quantum mechanics model is going to be the space
of all $X$-valued fields on $D$, or in other words
\be
\CX = {\rm Maps}(D \to X).
\ee
$\CX$ inherits a natural metric from the K\"ahler metric of $X$:
\be
\d \ell^2 = \half\left( g_{I\bar J}\d \phi^I\otimes \d\bar\phi{}^{\bar J} + {\rm complex~~conjugate} \right)
\ee
where $\phi^I$ are local holomorphic coordinates on $X$. Thus $\CX$ has metric:
\be |\delta\phi|^2=\half \int_D \d x \left( g_{I\bar J}\delta\phi^I\delta\bar\phi{}^{\bar J}+ {\rm complex~~conjugate} \right). \ee

Our motivating example is $X=\IC^n$ with a flat K\"ahler metric
\be\label{flatk}
\half \sum_K \left( \d\phi^K \otimes \d\bar\phi^{\bar K} +\d\bar\phi^{\bar K} \otimes  \d\phi^K\right).
\ee
In this case, the K\"ahler form $\omega=\frac{\I}{2} \sum_K\d\phi^K\wedge \d\bar{\phi}{}^{\bar K}$
is exact, $\omega=\d\lambda$ with $\lambda=\mathrm{Re}(\frac{\I}{2} \sum_K \phi^K\d\bar {\phi}{}^{\bar K})$.
The following construction applies whenever $\omega$ is exact.
When this is not the case, some slight modifications are needed since
the superpotential $h$ that we introduce momentarily is not single-valued.
In this case one should replace  $\CX$ by a suitable cover on which $h$ is
single-valued.

To put the $\sigma$-model in the supersymmetric quantum mechanics framework of section \ref{review},
all we need is to define the superpotential $h$.  We take this to be
\be\label{defh}
h := -\half \int_D\d x \,\Re\left({\I} \sum_I\phi^I\frac{\partial}{\partial x} \bar{\phi}{}^{\bar I}
-  \zeta^{-1} W \right).
\ee
for the case that $X=\IC^n$ with
$\omega=\frac{\I}{2}\sum_I\d\phi^I\wedge\d\bar \phi{}^{\bar I}$.  In general, one replaces $\Re(\frac{\I}{2}\sum\phi^I\d \bar\phi^{\bar I})$ with any
1-form $\lambda$ such that $\omega=\d\lambda$.  Parametrizing $X$ by an arbitrary set of real coordinates $u^a$, we write $\lambda=\lambda_a\d u^a$ and then
\be\label{defho}
h := - \int_D\d x  \left(  \lambda_a \frac{\d u^a}{\d x} - \half \Re\left(   \zeta^{-1} W \right)\right).
\ee
(If one transforms $\lambda$ to $\lambda+\d \alpha$ for some function $\alpha$ on $X$, $h$ is modified by boundary terms that we will discuss in
section \ref{boundary}.)

The construction reviewed in section \ref{review}, applied to any Riemannian manifold
(in this case $\CX$), with any superpotential
(in this case $h$), gives a quantum mechanical model with two supercharges $\CQ_{\zeta}$ and $\bar \CQ_{\zeta}$ of fermion number $\FF=1$ and $-1$,
respectively.  The supersymmetry algebra
of the quantum mechanical model includes time translations, but of course it does not include spatial translations,
which are not defined in the general quantum mechanical framework.  From eqn. (\ref{olt}), the kinetic energy of the quantum mechanical
model is $T=\frac{1}{2}g_{ab}\dot u^a\dot u^b$, which in the present context becomes
\be T=\int_D \d x \half g_{I\bar J}\frac{\d\phi^I}{\d t}\frac{\d\bar\phi{}^{\bar J}}{\d t}. \ee
The potential energy of the quantum mechanical system, again from (\ref{olt}), is $V=\frac{1}{2}g^{ab}\partial_a h\partial_b h$.
In the present case, this becomes
\be\label{pep}V=\half \int_D \d x \left|\frac{\d\phi^I}{\d  x}-\frac{\I \zeta}{2}g^{I\bar J}\frac{\partial \bar W}
{\partial \bar\phi{}^{\bar J}}\right|^2 , \end{equation}
 or,  after integration by parts,
\be\label{pe}V=\int_D \d x \half \left(g_{I\bar J}\frac{\d\phi^I}{\d  x}\frac{\d\bar\phi{}^{\bar J}}{\d  x}+\frac{1}{4}g^{I\bar J}\frac{\partial W}{\partial \phi^I}\frac{\partial \bar W}
{\partial \bar\phi{}^{\bar J}}\right) + \half
\biggl[\,\mathrm{Im}(\zeta^{-1}W)\biggr]_{\partial_\ell D}^{\partial_rD}~. \end{equation}
Here $\partial_r D$ and $\partial_\ell D$ are the left and right boundaries of $D$, which for the moment we assume to be at $x\to \pm\infty$.
(In case $D$ has boundaries at finite points $x_\ell$ and/or $x_r$, the same formula holds after imposing some further conditions
 that we discuss in section
\ref{boundary}.)

Still assuming that $D=\IR$, and assuming a reasonable behavior at infinity  as discussed in section \ref{boundary}),
the boundary terms
in (\ref{pe}) are just  constants that depend on the boundary conditions. These constants are responsible for the central charge term
in the small supersymmetry algebra (\ref{anifest}).   Apart from this constant term,
the potential energy $V$ of the $\sigma$-model is independent of $\zeta$.  Moreover, the sum $T+V$
is simply the bosonic part of the Hamiltonian of the standard LG model with superpotential $W$.  When one adds in the fermionic
terms in the Hamiltonian of the quantum mechanical model, one simply gets the full supersymmetric LG Hamiltonian.

What we have achieved via this construction of  the LG model  is to make manifest an arbitrary $\zeta$-dependent small subalgebra of the
supersymmetry algebra.  This is useful because we are primarily interested in branes and supersymmetric states
that are invariant under such a small subalgebra but not under the full $\N=2$ supersymmetry algebra.

As an immediate application, let us discuss the states that are annihilated by  $\CQ_{\zeta}$ and $\bar\CQ_{\zeta}$.
From the general quantum mechanical discussion of section \ref{review}, we know that such
states\footnote{States  annihilated by the full supersymmetry algebra with four supercharges, as opposed to the small subalgebra
generated by $\CQ_{\zeta}$ and $\bar\CQ_{\zeta}$, are the supersymmetric
vacua of the theory and correspond, of course, to critical points of $W$.}  correspond in the classical limit to critical points of $h$.
A simple computation shows that stationary  points of $h$ must satisfy
\be\label{eq:LG-flow}
 \frac{\d}{\d x} \phi^I = g^{I\bar J}\frac{\I \zeta}{2}
\frac{\p \bar W}{\p \bar \phi^{\bar J}}
\ee
We call this equation the \emph{$\zeta$-soliton equation.}  Not coincidentally, the potential energy is written in eqn. (\ref{pep}) as the integral of the square
of the left hand side of this equation.  For $D=\Bbb R$, with the fields  required
to approach specified vacua $i,j,\in \Bbb V$ at the two ends of $D$, a solution of this equation gives the classical approximation to an $ij$
BPS soliton \cite{Cecotti:1992rm}.

Another view of the $\zeta$-soliton equation is as follows.  We can think of $h$, as defined in eqn. (\ref{defho}), as the action of a classical
mechanical system in which the symplectic form is $\omega=\d\lambda$ and the Hamiltonian is
$H=-\half \mathrm{Re}(\zeta^{-1}W)$.
\footnote{ This $H$ is distinct from the Hamiltonian $\CH$ in equation
\eqref{anifest}. Note too that one often splits
the real coordinates $u^a$ into canonically conjugate pairs $p_i$ and $q^i$, with $\omega=\sum_i \d q^i \wedge \d p_i$ and $\lambda=-\sum_i p_i\d q^i$.
This might make $h=  \int (p_i\d q^i - H\d x)$ look more familiar.}
So the $\zeta$-soliton equation is a Hamiltonian flow equation
\begin{equation}\label{turmi} \omega_{ab}\frac{\d u^b}{\d x}+\frac{\partial H}{\partial u^a}=0. \end{equation}
Since the Hamiltonian is a conserved quantity in a Hamiltonian flow, an immediate consequence is that $H=-\half\mathrm{Re}(\zeta^{-1}W)$ is independent of
$x$ in a solution of the $\zeta$-soliton equation.
On a K\"ahler manifold, Hamiltonian flow for a Hamiltonian that is the real part of a  holomorphic function is the same as gradient flow with respect to the
imaginary part of the same holomorphic function.
So it is also possible to write the $\zeta$-soliton equation as a gradient flow equation:
\be
\frac{\d u^a}{\d  x} = g^{ab} \frac{\p}{\p u^b}\Im\biggl(\half \zeta^{-1} W\biggr).\label{turmix}\ee
(Concretely, the equivalence of these two forms of the $\zeta$-soliton equation is proved using the Cauchy-Riemann equation for the
holomorphic function $\zeta^{-1}W$.)
Hence $\Im(\zeta^{-1}W)$ is an increasing function of $x$ for any flow.  Combining these statements, it follows, for example, that an $ij$ soliton (interpolating from $\phi_i$ at $\tau=-\infty$ to $\phi_j$ at $\tau=+\infty$) can only exist for a particular value of $\zeta$:
\be\label{hopeful}
\I \zeta =
\I \zeta_{ji} := \frac{W_j - W_i}{\vert W_j - W_i \vert}
\ee
We recall that the central charge in this sector is $Z_{ji}=W_j-W_i$.

 In general, a solution of the $\zeta$-soliton equation gives only a classical approximation to a quantum BPS
state in the $ij$ sector.  To get the exact spectrum of quantum BPS states, one needs to modify the classical approximation
by instanton corrections.
We defer the details to section \ref{subsec:MorseComplexRealLine}, and for now merely remark that the
framework to compute instanton corrections is simply the standard framework for instanton corrections in supersymmetric
quantum mechanics, as described in section \ref{review}.  Thus, one needs to count (with signs) the solutions of the instanton
equation of the supersymmetric quantum mechanics.
The general instanton equation (\ref{olto})
of supersymmetric quantum mechanics, specialized to the case that the target space is K\"ahler, is
\be
\frac{\d \phi^I}{\d\tau} = 2g^{I\bar J}\frac{\delta h}{\delta \bar \phi{}^{\bar J}}.
\ee
For the case that the target space is $\CX$ and with our choice of $h$, this becomes
\be\label{eq:LG-INST}
\left(\frac{\p  }{\p x} +\I  \frac{\p }{\p \tau} \right)\phi^I = \frac{\I \zeta}{2}g^{I\bar J} \frac{\p \bar W}{\p \bar\phi{}^{\bar J}},
\ee
or alternatively
\be\label{eq:instanton}
\frac{\p \phi^I}{\p \bar s} =\frac{\I \zeta}{4}g^{I\bar J} \frac{\p \bar W}{\p \bar\phi{}^{\bar J}},
\ee
where $s= x + \I \tau$. We call this  the \emph{$\zeta$-instanton equation}, and we call its solutions $\zeta$-instantons.

\bigskip
\noindent
\textbf{Remark}: It is sometimes useful to have a clear idea about how the discrete spacetime
symmetries $P$ and $PT$ are implemented in Landau-Ginzburg theory. $PT$ corresponds to a rotation
by $\pi$ in Euclidean space and is therefore a symmetry of the theory. Under $PT$
the bosonic fields transform as $\phi(x,\tau)  \to \phi(-x,-\tau)$, and the
fermion fields transform as $\psi_\pm \to  \mp  i   \psi_\pm $. This is not
a symmetry of the $\zeta$-instanton equation but rather transforms it by $\zeta \to - \zeta$.
Parity, on the other hand, is in general not a symmetry of the Landau-Ginzburg theory.
Under parity we must have $\phi(x,\tau)  \to \bar\phi(-x, \tau)$, while the fermionic
fields transform as $\psi_\pm \to e^{\pm \I \alpha} \bar \psi_{\mp}$.  In general,
this transformation will map one Landau-Ginzburg model to another. A sufficient
criterion for parity invariance is $W(\bar\phi) =  (W(\phi))^*$.  In particular,
if $W$ is a polynomial it should have real coefficients.

\subsection{Boundary Conditions}\label{boundary}

\subsubsection{Generalities}\label{generalities}

The nature of the boundary conditions we impose
on \eqref{eq:LG-flow} depends on the domain $D$.

At an infinite end of $D$, to keep the energy of the LG model finite,
the fields must approach one of the critical points $\phi_i$, $i\in\IV$.
So if $D$ extends to $-\infty$,  then we require
\be\label{eq:left-infty-bc}
\lim_{x\to -\infty} \phi = \phi_i
\ee
where $\phi_i$ is some critical point of $W$. Similarly, if $D$
extends to $+\infty$ then we require
\be\label{eq:right-infty-bc}
 \lim_{x\to + \infty } \phi= \phi_j
\ee
where again $\phi_j$ is a critical point of $W$.\footnote{The integral (\ref{defh}) defining
 $h$ is infinite if $\Re(\zeta^{-1}W(\phi))$
is nonzero  at an infinite end of $D$.  However, in each  sector defined by the choices of critical points at infinity,
$h$ can be naturally defined up to an overall constant;
the variation of $h$ (under a local variation of
$\phi( x)$) and more generally the differences in the values of $h$ for different fields in the same  sector
are finite and well-defined.
Actually,  for $D=\Bbb R$, in a sector that contains BPS solitons, we can do better.  In such a sector,  $\Re(\zeta^{-1}W(\phi))$ has the same
value for $\phi=\phi_i$ or $\phi_j$, and the problem in defining $h$
can be eliminated by subtracting a constant from $W$ to ensure that this value is 0. If $D$ has only one infinite end, one can do the
same for each choice of vacuum at infinity.}

Let us now consider boundaries of $D$ at
finite distance.   We want to describe boundary conditions that preserve the small supersymmetry algebra.
Up to a certain point, the abstract quantum mechanical construction of the LG model tells us how to do that.
Formally, with an arbitrary target space and an arbitrary real superpotential, we can make a quantum mechanical model
with 2 supersymmetries.  So from that point of view, we can take the target space to be $\CX=
\mathrm{Maps}_*(D,X)$ where the notation $\mathrm{Maps}_*$ means that at the boundary of $D$, we place a
restriction of our choice on the map from $D$ to $X$.  For example, if $D=[x_\ell,x_r]$ is a compact interval (the analog if
$D\cong \IR_+$ has only one boundary point is obvious), we
can pick submanifolds $U_\ell,U_r\subset X$ and require that $x_\ell$ maps to $U_\ell$ and $x_r$ to $U_r$.  Similarly, from a
formal point of view, we can add any boundary terms that we want to the bulk superpotential $h$ defined in (\ref{defh}).
In the spirit of LG models, we will  take the boundary terms to be functions of the fields $\phi^I$ only and not their
derivatives.   So we pick arbitrary real-valued functions $k_\ell$ on $U_\ell$ and
 $k_r$ on $U_r$ and add the corresponding
boundary terms to $h$ to get:
%%
%%\be\label{defhi}
%%h := -\half \int_D\d x \,\Re\left(-i g_{I\bar J}\phi^I\frac{\partial}{\partial x} \bar{\phi}{}^{\bar J}+  \zeta^{-1} W %%\right)-\frac{k_\ell(\phi(x_\ell))}{2}+\frac{k_r(\phi(x_r))}{2}.
%%\ee
%%It is actually instructive to rewrite (\ref{defhi}), as in the bulk formula (\ref{defho}), in terms of an arbitrary 1-form $\lambda$ %%such that $\omega=\d\lambda$:
%%
\be\label{defhox}
h := -\half \int_D\d x  \left(2 \lambda_a \frac{\d u^a}{\d x} - \Re\left(   \zeta^{-1} W \right)\right)- k_\ell(u(x_\ell)) + k_r(u(x_r)) .
\ee
There is no natural choice of $\lambda$; we are always free to transform $\lambda\to\lambda+\d \alpha$ for any 0-form $\alpha$.
But in (\ref{defhox}), it is clear that such a redefinition of $\lambda$ can be absorbed in $k_\ell\to k_\ell+\alpha_\ell$,
 $k_r\to k_r+\alpha_r$.
So, since we allow any $k_\ell$, $k_r$, a shift of $\lambda$ by an exact form does not matter.

The general quantum mechanical construction gives us an action with target $\mathrm{Maps}_*(D,X)$ and superpotential
$h$; the action is invariant under  two supercharges $\CQ_{\zeta}$, $\bar\CQ_{\zeta}$ for any choices of $U_\ell,U_r,k_\ell$ and $k_r$.  However, here we have to be careful because
not every action function constructed from infinitely many variables can be quantized in a sensible way.  For example, if we simply drop the $\frac{du^a}{dx}$
term in (\ref{defhox}), we would lose the corresponding $|\partial_ x\phi|^2$ term in the Hamiltonian (or in the potential
energy of eqn. (\ref{pe})). The theory would then be ``ultra-local,'' with no energetic cost in fluctuations of short wavelength, and we would
not expect to be able to quantize it sensibly.

A more subtle variant of this problem arises in the present context unless $U_\ell$ and $U_r$ are middle-dimensional in $X$.  Quantization will only
be possible if the choices of $U_\ell$ and $U_r$ lead to elliptic boundary conditions on the Dirac equation of the two-dimensional $\sigma$-model.
A simple elliptic boundary condition on the Dirac equation on a two-manifold $\Sigma$ requires that one-half of the fermion components
vanish on the boundary of $\Sigma$.  In the present context, in the abstract quantum mechanical
model of section \ref{review}, the fermions $\psi,\bar\psi$ take values in (the pullback to the worldline of) the tangent space of the target space $M$.  In the present
context, with $M=\CX={\mathrm{Maps}}_*(D,X)$, this means that the restrictions of $\psi,\bar\psi$ to the boundaries of $D$ take values in (the pullbacks of)
the tangent bundles of $U_\ell$, $U_r$.  Thus the condition that the boundary values of the fermions take values in a middle-dimensional subspace means
that $U_\ell$, $U_r$ must be middle-dimensional.  If we do not obey this condition, we can write down a supersymmetric action, but we cannot quantize it
in a supersymmetric fashion.

Actually, $U_\ell$ and $U_r$ are subject to a much stronger constraint, which we can discover from the condition for a critical point of $h$.
Asking for $h$ to be stationary under variations of $\phi^i$ or $u^a$ that vanish at the boundary of $D$ will give the $\zeta$-soliton equation that
we have already discussed.   But there are also boundary terms to consider in the variation of $h$.  These terms are
\begin{equation}\label{zobbo}\frac{\delta h}{\delta u^a( x)}=\cdots +  \delta( x-x_\ell)\left( \lambda_a( x)-\frac{\partial k_\ell}{\partial u^a}\right)
- \delta( x-x_r)\left(\lambda_a( x)-\frac{\partial k_r}{\partial u^a}\right). \end{equation}
In general, in supersymmetric quantum mechanics, $h$ is certainly not stationary at a generic point in field space.  But in the particular
case of the infinite-dimensional target space $\mathrm{Maps}_*(D,X)$, to get a sensible model, we do need to work in a function space in which
the delta function terms in the variation of $h$ vanish.  Otherwise, when we compute the potential energy $\frac{1}{2}|\d h|^2$, we will find
terms proportional to $\delta(0)$.

In the present context, the only way to eliminate the delta function terms in the variation of $h$ is to constrain suitably $U_\ell$, $U_r$, $k_\ell$, and $k_r$.
Writing $\lambda|_U$ for the restriction of $\lambda$ to $U\subset X$, the conditions we need are
\begin{equation}\label{uruz} \lambda|_{U_\ell}=\d k_\ell,~~~~ \lambda|_{U_r}=\d k_r. \end{equation}
In particular,  $\lambda$ restricted to $U_\ell$ or $U_r$ is exact, and therefore $\omega=\d\lambda$ vanishes when restricted to $U_\ell$ or $U_r$.
Since $U_\ell$ and $U_r$ are middle-dimensional, this means that $U_\ell$ and $U_r$ are ``Lagrangian submanifolds'' of $X$.
To emphasize that they are Lagrangian, we will henceforth denote them as $\L_\ell$ and $\L_r$ rather than $U_\ell$ and $U_r$.
Moreover, eqn. (\ref{uruz})
imply that (up to inessential additive constants) $k_\ell$ and $k_r$ are uniquely determined by $\lambda$.

The condition on $U_\ell$ and $U_r$ that we have found is simply independent of $W$, so it must agree with what happens at $W=0$.  Indeed,
at $W=0$, a brane invariant under the small supersymmetry algebra is usually called an $A$-brane, and the usual $A$-branes are supported on Lagrangian
submanifolds.  $\zeta$ does not enter in standard discussions of $A$-branes; the reason for this is that  if $W=0$, there is an extra $U(1)$ $R$-symmetry
that can be used to rotate away $\zeta$. The $A$-model at $W=0$ in general may have ``coisotropic'' branes \cite{KO} as well as the usual Lagrangian branes,
but we will not try to generalize them in the presence of a superpotential.  The fact that the usual Lagrangian $A$-branes at $W=0$ can be generalized
so as to preserve the small supersymmetry algebra
also for $W\not=0$ has been shown before in a somewhat different way (see section 7.1 of \cite{HHP}).

Just as it is not strictly necessary to assume that $\omega$ is globally exact and therefore that $\lambda$ is globally-defined, similarly it is not
strictly necessary to assume that $k_\ell$ and $k_r$ are globally-defined.  Only their derivatives appear in the Lagrangian, and one can consider
the case that $k_\ell$ and $k_r$ are defined only up to additive constants.  In this case, one must develop the theory with a multivalued superpotential $h$
or else replace $\mathrm{Maps}_*(D,X)$ by a cover on which $h$ is single-valued.  However, the theory has particularly simple properties if one assumes
that $\lambda$, $k_\ell$, and $k_r$ are all single-valued, and this assumption is rather natural in the context of LG models.
Therefore, in this paper we will usually make that assumption.

Under this assumption,
eqn. (\ref{uruz}) says that $\lambda|_{\L_\ell}$ and $\lambda|_{\L_r}$ are globally exact.
On a symplectic manifold $X$ with exact symplectic form
$\omega=\d\lambda$, one says that a Lagrangian submanifold $\L\subset X$ is exact if $\lambda|_\L$ is exact.  One implication
of exactness in the standard $A$-model at $W=0$ is as follows.  In general, a (closed)
Lagrangian submanifold $\L\subset X$ determines a classical boundary condition in the
$A$-model, but because of disc instanton effects, this classical boundary condition might not really correspond to a supersymmetric $A$-brane.  In fact,
disc instanton effects can cause $\CQ_{\zeta}^2$ to be nonzero in the presence of such a brane, as we explain in section \ref{impanom}.
A disc instanton is a holomorphic map $\psi:H\to X$, where $H$ is a disc, such that $\psi(\partial H)\subset \L$.  Disc instantons do not exist if $\omega$
and $\L$ are exact, that is if $\omega=\d\lambda $ and $\lambda|_\L=\d k$ with $\lambda$ and $k$ globally defined, for then the area of a hypothetical
disc instanton would have to vanish:
\begin{equation}\label{ondog} \int_H\psi^*(\omega)=\int_{\partial H}\psi^*(\lambda)=\int_{\partial H}\psi^*(\d k)=\int_{\partial H}\d \psi^*(k)=0.\end{equation}
So in the usual $A$-model, an exact Lagrangian submanifold always does correspond to a supersymmetric brane.  The same is true in the
presence of a superpotential, since as we will explain in section \ref{impanom}, the disc instantons that are important here are ``small'' ones (localized
near a particular boundary point) that
are not affected by a superpotential.

We conclude this introductory discussion with some general remarks..
The $(2,2)$ supersymmetric sigma model with K\"ahler target $X$ and no superpotential
has two classical $R$-symmetries. $U(1)_{\rm Axial}$ rotates $Q_-, \bar Q_+$ with a phase and
$Q_+, \bar Q_-$ with the opposite phase, while $U(1)_{\rm Vector}$  rotates $Q_-, Q_+$ with
a phase and $\bar Q_-, \bar Q_+$ with the opposite phase. The topological $A$-model is obtained
by topological twisting using the current generating $U(1)_{\rm Vector}$. $A$-branes
are the branes in this topological field theory. When we turn on a generic\footnote{In the presence of
a quasihomogeneous superpotential, meaning that there is an action on $X$ of a group $U(1)_X$ under
which $W$ has ``charge 1,'' there is still an $R$-symmetry that acts on the supersymmetries
as $U(1)_{\rm Vector}$.  It is a diagonal combination of $U(1)_{\rm Vector}$ with $U(1)_X$.}
superpotential $W$, the $U(1)_{\rm Axial}$
is unbroken classically but the $U(1)_{\rm Vector}$ symmetry is broken classically
 and we cannot twist to make a topological $A$-model.

 Nevertheless, as we have seen, it is still possible
 to define branes that correspond rather closely to the usual $A$-branes at $W=0$.  We will just call them
 $A$-branes.  Moreover, as we will
 learn in section \ref{seidelfukaya}, it is also possible with $W\not=0$ to define tree-level amplitudes with many
 of the properties of standard $A$-model amplitudes (though not the usual cyclic symmetry).
 We will refer to the model that can be constructed with $W\not=0$
as the \emph{$A$-model with superpotential}.   This theory is not
a topological field theory, but shares many features of a topological
field theory.  In what follows, many statements are equally
 applicable to either a standard $A$-model (with a compact target space, or with branes required to be compact, or
 some other condition placed on branes for reasons explained in section \ref{noncompact})
 or to the  $A$-model with superpotential $W\not=0$.   When we refer loosely to the $A$-model,
 we are making statements that apply equally to the different cases.

%  One might
% refer to the model that can be constructed with $W\not=0$
%  as a partial $A$-model.  In what follows, many statements are equally
% applicable to either a standard $A$-model (with a compact target space, or with branes required to be compact, or
% some other condition placed on branes for reasons explained in section \ref{noncompact})
% or to the partial $A$-model that we will describe with $W\not=0$.   When we refer loosely to the $A$-model,
% we are making statements that apply equally to the different cases.
%

At $W=0$, the standard definition  of a Lagrangian $A$-brane involves specifying not just the support $\L$ of the
brane, but also a flat unitary Chan-Paton
vector bundle\footnote{This description is a little over-simplified, as one knows from the
K-theory interpretation of $D$-branes.  The Chan-Paton bundle on a brane is not quite a flat unitary vector bundle
but is twisted by a gerbe of order 2 associated to $w_2(\L)$.} over $\L$.  This part of the brane story is not affected
by introducing $W$, and does not interact in a very interesting way with what we will describe below.
In principle we should always denote a brane by $\B$ to distinguish it from its
support $\L$. Nevertheless, we will sometimes trust to the reader's indulgence
 and simply refer to a brane   by $\L$.

\subsubsection{Hamiltonian Symplectomorphisms}\label{hamilton}

In the conventional topological $A$-model -- with a compact target space, for example --
the brane determined by a Lagrangian submanifold $\L$ is
supposed to be invariant under deformations of $\L$ that are induced by  Hamiltonian symplectomorphisms of $X$,
provided the Hamiltonian symplectomorphisms are isotopic
to the identity.  The group of Hamiltonian symplectomorphisms is the group generated
by Hamiltonian flows with single-valued Hamiltonian functions $H$.  To a function $H$, we associate the Hamiltonian
vector field
\be \label{hamv} V_H^a=\omega^{ab}\partial_b H. \ee
The group of Hamiltonian symplectomorphisms is the group generated by these vector fields.

To determine the infinitesimal motion of a Lagrangian submanifold $\L$ generated by a given Hamiltonian
function $H$, we only care about the corresponding Hamiltonian vector field $V_H$ modulo vectors that are tangent
to $\L$, since a vector field tangent to $\L$ generates a reparametrization of $\L$, rather than a motion of $\L$ in $X$.
To determine $V_H$ modulo vector fields tangent to $\L$, we only need to know the first derivatives of $H$ along
$\L$ (rather than its derivatives in the normal direction).  So if we are given a function $H$ that is defined just on $\L$
(and not on all of $X$), this suffices to determine a motion of $\L$ in $X$ to first order, though of course only to first
order.

In the mathematical literature on the topological $A$-model, going all the way back to the work of A. Floer
in the mid-1980's, invariance of $A$-branes under Hamiltonian symplectomorphisms is one of the most central properties.
Yet this fact is relatively little-known among physicists.
The reason that the statement is not more familiar to string theorists
is the following.  In a class of Lagrangian submanifolds  that are equivalent under
Hamiltonian diffeomorphisms that are isotopic to the identity, there is
at most one special Lagrangian representative, and this is the representative that is
important in most physical applications of the $A$-model.
If there is no special Lagrangian representative in the given class, then the branes in question
are ``unstable,'' analogous to unstable
holomorphic bundles on the $B$-model side, and cannot be used in a superconformal construction.

Actually, invariance of the $A$-model under Hamiltonian
symplectomorphisms of branes is mirror dual to invariance of the $B$-model under complex
gauge transformations of the Chan-Paton gauge field of a brane.    One expects to be able to make Hamiltonian
symplectomorphisms independently for each $A$-brane,
just as in the $B$-model, one can make separate complex gauge transformations for each $B$-brane.

Here is a more detailed explanation.  First we consider the $B$-model, and then we will
consider the $A$-model in parallel.
In a $\sigma$-model, a brane $\B$ is described by a submanifold $Y\subset X$ that is equipped
with a Chan-Paton vector bundle $E\to Y$
that is endowed with a unitary connection $A$.  A unitary gauge transformation of $A$ just changes
the worldsheet action by a total
derivative, so it is trivially a symmetry.  However, the $B$-model is actually invariant under complex gauge
 transformations of $A$,
not just unitary ones.  The way that this happens is that the change in the action when $A$ is changed
by a gauge transformation with
an imaginary generator is $\Q$-exact.  This is a special case of the fact that a $D$-term in the action is
always $\Q$-exact.
For simplicity, in both the $A$-model and the $B$-model, we will consider only branes of rank 1; for
the $B$-model, this means that $A$
is a $U(1)$ gauge field.
 The boundary
$D$-terms
of lowest dimension take the form
\be\label{zollo}\int_{\partial\Sigma} \d t \d^2\theta \,k(u^a),\ee
where $X$ is parametrized locally by some functions $u^a$ and $k$ is a real-valued
function on the support $\L$ of a brane $\B$.
The integral runs over the portion of $\partial\Sigma$ that is labeled by a particular brane.
After performing the $\theta$ integrals, in the case of the $B$-model,
one can recognize (\ref{zollo}) as the change in the action under an infinitesimal gauge
transformation of $A$ with imaginary gauge parameter
$ik$.

To imitate this in the $A$-model, we proceed in exactly the same way, using the same $D$-term (\ref{zollo}).  But this
is actually not anything new.  We already allowed for such a boundary $D$-term in eqn. (\ref{defhox}), where we included
in the Morse function $h$ a boundary contribution involving an {\it a priori} arbitrary function $k$ on $\L$.  (In writing
this equation, we allowed for the possibility of separate Lagrangian submanifolds $\L_\ell$ and $\L_r$ at the two
ends, with separate functions $k_\ell$ and $k_r$.) The contribution of the Morse function to the action is the $D$-term
$\int \d t\d^2\theta\, h$, so the dependence of a brane on this interaction is $\Q_\zeta$-exact.  However, we learned in the
subsequent analysis that actually $k$ cannot be specified independently of the choice of $\L$.  Having specified
once and for all a one-form $\lambda$ with $\d\lambda=\omega$,  the restriction of $\lambda$ to $\L$ must be
related to $k$ by eqn. (\ref{uruz}):
\begin{equation}\label{noruz}\lambda|_\L=\d k. \end{equation}
This means that, up to an additive constant, $k$ cannot be varied independently of $\L$.  If we change $\L$ to
first order by a Hamiltonian vector field $V$, then the change in $\lambda|_\L$ is ${\bf{i}}_V\omega$, where ${\bf{i}}_V$
is the contraction operation (in coordinates, ${\bf{i}}_V\omega=V^a\omega_{ab}\d u^b$).  Thus eqn. (\ref{noruz})
tells us that if we want to change $k$ by an amount $\delta k$, then we also need to move $\L$ by the Hamiltonian
vector field
\begin{equation}\label{poruz} V^a=\omega^{ab}\partial_b(\delta k). \end{equation}

For a more complete picture, suppose  first
 that we are given a Hamiltonian function $H$ that is defined throughout $X$.  We consider making a change of
 variables in the theory with $\delta u^a=V^a_H$. In first order, this
  moves each Lagrangian submanifold $\L$ by the restriction to
 $\L$ of $V_H$.   To decide if this is an invariance of the $A$-model, we must
  see what happens to the Morse function $h$.  This is a sum of
 three terms
 \begin{align}\label{funn} h_1& = -\int_D \lambda_a\d u^a \cr
                                       h_2 & = \int_D \d x \frac{1}{2}\mathrm{Re}(\zeta^{-1}W) \cr
                                        h_3 & = -k_\ell(u_\ell)+k_r(u_r).  \end{align}
None of these terms is separately invariant under $\delta u^a=V^a_H$.  For example, $h_2$ changes
by a bulk integral
\begin{equation}\label{punn}    \delta h_2=\int_D\d x\frac{1}{2}\{H,\mathrm{Re}(\zeta^{-1}W)\}, \end{equation}
where $\{f,g\}$ is the Poisson bracket of two functions $f$ and $g$.  Because this is purely a bulk integral with
no delta function terms at the end, the corresponding contribution $\int\d t\d^2\theta \,\delta h_2$ is a harmless
$D$-term.  By contrast,  the change in $h_1$ is only a boundary term:
\be\label{wunn} \delta h_1=\int_D  \delta u^a\omega_{ab}\d u^b=\int_D \omega^{ac}\partial_c H ~ \omega_{ab}\d u^b
=-\int_D\d H=-H(x_r)+H(x_\ell).  \ee
A change in the Morse function by boundary terms will make  $\delta(0)$ contributions to the potential energy
$|\d h|^2$ of the $\sigma$-model.  So these terms must be canceled by the variation of $h_3$.  To do this, we compensate for the change of variables $\delta u^a=V^a_H$ by allowing variations in $k_r$ and $k_\ell$:
\begin{equation}\label{murn}\delta h_3=-\delta k_\ell(u_\ell)+\delta k_r(u_r). \end{equation}
To cancel the boundary terms in $\delta h$, we simply choose $\delta k_r$ and $\delta k_\ell$ to equal
$H(x_r)$ and $H(x_\ell)$, respectively, as was already explained in the last paragraph.

The result just described  is not general enough, since modulo $D$-terms, we are supposed
to be able to make  {\it separate}
Hamiltonian symplectomorphisms at the two ends of an open string.  Suppose that we want to transform $\L_\ell$
by a Hamiltonian function $H_\ell$ and $\L_r$ by another Hamiltonian function $H_r$.  We pick some more general
function $H(u^a;x)$ that depends explicitly on the point $x$ along a string, chosen to coincide with $H_\ell$
in a neighborhood of $x=x_\ell$ and with $H_r$ in a neighborhood of $x=x_r$.  The important boundary terms
in the preceeding analysis are unaffected.  The only change in the analysis is that $\delta h_1$ becomes more
complicated, with an additional bulk contribution that involves the explicit $x$-dependence of $H(u^a;x)$ and
contributes a harmless $D$-term.

There are still a few loose ends to tie up.  First, in addition to the term $|\d h|^2$, the $\sigma$-model action
contains another $D$-term, the kinetic energy of the $\sigma$-model.  This is not invariant under a generic Hamiltonian
symplectomorphism.  However, since it is  a $D$-term, its variation is also a $D$-term, and moreover a harmless
one, like $\delta h_2$, with no boundary contributions.  That is so simply because the kinetic energy has time derivatives
only, and no spatial derivatives, so there is no way for it to generate a boundary term.

\begin{figure}
 \begin{center}
   \includegraphics[width=1.8in]{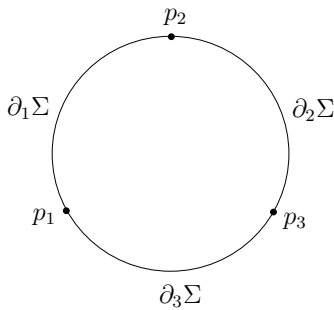}
 \end{center}
\caption{\small   A disc with its boundary divided in three segments $\partial_i\Sigma$, $i=1,2,3$, labeled by different
branes.}\label{cyclicends}
\end{figure}
Second, the $A$-model action also has a topological term that is not $\Q_\zeta$-exact:
\be\label{zunn}I'=\int_\Sigma \Phi^*(\omega)=\int_\Sigma \,\omega_{ab}\d u^a\wedge \d u^b. \end{equation}
In discussing this term, we can assume an arbitrary $\Sigma$, not necessarily a strip in the plane.
$I'$ is a topological invariant -- and hence in particular is $\Q_\zeta$-invariant --
if $\Phi(\partial\Sigma)$ is contained in a Lagrangian submanifold $\L$.  What happens under a Hamiltonian
symplectomorphism $\delta u^a=V^a_H$ that changes the map $\Phi$ and also changes $\L$?  The change in $I'$ is
\be\label{unn}\delta I'=\int_\Sigma\d(\delta u^a\omega_{ab}\d u^b)=\int_{\partial\Sigma}V^a_H\omega_{ab}\d u^b
= - \int_{\partial \Sigma}\d H.  \end{equation}
This vanishes if all of $\partial\Sigma$ is mapped to the same Lagrangian submanifold with the same $H$
(since $H$ is single-valued).  More generally,
 $\partial\Sigma$ may be  a union of segments $\partial_i\Sigma$ whose left and right  endpoints we call
 $p_i$ and  $p_{i+1}$ (Figure \ref{cyclicends}.  We assume that the $\partial_i\Sigma$ are mapped to different Lagrangian
submanifolds $\L_i$, which we want to deform using different Hamiltonian function $H_i$.  The generalization of eqn. (\ref{unn}) is
\be\label{zunno}\delta I'=- \sum_i\int_{\partial_i\Sigma}\d H_i=\sum_i (H_{i}(p_i)-H_{i-1}(p_{i})). \end{equation}
At each intersection point $p_i$, a vertex operator is inserted for an external string state, and the contribution
$H_{i}(p_i)-H_{i-1}(p_i)$ to the action can be absorbed in the normalization of this vertex operator.

Finally, in our discussion of the Morse function and kinetic energy of the $\sigma$-model, we considered
only a time-independent situation in which $\Sigma$ is a strip in the plane.  This makes it possible to consider
$A$-branes and strings that preserve two supersymmetries, namely $\Q_\zeta$ and its adjoint $\bar\Q_\zeta$.
In that context, we have shown that varying $\L_\ell$ and $\L_r$ by independent Hamiltonian symplectomorphisms
is equivalent to adding to the action a $D$-term -- a term that is both $\Q_\zeta$-exact and $\bar\Q_\zeta$-exact.
In a more general situation in which $\Sigma$ is not simply a strip (for example, in the study of tree-level
amplitudes described in section \ref{seidelfukaya}), there is no time-translation invariance and one cannot maintain
both $\Q_\zeta$ and $\bar\Q_\zeta$ symmetry.  Instead of writing the response to Hamiltonian symplectomorphisms of branes
as $\int_\Sigma \d t\d x\d\theta\d\bar \theta \,F$  (where $F$ was described in the above construction), we have to
write it simply as $\int_\Sigma\d^2x \int\d\theta\, \widehat F$, where in the time-independent case $\widehat F=\int \d\bar\theta F$.
In general, we would pick an $\widehat F$ that everywhere near the boundary of $\Sigma$ looks like the functional
the $\widehat F$ that we used in analyzing the problem on a strip.  In this way, we would establish the desired
invariance.

\subsubsection{Branes With Noncompact Target Spaces}\label{noncompact}

In the $A$-model with a compact symplectic manifold $X$ as target, one defines an $A$-brane supported on  any (closed) Lagrangian submanifold $\L$, subject
to some mild restrictions that are not pertinent at the moment.\footnote{One restriction is associated to the $K$-theory
interpretation of $D$-branes (the normal bundle to $\L$ in $X$ must admit a $\mathrm{Spin}_c$ structure \cite{FreedWitten}.  Another restriction, described in section \ref{impanom},
 involves disc instantons.} However, a compact $X$ is not relevant for the present paper, since if $X$ is
compact, it is not possible to introduce a nonconstant holomorphic superpotential $W$.

\begin{figure}
 \begin{center}
   \includegraphics[width=2.2in]{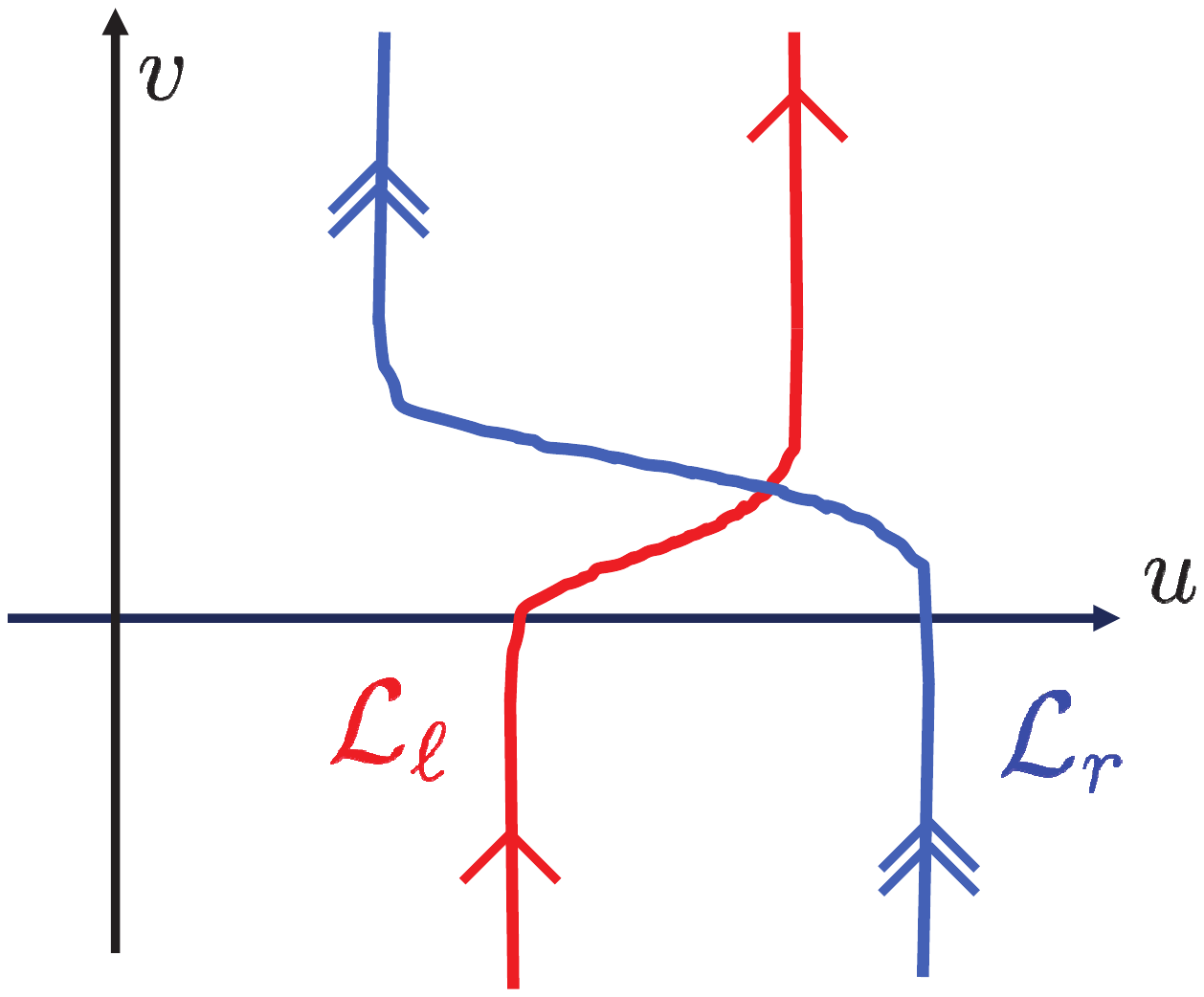}
 \end{center}
\caption{\small A pair of Lagrangian submanifolds $\L_\ell$, $\L_r$ embedded in the $u-v$ plane.  $\L_\ell$ and $\L_r$ intersect at the one
point indicated.  $u$ is plotted horizontally and we assume that $\L_\ell$, $\L_r$
are embedded in the half-plane $u>0$.}
 \label{Lagrangians}
\end{figure}
Once $X$ is not compact, one usually wants to impose some sort of condition
on the behavior of a Lagrangian submanifold at infinity.  The most basic reason is that otherwise the space of supersymmetric states in quantization
on a strip with boundary conditions set at the ends by a pair of Lagrangian submanifolds $\L_\ell$, $\L_r$ will not have the expected behavior.  To see what
will go wrong in general, consider the case that $X=\IR^2$ with the standard symplectic form $\omega=\d u\wedge \d v$, and with $\L_\ell$ and $\L_r$ as
depicted in  Figure \ref{Lagrangians}.  We assume that the $u$-axis runs horizontally in the figure, and that $\L_\ell$ and $\L_r$ are embedded in the half-plane
$u>0$.  Let us consider this system first in the ordinary $A$-model without a superpotential.  The classical approximation to a supersymmetric state of the $(\L_\ell,\L_r)$
system is given by an intersection point of the Lagrangian submanifolds $\L_\ell$ and $\L_r$.  In the example shown in the figure, there is precisely one such intersection point.
Since the MSW complex is thus of rank 1, its differential necessarily vanishes and the space of quantum supersymmetric states of the $(\L_\ell,\L_r)$ system is one-dimensional.

Now let us introduce a superpotential and quantize the theory on a strip of width $w=x_r-x_\ell$.  The classical approximation to a supersymmetric state of the $(\L_\ell,\L_r)$ system is given,
just as in the usual $A$-model, by a time-independent supersymmetric state, but now the condition for supersymmetry is the $\zeta$-soliton equation.
So supersymmetric states, in the classical approximation, correspond
to solutions of the $\zeta$-soliton equation that start somewhere on $\L_\ell$ at  $x=x_\ell$ and end somewhere on $\L_r$ at $x=x_r$. (For more on this, see
section \ref{subsec:MorseComplexHalfLine}.)    In this discussion, we consider only solutions that are independent of the usual time coordinate $\tau$
and it is convenient to refer to the usual spatial coordinate $x$ as
``time.''  Let $\L_\ell^w$ parametrize the points that can be reached by starting somewhere on $\L_\ell$ and
evolving for time $w$ via the $\zeta$-soliton equation.  Then $\L_\ell^w$ is a Lagrangian submanifold that is very close to $\L_\ell$ if $w$ is small.  ($\L_\ell^w$ is Lagrangian
because the $\zeta$-soliton equation describes Hamiltonian flow with the Hamiltionian $-\half \mathrm{Re}\,(\zeta^{-1}W)$.)  $\zeta$-soliton solutions that flow from $\L_\ell$ to $\L_r$ in
``time'' $w$ are simply intersections of $\L_\ell^w$ with $\L_r$.  For small enough $w$, the difference between $\L_\ell$ and $\L_\ell^w$ is unimportant and the classical supersymmetric
states with $W\not=0$ correspond naturally to those with $W=0$.

However, with everything being noncompact, intersection points of $\L_\ell^w$ and $\L_r$ can flow to infinity at finite $w$.  This will actually happen in the example
of the figure if we take $\zeta^{-1}W=\I \phi^2 $ (where $\phi=u+\I v$ with real $u,v$ and we take the Kahler metric of the $\phi$-plane to be $\d \ell^2=\d u^2+\d v^2$).
So $\mathrm{Im}(\zeta^{-1}W)=(u^2-v^2) $ and
 the $\zeta$-soliton equation is $\partial_x u=u$, $\partial_x v=-v$.
In particular, $\L_\ell^w$ is obtained from $\L_\ell$ by $(u,v)\to (e^wu,e^{-w}v)$.  In the figure (in which $u$ is plotted horizontally), we can assume that $\L_\ell$ and $\L_r$ are contained in a strip $u_0<u<u_1$ with $u_0>0$.  Then for large enough $w$, $\L_\ell^w$ is entirely to the right of the strip and has no intersections with $\L_r$.  Thus in this
example, for large enough $w$, supersymmetry is broken, even though for small $w$ it is unbroken (with precisely one supersymmetric ground state).
For some purposes this might be an interesting example of supersymmetry breaking. However in the present
context it is a problem. What has gone wrong is that the intersection of $\L_\ell^w$ with $\L_r$ goes to
infinity at a finite value of $w$.

A related problem  arises even in the absence of a superpotential if we consider the fact that the $A$-model is supposed to be invariant under Hamiltonian
symplectomorphisms applied separately to each brane, as we described in section \ref{hamilton}.
This fails if we consider branes and Hamiltonian symplectomorphisms with no restriction on their behavior at infinity
For instance, in our example, we can eliminate the intersection point of $\L_\ell$ with $\L_r$ by transforming $\L_\ell$
via a Hamiltonian symplectomorphism $(u,v)\to (u+c,v)$ for a large constant $c$ (the single-valued Hamiltonian that generates this
symplectomorphism is simply $v$).  So the space of supersymmetric
states of the $(\L_\ell,\L_r)$ system is not invariant under Hamiltonian symplectomorphisms
applied separately to $\L_\ell$ or $\L_r$.

\subsubsection{$W$-Dominated Branes}\label{goodclass}

To avoid both of these problems, we will place some conditions on $\L_\ell$ and $\L_r$, to prevent their intersections from going to infinity. In the present section, we describe a class of branes for which intersections are bounded and for which the machinery of the present paper applies naturally.
In defining this class of branes, we will make use of the superpotential $W$.   However, for Lagrangian submanifolds  $\L_\ell$ and $\L_r$ obeying the conditions
that we will state momentarily, the space of supersymmetric $(\L_\ell,\L_r)$ states makes sense in the ordinary $A$-model without a superpotential,
and is unchanged
when one turns on the superpotential $W$.

We simply require that $\mathrm{Im}(\zeta^{-1}W)$ goes to $+\infty$ at infinity along $\L_\ell$, and to $-\infty$
at infinity along $\L_r$.
We will refer to left- and right- branes obeying this condition as \emph{$W$-dominated branes}.
(The reversal of sign between $\L_\ell$ and $\L_r$, which might look peculiar at first sight, is natural in our formalism because a $\pi$ rotation of the plane,
which exchanges the left and right boundaries, reverses the sign of $\zeta$ and hence of $\mathrm{Im}(\zeta^{-1}W)$.)

One might think that
this condition is required for bounding the surface terms in eqn. (\ref{pe}).  But actually, that is not necessary;
the potential $V$ is positive-definite in any case since it can be written as
in eqn. (\ref{pep}).

The real virtue of the $W$-dominated branes is that the growth
condition on  $\mathrm{Im}(\zeta^{-1}W)$ prevents the intersections of left- and right- branes from going to infinity.
As one varies $\L_\ell$ and/or $\L_r$ to make an intersection point $p$ go to infinity, $\mathrm{Im}(\zeta^{-1}W)$ would have to go to $+\infty$
(since $p\in \L_\ell$) and to $-\infty$ (since $p\in \L_r$).   So intersection points do not go to infinity\footnote{We consider a family of
branes parametrized by a compact parameter space.  In that situation, the upper and lower bounds on $\mathrm{Im}(\zeta^{-1}W)  $
hold uniformly. The same comment is relevant at several points below.}
 and
the space of $(\L_\ell,\L_r)$ strings is well-defined in the ordinary $A$-model without a superpotential.  (In other words, the space
of $(\L_\ell,\L_r)$ strings is well-defined even if we only use $W$ for guidance in deciding what $\L_\ell$ and $\L_r$ to allow
and do not actually turn on $W$ as a contribution  to the Lagrangian.)

Also, if $\mathrm{Im}(\zeta^{-1}W)$ diverges at infinity on $\L_\ell$, then the same is true on $\L_\ell^w$ for any $w>0$,
since the $\zeta$-soliton equation is
ascending gradient flow for  $\mathrm{Im}(\zeta^{-1}W)$.  So intersection points of $\L_\ell^w$ with $\L_r$ do not go to infinity with increasing $w$.  This means
that intersection points cannot flow in from or out to infinity  when $w$ is turned on, so that, at the level of
 cohomology, the space of supersymmetric $(\L_\ell,\L_r)$ strings is unchanged when $W$ is
turned on and is independent of $w$.

A further virtue of $W$-dominated branes is that
the spaces of supersymmetric states are finite-dimensional spaces.
We will phrase our argument here for the standard $A$-model without a superpotential; including a superpotential
simply replaces $\L_\ell$ with $\L_\ell^w$  in what follows.
In quantization of the $(\L_\ell, \L_r)$ system, if  $\L_\ell$ and $\L_r$ intersect in a discrete set of points $\Bbb T(\L_\ell,\L_r)$,
then in the classical
approximation, there is one supersymmetric state $\Phi_\alpha$ for each $\alpha\in \Bbb T(\L_\ell,\L_r)$.   The theory
has a much simpler flavor if the sets $\Bbb T(\L_\ell,\L_r)$
are always finite, and  more generally the intersections $\L_\ell\cap \L_r$ are always compact.   Otherwise, one has to deal with infinite-dimensional
spaces of supersymmetric states, in the classical approximation and perhaps in the exact theory.
A simple example of what one would like to avoid is provided by again taking $X$ to be the $u-v$ plane, with
symplectic form $\d u\wedge\d v$, and
taking for $\L_\ell$ and $\L_r$ the $u$-axis and the curve $v=\sin u$.   Here the intersection $\L_\ell\cap \L_r$
consists of infinitely many points. It is easy
to construct wilder examples involving spirals in the plane.
For $W$-dominated branes, compactness of the intersection $\L_\ell\cap \L_r$ is insured since $\mathrm{Im}(\zeta^{-1}W)$
goes to $+\infty$ at infinity  on $\L_{\ell}$ and to $-\infty$ at infinity on $\L_r$.
In this case, the subset $\Delta\subset
\L_\ell$ on which $\mathrm{Im}(\zeta^{-1}W)$ is less than its upper bound on $\L_r$ is compact.\footnote{We consider a family of
branes parametrized by a compact parameter space.  In that situation, the upper and lower bounds on $\mathrm{Im}(\zeta^{-1}W)  $
hold uniformly and all intersections occur in a fixed compact set $\Delta\subset X$ that is independent of the parameters.}
 The intersection
$\L_\ell\cap \L_r$ is a closed and therefore
compact subspace of $\Delta$ and consequently the space of supersymmetric states of the $(\L_\ell,\L_r)$ system will
always be finite-dimensional.

 In our example of \S \ref{noncompact}  with $\mathrm{Im}(\zeta^{-1}W)=(u^2-v^2)$,
the brane $\L_r$ of Figure \ref{Lagrangians}
has the desired property, since $(u^2-v^2)$ goes to $-\infty$ at infinity along $\L_r$, but $\L_\ell$ does not obey
the appropriate condition at infinity. To ensure that $(u^2-v^2)$
goes to $+\infty$ at infinity along $\L_\ell$, we could rotate $\L_\ell$ by $\pm \pi/2$.
In this case, the pathologies noted in Section of \S \ref{noncompact} would disappear.
Once we restrict the class of branes so
 that $\mathrm{Im}(\zeta^{-1}W)$ goes to $+\infty$ or $-\infty$ at infinity on $\L_\ell$ or $\L_r$, we must also
  restrict the class of gauge transformations to include only those Hamiltonian symplectomorphisms that preserve these conditions. In particular,
a $\pi/2$ rotation, although symplectic, is not an allowed gauge transformation.

In this section, we have explained one natural answer to the question, ``For what kind of branes $\L_\ell,\L_r$ is the space of
supersymmetric $(\L_\ell,\L_r)$ states
well-defined?''  However, this question has at least one more interesting answer, which we return to in section \ref{morebranes}
after introducing the concept
of a thimble.

\subsubsection{Thimbles}\label{thimbles}

Generically, for every critical point $\phi_i$, there is  a canonical example of a left-brane $L_i^\zeta$ and
also of a right-brane $R_i^\zeta$ satisfying the conditions of section \ref{goodclass}.
To construct $L_i^\zeta$, we consider the $\zeta$-soliton equation on the
half-line $(-\infty,0]$ with the boundary condition that the solution approaches a critical point
$\phi_i$ at $x=-\infty$.   Regarding the $\zeta$-soliton equation as a gradient flow equation, the flows of this type are parametrized by the constants $c_i$  in eqn. (\ref{ubo}) with $f_i>0$, so the dimension of $L_i^\zeta$
is the Morse index $y$ of the function $\Im(\zeta^{-1}W)$ at its critical point, and $L_i^\zeta$ is a copy of $\R^y$.
Like the real or imaginary part of any holomorphic function that has a nondegenerate critical point,
this function has middle-dimensional Morse index, so in particular $y=\mathrm{dim}_\IC X$.
By mapping an ascending flow on $(-\infty,0]$ that starts at $\phi_i$
 to its value at $x=0$ (here we include the trivial flow line that sits at $\phi_i$ at all times), we can interpret the space of such  flows as a middle-dimensional submanifold $L_i^\zeta\subset X$.
Since the $\zeta$-soliton equation is translationally-invariant, the value $\phi(x_0)$ of an ascending flow
from $\phi_i$ at any point $x_0$ with $-\infty < x_0 < \infty$ is on the submanifold $L_i^\zeta$.
So $L_i^\zeta$ can be viewed as the union of all ascending flow lines that start at the critical point $\phi_i$ in the far past.
If all flow lines that start at $\phi_i$ in the past flow to infinity in the future, then $L_i^\zeta$ is a closed submanifold of $X$, and in this
case we call it a Lefschetz thimble.

There is an important circumstance in which this can fail.
Suppose that $\zeta=\zeta_{ji}$ for some $j$.  Then an $ij$ soliton may exist.  Such a soliton
is a flow line that starts arbitrarily close to $\phi_i$ in the past and flows arbitrarily close to $\phi_j$ in the future, but never reaches it.
So in  this case, the point $\phi_j$ is contained in the closure of $L_i^\zeta$ but not in $L_i^\zeta$ itself.
(One could of course replace $L_i^\zeta$ with its closure, but this does not work well; for example, for $\dim_\C X>1$, the
closure is generically not a manifold, and for $\dim_\IC X=1$ it may be a manifold with boundary.  The fundamental reason that there is
no good definition of a Lefschetz thimble at $\zeta=\zeta_{ji}$ is really that in crossing such a value, the topology of the Lefschetz thimble can jump.)

As long as $\zeta$ does not equal any of the $\zeta_{ij}$, we do not meet this problem and $L_i^\zeta$ is a closed submanifold of $X$.
 We have already seen that $L_i^\zeta$ is middle-dimensional and topologically $\R^y$.
  To show that $L_i^\zeta$ is Lagrangian, we use the fact that the $\zeta$-soliton equation is
Hamiltonian flow (eqn. (\ref{turmi})).  Using the identification of $L_i^\zeta$ with the value $u^a(0)$ of a flow at time $x=0$, and writing  the symplectic form
of $X$ as $\omega_{ab}\d u^a\d u^b$, the restriction of this form to
$L_i^\zeta$ is $\omega|_{L_i^\zeta}=\omega_{ab}\d u^a(0)\d u^b(0)$.  But Hamiltonian flow preserves the symplectic form,
so we can equally write $\omega|_{L_i^\zeta}=\omega_{ab}\d u^a(x)\d u^b(x)$ for any $x$.  Taking $x\to-\infty$, this vanishes for flows
that start at $\phi_i$, so $\omega|_{L_i^\zeta}=0$ and $L_i^\zeta$ is Lagrangian.
Since $L_i^\zeta$  is topologically $\IR^y$, any closed form on $L_i^\zeta$ is exact, and $L_i^\zeta$ is exact Lagrangian.
Finally, because $L_i^\zeta$ was defined by ascending gradient flow from $\phi_i$, $\mathrm{Im}(\zeta^{-1}W)$ is bounded below along $L_i^\zeta$
by its value at $\phi_i$.   If the K\"ahler metric of $X$ is complete, $\mathrm{Im}(\zeta^{-1}W)$ goes to infinity at infinity along $L_i^\zeta$.  (Completeness
of the metric and the fact that each ascending flow from $\phi_i$ goes to infinity in $X$ implies that each ascending flow line has infinite length;
this plus the
ascending flow equation implies that $\mathrm{Im}(\zeta^{-1}W)$ goes to infinity along each such line.) The corresponding right-brane $R_j^\zeta$, which we call a right Lefschetz thimble,
is defined in precisely the same way, with similar properties. It parametrizes ascending gradient flows
on the half-line $[0,\infty)$ that approach $\phi_j$ for $x\to\infty$,
and can be identified with the value of such a flow at $x=0$.  Clearly $\mathrm{Im}(\zeta^{-1}W)$ is
bounded above along $R_j^\zeta$ by its value at $\phi_j$.

As a simple example we return to $\zeta^{-1}W=\I \phi^2 $  so  $\mathrm{Im}(\zeta^{-1}W)=(u^2-v^2) $.
There is a single critical point at $\phi=0$. The left-Lefshetz thimble is the $u$-axis and the right-Lefshetz
thimble is the $v$-axis.

\subsubsection{Another Useful Class Of Branes: Class $T_{\kappa}$ }\label{morebranes}

 \begin{figure}
 \begin{center}
   \includegraphics[width=1.5in]{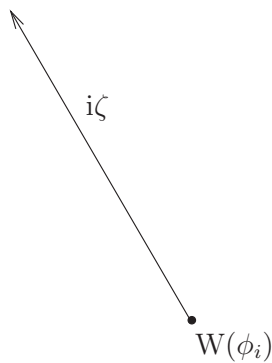}
 \end{center}
\caption{\small  A ray in the complex $W$-plane, starting at $W(\phi_i)$ and running in the $\i\zeta$ direction.} \label{ray}
\end{figure}
 In any Hamiltonian flow, the Hamiltonian is a conserved quantity.  So in particular, the
 Hamiltonian $H=-\half\mathrm{Re}( \zeta^{-1}W)$ is a conserved quantity for the $\zeta$-soliton equation.  Moreover, $\mathrm{Im}(\zeta^{-1}W)$ is
 an increasing function of $x$ along any non-constant solution of this equation.  Putting these facts together, the values of $W$ along the left thimble
 $L_i^\zeta$ lie on a ray that begins at the point $W(\phi_i)$ and extends in the direction $\i\zeta$ in the complex $W$-plane (Figure \ref{ray}).  (The following discussion
 could be presented in terms of right thimbles rather than left thimbles, but this would add nothing as it would be equivalent to replacing $\zeta$ by $-\zeta$.)

\begin{figure}
 \begin{center}
   \includegraphics[width=2.5in]{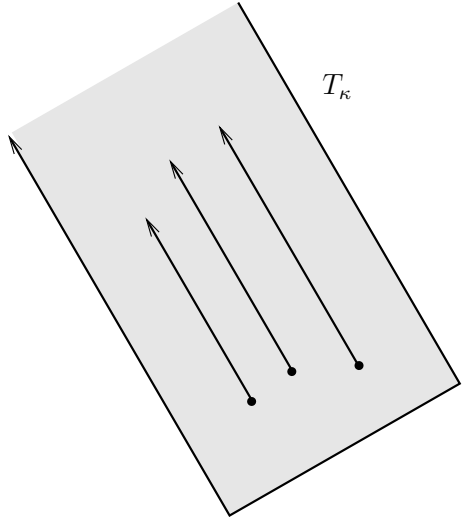}
 \end{center}
\caption{\small  The rays in the complex $W$-plane that start at critical points and all run in the $\i\kappa$ direction fit into the semi-infinite strip $T_\kappa$, which is shown as a shaded region. }\label{manyrays}
\end{figure}

 The images in the $W$-plane of all the left thimbles $L_j^\zeta$, $j\in \IV$, form a collection of parallel rays, starting at the critical values $W(\phi_j)$.  Assuming that $\IV$
 is a finite set, the images of these thimbles are all contained in a semi-infinite strip $T_\zeta$ of finite width in the $W$-plane.  (This is sketched in Figure \ref{manyrays}, except
 that, for reasons that will soon be apparent, in the figure $\zeta$ is replaced by a complex number $\kappa$ of modulus 1, not necessarily equal to $\zeta$.)
$T_\zeta$ is defined by
  \begin{align}\label{zelbo} |\mathrm{Re}\,(\zeta^{-1}W)| & \leq  c\cr
 \mathrm{Im}\,(\zeta^{-1}W)& \geq c', \end{align}
 for some constants $c,c'$.
We say that a Lagrangian submanifold $\L$ -- or a brane supported on $\L$ -- is of class $T_\zeta$ if $W$ restricted to $\L$ is valued in $T_\zeta$.
Branes of class $T_\zeta$ (or obeying an equivalent condition) are  considered in the mathematical theory of the Fukaya-Seidel
 category \cite{SeidelBook}; see section \ref{seidelfukaya}.  This is a mathematical theory related to what in physical terms is the $A$-model with a superpotential $W$.  In addition to
 some reasoning that is described below, the construction of the Fukaya-Seidel category is motivated by mirror symmetry.

 Let $\L$ and $\L'$ be Lagrangian submanifolds  of class $T_\zeta$.
 In the $A$-model without a superpotential, the space of supersymmetric $(\L,\L')$ states is not well-defined, because the strip $T_\zeta$ is not compact and intersections of
 $\L$ and $\L'$ can go off to infinity (if, for example, a Hamiltonian symplectomorphism is applied to $\L$ or $\L'$).  What happens if we turn on a superpotential $W$? By itself, this does not help.
 In studying supersymmetric $(\L,\L')$ states on a strip of width $w>0$, the effect of turning on the superpotential is that instead of looking at intersections
 $\L\cap \L'$, we have to look at intersections $\L^w\cap \L'$, where $\L^w$ is obtained
 from $\L$ by ascending gradient flow with respect to $\mathrm{Im}(\zeta^{-1}W)$.  The image under $W$ of
  $\L^w$ is contained in the strip $T_\zeta$   itself, so
 $\L^w$ is again of class $T_\zeta$ and the intersections
 $\L^w\cap \L'$ are not well-behaved.

 However, a simple variant of this idea does work.  We pick a complex number $\kappa$ of modulus 1, but not equal to $\pm\zeta$.  Then instead of branes
 of class $T_\zeta$, we consider branes of class $T_\kappa$.  These branes are characterized by the condition (\ref{zelbo}), but with $\zeta$ replaced by $\kappa$
 (as is actually shown in Figure \ref{manyrays}).
Omitting the points $\pm \zeta$ divides the unit circle into two connected components, and it is convenient to make a choice that $\kappa$ lies to the ``left'' of $\zeta$ (meaning
 that $\pi>\mathrm{Arg}(\zeta^{-1}\kappa)>0$).

 The classical approximation to a supersymmetric $(\L,\L')$ state on a strip of width $w$ is now given by a solution of the $\zeta$-soliton equation,
 starting somewhere on $\L$ at the left end of the strip and ending somewhere on $\L'$ at the right end.  Differently put, the classical approximation to such a supersymmetric
 state is an intersection point of $\L^w$ with $\L'$, where  $\L^w$ is obtained from $\L$ by evolving for a ``time'' $w$ via the $\zeta$-soliton equation.

 This evolution is in the direction of increasing $\mathrm{Im}\,(\zeta^{-1}W)$ (with $\mathrm{Re}\,(\zeta^{-1}W)$ fixed).  Because of our hypothesis that $\kappa\not=\pm\zeta$, this evolution tends to move the image of $\L^w$ out of the strip $T_\kappa$.
 As we will see momentarily,  under reasonable conditions on the growth of $W$ at infinity, the intersection $\L^w\cap \L'$ is
 bounded, so the space of supersymmetric $(\L,\L')$ states is well-defined.

\begin{figure}
 \begin{center}
   \includegraphics[width=3in]{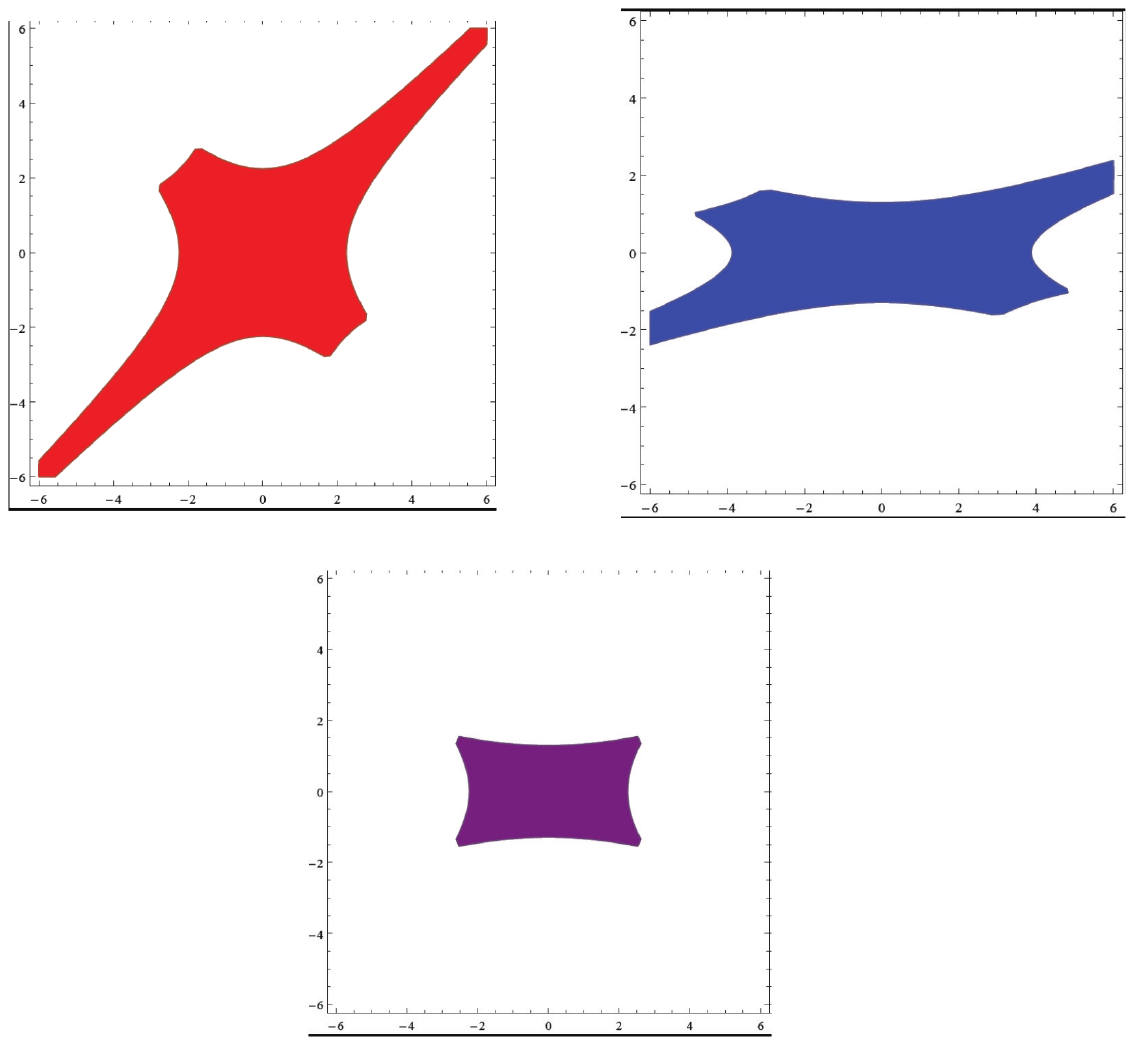}
 \end{center}
\caption{\small Illustrating the regions $X_{\kappa}$, $X_{\kappa}^w$ and their intersection.
Here we choose a single chiral superfield $\phi$ with $W=\i \phi^2$ and $\zeta=1$. The
region $X_{\kappa}$ for $\kappa=\i$, $c=5$, and $c'=-5$ is illustrated in the upper left figure. The region is noncompact,
with northeast and southwest boundaries asymptoting to the line $v=u$. Under the flow $u\to e^w u$,
$v\to e^{-w} v$ the region evolves (for $e^{2w}=3$) to the blue region shown in the upper right figure.
Again this region is noncompact.
The intersection, shown below in purple is compact for all $w>0$ and decompactifies as $w \to 0$.}
 \label{fig:XKappa-Regions}
\end{figure}

 To show the boundedness, we proceed as follows.
 Let $X_\kappa$ be the portion of the target space $X$ of the $\sigma$-model in which $W$ takes values in $T_\kappa$.
 And let $X_\kappa^w$
be the subset of $X$ that $X_\kappa$ flows to
 under $\zeta$-soliton flow for ``time'' $w$.  The key point is now to show that under suitable conditions, $X_\kappa^w\cap X_\kappa$ is compact
 for all $w>0$.  For if $\L$ and $\L'$ are any (closed) Lagrangian submanifolds  of class $T_\kappa$, then the intersection $\L^w\cap \L'$ is a closed subset of $X_\kappa^w\cap X_\kappa$, and so is compact if $X_\kappa^w\cap X_\kappa$ is compact.  As usual, the compactness of $\L^w\cap \L'$ will ensure that the space of supersymmetric states of the $(\L,\L')$ system is well-defined. See Figure \ref{fig:XKappa-Regions}.

 To understand the compactness of $X_\kappa^w\cap X_\kappa$ without any excessive clutter, let us take $\zeta=1$ and $\kappa=\I$.
  So the $\zeta$-instanton equation is
 Hamiltonian flow for the Hamiltonian $H=-\half\mathrm{Re}\,(\zeta^{-1}W)=-\half \mathrm{Re}\,W$.  The ``time''-dependence of $\mathrm{Re}(\kappa^{-1}W)=\mathrm{Im}\,W$
 along the flow is
 \begin{equation}\label{toro}
 \frac{\d }{\d x}\mathrm{Re}(\kappa^{-1}W)=  \frac{\d }{\d x}\mathrm{Im}\,W=\{H,\mathrm{Im}\,W\}=
 -\half \{\mathrm{Re}\,W,\mathrm{Im}\,W\}=- \frac{1}{4} |\d W|^2.\end{equation}
 Here $\{~,~\}$ is the Poisson bracket computed using the symplectic form of the target space $X$, and we have evaluated this Poisson bracket using the Cauchy-Riemann equations obeyed by the holomorphic function $W$. (In general we have
 $\frac{\d }{\d x}\mathrm{Re}(\kappa^{-1}W) = \frac{1}{4} \Im(\frac{\zeta}{\kappa}) |dW|^2$.)

 Usually, we are interested in models in which $|\d W|^2$ goes to infinity at infinity along $X$, and hence also along $X_\kappa$.  (For example, in the most standard
 Landau-Ginzburg model, $X=\IC^n$ for some $n$ and $W$ is a polynomial that is sufficiently generic so that $|\d W|^2$ grows polynomially at infinity.)  In this case,
 eqn. (\ref{toro}) implies that the rate at which  $\mathrm{Re}(\kappa^{-1}W)$ increases under $\zeta$-soliton flow increases near infinity in $X_\kappa$.  This means
 that $\zeta$-soliton flow for any positive ``time'' $w$ maps  a neighborhood of  infinity in $X_\kappa$  strictly outside of $X_\kappa$ (this neighborhood
 depends on $w$), and hence $X_\kappa^w\cap X_\kappa$ is indeed compact for all $w>0$.

 In general, we might not want to assume that $|\d W|^2$ goes to infinity at infinity along $X$, but it is always reasonable to assume that $|\d W|^2$ is bounded above
 0 near infinity.  (Otherwise, $W$ has a critical point at infinity and one should not expect to get a good description based only on the set $\IV$ of finite
 critical points.)  With $|\d W|^2$ bounded above 0, the same reasoning as before shows that if $w$ is sufficiently large, then $X_\kappa^w\cap X_\kappa$ is compact.

 Putting these statements together, under reasonable conditions,
 the space of supersymmetric $(\L,\L')$ states is well-defined for any $\L,\L'$ of class $T_\kappa$. In the above, we took $\kappa =i\zeta$, but the same reasoning applies as long as $\kappa\not=\pm \zeta$.
  Saying that the space of $(\L,\L')$ strings is well-defined means that it invariant under Hamiltonian symplectomorphisms of $X_\kappa$
  (applied separately to $\L$ and $\L'$), invariant under changes in the K\"ahler metric of $X$ (as long as this is not changed too drastically at infinity)
  and invariant under changes in $\kappa$ (as long as one keeps away from $\kappa=\pm \zeta$).

\subsection{The Fukaya-Seidel Category}\label{seidelfukaya}

 Having come this far, it is not too hard to understand how to go farther and define
 open-string tree amplitudes for branes of class $T_\kappa$.  (By contrast, one cannot do this for $W$-dominated branes, as we will soon explain.)

These open-string tree amplitudes -- when specialized to the case of just one string in the future, as discussed below -- give what would be
called mathematically an $A_\infty$ algebra if one considers just one brane of class $T_\kappa$, or an $A_\infty$ category
if one considers all of them.    The $A_\infty$ category that we obtain is presumably the Fukaya-Seidel
category \cite{SeidelBook,SeidelOne,SeidelTwo,SeidelOlder}, or its close cousin, the Fukaya category of the superpotential.  (These are expected\footnote{For example, see the end of section 2 of \cite{SeidelOlder}, where the Fukaya cateory of the superpotential is called $F(\pi)$ and the Fukaya-Seidel category is called $A$.  This and other matters described in the next paragraph were explained to us by N. Sheridan.}
to have the same derived categories of branes.)

It seems that, mathematically, it is understood that the Fukaya-Seidel category
 should have a definition along the lines of what we sketch below, but this has not yet appeared in the literature because of analytical details.
 The existing literature is thus based on alternative approaches that circumvent some analytical difficulties but will be less transparent to a quantum field theorist.
 Also, some of the details of the setup we use here seem fairly natural from a quantum field theory point
of view, but a  rigorous approach might use somewhat different definitions because
purely from the standpoint of  partial differential equations, one can make some more general choices and this freedom might be useful.  For example, instead of branes of class $T_\kappa$, one
could consider branes whose image in the $W$-plane is the union of a semi-infinite ray and a compact set.  Similarly, instead of using the global $\zeta$-instanton
equation, one can consider a more general equation that looks like the $\zeta$-instanton equation near the infinite ends of the worldsheet.   Despite some detailed differences in
approach, we expect the open-string
amplitudes that we define to have essentially the same content as the Fukaya-Seidel category.

To define open-string amplitudes, it is important to spell out a consequence of the restriction $\kappa\not=\pm\zeta$. Concretely, to compute the space of supersymmetric $(\L,\L')$ strings, we quantize the $\sigma$-model on a strip $S$ in the $x-\tau$ plane
 that is defined by $x_\ell\leq x\leq x_r$
 with $x_r-x_\ell=w$. This is a strip that runs in the $\tau$ direction.
  We construct an MSW complex as usual with a basis given by solutions of the $\zeta$-soliton equation and a differential found
 by counting solutions of the $\zeta$-instanton equation.   The reason that we specified that the strip $S$ runs in the $\tau$ direction
  is that, unlike the equation for a pseudoholomorphic curve that is usually considered in the $A$-model, the $\zeta$-instanton
 equation is not invariant under rotation of the $x-\tau$ plane.  If we rotate $S$ in the $x-\tau$ plane, so that $S$ is at an angle $\vartheta$ to
 the $\tau$-axis, this would be equivalent to replacing $\zeta$ by $\zeta e^{i\vartheta}$.  Since the only restriction on $\zeta$ is $\zeta\not=\pm\kappa$,
 we may rotate $S$ by an angle $\vartheta$ as long as $\zeta e^{i\vartheta}\not=\pm \kappa$.  For example, if $\kappa=i\zeta$,  we may take $S$ to be a strip
 propagating in any direction in the $x-\tau$ plane except the horizontal.
 This restriction on the slope of $S$ means that $S$ has a well-defined ``past'' (an end with $\tau\to -\infty$) and ``future'' (an end with $\tau\to +\infty$).

 The procedure for defining the open-string amplitudes in this situation is standard, except for a few key details.   Let us recall that in the usual $A$-model without a superpotential
 (or in physical string theory), to compute a tree-level amplitude with $q$ open strings,
 one takes the string worldsheet  to be a disc $H$ with $q$ marked points on its boundary.  The regions on the boundary
 between the marked points are labeled by branes and the marked points are labeled by vertex operators.   Modulo conformal transformations,
 $H$ depends on $q-3$ real moduli.  To compute the usual $A$-model amplitudes, we count (with signs) all pseudomolomorphic maps from $H$ to the target space
 $X$, obeying conditions determined by the choices of branes and vertex operators.  In the counting, we do not specify {\it a priori} the conformal
 structure of $H$ and we include pseudoholomorphic curves  with any values of their moduli.

 To define open-string amplitudes in the present context,
 roughly speaking, we do the same thing, with the equation for a pseudoholomorphic map replaced by the $\zeta$-instanton
 equation.  There are some changes because the $\zeta$-instanton equation is not conformally-invariant or even rotation-invariant.  We take $H$ to be a
 region in the complex plane, where we know how to define the $\zeta$-instanton equation.\footnote{It is possible but not necessary for our purposes here  to generalize this slightly -- $H$ could be a Riemann surface
with boundary with strip-like ends with a not necessarily holomorphic trivialization of its canonical line bundle and some
conditions on how the trivialization behaves at infinity and along the boundary.}
  Without conformal invariance, there is no direct equivalence between
 states and vertex operators, so we represent the external strings by semi-infinite strips of specified
 widths.  These strips are not allowed to be horizontal,
 because then in the case of branes of class $T_\kappa$, the external string states would not be well-defined, as was just explained.
 So in contrast to physical string theory (or the usual $A$-model), there is a well-defined
 distinction between string states that come in from the ``past'' ($\tau\to-\infty$) and those that go out to the future ($\tau\to+\infty$).  There are well-defined
 amplitudes with any number $n$ of strings coming in from the past and any number $m$ going out to the future.  However, in the context of the Fukaya-Seidel category or the Fukaya
 category of the superpotential, it is usual to consider only the case $m=1$.   (This case leads to amplitudes that can be put in a convenient algebraic framework -- an $A_\infty$
 algebra -- and whose counterparts under mirror symmetry are relatively well-understood.)

  \begin{figure}
 \begin{center}
   \includegraphics[width=3.9in]{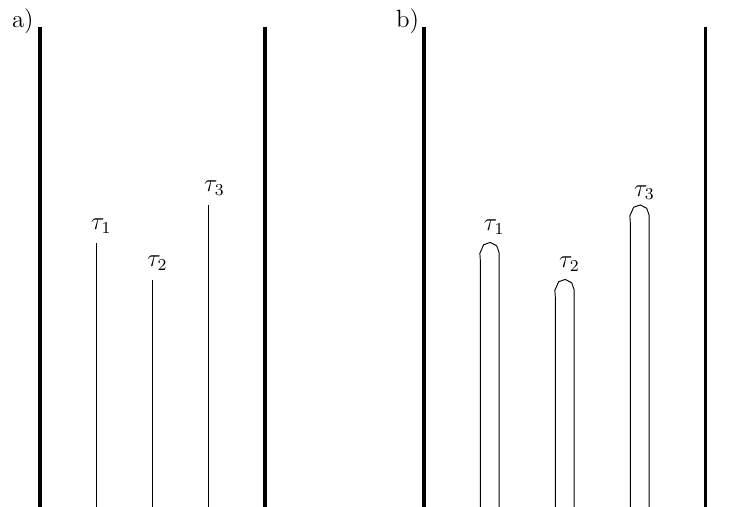}
 \end{center}
\caption{\small  (a) An open-string worldsheet $H$ in a form familiar in  light-cone gauge.  $n$ open strings all of width $w$ come in from the past ($\tau=-\infty$) and a single one of width $nw$
goes out to the future ($\tau=+\infty$).  There are $n-1$ values of $\tau$ at which two open strings combine to one.  The
 linearly independent differences between these critical
values of $\tau$ are the $n-2$ real moduli of this worldsheet.  (b)  The picture in (a) can be slightly modified in this fashion -- if one wishes -- so that $H$
becomes smooth. The moduli are still the differences between the critical values of $\tau$.} \label{openstrings}
\end{figure}
 For brevity, we will consider only the case $m=1$.  The total number of external string states is therefore $q=n+1$, and the number of real moduli is $q-3=n-2$.
 To define the $n\to 1$ amplitude, we need a family of regions $H$ in the $x-\tau$ plane that depend on the usual $n-2$ real moduli of a disc with $n+1$ marked
 points on its boundary.  These regions are far from being uniquely determined.  One convenient choice  is  the parametrization of the moduli
 space of a disc with $n+1$ punctures that  in ordinary string theory is used  in light cone gauge (for the case that all incoming particles
 have equal $p_+$).  The strings coming from the past all have equal width $w$ and the string going out to the future has width $nw$;
 as usual in light cone gauge, the moduli are the differences between
 the values of $\tau$ at which two strings join.  This is depicted in Figure \ref{openstrings}(a).
 If one prefers, one can use the less singular worldsheets of Figure \ref{openstrings}(b).

 As long as all branes considered are of class $T_\kappa$, the counting of $\zeta$-instanton solutions to define string amplitudes in this situation
 is well-defined, basically because the properties of the branes that make the external string states
 well-defined also ensure that solutions of the $\zeta$-instanton equation
 cannot go to infinity.   The resulting tree amplitudes have all the usual properties except cyclic symmetry.  Lack of cyclic symmetry means that
 these amplitudes cannot be derived in a natural way from a $\Q$-invariant effective action (except possibly by introducing separate fields to represent
 incoming and outgoing strings) but can be interpreted as constructing a nonlinear $\Q$ operator (which acts on a Fock space of open strings).
 Mathematically, lack of cyclic symmetry means that one gets an $A_\infty$ algebra without a trace.

This construction would not work for $W$-dominated branes.  $W$-dominated branes lead to well-defined
spaces of BPS states, but not to tree amplitudes. The reason is that in trying to define a tree amplitude for $W$-dominated branes,
there is no natural way
to decide if we should use a left-brane (with $\Im(\zeta^{-1} W) \to + \infty$ at infinity)
 or a right-brane  (with $\Im(\zeta^{-1} W) \to - \infty$ at infinity) on the intermediate boundaries
 in Figure \ref{openstrings}(b).  Moreover, neither choice leads to a well-controlled counting.

 On the other hand, for branes of class $T_\kappa$, the definition of  $n\to 1$ amplitudes by counting
 of $\zeta$-instantons works fine
for any $\kappa \in U(1) - \{ \pm \zeta \}$. If $\kappa$ and $\kappa'$ are in the
same connected component of $U(1) - \{ \pm \zeta \}$, the  categories associated to $\kappa$ and $\kappa'$
are naturally equivalent to each other via rotation. Thus, using branes of class $T_\kappa$ (with the same
${\kappa}$ on all connected components
of all boundaries), by counting $\zeta$-instantons in Figure \ref{openstrings}
we can define two $A_\infty$ categories $\fB\fr_{\pm \kappa}$.
In Section \S \ref{notif} below we will argue that the two categories $\fB\fr_{\pm \kappa}$
 are $A_\infty$-equivalent to the two categories of branes
 constructed in the abstract part of this paper using
 left- and right- thimbles
 of class $T_\zeta$  and positive or negative half-plane webs.

We should stress that even for branes of class $T_\kappa$, while we can define tree-level amplitudes
in parallel with standard tree-level $A$-model amplitudes, we cannot define analogs of higher genus $A$-model
amplitudes.  This is basically because with $W\not=0$, the standard $A$-model twisting is not available.  In the future,
to avoid repeating ourselves many times, we will use the phrase ``$A$-model'' to refer to either a standard $A$-model with
$W=0$, or a partial $A$-model with $W\not=0$ in which one considers only tree amplitudes.

\section{MSW Complex On The Real Line: Solitons And Instantons}\label{subsec:MorseComplexRealLine}

Up to this point, our preliminary discussion of BPS solitons has been purely classical.
We  now want to study these BPS solitons at the quantum level.  Everything about this problem will closely parallel the general quantum mechanical
analysis of section \ref{review}, except that we have to take into account
 certain zero-modes associated to broken bosonic and fermionic symmetries; these do not have
a close analog in the generic quantum mechanical case.

Let $\CS_{ij}$ denote the set of classical $ij$ solitons, that is solutions of the $\zeta$-soliton equation that
interpolate from $\phi_i$ at $x=-\infty$ to $\phi_j$ at $x=+\infty$.
From eqn. (\ref{hopeful}), we know that such solutions exist only for $\zeta=\zeta_{ji}$,
so in what follows we choose that value of $\zeta$.  The group $\IR$ of translations of the $x$-axis acts freely on the space of classical $ij$ solitons,
and generically an $ij$ soliton has no bosonic zero-mode except the one associated to translation invariance.  We will assume that we are in
this situation.  A classical $ij$ soliton always has a pair of fermionic zero-modes, one of fermion number $\FF=1$ and one of fermion number $-1$,
generated by the 2 supercharges that are not in the small subalgebra.  When the zero-mode associated to translation symmetry is the only
bosonic zero-mode, the 2 zero-modes associated to broken supersymmetries are the only fermionic ones.

In this situation, the quantization of a classical $ij$ soliton solution, in perturbation theory, is relatively straightforward.  The non-zero bosonic
and fermionic modes are simply placed in their ground state.  The only subtlety here is that one must determine the fermion number $f_0$ of
the ground state of the nonzero modes of the fermions.  We consider this question presently.  The quantization of the bosonic and fermionic zero-modes
is slightly subtle but is well-known.  If we write $a$
for the translational zero-mode of the soliton, then a wavefunction $e^{ipa}$ for this mode describes a soliton in a state of arbitrary momentum $p$.
A quantum BPS state invariant under the small supersymmetry algebra generated by $\Q_\zeta$ and $\bar \Q_{\zeta}$
 arises for $p=0$.  (A soliton in a momentum eigenstate
with $p\not=0$ is invariant under a boosted version of this algebra.)  The fermion zero-modes are a pair of operators $\chi_0$, $\bar\chi_0$ of $\FF=\pm 1$
generating a two-dimensional Clifford algebra; the representation of this algebra gives a pair of states of fermion number $\FF=\pm 1/2$.
So overall, the quantized soliton in perturbation theory can have any momentum and has fermion number $f_0\pm 1/2$.

\subsection{The Fermion Number}\label{ferminumber}

The fermion number $f_0$ of the Fock vacuum (of non-zero fermion modes) is formally the fermion number of the filled Fermi sea.  This of course
diverges and needs to be regularized.  The physically natural approach is to define first the renormalized, conserved, Lorentz-covariant fermion number current
(for example, via a process of point-splitting and normal-ordering) and then compute the matrix element in the soliton state, in perturbation theory,
of the integral of the fermion charge density.   The point-splitting and normal ordering deal with ultraviolet divergences in the definition of $f_0$.
And even if $D$ has infinite ends (for instance $D=\Bbb R$),
there is no infrared problem in defining the integrated fermion number
because, by virtue of  Lorentz invariance, the expectation value of the fermion charge density vanishes in the massive vacua  at $x\to \pm\infty$. (It turns out that there is a subtlety at
finite distance boundaries of $D$ rather than at ends at infinite distance; see section \ref{fnagain}.)

An alternative procedure is more convenient for some purposes.  In its standard form that
we describe first, this procedure
deals with ultraviolet divergences but not infrared ones, so it applies for the case that space is a compact 1-manifold $D=[x_\ell,x_r]$ (with boundary
conditions as in section \ref{boundary}).
 In what follows, by
 a ``fermion state,'' we mean an energy level of the single-particle Dirac equation that governs the fermions of $\FF=1$.  (We need not discuss separately
the single-particle $\FF=-1$ modes; they are canonically conjugate to the $\FF=1$ modes, and the Fock vacuum can be completely characterized by
saying which $\FF=1$ modes annihilate it.)  From $f_0$, which formally is the number of filled states of negative energy, formally we subtract  a constant, namely 1/2 the total number of fermion states.
Subtracting this constant can be thought of as measuring the fermion number of the soliton relative to the fermion number of the vacuum.  Still formally,
with this subtraction,
 $f_0$ is $1/2$ of the number of $\FF=1$ fermion modes of negative energy (the ones that are filled in the Fock vacuum)
minus $1/2$ the number of $\FF=1$ states of positive energy (the ones that are unfilled).   This result, whose quantum  mechanical
analog is eqn. (\ref{hoobo}),  still needs to be regulated.
 We weight a mode of
energy $E$ by a factor of $\exp(-\varepsilon |E|)$, for small positive $\varepsilon$, and take $\varepsilon\to 0$ at the end of the calculation.
Thus if $\Bbb T$ is the set of all {\it non-zero energy}  fermion modes of $\FF=1$, we define $f_0$ as
\begin{equation}\label{momzo} f_0=-\frac{1}{2}\lim_{\varepsilon\to 0}\sum_{i\in \Bbb T} \exp(-\varepsilon |E_i|) \mathrm{sign}(E_i).  \end{equation}
This formula is the general formula (\ref{hoobo}) of supersymmetric quantum mechanics for the fermion number of a state associated to a critical
point, except that in infinite dimensions we require a regulator, such as $\exp(-\varepsilon|E|)$, and in the finite-dimensional problem with a nondegenerate Morse function, there is no need to discuss fermion zero-modes.
The $\eta$-invariant of the single-particle Dirac Hamiltonian $\mathcal D$ (this is the operator
whose eigenvalues are the $E_i$; see eqn. (\ref{dolbo})), is usually defined as the ``limit''
\footnote{Unlike the sum in (\ref{momzo}), the one in (\ref{omzo}) is not absolutely convergent.  So we should specify that
the meaning of the ``limit''  in this equation is that
the sum over states on the right hand side  of eqn. (\ref{omzo}) converges for sufficiently large $\mathrm{Re}\,\veps$ and defines
an analytic function of $\veps$ that has an analytic continuation to $\veps=0$ where it is non-singular.  Also, $\eta$ is sometimes defined to include a contribution from the zero-modes,
but here it will be more convenient to omit them.}
\begin{equation}\label{omzo}\eta(\mathcal D)=\lim_{\varepsilon\to 0}\sum_{i\in \Bbb T}|E_i|^{-\varepsilon} \mathrm{sign}(E_i),\end{equation}
but the precise choice of regulator does not matter; one can here replace   $|E|^{-\varepsilon}=\exp(-\varepsilon\log|E|)$
with $\exp(-\varepsilon|E|)$, as in (\ref{momzo}).  So
\begin{equation}\label{mezzo}f_0=-\frac{\eta(\D)}{2}. \end{equation}
Including the contributions of the zero-modes, the soliton has two states of fermion number $f_0\pm 1/2$.
In other words, the two states, which we will call $\Psi_{ij}^f(p)$ and $\Psi_{ij}^{f+1}(p)$ (where $p\in \CS_{ij}$ labels a particular classical $ij$ soliton),
have fermion numbers $f$ and $f+1$, with
\begin{equation}\label{zelbor} f= f_0-\frac{1}{2}=-\frac{\eta(\D)+1}{2}=-\frac{\eta(\D+\varepsilon)}{2}.\end{equation}
This is the proper formula for $f$ on a compact manifold without boundary.  In our application, $D$ always has boundaries and/or infinite
ends.  In the presence of an infinite end, an infrared regularization of $\eta$ is required, as in eqn. (\ref{zonf}) below.  In the presence of a boundary,
the formula for $f$ requires
a boundary correction that is explained in section \ref{fnagain}.

A shortcut to find the appropriate Dirac operator $\D$ whose $\eta$-invariant enters this formula is to use the general formalism of supersymmetric
quantum mechanics, as reviewed in section \ref{review}.  In general, the Hamiltonian operator acting on the fermions (which in the quantum mechanical
context is the fermion mass matrix $\partial^2 h/\partial u^i\partial u^j$) is the  operator that arises in linearizing the equation ($\partial h/\partial u^i=0$)
for a critical point.   In our present context, the equation for a critical point is the $\zeta$-soliton equation and its linearization is the condition
\be\label{dolbo}
 \frac{\p  }{\p x}  \delta \phi^I -\frac{\I \zeta g^{I\bar J}}{2} \frac{\p^2 \bar W}{\p \bar\phi{}^{\bar J}\p\bar\phi{}^{\bar K}} \delta \bar \phi{}^{\bar K}=0,
\ee
together with the complex conjugate of this equation.  Writing the left hand side of (\ref{dolbo}) as a linear operator acting on the pair
$\begin{pmatrix}\delta\phi^I\cr \delta\bar\phi{}^{\bar J}\end{pmatrix}$ and including the complex conjugate equation,
we arrive at a formula for the appropriate Dirac operator:
\be\label{eq:NewD}
\D =  \sigma^3 \I \frac{\d}{\d x} + \begin{pmatrix} 0 & 0 \\ 1 & 0 \\ \end{pmatrix} \frac{\zeta^{-1}}{2}g^{J\bar I}\frac{\partial^2 W}{\partial \phi^J\partial\phi^K}
+ \begin{pmatrix} 0 & 1 \\ 0 & 0 \\ \end{pmatrix} \frac{\zeta}{2}g^{I\bar J}\frac{\p^2 \bar W}{\p \bar\phi{}^{\bar J}\p\bar\phi{}^{\bar K}} .
\ee
Here the Hessians of $W$ and $\bar W$ are evaluated on the soliton configuration and $\sigma^3$ is short for
$$
\begin{pmatrix} \delta^I_{~K} & 0 \\ 0 & - \delta^{\bar I}_{~\bar K} \\ \end{pmatrix}.
$$

The standard definition of the $\eta$-invariant that we have given above assumes that $\D$ has a discrete spectrum, so it applies for quantization on
$D=[x_\ell,x_r]$ (where, however, it does not quite give the complete answer for the fermion number, as we explain in section \ref{fnagain})
but not otherwise.  In the case of a BPS soliton with $D=\R$, because there is a mass gap at infinity, $\D$ has a  discrete spectrum near
zero energy, but it has a continuous spectrum above some threshold.  One needs to generalize slightly the definition of the $\eta$-invariant in this situation  (for a rigorous treatment, see \cite{Muller}).
The contribution of the discrete spectrum does not need any change (there are only finitely many states in the discrete spectrum, and their
contribution to the $\eta$-invariant is actually simply the number of positive energy normalizable eigenstates of $\D$ minus the number of negative
energy normalizable eigenstates of $\D$).  The contribution of the continuous spectrum needs to be defined more precisely.  If $P$ is the orthogonal
projector onto the continuous spectrum of $\D$, the contribution of the continuous spectrum to the $\eta$-invariant is
\begin{equation}\label{zonf}\eta^{\mathrm{cont}}(\D)=\lim_{\varepsilon\to 0}
\int_{-\infty}^\infty \d x\,\sum_{s=1}^{2d}\bigl\langle x,s| P \,\mathrm{sign}(\D)\exp(-\varepsilon|\D|)|x,s\bigr\rangle. \end{equation}
As usual, $|x\rangle$ is a state with delta-function support at a point $x\in \IR$, and $s$ parametrizes
the additional labels carried by such a state.\footnote{The operator $\D$ acts on a fermi field valued in the tensor product of a two-dimensional Clifford module with
the pullback of the complex tangent bundle of the target space $X$, which has rank $d=\mathrm{dim}_{\IC}\,X$.  So $s$ runs over
an orthonormal  basis of a $2d$-dimensional vector space.}
The definition of the operator $\mathrm{sign}(\D)$ is potentially troublesome because of zero-modes of $\D$, but the zero-modes are in the discrete spectrum
and so are annihilated by $P$, and hence  there is no problem in defining the product
$P\,\mathrm{sign}(\D)$.  The fact that is being generalized  in the formula (\ref{zonf}) is that  if $M$ is an operator
of finite rank  or more generally of ``trace class'' (represented by an $x$-space kernel that we also call $M$), then $\Tr\,M= \int_{-\infty}^\infty\d x\,\sum_s
 \langle x,s|M|x,s\rangle$.  The contribution of the continuous spectrum to the $\eta$-invariant is
supposed to be, naively, $\lim_{\varepsilon\to 0}\Tr\, P\,\mathrm{sign}(\D)\exp(-\varepsilon|\D|)$, but this trace is only conditionally-convergent
 (the operator
whose trace we are trying to take is not trace class) and is not well-defined.  The formula (\ref{zonf}) is a well-defined, regularized version of this trace.   It is well-defined because
the integrand $\sum_s\bigl\langle x,s| P \,\mathrm{sign}(\D)\exp(-\varepsilon|\D|)|x,s\bigr\rangle$ vanishes exponentially for $x\to \pm \infty$.  This is so for the same reason
that there is no infrared problem in the approach to defining $f_0$ via point-splitting and normal-ordering:  the fermion number density vanishes in the vacua
at $\pm \infty$.

\subsection{Properties Of The $\eta$-Invariant}\label{properties}

What can we say about the invariant $\eta(\D)$?  There is no simple formula for $\eta(\D)$ for a particular $ij$ soliton solution,
but there is a useful general statement comparing the values of $\eta(\D)$ for different $ij$ solitons.
(Since we have not yet analyzed boundary
contributions to the fermion number, the following analysis applies strictly for the case $D=\IR$, but
similar statements hold for other cases.
See section \ref{fnagain}.)

Suppose that $p,p'\in \CS_{ij}$ are two different $ij$ soliton solutions, interpolating between the same vacua at both ends.  They have
two different Dirac operators $\D^p$ and $\D^{p'}$ and so two different $\eta$-invariants $\eta(\D^p)$ and
$\eta(\D^{p'})$ and two different fermion
numbers $f^p$ and $f^{p'}$.  These can differ, but  their difference $f^{p'}-f^{p}$ is always an integer.
The proof can be expressed in either mathematical or physical language.  To
express the proof in physical language first, we observe that the bosonic fields of the LG model do not
carry fermion number and hence
$\FF$ is conserved in the propagation of the fermion fields in an arbitrary time-dependent background constructed from the bosons.
So in particular (here we assume that $X$ is simply-connected), we can construct a time-dependent
background that interpolates from the soliton solution
$p$ in the far past to the soliton solution $p'$ in the far future; moreover, we can do this with fields
that are time-independent at spatial infinity.
So time evolution gives an $\FF$-conserving mapping from the fermion Hilbert space $\mathcal H^p$
constructed in the past by expanding around soliton $p$ to
the corresponding Hilbert space $\mathcal H^{p'}$ constructed in the future by expanding around $p'$.
All states in $\mathcal H^{p}$ have $\FF=f^p$
mod $\Bbb Z$, since the modes of the fermion field, which act irreducibly in $\mathcal H^p$, carry $\FF=\pm 1$.
Likewise all states in $\mathcal H^{p'}$
have $\FF=f^{p'}$ mod $\Bbb Z$.  So the existence of an $\FF$-conserving map between these two Hilbert
spaces implies that $f^p=f^{p'}$ mod $\Bbb Z$.

For a more mathematical version of this argument, observe first that in general, changing only finitely many
eigenvalues of an operator $\D$
does not change $\eta(\D)$ mod 2, since for $\varepsilon\to 0$, each eigenvalue contributes $\pm 1$ to $\eta(\D)$.
In varying finitely many eigenvalues,
$\eta(\D)$ only changes when an eigenvalue changes sign, in which case it jumps by $\pm 2$.  The same is true if one
varies infinitely many eigenvalues
provided that the change in the $n^{th}$ eigenvalue vanishes rapidly enough for $n\to\infty$.  We are in this situation
if we change $\D$ by varying
the $x$-dependent matrix $\partial^2 W/\partial\phi^I\p \phi^J
$ in an arbitrary fashion (replacing it with an
arbitrary $m_{IJ}(x)$, not necessarily derived from an LG field), keeping fixed its behavior for $x\to \pm \infty$.
Since we do not change $\D$ at spatial infinity, we do not change the
energies of states at large $|x|$; since we do not change the term $\sigma^3\i \d/\d x$ in $\D$ that dominates at high
energies, we do not change
the eigenvalues of high energy.  So in such a variation, only a finite number of eigenvalues change substantially, and
  $\eta(\D)$ only changes when an eigenvalue passes through 0.  When that happens, $\eta(\D)$  jumps by $\pm 2$, so that
$f=-(1/2)\eta(\D)$ is constant mod 1.  Note that in this argument (as opposed to the physical
argument), we do not need to assume that $X$ is simply-connected;
we can interpolate between
the Dirac operators $\D^p$ and $\D^{p'}$ whether or not we can interpolate between the solitons $p$ and $p'$.

Clearly,  there is a topological invariant, namely the value of $-\half \eta(\D)$ mod $\Bbb \Z$, which only depends on the matrices
$m_{IJ}=\partial^2 W/\partial \phi^I\partial\phi^J$ at $x=\pm \infty$.   How can one compute this topological invariant?
One way to compute the value of $-\half \eta(\D)$ mod $\Bbb \Z$ is to deform
 to the case that $m_{IJ}(x)$ varies adiabatically as a function of $x$, between its given limiting values at $x=\pm \infty$.
 The adiabatic condition is
that  if $|m|$ is the smallest eigenvalue of $m_{IJ}$, then $|m|^{-2}\d m_{IJ}/\d x$ is everywhere small.
Under this condition, $\eta(\D)$ can be computed
as  the integral over $x$ of a universal local expression constructed from $m$ and its first derivative.
A straightforward computation using perturbation theory then yields the formula
\be\label{explained}
f  =   \frac{1}{2\pi} \left( \arg \det\left.\frac{\partial^2W}{\partial\phi^I\partial\phi^J}\right|_{\phi_j} - \arg\det \left.\frac{\partial^2W}{\partial\phi^I
\partial\phi^J}\right|_{\phi_i} \right) ~\mod  ~\IZ.
\ee
This agrees with the formulae  stated in \cite{Fendley:1992dm,Cecotti:1992qh,Cecotti:1992rm}.
Incidentally, in the case of solitons on the real line,
one might wonder if one can write an exact formula for $f$, not just a mod $\IZ$ formula,
as the integrated ``winding number'' of the matrix of second derivatives of $W$:
\be\label{zex}  f\overset{?}= \frac{1}{2\pi}\int_{-\infty}^\infty \d x \,\frac{\d }{\d x} \arg\det \left.\frac{\partial^2W}{\partial\phi^I
\partial\phi^J}\right|_{\phi_i}.  \ee
This formula actually does not make sense, since in general $\det \,\partial^2 W/\partial \phi^I\partial \phi^J$
can have zeroes; moreover,
as one varies the K\"ahler metric of $X$, a $\zeta$-instanton trajectory can cross such a zero, whereupon the
right hand side of (\ref{zex}) would jump,
 contradicting fermion number conservation if  eqn. (\ref{zex}) were valid.

The continuous variation between the soliton solutions $p$ and $p'$ can also be used, in principle,
to compute the integer $f^{p'}-f^p$.  This integer
is the ``spectral flow,'' the net number of eigenvalues of $\D$ that pass through 0 in the downward direction, in interpolating from $p$ to $p'$.  This spectral
flow is a regularized version of the difference in the Morse index of the soliton $p'$ and the soliton $p$.

A standard argument in index theory says that the spectral flow of a Dirac operator in $d$ dimensions gives the index of a Dirac operator in $d+1$ dimensions.
We actually explained this argument for $d=0$ in section \ref{fermanom}, and this particular argument is  independent of $d$.  The argument
uses the fact that the $d+1$-dimensional Dirac equation can be written $L\psi=0$, where $L=\partial_\tau+\D$; here $\D$ is a $d$-dimensional self-adjoint Dirac operator
(for $d=0$, as in section \ref{fermanom}, $\D$ is simply a finite rank matrix).  The
index is the number of normalizable solutions of the equation $L\psi=0$  minus the number of normalizable solutions of the adjoint equation $L^\dagger\bar\psi=0$,
where  $L^\dagger=-\partial_\tau+\D$.
By reducing to the case that the eigenvalues of $\D$ vary adiabatically with $\tau$ and performing the analysis of eqn. (\ref{ooby}) for each eigenvalue,
one finds that the index of $L$ is the spectral flow of $\D$ (defined as the net number of eigenvalues of $\D$ that pass from positive
to negative between $\tau=-\infty$ and $\tau=+\infty$).  In the context of Morse theory, the $d+1$-dimensional Dirac operator $L$
is the linearization of the gradient flow equation.    In two-dimensional LG theory, the gradient flow equation is the $\zeta$-instanton equation
(eqn. (\ref{eq:instanton})) and the analog of $L$ is the linearization of this equation.  As is the case in supersymmetric quantum mechanics in general,
$L$ is the kinetic operator for the fermions of $\FF=1$ (and its adjoint is the corresponding operator for the fermions of $\FF=-1$).
 In a process involving an instanton transition from the soliton $p$ in
its ground state of lower fermion number to a soliton $p'$ in the analogous state, the fermion number $\FF$ changes by $f^{p'}-f^p$; everything is consistent
because this number is the index of $L$, and therefore equals the fermion number of the operator insertions that must be made to get a nonzero amplitude for this transition.

\subsection{Quantum BPS States}\label{quantumbps}

The basic framework to study BPS states in the full quantum theory is the same as in section \ref{review}.  We construct a complex
which additively is given by the semiclassical spectrum of BPS solitons, and on this complex, we use instantons to define a differential
$\hat \Q_\zeta$
(a normalized version of $\Q_\zeta$ with the values of $h$ for the classical soliton solutions removed).
The cohomology of this differential gives the exact quantum spectrum of BPS solitons.

So additively, the complex describing $ij$ solitons is
\be\label{eq:MorseComplexR}
\IM_{ij}= \oplus_{p\in \CS_{ij} } \left( \IZ \Psi^f_{ij}(p)\oplus \IZ \Psi^{f+1}_{ij}(p) \right)
\ee
where $\CS_{ij}$ is the set of intersection points of a left thimble of type $i$ with a right thimble
of type $j$.
The grading of the complex is given by the fermion number $\FF$.  This grading is not really a $\Bbb Z$-grading,
since the values of $\FF$ differ from integers as explained in (\ref{explained}), but it is shifted from a $\Bbb Z$-grading by an overall constant
that depends only on $i$ and $j$.

Just as in section \ref{review},  the
differential on the complex (\ref{eq:MorseComplexR}) arises from counting
instantons interpolating between states whose fermion number differs
by $+1$. In other words we consider solutions of the   $\zeta$-instanton
equation with $\zeta = \zeta_{ji}$ and boundary conditions
\be \label{bcond}\lim_{x\to -\infty}\phi(x,\tau)=\phi_i \qquad \im_{x\to +\infty}\phi(x,\tau)=\phi_j\ee
 together with
\be\label{ccond}
\lim_{\tau\to - \infty} \phi(x,\tau) = \phi^{p_1}_{ij}(x) \qquad
\lim_{\tau\to + \infty} \phi(x,\tau) = \phi^{p_2}_{ij}(x)
\ee
where $\phi^{p_1}_{ij}(x)$ and $\phi^{p_2}_{ij}(x)$ are two $ij$ $\zeta$-solitons.
In section \ref{genex}, we give an example showing that in general there are $\zeta$-instantons obeying these boundary conditions
and contributing to the differential $\hat \Q_\zeta$. So not all BPS solitons give rise to true BPS states.

In the  general quantum mechanical analysis of section \ref{review}, there were only two supersymmetries, one of which was a symmetry
of the instanton.  The instanton therefore had just 1 fermion zero-mode, and this mode was responsible for the fact that the instanton amplitude
increases $\FF$ by 1.  In the present situation, the underlying two-dimensional model  altogether has
four supersymmetries, only two of which (the ones in the small subalgebra)
are symmetries of the initial and final states.
As a result, the initial and final states both represent a rank 2 Clifford algebra of broken supersymmetries. This
 is responsible for the doubling of the spectrum:
a classical soliton corresponds to two quantum states of fermion numbers $f, f+1$.
The instanton still preserves only one supersymmetry, so now
there are three fermion zero-modes, of which one is normalizable.  The two zero-modes that are generated by supersymmetries that are not in the
small subalgebra
 are localized in space but not in time, so they are not normalizable (and do not contribute to the index of the operator $L$). They
 go over in the far future or past to the fermion zero-modes
of the individual solitons.  The fact that the zero-modes of the individual solitons can be extended to time-dependent zero-modes in the instanton background
means that the instanton amplitudes commute with the action of the Clifford algebra on the initial and final states.   The third fermion zero-mode in the field
of the instanton is normalizable and is
the analog of the single zero-mode of the quantum mechanical analysis.  It is localized in space and time and ensures that $\zeta$-instantons
contribute to the matrix element of the differential $\hat\Q_\zeta$ only in the case $f^{p_2}-f^{p_1}=1$.

The fact that the instanton amplitude commutes with the Clifford algebra means that we can write the complex a little more economically:
\be\label{TC}\IM_{ij}=\Bbb{W}\otimes \IM'_{ij}.\ee
Here $\Bbb{W}\cong \IZ\oplus\IZ$ is an irreducible module for the Clifford algebra generated by two basis vectors $\vert -\half \rangle,
\vert +\half \rangle$ with $\FF=-\half,+\half$, respectively, and $\IM'_{ij}$ is a reduced complex. Indeed, we can write $\Psi^f_{ij}(p) = \vert -\half \rangle \otimes m^{f_0}_{ij}(p)$
and $\Psi^{f+1}_{ij}(p) = \vert +\half \rangle \otimes m^{f_0}_{ij}(p)$ where $f_0 = f+\half$. Thus,
\be\label{MC}
\IM'_{ij}= \oplus_{p\in \CS_{ij} } \IZ m^{f_0}_{ij}(p).
\ee
%%%%
%%
%%Actually,  $\Psi^f_{ij}(p)$ is of the form
%%and in Section \S \ref{collective} below we will drop the factor $\vert 0 \rangle$.  Similarly we set
%
In Section \S  \ref{collective} below, we show that in matching to the web-based formalism we should
take what there was called $R_{ij}$ to be the complex generated by the states of upper fermion number:
\be\label{MCdp}
R_{ij}= \oplus_{p\in \CS_{ij} } \IZ   \Psi^{f+1}_{ij}(p)
\ee
The instanton-generated differential $\hat\Q_\zeta$ acts on the reduced complexes $\IM'_{ij}$ and $R_{ij}$.

Once we get rid of the doubling of the spectrum in this way, the analogy with the general quantum mechanical analysis of section \ref{review} is much closer.
As in that analysis, each instanton solution  that interpolates from a soliton
$p_1$ in the past to a soliton $p_2$ in the future -- and has no moduli except the minimum possible number  -- contributes
$\pm 1$ to the relevant matrix element of the normalized differential $\hat\Q_\zeta$ acting on the reduced complex.
The sign of this contribution is given by the sign of the fermion determinant.
  The problem of defining
this sign -- and how it should be interpreted\footnote{The most powerful mathematical framework is provided by the theory of real determinant line bundles, as we discuss in the quantum mechanical case in Appendix \ref{sec:SQMSigns}.  Here we content ourselves with the following simple remarks
showing that as in the general quantum mechanical case, the theory is uniquely determined -- up to physically-understood choices -- if it works.} -- is very similar in the present
context to what it was in the general quantum mechanical discussion.  If $\pi_2(X)$ is trivial   then any two fields
obeying the boundary conditions (\ref{bcond}), (\ref{ccond})
are homotopic, so the fermion determinant is uniquely determined up to an overall sign.  Moreover, cluster decomposition
can be used to determine all the signs except for signs that can be absorbed in the definitions of the initial and final soliton states.
More generally, if $\pi_2(X)\not=0$, then the general analysis of the signs of the fermion determinants leads to the possibility of
discrete theta-angles, similarly to what happens in section \ref{review} when $\pi_1(M)\not=0$.

Now, let us consider an $ij$ soliton at rest with vacuum $i$ on the left and vacuum $j$ on the right.  If we turn the picture upside-down, rotating
by an angle $\pi$ in Euclidean signature, an $ij$ soliton becomes a $ji$ soliton.  The $\pi$ rotation, which is the Euclidean version of a CPT
transformation, also reverses the sign of the fermion
number current.  So if a multiplet of $ij$ solitons have fermion numbers $\FF=f,f+1$, then the rotation gives a pair of $ji$ solitons of fermion numbers $\FF=-f,-f-1$.
A static picture of an $ij$ soliton sitting at rest can be viewed as a pairing between an $ij$ soliton coming in from $\tau=-\infty$ and a $ji$
soliton coming \emph{in} from $\tau=+\infty$.  The path integral gives a non-degenerate and $\FF$-conserving pairing
\begin{equation}\label{ombo} \IM_{ij}\otimes \IM_{ji}\to \IZ. \end{equation}

Since the full complexes $\IM_{ij}$ admit this $\FF$-conserving pairing, the state $\Psi^{f+1}_{ij}$
must pair with the state $\Psi^{-f-1}_{ji}$ and the state $\Psi^{f}_{ij}$
must pair with the state $\Psi^{-f}_{ji}$. Therefore, the
 pairing on $\Bbb{W}$ is off-diagonal, pairing $\vert -\half \rangle $ with $\vert +\half \rangle$ to give
 (without loss of generality)
  $+1$.   Since the  pairing on the Clifford algebra has fermion number $0$ and the total pairing has fermion number $0$,
   the induced pairing on the reduced states $m_{ij}$ and $m_{ji}$ has fermion number
  $0$. That is, the induced pairing on the states of lower fermion number in each doublet,
  \begin{equation}\label{welz}K': \IM'_{ij}\otimes \IM'_{ji}\to \IZ,\end{equation}
  pairs $m_{ij}^{f_0}(p)$ with $m_{ji}^{-f_0}(p)$ and hence $K'$ has fermion number $0$.

When we explain the relation of the Landau-Ginzburg theory to the formalism of Section \S \ref{sec:RepWeb}
in Section \S \ref{collective} below it will turn out that we should not identify the
 reduced complex of solitons  $\IM'_{ij}$ with  what in the web treatment was called $R_{ij}$.  Rather,
$R_{ij}$ will actually be identified with the complex of soliton states of
``upper'' fermion number as in equation \eqref{MCdp}. The pairing $K'$ induces a similar pairing
\begin{equation}\label{elz}K:R_{ij}\otimes R_{ji}\to\IZ. \end{equation}
To define $K$ we simply drop the factor  $\vert +\half \rangle$ in both
$\Psi^{f+1}_{ij}(p) = \vert +\half \rangle \otimes m_{ij}^{f_0}(p)$
and $\Psi^{-f}_{ji}(p) = \vert +\half\rangle \otimes m_{ji}^{-f_0}(p)$,
and use the
pairing $K'$ on $m_{ij}^{f_0}(p)$ and  $m_{ji}^{-f_0}(p)$. Thus, $K$ is
a   nondegenerate pairing
of   fermion number $-1 +0 = -1$, in harmony with the definition of a web representation
in Section \S \ref{sec:RepWeb}. It is symmetric by CPT invariance.

\begin{figure}[htp]
\centering
\includegraphics[scale=0.3,angle=0,trim=0 0 0 0]{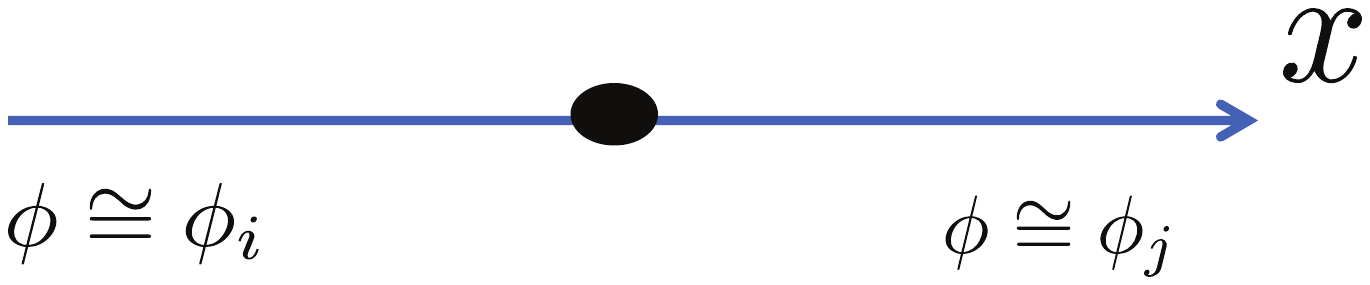}
\caption{In a massive theory, when viewed from long distance, or in the limit that the mass goes to
infinity, the soliton solution is well approximated by a discontinuous solution,
discontinuous at some point $x=x_0$.    }
\label{fig:SOLITON}
\end{figure}
\begin{figure}[htp]
\centering
\includegraphics[scale=0.5,angle=0,trim=0 0 0 0]{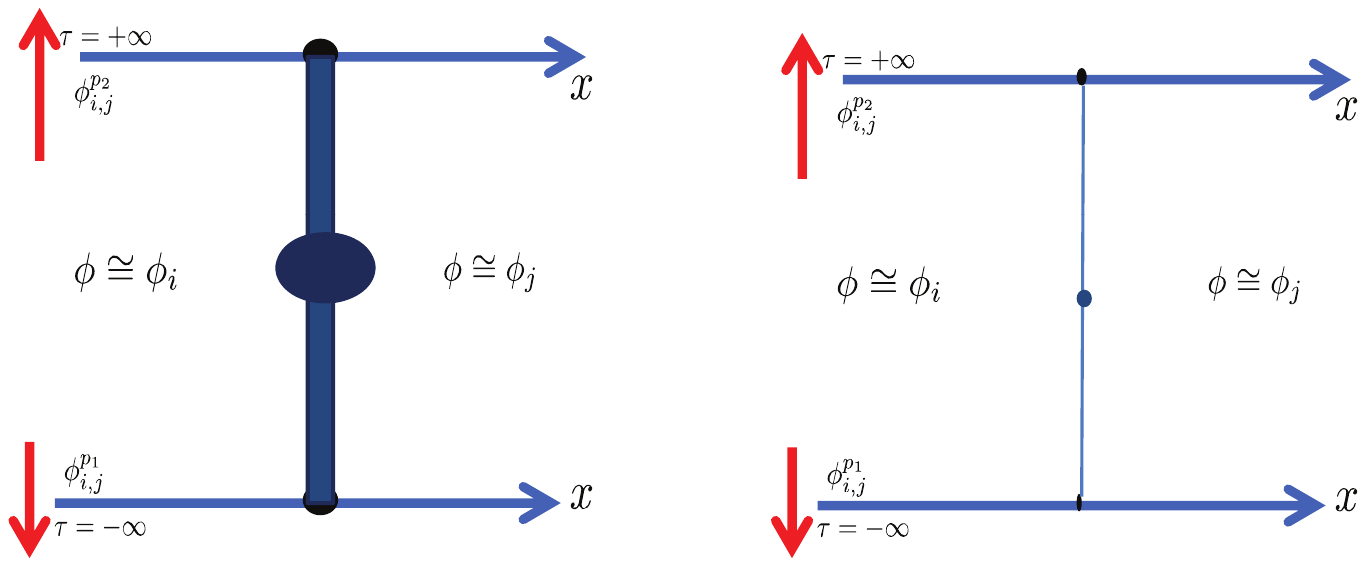}
\caption{Left: An instanton configuration contributing to the
differential on the   MSW complex. The black regions
indicate the locus where the field $\phi(x,\tau)$ varies vary significantly from the
vacuum configurations $\phi_i$ or $\phi_j$. The length scale here is $\ell_W$,
set by the superpotential.   Right: Viewed from a large
distance compared to the length scale $\ell_W$ the instanton looks like a
straight line $x=x_0$, where the vacuum changes discontinuously from vacuum $\phi_i$
to $\phi_j$. The nontrivial $\tau$-dependence of the instanton configuration,
interpolating from a   soliton $p_1$ to another soliton $p_2$ has been contracted to
a single vertex located at $\tau = \tau_0$. This illustrates the origin of the 2-valent
vertices of extended webs in the context of LG theory.  }
\label{fig:INSTANTON-ON-R}
\end{figure}

\textbf{Remarks}

\begin{enumerate}

\item
A soliton solution $\phi_{ij}(x)$ is very near $\phi_i$ or $\phi_j$ for ``most'' of
the values of $x$ and only shoots from vacuum $i$ to $j$ in a very short
interval $\Delta x$ set by the inverse mass scale $\ell_W$ of the theory.  Thus, ``in the infrared limit''
where we take the mass scale large the solution can be thought of as a discontinuous
function with vacuum $i$ at $x<x_0$ and $j$ at $x> x_0$.  See Figure \ref{fig:SOLITON}.
Similarly, the instantons can be viewed as stationary soliton worldlines with a
small dot inserted as in Figure \ref{fig:INSTANTON-ON-R}. This is the beginning of
the connection to the (extended) webs of previous sections.

\item We will show below that the counting of solitons and instantons leads to
web representations.  So, starting with the data defining a LG field theory, we can deduce a
mathematical structure that is defined over the integers, and this is how it was presented
in the first half of the paper. On the other hand,  the field theory is defined in terms of complex amplitudes
and vector spaces, and hence does not give an entirely natural explanation of why the mathematical
structures, such as equation \ref{ombo}, are in fact defined over $\Bbb Z$. (It is natural
for topological field theory path integrals to have integral values, but we are not
discussing topological field theory here.)

\item We can now introduce the Witten index, which in
this context is known as the  BPS   index $\mu_{ij}$. This is just the
Euler character of the complex $\IM_{ij}$ of \eqref{eq:MorseComplexR}
appropriately interpreted to take into account the fact that we are working
with a slightly degenerate Morse function. We should compute \cite{Cecotti:1992qh}
\be
\mu_{ij} := {\Tr}_{\IM_{ij}} F e^{i \pi F}  =  - \sum_{p \in L_i^{\zeta} \cap R_j^{\zeta  } \cap X_{W_0} } e^{\I \pi f(p)}
\ee
where $X_{W_0}$ is the preimage under $W$ of a regular value $W_0$ of the superpotential.
Here $  \zeta = \zeta_{ji}$ and $W_0$ lies on the interior of the line segment between
the critical values $W_i$ and $W_j$. As we have shown in equation \eqref{explained}
the fermion number of a classical soliton $\phi^p_{ij}$ has the form $f(p) = f_j - f_i + n_{ij}(p)$ where $n_{ij}(p)$
is an integer.  According to \cite{Cecotti:1992qh,Cecotti:1992rm,Hori:2000ck} the integer $n_{ij}(p)$, reduced mod 2, is the
contribution of $p$ to  the oriented intersection number of the Lefshetz thimbles, and hence
\be
\mu_{ij} =e^{\I \pi (f_j - f_i+1) } \# L_i^{\zeta} \cap R_j^{\zeta  }
 = e^{\I \pi (f_j - f_i+1) } \sum_{p \in L_i^{\zeta} \cap R_j^{\zeta  }\cap X_{W_0} } (-1)^{\iota(p)}
\ee
where $\iota(p)$ is the oriented intersection number.

\end{enumerate}

\subsection{Non-Triviality Of The Differential}\label{genex}

%%%
Some of the literature on BPS states in two-dimensional LG models
studies special cases which might give one the
impression that BPS solitons always lead to BPS states.
In this section we show that, in general, the differential $\Q_\zeta$ acting
on the space of classical $\zeta$-solitons is non-trivial.

The strategy will be to adapt a simple fact in ordinary Morse theory.  We consider a  family of Morse functions in one variable
$u$ that near $u=0$ look like
\begin{equation}\label{dolf} h_\varepsilon(u) = \frac{u^3}{3}-\varepsilon u ,\end{equation}
where $\varepsilon$ is a real parameter.
The equation for a critical point is $u^2=\varepsilon$. It has no real root for $\varepsilon<0$ but has a pair of real roots $u_\pm=\pm \sqrt{\varepsilon}$
for $\varepsilon>0$.  For $\varepsilon<0$, there are no classical vacuum states near $u=0$, but for $\varepsilon>0$, there are two of them.
However, extra quantum vacua cannot appear as $\varepsilon$ is varied, so there must be an instanton effect that lifts the two approximately
supersymmetric states that appear for $\varepsilon>0$.  Indeed, the portion of the $u$-axis between $u_+$ and $u_-$ is a gradient flow line
that connects the two critical points, and the contribution of this gradient flow line to the differential removes from the cohomology the states
supported at $u_+$ and $u_-$.

Notice that at $\varepsilon=0$, where the two critical points appear or disappear, $h_\varepsilon$ is not a Morse function, since its unique
critical point at $u=0$ is degenerate -- the second derivative of $h_0(u)$ vanishes at $u=0$.  Consider in any number of variables
a critical point that is degenerate in this way -- it is cubic  in one variable $u$ and quadratic  in any number of additional variables:
\begin{equation}\label{tofo} h(u,v_1,\dots, v_s, w_1,\dots, w_t)= \frac{u^3}{3}+\sum_{i=1}^s v_i^2- \sum_{j=1}^t w_j^2. \end{equation}
Under a generic perturbation, which will include a term linear in $u$,
such a critical point will either disappear or split into a pair of nondegenerate critical points, depending on the sign
of the perturbation.  If the sign the perturbation is such as to generate two new critical points,
there can be no exact quantum supersymmetric state associated
to them, since with the opposite perturbation, these critical points would be absent  even classically. So there will always be a gradient flow
line removing this pair of approximate ground states from the supersymmetric spectrum.

The $\zeta$-soliton equation on the real line is a problem of roughly this nature, with the role of the Morse function played by the functional
\be\label{defny}
h := -\half \int_{-\infty}^\infty\d x \,\Re\left( \frac{\I}{2} g_{I\bar J} \sum_I\phi^I\frac{\partial}{\partial x} \bar{\phi}{}^{\bar J} -  \zeta^{-1} W \right).
\ee
in the case that $g_{I\bar J}$ is constant.
$h$ is a function of infinitely many variables, and also we should factor out by translations of $x$ to think of $h$ precisely as a Morse function.
We will find a situation in which, near a certain critical point (and omitting the mode corresponding to spatial translations), $h$ will look like
(\ref{tofo}), but with infinitely many $v$'s and $w$'s.  By varying one parameter that will correspond to $\varepsilon$, we will be able to make
a pair of $\zeta$-solitons appear or disappear.  When these two classical solitons appear, for the same reasons explained above,
there will have to be a $\zeta$-instanton
interpolating between them and ensuring that the corresponding quantum states are not exactly supersymmetric.

The parameter playing the role of $\varepsilon$ can actually be either a parameter appearing in the K\"ahler metric of the target space $X$ or a parameter in the superpotential.
For $X=\IC$, varying the K\"ahler metric of $X$ will never cause a $\zeta$-soliton to appear or disappear,\footnote{\label{holfo}
For $X=\IC$ and $i,j\in\IV$, $ij$ $\zeta$-solitons are in 1-1 correspondence with paths in $X$ from $i$ to $j$ with $\mathrm{Re}(\zeta^{-1}W)$ constant.  Such a path, if properly
parametrized, becomes a $\zeta$-soliton.  The K\"ahler metric of $X$ enters only in determining the proper parametrization.}
but it is not difficult to give an example for $X=\IC^2$.  We parametrize $\IC^2$ with complex variables $Y,Z$.  To begin with, we take the K\"ahler metric
to be
\begin{equation}\label{dolof}\d \ell^2=|\d Y|^2+|\d Z|^2. \end{equation}
Later, we will make a perturbation of this K\"ahler metric.

We start with a 1-variable superpotential $W_0(Y)$, and for brevity we choose $\zeta^{-1}=\I$, so that the $\zeta$-soliton equation is ascending gradient flow
for $\mathrm{Re}\,W_0$ and a solution will have a fixed value of
$\mathrm{Im}\,W_0$.  We pick $W_0$ so that a non-trivial $ij$ $\zeta$-soliton $Y_0(x)$ does exist, for some $i,j\in\IV$.
After adding a constant to $W_0$,
we can assume that in this $\zeta$-soliton, $W_0(Y_0(x))$ is everywhere real and moreover that $W_0(Y_0(x))$ is negative for $x\to -\infty$
and positive for $x\to +\infty$.

In complex dimension 1, the operator $L$ obtained by linearizing around a $\zeta$-soliton has no zero-mode except the one associated to spatial translations.  (This is related to the
remark in footnote \ref{holfo}.)  To mimic the situation described in eqn. (\ref{tofo}), we need a $\zeta$-soliton that is ``degenerate,'' meaning
that the linearized operator $L$ has a zero-mode not associated to translation symmetry.   To achieve this, we include a second superfield $Z$
in the discussion, generalizing the superpotential to
\begin{equation}\label{monzo} W_1(Y,Z)=W_0(Y)(1+ Z^2).  \end{equation}
We have picked $W_1$ so that $\partial_ZW_1=0$ at $Z=0$, ensuring that (with the K\"ahler metric (\ref{dolof}))
the $\zeta$-soliton equation has the solution $Y=Y_0(x)$, $Z=0$.
Now let us expand around this solution.  The choice of $W_1$ ensures that the fluctuations in $Y$ and $Z$ obey separate equations. $Y$ has the
one zero-mode associated to spatial translations, and the linearized equation for $Z$ is
\begin{equation}\label{onzo}\frac{\d Z}{\d x}= W_0(Y_0(x)) \bar Z. \end{equation}
Setting $Z=z_1+\i z_2$ with $z_1,$ $z_2$ real and recalling that $W_0(Y_0(x))$ is real, we get the solutions
\begin{align}\label{tonzzo}  z_1(x) & = \exp\left( \int_0^x\d x'\, W_0(Y_0(x')) \right) \cr
                                          z_2(x) & = \exp\left(-\int_0^x\d x'\, W_0(Y_0(x')) \right) . \end{align}
Here, given our assumptions about $W_0(Y_0(x))$, $z_1(x)$ blows up exponentially at $x=\pm\infty$, but $z_2(x)$ vanishes
exponentially and thus we have found a $\zeta$-soliton with a zero-mode that is not associated to translation symmetry.

Now introduce a small parameter $u$ and consider  a family of approximate $\zeta$-solitons with
\begin{equation}\label{pilg} (Y,Z)=(Y_0(x),uz_2(x))+\O(u^2). \end{equation}
This obeys the $\zeta$-soliton equation to order $u$, but can we choose the $\O(u^2)$ terms to obey the equation in order $u^2$?
The case of interest is that this is not possible; that will  be so if and only if (after integrating out all the non-zero modes), the function $h$ of
eqn. (\ref{defny}), when expanded in powers of $u$, has a term of order $u^3$.  (The fact that $(Y,Z)$ defined in eqn. (\ref{pilg}) is a $\zeta$-soliton
up to order $u^2$ means that $h$ has no term linear or quadratic in $u$.)  The model as we have defined it so far has a symmetry $Z\to -Z$
which ensures that $h$ is an even function of $u$ and hence that there is no $u^3$ term.  However, we can easily modify the model (without
affecting the fact that $h$ has no term linear or quadratic in $u$) so
that $h$ acquires a $u^3$ term.  We simply add to the superpotential a term cubic in $Z$.  The superpotential
thus becomes
\begin{equation}\label{ilg} W_1(Y,Z)= W_0(Y)(1+Z^2) +A(Y)Z^3, \end{equation}
for a generic function $A(Y)$.

At this point, the solution $(Y,Z)=(Y_0,0)$ of the $\zeta$-soliton equation is a degenerate critical point of the functional $h$ that
can be modeled as in (\ref{tofo}) (with infinitely many $v$'s and $w$'s and an additional zero-mode - due to translation
invariance - that does not have a simple finite-dimensional analog).
To complete the story, we want to show that a perturbation of the K\"ahler metric or a further perturbation
of the superpotential can add to $h$ a term linear in $u$.
By varying the sign of this term, we can then make a pair of classical $\zeta$-solitons appear and disappear.  For the sign of the perturbation
for which the pair of classical solutions exists, there will have to be a $\zeta$-instanton interpolating between the two $\zeta$-solitons and
removing the corresponding quantum states from the supersymmetric spectrum.

A further perturbation of the superpotential that will have the desired effect is simply a term $\varepsilon B(Y)Z$,
with a generic function $B(Y)$ and a sufficiently small $\varepsilon$.  Thus the full superpotential becomes
\begin{equation}\label{nilg}W=W_0(Y)(1+Z^2) +A(Y)Z^3+\varepsilon B(Y)Z. \end{equation}

Alternatively, we can induce the desired linear term in $h$ by a small  perturbation of  the K\"ahler metric (\ref{dolof}) that breaks the symmetry $Z\to -Z$ of this metric.
We can choose a simple perturbation that preserves the flatness of the metric:
\begin{equation}\label{omolf}\d \ell^2\to \d \ell^2+ \varepsilon \left(\d Y \d\bar Z+ \d Z\d\bar Y\right). \end{equation}
For a generic superpotential $W_1(Y,Z)$ as constructed above, this perturbation adds to $h$ a term $\varepsilon u$ and ensures
that all of the desired conditions are satisfied.

\section{MSW Complex On The Half-Line And The Interval}\label{subsec:MorseComplexHalfLine}

Now we will consider the analogous questions when  the LG model is formulated  on  a two-manifold that is either a half-plane or a strip, in other
words it is  $\R\times D$ where $\R$ parametrizes time and $D$ is a half-line
$[x_\ell,\infty)$ or $(-\infty,x_r]$ or a compact interval $[x_\ell,x_r]$.
An important difference from the discussion of BPS solitons on the real line is that in the discussion of $ij$ solitons,
we set $\zeta$ to the unique value $\zeta_{ji}$ at which such solitons may exist.  By contrast, when we quantize
the LG model on a strip or a half-plane, we pick a generic value of $\zeta$ at which there are no $ij$ solitons
for any $i,j$.
\footnote{Using equation \eqref{hopeful} and Remark 9 of Section
   \S \ref{planewebs} this statement is equivalent to the criterion, used in the
   abstract part of the paper, that no difference of vacuum weights $z_{ij}$ is parallel
   to the boundary $\p \CH$.}
Wall-crossing phenomena can occur when $\zeta$ crosses the special values at which $ij$ solitons do
exist, and it is simplest to work away from these walls.

In quantization on either a half-line or an interval, we need a boundary condition at each finite end of
$D$. Any such boundary condition will explicitly break the $\N=(2,2)$ supersymmetry of the bulk
LG model to a subalgebra with at most two  supercharges,
and we want this to be the small supersymmetry algebra generated by $\Q_\zeta$ and its adjoint.  Since the
spatial translations and supersymmetries that are not in the small subalgebra are explicitly broken by the boundary
conditions, they do not lead to any zero-modes.
This makes the analogy with the generic quantum mechanical analysis of section \ref{review} much
more straightforward.  Generically, for each choice of  boundary condition,
the $\zeta$-soliton equation has only a finite set of solutions, none of which admit any bosonic
or fermionic moduli.  Each such solution corresponds to a single approximately supersymmetric state
of some fermion number $f$ (whose definition is analyzed in  section \ref{fnagain}).
As usual, we make a complex with a basis corresponding to the solutions of the $\zeta$-soliton equation,
and on this complex we define a differential $\hat\Q_\zeta$ whose matrix elements are found by counting
(with signs that come from the sign of the fermion determinant) the $\zeta$-instantons that interpolate between
two given $\zeta$-solitons.

We now make a few more detailed remarks about  the cases of a half-line or a strip.

\subsection{The Half-Line}\label{halfline}

If we do not impose a boundary condition at $x=0$, then the space of $\zeta$-solitons on the half-line $[0,\infty)$ that approach
$\phi_j$ at infinity is the exact
Lagrangian submanifold $R^\zeta_j$ that was described at the end of section \ref{boundary}.  $R^\zeta_j$ is naturally embedded
in $X$ by identifying a $\zeta$-soliton with the value $\phi^I(0)$.  To quantize the LG model on $[0,\infty)$, we require
a boundary condition at $x=0$ stating that $\phi^I(0)$ must lie in a specified Lagrangian submanifold $\L$.
$\zeta$-solitons on $[0,\infty)$ that obey this condition at $x=0$ and also approach $\phi_j$ at $x=\infty$ are simply
in one-one correspondence with the intersection points $\L\cap R^\zeta_j$.

\begin{figure}[htp]
\centering
\includegraphics[scale=0.3,angle=0,trim=0 0 0 0]{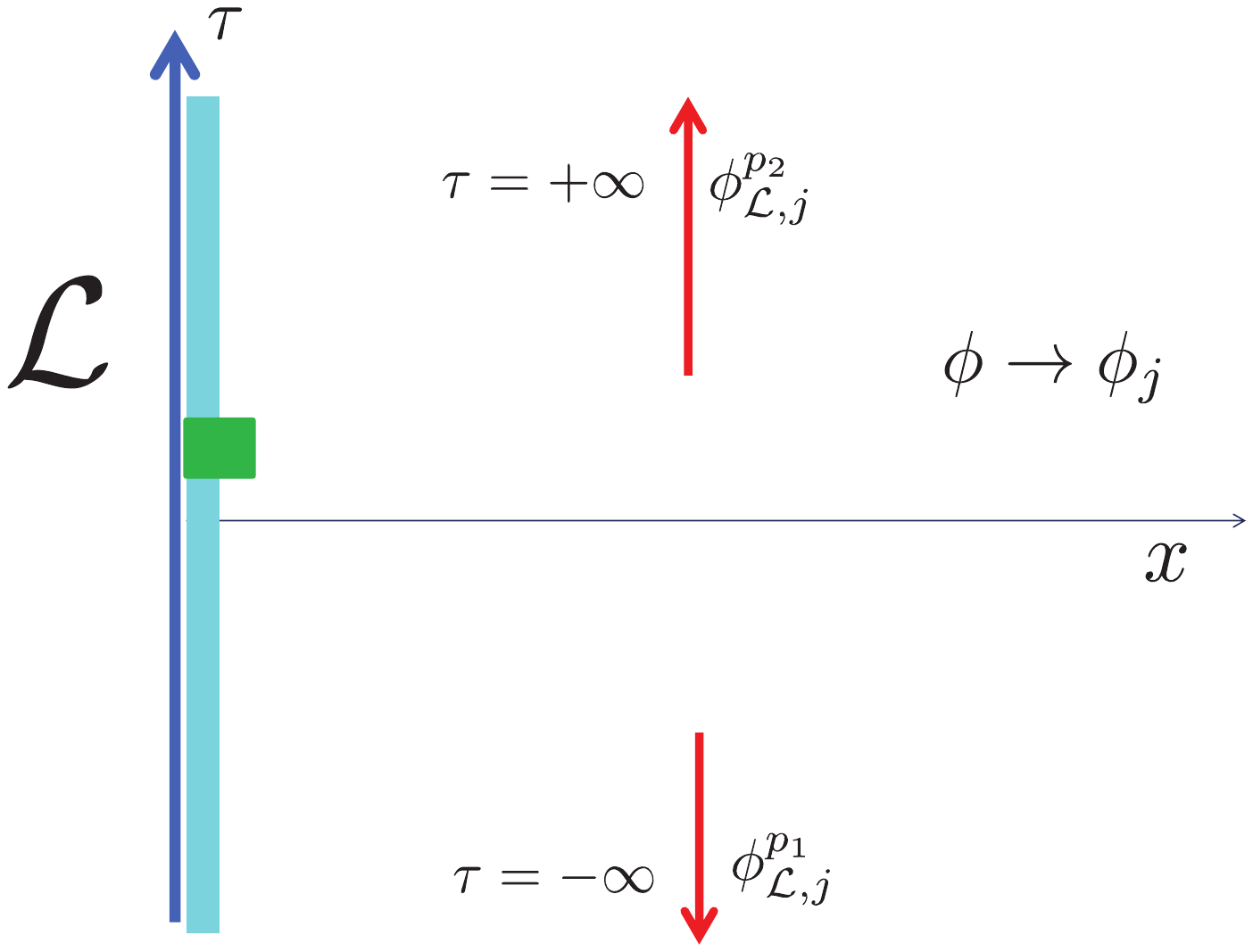}
\caption{An instanton in the complex $\IM_{\CL,j}$.  The solitons corresponding to
$p_1,p_2 \in \CL\cap R^\zeta_j$ are exponentially close to the vacuum $\phi_j$ except
for a small region, shown in turquoise,  of width $\ell_W$. In addition, the instanton
transitions from one soliton to another in a time interval of length $\ell_W$,
indicated by the green square. At large distances the green square becomes the
$0$-valent vertex used in extended half-plane webs. }
\label{fig:HALFPLANE-INSTANTON-1}
\end{figure}

As usual, to find the quantum BPS states, we construct a complex $\IM_{\L,j}$ that additively has a basis corresponding to the intersection
points just mentioned, with a differential $\hat \Q_\zeta$ that can be found by counting $\zeta$-instantons.  The instantons  are schematically illustrated
in Figure \ref{fig:HALFPLANE-INSTANTON-1}.

To study BPS states on
the half-line $(-\infty,0]$, again with boundary conditions set by $\L$, we proceed in just the same way.  The $\zeta$-solitons that obey
the boundary conditions are the intersections $L^\zeta_i\cap \L$, where as in section \ref{boundary}, $L^\zeta_i$ is
defined by ascending gradient flows on $(-\infty,0]$ that start at $\phi_i$ at $x=-\infty$.  This gives a basis for a complex $\IM_{i,\L}$,
whose normalized differential $\h\Q_\zeta$ is found by counting $\zeta$-instanton solutions.

In general, a Lagrangian submanifold $\L$ may intersect $R^\zeta_j$ (or $L^\zeta_i$) in many points.  Moreover, if $\L$ is varied by
an exact symplectomorphism of $X$ (and thus without changing it as an $A$-brane), the number of these intersection
points may vary.  Since the number of exactly supersymmetric states on the half-line does not change if $\L$ is varied
in this way, but the dimension of the complex $\IM_{\L,j}$ can jump, the differential $\h\Q_\zeta$ is certainly nonzero in general
and $\zeta$-instantons
are important.  However, it is difficult to give explicit examples.

We will describe an important example in which it is straightforward to identify the intersections $\L\cap R^\zeta_j$.  We simply take $\L$ to be
one of the Lagrangian submanifolds $L^\zeta_i$.
So $L\cap R^\zeta_j=L^\zeta_i\cap R^\zeta_j$ consists of points in $X$ that
are boundary values of flows on $(-\infty,0]$ that start at $\phi_i$ and also are boundary values of flows on $[0,\infty)$ that end at $\phi_j$.
Gluing these flows together, we get an $ij$ soliton, but by our hypothesis there are no such solitons for $i\not=j$.  So $L^\zeta_i\cap R^\zeta_j$
is empty for $i\not=j$.
  On the other hand, the only solution of the $\zeta$-soliton
equation interpolating
from a vacuum $i$ to itself is the trivial constant solution (in a non-trivial solution, the function $\mathrm{Im}(\zeta^{-1}W)$ is strictly increasing), 	
so $L^\zeta_i\cap R^\zeta_i$ consists of a single point, the critical point $\phi_i$.   With only a single approximately supersymmetric
state, there is no possibility of an instanton transition.

Of course, the analysis of $L^\zeta_i\cap\L$ is the same if $\L=R^\zeta_j$.  So
\begin{equation}\label{menod} \IM_{L^\zeta_i, \,j}=\IM_{i ,\,R^\zeta_j}=\begin{cases}0 & \mathrm{if}~ j\not=i\cr  \Bbb Z & \mathrm{if}~ j=i ,
\end{cases}\end{equation}
in each case with trivial differential.

\textbf{Remark}: When comparing with the abstract discussion
of Section \S \ref{subsubsec:WebRep-Halfplane} we should identify the Chan-Paton factors
for a brane defined by a Lagrangian subvariety $\CL$ with
\be\label{eq:CP-realized}
\CE_j := \IM_{\CL,j} \qquad\qquad \& \qquad\qquad \tilde \CE_i := \IM_{i,\CL}
\ee
for  left- and right- boundaries, respectively.

\subsection{The Strip}\label{thestrip}

In quantization on the closed interval  $D=[x_\ell, x_r]$,
with boundary conditions set at the ends
by two Lagrangian submanifolds  $\CL_\ell, \CL_r$,
 a classical ground state is a  solution of the $\zeta$-soliton equation with
$\phi(x_\ell)\in \CL_\ell$, $\phi(x_r)\in \CL_r$.
For the case that the strip is very long compared to the longest Compton
wavelength of any massive particle in the theory, we can give a simple
counting of such solutions.   At long distances from the boundaries, the theory
will be close to one of the vacuum states, which correspond to the critical points
$\phi_i$, $i\in \IV$.  Given this, near  the left boundary, the $\zeta$-soliton solution will approximate
one of the half-line solutions that contribute basis vectors of $\IM_{\CL_\ell,i}$;
and near the right boundary, it will approximate one of the half-line solutions that contribute
basis vectors of $\IM_{i,\CL_r}$.  Conversely, if  the strip is wide, then generalities of
index theory and elliptic operators imply that every pair of left- and right- half-line solutions arises
in this way from a solution on the strip.\footnote{If the strip is wide, then a simple guing starting with
a left- and a right- half-line solution comes exponentially close to an exact solution on the strip.  Using the
assumed nondegeneracy of the solutions and the fact that the relevant index vanishes because $\L_\ell$ and $\L_r$
are both middle-dimenisonal, this approximate solution can be corrected to an exact solution in a unique way.}
So additively, the complex  $\IM_{\CL_\ell, \CL_r}$ of the two-sided problem
has a simple description:
\be\label{eq:appxt-complex-ii}
 \IM_{\CL_\ell, \CL_r} \cong \oplus_{i\in
 \IV}  \IM_{\CL_\ell,i} \otimes \IM_{i, \CL_r}
\ee
At the end of section \ref{fnagain}, we explain why the grading of $\IM_{\CL_\ell,\CL_r}$ by
fermion number is related to those of $\IM_{\CL_\ell,i}$ and $\IM_{i,\CL_r}$ in the way suggested
by this decomposition.

As usual, the normalized differential $\hat\Q_\zeta$ on this complex is computed
by counting $\zeta$-instanton solutions.   Some such instantons are localized
at one end of the strip or the other.   These instantons  contribute the differentials
of individual factors  $ \IM_{\CL_\ell,i} $ and $\IM_{i, \CL_r}$ in the sum over $i\in\IV$
in (\ref{eq:appxt-complex-ii}).  If they were the only instantons on the strip, then the cohomology of the
complex $ \IM_{\CL_\ell, \CL_r}$ would have the same sort of decomposition as the complex itself:
\be\label{wrong}
H^*( \IM_{\CL_\ell, \CL_r})\overset{?}{\cong} \oplus_{i\in \IV}  H^*(\IM_{\CL_\ell,i}) \otimes H^*(\IM_{i, \CL_r}).
\ee

As discussed qualitatively in the introductory section \ref{sec:Introduction}, in general
the differential for the problem on the strip is much more complicated and this simple formula is not correct.
The differential can receive corrections from instantons that cannot be localized at one end of the strip
or the other.
 To a large extent, the goal of this paper is to understand
these unlocalized contributions to the differential, which prevent
an elementary description of the space of physical states on the strip.
Starting in Section \S \ref{zetawebs},
we will seek to understand these contributions in the context of LG models.
But first we pause for another interlude on the fermion
number (Section \S \ref{fnagain}).

Although there is no simple statement about the cohomology on the strip, there is a simple
statement about
the Euler characteristic of the cohomology.  The Euler characteristic of the cohomology of a complex does not depend on the differential (the effect
of the differential is eliminate pairs of states of opposite statistics that anyway make no net contribution to the Euler charactistic),
so the supersymmetric index of this problem factorizes nicely in the fashion suggested  by \eqref{eq:appxt-complex-ii}.
If we define the  index
\be
\mu_{\CL_\ell,i} = {\Tr}_{\IM_{\CL,i}} e^{i \pi \FF}
\ee
on the half-line, with a similar definition
for $\mu_{i, \CL_r}$ and $\mu_{\CL_\ell, \CL_r}$,
then, provided the fermion numbers are coherently related,
 it follows from equation \eqref{eq:appxt-complex-ii}
that  these  indices nicely factorize:
\be\label{eq:WittenIndexFactorize}
\mu_{\CL_\ell, \CL_r} = \sum_{i \in \IV} \mu_{\CL_\ell,i} \mu_{i, \CL_r}
\ee

\subsection{The Fermion Number Revisited}\label{fnagain}

Much of our analysis of the fermion number and the $\eta$-invariant, starting
in section \ref{ferminumber}, was equally valid whether the spatial manifold $D$ on which
we quantize is the real line, a half-line, or a compact interval.  However, we will now describe  some special
features that occur when $D$ has a boundary (as opposed to an infinite end).     In doing so, it is convenient to assume that
$D$ has only boundaries and no infinite ends -- the general
case is a mixture of the two problems.

We begin by considering, in the absence of a superpotential, the standard $A$-model of maps $\phi:\Sigma\to X$,
where $\Sigma$ is a compact  Riemann surface without boundary of Euler characteristic $\chi(\Sigma)$ and $X$ is a compact K\"ahler manifold.  In general,
fermion number is not conserved in this model.  The net violation of fermion number is given by the index of the
two-dimensional Dirac
 operator coupled to $\phi^*(T^{1,0}X)$. After topological twisting\footnote{After this twisting, the relevant Dirac operator is naturally regarded as a $\bar\partial$ or Dolbeault operator.} the resulting Dirac operator $L$
has index:
\be\label{zelb}\iota(L)=\chi(\Sigma)\dim_{\IC}X-2\int_\Sigma c_1(\phi^*(K_X)), \end{equation}
where $K_X$ is the canonical bundle of $X$ -- the bundle of $(n,0)$ forms on $X$, where $n=
\dim_\C X$. (In the untwisted $\sigma$-model, one would drop the first term on the right-hand side of \eqref{zelb}.)
The index \eqref{zelb} is the expected real dimension of the moduli space of holomorphic maps $\phi:\SIgma\to X$.
Equivalently, it is the net fermion number $\FF$ of operators that must be inserted to get a non-zero $A$-model
amplitude.  For our purposes, the important term in the index formula  is the one involving $c_1(\phi^*(K_X))$.  This term depends on $\phi$,
so when it is nonzero, different $A$-model amplitudes on the same $\Sigma$ violate the fermion number $\FF$ by different amounts and
we say that the $A$-model does not have fermion number symmetry.   The $\chi(\Sigma)\dim_\C(X)$ term in the index formula, since it is independent of
$\phi$, is a sort of $c$-number anomaly, and actually, in our eventual applications, we will have $\chi(\Sigma)=0$.     (The $A$-model still makes
sense -- and is much studied -- as a topological quantum field theory even when it does not conserve an integer-valued fermion number.  Because of the factor of 2 multiplying the $\phi$-dependent
term, the $A$-model always conserves fermion number modulo $2k$ for some integer $k$, and similarly in the analysis below, even when there
is no $\Z$-graded MSW complex, there is always a $\Z_{2k}$-graded one.)

What happens if we turn on a superpotential?  Then the equation $\partial_{\bar s}\phi^I=0$ for a holomorphic map  is deformed to the $\zeta$-instanton equation
\be\label{eq:instant}
\frac{\p \phi^I}{\p \bar s} =\frac{\I \zeta}{4}g^{I\bar J} \frac{\p \bar W}{\p \bar\phi{}^{\bar J}}.
\ee
This equation depends on the choice of local complex parameter $s$ on $\Sigma$ and does not make sense on an arbitrary $\Sigma$.
The left hand side of (\ref{eq:instant}) is a  $(0,1)$-form and the right hand side is a $0$-form;
to set these equal implies picking a trivialization of the bundle of $(0,1)$-forms.  For $\Sigma$ of genus 1, this is given by the globally-defined
$(0,1)$-form $\d\bar s$, but for $\chi(\Sigma)\not=0$,  there is a topological obstruction to the definition.  However, whenever the deformation from the usual
$A$-model equation to the $\zeta$-instanton equation is possible, this deformation
certainly does not affect the index, since the superpotential $W$ is a lower
order term in the $\zeta$-instanton equation, or alternatively since the index is an integer and (for compact $\Sigma$) the spectrum of $L$ varies smoothly when
$W$ is turned on.

Now let us allow $\Sigma$ to have a boundary, labeled by a brane.  To define a boundary condition on the equation for a holomorphic map, or on the $\zeta$-instanton equation, we pick a Lagrangian submanifold $\CL\subset X$ and require that $\phi(\partial\Sigma)\subset \CL$, as discussed in section \ref{boundary}. In addition we choose local boundary
conditions on the fermions which then follow from the preservation of $\CQ_{\zeta}$-supersymmetry.
In this case, the index of the Dirac/Dolbeault
 operator $L$ still makes sense and still determines the expected dimension of the moduli space and the
net violation of fermion number in amplitudes.  Moreover, if $\Sigma$ is compact,
 and if it makes sense to turn on a superpotential $W$, the index is independent of $W$, since $W$ contributes a lower order term in the $\zeta$-instanton equation and  does
not affect the boundary condition.  For similar reasons, when we discuss later the $\eta$-invariant on a one-manifold $D$ with boundary (but no
infinite ends), $W$ will be irrelevant and can be set to 0.  Accordingly, the following discussion of the index theorem when $\Sigma$ has a boundary,
and later of the $\eta$-invariant when $D$ has a boundary, can be carried out at $W=0$ and is simply part of the analysis of the usual $A$-model.
By contrast, if $\Sigma$ or $D$ has an infinite end, then $W$ affects the asymptotic behavior at the end and does affect the index theory
and the $\eta$-invariant, as for instance in eqn. (\ref{explained}).

So setting $W=0$, we now consider
 the formula for the index when $\Sigma$ has a boundary.  The formula (\ref{zelb}) does not quite make sense without some explanation
 because to define a first Chern class of the line bundle $\phi^*(K_X)$ over $\Sigma$, one needs a trivialization of $\phi^*(K_X)$ on $\partial\Sigma$.
 The proper interpretation is as follows.  For simplicity, we assume that the Lagrangian submanifold $\CL$ is orientable
and pick an orientation.  $\CL$ acquires a Riemannian metric from its embedding in the K\"ahler manifold $X$, so it has a volume form $\mathbf{vol}$.
Setting $n=\dim_\C X$, $\mathbf{vol}$ is a real-valued $n$-form on $\CL$.  On the other hand, a section of  $K_X$  is an $(n,0)$-form
whose restriction to $\CL$ is a complex-valued $n$-form on $\CL$.  A real-valued $n$-form is a special case of a complex-valued one, so we can regard $\mathbf{vol}$ as a trivialization of $K_X|_\CL$ (the restriction of
$K_X$ to $\CL$).  Given a map $\phi:\Sigma\to X$ that maps $\partial\Sigma$ to $\CL$, the trivialization of $K_X|_\CL$ pulls back to a trivialization of
$\phi^*(K_X)|_{\partial\Sigma}$.  We define $\int_\Sigma c_1(\phi^*(K_X))$ using this trivialization of $\phi^*(K_X)$ on $\partial\Sigma$. With this
interpretation, the index formula (\ref{zelb}) remains valid. (For mathematical background on this and related statements,
see \cite{Gromov}, especially section 2.1.D, and \cite{RS}, especially
 section 5.)
 To emphasize the role of the trivialization, we write the index formula as
\be\label{welb}\iota(L)=\chi(\Sigma)\dim_{\IC}X-2\int_\Sigma \left.c_1(\phi^*(K_X))\right|_{\mathbf{vol}}, \end{equation}
where the notation is meant to remind us that the first Chern class is defined using the trivalization via $\mathbf{vol}$ on the boundary.  If one asks ``why''
the trivialization we have used is the right one in the index formula, one answer is that this trivialization is the only one that can be described in a universal
local way, and  the heat kernel proof of the index theorem makes clear that
there must be a universal local formula.

Now let us ask under what conditions the $A$-model has fermion number symmetry, in the sense that the index $\iota(L)$ does not depend on $\phi$.  A
necessary condition is certainly that $c_1(K_X)=0$.  Otherwise, the $A$-model  has no fermion number symmetry
even in the absence of a boundary.  However, even if $c_1(K_X)=0$, it does not necessarily follow that $c_1(\phi^*(K_X))|_{\mathbf{vol}}$ vanishes.
This involves a condition on $\CL$, and the $A$-model has fermion number symmetry only when this condition is obeyed.

Let us consider an example.  We take $X$ to be the complex $\phi$-plane, and we take $\CL$ to be the unit circle
in the $\phi$-plane.\footnote{This particular
Lagrangian submanifold does not actually correspond to a quantum $A$-brane, because of disc instanton effects.  But we can still use it to illustrate the index theorem.}
We take $\Sigma$ to
be the unit disc in the complex $z$-plane.  A degree 0 map from $\Sigma$ to $X$, mapping $\partial \Sigma$ to $\CL$, is a constant map, depending
on 1 real  parameter (the choice of a point in $\CL$).   A degree 1 map from $\Sigma$ to $X$, mapping $\partial\Sigma$ to $\CL$, is a fractional
linear transformation $\phi=(az+b)/(\bar b z+ \bar a)$,  depending on 3 real parameters (the real and imaginary parts of  $a,b$ with the equivalence
$a,b\cong \lambda a,\lambda b$ for $\lambda>0$; one also requires $|b|<|a|$).
More generally, a degree $r$ map is $\phi=p(z)/z^r \bar p(1/z)$,
where $p$ is a degree $r$ polynomial which depends on $2r+1$
parameters (after allowing for an equivalence $p\cong \lambda p$, $\lambda\in\IR$; $p(z)$ must have all its zeroes inside the unit
disc).  So the index $\iota(L)$ equals $2r+1$.  Since $\iota(L)$ depends on $\phi$ in this example, this means that with this choice of $\CL$,
the $A$-model does not have fermion number symmetry.  (In fact, $\FF$ is conserved
only mod 2 in this situation, since by changing $r$ one changes $\iota(L)$ be an arbitrary
multiple of 2.)

Now let us try to understand  the dimension of the moduli space from the index formula (\ref{welb}).  Since $\chi(\Sigma)=\dim_{\IC}X=1$,
to  get $\iota(\D)=2r+1$,
we need $\int_\Sigma \left.c_1(\phi^*(K_X))\right|_{\mathbf{vol}}=-r$.  The basic case to understand is $r=1$, as the general case will then follow by
taking an $r$-fold cover.  For $r=1$, we can take the map from $\Sigma$ to $X$ to be $\phi=z$, so we identify $\SIgma$ with the unit disc $Y\subset X$
and perform the calculation there.  In terms of polar coordinates $\phi=r e^{\i\alpha}$, the volume form of $\CL$ is $\mathbf{vol}=\d \alpha$.  We want
to calculate $\int_Y c_1(K_X)|_{\mathbf{vol}}$, where we interpret $\mathbf{vol}$ as a trivialization of $K_X|_{\CL}$. One way to do the computation
is to extend $\mathbf{vol}=\d\alpha$ to a meromorphic section $\Upsilon$ of $K_X$ over $Y$.  Then the difference between the number of zeroes
and poles of $\Upsilon$ in $Y$  is $\int_Y c_1(K_X)|_{\mathbf{vol}}$.  The formula $(\d z/z)|_{|z|=1}=\I \d\alpha$ shows that we can take $\Upsilon=-\I \d z/z$,
with one pole and no zeroes in the unit disc, so $\int_Y c_1(K_X)|_{\mathbf{vol}}=-1$, as expected.

When $c_1(K_X)=0$, the relevant part of the index formula can be written as follows as an integral over $\partial\Sigma$.  Let $\Omega$ be
a holomorphic $n$-form on $X$, normalized so that $|\Omega\wedge\bar \Omega|$ agrees with the Riemannian volume form of $X$.  Then when
restricted to $\CL$, the expression
\begin{equation}\label{dommy}e^{\i\vartheta}=\frac{\mathbf{vol}}{\Omega|_{\CL}} \end{equation}
is a $U(1)$-valued function on $\L$.  ($\L$ is said to be special Lagrangian if this function is constant.)
If $\Sigma$ has several boundary components $\Sigma_s$ mapping to Lagrangian submanifolds
$\CL_s$, then each has its own volume-form $\mathbf{vol}_s$ and we set $e^{\i\vartheta_s}=\mathbf{vol}_s/\Omega|_{\CL_s}$.
The index formula is then
\begin{equation}\label{ommy}\iota(L)=\chi(\Sigma)\dim_{\IC}X-2\sum_s \oint_{\partial_s\Sigma}\frac{\d\varphi_s}{2\pi}. \end{equation}
where we define $\varphi_s := \phi_s^*(\vartheta_s)$ and  $\phi_s$ is the restriction of $\phi$ to  $\partial_s\Sigma$.
To get from (\ref{welb}) to this formula, we observe that if we were to trivialize $\phi^*(K_X)$ via the everywhere nonzero section $\phi^*(\Omega)$,
then $\int_\Sigma c_1(\phi^*(K_X))$ would vanish.  The actual definition involves a trivialization on each $\partial_s\Sigma$ via $\phi^*(\mathbf{vol}_s)$,
and then $\int_\Sigma c_1(\phi^*(K_X))$ is a sum of boundary contributions, where the contribution of $\partial_s\Sigma$ is  made by comparing the
two trivializations of $\phi^*(K_X)|_{\partial_s\Sigma}$.

Finally this gives a necessary and sufficient condition for the $A$-model with branes defined by Lagrangian submanifolds $\CL_s$ to
have conserved fermion number.   The condition under which the index does not depend on the map $\phi$ is that on each $\CL_s$,
it is possible to define $\vartheta_s$ as a real-valued function, not just an angle-valued function.

Now we turn to our real interest, which is to analyze the fermion number of physical states in quantization of the model.  For this, we take
$\Sigma=\IR\times D$, where $D$ is a compact interval, for instance the interval\footnote{For $W=0$, conformal invariance
ensures that the width of interval $D$ does not matter.  It is convenient to take an interval of width $\pi$.}  $[0,\pi]$.  We label the two ends of the interval
with Lagrangian submanifolds  $\CL_\ell$ and $\CL_r$.  Assuming $c_1(X)=0$, we want to define the conserved fermion number of the $A$-model.  From
section \ref{ferminumber}, we certainly expect that the $\eta$-invariant of the one-dimensional Dirac operator will be part of the answer.  But since
there is a boundary correction in the index formula (\ref{ommy}), it is not surprising that a boundary contribution is needed in the definition of a conserved
fermion number.

Rather than developing a general theory, we will continue with the example $X=\C$.  (The interested reader should be able to generalize the following
computation.)  We parametrize $X$ by a single
complex field $\phi$.  To start with, we take $\CL_r$ to be a straight line in $X$ at an angle $-\vartheta_r$ to the $\mathrm{Re}\,\phi$ axis.
The equation defining this straight line is
\begin{equation}\label{xoom}\bar\phi=e^{ 2\i\vartheta_r}\phi. \end{equation}
We similarly take $\CL_\ell$ to be a straight line in $\C$ at an angle $\varphi_\ell$ to the real axis, and so described by
\begin{equation}\label{oom}\bar\phi =e^{2\i\vartheta_\ell}\phi. \end{equation}

For compact $D$, we can set $W$ to 0 without modifying the fermion number.  When we do this,  the Dirac operator (\ref{eq:NewD}) is simply
\be\label{nux} \D=\sigma^3\i \frac{\d}{\d x}. \ee
We recall that this was written in a basis $\begin{pmatrix}\delta\phi\cr \delta\bar\phi\end{pmatrix}$.   We write $\psi_+$, $\psi_-$ for fermions corresponding
respectively to $\delta\phi$, $\delta\bar\phi$,
\footnote{These are actually linear combinations of the fields $\psi_\pm$ and
$\bar\psi_\pm$ used in the standard Lagrangian.}
so $\D$ acts on the fermions as
\begin{equation}\label{ux}\D\psi_\pm=\pm\i\frac{\d}{\d x}\psi_\pm. \end{equation}
The boundary conditions (\ref{xoom}) and (\ref{oom}) imply that variations of $\phi$ obey
$\delta\bar\phi=e^{ 2\i\vartheta_r}\delta\phi$ at $x=\pi$ and
$\delta\bar\phi=e^{2\i\vartheta_\ell}\delta\phi$ at $x=0$.
So the boundary conditions on the fermions are
\be \label{fbc} \psi_-=\psi_+\cdot \begin{cases} e^{ 2\i\vartheta_r},& x=\pi\cr e^{2\i\vartheta_\ell},& x=0.\end{cases}\ee
With these boundary conditions, and the specific form of the Dirac operator $\D$ in (\ref{ux}),
we can glue $\psi_+$, $\psi_-$ to a single fermi field
$$
\psi(x) = \begin{cases}
\psi_+(x) & 0 \leq x \leq \pi \\
e^{ - 2\i\vartheta_r}\psi_-(2\pi-x) & \pi \leq x \leq 2\pi \\
\end{cases}
$$
that obeys
\be \label{noxt}\D\psi=\i\frac{\d}{\d x}\psi \end{equation}
and
\be \label{flox}\psi(x+2\pi)=e^{-2\i(\vartheta_r-\vartheta_\ell)}\psi(x). \end{equation}
The eigenvalues of the Dirac operator acting on states with this periodicity condition are
\be \label{zox}\lambda_n =n+\frac{\vartheta_r-\vartheta_\ell}{\pi},~~ n\in \IZ. \ee
Since this spectrum is a periodic function of $\Delta = \frac{\vartheta_r-\vartheta_\ell}{\pi}$ with period 1,
$\eta(\D)$ is also a periodic function of $\Delta $ with that period. Using the Hurwitz
zeta function $\zeta(s,\alpha) = \sum_{n=0}^\infty \frac{1}{(n+\alpha)^s}$ and its analytic
continuation to $s=0$, given by $\zeta(0,\alpha) = \half - \alpha$ (and valid for
$\Re(\alpha)>0$) one easily shows
that for $0<\Delta<1$, $\eta(\D)=1-2\Delta$.  In general this is correct mod 2,
\begin{equation}\label{poly}\eta(\D)=1-4\frac{\vartheta_r-\vartheta_\ell}{2\pi}~{\mathrm{mod}}\,2\Bbb Z,  \end{equation}
with even integer jumps so that $\eta(\D)$ is periodic.

Let us write $f^\psi$ for the fermion number of the filled fermi sea.  By equation
\eqref{mezzo}, this is $-\eta(\D)/2$, which in view of eqn. (\ref{poly}) leads to
\begin{equation}\label{fgt} f^\psi=-\frac{1}{2}+2\frac{\vartheta_r-\vartheta_\ell}{2\pi}~~\mathrm{mod}\,\Bbb Z. \end{equation}

To see why the fermion number needs an additional contribution, let $\CL_\ell$ and $\CL_r$ be more general Lagrangian submanifolds of $X$,
not necessarily straight lines.  Pick a time-independent bosonic field of the LG model; this is just a map $\phi:[0,\pi]\to X$ with $\phi(0)\in \CL_\ell$
and $\phi(\pi)\in \CL_r$.  Since the fermions are free for the case that $X=\C$, the computation of the $\eta$-invariant is the same as above,
with $\varphi_\ell=\vartheta_\ell(\phi(0))$ and $\varphi_r= \vartheta_r(\phi(\pi))$ where
 $\vartheta$ is  defined by (\ref{dommy}).  Another way to say this is that for any choice of $\phi:[0,\pi]\to X$,
the calculation is the same as above, with $\CL_\ell$ and $\CL_r$ replaced by their tangent lines at the points $\phi(0)\in  \CL_\ell$ and $\phi(\pi)\in \CL_r$.  Now
give $\phi$ a slow time dependence.  The formula for $f^\psi$ remains the same, but now it is time-dependent, since $\varphi_r$ and $\varphi_\ell$
are time-dependent.    To define a conserved fermion number, we have to add to $f^\psi$ a boundary correction that cancels this time-dependence.
As one might guess from the analysis of the index, this is only possible if one can define $\varphi_\ell$ and $\varphi_r$ as real-valued functions
(rather than angle-valued functions) and in that case one can take the conserved fermion number to be
\begin{equation}\label{cgt} f=f^\psi -2\frac{\varphi_r-\varphi_\ell}{2\pi}=-\frac{1}{2}\eta(\D) -2\frac{\varphi_r-\varphi_\ell}{2\pi}.  \end{equation}
In view of (\ref{fgt}), $f$ takes values in $-1/2+\Bbb{Z}$.  The generalization of this to higher complex dimension is $-\frac{1}{2}\dim_\C X+\Bbb Z$.
From the point of view of the physical approach described at the beginning of section \ref{ferminumber}, the reason that a boundary correction to the fermion
number is needed is that although one can define a conserved current that is bilinear in the fermion fields, its flux through the boundary may be nonzero.

We note, however, that for a Lagrangian submanifold $\CL$, if the angle-valued function $\vartheta$ can be lifted to an integer-valued function,
then this lift is only uniquely determined mod $2\pi\Bbb Z$ and the choice of lift (which in (\ref{cgt}) is made independently for $\CL_\ell$ and $\CL_r$)
affects the definition of the fermion number.
  Moreover, if we change the orientation of $\CL$, this changes the sign of $\mathbf{vol}_s$
and hence shifts $\vartheta_s$ by $\pi$, again changing the fermion number.
 So an orientation of $\CL$ and a choice of lift of $\vartheta$ have to be regarded as part of the definition of an $A$-brane.  A change in orientation or
 lift of either $\CL_\ell$ or $\CL_r$ changes $f$ by an integer.
(We do not need to worry about the dependence of $\vartheta$ on the choice of $\Omega$, because, assuming we use the same $\Omega$ to define
$\vartheta_s$ for all $\CL_s$, a change in $\Omega$ shifts all $\vartheta_s$ by the same amount and never contributes to $\varphi_r-\varphi_\ell$.)

The physical application of this is as usual.  In quantization on $D=[0,\pi]$, in a transition between two $\zeta$-solitons, the integer-valued
conserved fermion number will in general change by an integer.  This integer will match the index of the two-dimensional Dirac operator $L$,
and this index gives the net fermion number of operators that must be inserted to make this transition possible.

Now let us restore $W$ and take the interval $D=[x_\ell,x_r]$ to be much longer than the Compton wavelength of any massive particle.
Then for each $\zeta$-soliton, the fermion number can be written as a sum of contributions
from the two ends of the strip, as we assumed  in discussing
(\ref{eq:appxt-complex-ii}).  The boundary corrections are manifestly a sum of contributions
from the two ends, and the $\eta$-invariant has the same property, since it can be written rather
as in (\ref{zonf}) as an integral over $D$:
\begin{equation}\label{ponf}\eta^{\mathrm{cont}}(\D)=\lim_{\varepsilon\to 0}
\int_D \d x\,\sum_{s=1}^{2d}\bigl\langle x,s|  \mathrm{sign}(\D)\exp(-\varepsilon|\D|)|x,s\bigr\rangle. \end{equation}  The integral receives a contribution
only from the regions near the boundaries of $D$, because the
integrand vanishes exponentially fast away from the boundaries.  The last statement just reflects the fact that the expectation value of the Lorentz-invariant fermion
number current is 0 in a Lorentz-invariant vacuum, and moreover, in a massive theory, the approach to the vacuum is exponentially fast.

Thus far, we have assumed $D$ is compact.
If we take $D$ to be a half-line $[0,\infty)$ or $(-\infty,0]$, then because of the infinite end of $D$,
we cannot simply set $W$ to 0.  However, $W$ does not affect the boundary contributions at the finite end of $D$
to the index or the fermion number.

In summary, although we have based our discussion on a representative example, in general
the fermion number of the MSW complex on the interval is defined by
\be
f= - \half \eta(\CD) -2 \frac{\varphi_r - \varphi_{\ell}}{2\pi}
\ee
where $\CD$ is the Dirac operator obtained from linearizing the $\zeta$-soliton equation. We choose
local $\CQ_\zeta$-preserving boundary conditions for the fermions and $\varphi_{\ell} = \vartheta_{\ell}(\phi(x_\ell))$
while $\varphi_r = \vartheta_r(\phi(x_r))$. On the interval we can simplify the Dirac operator
by setting $W=0$. On the half-line we drop $\varphi_r$ or $\varphi_\ell$, as appropriate, and
we cannot set $W=0$.

\subsection{Implications Of The Bulk Anomaly}\label{panom}

What are the implications  of the fermion number anomaly?  Our next goal is to explain that in general, the
framework of the present paper only applies when the fermion number anomaly vanishes.
We discuss first the bulk anomaly and then (in section \ref{impanom}) the boundary anomaly.

The bulk anomaly is proportional to the first Chern class $c_1(X)$ of the target space $X$ of the $\sigma$-model, so
it is absent for simple Landau-Ginzburg models with $X=\IC^n$.  We pause to give a few illustrative examples of massive
$\sigma$-models with more complicated target spaces $X$, constructed from  a $U(1)$ gauge theory
coupled to chiral superfields.  Here are two examples:

{\it {(A)}}  For our first model, we consider three  chiral superfields $u,v,b$ with $U(1)$ charges $1,1,-2$.  The $D$-term constraint
coming from the $U(1)$ gauge-invariance is $\vert u \vert^2 + \vert v \vert^2 -2|b|^2=r$.  If the constant $r$ is large and positive, the model is equivalent
at low energies to a $\sigma$-model in which the target space $X$
is the Eguchi-Hansen manifold,
the total space of the line bundle $\O(-2)\to \Bbb{CP}^{1}$, where $\Bbb{CP}^{1}$ is embedded as the locus $b=0$.
We take the superpotential to be $W=buv+b^2 G(u,v)$ where $G$ is a generic homogeneous polynomial of degree 4.
This gives a massive model with $2$ vacua at $b=0$ (with $u=0$ or $v=0$) and more at $b\not=0$.

{\it {(B)}} For our second model, we consider three chiral superfields $u,v,b$ with $U(1)$ charges $1,1,-1$.  The $D$-term constraint
is $|u|^2+|v|^2-|b|^2=r$, and if $r>>0$, the target space $X$ is the total space of the line bundle ${\mathcal O}(-1)\to \Bbb{CP}^1$,
where again $\Bbb{CP}^1$ is embedded in $X$ as the locus $b=0$.  We take the superpotential to be $W=bu$. There is a unique massive vacuum at $b=u=0$.

In general, in a theory of a $U(1)$ vector multiplet coupled to chiral multiplets of charge $q_i$, the condition for a fermion number anomaly is
$\sum_i q_i\not=0$.  Accordingly,
model {\it {(A)}} actually does not have a fermion number anomaly, since the sum of the chosen $U(1)$ charges vanishes, while model
 {\it {(B)}} does have an anomaly.

Parenthetically, we remark that while model {\it {(A)}} does not have an anomaly, nevertheless it does
illustrate another point of interest in the present paper: the superpotential $W$ has equal values at both of the $b=0$ vacua.  This
does not reflect global symmetries: for generic $G(u,v)$, there are none.  Rather, it reflects the fact that the two vacua are contained
in a common holomorphically embedded $\Bbb{CP}^1$. In general, if $W$ is restricted to any \emph{compact and holomorphic }
 subvariety of $X$ it must be constant there, and hence the critical values of any two critical points
 on such a subvariety are equal.
\footnote{In the particular case of a $(2,2)$ supersymmetric $\sigma$-model
with a hyperk\"ahler target $X$, there is a possible modulus that is not easily visible in the linear $\sigma$-model language and that
removes holomorphic subvarieties. The modulus in question is a rotation of the complex structure of $X$
away from the one that is visible in the linear $\sigma$-model. In particular, in model {\it {(A)}},
the target has the Eguchi-Hanson metric and is hyperk\"ahler. A rotation of complex structure removes
the holomorphically embedded $\IC\IP^1$. For such a rotation (with a corresponding change of $W$ to remain
holomorphic) the values of $W$ in the two vacua will in general become unequal. We can still illustrate the main point since
 a more elaborate model of this type  has no such modulus. An example is a
$U(1)$ gauge theory with chiral superfields $u,v,w,b,b'$ of charges $1,1,1,-1,-2$
and superpotential $W=bu+b'vw$. There are two massive vacua with $b=b'=u=0$ and $v$ or $w$ vanishing, and $W=0$ in each vacuum. }
So  in general in a massive $(2,2)$ model in two dimensions, it is not true that parameters can be varied so that  the superpotential takes generic values in the different vacua.  Hence the genericity assumption that we have made throughout this paper needs to be treated with care.  However,
the case that everything is generic except that two vacua unavoidably have the same value of $W$ fits into our framework
with a minor modification:  precisely because the two vacua have the same value of $W$,
there are no BPS solitons connecting them, and in many statements we can simply conflate the two vacua in question.

Returning to the consequences of a nonvanishing fermion anomaly, as we have noted,
model {\it {(B)}} actually does have a fermion number anomaly, since in this model $q:=\sum_iq_i=1$.  To understand the
consequences, recall that in the theory of $\N=(2,2)$ vector multiplets coupled to chiral multiplets in two dimensions, the vector multiplet
field strength is a twisted chiral superfield $\Sigma$.  Classically, the twisted chiral superpotential for this field is $\t W=t \Sigma$.  Here $t$ is
a complex constant, the K\"ahler modulus, whose imaginary part is the constant $r$ that enters the classical $D$-term constraint;
its real part is $\theta/2\pi$, with $\theta$ the quantum
mechanical $\theta$-angle.  But at 1-loop order, when $q \not=0$, the twisted chiral superpotential becomes
 $\t W=t\Sigma+q \Sigma\log \Sigma/2\pi \i$.
In our model {\it {(B)}}, with  $q = 1$, this  $\t W$ turns out to have  a unique critical point, which corresponds to the unique massive vacuum of the
model.\footnote{We are eliding a few points, which are explained for example in section 13.6 of \cite{Deligne:1999qp}.
The semiclassical method described in the text to evaluate $\t W$ in a massive vacuum is valid when $|r|$ is large and the signs of $\sum_i q_i$ and
of $r$ are such that the value of $\Sigma$ is large in this vacuum. Even when this derivation is not valid, the answer it gives for the value
of $\t W$ in a massive vacuum can be justified using holomorphy.  }  The value of $\t W$ at the critical point depends on the K\"ahler modulus $t$.
If we take model {\it {(B)}} as it stands, since it has only one vacuum, the value of $\t W$ in this vacuum does not really matter (an additive constant
in either $W$ or $\t W$ is irrelevant),
and moreover with only one vacuum, all considerations of the present paper are trivial.  To make a more interesting variant of model {\it (B)}, we first observe
that  model {\it {(B)}} can be interpreted as a $\sigma$-model whose target $X$ is $\IC^2$ with a point blown up.  (As a $\sigma$-model, this model
is asymptotically free, since $\sum_iq_i>0$, so this description is good even in the ultraviolet.)  Here $\IC^2$ is parametrized by
$x=ub$ and $y=vb$, and the point $x=y=0$ is blown up in $X$ to the copy of $\Bbb{CP}^1$ defined by $b=0$.  In this description, the
superpotential is $W=x$.  Though the function $x$ has no critical point on $\IC^2$, when one blows up a point, it acquires a nondegenerate critical
point on the resulting exceptional divisor $\Bbb{CP}^1$.  Now let us blow up $k$ points $p_1,\dots, p_k\in \IC^2$.  After the blowup, each $p_i$ is replaced by
a copy of $\Bbb{CP}^1$, and on each of these $\Bbb{CP}^1$ the superpotential $W$ has a nondegenerate critical point $\phi_i$, with the value
of $W(\phi_i)$ being simply the $x$-coordinate of the point $p_i$; in particular, these values can be generic.  Each $\Bbb{CP}^1$ also has its
own K\"ahler modulus $t_i$.  The value of the twisted chiral superpotential $\t W$ at the critical point $\phi_i$ depends on $t_i$ just as if
the other $\Bbb{CP}^1$'s did not exist (it can be computed in $A$-model terms by counting holomorphic maps to the given $\Bbb{CP}^1$,
and this counting does not ``know'' whether other points have been blown up).  In particular, in this model, both the ordinary superpotential
$W$ and the twisted superpotential $\t W$ take generic values in the massive vacua.\footnote{  The techniques
described in \cite{Martinec:2002wg} could be used to make the above construction quite concrete.}

The significance of this arises when we consider solitons in the $ij$ sector.  In addition to the familiar central charge $W(\phi_i)-W(\phi_j)$,
the supersymmetry algebra acquires another central charge term $\t W(\phi_i)-\t W(\phi_j)$.  The supercharge $\CQ_{\zeta}$ whose cohomology we have
been analyzing is still conserved, but it is no longer nilpotent in a soliton sector; rather
$ \CQ_{\zeta}^2$ is proportional to $\zeta^{-1} \left( \t W(\phi_i)-\t W(\phi_j)\right) $.  The supersymmetric model still exists,
but it cannot be studied using the methods of this paper, which assume that $\CQ_{\zeta}^2=0$. Generically, fermion number conservation is the only obvious mechanism to ensure that
 the $\FF=1$ supercharge $\CQ_{\zeta}$ obeys $\CQ_{\zeta}^2=0$ in a soliton sector. That is why we require fermion number conservation.

  As usual, the relation between $W$ and $\t W$ is actually symmetrical.
{\it A priori}, an $\N=(2,2)$ model in two dimensions can have separate $R$-symmetry groups acting on positive chirality and negative
chirality supercharges.   Taking $W\not=0$, $\t W=0$  leaves one linear combination of these as a possible symmetry, and this is what we have called fermion number.
Taking  $W=0$, $\t W\not=0$
leaves a different symmetry and a different supercharge $\t\Q$ that obeys $\t\Q^2=0$ in soliton sectors. The reasoning of this present paper applies
 equally well to a model with $W=0$, $\t W\not=0$.

\subsection{Implications Of The Boundary Anomaly}\label{impanom}

Even if  the bulk fermion number anomaly vanishes, it is natural now to suspect that a similar phenomenon can arise from
the boundary contribution to the fermion number anomaly.  Indeed, this is the case.  As an example, we consider the model that we already
used to illustrate the boundary  anomaly, with  $X$ being the complex $\phi$-plane.
\footnote{This example is closely related to Example 1.11 discussed in \cite{Auroux}.}
We take the
superpotential to be $\zeta^{-1} W=\I \phi^2$, which has just one critical point at $\phi=0$.
The right thimble $R^\zeta$ is the imaginary $\phi$ axis.  We consider the $\sigma$-model on the
half-plane $x\geq 0$ in the $x-\tau$ plane with boundary condition at $x=0$ set by a Lagrangian submanifold $\CL$.  As in section \ref{halfline},
the classical approximation to a supersymmetric state on the half-plane is given by an intersection $\CL\cap R^\zeta$.  We take $\CL$
to be the unit circle in the $\phi$-plane. This is an example where the phase function $e^{\I\vartheta}$
 defined in \eqref{dommy} does not have a well-defined logarithm, so we should expect  trouble.
The Lagrangian $\CL$  intersects $R^\zeta$ in the two points $\phi=\pm i$, which we call $p$ and $p'$.  So in the classical approximation,
there are two supersymmetric
half-plane states, $|p\rangle$ and $|p'\rangle$.  The state $|p\rangle$ corresponds
to a $\zeta$-soliton $\Phi_p$ on the half-line that interpolates from $\phi=i$ at $x=0$ to $\phi=0$ at $x=\infty$, and similarly
$|p'\rangle$ corresponds to a $\zeta$-soliton $\Phi_{p'}$ that interpolates from $\phi=-i$ at $x=0$ to $\phi=0$ at $x=\infty$.
The model has a classical symmetry
$\kappa:\phi\to -\phi$ that exchanges the points $p$ and $p'$, and also the states $|p\rangle$ and $|p'\rangle$.   What does
$\kappa$ do to the fermion number of  a half-plane state?  The $\eta$-invariant is invariant under $\kappa$, but $\kappa$ increases
the boundary contribution to the fermion number by 1.  (This is clear from eqn. (\ref{cgt}): the boundary contribution at $x=0$
is $\varphi_\ell/\pi$
where $\varphi_\ell$ differs by $\pi$ between the points $p$ and $p'$.) The fact that $\kappa$ increases $\FF$ by 1 is compatible with $\kappa^2=1$,
since $\FF$ is only conserved modulo 2.  There is no natural way to define a zero of the fermion number and say which of $|p\rangle$ and
$|p'\rangle$ is even and  which is odd.

Quantum mechanically, there must be no supersymmetric states on the half-plane in this problem.
This follows from the fact that the space of supersymmetric
states must be invariant under Hamiltonian symplectomorphisms acting on the Lagrangian submanifold $\CL$.  We can use such a symplectomorphism
to move $\CL$ so that it does not intersect $R^\zeta$ at all.  (This is possible, in part, because the algebraic intersection number of $\CL$ and
$R^\zeta$ vanishes in this example.)  Accordingly, it must be that quantum mechanically $\CQ_{\zeta}$ does
not annihilate the states $|p\rangle$ and $|p'\rangle$.  In fact, $\CQ_{\zeta}$ must exchange these two states, as $\CQ_{\zeta}$ reverses the $\IZ_2$-valued fermion
number and the two states have opposite fermion number.  So $\CQ_{\zeta}$ must act by
\begin{align}\label{zonk} \CQ_{\zeta}|p\rangle & =\lambda|p'\rangle \cr \CQ_{\zeta}|p'\rangle & = \lambda'|p\rangle, \end{align}
where $\lambda$ and $\lambda'$ cannot both be zero.  However, the symmetry $\kappa$ ensures that $\lambda$ is nonzero if and only if $\lambda'$
is nonzero.\footnote{One might be surprised to have both $\lambda$ and $\lambda'$ nonzero, since one counts ascending gradient flows
from $p$ to $p'$ and the other counts ascending flows from $p'$ to $p$.  How can there be ascending flows in both directions?  The point is
that we are in a situation in which the superpotential is not single-valued.}  Hence in this space of states $\CQ_{\zeta}^2=\lambda\lambda'\not=0$.

In the topological $A$-model, one would describe the fact that $\CQ_{\zeta}^2\not=0$ by saying that the brane under consideration is not a valid $A$-brane.
It appears that in the physical supersymmetric model, this brane makes sense but that for strings ending on this brane, there is a central
charge such that $\CQ_{\zeta}^2$ is a nonzero constant.  In any event, the fact that $\CQ_{\zeta}^2\not=0$ in the presence of this brane means that supersymmetric
states involving this brane cannot be studied using the methods of the present paper.

But what goes wrong with the proof that $\CQ_{\zeta}^2=0$?  We view the $\sigma$-model on the half-line as an infinite-dimensional version of Morse theory,
with the Morse function of eqn. (\ref{defhox}).  As explained in section \ref{whyindeed}, to try to prove that $\CQ_{\zeta}^2=0$, we are supposed
to look at two-dimensional moduli spaces of the flow equation, which in the present context is the $\zeta$-instanton equation on the half-plane
$x\geq 0$.  To be more precise, to try to prove that $\CQ_{\zeta}^2$ annihilates the state $|p\rangle$, we
must study solutions of the $\zeta$-instanton equation that approach the $\zeta$-soliton
$\Phi_p$ for $\tau\to\pm \infty$, approach $\phi=0$ for $x\to\infty$, and obey $|\phi|=1$ at $x=0$.
To get a two-dimensional moduli space, we need a fermion number anomaly of 2, and this means, since the fermion number
anomaly is twice the winding number of the boundary, that along the boundary of the half-plane (which is the line $x=0$),  $\phi$ wraps once around the unit circle.

If
$\lambda$, $\lambda'\not=0$, then solutions obeying the appropriate conditions do exist.  For to get a non-zero matrix element $\CQ_{\zeta}|p\rangle=\lambda|p'\rangle$, there is a solution
of the $\zeta$-instanton equation of the half-plane that describes a flow from $\Phi_p$ in the past to $\Phi_{p'}$ in the future; and this
solution has only the one modulus that results from time-translation symmetry.  Similarly, a non-zero matrix element $\CQ_{\zeta}|p'\rangle=\lambda'|p\rangle$
means that there are $\zeta$-instantons describing flows from $\Phi_{p'}$ back to $\Phi_p$, again with only one modulus associated to
time translations.  As always in Morse theory, by combining these solutions to a broken path $\Phi_p\to \Phi_{p'}\to \Phi_p$ and
correcting the broken path slightly to get a family of exact flows, we get a family of solutions of the $\zeta$-instanton equation with
a two-dimensional moduli space $\M$ and hence a one-dimensional reduced moduli space $\M_\red$.  $\M_\red$ has one end corresponding
to the broken path $\Phi_p\to \Phi_{p'}\to \Phi_p$.  This broken path by itself makes a nonzero contribution to the matrix element of $\CQ_{\zeta}^2$
from $|p\rangle$ to itself.

Usually, in Morse theory, the proof that $\CQ_{\zeta}^2=0$ comes by arguing that any such $\M_\red$ must have a second end corresponding to another
broken path starting and ending at $\Phi_p$, and that the two broken paths make canceling contributions to the matrix element of $\CQ_{\zeta}^2$.
However, in the present context, the second end of $\M_\red$ is not another broken path; it is a sort of  ultraviolet end that is possible because
of the fermion number anomaly.  It seems to be difficult to find exactly the relevant two-parameter family of $\zeta$-instantons (with a quadratic
superpotential, the $\zeta$-instanton equation is linear but it is difficult to satisfy the boundary conditions).  However, we can easily find the
end of $\M_\red$ that is not a broken path and so leads to $\CQ_{\zeta}^2$ being non-zero.  This end is described by a $\zeta$-instanton that
coincides with the $\zeta$-soliton $\Phi_p$ except very near a boundary point $x=0$, $\tau=\tau_0$ (for some $\tau_0$).
At short distances, the $\zeta$-instanton equation can be approximated by the equation $\partial\phi/\partial\bar s=0$ for a holomorphic map;
 as usual $s=x+i\tau$.  A simple
family of such holomorphic maps from the right-half plane to the unit disc, mapping the boundary of the right half plane to the boundary of the disc
with winding number 1 is given by a fractional linear transformation
\begin{equation}\label{fraclin}\phi_{a,\tau_0}(s)=i\frac{(s-i\tau_0)-a}{(s-i\tau_0)+a},~~\tau_0\in\IR,~~a>0. \end{equation}
This is a family of holomorphic maps from the right half-plane to the unit disc, mapping the boundary of the half-plane to the
boundary of the disc, and depending on the two real parameters $\tau_0$ and $a>0$.
 (A third parameter has been fixed by requiring that $\phi_{a,\tau_0}(\pm i\infty)=i$.)  For very small $a$, $\phi(s)$ differs substantially from $i$
 only for $|s-i\tau_0|\lesssim a$.  In the limit $a\to 0$, the reduced moduli space of holomorphic maps has an ``end.''  For small $a$,
 we can find a family of $\zeta$-instantons that coincide with $\Phi_p$ except very near $s=i\tau_0$ and with $\phi_{a,\tau_0}$ near
 $s=i\tau_0$; the asymptotic behavior of $\phi_{a,\tau_0}$ has been chosen to make this possible. (As usual, gluing gives a family of
  approximate solutions
 and index theory and general properties of nonlinear partial differential equations predict that this can be slightly corrected to a family
 of exact solutions.) Thus in this example, the $\zeta$-instanton
 moduli space has an end which is not a broken path and  which spoils the usual strategy to show that $\CQ_{\zeta}^2=0$.

\section{$\zeta$-Instantons And $\zeta$-Webs}\label{zetawebs}

\subsection{Preliminaries}\label{prelims}

One of the most important properties of the $\zeta$-instanton equation is that in any massive theory, it has no pointlike
solutions.   To be more precise, the  only solution on $\R^2$
that approaches a prescribed critical point $\phi=\phi_i$ at infinity is the constant solution with $\phi=\phi_i$ everywhere.

This is proved by a standard type of argument.   Suppose we are given a solution of  the $\zeta$-instanton equation
\begin{equation}\label{tox}\frac{\partial \phi^I}{\partial \bar s}-\frac{\i \zeta}{4}g^{I\bar J}
\frac{\partial \bar W}{\partial \bar\phi{}^{\bar J}}=0\end{equation}
on the whole complex $s$-plane. Then
\begin{equation}\label{mofox}0=\int\d^2s\left(
\frac{\partial \phi^I}{\partial \bar s}-\frac{\i \zeta}{4}
g^{I\bar K}\frac{\partial \bar W}{\partial \bar\phi^{\bar K} }\right)
\left( \frac{\partial \bar \phi^{\bar J} }{\partial  s}+\frac{\i \zeta^{-1}}{4}
g^{L\bar J}\frac{\partial   W}{\partial  \phi^L }\right)g_{I\bar J}
. \end{equation}
Expanding this out, we obtain
\be\label{rofox}
0 = \int \d^2 s \left( \left\vert \frac{\p\phi }{\p  \bar s}  \right \vert^2 + \frac{1}{16}\left \vert \frac{\p   W}{\p   \phi}\right \vert^2 \right)
+\half \Im \left[ \zeta \int \d^2 s     \frac{\p }{\p   s}  \overline{W} \right] .  \ee
Alternatively, for any $R>0$, after integrating by parts,  we have
\be\label{tofox}
0 = \int_{|s|\leq R} \d^2 s \left( \left \vert \frac{\p\phi }{\p  \bar s}  \right \vert^2 + \frac{1}{16} \left\vert \frac{\p   W}{\p   \phi} \right
\vert^2 \right)
+ \frac{R}{4}\, \Im\left[ \zeta  \int_{0}^{2\pi} e^{-\I \theta} \bar W d \theta \right]_{|s|=R},
\ee
where $\d^2 s = \frac{\I}{2}  \d s\wedge \d \bar s $, $\theta = \arg(s)$.

After possibly adding a constant to $W$ (which does not change the $\zeta$-instanton equation),
we can assume that $W(\phi_i)=0$.  In any massive theory, a solution in which $\phi\to \phi_i$
at infinity has the property that $\phi$ approaches $\phi_i$ exponentially fast.  Hence $W(\phi)$ approaches its limiting value $W(\phi_i)=0$
exponentially fast, and therefore the surface term in (\ref{tofox})
vanishes for $R\to\infty$.  Accordingly, these formulas imply that
$\partial W(\phi)/\partial\phi^I$ identically vanishes in any such solution of the $\zeta$-instanton equation,
so that $\phi$ is everywhere equal to a critical
point of $W$.  In a massive theory, the critical points are
a discrete set, and therefore  $\phi$ must be identically equal to its limiting value at infinity.

\begin{figure}[htp]
\centering
\includegraphics[scale=0.3,angle=0,trim=0 0 0 0]{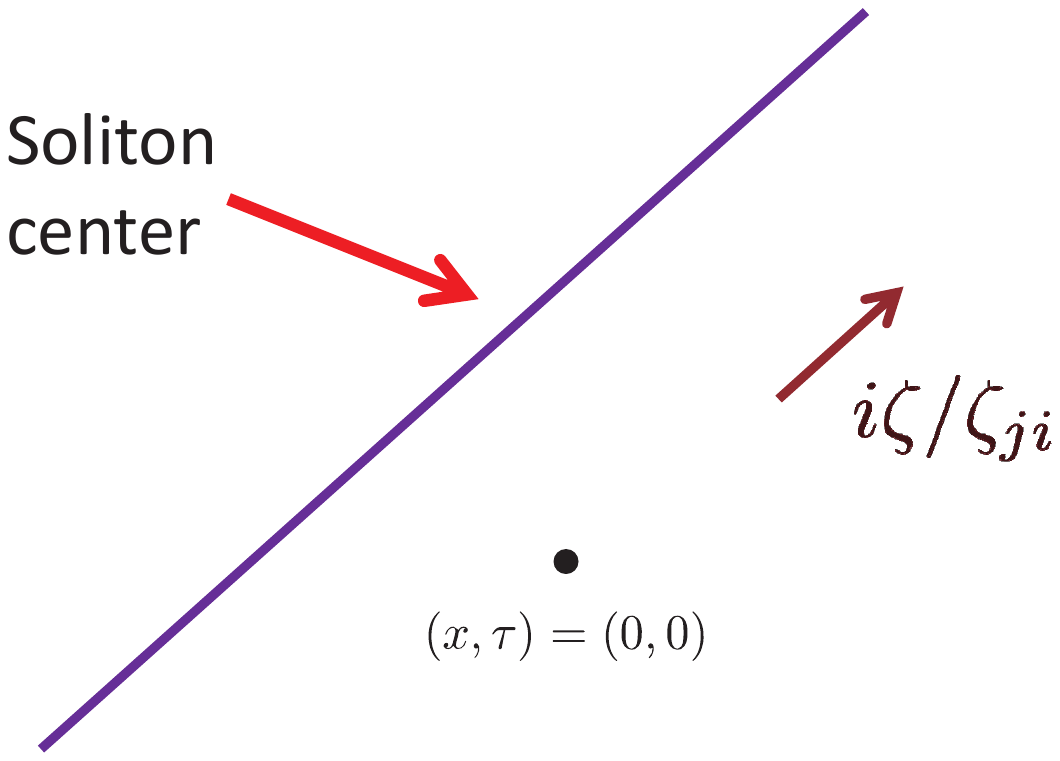}
\caption{A ``boosted'' $ij$ soliton defines a $\zeta$-instanton with ``core'' along
an oriented line $\ell$ in the $(x,\tau)$ plane parallel to the phase $\pm \I \zeta/\zeta_{ji}$. The same line with opposite orientation determines
a $ji$ soliton with opposite impact parameter.  }
\label{fig:BOOSTEDSOLITON}
\end{figure}

Although the $\zeta$-instanton equation does not have point-like solutions, it does have solutions localized on lines,
as we explained qualitatively in the introduction.  We have chosen $\zeta$ so that for any $i,j\in\IV$, there does not exist
a time-independent solution of the $\zeta$-instanton equation interpolating between $\phi=\phi_i$ at $x=-\infty$ and $\phi=\phi_j$
at $x=+\infty$.  However, such a solution $\phi_{ij}(x)$ might exist if we replace $\zeta$ in the $\zeta$-soliton equation by an appropriate
value $\zeta_{ji}$.  If $\phi_{ij}(x)$ obeys the $\zeta_{ji}$-soliton equation
\begin{equation}\label{mondox}\frac{\d}{\d x}\phi^I_{ij}(x)=\frac{\I \zeta_{ji}}{2}g^{I\bar J}\partial_{\bar J}\bar W(\phi_{ij}(x)),\end{equation}
then clearly its rotated counterpart
\be
\phi_{ij}^{\rm euc}(x,\tau) := \phi_{ij}(\cos\trho x + \sin\trho \tau),
\ee
obeys
\be\label{eq:boosted-soliton}
\left(\frac{\p  }{\p x} + \I \frac{\p }{\p \tau} \right)\phi_{ij}^{{\rm euc},I}(x,\tau)
=  \frac{\I e^{\I \trho}\zeta_{ji} }{2}g^{I\bar J}\partial_{\bar J}\bar W(\phi_{ij}(x))
\ee
We call this a \emph{boosted soliton}, although in Euclidean signature it might be more apt to speak of a \emph{rotated soliton}.
This construction gives a solution to
the $\zeta$-instanton equation
\eqref{eq:LG-INST} provided we choose $\trho$ so that
\be\label{eq:xi-to-zeta}
e^{\I \trho}  \zeta_{ji} = \zeta
\ee
%
%{\it I actually do not understand the factors of $i$ in these formulas.  If there is an $i$ in the $\zeta$-instanton
%eqn, isn't there one in the $\zeta$-soliton equation?}
%
The boosted soliton is of course not at rest; rather, it is localized along a line $L$ in the complex $s$-plane.
This line is parallel to the complex number $ \I e^{\I\trho}$ as in Figure
\ref{fig:BOOSTEDSOLITON}. Comparing this with the definition of plane webs in Section \S \ref{planewebs} and
using equation \eqref{hopeful}
we deduce that the relation between plane web vacuum weights and the critical values of the superpotential
is $z_j = \zeta \bar W_j$.

In saying that the solution is localized on a line,
we ignore the width of the BPS solitons.  When one looks more closely, these solitons have a width no greater than
 $\ell_W = 1/m$, where $m$ is the mass of the lightest particle of the theory.  Because of this width, though $L$ has a precisely
defined angle, its ``impact parameter'' is only naturally defined to within a precision of $\ell_W$.
A similar remark holds in many statements below about the relation between $\zeta$-instantons and webs.

Though there is no natural definition of  the impact parameter of a $\zeta$-soliton, it is convenient for some purposes
to pick a specific definition. First we define for each $ij$ soliton precisely what we mean by its center of mass.  There is no completely
natural choice, and we pick some definition.  (For instance, for a soliton at rest, we can pick the unique point with equal integrated energy to its
left and right.)  Having done this, a possibly boosted soliton
  is centered on a well-defined oriented line $\ell$.
 We pick an arbitrary origin $\v\in \R^2$ and use the orientation of $\R^2$ that is built into the $\zeta$-instanton equation. Then we
 define the impact parameter of the given soliton as the signed distance by which $\ell$ passes
 to the right of $\v$ (Figure\ref{fig:BOOSTEDSOLITON}).

\subsection{Fan-Like Asymptotics and $\zeta$-Webs}\label{zetafan}

The $\zeta$-instanton equation might also have solutions of a more complicated sort that again was schematically described in the introduction.
At large values of $\vert s\vert$ the solutions approach piecewise constant functions on the complement of certain rays in the complex $s$-plane.
(More precisely the rays should be replaced by   certain strips with a width of order $\ell_W$.) The constant values of $\phi$ are
a cyclically ordered sequence of vacua $i_1,\dots,i_p\in \IV$, for some $p$. Across a ray separating
 vacuum $i_k$ from vacuum $i_{k+1}$ the field is well approximated by a boosted $i_ki_{k+1}$ soliton of the type just
discussed.  The picture is schematically depicted  in Figure  \ref{fig:WEDGES} of Appendix \ref{subsec:DefFanBC} below
(where the boundary conditions are spelled out a bit more precisely).
In the picture, in crossing from vacuum $i_k$ to vacuum $i_{k+1}$, the central charge jumps from $W(\phi_{i_k})$ to $W(\phi_{i_{k+1}})$,
and the angle $\trho$ at which the boosted $i_ki_{k+1}$ soliton emerges is described
by $e^{\I \trho}  \zeta_{ji} = \zeta$, as  in eqn. (\ref{eq:xi-to-zeta}).  This angle must be a decreasing function of
$k$ in order for the picture to make sense,
and therefore $i_1,\dots,i_p$ must be a cyclic fan of vacua in the sense defined in section
\ref{planewebs}.\footnote{Recall our convention that for a fan of vacua $i_1,\dots,i_p,$
   reading left to right the vacua are located in regions in the clockwise
 direction.}   An important detail  is that although the
angles of the outgoing solitons are directly determined by  $\zeta$ and the critical values $W_i$,
the impact parameters of the outgoing solitons (defined by some specific procedure like that of section \ref{prelims})
are not; they must be found by solving the $\zeta$-instanton equation.   The outgoing rays are only defined near infinity on the $s$-plane, and if continued into the interior of the picture,
they do not necessarily meet. Motivated by this picture, we define a
 a \emph{fan of solitons} to be a cyclically-ordered collection of solitons of type\footnote{We will be more precise in section \ref{collective}
 about how to treat the fact that each classical soliton
corresponds to two quantum states.}
 $\{ \phi_{i_1,i_2}, \dots, \phi_{i_p,i_1} \}$.

As was discussed qualitatively in the introduction, low energy field theory (as opposed to a full study of the nonlinear $\zeta$-instanton equation) is not powerful enough to determine
whether solutions with such fan-like asymptotics at infinity actually exist.
The considerations of the present paper are most interesting if they do exist and we proceed assuming this
is the case. In fact, the $\zeta$-instanton equations with such boundary conditions have been studied
previously in the context of domain wall junctions in \cite{Carroll:1999wr,Gibbons:1999np} where, in some
special cases, existence proofs are given.

Although we refer to solutions of the $\zeta$-instanton equation with fan asymptotics as ``$\zeta$-instantons,'' the fact that such solutions have
outgoing solitons means
that they are not localized in spacetime and are not instantons in the usual sense.
 In fact they have infinite action if we use the
   standard Landau-Ginzburg action discussed (implicitly) in Section
   \S \ref{subsec:LG-Theory} above. More precisely, using
   \eqref{tofox} one easily shows that the contribution to the standard Landau-Ginzburg action
   from integrating over a   disk of radius $R$ approaches
\be\label{eq:WaveFunRenorm}
 R  \sum_k \vert W_{i_{k+1}} - W_{i_k} \vert + \rho + 2 \int_{\IR^2} \phi^*(\omega)
\ee
as $R \to \infty$. Here $\omega$ is the K\"ahler form on $X$
and one can show, using \eqref{tofox} that $\rho$ has a finite limit as $R\to \infty$; it is an
interesting function of the fan of solitons. The term linear in $R$ is just $R$ times the sum of the masses of the
solitons. One could remove this term by a  suitable ``wavefunction renormalization.''
We will, actually, consider a slightly different action for the path integral in our discussion
below. Namely, we take the action given by squaring the $\zeta$-instanton equation,
as in equation \eqref{mofox}, for the bosonic fields, and then adding the supersymmetric
completion. In this theory, the $\zeta$-instantons have zero action.

In a model in which solutions with fan-like asymptotics
do exist, the most important ones in a certain sense are the ``vertices,'' the families of irreducible  solutions.  This concept
requires some explanation.  Suppose that
for a specified fan $i_1i_2\dots i_p$ of vacua, there is a nonempty moduli space $\M=\M_{i_1i_2\dots i_p}$ of solutions of the $\zeta$-instanton
equation. (In defining $\M$, we do not specify the impact parameters of external solitons.  See section
\ref{tonzo} for more on this.)
$\M$ is not necessarily connected and might  have components of different dimension.
Our considerations are most natural in an
$\FF$-conserving theory, as we have explained in section \ref{impanom}.
The conservation of the fermion number $\FF$ together with
the analysis of Sections \S\ref{tonzo} and \S\ref{bottom} implies a relation between the expected
dimension of $\M$ and the fermion numbers $(f_a,f_a+1)$ of the solitons $\phi_{i_a,i_{a+1}}$ in the fan:
\begin{equation}\label{eq:dimformula}
\mathrm{dim}\, \M = \sum_a (f_a+1)
\end{equation}
In other words, the expected dimension equals the sum of the upper fermion numbers of the outgoing solitons in the fan. See
Appendix \S \ref{subsec:DefFanBC} for further details.

The $\zeta$-instanton equation has translation symmetry but no additional relevant symmetries.  The quotient
of $\M$ by translations is a reduced moduli space $\M_{\red}$ of dimension $d-2$.  A basic question now is whether $\M_\red$ is compact,
and if not, what are its ``ends''?  In other words, after dividing by spatial translations to eliminate a trivial ``end'' of the moduli space in which a solution moves off to spatial infinity, in what ways can a sequence of solutions with  given fan-like asymptotics blow up or diverge?

We make use of the trichotomy that was explained at the end of section \ref{whyindeed}.
An end of the moduli space is either an ultraviolet effect (something blows up at short distances),
a large field effect (some fields go to infinity), or an infrared effect (something happens at large distances).
In a massive theory, we do not anticipate a large field effect, since the potential $|\d W|^2$ blows up if the fields become large.
Before discussing ultraviolet effects, we recall some examples. A typical example of an ultraviolet
``end'' of a moduli space of solutions of a partial differential equation  is the small instanton
singularity in four-dimensional Yang-Mills theory, or the analogous small instanton
singularity in a two-dimensional $\sigma$-model with a suitable K\"ahler target $X$.  The latter example is more relevant to us,
since at short distances the $\zeta$-instanton equation can be approximated by the equation $\partial_{\bar s}\phi^I=0$ for a holomorphic
map to $X$.  In the case of a massive LG model with target $X=\IC^n$, we do not have to worry about ultraviolet ends of the moduli
space, because the equation $\partial_{\bar s}\phi^I=0$ is linear and does not have an analog of the small-instanton singularity
(paying proper attention to domains, a sequence of holomorphic functions does not converge to a
meromorphic function with a pole).  The considerations of the present paper also apply
for more general $X$'s, on which small instanton singularities might in general occur.  However,
we can assume that  the $A$-model of $X$ exists without a superpotential (otherwise turning on a superpotential will hardly help), and
this means that the small instanton ends of the moduli space do not contribute to supersymmetric Ward identities.
So even if $X$ is not $\IC^n$, we do not have to worry about
ultraviolet ends of the moduli space.\footnote{Even for $X=\IC^n$, if the $\sigma$-model is formulated on a two-manifold with boundary,
in general the $\zeta$-instanton moduli space has ultraviolet ends, as we have explained in section \ref{impanom}.  But as was also explained
there, such effects are not relevant if the boundary conditions are set by a valid $A$-brane -- or are
such that the methods of the present paper are applicable.  In that analysis, we made use of the fact that ultraviolet ends of the moduli
space are not affected by a superpotential.}

Just as in the Morse theory problem of section \ref{warmup}, we \underline{do} have to worry about infrared effects. Morse theory involves a massive
theory in 1 dimension, and in this case the only interesting infrared effect is a ``broken path'' involving successive jumps
from one critical point to another.  The basic example was a two-step jump  $i_1\to i_2\to i_3$ that was crucial in understanding the
MSW complex.   To construct a solution of the gradient flow equation representing this two-step jump, we glue together two widely separated
solutions representing transitions $i_1\to i_2$ and  $i_2\to i_3$, and deform slightly to make an exact solution corresponding to
a broken path.  The ends of moduli spaces of gradient flows are given by such broken paths.

\begin{figure}
 \begin{center}
   \includegraphics[width=4.5in]{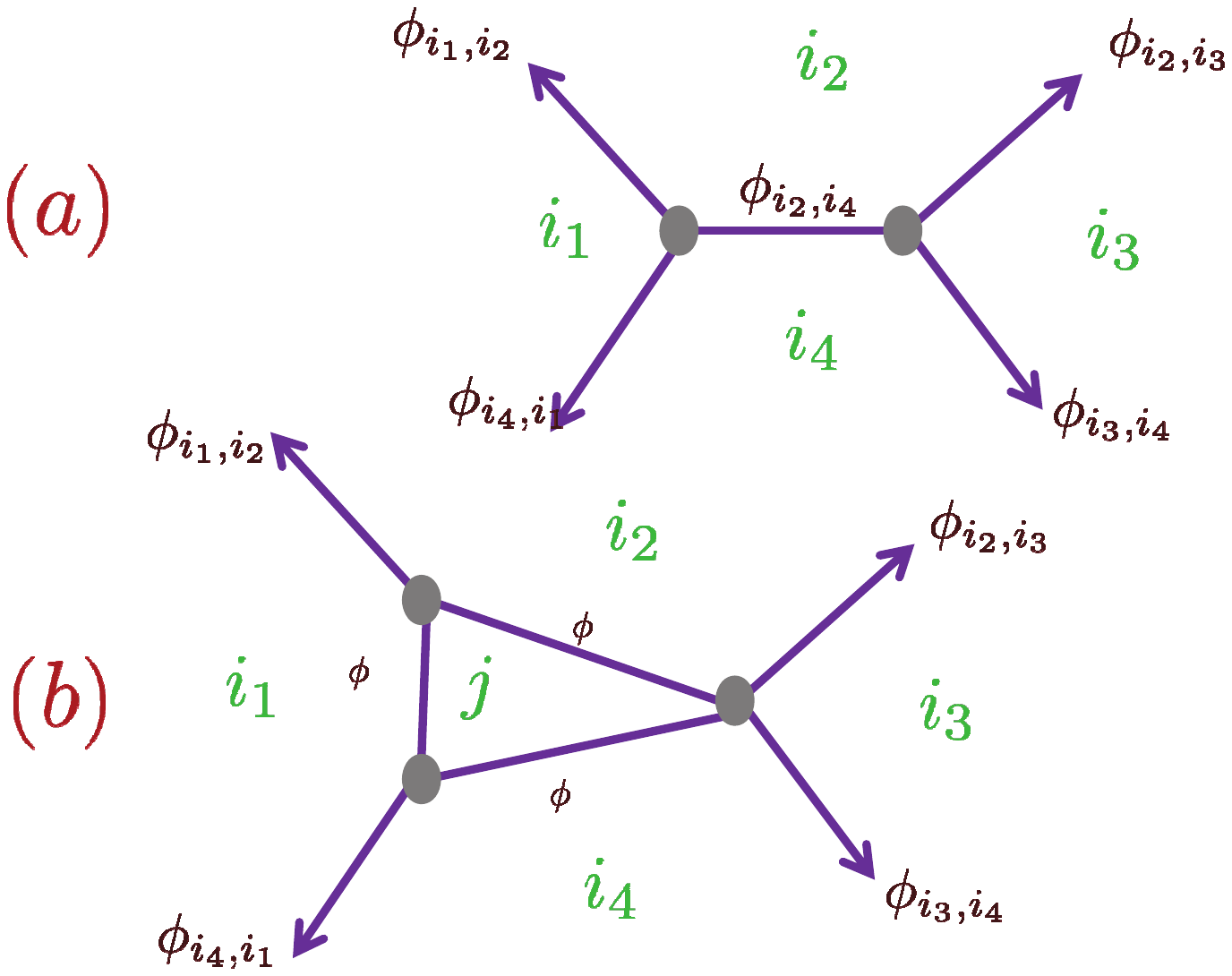}
 \end{center}
\caption{\small  Sketched here are some of the possible ``ends'' of the moduli space $\M$ of
$\zeta$-instantons with fan boundary conditions $\{ \phi_{i_1,i_2}, \phi_{i_2,i_3}, \phi_{i_3,i_4}, \phi_{i_4,i_1} \}$.
An end of the moduli space can be described by a web-like picture in which the lines represent BPS solitons that propagate
for long distances (compared to $\ell_W$), the regions between the lines are labeled by vacua, and the ``blobs'' where lines meet
represent in their own right moduli spaces of solutions of the $\zeta$-instanton equation with fan-like boundary conditions.  Thus
an end of $\M$ is associated to a gluing of moduli spaces $\M_i$ of $\zeta$-instantons (the $\M_i$ have smaller
dimension than $\M$) which themselves satisfy fan-like boundary conditions. In subsequent figures of $\zeta$-webs, we will
generally drop the explicit labeling of the edges by solitons, but they are implicitly there.  }
 \label{webend}
\end{figure}

In two dimensions, there is a similar operation of combining solutions to make a more complicated solution, but,
once we have introduced fan boundary conditions,  this operation is much
more complicated because there are many more possible gluing operations in two dimensions than in one dimension.
The basic gluing operation is made by
embedding in $\R^2$ widely separated $\zeta$-instanton sub-solutions each of which has fan-like asymptotics,  and which are positioned  in such
a way that the various soliton lines that are outgoing from the various individual sub-solutions
fit into a web.  Some examples are sketched in Figure \ref{webend}.
Such a picture represents something that is exponentially close to a solution of the $\zeta$-instanton equation, and (modulo technicalities
that we will certainly not resolve in the present paper) index theory can be used to predict that this approximate solution can be slightly
corrected to make an exact solution.  We say a little more on this in section \ref{bottom}.
We will call the picture representing such a gluing a \emph{ $\zeta$-web}.
$\zeta$-webs are analogous
to the abstract webs of section \ref{planewebs}, but  differ in two key ways:
First, an edge separating a vacuum $i$ from vacuum $j$ is labeled by a choice of
classical $ij$ soliton, as shown in Figure \ref{webend}. Second,    the vertices in a $\zeta$-web
are not simple points but are rather regions in which the approximation to a boosted soliton
breaks down. Within this region the solution should be a solution of the $\zeta$-instanton equation
with boundary conditions given by the fan of solutions defined by the lines coming out of the vertex.
Thus, we may think of the vertex-regions as representing moduli spaces of solutions with fan-like
boundary conditions.
Since   {\it a priori} we know little about these moduli spaces,
 a $\zeta$-web is a concept that only makes  sense in the limit that the vertex regions are small
 relative to the lengths of the  the internal lines in the $\zeta$-web.

A  reduced moduli space of $\zeta$-instanton solutions
corresponding to a given fan of solitons can have an ``end'' corresponding to a $\zeta$-web.
  Sketched in Figure \ref{webend} are some
examples of possible ends of the moduli space $\M_\red$ corresponding to fan boundary
conditions $\{ \phi_{i_1,i_2}, \phi_{i_2,i_3}, \phi_{i_3,i_4}, \phi_{i_4,i_1} \}$.
We claim that the only ends of reduced $\zeta$-instanton moduli spaces with fan-like asymptotics correspond to such $\zeta$-webs.
From a physical point of view, we expect this claim to be valid since degeneration to a $\zeta$-web with widely separated vertices
is the only natural infrared effect in a massive theory.  Mathematically, our claim, and even the existence of $\zeta$-instanton moduli
spaces with fan-like asymptotics, is almost certainly not a direct consequence of any standard theorem and will involve new analysis.

\begin{figure}
 \begin{center}
   \includegraphics[width=3in]{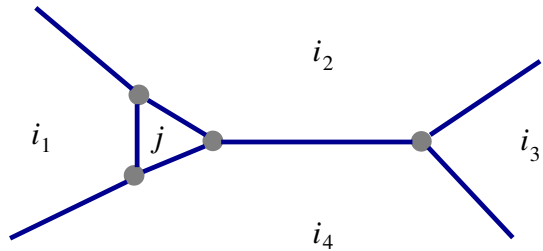}
 \end{center}
\caption{\small The ends of $\M$ depicted in Figure \protect\ref{webend}.
%
%  \ref{webend}
%
 may themselves have ends; these ends would arise by further resolving
one of the blobs of that figure as a convolution of $\zeta$-instanton moduli spaces of yet smaller dimension.  (The particular
example shown here can arise by resolving a vertex in either Figure \protect\ref{webend}(a)
%
%\ref{webend}(a)
%
of Figure \protect\ref{webend}(b)
%
% \ref{webend}(b).
%
)
This process only ends
when the blobs are $\zeta$-vertices, which by definition cannot be further resolved. }
 \label{webmore}
\end{figure}

If $\M_\red$ does have an end corresponding to a $\zeta$-web such as one of those in Figure \ref{webend}, then we can repeat the question.
Pick one of the blobs corresponding to a sub-solution in this $\zeta$-web.
 This blob
is associated to a new fan of vacua $j_1,\dots,j_q$ and has its own reduced moduli space $\M'_\red$.
We should ask the same question about $\M'_\red$ that we originally asked about $\M_\red$.  Is $\M'_\red$ compact, and if not what are
its ends?  The same reasoning as before makes us  expect
 that ends of $\M'_\red$ arise just like the ends of $\M_\red$, by resolving the chosen blob
  into a new $\zeta$-web. This is illustrated in Figure \eqref{webmore}.  The process of finding an end of a $\zeta$-web moduli space
by resolving one of the sub-solutions from which this  $\zeta$-web is constructed into a new $\zeta$-web
might be called $\zeta$-convolution as it is
fairly analogous to the ordinary convolution of webs as introduced in section \ref{sec:Webs}.

This process cannot go on indefinitely, because at each step the dimension of the moduli spaces represented by the remaining blobs
 becomes smaller.
(We will be more precise about this in section \ref{zetafan}.)  So eventually we arrive at a web that is made from ``$\zeta$-vertices.''
By a $\zeta$-vertex $\V$ we mean a fan of solitons together with a {\it compact} and \emph{connected}
component of its reduced moduli space. We denote it by $\M_\red(\V)$.
Thus a $\zeta$-vertex
represents a family of solutions of the $\zeta$-instanton equation that does not have any ends at which it can be resolved into
a non-trivial $\zeta$-web (a $\zeta$-web with more than one vertex).

There can be several $\zeta$-vertices for a given fan of
solitons. That is, there can be several compact connected components of the reduced moduli space of $\zeta$-instantons with a fixed
fan of solitons at infinity. More generally,  for a given  fan at infinity, one expects the moduli space $\M$
of $\zeta$-instantons to have only finitely many components.  This is actually a special case of the statement that $\M$
is compact except for ends arising from gluings of sub-solutions.

The $\zeta$-vertices of reduced dimension 0 play a special role in our construction.
We will call them rigid $\zeta$-vertices.  The reduced moduli space of a rigid $\zeta$-vertex is by definition a point.
The algebraic structures constructed in this paper can be understood entirely in terms of rigid  $\zeta$-vertices.
In particular, the rigid $\zeta$-vertices suffice for answering one of the basic questions in the present paper,
which is to understand in terms of webs the space of supersymmetric states of the $(\B_\ell,\B_r)$ system, where $\B_\ell$ and $\B_r$
are left and right $W$-dominated branes  described in section \ref{goodclass}.   This statement reflects the following considerations.

A $\zeta$-vertex $\V$,
if we forget the $\zeta$-instanton equation, determines in particular a fan of vacua to which we can associate a web $\frak w_\V$
with only one vertex.  In the language of section \ref{planewebs}, since $\frak w_\V$ only has one vertex, its reduced moduli space  is
a point, of dimension 0.  This coincides with $\M_\red(\V)$ if $\V$ is rigid, but if
$\V$ is a non-rigid $\zeta$-vertex with $\M_{\red}(\V)$ of dimension $\varepsilon_\V>0$, then $\M_\red(\V)$ parametrizes
internal degrees of freedom of a family of $\zeta$-instanton solutions that cannot be described in terms of webs.
We call $\varepsilon_\V$
the \emph{excess dimension} of $\M_\red(\V)$.

Our definitions and assumptions imply that if $\SSS$ is any component of the moduli space of   $\zeta$-instantons  on
the $s$-plane with fan-like behavior at infinity, then either $\SSS$ is the moduli space associated to a $\zeta$-vertex $\V$ or $\SSS$ has an
end  corresponding to  a $\zeta$-web  made from $\zeta$-vertices $\V_i$.
In the latter case, we define $\varepsilon(\SSS)=\sum_i\varepsilon(\V_i)$.  (We claim that this sum does not depend on the choice of a particular
end of $\SSS$.)  As we explain in section \ref{bottom},
$\varepsilon(\SSS)$ is the excess dimension of the moduli space $\M(\SSS)$ associated to $\SSS$, relative to the dimension of the moduli
space of the web $\frak w_\SSS$.   This  excess dimension is always non-negative and vanishes precisely if  the ends of $\SSS$ are
built from
rigid $\zeta$-vertices.

Finally we can explain the importance of the case that the excess dimension is zero.
For the same reasons
as in the supersymmetric approach to Morse theory, the dimension of a moduli space of solutions of the $\zeta$-instanton
equation determines the net violation of fermion number in an amplitude derived from that moduli space. (As in Morse theory, the relevant
fact is that the Dirac equation for the fermions of fermion number 1 is the
linearization of the $\zeta$-instanton equation.) By considering only moduli
spaces of zero excess dimension, we can answer questions that involve the minimum violation of fermion number.  Such a question
is to determine the space of supersymmetric $(\B_\ell,\B_r)$ states on a strip; the differential of the MSW complex
comes from $\zeta$-instanton solutions on the strip with no
 reduced moduli (except the one that is required by time-translation invariance), so that the excess dimension of the moduli space must
 vanish. Since excess dimensions are non-negative and add under natural operations of combining webs or vertices, to construct the differential
 of the MSW complex, we only need to study $\zeta$-vertices of excess dimension 0, in other words the rigid ones.
However, the $A$-model can have local observables of positive fermion number, and to compute their matrix elements requires
considering moduli spaces with a positive excess dimension; see section \ref{sec:RemarksLocalObs}.  For this application,
 we do need to consider the non-rigid $\zeta$-vertices.

We stress again that because a BPS soliton has a nonzero width of order $\ell_W$, the solitons emanating from a rigid $\zeta$-vertex
cannot literally be identified with rays that emanate from a point in the plane.   To define  rays associated to solitons, one must
choose  a ``center of mass'' for each BPS soliton, as in section \ref{prelims}.  Regardless of those choices, one should not
expect the solitons emanating from an arbitrary
$\zeta$-vertex to correspond to rays that emanate from a point in the plane.
This will only work to within a precision of order $\ell_W$. In the case of a non-rigid $\zeta$-vertex, after making a
precise definition of the centers of masses of the solitons,
the offsets of the rays representing the solitons will depend on the excess moduli of the vertex, though only by an amount of order $\ell_W$.
The relationship between $\zeta$-solitons and webs is a statement about the infrared limit and
always involves ignoring a discrepancy of order $\ell_W$.

\subsection{The Index Of The Dirac Operator}\label{tonzo}

Let $L$ be the Dirac operator in this theory for fermions of fermion number $\FF=1$.
The Dirac equation is
\be\label{eq:Inst-L}
\biggl[ \frac{\p}{\p r} + \I \sigma^3 \frac{1}{r} \frac{\p}{\p \theta}
+ \frac{\I}{2}
\begin{pmatrix}
0 & - \zeta e^{-\I\theta} \bar W'' \\
 \zeta^{-1}  e^{\I\theta}   W'' & 0  \\
\end{pmatrix} \biggr]
\begin{pmatrix} \delta\phi \\  \delta\bar\phi \\ \end{pmatrix} = 0
\ee
where $s = x+ \I \tau = r e^{\I \theta}$ and for simplicity we take $X=\IC$ with standard Euclidean metric.
For the same
reasons as in supersymmetric quantum mechanics, the index $\iota(L)$ is the expected dimension of the moduli
space $\M$ of $\zeta$-instantons. This happens because the operator $L$ is the linearization of the $\zeta$-instanton
equation.

 However, there is a subtlety that does not have an analog in supersymmetric quantum mechanics. When we consider
 a  $\zeta$-instanton that is asymptotic at infinity to a fan consisting of $p\geq 3$ solitons, each soliton has its own ``impact parameter.''
 We have to decide whether in defining a moduli space $\M$ of such $\zeta$-instantons, we want to allow deformations
 in which these impact parameters change.  Such deformations correspond to zero-modes of $L$ that are asymptotically constant
 along each outgoing soliton;
 \footnote{That is,  one could make the ansatz for the Dirac equation $\delta\phi(x,\tau) = f(r,\theta) \phi_{ij}'(r(\psi_{ij}-\theta))$ in the neighborhood of an outgoing boosted $ij$ soliton where the soliton ray is parallel to $e^{\I\psi_{ij}} = \I \zeta/\zeta_{ji}$. As $r\to \infty$ the Dirac equation rapidly approaches the free Dirac equation for $f(r,\theta)$. We require that $f(r,\theta)$ approaches a constant for $r\to \infty$.}
these represent an asymptotically constant displacement of the soliton in question. They are certainly not square-integrable.

 For most purposes, it is more natural for us to define a moduli space $\M$ of $\zeta$-instantons in which the impact
 parameters of outgoing solitons are {\it not} specified.  The dimension of $\M$ is then the index $\iota(L)$ of the operator
 $L$, acting on a space of $\FF=1$ fermion states that are allowed to be asymptotically constant along each outgoing soliton.
 This motivates our conjectural dimension formula \ref{eq:dimformula}.

 An important detail is that $\iota(L)$ is the difference in dimension between the kernel of $L$, acting on $\FF=1$ fermions, and the
 kernel of the adjoint operator $L^\dagger$, acting on $\FF=-1$ fermions.  The adjoint condition to ``constant at infinity'' is
 ``vanishing at infinity'' so in computing $\iota(L)$, one counts $\FF=-1$ zero modes that vanish at infinity.

 Although we will not often find this useful, we could alternatively define a moduli space $\M^\diamond_{x_1,\dots,x_p}$ in which
 the impact parameters of the solitons are required to take specified values $x_1,\dots, x_p$.  To make sense of this, as described at the end of section \ref{prelims},
 we first define for each $ij$ soliton  precisely what we mean by its impact parameter.  Then we denote as
  $\M^\diamond_{x_1,\dots,x_p}$ the $\zeta$-instanton moduli space in which the impact parameters take specified values $x_1,\dots,x_p$.
  The expected dimension of
 $\M^\diamond_{x_1,\dots,x_p}$ is $\iota^\diamond(L)$, where $\iota^\diamond(L)$ is the index of $L$ acting on $\FF=1$ fermions that vanish at infinity
 (and dually, on $\FF=-1$ fermions that are allowed to be constant at infinity).  The relation between the two
notions of the index of $L$ is
\begin{equation}\label{zota}\iota^\diamond(L)=\iota(L)-p. \end{equation}
Indeed, each time we constrain the $\FF=1$ fermions to vanish at infinity along a particular outgoing soliton, and drop a dual requirement
for $\FF=-1$ fermions to vanish at infinity along that soliton, we  remove an $\FF=1$ zero-mode
or add an $\FF=-1$ zero-mode, in either case reducing the index by 1.

If $\iota^\diamond(L)>0$, this implies that there are normalizable fermion zero-modes of $\FF=1$.  All the amplitudes we compute below will then vanish
unless we insert local $A$-model observables to absorb those zero-modes.
 In a Landau-Ginzburg model with target $X=\IC^n$, there are no
such observables in bulk, but more general models can have such observables. (Even for target $\IC^n$, there can be boundary
$A$-model observables for suitable choices of brane.)   Dually, if $\iota(L)<0$, there are normalizable fermion zero-modes
of $\FF=-1$, which will also ensure vanishing contributions to $A$-model amplitudes.  However, $\zeta$-instanton moduli spaces
with $\iota(L)<0$ have negative expected dimension and generically do not exist.  In fact, translation invariance implies that an actual
$\zeta$-instanton moduli space has dimension at least 2, so $\zeta$-instanton moduli spaces with $\iota(L)<2$ generically do not exist.
Hence we are primarily interested in the case that $\iota(L)\geq 2$ and (if local $A$-model observables are not relevant) $\iota^\diamond(L)\leq 0$.
The last condition tends to be violated if excess dimensions are too large, so it is part of the reason that $\zeta$-instanton moduli spaces with
positive excess dimension are not relevant unless we consider local $A$-model observables.

When $\iota(L)=d\geq 2$ for a given component $\M$ of $\zeta$-instanton moduli space, we expect  $\M$ to be generically a smooth
manifold of dimension $d$.    When this is so, if $\iota^\diamond(L)=-k$ is negative, what this means generically is that $\M^\diamond_{x_1\dots x_p}$ is
nonempty only if $k$ conditions are placed on the impact parameters $x_1,\dots,x_p$.

\subsection{More On $\zeta$-Gluing}\label{bottom}

Suppose that we are given a fan of solitons corresponding to vacua $i_1,\dots,i_p$, $p\geq 3$.
  The sum of the fermion numbers of the chosen solitons is an integer.  Indeed, the fermion
numbers of the individual solitons in the fan are given mod $\IZ$ by certain boundary terms (eqn. (\ref{explained})), and these boundary
terms cancel when we add up the fermion numbers of all the solitons in the fan.

Let us suppose that there exist solutions of the $\zeta$-instanton equation on the $s$-plane that are asymptotic at infinity to the chosen
fan of solitons and let $\M$ be a component of the corresponding moduli space. As already stressed in section \ref{tonzo}, in defining $\M$ we do not
specify the ``impact parameters'' of the outgoing solitons.  In the notation of that section, when
 $\iota(L)=d$, this means that in trying to solve the $\zeta$-instanton equation (without specifying the impact parameters of outgoing
 solitons), there are in effect $d$ more unknowns than equations.
That is why the moduli space has dimension $d$.  Now let us consider the gluing operation of section \ref{zetafan} from the point of
view of index theory.  We try to glue various sub-solutions of the $\zeta$-instanton equation corresponding to moduli spaces $\M_i$
of dimensions $d_i\geq 2$.  We make this gluing via a web $\frak w$.  The corresponding web moduli space $\D(\frak w)$
 was studied in section \ref{planewebs} and (for generic vacuum data) has dimension
 \be \label{webdim}d(\frak w)=2V-E,\end{equation}
 where $V$ and $E$ are respectively the numbers of vertices and internal edges in the web $\frak w$.  Because the three-dimensional
 group of translations and scalings of $\IR^2$ acts on $\D(\frak w)$, we are usually only interested in gluings
 leading to webs such that $d(\frak w)\geq 3$; other gluings cannot be realized by webs of BPS solitons unless the
 central charges of the solitons take special values.

 The analog of eqn. \ref{webdim} for the dimension of a hypothetical moduli space $\M^*$ of $\zeta$-instantons that arises by gluing
 of sub-solutions with moduli spaces $\M_i$ is
 \begin{equation}\label{zetadim} d(\M^*)=\sum_i d(\M_i)-E. \end{equation}
 For webs, each vertex carries 2 moduli, accounting for the contribution $2V$ in (\ref{webdim}).  For $\zeta$-webs,
 the contribution of each sub-solution is instead $d(\M_i)$, accounting for the formula (\ref{zetadim}).  The $-E$ in (\ref{zetadim})
has the same origin as in (\ref{webdim}):  for every edge in a web or $\zeta$-web, there is one constraint so that the
two vertices or sub-solutions can be connected by a line or a $\zeta$-soliton at the appropriate angle.
This result is compatible with our conjectural dimension formula \ref{eq:dimformula}: each internal edge removes
the upper fermion numbers of two CPT-conjugate solitons, which add to $1$.

{\it A priori}, when we  glue together widely separated sub-solutions -- one for each vertex in the web $\frak w$ --  we do not get an exact $\zeta$-instanton solution, but we come exponentially close to one. However, index theory and
generalities about nonlinear equations make one expect that when the expected dimension $d(\M^*)$ is positive and the individual
moduli spaces that are being glued are smooth (no fermion zero-modes of $\FF=-1$), a very good approximate
solution can be corrected to a nearby exact solution.   The condition on the expected dimension is satisfied, since $d(\M^*)\geq d(\frak w)\geq 3$.
The formulas (\ref{webdim}) and (\ref{zetadim}) and
the conditions $d(\frak w)\geq 3$, $d(\M_j)\geq 2$ for all $j$,
  imply that $d(\M^*)>d(\M_i)$ for all $i$.  This inequality ensures that the process of repeatedly resolving a $\zeta$-moduli
space by blowing up a sub-solution into a $\zeta$-web  must terminate, as claimed in section \ref{zetafan}.
By definition, when it terminates, the $\M_i$ are all $\zeta$-vertices.
Then $d(\M_i)=2+\varepsilon_i$, where $\varepsilon_i$ is the excess dimension defined in section \ref{zetafan}.
So the excess dimension of the $\zeta$-web whose moduli space is $\M^*$ (defined as the difference between $d(\M^*)$
and the corresponding dimension $d(\frak w)$ of the ordinary web moduli space) is
\begin{equation}\label{ongoing} d(\M^*)-d(\frak w)=\sum_i\varepsilon_i. \end{equation}
This is the additivity of the excess dimension claimed in section \ref{zetafan}.

Since $d(\M_i)$ might be greater than 2, $d(\M^*)$ might be large even if $d(\frak w)<3$.  This may make one wonder if
gluing of $\zeta$-instantons can produce a $\zeta$-web that does not have an analog (for generic central charges) in ordinary  webs.
However, if a given web $\frak w$ cannot be constructed (with a given set of central charges) if the vertices are points,
then it also cannot be constructed if the vertices are pointlike within an error of order $1/\ell_W$.
To construct such a web with vertices that are $\zeta$-instanton sub-solutions, one would really have to resolve
some of the sub-solutions into webs (so that they become far from point-like),
taking advantage of their moduli spaces $\M_i$.  What would arise this way is an end of $\M$
that we would associate not to the web $\frak w$, but to some other web $\frak w'$ obtained by resolving some of the vertices in $\frak w$.
Thus ends of a moduli space $\M$ of $\zeta$-instantons with fan-like asymptotics can be put in correspondence with  ordinary webs with point vertices,
the same objects studied in Sections \S\S\ref{sec:Webs}-\S\ref{sec:LocalOpsWebs} of this paper.

\begin{figure}
 \begin{center}
   \includegraphics[width=4.5in]{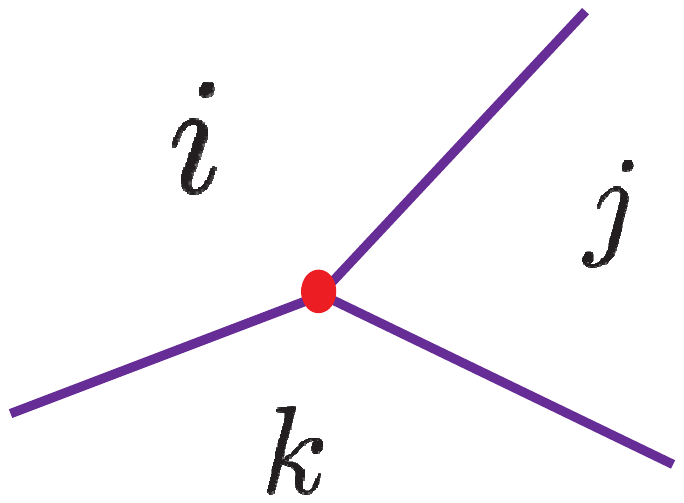}
 \end{center}
\caption{\small This $\zeta$-web has only two moduli (assuming the vertex is rigid) so the ``impact parameters'' $x_1,x_2,x_3$ of the three outgoing solitons cannot
be varied independently.  They obey a linear relation $a_1x_1+a_2x_2+a_3x_3=0$, for some constants $a_i$.  }
 \label{fourthly}
\end{figure}

\begin{figure}
 \begin{center}
   \includegraphics[width=2.7in]{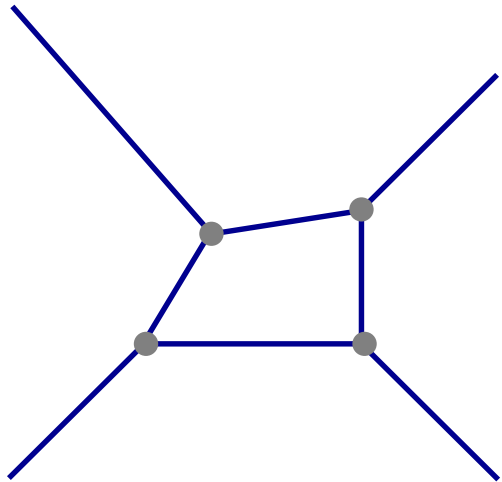}
 \end{center}
\caption{\small A $\zeta$-web with four external solitons and a four-dimensional moduli space.
The impact parameters can be varied independently but only in  a certain region in
$\IR^4$.  }
 \label{instructive}
\end{figure}

\subsection{The Collective Coordinates}\label{collective}
Now let us examine in this context the consequences of
the fact that a classical $\zeta$-soliton solution actually corresponds to a pair of quantum states
of fermion numbers $f,\, f+1$ for some $f$.  The doubling of the spectrum
arises  from quantizing the center of mass motion of the soliton and its supersymmetric
counterpart.  It is convenient here, as in section \ref{prelims},  to choose a precise definition of the impact parameter $x$ of each soliton,
even though this concept is
really only naturally defined up to an additive constant.
The fermionic collective coordinate of the soliton is $\{\CQ_{\zeta},x\}$.
 It is evocative to write this as $\d x$.  Now consider a $\zeta$-instanton
asymptotic to a fan of, say, $n$ solitons.  Each of the
 $n$ outgoing solitons has a bosonic collective coordinate $x_1, \dots, x_n$, together with
 corresponding fermionic collective coordinates $\d x_1, \dots, \d x_n$.
  The $\zeta$-instanton amplitude to create the outgoing solitons  is
a state
$\Psi(x_1,\d x_1;\dots;x_n,\d x_n)$; we can think of this state as a differential form on a copy of $\IR^n$ parametrized by the $x_i$.
This differential form is valued in the tensor product
\be
\IM'_{I}=\IM'_{i_1 i_2}\otimes \IM'_{i_2 i_3 }\otimes \cdots \IM'_{i_n i_1}
\ee
of the reduced complexes
\eqref{MC}
that describe BPS solitons without their zero-modes; $\Psi$ is the product of an ordinary differential form and an element
$m_{i_1 i_2}\otimes m_{i_2 i_3 }\otimes \cdots m_{i_n i_1}$  of this tensor product
\footnote{We abbreviate the expression $m_{ij}^{f_0}(p)$ of equation \eqref{MC} by $m_{ij}$
and also shift its fermion number by $-\half$ so that two Clifford module generators of
the module $\Bbb{W}$ in equation \eqref{TC} have fermion numbers $0$ and $1$.}
\be\label{eq:WaveDecomp}
\Psi = \psi(x_i,\d x_i)\cdot m_{i_1 i_2}\otimes m_{i_2 i_3 }\otimes \cdots m_{i_n i_1}
\ee
$\CQ_{\zeta}$ acts on this $\Psi$ as the exterior derivative $\d$ on $\psi$. As in any instanton calculation, in discussing the state $\Psi$ generated by a given $\zeta$-instanton moduli space,
we need not consider higher order $\zeta$-instanton effects that generate the differentials for the individual soliton states that appear on the right hand side of the formula for $\Psi$.  (How they enter will be explained in footnote \ref{tautextended}.)
Now  $\CQ_{\zeta}$-invariance of the path integral is the statement that $\d \Psi=0$.

To understand $\Psi$, we need to know whether, as we vary the $\zeta$-instanton moduli, the $x_i$ can be varied independently.
A typical example in which they cannot be varied independently arises from a $\zeta$-web with a single  trilinear vertex (Figure \ref{fourthly}).
Assuming the vertex is rigid, it has only two moduli so the $x_i$ are not independent, but obey a linear relation $a_1x_1+a_2x_2+a_3x_3=0$,
with some constants $a_i$.
(With natural outgoing normalizations for all solitons, we can take the $a_i$ to be all positive.
Even if the vertex is not rigid, the relation is obeyed to within an error of order $\ell_W$.)
The wavefunction $\Psi$ is therefore supported on the locus $a_1x_1+a_2x_2+a_sx_3=0$.
It  takes the form
\begin{equation}\label{zob} \Psi=\pm\delta(a_1x_1+a_2x_2+a_3x_3)
(a_1\d x_1+ a_2 \d x_2+ a_3 \d x_3)\cdot m_{ij}\otimes m_{jk}\otimes m_{ki} , \end{equation}
where as usual in the $A$-model, the signs of boson and fermion determinants cancel, leaving a ``constant'' wavefunction with a
sign $\pm$ that comes from the sign of the fermion
determinant (as usual, this sign depends on the signs chosen for the soliton states $m_{ij}$, $m_{jk}$, $m_{ki}$).

To explain the above formula better, we note that
$A$-model results are usually expressed as forms on the appropriate moduli space, which here is defined by $\jmath=0$ where
$\jmath=a_1x_1+a_2x_2+ a_3x_3$. If one does this, the wavefunction would just be $\pm 1$ (times the tensor product of symbols $m_{ij}$, etc., representing the external
states). However, in the present context, it is much more natural to express $\Psi$ as a wavefunction on the space $\IR^3$
that parametrizes the three centers of mass.  We do this by multiplying by the form $\delta(\jmath)\d \jmath$. In general, for any manifold $X$ with submanifold $Y$ whose interior points are
defined locally by equations $\jmath_1=\dots=\jmath_k=0$, where an orientation of the normal bundle to $Y$ is defined by $\d \jmath_1\wedge\dots\wedge \d \jmath_k$, one
defines the $k$-form Poincar\'e dual to $Y$ by
\begin{equation}\label{dorb} \Theta_Y=\delta(\jmath_1)\dots\delta(\jmath_k)\d \jmath_1\dots\d \jmath_k ,\end{equation}
which depends only on $Y$ and not on the choices of the functions $\jmath_i$.  In our discussion it is important to
consider the case where $Y$ has a boundary. If $Y$ has a boundary $\partial Y$, and near $\partial Y$ it is defined locally by equations
$\jmath_1=\dots=\jmath_k=0$ together with an inequality
$h\geq 0$,  then \eqref{dorb} should be modified to
\begin{equation}\label{dorbo} \Theta_Y=\Theta(h)\delta(\jmath_1)\dots\delta(\jmath_k)\d \jmath_1\dots\d \jmath_k ,\end{equation}
where $\Theta(\alpha)$ for $\alpha$ real
is the standard Heaviside function given by $0$ for $\alpha <0$ and $1$ for $\alpha>0$.
In this situation, $\partial Y$ is defined by equations $\jmath_1=\dots \jmath_k=0=h$, so the definitions just given immediately imply that
\begin{equation}\label{worb} \d\Theta_Y=\Theta_{\partial Y}.\end{equation}
In our example, $\Psi$ is Poincar\'e dual to the submanifold $Y$ defined by   $a_1x_1+a_2x_2+a_3x_3=0$; $Y$
has no boundary, so $\d\Psi=0$ in eqn. (\ref{zob}).  This is
a consequence of the underlying $\CQ_{\zeta}$-invariance, since as usual in the $A$-model, $\CQ_{\zeta}$ behaves as the exterior derivative $\d$ on  moduli spaces
of classical solutions.

We note from eqn. (\ref{zob}) that although the three-soliton state created by the given $\zeta$-instanton has definite fermion number, the individual outgoing
solitons are not created in states of definite fermion number: the factors $\d x_1$, $\d x_2$, and $\d x_3$ each have fermion number 1.  Another
interesting point concerns the total fermion number of the state $\Psi$. In an $\FF$-conserving theory, the outgoing state created by a $\zeta$-instanton
moduli space must have fermion number 0.  So as the explicit factor $a_1\d x_1+a_2\d x_2+a_3\d x_3$ has $\FF=1$, the factor
$m_{ij}\otimes m_{jk}\otimes m_{ki}$ must have $\FF=-1$.  Similarly, in eqn. (\ref{buggy}) below, the factor $m_{i_1i_2}\otimes m_{i_2i_3}\otimes m_{i_3i_4}
\otimes m_{i_4i_1}$ has $\FF=0$.

If we want to get a number from this $\zeta$-instanton amplitude, we have to pair $\Psi$ with an external state of the outgoing solitons.
Such a state would have to be a two-form, and we want a closed two-form as we want a $\CQ_{\zeta}$-invariant state.  An example would
be the closed two-form $\Psi'=\delta(x_1-c_1)\d x_1\,\delta(x_2-c_2)\d x_2\cdot m_{ik}\otimes m_{kj}\otimes m_{ji}$, where $c_1,c_2$ are
constants and, for example, $m_{ji}$ is the state in the reduced complex $\IM'_{ji}$ that is obtained from $m_{ij}$ by a $\pi$
rotation (and so is dual to $m_{ij}$ under the pairing (\ref{welz})).   The pairing
\begin{equation}\label{tomox}Z_{\Psi'}=\int_{\R^3}
{\mathcal D}(x_i,\d x_i)\, \bigl( \Psi(x_i,\d x_i) ,\Psi'(x_i,\d x_i)\bigr) \end{equation}
 is clearly nonzero.
To evaluate it, we contract out the internal states $m_{ij}$, $m_{ji}$, etc., using the pairing (\ref{welz})
(this gives a factor $\pm  K'(m_{ij}, m_{ji} ) K'(m_{jk}, m_{kj}) K'( m_{ki}, m_{ik})$)
and integrate over the other variables using the natural  measure
$\mathcal{D}(x_i,\,\d x_i)$  for integration over the pairs of bosonic and fermionic variables $x_i$ and $\d x_i$.
(So for a function $f(x, \, \d x)$ defined on $\IR$,
 the integral $\int_\IR{\mathcal D}(x, \d x)\,f(x, \d x)$ is simply the integral over $\IR$ of the differential
form $f(x, \,\d x)$.)
$Z_{\Psi'}$ is the $\zeta$-instanton amplitude to create the outgoing solitons in the state $\Psi'$.  As usual in an $A$-model
(with or without a superpotential) this amplitude is the answer to a counting question ($A$-model ``counting'' is always weighted by the sign
of a fermion determinant).  In the present example, what we have counted (the answer being  $\pm 1$)
is the number of $\zeta$-instantons with the given fan asymptotics that create outgoing solitons with impact parameters obeying the
constraints $x_1=c_1$ and $x_2=c_2$.

A final comment is that the relation
\begin{equation}\label{molg}\d\Psi=0, \end{equation}
which expresses the underlying $\CQ_{\zeta}$-invariance, is needed to ensure that the amplitude (\ref{tomox}) is invariant under
$\Psi'\to \Psi'+\d\Chi$ for any $\Chi$.

Another instructive example is sketched in Figure \ref{instructive}.  Here we consider a $\zeta$-web with four external soliton lines and
four vertices that we will assume rigid; hence this $\zeta$-web has
a four-dimensional moduli space.  The impact parameters $x_1,\dots, x_4$ of the external solitons
can be varied independently, but subject to inequalities that come from the fact that the
lengths of the edges in the $\zeta$-web must all be positive.  Thus the wavefunction $\Psi$ is nonzero
only in a certain (noncompact) polytope $U\subset \IR^4$.
Within $U$, $\Psi=\pm 1$ (times a state $\otimes_{j=1}^4m_{i_ji_{j+1}}$ in the
appropriate reduced complex), expressing the fact that if we specify
the desired values of the $x_i$, there is a unique point in the given $\zeta$-web moduli space  that yields these values.   (This statement
assumes that we keep away from the boundaries of $U$ by an amount much greater than $1/m$, the inevitable error in any statement
based on a $\zeta$-web.)  So the wavefunction of the outgoing solitons is
\begin{equation}\label{buggy}\Psi=\pm \Theta_U\otimes_{j=1}^4m_{i_ji_{j+1}},\end{equation}
where $\Theta_U$ is the characteristic function of the region $U$,  the sign depends on the sign of a fermion determinant, and
$\otimes_j m_{i_j i_{j+1}}$ is the $\CQ$-invariant
state in the reduced complex $\IM'_{i_ji_{j+1}}$ corresponding to the specific outgoing solitons. We immediately
see that this wavefunction is not closed: $\d\Psi$ is a delta function supported on the boundaries of $U$.
In fact, as $U$ has codimension 0 in $\IR^4$, the characteristic function $\Theta_U$ is the Poincar\'e dual to $U$ as defined in eqn.
(\ref{dorb}), and a special case of eqn. (\ref{worb}) tells us that
\begin{equation}\label{ulu}\d\Theta_U=\Theta_{\partial U},\end{equation}
where now $\partial U$ has codimension 1 so $\Theta_{\partial U}$ is a 1-form.
Even though the formula
(\ref{buggy}) is precisely valid only away from the boundaries of $U$, modifying $\Psi$ only near those boundaries will not help in achieving
$\d\Psi=0$. After all, for a zero-form $\Psi$ to be annihilated by $\d$, it must be constant throughout $\IR^4$.

What is going on here is that the $\zeta$-web in question really represents one end of a four-dimensional moduli space $\M$ of
$\zeta$-instantons with fan-like asymptotics.  If $\M$ has only one end corresponding to the $\zeta$-web in the figure,   we will
indeed get a contradiction.  There must then be other ends of the moduli space that either cancel $\Psi$ or complete it to a closed 0-form on
$\R^4$.  This paper is really based on the possibility that this cancellation occurs in an interesting way,  such that the interior amplitudes defined and studied
in section (\ref{webscot}) are not identically zero.   A less interesting but logical possibility is
that one of the rigid $\zeta$-vertices in Figure \ref{instructive}
can be replaced by a second $\zeta$-vertex with the same asymptotics but contributing with the opposite sign of the fermion determinant.
(Remember that a choice of a component of the moduli space is part of the definition of a $\zeta$-vertex.)  This case is less interesting,
in the sense that if it occurs, the two $\zeta$-vertices in question will always make canceling contributions in all of our considerations.
A final comment is that hypothetical ultraviolet ends of $\M$ -- ends supported on a region in which one or more of the internal
lines in Figure \ref{instructive} collapses to a length of order $1/m$ -- could not help in  correcting $\Psi$ to satisfy $\d\Psi=0$.  But as
explained in section \ref{zetafan}, as long as the corresponding $A$-model under discussion exists
as a topological field theory, we do not expect contributions from
such ultraviolet ends.

\begin{figure}
 \begin{center}
   \includegraphics[width=3.5in]{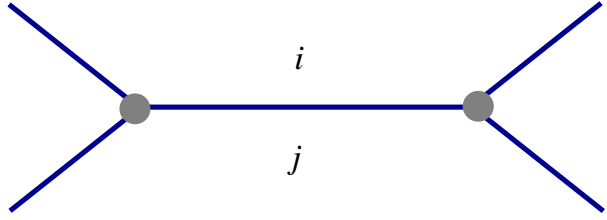}
 \end{center}
\caption{\small A $\zeta$-web with four external solitons and a four-dimensional moduli space.
The impact parameters can be varied independently but only in  a certain region in
$\IR^4$.  }
 \label{stillmore}
\end{figure}

Finally, we discuss what happens to the collective coordinates of a given soliton line under gluing.
To be specific, consider the $\zeta$-web in Figure  \ref{stillmore} showing a solution made by convolution of two sub-solutions connected
by a single soliton.  This soliton is of type $ij$ if viewed as propagating from left to right in the figure, or of type $ji$ if we consider
it to be propagating from right to left.  It has collective coordinates $x,\,\d x$.  Write generically $y,\,\d y$ for the collective
coordinates of solitons emerging from the subsolution on the left and $z,\,\d z$ for the collective coordinates of solitons
emerging on the right.  The subsolution on the left creates a fan of solitons in a state that we write generically as $\Psi_\ell(x,\d x; y,\d y)$
and the subsolution on the right creates a fan of solitons in a state that we write generically as $\Psi_r(x,\d x; z,\d z)$.
The $\zeta$-web in the figure has fan-like asymptotics with the outgoing solitons described by the whole collection of variables
$y,\,\d y; z\,\d z$.   Cluster decomposition in a massive theory tells us that the wavefunction of the solitons emerging from the whole
web is the product of the left and right wavefunctions with the collective coordinates $x, \, \d x$ of the internal soliton integrated out:
\begin{equation}\label{xelbozz} \Psi(y,\d y;z,\d z)=\int_{\IR}{\mathcal D}(x, \d x) \,  \,
\biggl( \Psi_\ell(x, \d x; y,\d y), \Psi_r(x,\d x;z,\d z)\biggr). \end{equation}
where, again, the pairing on the integrand uses $K'$.

Let us formalize this result.  The variables $x$ and $\d x$ are collective coordinates of an $ij$ soliton emerging from the subsolution on the
left of the figure.  This $ij$ soliton is a vector in the reduced complex $\IM'_{ij}$ of classical $ij$ solitons.  Similarly, the $ji$ soliton emerging
from the subsolution on the right of the figure is a vector in the space $ \IM'_{ji}$ of classical $ji$ solitons.
The pairing between these states is the nondegenerate pairing of eqn. (\ref{welz}).
We recall that this pairing is nondegenerate because whatever  $ij$ $\zeta$-soliton  emerges
from the left subsolution in the figure can be paired with a unique $ji$ $\zeta$-soliton, namely the ``same''
classical solution rotated by an angle $\pi$.

It remains to discuss the integration over the bosonic collective coordinate $x$.
Let $\M_\ell$ and $\M_r$ be the moduli of the sub-solutions on the left and right of the figure and let $\M$ be the moduli space
of the overall $\zeta$-web.  For a given choice of a point in $\M_\ell$,  the $ij$ soliton connecting the two parts of the figure has
a definite impact parameter, so it has a wavefunction $\delta(x-a)$, where $a$ is a function on $\M_\ell.$  Similarly, the wavefunction
of the $ji$ soliton emerging from the right is proportional to $\delta(x-b)$, where $b$ is a function on $\M_r$.  The integral over $x$
gives
\begin{equation}\label{yorb} \int_{\IR}\d x\,\delta(x-a)\delta(x-b)=\delta(a-b).\end{equation}
The constraint that $a=b$ in order for the $\zeta$-web in the figure to exist means that ${\mathrm{dim}}\,\M={\mathrm{dim}}\,\M_\ell+
{\mathrm{dim}}\,\M_r-1$.  The term $-E$ in eqn. (\ref{zetadim}) arises because each internal line in a $\zeta$-web makes
in this way a contribution $-1$.

One last comment is that in all of these examples, because we have assumed all $\zeta$-vertices  to be rigid, there are no normalizable
moduli.  All deformations of the $\zeta$-webs that we have considered change the impact parameters of the external solitons.
In the case of a $\zeta$-web that has a modulus that can be varied without changing the impact parameters -- for example, a non-rigid
$\zeta$-vertex -- there is  a normalizable fermion zero-mode.  Unless we insert a suitable vertex operator to absorb this zero-mode (and
Landau-Ginzburg models with target $\IC^n$ do not have such operators),
an amplitude derived from a web with a normalizable zero-mode vanishes. That is why we
can restrict our attention here to $\zeta$-vertices of zero excess dimension.

\subsection{Interior Amplitudes And The Relations They Obey}\label{webscot}

The goal of this section is to explain the origin of the fundamental relation for representations of plane
webs that was presented in eqn. (\ref{eq:bulk-amp}).  Rather like the proof that $\CQ_{\zeta}^2=0$ in the context of the MSW
complex (section \ref{whyindeed}), the basic idea is to consider the two ends of 1-dimensional reduced moduli spaces.

\begin{figure}
 \begin{center}
   \includegraphics[width=4.5in]{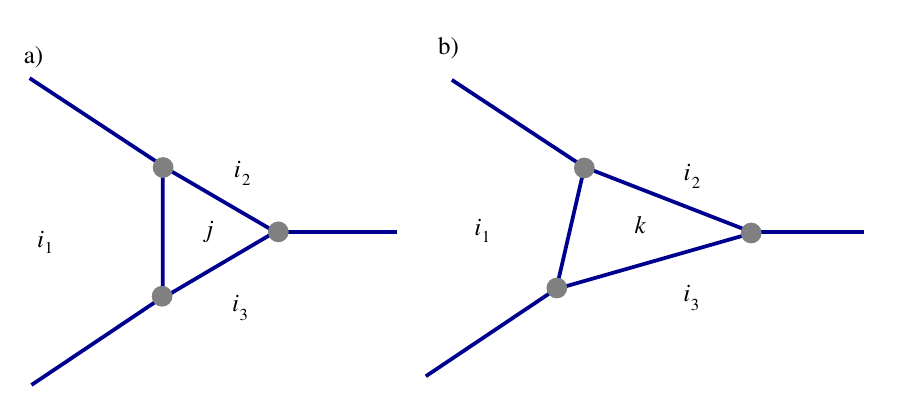}
 \end{center}
\caption{\small These two taut $\zeta$-webs might arise as the ends of the same 1-dimensional reduced moduli space $\M_\red$.  (In general,
the ends of $\M_\red$ might not be the same topologically, but in this example they are.) }
 \label{taut}
\end{figure}
\def\red{{\mathrm{red}}}
\def\fr{{\mathfrak r}}
\def\o{{\mathbf o}}

Let $i_1i_2\dots i_n$ be a cyclic fan of vacua and let $\IM'_{i_1\dots i_n}=\IM'_{i_1i_2}\otimes\dots \otimes \IM'_{i_ni_1}$ be
the tensor product of the reduced complexes of the outgoing solitons.  The impact parameters of the outgoing solitons define a point in
$\IR^n$, with one copy of $\IR$ for each soliton.  We write $\Omega^*(\IR^n)$ for the space of differential forms on $\IR^n$.

If $\V$ is a rigid $\zeta$-vertex that is asymptotic to the given fan of vacua,
then the path integral associated to this family of $\zeta$-instantons
determines an element $B(\V)\in\Omega^*(\IR^n)\otimes  \IM'_{i_1,\dots,i_n}$.  Examples were discussed in section \ref{collective}.
A small simplification is that, dividing by overall translations, $\IR^n$ projects to a reduced parameter space $\IR^{n-2}$ and $B(\V)$
is actually always the pullback of an element $B_\red(\V)\in \Omega^*(\IR^{n-2})\otimes \IM'_{i_1,\dots,i_n}$, which we call the reduced state.
Some statements are more
transparent in terms of $B_\red(\V)$.   We note that $\IR^{n-2}$ has a natural origin $\o$, corresponding
to solitons that all emanate from a common point in $\IR^2$.  Since $\V$ is rigid, its reduced moduli space is 0-dimensional
and $B_\red(\V)$ is actually a simple product
\begin{equation}\label{hopf}B_\red(\V)=\Theta_\o\cdot B_\red'(\V),  \end{equation}
where $\Theta_\o$ is an $n-2$-form with delta function support at $\o\in \IR^{n-2}$ and $B_\red'(\V)\in \IM'_{i_1,\dots,i_n}$.
To be more precise, $B_\red(\V)$ takes this form
  within the usual error of order $1/m$.  $B_\red(\V)$ differs from the expression just given by an exact
form $\d\Chi$, where $\Chi$ vanishes exponentially fast at distances greater than $1/m$ from the point $\o$.

We will say that a component $\M$ of the moduli space of $\zeta$-instantons is ``taut''  if {\it (i)} it is three-dimensional, so that
 the corresponding reduced dimension $\M_\red=\M/\IR^2$ has dimension 1; and {\it (ii)} this reduced space is a copy of $\IR$.
 The reason for the definition is that $\zeta$-instanton moduli spaces that are taut in this sense will play a role
 somewhat analogous to the role played by taut webs.
  Indeed, if  $\M$ is taut, then $\M_\red\cong\IR$
 has two ends (Figure \ref{taut}).  Each of these ends corresponds to what we will call a taut $\zeta$-web, that is a $\zeta$-web
with a 1-dimensional reduced moduli space.    (If $\M$ obeys condition {\it (i)}  but not condition {\it (ii)},  then $\M_\red$
 is a copy of $S^1$.  In this case, $\M$ is a non-rigid $\zeta$-vertex with an excess dimension of 1.  Such
 components of $\zeta$-instanton moduli space will play no role  in the present section.)

Suppose that $\ZZ$ is a taut $\zeta$-web asymptotic to the fan of vacua $i_1\dots i_n$. ($\ZZ$ is actually asymptotic at infinity to a specific
fan of solitons, but it is more convenient here  simply to specify the fan of vacua.) Then the path integral for this family of $\zeta$-instantons
determines a state $\Psi_\ZZ\in \Omega^*(\IR^n)\otimes \IM'_{i_1\dots i_n}$, which as in the case of a rigid $\zeta$-web is a pullback
of a reduced state
 $\Psi_{\ZZ,\red}\in \Omega^*(\IR^{n-2})\otimes \IM'_{i_1\dots i_n}$.   Moreover this state is supported on the 1-dimensional reduced moduli
space of the web $\ZZ$.  By scale-invariance of webs, this reduced moduli space is a ray $\fr$ starting at the origin  $\o\in\IR^{n-2}$ (to within
the usual error of order $1/m$).
As in section \ref{collective}, the state determined by $\ZZ$ is the Poincar\'e dual of $\fr$ times a state $\Psi_\ZZ^*\in \IM'_{i_1\dots i_n}$:
\begin{equation}\label{nozo} \Psi_{\ZZ,\red}=\Theta_\fr\cdot \Psi_\ZZ^*. \end{equation}
It immediately follows from this that $\Psi_{\ZZ,\red}$ is not closed.  We have $\d\Theta_\fr=\Theta_{\partial\fr}=\Theta_\o$ (since the boundary of
the ray $\fr$ consists of its endpoint $\o$), so
\begin{equation}\label{ozo}\d\Psi_{\ZZ,\red}=\Theta_\o\cdot \Psi_\ZZ^*.\end{equation}
(We have  $\d \Psi_\ZZ^*=0$ since $\Psi_\ZZ^*$ is just a fixed state in $ \IM'_{i_1\dots i_n}$ corresponding
to the relevant fan of solitons.)
The reduced state associated to the whole moduli space $\M$ will be closed because of the underlying $\CQ_{\zeta}$-invariance; $\Psi_{\ZZ,\red}$,
which is not closed, is  the contribution of just one end of the moduli space.
$\M_\red$ has precisely  1 additional end, corresponding to another taut\footnote{\label{tautextended}
Actually, this taut web could be what in the
abstract part of this paper was called a taut extended web: that is, it could be constructed from a rigid $\zeta$-vertex with
a $\zeta$-instanton correction to the MSW complex on one of the external lines.} web $\ZZ'$.  The
state associated to this second taut web is supported on another ray $\fr'$ from the origin, and has the form
\begin{equation}\label{nozzo} \Psi_{\ZZ',\red}=\Theta_{\fr'}\cdot \Psi_{\ZZ'}^*, \end{equation}
for some $\Psi_{\ZZ'}^*\in \IM'_{i_1\dots i_n}$.  Hence
\begin{equation}\label{zozzo}\d\Psi_{\ZZ',\red}=\Theta_{\o}\cdot \Psi_{\ZZ'}^*. \end{equation}
The condition that $\d(\Psi_{\ZZ,\red}+\Psi_{\ZZ',\red}^*)=0$ is thus simply
\begin{equation}\label{funda} \Psi_{\ZZ}^*+\Psi_{\ZZ'}^*=0. \end{equation}
This is the  identity that leads to the basic algebraic relation for plane webs that was proposed in eqn. (\ref{eq:bulk-amp}).

Before explaining this claim, we pause to point out that it is oversimplified to expect that the reduced state produced by the moduli space
$\M$ is precisely $\Psi_{\ZZ,\red}+\Psi_{\ZZ',\red}$.  The $\zeta$-webs
$\ZZ$ and $\ZZ'$ are really only well-defined when they are large, in other words far from the origin $\o\in \IR^{n-2}$.  The state
$\hat\Psi_\M$ is not just the pullback of $\Psi_{\ZZ,\red}+\Psi_{\ZZ',\red}$; it is the pullback of a state $\hat\Psi_{\M,\red}$ that
differs from $\Psi_{\ZZ,\red}+\Psi_{\ZZ',\red}$ by a state $\Chi$ supported near the point $\o$.  The reason for the identity (\ref{funda}) is
that without this identity,
 the condition  $\d(\Psi_{\ZZ,\red}+\Psi_{\ZZ',\red}+\Chi)=0$ is not satisfied for any compactly supported $\Chi$.  (The obstruction comes from
 the fact that $\Theta_\o$, since its integral over $\IR^{n-2}$ is non-zero, represents a non-zero element in the compactly supported cohomology
 of $\IR^{n-2}$, so it is not $\d\Chi$ for any compactly supported $\Chi$.  Here it does not matter if
 ``compact support'' is replaced by ``exponential decay at
 distances large compared to $1/m$.'')

To understand the identity (\ref{funda}), let us first note that it trivially implies a relation for the sum of contributions of
all taut webs asymptotic to the fan $i_1i_2\dots i_n$ of vacua.  Let $\SSS_{i_1i_2\dots i_n}$ be the set of all taut $\zeta$-webs
asymptotic to the given fan.  Then
by virtue of (\ref{funda}),
\begin{equation}\label{boffo}\sum_{\ZZ\in\SSS_{i_1\dots i_n}} \Psi_\ZZ^*=0.  \end{equation}
Indeed, each taut $\zeta$-web with the given asymptotics is an end of a unique
1-dimensional reduced moduli  space of $\zeta$-instantons, and, because of (\ref{funda}), the two taut $\zeta$-webs that are the ends of this reduced moduli
space make canceling contributions in (\ref{boffo}).

The sum in eqn. (\ref{boffo}) is reminiscent of the sum over ordinary taut webs with given asymptotics in eqn. (\ref{eq:bulk-amp}).
To understand the relationship,
we have to analyze the state $\Psi_{\ZZ}^*\in \IM'_{i_1\dots i_n}$.  Here the moduli
$\ZZ$ are not relevant; by definition, $\Psi_{\ZZ}^*$ is a state in the product $\IM'_{i_1i_2}\otimes \dots\otimes \IM'_{i_ni_1}$ of the reduced
complexes.  To determine this state, we can work near infinity in the reduced moduli space of $\ZZ$: thus $\ZZ$ is built from widely
separated rigid $\zeta$-vertices $\V_a$, $a=1,\dots, t$, connected by soliton lines.  The path
integral for each $\zeta$-vertex $\V_a$ determines a state
$B(\V_a)$ of the solitons emanating from this vertex.  When two $\zeta$-vertices $\V_a$ and $\V_b$ are connected by a soliton
line in the $\zeta$-web $\ZZ$, this means that the corresponding soliton states emanating from $\V_a $ and $\V_b$ have to be
contracted via the nondegenerate degree 1 pairing (\ref{welz}).  The combined operation of computing a state $B(\V_a)$
and then contracting these states whenever two vertices are joined by lines is precisely the operation $\rho$
 defined in \eqref{eq:web-rep-1} et. seq., used in formulating the fundamental
algebraic identity  \eqref{eq:bulk-amp} (this interpretation of $\rho$ was the motivation for the way it was defined).
The terms $\Psi_\ZZ^*$ that contribute in the identity \eqref{boffo} are in natural corresponence with contributions to the identity
\eqref{eq:bulk-amp} of the abstract part of the paper: in either case, a contribution is made by arranging rigid vertices as the vertices of a taut web,
computing states associated to the vertices and contracting those states whenever two vertices are joined by a line.

In relating our present analysis to  the abstract discussion, the object called $\beta$ in the abstract discussion should
be defined as a sum over all rigid $\zeta$-vertices $\V_a$.  If $\V_a$ is
asymptotic to the fan of vacua $j_1^a\dots j_{s_a}^a$, and the corresponding solitons have center of mass coordinates $x_{j_1^a},\dots,
x_{j_s^a}$, then
\begin{equation}\label{torrix} \beta=\sum_a\d x_{j_1^a}\dots \d x_{j_s^a} B'_\red(\V_a). \end{equation}
See eqn. (\ref{hopf}) for the definition of $B'_\red(\V_a)$. Now, $B_\red$, being the state produced by the path integral in a fermion-number conserving
theory, has $\FF=0$; since $\Theta_\o$ is an $(n-2)$-form, $B'_\red$ has fermion number $-(n-2)$.  Including the factors of $\d x_{j_k^a}$
increases the fermion number by $n$, so $\beta$ has fermion number 2, as in the definition   (\ref{eq:bulk-amp})
in Section \ref{subsec:WebRepPlane}.
The factors $\d x_{j_k^a} $ have been included in the definition of $\beta$ for the following reason.
In the abstract algebraic treatment, it is convenient to imagine that a classical soliton that has two states
of fermion number $f,f+1$ is always created in the state of fermion number $f+1$.  In the more microscopic path integral treatment,
the truth is more complex, as we see in eqn. (\ref{zob}), and in the above derivation, it was actually
more convenient to use reduced states of fermion number $f$. To compensate for this, we include a factor of $\d x$ for each soliton
in the formula for $\beta$. This also means that the pairing
used in the abstract description is not the degree 1 pairing (\ref{welz}) but is the degree $-1$ pairing
$K$ that is obtained by composing (\ref{welz}) with the operation that removes the fermion zero-mode
$\d x$ from each soliton state.

At this point, using the Landau-Ginzburg theory based on $(X,W)$, we have
determined vacuum data $(\IV,z)$ (from the superpotential),
  web representations $\CR$ (from equations \eqref{MCdp} and \eqref{elz}),
 and an  interior amplitude $\beta$ (from equation  \eqref{torrix}), thus
 defining a Theory in the formal sense of Section \S \ref{subsec:WebRepPlane}. We have accordingly
   recovered all the data needed to form the vacuum
category $\fVac$ using the construction of Section \S  \ref{subsec:VacCategory}, once we choose a half-plane $\CH$.
We let $\fVac(X,W)$ denote the category for the positive half-plane so that  $\fVac^{\rm opp}(X,W)$
is the category for the negative half-plane.  It now follows from the purely formal constructions of
Section \S \ref{subsec:BraneCat} above that there is a
corresponding \afty\ category of Branes $\fB\fr(X,W)$,
and it is natural to wonder if the corresponding $A_\infty$ amplitudes are equivalent to those
that were constructed in section \ref{seidelfukaya} for branes of class $T_\kappa$.
We will discuss this question in detail in Section \S \ref{notif}.

\begin{figure}[htp]
\centering
\includegraphics[scale=0.3,angle=0,trim=0 0 0 0]{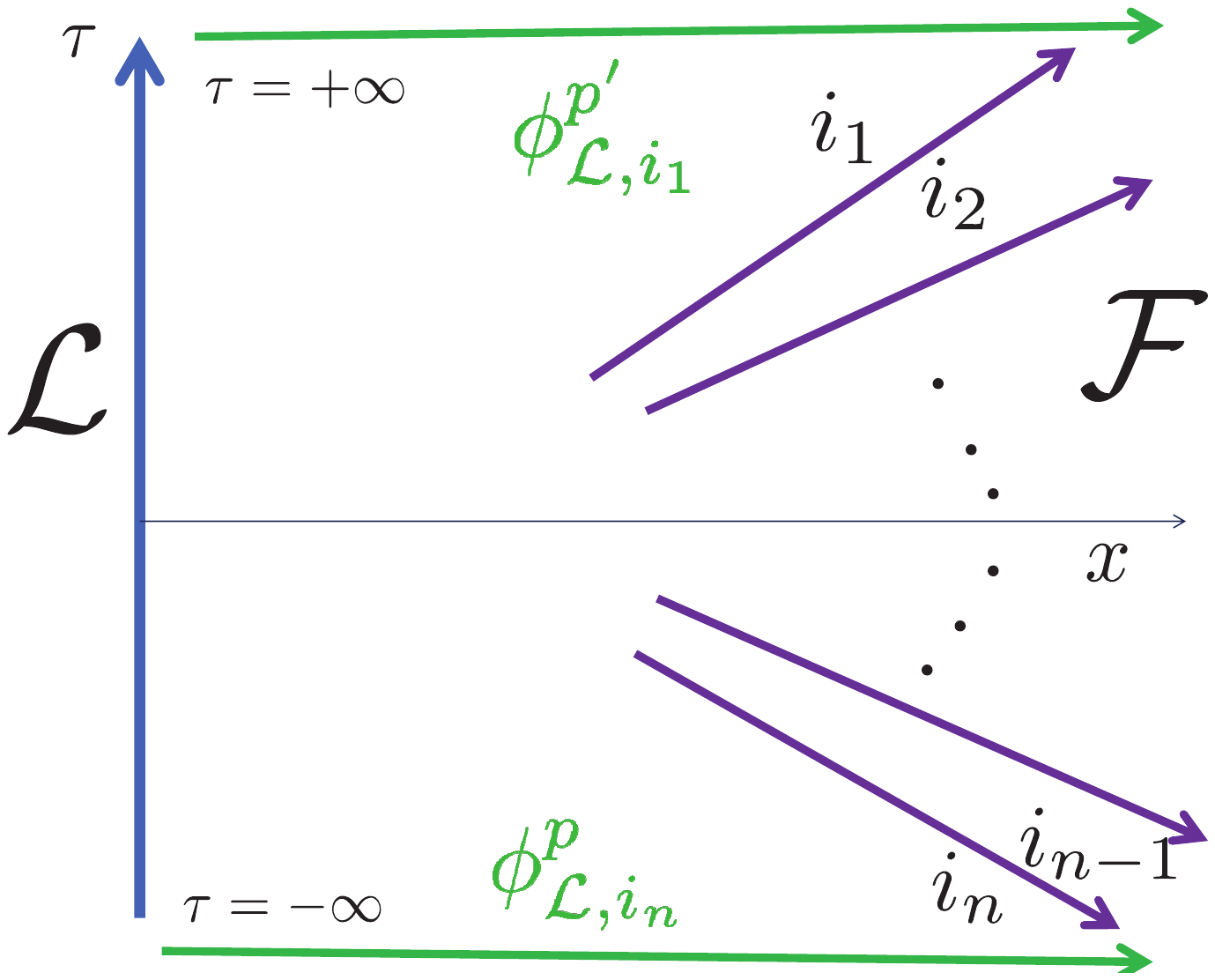}
\caption{Boundary conditions for general half-plane instantons with
fan boundary conditions at $x\to + \infty$ and solitons at $\tau \to \pm \infty$.
   }
\label{fig:HALFPLANEBC}
\end{figure}

\subsection{$\zeta$-Instantons On A Half-Space Or A Strip}\label{halfspace}

\subsubsection{Preliminaries}
In this section, we will first analyze $\zeta$-instantons on the half-plane $x\geq 0$ in the $x-\tau$ plane,
with boundary conditions set by a Lagrangian submanifold
$\L$.  After some preliminaries, we  explain in section \ref{longwinded} how $\zeta$-instantons
can be used to define a boundary amplitude $\mathcal B$ as introduced in section \ref{subsubsec:WebRep-Halfplane} and obeying the key
identity \ref{eq:boundary-amp}.   The considerations here are very similar to those of sections \ref{zetawebs} and \ref{webscot}.  Then we consider
$\zeta$-instantons on a strip in section \ref{zetastrip}.

We began section \ref{zetawebs} by showing that in any massive theory, there are no localized $\zeta$-instantons on $\R^2$.   The analog
in a half-plane is more subtle.  To formulate the question, we pick a critical point $\phi_i$ corresponding to a
vacuum $i\in\IV$ at $x\to\infty$, and we also choose a half-line
$\zeta$-soliton $\phi_{\L,i}$ that interpolates from $\L$ at $x=0$ to vacuum $i$ at $x=\infty$.  Then we ask if there are $\zeta$-instantons on the half-plane
that map the $\tau$-axis to $\L$, approach $\phi_{\L,i}$  for $\tau\to\pm\infty$, and approach $\phi_i$ for $x\to\infty$.  Such a $\zeta$-instanton
is localized in the sense that it differs substantially from the chosen $\zeta$-soliton only in a localized region of the half-plane.  In general,
depending on $\L$, there may be such localized half-plane $\zeta$-instantons.  We have seen an example in section  \ref{impanom}. As we learned there,
when localized half-plane instantons exist, it often means that the brane in question is not a valid $A$-brane and our methods do not apply
in its presence.

In this section,
we will not consider such localized half-plane $\zeta$-instantons.  This does not mean that we have to assume they do not exist;  they are not
relevant to the questions we will consider here.  When moduli spaces of localized half-plane $\zeta$-instantons
exist (but the brane in question is a valid $A$-brane),  they are somewhat like $\zeta$-instanton moduli
spaces with positive excess dimension, relevant only if one inserts local $A$-model observables.
A simple criterion  that ensures that there are no localized half-plane $\zeta$-instantons is that the superpotential $h$ of eqn.
(\ref{defhox}) is single-valued.  In particular, this is so if
the symplectic form $\omega$ of $X$ is exact and the Lagrangian submanifold $\L$ is also exact (for these notions, and their
significance, see the discussion of eqn. (\ref{ondog}); the assumptions of exactness are
often made mathematically in the Fukaya-Seidel category).
Since the $\zeta$-instanton equation is gradient flow for $h$,  $h$ strictly increases with $\tau$ in any non-trivial $\zeta$-instanton.
So single-valuedness of $h$ implies
that there are no non-trivial $\zeta$-instantons beginning and ending at the same  $\zeta$-soliton $\phi_{\L,i}$.

On the other hand, if $p$ and $p'$ are two {\underline {distinct}} intersection points  of $\L$ with the right thimble of type $i$, corresponding to
two different half-line solitions, then there can be
nontrivial $\zeta$-instantons with the boundary condition that $\phi \to \phi_{\CL,i}^{p}$ for $\tau \to -\infty$ and
$\phi \to \phi_{\CL,i}^{p'}$ for $\tau\to +\infty$. Indeed, such instantons of fermion number one
define the differential on the MSW complex $\IM_{\CL,i}$.
They are localized both in $\tau$ and near the boundary and correspond to boundary vertices of valence zero in the extended webs of the
web-based formalism. Recall Figure \ref{fig:HALFPLANE-INSTANTON-1}.

In analogy with Section \S \ref{zetawebs}, our main interest is in $\zeta$-instantons on the half-space that are asymptotic to a half-plane fan of solitons,
as sketched in Figure \ref{fig:HALFPLANEBC}, with $i_1\not= i_n$.  We trust that this notion is clear: as in the figure, we start with a half-plane fan of vacua $i_1,\dots,i_n$
and then choose suitable half-line $\zeta$-solitons at $\tau\to \pm \infty$, and suitable boosted
$\zeta$-solitons along outgoing lines separating the vacua
in the fan.   Generally speaking, all notions of Section \S \ref{zetafan} have analogs in this situation.   If $\M$ is a component of the moduli space\footnote{As
discussed most fully in section \ref{tonzo}, in defining $\M$, we do not specify the impact parameters of outgoing solitons.}
of half-space $\zeta$-instantons with fan-like asymptotics, then the group $\IR$ of time translations acts freely on $\M$; we define the reduced
moduli space $\M_\red=\M/\IR$.  We call a connected component of $\M$ a half-space $\zeta$-vertex if the corresponding reduced component
in $\M_\red$ is compact.  If in addition that component is a point,
we call it a rigid $\zeta$-vertex; if $\M_\red$ has positive dimension, we call its dimension the excess dimension of $\M$.
We can write again a formula for the expected dimension of $\M$ in an $\FF$-conserving system in terms of the upper fermion numbers
$f_a$ of the solitons in the fan and $f^{p'}$, $f^p$ of the half-line soliton states $|p\rangle$ and $|p'\rangle$:
\begin{equation}\label{eq:hdimformula}
\mathrm{dim} \,\M = f^{p'} - f^p + \sum_a (f_a+1)
\end{equation}
 In general, if $\M_\red$ is not compact,
it has ``ends'' that can be constructed by gluing of sub-solutions, as we discussed for $\zeta$-instantons with fan-like asymptotics
on $\IR^2$ starting in  Section \S \ref{zetafan}.  The sub-solutions can now be  $\zeta$-instantons with fan-like asymptotics
on either $\IR^2$ or the half-space.  Thus, in general the ``ends'' of $\M$ correspond to half-space $\zeta$-webs.  After repeatedly
resolving the sub-solutions into $\zeta$-webs, we eventually learn that any end of $\M$ can be built by gluing of $\zeta$-vertices $\V_J$, arranged in a $\zeta$-web $\frak u$.
As in eqn. (\ref{ongoing}), the dimension of $\M$ exceeds the dimension of the moduli space of the half-space web $\fu$ (for any $\fu$ associated to
 an end of $\M$) by the sum of the excess dimensions of the $\zeta$-vertices $\V_J$.

\subsubsection{Boundary Amplitudes And The Relations They Obey}\label{longwinded}

The general analysis of boundary amplitudes and the relations they obey
is very similar to what we said in action \ref{webscot} concerning interior
amplitudes -- so similar that we will be brief.

We start with a boundary condition associated to a Lagrangian submanifold $\L$,
with a half-plane fan of vacua $i_1,\dots, i_n$, as in Figure \ref{fig:HALFPLANEBC}.
To this data, we have complexes $\IM_{\L,i_1}$ and $\IM_{ \L,i_n}$ of initial and final classical
 half-plane states, and reduced complexes $\IM'_{i_1i_2},\dots,\IM'_{i_{n-1}i_n}$
of classical outgoing solitons.  The tensor product
\be
\IM^{\L}_{i_1\dots i_n}=\IM_{\L,i_1}\otimes
\IM'_{i_1i_2}\otimes \dots\otimes \IM'_{i_{n-1}i_n}\otimes (\IM_{\L,i_n})^*
\ee
has a basis corresponding
to choices of classical BPS half-line and soliton states.  The spaces $\IM_{\L,i}$   of
half-line solitons play the role of the Chan-Paton spaces $\mathcal E_i$  of the abstract discussion.
The impact parameters of the $n-1$ outgoing
solitons define a point in $\IR^{n-1}$, and we write $\Omega^*(\IR^{n-1})$ for the corresponding
space of differential forms.

If $\V$ is a rigid half-plane $\zeta$-vertex that is asymptotic to the given
fan of vacua, then the corresponding path integral defines an element
$B(\V)\in\Omega^*(\IR^{n-1})\otimes \IM^{\L}_{i_1\dots i_n}$.  Dividing by time
translations, $\IR^{n-1}$ projects to $\IR^{n-2}$ and $B(\V)$ is the pullback of an element
$B_\red(\V)\in \Omega^*(\IR^{n-2})\otimes \IM^{\L}_{i_1\dots i_n}$. This copy of $\IR^{n-2}$ has a natural
origin $\o$, corresponding to a collection of $n-1$ solitons that all emanate from a common point on $\IR$, the boundary
of the half-plane.  Since $\V$ is rigid, its reduced moduli space is 0-dimensional and $B_\red(\V)$ is
a simple product
\begin{equation}\label{merf} B_\red(\V)=\Theta_\o\cdot B'_\red(\V), \end{equation}
where $\Theta_\o$ is Poincar\'e dual to the point $\o\in\IR^{n-2}$ and $B'_\red(\V)\in \IM^{\L}_{i_1\dots i_n}$.
(As usual, such a statement holds modulo a correction $\d\chi$, where $\chi$ vanishes rapidly at infinity.)

We say that a component $\M$ of the moduli space of half-plane $\zeta$-instantons is ``taut'' if
{\it (i)} it is two-dimensional, so that the reduced space $\M_\red=\M/\IR$ has dimension 1; {\it (ii)} this
reduced space is a copy of $\IR$.  Such components play a role analogous to that played by
taut half-plane webs in the abstract discussion of section \ref{subsubsec:WebRep-Halfplane}.
In this situation, each of the two ends of
$\M_\red\cong \IR$ corresponds to what we will call a taut half-plane $\zeta$-web, by which we mean
a half-plane $\zeta$-web with a 1-dimensional reduced moduli space.

From here, the goal is to show that the objects $B'_\red(\V)$ (after dressing as in eqn. (\ref{torrix}) with
1-forms for outgoing solitons) satisfy the fundamental identity (\ref{eq:boundary-amp}) or (\ref{eq:boundary-amp2}) of a boundary
amplitude, as formulated abstractly in section   \ref{subsec:WebRepPlane}.
The argument will simply follow what we said for interior amplitudes in section \ref{webscot}.

Let $\M$ be a taut family of half-plane $\zeta$-instantons, and
let  $\ZZ$, $\ZZ'$ be  taut $\zeta$-webs representing the two ends of $\M_\red$.
The path integral associated to the  family $\ZZ$ of $\zeta$-instantons
determines a state $\Psi_\ZZ\in \Omega^*(\IR^{n-1})\otimes \IM^{\L}_{i_1\dots i_n}$, which is the pullback
of a reduced state $\Psi_{\ZZ,\red}\in\Omega^*(\IR^{n-2})\otimes \IM^{\L}_{i_1\dots i_n}$.  Just as in
(\ref{nozo}), this state is supported on a ray $\fr$ emanating from the origin $\o\in\IR^{n-2}$:
\begin{equation}\label{mozo}\Psi_{\ZZ,\red}=\Theta_\fr\cdot \Psi^*_\ZZ,~~\Psi^*_\ZZ\in \IM^{\L}_{i_1\dots i_n}.\end{equation}
Hence
\begin{equation}\label{moxo}\d\Psi_{\ZZ,\red}=\Theta_\o\cdot \Psi^*_\ZZ. \end{equation}
We can make the same construction for $\ZZ'$; $\Psi_{\ZZ',\red}$ is supported on another
ray $\fr'$ from $\o$, and again
\begin{equation}\label{oxo}\d\Psi_{\ZZ',\red}=\Theta_\o\cdot \Psi^*_{\ZZ'}, \end{equation}
for some $\Psi^*_\ZZ\in \IM^{\L}_{i_1\dots i_n}$.
The state determined by the full moduli space $\M$ must be closed, and this tells us that
\begin{equation}\label{boroxo} \Psi^*_\ZZ+\Psi^*_{\ZZ'}=0,\end{equation}
just as in eqn. (\ref{funda}).

From eqn. (\ref{boroxo}), we can deduce a result analogous to eqn. (\ref{boffo}).
Let $\SSS^\L_{i_1\dots i_n}$ be the set of all taut $\zeta$-webs asymptotic to the given
half-plane fan of vacua.  Then
\begin{equation}\label{mooxo}\sum_{\ZZ\in\SSS^\L_{i_1\dots i_n}}\Psi^*_\ZZ=0. \end{equation}
The proof mimics the proof of eqn. (\ref{boffo}).  Each taut half-plane $\zeta$-web is associated to an end of
a 1-dimensional reduced $\zeta$-instanton moduli space; each such moduli space has two ends; and
by virtue of (\ref{boroxo}), these two ends cancel in pairs in eqn. (\ref{mooxo}).

The relation between (\ref{mooxo}) and the identity (\ref{eq:boundary-amp})
or (\ref{eq:boundary-amp2})
of the abstract
discussion of boundary amplitudes
 is quite analogous to the corresponding relation between eqn. (\ref{boffo}) and
the identity (\ref{eq:bulk-amp}) of the abstract discussion of bulk amplitudes.  $\Psi_\ZZ^*$ can be
computed by working at infinity in the reduced moduli space of $\ZZ$, where it is built from
widely separated bulk and boundary rigid $\zeta$-vertices, connected in general by classical
solitons that propagate for long distances.   Let us denote the bulk and boundary $\zeta$-vertices
as $\V_a$ and $\V_\alpha$, respectively.  To compute $\Psi_\ZZ^*$, we take the tensor
products of all states $B(\V_a)$ and $B(\V_\alpha)$, and then, whenever two vertices
are connected by a soliton line in the $\zeta$-web $\ZZ$, we contract out the corresponding
soliton states using the degree 1 pairing (\ref{welz}).
Thus, the left hand side of (\ref{mooxo}) is obtained by summing over all taut half-plane $\zeta$-webs $\ZZ$,
and for each such web,
 computing a state
$B(\V_a)$ or $B(\V_\alpha)$ for each bulk or boundary vertex and then contracting these states
whenever two vertices are joined by lines.  All this matches  precisely the operation $\rho_\beta$ used in
the abstract statements (\ref{eq:boundary-amp})
or (\ref{eq:boundary-amp2}) and therefore  (\ref{mooxo}) essentially matches those identities.

To match the notation used in the abstract discussion, we proceed as in
eqn. (\ref{torrix}).  Let $\V_\alpha$ be a rigid half-plane $\zeta$-vertex, asymptotic to a half-plane
fan with $s_\alpha$ vacua and $s_\alpha-1$ outgoing solitons with impact parameters
$x^\alpha_{1},\dots x^\alpha_{s_\alpha-1}$.
In a fermion-number conserving theory, the state $B(\V_\alpha)$ produced by the path integral
has fermion number 0.  However, $\Theta_\o$, being Poincar\'e dual to a point $\o\in\IR^{s_\alpha-2}$,
is an $(s_\alpha-2)$-form, and hence the reduced state $B'_\red(\V_\alpha)$ has fermion number $-(s_\alpha-2)$.
To match the abstract discussion, we should define
\begin{equation}\label{horf} {\mathcal B}=\sum_\alpha\d x^\alpha_{1}\dots \d x^\alpha_{s_\alpha-1}
B'_\red(\V_\alpha), \end{equation}
where the sum runs over rigid half-plane $\zeta$ vertices.
Clearly, ${\mathcal B}$ is a sum of terms all of fermion number 1.  This is the object that appears in the identity
(\ref{eq:boundary-amp}) or (\ref{eq:boundary-amp2}) of the abstract discussion.

\begin{figure}[htp]
\centering
\includegraphics[scale=0.3,angle=0,trim=0 0 0 0]{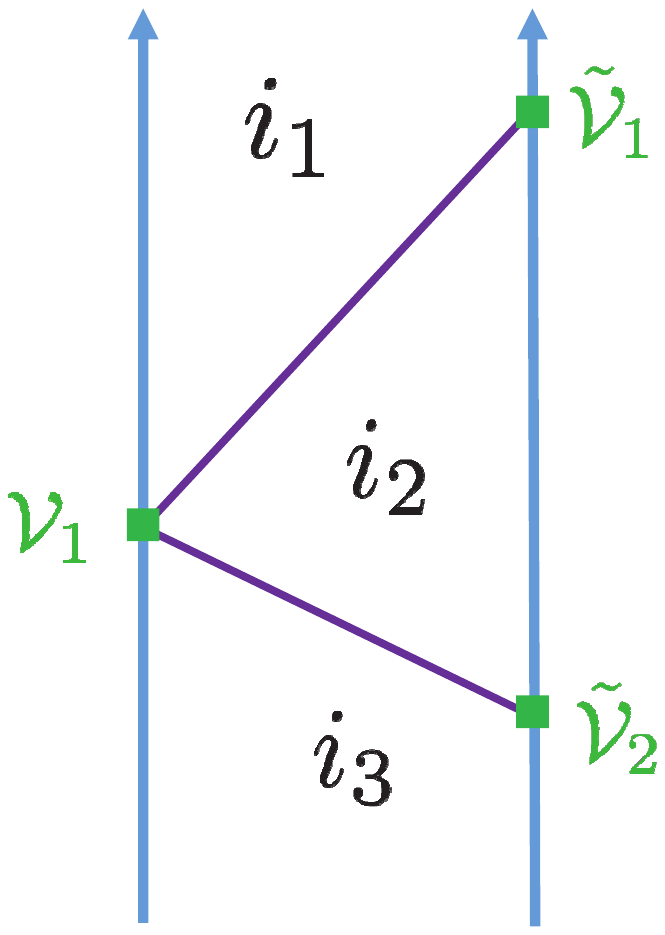}
\caption{A rigid strip web $\fs$.   }
\label{fig:RIGIDSTRIPWEB}
\end{figure}

\subsubsection{$\zeta$-Instantons On A Strip}\label{zetastrip}

Now we return to the problem formulated in section \ref{thestrip} of finding physical states when the theory
is quantized on an interval $D=[x_\ell,x_r]$, with boundary conditions at the two ends of the strip set by Lagrangian
submanifolds $\L_\ell$, $\L_r$.  We recall from eqn. (\ref{eq:appxt-complex-ii}) that, if the width of the strip is much greater than the largest length
scale in the theory, there is no problem to describe the classical approximation to the space of physical
states on the strip:
\be\label{goodapprox}
 \IM_{\CL_\ell, \CL_r} \cong \oplus_{i\in
 \IV}  \IM_{\CL_\ell,i} \otimes \IM_{i, \CL_r}
\ee
The problem raised in section \ref{thestrip} was to compute the differential $\CQ_{\zeta}$ that acts on this complex.
Matrix elements of this differential are supposed to be computed by counting $\zeta$-instantons on the strip.
To be more precise,
we are supposed to count families of rigid $\zeta$-instantons on the strip, that is families of $\zeta$-instantons  that have
no modulus except the inevitable modulus associated to time translations.   To compute a matrix element of $\CQ_{\zeta}$ between
specified initial and final states, we have to count families of rigid $\zeta$-instantons that interpolate between
specified initial and final $\zeta$-solitons -- that is elements in the complex $ \IM_{\CL_\ell, \CL_r}$ of
(\ref{goodapprox}) -- in the far past and the far future.

We hope it is now clear how to do the counting.  On a very wide strip, the $\zeta$-instantons can be represented
by webs -- or $\zeta$-webs -- in which the lines are classical solitons and the vertices are bulk and boundary
$\zeta$-vertices.  So if $\M$ is a component of $\zeta$-instanton moduli space on the strip, we can associate
to it a strip web $\fs$ in the sense of section \ref{subsec:Strip-Webs}.  The $\zeta$-instantons parametrized
by $\M$ are built by gluing bulk and boundary $\zeta$-vertices $\V_a$, $a=1,\dots,s$ and $\V_\alpha$, $\alpha=1,\dots,t$
 via the strip web $\fs$.  The web dimension $d(\fs)$ is always at least 1, because
of time translations, and $\fs$ is called a rigid strip web if $d(\fs)=1$.  The dimension of $\M$ exceeds
$d(\fs)$ by the sum of the excess dimensions of the $\zeta$-vertices $\V_a$ and $\V_\alpha$.  (See
eqn. (\ref{ongoing}) for a formula of this type.)  So for $\M$ to be 1-dimensional, $\fs$ must be rigid
and the $\zeta$-vertices $\V_a$ and $\V_\alpha$ must have 0 excess dimension, that is they must also be rigid.
Given a rigid strip web $\fs$, such as the one sketched in Figure \ref{fig:RIGIDSTRIPWEB}, to count the corresponding
$\zeta$-instantons, we just sum over all possible labelings of the internal lines in the web by  solitons,
the boundary segments by half-plane states, and the bulk and boundary vertices $\V_a$ and $\V_\alpha$
by the corresponding bulk and boundary
amplitudes $B(\V_a)$ and $B(\V_\alpha)$.    We have shown that $B(\V_a)$ and $B(\V_\alpha)$
obey the algebraic identities that were assumed in section \ref{subsec:WebRepStrip}, and the  counting procedure
for rigid $\zeta$-instantons on the strip
coincides with what was used in that section to define a differential.  So we conclude that
the procedure of that abstract discussion can indeed be used to determine the differential whose
cohomology gives the exact supersymmetric ground states on the strip.

\section{Webs And The Fukaya-Seidel Category}\label{notif}

\subsection{Preliminaries}\label{realprelims}

In this paper, we have described two algebraic structures associated to open strings in the same massive Landau-Ginzburg model:

{\it (i)}  In the abstract part of this paper, Sections \S\S \ref{sec:Webs}-\ref{sec:CategoriesBranes}, we described
an algebraic structure that is built, roughly speaking, by multiplying boundary web vertices.   This algebraic
structure is an $A_\infty$ algebra, something that in physical applications is usually associated to  open-string amplitudes at tree level.

{\it (ii)}  In section \ref{seidelfukaya}, we described the Fukaya-Seidel category, or more precisely in our context the Fukaya category of
the superpotential, in which an $A_\infty$ algebra
is actually defined from open-string amplitudes.

It would be surprising for two essentially different $A_\infty$ algebras to arise in the same model, and indeed in the present
section we will argue that, once one restricts the web-based construction to a smaller class of branes in a way that we will  explain, the two $A_\infty$ algebras are indeed equivalent. Roughly speaking, $(i)$ and $(ii)$ correspond, respectively, to methods of describing the same category
in the infrared (long distances) and in the ultraviolet (short distances).  Naively, the superpotential $W$ is very
important in the infrared and not important in the ultraviolet. Unfortunately, this is a little oversimplified.
It is true that in method $(i)$, the superpotential plays a very important role, and we will explain below in what sense
webs emerge from the $\zeta$-instanton equation in the infrared limit.  But, unless the
branes considered are all compact, there is no ultraviolet limit in which one completely forgets the superpotential;
as explained in Section \S \ref{boundary}, it plays a role in controlling the classes of branes that one can consider and in ensuring that
counting of solutions is always well-defined.

%The equivalence of these categories can be motivated physically
% since $(i)$ and $(ii)$ above correspond, roughly speaking, to the infrared and ultraviolet description of the category
% of branes in the Landau-Ginzburg model, respectively. The ultraviolet description corresponds
%to the Fukaya-Seidel category, based as it sometimes is, on the pseudoholomorphic map equation,
%corresponding to the leading UV terms in the $\zeta$-instanton equation. The superpotential is essential in delimiting the
%appropriate class of Lagrangians we allow as objects, but $W(\phi)$ is needed to define the theory in the UV
%and should be considered as UV data.
% On the other hand, the infrared description corresponds to the web-based formalism, based as it is
%on the triviality of the massive theory in the far infrared, except for the solitons which lead to the edges of the webs.
%Note that the webs depend only on the \emph{critical values} $W_i$ of the superpotential, this, together with the
%soliton spaces $R_{ij}$ can be considered as purely IR data.
%

Approach {\it (i)} was part of a larger discussion in which the largest class of branes that one can consider are
the $W$-dominated branes described in section \ref{goodclass}. Recall this means that on
 left branes   $\Im(\zeta^{-1}W)$
goes to $+\infty$ at infinity, and on right branes  $\Im(\zeta^{-1}W)$ goes to $-\infty$ at $\infty$.
%%
%Let us call these the \emph{minimal conditions} on the behavior of $W$ along a brane.
%%
If $\B_\ell$ is a left brane and $\B_r$ is a right brane, then there is a well-defined space of $(\B_\ell,\B_r)$
strings.  This space can be computed by solving the $\zeta$-instanton equation on a vertical strip in the $s=x+i\tau$ plane.
To ensure that the space of $(\B_\ell,\B_r)$ strings is well-defined, the strip has to be vertical (as drawn, for example, in Figure
\ref{thirdone} from the introduction, and in many other figures in this paper),
assuming no restriction except that the branes are $W$-dominated.
We always think of a left brane as attached to a left vertical boundary of a region in the $s$-plane, and a right brane
as attached to a right vertical boundary of such a region.

Manipulating boundary vertices by the procedure of Section \S \ref{sec:CategoriesBranes}
gives one $A_\infty$ algebra for the left branes, and another for the right branes.
Similarly, there are really two versions of the Fukaya-Seidel
category.  That follows because in Section \S \ref{seidelfukaya} we considered branes of class $T_\kappa$ where $\kappa$ is a
complex number of modulus 1 that is required to differ from $\pm \zeta$.  Removing the points $\pm \zeta$ from the unit
circle divides it into two components, and there are really two Fukaya-Seidel categories depending on which component contains
$\kappa$.  We will match the two Fukaya-Seidel categories with the web-based categories for left and right branes.

However, to make contact with the Fukaya-Seidel category, it is not sufficient simply
 to require that the left and right branes be $W$-dominated.
 Let us therefore discuss what is the smallest class of branes that we could reasonably consider.
In the abstract discussion, a brane $\B$ has Chan-Paton factors $\E_i(\B)$ for each vacuum state $i\in\IV$.
$\E_i(\B)$ is a complex whose cohomology is
the space of supersymmetric physical states when the theory is formulated on a half-line with $\B$ at the finite
end and vacuum $i$ at infinity.  (In the case of a left or right brane, the finite end of the half-line is taken to be on the left or right.)
The axioms of the abstract discussion ensure the existence, for each vacuum $i\in \IV$,
 of a distinguished left brane $\fT_i$ whose Chan-Paton
factors are $\delta_{ij}\IZ$ (or $\delta_{ij}\IC$ if one considers physical states to be complex vector spaces rather than
$\IZ$-modules); in other words, there are no physical states with brane $\fT_i$ at the finite end and vacuum $j$ at infinity,
unless $i=j$, in which case the space of such states has rank (or dimension) 1.    In terms of Landau-Ginzburg models,
the $\fT_i$ are the left thimbles $L_i^\zeta$ that were introduced in section \ref{thimbles}.   The Chan-Paton factors of these
thimbles
were described in eqn. (\ref{menod}) and are as desired.  Thus, certainly, in a Landau-Ginzburg construction
that is supposed to illustrate the abstract part of this paper, among the left branes we must at least include the left thimbles
$L_i^\zeta$. (And similarly we must include the right thimbles $R_i^\zeta$ among the right branes.)
The abstract part of this paper also includes a criterion (section \ref{subsubsec:WebRep-Halfplane})
for what is a brane with more general Chan-Paton factors,  and to match this discussion, we need to allow
more general branes built from the thimbles.
 We therefore need a reasonable class of left branes that includes these thimbles, and is closed under the relevant
 operations (which basically involve building a new brane as an extension of one brane by another) yet is
  sufficiently small to allow a construction we will come to momentarily.
\footnote{Mathematically we would certainly want to include the smallest triangulated category
containing the thimbles.}
It turns out that a suitable condition is that a left brane must be of class $T_\zeta$, as described in section
\ref{morebranes}.  (We recall that the $T_\zeta$ condition means that, along the support of the brane, $W$ must take values in a certain
semi-infinite strip that contains the images in the $W$-plane of the $L_i^\zeta$, as in Figure (\ref{manyrays}).)  The analogous
condition on right branes is that they should be of class $T_{-\zeta}$.

To compare the web-based approach to the Fukaya-Seidel category, we want to interpret left branes of class $T_\zeta$
as objects in the Fukaya-Seidel category.  At first sight there is some tension here.
In section \ref{seidelfukaya}, we found that to describe the Fukaya-Seidel category using the $\zeta$-instanton equation,
we had to consider not branes of class $T_{\pm \zeta}$, but branes of class $T_\kappa$ where $\kappa$ is a complex number
of modulus 1 and not equal to $\pm \zeta$.  But  in the web-based procedure using
the $\zeta$-instanton equation, we definitely want left branes of class $T_\zeta$, not of class $T_\kappa$ with any other $\kappa$.
(And the right branes should definitely be of class $T_{-\zeta}$.)       However, there is  a simple way to reconcile the different
statements.  We use the fact that the $\zeta$-instanton equation is not invariant under rotations of the $s$-plane, and that
such a rotation is equivalent to a rotation of the complex number $\zeta$.  In section \ref{seidelfukaya}, we assumed
that open-string amplitudes relevant to the Fukaya-Seidel category are to be computed with a quantum field theory defined
in a region of the $s$-plane whose boundaries are asymptotically vertical, as in Figure \ref{openstrings}.  In this case,
the branes must be of class $T_\kappa$ with $\kappa\not=\pm\zeta$.  If we want instead to use branes of class $T_\zeta$,
we simply have to rotate  Figure \ref{openstrings} so that the boundaries of incoming and outgoing open strings are not vertical.
For example, it is convenient to rotate the figure by an angle of $\pm \pi/2$ so that incoming open strings come in from the left
of the $s$-plane or from the right.  (The angle $\pi/2$ is not essential; any angle other than 0 or $\pi$ will do.)

As we explained above, there are really two Fukaya-Seidel categories, depending on which component of $S^1\backslash
\{\pm\zeta\}$ contains $\kappa$.  After we rotate Figure \ref{openstrings} so that the open strings no longer come in from the  bottom
(or top)  of the figure, the two categories differ by whether the open
strings come in from the left half of the $s$-plane or from the right half.
We will match the Fukaya-Seidel category constructed with open strings that come in from the left to the web-based category
of left branes.  Similarly, the Fukaya-Seidel category with open strings that come in from the right matches the web-based
category of right branes.

Some of the ideas in the following analysis have been explained in a simpler context in Section \S \ref{whynot}.

\subsection{Morphisms}\label{morphisms}

 \begin{figure}
 \begin{center}
   \includegraphics[width=2.5in]{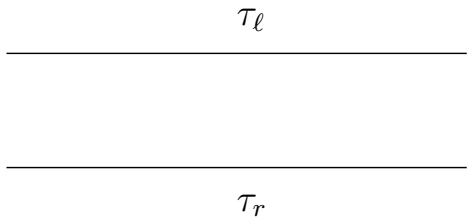}
 \end{center}
\caption{\small A horizontal strip $\tau_\ell>\tau>\tau_r$ in the $s$-plane.
}
 \label{horizontal}
\end{figure}

 \begin{figure}
 \begin{center}
   \includegraphics[width=4.5in]{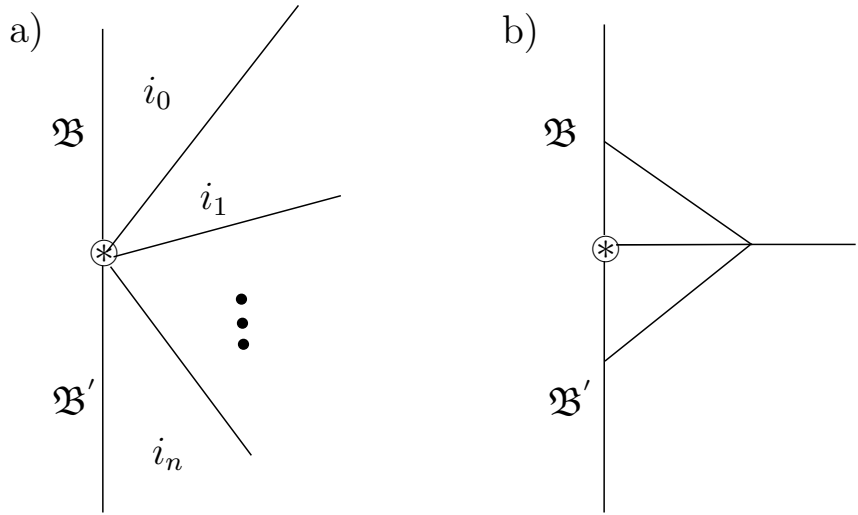}
 \end{center}
\caption{\small (a) The symbol $\circledast$ represents a half-space fan of solitons, as indicated, or
equivalently an element
$\delta \in \Hop(\B,\B')$ in the web-based formalism.  The symbol $\circledast$
with the indicated fan of solitons emerging from it
represents a possible asymptotic behavior at infinity of a solution of the $\zeta$-instanton equation with no knowledge of how the solution behaves in the interior.
(b) This
picture, which is our first example of a ``hybrid'' $\zeta$-web, would represent a
$\zeta$-web with a 1-dimensional reduced moduli space, except that one of the vertices is not a  $\zeta$-vertex but rather is
an abstract vertex labeled $\circledast$, which represents a fan of solitons rather than a solution of the $\zeta$-instanton equation.
All other bulk and boundary vertices in the picture are conventional $\zeta$-vertices, contracted together by ``propagators'' in the usual way.  }
\label{labeledfan}
\end{figure}

If $\B$ and $\B'$ are two branes of class $T_\zeta$, then we can calculate the space of supersymmetric $(\B,\B')$ states
on the interval, which we will call $\H_{\B,\B'}$,
in the Fukaya-Seidel category, and we can also compute $\H^{\rm web}_{\B,\B'}:=H^*(\Hop(\B,\B'),M_1)$
in the web-based construction of an $A_\infty$
category.
Our goal in this section is to sketch the construction of a natural isomorphism $\H_{\B,\B'}\cong \H^{\rm web}_{\B,\B'}$.
(Of course, like most claims in this paper about the $\zeta$-instanton equation, the arguments presented here are not complete
mathematically.)
When combined with a similar analysis that we will make of the multiplication of string states in section \ref{multiplication},
and of the higher order operations in section \ref{higherones}, this will show the equivalence between the two constructions
of an $A_\infty$ category.
\footnote{In the abstract part of this paper, we introduced (1) a space of $(\B,\B')$ ground states, where $\B$ is a left-brane
and $\B'$ is a right-brane, and also  (2) a space
 $\H^{\rm {web}}_{\B,\B'}$ of morphisms between two left-branes (or two right-branes).  What is relevant to the present discussion
 is definitely construction (2).  From the $\sigma$-model point of view, the definition (1) assumes that $\B$ and $\B'$ are $W$-dominated
 branes of opposite type, and the definition (2) applies to branes of class $T_\zeta$.    $\H^{\rm web}_{\B,\B'}$ can be interpreted as a space of open-string
 ground states in quantization on a horizontal strip as in Figure \ref{horizontal},
 but since the class of branes is different and both branes are of the same type, these are not the open-string states studied in the abstract part of the paper. }

First we review the definition of the two spaces that are supposed to be isomorphic.
The definition of $\H_{\B,\B'}$ involves two steps:

{\it (a)} We consider the LG model of interest on a strip $\IR\times I$, where $I$ is a closed interval, with boundary
conditions at the two ends set by $\B$ and $\B'$, where $\B,\B'$ are both of class $T_\zeta$.  For reasons explained in section \ref{realprelims}, we take this
strip to run horizontally, rather than vertically, in the $s=x+i\tau$ plane (Figure \ref{horizontal}).  Thus the strip is defined by
$\tau_\ell\geq \tau\geq \tau_r$ (for some $\tau_\ell$, $\tau_r$), and $x$ plays the role of ``time.''
We define a vector space   $\IM_{\B,\B'}$ (actually a $\IZ$-module)
that has a basis vector for every $\zeta$-soliton -- that is, for every
solution of the $\zeta$-instanton equation \eqref{eq:LG-INST} on the strip that depends only on $\tau$.  This complex is graded
by fermion number in the usual way.

{\it (b)} On $\IM_{\B,\B'}$, we define a differential $\h\CQ_{\zeta}$ by counting suitable 1-parameter families of
$\zeta$-instantons on the strip.  The solutions are required to be independent of $x$ for $x<<0$ and also for
$x>>0$.  This means that the ``initial data'' (for $x<<0$) are associated to a basis vector $\vert \phi_{\ell} \rangle$ of
$\IM_{\B,\B'}$, corresponding to some $\zeta$-soliton that satisfies the boundary conditions, and the ``final state''
(for $x>>0$) is similarly associated to a possibly different $\zeta$-soliton $\vert \phi_{r} \rangle$.
The one-parameter family corresponds to translation in the $x$-direction.
Counting 1-parameter families
of $\zeta$-instantons with these boundary conditions gives the matrix element of $\h\Q_{\zeta}$ from $\vert \phi_{\ell} \rangle$ to $\vert \phi_{r} \rangle$.
(As always, in this ``counting,'' a $\zeta$-instanton is weighted with a factor $\pm 1$ coming from the sign of the fermion
determinant.)  The space $\H_{\B,\B'}$ is the cohomology of the differential $\h\Q_{\zeta}$.

\def\w{{\frak w}}
The definition of $\H^{\rm web}_{\B,\B'}$ in the web-based procedure of section \ref{sec:CategoriesBranes}
involves two analogous steps:

{\it ($a'$)} We first choose a half-plane, $\CH$, and here we choose it to be the
 positive half $s$-plane as in Figure \ref{labeledfan}(a)). Next we
 define a complex $\IM^{\rm web}_{\B,\B'}$  with a basis vector associated the
 the following data: As in Figure \ref{labeledfan}(a)) we choose a half-plane
 fan of vacua $J= \{i_0,i_1,\dots, i_n \}$.     The regions between
the lines in the figure are labeled by vacua $i_0,\dots,i_n\in\IV$, representing constant solutions of the $\zeta$-instanton
equation.  The line separating any two consecutive vacua $i_{k}$ and $i_{k+1}$  is labeled by an $i_k i_{k+1}$ $\zeta$-soliton.
When restricted to large $|s|$, $\CH$ has both an upper and a lower boundary.  The upper boundary is labeled by
a half-line $\zeta$-soliton interpolating from brane $\B$ to vacuum $i_0$; the lower boundary
is labeled by a half-line $\zeta$-soliton   set by the brane $\B'$ and the vacuum $i_n$. If we compare with
the definitions in Section \S \ref{subsec:BraneCat} then we should
identify
\be
\IM^{\rm web}_{\B,\B'}=  \Hop(\B,\B') = \Hom( \B',\B) = \oplus_{i,j\in \IV} \CE(\B)_i \otimes
\hat R_{ij} \otimes (\CE(\B')_j)^*.
\ee

In short, $\IM^{\rm web}_{\B,\B'}$ has a basis in which the basis vectors are half-plane fans of solitons, interpolating between branes $\B$ and $\B'$.
Such a fan is indicated in Figure \ref{labeledfan}(a).  It represents
the asymptotic behavior at infinity of a possible solution of the $\zeta$-instanton equation on the half-plane,
with no knowledge concerning the behavior in the interior.

In the abstract part of this paper, we studied webs in which the vertices represented elements of $\IM^{\rm web}_{\B,\B'}$ and its bulk analog.
 Starting in section \ref{zetafan}, we have studied $\zeta$-webs, in which the vertices are $\zeta$-vertices, which represent solutions of the $\zeta$-instanton
 equation with fan-like asymptotics.  It will now be useful to consider webs with ingredients of both kinds.  We will call these {\it hybrid} $\zeta$-webs.
 In a hybrid $\zeta$-web -- our first example is in Figure \ref{labeledfan}(b) -- a vertex labeled by $\circledast$ represents a fan of solitons, and a vertex
 not so labeled is a conventional $\zeta$-vertex.   Thus, the symbol $\circledast$, which we call an abstract vertex (in homage to the abstract nature
 of the web-based construction) represents an element of $\IM^{\rm web}_{\B,\B'}$ associated to a given fan of solitons,
 while the other vertices are the ones that we have used until
 now in the $\sigma$-model approach.  Consideration of these hybrid $\zeta$-webs will be helpful in bridging the gap between the two approaches to open-string amplitudes.

The next step is to define a differential on the space $\IM^{\rm web}_{\B,\B'}$,
making it a complex whose cohomology is $\H^{\rm web}_{\B,\B'}$.   The differential is called $M_1$ in eqn. (\ref{eq:BraneMultiplications}) and may be described as follows:

{\it ($b'$)} In the web-based treatment, the differential acting on $ \delta\in \IM^{\rm web}_{\B,\B'}$ is defined by inserting $\delta$ (or the associated
half-space fan) at one boundary vertex of a taut half-plane web, and weighting all other interior or boundary vertices by the appropriate interior
or boundary amplitudes.  We follow the same procedure now, with one difference: we interpret the interior or boundary amplitudes as the ones
that are computed in the $\sigma$-model, by counting solutions of the $\zeta$-instanton equation.  This means that the bulk and boundary vertices
-- other than the one associated to $\delta$ -- are now going to be $\zeta$-vertices.\footnote{Likewise in discussing below other constructions of the web-based
approach, all bulk or boundary vertices not labeled by $\circledast$ will be $\zeta$-vertices.}

Accordingly, we  consider a hybrid $\zeta$-web (Figure \ref{labeledfan}(b)), of the type just described.
We use such hybrid $\zeta$-webs, for the time being, merely to give a pictorial representation of
some of the abstract web-based constructions, with the one change that bulk and interior amplitudes are now derived by counting $\zeta$-instantons.
The hybrid $\zeta$-webs in Figure \ref{labeledfan}(b) are required to be taut; that is, they have precisely one reduced modulus, derived from
an overall scaling of the half-plane, keeping the abstract vertex fixed.

To a hybrid $\zeta$-web with one abstract vertex $\circledast$ and with fan-like asymptotics at infinity, we can associate a pair of elements
of the web-based complex $\IM^{\rm web}_{\B,\B'}$.   By restricting the hybrid $\zeta$-web to a neighborhood of the abstract vertex
$v$ we find a labeled fan of vacua $J_v$ and a corresponding
basis vector $|\phi_F^{\rm web}(\circledast)\rangle$ of $\IM^{\rm web}_{\B,\B'}$. (The subscript $F$ is meant to
 remind us that $\phi_F$ depends on a fan of solitons, rather than a single soliton.)
 Similarly, by   restricting the web to a neighborhood of $|s|=\infty$, we get a second labeled fan of vacua $J_\infty$
and correspondingly a second
basis vector $|\phi_F^{\rm web}(\infty)\rangle$. It is convenient to refer to a neighborhood of the abstract vertex and a neighborhood of infinity as the two
ends of the hybrid $\zeta$-web in Figure \ref{labeledfan}(b).  The matrix element of the web-based differential $M_1=\h\Q_{\zeta}^{\rm web}$ from $|\phi_F^{\rm web}(\circledast)\rangle$ to
$|\phi_F^{\rm web}(\infty)\rangle$
is computed by counting the number of hybrid $\zeta$-webs, as in Figure \ref{labeledfan}(b), with precisely one reduced modulus
associated to scaling, and with $|\phi_F^{\rm web}(\circledast)\rangle $ and $|\phi_F^{\rm web}(\infty)\rangle$ as the restrictions to the two ends. The webs are weighted with the signs carefully spelled out in equation \eqref{eq:bdy-rho-signs} above.
Physically these signs come from the fermion determinants (and they agree because fermion determinants satisfy the gluing laws that were built
into the other approach).  The algorithm we have just described is precisely the
definition of $M_1(\delta)$ as defined in equation \eqref{eq:BraneMultiplications}, where the morphism $\delta$
is inserted at the abstract vertex $\circledast$.

\def\IW{{\Bbb W}}
 \begin{figure}
 \begin{center}
   \includegraphics[width=4.5in]{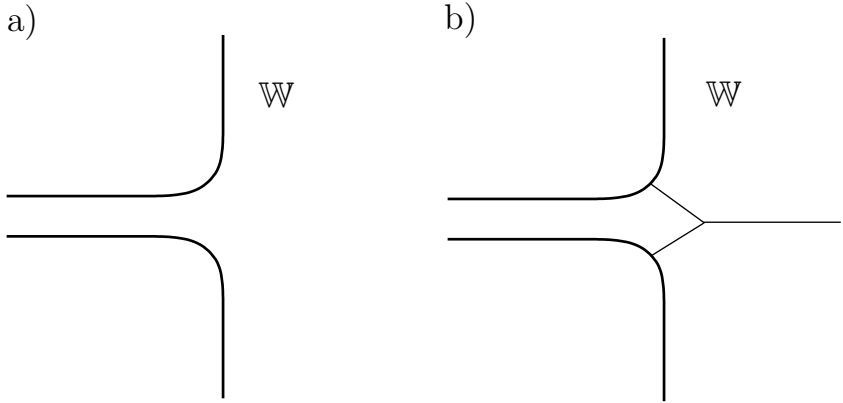}
 \end{center}
\caption{\small    (a) A region $\IW$ in the $s$-plane; a strip coming in from the left opens out to the half-plane $\CH$.
(b) A $\zeta$-web in the region $\IW$.   The web sketched here is a simple one in which two BPS solitons emitted from the boundary join into one.  This web corresponds to a moduli space of zero dimension if and only if the indicated boundary vertices are possible only at specific
points along the boundary, because the $\zeta$-solitons in question can be emitted only when the boundary slopes at a favored angle. }
 \label{duplex}
\end{figure}

To establish a natural isomorphism between $\H_{\B,\B'}$ and $\H^{\rm web}_{\B,\B'}$, we will imitate the procedure explained in
section \ref{whynot} for showing that the cohomology of the MSW complex does not depend on the metric or superpotential on $X$.
First we define a linear map
\begin{equation}\label{diflo}\U:\IM_{\B,\B'}\to \IM^{\rm web}_{\B,\B'}, \end{equation}  as follows.
We consider a region $\IW$ of the complex $s$-plane in which a semi-infinite strip comes in from $x=-\infty$
and fans out to the positive half plane (Figure \ref{duplex}(a)).  We consider solutions of the $\zeta$-instanton equation on $\IW$
with boundary conditions determined by $\B$ and $\B'$ on the upper and lower boundaries, respectively, and which
moreover are independent of $x$ for $x\to -\infty$ and finally  have fan-like asymptotics for $|s|\to\infty$ for large $\Re(s)$
in the positive half-plane.
The behavior of such a solution for $x\to-\infty$ gives a basis vector $\vert \phi_{\ell} \rangle$ of $\IM_{\B,\B'}$, and its behavior for $|s|\to\infty,
~x\gg 0$ gives a basis vector $|\phi_F^{\rm web}(\infty)\rangle$ of $\IM^{\rm web}_{\B,\B'}$.  We consider only the components of the moduli space of such
solutions with an expected dimension of 0.  The actual dimension is then also 0 if the K\"ahler metric of the target space $X$ is
generic.   Counting (with signs, as always) the solutions that are in zero-dimensional moduli spaces and with the asymptotics that we have
specified
gives the matrix element of the desired operator $\U$ from $\vert \phi_{\ell} \rangle $ to $|\phi_F^{\rm web}(\infty)\rangle.$  The map $\U$ that is defined this way
preserves the fermion
number since it comes from components of moduli space with expected dimension 0.

In this construction, the width of the strip in the left half plane is arbitrary.  In particular, we are free to choose
this strip to be much wider than the natural length scale of the massive LG theory under study.   If we do this, then a
$\zeta$-instanton contributing to $\U$ can everywhere be represented by a $\zeta$-web.  A possible example is sketched in Figure \ref{duplex}(b).
(The $\zeta$-web associated to a $\zeta$-instanton without moduli extends only a bounded distance to the left down the strip,
but given the desired fan-like asymptotics for $x\gg 0$, it is unbounded to the right.)  Computing the dimension of the
moduli space associated to such a web involves some new ingredients, which we will not explore,
since some boundary vertices might exist only when the boundary is oriented
at a favored angle.
If we work with wide strips, then other solutions that we encounter presently are similarly web-like.

 \begin{figure}
 \begin{center}
   \includegraphics[width=4.5in]{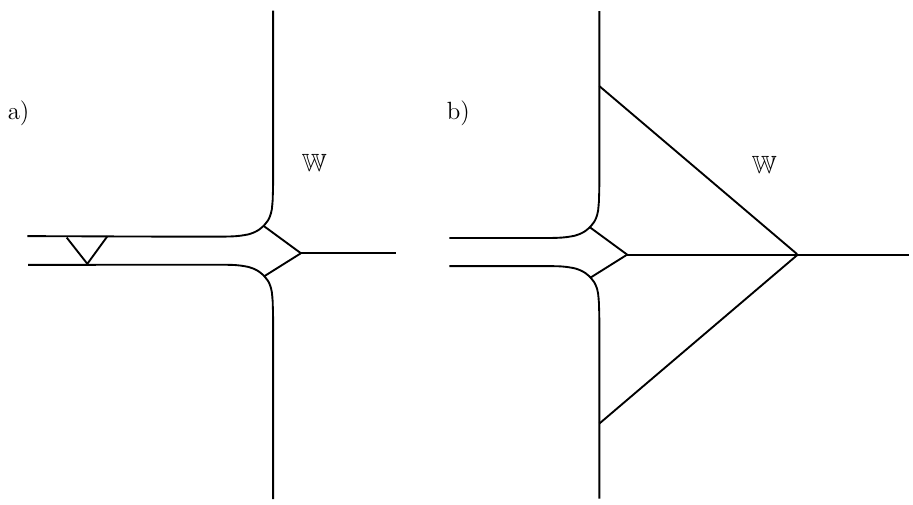}
 \end{center}
\caption{\small  The two types of end of a 1-dimensional moduli space of $\zeta$-instantons on the region $\IW$
are depicted here.  (These webs correspond to 1-dimensional moduli spaces if the web in Figure \protect\ref{duplex}(b)
%
% \ref{duplex}(b)
%
has a zero-dimensional moduli space.  As observed in the caption to that figure, this depends on some assumptions about the boundary
vertices.)  The end in (a)  correponds to the $\zeta$-instanton of Figure \protect\ref{duplex}(b),
 %
 %  \ref{duplex}(b),
 %
  which contributes to $\U$, concatenated with
 a $\zeta$-instanton on the strip, which contributes to $\h\Q_{\zeta}$.  An end of this type contributes
to the product $\U\h\Q_{\zeta}$.  The end in (b)  is made by replacing the abstract vertex in Figure \protect\ref{labeledfan}(b),
%
%  \ref{labeledfan}(b),
%
which contributes to $\h\Q_{\zeta}^{\rm web}$, with the $\zeta$-instanton of Figure \protect\ref{duplex}(b),
 %
 % \ref{duplex}(b),
 %
which contributes to $\U$.  An end
of this type contributes to the product $\h\Q_{\zeta}^{\rm web}\U$. }
 \label{wends}
\end{figure}

\def\IY{{\Bbb Y}}
\def\IW{{\Bbb W}}
As in eqn. (\ref{suffo}), $\U$ induces a map on cohomology, in other words a linear transformation
$\h\U:\H_{\B,\B'}\to \H_{\B,\B'}^{\rm web}$, because it obeys (up to sign)
\begin{equation}\label{mofoxp} \h\Q_{\zeta}^{\rm web}\,\U=\U\h\Q_{\zeta}.\end{equation}
The proof of this formula proceeds like the proof of eqn. (\ref{suffo}).  We consider the 1-dimensional
moduli space $\M$ of solutions of the $\zeta$-instanton equation
on the region $\IW $, which we require to have asymptotic behavior corresponding to basis vectors $\vert \phi_{\ell} \rangle $ and $|\phi_F^{\rm web}(\infty)\rangle$
(whose fermion numbers now differ by 1, since $\M$ is 1-dimensional).  The moduli space $\M$ in general
has several connected components. As usual the compact components do not affect the following discussion.  Alternatively, a given
component $\M_\alpha$ of $\M$ could instead
be a copy of $\R$, with two ends.  Such ends are of two possible types as illustrated in
Figure \ref{wends}.  One type of end contributes
to the left hand side of eqn. (\ref{mofoxp}), and one contributes to the right hand side.
If a given component $\M_\alpha$ has two ends of the same type, these ends both contribute to the same side of eqn. (\ref{mofoxp}) but
with opposite signs.  If the ends are of opposite types, one end contributes $\pm 1$ to the left hand side of the identity and
one makes an equal contribution to the right hand side.
 After summing the contributions of all
1-dimensional components $\M_\alpha$ of $\M$, we arrive at the identity \eqref{mofoxp}.

The interpolation we have made to define the map $\U$  involved a number of arbitrary choices: the region $\IW$ was not
uniquely determined (we only specified its asymptotic form), the K\"ahler metric on the target space $X$ is supposed to be
irrelevant, and for that matter the K\"ahler metric on $\IW$ should likewise be unimportant, as long as its asymptotic behavior is kept fixed.
To show that the induced map $\h\U$ on cohomology does not depend on the choices, one needs an identity of the same
form as eqn. (\ref{mizzo}).  If $\U_0,\U_1:\IM_{\B,\B'}\to \IM^{\rm web}_{\B,\B'}$ are computed with two different choices
related by homotopy, we need
a map $\EE:\IM_{\B,\B'}\to \IM^{\rm web}_{\B,\B'}$ of fermion number $-1$ obeying
\begin{equation}\label{zonfo}\U_1-\U_0=\h\Q_{\zeta}^{\rm web}\EE-\EE\h\Q_{\zeta}. \end{equation}
This is established by imitating the proof of eqn. (\ref{mizzo}): one interpolates between the choices made to define $\U_0$
and $\U_1$, and then looks at certain 1-dimensional moduli spaces.  These moduli spaces have ends of four types that
correspond to the four terms in eqn. (\ref{zonfo}). The map $\EE$ is defined
by counting exceptional solutions of the $\zeta$-instanton equation that exist only at specific points during the interpolation
from $\U_0$ to $\U_1$.

 \begin{figure}
 \begin{center}
   \includegraphics[width=6in]{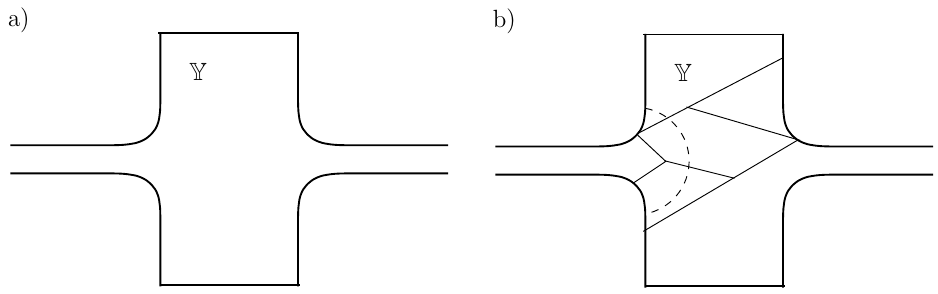}
 \end{center}
\caption{\small   (a)  Since the region $\IY$ is homotopic to an infinite strip, counting of $\zeta$-instantons on $\IY$ gives a map $\mathcal V$ that induces an isomorphism on cohomology.
(b)  Restriction of a $\zeta$-instanton on $\IY$ to the dotted semi-circle reveals a fan of solitons that determines
a basis vector  of $\IM^{\rm web}_{\B,\B'}$.  This implies that the map $\V$ determined by $\zeta$-instantons on $\IY$
can be factored through $\U$, implying that $\U$ also induces an isomorphism on cohomology.}
 \label{hoperegion}
\end{figure}
Finally, we would like to establish that the induced map $\h\U$ on cohomology is an isomorphism.  As in section \ref{whynot},
this is most naturally done by finding an inverse map.  For this, we consider $\zeta$-instantons on the region $\IY$ sketched
in Figure \ref{hoperegion}(a).   A strip coming in from $x=-\infty$ opens out to a very large portion of the right half-plane (compared
to the width of the strip) which is part of
$\IY$, but eventually this is cut off and the region $\IY$ reduces back to a strip for
$x\to +\infty$.  By counting $\zeta$-instantons on $\IY$ that are independent of $x$ for $x\to\pm\infty$, we get a linear
map $\V:\IM_{\B,\B'}\to \IM_{\B,\B'}$.  The induced map on cohomology is the identity.
(If $\IY$ were replaced by a simple product $\R\times I$, rather than a region asymptotic to such a product, then $\V$
would be the identity even before passing to cohomology; a special case of what is explained in the last paragraph is
that the induced map on cohomology is unaffected if $\R\times I$
is replaced by $\IY$.)  On the other hand, the $\zeta$-instantons contributing
to $\V$ have a web-like description, as illustrated in Figure \ref{hoperegion}(b).  If one of these solutions is restricted to the
semi-circle indicated in that figure, it determines a labeled fan of vacua or in other words a basis vector $|\phi_F^{\rm web}(\infty)\rangle$ of
$\IM^{\rm web}_{\B,\B'}$.  Accordingly, the map $\V$ can be factored as $\V'\circ \U$, where $\U:\IM_{\B,\B'}\to \IM^{\rm web}_{\B,\B'}$
was defined above, and $\V'$ is a map in the other direction.  It follows that the map $\h\U$ on cohomology
is an isomorphism.

\subsection{Multiplication}\label{multiplication}

\def\IX{{\Bbb X}}
 \begin{figure}
 \begin{center}
   \includegraphics[width=3.5in]{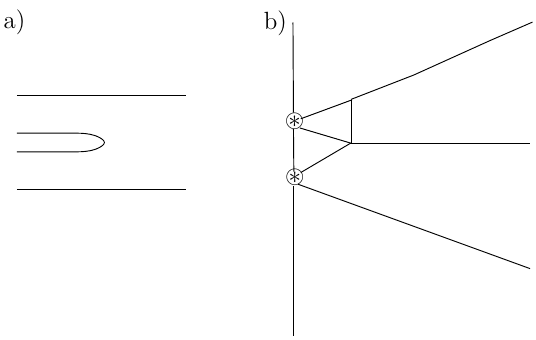}
 \end{center}
\caption{\small   (a)  In the $\sigma$-model approach, the multiplication $\IM_{\B,\B'}\otimes \IM_{\B',\B''}\to \IM_{\B,\B''}$ is defined
by joining open strings.  (b)  In the web-based approach, the multiplication $\Hom(\B,\B')\otimes \Hom(\B',\B'')\to \Hom(\B,\B'')$ is defined by
configurations similar to rigid half-plane $\zeta$-webs but with  abstract vertices (depicted as $\circledast$ in the figure ) at two specified points
on the boundary.}
 \label{joining}
\end{figure}

 \begin{figure}
 \begin{center}
   \includegraphics[width=4.5in]{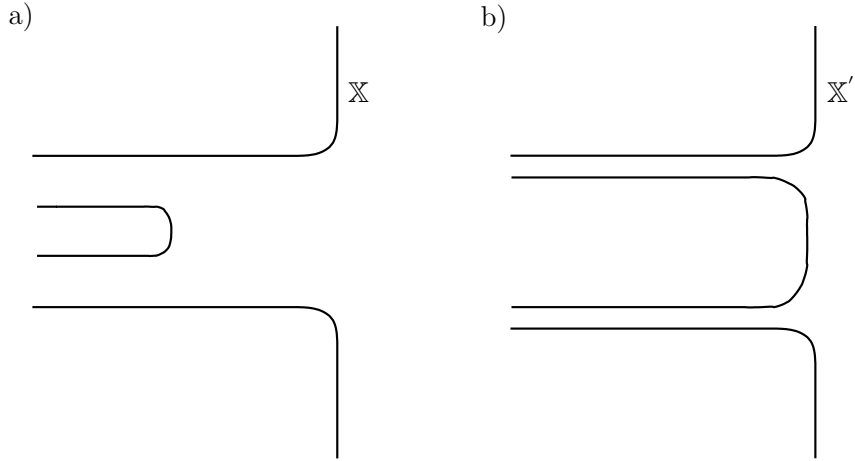}
 \end{center}
\caption{\small   In (a), we multiply two string states within the $\sigma$-model approach, and then apply the map $\U$
 to the web-based category.  The string states come in from the left, are multiplied when the two strings join,
and are mapped to the web picture when the strip widens out into a half-plane. In (b), we first map each state separately to the web-based category, and then multiply
them using the multiplication law of that category.  The two results are equivalent modulo a sum of terms in which the differential $\h\Q_{\zeta}$ or $\h\Q_{\zeta}^{\rm web}$
acts on the initial or final states, because the regions $\IX$ and $\IX'$ of the $s$-plane can be deformed into each other without
altering anything at infinity.
Under this deformation, the  $A$-model action changes by an exact term, which after integration by parts is equivalent  to a sum of contributions in which the differential acts
on external states. In particular, these contributions vanish if we consider the multiplication law on cohomology.}
 \label{comparing}
\end{figure}
Let $\B, \B'$, and $\B''$ be three branes.
In the $\sigma$-model approach,
a string state $x\in \IM_{\B,\B'}$ and a string state $y\in \IM_{\B',\B''}$ are multiplied by joining the two strings (Figure \ref{joining}(a)).  This
operation gives a bilinear map $m_2:\IM_{\B,\B'}\otimes \IM_{\B',\B''}\to \IM_{\B,\B''}$.

On the other hand, there is also a bilinear product $M_2:\IM^{\rm web}_{\B,\B'}\otimes \IM^{\rm web}_{\B'.\B''}\to \IM^{\rm web}_{\B,\B''}$, defined in eqn.  (\ref{eq:BraneMultiplications}).
The definition is as follows.  We recall that an element of $\IM^{\rm web}_{\B,\B'}$ or $\IM^{\rm web}_{\B',\B''}$ represents a half-plane fan of solitons emanating from an abstract
vertex.  To multiply two such fans, we place the two abstract vertices at chosen points on the boundary of the half-plane $\H$  and draw  a picture that
is analogous to a rigid $\zeta$-web  except that two of the vertices are abstract vertices rather than $\zeta$-vertices.
 (Choosing the points at which the two abstract vertices are inserted on the boundary of $\H$ can be understood as dividing by
the translation and scaling symmetries of $\H$.)   This procedure is illustrated in Figure \ref{joining}(b); every line or vertex in this figure,
 except the two abstract vertices
that are denoted as $\circledast$,  represents a boosted $\zeta$-soliton or $\zeta$-instanton solution with  appropriate asymptotics.  Then for
$\delta_1^{\rm web}\in \IM^{\rm web}_{\B,\B'}$, $\delta_2^{\rm web}\in \IM^{\rm web}_{\B',\B''}$, we want to define $M_2(\delta_1^{\rm web}\otimes \delta_2^{\rm web})\in \IM^{\rm web}_{\B,\B''}$.  The definition is as follows: the coefficient with which
a given basis vector of $\IM^{\rm web}_{\B,\B''}$ -- corresponding to an outgoing fan of solitons at infinity -- appears in $M_2(\delta_1^{\rm web}\otimes \delta_2^{\rm web})$ is given
by counting (with signs, as usual) the possible rigid pictures of the type sketched in Figure \ref{joining}(b).

We now should recall that we also have, for any branes $\B_1,\B_2$, the natural map $\U:\IM_{\B_1,\B_2}\to \IM^{\rm web}_{\B_1,\B_2}$.  So given $\delta_1\in \IM_{\B,\B'}$,
$\delta_2\in \IM_{\B',\B''}$, we have two ways to produce an element of $\IM^{\rm web}_{\B,\B''}$:

{\it (i)} We first multiply $\delta_1$ and $\delta_2$ in the $\sigma$-model approach, and use $\U$ to map the result to the web-based category, to get
$\U(m_2(\delta_1\otimes \delta_2))\subset \IM^{\rm web}_{\B,\B''}$.

{\it (ii)} Alternatively, we first map $\delta_1$ and $\delta_2$ to $\U \delta_1$ and $\U \delta_2$ in the web-based category, and then use the multiplication $M_2$
in that category, to get $M_2(\U \delta_1\otimes \U \delta_2)$.

The relation between these two procedures is that there is a map $\EE:\IM_{\B,\B'}\otimes \IM_{\B',\B''}\to \IM^{\rm web}_{\B,\B''}$,
reducing the fermion number by 1,
such that
\begin{equation}\label{zelfox} M_2\circ (\U\otimes \U)-\U\circ m_2=\h\Q_{\zeta}^{\rm web}\EE-\EE\h\Q_{\zeta}. \end{equation}
This formula can be understood by considering Figure \ref{comparing}.  In part (a) of this figure, two open strings join to
a single open string -- corresponding to the operation $m_2$ in the $\sigma$-model approach -- and then the worldsheet
of that string opens out to a half-plane -- giving the map $\U$ to the web-based category.  Solving the $\zeta$-instanton
equation in the region $\IX$ of Figure \ref{comparing}(a), with  asymptotic conditions of the standard type  at the ends, gives
a matrix element of
$\U\circ m_2$.  In part (b),  each open-string worldsheet fans out into a region of a half-plane, with opening
angle $\pi$, to describe the separate action of $\U$ on $\delta_1$ and $\delta_2$ (in other words to describe $\U\otimes \U$),
and then, on a larger length scale, the two regions fit into a single half-plane,
to describe $M_2(\U\otimes \U)$.  Solving the
$\zeta$-instanton equation on the resulting region $\IX'$, with the standard type of asymptotic condition, gives
a matrix element of $M_2\circ (\U\otimes \U)$.   To understand the relation between corresponding matrix elements of
$M_2\circ (\U\otimes \U)$ and of $\U\circ m_2$, we need to compare the counting of $\zeta$-instantons in the regions $\IX$
and $\IX'$ of figures \ref{comparing}(a) and (b).

Those regions are topologically the same, and differ only by a change of metric; moreover, this change of metric is
trivial near infinity.  As in the A-model the stress tensor is $\Q$-exact, so the change in the action resulting from
a change in the metric is
$\Q$-exact.  After integration by parts, a $\Q$-exact term in the action leads to contributions in which the differential
acts on initial or final states; these contributions make up the right hand side of eqn. (\ref{zelfox}).  Thus the origin of the
right hand side of this equation is precisely analogous to the origin of the right hand side of eqn. (\ref{mizzo})
of section \ref{whynot}.

A detailed explanation of the right hand side of eqn. (\ref{zelfox}) -- by
describing moduli spaces of solutions of the $\zeta$-instanton equation, rather than by a quantum
field theory argument using $\Q$-exactness -- would proceed much like the explanation of
eqn. (\ref{zonfo}).  One would pick an interpolation between the two pictures of Figure
\ref{comparing}(a,b), involving a new parameter $u$.  Including $u$ in the description, one would look
at one-parameter families of solutions of the $\zeta$-instanton equations.  Such families would have ends
of four possible types, which would contribute to the four terms in eqn. (\ref{zelfox}).  In particular, $\EE$
would be computed by counting exceptional solutions of the $\zeta$-instanton equation that exist
at special values of $u$.

The map $m_2$ induces a map on cohomology groups, $\h m_2:\H_{\B,\B'}\otimes\H_{\B',\B''}\to\H_{\B,\B''}$,
and likewise $M_2$ induces a map $\h M_2:\H^{\rm web}_{\B,\B'} \otimes\H^{\rm web}_{\B',\B''} \to \H^{\rm web}_{\B,\B''} $.
When we pass to cohomology, the right hand side of eqn. (\ref{zelfox}) drops out and we get
$\h\U\circ \h m_2=\h M_2\circ (\h\U\otimes \h\U)$.  In other words, the isomorphism between the cohomology
groups of the two categories coming from $\h\U$ extends to an isomorphism between the two
multiplication laws.

\subsection{The Higher $A_\infty$ Operations}\label{higherones}

The $\sigma$-model and web-based categories also have higher multiplication
laws, which we have described in sections \ref{seidelfukaya} and \ref{subsec:BraneCat}.
In either case, one is given branes $\B_0,\dots,\B_n$ of class $T_\zeta$ and elements
$\delta_i\in \IM_{\B_{i-1},\B_i}$ or $\delta_i^{\rm web}\in \IM^{\rm web}_{\B_{i-1},\B_i}$, $i=1,\dots,n$ which can be multiplied to get
$m_n(\delta_1\otimes \delta_2\otimes\dots\otimes  \delta_n)\in \IM_{\B_0,\B_n}$ or $M_n(\delta_1^{\rm web}\otimes \delta_2^{\rm web}\otimes \dots\otimes
\delta_n^{\rm web})\in \IM^{\rm web}_{\B_0,\B_n}$.
We want to compare these operations.  First let us summarize the two definitions.

 \begin{figure}
 \begin{center}
   \includegraphics[width=4.5in]{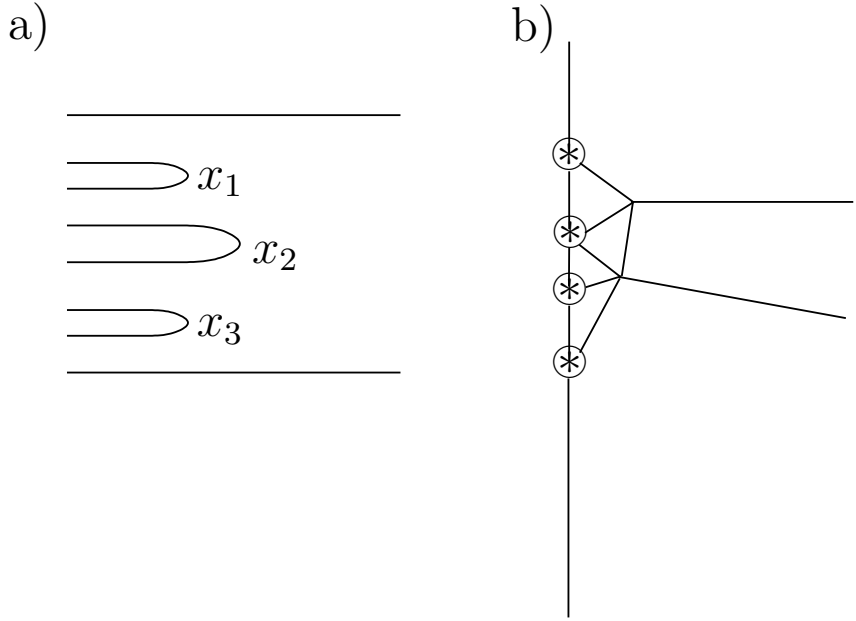}
 \end{center}
\caption{\small  The purpose of this picture is to sketch the higher $A_\infty$ operations
$m_n$ and $M_n$ in the $\sigma$-model approach and in the web-based approach.  (We take $n=4$ for illustration.)
(a)  In the $\sigma$-model approach, $m_n$ is defined by joining $n$ open strings and ``integrating'' over  the differences
between the horizontal positions $x_i$ at which this joining takes place.  (This is simply
Figure \protect\ref{openstrings}(b)
 of \S \protect\ref{seidelfukaya},
 but rotated by an angle $\pi/2$.) (b) In the web-based approach, the multiplication is defined
by counting rigid pictures such as that shown here.  In this picture, there are $n$ abstract vertices labeled by the symbol
$\circledast$.
One can use the scaling
and translation symmetries of the right half plane to place the first and last of them at specified values of $\tau$, say
$1$ and 0;  otherwise, the positions of the   abstract vertices are unspecified.  As usual, in the web-based approach, the abstract
vertices represent morphisms $\delta_i \in \Hop(\B_{i-1},\B_i)$.   All lines and all  other vertices in the figure
represent $\zeta$-solitons or $\zeta$-instantons with appropriate asymptotics. }
 \label{higherjoin}
\end{figure}

In the $\sigma$-model approach,  $m_n$ is defined by joining
together $n$ open strings (via a worldsheet with disc topology) to make a single string; one has to integrate over the $n-2$ real
moduli that are involved in this gluing.
Since we depict strings in the $\sigma$-model approach as propagating from left to right, we depict
this gluing as in Figure \ref{higherjoin}(a).  The moduli in the gluing are the differences between the horizontal positions
$x_i$ of the joining events.
 We should point out, however, that in an $A$-model, ``integrating''
over the moduli has a special meaning.  Amplitudes are defined in the $A$-model by counting (with signs) solutions
of the $\zeta$-instanton equation that have appropriate asymptotics.  If the fermion numbers of external states are such that  the matrix element of $m_n$ that we are trying to compute is not trivially zero, then the expected dimension of the moduli
space $\M$ of $\zeta$-instantons is 0.  Generically,
$\M$ then actually consists of finitely many (nondegenerate) points, which correspond to solutions that  exist only for particular values of the $n-2$
moduli of Figure \ref{higherjoin}(a).  Given this, the ``integration'' that is involved in evaluating a matrix element of $m_n$ reduces to
the sum of finitely many delta-function contributions, corresponding to these solutions.

In the web-based description, the higher multiplication operation $M_n$ of eqn. (\ref{eq:BraneMultiplications}) can be described
as follows.  The elements $\delta_i^{\rm web}$ that we wish to multiply correspond to morphisms attached to the $\circledast$
symbols in Figure \ref{higherjoin}(b).  To compute a matrix element of $M_n$ from $\delta_1^{\rm web}\otimes\dots\otimes \delta_n^{\rm web}$ to
a given fan at infinity, one counts certain pictures of the form shown in the figure. (Recall the morphisms
dictate a particular fan of solitons at each abstract vertex $\circledast$.) In the figure, as usual, all vertices are $\zeta$-vertices,
except that the $\circledast$ symbols
represent abstract vertices.
We only care about pictures like that of Figure \ref{higherjoin}(b)
modulo the translation and scaling symmetries of the right half plane.  We can use those symmetries, for example,
to fix the top and bottom abstract vertices in the figure to specified positions, say $\tau=1$ and $\tau=0$.  However,
the locations of the other abstract vertices are then ``moduli,'' over which we have to integrate.  An integration over this
moduli space is hidden in eqn.  (\ref{eq:BraneMultiplications}), in the following sense.
  If the fermion numbers of the initial and final fans are chosen so that the
matrix element of $M_n$ is not trivially zero, then (for generic superpotential) pictures of the form sketched in Figure \ref{higherjoin}(b)
are rigid up to the translation and scaling symmetries of the right half plane, and each such picture determines the moduli -- the
positions of the abstract vertices --
uniquely.  In eqn. (\ref{eq:BraneMultiplications}), no restriction is placed on the moduli and a given matrix element of $M_n$
is determined by counting (in our present language)  all rigid $\zeta$-webs with specified fans at the abstract vertices and at infinity.
Since the moduli are unspecified, one can think of this procedure as an instruction to integrate over moduli space,
where the integrand has a delta function contribution (with coefficient $\pm 1$) at any point in moduli space at which an
appropriate picture exists.  This ``integration'' is precisely in parallel with the integration in the $\sigma$-model approach to the Fukaya-Seidel category.

 \begin{figure}
 \begin{center}
   \includegraphics[width=5in]{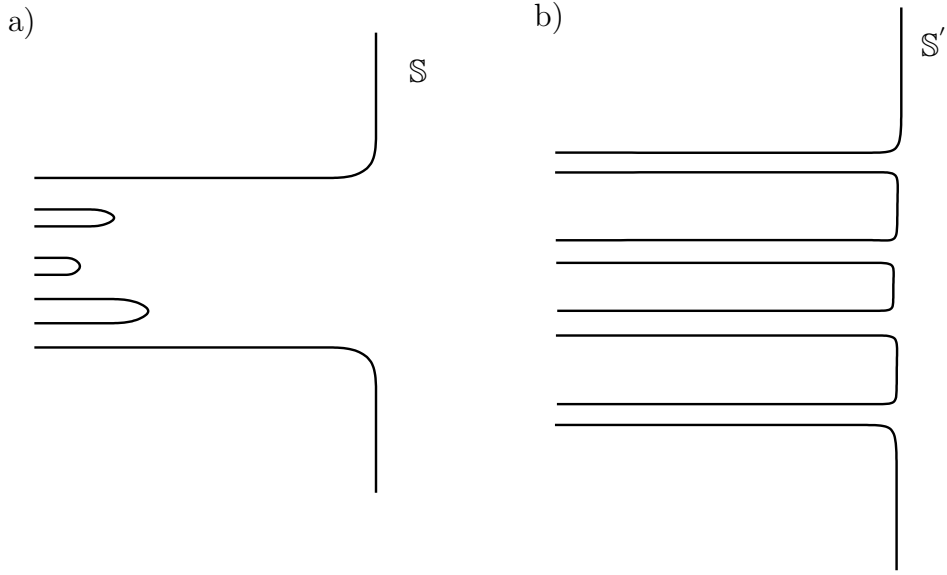}
 \end{center}
\caption{\small   In (a), we sketch the worldsheet that can be used to multiply $n$ open strings in the $\sigma$-model approach
 (sketched here for $n=4$)  and then map them to the web-based category.  To instead map the open strings to the web-based category
before multiplying them, we should use the worldsheet (b).  The topological equivalence between the two pictures
leads to the $A_\infty$ equivalence between the two categories.}
 \label{highercompare}
\end{figure}

Now we would like to compare the result of first multiplying $n$ open strings in the $\sigma$-model approach and then
mapping to the web description, on the one hand, with the result of first mapping to the web description and then multiplying,
on the other hand.
In other words we want to compare $\U\circ m_n$ to $M_n\circ(\U\otimes \dots\U)$.  The basis for the comparison is
shown in Figure \ref{highercompare}, in which part (a) shows the worldsheets that must be used to compute
$\U\circ m_n$, and part (b) shows the worldsheets that must be used to compute
$M_n\circ(\U\otimes \dots\otimes\U)$.  As in section \ref{multiplication}, these worldsheets differ only by having
different K\"ahler metrics (the worldsheets can   actually be identified in such a way that the metrics are the same at infinity).
Since the K\"ahler metric is irrelevant in the $A$-model, in some sense there is an equivalence between
$\U\circ m_n$ and $M_n\circ(\U\otimes \dots\otimes\U)$.  Technically the equivalence is a morphism of $A_\infty$ algebras,
a fact that can be understood as follows.

A change in K\"ahler metric changes the $A$-model action by a $\h\Q_\zeta$-exact term, which can be analyzed by integration by parts on the worldsheet.
In the case of the multiplication law, this integration by parts leads to the terms $\h\Q_{\zeta}^{\rm web}\EE-\EE\h\Q_{\zeta}$ on the right hand side
of (\ref{zelfox}), and certainly a term of this kind also appears in the difference $M_n\circ(\U\otimes \dots\otimes \U)-\U\circ m_n$.
What is new for $n>2$ is that there are also moduli, and $\Q$-exact terms can contribute total derivatives on moduli
space, leading to surface terms on moduli space. Thus, $\CU$ is extended to include maps
$\CU: \IM_{\B_0,\B_1}\otimes \cdots \otimes \IM_{\B_{n-1},\B_n} \to \IM^{\rm web}_{\B_0,\B_n}$
(the map $\EE$ in equation \eqref{zelfox} is already the case $n=2$), and the result of these surface terms is that
there will be extra terms in the identity:

\begin{enumerate}

\item  The region where $x_{k+1}, \dots, x_{k+k'} \ll  x_1,\dots, x_k, x_{k+k'+1},\dots, x_n$
 in Figure \ref{highercompare}(a)  leads to terms
 of the form
\be
\U\left( \delta_1 , \cdots , \delta_{k} ,  m_{k'}(\delta_{k+1} ,  \cdots ,  \delta_{k+k'}),  \delta_{k+k'+1}
  ,  \cdots , \delta_n \right)
\ee
with $k'<n$.

\item  The collision of abstract vertices in  Figure \ref{highercompare}(b) leads to
operations $\U(\delta_i, \dots, \delta_j)$ and terms of the form
\be
M_s\left( \U(\delta_1, \cdots \delta_{k_1}), \U(\delta_{k_1+1}, \cdots \delta_{k_1+k_2}), \dots, \U(\delta_{k_1+k_2+\cdots + k_{s-1}+1}, \cdots \delta_{n}) \right)
\ee

\end{enumerate}

Taking all boundaries into account and recalling that $\h\Q_{\zeta} = m_1$ and $\h \Q^{\rm web} = M_1$ we find
that $\U$ is simply an \afty-functor:
\be
  \sum_s \sum_{{\rm Pa}_s(P)} M_s\left( \U(P_1),\dots,  \U(P_s)\right)
  = \sum_{{\rm Pa}_3(P)} \epsilon_{P_1,P_2,P_3}  \U\left( P_1, m_{p_2} (P_2), P_3 \right)
\ee
where $P=\{\delta_1,\dots, \delta_n\}$ and $p_k = \vert P_k \vert$.

From a physical point of view,
one would simply summarize the above discussion as follows (in the context of the $A$-model):
 An $A_\infty$ algebra or category is a way of describing the tree level
approximation to an open-string theory.  Two open-string theories whose construction differs by $\h\Q_\zeta$-exact terms
should be equivalent.  A map from one open-string theory to another that maps the operations of one to the operations
of the other, modulo $\h\Q_\zeta$-exact terms, should be an equivalence.

\section{Local Observables}\label{sec:RemarksLocalObs}

\subsection{The Need For Unfamiliar Local Observables}\label{need}

The familiar local observables of the $A$-model with target $X$ are associated to the cohomology of $X$.
One way to describe the local operator corresponding to a given cohomology class makes use of a dual homology cycle.
Indeed, if  $H\subset X$ is any oriented submanifold, one defines in the $A$-model a local
operator $\O_H$ as follows.  In general, $A$-model observables on a two-manifold $\Sigma$ are defined by counting holomorphic maps $\Phi:\Sigma\to X$
(or solutions of a corresponding $\zeta$-instanton equation) that obey
suitable conditions.  An insertion of $\O_H$ at a point $p\in \Sigma$ imposes the condition that $\Phi(p)$ must lie in $H$.  The fermion number of $\O_H$ is the codimension of $H$; it is also the degree of the cohomology class dual to $H$.
It is also possible, in a standard way, to define one-form and two-form descendants of $\O_H$.

However, even in the absence of a superpotential, these familiar $A$-model observables are not the whole story.  To see this,
let us consider the $B$-model of $\C^*=\IR\times \t S^1$, where $\t S^1$ is a circle
and we regard $\C^*$  as the quotient of the complex $\tilde \phi$-plane by
$\tilde \phi \sim \tilde \phi+2\pi\i$.
We recall that in general the most familiar local observables of the $B$-model with target $X$ correspond to elements of $H^p(X,\wedge^q TX)$,
where $TX$ is the tangent bundle of $X$. To be more precise, an element of $H^p(X,\wedge^q TX)$ corresponds to a local $B$-model observable
of fermion number $p+q$.  In particular, setting $p=q=0$, we have the holomorphic functions  $\CO_n=e^{n\tilde\phi}$, $n\in\IZ$,
and these correspond to observables of fermion number 0.  $H^0(\C^*,T\C^*)$ is also infinite-dimensional, with sections $\widehat{\CO}_n=e^{n\tilde\phi}\partial_{\tilde\phi}$. These
correspond to observables of fermion number 1.   (This completes the story, since $H^1(\C^*,\wedge^q\T \C^*)=0 $ for all $q$.)

By $T$-duality on the second factor of $\C^*=\IR\times \t S^1$, we can map the $B$-model of $\C^*$ to the $A$-model of a dual $\C^*$, namely $\IR\times S^1$,
where $ S^1$ is the circle dual to the original $\t S^1$.  So the $A$-model of $\C^*$ must also have infinitely many local operators of fermion number 0 or 1.
The standard construction gives only one local operator of degree 0 and one of degree 1, since the cohomology of $\C^*$ is of rank 1 in degree 0 or 1 and
vanishes otherwise.  What are all the other local observables of the $A$-model of $\C^*$?

Going back to the $B$-model, the observables corresponding to $\CO_0$ and $\widehat{\CO}_0$ have
momentum $0$ around the circle in $\C^*=\IR\times \t S^1$,
and they correspond to the standard $A$-model observables, i.e. the cohomology classes mentioned in the first paragraph
of this section.  But the operators $\CO_n$, $\widehat{\CO}_n$
with $n\not=0$ carry momentum around the circle.  Their duals will have to correspond to $A$-model observables with winding number.

What is an $A$-model observable with winding number?  It is a disorder operator, which creates a certain type of singularity in the holomorphic map
(or solution of the $\zeta$-instanton equation) that is counted in the $A$-model.  Let us take the worldsheet $\Sigma$ to be the complex $s$-plane
and let us parametrize the dual $\C^*$ of the $A$-model by a complex variable $\phi$ with $\phi\sim \phi+ 2\pi \i$.  An $A$-model observable of winding number $n$,
inserted at a point $ s=s_0$, imposes a constraint that the holomorphic map $\phi(s)$ should have a singularity at $s\to s_0$
\begin{equation}\label{onog}\phi(s)\sim n \log(s-s_0), \end{equation}
where $n$ is the winding number.  Indeed, this singularity of $\phi(s)$ is characterized by the fact that as $s$ loops around $s_0$, $\phi(s)$ loops $n$ times
around the second factor in $\C^*=\IR\times  S^1$.

Actually, the concept of a disorder operator in the $A$-model, which creates some sort of singularity in a holomorphic map, is much more general than
the concept of an $A$-model observable that carries winding number.  It is not difficult to give a more sophisticated example of a $B$-model with infinitely many observables
such that the dual $A$-model has a target space with finite rank cohomology that moreover has a trivial or  finite fundamental group.
\footnote{A good example is the B-model on Hitchin moduli space in a complex structure other than $\pm I$.
Consider the mirror duals of the characters of holonomies of the flat gauge field parametrized by the moduli space.}
In such a situation, most of the $B$-model observables will correspond to disorder operators of the $A$-model, but these operators cannot be characterized
by their winding numbers.  To describe them requires more insight about what sort of condition should be placed on the singularity of a holomorphic map.

We will not pursue this issue here.  Instead, we will consider a more sophisticated example of an $A$-model that must have local observables of an unfamiliar
sort.  This example will provide a good illustration of ideas we develop later.

Our example is the $B$-model of $\Bbb{CP}^1$.  In this example, $H^0(\Bbb{CP}^1,\O)\cong \IC$; $H^0(\Bbb{CP}^1,T\Bbb{CP}^1)$ is of rank 3 (and corresponds to the Lie algebra
of $SL(2,\IC)$); and $H^1(\Bbb{CP}^1,\wedge^q T\Bbb{CP}^1)=0$, $q=0,1$.  So the space of local observables has rank 1 in fermion number 0 and rank 3 in fermion
number 1.

The $B$-model of $\Bbb{CP}^1$ has a mirror that can be obtained
\cite{Hori:2000kt}  by $T$-duality on the orbits of a $U(1)$ subgroup of the $SL(2,\IC)$ symmetry
group of $\Bbb{CP}^1$.
This dual has target space $\C^*$, which we again parametrize by a complex variable $\phi$ with $\phi\sim \phi+2\pi \i$.  In the $\sigma$-model,
$\phi$ is promoted to a chiral superfield.  The difference between the mirror of $\IC^*$ and the mirror of $\Bbb{CP}^1$ is that in the latter case,
there is a superpotential
\be\label{twops} W=e^\phi+e^{-\phi}. \ee

The $U(1)$ symmetry of the $B$-model that is used in the $T$-duality is converted to the winding number symmetry of the $A$-model of $\C^*$.
(Except for the Weyl symmetry $\phi\leftrightarrow -\phi$, which will play an important role, the
 rest of the $SL(2,\C)$ symmetry of the $B$-model is not manifest in the $A$-model mirror.)
So let us enumerate the $U(1)$ charges of the $B$-model observables.  This will tell us something about what to expect in the $A$-model.
The observable of the $B$-model of $\Bbb{CP}^1$ of fermion number 0 corresponds to the constant function $1$; it is $U(1)$-invariant, so its dual will
be an $A$-model observable of winding number 0.  The three observables of the $B$-model of $\Bbb{CP}^1$ with fermion number 1 correspond to the Lie
algebra of $SL(2,\C)$, so (if we normalize the $U(1)$ generator to assign charges $\pm 1/2$ in the two-dimensional representation of $SL(2,\C)$)
they have $U(1)$ charges $1,0,-1$.

This then gives our expectations for the $A$-model of $\C^*$ with the superpotential of eqn. (\ref{twops}): there should be one observable of
fermion number 0 and winding number 0, and three observables of fermion number 1 and winding number $-1,0,1$.   We return to this example
after developing the necessary machinery.

We have explained here one reason, based on $T$-duality, that in general closed-string
disorder operators must be considered in the $A$-model with a noncompact target space. This argument could be adapted for
open-string observables, but instead in section \ref{openlocal}, we give a different explanation
of why in the $A$-model with a superpotential, the natural class of local open-string observables includes disorder operators.

\subsection{Local Open-String Observables}\label{openlocal}

 \begin{figure}
 \begin{center}
   \includegraphics[width=3.5in]{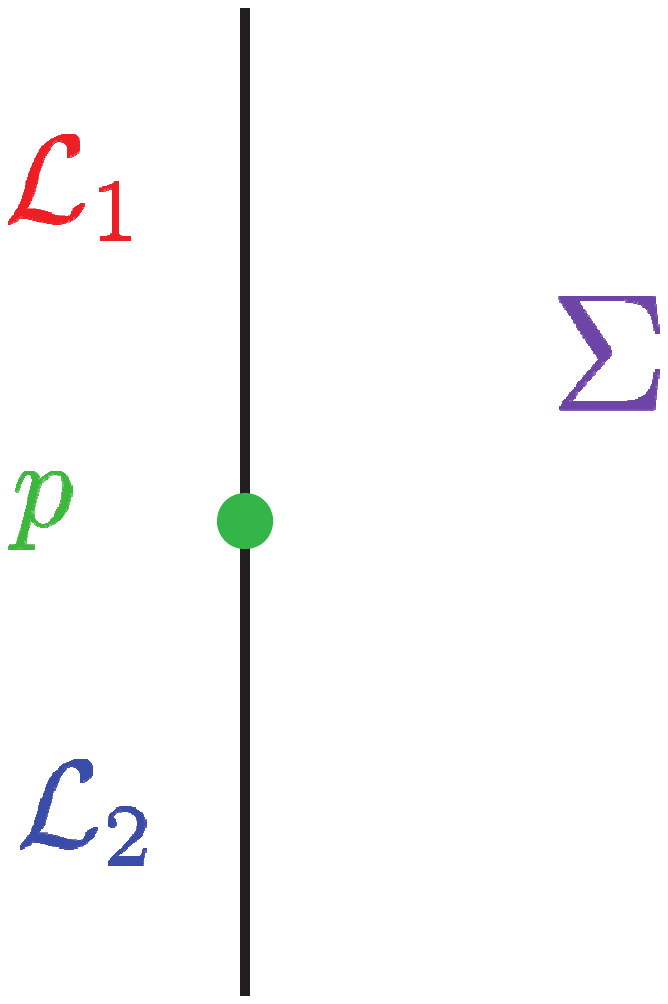}
 \end{center}
\caption{\small   The vertical line is the boundary of a string worldsheet $\Sigma$.  A point $p\in\Sigma$ divides the boundary in portions
that are mapped to two different Lagrangian submanifolds $\L_1$ and $\L_2$, which are the supports of branes $\B_1$ and $\B_2$.}
 \label{Division}
\end{figure}
Local open-string observables in the presence of a superpotential
can be studied rather directly using the construction already described in  section \ref{morphisms}.  (To adapt this for closed strings will require
some further ideas that are explained in section \ref{closedstrings}.)

An open-string observable is inserted at a point $p$ in the boundary of a string world-sheet $\Sigma$.  The point $p$ locally divides the boundary into
two pieces which in general are labeled by branes $\B_1$ and $\B_2$ that are supported on Lagrangian submanifolds $\L_1$ and $\L_2$ (Figure \ref{Division}).
In the ordinary $A$-model with compact target space $X$, the space of local operators that can be inserted at $p$
is the same as the space of $(\L_1,\L_2)$ strings.  To argue that this is also true in the presence of a superpotential, we return to Figure \ref{duplex}(a)
of section \ref{morphisms}, where a strip comes in from the left of the $s$-plane and attaches to the right half-plane.  In the limit that the strip
is very narrow, the incoming string can be replaced by a local operator of some sort inserted at a point on the boundary, as in Figure \ref{Division}.
However, there is some subtlety in describing this local operator.

$A$-model observables are computed by summing contributions of solutions of the $\zeta$-instanton equation (or simply the equation for a holomorphic
map if the superpotential vanishes).  Thus, whatever we compute in the picture of Figure \ref{duplex}(a) involves summing over solutions of the $\zeta$-instanton
equation in the geometry $\Bbb{W}$ of that picture. In the limit that the strip that comes in from the left in Figure \ref{duplex}(a) is very narrow,
$\Bbb{W}$ can be replaced by the right half-plane $\Sigma$ of Figure \ref{Division} and a solution of the $\zeta$-instanton equation on $\Bbb{W}$ converges
to a solution on $\Sigma$.  However, the limiting solution on $\Sigma$ might have a singularity at $p$.

If the limiting solution has no singularity at $p$, we call the local operator $\O$ that is inserted at $p$ an order operator.  In this case, the role of $\O$ in
the $A$-model is to put a constraint on the solution of the $\zeta$-instanton equation.  We will explain this more precisely in a moment.  Alternatively,
the limiting solution might have a singularity at $p$.  In that case, we call $\O$ a disorder operator.  Such a disorder operator
is characterized by the precise type of singularity
that occurs at $p$; this is determined by the choice of the open-string state that comes in from the left in the original description on $\Bbb{W}$.

In the conventional $A$-model with compact target space (and therefore no superpotential),
all  local open-string (and closed-string) observables are order operators.
We will recall why this is true and explain  why it is not true in the presence of a superpotential.  More fundamentally, we will see that in the presence of a superpotential,
 the distinction between order and disorder
operators is not really natural for open-string observables of the $A$-model, in the sense that it is not invariant under independent
Hamiltonian symplectomorphisms of $\L_1$ and $\L_2$.   This fact  will provide  useful background for our study of the closed-string case
in section \ref{closedstrings}.

Let us look at Figure \ref{Division} and ask what an order operator might be.  In doing this, we will assume for simplicity that
$\L_1$ and $\L_2$ intersect transversely at finitely many points $r_1,\dots,r_k$.  We can expect to reduce to this case\footnote{We can expect to do this even if
$\B_1=\B_2$.  This case arises in mathematical work on the Arnold-Givental conjecture, and its special case, the Arnold conjecture.  The work of A. Floer
on the Arnold conjecture in the 1980's was one of the original mathematical applications of the $A$-model.} by applying suitable Hamiltonian symplectomorphisms
to $\L_1$ and $\L_2$. What can a solution of the $\zeta$-instanton equation on the half-plane $\Sigma$
look like in this situation, assuming that it has no singularity at $p$?
The upper half of the boundary is mapped to $\L_1$ and the lower half to $\L_2$.  So the point $p$ must be mapped to one of the intersection points $r_1,\dots,
r_k\in\L_1\cap \L_2$.  For each such point, we can define a local operator $\O_{r_k}$ that imposes a constraint that the point $p$ must be mapped
to $r_k$.  It is not true necessarily true that the $\O_{r_k}$ are $A$-model observables, because there might be a nontrivial differential acting on the space
spanned by the $\O_{r_k}$.  (This will happen, in particular,  if $\L_1$ and $\L_2$ were chosen to have unnecessary intersections that could be deformed
away.)   Rather, one can define a complex with basis given by the $\O_{r_k}$ such
that the cohomology of this complex is the space of local $A$-model open-string
observables.

To see that all $A$-model open-string observables in the absence of a superpotential are order operators, let us just ask what is the space of $(\B_1,\B_2)$
strings in the ordinary $A$-model without a superpotential. A classical zero energy state of an open-string, with the left endpoint mapped to $\L_1$ and
the right endpoint mapped to $\L_2$, is given by a constant map of the string to one of the intersection points $r_k\in \L_1\cap \L_2$.   So the MSW
complex of the open strings has a basis corresponding to the $r_k$.  The differential acting on this complex is the same as the differential acting on the
corresponding space of operators $\O_{r_k}$. This follows by
  a standard argument involving a conformal mapping from a strip $1\geq \mathrm{Im}\,s \geq 0$ in the complex
$s$-plane,
on which one defines the MSW complex, to the half-plane $\Sigma$ of Figure \ref{Division}.  In this mapping, the left end of the strip (where an initial string
state comes in from $\mathrm{Re}\,s=-\infty$, as in Figure \ref{duplex}(a)) is mapped to a boundary point $p$ at which a local operator is inserted.

The reason that this argument does not apply in the presence of a superpotential is that
when there is a nontrivial superpotential the MSW complex does not
have a basis corresponding to the intersections of $\L_1$ and $\L_2$.  Rather, as we explained in section \ref{noncompact}, it has a basis
corresponding to intersection points of $\L_1^w$ with $\L_2$, where $w$ is the width of the strip on which we quantize
and $\L_1^w$ is obtained from $\L_1$ by Hamiltonian flow for ``time'' $w$ with
Hamiltonian $\frac{1}{2}\mathrm{Re}\,((\i\zeta)^{-1}W)$. The reason for the ``extra'' factor of $-\i$ is that
far down the funnel of Figure \ref{duplex}(a) the relevant $\zeta$-instantons are
(to exponentially good accuracy)  $x$-independent, and therefore solve the soliton
equation in the $\tau$-direction with phase $\zeta \to - \i \zeta$.
%
%Thus $\L_1^w=
%\varphi(\L_1)$ for some Hamiltonian symplectomorphism $\varphi$.
%
There is no simple correspondence in general between $\L_1^w\cap \L_2$ and $\L_1\cap \L_2$.

In order to define a local operator we need to take the limit $w\to 0$ of the solution to the
$\zeta$-instanton equation in the geometry of Figure \ref{duplex}(a). We consider
$(\B_1,\B_2)$ strings such that the corresponding supports
$\L_1,\L_2$ are both of class $T_\zeta$, as defined in
Section \S \ref{morebranes} above. As long as $w>0$ there will be a finite number of
points in $\L_1^w\cap \L_2$. However, it can happen that as $w\to 0$ some of those
points move off to infinity. If the intersection
point moves off to infinity as $w \to 0$ then the
limiting solution $\phi_{\rm lim}$ on the positive half-plane must be
such that $ \phi_{\rm lim}(q) \to \infty$ as $q\to p$. Such solutions correspond to disorder operators.
The remaining points in the intersection $\L_1^w\cap \L_2$ will have smooth limits
to intersection points in $\L_1 \cap \L_2$. Such solutions correspond to order operators,
as in the standard $A$-model without superpotential.  It can well happen that $\L_1 \cap \L_2= \emptyset$
and yet $\L_1^w \cap \L_2$ is nonempty for all positive $w$.
(For an example, consider the two Lagrangians $v=u\pm \epsilon$ in the upper left region
of Figure \ref{fig:XKappa-Regions}.) In this case all the
local operators in $\Hop(\fB_1,\fB_2)$ are disorder operators.

We have just explained that the local operators in $\Hop(\fB_1,\fB_2)$ will,
in general, consist of both order and disorder operators. Actually,
this distinction is not invariant under
Hamiltonian symplectomorphisms applied separately to $\L_1$ and $\L_2$.
The relevant symplectomorphisms $\varphi$ in this case are those
 that preserve the region $X_{\zeta}$ (and which are isotopic to the identity).
Recall from Section \S \ref{morebranes} that $X_\zeta$ is  the preimage
under $W$ of the rectangle $T_\zeta$. As we saw in the example of
Figure \ref{Lagrangians}, such symplectomorphisms can map a pair
$\L_1, \L_2$ with nontrivial intersections to a pair $\varphi(\L_1), \L_2$
with no intersections. Hence, while there is a well-defined space
$\Hop(\fB_1,\fB_2)$ of boundary-condition-changing local operators,
there is no invariant distinction between which operators are
order operators and which operators are disorder operators.
This fact will be useful background for our study of the closed strings.
In that study, we will consider all local operators together, and it will not be very apparent
which are order operators and which are disorder operators.

\subsection{Closed-String Observables}\label{closedstrings}

\subsubsection{Twisted Closed-String States}\label{twclosed}

For closed-string observables, there is an essentially new ingredient.  We associated open-string observables to ordinary open-string states,
but closed-string observables will be associated to what one might call twisted closed-string states.

 \begin{figure}
 \begin{center}
   \includegraphics[width=3.5in]{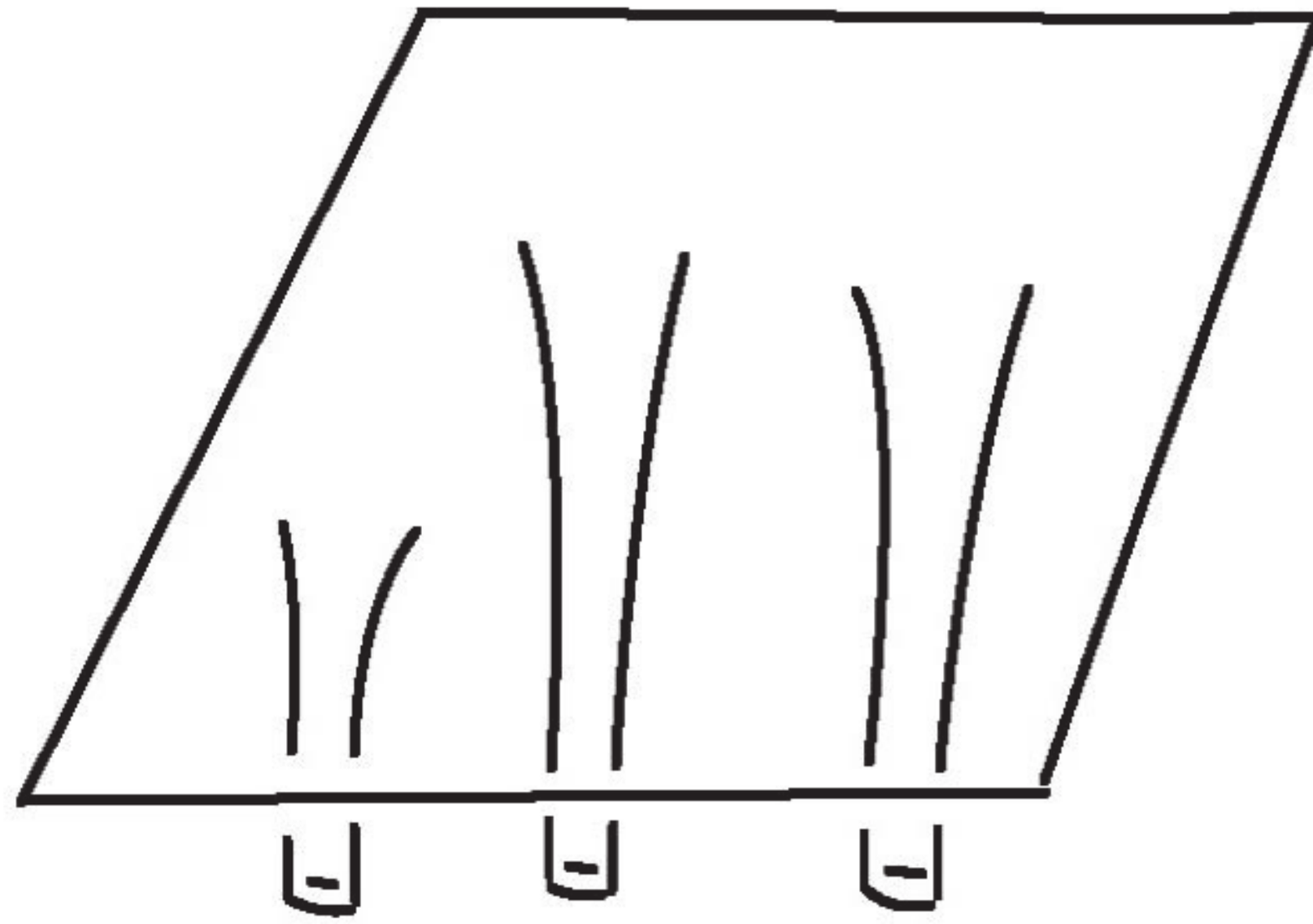}
 \end{center}
\caption{\small   Three tubes have been attached to the complex $s$-plane.  Closed-string states will propagate in from these tubes.  In the limit
that the circumference of the tubes goes to zero, these closed-string insertions will converge to local closed-string operators of the $A$-model. }
 \label{Tubers}
\end{figure}
By analogy with Figure \ref{duplex}(a), which was the starting point to define the map from open-string states to the corresponding local observables,
we can start by considering a worldsheet $\Sigma$ -- we will take it to be the complex $s$-plane -- to which semi-infinite tubes are attached (Figure \ref{Tubers}).
Closed-string states of some kind will be inserted at the ends of these tubes.  In the limit that the tubes shrink to zero circumference, the tubes and the incoming
closed-string states can be replaced by local operator insertions.  The question is what kind of closed-string states we will have to use in this process.

The complex $s$-plane $\Sigma$ with the tubes attached is conformally equivalent to the $s$-plane with finitely many points $s_1,\dots,s_k$ removed.
However, as the $A$-model with a superpotential is not conformally-invariant, we have to specify a Kahler metric on the punctured $s$-plane.  We want
this metric to be tubelike, so we choose something like
\be\label{zomm}\d \ell^2=|\d s|^2\left(1+\sum_{i=1}^k\frac{f_i^2}{|s-s_i|^2}\right) \ee
with small positive constants $f_i$.

%
%\cg{Perhaps rewrite this paragraph a bit. March 4, 2015. }
%
To formulate the $A$-model with superpotential on the punctured $s$-plane, we need to study the $\zeta$-instanton equation.  It will
have the form
\be\label{rom}\bar\partial\phi^I-\bar{\xi} \frac{\i\zeta}{4} g^{I\bar J}\frac{\partial\bar W}{\partial \bar\phi^{\bar J}}=0, \ee
where $\bar{\xi}$ will be a $(0,1)$-form that should be everywhere non-zero. In fact $\bar{\xi}$ should be such that
$\vol:=\frac{\i}{2}\xi \wedge \bar \xi$ is the positive $(1,1)$ form associated with the metric $d\ell^2$ (where
$\xi$ is the complex conjugate of $\bar \xi$).
Indeed, let $\bar \CE^{I}$ denote the left-hand-side of \eqref{rom}. Then one can show that on
any Riemann surface $\Sigma$ with boundary:
\be\label{eq:genaction}
\begin{split}
-2\i \int_{\Sigma} g_{I\bar J} \bar \CE^{I} \CE^{\bar J} & = \int_{\Sigma} g_{I\bar J} d\phi^I * d \bar\phi^{\bar J} +
\frac{1}{4} \vol g^{I\bar J}\p_I W \p_{\bar J} \bar W - 2\int_{\Sigma} \phi^*(\omega) \\
& - \int_{\Sigma} \Re\left(\zeta \bar W \p \bar \xi\right) + 2 \oint_{\p \Sigma} \Re\left(\zeta \bar W \bar \xi\right)\
\end{split}
\ee
where the Hodge $*$ in line one is computed with the metric whose $(1,1)$-form is $\vol = \frac{\i}{2} \xi \bar\xi$.

%In the absence of the tubes, the standard choice is
%$\bar{\xi} =\d\bar s$, and one might be tempted to make the same choice on the punctured $s$-plane.  This, however, is not a good choice
%because the norm of $\bar{\xi}$ computed with the Kahler metric (\ref{zomm}) vanishes for $s\to s_i$.  So if we take $\bar{\xi}=\d\bar s$,
%the superpotential is effectively turned off as we go down the tube, and we would be trying to couple closed-string states without a superpotential
%to the $A$-model with a superpotential.  This might make sense, but it cannot be the whole story, as we know from section \ref{need}.
%
%To ensure that the closed-string states at the ends of the tubes {\it do} know about the superpotential, we have to choose $\bar{\xi}$
%to have a constant norm in the tube metric (\ref{zomm}) as $s\to s_i$.  This condition would be satisfied
%(in the limit) by, for example,
%$\bar{\xi}=\d\bar  s ( 1+\sum_{i=1}^k \frac{f_i}{\bar s-\bar s_i} )$,
%but this expression has unwanted zeroes away from the punctures.  To avoid any zeroes, we take

Returning to the metric \eqref{zomm}, a  simple choice for $\bar{\xi}$ is:
\be\label{zommi}\bar{\xi}=\d\bar  s\left( 1+\sum_{i=1}^k\frac{f_i^2}{|s-s_i|^2} \right)^{1/2}. \ee
This expression has norm one near the punctures and is free of zeroes.  But it leads to ``twisted closed-string states,''
as we now explain.

To study the closed-string state that is inserted at $s=s_j$,
we set $z=\i\log (s-s_j)$, and write
\be\label{huffing}z=x+\i\tau,~~~0\leq x\leq 2\pi, \ee
so
\be\label{puffing}s-s_j=\exp(-\i z)=e^\tau e^{-\i x}.\ee
Conventions have been chosen to match with Section \S \ref{lgassuper}, with $x$ and $\tau$ understood as ``space'' and ``time'' coordinates
for the closed string and with Section \S \ref{sec:CatTransSmpl} for wedge webs.
In the limit that $s \to s_i$, that is, such that $\tau \to -\infty$, the   $\zeta$-instanton equation (\ref{rom}) becomes
\be\label{romm}\frac{\partial \phi^I}{\partial\bar z}= -\frac{\zeta e^{\i x}}{4}f_i g^{I\bar J}\frac{\partial\bar W}{\partial \bar\phi^{\bar J}}. \ee
Apart from a factor of $f_i$ this
  differs from the standard form of the $\zeta$-instanton equation  on the flat $z$-plane (i.e., with the substitution $s\to z$ in
eqn. (\ref{eq:instanton})) in one important way:  $\zeta$ is replaced by $\zeta_{\mathrm{eff}}=\i
\zeta e^{\i x}$.  Thus, effectively all values
of $\zeta$ are realized at a unique point along the circle parametrized by $x=\mathrm{Re}\,z$.

This twisted version of the $\zeta$-instanton equation describes gradient flow in the $\tau=\mathrm{Im}\,z$ direction,
with a superpotential that is the obvious generalization of eqn. (\ref{defho}):
\be\label{nefho}
h := - \int_0^{2\pi}\d x  \left(  \lambda_a \frac{\d u^a}{\d x} - \half \Re\left(-\i   \zeta^{-1}e^{-\i x}f_i W \right)\right).
\ee
We call this a ``twisted'' version of the usual superpotential, where what is ``twisted'' is $\zeta$ as a function of $x$.

The standard machinery of supersymmetric quantum mechanics and the MSW complex can be used with this superpotential
to determine what we will call the twisted closed-string states.
Twisted closed-string states are the ones that correspond to local
closed-string observables of the $A$-model with a superpotential.
In the absence of a superpotential, the twisted $\zeta$-instanton equation, just like the ordinary one,
would reduce to the equation for
a holomorphic map, and the twisted closed-string states would reduce to the standard closed-string states of the $A$-model.

For the twisted theory to be satisfactory, and to give a space of twisted closed-string states
that has the expected invariances of the $A$-model we need to know that the critical points of $h$ and the flows between them cannot
go to or from infinity as one varies the Kahler metric
of the target space $X$ or of the tube parametrized by $z$.  In later applications,
we need an analogous compactness result for solutions
of the $\zeta$-instanton equation on a more general worldsheet $\Sigma$, such as that of Figure \ref{Tubers}, with tubes attached, and possibly with
boundaries.  A sufficient
condition for such compactness arguments is that
\be\label{zelix} |\d W|^2 \gg |W| \ee
at infinity on $X$.  This condition  is satisfied for most interesting choices of $X$ (and its Kahler metric) and $W$.
To show the relevance of the condition (\ref{zelix}), we refer to equation \eqref{eq:genaction} above.
%
%The  replacement $\zeta\to \i\zeta e^{\i x}$ (or more generally $\d \bar s\to \bar{\xi}$)
%in that equation leads, after the integration by parts that is used in arriving at eqn. (\ref{tofox}), to
%
On the second line there is an extra term proportional to $ W $.  The condition
(\ref{zelix}) means that near infinity on $X$, this extra term is subdominant compared to the $|\d W|^2$ term that
appears in the first line of  equation \eqref{eq:genaction}.
One expects that in a rigorous mathematial theory, this will give the compactness of moduli spaces of twisted $\zeta$-solitons and
$\zeta$-instantons that is needed to justify the reasoning that follows.

\subsubsection{Fans Of Solitons}\label{statefans}

In the above construction, the Kahler metric on the tube is $R^2|\d z|^2$, where for the $j^{th}$ closed-string insertion,
 $R=f_j$.  Though the $\sigma$-model
is not conformally-invariant, as always (given the compactness that was just explained) the cohomology of the MSW
complex does not depend on the Kahler metric.  The connection to local operators arises if $R$ is small.
If $R$ is large, we can get a useful alternative description of the MSW complex.
Almost everywhere along the circle, the $\sigma$-model fields will sit at a critical point $i\in \IV$,
corresponding to one of the vacua of the theory.   An $ij$ soliton involving a transition from vacuum
$i$ to vacuum $j$ can occur at a unique  angular position $x$ along the circle, namely
the position at which the effective value $\zeta_{\mathrm{eff}}=\i\zeta e^{\i x}$ of $\zeta$ coincides with the value
$\zeta_{ji}=-\i (W_j-W_i)/|W_j-W_i|$ (eqn. (\ref{hopeful})) at which an $ij$ soliton can actually occur.  The condition
for this is $x=x_{ji}$ with
\be\label{murf}\frac{W_j-W_i}{|W_j-W_i|} =-\zeta e^{\i x_{ji}}. \ee
Using the relation between the vacuum weights of the web formalism and the
critical values of $W$, which in this case becomes $z_i(x) = \zeta_{\rm eff} \bar W_i$,
we see that the $x_{ji}$ of equation \eqref{murf} correspond precisely to
the binding points of type $ij$ of Section \S \ref{sec:CatTransSmpl}. (See
equation \eqref{eq:Sij-Ray-def}.)

Thus, when the radius of the tube is large compared to $\ell_W$,
a basis of the MSW complex is given by a cyclic fan of solitons, arranged at preferred
locations around the circle.  This might not be a complete surprise, since in the abstract discussion of local operators
in section \ref{sec:LocalOpsWebs}, such a cyclic fan of solitons gave a basis for the complex that was used
to describe local operators.  (Vacua in such a fan were enumerated in clockwise order, which in view of eqn. (\ref{puffing}) means
in the order of increasing $x$.)

The fact that the solitons have a preferred location along the circle has an important implication for the fermion number
of a twisted closed-string state.  Usually, a soliton along the real line has an arbitrary position.  That is so because the appropriate
superpotential (\ref{defh}) does not depend on the position of the soliton.  The position of the soliton is a bosonic
collective coordinate that has two fermionic partners.  Quantization of the fermion zero-modes gives a pair of states
differing in fermion number by 1, and accordingly the quantum soliton really represents such a pair  states.

Things are different for twisted closed-string states.  The appropriate superpotential $h$ of eqn.
(\ref{nefho}) depends on the angular
position $x_*$ of a soliton.  For an $ij$ soliton,  it is stationary at $x_*=x_{ji}$.  Moreover, as a function of $x$,  $h$ has a local maximum
at $x_* =x_{ji}$ if $R$ is large enough. This splits the degeneracy between the two states of the soliton.  In general, in Morse theory, an unstable direction
contributes $+1$ to the fermion number of the quantum state associated to a critical point.  So a twisted closed-string state
in the classical approximation corresponds to a cyclic fan of solitons, with the upper value of the fermion number
chosen for each soliton.  This coincides with the recipe of section \ref{sec:LocalOpsWebs}, once we take
into account \eqref{MCdp}.

To verify that $h$ has a local maximum at $x_*=x_{ji}$, we will literally evaluate $h$ as a function of the assumed position $x_*$ of the soliton
along the circle.  Since the first term in $h$ (namely $-\oint \lambda_a \d u^a$) is rotation-invariant,
we only have to examine the $x_*$-dependence of
\be\label{ziff}\Delta h=\frac{1}{2}\int_0^{2\pi}\d x\, \mathrm{Re}\left(-\i\zeta^{-1} e^{-\i x}W\right). \ee
In this computation, assuming that the circle is very large compared to the internal structure of the soliton, we can ignore that internal
structure and think of the soliton as a discontinuity in the fields, which we assume jump from one vacuum $i\in \IV$ to another vacuum $j\in\IV$
at  $x=x_*$ along the circle.  We work in  an interval $I:x_0\leq x\leq x_1$ that we assume contains only one soliton.
More specifically, we assume the fields are in vacuum $i$ for
$x_0\leq x<x_*$ and in vacuum $j$ for $x_*<x\leq x_1$ for some point $x_*\in I$.
As usual, we also write $W_i$ and $W_j$ for the values of $W$ in vacua $i$ and $j$.
The contribution to $\Delta h$ from the interval $I$ is
\be\label{miff}\Delta h_I=\int_{x_0}^{x_*} \d x \, \mathrm{Re}\left(-\i\zeta^{-1} e^{-\i x}W_i\right) +\int_{x^*}^{x_1}\d x\, \mathrm{Re}\left(-\i\zeta^{-1} e^{-\i x}W_j\right). \ee
The second derivative of this with respect to $x_*$ is
\be\label{wiff}\frac{\partial^2\Delta h_I}{\partial x_*^2}=\mathrm{Re}\left(\zeta^{-1} e^{-\i x_*}(W_j-W_i)\right).\ee
We want to evaluate this at $x_*=x_{ji}$, where $x_{ji}$ satisfies eqn. (\ref{murf}).  We get simply
\be\label{pliff}\left.\frac{\partial^2 \Delta h_I}{\partial x_*^2}\right|_{x_*=x_{ji}}=-|W_j-W_i|<0.\ee
So  $h$ has a local maximum as a function of the assumed position of the soliton, and hence the quantum soliton must be placed in the
state of upper soliton number.

% \begin{figure}
% \begin{center}
%   \includegraphics[width=3.5in]{fig:BULKOBS-eps-converted-to.pdf}
% \end{center}
%\caption{\small   A tube is attached to the complex $s$-plane.  Sketched is a solution of the $\zeta$-instanton equation
%on the $s$-plane with fanlike asymptotics at infinity on the $s$-plane and asymptotic to a constant twisted soliton
%at the end of
%the tube. The solution has no simple description in the tube (unless the radius
%of the tube is large), but asymptotically it is fanlike.  }
% \label{fig:BULKOBS}
%\end{figure}
%

%
\begin{figure}
 \begin{center}
   \includegraphics[width=2.5in]{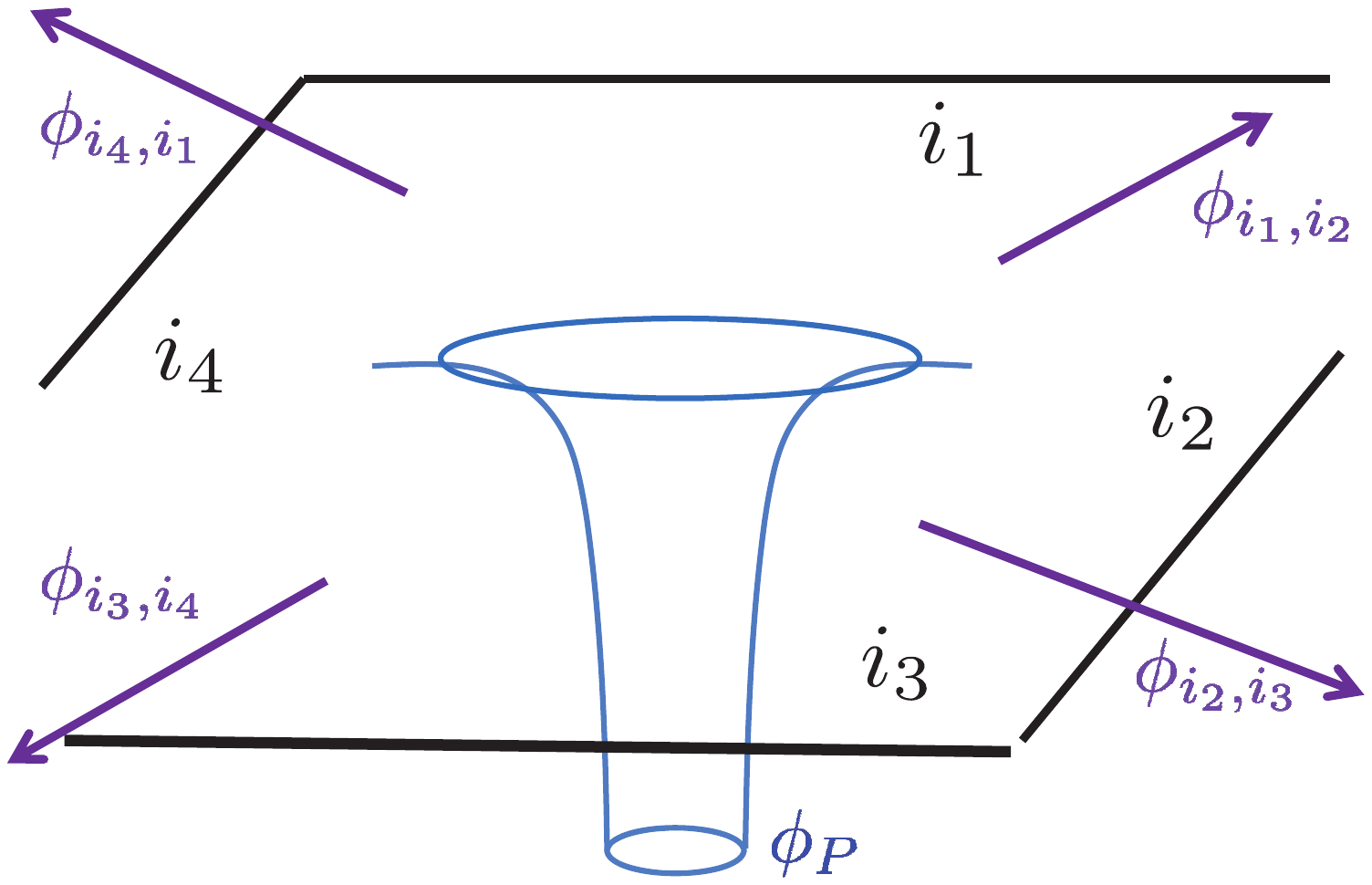}
 \end{center}
\caption{\small   A tube is attached to the complex $s$-plane.  Sketched is a solution of the $\zeta$-instanton equation
on the $s$-plane with fanlike asymptotics at infinity on the $s$-plane and with a solution asymptotic to a  twisted soliton
at the end of
the tube. The solution has no simple description in the tube (unless the radius
of the tube is large), but asymptotically it is fanlike.  }
 \label{fig:BULKOBS}
\end{figure}

\begin{figure}
 \begin{center}
   \includegraphics[width=3.5in]{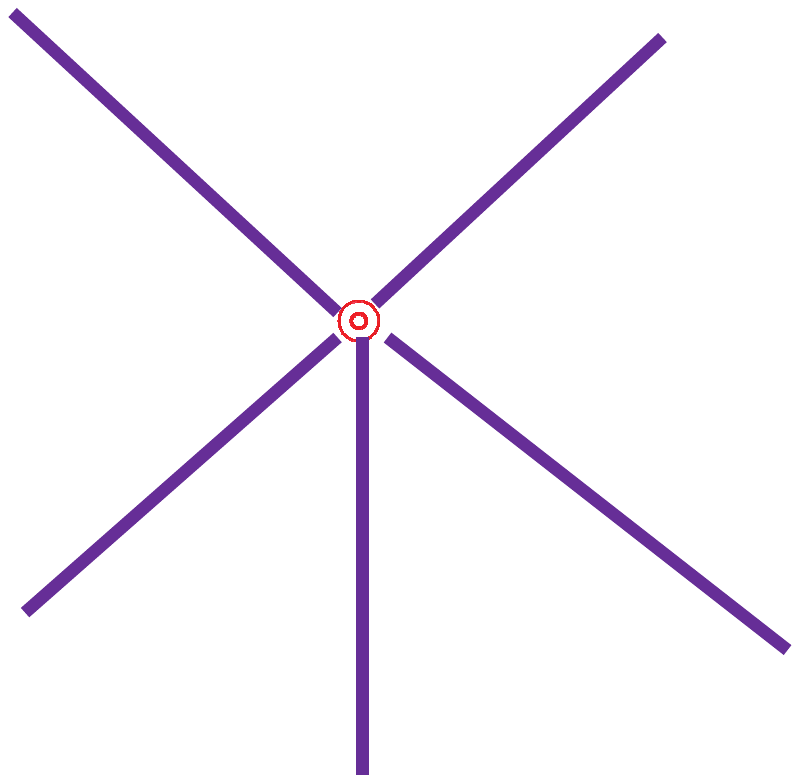}
 \end{center}
\caption{\small A cyclic fan of solitons.  The solitons emanate from a point that has been labeled $\circledcirc$.
This is an abstract symbol that does not necessarily have an interpretation in terms of a solution of the $\zeta$-instanton
equation on the $s$-plane, even a singular one.    }
 \label{FanBasis}
\end{figure}

To map the MSW complex of twisted closed strings to a complex of cyclic fans of solitons, we imitate the
open-string construction of Figure \ref{duplex} and equation \eqref{diflo}.
Thus, we seek to define a chain map
\be\label{eq:Utilmorph}
\tilde\CU: \IM_{S^1} \to \IM^{\rm web}_{S^1}.
\ee
Here $\IM_{S^1}$ is the MSW complex of twisted closed strings, that is,
periodic solutions of the soliton equation \eqref{romm}, and
\be
\IM^{\rm web}_{S^1} = R_c
\ee
where $R_c$ is the complex defined in equation \eqref{eq:LocalOpComplex}: It is the
sum over all cyclic fans of  solitons (where the singleton fan $\{i\}$
corresponds to the vacuum $\phi(x) = \phi_i$). In analogy to equation \eqref{diflo}
we define $\tilde\CU$ by defining its matrix elements using the counting of
$\zeta$-instantons. If $\phi_P(x)$ is a twisted closed-string state, and
$\phi_{\CF}$ is a cyclic fan of solitons then the matrix element
between $\phi_P(x)$ and $\phi_{\CF}$ is given by counting, with signs,
the zero-dimensional components of the moduli space of the solutions
to the following $\zeta$-instanton equation:  We attach a tube to the
$s$-plane as in Figure \ref{fig:BULKOBS}. We consider the $\zeta$-instanton
equation \eqref{rom} with the choice of $\bar\xi$ in \eqref{zommi} (with $k=1$).
The boundary conditions for $s\to s_i$ (i.e. $\tau \to -\infty$ in the
coordinates \eqref{puffing} ) are specified by the choice of twisted closed
string state $\phi_P(x)$, while the boundary conditions for $s\to \infty$
are specified by fan boundary conditions with data $\phi_{\CF}$. We argue
below equation \eqref{zelib} that $\t\U$ is a chain map.

%
%We attach a tube to the $s$-plane (Figure \ref{fig:BULKOBS}) and consider
%a solution of the $\zeta$-instanton equation that is asymptotic to a twisted soliton at the end
%of the tube (i.e. it is asymptotically constant as a function of the
%coordinate $\tau$ defined in equation \eqref{puffing})
%and has fanlike asymptotics at infinity on the $s$-plane. In analogy to the
%discussion below equation \eqref{diflo},  we get a map $\U$
%from the MSW complex of twisted closed strings to a complex that has a basis labeled by closed-string soliton fans,
%that is, if we define $\IM_{S^1}$ to be the MSW complex on the circle with and we recall
%the definition of
%We have defined a map
%
%by defining the matrix elements of $\t \CU$ to be given by counting, with signs, the
%zero-dimensional moduli spaces of $\zeta$-instantons with the boundary conditions specified by the
%in and out vectors.
%

If the radius of the tube is large compared to $\ell_W$, a twisted closed-string state in the tube is itself labeled by a fan of solitons, as explained above. The MSW differential $\h\Q^{\rm MSW}_{\zeta}$ itself can be represented by counting webs on the large cylinder.
The solution in Figure \ref{fig:BULKOBS} has a simple description everywhere in terms of solitons:
the solitons are located at fixed positions in the tube, and when the tube opens up to the $s$-plane,
they remain at the corresponding angular positions in the $s$-plane.

On the other hand, to understand local closed-string operators, we want to consider the tube to have a small circumference.  A solution
of the $\zeta$-instanton equation in this situation can then  be represented by a fan only at large distances on the $s$-plane.
We consider all such fans and label the vertex from which a fan emanates as $\circledcirc$
%$\odot$  \textcircled{$\bullet$}
%
%\cg{Need to replace \textcircled{$\bullet$} by $\circledcirc$ below.}
%
as shown in (Figure \ref{FanBasis}).
We think of this
as an abstract vertex analogous to the abstract vertices that were labeled $\circledast$ in the open-string discussion.

The abstract vertex does not necessarily represent a $\zeta$-instanton solution on the $s$-plane, not even a singular one.
The reason for this statement is that
 the map $\t\U$ from the twisted MSW complex to the set of cyclic fans is not necessarily injective or surjective
if the tube has a small circumference.  The only general statement, which we will justify in a moment,
 is that  $\t\U$ induces an isomorphism on
cohomology; there is no simple general statement about how it acts on the complex.   In the limit that the tube in Figure \ref{fig:BULKOBS}
has a small radius the $\zeta$-instanton sketched in that figure plausibly converges to a singular $\zeta$-instanton on the $s$-plane.
But this does not give a natural way to interpret the $\circledcirc$ solution in terms of singular $\zeta$-instantons;
when the radius is small, some cyclic fans of solitons might arise from more than one solution in Figure \ref{fig:BULKOBS}
and some might not arise at all. Just as $\circledast$ is can be viewed as a receptacle for open string local
operators, so too $\circledcirc$ can be viewed as a receptacle for local closed-string operators.

To prove that the map $\t\U$ induces an isomorphism on cohomology, we simply repeat the reasoning from the open
string case.  First we show that
\be\label{zelib}\Q^{\mathrm {web}}_\zeta\t\U=\t\U\h\Q^{\mathrm{MSW}}_\zeta,\ee
where $\h\Q^{\mathrm{MSW}}_\zeta$ is the differential of the MSW complex, and we will recall in a
moment the definition of the corresponding
web differential $\Q^{\mathrm{web}}_\zeta$.  This means that the map $\t\U$ between the two complexes induces a map on
cohomology.  To show eqn. (\ref{zelib}),
we consider 1-parameter families of $\zeta$-instantons in the
geometry of Figure \ref{fig:BULKOBS}.

\begin{figure}
 \begin{center}
   \includegraphics[width=3.5in]{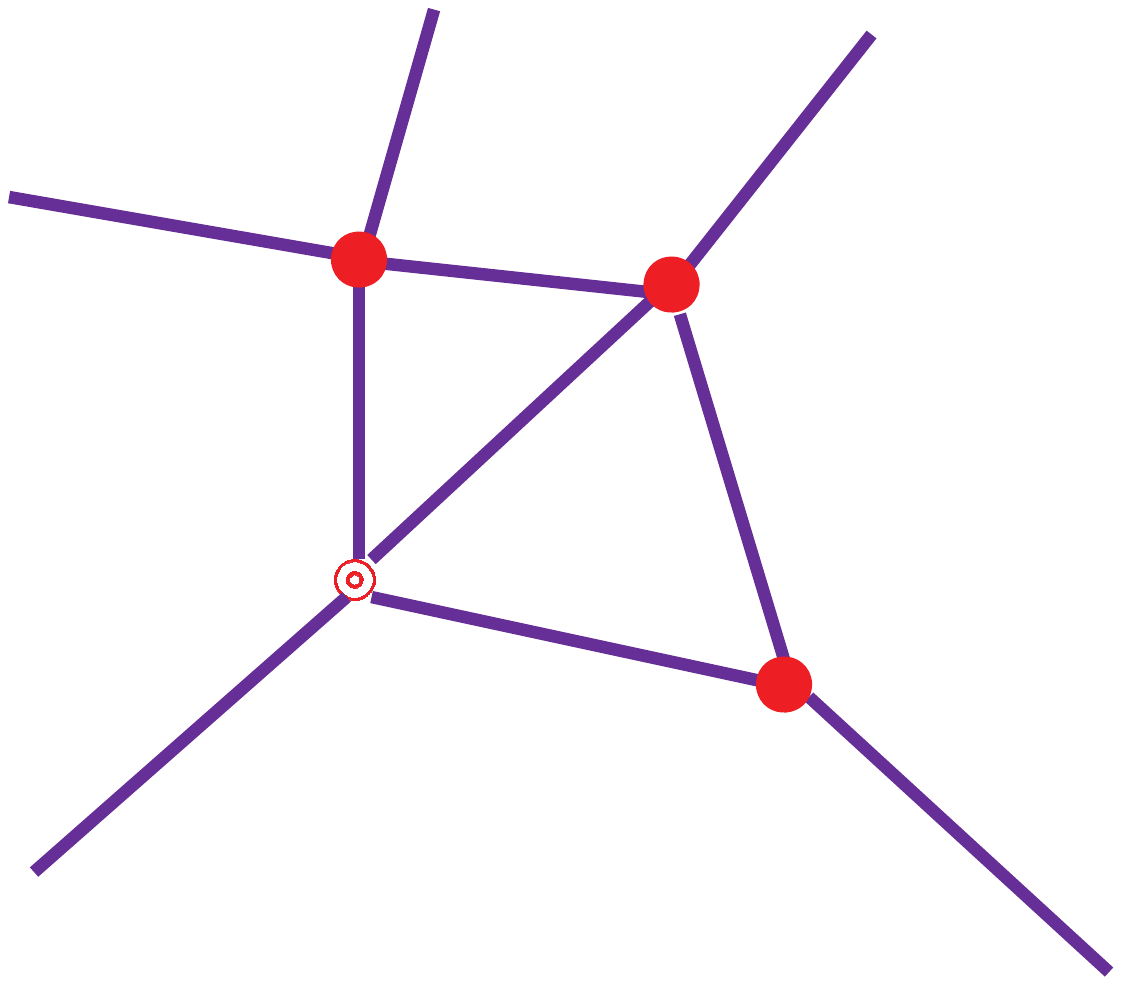}
 \end{center}
\caption{\small A cyclic fan of solitons.  The solitons emanate from a point that has been labeled $\circledcirc$.
This is an abstract symbol that does not necessarily have an interpretation in terms of a solution of the $\zeta$-instanton
equation on the $s$-plane, even a singular one.    }
 \label{SuperFan}
\end{figure}

We now make an argument of a type that should be familiar; accordingly, we will be brief.
The compact components of $\M$ make no contribution. The noncompact components of $\M$ will have
two ends. Each of these ends be one of two types. One type of end arises from convolving a $\zeta$-instanton solution
on the punctured plane with  a tunneling event between
two closed-string solitons far down the tube.   Such a tunneling event represents a contribution to $\h\Q_\zeta^{\mathrm{MSW}}$.
When it occurs far down the tube in the geometry of Figure \ref{fig:BULKOBS}, this gives a contribution
to $\t\U\h\Q^{\mathrm{MSW}}_\zeta$, the right hand side of eqn. (\ref{zelib}).  The other possible type of end of $\M$  arises
from convolving a $\zeta$-instanton solution on the punctured plane with a taut $\zeta$-web, as pictured
in  Figure \ref{SuperFan}.  In this picture, all vertices
except the one labeled $\circledcirc$ are $\zeta$-vertices (i.e. interior amplitudes)
and represent solutions of the $\zeta$-instanton equation with the indicated asymptotics.
(The $\zeta$-web in the figure is taut if its only modulus, keeping fixed the position of the special
vertex $\circledcirc$, is the one derived from scaling.)
The vertex labeled $\circledcirc$ simply represents a cyclic fan of solitons emanating from a $\zeta$-instanton solution
on the punctured plane.  The web differential
$\Q^{\mathrm{web}}_\zeta$ in the abstract treatment of local operators
(described in Section \S \ref{sec:LocalOpsWebs}) was defined as follows.  Its matrix
element from a given ``initial'' fan of solitons to a given ``final'' one is obtained by counting taut webs that interpolate
from the initial fan at the origin to the final fan at infinity.  In this counting, one considers only taut webs and one weights each vertex by the corresponding interior amplitude.
With this definition of $\Q^{\mathrm{web}}_\zeta$, an end of $\M$ of the second type represents a contribution to
$\Q^{\mathrm{web}}_\zeta\t\U$.

A given component $\M$ might have two ends of the same type.  If so, they make canceling contributions to the left
or to the right of eqn. (\ref{zelib}).  A component with ends of opposite types makes equal contributions to the left and to the
right.
As usual, any matrix element of either the left or the right can be interpreted as a sum of ends of one-parameter
families of $\zeta$-instantons.\footnote{This is proved by the usual type of gluing argument.  The convolution of a $\zeta$-instanton
on the punctured plane with a tunneling event far down the tube or a very large $\zeta$-web is exponentially close to a $\zeta$-instanton
solution and can be slightly perturbed to make an exact solution.}  The identity of eqn. (\ref{zelib})
comes from summing the contributions of ends of 1-parameter
families of $\zeta$-instantons.

So $\t\U$ induces a map,   $\h\U$, on cohomology.  The fact that $\h\U$ is an isomorphism is proved by imitating the argument
of Figure \ref{hoperegion}:  one lets a tube fan out into a plane that then closes back into a tube, to give an inverse to $\h\U$.
The fact that $\h\U$ does not depend on arbitrary choices made in the construction
 is shown by an argument similar to the one discussed
in relation to eqn. (\ref{zonfo}).

\subsubsection{More On The Mirror of $\Bbb{CP}^1$}\label{moremirror}

We now return to an example described in section \ref{need}: the $A$-model mirror of the $B$-model of $\Bbb{CP}^1$.
This is a model with a single chiral superfield $\phi$, with an equivalence $\phi\cong \phi+2\pi\i$ so that it parametrizes
$X=\IC^*$, and with superpotential $W(\phi)=e^\phi+e^{-\phi}$.

The Weyl group of the original $SL(2,\IC)$ symmetry of the $\Bbb{CP}^1$ acts by $\phi\to -\phi$.  This symmetry
plays a significant role, so we want to pick a Kahler metric on $X$ that respects this symmetry.  For example, we can
use the flat Kahler metric $\d \ell^2=|\d \phi|^2$.

\begin{figure}
 \begin{center}
   \includegraphics[width=3.5in]{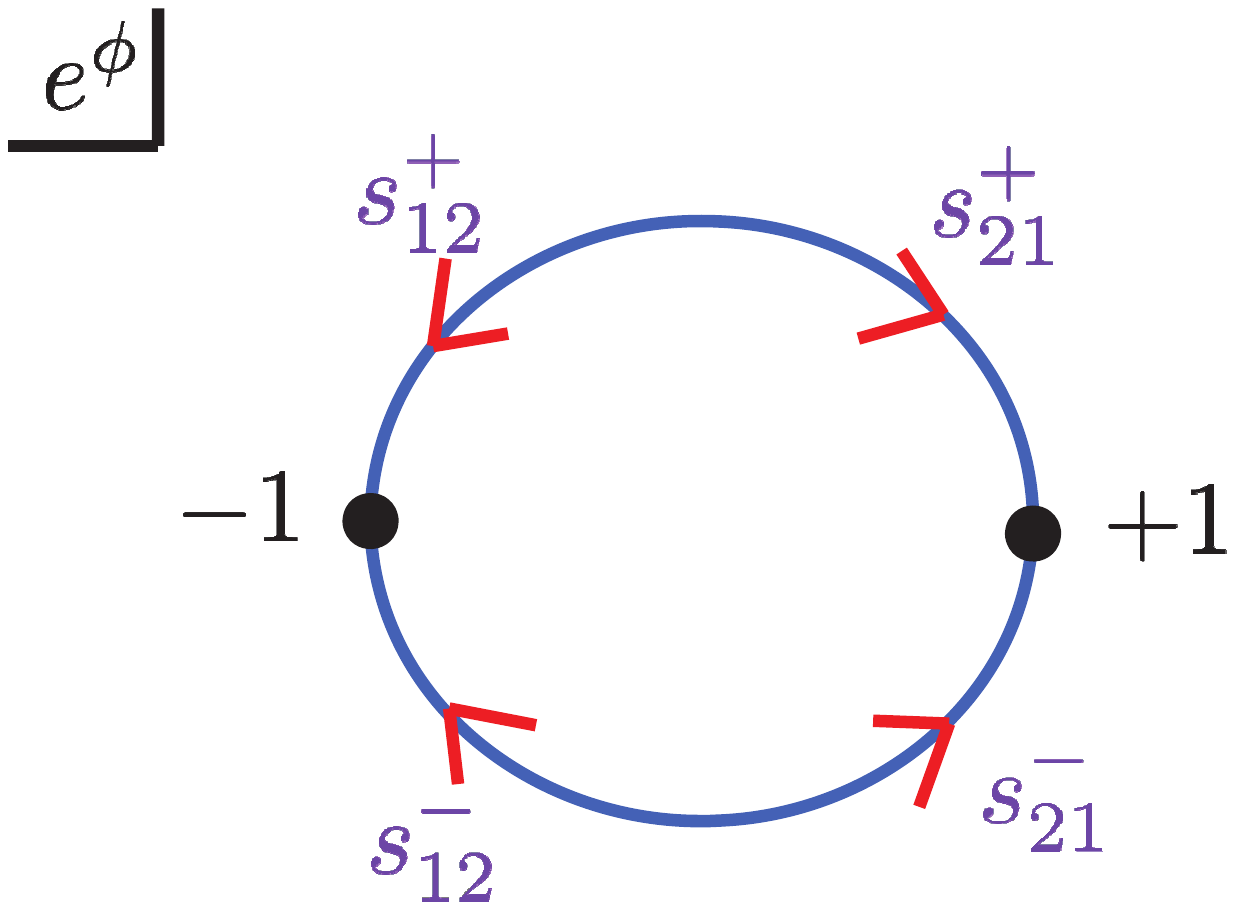}
 \end{center}
\caption{\small  Illustrating the path of $e^{\phi(x)}$ for the four nontrivial
solitons.  }
 \label{fig:FOUR-CP1-SOLITONS}
\end{figure}

The superpotential has two critical points at $\phi=\phi_1=0$ and $\phi=\phi_2=\pi\i$.
 The values of the superpotential at the two critical points are $W_1=2$
and $W_2=-2$.  In particular, $W_1-W_2$ is real, and this means that 12 and 21 solitons only occur when $\zeta=\pm\i$.
In turn this means that $\mathrm{Im}\,W$ will be a conserved quantity in the $\zeta$-soliton equation.
Since $\mathrm{Im}\,W=0$ at each critical point, the soliton trajectory will be one along which $\mathrm{Im}\,W$ is
identically 0.
The locus $\mathrm{Im}\,W=0$ is the union of the real $\phi$ axis, the line $\mathrm{Im}\,\phi=\pi$, and
  the circle $\mathrm{Re}\,\phi=0$.
(This is a circle because $\phi\cong \phi+2\pi\i$.)  A trajectory on the real $\phi$ axis or on the line $\mathrm{Im}\,\phi=\pi$
cannot connect the two critical points at $\phi=0$ and $\phi=\pi \i$, but we can connect them by a trajectory on the
circle $\mathrm{Re}\,\phi=0$.  In fact, we can do this in two different ways: the soliton trajectory can be the ``upper'' segment
$0\leq \mathrm{Im}\,\phi\leq \pi$ or ``lower'' segment $-\pi\leq \mathrm{Im}\,\phi\leq 0$ (in each case with $\mathrm{Re}\,\phi=0$).
The $\zeta$-soliton equation on the real $x$-line
determines a unique parametrization of these segments (as a function of
$x$, and up to an additive constant) to make a 12 or 21 solition.  So there are two 12 solitons, say
$s^+_{12}$ and $s^-_{12}$,
where the superscript refers to the upper or lower soliton trajectory, and likewise two 21 solitons, $s^\pm_{21}$.
See Figure \ref{fig:FOUR-CP1-SOLITONS}.

Now we can enumerate a basis for the MSW complex.  There are two states consisting of trivial fans that simply
sit at critical point 1 or 2 on the whole circle.  And there are four states consisting of a fan of two solitons
$s^\pm_{12}$ and $s^\pm_{21}$ that sit at antipodal points $x_{12},x_{21}$
on the circle, chosen to give the right values of $\zeta_{\mathrm{eff}}$.

We would like to know the winding numbers and fermion numbers of these six states.
For the winding numbers, this is immediate.  The trivial fans have winding number 0.  For the four
fans $s_{12}^\pm s_{21}^{\pm}$,
we observe that two of these four go forwards and then back along the upper or lower half of the circle, producing a net
winding of 0.  The other two states go all the way around the circle in one direction or the other, traversing first the upper
half of the circle and then the lower half, or vice-versa.  These two states have a net winding number of 1 and $-1$,
respectively.

It is almost equally easy to determine the fermion numbers.   The trivial fans correspond to states with fermion number 0.
This is a universal statement about a massive  $A$-model with a superpotential.
A trivial fan corresponds to a closed-string state that lives in a particular vacuum $i\in \IV$.
Since the fermion number current
is a Lorentz vector, its expectation value in any vacuum is $0$ in infinite volume.  In a massive theory, this statement
is not significantly affected by compactification on a very large circle to get a closed-string state, so the state associated
to a trivial fan always has fermion number $0$.

What about the non-trivial fans of solitons?   The fermion number of a fan is the sum of the fermion numbers
of the individual solitons.  The fermion number of an individual solition is an $\eta$-invariant computed from the
non-zero eigenvalues of the Dirac operator, plus a contribution $\pm 1/2$ from the zero-modes.  For the particular
model under study here, with $W=e^\phi+e^{-\phi}$, $\phi$ imaginary in a soliton trajectory, and $\zeta_{\mathrm{eff}}$ imaginary, the
appropriate Dirac operator $\mathcal D$ of eqn. (\ref{eq:NewD}) is odd under complex conjugation
(or alternatively under conjugation with the Pauli matrix $\sigma^1$).  This implies
the spectrum is invariant under $E\to - E$,   and accordingly the
$\eta$-invariant computed from the non-zero modes vanishes.
\footnote{Since the real line on which the soliton is defined is not compact, the $\eta$-invariant is defined not just in terms of
 the eigenvalues of $\mathcal D$ but also requires a regularization, such as the one in eqn. (\ref{zonf}).  With such a regularization,
 the fact that $\mathcal D$ is odd under complex conjugation does imply that the nonzero eigenvalues do not contribute to $\eta(\mathcal D)$.}
Hence each soliton has two states of fermion number
$\pm 1/2$ from quantization of the zero-modes. As explained in section \ref{statefans}, to make a fan we take for each
soliton  the state
of upper fermion number.  In the present context, this means that the solitons all have fermion number $1/2$ and
hence that
each of the four fans $s_{12}^\pm s_{21}^\pm$ has a fermion
number of 1.

In summary, the MSW complex for this problem has a basis consisting of six states, as follows:
(1)  there are two states of fermion number 0 and winding 0; (2) there are four states of fermion number 1 and
winding number $0,0,1,-1$.

Since the differential of the MSW complex commutes with winding number, the two states of fermion number 1
and winding number $\pm 1$ will certainly survive in the cohomology.  However, there could be
a nonzero differential acting on the
states of fermion number 0 and winding number 0, mapping them to fermion number 1 and winding number 0.
From the point of view of the MSW complex on the cylinder, it is not obvious what this differential would be.  However,
this differential was determined in the web-based formalism below equation \eqref{eq:ExtDiff},  and this answer carries over to the
MSW complex, since we identified the two complexes in section \ref{statefans}.  Equation \eqref{eq:ExtDiff} shows that
when there are only two vacua, the two states corresponding to trivial fans, which in that language are $R_1$ and $R_2$,
have up to sign the same non-zero image $R_{12}\otimes R_{21}$ in fermion number 1.  So precisely one linear
combination of states of fermion number 0 survives in the cohomology.  This corresponds to the identity operator
of the field theory.  Dually, the cohomology in fermion number 1 and winding 0 is one-dimensional, generated by any
state of winding number zero that is not in the image of the one-dimensional space of winding number zero
states obtained from the differential acting on $R_1\oplus R_2$.

Accordingly, the cohomology of the MSW complex is as follows: (1) there is one state of fermion number 0 and winding
0; (2) there are three states of fermion number 1 and winding $1,0,-1$.  This agrees with the prediction from mirror symmetry.

\subsubsection{Closed-String Amplitudes}\label{closedamp}

Having come this far, it is not hard to see how to define closed-string amplitudes, making up what mathematically is called
 an $L_\infty$ algebra.  We simply imitate what we did in section \ref{seidelfukaya} for open strings.

\begin{figure}
 \begin{center}
   \includegraphics[width=3.5in]{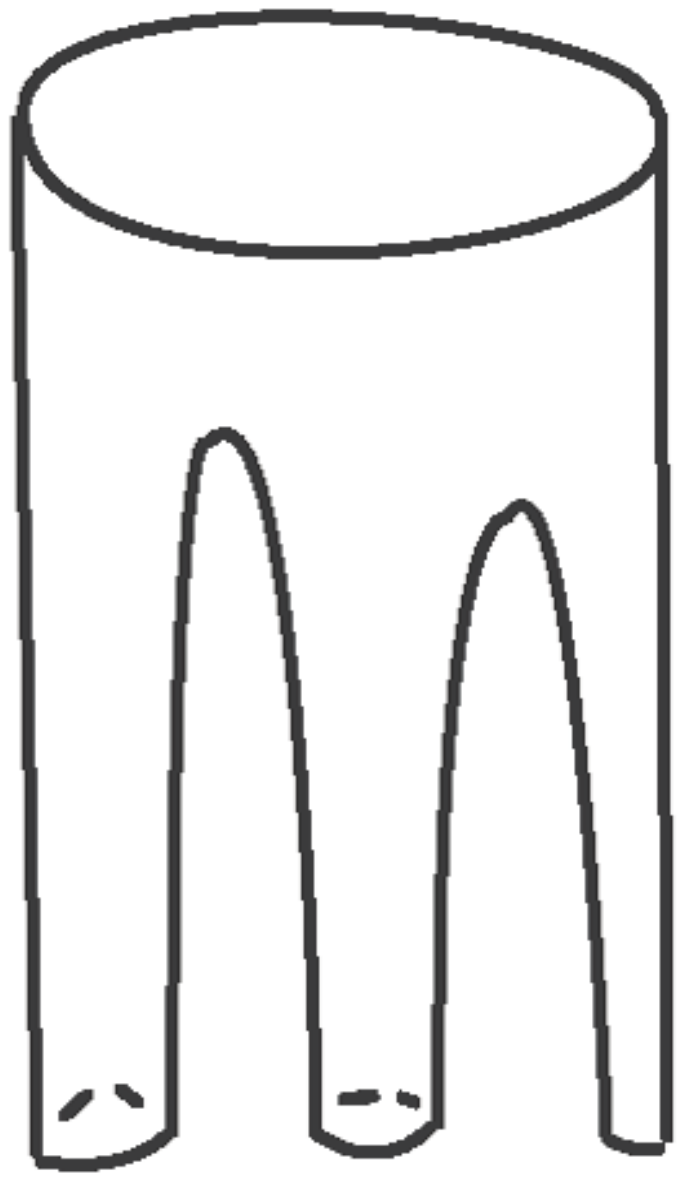}
 \end{center}
\caption{\small $k$ tubes joining to a single one, sketched here for $k=3$.    }
 \label{MultiTubes}
\end{figure}

Imitating the open-string worldsheets of  Figure \ref{openstrings}, we need a suitable family of closed-string worldsheets
that describes a transition from $k$ twisted closed strings to a single twisted closed string.\footnote{In contrast to
the open-string case, it appears difficult to define $k\to m$ amplitudes with $m>1$.  The singularity of $\bar{\xi}$ (eqn. (\ref{zommim}) near
an incoming string at, say, $s_j=0$ is $\bar{\xi}\sim \d\bar s/|s|$. We call this an incoming singularity.
  The singularity near $s=\infty$, in terms of $t=1/s$, is $\bar{\xi}\sim |t|\d\bar t/\bar t^2$.  We call this an outgoing singularity.  For topological
  reasons, on a surface of genus 0,  if $\bar{\xi}$ has no zeroes and has only incoming and outgoing singularities, the number of outgoing singularities is precisely 1,
  though there may be any number of incoming singularities. To see this note that a nonzero $\bar\xi$ defines a
   trivialization of $T^{(0,1),*}\Sigma$, where $\Sigma$ is the surface without the singular points.
   At incoming singularities $\bar \xi$ has zero winding number, and
   the trivialization may be extended over these points. On the other hand, at outgoing
    singularities $\bar \xi$  has a winding number of
   $-2$. The sum of the winding numbers must equal the first Chern class of $T^{(0,1),*}\bar\Sigma$
   where $\bar\Sigma$ is the surface with the singular points filled in.
   This is given by minus the Euler character of $\bar\Sigma$. Therefore there is precisely one outgoing puncture for $g=0$,
   there are no outgoing punctures for $g=1$, and for $g>1$   a $\bar{\xi}$ with the assumed properties does not exist.}    For the necessary
worldsheet $\Sigma$, we simply use the $s$-plane with punctures at $s_1,\dots, s_k$, but we omit the ``1'' from the metric
and from the $(0,1)$-form $\bar{\xi}$.  Thus the metric becomes
\be\label{zommm}\d \ell^2=|\d s|^2\sum_{i=1}^k\frac{f_i^2}{|s-s_i|^2} \ee
and $\bar{\xi}$ becomes
\be\label{zommim}\bar{\xi}=\d\bar  s\left(\sum_{i=1}^k\frac{f_i^2}{|s-s_i|^2} \right)^{1/2}. \ee
The $s$-plane with the metric (\ref{zommm}) is sketched in Figure \ref{MultiTubes}. The regions near $s=s_i$, $i=1,\dots,k$
make up $k$ tubes that we consider ``incoming''; they  join into a single ``outgoing''
tube, represented by the region near $s=\infty$.

We define a $k$-fold product that maps $k$ states in the twisted closed-string MSW complex  to a single one as follows.
We choose $k$ incoming states in the twisted MSW complex of the closed strings, and one such outgoing state.
These $k+1$ states determine the desired asymptotic behavior of a solution on $\Sigma$ of the $\zeta$-instanton
equation.  We count the number of solutions (with signs, as always)
of that equation, with the desired asymptotics.  In the counting,
we use the translation and scaling symmetries of the $s$-plane to set $s_1=0$, $s_2=1$, but we leave $s_3,\dots,s_k$
unspecified.  The number of such solutions, with the stated asymptotics, and any values of $s_3,\dots,s_k$,
gives the chosen matrix element of the $k^{th}$ $L_\infty$ operation.
The procedure is sketched in Figure \ref{MultiTubes}.

What we have stated is the standard procedure
for defining closed-string amplitudes in topological string theory, adapted to a situation in which we need to specify
more carefully what worldsheets should be used, since some of the symmetry is missing.
This procedure nevertheless defines an $L_\infty$ algebra structure on the MSW complex $\IM_{S^1}$
of twisted closed string states.

\begin{figure}
 \begin{center}
   \includegraphics[width=3.5in]{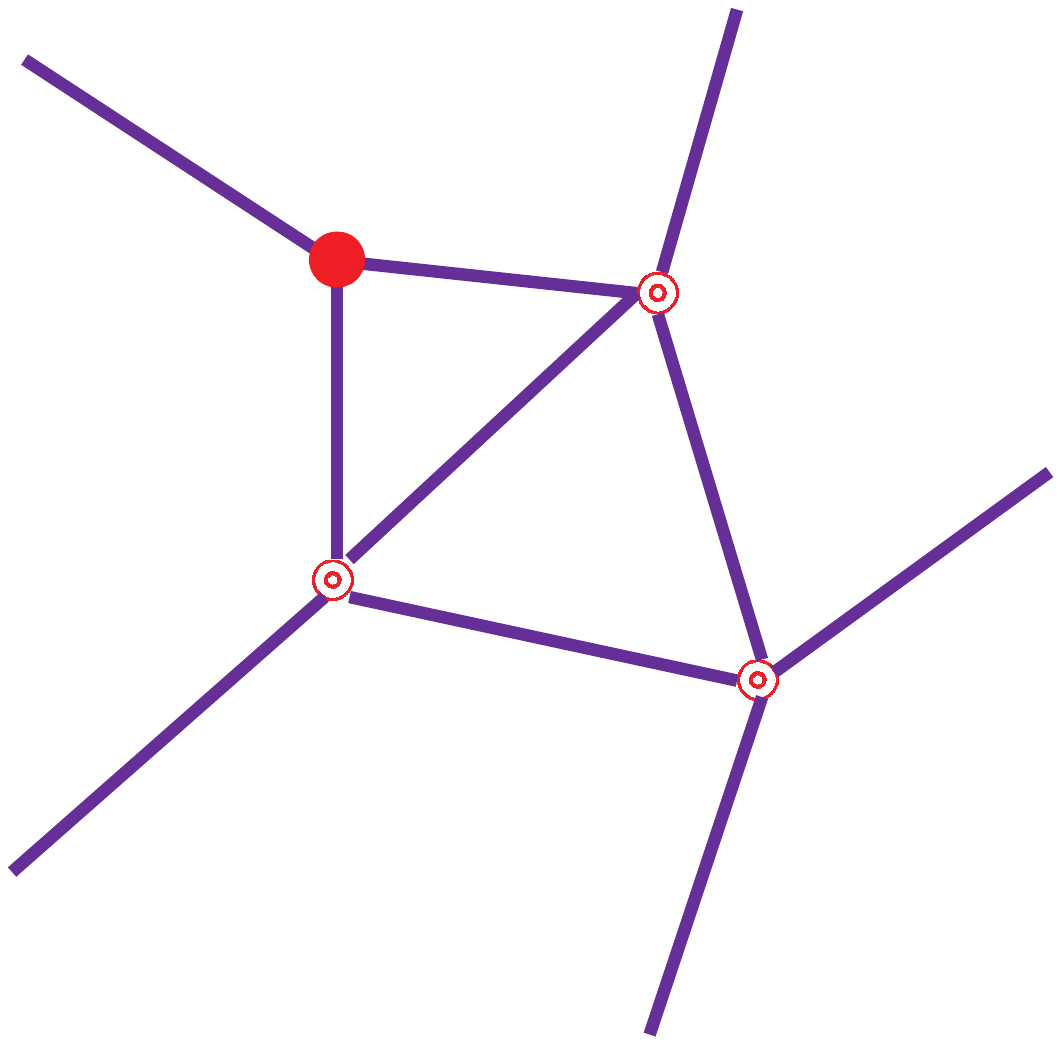}
 \end{center}
\caption{\small The $s$-plane with a web containing $k$ abstract vertices denoted $\circledcirc$ (here $k=3$).
Additional vertices are $\zeta$-vertices.  In this example, only one such $\zeta$-vertex is shown.  }
 \label{LongWeb}
\end{figure}

To relate what we have just described
 to the web-based  definition in section \ref{sec:LocalOpsWebs}, we imitate what we did in the open-string
case in section \ref{notif}. First, if we deform the metric on $\Sigma$ so that the tube in
Figure \ref{MultiTubes} joins on a flat $s$-plane then we obtain a sequence of maps
\be
\t \CU_k: \IM_{S^1}^{\otimes k} \rightarrow \IM^{\rm web}_{S^1}  \qquad \qquad k\geq 1
\ee
Next we claim that these maps define an $L_\infty$ morphism. (See Section \S \ref{subsec:LINF-ALG}.)
To see this we now deform the metric to
a different picture in which $k$ small tubes join separately to the asyptotically flat $s$-plane (this picture was sketched in Figure \ref{Tubers}).
In a limit in which the tubes are all widely separated compared to their sizes and to the natural length scale of the theory,
a solution of the $\zeta$-instanton
equation becomes weblike, but now with $k$ abstract vertices as well as possible  $\zeta$-vertices, as shown
in Figure \ref{LongWeb}.   In this description, to compute the $k$-fold $L_\infty$ product, we have to count such
webs, where two of the abstract vertices are placed at $s_1=0$ and $s_2=1$, and the others are at unspecified points
$s_3,\dots,s_k$.  In the counting, each $\zeta$-vertex is weighted by the corresponding interior amplitude.
But this is precisely the definition of $\rho(\ft_p)$ used in defining the planar $L_\infty$ algebra
 in Section \S \ref{subsec:WebRepPlane} and the $L_\infty$ algebra $R_c$ in Section \S \ref{sec:LocalOpsWebs}. The argument of
 Section \S \ref{higherones} can now be imitated to establish that the $\t\CU_k$ define an $L_\infty$-morphism.

\begin{figure}
 \begin{center}
   \includegraphics[width=2.5in]{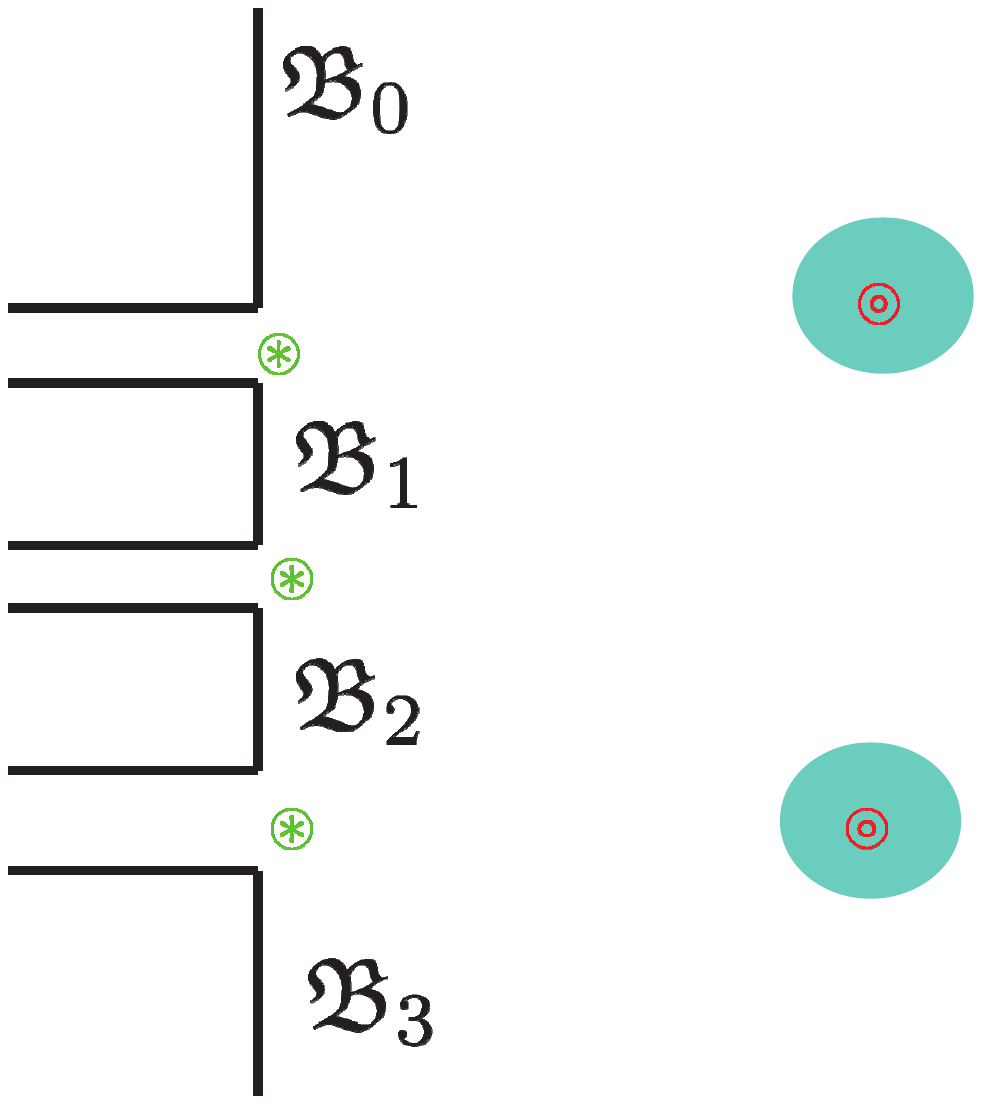}
 \end{center}
\caption{\small The web-based $LA_\infty$ algebra is reproduced by counting
$\zeta$-instantons with abstract vertices (representing insertions of local operators)
 Each turquoise disk is removed and the boundary is smoothly attached to a semi-infinite cylinder. }
 \label{fig:LA-MORPHISM}
\end{figure}

We can extend further the maps $\tilde \CU_k$ defined above to give an $LA_\infty$ morphism
of open-closed $LA_\infty$ algebras. (See Section \S \ref{subsec:LA-ALG}.)
We now attach tubes to the geometries of the form shown in Figure \ref{openstrings}
to define $LA_\infty$ products on the Fukaya-Seidel morphisms $\IM_{\fB,\fB'}$
with $\IM_{S^{1}}$. Then by attaching a geometry such as Figure \ref{MultiTubes}
to, say, Figure \ref{higherjoin}(a) and counting $\zeta$-instantons we obtain
a sequence of linear maps
\be
\tilde\CU_{n,m}: \IM_{\B_0,\B_1} \otimes \cdots \IM_{\B_{n-1},\B_n}\otimes \IM_{S^1}^{\otimes m}
\rightarrow \IM^{\rm web}_{\B_0,\B_n} \qquad n\geq 1, m\geq 0 .
\ee
Then, by deforming these geometries to those such as the ones shown in
Figure \ref{fig:LA-MORPHISM}, (corresponding to the case $n=3$
and $m=2$) we can prove that the $\tilde\CU_{n,m}$ satisfy the axioms
of an $LA_\infty$-morphism.

\subsection{Direct Treatment Of Local $A$-Model Observables}\label{localobs}

This section is somewhat outside our main line of development.  We will ask what we can learn by studying
standard local $A$-model operators in a web framework.   This is a little unnatural, for we have learned in sections \ref{need}
and \ref{openlocal} that in the presence of a superpotential,
the standard local operators of the  $A$-model  are not the whole story.  Still
it seemed interesting to ask what we could say by directly studying the standard operators.

We will do this just in the special case of boundary
 local operators, associated to a particular brane $\B$ that is supported on a Lagrangian submanifold $\L$.  These operators
 are associated to the cohomology
of  $\L$.   Let $H\subset \L$ be a homology class of
real codimension $r$. The $A$-model
has a corresponding boundary observable $\O_H(\textbf{w})$ which should be inserted at a point $\textbf{w}$ in a boundary $\partial_\L
\Sigma$ (of the Riemann surface $\Sigma$ on which the $A$-model is defined)
that maps to $\L$: this operator imposes a constraint that the point $\textbf{w}$ should map to $H\subset \L$. The descent procedure gives a corresponding 1-form operator $\O^{(1)}_H$, characterized by the
condition
\begin{equation}\label{cond} \{\CQ_{\zeta},\O_H^{(1))}\}=\d \O_H(\textbf{w}). \end{equation}
The insertion $\int_{\partial_\L\Sigma}\O_H^{(1)}$ imposes a constraint that some point on $\partial_\L\Sigma$
should map to $H$.   Modulo
$\{\CQ_{\zeta},\cdot\}$, the operators $\O_H$ and $\int_{\partial_\L\Sigma}\O_H^{(1)}$ depend only on the homology
class of $H$.  The fermion number of the operator $\O_H$ is the codimension $r$ of $H$, and that of $\O_H^{(1)}$ is
$r-1$.

For a simple example, we will take $\Sigma$ to be the strip $[x_\ell,x_r]\times \IR$, and as usual
 define boundary conditions at the two ends of the strip by Lagrangian submanifolds
$\L_\ell$, $\L_r$.  We write
$\Sigma_\ell$ for the left
boundary at $x=x_\ell$.
For a given $H\subset \L_\ell$, and a choice of the point $\textbf{w}\in \Sigma_\ell$,
 we attempt to calculate a matrix element
\begin{equation}\label{matemt} \langle f|\O_H(\textbf{w})| i \rangle. \end{equation}
where $| i \rangle$ and $\langle f$ are some initial and final states, whose details
will not be important in what follows.
First, we assume that $H$ is a hypersurface in $\L_\ell$, of codimension 1, so that $\O_H$ has fermion number 1.
In this case, a nonzero contribution to the matrix element in (\ref{matemt}) must come from a component $\M$
of $\zeta$-instanton moduli space of dimension 1.  Thus $\M$ parametrizes a family of rigid $\zeta$-instantons,
with time translation as the only modulus.  To get a non-zero contribution to the given matrix element,
we must adjust this modulus so that the $\zeta$-instanton maps the point $\textbf{w}$ to $H$.

\begin{figure}
 \begin{center}
   \includegraphics[width=2in]{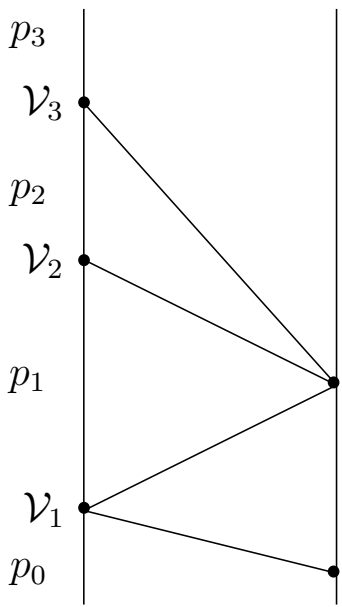}
 \end{center}
\caption{\small  A rigid strip web with left-boundary Lagrangian $\L_{\ell}$.
 The $\zeta$-vertices on the left boundary have been
labeled as $\mathcal V_1,\dots, \mathcal \V_s$ (where $s=3$ in this example). A particular
 $\zeta$-instanton will be exponentially close
 to points    $p_0,\dots,p_s\in \L_{\ell}$ on the boundary segments between vertices.}
 \label{rigidweb}
\end{figure}

A family of rigid $\zeta$-instantons is associated to a $\zeta$-web with rigid vertices as in
figure \ref{rigidweb}.
Let us discuss how $\SIgma_\ell$ is mapped to the target space $X$ by such a $\zeta$-instanton.
 There is a finite set $P$ of points in $\L_\ell$  that are boundary values of some half-line $\zeta$-soliton
that interpolates between $\L_\ell$ and one of the critical points $i\in\IV$.    Suppose that along $\Sigma_\ell$,
there are a total of $s$ rigid $\zeta$-vertices $\V_\alpha$, $\alpha=1\dots s$.  The intervals before and
after these vertices (including the semi-infinite intervals with $\tau\to \pm\infty$) are mapped to points in $P$
(within an exponentially small error).  We call these points $p_0,\dots,p_s$
where $p_{\alpha-1}$ is just
before $\V_\alpha$ (in time) and $p_\alpha$ is just after.

 Let us choose $H$, within its homology class, so that $H$ does not
intersect $P$.  Then it is only within the vertices $\V_\alpha$ that we may find a point on $\Sigma_\ell$ that maps
to $H$.  Each $\V_\alpha$ defines a half-plane $\zeta$-instanton that, when restricted to $\Sigma_\ell$, determines
a path $\rho_\alpha$ from $p_{\alpha-1}$ to $p_\alpha$.  Let $n_\alpha^H$ be the intersection number of $H$
with the path $\rho_\alpha$.  (This intersection number is defined as usual by counting intersections with signs
that depend on relative orientations.)
This number depends not only on the homology class of $H$ in $\L$ but on its homology
class in $\L\backslash P$, in other words it depends on how we chose $H$ so as not to intersect $P$.
To evaluate
the contribution of a family $\M$ of rigid $\zeta$-instantons to the matrix element (\ref{matemt}), we have to count
(with signs) the points in $\M$ that parametrize $\zeta$-instantons that map the point $\textbf{w}\in \Sigma_\ell$ to $H$.
This counting is the same as the counting of the intersections $H\cap \rho_\alpha$ (for all possible $\alpha$),
since any point  $y\in H\cap \rho_\alpha$ is the image of the point $\textbf{w}\in \Sigma_\ell$ for a unique $\zeta$-instanton
in the family $\M$.  (One simply slides the $\zeta$-instanton forwards and backwards in time until the point
in $\partial_\ell\Sigma$ that maps to $y$ is $\textbf{w}$.)  So the contribution of $\M$ to the matrix
element (\ref{matemt}) is given by
\begin{equation}\label{urz}\pm\sum_{\alpha=1}^s n_\alpha^H.\end{equation}
As usual, the overall sign here is determined by the sign of a fermion determinant (and depends on the
signs of the initial and final states $|i\rangle$ and $|f\rangle$).
The full matrix element is obtained by summing
this expression over all families $\M$ of rigid $\zeta$-instantons that satisfy the boundary conditions.

An attentive reader might notice a puzzle here.  If $H$ is moved across one of the points $p_\alpha$,
$\alpha=1,\dots,s-1$, which label an interval of  $\partial_\ell\Sigma$ between adjacent vertices $\V_{\alpha}$ and $\V_{\alpha+1}$,
then the integers $n_\alpha^H$ and $n_{\alpha+1}^H$ make equal and opposite jumps and the sum (\ref{urz}) does not change.
However, this sum is not invariant if $H$ crosses one of the points $p_0$ or $p_s$ at the ends of the chain, for then there is jumping
only of $n_1^H$ or $n_s^H$.  The interpretation is as follows.  In general, when $H$ is replaced by another cycle in its homology
class, $\O_H(\textbf{w})$ changes by $\{\CQ_{\zeta},\mathcal X(\textbf{w})\}$ for some $\mathcal X$.  The matrix element $\langle f|\O_H(\textbf{w})|i\rangle$ shifts
by $\langle f|\{\CQ_{\zeta},\mathcal X(\textbf{w})\}|i\rangle$, and for this to vanish, the states $|i\rangle$ and $|f\rangle$ must be $\CQ_{\zeta}$-invariant.  However, the
family $\M$ of $\zeta$-instantons whose contribution we have been evaluating would by itself contribute $\pm 1$ to the matrix element
$\langle f|\CQ_{\zeta}|i\rangle$, since it parametrizes a one-parameter family of $\zeta$-instantons that interpolates between the state $|i\rangle$
in the past and the state $|f\rangle$ in the future.  If it is true that $\CQ_{\zeta}|i\rangle=0$, then there must be a second one-parameter family $\M'$
of $\zeta$-instantons also interpolating from $|i\rangle$ in the past to $|f\rangle$ in the future, and contributing to the matrix element
of interest by a formula similar to (\ref{urz}) (in general with $s$ replaced by some $s'$ and the  $\V_1,\dots,\V_s$
replaced by another set of rigid $\zeta$-vertices $\V'_1,\dots, \V'_{s'}$) but with
a relative minus sign because of an opposite sign of the fermion determinant.   In such a situation, the jumping when $H$ moves across
an endpoint of the chain cancels between $\M$ and $\M'$, and the matrix element of $\O_H(\textbf{w})$ depends only on the homology class of $H$.

The moral of the story is that in
evaluating the matrix element $\langle f|\O_H(\textbf{w})|i\rangle$, the operator $\O_H(\textbf{w})$ can be replaced by an instruction to count only
$\zeta$-instantons in which one of the vertices $\V_\alpha$ is located at the point $\textbf{w}$, and to weight every such $\zeta$-instanton
by a factor $n_\alpha^H$.   Here the coefficients $n_\alpha$ are
not completely natural but depend on how $H$ is chosen to not intersect the finite set $P$.

Now let us go
on and consider the case that $H\subset \L_\ell$ is of codimension 2. In this case, the matrix element (\ref{matemt}) will be a sum of contributions of two-parameter moduli spaces
$\M$ of $\zeta$-instantons.  There are two types of two-parameter moduli spaces on a strip.
 (1) In one case, we construct $\zeta$-instantons on the strip by gluing $\zeta$-vertices, one of which has
 an excess dimension 1, and the rest of which are rigid, using a rigid web $\fs$. (2) In the second case,
 we construct $\zeta$-instantons on the strip by gluing rigid $\zeta$-vertices via a strip web $\fs$ that
 has a two-dimensional moduli space. (This gives a description of one region in a two-parameter moduli
 space $\M$, and a full explanation will involve considering some of the other regions, as discussed below.)
  Each of the two cases can be relevant, in general, to evaluating the
 matrix element (\ref{matemt}).

 We consider first the case of a moduli space $\M$ of type (1).  This case is particularly interesting as it gives
 an example (the only example that will be studied in the present paper) in which a non-rigid $\zeta$-vertex is
 relevant.  The relevant picture is again that of Figure \ref{rigidweb}, but now one of the $\zeta$-vertices on the left boundary,
 say $\V_*$, has an excess dimension of 1.  Since  $H$ has codimension 2 in $\L$, we can
 choose it   not to intersect any of the paths $\rho_\alpha$
 associated to the rigid $\zeta$-vertices $\V_\alpha$ on $\SIgma_\ell$.  With such a choice, the contribution
 of $\M$ to the desired matrix element comes entirely from the non-rigid vertex $\V_*$.
  The part of $\Sigma_\ell$ just before or after $\V_*$ is mapped
 to points $p,p'\in P$, and $\V_*$ parametrizes a 1-parameter family of paths from $p$ to $p'$, sweeping
 out a two-manifold $D\subset \L$.  Let $n_*$ be the intersection number $H\cap D$.   The contribution of $\M$
 to the matrix element (\ref{matemt}) is then $\pm n_*$, where as usual the sign comes from a fermion determinant.
 This sort of contribution to the matrix element can be described by saying that the insertion of $\O_H(\textbf{w})$ receives
 a contribution in which this operator is replaced by an insertion of the non-rigid $\zeta$-vertex $\V_*$ at $\textbf{w}$,
 with amplitude $n_*$.

 \begin{figure}
 \begin{center}
   \includegraphics[width=5in]{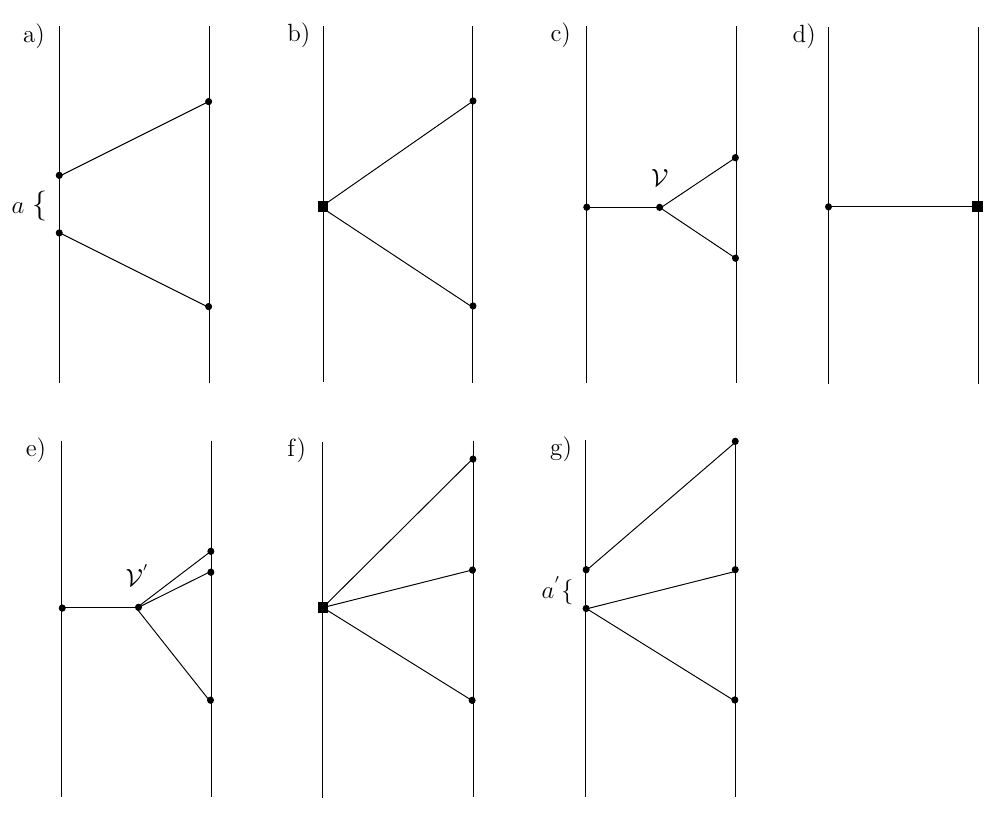}
 \end{center}
\caption{\small     A
 two-parameter $\zeta$-instanton moduli space $\M$
of type (2) is sketched here.  (a) is a weblike region of this moduli space, with a single reduced modulus $a$, as indicated in the picture.  For $a\to 0$, the web description
breaks down and we get a region of the moduli space indicated in (b); the reduced modulus is hidden in  the small square, which represents a family
of half-plane $\zeta$-instantons with a one-dimensional reduced moduli space (a copy of $\IR$) that interpolates between  (a) and (c).  Here (c) is another web-like
region, the reduced modulus being the horizontal position of the vertex labeled $\mathcal V$.  This web-like description breaks down when $\mathcal V$
reaches the right boundary; after another transition region (d), the reduced moduli space continues with still  another web-like region (e), in which the reduced
modulus is the horizontal position of the vertex $\mathcal V'$.  This description breaks down when $\mathcal V'$ reaches the left boundary.
After one  more transition region (f), the moduli space ends in one last
web-like region (g), the reduced moduli space here being the distance $a'$. The reduced moduli space $\M$ has two ends, corresponding to $a\to\infty$ in (a)
and $a'\to\infty$ in (g).   In the transition regions (b), (d), and (f), the small square on the left or right boundary does not represent
a $\zeta$-vertex (whose reduced moduli space by definition has no ends) but rather a family of half-plane $\zeta$-instantons whose reduced moduli
space is a copy of $\IR$, with two ends, related to two different webs.}
 \label{hallustrip}
\end{figure}

 Now let us consider the contribution of a two-parameter moduli space $\M$ of type (2). The reduced
 space $\M_\red$ of $\M$ is 1-dimensional.   When the strip $[x_\ell,x_r]$ is very wide, $\M_\red$ is divided
 into regions each of which can be constructed by gluing of $\zeta$-vertices using a strip web $\fs$ that has a 1-dimensional
 reduced moduli space.
  The reduced moduli space of such a $\fs$ has two ends.  At  an end, the description of
 $\M$ by the strip web $\fs$ breaks down, but $\M$ continues past this breakdown, with a new region that
 is described by a generically different web $\fs'$.  So overall, $\M$ has regions
  related to various webs $\fs_\sigma$.  See Figure \ref{hallustrip} for an example.  In this example, $\M$ has
  four web-like regions, with three transition regions between them.  The transition regions are represented by
  pictures that look web-like, but they do not really represent $\zeta$-webs, since they each contain one ``vertex'' (labeled
  in the figure by a small square)
  that is not what we usually call a $\zeta$-vertex (it represents a family of $\zeta$-instantons with a one-dimensional
  reduced moduli space with ends that correspond to $\zeta$-webs).  The reduced moduli space of $\M$ might be
  compact or, as shown in the figure, it might have ends that correspond to time-convolution of webs, an operation
  considered in section \ref{subsec:Strip-Webs}.

  Assuming that $H$ does not intersect any of the paths $\rho_\alpha$ associated to $\zeta$-vertices,
  the web-like regions in a type (2) moduli space $\M$ do not contribute to the matrix element of eqn (\ref{matemt}).
  However, in general, the transition regions may contribute.  The web-like regions represent $\zeta$-instantons
  that map $\partial_\ell\Sigma$ to $\L$ via a sequence of paths $\rho_s\star\rho_{s-1}\star\dots\star\rho_1$ (where
  $\star$ represents amalgamation of paths).  A transition region of the type shown in Figure \ref{hallustrip}(b,f) (where
  the transition occurs along $\Sigma_\ell$) involves a transition from one sequence $\rho_k\star \rho_{k-1}$
  (for some $k$)
  between points $p,p'\in P$ to another sequence $\rho'_k\star\rho'_{k-1}$.  As the transition is made, the relevant family
  of $\zeta$-instantons sweeps out a two-dimensional surface $D'$ that interpolates between $\rho_k\star\rho_{k-1}$
  and $\rho'_k\star\rho'_{k-1}$.   The contribution of the transition region in question to the desired matrix
  element (\ref{matemt}) is $\pm H\cap D'$, where as usual the sign is the sign of the fermion determinant.  The contribution
  of $\M$ to the matrix element is the sum of these contributions, for all the relevant transition regions.

  Combining what we have said about moduli spaces of types (1) and (2), an insertion of $\O_H(\textbf{w})$ can be represented
  in a weblike picture as a sum of insertions of effective $\zeta$-vertices.  The effective $\zeta$-vertices in question
  can be either (1) non-rigid $\zeta$-vertices whose excess dimension is 1 and whose  internal modulus is fixed by requiring that
  the point $\textbf{w}$ is mapped to $H$, or (2) effective $\zeta$-vertices that arise by using this constraint that $\textbf{w}$ maps to $H$ to fix
   a modulus in a transition region
  between two $\zeta$-webs.

  Though we will not try to develop a systematic theory in the present paper, hopefully we have said enough
  to convince the reader that it is possible to give a recipe to compute matrix elements of standard local observables of the
  $A$-model in terms of $\zeta$-webs.  The intrepid reader can consider in a similar spirit the case that the codimension
  of $H$ is greater than 2, the case that $\O_H(\textbf{w})$ is replaced by insertion of an integrated  descendant $\int \O_H^{(1)}$,
  or the case that the boundary operator $\O_H(\textbf{w})$ is replaced by a bulk local observable of the $A$-model.

  In this analysis, we found a role for $\zeta$-vertices with positive excess dimension.  One may be puzzled, as they did not
  enter the seemingly more general analysis of section \ref{openlocal}.    This has happened because here we considered order operators
  only and labeled the entire left boundary of the worldsheet by a single Lagrangian submanifold $\L$. An order operator then places a constraint
  on the solution of the $\zeta$-instanton equation and can remove the excess moduli associated to a non-rigid $\zeta$-vertex.
   By contrast, in section \ref{openlocal},
  we always assumed
   that if an open-string observable
  is inserted at a boundary point $p$ that separates regions labeled by Lagrangians $\L_1$ and $\L_2$ (Figure \ref{Division}), then $\L_1$ and $\L_2$ intersect  transversely,
  even if they are equivalent under Hamiltonian symplectomorphisms.
  It is then still true that an order operator places a constraint on the solution of the $\zeta$-instanton equation, but the interpretation in terms of non-rigid
  $\zeta$-vertices is hidden.

\section{ Interfaces And Forced Flows}\label{subsec:LG-Susy-Interface}

Now we study a \emph{family} of massive superpotentials (i.e. holomorphic Morse
functions), all defined on a fixed target space $X$. Given such a family we
can define a set of interesting supersymmetric interfaces between Landau-Ginzburg theories,
thus illustrating the abstract ideas of Sections \S \ref{sec:Interfaces} - \S \ref{sec:GeneralParameter}
in the concrete example of Landau-Ginzburg models.

We denote a typical superpotential by $W(\phi;z)$ where in this section $z$ is in a parameter
space $C$. Some  notable examples of $C$
include the case where $C$ is a  Riemann surface, as in
\cite{Gaiotto:2009fs,Gaiotto:2011tf,Gaiotto:2012rg},
or   $C= {\rm Sym}^n(C_{uv})$, with $C_{uv}$ a Riemann surface,  as in \cite{Gaiotto:2011nm}
and in Section \S \ref{subsec:KnotHomology} below.
The parameter   $z$ is not to be confused with a vacuum weight. In fact, for
a fixed $z\in C$ the superpotential $W(\phi;z)$  has critical points $\phi_{i,z}$,
labeled by $i\in \IV$ and, introducing a phase $\zeta$, the critical values
\be
z_i := \zeta \bar W(\phi_{i,z} ;z)
\ee
define the vacuum weights of the corresponding Theory. The set of vacuum weights is the fiber
of an $N:1$ covering space of $C$ which we denote by $\pi: \Sigma \to C$.
In general this covering space will have nontrivial monodromy.
In the examples discussed in \cite{Gaiotto:2009fs,Gaiotto:2011tf,Gaiotto:2012rg}
the covering extends to a branched covering $\pi: \bar \Sigma \to \bar C$,
where $\bar C$ is a punctured Riemann surface.
The superpotential is not massive at the branch points of this covering,
so in this Section we avoid those points and just work on $C$. Some preliminary remarks on
extending our considerations to the full branched covering are deferred to Section \S \ref{subsec:partialRG}
below.  \footnote{In the papers just cited, what is here called $C$ would be
called $C'$, while what is here called $\bar C$ would simply be called $C$.  }

Now let $D\subset \IR $ be a spatial domain and consider
a continuously differentiable map   $z:D \to C$ with compact support for $\frac{d}{dx} z(x) $ contained
within the interval $[x_-, x_+]$.
Just as in Section \S \ref{lgassuper} we can define a $1+1$ dimensional QFT
by considering the supersymmetric quantum mechanics with real superpotential
\be\label{eq:Interface-h}
h = - \half \int_{D}
\left[2\phi^*(\lambda) -  {\Re}(\zeta^{-1} W(\phi;z(x) )dx \right]
\ee
where $\lambda = pdq$ is a Liouville one-form for the symplectic form on $X$. The resulting $1+1$
dimensional QFT manifestly has the supersymmetries $\CQ_\zeta, \bar \CQ_\zeta$
of equation \eqref{manifest}. For $x \leq x_-$ the integrand of $h$ coincides with that associated
to the LG theory determined by $W(\phi;z_{-})$ and for $x \geq x_+$ it coincides with that associated to $W(\phi;z_+)$.
Therefore we can consider this QFT to be the theory of a supersymmetric interface between
the theories at $z=z_-$ and $z=z_+$.

To make contact with the abstract part of the paper we   denote the   path
traced out by $z(x)$ in $C$ by $\wp$. Along $\wp$ there is a parallel
transport of the vacua $\phi_{i,z}$, allowing us to define ``local vacua'' $\phi_{i,x}$
for the theory at $z=z(x)$. The corresponding   critical values are denoted $W_{i,x}$.
Now, defining
\be\label{eq:LG-vac-hmtpy}
z_i(x) = \zeta \bar W_{i,x}
\ee
we obtain a ``vacuum homotopy,''  in the language of
Sections \ref{sec:CatTransSmpl} and \ref{sec:GeneralParameter}.
It now follows from the results of Sections \S\S \ref{lgassuper}-\ref{zetawebs}
that we have a family of Theories. The representation of webs for the Theory
at $x$ is provided by the MSW complex for the superpotential at $z(x)$
 and the local interior amplitude $\beta(x)$ for the Theory $\CT(x)$ is provided by the amplitude
\eqref{torrix} for the LG theory $W(\phi;z(x))$.  From the abstract discussion
it follows that the
resulting family of Theories defined by \eqref{eq:LG-vac-hmtpy}  have a corresponding Interface
$\fI[\wp]$. We claim that this Interface is precisely the theory defined by
\eqref{eq:Interface-h}. Our goal is now to describe this Interface in conventional Landau-Ginzburg terms.

To this end we  follow once again the standard SQM interpretation of Morse theory.
 When   $D=\IR$ we choose boundary conditions
\be\label{eq:FFlow-bc1}
\lim_{x\to + \infty} \phi(x) = \phi_{j',z_+}
\ee
\be\label{eq:FFlow-bc2}
\lim_{x\to - \infty} \phi(x) = \phi_{i,z_-}
\ee
where $z_\pm := \lim_{x\to \pm\infty} z(x)$ and the vacua $j'$ for $x \geq x_+$
are to be compared to the vacua $j$ for $x\leq x_-$ by
parallel transport on the covering space $\Sigma$. The prime on $j'$ is meant to remind us that $\phi_{j',z_+}$
are vacua in a Theory $\CT^+$ different from the vacua $\phi_{i,z_-}$ of the Theory $\CT^-$.

The stationary points of \eqref{eq:Interface-h} are  given by solutions to the
differential equation
\be\label{eq:LG-forced-flow}
 \frac{d}{dx} \phi^I = \frac{\I \zeta}{2} g^{I \bar J}
\frac{\p \bar W}{\p \bar \phi^{\bar J}}(\bar\phi;z(x))
\ee
but now there are some important differences from the $\zeta$-soliton equation. Compared
to the old equation there is
extra $x$-dependence on the right hand side of \eqref{eq:LG-forced-flow}
due to the explicit $x$-dependence of $z(x)$. Moreover,
the phase $\zeta$ is a choice fixed from the beginning and
we do not take it to be related to the phase $\zeta_{j'i}$.
We call \eqref{eq:LG-forced-flow} the \emph{$\zeta$-forced flow equation}.
\footnote{The local vacua $\phi_{i,x}$ are   \emph{not} to be confused with solutions to the forced flow
equation \eqref{eq:LG-forced-flow}.}

Following the usual Morse-theoretic interpretation of SQM we define
the MSW complexes:
\be
\IM_{ij'}^\bullet(\wp) = \oplus_p  \Psi^f_{ij'} (p) \IZ
\ee
Once again, $p$ enumerates the solutions $\phi^p_{ij'}(x)$ of the forced flow
equation with boundary conditions \eqref{eq:FFlow-bc1}, \eqref{eq:FFlow-bc2}.
If $z(x)$ has nontrivial $x$-dependence the equation \eqref{eq:LG-forced-flow} is
no longer translation invariant and hence there will in general be a unique BPS groundstate $\Psi^f_{ij'} (p)$. The
grading/fermion number will be given once again by the eta invariant:
$f = - \half \eta(\CD)$ with $\CD$ given
by \eqref{eq:NewD}. The differential on the complex is again
given by computing solutions to the $\zeta$-instanton equation
\be\label{eq:LG-INST-forced}
\left(\frac{\p  }{\p x} +\I  \frac{\p }{\p \tau} \right)\phi = \frac{\I \zeta}{2} \frac{\p \bar W}{\p \bar\phi}(\bar\phi(x,\tau); z(x))
\ee
interpolating between solutions whose fermion number differs by $1$.
The Chan-Paton complex of the Interface $\fI[\wp]$ is now provided by
the MSW complex:
\be
\CE(\fI[\wp])_{ij'} = \IM^\bullet_{ij'}(\wp).
\ee
\begin{figure}[htp]
\centering
\includegraphics[scale=0.3,angle=0,trim=0 0 0 0]{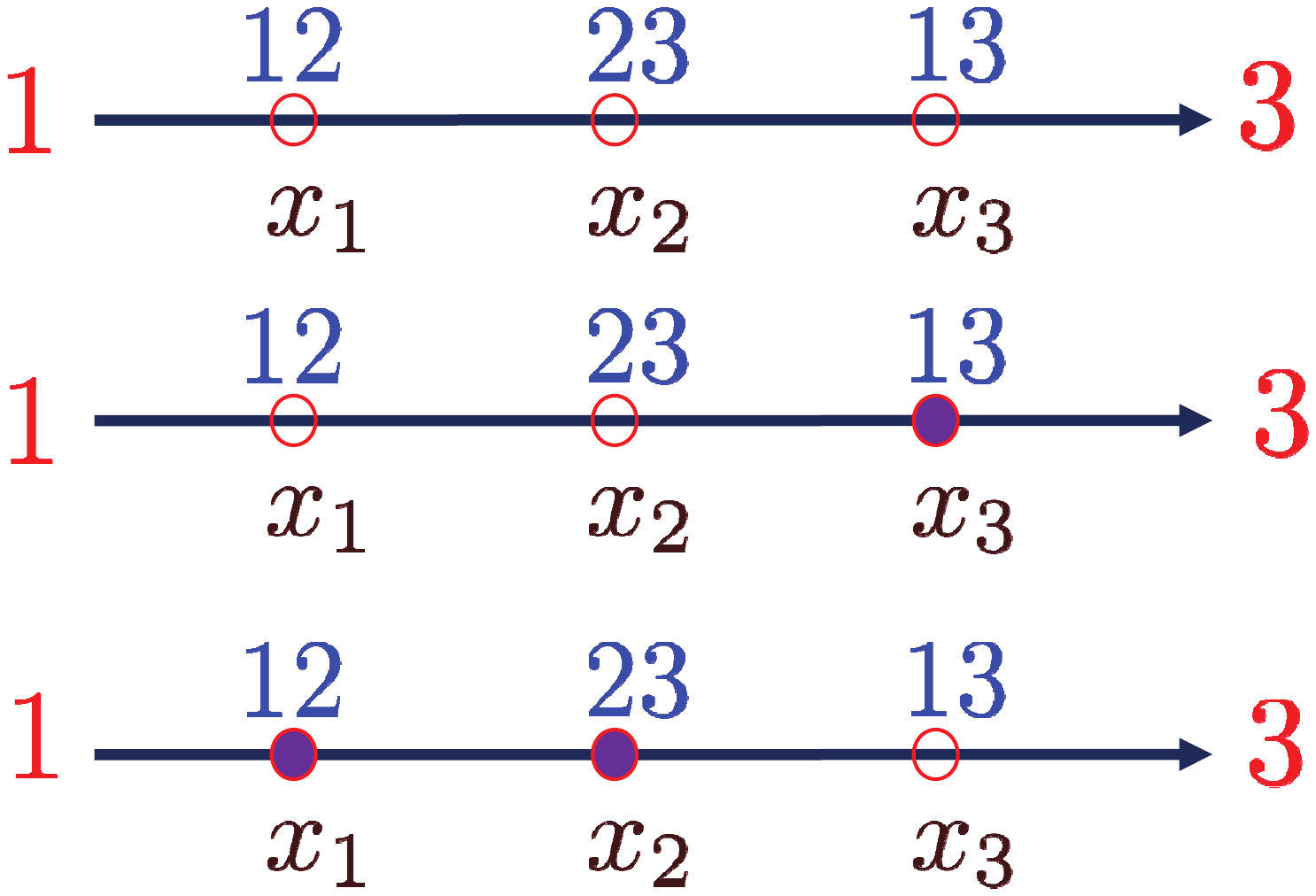}
\caption{An example of the bound-soliton basis for the complex $\IM^\bullet_{13'}(\wp)$.
In the first line the empty circles indicate potential positions of solitons at
binding points $x_1, x_2, x_3$. We could potentially glue in a soliton  of
type $12$ at $x_1$, $23'$ at $x_2$ and $13'$ at $x_3$.
In the second line, the $13'$ soliton is filled at $x_3$ and the solutions is
approximately in the vacuum $\phi_{1,x}$ for $x< x_3$. This represents one basis
vector for the complex. In the third line the soliton at $x_1$ of type $12$
and that at $x_2$ of type $23$ are filled and the soliton is approximately in the
vacuum $\phi_{3',x}$ for $x> x_2$. This represents a second basis vector for
the complex. We claim that in this example $\IM^\bullet_{13'}(\wp)$ will have
precisely two generators of the space of states that undergo framed
wall-crossing. They are   represented by the above approximate solutions because there
are no other ways the vacua of bound solitons can be compatible with the boundary conditions. }
\label{fig:FORCEDFLOWCOMPLEX-1}
\end{figure}

We now describe the origin of curved webs in the present context.
We begin with  a very useful picture of some elements of the Morse
complex $\IM_{ij'}^\bullet(\wp)$. Starting with the vacua of
$W(\phi;z_-)$ at $x=-\infty$ we parallel transport the
vacua along $\wp$ to produce $\phi_{i,x}$ with corresponding
critical values $W_{i,x}$. Now, near values of $x=x_0$ such that
\be\label{eq:BindingPoints}
\frac{W_{j_2,x_0} - W_{j_1,x_0} }{\I \zeta} \in \IR_+
\ee
we can produce an approximate solution of the
$\zeta$-forced flow equation by using a soliton solution of type
$\phi_{j_1, j_2}(x)$  for the
superpotential $W(\phi; z(x_0))$. Moreover, we can choose the solution
to be ``centered'' near $x_0$. (Such a center is only well-defined within
a range of order $1/m$, where $m$ is the mass scale of the interface.)   Such solutions
can be glued together to produce good approximations
to true solutions satisfying the boundary conditions
\eqref{eq:FFlow-bc1},\eqref{eq:FFlow-bc2} provided
the intermediate vacua of subsequent solitons agree.
Physically, such solutions correspond to boundstates of the
solitons to the interface, binding near the special point $x=x_0$.
This is the origin of the terminology \emph{binding points} used
in Section \S \ref{subsec:BindPoints} and indeed the condition
\eqref{eq:BindingPoints} is equivalent to the definition used in
Section \S \ref{subsec:BindPoints}, given the vacuum homotopy \eqref{eq:LG-vac-hmtpy}.

A basis for the space of states in the MSW complex $\IM^\bullet_{ij'}(\wp)$
that can undergo framed wall-crossing can be obtained by
attaching solitons to the binding points in all ways consistent with
boundary conditions as described above. (It is clear that we can produce solutions to
\eqref{eq:LG-forced-flow} by gluing together such
solitons, but it is not self-evident that these are the only solutions.)
There is a simple pictorial formalism
for these generatoris,  illustrated in   Figure \ref{fig:FORCEDFLOWCOMPLEX-1}.
\begin{figure}[htp]
\centering
\includegraphics[scale=0.3,angle=0,trim=0 0 0 0]{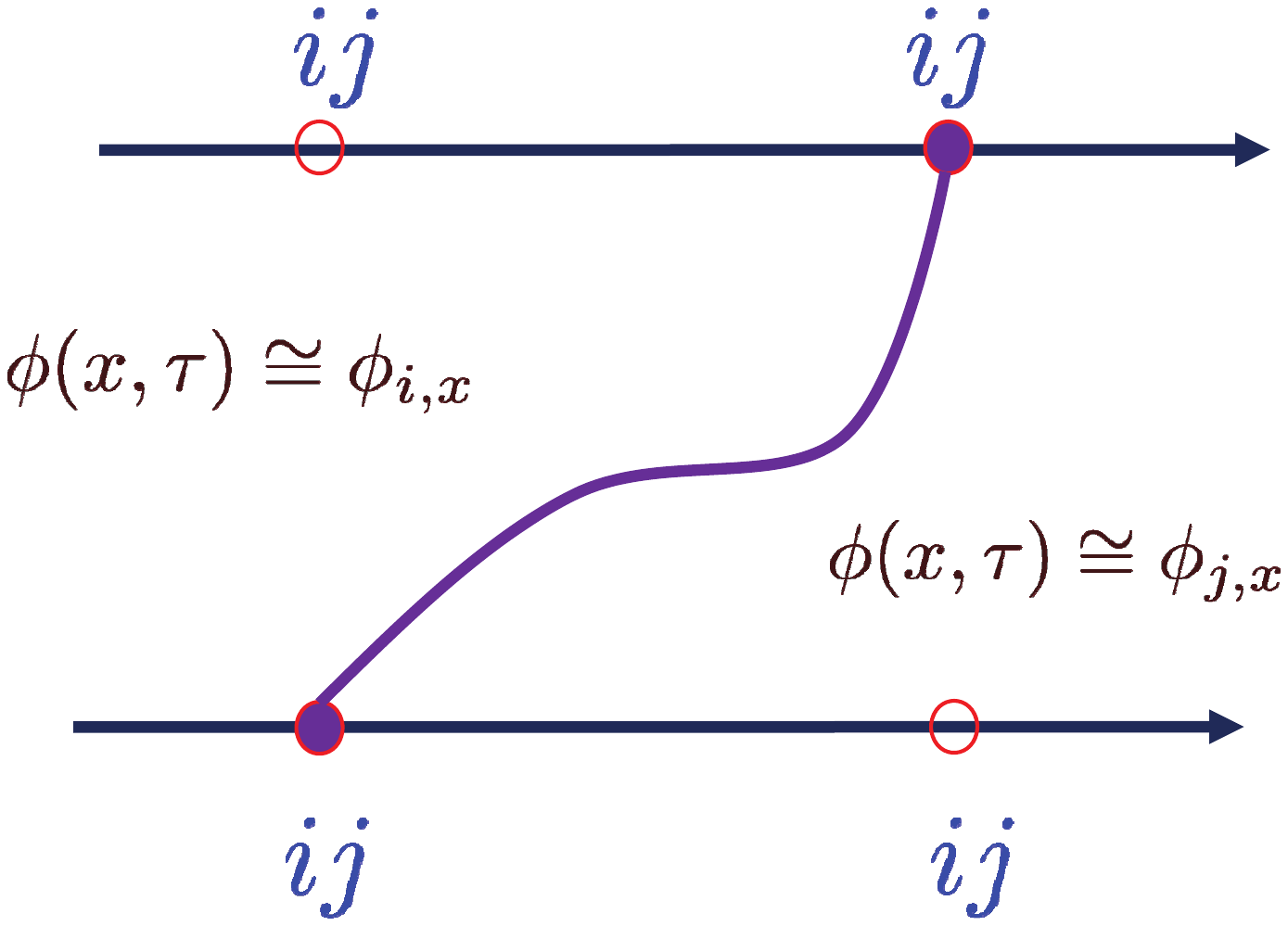}
\caption{An analog of the boosted soliton for the case of a supersymmetric
interface.   }
\label{fig:FORCEDFLOWBOOSTEDSOLITON}
\end{figure}

Now we are in a position to describe the physical origin of curved webs.
There will be time-independent solutions to the forced $\zeta$-instanton equation
which, at long distances, have vertical worldlines of solitons
with a single dot inserted at some $\tau$, analogous to
Figure \ref{fig:INSTANTON-ON-R}. In addition, there will be solutions
analogous to the boosted solitons of Section \S \ref{prelims}.
To describe these suppose we have (in some region of $\wp$ in $C$)
a family of solitons $\phi_{ij}(x;z)$ satisfying the (ordinary)
soliton equation for $W(\phi;z)$ with phase $\zeta_{ji}(z)$ given
by the phase of   $W_{j}(z)  - W_{i}(z)$.  Due to translation invariance
of the soliton equation we can, moreover, assume that $\frac{d}{dx}\phi_{ij}(x;z)$
has its support near $x\cong 0$. We now make an ansatz for the
forced $\zeta$-instanton equation of the form:
\be\label{eq:ForceZetaAnsatz}
\phi_{ij}(x-x(\tau); z(x(\tau)))
\ee
for some trajectory $s(\tau) = x(\tau) + i \tau$ in the complex $s$-plane.
If
\be
\zeta^{-1} \frac{d s(\tau) }{d\tau} \frac{ W_{j, x(\tau)} - W_{i, x(\tau)} }{\vert W_{j, x(\tau)} - W_{i, x(\tau)} \vert} = -1
\ee
and at the same time
\be
\vert \dot x(\tau) \left( \p_z \phi_{ij} z' + \p_{\bar z} \phi_{ij} \bar z' \right) \vert
\ee
is small compared to other terms in the $\zeta$-instanton equation (as will be guaranteed if
$z(x)$ evolves adiabatically in $x$)  then the ansatz \eqref{eq:ForceZetaAnsatz} will be a good
approximation to the forced $\zeta$-instanton equation. The slope of the curve $s(\lambda)$
defined by the center of the soliton will be parallel to $z_{i,x(\lambda)}- z_{j,x(\lambda)}$ where
$z_{i,x}$ are the ``local'' values of the LG vacuum weights.
These instantons may be depicted as in Figure \ref{fig:FORCEDFLOWBOOSTEDSOLITON}  and are the main motivation for the
definition of curved webs in Section \S \ref{subsec:CurvedWebs}.
Of course, when there are three or more vacua we will have local versions of the
vertices used to construct $\zeta$-webs. These define $\beta(x)$, as noted above.

We can now follow the general
discussion of Sections \S \ref{sec:CatTransSmpl} and \S \ref{sec:GeneralParameter}.
The cohomology of the complex
$H^*(\IM_{ij'}(\wp))$  is the space of  ``framed BPS states,''
and its Witten index is the framed BPS index $\fro(\fI[\wp], ij')$,
in the language of \cite{Gaiotto:2010be,Gaiotto:2011tf}. Thus the complex
$\IM_{ij'}(\wp)$   ``categorifies'' the framed BPS indices.

\begin{figure}[htp]
\centering
\includegraphics[scale=0.3,angle=0,trim=0 0 0 0]{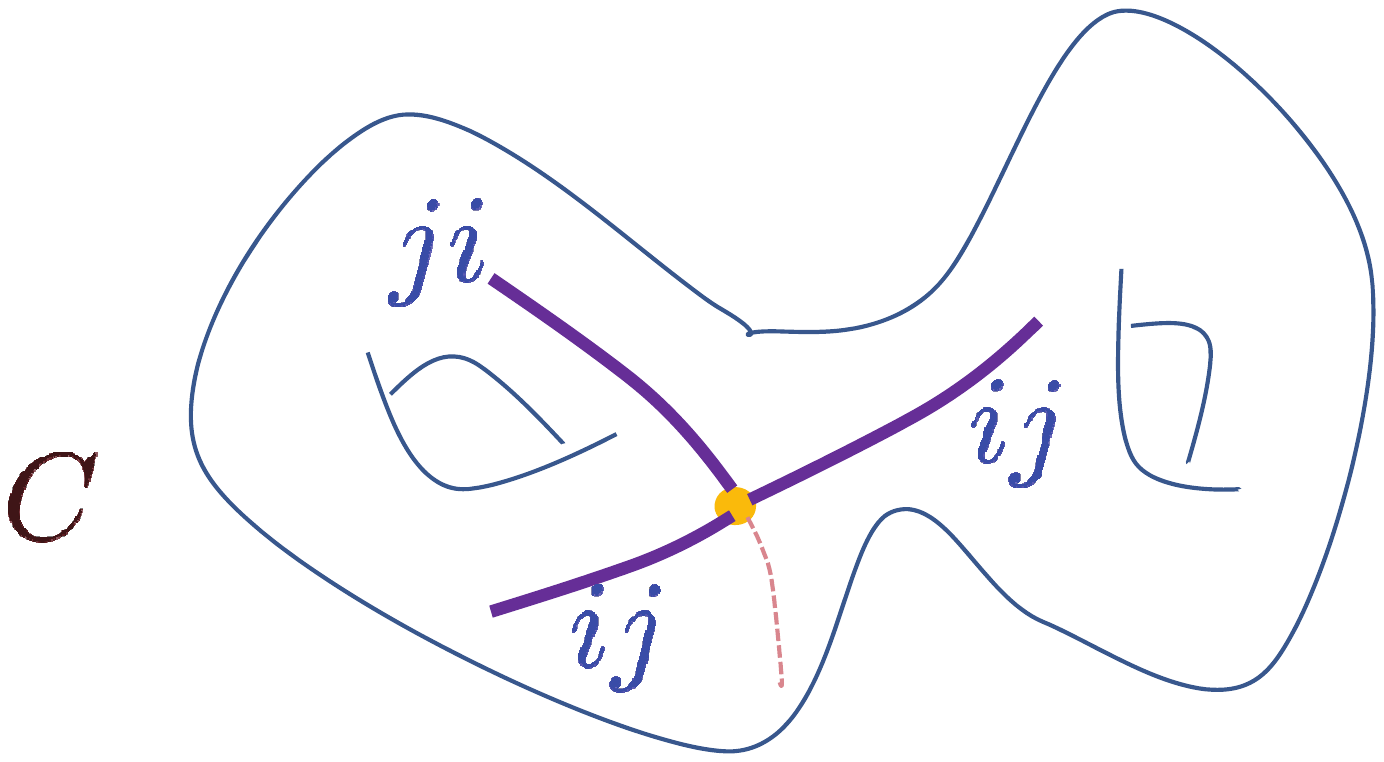}
\caption{When $\bar C$ is the base of a branched cover of vacua there will be a real
codimension two branch locus, indicated here by the orange dot. Three $S$-walls of
type $ij$ can terminate on a simple branch point of type $ij$ as shown. There will be a monodromy exchanging vacua $i$
and $j$ around the branch locus and the pink dashed line indicates a cut for the
trivialization of the cover we have chosen.  }
\label{fig:BRANCHPOINT}
\end{figure}

\textbf{Remarks}:

\begin{enumerate}

\item
There is a nice interpretation of the meaning
of the special binding points defined by
\eqref{eq:BindingPoints}. On the space $C$ we
can define  ``S-walls of type $ij$'' and phase $\vartheta$ by the equation:
\be
\{ z: \frac{W_i(z) - W_j(z)}{\I \zeta} \in \IR_+ \}
\ee
where $\zeta=e^{\I \vartheta}$ and the meaning of $i,j$ depends on choosing a
local trivialization of the cover $\pi: \Sigma  \to C $.
Then the binding points correspond to the values of $x$ where
the path $\wp$ crosses the various S-walls. In fact, as we have noted,
in the physical examples motivating this construction it is
useful to consider $C$ embedded into a family $\bar C$ such
that the cover $\pi: \bar \Sigma \to \bar C$ is a branched cover.
In this case the $S$-walls can end on a branch locus
as in Figure \ref{fig:BRANCHPOINT}. What we are describing here is
a piece of a spectral network. See Section \S \ref{subsec:CatSpecNet}
below for further discussion.

\item  If we now consider a family of paths $\wp_s$ with variable endpoint $z_f(s)$
 then when that endpoint    passes across an $S$-wall there will be wall-crossing of the
framed BPS states: Physically, as the parameters of the  supersymmetric interface
are changed, it will emit or absorb some of the solitons associated to the binding point.
This physical picture was described in
\cite{Gaiotto:2011tf}, but now we can describe it at the level of complexes -
so again we have, in some sense, categorified the wall-crossing story.
We did this in the abstract half of the paper when we defined the
Interfaces $\fS^{p,f}_{ij}$ in Section \S \ref{subsec:SWallIntfc} to describe
a categorified notion of ``$S$-wall crossing.''

\item As a simple example, consider the family of theories with superpotential
\be
W = \frac{1}{3}\phi^3 - z \phi
\ee
corresponding to a family of theories of the type $\CT^N$, with $N=2$,
discussed in Section \S \ref{subsec:CyclicVacWt}.
The family is parametrized by $z\in C$ with $C=\IC^*$.  There are two massive vacua
at $\phi_{\pm } = \pm z^{1/2}$ where we choose the principal branch of the logarithm
and $z$ is not a negative real number. We choose a path $\wp$ defined by
$z(x)$ in $\IC^*$ where $x\in [\epsilon, 1-\epsilon]$
for $\epsilon$ infinitesimally small and positive with $z(x) = e^{\I (1-2x) \pi}$. The spinning vacuum weights
satisfy $z_{-+}(x) = \frac{4}{3} \zeta e^{\I (2x-1) 3 \pi/2}$. If we also take $\zeta$ to have a
small and positive phase then,  applying the criterion
of \eqref{eq:BindingPoint-def} we find that there are two binding points of type $+-$
at $x= 1/3-0^+ , 1-0^+$ and one binding point of type $-+$ at $x=2/3-0^+$. They are all
future stable. The two $S_{+-}$ and one $S_{-+}$ walls emanate from $z=0$ with angle $2\pi/3$
between them. There is a corresponding family of Interfaces $\fI_x=\fI[\wp_x]$ given by the
path $\wp_x$ that evolves along $\wp$ from $\epsilon$ to $x$.
The wall-crossing formula for the framed BPS indices amounts to a simple matrix identity:
\be\label{eq:Spl-Fr-WC}
\begin{pmatrix} 1 & 0 \\  1 & 1 \\ \end{pmatrix}
\begin{pmatrix} 1 & -1 \\  0 & 1 \\ \end{pmatrix}
\begin{pmatrix} 1 & 0 \\  1 & 1 \\ \end{pmatrix}
=
\begin{pmatrix} 0 & -1 \\  1 & 0 \\ \end{pmatrix}
\ee
where the three factors on the LHS reflect the wall-crossing across the three $S_{ij}$-rays,
and the matrix on the right accounts for the monodromy of the vacua.
(See Section 8.1.1 of \cite{Gaiotto:2011tf} for an extended discussion.)
We now use equations \eqref{eq:SijFactor-def-fs} and \eqref{eq:TautCurvedCP} together with
the web representation provided by \eqref{eq:ExpleTN-webrep}:
\be
\begin{split}
R_{-+} & = \IZ[f_1]\\
R_{+-} & = \IZ[f_2]\\
\end{split}
\ee
where the fermion number shifts $f_1,f_2$ must
satisfy $f_1 + f_2 = 1$, since $K$ must have degree minus one.
The categorification of the wall-crossing identity \eqref{eq:Spl-Fr-WC},
at least at the level of Chan-Paton complexes,  is obtained by generalizing the
left-hand-side of \eqref{eq:Spl-Fr-WC} to:
\be\label{eq:CP-prod}
\begin{split}
\begin{pmatrix}\IZ & 0 \\  \IZ[f_2] ~~ & \IZ \\ \end{pmatrix}
\begin{pmatrix} \IZ & ~~ \IZ[f_1] \\  0 & \IZ \\ \end{pmatrix}
\begin{pmatrix}\IZ & 0 \\  \IZ[f_2]~~ & \IZ \\ \end{pmatrix}
& =
\begin{pmatrix}\CE_{--}   & \CE_{-+} \\
\CE_{+-}  & \CE_{++}  \\ \end{pmatrix}\\
\end{split}
\ee
Here $\CE_{-+}= \IZ[f_1]$, while
\be
\CE_{--} = \CE_{++}= \IZ \oplus \IZ[f_1+f_2]
\ee
is a complex with a degree one differential (note that $f_1+f_2 =1$) and
\be
\CE_{+-} =  \IZ[f_2]\oplus \IZ[f_2]\oplus \IZ[f_2+1]
\ee
is another complex with a degree one differential.
The differential comes from combining two boundary
amplitudes of the form in Figure \ref{fig:SWALL-AMP-MINUS}
in a way similar to what happens in Figure
\ref{fig:IPM-MP-HE}. The matrix of complexes \eqref{eq:CP-prod} is
quasi-isomorphic to the categorified version of the monodromy:
\be
\begin{pmatrix} 0 & \IZ[ 1-f_2 ] \\  \IZ[f_2] & 0 \\ \end{pmatrix}.
\ee
The identity \eqref{eq:Spl-Fr-WC} is important in the nonabelianization
map of \cite{Gaiotto:2012rg} when extending the construction across
branch points of a spectral covering. We expect the above identity to
be important in the extension of this construction to the categorified
context.

\end{enumerate}

\section{Generalizations, Potential Applications, And Open Problems}\label{sec:PotentialAppsOpenProb}

\subsection{Generalization 1: The Effect Of Twisted Masses}\label{subsec:TwistedMasses}
In the main text of the paper we have encountered examples of Theories
enriched by conserved global symmetries, namely the $\CT^{SU(N)}_\vartheta$ Theories.
These global symmetries could be identified in their physical counterparts
(see Section \S \ref{subsubsec:RelationTSUN-Physical})
either as winding symmetries of an LG theory with a non simply connected
target space $X \subset (\C^*)^N$, or as isometries of the target space
in a mirror description as a $\C P^{N-1}$ sigma model.

In either case, the underlying physical theories admit a special class of
relevant deformations which deform the supersymmetry algebra and are mirror to each other.
The LG superpotential can be deformed to a multivalued function with single-valued first derivatives:
\be
W = \sum_a Y_a + m_a \log Y_a
\ee
The $\C P^{N-1}$ sigma model can be deformed by twisted masses $m_a$ sitting in the Cartan
subalgebra of the $SU(N)$ global symmetry. In either case, the mass parameters $m_a$ can be interpreted as
the expectation value of scalar components of background gauge supermultiplets coupled to the corresponding
global symmetries: twisted vector multiplets \cite{Gates:1984nk} in the A-type description and vectormultiplets in the mirror B-type description.

Abstractly, the $\N=2$ algebra in two dimensions allows a central charge $Z=\{\bar Q_+,\bar Q_-\}$ and a twisted central charge
$\t Z=\{Q_+,\bar Q_-\}$.   Either one of these breaks one of the two $U(1)$ $R$-symmetries of the theory.  As the framework of the present
paper requires such a $U(1)$ $R$-symmetry, we can include one or the other of these but not both; because of the mirror symmetry of the $\N=2$
algebra, it does not matter which we include.  By convention, we have assumed in this paper that $\t Z=0$, $Z\not=0$.

In general, every $(2,2)$ theory equipped with global symmetries which can be coupled to background vectormultiplets
will admit twisted mass deformations, which modify the twisted central charge $\t Z=\{Q_+,\bar Q_-\}$ to include
contributions proportional to the corresponding global charges. Dually, every $(2,2)$ theory equipped with global symmetries
which can be coupled to background twisted vectormultiplets will admit mass deformations which modify the central charge
$Z=\{\bar Q_+,\bar Q_-\}$ to include contributions proportional to the corresponding global charges. We are not aware of a
specific pre-existing naming convention for the mirror notion to twisted masses. In the context of this paper,
it seems reasonable to dub them A-twisted masses and denote the usual twisted masses as B-twisted masses.

In order to keep $\t Z=0$, we will only allow A-twisted mass deformations. The general expression for the central charge in a massive
$(2,2)$ theory with A-twisted masses $M$, for states which interpolate between vacua $i$ and $j$ and carry charge global $\gamma$ is
\be
Z = W_i-W_j + M \cdot \gamma
\ee
Here the charge $\gamma$ is valued in a lattice $\Gamma$ of global charges and $M$ is valued in the Cartan subalgebra of the global symmetry group.
This expression is a bit ambiguous: the superpotential vevs $W_i$ and $W_j$ are defined up to integral shifts of
$M$, which are related to the possibility to re-define the global symmetry charge by some $\gamma \to \gamma + \gamma^{(i)} - \gamma^{(j)}$.
We can fix the ambiguity by selecting specific values for $W_i$ and $W_j$. \footnote{Even better, we can embrace the ambiguity
by taking the charges of solitons to live in a torsor for the lattice of global charges, see i.e. \cite{Gaiotto:2011tf}}

A crucial new physical phenomenon in the presence of A-twisted masses is the existence of charged BPS particles which live in a specific vacuum,
rather than interpolating between two vacua. In particular, these charged BPS particles modify the standard
Cecotti-Vafa wall-crossing formalism. The correct wall-crossing formula is a specialization of the 2d-4d wall-crossing formula of
\cite{Gaiotto:2011tf} where the charge lattice is restricted to contain flavor charges only. Intuitively, the $S_{ij}$ factors for
standard BPS solitons are refined to keep track of global charges and new $K_\gamma$ factors are introduced to account for
the contribution to bound states of whole Fock spaces of BPS particles of charge $\gamma$.

We are thus presented with the natural problem of extending our formalism to theories with A-twisted masses.
In the context of LG theories, there is a simple way to recast the problem which allows one to employ directly
much of our standard formalism: one can replace the target space $X$ with some minimal cover $\h X$ on which $W$ is single-valued.

Generically, the fiber of such cover is naturally a torsor for the lattice $\Gamma^f$ of global charges and $\Gamma^f$ acts by
deck transformations. The theory has a finite set $\IV$ of vacua corresponding to the critical points of $W$
on $X$ (these critical points are well-defined even though $W$ is only single-valued up to an additive constant).
The set $\IV$ of vacua is covered by the set $\h \IV$ of critical points of $W$ regarded as a function on $\h X$.  A point in $\h \IV$
is a point in $\IV$ -- labeling a critical point $p\in X$ -- together with a choice of lift of $p$ to $\h X$. The weights attached
to elements of $\h \IV$ which cover a given vacuum $p$ will take the schematic form $z_p + m \cdot \gamma$, where
we use the charge $\gamma$ to label the possible lifts of $p$. \footnote{If we pick a reference sheet, $\gamma$ will be an element
of $\Gamma^f$. If not, $\gamma$ will be an element of some torsor $\Gamma_p^f$.}

The first obstruction one encounters in applying our formalism to build a Theory associated to the infinite set of vacua
$\h \IV$ is the infinite proliferation of possible webs. This obstruction, though, must be purely formal. It is clear that any physical
calculation, say of a strip differential in the underlying LG theory, must only involve a finite set of terms in each matrix element. After all,
the complexes may even be defined over the integer numbers. Some finiteness principle must limit the number of
webs which may occur as $\zeta$-webs in a given physical model.

We expect that such physical restrictions will be encoded in the abstract web formalism by adding some extra selection rules to
the vacuum data which select some subset of all possible webs which still satisfies all the required convolution identities
but has better finiteness properties. A simple example could be a list of allowed pairs of elements in $\h \IV$ which restricts which edges are allowed in the webs.
Such a restriction is compatible with convolution identities and preserves our algebraic structures.

A second, more significant obstruction is the need to accommodate charged BPS particles within the web representation data.
In principle, an abstract edge of slope $m \cdot \gamma$ may represent
the trajectories of multiple particles of charges proportional to $\gamma$. Thus the representation data will have to include
both $R_{p, p',\gamma}$ spaces encoding standard BPS solitons and extra Fock spaces encoding charged BPS particles.
The notions of pairing, representations of fans, etc. will have to be adjusted accordingly.

A categorical wall-crossing formula adapted to this deformed context will have to include a categorification of $K_\gamma$ factors.
It would be interesting to find out the corresponding generalization of the notions of exceptional collections and mutations.
We can sketch here some basic idea for such a generalization, which we can dub ``flavoured exceptional collection''.
\begin{itemize}
\item We still expect to have a collection of basic objects $\fT_p$ attached to vacua in $\IV$.
The spaces of morphisms between these objects will be graded by the lattice $\Gamma^f$.
\item The $\Hop$ spaces will have a graded triangular structure, i.e. the $\gamma$-graded subspace $\Hop_\gamma(\fT_p, \fT_{p'})$ vanishes if $z_{pp'}+ m \cdot \gamma \notin \CH$.
\item The categorification of $S$ walls will be encoded by ``partial mutations'', involving only the $\gamma$-graded subspace $\Hop_\gamma(\fT_p, \fT_{p'})$
associated to the weight which is entering/exiting $\CH$.
\item The categorification of $K_\gamma$, walls will be encoded by ``categorical reflections'' which completely reorganize the triangular structure
of the collection, as $m \cdot \gamma$ enters/exits $\CH$.
\end{itemize}

\subsection{Generalization 2: Surface Defects, Spectral Networks And Hitchin Systems}\label{subsec:CatSpecNet}

One of the main motivations for the present work was the desire to categorify the
2d/4d wall-crossing formula for BPS states associated surface defects in four-dimensional $\CN=2$
theories. For background see \cite{Gaiotto:2011tf,MOORE_FELIX}.
One way to produce such defects is to consider an embedding of two-dimensional Minkowski space $\IM^{1,1}$
into $\IM^{1,3}$ and to couple a $1+1$ dimensional field theory
with $(2,2)$ supersymmetry, supported on the embedded $\IM^{1,1}$, to the ambient four-dimensional
theory.

A $(2,2)$ defect with a $U(1)$ $R$-symmetry, leftover from the bulk $SU(2)_R$ symmetry,
has much in common with a massive $(2,2)$ theory deformed by A-twisted masses.
The low energy bulk gauge symmetries play a similar role to the global symmetries
and the central charge includes a contribution from the bulk gauge charges. Schematically,
\be \label{eq:ccharge}
Z = W_i-W_j + Z_\gamma
\ee
where $Z_\gamma$ is the bulk central charge for a particle of gauge and flavor charge $\gamma$.

A simple $\Omega$-deformation in the plane orthogonal to the defect
(even in the absence of an actual $(2,2)$ defect) is expected to reduce the system to
an effective $(2,2)$ theory \cite{Nekrasov:2009rc}, breaking down the bulk BPS particles to infinite towers
of angular momentum modes, each behaving as a 2d BPS particle.

Based on such an analogy, we expect it should be possible to develop a consistent web formalism
to study the space of ground states of the system in the presence of boundary conditions
or interfaces for the 2d defect, or even for the bulk theory. Compared to the 2d setup with A-twisted masses
of Section \S \ref{subsec:TwistedMasses}, the new ingredient will be the presence of bulk Abelian gauge fields.
Even at the level of the 2d/4d wall-crossing formula the effect of the bulk Abelian gauge fields is rather dramatic:
the $K_\gamma$ factors commute in the 2d setup, but not in the full 2d/4d setup. We expect the effect to be equally dramatic
in the full categorical setup. Although we do not know how to construct such a generalized 2d/4d web formalism, we can
describe some possible applications, especially those which only involve the categorification of $S_{ij}$ transformations.

In theories of class S, characterized by a triplet of data $(\fg, \bar{C},   D)$,  where $\fg$ is a
Lie algebra of ADE type, $\bar{C}$ is a punctured Riemann surface, and $D$ is a collection of codimension two
defects located at the punctures of $\bar{C}$,  there is a canonical surface defect $\IS_z$ associated to a point $z$ on
the ultraviolet curve $\bar{C}$. Its origin in $M$ theory is a semi-infinite $M2$ brane whose boundary
is $\IM^{1,1} \times \{z \}$.  In some regions of parameters this surface defect can be viewed as an
LG model coupled to the ambient four-dimensional theory.

Let us recall the basic mathematical setup for the theory of canonical surface defects
in theories of class $S$ \cite{Gukov:2006jk,Alday:2009fs,Gaiotto:2009fs,Gaiotto:2011tf}.
We begin with the data of an $N:1$ branched cover $\pi: \bar{\Sigma} \to \bar{C}$.
As before we let $C$ be $\bar{C}$ minus the branch points.
Physically, $\pi: \bar{\Sigma} \to \bar{C}$ is the covering of the Seiberg-Witten curve
over the UV curve $\bar{C}$ and mathematically $\bar{\Sigma} $ is  the spectral cover associated with
a Hitchin system.
Families of 1+1-dimensional LG models $\IS_z$ parametrized by $z\in C$ also fit into
this framework.   In such cases, the ambient four-dimensional
theory is trivial.  Whether or not the ambient theory is trivial,
vacua of the $1+1$ dimensional defect theory $\IS_z$ are  labeled by the sheets  $z^{(i)}$ of the covering,
$i=1,\dots, N,$ and hence we identify
\be
\IV(\IS_z) = \pi^{-1}(z).
\ee

In the present paper - with the notable exception of the   Section
\S \ref{subsec:TwistedMasses} - an ordered  pair of vacua $(i,j)$ with $i,j\in \IV$ uniquely determines a soliton
sector for the theory on a spatial domain $\IR$. By contrast, in the theory of
the surface defect $\IS_z$, the soliton sectors are labeled, roughly speaking,
by homology classes of paths on $\bar \Sigma$ connecting the vacua $z^{(i)}$
and $z^{(j)}$. To be slightly more precise, they are labeled by equivalence classes of
open chains $\fc \subset \bar{\Sigma}$ with the constraint that $\p \fc = z^{(j)}- z^{(i)}$.
The set of these ``charges'' is a sublattice of a relative homology lattice
\be
\Gamma(z^{(i)}, z^{(j)})\subset H_1(\bar\Sigma, \{ z^{(i)}, z^{(j)} \};\IZ),
\ee
and is a torsor for a sublattice $\Gamma$ in  $H_1(\bar{\Sigma}; \IZ)$. For example,
when $\fg = A_1$, and the Seiberg-Witten curve is a two-fold cover, $\Gamma$ is
the
anti-invariant sublattice of $H_1(\bar\Sigma;\IZ)$ under the deck transformation.
\footnote{More generally, $\Gamma$ might be a subquotient of $H_1(\bar\Sigma;\IZ)$.}

The central charge associated to a soliton sector $\gamma_{ij} \in \Gamma(z^{(i)}, z^{(j)})$ is just
\be
Z_{\gamma_{i,j}}  = \frac{1}{\pi} \int_{\gamma_{i,j}} \lambda
\ee
where   $\lambda$ is the
Seiberg-Witten differential (the canonical Liouville form for the natural holomorphic
symplectic structure on $T^*\bar{C} $).

The lattice $\Gamma$ is, physically, the character lattice of the group which is the
product of the unbroken gauge symmetry and the continuous global flavor symmetry
of the four-dimensional theory, i.e. the lattice of gauge and global charges of the bulk theory.
The central charge $Z_\gamma$ of the bulk theory is simply given by the contour integral of $\lambda$ along $\Gamma$.

The expression for $Z_{\gamma_{i,j}}$ is a slightly more canonical version of equation \ref{eq:ccharge}.
Although it does not look like a difference of two weights, all the essential constructions in the paper only involve differences
of vacuum weights, and not the vacuum weights themselves, it is not,
 strictly speaking, necessary to identify particular vacuum weights. The edges of webs
can simply be labeled by $\gamma_{i,j}$ (so that the cyclic sum of charges
around a vertex is zero). The phases of $Z_{\gamma_{i,j}} $
suffice to define the slopes of the edges. Similarly, the generalization of
the complexes $R_{ij}$ used in a representation of webs is a set of
 complexes  $R_{\gamma_{ij}}$. The contraction $K: R_{\gamma_{ij}} \otimes R_{\gamma'_{ji}} \to \IZ$
is a symmetric degree $-1$ perfect pairing when $\gamma_{ij} + \gamma'_{ji} =0$. , and so on.

The theory of Interfaces has a very natural formulation in theories of class S
and this was, in fact, one of the main motivations for the discussion in
Section \S \ref{subsec:LG-Susy-Interface} above. To each  path $\wp$ in $C$
connecting $z_1$ to $z_2$ and a choice of phase $\zeta$ one can
define a supersymmetric interface $\fI[\wp,\zeta]$ between the theories $\IS_{z_1}$
and $\IS_{z_2}$. The framed BPS states associated with vacua $z_1^{(i)}$ and
$z_2^{(j')}$ have a ``charge'' in the relative homology lattice $\Gamma(z_1^{(i)}, z_2^{(j')})$.
Just as for the soliton sectors $\Gamma(z_1^{(i)}, z_2^{(j)})$ the lattice
$\Gamma(z_1^{(i)}, z_2^{(j')})$  is a $\Gamma$-torsor
of chains with $\p \fc = z_2^{(j')}- z_1^{(i)}$, up to homology. The interfaces
support framed BPS states and the central charge of these framed BPS states is given by
\be
Z_{\gamma_{i,j'}}  = \frac{1}{\pi} \int_{\gamma_{i,j'}} \lambda
\ee
If we label vacua by $z_2^{(j')}$ and $z_1^{(i)}$ then the  Witten index
of framed BPS states $\fro(\fI[\wp,\zeta], ij')$
will in general be infinite. However, we can
grade the cohomology by characters of a global symmetry
group with character lattice $\Gamma$.   Then we can
 consider the   indices $\fro(\fI[\wp,\zeta], \gamma_{ij'} )$ of the subspace transforming
 in a given representation. From physical reasoning the indices with fixed charge $\gamma_{ij'}$ are expected to be finite.

One of the interesting aspects of the theory of surface defects is that one
can construct a ``nonabelianization map'' which is a converse to the standard
abelianization map of the theory of Hitchin systems. (See
\cite{Gaiotto:2012rg} Section 10 for the definition of the nonabelianization map
and \cite{Gaiotto:2012db} for an extended
example. See also \cite{MOORE_FELIX} for more expository remarks.) We now describe
how that is related to our theory of Interfaces.

Given a phase
$\zeta = e^{\I \vartheta}$ one can construct (WKB) $S_{ij}$-walls on $\bar{C}$ which are
essentially the same as the $S_{ij}$ walls used in this paper.
A suitable collection of such walls forms a graph on $\bar{C}$ known as   a ``spectral
network'' $\CW$, so-called because the combinatorics of the network allow one to construct the
spectrum of BPS degeneracies of the 2d4d system \cite{Gaiotto:2012rg,Gaiotto:2012db,Galakhov:2013oja,Galakhov:2014xba}.
In the theory of
Hitchin systems  the spectral curve   $\bar{\Sigma}$ is equipped
with a holomorphic line bundle $L$ with a  flat connection $\nabla^{\rm ab}$
\cite{Hitchin,Frenkel:2007tx}. The map from the nonabelian Hitchin system on $\bar{C}$ to
the flat abelian connection on $\bar{\Sigma}$ is known as the ``abelianization map.''
Conversely, given a line bundle $L$ with flat connection $\nabla^{\rm ab}$
on $\bar{\Sigma}$  we can define the parallel transport $F(\wp)$ of a flat nonabelian connection on a
certain rank $N$ bundle $E\to \bar{C}$ along a path $\wp \subset \bar{C}$ using
\be\label{eq:NonabMap}
F(\wp) = \sum_{\gamma_{ij'}}  \fro(\fI[\wp,\zeta],\gamma_{ij'}) \CY_{\gamma_{ij'}}.
\ee
We have $E \cong \pi_*(L)$ away from the network $\CW$ and $\CY_{\gamma_{ij'}}$ are the parallel transports using the connection
$\nabla^{\rm ab}$ on   $L\to \bar{\Sigma}$.

If we assume the existence of a categorification of $S_{ij}$ walls in the full 2d/4d setup, either
defined by a direct web construction or through the abstract categorification sketched at the end of Section \S  \ref{subsec:TwistedMasses},
then one will obtain directly a categorification of equation \eqref{eq:NonabMap}. $F(\wp)$ is generalized
from the parallel transport operator associated with a flat nonabelian connection
 to an $A_\infty$-functor between Brane categories for the surface defect theories $\IS_{z_1}$ and $\IS_{z_2}$,
 implemented via an Interface $\fI[\wp]$ as in Section \S \ref{sec:CatTransSmpl}. The Chan-Paton factors of this Interface
 provide a ``lift'' of the framed BPS degeneracies $\fro(\fI[\wp,\zeta],\gamma_{ij'})$
 to complexes $\CE_{\gamma_{ij'} }$. Our rules for the composition of
 rotation Interfaces $\fR[\vartheta(x)]$ which do not cross $S$-walls, as well as those for the wall-crossing
 Interfaces $\fS^{p,f}_{ij}$,  can be recognized as
  a categorification of the ``detour rules'' of \cite{Gaiotto:2012rg,Gaiotto:2012db,MOORE_FELIX}.

The parallel transport $F(\wp)$ is that of a flat connection on $\bar C$, not just $C$. That is, it smoothly extends over the branch points of the covering. This is its claim to fame!
 Now, we discussed the sense in which the corresponding functor $F[\wp]$ on Brane categories
 is homotopy invariant in Sections \S \ref{sec:CatTransSmpl}
and \S \ref{sec:GeneralParameter}, but we did not discuss the crucial notion of homotopy invariance for deforming
 $\wp$ across a branch point of the cover $\pi: \bar\Sigma\to \bar C$. This will involve extending our formalism to
 theories which are not completely massive, because at a branch point
 two vacua have coinciding values of $W_i$ and hence some solitons become massless. Some
 preliminary remarks about this issue can be found in Section \S \ref{subsec:partialRG}.
 We leave the matter here for the present paper. Clearly, it will be an interesting project to generalize
 the considerations of this paper to the case where $\pi: \bar\Sigma\to \bar C$ is a branched cover with nontrivial
 monodromy and where $\lambda$ has periods densely filling the complex plane.

One generalization of the surface defects $\IS_{z}$ studied in the literature
is of great significance for the applications to
knot homology. In the M-theory context we can imagine several parallel semi-infinite $M2$ branes ending
on the $M5$-brane. So the boundary is now $\IM^{1,1} \times \{ z_1, z_2, \dots , z_n \}$
where the $z_i$ are distinct points of $C$. Naively, these M2 branes are mutually BPS and
would seem to have no effect on each other. However there are in fact interesting ``topological interactions''
and, as we will see below,  these are responsible for nontrivial knot homologies. Indeed,
the knot homologies are closely related to the spaces of framed BPS states for the generalized
surface defect theories $\IS_{z_1, \dots, z_n }$. We turn to a more detailed
 discussion of potential  applications to knot homology in Section \S \ref{subsec:KnotHomology}.

It is likely that a  study of categorical wall-crossing for surface defects would have other
interesting applications.  It might  provide new insights in a variety of
interesting subjects, such as quantum Teichm\"uller theory and the Stokes theory of asymptotics of holomorphic functions on Hitchin moduli space.

\subsection{Generalization 3: Hierarchies Of Scales And Cluster-Induced Webs}\label{subsec:partialRG}

In this Section we sketch an interesting construction that becomes
available when there is a hierarchy of scales among sets of vacuum
weights. There are several potential applications of the construction
described at the end of this Section.

By a hierarchy of scales we mean that we consider   Theories
in which  we can divide up the vacua into a
disjoint union:
\be
\IV = \amalg_{\mu}\IV^{(\mu)},
\ee
where the vacuum weights form well-separated clusters. Here
the labels $\mu$ run over some finite set. (As we will soon see,
it is a set of Theories.)
The vacua will be denoted as $(\mu,i)$, $i=1,\dots, \vert \IV^{(\mu)}\vert$.
The vacuum weight functions are denoted $z^{(\mu)}: \IV^{(\mu)}\to \IC$
 and specific vacuum weights are denoted by $z^{(\mu)}_{i}$,  $i=1,\dots, \vert \IV^{(\mu)}\vert$.
We  let $Z_\mu$  be the center of mass of the $z^{(\mu)}_{i}$ for fixed $\mu$.
Thus we can write
\be
z^{(\mu)}_{i} = Z_\mu +   \delta z^{(\mu)}_{i}
\ee
To be precise, by a \emph{hierarchy of scales} we mean that, for all $\mu,i, \nu, \lambda$
\be
\vert \delta z^{(\mu)}_{i}\vert \ll \vert Z_{\nu\lambda}\vert
\ee
where $Z_{\nu\lambda}:= Z_{\nu}- Z_{\lambda}$. Note, in particular, that
the convex hulls of the images of the $z^{(\mu)}$ for different $\mu$ do not intersect.

Let us now assume that we are given a Theory $\CT$, in the sense of Section \S \ref{subsec:WebRepPlane}.
If the convex hulls of $z^{(\mu)}_i$ do not intersect (for fixed $\mu$), then by the dual interpretation
of webs through convex polygons  (Section \S \ref{planewebs}, Remark 5)
it follows that any web $\fw$ with a fan $I_\infty$ which involves only vacuum weights of type $\mu$
will also only have pairs of weights of type $\mu$ on internal edges as well. Therefore,
if we consider the restriction of  the web representation $\CR$ and the interior amplitude  $\beta$ to vacua purely of
type $\mu$ then the data $(\IV^{(\mu)}, z^{(\mu)}, \CR^{(\mu)}, \beta^{(\mu)} )$ by themselves
define a Theory, which we will denote $\CT^{(\mu)}$. In particular, $\beta^{(\mu)}$ satisfies
the $L_\infty$ Maurer-Cartan equation. Note that if $\vert \IV^{(\mu)}\vert = 1$
then the Theory $\CT^{(\mu)}$ is trivial.

Let us now consider two sub-Theories $\CT^{(\mu)}$ and $\CT^{(\nu)}$ with $\mu\not=\nu$.
We claim that the data of the parent Theory $\CT$ allows us to construct an Interface
$\fI^{(\mu,\nu)} \in \fB\fr( \CT^{(\mu)}, \CT^{(\nu)})$ between the sub-Theories,
where the domain wall $D_{\mu\nu}$ is parallel to $Z_{\mu\nu}$.
The Chan-Paton factors of the Interface are given by
\be\label{eq:ThryThry-CP-data}
\CE(\fI^{(\mu,\nu)})_{(\mu,i),(\nu,j')} := R_{(\mu,i),(\nu,j')}
\ee
\begin{figure}[htp]
\centering
\includegraphics[scale=0.3,angle=0,trim=0 0 0 0]{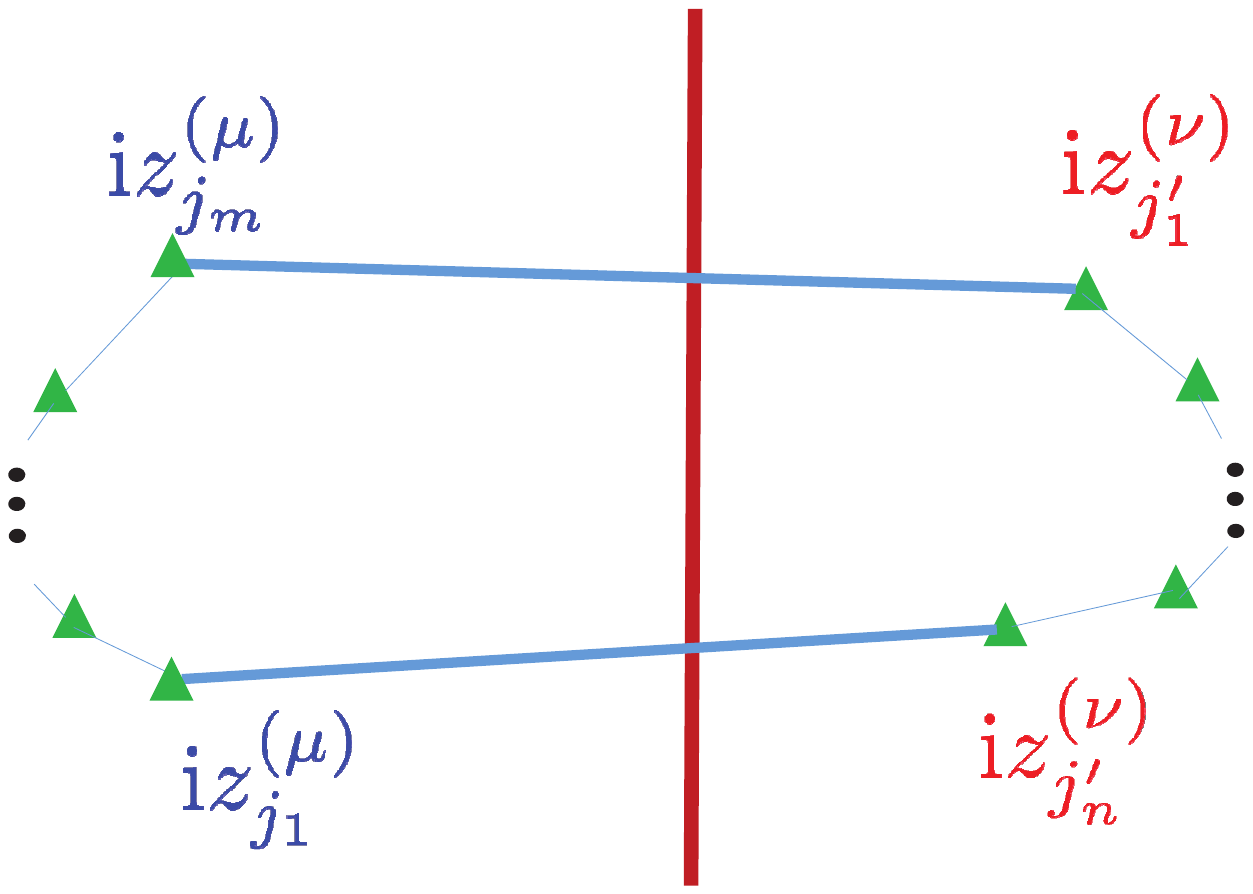}
\caption{The dual of a fan in the Theory $\CT$ which involves two half-plane fans in
the sub-Theories $\CT^{(\mu)}$ and $\CT^{(\nu)}$. This can be interpreted as
defining an interface fan between the sub-Theories.   Heavy blue
lines connect vacua of type $\mu,\nu$. Light blue lines connect vacua of the
same type.  The vertical maroon line is parallel to $Z_{\mu\nu}$.     }
\label{fig:THEORYINTERFACE-1}
\end{figure}

The absorption/emission amplitudes $\CB$ of $\fI^{(\mu,\nu)}$ are derived from $K$ and
$\beta$ of the parent theory as follows. Consider a fan of vacua $I_\infty$ in $\CT$ which involves
only the $\mu$ and $\nu$-type vacua. Because the clusters are well-separated, and
the fan must involve a convex set of weights it will be of the form $I_\infty=\{ J^+, J^- \}$
where
\be
\begin{split}
J^+ & = \{ (\nu,j_1'), (\nu, j_2'), \dots, (\nu, j_n') \}\\
J^- & = \{(\mu,j_1),(\mu,j_2),\dots, (\mu,j_m) \} . \\
\end{split}
\ee
See Figure \ref{fig:THEORYINTERFACE-1}, which should be compared to Figure
\ref{fig:DOMAINWALL-CHANPATON}. Now, the component of the interior amplitude
$\beta$ of the parent Theory $\CT$ with this fan at infinity is a degree two
element
\be
\beta_{I_\infty} \in R_{(\mu,j_m), (\nu,j_1')} \otimes  R^+_{J^+}  \otimes R_{(\nu, j_n'), (\mu,j_1) }
\otimes  R^-_{J^-}
\ee
where
\be
\begin{split}
R^+_{J^+} & =   R_{ (\nu,j_1'), (\nu, j_2')  } \otimes
\cdots \otimes R_{ (\nu,j_{n-1}'), (\nu, j_n')  } \\
R^-_{J^-} & =    R_{ (\mu,j_1),(\mu,j_2) } \otimes
\cdots \otimes R_{ (\mu,j_{m-1}), (\mu,j_m) } . \\
\end{split}
\ee
If we apply $\check K_{(\nu, j_n'), (\mu,j_1) }: R_{(\nu, j_n'), (\mu,j_1) }
\rightarrow R_{(\mu,j_1),(\nu, j_n') }^*$ then we get a degree one element
$\CB_I$ for an interface amplitude appropriate to the Chan-Paton data
\eqref{eq:ThryThry-CP-data}. (Compare equation \eqref{eq:RJ-intfc} above.)
We claim that in fact
\be\label{eq:ThryThry-intfc-amp}
\CB_{J^+,J^-} :=  \check K_{(\nu, j_n'), (\mu,j_1) }(\beta_{I_\infty} ).
\ee
is an interface amplitude. (We are not being careful about signs here.)

\begin{figure}[htp]
\centering
\includegraphics[scale=0.3,angle=0,trim=0 0 0 0]{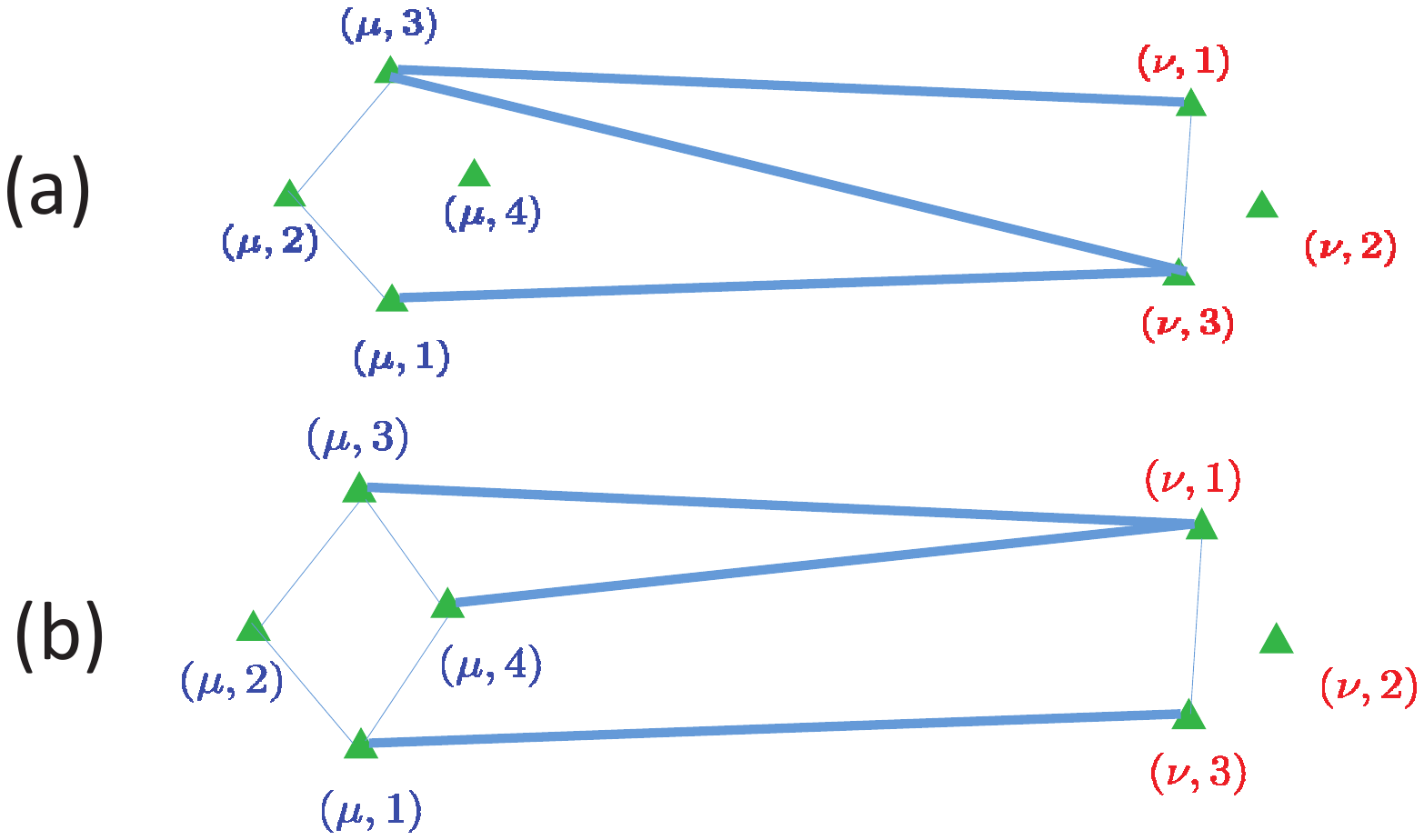}
\caption{Two representative decompositions of a dual polygon giving a web with
$I_\infty = \{ (\nu,1), (\nu,3),(\mu,1),(\mu,2),(\mu,3) \} $.    }
\label{fig:THEORYINTERFACE-7}
\end{figure}
\begin{figure}[htp]
\centering
\includegraphics[scale=0.3,angle=0,trim=0 0 0 0]{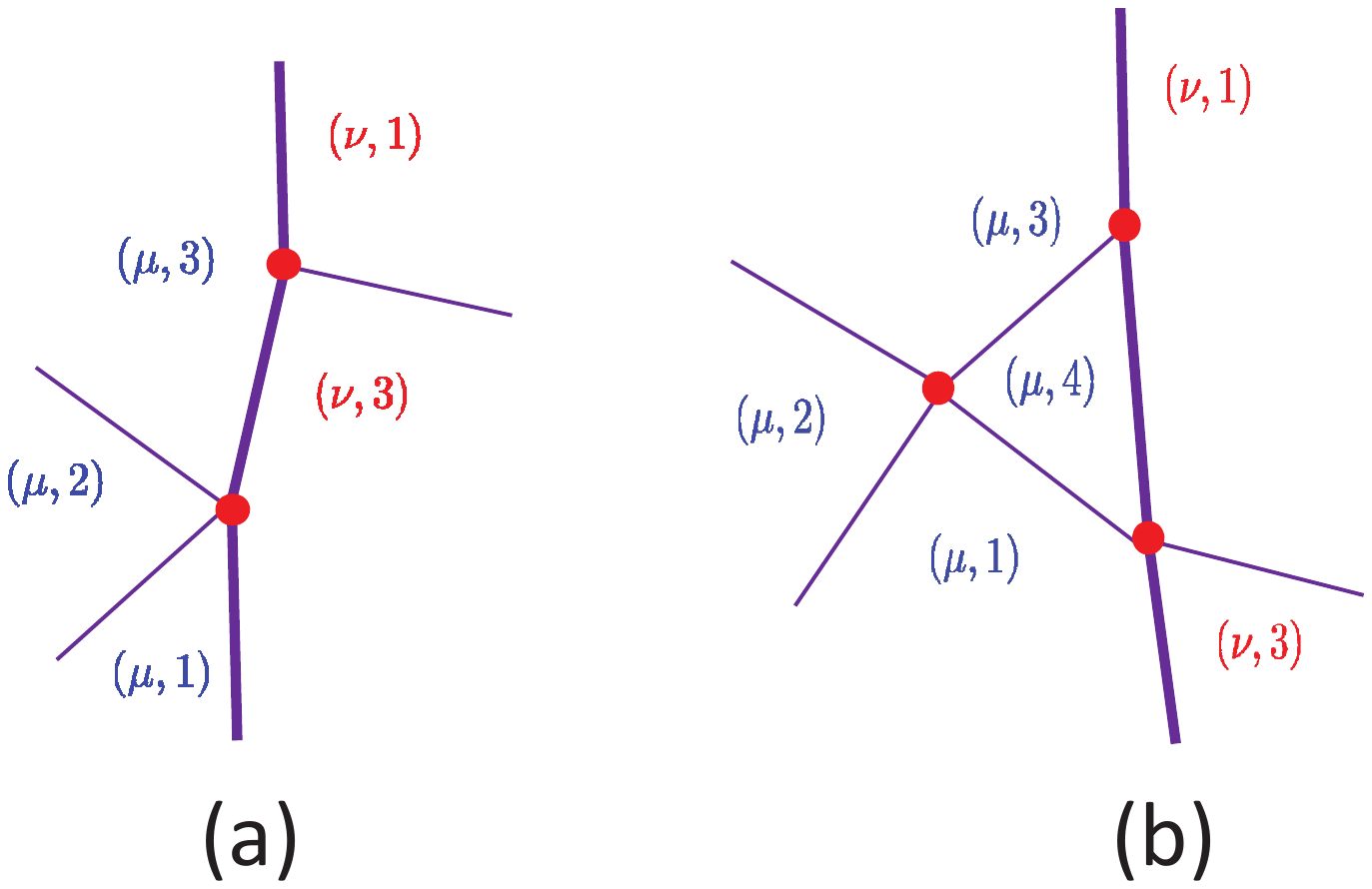}
\caption{Two webs with
$I_\infty = \{ (\nu,1), (\nu,3),(\mu,1),(\mu,2),(\mu,3) \} $. Heavy purple lines
separate vacua of type $\mu,\nu$. They are approximately vertical and correspond to
the domain wall of a corresponding interface. Light purple lines separate vacua of the same type.
They correspond to edges of webs in the corresponding sub-Theories.       }
\label{fig:THEORYINTERFACE-6}
\end{figure}

To prove this, let us consider the plane webs $\fw$ contributing to the $L_\infty$ identity
satisfied by $\beta$ with fan of vacua $I_\infty=\{ J^+, J^- \}$.
These are in one-one correspondence with decompositions of the convex polygon with
vertices
\be
\I z^{(\mu)}_{j_m},\I  z^{(\nu)}_{j_1'}, \dots, \I z^{(\nu)}_{j_n'},\I  z^{(\mu)}_{j_1}, \dots, \I z^{(\mu)}_{j_m}
\ee
into convex polygons with vacuum weights as vertices. We can separate the edges into two types.
Edges connecting vacua of different types are called ``heavy,'' because in the Landau-Ginzburg
incarnation the corresponding classical solitons will be heavy.
\footnote{There is a potential for confusion here. A given heavy classical soliton which
connects vacua of different types can function as a domain wall between the Theories. But
our domain walls are labeled by pairs of Theories, and not by specific solitons between
the vacua of different types. }
In the limit that the clusters
become infinitely separated all the heavy lines will   be parallel to $Z_{\mu\nu}$. Edges connecting vacua
of the same type are called ``light.'' The heavy lines are all nearly parallel and their
common parallel serves as a locus for the domain wall $D_{\mu\nu}$. Note that fixing the
transverse position of the domain wall eliminates one degree of freedom so in this mapping
of plane webs for $\CT$ to interface webs between $\CT^{(\mu)}$ and $\CT^{(\nu)}$, taut webs
are mapped to taut webs. Conversely, every taut interface web between $\CT^{(\mu)}$ and $\CT^{(\nu)}$
can be ``lifted'' to a taut plane web for $\CT$ .

The extra factor $K$ used in defining $\CB_{J^+,J^-}$ in equation
\eqref{eq:ThryThry-intfc-amp} is precisely what is needed in order
to convert the contraction of interior amplitudes for
 a plane web $\rho(\fw)$ in $\CT$ to the contraction
of interface webs $\rho(\fd)$ between Theories $\CT^{(\mu)}$ and $\CT^{(\nu)}$.
Thus the $L_\infty$ equations satisfied by $\beta$ in the parent theory become
the $A_\infty$ equations satisfied by the interface amplitude of equation \eqref{eq:ThryThry-intfc-amp}.
The ordering of   vertices  along the heavy edges nearly parallel   to $Z_{\mu\nu}$ collapses
the $L_\infty$ combinatorics to $A_\infty$ combinatorics.
In conclusion, a pair of far separated clusters of vacua canonically defines
an  Interface $\fI^{(\mu,\nu)}$ as claimed above.

We can now envision a nontrivial generalization of our entire formalism, where
vacua are replaced by Theories, and edges of webs support Interfaces. These
Interfaces will themselves interact at junctions, thus we again have a system of
webs, which we will call \emph{cluster-induced webs}, whose  edges are parallel to $Z_{\mu\nu}$.
Moreover, on a half-plane the Interfaces can have junctions at the boundary of
the half-plane, leading to cluster-induced half-plane webs where again edges are associated to Interfaces
and segments on the boundaries are associated with Branes within the various Theories.
We will next sketch how this idea can be made more precise, but we will leave detailed
verification of the full picture for future work.

\def\fW{{\mathfrak{W}}}

\begin{figure}[htp]
\centering
\includegraphics[scale=0.3,angle=0,trim=0 0 0 0]{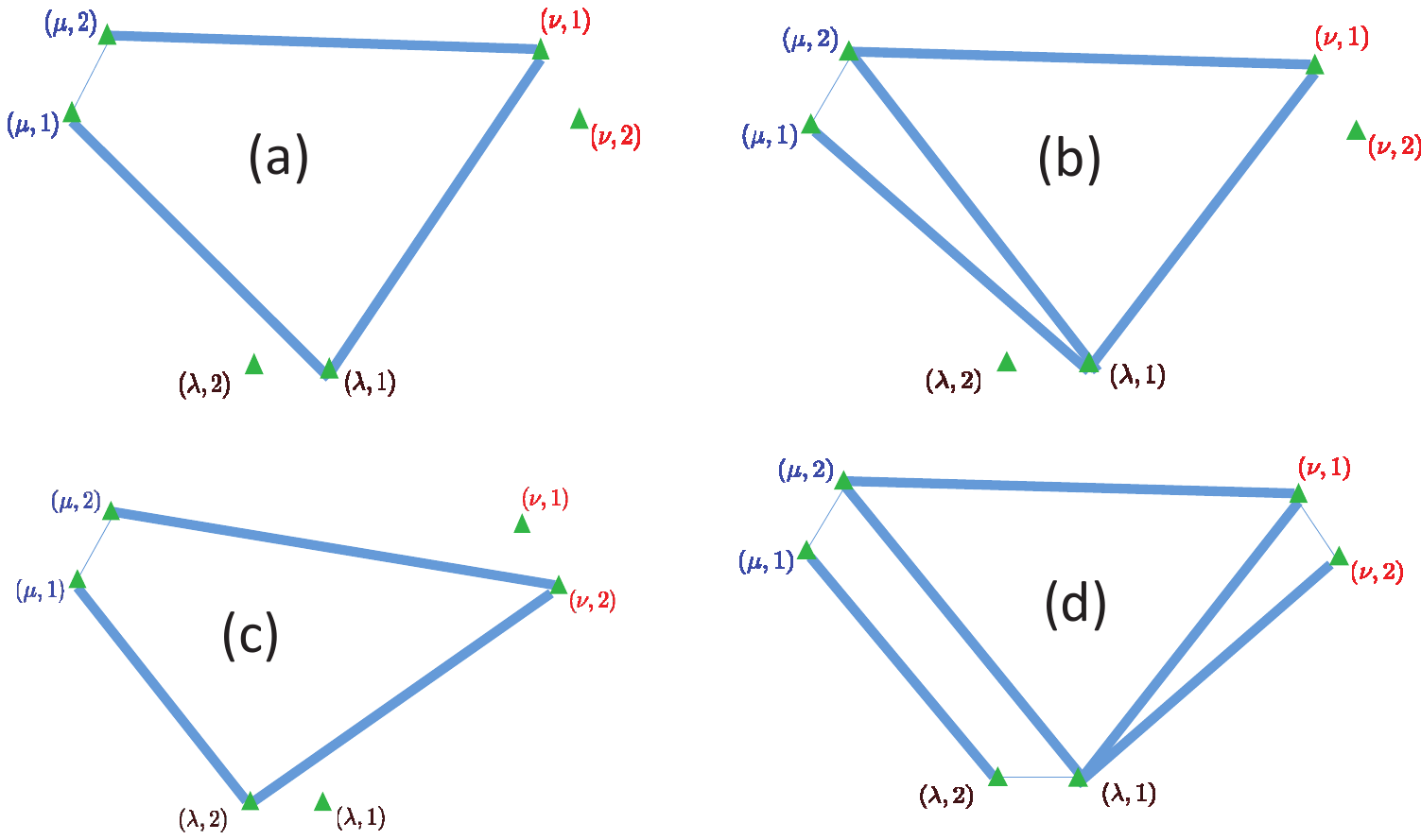}
\caption{We illustrate three clusters of vacua of type $\mu,\nu,\lambda$. Associated to this
``fan of Theories'' are several fans of vacua $I_\infty$ in the parent theory as well as several
webs in the parent Theory with fixed $I_\infty$. A few examples are shown here.      }
\label{fig:THREE-THRY-VTX-1}
\end{figure}

\begin{figure}[htp]
\centering
\includegraphics[scale=0.3,angle=0,trim=0 0 0 0]{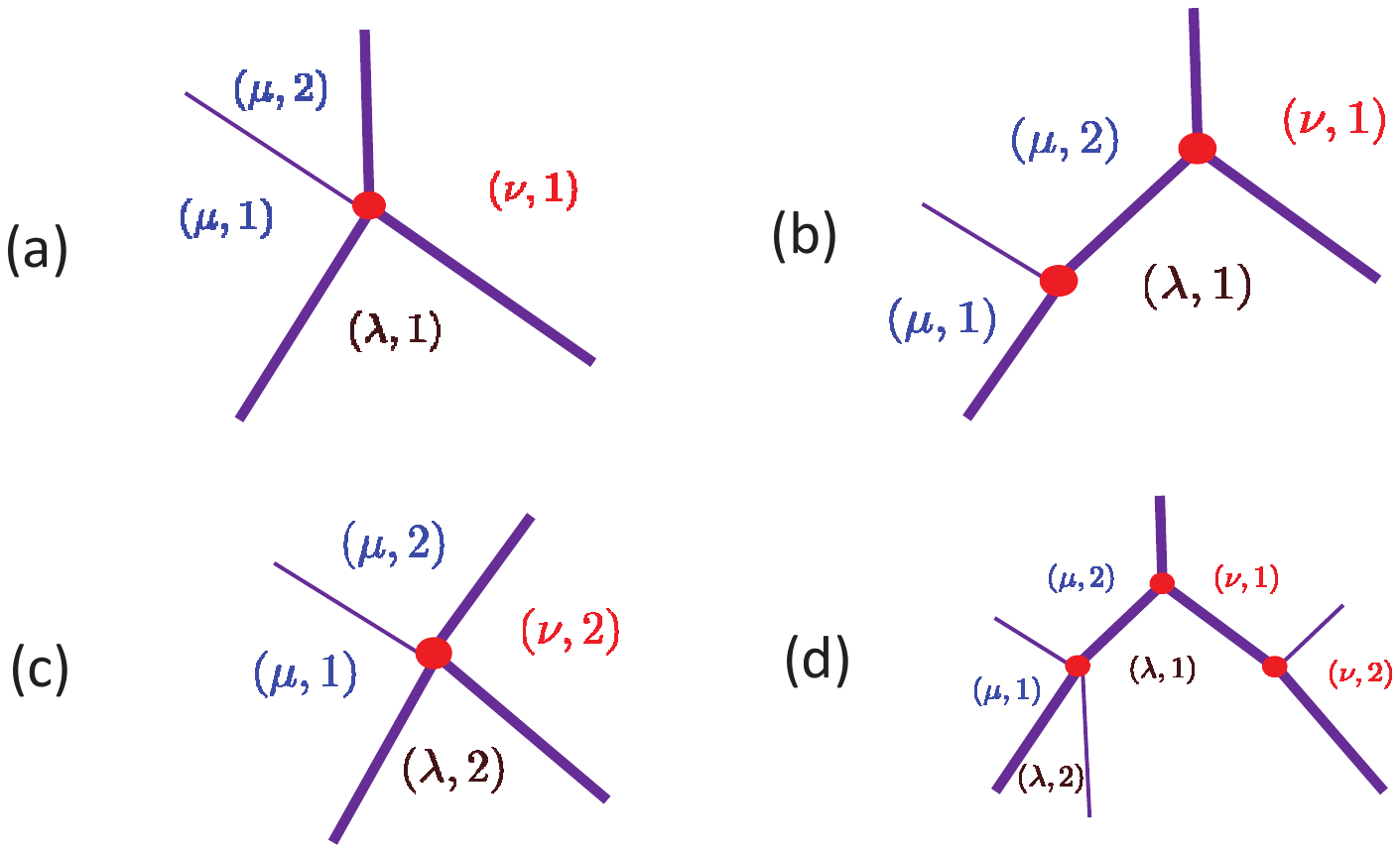}
\caption{This figure shows some of the webs associated with the fan of Theories
$\{ \CT^{(\mu)}, \CT^{(\nu)}, \CT^{(\lambda)} \} $.      }
\label{fig:THREE-THRY-VTX-2}
\end{figure}

\begin{figure}[htp]
\centering
\includegraphics[scale=0.3,angle=0,trim=0 0 0 0]{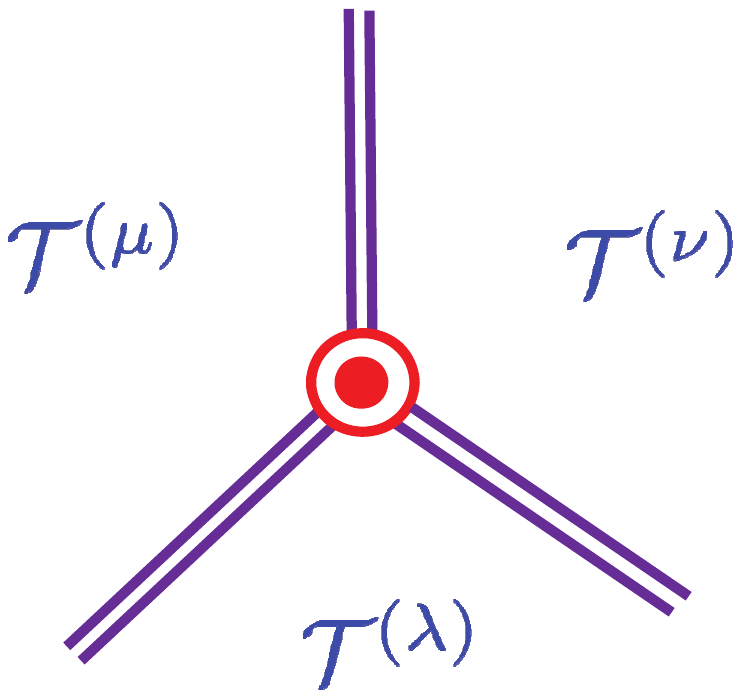}
\caption{A junction of Interfaces $\CT^{(\mu,\nu)}, \CT^{(\nu,\lambda)}, \CT^{(\lambda,\mu)}$.  This serves as  a
vertex for the cluster-induced webs.      }
\label{fig:THREE-THRY-VTX-3}
\end{figure}

We first use a key idea from the construction of the Interfaces $\fI^{(\mu,\nu)}$.
We consider webs $\fW$ whose set of vacuum labels   are the Theories $\CT^{(\mu)}$ and whose
vacuum weights are the center of mass coordinates $Z_\mu$. These are the
\emph{cluster-induced webs} mentioned above. In the limit that the clusters of
vacua $z^{(\mu)}_i$ are well-separated for different $\mu$, to every web $\fw$ of the parent
Theory we can associate a cluster-induced web $\fW[\fw]$, called the \emph{skeleton of $\fw$}. It is defined by keeping
the heavy lines in $\fw$ and collapsing the light lines. Conversely, given a web of type $\fW$
there will be several webs $\fw_1, \fw_2,\dots $ in the parent Theory which have the same
skeleton $\fW$. Instead of web representations, to the edges of $\fW$ we associate the
Interfaces $\fI^{(\mu,\nu)}$.
See, for example Figures \ref{fig:THREE-THRY-VTX-1}, \ref{fig:THREE-THRY-VTX-2}, and
\ref{fig:THREE-THRY-VTX-3}.


The vertices of $\fW$ correspond now to junctions of Interfaces. See again Figure \ref{fig:THREE-THRY-VTX-3}.
Instead of a representation of a fan, the natural object to associate to a cyclic junction of
Interfaces $\fI^{(\mu_1,\mu_2)}, \fI^{(\mu_2,\mu_3)}, \dots, \fI^{(\mu_n,\mu_1)}$ is the
chain complex defined by the trace of the composite Interface. Recall that each Interface
$\fI^{(\mu,\nu)}$ is associated with a domain wall parallel to $  Z_{\mu\nu}$, so let $\fR_{\mu,\nu}$
be the rotation Interface of Section \S \ref{sec:CatTransSmpl} (defined in the parent Theory $\CT$)
that rotates $Z_{\mu\nu}$ through an angle less than $\pi$ to be vertical (so the Interface domain wall is vertical).
Then, instead of associating a fan representation $R_I$ to a vertex, as we do in
the parent theory $\CT$,   to a junction of Interfaces we now associate the chain complex:
\be
\underline{R}_{\mu_1, \dots, \mu_n} :=
\Tr \biggl[ \left( \fI^{(\mu_1,\mu_2)}\IntfcTimes \fR_{\mu_1,\mu_2} \right) \IntfcTimes
 \left( \fI^{(\mu_2,\mu_3)}\IntfcTimes \fR_{\mu_2,\mu_3} \right) \IntfcTimes \cdots
 \IntfcTimes  \left( \fI^{(\mu_n,\mu_1)}\IntfcTimes \fR_{\mu_n,\mu_1} \right) \biggr]
\ee
where the trace of an Interface was defined in Section \S \ref{sec:LocalOpsWebs} above.

Now, we conjecture that the data of the interior amplitude $\beta$ of the parent Theory
allows us to construct distinguished elements in the complexes associated to the
vertices of $\fW$, say $\underline{\beta}_{\mu_1, \dots, \mu_n } \in \underline{R}_{\mu_1, \dots, \mu_n}$.
Then, the direct sum over all fans of Theories defines an analog $\underline{R}^{\rm intfc}$ of
$\Rvtx$. We further conjecture that $\underline{R}^{\rm intfc}$ carries the structure
of an $L_\infty$ algebra and $\underline{\beta}_{\mu_1, \dots, \mu_n } $ define a solution to the
Maurer-Cartan equation of that $L_\infty$ algebra.

We can be more specific. The notion of amplitude for a ``web of interfaces'' $\fW$ makes perfect sense in our general algebraic setup even outside the context of cluster-induced webs.
Given a general collection of Theories $\CT^{(\mu)}$ attached to the faces of the web, of Interfaces $\fI^{(\mu,\nu)}$ attached to the edges of the web, each belonging to the category of interfaces with the slope of the corresponding edge,
and a collection of elements $\underline r_a$ in the chain complexes associated to the vertices of the web, we can {\it define} the amplitude $\underline \rho(\fW)[\underline r_a]$ of such a web in a straightforward manner.
We can identify $\fW$ as a sum over composite webs (including edges going into the vertices of $\fW$, as for wedge webs)
of the underlying $\CT^{(\mu)}$ Theories defined within the corresponding faces, and define $\underline \rho(\fW)[\underline r_a]$ by inserting the appropriate interior and boundary amplitudes in
$\rho(\fW)[\cdots; \cdots; \underline r_a]$.

The above conjecture can be stated as the claim that for the Theories and Interfaces defined in this section, there exist a
sum of cluster-induced webs $\ft_{\mathrm{cluster}}$ such that $\underline \rho(\ft_{\mathrm{cluster}})[\underline r_a]$
defines an $L_\infty$ algebra and $\underline{\beta}_{\mu_1, \dots, \mu_n } $ is a solution to the
Maurer-Cartan equation of that $L_\infty$ algebra. We expect $\ft_{\mathrm{cluster}}$ to arise from the decomposition into cluster-induced webs of the taut element
$\ft$ in the underlying theory.

The plane webs $\fW$ described above should have half-plane analogs.  Let us fix the positive half-plane.
Then, in close analogy to the definitions of the Theories $\CT^{(\mu)}$, a Brane $\fB$ with Chan-Paton
spaces $\CE_{(\mu,i)}$ and boundary amplitudes $\CB_{J}$ in the parent Theory $\CT$ defines a collection
of Branes $\fB^{(\mu)}$, one for each Theory  $\CT^{(\mu)}$. The key observation is again that if
$J_\infty$ involves only vacua of type $\mu$ and all the emission amplitudes only involve vacua
of type $\mu$ then all the interior vertices must also correspond to fans of
 vacua solely of type $\mu$. \footnote{Intuitively, heavy lines cannot end away from the boundary and cannot turn back.  }
Consequently,  for fixed $\mu$, we
can define $\fB^{(\mu)}$ to have Chan-Paton spaces $\CE_{(\mu,i)}$, $i=1,\dots, \vert \IV^{(\mu)} \vert$,
with amplitudes $\CB_J$ where $J$ is a half-plane fan of vacua all of type $\mu$. If we contract interior
vertices with $\beta^{(\mu)}$ then these will satisfy the $A_\infty$ identities by themselves, and hence
define Branes in $\CT^{(\mu)}$.

\begin{figure}[htp]
\centering
\includegraphics[scale=0.3,angle=0,trim=0 0 0 0]{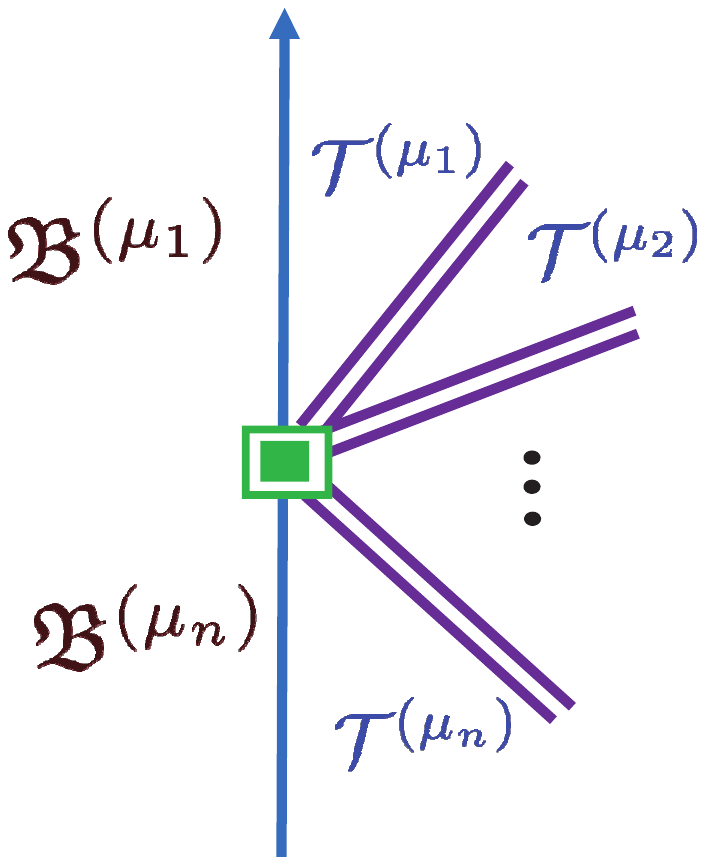}
\caption{Supersymmetric Interfaces can end on boundaries. Shown here is a cluster-induced
half-plane web $\fU$.  The ``emission amplitude'' that
interpolates between Branes of two Theories and joins several interfaces is an amplitude $\underline{\CB}$
constructed from the data of the boundary amplitudes of the underlying parent Theory $\CT$.       }
\label{fig:THRY-BR-JCTN}
\end{figure}

Now, if we consider half-plane fans in the parent theory of type
\be
J_\infty = \{ (\mu,j_1), \dots, (\mu,j_n), (\nu,j_1'), \dots, (\nu,j_m') \}
\ee
the corresponding half-plane webs will involve heavy edges, approximately parallel
to $Z_{\mu\nu}$ (which we assume points into the positive half-plane) and
terminating at the boundary of $\CH$. Thus,
supersymmetric Interfaces $\fI^{(\mu,\nu)}$ should also have boundary junctions.
Instead of a representation $R_J$ associated with an emission vertex in the parent
Theory $\CT$, now, to a junction such as that shown in Figure \ref{fig:THRY-BR-JCTN}  we associate again a complex
\be
\begin{split}
\underline{R} & = \fB^{(\mu_1)} \IntfcTimes  \left( \fI^{(\mu_1,\mu_2)}\IntfcTimes \fR_{\mu_1,\mu_2} \right) \IntfcTimes
 \left( \fI^{(\mu_2,\mu_3)}\IntfcTimes \fR_{\mu_2,\mu_3} \right) \IntfcTimes \cdots
 \IntfcTimes  \left( \fI^{(\mu_n,\mu_1)}\IntfcTimes \fR_{\mu_n,\mu_1} \right)\IntfcTimes \fB^{(\mu_n)}[\pi] \\
\end{split}
\ee
The analog of the Chan-Paton factors are now the Branes $\fB^{(\mu)}$ in the Theories $\CT^{(\mu)}$.
A conjecture analogous to that we made above states that the data of the boundary amplitudes $\CB_J$
of the parent theory allow us to define elements $\underline{\CB}$ of these complexes so that the $A_\infty$ relations
of the half-plane webs $\mathfrak{U}$ associated to data $(\CT^{(\mu)}, Z_\mu, \underline{R}, \underline{\beta})$
will be satisfied. Again, the operations of the $A_\infty$ category will be defined from the
amplitudes of appropriate webs of Interfaces drawn in the half-plane, including both the  $\fB^{(\mu)}$ Branes and
the $\fI^{(\mu,\nu)}$ Interfaces.

In conclusion, we have sketched how, in the limit that vacuum weights form well-separated clusters we can build
a new set of webs, called cluster-induced webs  and a ``representations of webs'' (the chain complexes $\underline{R}$)
with elements $\underline{\beta}$ and $\underline{\CB}$ satisfying the $LA_\infty$ relations. If, then,
there are hierarchies of clusters within clusters there will be a corresponding hierarchy of these cluster-induced
structures. This might be a way to generalize our formalism to infinite collections of vacuum weights with accumulation
points. But we leave that for the future.

Finally, let us sketch some of the physical motivations and potential applications of the above
construction:

\begin{enumerate}

\item In this paper we have heavily used  the assumption that
the IR vacua of the   field theory under study are all massive.
It is natural to ask whether the formalism can be extended
to include interacting massless vacua, and in particular
nontrivial interacting CFT's. We believe the the above
construction can be used to define a web-formalism for
such interacting CFT's. In the context of LG theories, massless vacua
appear when we consider families of superpotentials
such that one or more Morse critical points approach a
common (non-Morse) critical point, and the cluster-webs
should allow a description of the physics of these models.
An interesting open problem is whether the formalism then
applies to non-Morsifiable singularities.

\item A closely related application is the extension of
the application to categorified spectral networks to include the
branch points of the spectral cover $\pi: \bar \Sigma \to \bar C$,
as mentioned in Section \S \ref{subsec:CatSpecNet} above.

\item A third application is to the construction of creation and fusion
Interfaces  in the knot homology application of our formalism sketched in
 Section \S \ref{subsec:KnotHomology} below. See, in particular
 Section \S  \ref{subsec:CategoryAppl}.  We turn to these matters next.

\end{enumerate}

\subsection{Potential Application: Knot Homology}\label{subsec:KnotHomology}

\subsubsection{The Main Point}

\def\kwc{ {\rm \textbf{c}} }
\def\kwt{ {\rm \textbf{t}} }
\def\Ad{{\rm Ad}}
\def\tvt{{\tilde \vartheta}}

Knot homology is an important topic in low-dimensional topology. It has
interesting relations to string theory and gauge theory and there
have been several interpretations of knot homology in the physics
literature.  The gauge-theoretic definition of knot homology proposed  
in \cite{Witten:2011zz} can be deformed to a setup which has many properties in common
with a massive 2d theory with $\CN=(2,2)$ supersymmetry compactified on a segment \cite{Gaiotto:2011nm}.

The objective of this Section is to describe how one might possibly employ the machinery developed in the present paper
in order to express the gauge-theoretic definition of knot homology in the language of webs, web representations,
Theories and Interfaces.

The immediate payoff of such a translation would be to bridge the conceptual gap
between the gauge-theoretic definition of knot homology, which has a direct relationship to the
three-dimensional geometry of the knot, and the standard combinatorial definitions of
Khovanov cohomology, which lack such a relationship. Ultimately, it should be possible to establish a direct equivalence between some
Vacuum $A_\infty$ categories associated to the gauge-theoretic construction and
the categories employed in the combinatorial definitions of Khovanov cohomology.

The gauge-theoretic definition of knot homology employs an MSW complex built from solutions of certain four and five-dimensional
$\CQ$-fixed point equations for a five-dimensional supersymmetric gauge theory on a five-manifold with boundary:
\be
M_5 = \IR \times M_3\times \IR_+,
\ee
where $M_3$ is a three-manifold. The knot resides in $M_3$ on the boundary and is
used to formulate the crucial boundary conditions for the instanton equations of the gauge theory.
Referring to the first factor as ``time'' the   solutions of time-independent, four-dimensional, ``soliton'' equations provide the complex of approximate ground states and the solutions of the
five-dimensional ``instanton'' equations with time-independent boundary conditions provide the differential on the complex.
We will review the setup in full detail in Section \S \ref{subsec:LightningReview}.
The relation to this paper begins to emerge when we realize that the 
 five-dimensional instanton equations are equivalent to the $\zeta$-instanton
equations for a gauged Landau-Ginzburg model whose ``worldsheet'' is $\IR\times \IR_+$ and 
whose target space is a space of complexified gauge connections on $M_3$, as 
described in Section \S \ref{subsec:LightningReview}.
The superpotential is the Chern-Simons functional. (This gauged LG model is referred to as CSLG1 below.) 
In the case when $M_3 = \IR \times C$, with $C$ a Riemann surface,
the equations are also the $\zeta$-instanton equations for another gauged Landau-Ginzburg model
(referred to as CSLG2 below). 
Referring to the first two coordinates of $M_5 = \IR\times\IR\times C \times \IR_+$ 
as $(x^0,x^1)$ the ``worldsheet'' of CSLG2 is the $(x^0,x^1)$ plane, while the target 
  is a space of complexified gauge fields on $\tilde M_3 = C \times \IR_+$. Again, the 
superpotential is a Chern-Simons functional. Since the Chern-Simons functional is 
multivalued, $dW$ can have periods, and the considerations of Section \S \ref{subsec:TwistedMasses}
become important. 
In either case, the knot complex is the MSW complex for the Landau-Ginzburg theory.

In order to illustrate the relation to the web formalism, we focus on the case when $M_3 = \IR\times C$ 
and use the formulation CSLG2. We should stretch the 
link along one spatial direction $x^1$, and introduce the
deformation of the boundary conditions proposed in \cite{Gaiotto:2011nm}. In the approximation that the link's strands are
parallel to the $x^1$ direction, the boundary conditions have 2d translation symmetry and the 5d equations have much in common with the $\zeta$-instanton equations of an ungauged massive LG model. 
%
%Indeed, they may be recast as $\zeta$-instanton equations
%for a massive LG model with an infinite-dimensional target space and a superpotential which is defined up to integer multiples of a %constant number.
%\footnote{The multi-valuedness of the superpotential is associated to the instanton number global symmetry in the same way as
%described in Section \S \ref{subsec:TwistedMasses}}
%
In particular, they admit isolated solutions akin to 2d vacua, which are independent of time and the $x^1$ space direction and only depend on the three remaining space directions. Conjecturally, these vacua are {\it massive}, with a mass scale controlled by the deformation parameters and the transverse separation between the
strands. Such a conjecture implies a familiar structure for solutions of the five-dimensional instanton equations: as long as the
 strands of the link are approximately parallel away from co-dimension one loci (``interfaces''), which are in turn 
 well-separated from each other, 
the solutions will be almost everywhere exponentially close to the 2d vacuum solutions, except in the neighbourhood
of a BPS web.

Assuming that this conjecture holds true, it should be possible to employ the 5d instanton equations to define the same variety of
counting problems as we did for the $\zeta$-instanton equation in standard LG theories in Sections \S\S \ref{lgassuper}-\ref{subsec:LG-Susy-Interface}, as long as we deal with the  non-single-valued
vacuum weights as sketched in Section \S \ref{subsec:TwistedMasses}. Thus, we expect that for any collection $S$ of parallel strands:
\begin{itemize}
\item Solutions of the 5d instanton equation which do not depend on $(x^0,x^1)$ will give the vacuum data $\IV_S$.
\item Solutions of the 5d instanton equation which depend only on the combination $x^1 \cos \mu + x^0 \sin \mu$ will provide the spaces of
solitons which can interpolate between any two given vacua and thus web representations for the vacuum data $\IV_S$.
\item Solutions of the 5d instanton equation with fan-like asymptotics in the $(x^0,x^1)$ plane will provide interior amplitudes $\beta_S$
and thus Theories $\CT_S$.

\item If $S$ is an empty collection the theory $\CT_S$ will be trivial. That is, it will have a unique vacuum.
\end{itemize}

Similarly, for any ``supersymmetric interface'' $\CI$, i.e. a time-independent boundary condition for the 5d equations which involves
a set of parallel strands $S^-$ for $x^1\ll -L$ and a set of parallel strands $S^+$ for $x^1\gg L$
\begin{itemize}
\item Solutions of the 5d instanton equation which do not depend on time will give Chan-Paton data $\CE^\CI_{j,j'}$.
\item Solutions of the 5d instanton equation with fan-like asymptotics in the $(x^0,x^1)$ plane will provide boundary amplitudes $\CB^\CI$
and thus an Interface $\fI[\CI]$ between Theories $\CT_{S^-}$ and $\CT_{S^+}$.
\end{itemize}

We can assume that the stretched link is approximated by a sequence of collections of strands $S_a$, starting and ending with the empty collection $S_0 = S_n = 0$,
separated by interfaces $\CI_{a,a+1}$. The approximate ground states and instantons of the knot homology complex
will literally coincide with the chain complex of the Interface $\fI({\rm Link})$ between the trivial Theory and
itself, defined as the composition of the Interfaces $\fI[\CI_{a,a+1}]$
\be\label{eq:Link-Interface0}
\fI({\rm Link}) :=  \fI[\CI_{0,1}]\boxtimes \cdots \boxtimes  \fI[\CI_{n-1,n}].
\ee
This complex is bigraded. One grading is the fermion number used throughout this paper. The second grading 
is related to the instanton number current in the five-dimensional theory and hence to the multi-valuedness 
of the Chern-Simons functional and the considerations of   Section \S \ref{subsec:TwistedMasses}.

Furthermore, if we allow the transverse position of the strands to evolve adiabatically in between discrete events such as recombination of strands,
according to some profile $S_a(x^1)$, we expect the knot homology complex to coincide with the chain complex of an Interface $\fI({\rm Link})$
which now includes the insertion of the corresponding categorical parallel transport interfaces:
\be\label{eq:Link-Interface0}
\fI({\rm Link}) :=  \fI[\CI_{0,1}]\boxtimes \fI[\CT_{S_1(x^1)}] \boxtimes \cdots \cdots \boxtimes \fI[\CT_{S_{n-1}(x^1)}] \boxtimes  \fI[\CI_{n-1,n}].
\ee

In the remainder of this Section, we will review in a little more detail the gauge theory definition of knot homology and the
relation between the five-dimensional instanton equations
and the $\zeta$-instanton equations. We will also review the definitions of a collection of finite-dimensional
auxiliary ungauged Landau-Ginzburg models introduced in \cite{Gaiotto:2011nm}. (These are the ``monopole'' 
 and   ``Yang-Yang'' models described below.) These models are expected to provide a low-energy effective description
for the full gauge theory model in the case when the collection $S$ consists of parallel strands. It might be possible to prove the equivalence of the
Theories $\CT_S$ and Interfaces $\fI[\CI]$ with the Theories and Interfaces computed from these finite-dimensional
ungauged LG models.

\subsubsection{Preliminary Reminder On Gauged Landau-Ginzburg Models}\label{subsec:Remind-GLG}

In Sections \S\S \ref{lgassuper}-\ref{subsec:LG-Susy-Interface}
 above we discussed at length $\CN=(2,2)$ Landau-Ginzburg
sigma models with K\"ahler target $X$ and holomorphic superpotential $W$.
In Sections \S\S \ref{subsec:LightningReview}-\ref{subsec:Reformulation}
 we will reformulate the gauge theoretic
approach to knot homology in terms of Landau-Ginzburg models.
The relevant superpotential, which will be a Chern-Simons functional,
will actually be a degenerate superpotential due to gauge invariance and hence
we need to generalize the discussion of Section \S \ref{lgassuper} slightly to include the case of
\emph{gauged} Landau-Ginzburg models. This is easily done. We briefly summarize
the generalization here. (For background see Section 5.1.1 of \cite{NewLook}. We reduce the
$d=4$, $\CN=1$ gauged nonlinear model of \cite{Wess:1992cp}, ch.24 following
the general procedure of \cite{Witten:1993yc}.)

Suppose the K\"ahler manifold $X$ has a continuous group $S$ of isometries
and suppose moreover that $S$ is a symmetry of the superpotential $W$.
\footnote{Or rather, a symmetry of the current generated by the pullback
of $dW$. The superpotential is allowed to shift by a constant under
symmetry transformations.}
The LG model then has a global symmetry and we can gauge it. We do so
in a supersymmetric way, coupling to a $(2,2)$ vectormultiplet with
bosonic fields $(\sigma, \bar\sigma,B , D)$. All fields are locally valued
in the Lie algebra of $S$.
$B$ is a gauge field for an $S$-bundle on the ``worldsheet.'' The remaining
fields are scalars. $D$ is an auxiliary field. In order to couple the gauge fields
supersymmetrically we assume furthermore that   $S$ acts symplectically on
$X$ so that there is a moment map $\mu: X \to {\rm Lie}(S)^*$.

We can define supersymmetries $\CQ_\zeta$ as before. The fixed point equations
of the topologically twisted theory
are an interesting combination of vortex and $\zeta$-instanton equations:
\begin{subequations}\label{eq:LG-gauged-Qfx}
\begin{align}
\bar\p_B \phi^I & = \frac{\I \zeta}{4} g^{I\bar J} \frac{\p \bar W}{\p \bar \phi^{\bar J}}d\bar w \label{eq:LG-gauged-F}\\
*_2 F_B + \mu & = 0 \label{eq:LG-gauged-D}
\end{align}
\end{subequations}
where $d\bar w = dx^1 - \I dx^0$ is a $(0,1)$ form on the worldsheet and
 the covariant derivatives on the scalars can be written:
\be\label{eq:GSM-CovDer}
d_B \phi^I =  d \phi^I + \langle B, V^I \rangle.
\ee
Here the action of $S$ on $X$ defines vector fields $V$ on $X$ valued in ${\rm Lie}(S)^*$,
that is, an  element $\fx$ of the Lie algebra of $S$ generates a vector field
\be
V(\fx) = V^I(\fx) \frac{\p}{\p \phi^I } + V^{\bar I}(\fx) \frac{\p}{\p \bar\phi^{\bar I}  }.
\ee
In \eqref{eq:GSM-CovDer} the angle brackets denote
 contraction with the ${\rm Lie}(S)$-valued gauge field $B$.
The second set of equations, \eqref{eq:LG-gauged-D}, minimize the kinetic energy
terms of the gauge fields in the action. In the twisted theory one adds a boundary
term $\sim \oint \langle B, \mu \rangle$
to the standard physical action allowing one to complete a square and
write the kinetic energy term for the gauge fields as $\sim \int (*F_B - \mu)^2$.
The final parts of the $\CQ_\zeta$-fixed point equations require that
the   Lie algebra valued field $\sigma$ is covariantly constant and that all the
other fields should be invariant under gauge transformation by $\sigma$.

An important simplification occurs when $S$ acts freely and the critical
points of $W$ correspond to isolated, nondegenerate $S^c$ orbits.  In this case the
low energy behavior of the gauged model is described by an ordinary LG
model on the symplectic quotient. Since the $S$-action  is fixed point free,
$\sigma=0$. Moreover,  since $\mu=0$ in the symplectic quotient we can gauge away
$B$ and the equation reduces to the ordinary $\zeta$-instanton equation.
By geometric invariant theory we know that, with a
 suitable stability condition, the symplectic quotient can be identified,
as a complex manifold, with $X/S^c$ and $W$ descends to a nondegenerate
Morse function on this space.
These conditions will hold in our application below since the Nahm pole boundary
conditions guarantee that the $\CQ_\zeta$-fixed points have no symmetries.

Finally, it is interesting to generalize the formulation of a LG model
as supersymmetric quantum mechanics, as explained in Section
\S \ref{lgassuper} above, to the case of a gauged model. Equation
\eqref{defho} is now generalized to
\be\label{eq:equi-h}
h = - \int_D \biggl[  \phi^*(\lambda)- \langle B, \mu\rangle  - \half \Re(\zeta^{-1} W)dx \biggr]
\ee
The equation for upwards gradient flow for $h$ becomes the pair of equations
\eqref{eq:LG-gauged-D} and \eqref{eq:LG-gauged-F}.

\subsubsection{Lightning Review: A Gauge-Theoretic Formulation Of Knot Homology}\label{subsec:LightningReview}

The gauge theoretic formulation of knot homology given in    \cite{Witten:2011zz}
has its origins in the theory of supersymmetric branes in string theory, or alternatively,
in the six-dimensional (2,0) superconformal field theory. These motivations are
explained at length in    \cite{Witten:2011zz}  and will not be repeated here.
 See \cite{WItten:2011pz,Witten:2014xwa} for brief reviews of \cite{Witten:2011zz}
and related papers.
Here we summarize   the final mathematical statement arrived at in
\cite{Witten:2011zz} but approaching the subject via Landau-Ginzburg theory and Morse
theory, the topics of such importance in this paper.

Let $L\subset M_3$ be a knot (or link) in an oriented and framed three-manifold $M_3$.
We wish to formulate a doubly-graded homology theory $\CK(L)$. This will be
the homology of a complex $\widehat{\CK(L)}$, which in turn will be a certain MSW complex.
In order to formulate the MSW complex and and its differential we introduce a metric $g_{ij}dx^i dx^j$
on $M_3$.
We also introduce a compact simple Lie group $G$ with real Lie algebra $\fg$, and a
principal $G$ bundle $E\to M_3$.
We further let $\CU$ be the space of connections on $E$
and $\CU^c$ its natural complexification. A generic element $\CA$ of $\CU^c$
 can be decomposed into ``real and imaginary
parts'' as $\CA = A+ \I \phi$ where $A$ is connection on $E$ and $\phi$ is a one-form
valued in the adjoint bundle. Locally, they are one-forms valued in the compact
real Lie algebra $\fg$.
\footnote{In our conventions $\fg$ is a real subalgebra of a Lie algebra of anti-hermitian matrices.}
The space $\CU^c$ is an infinite-dimensional
K\"ahler manifold with metric
\be\label{eq:Uc-metric}
\d \ell^2 =  \int_{  M_3} \Tr(\delta \CA * \delta \overline{\CA} )
\ee
where $\Tr$ is a a positive-definite Killing form on $\fg$. The
normalization of the Killing form does not affect the flow equations.
The symplectic structure is
\be\label{eq:Uc-Symp}
\omega = \int_{  M_3} \Tr( \delta A *  \delta \phi).
\ee
and the complex structure maps $\delta A$ to $\delta \phi$.

We will consider a Landau-Ginzburg theory on the half-line $\IR_+$, parametrized by $y$,
with target space $\CU^c$, with certain boundary conditions at $y=0, \infty$ that
will be sketched below. The data of the knot enters in the boundary conditions at $y=0$.
\footnote{This LG theory provides a convenient way to derive the
gauge theory BPS equations. It should not be confused with the LG theory
in the $(x^0,x^1)$ plane which will be used in Section \S \ref{subsec:Reformulation}
to establish the relation to the web formalism}

The superpotential of the model will be the Chern-Simons term
\be\label{eq:CS-kappa}
W^{cs}(\CA) =  \int_{ M_3} \Tr\left(\CA d \CA + \frac{2}{3} \CA^3 \right)
\ee
Of course this is not single-valued, but $dW^{cs}$ is single-valued, and
this is all that is needed for the construction.
Note that we have not chosen a normalization
of the Killing form, so the periods of $W^{cs}(\CA)$ have not yet been specified.
The superpotential $W^{cs}(\CA)$ is a degenerate holomorphic Morse function on $\CU^c(\CB\CC)$
due to the gauge invariance   of $dW^{cs}$. Introduce
the group $\CG$ of unitary automorphisms of $E$. When $E$ is trivializable, as we will
assume, $\CG$ is just the group   $\CG = {\rm Map}(M_3, G)$.  The group $\CG$
acts as a group of isometries preserving the symplectic structure as well as $dW^{cs}$.  We are therefore in a
position to consider - at least formally -
the gauged Landau-Ginzburg model, as described in Section \S \ref{subsec:Remind-GLG},
with symmetry group  $S = \CG$. In particular
the group of gauge transformations of this gauged LG
model consists of maps from $\IR\times \IR_+ $, parametrized by $(x^0,y)$,  into $\CG$.

Viewed as a problem in equivariant Morse theory on $X={\rm Map}(\IR_+, \CU^c)$
(or rather, on a cover on which $W^{cs}$ is single-valued)  the Morse function is, according
to \eqref{eq:equi-h}
\be\label{eq:GgeMrse-CS-1}
h = -  \int_{\IR_+} dy \int_{M_3}  \vol(g)  g^{ij} \left(  \phi_i \p_y A_j - \phi_i D_{j} B_y\right) -
\half {\rm Re}\biggl[ \I e^{-\I \tilde\vartheta} CS(\CA) \biggr]
\ee
where $D_j$ is the covariant derivative with respect to $A_j$.

The flow equations, or, equivalently, the $\CQ_{\zeta}$-fixed point equations (with $\zeta = - \I e^{\I \tilde\vartheta}$)
are now easily written.
When covariantized in the time direction they become
\begin{subequations}\label{eq:CSLG-FP-1}
\begin{align}
 [(D_y - i D_0) , \CD] & = e^{\I\tilde \vartheta}    *_{ M_3}  \CF^*   \\
[D_0,D_y] +  D_A * \phi & = 0
\end{align}
\end{subequations}
The conventions here are the following: $\CD$ is the covariant derivative with respect to $\CA$ on $M_3$.
In local coordinates $\CD=\sum_{i=1}^3 dx^i(\p_i + A_i + \I \phi_i)$, and the fieldstrength is
$\CF_{ij}= [\CD_i,\CD_j]$. The complex conjugation $*$ is an anti-linear involution acting as $-1$ on $\lieg$.
The covariant derivatives   $D_0,D_y$
on the ``worldsheet'' $\IR\times \IR_+$ have gauge field $B = dx^0 B_0 + dy B_y$ valued in the Lie algebra
${\rm Map}(M_3, \fg)$.

These equations can be rewritten as equations on the   five-dimensional space of the form
\be
M_5 = \IR \times M_3 \times \IR_+
\ee
(with local coordinates $(x^0, x^i, y)$)  for a \underline{five-dimensional} gauge
field, locally a $\lieg$-valued one-form:
\be
A^{\rm 5d} = B_0 dx^0 + A_i dx^i + B_y dy,
\ee
together with a $\lieg$-valued field $\phi$ on $M_5$ that is cotangent to $M_3$. Locally
$\phi = \sum_{i=1}^3 \phi_i dx^i$. Put more formally, $\phi\in \Gamma(M_5, \pi^*(T^*M_3)\otimes \Ad(E))$
where $\pi: M_5 \to M_3$ is the projection. Note that the first term in the expression for
$h$ in \eqref{eq:GgeMrse-CS-1} can be written as $\int_{\IR_+ \times M_3} \Tr(\phi * F^{\rm 5d})$,
showing that the interpretation of $A^{\rm 5d}$ as a five-dimensional gauge field is natural.
(Compare with equation (5.42) of \cite{Witten:2011zz}.)

In local orthonormal coordinates, the $\zeta$-instanton equations of the CSLG model
are  seven ``real'' equations for $8$ real fields on $M_5$:
\begin{subequations}\label{eq:KW5}
\begin{align}
F_{yi} + D_0\phi_i    + \half \epsilon_{ijk}  (F - \phi^2)_{jk}
+  \kwt \left(  D_y \phi_i  + F_{i0}  + \half \epsilon_{ijk}  (D_A \phi)_{jk} \right)   & = 0  \qquad \forall i=1,2,3 \\
F_{yi} + D_0\phi_i  - \half \epsilon_{ijk}  (F - \phi^2)_{jk}
-  \kwt^{-1}  \left(   D_y \phi_i  + F_{i0}   -  \half \epsilon_{ijk}  (D_A \phi)_{jk} \right)   & = 0 \qquad \forall i=1,2,3  \\
F_{0y}  + D_i \phi_i  & = 0
\end{align}
\end{subequations}
where it is useful to introduce the parameter:
\be
\kwt = \frac{\sin\tilde\vartheta}{1+\cos\tilde\vartheta} = \tan(\half \tilde\vartheta) .
\ee
We should regard $\kwt$ as the stereographic projection of $e^{\I \tilde\vartheta}$ on
the unit circle to the real line.
These equations have some remarkable properties:

\begin{enumerate}

\item When $\kwt=1$ they become covariant equations on the four-manifold
$\widehat M_4 = \IR \times M_3$, with local coordinates $(x^0,x^i)$,
where $\phi_i$ are reinterpreted as the three components of a self-dual
$2$-form on $\widehat M_4$. These equations were written in equation (5.36)
of \cite{WItten:2011pz}. Similar equations were written in \cite{Haydys},
and hence we will refer to the equations \eqref{eq:KW5} as the $HW(\kwt)$
equations.

\item For general $\kwt$, if the fields are taken to be time-independent, i.e.
if they are pulled
back from the four-manifold $M_4 = M_3 \times \IR_+$, with local coordinates
$(x^i, y)$, then, if we rename $B_0 \to -\phi_y$, and introduce a new
adjoint-valued 1-form:
$\phi^{\rm kw} = \phi_i dx^i + \phi_y dy$  the equations become
the Kapustin-Witten equations with parameter $\kwt$:
\begin{subequations}\label{eq:KW}
\begin{align}
F - (\phi^{\rm kw})^2 + {\rm \textbf{t}}  (d_A \phi^{\rm kw} )^+ - {\rm \textbf{t}}^{-1}  (d_A \phi^{\rm kw} )^- & = 0   \label{eq:KW-FTERM} \\
d_A * \phi^{\rm kw}  & = 0  \label{eq:KW-DTERM}
\end{align}
\end{subequations}
Here the superscript $\pm$ refer to the self-dual
projections with the product metric $g_{ij} dx^i dx^j + dy^2 $ on $M_4$ and
the orientation $dy dx^1 dx^2 dx^3$. We will refer to the equations
\eqref{eq:KW} as the $KW(\kwt)$ equations.

\end{enumerate}

The desired knot homology complex $\widehat{\CK(L)}$ will be an MSW complex for the above gauged LG model
on the half-plane with coordinates $(x^0,y)$. The vacua are solutions of the $KW(\kwt)$ equations
and the instantons between them are the solutions of the $HW(\kwt)$ equations.
That is, the differential on the complex $\widehat{\CK(L)}$ is obtained
by counting solutions to the five-dimensional  equations.
However, to specify the complex precisely we need to specify boundary
conditions for $y\to 0$ and $y\to \infty$. We now turn to these boundary conditions. They should be viewed
as part of the specification of the gauged LG model, including a brane at $y=0$.

In our main application below we will take $M_3 = \IR \times C\cong \IR^3$ where $C$ is the complex plane. In this case, a suitable boundary condition for $y\to\infty$ is simply to require
that $A\to 0$, and $\phi\to \vec c\cdot \d \vec x$, where $\vec c$ is a chosen triple of commuting elements of $\frak g$.  Thus we can conjugate
$\vec c$ to a Cartan subalgebra $\frak t\subset \frak g$.  For a more general $M_3$, a suitable condition for $y\to \infty$ is to require
the fields to approach a specified, $y$-independent, solution of the $KW(\kwt)$ equations.

The boundary conditions at $y=0$ are more subtle ones and involve specifying a singularity that the fields are supposed to have.  One chooses
the following data:

\begin{enumerate}

\item A homomorphism of Lie algebras $\rho: \fs\fu(2) \to \fg$.  In the usual application
to knot homology, this is taken to be a principal embedding.

\item A representation $R^\vee$ of the Langlands or GNO dual group $G^\vee$
associated to each connected component of $L$. In general different components
are associated to different representations.

\end{enumerate}

We describe the boundary condition first for the case that $M_3=\IR^3$ with Euclidean metric $\sum_i \d x_i^2$, and without knots,
and that  $G=SU(2)$.  Also, we take $\rho$ to be the principal embedding, which for $\frak g=\frak{su}(2)$
is just the identity map $\frak{su}(2)\to\frak{su}(2)$.
In this situation, we impose the following\footnote{The name reflects the fact that the singular behavior that we are about to specify for $y\to 0$
was originally introduced by W. Nahm in his study of Nahm's equations associated to magnetic monopoles.}  ``Nahm pole boundary condition.''
We require that for $y\to 0$, $A$ vanishes and
\be\label{eq:NP-1}
\phi \sim  \frac{1}{y}\sum_{i=1}^3 \rho(\ft_i) \d x^i + \CO(y).
\ee
This boundary condition has a natural generalization on a general Riemannian three-manifold $M_3$.  We state this
first in the absence of knots.  We fix a spin bundle
\footnote{In fact, for the principal embedding the choice of spin structure does not matter since the Lie algebra $\frak g$
pulls back to an integer spin representation. }
%
%
%We fix a spin bundle
%
%\footnote{For certain $G$, this construction does not really depend on the choice
%of the spin structure of $M_3$.  This is the case if the principal embedding $\rho:\frak{su}(2)\to \frak g$ corresponds, at the level %of groups,
%to an embedding in $G$ of $SO(3)$ rather than $SU(2)$.}   
%
${\mathcal S}\to M_3$, with structure group $SU(2)$, and write $P_{\mathcal S}$ for
the corresponding principal bundle.
A $G$-bundle  $E= M_3$ is then defined by $E=P_{\mathcal S}\times_{SU(2)} G$, where $SU(2)$ is embedded in $G$ via $\rho$
and acts on $G$ on the left.  Eqn. (\ref{eq:NP-1}) then makes sense for a section $\phi$ of $T^*M_3\otimes \mathrm{ad}(E)$.
The gauge field $A$ is  required to approach the
spin connection on $TM_3$ as $y\to 0$, where the
spin connection is  embedded in $G$ via
 $\rho:\frak{su}(2)\to \frak g$. 
   The point  is  that with this choice, the singular part of the $KW$ equations is obeyed near $y=0$, assuming
 that  $\kwt=1$ (a minor
 modification is needed for more general $\kwt$).
 To generalize the Nahm pole boundary condition in the presence of a knot, one
``embeds a Dirac monopole singularity in the Nahm pole singularity.''
What this means is that the
solution is required to asymptote along $L$ to a certain model solution of the
$KW(\kwt)$ equations that informally is a combination of a Dirac monopole with a Nahm pole; the model solution depends on the choice of a representation of the dual group $G^\vee$.
The basic model solution for $G=SU(2)$, $\rho={\rm Id}$,
$R^\vee$ the fundamental representation,
and $\kwt=1$ is given in Section 3.6 of \cite{Witten:2011zz}.   Near $y=\infty$
the gauge field $A$ vanishes and $\phi_i$ approaches a constant determined, up to
conjugacy, by the $c_i$. In particular for $M_3 = \IR^3$,
\be
\phi \to {\rm Ad}(g)\left( \sum_{i=1}^3   c_i    dx^i \right) + \CO(y^{-\delta})
\ee
for some $g \in G$ and some positive $\delta$.
In the presence of a Nahm pole, a gauge transformation is required to be trivial at $y=0$, and this ensures
that the gauge group acts freely.

A demonstration that the Nahm pole boundary conditions for $KW(\kwt)$ and $HW(\kwt)$
at $\kwt=1$ are elliptic is given in \cite{Witten:2011zz,Mazzeo:2013zga}
and the generalization to $\kwt\not=\pm 1$ and to include singular monopoles is
fully expected to hold. 
%
%We will simply
%denote the boundary conditions as $\CB\CC(\rho,R^\vee; \vec c)$.  Note that this notation
%combines the boundary conditions at the two ends $y\to 0$ and $y\to\infty$.
%

%
\begin{figure}[htp]
\centering
\includegraphics[scale=0.3,angle=0,trim=0 0 0 0]{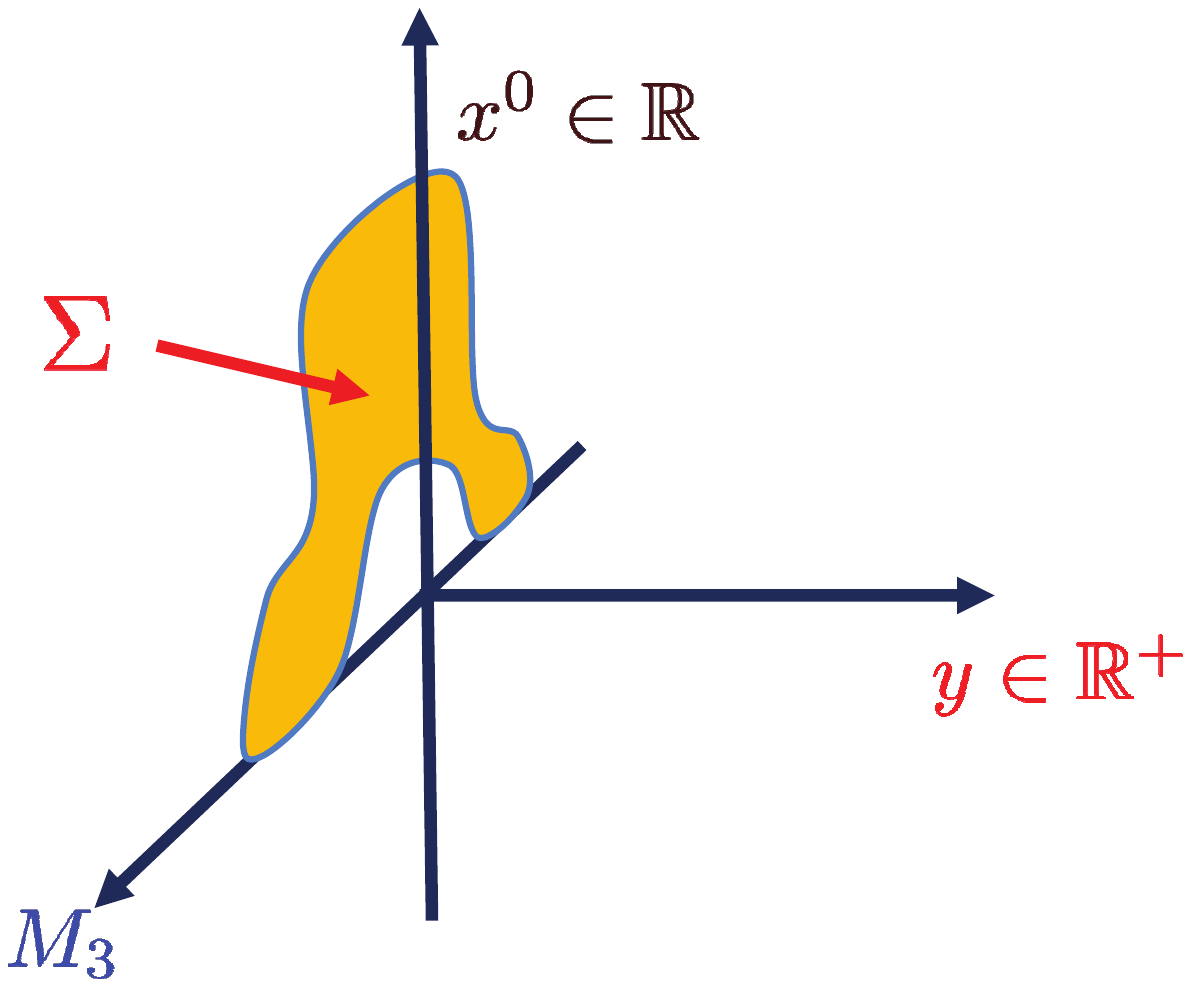}
\caption{The basic setup for the gauge-theoretic approach to knot homology.
One considers 5 dimensional SYM on a space $\IR \times M_3 \times \IR_+$ where
$M_3$ is in an oriented Riemannian 3-fold. For fixed $x^0$, $M_3$ contains a
link $L$ located at $y=0$. The link evolves as a function of $x^0$ to
produce a knot bordism $\Sigma$.    }
\label{fig:KNOT-HOM-3}
\end{figure}

\bigskip
\textbf{Remarks}:

\begin{enumerate}

\item In knot homology an important role is played by morphisms $\Phi(\Sigma): \CK(L_1) \to \CK(L_2)$
associated with knot bordisms in $\IR \times M_3$ such that $\p \Sigma = L_2 - L_1$.
In the gauge theoretic approach these, too, are located at $y=0$ as illustrated in Figure
\ref{fig:KNOT-HOM-3}. There are corresponding boundary conditions on the $\zeta$-instantons,
that is, on the $HW(\kwt)$ equations in the presence of such bordisms.

\item The advantage of the gauge-theoretic approach to knot homology
is that the definition of the homology groups $\CK(L)$ does not employ
a special choice of direction in $M_3$ unlike, say, combinatorial definitions based on
knot projections (such as Khovanov's original definition \cite{Kh,BarNatan}).
  The origin of the   equations \eqref{eq:KW} and their 5d counterparts
in topological field theory
lead us to expect that the knot homology will be independent
of the metric $g_{ij}$,   the parameter $\kwt$, and the symmetry-breaking
boundary conditions $\vec c$.

\item The knot homologies do, of course, depend on the data $G, R^\vee, \rho$.  In \cite{Witten:2011zz} many of the expected properties
 of knot homologies such as a $\IZ \times \IZ$ grading and its behavior with
 respect to change of framing and knot bordisms   were established.  (The second factor in the $\IZ\times \IZ$ framing
depends also on a choice of framing of the manifold $M_3$ and of the link $L$.)

\end{enumerate}

\begin{figure}[htp]
\centering
\includegraphics[scale=0.3,angle=0,trim=0 0 0 0]{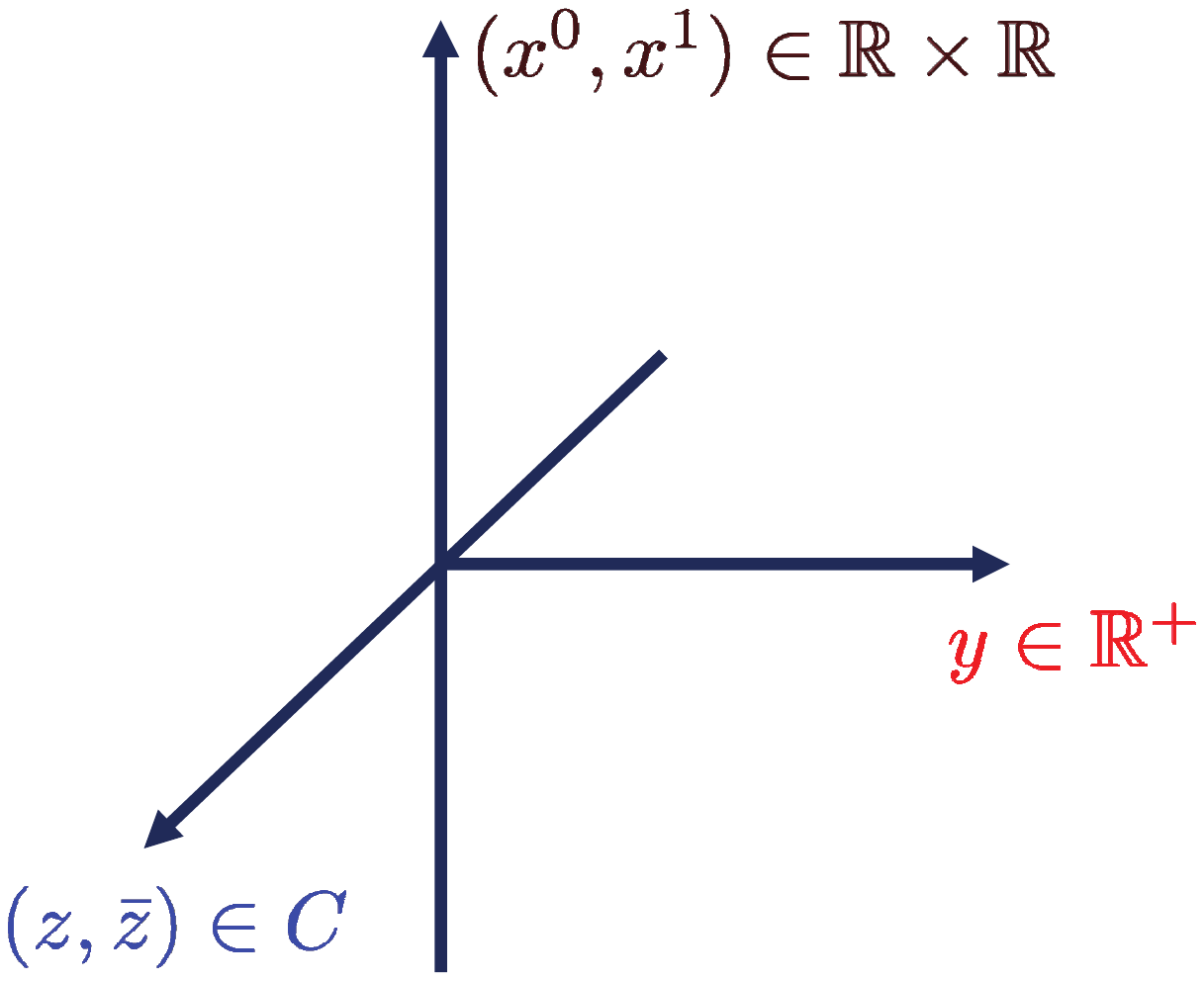}
\caption{In the second formulation of $\widehat{\CK(L})$ in terms of a gauged Landau-Ginzburg
model, we specialize to $M_3 = \IR \times C$, with $C$ a Riemann
surface. Thus, we consider five-dimensional SYM on a space $\IR\times \IR \times C \times \IR_+$.
We will eventually take $C$ to be the complex plane, but the basic picture
should hold for a general Riemann surface. (The surface $C$ must have at least one
puncture so that $M_3$ admits a framing.)  The link bordism at $y=0$ is not shown here.
We use local complex coordinates $z = x^2 + \I x^3$ on $C$.    }
\label{fig:KNOT-HOM-1}
\end{figure}

\subsubsection{Reformulation For $M_3 = \IR\times C$}\label{subsec:Reformulation}

In order to make contact with the web formalism, we need an alternative formulation of the
gauge theory equations which is available in the special case when the three-manifold factorizes
$M_3 = \IR \times C$, where $C$ is a Riemann surface.
Remarkably, the full set of equations $HW(\kwt)$ in this case can also
be described as the $\zeta$-instanton equations in a
\underline{second} gauged Landau-Ginzburg model.
Since there are now two gauged Chern-Simons-Landau-Ginzburg models in play
we will refer to them as CSLG1 and CSLG2.

The model CSLG2 is again a  gauged LG model with a target space  of complexified
gauge fields, but now they are gauge fields on the three-manifold
\be
\tilde M_3 = C \times \IR_+ ,
\ee
where $\IR_+$ is the $y$-direction. We must also assume the link $L$ is translation invariant
in the $x^1$-direction. Configurations involving a slow $x^1$ dependence of the link $L$ or even the Riemann surface
$C$ can be included  as in Section \S \ref{subsec:LG-Susy-Interface}

See Figure \ref{fig:KNOT-HOM-1}. We consider a sigma model of maps
from $\IR^2$, with coordinates $(x^0,x^1)$, to a space of complexified $G^c$ gauge fields
on $\tilde M_3 = C \times \IR_+$. We denote this space of gauge fields by
$\tilde\CU^c(\CB\CC)$, where, now, the boundary conditions  $\CB\CC$ serve to
define the  target space of the LG model, rather than the boundary conditions of the LG model.
 The conditions $\CB\CC$
will be specified in Section \S \ref{subsec:CSLG-Model} below for the case of $G=SU(2)$ or $G=SO(3)$.

In some more detail, we denote the complexified gauge field by
\be
\tilde \CA = \tilde\CA_2 dx^2 + \tilde\CA_3 dx^3 + \tilde\CA_y dy.
\ee
 We use the same formulae \eqref{eq:Uc-metric} for the metric,
 \eqref{eq:Uc-Symp} for the symplectic form and \eqref{eq:CS-kappa} for the superpotential,
 but now with the replacement   $M_3 \to \tilde M_3$ and $\CA \to \tilde \CA$.
 Viewed as equivariant Morse theory on ${\rm Map}(\IR, \tilde\CU^c(\CB\CC) )$, the Morse function is now
\be\label{eq:GgeMrse-CS-2}
h = -  \int_{\IR} dx^1 \int_{\tilde M_3}  \vol(g)  g^{ij} \left(  \tilde \phi_i \p_{x^1}  \tilde A_j
-  \tilde \phi_i \tilde D_{j}  \tilde B_1 \right) -
\half {\rm Re}\biggl[ \I e^{-\I \tilde\vartheta} CS(\tilde \CA) \biggr]
\ee
and the $\CQ_{\zeta}$-fixed point equations (again with $\zeta = - \I e^{\I \tilde\vartheta}$)
 are again of the form \eqref{eq:CSLG-FP-1}, but with all covariant
derivatives replaced by their corresponding version with a tilde:
\begin{subequations}\label{eq:CSLG-FP-2}
\begin{align}
 [(\tilde D_1 - \I \tilde D_0) , \tilde \CD ] & = e^{\I\tilde \vartheta}  \left( *_{ M_3} \tilde\CF^* \right) \\
[\tilde D_0,\tilde D_1] +  \tilde D  * \tilde \phi & = 0
\end{align}
\end{subequations}

We stress that the equivalence of the equations \eqref{eq:CSLG-FP-1} and \eqref{eq:CSLG-FP-2} is not entirely trivial.
The fields of the two CSLG models are different. In CSLG1, which applies for general three-manifolds $M_3$,
we use one-form-valued fields with components $\phi_1, \phi_2, \phi_3$. In CSLG2, which applies
for $M_3 = \IR \times C$, we use one-form-valued fields with components $\phi_2, \phi_3,\phi_y$.
Nevertheless, the new flow equations  \eqref{eq:CSLG-FP-2} are in fact \underline{equivalent}
to the original flow equations \eqref{eq:CSLG-FP-1}. To see this, let $D_{i'} = \Ad(\phi_i)$, $i=1,2,3$
and   $\tilde D_{a'} = \Ad(\tilde\phi_a)$, $a=2,3,y$. If  we make the replacement:
\be\label{eq:HW-ROT}
\begin{split}
D_{1'} & \rightarrow \tilde D_0 \\
D_{0} & \rightarrow - \tilde D_{y'} \\
\end{split}
\ee
as well as  $D_{2',3'} \rightarrow \tilde D_{2',3'} $ and $D_{1,2,3,y} \rightarrow \tilde D_{1,2,3,y}$
in \eqref{eq:CSLG-FP-1}, then, after some nontrivial rearrangement we obtain precisely the set of
equations \eqref{eq:CSLG-FP-2}.

The great advantage of the second formulation is that the equations for the vacua of the model
are much simpler than the $KW(\kwt)$ equations. Indeed, they simply say that the gauge field
is flat and satisfies a moment map equation:
\footnote{In our conventions $A_j,\phi_j\in \lieg$ are regarded as anti-hermitian
in real local coordinates, so, for example, $(\p_j + A_j + \I \phi_j)^\dagger = - (\p_j + A_j - \I \phi_j )$. }
\begin{subequations}\label{eq:CS2-vac}
\begin{align}
[\tilde \CD_a, \tilde \CD_b ] &  = 0  \\
\sum_{a=2,3,y} [\tilde \CD_a, \tilde \CD_a^\dagger ] & = 0
\end{align}
\end{subequations}
Indeed, in CSLG1, the vacua cannot be ``spatially independent'' (i.e. $y$-independent)
because of the boundary conditions on the LG model. In CSLG2, the vacua can be translationally
invariant in the $x^1$-direction.

\begin{figure}[htp]
\centering
\includegraphics[scale=0.3,angle=0,trim=0 0 0 0]{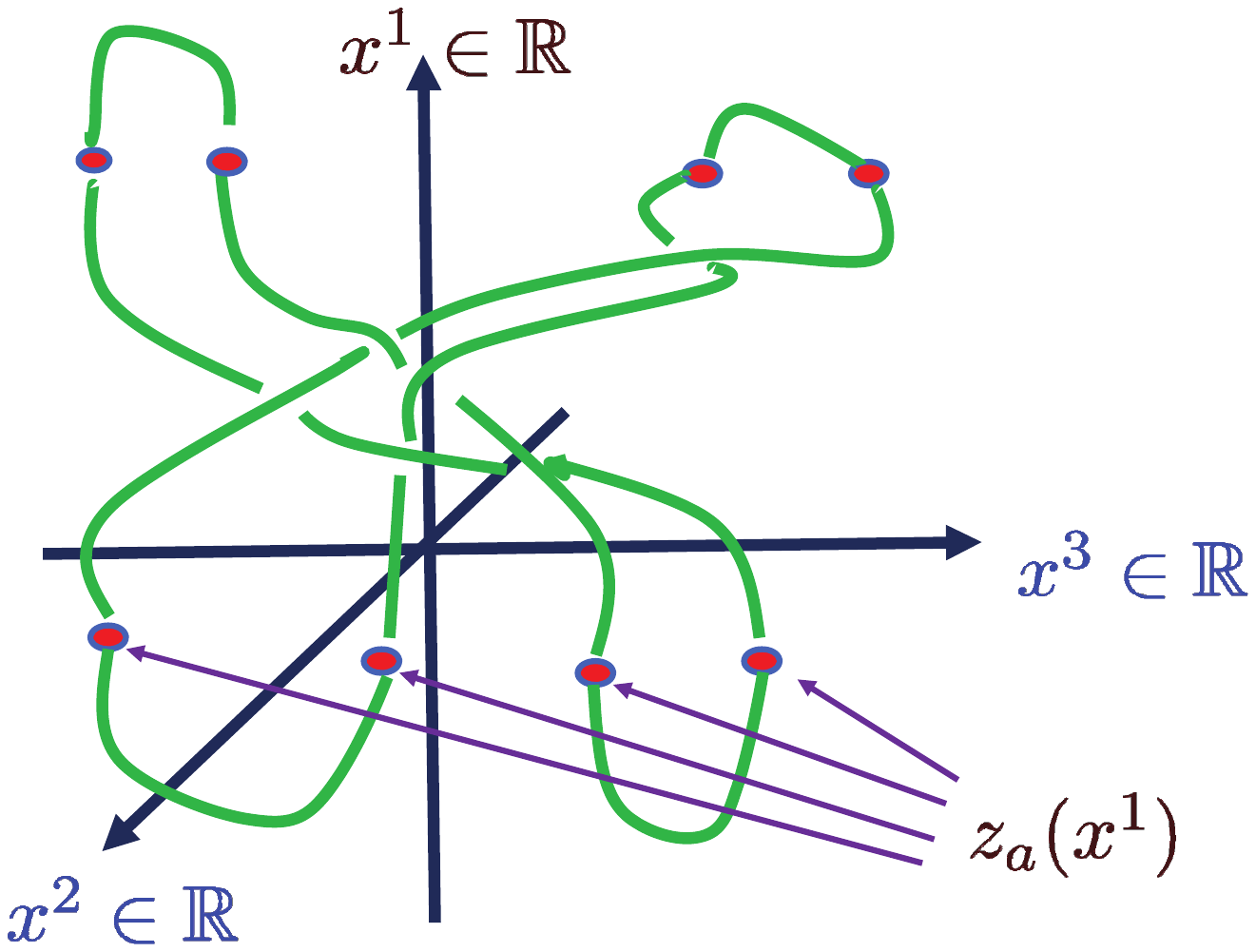}
\caption{This figure depicts the link $L$ in the boundary at $y=0$ at a fixed value of $x_0$.
It is presented as a tangle evolving in the $x^1$ direction and therefore can be characterized
as a trajectory   of points $z_a(x^1)$ in the complex $z=x^2 + \I x^3$ plane.  The points
$z_a$ are decorated with irreducible representations $R_a$ of $G^\vee$. The tangle is closed
by ``creation'' and ``annihilation'' of the points $z_a$  in pairs (decorated with dual representations).     }
\label{fig:KNOT-HOM-4}
\end{figure}

As mentioned above, the model CSLG2 is defined when the link $L$ is translation-invariant in the $x^1$ direction.
Of course most links $L\subset \IR\times C$ do not have this property.
An important part of the argument of \cite{Gaiotto:2011nm} is to employ
topological invariance of the underlying topologically twisted 5d SYM theory  to present the link $L$
as the closure of a tangle, that is, as
an adiabatically evolving collection of points in $C$. They will be
denoted $z_a(x^1)$, $a=1,\dots, n$.  The evolution of $z_a(x^1)$ defines a
braid. We may take the evolution to be adiabatic, thus justifying various
low-energy approximations used in the physical arguments. However, since we
ultimately use a topological field theory this is not strictly necessary.
At some critical values of $x^1$ the
number of points $n$ jumps by $n \to n \pm 2$ due to   ``creation'' and ``annihilation'' processes.
Indeed, we assume that for sufficiently large $\vert x^1\vert$ there are no such points
at all. That is, all the strands have been closed off in the far future and past.
 See Figure \ref{fig:KNOT-HOM-4}.

Therefore, our strategy will be first, to understand the model CSLG2  (and its finite-dimensional reductions)
for the case where the link $L$ is   just a disjoint
union of lines $L = \amalg_{a=1}^n\IR \times \{ z_a \} $, where $z_a$ are a collection of
points on the Riemann surface $C$.
Then we will understand the $x^1$-evolution of the points $z_a$ as defining a
path of such theories. Accordingly, we can apply our general theory of Interfaces developed in
Sections \S\S \ref{sec:Interfaces}-\ref{sec:CatTransSmpl}.
Ultimately we will have an Interface between the trivial theory and itself. As we saw in
Section \ref{subsubsec:TrivialTheories} above, such an Interface is a complex. This will be our
proposal for a knot homology
complex $\widehat{\CK}(L)$. It is meant to be homotopy equivalent to the original MSW complex of CSLG1.

\begin{figure}[htp]
\centering
\includegraphics[scale=0.3,angle=0,trim=0 0 0 0]{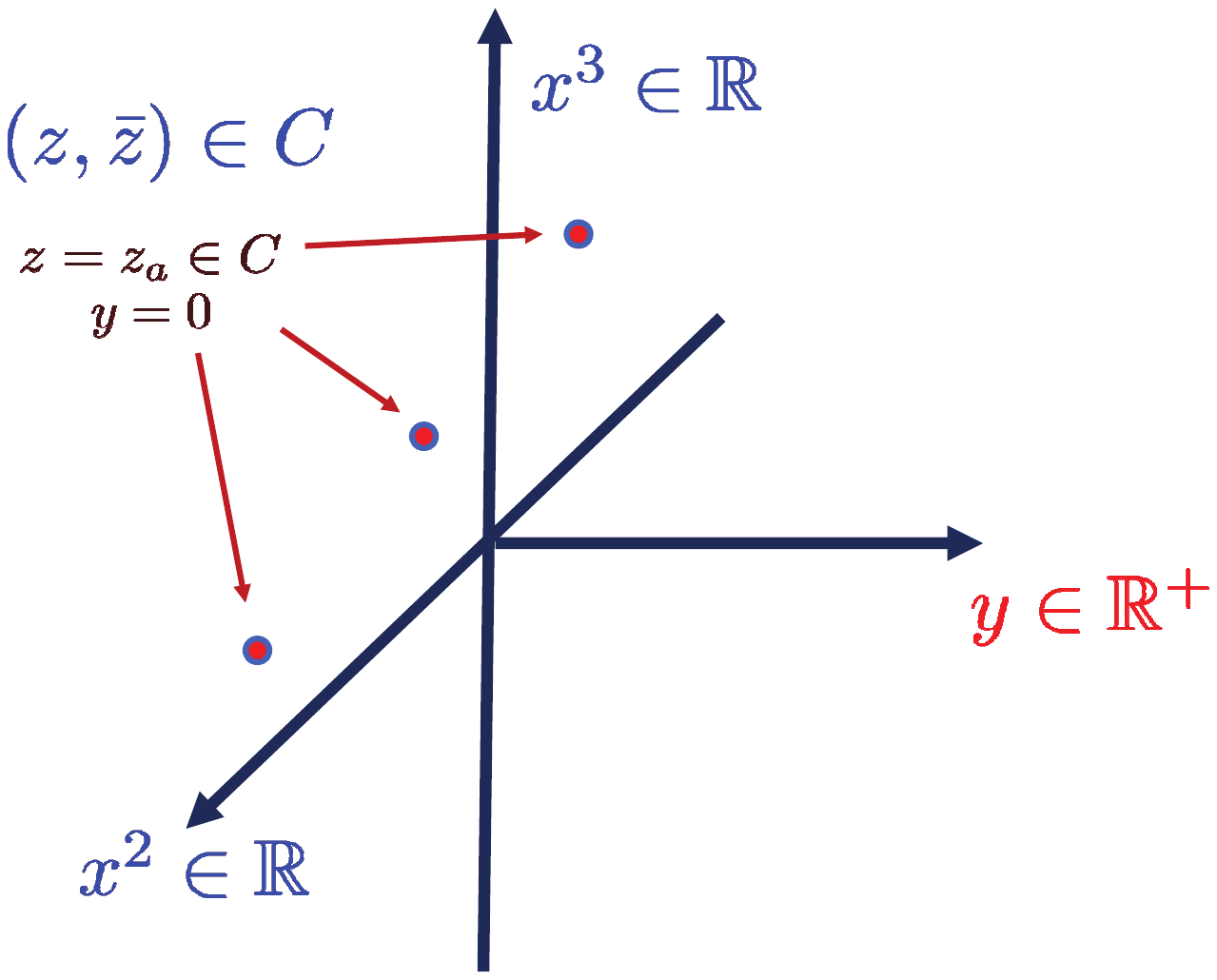}
\caption{At a fixed value of $x^0, x^1$ we have a 3-fold $\tilde M_3=C\times \IR_+$ shown here.
The equivalent gauged Landau-Ginzburg model is formulated in terms of target space
of complexified gauge fields $\CA$ defined on this space and satisfying suitable
boundary conditions $\CB\CC$ at $y\to 0,\infty$, together with extra conditions
at   $z=z_a$.}
\label{fig:KNOT-HOM-2}
\end{figure}

\subsubsection{Boundary Conditions Defining The Fields Of The CSLG  Model}\label{subsec:CSLG-Model}

Now we must discuss the boundary conditions $\CB\CC$ used to define the target space $\tilde\CU^c(\CB\CC)$
of the model CSLG2. The link $L$ is encoded in a collection of points $z_a\in C$
at $y=0$, as shown in Figure \ref{fig:KNOT-HOM-2}.

The rotation \eqref{eq:HW-ROT} is incompatible with Nahm pole boundary conditions on $\phi$.
Therefore, we must find a new set of boundary conditions which correctly implements the
presence of the knot at $y\to 0$. It is also useful to include boundary conditions
at $y\to \infty$ that reduce the structure group $G$ of $E$ to a maximal torus. Physically,
this corresponds to moving onto the Coulomb branch of a gauge theory and resolves some confusing
singularities in the moduli space of vacua.

 While the papers \cite{Witten:2011zz,WItten:2011pz,Witten:2014xwa}
emphasize the formulation reviewed in Section \S \ref{subsec:LightningReview}, in fact, there is some freedom in
the choice of which supersymmetry to use in deriving the $Q$-fixed point equations,
as well as the precise boundary conditions which are imposed on the fields in
the formulation of  $\widehat{\CK(L)}$. It should be possible to deform the boundary conditions
and still define an equivalent knot homology.
Some of this freedom was employed in \cite{Gaiotto:2011nm}. See Appendices A and B of
\cite{Gaiotto:2011nm} for a detailed discussion.

For technical simplicity, the discussion of \cite{Gaiotto:2011nm} is restricted to
$G=SU(2)$ or $G=SO(3)$. Moreover, the Riemann surface $C$ is taken to be the
complex plane with Euclidean metric and trivial framing. \footnote{It would be
interesting to generalize the following considerations to an arbitrary Riemann surface.}
The irreducible representations $R_a^\vee$ of $G^\vee$ at $z_a$ are described
by their dimensions $k_a + 1 $, where $k_a$ is a positive integer. If $G=SU(2)$ then $k_a$
must be even, and if $G=SO(3)$ then $k_a$ can have either parity.\footnote{In the example relevant to a knot or link, presented as in Figure
\ref{fig:KNOT-HOM-4}, the $k_a$  are not independent but always appear in pairs.}
It is convenient to package these data into a monic polynomial
\be\label{eq:defK}
K(z):= \prod_{a=1}^n (z-z_a)^{k_a}.
\ee

To define the space $\tilde\CU^c(\CB\CC)$ we consider all $G^c$ gauge fields on
$C \times \IR_+$ that satisfy the following boundary conditions: For $z\not = z_a$ and
$y \to 0$ we require that there is a gauge in which
\be\label{eq:NP-2}
\tilde \CA \to \frac{1}{2y} \begin{pmatrix} dy & 2 dz  \\  0 &  - dy \\ \end{pmatrix} + \cdots \qquad z\not=z_a, y\to 0
\ee
This is the complex gauge field analog of the Nahm pole boundary condition.

For $y \to \infty$ at fixed $z$ we require that in some gauge
\be\label{eq:y-infty-bc}
\tilde \CA \to  \frac{dz}{\xi} \begin{pmatrix} \kwc & 0 \\  0 & - \kwc \\ \end{pmatrix} + \xi d \bar z \begin{pmatrix} \bar \kwc & 0 \\  0 & - \bar \kwc \\
\end{pmatrix} + dy \begin{pmatrix} c_1 & 0 \\  0 & - c_1 \\ \end{pmatrix}
\ee
where the symmetry-breaking data $\vec c$ is encoded here as three real numbers $c_i$ and we define
 $\kwc :=\half( c_2 - \I c_3)$. The parameter $\xi$
that enters here is a complex number $\xi \not= 0, \infty$ and is used to deform the original Nahm pole
boundary conditions of CSLG1.
\footnote{In  \cite{Gaiotto:2011nm} the parameter $\xi$ was called $\zeta$, but
we have renamed it here since it should not be confused with the $\zeta$ used for the Landau-Ginzburg model.}
Away from knots, the Nahm pole boundary condition implies the existence of an everywhere nonzero
``flat section''  $s$ of the rank 2 associated bundle to $E^c$
which solves
\be\label{eq:Flt-Sec}
\begin{split}
\tilde \CD_{\bar z} s  & = 0 \\
\tilde \CD_{y} s & = 0 \\
\end{split}
\ee
Moreover, one can choose $s$ to grow only polynomially for $z\to\infty$ (without this condition, one would be free to multiply $s$ by
an entire function such as $e^z$).
If $\CF=0$ and $s$ satisfies \eqref{eq:Flt-Sec}, then the quantity $s \wedge \CD_z s$ is $y$-independent.
Although we refer to $s$ as a ``flat section'' in fact $\CD_z s $
is \underline{not} zero. Knots are most succinctly incorporated by saying that $s$ has a zero of order $k_a$ along a knot, leading to
\be\label{eq:Flt-Sec-bc-1}
\begin{split}
s \wedge \CD_z s & = K(z)\vol(E)  \\
\end{split}
\ee
where $\vol(E)$ is a fixed constant volume form on the rank two associated bundle.
The boundary condition \eqref{eq:Flt-Sec-bc-1}
 encodes the presence of the knots at $z=z_a$ and $y=0$.

This completes the formulation of the boundary conditions, and hence the specification
of the target space $\tilde\CU^c(\CB\CC)$ of the model CSLG2.

\textbf{Remarks}

\begin{enumerate}

\item We will henceforth write the space of gauge fields $\tilde \CA$ determined by these
boundary conditions as $\tilde\CU^c(z_a,k_a; \xi, \vec c)$.

%\item The boundary condition \eqref{eq:Flt-Sec-bc-2} means that $\CA$ has an irregular
%singular point at $z\to \infty$.

\item In the gauge \eqref{eq:NP-2} the section $s$ is the solution of $\CD_y s =0$
which vanishes for $y\to 0$. Therefore it is called the ``small flat section.''

\item
Note that even though $\tilde \CA$ is a flat connection on the simply connected domain $\IR^2 \times \IR_+$
we cannot write a solution to both \eqref{eq:Flt-Sec} and $\CD_z s=0$ which satisfies the above
boundary conditions. There can therefore be an interesting moduli space of such connections.

\end{enumerate}

  Reference \cite{Gaiotto:2011nm} argues that one can parametrize the moduli space of solutions
to the vacuum equations by making a complex gauge transformation, growing at most polynomially
for $z\to\infty$,  such that, away from $y=0$,  $\CA$ is given by \eqref{eq:y-infty-bc}  \underline{exactly}, not just asymptotically, on $\tilde M_3$.
The moduli space is then parametrized by the data of the small flat section, which, in this gauge,
must have the form
\be
s = \exp[ - (\xi \bar \kwc \bar z + c_1 y ) \sigma^3] \begin{pmatrix} P(z) \\ Q(z) \\ \end{pmatrix}
\ee
where $P(z),Q(z)$ are \underline{polynomials},
 unique up to rescaling $(P,Q) \to (\lambda P, \lambda^{-1} Q)$.
Then equation \eqref{eq:Flt-Sec-bc-1} implies that
\be\label{eq:PQK}
P Q' - Q P' - c_0  PQ = K
\ee
where $c_0 := - \frac{2\kwc}{\xi}$. Following \cite{Gaiotto:2011nm} we
rewrite this as   $ e^{-c_0 z} \frac{K}{Q^2} =- \p_z (e^{-c_0 z} \frac{P}{Q}) $
and since $P,Q$ are polynomials it follows that $e^{-c_0 z} \frac{K}{Q^2}$
must have
zero residue at all the zeroes of $Q$. Using the scaling freedom we can
assume that $Q(z)$ is monic, so it must have the form
\be
Q(z) = \prod_{i=1}^q (z-w_i)
\ee
and moreover the roots $w_i$ of $Q$ are constrained to satisfy
\be\label{eq:BetheAnsatz}
\sum_{a=1}^n \frac{k_a}{w_i - z_a} = c_0 + \sum_{j\not=i} \frac{2}{w_i - w_j}  \qquad i=1,\dots, q
\ee
These equations are of course the critical points of the function
\be\label{eq:YangYangW}
W =  \sum_{i,a} k_a \log(w_i-z_a) - \sum_{i\not=j} \log(w_i - w_j) - c_0 \sum_i w_i
\ee

As discussed at length in \cite{Gaiotto:2011nm}, the equations \eqref{eq:BetheAnsatz} define
an oper with monodromy-free singularities. Consequently, there
are close connections to integrable systems such as the Gaudin model
(with an irregular singular point at infinity when $c_0\not=0$) \cite{Frenkel:2004qy}. In particular,
the equations \eqref{eq:BetheAnsatz} are just the Bethe ansatz equations for the Gaudin model.
Moreover, after adding a term depending only on $k_a$ and $z_a$:
\be
\Delta W =   \frac{c_0}{2} \sum_a k_a z_a -
\frac{1}{4} \sum_{a\not=b} \log(z_a - z_b)
\ee
the function $W+\Delta W$ is the Yang-Yang function of the model.
It is shown in
\cite{GaudinIrregular}   that there are generically $\prod_a (k_a+1)$ solutions to equation \eqref{eq:BetheAnsatz},
provided we consider all $0\leq q \leq k$. (This is also
the dimension of $\otimes_a R_a^\vee$, a statement which is important in the theory of the Bethe
ansatz.)  In particular the space of vacua generically consists of
a finite collection of distinct flat connections.

Moreover,  the integral of  $\exp[(W + \Delta W)/b^2]$
over the Lefshetz thimbles associated to \eqref{eq:YangYangW} are the renowned free-field representations of
conformal blocks of degenerate representations of the Viraroso algebra. This in turn leads to a
demonstration that, in the case of $M_3 = \IR \times \IC$
and $G=SU(2)$,  the Euler character of the MSW complex
built on the $KW(\kwt)$ equations is indeed the Jones polynomial \cite{Gaiotto:2011nm},
thus providing substantial
evidence that $\CK(L)$ is equivalent to Khovanov homology.

 All of this strongly suggests
that there is a low energy effective description of the model CSLG2 -- equivalent at the level of $LA_\infty$ algebras
and the $A_\infty$ category of Interfaces -- in terms of a simpler \underline{ungauged} LG model with
a finite-dimensional target space and with superpotential \eqref{eq:YangYangW}. We next
proceed with an argument that this is indeed the case.

\subsubsection{Finite-Dimensional LG Models: The Monopole Model}\label{subsec:Monopole-Model}

The first step in simplifying CSLG2, again described in \cite{Gaiotto:2011nm}, is
to reduce it to an ungauged LG model whose target space is a moduli space of
magnetic monopoles. We will refer to that as the \emph{monopole model}.
We now briefly sketch how this is done. A full derivation, taking careful account
of $D$-terms, remains to be done.

A quick route to the monopole model is provided by applying the remark at the end of
Section \S \ref{subsec:Remind-GLG}. Since $S$ acts on $X$ without fixed points,
we can consider an equivalent ungauged Landau-Ginzburg model with target space the
symplectic quotient. As a complex manifold this is
 $X/S^c = \tilde\CU^c(z_a, k_a;\xi; \vec c)/\CG^c$ and $W^{cs}$ (modulo periods) descends to a nondegenerate
Morse function on this space. The  critical points are, of course, just the
gauge equivalence classes of flat connections $\tilde\CA$ on $\tilde M_3$ satisfying the
boundary conditions \eqref{eq:NP-2} - \eqref{eq:Flt-Sec-bc-1}.
We claim that, at least when all three components $c_i\not=0$
and $\xi\not= 0,\infty$, there is a finite set of critical points which can identified as a finite
set of points in a moduli space of smooth magnetic monopoles for $SU(2)$ of charge
\be
m := \half \sum_a k_a.
\ee
(For $G=SU(2)$, the integers $k_a$ are all even. For
 $G=SO(3)$, the integers $k_a$ can have any parity, but the sum is
constrained to be even.)

To explain this claim in some more detail, we begin with  an important preliminary remark.
Due to the Nahm pole, $\tilde\varphi := \half (\tilde\phi_2 - \I \tilde\phi_3)$
is nonzero and hence $\CA_z$ is not unitary. For the moment take the complex symmetry breaking
parameter $c_0=0$ but $c_1 \not= 0$. Since $c_1$ is nonzero the gauge symmetry is spontaneously broken to
$U(1)$ and hence   charged fields, such as $\tilde\varphi$, will decay exponentially fast for
$y\to \infty$   on a scale set by $c_1$.
On the other hand, when  $\tilde\varphi=0$ the equations \eqref{eq:Flt-Sec} are equivalent
to the ``holomorphic part'' of the standard Bogomolnyi equations for ordinary magnetic monopoles
in a Yang-Mills-Higgs theory.

We are therefore led to consider the space $ \CM(c_1,m)$ of smooth $SU(2)$ magnetic
monopoles with asymptotic Higgs field
\be\label{eq:MonopoleAsympt}
\tilde\phi_y \to  c_1 \fh  - \frac{m}{2r} \fh + \cdots \qquad  \tilde F \to \half m \fh \sin\theta d\theta d\phi + \cdots
\ee
where $\fh =\I \sigma^3$ is a simple coroot in a standard Cartan subalgebra of $su(2)$
and we chose a gauge with $\tilde\phi_y$ constant at infinity. Moreover, we choose the gauge so that $c_1$ is
positive. (This still leaves an $SO(2)$ subgroup of global gauge transformations unfixed. We will fix it below.)
Although $\CM(c_1,m)$  is hyperk\"ahler, the choice of a distinguished direction $y$
selects a distinguished complex structure on $\CM(c_1,m)$  and we will simply regard
it as a K\"ahler manifold in this complex structure.
In order to incorporate the other flatness equations from $W^{cs}$, and the
fact that $\tilde\varphi$ is not exactly zero we will introduce an effective superpotential on the
space $\CM(c_1,m)$. It will be holomorphic in the complex structure selected by $y$.

In order to write the effective superpotential, we
will make use of a well-known  presentation of the monopole moduli space $\CM(c_1,m)$
\cite{HitchinMonopoles,DonaldsonRatMap,HurtubiseRatMap,AtiyahHitchinBook}.
We consider the scattering problem along the $y$ axis at fixed $(z,\bar z)$ for the operator
$\CD_y$. The  evolution operator along the $y$-axis is just
\be\label{eq:Evolution}
P\exp \left[ -  \int_{y_-}^{y_+}  ( \tilde A_y + \I\tilde \phi_y) dy\right]
\ee
Using \eqref{eq:MonopoleAsympt}, it follows that there exists a basis of covariantly constant sections
with $y\to +\infty$ asymptotics
\be\label{eq:flatlim-1}
\begin{split}
s^{(+\infty,+)} & \sim e^{c_1 y} y^{-m/2} \begin{pmatrix} 1\\ 0 \\ \end{pmatrix}\left( 1+ \CO(1/y) \right) \\
s^{(+\infty,-)} & \sim e^{-c_1 y} y^{m/2} \begin{pmatrix} 0\\ 1\\ \end{pmatrix}\left( 1+ \CO(1/y) \right) \\
\end{split}
\ee
 Similarly,   there is a basis of such sections with $y\to -\infty$ asymptotics
\be\label{eq:flatlim-2}
\begin{split}
s^{(-\infty,-)} & \sim e^{c_1 y} \vert y\vert^{m/2} \begin{pmatrix} 1 \\ 0 \\ \end{pmatrix}\left( 1+ \CO(1/y) \right) \\
s^{(-\infty,+)} & \sim e^{-c_1 y} \vert y\vert^{ - m/2} \begin{pmatrix} 0\\ 1 \\ \end{pmatrix}\left( 1+ \CO(1/y) \right) \\
\end{split}
\ee
Since the space of flat sections is two-dimensional we have
\be\label{eq:Def-PQR}
\begin{split}
s^{(-\infty,-)} & = Q s^{(+\infty,+)} - \tilde P s^{(+\infty,-)}\\
s^{(-\infty,+)} & = P s^{(+\infty,+)} + R s^{(+\infty,-)}\\
\end{split}
\ee
where the scattering matrix
\be\label{eq:ScattMtrx}
  \begin{pmatrix} Q & P \\  - \tilde P &  R \\ \end{pmatrix}
\ee
satisfies $QR + P \tilde P = 1$ since the field $\tilde\CA_y$ is traceless.
The peculiar sign choice in front of $\tilde P$ will be convenient later.

Now, the holomorphic part of the Bogomolnyi equations can be written as
\be\label{eq:MonopoleHE}
 [\CD_{\bar z} , \CD_y]=0.
\ee
We may therefore choose the sections $s^{(\pm \infty,\pm)}$ to be annihilated by $\CD_{\bar z}$
as well as $\CD_y$ and therefore the ``S-matrix'' \eqref{eq:ScattMtrx}  is
holomorphic in $z$. The asymptotics \eqref{eq:flatlim-1}
and \eqref{eq:flatlim-2} only determine the bases up to a shift of the growing
solution by a multiple of the decaying solution. This multiple can be a holomorphic
function of $z$ and therefore the S-matrix is only determined up to multiplication
\be\label{eq:ShiftFreedom}
  \begin{pmatrix} Q & P \\  -\tilde P &  R \\ \end{pmatrix} \rightarrow
  \begin{pmatrix} 1 & 0  \\ U_+(z)  & 1 \end{pmatrix}
    \begin{pmatrix} Q & P \\  -\tilde P &  R \\ \end{pmatrix}     \begin{pmatrix} 1 & U_-(z) \\ 0 & 1 \end{pmatrix}
\ee

Note that if the monopoles are all uniformly translated by $\Delta y$ in the
$y$-direction then, from the asymptotics \eqref{eq:flatlim-1} and \eqref{eq:flatlim-2}
we see that the scattering matrix is transformed to
\be\label{eq:TranslateMon}
  \begin{pmatrix} Q & P \\  -\tilde P &  R \\ \end{pmatrix} \rightarrow
  \begin{pmatrix} e^{c_1 \Delta y}  & 0  \\   &  e^{-c_1 \Delta y}  \end{pmatrix}
    \begin{pmatrix} Q & P \\ - \tilde P &  R \\ \end{pmatrix}     \begin{pmatrix} e^{-c_1 \Delta y}  & 0  \\   &  e^{c_1 \Delta y}  \end{pmatrix}
\ee
and in particular $P \to e^{2c_1 \Delta y} P$.
From equation \eqref{eq:Def-PQR} it is clear that the
zeroes of $Q$ correspond to the points $z=w_i$ where there is a
boundstate for the scattering problem, with
exponential decay at both ends $y\to \pm \infty$.  In
the asymptotic region of moduli space with well-separated monopoles
these zeroes represent the positions of the monopoles in the
complex plane, and the $S$-matrix approaches a product of
factors $S_1 S_2 \cdots S_m$ where $S_i$ are the scattering matrices
computed from the one monopole problem
\be\label{eq:OneMonS}
 S_i = \begin{pmatrix} (z-w_i) & e^{\CY_i}  \\  - e^{-\CY_i}   & 0 \\ \end{pmatrix}
\ee
and
\be
\Re(\CY_1) \ll \Re(\CY_2) \ll \cdots \ll  \Re(\CY_m) .
\ee
Again,  $\Re(\CY)$ shifts by $2c_1 \Delta y$ under translation of the
monopole by $\Delta y$ in the $y$-direction, and hence represents the $y$-position
of the monopole.

Given the asymptotic factorization of the scattering matrix into factors
of the form \eqref{eq:OneMonS} it follows that the components of
the matrix \eqref{eq:ScattMtrx} are polynomial functions of $z$.
In particular, $Q$ has degree $m$. Using the residual $SO(2)$ gauge freedom
left unfixed from \eqref{eq:MonopoleAsympt} we can take $Q(z)$ to be monic:
\be
Q(z) = \prod_{i=1}^m(z-w_i).
\ee
Now we can fix the ambiguity \eqref{eq:ShiftFreedom}
by requiring $P,\tilde P$ to have degree $m-1$ and $R$ to have
degree $m-2$.
%A simple way to distinguish $P$ from $\tilde P$
%is to consider the
%asymptotic region of moduli space with well-separated monopoles.
%Under a large translation in the positive $y$ direction
%$\tilde P$ becomes exponentially smaller and $P$ becomes exponentially larger.
%
Thus, we can uniquely associate to a point in monopole moduli space a
rational map $\tilde P/Q$ and, by \cite{DonaldsonRatMap} this is in fact
a diffeomorphism with the space of rational maps.

Given the above rational map presentation of
$\CM(c_1,m)$ we are now ready to introduce the effective superpotential
 on the space $\CM(c_1,m)$. By combining physical arguments with
qualitative features of the moduli space of opers,
reference \cite{Gaiotto:2011nm} proposed the effective superpotential
on monopole moduli space to be:
\be\label{eq:MonopoleW}
W = \sum_{i=1}^m \left( {\rm Res}_{z=w_i} \frac{K(z)  \tilde P(z) dz}{Q(z)} -  \log  \tilde P(w_i) - c_0 w_i  \right) +
\Delta W(z_a)
\ee
where $K(z)$ was defined in \eqref{eq:defK} above and the last term $\Delta W(z_a)$ is independent of the $w_i$,
and cannot be determined by the above arguments.
\footnote{In principle, $\Delta W(z_a)$ could be determined by evaluation of the Chern-Simons
functional on the critical flat connections. In \cite{Gaiotto:2011nm}, it was determined
by a relation to conformal field theory. In any case, in the application to webs below, it
is a constant shift in the superpotential and will not affect the webs.}
Note that we have taken into account the complex symmetry breaking by $c_0$ through the superpotential.

The justification for \eqref{eq:MonopoleW} is given by matching the critical points to
the moduli space of opers with monodromy-free singularities. Assuming that the points $w_i$ are
all distinct we can change coordinates to $\tilde P(w_i) := e^{-\CY_i}$.
As we have seen, in the asymptotic regions of moduli space the
parameters $\Re(\CY_i)$ measure the positions of $m$ basic 't Hooft-Polyakov
monopoles in the $y$ direction so long as these values are large and positive. In this
limit the fields are heavy and it is justified to integrate them out leaving an
effective superpotential for a collection of chiral superfields $W_i$, whose leading term is $w_i$.
The critical points are determined by
\be
e^{\CY_i} =  \frac{K(w_i)}{Q'(w_i)}
\ee
But, since $QR + P \tilde P = 1$, we have $P(w_i) = e^{\CY_i}$. This matches
beautifully with \eqref{eq:PQK}. Moreover, the effective superpotential for the
remaining fields $w_i$ is precisely \eqref{eq:YangYangW}.

Of course, $W$ is not single-valued due to the terms $\log\tilde P(w_i)$. In fact $\pi_1(\CM)=\IZ$ and the
 universal cover of the moduli space is
$\widehat{\CM} = \IR^4 \times \CM_0$, where $\CM_0$ is the simply connected
moduli space of centered monopoles. The superpotential will be
single-valued on this cover. Note that deck transformations shift $W$
by elements of $2\pi \I \IZ$. Until now we have not said how the trace $\Tr$
on $\fg$ used in \eqref{eq:Uc-metric}, \eqref{eq:Uc-Symp}, and \eqref{eq:CS-kappa}
is normalized, but at this point it should be normalized so that the periods of
the Chern-Simons form for the unitary gauge field $A$ are in $2\pi \I \IZ$.

In summary we have the  \emph{monopole model}:

\begin{enumerate}

\item   The K\"ahler  target space is the space of \emph{smooth} monopoles $\CM(c_1,m)$ on $\IR^3$ with
magnetic charge $ m \fh$  and Higgs vev at infinity $c_1 \fh$. Using the standard hyperk\"ahler
metric we  consider it to be K\"ahler in the complex structure determined by $y$.

\item The superpotential is given by \eqref{eq:MonopoleW}.

\end{enumerate}

%
%\cg{Aren't the vacuum weights - defined by the \emph{values} of the Chern-Simons functional
%on the special flat connections supposed to match the vacuum weights of the monopole
%model and the Yang-Yang model? this will fix the normalization of the CS functional
%and maybe determines $\Delta W(z_a)$.}
%

\subsubsection{Finite-Dimensional Models: The Yang-Yang Model}\label{subsubsec:YangYang-Model}

The procedure of integrating out the fields $\CY_i$ in the
 monopole model leads to a model we call the
\emph{Yang-Yang model}. Integrating out heavy fields is expected to
produce a LG model whose associated $LA_\infty$ algebra and categories of
Branes and Interfaces is equivalent to the original one.

To define the  \emph{Yang-Yang model},  we choose
   a collection of $n$ distinct points $z_a\in \IC$ and label them with
 positive integers $k_a$ such that $k = \sum_a k_a $ is even.  Fix an integer $1\leq q \leq k$.
(For $q$ outside this range, the Bethe equations that we arrive at will have no solutions.)
The target space of the model is a covering space of the configuration space $\CC(q,S)$
of $q$ distinct, but indistinguishable points $w_i$, $i=1, \dots, q$ on $S := \IC - \{ z_1, \dots, z_n \}$.
To define the covering space we introduce the superpotential:
\be\label{eq:YangYangW-1}
W = \sum_{i,a} k_a \log(w_i-z_a)  - \sum_{i\not=j} \log(w_i - w_j) -c_0   \sum_i w_i
\ee
The target space $X$ should be the smallest cover of  $\CC(q,S)$ on which $W$ is single-valued as a function of the $w_i$.
Thus, explicitly,  $X=\widehat{\CC}(q,s)/H $ where  $\widehat{\CC}(q,s)$ is the universal cover
and $H$ is the subgroup of $\pi_1$ given by the kernel of the homomorphism  $\oint dW:\pi_1 \to 2\pi \I \IZ$.
$X$ is thus a Galois cover with
covering group $\pi_1/H\cong \IZ $.  We are primarily interested in the case $q = \frac{k}{2}$,
since this is the case that arises from the monopole model.
The derivation of the model from the monopole model suggests that the  K\"ahler metric
should be taken to be the metric induced from the hyperk\"ahler metric
on $\CM(c_1,m)$. Using the discussion of
Section \ref{whynot} above, the algebraic structures of concern to us will be unaffected
if we replace that metric by the much simpler Euclidean metric $\sum_i \vert dw_i \vert^2$
(pulled back to $X$) and we make this choice. In particular, with this choice we
can define the model for any value of $q$. We denote the Yang-Yang model by
$\CT(q,\{k_a,z_a\})$.

\subsubsection{Knot Homology From Interfaces Between Landau-Ginzburg Models}\label{subsec:CategoryAppl}

We can now use the Yang-Yang model to formulate a knot homology complex.
The vacua $\IV$ are in 1-1 correspondence with the (lifts of) the Bethe roots of
\eqref{eq:BetheAnsatz}. They can be labeled (noncanonically)
as  $\vec w^{(r,n)}$, with $r=1,\dots, \prod_{a}(k_a+1)$
and $n\in \IZ$. For each Bethe root $\vec w^{(r,n)}$ the vacuum weight is
determined by the value of the
Yang-Yang function $W^{r,n}= W(\vec w^{(r,0)}) + 2\pi \I n $.
In addition the web representation $\CR$, the interior amplitude $\beta$,
and, once we choose a phase $\zeta$ and a half-plane, the category of branes
are all determined,  in principle, by the physical model.

Now we consider again the link presented as a tangle determined by
the functions $z_a(x^1)$. The vacuum data evolves with $x^1$, as does
the web representation and interior amplitude. When the number of strands
is conserved as a function of $x^1$ the theory of Sections
\S  \ref{sec:Interfaces}-\ref{sec:GeneralParameter} determines an
Interface $\fI[z_a(x)]$ between the initial and final Theories
$\CT(q,\{k_a,z_a^{\rm in} \})$ and $\CT(q,\{k_a,z_a^{\rm out}\})$.
As we have explained, this can be used to construct an \afty\ functor
between the initial and final \afty\ categories of Branes.

In addition, if two points, say $z_{a_1}(x)$ and $z_{a_2}(x)$, carry the same
integer, $k_{a_1}=k_{a_2}=k_1$, then they can be ``annihilated,'' changing $n+2 \to n$.
The ``time-reversed'' process creates two points and changes $n \to n+2$.
Taking, for simplicity $q = \half k$, the annihilation must lead to
an Interface  $\fI^{n+2 \to n}(a_1,a_2) \in \fB\fr\bigl( \CT(q,\{k_a,z_a\}), \CT(q',\{k_a,z_a\}')\bigr) $,
where $q' = q-k_1$ and $\{k_a,z_a\}'= \{k_a,z_a\} - \{ k_{a_1}, k_{a_2}, z_{a_1},z_{a_2} \}$.
Similarly, the creation process must lead to an interface
$\fI^{n\to n+2}(a_1,a_2) \in \fB\fr\bigl(\CT(q',\{k_a,z_a\}') ,\CT(q,\{k_a,z_a\}) \bigr)$.
Moreover, these interfaces should be related by parity-reversal.

Up to homotopy equivalence of Interfaces the Interface associated with
a tangle can be decomposed into products of Interfaces
associated with elementary positive or negative braidings
of single pairs of points and associated to creation and
annihilation of pairs of points. We are thus led to consider
four types of basic interfaces:

\begin{enumerate}

\item If the path $\wp^\pm_{a_1,a_2}$ braids two points
$z_{a_1}(x)$ and $z_{a_2}(x)$ while all other points $z_a(x)$, for $a\not= a_1,a_2$
are fixed (on some small interval in $x$) then there will be \emph{braiding
Interfaces} $\fI^{\pm}(\wp^\pm_{a_1,a_2})$ between the theory
$\CT(\{ z_a \}, \{ k_a \})$ and the theory with $k_{a_1} \leftrightarrow k_{a_2}$.  The superscript indicates whether
the braiding is clockwise or counterclockwise. These will be very similar to the S-wall
interfaces discussed above.

\item If two points $z_{a_1}(x)$ and $z_{a_2}(x)$ annihilate
then there will be an \emph{annihilation  Interface} $\fI^{n\to n+2}(a_1,a_2)$
as described above. Similarly, there will be creation Interfaces
$\fI^{n+2\to n}(a_1,a_2)$.

\end{enumerate}

Now, a tangle such as shown in Figure \ref{fig:KNOT-HOM-4} is an $x^1$-ordered instruction of
creation of pairs of points, braidings of points, and annihilations of pairs of points.
Let us denote the corresponding ordered set of Interfaces for the tangle as $\fI_1,\dots, \fI_N$,
for some $N$, where each $\fI_s$ is one of the four types of basic interfaces described
above. Then we can use the interface product $\boxtimes$ described above to
construct
\be\label{eq:Link-Interface}
\fI({\rm Tangle}) :=  \fI_1\boxtimes \cdots \boxtimes  \fI_N.
\ee
The Interface \eqref{eq:Link-Interface} is an Interface between the trivial Theory
and itself.  As we saw in Section
\S \ref{subsubsec:TrivialTheories} an Interface between the trivial Theory and
itself is a chain complex. We propose that this chain complex defines a
 knot homology complex $\widehat{\CK(L)}$. Moreover, in the case of $G=SO(3)$ and all $k_a = 1$
this should give a theory equivalent to Khovanov homology.

The required double-grading on $\widehat{\CK(L)}$ comes about as follows: The $R_{ij}$ and Chan-Paton
data have the usual grading by Fermion number $\CF$. The second grading comes from the
fact that $dW$ has periods. As we have seen, the vacua $\vec w^{r,n}$ are labeled
(noncanonically) by a sheet index $n\in \IZ$. It is natural to assign $q$-grading
$(n_2 - n_1 )$ to the Morse complexes $R_{(r_1,n_1),(r_2,n_2)}$ and $q$-grading
$n$ to the Chan-Paton spaces $\CE_{(r,n)}$ of the Branes and Interfaces.
One important statement which is expected, but should be proven, is that the Interface
\eqref{eq:Link-Interface} does not depend on the tangle presentation of $L$, up to
homotopy equivalence of Interfaces.

It remains to construct the elementary interfaces. This is beyond
the scope of the present work and will be addressed elsewhere.
We simply mention that one can get a good analytic understanding of the Bethe
roots and the critical values of $W$ in the limit that $c_0$ is large but $\delta = c_0(z_1-z_2) \to 0$
as a double expansion in $1/c_0$ and $\delta$. Moreover, the critical values form
two clusters of vacua
\footnote{These clusters can be nicely understood in
terms of bases of conformal blocks for degenerate representations of the Viraroso
algebra.}
and hence the results on cluster webs from Section
\S \ref{subsec:partialRG} will be relevant to the construction.

\bigskip
\noindent
\textbf{Example} In order to illustrate some of the issues which must be overcome
to make this proposal computationally effective let us consider the construction of
the complex  $\widehat{\CK(L)}$  for the unknot with $k_1=k_2=1$.
In this case two points $z_a$, $a=1,2$ are simply
created and then annihilated. The superpotential, after creation,  is just
\be
W = \log(w - z_1) + \log(w-z_2) - c_0 w
\ee
The target space $X$ is therefore a cyclic cover of the plane with two punctures.
%
%\footnote{In more detail: The universal cover of the plane with two punctures
%has a fundamental group based, say, at $w_*$. It is a free group, and two
%generators can be taken to be two simple loops $\gamma_1,\gamma_2$ based at $w_*$ and winding
%once in the clockwise direction around $z_1,z_2$. Therefore, $X$ is the quotient of the universal cover
%by the subgroup generated by $\gamma_1\gamma_2^{-1}$. }
%
Writing $z_1 = \delta/c_0$ and $z_2 = -\delta/c_0$, the two Bethe roots are
\be
w^\epsilon =  \frac{1}{c_0} + \epsilon \frac{\sqrt{ 1+ \delta^2 }}{c_0}
\ee
where $\epsilon \in \{ \pm 1 \}$.
Hence the vacua on $X$ can be denoted by  $w^{\epsilon,n}$, $n\in \IZ$. The
critical values of $W$ at these vacua are
\be\label{eq:CritExact-1}
W^{\epsilon,n} =    2\log \frac{1}{c_0} + \log[ 2(1 + \epsilon  \sqrt{ 1+ \delta^2 } ) ]   -2  + 2\pi \I n
\ee
The soliton spaces $R_{ (\epsilon_1, n_1), (\epsilon_2, n_2) }$ clearly only depend on
$n=n_1-n_2$ so denote we denote them simply by $R_{\epsilon_1,\epsilon_2,n}$. This space
should have $q$ grading $q^n$.
For $\epsilon_1\epsilon_2=+1$ the spaces are nonzero only for $n=\pm 1$.
When $\epsilon_1 \epsilon_2 = -1$ there will be an infinite number of nonzero spaces $R_{\epsilon_1,\epsilon_2,n}$.
\footnote{Whether or not the cohomology of $R_{\epsilon_1,\epsilon_2,n}$ is nonzero is more subtle.
Experience with the closely related $\IC\IP^1$ model with a twisted mass parameter suggests that
this depends on $c_0$ and $\delta$. }
Hence there is an infinite number of soliton slopes, with an accumulation slope along the vertical axis.
We therefore choose
boundary conditions to preserve $\zeta$-supersymmetry with $\zeta \not=\pm 1$. There are in
principle an infinite number of cyclic fans and interior amplitudes, but the $L_\infty$
Maurer-Cartan equations will be well-defined, as discussed above in Section \S \ref{subsec:TwistedMasses}.
Specifying the creation
Interface $\fI^{>}$ requires specifying the Chan-Paton spaces $\CE(\fI^{>})_{\epsilon,n}$
together with the boundary amplitudes. Once this is determined the annihilation Interface
$\fI^{<} $ is just the parity reverse. Thus, altogether, the complex for the unknot
is
\be
\oplus_{\epsilon,n}  \CE(\fI^{>})_{\epsilon,n}\otimes   \CE(\fI^{<})_{\epsilon,n}^*
\ee
with a differential obtained by our formalism using the taut strip-webs and the boundary and interior amplitudes.
We hope to return to a more complete analysis on another occasion.

\textbf{Remarks}

\begin{enumerate}

\item The generalization where $C$ is a Riemann surface makes contact with
the theory of surface defects in class S theories. The surface defect theories
for  $\IS_{z_1,\dots, z_n}$ mentioned above should be closely related
to the Landau-Ginzburg models we have just discussed based on the
data $z_a$. An important lesson we learn is that there are topological interactions between the
distinct M2 branes ending on the UV curve, and they cannot be treated independently,
even when they are far separated.

\item Knot bordisms can be incorporated into our formalism by using
the $(x^0,x^1)$-dependent data of Theories discussed in Section
\S \ref{subsubsec:Homotopy-homotopy} and \S \ref{sec:GeneralParameter}.

\item  The
categorified version of the skein relations should translate into some
interesting relations between
the basic Interfaces described above. In order to prove, for example, that
the Interface \eqref{eq:Link-Interface} is independent of the tangle
presentation up to homotopy equivalence it would then
 suffice to prove these relations on the Interfaces.

\end{enumerate}

\section*{Acknowledgements}

We would especially like to thank Nick Sheridan for many useful
discussions, especially concerning the Fukaya-Seidel category. We would also like to thank M. Abouzaid, K. Costello, T. Dimofte,
D. Galakhov, E. Getzler, M. Kapranov, L. Katzarkov, M. Kontsevich,  Kimyeong Lee,  S. Lukyanov, Y. Soibelman, and A. Zamolodchikov
for useful discussions and  correspondence.
The research of DG was supported by the
Perimeter Institute for Theoretical Physics. Research at Perimeter Institute is supported
by the Government of Canada through Industry Canada and by the Province of Ontario
through the Ministry of Economic Development and Innovation.
The work of   GM is supported by the DOE under grant
DOE-SC0010008 to Rutgers and NSF Focused Research Group award DMS-1160591.  GM also gratefully acknowledges
the hospitality of  the Institute for Advanced Study, the Perimeter Institute for Theoretical Physics,
the Aspen Center for Physics  (under
NSF Grant No. PHY-1066293), the
KITP at UCSB (NSF Grant No. NSF PHY11-25915) and the
Simons Center for Geometry and Physics. 
The work of EW is supported in part
by NSF Grant PHY-1314311.

\appendix

\section{Summary Of Some Homotopical Algebra }\label{App:HomotopicalAlgebra}

This subject is well-reviewed. See, for examples, \cite{ClayBook2,KellerIntroduction,SeidelBook}.
We briefly summarize some material to establish our notation and conventions and, in
some places, to emphasize a slightly nonstandard viewpoint on this standard material.

Throughout this appendix and the next, the term ``module'' refers to either a $\IZ$-module,
i.e., an abelian group, or a vector space over a field. All modules
are assumed to be $\IZ$-graded. Moreover,
all infinite sums are assumed to be convergent.

\subsection{Shuffles And Partitions}

If $P$ is any ordered set we define an ordered $n$-partition of $P$ to be an ordered disjoint
decomposition into $n$ ordered subsets
\be
P = P_1 \amalg P_2 \amalg \cdots \amalg P_n
\ee
where the ordering of each summand $P_\alpha$ is inherited
from the ordering of $P$ and all the elements of $P_\alpha$ precede all elements of $P_{\alpha+1}$  inside $P$.
We allow the $P_\alpha$ to be the empty set.
For an ordered set $P$ we let ${\rm Pa}_n(P)$ denote
the set of distinct $n$-partitions of $P$. If $p=\vert P\vert$ there are
${n+p-1 \choose p}$ such partitions. For example, if $n=2$ there are $p+1$ different 2-partitions.
Each 2-partition is completely determined by specifying the number of elements in $P_1$.
This number can be any integer from $0$ to $p$.

If $S$ is an ordered set then an $n$-shuffle of $S$ is an ordered disjoint
decomposition into $n$ ordered subsets
\be\label{eq:S-shuff}
S = S_1 \amalg S_2 \amalg \cdots \amalg S_n
\ee
where the ordering of each summand $S_\alpha$ is inherited
from the ordering of $S$ and the $S_\alpha$ are allowed
to be empty. Note that the ordering of the
sets $S_\alpha$ also matters so that $S_1\amalg S_2$
and $S_2 \amalg S_1$ are distinct 2-shuffles of $S$.
For an ordered set $S$ we let ${\rm Sh}_n(S)$ denote
the set of distinct $n$-shuffles of $S$.
We can count $n$-shuffles by successively asking
each element of $S$  which set $S_\alpha$ it belongs to.
Hence there are  $n^{\vert S\vert}$ such shuffles.

\subsection{$A_\infty$ Algebras}

\def\Pa{{\rm Pa}}

An $A_\infty$ structure $\mu$ on  a module $\CA$ is defined by a collection of multilinear maps
\be
\mu_n: \CA^{\otimes n} \to \CA
\ee
of degree $2-n$, which satisfies
the following $A_\infty$ associativity relation:
\be\label{eq:Aass}
\sum_{P_{1,2,3} \in {\rm Pa}_3(P)} \epsilon_P(P_{1,2,3}) \, \mu\left(P_1,\mu(P_2), P_3 \right) = 0
\ee
for all ordered sets $P$ of elements in $\CA$. (We identify ordered sets of elements in $\CA$ with
monomials in $\CA^{\otimes n}$.)  We define  for convenience
$\mu(P_\alpha) := \mu_{|P_\alpha|}(P_\alpha)$ and we take $\mu(\emptyset) := 0$.
\footnote{It is straightforward, if needed, to relax the axioms
for $A_\infty$ structures by allowing a choice of nonvanishing ``source'' $\mu(\emptyset)\in \CA$.
Such an algebra is called a ``curved \afty-algebra'' in the math literature.
}
%
%\cg{It might be nice to explain that this occurs in $\Hop(\fB,\fB)$ for
%boundary conditions where $Q^2\not=0$. If we can't understand that, just
%drop this marginalia.}
%
We will define the sign $\epsilon_P(P_{1,2,3}) $ momentarily. The pair $(\CA, \mu)$ is called an $A_\infty$ algebra.

In this paper, we choose the sign convention where $\epsilon_P(P_{1,2,3})= (-1)^{\deg_r(P_1)}$,
where the reduced degree $\deg_r$ for a set $\{ a_1, \cdots a_n\}$ of elements in $\CA$
is
\be \label{eq:degr}
\deg_r(\{ a_1, \cdots a_n\}) = \sum_{k=1}^n \left(\deg(a_k) -1 \right)
\ee
This is an example of the {\it reduced Koszul rule} we use throughout this paper:
in order to determine the sign $\epsilon_P(P_{1,2,3})$ we assign {\it reduced degree} $\deg_r(a) = \deg(a)-1$
to arguments in $P$ and $\deg_r(\mu)=1$ to $\mu$ and use the Koszul rule
with the reduced degree in order to bring the symbols in \eqref{eq:Aass} from a canonical
order $\mu \mu P$ to the order they appear with in the equation.
For example, the first few associativity relations are
\begin{align} \label{eq:Aex}
& \mu_1\left(\mu_1(a)\right)=0 \cr
& \mu_1\left(\mu_2(a_1,a_2)\right) + \mu_2\left(\mu_1(a_1),a_2\right) +(-1)^{\deg(a_1)-1} \mu_2\left(a_1,\mu_1(a_2)\right) =0 \cr
& \mu_1\left(\mu_3(a_1,a_2,a_3)\right) + \mu_2\left(\mu_2(a_1,a_2),a_3\right) +(-1)^{\deg(a_1)-1} \mu_2\left(a_1,\mu_2(a_2,a_3)\right)
+ \mu_3\left(\mu_1(a_1),a_2,a_3\right) \cr& +(-1)^{\deg(a_1)-1} \mu_3\left(a_1,\mu_1(a_2),a_3\right) +(-1)^{\deg(a_1)+\deg(a_2)}\mu_3\left(a_1,a_2,\mu_1(a_3)\right)  =0 \cr
& \cdots
\end{align}

The choice of signs   based on the reduced Koszul rule may appear surprising at first sight.
In order to understand why it is the ``correct'' choice, as opposed to naive alternatives,
such as a sign based on a standard Koszul rule, it is useful to point out a simple consistency condition on
the associativity constraint \eqref{eq:Aass}. Choose an ordered set of elements $P$ in $\CA$ and, for
any $P_{1,2,3}\in \Pa_3(P)$ let $\widehat{P_{1,2,3}}$ denote the ordered set of objects $P_1 \amalg \{ \mu(P_2) \} \amalg P_3 $.
We consider the associativity constraint \eqref{eq:Aass} for the ordered set $\widehat{P_{1,2,3}}$,
and then sum that over $\Pa_3(P)$, weighted by $\epsilon_P(P_{1,2,3})$:
\be\label{eq:DoubSum}
\sum_{P_{1,2,3}\in \Pa_3(P)} \epsilon_P(P_{1,2,3}) \sum_{Q_{1,2,3}\in \Pa_3(\widehat{P_{1,2,3}})}
\epsilon_{ \widehat{P_{1,2,3}} }(Q_{1,2,3}) \mu(Q_1, \mu(Q_2), Q_3) = 0
\ee
Now, to each nested partition $Q_{1,2,3}$ we can associate a 5-partition of $P$.
There are three cases, according to whether $\mu_2(P_2)$ is in $Q_1$, $Q_2$, or $Q_3$.
In each case, if   $\mu(P_2) \in Q_a$ then we write $Q_a$ as a disjoint union $Q_a = Q_a' \amalg \{ \mu(P_2) \} \amalg Q_a'' $
and to the partition $Q_{1,2,3}$ we associate the 5-partition of $P$ given by
\be
 Q_1' \amalg P_2 \amalg Q_1'' \amalg Q_2 \amalg Q_3  \qquad  \qquad  \mu_2(P) \in Q_1
\ee
\be
 Q_1 \amalg Q_2' \amalg P_2 \amalg Q_2'' \amalg Q_3  \qquad  \qquad  \mu_2(P) \in Q_2
\ee
\be
 Q_1 \amalg Q_2  \amalg Q_3' \amalg P_2 \amalg Q_3''  \qquad  \qquad  \mu_2(P) \in Q_3
\ee
We now decompose the sum \eqref{eq:DoubSum} into three terms corresponding to these three cases:
\be\label{eq:Aa3case}
\begin{split}
0 = & \sum_{\mu_2(P_2)\in Q_1}   \epsilon_P(P_{1,2,3})\epsilon_{ \widehat{P_{1,2,3}} }(Q_{1,2,3})
\mu( Q_1', \mu(P_2) , Q_1'', \mu(Q_2), Q_3) \\
 & + \sum_{\mu_2(P_2)\in Q_2}   \epsilon_P(P_{1,2,3})\epsilon_{ \widehat{P_{1,2,3}} }(Q_{1,2,3})
\mu( Q_1 , \mu(Q_2', \mu(P_2), Q_2''), Q_3) \\
& + \sum_{\mu_2(P_2)\in Q_3}   \epsilon_P(P_{1,2,3})\epsilon_{ \widehat{P_{1,2,3}} }(Q_{1,2,3})
\mu( Q_1 , \mu(Q_2), Q_3', \mu(P_2), Q_3'') \\
\end{split}
\ee
The middle sum can be rearranged with an inner sum over partitions of the fixed set $R= Q_2' \amalg P_2 \amalg Q_2''$.
Since $\epsilon_P(P_{1,2,3})\epsilon_{ \widehat{P_{1,2,3}} }(Q_{1,2,3})= (-1)^{\deg_r(Q_2')}$ the signs
are such that the middle sum is zero by \eqref{eq:Aass}. The first and third lines of \eqref{eq:Aa3case}
will cancel each other by exchanging the roles of $\mu(P_2)$ and $\mu(Q_2)$, but the cancelation
requires that we use the reduceed Koszul rule in formulating the equations. If we had used a standard
Koszul rule, rather than the reduced Koszul rule, the terms would not have cancelled out,
and the associativity constraints would have been over-constraining.

%Pick an ordered set $P'$ of elements in $\CA$ and sum over all ordered $3$-partitions
%$P'_1 \amalg P'_2 \amalg P'_3$ the associativity constraint with argument $P[P'] = P'_1 \amalg \{ \mu(P'_2) \} \amalg P'_3$:
%\be\label{eq:Acons}
%\sum_{{\rm Pa}_3(P')} \epsilon_{P'} \left[ \sum_{{\rm Pa}_3(P[P'])}  \epsilon_{P[P']} \, \mu\left(P_1,\mu(P_2), P_3 \right) \right]= 0
%\ee
%We can regroup the terms in this sum depending on $\mu(P'_2)$ being contained in the first, second or third subsets.
%The terms within each group are naturally in one-to-one correspondence with ordered $5$-partition of $P'$:
%\begin{align}\label{eq:Acons2}
%\sum_{{\rm Pa}_5(P')} &\epsilon_{P'} \, \mu\left(P'_1,\hat \mu(P'_2), P'_3,\mu(P'_4), P'_5 \right) +\epsilon'_{P'} \, \mu\left(P'_1,\mu(P'_2,\hat %\mu(P'_3),P'_4), P'_5 \right) + \cr + &\epsilon''_{P'} \, \mu\left(P'_1,\mu(P'_2), P'_3,\hat \mu(P'_4), P'_5 \right) = 0
%\end{align}
%%We marked with a hat the $\mu(P_\alpha)$ argument which arises from $\mu(P_2)$ in \ref{eq:Acons}.
%The signs are determined by the reduced Koszul rule, starting from a $\mu \mu \hat \mu P$ order of the symbol.
%
%The middle term in the sum cancels out because of the associativity constraint applied to the inner pair of $\mu$
%operations, with argument $P'_2 \amalg P'_3 \amalg P'_4$. The first and last term cancel against each other, because they
%differ only by the relative order of the inner $\mu$ and $\hat \mu$ symbols, which have reduced degree $1$.
%

It is possible to seek for sign redefinitions of the $\mu_n$ maps which give the axioms a more familiar form. For example, we could define maps
$\tilde \mu_n$ by
\be
\left[ \prod_{m=1}^n \theta_m\right] \mu_n(a_1, \cdots a_n)  =\tilde \mu_n( \theta_1 a_1, \cdots \theta_n a_n),
\ee
where $\theta_m$ are formal variables of degree $1$ which are manipulated on the right hand side using a standard Koszul relation. Concretely,
\begin{align}
\mu_1(a_1) &= - \tilde \mu_1(a_1) \cr
\mu_2(a_1, a_2) &= (-1)^{\deg(a_1)} \tilde \mu_2(a_1, a_2) \cr
\mu_3(a_1, a_2,a_3) &= - (-1)^{2 \deg(a_1)+\deg(a_2)} \tilde \mu_3(a_1, a_2,a_3) \cr
\cdots &
\end{align}
Indeed, plugging this into the associativity relations \ref{eq:Aex}, we get conventional-looking graded associativity relations
\begin{align}
& \tilde \mu_1\left(\tilde \mu_1(a)\right)=0 \cr
& \tilde \mu_1\left(\tilde \mu_2(a_1,a_2)\right) - \tilde \mu_2\left(\tilde \mu_1(a_1),a_2\right) -(-1)^{\deg(a_1)} \tilde \mu_2\left(a_1,\tilde \mu_1(a_2)\right) =0 \cr
& \tilde \mu_1\left(\tilde \mu_3(a_1,a_2,a_3)\right) +\tilde \mu_2\left(\tilde \mu_2(a_1,a_2),a_3\right) - \tilde \mu_2\left(a_1, \tilde \mu_2(a_2,a_3)\right)
+ \tilde \mu_3\left(\tilde \mu_1(a_1),a_2,a_3\right) \cr& +(-1)^{\deg(a_1)} \tilde \mu_3\left(a_1,\tilde \mu_1(a_2),a_3\right) +(-1)^{\deg(a_1)+\deg(a_2)}\tilde \mu_3\left(a_1,a_2,\tilde \mu_1(a_3)\right)  =0 \cr
& \cdots
\end{align}

Although homotopical algebra has its origins in the homotopy theory of H-spaces, it has a highly
abstract and algebraic character. It is useful to give the equations a geometrical interpretation.
One such interpretation involves odd vector fields on non-commutative manifolds.
Here we will emphasize the physical interpretation according to which  $A_\infty$ algebras
encode general non-linear gauge symmetries. This interpretation arises naturally in the applications to
string field theory  (see, for examples,
\cite{Gaberdiel:1997ia,Zwiebach:1997fe,Zwiebach:1992ie})  and constitutes a particularly interesting
class of such odd vector fields.
In order to make this relation explicit, we should identify the degree $1$ elements $a \in \CA$ as ``connections,'' and define a ``covariantized'' differential $\CA \to \CA$
\be
\mu_a(\tilde a) = \sum_{k=0}^\infty \sum_{n=0}^\infty \mu(a^{\otimes k}, \tilde a, a^{\otimes n}).
\ee
The ``field strength'' for such an $\CA$-connection is defined naturally as
\be
\CF_a = \sum_{n=1}^\infty \mu( a^{\otimes n} )
\ee
and has degree two.

If we plug $a^{\otimes n}$ into the $A_\infty$ associativity relations and sum over $n$ we find that the field strength is covariantly closed
\be
\mu_a( \CF_a) =0.
\ee
If we plug $a^{\otimes k} \otimes \tilde a \otimes a^{\otimes n}$ and sum over $k,n$,   we find that the square of the covariantized differential
is proportional to the field strength:
\be \label{eq:dda}
\mu_a\left( \mu_a( \tilde a) \right) + \mu_a(\CF_a, \tilde a) + (-1)^{\deg_r(\tilde a)} \mu_a(\tilde a,\CF_a) =0
\ee
where we defined the second deformed operation
\be
\mu_a(\tilde a_1, \tilde a_2) = \sum_{n_0=0}^\infty \sum_{n_1=0}^\infty \sum_{n_2=0}^\infty\mu(a^{\otimes n_0}, \tilde a_1, a^{\otimes n_1},\tilde a_2, a^{\otimes n_2}).
\ee
The first term in \ref{eq:dda} collects all terms in the associativity relation \ref{eq:Aass} such that $\tilde a \in P_2$, the second and third collect all terms where
$\tilde a \in P_3$ and $\tilde a \in P_1$ respectively.
In particular, the field strength transforms covariantly under the infinitesimal ``gauge transformations'' $a \to a + \mu_a(\tilde a)$ (where $\tilde a$ has degree zero):
\be
\CF_a \to \CF_a + \mu_a(\mu_a(\tilde a)) = \CF_a + \mu_a(\tilde a,\CF_a) - \mu_a(\CF_a, \tilde a).
\ee

Thus ``flat connections'' coincide with solutions of the Maurer-Cartan equation for the $A_\infty$ algebra:
\be \label{eq:MC}
\sum_{n=1}^\infty \mu( a^{\otimes n} ) =0.
\ee
Given such a flat $\CA$-connection $a$, we can actually define a
full finite deformation of all the original $A_\infty$ operations on $\CA$:
\be
\mu_a(\tilde a_1, \cdots, \tilde a_k) = \sum_{n_0=0}^\infty \cdots \sum_{n_k=0}^\infty \mu(a^{\otimes n_0}, \tilde a_1, a^{\otimes n_1},\cdots,a^{\otimes n_{k-1}}, \tilde a_k, a^{\otimes n_k})
\ee
The $A_\infty$ associativity relation for $\mu_a$ can be easily rewritten as a linear combination of associativity relations for $\mu$.\footnote{The proof of this is a special case of the discussion below
\eqref{eq:BrCatDef} below. In $a$ is not flat, we find instead the weaker version of the $A_\infty$ relations, with non-zero source $\mu_a(\emptyset) = \CF_a$}
If $\tilde a$ satisfies the MC equations for $\mu_a$, then $a + \tilde a$ satisfies the MC equations for $\mu$.

It is straightforward to extend this analogy to ``matrix-valued connections''. Given any differential graded module
$\CE$, we can extend the $\mu$ operations to multilinear maps $\mu^\CE$ on $\CA^\CE := \CE \otimes \CA \otimes \CE^*$ by matrix multiplication, i.e.
simply by contracting the $\CE^*$ and $\CE$ factors of consecutive arguments and acting on the $\CA$ factors with the vanilla $\mu$ maps. The sign conventions needed to to satisfy \eqref{eq:Aex} are a little tricky, so we spell them out.
The differential $\mu_1$ on $\CA^\CE$ is the natural one induced by that on the three factors.
It is crucial to use the convention that the differentials on $\CE$ and $\CE^*$ are related
by the sign rule $\mu_{1,\CE^*} (e_1^*)\cdot e_2 =  (-1)^{\deg(e_1^*)  } e_1^* \cdot \mu_{1,\CE}(e_2)$
when checking \eqref{eq:Aex}.
Thus, for example,
if $a\in \CA$, $e\in \CE$ and $e^*\in \CE^*$ are three independent homogeneous elements then
\be\label{eq:MatEx-1}
\mu_1(  e a e^*) = \mu_{1,\CE}(e) a e^* + (-1)^{\deg(e)}  e\mu_{1,\CA}(a) e^* + (-1)^{\deg(e) + \deg(a) } e a \mu_{1,\CE^*}(e^*)
\ee
(Notice we use the standard Koszul rule for computations with $\mu_1$.)
For $n>1$ the multiplication is defined by
\be\label{eq:MatEx-2}
\mu_n(  (e_1 a_1 e_1^*) \cdots (e_n a_n e_n^*) ) = (-1)^{\deg(e_1) } ~ \sigma ~~ e_1 \mu_n(a_1, \dots, a_n) e_n^*
\ee
where $\sigma $ is the scalar:
\be\label{eq:MatEx-3}
\sigma = (e_1^* \cdot e_2) \cdots ( e_{n-1}^* \cdot e_n)
\ee
Again, the sign in \eqref{eq:MatEx-1} is crucial to checking \eqref{eq:Aex}.

A (flat) matrix-valued $\CA$ connection is defined simply as a (flat) $\CA^\CE$-connection.

\subsection{$A_\infty$ Morphisms}

Given two $A_\infty$ algebras $\CA$ and $\CB$, with operations $\mu_{\CA}$ and $\mu_{\CB}$, we can define an $A_\infty$ morphism from $\CA$ to $\CB$
as a collection of multi-linear maps
\be
\phi_n: \CA^{\otimes n} \to \CB  \qquad\qquad n\geq 1.
\ee
The degree of $\phi$ is   $1-n$ and hence it has reduced degree $\deg_r(\phi)=0$. It must satisfy
the following relations:
\be \label{eq:Amorph}
\sum_{{\rm Pa}_3(P)} \epsilon_P(P_{1,2,3})
 \, \phi\left(P_1,\mu_\CA(P_2), P_3 \right) = \sum_{n\geq 1} \sum_{{\rm Pa}_n(P)} \mu_\CB(\phi(P_1), \cdots, \phi(P_n)),
\ee
for all ordered sets $P$ of elements in $\CA$, where we defined for convenience $\phi(P_\alpha) := \phi_{|P_\alpha|}(P_\alpha)$ and $\phi(\emptyset) := 0$. The identity morphism, $\CB=\CA$,  with $\phi_n=0$ for $n>1$ and $\phi_1(a)=a$, is an \afty-morphism.

Given three $A_\infty$ algebras $\CA$ and $\CB$ and $\CC$, and $A_\infty$ morphisms $\phi$ from $\CA$ to $\CB$ and $\phi'$ from $\CB$ to $\CC$,
we can define the composition of the two $A_\infty$ morphisms by:
\be
\left[ \phi' \circ \phi\right](P) :=  \sum_{n\geq 1} \sum_{{\rm Pa}_n(P)} \phi'(\phi(P_1), \cdots, \phi(P_n)).
\ee
Composition with the identity morphism is a left- and right- identity element for this composition.
One can verify that this composition is associative as follows:
The composition $\phi''\circ (\phi'\circ\phi)(P)$ involves a sum over nested
partitions
\be\label{eq:NstP-1}
P = \bigl[ P_1^{(1)} \amalg \cdots \amalg P_1^{(k_1)}\bigr] \amalg \bigl[ P_2^{(1)} \amalg \cdots \amalg P_2^{(k_1)}\bigr]
\amalg \cdots \amalg \bigl[ P_n^{(1)} \amalg \cdots \amalg P_n^{(k_n)}\bigr]
\ee
where $n, k_1, k_2 , \dots \geq 1$. On the other hand, the composition $(\phi''\circ \phi')\circ\phi  (P)$
is a sum over all double partitions, where we first partition $P$ into $P_{1,\dots, n}$ and then consider
partitions:
\be\label{eq:NstP-2}
\{ \phi(P_1), \dots , \phi(P_n)\} =  Q_1 \amalg \cdots \amalg Q_N.
\ee
But there is a 1-1 correspondence between these two kinds of partitions.  Similar manipulations
confirm that in fact $\phi'\circ\phi$ does define an \afty-morphism from $\CA$ to $\CC$.
It is simplest to start with the right-hand-side of the identity, identify it as a sum over
nested partitions and rearrange that sum as a sum over partitions of the form \eqref{eq:NstP-2}:
\be
\sum_{n\geq 1} \sum_{\Pa_n(P)} \sum_{N\geq 1} \sum_{\Pa_N(\{\phi(P_1),\dots, \phi(P_n)\} )}
\mu_{\CC}(\phi'(Q_1), \dots, \phi'(Q_N))
\ee
Next use the fact that $\phi'$ is an \afty-morphism. The resulting sum can be rearranged
to give the left-hand-side of the desired identity. Thus, the set of \afty-morphisms
forms a standard category.

A simple observation is that an $A_\infty$ morphism $\phi$ from $\CA$ to $\CB$ can be used to map flat $\CA$-connections to flat $\CB$-connections:
if $a$ satisfies the MC equations for $\CA$, then the degree one element
\be \label{eq:pull}
b:= \phi_a(\emptyset) := \sum_{n=1}^\infty \phi( a^{\otimes n} )
\ee
satisfies the MC equations for $\CB$. This can be shown simply by plugging $a^{\otimes n}$ in \ref{eq:Amorph},
recalling that $\deg_r(\phi)=0$,  and summing over $n$.
Similarly, the morphism maps gauge transformations to gauge transformations: If we define
\be \label{eq:pull2}
\tilde b:= \phi_a(\tilde a)= \sum_{k=0}^\infty \sum_{n=0}^\infty \phi( a^{\otimes k}, \tilde a, a^{\otimes n})
\ee
then the gauge transformation $a \to a + \mu_a(\tilde a)$ induces the transformation
 \be \label{eq:pull3}
\phi_a(\emptyset) \to \phi_a(\emptyset) + \phi_a(\mu_a(\tilde a)) = \phi_a(\emptyset) + \mu_{\phi_a(\emptyset)}(\phi_a(\tilde a))
= b + \mu_{\CB,b}(\tilde b).
\ee
Similarly multilinear maps $\phi_a: \CA^{\otimes n} \to \CB$ give an $A_\infty$ morphism from the finite deformation of $\CA$ by $a$ to the finite deformation of $\CB$ by $b=\phi_a(\emptyset)$.

A more surprising observation is that the equations \ref{eq:Amorph} which define $A_\infty$ morphisms from $\CA$ to $\CB$
can be reinterpreted as the MC equations for an $A_\infty$ algebra $\Hop(\CA,\CB)$. A degree $k$ element of the algebra $\Hop(\CA,\CB)$ is
a collection $\alpha$ of multi-linear maps $\alpha_n: \CA^{\otimes n} \to \CB$ of degree $k-n$ (i.e. reduced degree $k-1$). The $A_\infty$ operations are
\begin{align}
\left[ \mu_{\Hop(\CA,\CB)}(\alpha) \right](P) &= \mu_\CB(\alpha(P)) + (-1)^{\deg(\alpha)} \sum_{{\rm Pa}_3(P)} \epsilon_P \, \alpha\left(P_1,\mu_\CA(P_2), P_3 \right) \cr
\left[ \mu_{\Hop(\CA,\CB)}(\alpha_1, \cdots, \alpha_n) \right](P) &= \sum_{{\rm Pa}_n(P)} \mu_\CB(\alpha_1(P_1), \cdots, \alpha_n(P_n))
\end{align}
The $A_\infty$ associativity relations for $\mu_{\Hop(\CA,\CB)}$ can be reduced to the relations for $\CA$ and $\CB$ in a straightforward, if tedious, way.
The three groups of terms involving two $\mu_\CB$, a $\mu_\CB$ and a $\mu_\CA$ or two $\mu_\CA$ respectively cancel out separately.
Thus $A_\infty$ morphisms from $\CA$ to $\CB$ coincide with flat $\Hop(\CA,\CB)$-connections. In particular they inherit the $A_\infty$ category structures discussed in appendix \ref{app:cat}.

\subsection{$A_\infty$ Modules}

A (left) $A_\infty$-module for an \afty-algebra $\CA$
is a module $\CM$ equipped with a collection of multi-linear maps
\be\label{eq:ModMaps}
\nu_n: \CA^{\otimes n} \otimes \CM \to \CM \qquad n\geq 0
\ee
of degree $1-n$. These maps must satisfy
the following relations:
\be \label{eq:Amod}
\sum_{P_{1,2,3}\in{\rm Pa}_3(P)} \epsilon_P(P_{1,2,3}) \, \nu\left(P_1,\mu_\CA(P_2), P_3;m\right) +
\sum_{P_{1,2}\in{\rm Pa}_2(P)} \epsilon_{P}(P_{1,2}) \nu(P_1; \nu(P_2;m)) =0,
\ee
for all ordered sets $P$ of elements in $\CA$ and any element $m$ in $\CM$.
Here we defined for convenience $\nu(P_\alpha) := \nu_{|P_\alpha|}(P_\alpha)$.
We denote the map for $n=0$ in \eqref{eq:ModMaps} by $\nu(m)$  and
define  $\nu(\emptyset;m) := \nu(m)$. The map $\nu:\CM\to \CM $ provides a differential on $\CM$.
The sign $\epsilon_{P}(P_{1,2})$ is obtained by starting with the order $\nu\nu P$ and rearranging
using the reduced Koszul rule, where $\deg_r(\nu_n) = 1$. As a check on  the sign in the second
term note that an \afty-algebra should be a left-\afty-module over itself with $\nu_n = \mu_{n+1}$.
With our sign convention this is indeed the case: Apply the relations \eqref{eq:Aass} to partitions
whose last element is $m$. The second term in \eqref{eq:Amod} corresponds to
the partitions with $P_3 = \emptyset$.

Following the analogy between $A_\infty$ algebras and gauge connections, we can
pick a ``connection'' $a \in \CA$ and define the ``covariantized'' differential $\nu_a$ on $\CM$
\begin{equation}
\nu_a(m) := \sum_{n=0}^\infty \nu(a^{\otimes n};m)
\end{equation}
and ``gauge transformations'' of parameter $\tilde a$:
\begin{equation}
m \to m + \nu_a(\tilde a;m)
\end{equation}
with
\begin{equation}
\nu_a(\tilde a;m) := \sum_{n=0}^\infty \sum_{n'=0}^\infty \nu(a^{\otimes n},\tilde a,a^{\otimes n'};m).
\end{equation}
The defining relations for the $A_\infty$ morphism insure that the differential $\nu_a$ transforms covariantly under
gauge transformations of $a$ and $m$ with parameter $\tilde a$.
\be
\nu_a(m) \to \nu_a(m) + \nu_a(\nu_a(\tilde a;m)) + \nu_a(\mu_a(\tilde a);m)= \nu_a(m)+ \nu_a(\tilde a;\nu_a(m))
\ee
To check the signs here recall that the gauge field $a$ has reduced degree zero and the gauge parameter $\tilde a$
has reduced degree $-1$.  The defining equations \eqref{eq:Amod}  also insure that a flat $\CA$-connection gives a nilpotent differential $\nu_a$.
Similarly defined multilinear maps $\nu_a: \CA^{\otimes n} \otimes \CM \to \CM$ give $\CM$ the structure of a left $A_\infty$ module for the
finite deformation of $\CA$ by $a$.

There are natural relations between $A_\infty$ morphisms and $A_\infty$ modules. As noted above,
$\CA$ is an \afty-module over itself. Moreover, modules pull back:
That is,  if $\nu_{\CB,n}: \CB^{\otimes n} \otimes \CM \to \CM$
defines the structure of an \afty-module on $\CM$ for an \afty-algebra $\CB$, and if $\phi$
is an \afty-morphism from $\CA$ to $\CB$ then $\nu_{\CA,n}:\CA^{\otimes n}\otimes \CM \to \CM $
defined by
\be
\nu_{\CA,p}(P;m):= \sum_{n\geq 1} \sum_{\Pa_n(P)} \nu_{\CB}(\phi(P_1),\dots, \phi(P_n);m)
\ee
where $p= \vert P \vert$,
defines the structure of an \afty-$\CA$-module on $\CM$. To prove this assertion it is
easiest to begin with the $\nu_\CA \nu_\CA P m$ term, and use the definition
to write it in the form $\nu_{\CB} \phi ; \nu_{\CB} \phi m $. The sum can be rearranged as
a sum over 2-partitions of sets of the form $\phi(P_1), \dots, \phi(P_n)$. One then
applies the module axiom for $\nu_{\CB}$ and rearranges the sum to be of the form of the first term
in the $\nu_{\CA}$-module axiom. As a corollary, if $\phi$ is an \afty-morphism from $\CA$ to $\CB$,
then $\CB$ is canonically an  $\CA$-module.

A second relation between \afty morphisms and modules is the following.
An  $A_\infty$ module $\CM$ for $\CA$ can be reinterpreted as an \afty-morphism into
a very simple \afty algebra $\CB$. As a module $\CB = \CM \otimes \CM^* \cong {\rm End}(\CM)$.
There are only two nontrivial operations on $\CB$, first,
\be
\mu_{\CB,1}(b) = - (\nu b + (-1)^{\deg_r(b) } b \nu ) = - [\nu, b]
\ee
where $\nu$ is the differential on $\CM$  and second,
\be
\mu_{\CB,2}(b_1,b_2) = (-1)^{\deg_r(b_1)} b_1 \circ b_2
\ee
where on the right hand side we have ordinary composition of endomorphisms.
Again we see that,  given an $A_\infty$ morphism from $\CA$ to $\CB$, and a $\CB$-module $\CM$
we get a $\CA$-module. This is simply a composition of $A_\infty$ morphisms.
%
%\cg{We ought to check that the remarks in the previous two paragraphs are consistent. }
%

A right $A_\infty$ module $\CM$ is a vector space equipped with a collection of multilinear maps $\nu_n:  \CM \otimes \CA^{\otimes n} \to \CM$ of degree $1-n$, which satisfies
the following relations:
\be \label{eq:Amod2}
\sum_{P_{1,2,3}\in {\rm Pa}_3(P)} \tilde \epsilon_P(P_{1,2,3}) \, \nu\left(m;P_1,\mu_\CA(P_2), P_3\right) + \sum_{{\rm Pa}_2(P)} \nu(\nu(m;P_1);P_2) =0.
\ee
The symbols and reduced degree are defined as before. The sign $ \tilde \epsilon_P(P_{1,2,3})$ is computed by starting
with the order $\nu m P \mu$ and then distributing the factors with the reduced Koszul sign rule.

Finally an $A_\infty$ bimodule $\CM$ is a vector space equipped with a collection of multi-linear maps $\nu_{m,n}: \CA^{\otimes m} \otimes \CM \otimes \CB^{\otimes n} \to \CM$ of degree $1-n-m$, which satisfies
the following relations:
\be \label{eq:Abimod}
\begin{split}
\sum_{{\rm Pa}_3(P)} \epsilon \, \nu\left(P_1,\mu_\CA(P_2), P_3;m;P'\right) & +\sum_{{\rm Pa}_3(P')} \epsilon' \, \nu\left(P;m;P'_1,\mu_\CB(P'_2), P'_3\right)\\
&  + \sum_{{\rm Pa}_2(P)}\sum_{{\rm Pa}_2(P')} \epsilon'' \nu(P_1; \nu(P_2;m;P'_2);P_2) =0.\\
\end{split}
\ee
The symbols and reduced degree are defined as before. The signs are given by combining
those of the previous two cases.

Notice that an $A_\infty$ bimodule maps an $\CA$-flat connection $a$ to a right $\CB$ module
\begin{equation}
\nu_a(m;P) := \sum_{n=0}^\infty \nu(a^{\otimes n};m;P).
\end{equation}

\subsection{$L_\infty$ Algebras, Morphisms, And Modules}\label{subsec:LINF-ALG}

Roughly speaking, an $L_\infty$ algebra is a graded-commutative version of an $A_\infty$ algebra.
In the context of the present paper, and in other physical contexts in which $A_\infty$ and $L_\infty$ algebras occur,
the former are associated to correlation functions of operators located on the boundary of a two-dimensional region,
the latter to correlation functions of operators located in the interior. As a consequence, the natural degree assignment
for the operations of $A_\infty$ and $L_\infty$ algebras in a physical context differ: the former are maps of degree $2-n$
acting on $n$ arguments, while the latter are maps of degree $3-2n$. Thus in standard mathematical notation
our $L_\infty$ algebras would be denoted as $L_\infty[-1]$ algebras. Correspondingly, the reduced degree of elements in $L_\infty$
algebras will coincide with the degree minus $2$.

We are   ready for our definition. Consider a module $\CL$.
An $L_\infty$ structure $\lambda$ on $\CL$ is defined by a collection of multi-linear, graded-commutative maps
$\lambda_n: \CL^{\otimes n} \to \CL$ of degree $3-2n$, which satisfies
the following $L_\infty$ associativity relation:
\be\label{eq:Lin-Def}
\sum_{S_{1,2}\in {\rm Sh}_2(S)} \epsilon_S(S_{1,2}) \lambda(\lambda(S_1), S_2) = 0
\ee
for all sets $S$ of elements in $\CL$, where we defined for convenience
$\lambda(S_\alpha) := \lambda_{|S_\alpha|}(S_\alpha)$ and $\lambda(\emptyset) := 0$. \footnote{It is straightforward, if needed, to relax the axioms
for $L_\infty$ structures by allowing a choice of ``source'' $\lambda(\emptyset) \neq 0$.}
The sign $\epsilon_S(S_{1,2}) $ is given again by the reduced Koszul rule, with the symbol $\lambda$ having degree $1$.
Of course, in this case this is the same as the standard Koszul rule. The pair $(\CL, \lambda)$ is called an $L_\infty$ algebra.

The counterpart to the equations \eqref{eq:Aex} are
\be\label{eq:LinftyExpl}
\begin{split}
\lambda_1(\lambda_1(s) & = 0 \\
\lambda_1(\lambda_2(s_1,s_2)) + \lambda_2(\lambda_1(s_1),s_2) + (-1)^{s_1 s_2 } \lambda_2(\lambda_1(s_2),s_1) & = 0 \\
\lambda_3(\lambda_1(s_1),s_2,s_3) + (-1)^{s_1 s_2} \lambda_3(\lambda_1(s_2),s_1,s_3) + (-1)^{s_3(s_1 + s_2)} \lambda_3(\lambda_1(s_3),s_1,s_2) & + \\
\lambda_2(\lambda_2(s_1,s_2),s_3) + (-1)^{s_2 s_3} \lambda_2(\lambda_2(s_1,s_3),s_2) + (-1)^{s_1(s_2 + s_3)}\lambda_2(\lambda_2(s_2,s_3),s_1) & + \\
+\lambda_1(\lambda_3(s_1,s_2,s_3)) & = 0 \\
\cdots & \cdots \\
\end{split}
\ee
The third equation is a version of the Jacobi identity, up to homotopy.

The natural MC equation associated to an $L_\infty$ algebra can be written compactly as
\be
\lambda(e^\beta) := \sum_{n=0}^\infty \frac{1}{n!} \lambda_n(\beta^{\otimes n}) =0
\ee
where $\beta$ has $\deg(\beta)=2$.
In analogy to the main text of the paper, we can denote a solution of such MC equation as an ``interior amplitude''.
We could pursue an analogy between interior amplitudes and flat 2-form connections, but we will not do so.
Again, an interior amplitude gives a finite deformation of an $L_\infty$ algebra $\CL$, with operations
\be
\lambda_\beta(S) = \lambda(e^\beta,S).
\ee
See equation \eqref{eq:L-infty-rhoz} for a detailed proof.

We can define $L_\infty$ morphisms from an $L_\infty$ algebra $\CL$ to an $L_\infty$ algebra $\tilde \CL$  in an obvious way,
as collection of maps $\varphi_n: \CL^{\otimes n} \to \tilde \CL$ of degree $2-2n$ (hence $\deg_r(\varphi)=0$)  which satisfy
\be\label{eq:Lin-Morph}
\sum_{S_{1,2}\in {\rm Sh}_2(S)} \epsilon_S(S_{1,2}) \varphi(\lambda_\CL(S_1), S_2) = \sum_{n>0}\frac{1}{n!} \sum_{S_{1,\dots, n}\in {\rm Sh}_n(S)} \epsilon_S(S_{1,\dots, n}) \lambda_{\tilde \CL}(\varphi(S_1), \cdots \varphi(S_n))
\ee
where the signs are given by the Koszul sign required for rearranging the elements of $S$.
Again $\varphi(\emptyset)=0$.
Such morphisms map any interior amplitude $\beta$ to an interior amplitude $\varphi(e^\beta)$.
Two $L_\infty$ morphisms can also be composed
\be
\left[ \varphi \circ \tilde \varphi \right](S)
= \sum_{n>0}\frac{1}{n!} \sum_{{\rm Sh}_n(S)} \epsilon_S(S_{1,\dots,n}) \varphi(\tilde \varphi(S_1), \cdots \tilde \varphi(S_n))
\ee
and composition is associative. Once again the identity morphism is a left- and right- identity
for this composition.

We can also define an $L_\infty$ module $\CM$. It is a module $\CM$
together with a collection of maps $\nu_n$ $\CL^{\otimes n} \otimes \CM \to \CM$
of degree $1-2n$ with
\be\label{eq:Lin-Mod}
\sum_{S_{1,2}\in {\rm Sh}_2(S)} \epsilon_S(S_{1,2}) \nu(\lambda(S_1), S_2;m)+
\sum_{S_{1,2}\in {\rm Sh}_2(S)} \epsilon_S'(S_{1,2}) \nu(S_1;\nu(S_2;m))= 0
\ee
The sign rule $\epsilon_S'(S_{1,2})$ can be deduced by requiring that
an $L_\infty$ algebra $\CL$ be a left-module over itself. Then
equation \eqref{eq:Lin-Mod} is the $L_\infty$-relation for a set $S$
whose last element is $m$. The first term in \eqref{eq:Lin-Mod} corresponds to
the term in \eqref{eq:Lin-Def} where $m$ is not an element of $S_1$ and
the second term in
\eqref{eq:Lin-Mod} corresponds to
the term in \eqref{eq:Lin-Def} where $m$ is  an element of $S_1$.

\subsection{$LA_\infty$ Algebras, Morphisms, And Modules}\label{subsec:LA-ALG}

In this paper we encounter a neat structure which allows an $L_\infty$ algebra to control deformations of an $A_\infty$ algebra.
This structure, which we dub an $LA_\infty$ algebra, is defined by an $L_\infty$ algebra $\CL$, together with a
module $\CA$ equipped with multilinear operations $\mu_{k,n}: \CL^{\otimes k}\otimes \CA^{\otimes n}    \to \CA$, $k\geq 0$, $n>0$,
graded symmetric in the first set of arguments, and of degree $2-n-2k$. We abbreviate the operation on monomials as
$\mu(S;P)$. The expression $\mu(S;P)$ is zero if $P=\emptyset$.
The operations satisfy the relations
\be\label{eq:la}
 \sum_{{\rm Sh}_2(S), {\rm Pa}_3(P)} \epsilon ~~ \mu(S_1; P_1,
\mu(S_2;P_2), P_3)   +
\sum_{{\rm Sh}_2(S)}  \epsilon_S(S_{1,2})~~ \mu(\lambda(S_1), S_2;P)   = 0  .
\ee
Here $\epsilon$ is the reduced Koszul sign computed from the order $\mu  \mu S P$  and
the sign in the second term comes   from $\mu  \lambda S P $. Therefore
\be\label{eq:la-sign}
\epsilon = \epsilon_S(S_{1,2}) (-1)^{d_r(S_1) + d_r(P_1) + d_r(P_1)d_r(S_2)}.
\ee

The maps $\mu_{0,n}$ endow $\CA$ with an $A_\infty$ algebra structure.
The maps $\mu_{k,1}$ alone give multilinear maps $\CL^{\otimes k} \otimes \CA \to \CA$ which endow $\CA$ with the structure of an $L_\infty$ module.
%
%In a sense, one can interpret the $L_\infty$ structure on $\CL$ as maps $\mu_{0,k}$ (but note
%that $\mu_{0,k}$ has degree $3-2k$).
%
Furthermore,
given any interior amplitude $\beta$ for $\CL$,
\begin{equation}
\mu_\beta(P) = \mu( e^\beta;P)
\end{equation}
endow $\CA$ with an $A_\infty$ algebra structure controlled by $\beta$.

The relations involving a non-empty set $S$ can be expressed as the statement of an $L_\infty$ morphism
from $\CL$ to the {\it Hochschild complex} of $\CA$, denoted $CC^*(\CA)$.
This complex  can be thought of as an $L_\infty$ algebra whose interior amplitudes
are deformations of the $\CA$ operations.

We define a degree $k$ element in $CC^k(\CA)$ as a collection of multilinear maps $\delta_n: \CA^{\otimes n} \to \CA$ of degree $k-n$.
The Hochschild complex is   equipped with  linear and   quadratic operations, and all higher operations can be taken to be $0$
when giving it the structure of an $L_\infty$ algebra.
The first operation on the complex is
\be\label{eq:Hoc1}
 \lambda_{\tilde \CL}(\delta)(P)
= \sum_{\Pa_3(P)} \epsilon_1  \delta(P_1, \mu(P_2), P_3)   - \sum_{\Pa_3(P)} \epsilon_2 \mu(P_1, \delta(P_2) , P_3)
\ee
where both $\epsilon_1$ and $\epsilon_2$ are determined by using the reduced Koszul rule starting from
the order $ \mu \delta   P$, so
\be
\epsilon_1 = (-1)^{ d_r(\delta) + d_r(P_1)} \qquad  \epsilon_2 = (-1)^{d_r(\delta) d_r(P_1) }
\ee
%
%
%%\be\label{eq:Hoc1}
%\left[\lambda_1(\delta) \right](P) = \sum_{{\rm Pa}_3(P)} \, \left[ \epsilon \mu \left(P_1,\delta(P_2), P_3 \right) + \epsilon' \delta %\left(P_1,\mu(P_2), P_3 \right) \right]
%\ee
The second operation on the Hochschild complex is the graded-symmetric operation:
\be\label{eq:Hoc2}
\lambda_{\CL,2}(\delta_1,\delta_2)(P)  = \sum_{{\rm Pa}_3(P)} \epsilon_3  \delta_1(P_1, \delta_2(P_2), P_3) +
\epsilon_4 \delta_2(P_1, \delta_1(P_2), P_3)
\ee
%
%
%The second operation is
%\be\label{eq:Hoc2}
%\left[\lambda_2(\delta,\delta') \right](P) = \sum_{{\rm Pa}_3(P)} \, \left[ \epsilon'' \delta \left(P_1,\delta'(P_2), P_3 \right) + \epsilon''' %\delta' \left(P_1,\delta(P_2), P_3 \right) \right]
%\ee
%
%The signs $\epsilon,\epsilon',\epsilon'',\epsilon'''$
%are determined using the reduced Koszul rule starting from the orders $\mu\delta P$, $\mu\delta P$,
%$\delta\delta' P$, and $\delta \delta' P$, respectively.
%
%\cg{We ought to check that these sign rules really work as claimed.}
%
%
where
\be
\epsilon_3 =  (-1)^{d_r(\delta_1)} (-1)^{ d_r(P_1)d_r(\delta_2) }
\ee
and
\be
\epsilon_4 = (-1)^{d_r(\delta_2)} (-1)^{d_r(\delta_1) d_r(\delta_2)  + d_r(P_1)d_r(\delta_1) }
\ee
These are almost but not quite what we would get from the reduced Koszul rule starting
from the ordering $\delta_1 \delta_2 P$. (The   factor $ (-1)^{d_r(\delta_1)}$ in $\epsilon_3$
violates the rule, but is
needed for the $L_\infty$ relations. We can restore the ordering rule by
shifting $d_r(P) \to d_r(P) + 1$.)

Now, the $L_\infty$ morphism $\varphi: \CL \to CC^*(\CA)$ takes a monomial $S$ to $\varphi(S)$ where
 $\varphi(S)$ is the Hochschild cochain taking $P \in \CA^{\otimes n}$ to
\be
\varphi(S)(P):= \mu(S;P)
\ee
Note that
\be
\deg\mu(S;P) = 2 - p - 2s + \deg(P) + \deg(S) = (2 + \deg_r(S)) + \deg_r(P)
\ee
so the Hochschild cochain $\delta = \varphi(S)$ has degree $k= 2 + \deg_r(S)$ as we
expect if $\varphi$ is to be an $L_\infty$ morphism.

Now we interpret the equation  \eqref{eq:la} as the condition for $\varphi$ to be an $L_\infty$ morphism.
The second term of \eqref{eq:la}
is   identified with the left hand side of equation \ref{eq:Lin-Morph}. The first term of
equation \eqref{eq:la} can be  identified with (minus) the right hand side of equation \ref{eq:Lin-Morph}.
It can be decomposed into terms with $S_1$ and $S_2$ both not empty,
which give terms quadratic in $\varphi$, and terms with $S_1$ or $S_2$ empty, which give terms linear in $\varphi$.
A special case of the relation of $LA_\infty$ algebras and the Hochschild complex is
worked out in detail in Section \S \ref{subsubsec:HalfPlane-CLA} above.

Given an $LA_\infty$ structure $\mu$ on $\CL$ and $\CA$, and an $L_\infty$ morphism $\varphi$ from $\tilde \CL$ to $\CL$, one gets an $LA_\infty$ structure
$\mu\circ \varphi $ by composing $\mu$ and $\varphi$,  interpreted as an $L_\infty$ morphism. Concretely,
\be
\left[ \mu\circ \varphi \right](S;P)= \sum_{n>0}\frac{1}{n!} \sum_{{\rm Sh}_n(S)} \epsilon_S(S_{1,\dots, n})
 \mu( \varphi(S_1), \cdots \varphi(S_n);P)
\ee

An $LA_\infty$ morphism between  $LA_\infty$ algebras $(\CL_1,\CA_1)$ and
$(\CL_2,\CA_2)$ can be defined as a collection of maps
\be
\begin{split}
\phi_{k,n}: \CL_1^{\otimes k}\otimes  \CA_1^{\otimes n}   &  \to  \CA_2 \qquad \qquad  n>0, k\geq 0 \\
\phi_{k,0}:  \CL_1^{\otimes k} &  \to \CL_2      \qquad\qquad k >0 \\
\end{split}
\ee
The degree of $\phi_{k,n}$ is $1-n-2k$ for $n>0,k\geq 0$ and $2-2k$ for $n=0$.
 Moreover the $\phi_{k,n}$ must satisfy:
\be
\begin{split}
\sum_{\Pa_3(P),  \Sh_2(S) }  \epsilon ~~ \phi(S_1; P_1,  \mu_{\CL_1,\CA_1} (S_2;P_2) , P_3
) & +   \sum_{ \Sh_2(S) }  \epsilon' ~~ \phi(S_1;\mu_{\CL_1 }(S_2;P) )  \\
= \sum_{n,m\geq 0}  \sum_{ \Pa_n(P)} \sum_{  \Sh_{n+m}(S) } \epsilon''  ~~
  \mu_{\CL_2,\CA_2}( \phi(S_{1}), ...  \phi(S_m) ;  \phi(S_{m+1};P_1),
&
...  , \phi(S_{m+n};P_n)).\\
\end{split}
\ee
Where (we didn't check) the signs should be determined from the reduced Koszul rule
starting with canonical orderings. This equation can be understood as a morphism
between two $L_\infty$ morphisms $\varphi_i: \CL_i \to CC^*(\CA_i)$.

A left $LA_\infty$ module can be defined as a collection of multi-linear maps
\be
\nu_{k,n} : \CL^{\otimes k} \otimes  \CA^{\otimes n} \otimes\CM \to \CM \qquad\qquad n,k \geq 0
\ee
We write $\nu(S;P;m)$ on monomials.  The structure of an $LA_\infty$ module
allows one to associate to any choice of interior amplitude $\beta$ an $A_\infty$ module for the corresponding
$A_\infty$ structure on $\CA$. The defining relations can be written as:
\be \label{eq:LAmod}
\begin{split}
0 =  \sum_{{\rm Sh}_2(S), {\rm Pa}_3(P)} \epsilon ~ \, \nu\left(S_1; P_1,\mu(S_2;P_2), P_3;m\right)
  &  \\
+ \sum_{{\rm Sh}_2(S)} \epsilon' ~ \nu\left(\lambda(S_1),S_2;P;m\right) &  + \sum_{{\rm Sh}_2(S),{\rm Pa}_2(P)} \epsilon'' ~
\nu(S_1; P_1; \nu(S_2;P_2)  ;m)).\\
\end{split}
\ee
The signs $\epsilon,\epsilon',\epsilon''$ are determined by using the reduced Koszul rule
starting with the orders $\nu\mu S P  m$, $\nu\lambda S P  m$ and $\nu\nu S P  m$, respectively.

Clearly, the $\nu_{0,m}$ maps give $\CM$ the structure of an $A_\infty$ module.
It would be interesting to interpret the remaining equations as some morphism from $\CL$ to some kind of
Hochshild complex for the $A_\infty$ module $\CM$, which encodes the possible deformations of
the $A_\infty$ module structure which possibly accompany deformations of the $A_\infty$ algebra.

Right $LA_\infty$ modules and $LA_\infty$ bimodules can be defined in a similar manner.

\section{$A_\infty$ Categories And Mutations } \label{app:cat}

\subsection{$A_\infty$ Categories And Exceptional Categories}

An $A_\infty$ category $\mathfrak{A}$ consists of the following data:
\begin{itemize}
\item A set of objects, which we   denote as $\textbf{Ob}(\mathfrak{A})$
\item For each pair of objects $A,B\in \textbf{Ob}(\mathfrak{A})$ a module
\footnote{See the preface to Appendix \ref{App:HomotopicalAlgebra}
for our use of the word ``module.'' } $\Hom(A,B) := \Hop(B,A)$
\item Multilinear composition maps
\be
\mu_n: \Hop(A_0,A_1) \otimes \cdots \otimes \Hop(A_{n-1},A_n) \to \Hop(A_0,A_n)
\ee
of degree $2-n$
which satisfy the $A_\infty$ associativity axiom.
\end{itemize}

Note that an $A_\infty$ algebra $\CA$ is a special case of an $A_\infty$ category, with a single object $A$ and $\CA = \Hop(A,A)$.
It is straightforward to refine the definitions we gave for $A_\infty$ morphisms, modules, etc. to corresponding categorical notions such as
$A_\infty$ functors, etc. The defining relations take an identical form, but the arguments live in compatible sequences of $\Hop$ spaces.

As a first example,
an $A_\infty$ functor from $\fA$ to $\fB$ is given by a map ${\cal F}$ from objects of $\fA$ to objects of $\fB$ together with a collection of multilinear maps
\be
\CF_n: \Hop(A_0,A_1) \otimes \cdots \otimes \Hop(A_{n-1},A_n) \to \Hop({\cal F}(A_0),{\cal F}(A_n))
\ee
of degree $1-n$ satisfying relations formally identical to
the those for morphisms, equation \eqref{eq:Amorph}. As a second example,
one can define a categorical analogue to an $A_\infty$ module by a map $\fM$ from objects $A_i$ of $\fA$ to modules $\fM(A_i)$ and collections of multilinear maps
\be
\nu_n: \Hop(A_0,A_1) \otimes \cdots \otimes \Hop(A_{n-1},A_n) \otimes \fM(A_n) \to \fM(A_0)
\ee
of degree $1-n$ and satisfying the analogue of equation \eqref{eq:Amod}. And so forth.

A somewhat surprising observation is that given an $A_\infty$ algebra $\CA$, the set of $\CA$-flat connections forms an
$A_\infty$ category.
The spaces $\Hop(a,a')$ from a Maurer-Cartan element $a'$ to a Maurer-Cartan element $a$
 all coincide with $\CA$ as a module,  and the multilinear operations
take the form
\be\label{eq:BrCatDef}
M(\tilde a_1, \cdots, \tilde a_n) = \sum_{k_i\geq 0 } \mu(a_0^{\otimes k_0}, \tilde a_1, a_1^{\otimes k_1},\cdots, \tilde a_n, a_n^{\otimes k_n}).
\ee
This fact was used in defining the category of Branes using the multiplications
 in equation \eqref{eq:BraneMultiplications} above. Here we give a simple proof that \eqref{eq:BrCatDef} satisfy
the \afty-relations.
Given $P = \{ \tilde a_1, \cdots, \tilde a_n \}$ we wish
to show that
\be
\sum_{P_{1,2,3}\in {\rm Pa}_3(P)} \epsilon_P(P_{1,2,3}) M(P_1, M (P_2), P_3) = 0.
\ee
For each vector $\vec k  = (k_0, \dots, k_n) \in \IZ_+^{n+1}$ define the ordered set:
\be
P(\vec k) = \{ a_0^{k_0}, \tilde a_1, a_1^{k_1}, \tilde a_2, \dots, \tilde a_n, a_n^{k_n} \}.
\ee
We then apply the \afty-relations for $\mu_{\CA}$ to the set $P(\vec k)$. Then we sum these over all
$\vec k$. Now we reorganize the terms in the disjoint union
\be
\amalg_{\vec k \in \IZ_+^{n+1} } {\rm Pa}_3(P(\vec k))
\ee
according to
\be
\amalg_{P_{1,2,3}\in {\rm Pa}_3(P) } \Biggl\{ \amalg_{\vec k_1 \in \IZ_+^{p_1+1}} \amalg_{\vec k_2 \in \IZ_+^{p_2+1}}
\amalg_{\vec k_3 \in \IZ_+^{p_3+1}} \{ P_1(\vec k_1)\} \amalg \{ P_2(\vec k_2)\} \amalg \{ P_3(\vec k_3) \} \Biggr\}
\ee
This gives precisely the expression
\be
\sum_{{\rm Pa}_3(P)} \epsilon_P(P_{1,2,3})  M_{p_1 + p_3 + 1}(P_1, M_{p_2}(P_2), P_3) = 0.
\ee
where we include $P_2=\emptyset$. The one place we use the property that the $a_i$ solve
the Maurer-Cartan equation is that
the $P_2 = \emptyset $ terms can be dropped.

This statement can be extended to the set of matrix-valued $\CA$-flat connections, given by pairs $\CB := (\CE, a\in \CE \otimes \CA \otimes \CE^*)$.
Then $\Hop(\CB_1, \CB_2) := \CE_1 \otimes \CA \otimes \CE_2^*$ as a module, with compositions $\mu$ including the contractions
of $\CE_i^*$ and $\CE_i$ factors in consecutive arguments. We use the sign conventions of
equation \eqref{eq:MatEx-1} et. seq. Taking together all modules $\CE$ (with suitable finiteness properties)
 the resulting $A_\infty$ category is denoted as $\fB\fr[\CA]$.

Given any $A_\infty$ category $\fA$, we can define a flat $\fA$ connection $\CB$ starting from a collection of pairs of objects and vector spaces $(A_i,\CE_i)$
and solving the MC equation for an element $a \in \oplus_{i,j} \CE_i \otimes \Hop(A_i,A_j) \otimes \CE_i^*$. Flat $\fA$ connections
also form a larger $A_\infty$ category $\fB\fr[\fA]$ with $\Hop(\CB,\CB') := \CE_i \otimes \Hop(A_i,A_j) \otimes (\CE'_i)^*$.

In general, spaces of flat $\CA$ or $\fA$ connections can be very intricate, as they are defined by intricate, possibly non-polynomial,
MC equations. The equations greatly simplify if the there is an ordering on the set of objects $\textbf{Ob}(\fA) = \{ T_i \}$   such that
there is a triangular structure on the hom-sets. This motivates the

\bigskip
\noindent
\textbf{Definition:} An \emph{exceptional category} $\fE$ is an \afty-category
such that there is an ordering on the (countable) set of objects  $\{ T_i \}$
such that
\be
\begin{split}
\Hop(T_i,T_j) & = 0 \qquad  i> j \\
%\Hop(T_i,T_i) & = \IZ \\
\end{split}
\ee
Moreover $\Hop(T_i,T_i) \cong \IZ$ is concentrated in
degree zero and the generator $\Id_i \cong 1$ is a graded identity
for $\mu_2$:
\be
\mu_2(\Id,a) = a  \qquad  \mu_2(a,\Id) = (-1)^{\deg(a)} a
\ee
where $\Id = \oplus_i \Id_i$ and   $a$ is homogeneous. Moreover,
$\mu_n(P)=0$ whenever
$n\not=2$ and $P$ contains a multiple of $\Id$.

The MC equations in an exceptional category have a triangular structure
and are rather tractable. \footnote{In this situation, the definition of flat $\fA$-connections coincides with the mathematical notion of
twisted complex. }
This is precisely the case which occurs in our paper. Physically, the objects in the exceptional collection coincide with the thimble branes.
General branes in a given theory appear as $\fA$-flat connections for the exceptional category.

\subsection{Mutations Of Exceptional Categories}\label{subsec:CatMut}

\subsubsection{Exceptional Pairs And Two Distinguished Branes}

An \emph{exceptional pair} is a pair of objects $T_1,T_2$ in a
category such that
\be
\begin{split}
\Hop(T_1,T_1) \cong \IZ  \qquad &  \qquad \Hop(T_1,T_2) := \CH \\
\Hop(T_2,T_1) = 0 \qquad & \qquad \Hop(T_2,T_2) \cong \IZ \\
\end{split}
\ee
where $\CH $ is in general nonzero,  and is simply
some differential graded module.

Given an exceptional pair $(T_1,T_2)$ and any  given differential
graded modules $\CE_1 = \CE_{T_1}$ and $\CE_2 = \CE_{T_2}$ we can
put a differential graded associative algebra structure on the set of
matrices with elements in
\be
\begin{pmatrix}
\CE_1 \CE_1^* ~~  &  ~~ \CE_1 \Hop(T_1,T_2) \CE_2^* \\
0 & \CE_2 \CE_2^* \\
\end{pmatrix}
\ee
To define $\mu_1$ we simply apply it to each of the matrix elements using the natural
induced differential.
The definition of $\mu_2$ requires some care, given the conventions above.
If we consider homogeneous matrices of monomials like
\be
X = \begin{pmatrix}  ee^*   &  \tilde e a \tilde f^* \\
0 & ff^* \\ \end{pmatrix}
\ee
then $\mu_2(X_1,X_2) $ is given by
\be
\begin{pmatrix}  (-1)^{e_1} (e_1^*\cdot e_2) e_1 e_2^* ~~~  &
~~~~ (-1)^{e_1} (e_1^*\cdot \tilde e_2) e_1 a_2 \tilde f_2^* + (-1)^{\tilde e_1 + a_1} (\tilde f_1^*\cdot f_2) \tilde e_1 a_1 f_2^*  \\
0 & (-1)^{f_1} (f_1^*\cdot f_2) f_1 f_2^* \\ \end{pmatrix}
\ee
where $(-1)^v$ is short for $(-1)^{\deg(v)}$ for a homogeneous vector $v$.
The extra sign $(-1)^{a_1}$ in the $12$ element comes about because the identity in $\Hop(T_1,T_1)$ and $\Hop(T_2,T_2)$ is
a graded identity, so $\mu_2(1,a) = a $ but $\mu_2(a,1) = (-1)^{a} a$. Taking $\mu_n =0 $ for $n>2$ one can
check that the equations  \eqref{eq:Aex} are indeed satisfied.

Now, using this construction we can construct two distinguished Maurer-Cartan elements
in the above differential graded algebra. They will therefore define  objects in $\fB\fr(\fA)$ in any
category $\fA$ which contains the exceptional pair $(T_1,T_2)$. We will therefore refer to them as ``Branes.''

The first Brane, denoted $L(T_1,T_2)$  may be denoted
\be
L(T_1,T_{2}) =\CE_1 T_1 \oplus \CE_{2} T_2 = T_1 \oplus \CH^{[-1]} T_{2}
\ee
where we recall that $\CH = \Hop(T_1,T_2)$.
\footnote{ We adopt the convention that for a degree shift by $s$, $\deg( v[s]) = \deg(v) + s $.}
The Maurer-Cartan element (``boundary amplitude'') is
\be\label{eq:Brane1}
\CB =  \begin{pmatrix} 0 &  \CB_{12}  \\     0 &  0 \\    \end{pmatrix}
\ee
with
\be
\CB_{12}  := \sum_s f_s \otimes (f_s[-1])^*
\ee
where we sum over a basis $\{ f_s \}$ for $\CH$. It is not difficult
to check that $\mu_1(\CB_{12})=0$ and hence the MC equation is satisfied.

The second Brane, denoted $R(T_1,T_2)$, has Chan-Paton factors:
\be
R(T_1,T_{2}) =\CE_1 T_1 \oplus \CE_{2} T_2 = (\CH^{[-1]})^* T_1 \oplus   T_{2}.
\ee
The Maurer-Cartan element (``boundary amplitude'') is again of the form
\be\label{eq:Brane1}
\CB =  \begin{pmatrix} 0 &  \CB_{12}  \\     0 &  0 \\    \end{pmatrix}
\ee
now with
\be
\CB_{12}  := \sum_s  (f_s[-1])^*\otimes f_s
\ee
where we sum over a basis $\{ f_s \}$ for $\CH$. It is not difficult
to check that $\mu_1(\CB_{12})=0$ and hence the MC equation is satisfied.

\subsubsection{Left And Right Mutations}

Now suppose that $\fE$ is an exceptional category. We begin with a definition:

\bigskip
\noindent
\textbf{Definition:} A \emph{left mutation at $j$ of an exceptional category $\fE$} is
another exceptional category $\fF$, with an ordering on its objects $\{S_i \}$
such that there is an \afty-functor
\be
\CF: \fF \rightarrow \fB\fr(\fE)
\ee
with
\be
\CF(S_i) =
\begin{cases}
T_i  &  i \not= j,j+1 \\
T_{j+1} & i = j \\
L(T_j,T_{j+1})  & i = j+1 \\
\end{cases}
\ee
and
\be
\CF_1: \Hop(S_i,S_{k}) \rightarrow \Hop(\CF(S_i) , \CF(S_k) )
\ee
is a quasi-isomorphism for all $i,k$.

We now show that left mutations in fact do exist. To do this, we first note
that it is not immediately obvious that quasi-isomorphisms of the desired type
in fact do exist. Note that if we consider the full subcategory $L_j(\fE)$
of $\fB\fr(\fE)$ whose set of objects is just $\{ \tilde T_i \}$ with
\be\label{eq:TildeTees}
\tilde T_i =
\begin{cases}
T_i  &  i \not= j,j+1 \\
T_{j+1} & i = j \\
L(T_j,T_{j+1})  & i = j+1 \\
\end{cases}
\ee
The resulting subcategory $L_j(\fE)$  is \emph{not} an exceptional category. It does not
satisfy the definition because $\Hop(\tilde T_{j+1}, \tilde T_{j+1})$ is not
isomorphic to $\IZ$ and $\Hop(\tilde T_{j+1},\tilde T_j)$ is not isomorphic
to zero. We therefore should check that they are at least quasi-isomorphic to $\IZ$ and zero, respectively.

We first show there is a quasi-isomorphism of $\Hop(\tilde T_{j+1}, \tilde T_{j+1})$  with $\IZ$.
Now,
\be
\Hop(L(T_j,T_{j+1}), L(T_j, T_{j+1}))
\ee
is the associative matrix algebra
associated with the  exceptional pair $(T_j, T_{j+1})$.  So, our
definition only makes sense if this matrix algebra is  quasi-isomorphic
to $\IZ$ with the differential $M_1$ in the brane category. Let us check that
this is indeed the case:

Morphisms $\delta \in \Hop(L(T_j,T_{j+1}), L(T_j, T_{j+1})$ can be thought of as being in
\be
\begin{pmatrix}
\IZ  ~~  &  ~~ \CH_j\otimes (\CH_j^{[-1]})^* \\
0 &  (\CH_j^{[-1]})\otimes (\CH_j^{[-1]})^* \\
\end{pmatrix}
\ee
where $\CH_j = \Hop(T_j, T_{j+1})$.
We write these as
\be
\delta = \begin{pmatrix} \delta_{11} & \delta_{12} \\  0 & \delta_{22} \\ \end{pmatrix}
\ee
and, using $\CB^2=0$ we compute
\be
M_1(\delta) = \mu_1(\delta) + \mu_2(\CB,\delta) + \mu_2(\delta,\CB)
\ee
So
\be
M_1(\delta) = \begin{pmatrix} 0 ~~~~ &  ~~ \mu_1(\delta_{12}) + \mu_2(\CB_{12}, \delta_{22}) + \delta_{11} \mu_2(1,  \CB_{12}) \\
0 & \mu_1(\delta_{22}) \\   \end{pmatrix}
\ee
It is not difficult to show that the cohomology of $M_1$ on the subspace of morphisms
of the form
\be\label{eq:BigKer}
\delta = \begin{pmatrix} 0 & \delta_{12} \\  0 & \delta_{22} \\ \end{pmatrix}
\ee
is zero, precisely because the operation $x \to \mu_2(\CB_{12},x)$ acts as a
``twisted degree shift.'' The projection of the kernel of $M_1$ to $\delta_{11}\in \IZ$
is then the desired quasi-isomorphism.

Now consider $(\alpha,\beta) = (j+1,j)$ then, on the one hand,   $\Hop(S_{j+1}, S_{j})=0 $,
and therefore
\be\label{eq:SumTwo}
\begin{split}
\Hop( L(T_j,T_{j+1}), T_{j+1}  ) & = \Hop(T_j, T_{j+1}) \oplus \CE_{j+1}\otimes \Hop(T_{j+1},T_{j+1}) \\
& \cong \CH_j \oplus \CH_j^{[-1]}\\
\end{split}
\ee
must be quasi-isomorphic to zero with the differential $M_1$ in the Brane category.
 Indeed this is the case. Once  again
\be
M_1(\delta) = \mu_1(\delta) + \mu_2(\CB,\delta) + \mu_2(\delta,\CB).
\ee
Writing an element of
\eqref{eq:SumTwo} as $\delta = \delta_1 \oplus \delta_2 $ we compute that
\be
M_1(\delta) = \biggl( \mu_1(\delta_1) + \mu_2(\CB_{12}, \delta_2) \biggr) \oplus \mu_1(\delta_2)
\ee
and again $\mu_2(\CB_{12}, \delta_2)$ acts as a twisted degree shift. Using this property it is
not difficult to show that  the cohomology is zero.

If we consider the fourteen remaining cases of $\CF_1: \Hop(S_\alpha, S_\beta) \to \Hop(\tilde T_{\alpha}, \tilde T_{\beta})$
then many cases are trivially quasi-isomorphisms, and the remaining ones simply constrain the relation
of the hom-sets in interesting ways. We will comment on those below.

We can now show the existence of left-mutations, in the sense of our definition. We use the result of
Kadeishvili, and of Kontsevich and Soibelman (see \cite{ClayBook2}, pp. 587-593 for a detailed exposition)
that there is always a quasi-isomorphism of any \afty-category with another \afty-category with the
same set of objects, but whose morphism spaces are the cohomologies of the original category.
If we apply the construction of Kadeishvili-Kontsevich-Soibelman to $L_j(\fE)$ we obtain the required
exceptional category $\fF$.

In an entirely analogous fashion we can give the:

\bigskip
\noindent
\textbf{Definition} A \emph{right mutation at $j$ of an exceptional category $\fE$} is
another exceptional category $\fF$, with an ordering on its objects $\{S_i \}$
such that there is an \afty-functor
\be
\CF: \fF \rightarrow \fB\fr(\fE)
\ee
such that
\be
\CF(S_i) =
\begin{cases}
T_i  &  i \not= j,j+1 \\
R(T_j,T_{j+1})   & i = j \\
 T_{j} & i = j+1 \\
\end{cases}
\ee
and
\be
\CF_1: \Hop(S_i,S_{k}) \rightarrow \Hop(\CF(S_i) , \CF(S_k) )
\ee
is a quasi-isomorphism for all $i,k$.

We would like to conclude with a number of remarks:

\begin{enumerate}

\item As a general rule, in a mutation, if we ``add'' a new brane of the
type $C=A + \CE B$ where $\CE \sim \Hop(A,B)$ (up to degree shift and duals) then
the new set of objects contains $C$ and $B$ but \emph{not} $A$.

\item The categorical mutations described here arise from framed-wall-crossing
on $S_{ij}$-walls, as described in Section \S \ref{subsec:Mutations} above.

\item As we remarked, the existence of a left- or right-mutation implies some interesting
relations between the spaces $\Hop(S_\alpha, S_{\beta} )$ and $\Hop(T_\alpha, T_\beta)$.
For example, considering the case $(\alpha, \beta) = (j, j+1)$ for a left-mutation at $j$ we find that
there must be a quasi-isomorphism
\be\label{eq:HopSS}
\CF_1: \Hop(S_{j}, S_{j+1}) \rightarrow (\Hop(T_j, T_{j+1})^{[-1]})^*
\ee
This is in harmony with the expectations of $S$-wall-crossing. Similarly,
the case $(\alpha,j+1)$ for   $\alpha < j$ implies there are quasi-isomorphisms
\be\label{eq:ExtraCoh1}
\CF_1: \Hop(S_\alpha, S_{j+1}) \rightarrow
\Hop(T_\alpha, L(T_j,T_{j+1}) ) \cong \Hop(T_\alpha, T_j) \oplus \Hop(T_\alpha, T_{j+1}) \otimes (\CH_j^{[-1]})^*
\ee
while $(j+1, b)$   for $\beta > j+1 $ implies there are quasi-isomorphisms
\be\label{eq:ExtraCoh2}
\CF_1: \Hop( S_{j+1},S_\beta) \rightarrow \Hop( L(T_j,T_{j+1}) , T_\beta) = \Hop(T_j, T_\beta) \oplus \CH_j^{[-1]}\otimes \Hop(T_{j+1},T_{\beta})
\ee
Again, in harmony with $S$-wall-crossing.

\item There is a sense in which left and right mutations are inverse to each other:
\emph{If $\fF$ is a left mutation at $j$ of $\fE$, then $\fE$ is a right mutation
at $j$ of $\fF$}. As a check on this assertion note that $\CF_j^{(R)}\circ \CF_j^{(L)}$
takes $S_j \to S_j$ but
\be
\CF_j^{(R)}\circ \CF_j^{(L)}: S_{j+1} \rightarrow
 \CE S_j \oplus S_{j+1}
\ee
where
\be
\CE = (\Hop(S_j, S_{j+1})^{[-1]})^* \oplus \Hop(T_j, T_{j+1})^{[-1]}
\ee
and by equation \eqref{eq:HopSS} above this space can indeed admit a
differential making it quasi-isomorphic to zero.
Similarly, $\CF_j^{(L)}\circ \CF_j^{(R)}$ takes $S_{j+1} \to S_{j+1}$ but
\be
\CF_j^{(L)}\circ \CF_j^{(R)}: S_{j} \rightarrow
 S_j \oplus \CE S_{j+1}
\ee
where
\be
\CE =  \Hop(S_j, S_{j+1})^{[-1]}  \oplus (\Hop(T_j, T_{j+1})^{[-1]})^{[-1]}
\ee
and again \eqref{eq:HopSS} shows this space can indeed admit a
differential making it quasi-isomorphic to zero.

\item Moreover, we expect the left and right mutations to define a
braid group action in the following sense. Suppose we have left-mutations:
\be
\begin{split}
\CF_j^{(L)}: \fE_1 & \rightarrow \fB\fr(\fE_2)\\
\CF_{j+1}^{(L)}: \fE_2 & \rightarrow \fB\fr(\fE_3)\\
\CF_j^{(L)}: \fE_3 & \rightarrow \fB\fr(\fE_4)\\
\end{split}
\ee
and similarly,
\be
\begin{split}
\CF_{j+1}^{(L)}: \fE_1 & \rightarrow \fB\fr(\fE_2')\\
\CF_{j}^{(L)}: \fE_2' & \rightarrow \fB\fr(\fE_3')\\
\CF_{j+1}^{(L)}: \fE_3' & \rightarrow \fB\fr(\fE_4')\\
\end{split}
\ee
then we expect that there is an equivalence of \afty-categories
$\CG : \fB\fr(\fE_4)\to \fB\fr(\fE_4')$ such that there is
an isomorphism of \afty-functors:
\be
\CG\circ \CF^{(L)}_j \circ \CF^{(L)}_{j+1}\circ \CF^{(L)}_j
\cong
\CF^{(L)}_{j+1} \circ \CF^{(L)}_{j }\circ \CF^{(L)}_{j+1}
\ee
That is, there is an invertible \afty-natural transformation
between these two functors. We have not checked the details
of these last assertions, so we'll leave it here for now.

\end{enumerate}

\section{Examples Of Categories Of Branes}\label{sec:ExamplesBraneCat}

For small numbers of vacua it is possible to write out in some generality
the full structure of the web-formalism of Sections \S \ref{sec:Webs}-\ref{sec:LocalOpsWebs}.
Of course the complexity increases rapidly as the number of vacua is increased.

\subsection{One Vacuum}

In this case there are no planar webs. There are no unextended half-plane webs but
there are extended half-plane and interface webs. As noted in
Section \S \ref{subsubsec:TrivialTheories}, the category of Branes,
and also the category of Interfaces between the trivial Theory and
itself is precisely the category of chain complexes.

\subsection{Two Vacua}

Suppose that $\IV$ has cardinality $2$. Then the entire web formalism can
still be written out quite explicitly.  We will describe the category of
Branes for the positive and negative half-plane and the differential
on the strip complex. Therefore we must assume that $\Re(z_{12})\not=0$.
By reordering vacua may assume without loss of generality that $\Re(z_{12})>0$.

There are no unextended plane webs. In particular there are no vertices
of valence three or higher.
The only plane webs consist of a single line with arbitrarily many 2-valent
vertices on it. For a web representation we are free to choose any pair of
$\IZ$-graded modules $R_{12}, R_{21}$ together with a perfect degree $-1$ pairing
$K: R_{12}\otimes R_{21} \to \IZ$. There is one taut web, illustrated in
Figure \ref{fig:EXTENDED-DIFFERENTIAL}(a). As explained in the text, the $L_\infty$
Maurer-Cartan equation implies that the interior amplitude defines a differential
$\Q$ on $R_{12}$ and $R_{21}$ so that the pairing is $Q$-invariant. Thus, the
Theory $\CT$ is entirely characterized by a choice of a complex $R_{12}$ and
a dual complex $R_{21}$ with a perfect pairing.

The complex $(R_c, d_c)$ of local operators is
\be
0 \rightarrow R_1 \oplus R_2 \rightarrow R_{12}\otimes R_{21} \rightarrow 0
\ee
where $R_1,R_2 \cong \IZ$ with generators $\phi_1,\phi_2$ and
\be
\begin{split}
d_c(\phi_1) & = K_{12}^{-1} \\
d_c(\phi_2) & = K_{21}^{-1} \\
\end{split}
\ee
so that the $\Q$-invariant local operators on the plane consist of the identity $\textbf{1}=\phi_1 + \phi_2$ in
degree zero
and a space of operators isomorphic to $\left( R_{12}\otimes R_{21}\right)/K_{12}^{-1}\IZ$. The $\IC\IP^1$ model discussed in
Section \S \ref{moremirror} above is a special case of this situation.

\begin{figure}[htp]
\centering
\includegraphics[scale=0.3,angle=0,trim=0 0 0 0]{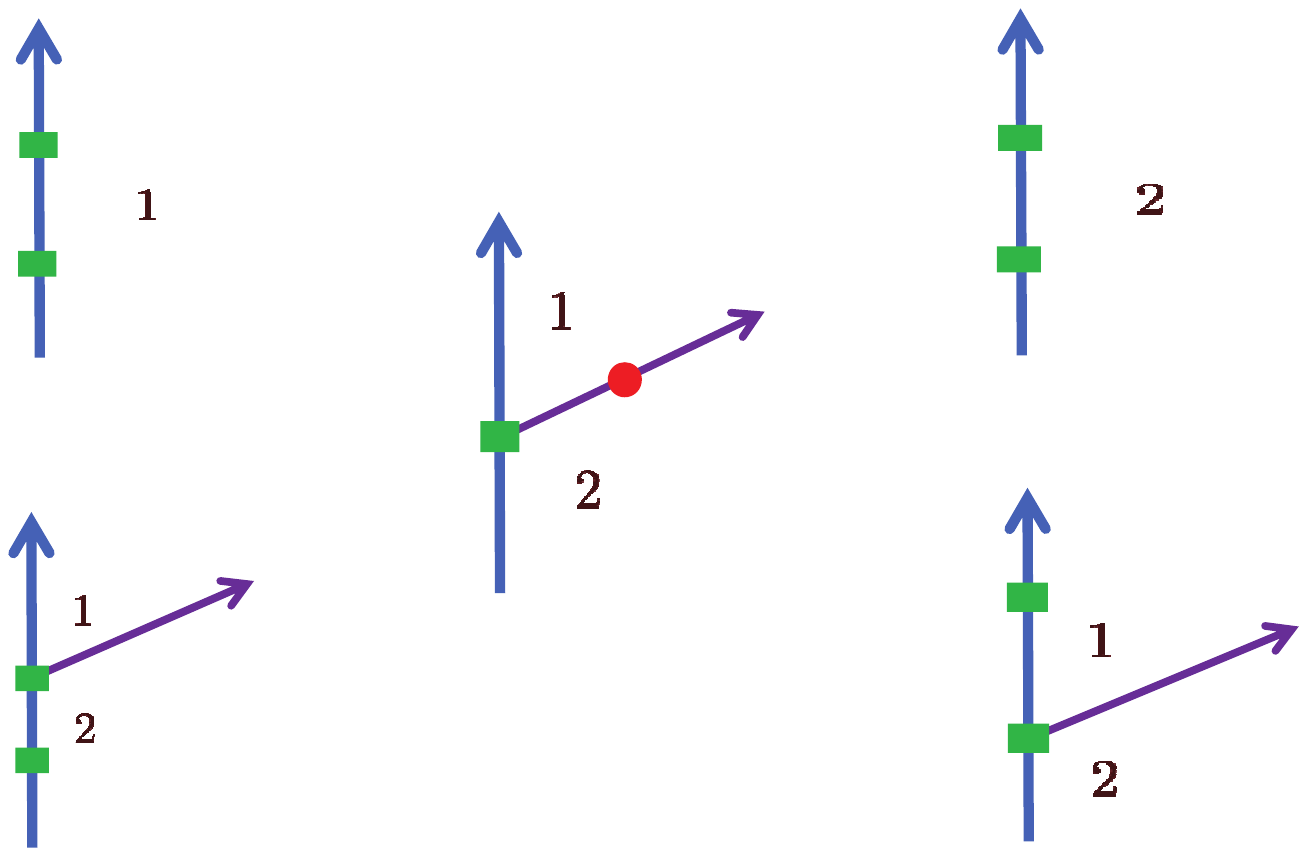}
\caption{In a Theory with two vacua, with $\Re(z_{12})>0$, there are five
taut positive half-plane webs, illustrated here.  }
\label{fig:FIVETAUT-2VACB}
\end{figure}

Now let us consider the category $\fB\fr(\CT,\CH^+)$ in the positive half-plane
for the above Theory. The half-plane webs consist of an arbitrary number of
zero-valent boundary vertices with one or zero emission lines separating vacua $1$
and $2$. If there is such a line it can have an arbitrary number of $2$-valent vertices
on it. In particular, there are five taut half-plane webs shown in Figure
\ref{fig:FIVETAUT-2VACB}.

Now we construct positive half-plane Branes. We choose two $\IZ$-graded Chan-Paton
modules $\CE_1,\CE_2$. The boundary amplitude is an element
\be
\CB \in \oplus_{i,j} \CE_i \otimes \Hop(i,j) \otimes \CE_j^*
\ee
and may be thought of as a $2\times 2$ matrix:
\be
\CB = \begin{pmatrix} \CB_{11} & \CB_{12} \\  0 & \CB_{22} \\  \end{pmatrix}
\ee
As in the discussion of Section \S \ref{subsec:RepExtendedWebs} the Maurer-Cartan
equation following from the two taut webs on the top of Figure \ref{fig:FIVETAUT-2VACB}
imply that $\CB_{ii} \in \CE_i\otimes \CE_i^* \cong \End(\CE_i)$ are differentials,
while the remaining three taut webs imply that
\be\label{eq:2Vac-1}
\CB_{12} \in \Hom(\CE_2,\CE_1\otimes R_{12})
\ee
is annihilated by the differential induced from that on $\CE_1,\CE_2$ and $R_{12}$.
We conclude that the objects in the category $\fB\fr(\CT,\CH^+)$ consist of a pair of $\IZ$-graded
complexes $(\CE_1,\CE_2)$ together with an arbirary degree one, $\Q$-invariant
morphism in \eqref{eq:2Vac-1}.

The space $\Hop(\fB,\fB')$ between two Branes $\fB,\fB'$ in $\fB\fr(\CT,\CH^+)$ can be thought of
as $2\times 2$ matrices valued in
\be\label{eq:2Vac-2}
\begin{pmatrix}  \Hom(\CE_1',\CE_1) & \Hom(\CE_2', \CE_1\otimes R_{12} )  \\  0 & \Hom(\CE_2',\CE_2)   \\  \end{pmatrix}
\ee
The differential $M_1$ on $\Hop(\fB,\fB')$ computed using equation \eqref{eq:BraneMultiplications}
with the taut element shown in Figure \ref{fig:FIVETAUT-2VACB} is the natural differential acting
on \eqref{eq:2Vac-2}, so the local operators between $\fB$ and $\fB'$ can be identified with the
$\Q$-cohomology of  \eqref{eq:2Vac-2}. The multiplication $M_2$ on the category is given by
the naive multiplication of elements of the form \eqref{eq:2Vac-2} since all taut half-plane
webs have at most two boundary vertices. Moreover, for this reason, the higher multiplications
$M_n$, $n\geq 3$ all vanish.

Similarly, the objects in the category $\fB\fr(\CT,\CH^-)$ associated with the negative
half-plane consist of a pair of $\IZ$-graded
complexes $(\tilde\CE_1,\tilde\CE_2)$ together with an arbirary degree one, $\Q$-invariant
morphism
\be\label{eq:2Vac-3}
\tilde\CB_{12} \in \Hom(\tilde \CE_2, R_{21}\otimes \tilde \CE_1)
\ee
\begin{figure}[htp]
\centering
\includegraphics[scale=0.3,angle=0,trim=0 0 0 0]{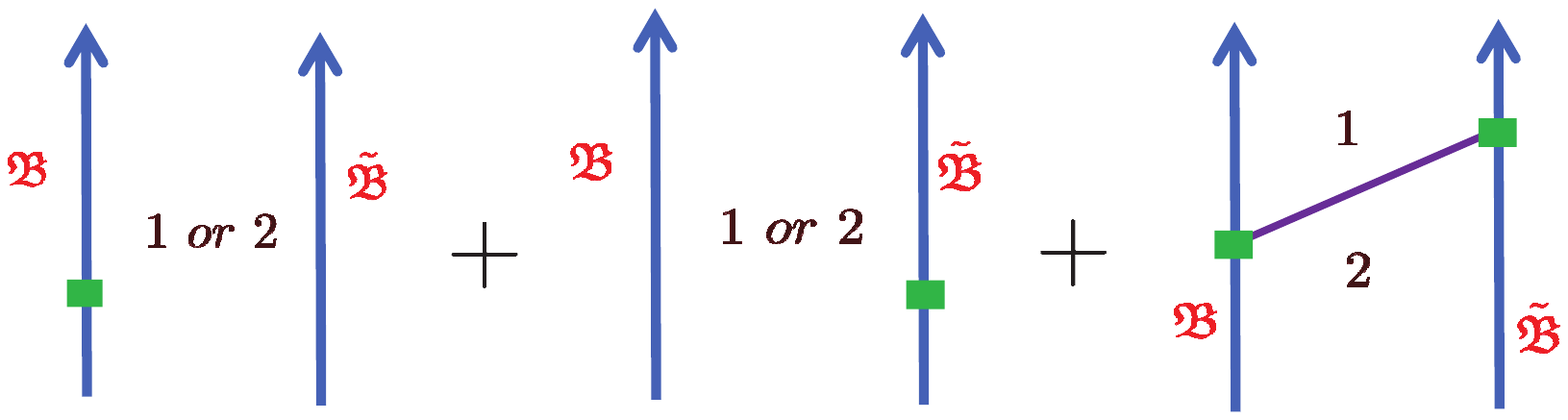}
\caption{In a Theory with two vacua, with $\Re(z_{12})>0$, there are three
taut webs on the strip.  }
\label{fig:FIVETAUT-2VAC-3}
\end{figure}

Now, consider the strip with a Brane  $\fB\in \fB\fr(\CT,\CH^+)$
on the left-boundary and $\tilde\fB\in \fB\fr(\CT,\CH^-)$ on the right boundary.
The complex of approximate groundstates is
\be\label{eq:2Vac-4}
\CE_1 \otimes \tilde\CE_1 \oplus \CE_2 \otimes \tilde\CE_2
\ee
with a differential induced from the taut strip-webs shown in Figure \ref{fig:FIVETAUT-2VAC-3}.
The first two types of webs give the naive differential on \eqref{eq:2Vac-4}. Using the
block form corresponding to the direct sum decomposition in \eqref{eq:2Vac-4} the differential
has the form
\be\label{eq:2Vac-5}
d_{LR} = \begin{pmatrix} \Q_1\otimes 1 + 1\otimes \tilde\Q_1  & \Q_{12}  \\  0 & \Q_2\otimes 1 + 1\otimes \tilde\Q_2  \\  \end{pmatrix}
\ee
where, up to sign
\be
\Q_{12}=K_{12}(\CB_{12}\otimes\tilde\CB_{12}) \in \Hom(\CE_2\otimes\tilde\CE_2, \CE_1\otimes\tilde\CE_1)
\ee
We can compute the cohomology of $d_{LR}$ by first passing to the naive cohomology of \eqref{eq:2Vac-4}
using the diagonal elements of \eqref{eq:2Vac-5}. The operator $\Q_{12}$ passes to an operator
$\widehat\Q_{12}$ on the this cohomology.  The ``space of exact ground states''
in the sense of Section \S \ref{subsec:WebRepStrip} is therefore
\be
{\rm Ker}(\widehat\Q_{12}) \oplus {\rm Cok}(\widehat\Q_{12}).
\ee
Note that if we work over the integers the cokernel can be a finite abelian group,
and therefore the space of exact ground states can have torsion. It would be interesting
to know if this has a physical interpretation.

One can similarly work out the general category of Interfaces, but to do this one
must drop the assumption that $\Re(z_{12})>0$ since there are now two Theories
associated (up to locally trivial parallel transport) with $\Re(z_{12})>0$ and
$\Re(z_{12})<0$. It is then possible to write out in full detail the Interfaces
$\fI[\wp]$ for a vacuum homotopy $z_{12}(x)$ and the $S$-wall Interfaces.
This is a good exercise that we will leave to the reader.

Similarly, one could move on and write out in full generality the web formalism
when $\IV$ has three vacua. Again, we leave this as an extensive exercise to the
reader.

%\section{Proof Of Equation \eqref{eq:CharPoly} }\label{app:ProveCharPol}
\section{Proof Of Equation (7.181) }\label{app:ProveCharPol}

In this appendix we prove that the characteristic polynomial
of the $N\times N$ matrix
\be\label{eq:CPpmp-2}
\begin{split}
\CE & = - R_{N-1}   e_{0,0} + \sum_{j=0}^{N-2}   e_{j+1,j} -  e_{0,N-1} \\
&  + \sum_{1\leq j < (N-1)/2} R_{N-1-2j} e_{N-j,j} + \sum_{0\leq j < (N-2)/2} R_{N-2- 2j } e_{N-1-j,j} \\
\end{split}
\ee
(where we treat $R_n$ as scalars)
is simply
\be\label{eq:CharPoly-2}
\det( x \textbf{1}_N - \CE)=x^N + \sum_{j=1}^{N-1} R_{j} x^j + 1
\ee

We prove equation \eqref{eq:CharPoly-2} by expansion by minors.
We use the last column of $x \textbf{1}_N - \CE$ because it has only
two nonzero entries. The minor of the $(N-1,N-1)$ matrix element
is lower triangular and immediately gives
\be
x^{N-1}(x+R_{N-1})
\ee
The minor of the $(0,N-1)$ matrix element is $(-1)^{N+1}$ times the determinant
of the $(N-1)\times (N-1)$ dimensional matrix:
\be\label{eq:CPpmp-3}
\begin{split}
 \tilde M &  =  x\sum_{j=0}^{N-3} e_{j,j+1} -  \sum_{j=0}^{N-2}  e_{j,j}  \\
& - \sum_{1\leq j < (N-1)/2} R_{N-1-2j} e_{N-1-j,j} - \sum_{0\leq j < (N-2)/2} R_{N-2- 2j } e_{N-2-j,j} \\
\end{split}
\ee
Now we use row and column operations to eliminate the $x$'s above the diagonal.
For simplicity assume that $N$ is even. Then, adding $x$ times row $(N-2)$ to row $(N-3)$,
then $x$ times row $(N-3)$ to row $(N-4)$ and so forth up to (but not including) row $N/2$ gives
a matrix of the form
\be\label{eq:CPpmp-4}
\begin{pmatrix} A & 0 \\ * & - \textbf{1}_{\frac{(N-2)}{2} } \\  \end{pmatrix}
\ee
where
\be\label{eq:CPpmp-5}
\begin{split}
 A &  =  x\sum_{j=0}^{N/2 -1 } e_{j,j+1} -  \sum_{j=0}^{N/2 -1 }  e_{j,j}  \\
  &  - x^{(N-2)/2} R_{N-2} e_{N/2-1, 0} - \left( x^{(N-2)/2}  R_{N-3}+ x^{(N-4)/2}  R_{N-4}\right)  \\
  & - \cdots    - (x^2 R_3 + x R_2) e_{N/2-1, N/2-2} - x R_1 e_{N/2-1, N/2-1 } \\
\end{split}
\ee
Now to determine $\det A$ use column operations to eliminate the $x$'s above the diagonal:
First add $x$ times column $0$ to column $1$,
then $x$ times column $1$ to column $2$ and so forth to produce a lower triangular matrix with diagonal
with $-1$ in every element except the $(\frac{N}{2}-1, \frac{N}{2}-1)$ matrix element which is
\be
-1- x R_1 - x^2 R_2 - \cdots - x^{N-2} R_{N-2}
\ee
Thus the contribution of the $(0,N-1)$ minor in $x \textbf{1}_N - \CE$
is (recall $N$ is even):
\be
\begin{split}
- \det M & =   (-1)^{N/2} \det A \\
& = - (-1)^{N/2}(-1)^{N/2-1} \left(1+ x R_1 + x^2 R_2 + \cdots + x^{N-2} R_{N-2}\right)\\
& = 1+ x R_1 + x^2 R_2 + \cdots + x^{N-2} R_{N-2}
\end{split}
\ee
completing the proof of \eqref{eq:CharPoly-2}.

\begin{figure}[htp]
\centering
\includegraphics[scale=0.3,angle=0,trim=0 0 0 0]{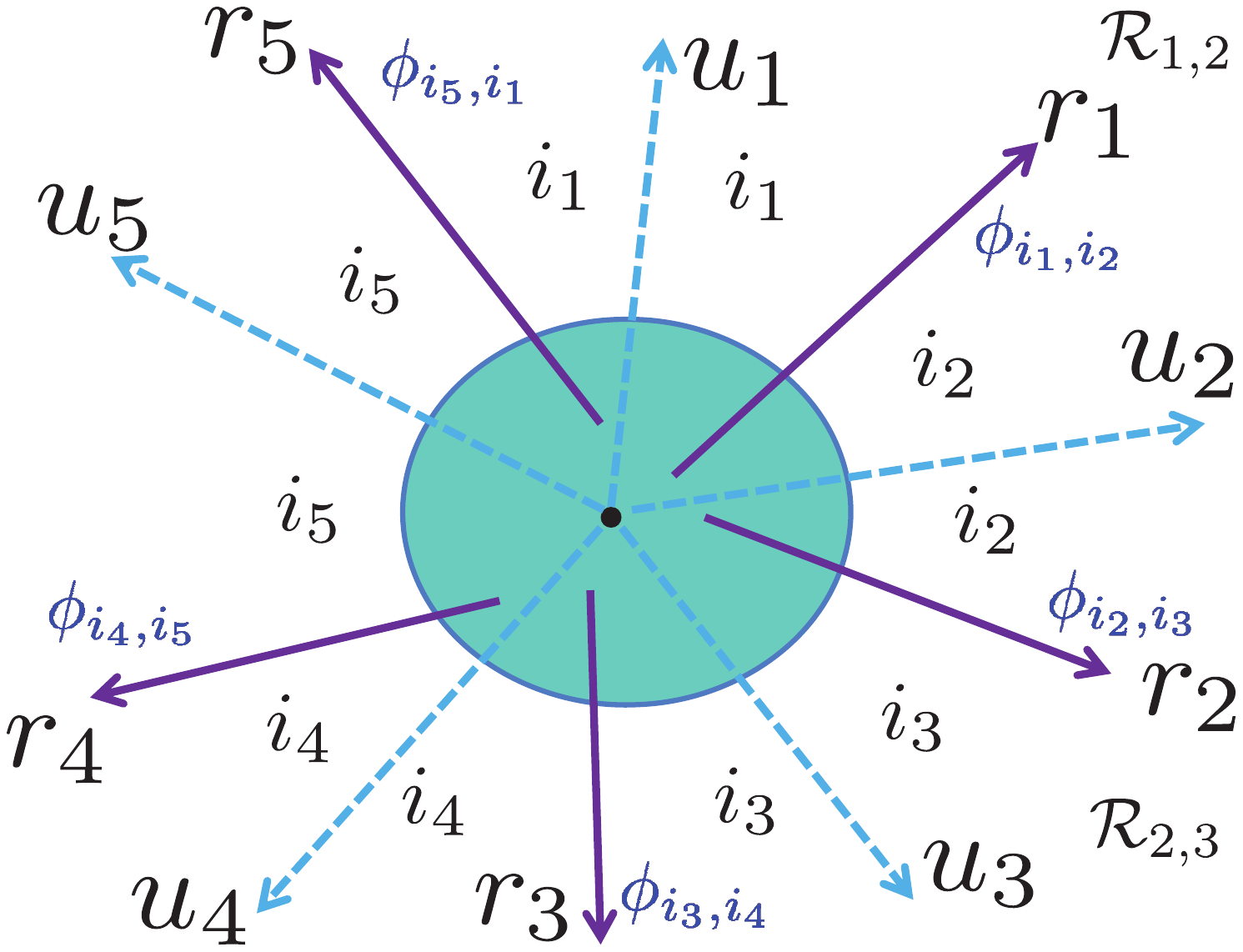}
\caption{In the regions $\CR_{k,k+1}$ which contain the boosted soliton
core rays $r_k$ and are at $\vert s\vert > R$ (the region outside the
blue circle) we can write approximate solutions to the instanton equations
to exponentially good accuracy.  }
\label{fig:WEDGES}
\end{figure}

\section{A More Technical Definition Of Fan Boundary Conditions }\label{subsec:DefFanBC}

In this appendix we describe how to define  a set of boundary conditions
for the $\zeta$-instanton equation \eqref{eq:instanton} associated with
a   fan of solitons
\be\label{eq:solseq}
\CF = \{ \phi_{i_1, i_2}^{p_1}, \dots, \phi_{i_n, i_1}^{p_n} \}.
\ee

To begin, we need to define some notation.
We   choose a set of points $s_1,\dots, s_n$ in the complex $s=x+\I \tau$
 plane. To these points we attach a
set of rays   $\fr_k= s_k+ z_{i_k,i_{k+1}}\IR_+$.
The index $k$ is now considered modulo $n$. Now we also choose an $R\gg \vert s_k\vert$.
Now consider the connected components of
  the complement of the rays $\fr_k$ in the region  $\vert s\vert>R$.
  The region   between $\fr_{k-1}$ and $\fr_k$ is labeled with a vacuum $i_k$.
  Now, in each such component write the ray $\fv_k$ with slope bisecting
  $\fr_{k-1}$ and $\fr_k$.  Finally, define a region
\be
\CR_{k,k+1}(R):= \{ s \vert  \arg \fv_k >  \arg s > \arg \fv_{k+1} \qquad \vert s \vert \geq R \}
\ee

In the region $\CR_{k-1,k}(R)$ we use a  boosted
soliton solution $\phi^{p_{k-1}}_{i_{k-1},i_{k}}$ with core centered
on the ray $\fr_{k-1}$. These are true solutions of the
$\zeta$-instanton equation,
but only in this region. On the boundary $\fv_k$
between $\CR_{k-1,k}(R)$ and $\CR_{k,k+1}(R)$ there is a discontinuity.
However, for large $R$ and for large mass scale $m$ of the LG theory
both solutions $\phi^{p_{k-1}}_{i_{k-1},i_{k}}$ and $\phi^{p_{k}}_{i_{k},i_{k+1}}$
are within order $\CO(e^{-m R})$ of the vacuum $\phi_{i_k}$.
Therefore, as $R\to \infty$ our solution, which is only defined on the
 open region $\cup_k \CR_{k-1,k}$ has exponentially small discontinuities
 at $\fv_k$. Let us denote these discontinuous
solutions by $\phi_{\CF,\vec s, R}$, where $\CF$ stands for the fan data \eqref{eq:solseq},
$\vec s$, is the vector of origins of the rays $\fr_k$, and $R$ is large.

Now we  state the the boundary conditions on the instanton
equation \eqref{eq:instanton}. First, our solutions $\phi(x,\tau)$
should be continuously differentiable. Next, we  require that there is some $\vec s$,
and some
$R_0$, sufficiently large, and some constant $C$
 so that admissible solutions satisfy
\be\label{eq:Fan-BC}
\vert \phi(x,\tau)  - \phi_{\CF,\vec s, R}\vert \leq C e^{-m R}  \qquad \forall \vert s \vert \geq R
\ee
when $R > R_0$.

Given a fan of solitons $\CF$ we let $\CM(\CF)$ denote the
moduli space of smooth solutions to the $\zeta$-instanton equation
\eqref{eq:instanton} with fan boundary conditions \eqref{eq:Fan-BC}.
As explained in Sections \S\S \ref{zetafan}-\ref{bottom}  when the
excess dimension dimension of all zeta vertices vanishes there is a physical expectation that
the component of $\CM(\CF)$ of maximal dimension has dimension given by:
\be\label{eq:DimensionM}
\dim \CM(\CF) = - \half \sum_{k} \eta(D_k-\epsilon)
\ee
where $D_k$ is the Dirac operator  \eqref{eq:NewD} on $\IR$ with $W''$ evaluated
for the soliton $\phi^{p_k}_{i_k,i_{k+1}}$.
In particular, it is physically reasonable to expect that the excess dimension
vanishes for $X = \IC^n$.
Giving a rigorous proof of \eqref{eq:DimensionM}, even in this case,  would seem
to be a challenging task. One needs to compute the index of the
Dirac operator corresponding to the first order variation
of the $\zeta$-instanton equation
\be\label{eq:2d-dirac}
\widehat{D} = \frac{\p}{\p \tau} - i \sigma^3 \frac{\p}{\p x}- \half
\begin{pmatrix} 0 & \zeta \bar W'' \\  \bar\zeta W'' &  0 \\ \end{pmatrix}
\ee
Superficially this would appear to be a problem of the kind studied in
\cite{Callias:1977kg,BottSeeley}. However, the presence of the center of mass
collective coordinates discussed at length in \S \ref{collective} implies that
the Dirac operator with our boundary conditions is \emph{not} Fredholm,
so standard theorems will not apply.  Nevertheless, there should be a good
theory of the Dirac operator $\widehat{D}$ with our boundary conditions.
In particular, we need to orient the moduli spaces $\CM(\CF)$. The relative
orientation for different fan boundary conditions should  be fixed from
a trivialization of the determinant line of $\widehat{D}$ on the space
of all LG fields.

\section{Signs In The Supersymmetric Quantum Mechanics Formulation Of Morse Theory}\label{sec:SQMSigns}

In this Appendix we briefly remark on the signs by which instantons are weighted in
the approach to Morse theory reviewed in Section \S \ref{review} above. In particular
we comment on how to choose the signs for instanton amplitudes in \eqref{bloto}
and \eqref{zelf}.  This is closely related to the proper interpretation of the fermion determinant
\eqref{molo}. The sign conventions for $m_{qp}$ were already discussed in footnote 3 of the
original article \cite{Witten:1982im} (see the end of Section 10.5.3 of \cite{Hori:2003ic}
for an elaboration of that discussion). We will take our cue from the description of signs on
p. 106 of \cite{BottIndomitable} and Section 2 of \cite{Hutchings}.

\subsection{Preliminaries}

\def\OR{{\rm \textbf{Or} }}

We first need a little notation. It will be useful to speak of determinant line bundles.
If $V$ is a finite-dimensional real vector space then $\DET(V)$
is the real line given by the highest exterior power. The set of nonzero vectors in $\DET(V)$
has two connected components. A choice of a component is, by definition, an orientation of $V$.
We can thus identify an orientation of $V$ with a choice of a nonzero vector in $\DET(V)$ up to
rescaling by positive real numbers.
If $V$ has a metric then a volume form on $V$ is a nonzero product of orthonormal vectors,
and corresponds to a vector of norm one in $\DET(V)$.
When $V$ is a vector bundle over a manifold a choice of metric defines a reduction of
structure group of $\DET(V)$ from $GL(1,\IR)\cong \IR^*$ to $O(1)\cong \IZ_2$.
%We refer to the
%real line bundle with $\IZ_2$-structure group as $\OR(V)$.
%
%\cg{Not sure we need the orientation bundle}
%

 The set of descending flows from a critical point $p\in M$
is the set of trajectories $u(\tau)$ satisfying \eqref{zoobo} with $\lim_{\tau\to - \infty} u(\tau) = p$.
The union of the descending flows is a cell $\CD_p\subset M$.
Similarly, the set of ascending flows
from $p$ is the set of trajectories satisfying \eqref{zobb} with $\lim_{\tau\to - \infty} u(\tau) = p$.
Their union is also a cell in $M$ denoted $\CU_p$.
\footnote{
Recall that a cell of dimension $n$ is
a topological space homeomorphic to the open $n$-dimensional ball in $\IR^n$. Using the Morse lemma
it is easy to see that the union of flows defines a cell in the neighborhood of a critical point.
For example, for ascending flows
the coefficients $c_i$ in \eqref{ubo} with $f_i>0$ provide a system of coordinates. Since we have
a first order differential equation the flows will not intersect even when we follow them
 beyond the neighborhood where the Morse lemma applies. See
\cite{AbrahamRobbin} for the rigorous proof that $\CD_p$ and $\CU_p$ are cells. One of the very useful
aspects of Morse theory in topology is that it provides explicit cell decompositions of topological
spaces.}
At a nondegenerate critical point the Hessian, or fermion
mass matrix:
\be
m_{ij} = \frac{D^2 h}{D u^i Du^j}
\ee
is a symmetric, real, nondegenerate form on $T_pM$. Let $P_+(p)$ be the projector onto
the subspace of $T_p M$ with positive fermion masses and $P_-(p)$ the projector onto the space
with negative fermion masses. We can identity
\be
P_+(p)T_pM  = T_p\CU_p \qquad \& \qquad  P_-(p) T_pM = T_p \CD_p
\ee
Thus, assuming $M$ is finite-dimensional, we can say that
 $\CD_p$ is a cell of dimension $n_p$ and $\CU_p$ is a cell of
 dimension $d-n_p$, where $d=\dim M$.

If  $(g,h)$ are sufficiently generic then the ascending and descending cells from
two critical points $p,q$ will intersect \emph{transversally} in the following
sense: We suppose $h(p)>h(q)$ so there is a moduli space $\M_{qp}$ of ascending
flows from $q$ to $p$. It will be a manifold, perhaps with many connected components,
of dimension $n_p - n_q$. We let $\Gamma_{qp}\subset M$ be the the image of
these flows in $M$. Of course, $\Gamma_{qp} \subset \CD_p \cap \CU_q$.
  Our primary example is when  $n_p = n_q+1$ so that
$\M_{qp}$ (generically) consists of an isolated set of instanton trajectories.
The image $\Gamma_{qp}$ is a union of disjoint one-manifolds in $M$ and the closure
is a graph with two vertices at $q$ and $p$.
We will  also need the
case $n_p = n_q+2$ below. We say that $(g,h)$ determine transversal flows if there is an
exact sequence
\be
0 \rightarrow T_{\CP} \Gamma_{qp} \rightarrow T_{\CP} \CU_q \oplus T_{\CP} \CD_p \rightarrow T_{\CP} M \rightarrow 0
\ee
for any point $\CP \in \Gamma_{qp}$. Linear algebra now guarantees the existence of a
\emph{canonical} isomorphism
\be\label{eq:DetExtSeq}
\DET(T_{\CP} \Gamma_{qp}) \otimes \DET( T_{\CP} M ) \cong \DET(T_{\CP} \CU_q) \otimes \DET(T_{\CP} \CD_p).
\ee
This will be the crucial identity in our determination of sign rules.

\subsection{A Mathematical Sign Rule}\label{AppE:MathSigns}

In many mathematical treatments of the MSW complex (see, for examples \cite{BottIndomitable}, \cite{Hutchings} Section 2,
or \cite{ClayBook2} Section 8.3) one chooses generators $[\CU_p]$ of the complex to be
orientations of $T_p\CU_p$, that is, a nonzero vector in $\DET(T_p\CU_p)$, up to positive scaling.
The choice of orientation at each critical point $p$ is made arbitrarily.
Since $\CU_p$ is a cell, such a choice determines an orientation of $T_\CP \CU_p$ for all $\CP \in \CU_p$,
say, by parallel transport.
Similarly, we have a canonical isomorphism $\DET(T_q\CU_q)\otimes \DET(T_q \CD_q) \cong \DET(T_qM)$
and hence an orientation of $T_q \CU_q$ also determines a nonzero vector (up to positive scaling)
of $\DET(T_qM)\otimes (\DET(T_q \CD_q))^{-1}$. Again, the orientation of this
 line can be extended continuously to an orientation of $\DET(T_\CP M)\otimes (\DET(T_\CP \CD_q))^{-1}$
for all $\CP \in \CU_q$. Now, rewrite the canonical isomorphism \eqref{eq:DetExtSeq} as
\be\label{eq:DetExtSeq2}
 \DET(T_{\CP} \CU_q) \cong \DET(T_{\CP} \Gamma_{qp}) \otimes \left( \DET( T_{\CP} M )\otimes ( \DET(T_{\CP} \CD_p))^{-1} \right)
\ee

Now suppose that $n_p = n_q+1$. Then $\CM_{qp}$ is a disjoint union over instantons $\ell$.
The matrix element $m_{qp}$ in \eqref{bloto} is a sum of contributions   $m_{qp}(\ell)\in \{ \pm 1 \}$
where, in this approach, one identifies $\Phi_p = [\CU_p]$ in \eqref{bloto}. For each instanton $\ell$
choose a point $\CP\not= p,q$ on $\ell$ and choose the upwards-flowing orientation of   $T_{\CP} \Gamma_{qp}$.
Then  the arbitrary choices
of orientations of $[\CU_p]$ and $[\CU_q]$ will either agree or disagree with the canonical isomorphism
\eqref{eq:DetExtSeq2} for a trajectory $\ell$ containing a point $\CP$. If the orientations agree
then  $m_{qp}(\ell)$ is +$1$ and if they disagree it is $-1$. Note, incidentally, that it was
not necessary to orient $M$. This is the sign rule given in \cite{BottIndomitable} and \cite{Hutchings} and it is
equivalent to that given in \cite{Witten:1982im}.

As an example let us check that the two upward flows in Figure \ref{Morse}(a) indeed cancel.
The ascending cell $\CU_q$ is the circle minus $p$, while $\CD_q$ is the $0$-cell
$q$ itself. The descending cell $\CD_p$
is the circle minus $q$ while $\CU_p$ is the $0$-cell $p$ itself.
 We choose the orientation of $\CU_q$ to be defined by
a nonzero horizontal tangent vector pointing left at $q$ and the orientation
of $\CD_p$ to be a nonzero horizontal tangent vector pointing right at $p$.
We choose the clockwise orientation for $TM$. The orientations of the two
ascending flows will be given by a nonzero tangent vector in the ascending
direction. With these choices we compute from the above sign rule that
 the contribution to $m_{qp}$
from the ascending flow on the left is $+1$ and that of the ascending flow
on the right is $-1$.

\subsection{Approach Via Supersymmetric Quantum Mechanics}\label{AppE:SQM}

Our goal here is to give a slight reformulation of the above sign rule which
makes more direct contact with the physical approach from supersymmetric quantum mechanics.
This approach provides a
framework for generalization to the case when $M$ is an infinite-dimensional function space,
such as is necessary in applications to quantum field theory (such as Landau-Ginzburg theory).

We must first return 
to the description of the Fermion state associated with a critical point $p\in M$
(See the discussion from equations \eqref{heffalo} to \eqref{hoobo} above.)

 At any point $\CP \in M$, the
 span of the Fermions $\psi^i$ is $\Pi T_\CP^* M$ and their
canonical conjugates $\bar \psi_i = g_{ij} \bar\psi^j$ span $\Pi T_\CP M$. (Here $\Pi$ indicates
parity reversal - these are odd real vector spaces.) The canonical quantization relations
define the natural Clifford algebra of $T_\CP^*M \oplus T_\CP M$ associated with the
natural signature $(d,d)$ form given by contraction. In order to write a definite
Fermionic Hilbert space we must choose a particular Clifford module and this is
done by choosing a maximal isotropic subspace of $T_\CP^*M \oplus T_\CP M$.
 If we quantize by choosing the vacuum line
to be determined by the maximal isotropic subspace $T_\CP M$ then the Hilbert space of Fermionic
states at $\CP \in M$ is the Clifford module
\be\label{eq:Fermi-HS-1}
\CH^{\rm Fermi}_\CP =  \Lambda^*(T_\CP^* M) \otimes \fM_{\CP}.
\ee
where, for any point $\CP\in M$ we have introduced the notation
$\fM_{\CP} := \DET(T_\CP M)$. Heuristically, this is the real span
of the product of the $\bar \psi_i$.
This quantization applies to general manifolds $M$, orientable or not.
When $M$ is orientable, there exist smooth trivializations
of $\DET(TM)$. Since $M$ is Riemannian there are two natural trivializations
of unit norm, corresponding to the two possible orientations of $M$.
Relative to such a trivialization a vector in
\eqref{eq:Fermi-HS-1} can be identified with a differential
form at $\CP$, and in this way the entire Hilbert space of the theory
is identified with $\Omega^*(M)$.  When $M$ is unorientable
there is no continuous trivialization of $\fM = \DET(TM)$. Rather
wavefunctions are sections of $\Omega^*(M;\fM)$. These are
known as densities, and can be integrated on any manifold, oriented or not.

\textbf{Remarks}

\begin{enumerate}

\item It is useful
to note that the supersymmetric quantum mechanics can be coupled to a
flat line bundle $\fL$. When we do this equation \eqref{eq:Fermi-HS-1} is
generalized to
\be\label{eq:Fermi-HS-2}
\CH^{\rm Fermi}_\CP =  \Lambda^*(T_\CP^* M) \otimes \fM_{\CP}\otimes \fL_{\CP}.
\ee
The instanton amplitudes $m_{qp}$ discussed below now have an extra factor
corresponding to the parallel transport from $\fL_q$ to $\fL_p$. This leads to
the MSW complex twisted by a flat line bundle. If $\fL$ is a real line bundle
then $\CH^{\rm Fermi}_\CP$ is a real Hilbert space.

\item There is a
dual quantization where the Clifford vacuum is based on the maximal isotropic
subspace $T^*_\CP M$. This corresponds to an exchange of $\psi^i$ for $\bar \psi_i$.
 Equation \eqref{eq:Fermi-HS-1} treats $\psi^i$ and $\bar \psi_i$ asymmetrically
and moreover differs in our conventions for the relation of Fermion number to the
degree of a form. In \eqref{eq:Fermi-HS-1} a differential
form of degree $n$ corresponds naturally to a Fermion state of Fermion number $n-d$.
If $M$ is orientable
then one can maintain the (Hodge) symmetry between $\psi^i$ and $\bar \psi_i$
by choosing $\fL^{-1}$ to be a real line bundle which squares to $\DET(TM) $.
We can use the metric to reduce the structure group of $\fM$ to $O(1)\cong \IZ_2$. If
we do so, then a choice of $\fL$ is a  choice of a flat order two real line bundle.
If we twist by such a line $\fL$ then a differential form of degree $n$ corresponds to a
Fermionic state in \eqref{eq:Fermi-HS-2} of  Fermion number $n-d/2$.

\end{enumerate}

Now we wish to define a (perturbative) ground state associated with a critical point $p$ in the
supersymmetric quantum mechanics.

In general, in quantum mechanics, a (pure) state is a one-dimensional complex line
in a complex Hilbert space. (In the physics literature such a line is often
called a ``ray.'')  In our problem the Hilbert space is naturally real.
The perturbative ground state  is a product of real lines for the
bosonic and Fermionic degrees of freedom. The bosonic groundstate, which has
Gaussian support near $p$, poses no sign difficulties and will henceforth be ignored.
The Fermionic ground state is therefore considered to be a real line in
the real Hilbert space \eqref{eq:Fermi-HS-1}. Physicists often use the word
``state'' to refer to either a line or a norm one vector in that line. Since
the distinction will be important in what follows, and in order to avoid confusion, we will use the term
\emph{Fermion vector} to refer to an actual vector in the line. In Morse theory we are interested
in transition matrix elements and not their squares so we must choose a Fermion
vector $\Psi(p) \in \CH^{\rm Fermi}_p$. Since the Fermion line is a real line
there are two normalized vectors $\pm \Psi(p)$. The vectors $\Phi_p$ used in
equations \eqref{mox} et. seq. of Section \S \ref{zyindeed} are then obtained
by taking a product with the (unproblematic) bosonic vector.

Now, for any critical point $p$,  let us determine the Fermion line in which $\Psi(p)$ must live.
Recall from the analysis above \eqref{hoobo} that if
 we choose a coordinate frame diagonalizing $m_{ij}(p)$ then the ground state
is made by acting on the vacuum line of \eqref{eq:Fermi-HS-1}  with the
wedge product of  $\psi^i$ for $i$ running over the
downward directions. Thus, the Fermion ground vector  at $p$ is an element of the real line
\be
\CS_p := \DET(T^*_p \CD_p) \otimes \fM_p .
\ee
Using the canonical isomorphisms
\be
\DET (T_p^* \CD_p) \cong (\DET (T_p \CD_p) )^{-1}
\ee
and
\be
\fM_p \cong \DET (T_p \CD_p) \otimes \DET(T_p \CU_p)
\ee
we can equally well write
\be\label{eq:Fermi-Crit}
\CS_p \cong \DET(T_p \CU_p)
\ee
This will be the form of most use to us, and we will find it useful to introduce the
notation
\be
\fU_p :=  \DET(T_p \CU_p)  \qquad\qquad \& \qquad\qquad \fD_p:= \DET(T_p \CD_p).
\ee

Now we can start to make contact with the mathematical formulation of Section \S \ref{AppE:MathSigns}
above. A choice at each critical point $p$ of $\Psi(p)$ determines a set of nonzero vectors
in $\CS_p$ up to \emph{positive} scaling. That is equivalent to a choice of orientation
of $\CU_p$. The difference is that in physics $\Psi(p)$ is vector in a Hilbert space and
in the mathematical approach $[\CU_p]$ is just an abstract generator of a $\IZ$-module.

Let us next turn to the matrix element $m_{qp}$ of equation \eqref{bloto}. Here $p,q$ are
critical points with $n_p = n_q +1$. The operator $\hat\Q$ is not a section of a line
bundle so we regard $m_{qp}$ as an element of a line
\be\label{eq:mqp-line}
m_{qp} \in \frac{\fU_q}{\fU_p}
\ee
The matrix element $m_{qp}$ is in the dual line to $\Hom (\CS_q , \CS_p)$.
The localization of the path integral states that $m_{qp} = \sum_{\ell \in \M_{qp}} m_{qp}(\ell)$,
and each $m_{qp}(\ell)$ is given by the one-loop path integral expanded around the instanton $\ell$.
This is the Fermionic path integral normalized by the (unproblematic) one loop bosonic
determinant. The Fermionic path integral is a section, $\det(L)$ of the determinant line bundle
$\DET(L)$. In general, if $T:\CH_1 \to \CH_2$ is a Fredholm operator between two Hilbert spaces
we can define $\DET(T)\cong \DET(\ker L)^{-1} \otimes \DET(\ker L^\dagger)$. As explained in
Section \S \ref{fermanom}, since $\iota(L) = n_p-n_q = 1$ is positive, we generically have
$\ker L^\dagger = \{ 0 \}$ and for simplicity we will assume that this is the case. Thus,
$\det(L) \in \DET(\ker L)^{-1}$. Indeed, this is where the measure of the Fermion zeromodes should live.
On the other hand, for any point $\CP\in \ell$, we can also identify $\ker L \cong T_{\CP} \ell$,
since the zeromode is just $\dot u^i(\tau)$, defining a nonvanishing vector field along $\ell$. Thus,
we can choose any point $\CP \in \ell$ and use this isomorphism to identify the path integral as an
nonzero vector in  $ \DET(T_\CP \ell )$.
Now we have   isomorphisms
\be\label{eq:MtxEl-Line}
\frac{\fU_q}{\fU_p} \cong \frac{\fU_q \fD_p}{\fM_p} \cong \frac{\DET(T_\CP \CU_q) \DET(T_\CP \CD_p) }{\fM_{\CP} }\cong\DET(T_\CP \Gamma_{qp} )
=\DET(T_\CP \ell ).
\ee
where the first isomorphism is canonical, the  second  uses parallel transport along $\ell$ to a point $\CP\in \ell$,
and the third is the canonical isomorphism \eqref{eq:DetExtSeq}. In this way, for
each $\ell$ we can map the amplitude to an element  $m_{qp}(\ell)\in \frac{\fU_q}{\fU_p}$ in a common line,
where the amplitudes can be sensibly added.

As explained in Section \S \ref{zyindeed}, $\det(L)$ is a product of the zeromode measure
(valued in $\DET(T_\CP\ell)$, for any choice of $\CP\in \ell$) and the ratio of determinants
in equation \eqref{molo}:
\be\label{eq:molop}
\frac{\det' L }{(\det' (L^\dagger L))^{1/2}}
\ee
The ratio of determinants is a number (in general complex) not a section of a line bundle
and can be defined, say, by $\zeta$-function regularization. In our case, since $L$ is real
this number is $\pm 1$. Once we have chosen a trivialization $\Psi(p), \Psi(q)$ of $\fU_q/\fU_p$
the contributions of the amplitudes $m_{qp}(\ell)$ for different instantons differ by $\pm 1$
where this sign is given by both the ratio of determinants and the ratio of orientations of
Fermion measures. This makes contact with the mathematical sign rule of Section
\S \ref{AppE:MathSigns} above.

As an example, consider again Figure \ref{Morse}(a). Let $\phi \sim \phi + 2\pi$ parametrize
the circle and let $h=\cos(\phi)$. The two instanton flows are given by
$\tan(\phi/2) = \pm e^{-(\tau-\tau_0)}$ and  $L = - \frac{d}{d\tau} + \cos\phi$.   In this
case \eqref{eq:molop} is clearly the same for the two flows, but the Fermion measure is oriented
in opposite directions for the two flows, and hence the two instanton amplitudes cancel.

%On the other hand, at a critical point $p$ it follows from
%
%\be\label{eq:Fermi-Crit}
%\Psi(p) \in \CS_p = \left( \DET(T_p \CU_p) \otimes (\DET (T_p^* \CD_p) \right)^{1/2}
%\ee
%
%
%The need for the square-root can be seen since the state should have Fermion number $n_p  - d/2$.
%To make precise sense of the square-root note that we have a canonical isomorphism
%
%\be
%\DET(T_p \CU_p) \otimes \DET (T_p^* \CD_p) \cong  (\DET (T_p^* \CD_p))^2\otimes \DET(T_p M)
%\ee
%
%and hence, given a choice of $(\DET T_p M)^{1/2}$ we can choose the precise definition:
%
%\be
%\CS_p :=   \DET (T_p^* \CD_p)\otimes \fM_{p}^{1/2}.
%\ee
%
%In particular there is a canonical identification of the state $\Psi(p)$ with a state in \eqref{eq:Fermi-HS}.
%In what follows it will be more useful to invoke the canonical isomorphism
%
%and write instead
%
%\be
%\CS_p = (\DET (T_p \CD_p) )^{-1}\otimes (\DET (T_p M) )^{1/2}.
%\ee
%
%In the unorientable case it is not natural to insist on the symmetry between $\psi^i$
%and $\bar \psi_i$ and, if we choose the quantization \eqref{eq:Fermi-HS-1},
%then we must take instead   $\Psi(p) \in \DET(T_p\CU_p)$ (as one obtains by
%formally multiplying $\CS_p$ by $(\DET (T_p M))^{1/2}$).
%

More generally,
it is instructive to go back to the localization formula for the vev of some generic operator of the form
\begin{equation}\label{odef}
\hat O = O_{i_1\cdots i_n} \psi^{i_1}\cdots \psi^{i_n}
\end{equation}
which is $\Q$-closed if the degree $n$ form $O = O_{i_1\cdots i_n} \d \uu^{i_1}\cdots \d \uu^{i_n}$ on $M$ is closed.
The same analysis as for $[\Q,f]$ given in Section \S \ref{zyindeed} shows that
\begin{equation}\label{geno}\langle \Phi_p|\hat O(\tau) |\Phi_q\rangle=  \int_{\Gamma_{qp}}   O \end{equation}
Here $n= n_p-n_q$, otherwise both sides are zero.  The right hand side is a sum over
the components  $\Gamma^\alpha_{qp}$ of $\Gamma_{qp}$, but how should these components be oriented?
  Having made a choice of $\Psi(p)$ and $\Psi(q)$
to define the left hand side, we can again use \eqref{eq:MtxEl-Line} to determine the orientations of
the components $\Gamma^\alpha_{qp}$. We define the sign of the integral $ \int_{\Gamma^\alpha_{qp}} O$
using this orientation.

\begin{figure}[htp]
\centering
\includegraphics[scale=0.3,angle=0,trim=0 0 0 0]{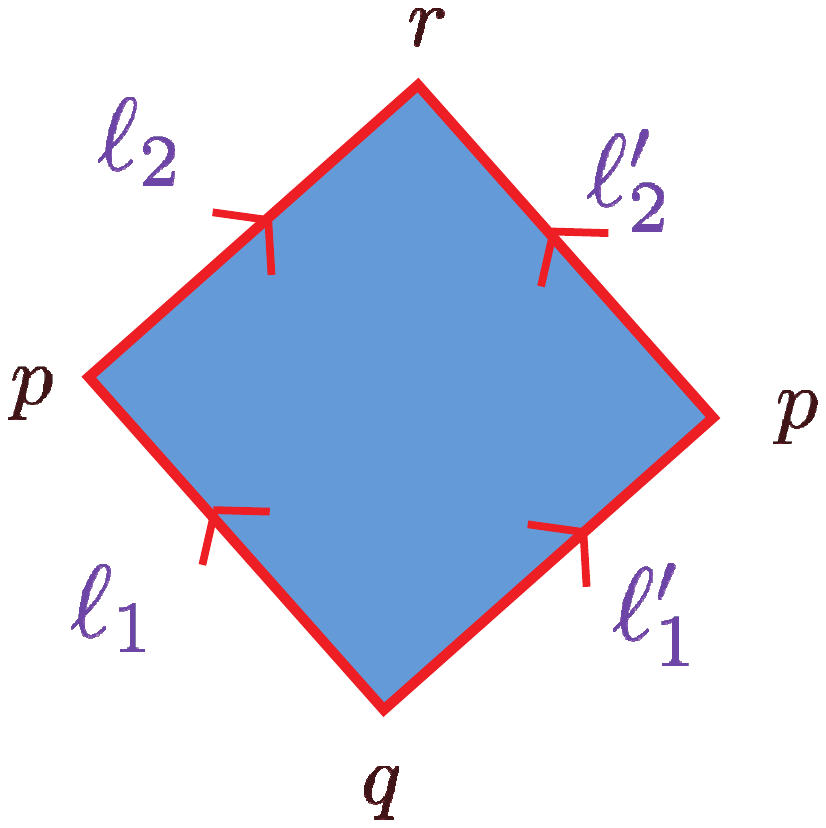}
\caption{A component of $\CM_{qr}$ when $n_r-n_q=2$ corresponds to a two-cell
$\Gamma_{qr}\subset M$ depicted schematically here. There are two one-dimensional
boundaries of this cell, corresponding two broken paths $\ell_1*\ell_2$ and
$\ell_1'*\ell_2'$ interpolating   $q\to p\to r$, where $n_p = n_q+1$.    }
\label{fig:TWOCELL-BROKENPATH}
\end{figure}

Let us now address the cancellations required for equation \eqref{wonox} to hold.
Suppose that $n_r = n_p +1 = n_q +2 $ and focus on a particular component of the
moduli space $\M_{qr}$. This component will map to a two-dimensional cell $\Gamma_{qp}\subset M$ with
boundaries corresponding to broken paths $\ell_1 * \ell_2$  and $\ell_1' * \ell_2'$
as in Figure  \ref{fig:TWOCELL-BROKENPATH}.  The product of instanton amplitudes
$m_{qp} m_{pr}$ for both broken paths is, according to \eqref{eq:mqp-line}, valued in
\be
m_{qp} m_{pr}  \in  \frac{\fU_q}{\fU_p} \otimes \frac{\fU_p}{\fU_r} \cong \frac{\fU_q}{\fU_r}
\ee
Now, once again, if we choose a particular two-cell in  $\Gamma_{qr}$
as in Figure  \ref{fig:TWOCELL-BROKENPATH}   we can use the   isomorphism \eqref{eq:MtxEl-Line}
to map to an element
\be
m_{qp}m_{pr} \in   \DET(T_\CP \CM_{qr} )
\ee
where $\CP \in \Gamma_{qr}$ is any point in the cell.  Thus the
contributions of the two bounding broken paths  $\ell_1 * \ell_2$  and $\ell_1' * \ell_2'$
lie in the same line and can meaningfully be compared. On the other hand, from \eqref{eq:MtxEl-Line}
we see that $m_{qp}(\ell_1)$ and $m_{pr}(\ell_2)$ determine orientations of $\ell_1$ and
$\ell_2$. Since the broken path $\ell_1*\ell_2$ is a limit of smooth paths, all of which
can be oriented, there is a natural correlation between the orientation of $\ell_1$
and $\ell_2$ so that there is a well-defined orientation of the boundary $\ell_1*\ell_2$.
Thus, $m_{qp}(\ell_1)m_{pr}(\ell_2)$ determines an orientation of the boundary one-cell
$\ell_1*\ell_2$. There are only two natural maps  into $  \DET(T_\CP \CM_{qr}) $,
namely taking a wedge product with the outward or the inward normal to the two-cell
$\Gamma_{qr}$.
Making the same choice for both $\ell_1*\ell_2$ and $\ell_1'*\ell_2'$ the
contributions from the two boundaries cancel. The reason we make the same choice is that
we then have a rule analogous to  \eqref{geno} for matrix elements of operators of
Fermion number $2$.

Finally, we discuss the sign choices for the exceptional instantons counted by $\EE$
in equation \eqref{zelf}. Let $s_0$ be an isolated value of $s$ such that
equation \eqref{luffo-p} has a solution. The relevant linear operator acts on
a Fermion field with one extra component
\be
 \psi(\tau) = (\psi^i(\tau), \psi^s(\tau))
\ee
where $\psi^s(\tau)$ takes into account the first order variation in the $s$
direction. The relevant Dirac operator  is given by
\be
(\tilde L \psi)^i = \frac{D\psi^i}{D\tau}-g^{ij}\frac{D^2h}{D \uu ^j D \uu ^k}\psi^k -
\frac{\p}{\p s}\biggr\vert_{s_0} \biggl[ g^{ij} \frac{\p h}{\p u^j} \biggr]\psi^s
\ee
where $g,h$ and their derivatives are evaluated on the exceptional instanton $u^i(\tau;s_0)$.
The contribution of the exceptional instanton at $s_0$  to $e_{pp'}$ in equation \eqref{zelf} is, as usual,
an element
\be
e_{pp'}(s_0) \in  \frac{ \fU_p}{\fU_{p'} }
\otimes T_{s_0}^*I
\ee
where the last factor $T_{s_0}^*I$ on the right hand side
comes from the extra component of the Fermi field, $\psi^s$,  in the domain of the Dirac operator $\tilde L$,
and $I=[0,1]$ is the space parametrized by $s$ and  describing the homotopy between
$(g(\tau), h(\tau))$ and $(g'(\tau), h'(\tau))$. Accordingly, once we choose an orientation for $I$
we can add the contributions of the
different exceptional instantons. Reversing the orientation changes the sign of the operator
$\EE$. From equation \eqref{mizzo} it is clear this must be the case.

\end{document}